13th International Conference on

**Elastic and Diffractive Scattering**

(Blois Workshop)

# Moving Forward into the LHC Era

**29th June - 3rd July 2009**
**CERN – Geneva**

**M. Deile, D. d'Enterria, A. De Roeck (eds.)**

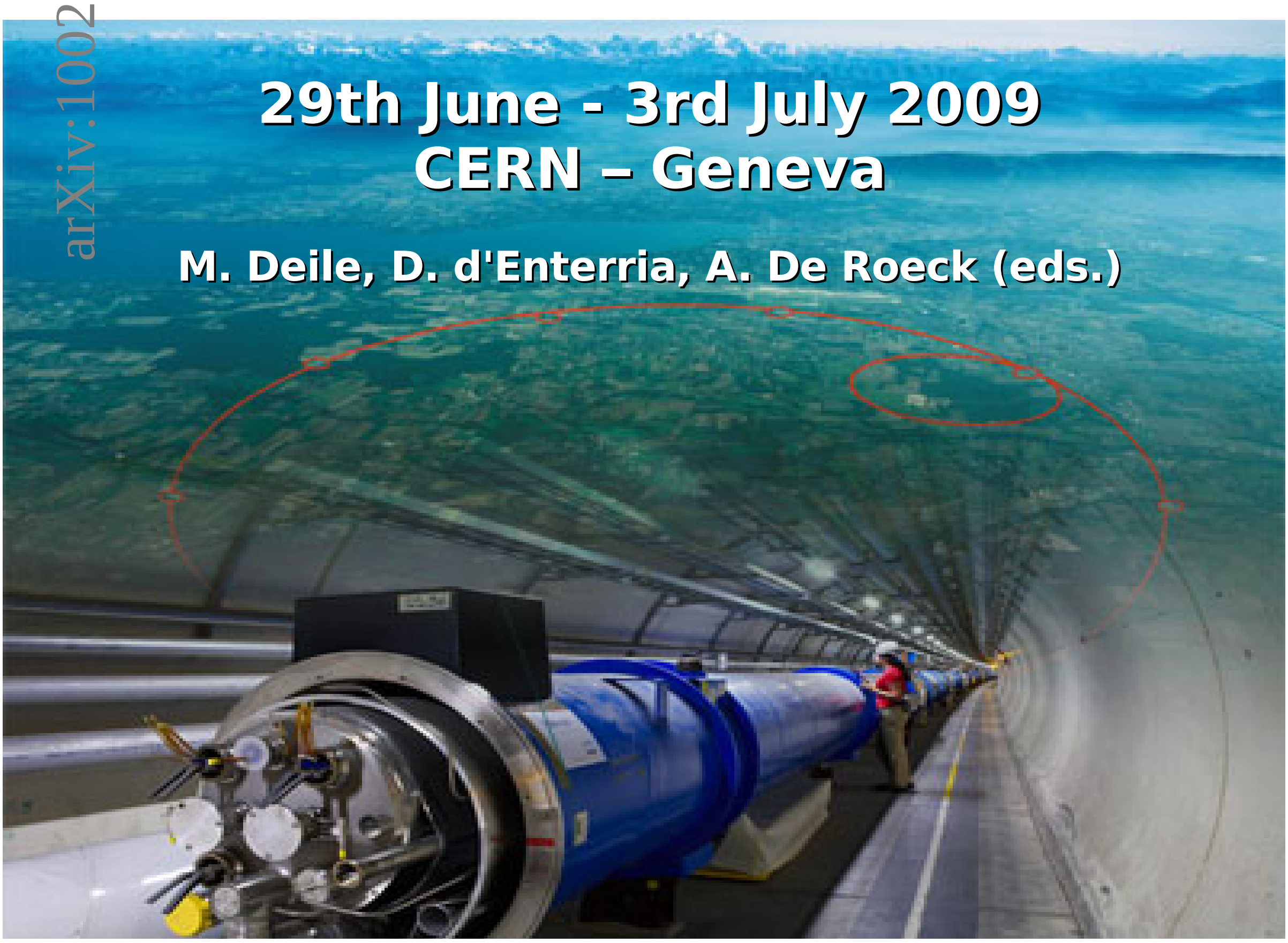

Proceedings of the

# 13th International Conference on Elastic and Diffractive Scattering ("Blois Workshop")

# EDS 2009

29th June - 3rd July 2009
CERN,Geneva, Switzerland

Editors: Mario Deile, David d'Enterria, and Albert De Roeck

## Impressum

**Proceedings of the 13th International Conference on Elastic and Diffractive Scattering 2009**

Conference homepage
`http://www.cern.ch/eds09/`

Slides at
`http://indico.cern.ch/conferenceDisplay.py?confId=41547`

Online proceedings at
`http://arxiv.org/abs/1002.3527`

## Organizing Committee:

M. Arneodo (U. Piemonte Orientale, Novara & INFN Torino), M. Bruschi (U. & INFN Bologna), M. Deile* (CERN), D. d'Enterria (U. Barcelona), A. De Roeck (CERN), B. Di Girolamo (CERN), M. Grothe (U. Wisconsin), R. Orava (HIP Helsinki).
* chair

## Scientific conveners:

**Elastic scattering and total cross-section:** M. Haguenauer (LLR Palaiseau), M. Islam (U. Connecticut), L. Jenkovszky (BITP Kiev), A. Krisch (U. Michigan), V. Kundrat (ASCR Prague)
**Soft and hard diffraction, diffraction at the LHC:** G. Alves (CBPF Rio de Janeiro), A. Kaidalov (ITEP Moscow), K. Oesterberg (HIP Helsinki), M. Ruspa (U. & INFN Torino), R. Schicker (U. Heidelberg), G. Watt (UCL London)
**Central exclusive production:** V. Khoze (U. Durham), A. D. Pilkington (U. Manchester)
**Heavy-ion physics:** F. Arleo (LAPP Annecy), D. d'Enterria (U. Barcelona)
**Photon-induced physics:** J. Gronberg (LLNL), J. Nystrand (U. Bergen)
**Cosmic rays and related studies at the LHC:** J. Hörandel (U. Nijmegen), G. Navarra† (U. Torino)
**Forward physics and low-x QCD:** F. Gelis (CEA Saclay), H. Jung (DESY)
**Past and future of forward physics experiments**: M. Bruschi (U. & INFN Bologna), B. Di Girolamo (CERN)

† deceased.

## Advisory Committee:

H. Abramowicz (U. Tel Aviv), J. Bartels (DESY), G. Bellettini (INFN Pisa), K. Borras (DESY), M. Diehl (DESY), G. Giacomelli (U. & INFN Bologna), K. Goulianos (Rockefeller U.), M. Islam (U. Connecticut), H. Jung (DESY), N. Khuri (Rockefeller U.), A. Krisch (U. Michigan), V. Kundrat (ASCR Prague), An. Martin (CERN), G. Matthiae (U. Roma T. V.), P. Newman (U. Birmingham), B. Nicolescu (IN2P3 Paris), R. Orava (HIP Helsinki), E. Predazzi (INFN Torino), M. Ryskin (U. Durham & PNPI St. Petersburg), J. Soffer (Temple U.), M. Strikman (Penn State U.), C.-I. Tan (Brown U.), J. Trân Thanh Vân (U. Paris XI, Orsay), T. Tsun Wu (Harvard U.).

## Permanent Committee of the EDS Conferences:

## Supported by:

European Organization for Nuclear Research, CERN
Istituto Nazionale di Fisica Nucleare, INFN, Sezione di Bologna

# Preface

The *13th International Conference on Elastic and Diffractive Scattering, EDS'09* – also known as "Blois Workshop" from its first venue at the Château de Blois back in 1985 – was held at CERN from 29 June – 3rd July 2009. The meeting focussed on experimental and theoretical studies of the QCD sector of the Standard Model probed in hadronic interactions at high energy. More than 100 participants from 18 countries attended the meeting. There were 70 talks presented in 8 sessions convened by respected experts on the following fields: *the total proton-proton (pp) cross section; elastic pp scattering; inelastic diffractive scattering in electron-proton (ep), pp and heavy-ion collisions; central exclusive production; photon-induced processes; forward physics and low-x QCD; cosmic-ray physics and related studies at the LHC*. These proceedings collect write-ups of most of the talks presented at the conference. Detailed information on the conference programme and presentation slides can be found at `http://www.cern.ch/eds09/`.

We thank all participants for their valuable contributions, the session conveners for defining the details of the programme and for proof-reading these proceedings. We are grateful to Patricia Mage-Granados (CERN) and to the organizers of the previous edition of this conference series, EDS'07 at DESY, for their advice and help in the preparations. The DESY management also contributed by kindly agreeing to have these proceedings printed in their printshop. The financial support from CERN and INFN is warmly acknowledged, as well as Rossella Magli's (INFN Bologna) very helpful administrative support.

The EDS'09 Organizing Committee:
Michele Arneodo, Marco Bruschi, Mario Deile (chair), David d'Enterria,
Albert De Roeck, Beniamino Di Girolamo, Monika Grothe, Risto Orava

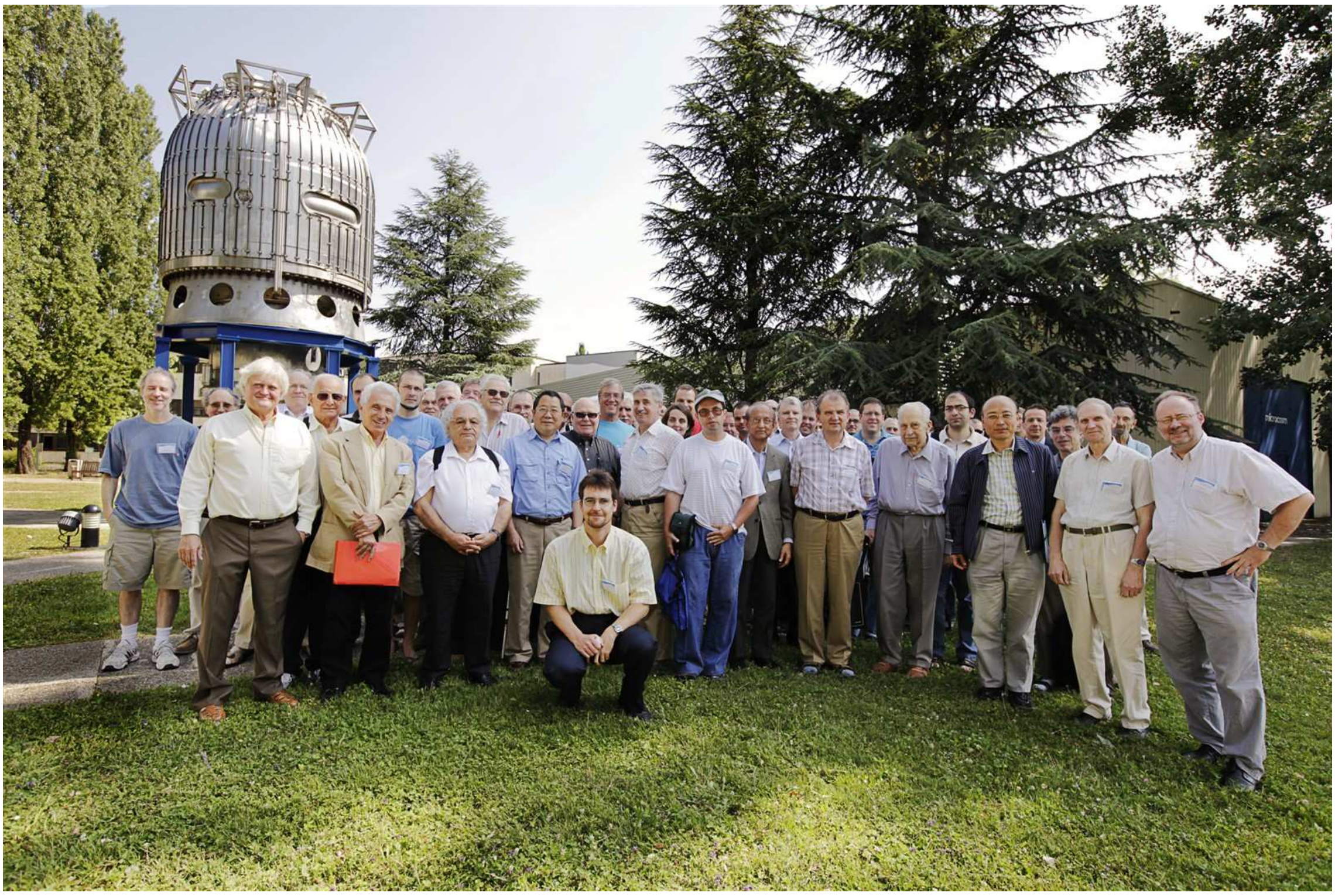

# Obituary

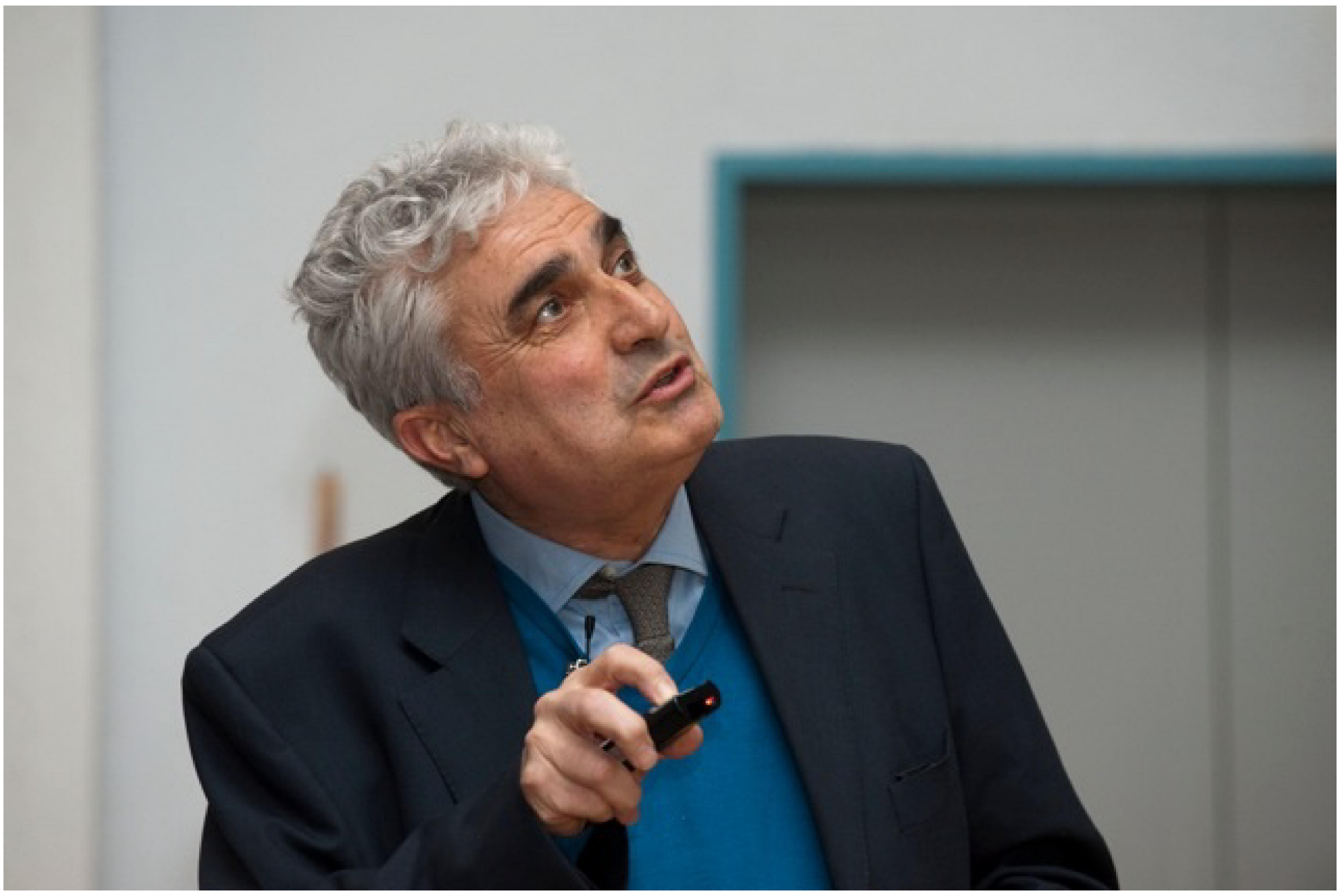

Professor Dr. Gianni Navarra 1945 – 2009

The session on cosmic-ray physics and related studies at LHC was jointly organized by Gianni Navarra and myself. Sadly, Professor Navarra passed away on August 24th, 2009. We lost a best friend. He was a scientist of amazing understanding and creativity. Even more remarkable was his modest deference to the opinions of others.

Gianni Navarra was born on September 12th, 1945. He graduated in physics in 1967 and specialized later in cosmic-ray physics (1974). The stations in his scientific career include: Researcher of Istituto di Cosmogeofisica (C.N.R.) from 1970 to 1983; Associated Professor of Physics in Torino from 1983 to 2000; Professor of Experimental Physics from 2000 to 2003; Full Professor at Torino University since 2003. He performed experimental research in the field of high-energy cosmic rays concerning their astrophysical origin and high-energy hadronic interactions.

His research activity took place in Italian and international laboratories. Among them are: (i) Monte dei Cappuccini (Torino, 1967-1970): study of the muon energy spectrum and of the properties of strong interacting particles at 70 m.w.e. depth. (ii) Pic du Midi (France, 1972-76): study of air-shower characteristics by means of atmospheric Cherenkov light detection in coincidence with a small particle array (run by the University of Kiel). (iii) Testa Grigia (Italy, 1979-89): gamma-ray astronomy, 10 years observation of the Cygnus region; experimental test of an idea (by Professor Chudakov) to measure the primary energy spectrum of cosmic rays with indirect Cherenkov light scattered from the snow. (iv) Chacaltaya (Bolivia): gamma-ray astronomy from the Southern Hemisphere, observation of gamma-ray emission from the region of SN1987A. (v) Baksan (URSS, 1984-85): search for gamma-ray point sources with the "carpet" air shower array. (vi) Gran Sasso Laboratory (Italy, 1985-2000): investigation of the properties of high-energy cosmic rays (energy spectrum, mass composition, arrival direction) in the knee region with the EAS-TOP experiment. (vii) Forschungszentrum Karlsruhe (Germany, 2001-2009): measurement of the energy spectrum and mass composition of cosmic rays from the knee to the transition region (from galactic to extragalactic cosmic rays) with the KASCADE-Grande experiment. (viii) Malargüe (Argentina, 2002-2009): investigation of the properties of the highest-energy cosmic rays with the Pierre Auger Observatory.

He was P.I. of the EAS-TOP experiment at Gran Sasso Laboratories, responsible for the Italian group in the Kascade-Grande experiment, and for the Surface Detector Group at Torino in the Pierre Auger Observatory.

We thoroughly regret this big loss. Our thoughts are with his family.

*Jörg R. Hörandel*

# Contents

# Part I

# Introductory Session

# Startup Planning for the LHC and Operation Scenario for Forward Physics

*Helmut Burkhardt*[1]

[1]CERN, 1211 Geneve 23, Switzerland

This contribution describes the status of the LHC and operational scenarios for forward physics.

## Short Summary with References

Commissioning of the LHC with beams started in September 2008. Initial progress was excellent and a lot of very useful information obtained. Details can be found in a series of LHC-Performance-Notes. In particular, it was possible to perform detailed optics checks in the LHC and to determine and correct the $\beta$-beating [1].

Unfortunately, the commissioning stopped after few days with an incident which required a major repair, resulting in a longer shutdown [2]. The LHC is scheduled to restart for operation with beams in November 2009.

Special high-$\beta$ optics have been prepared for forward physics for the TOTEM experiment [3] and the ALFA option [4] of the ATLAS experiment. The very high $\beta^* = 2625\,\mathrm{m}$ optics for ATLAS is described in [5].

Both an intermediate $\beta^* = 90\,\mathrm{m}$ and a high $\beta^* = 1535\,\mathrm{m}$ optics have been developed for TOTEM [6]. The 90 m option is designed such, that it is reachable from normal physics operation by an un-squeeze to $\beta^* = 90\,\mathrm{m}$. This makes it suitable for tests in earlier physics operation.

Details of both the TOTEM and ATLAS high-$\beta$ optics are described in [7].

## References

[1] M. Aiba *et al.*, "First beta-beating measurement and optics analysis for the CERN Large Hadron Collider", *Phys. Rev. ST Accel. Beams* 12 (2009) 081002.

[2] M. Lamont, "LHC: status and commissioning plans", 0906.0347.

[3] **TOTEM** Collaboration, G. Anelli *et al.*, "The TOTEM experiment at the CERN Large Hadron Collider", *JINST* 3 (2008) S08007.

[4] ATLAS collaboration, "ATLAS Dectectors for Measurement of Elastic Scattering and Luminosity", CERN LHCC-07, 2007.

[5] S. M. White, H. Burkhardt, P. M. Puzo, S. Cavalier, and M. Heller, “Overall Optics Solutions for Very High Beta in Atlas”, Proc. EPAC 2008.

[6] H. Burkhardt, S. White, and Y. Levinsen, “Study of High Beta Optics Solution for TOTEM”, Proc. PAC 2009, CERN-ATS-2009-034.

[7] H. Burkhardt and S. White, “High-beta Optics for the LHC”, LHC Project Note, in preparation.

# Part II

# Elastic Scattering and Total Cross-Section

# Diffractive and Total pp Cross Sections at LHC

*Konstantin Goulianos*

The Rockefeller University, 1230 York Avenue, New York, NY 10065-9965, USA

The single-diffractive and total *pp* cross sections at the LHC are predicted in a phenomenological approach that obeys all unitarity constraints. The approach is based on the renormalization model of diffraction and a saturated Froissart bound for the total cross section yielding $\sigma_t = (\pi/s_o) \cdot \ln^2(s/s_F)$ for $s > s_F$, where the parameters $s_o$ and $s_F$ are experimentally determined from the $\sqrt{s}$-dependence of the single-diffractive cross section.

## 1 Single Diffraction

The measurements of the elastic [1] ($\sigma_{el}$), total [2] ($\sigma_t$), and single-diffractive [3] ($\sigma_{sd}$) cross sections by the Collider Detector at Fermilab (CDF), published in 1994, brought into sharp focus the unitarity problems inherent in the traditional Regge theory description of soft cross sections (see [4]). According to the theory, the cross sections at high energies are dominated by Pomeron ($I\!P$) exchange, and with a Pomeron trajectory of intercept $\alpha(0) = 1 + \epsilon$ the $s$-dependence is given by:

$$\left.\frac{d\sigma_{el}}{dt}\right|_{t=0} \sim \left(\frac{s}{s_o}\right)^{2\epsilon},$$

$$\sigma_t \sim \left(\frac{s}{s_o}\right)^{\epsilon},$$

$$\sigma_{sd} \sim \left(\frac{s}{s_o}\right)^{2\epsilon}.$$

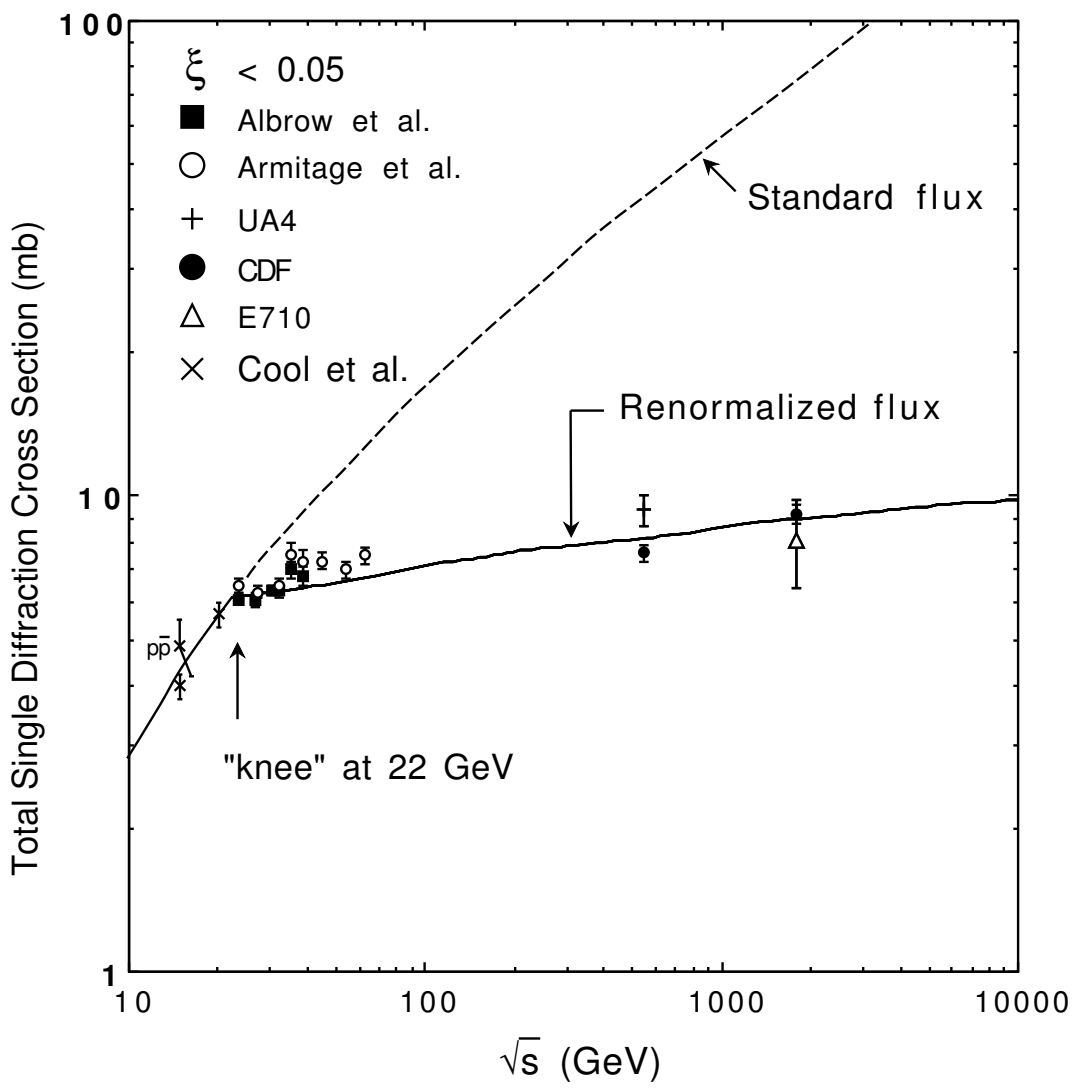

Figure 1: Total $pp/\bar{p}p$ single diffraction dissociation cross section data (*both* $\bar{p}$ and $p$ sides) for a forward $\bar{p}$ or $p$ momentum loss fraction $\xi < 0.05$ compared with Regge theory predictions based on the standard and the renormalized Pomeron flux (from Ref. [5]).

Such behaviour would violate unitarity at high energies with the elastic and/or the single-diffractive cross section(s) becoming larger than the total cross section. Unitarity, of course, should be obeyed in nature, and the CDF measurements of $\sigma_{sd}$ at $\sqrt{s} = 540$ and 1800 GeV showed that $\sigma_{sd}$ is suppressed at high energies relative to Regge predictions preserving unitarity. This result is spectacularly displayed in Fig. 1 from Ref. [5].

The Pomeron exchange contribution to the cross sections can be written as:

$$\sigma_t(s) = \beta^2_{I\!Ppp}(0)\left(\frac{s}{s_\circ}\right)^{\alpha(0)-1} \Rightarrow \sigma_\circ\left(\frac{s}{s_\circ}\right)^{\epsilon}, \tag{1}$$

$$\frac{d\sigma_{el}(s,t)}{dt} = \frac{\beta^4_{I\!Ppp}(t)}{16\pi}\left(\frac{s}{s_\circ}\right)^{2[\alpha(t)-1]}, \tag{2}$$

$$\frac{d^2\sigma_{sd}(s,\xi,t)}{d\xi dt} = \underbrace{\frac{\beta^2_{I\!Ppp}(t)}{16\pi}\,\xi^{1-2\alpha(t)}}_{f_{I\!P/p}(\xi,t)}\ \underbrace{\beta_{I\!Ppp}(0)\,g(t)\left(\frac{s'}{s_\circ}\right)^{\alpha(0)-1}}_{\sigma^{I\!Pp}(s',t)}. \tag{3}$$

The two terms in the diffractive cross section, Eq. (3), are the *Pomeron flux*, $f_{I\!P/p}(\xi,t)$, presumed to be emitted by the diffractively scattered proton, and the $I\!P$-$p$ total cross section, $\sigma^{I\!Pp}(s',t)$. The parameters appearing in Eq. (3) are identified as follows:

- $\alpha(t) = \alpha(0) + \alpha' t = (1+\epsilon) + \alpha' t = (1+0.08) + 0.25\,t$ is the Pomeron trajectory;
- $\beta_{I\!Ppp}(t)$ is the coupling of the Pomeron to the proton, $\beta^2_{I\!Ppp}(t) = \sigma_\circ \cdot e^{b_\circ t}$, where $\sigma_\circ \equiv \beta^2_{I\!Ppp}(0)$ and $e^{b_\circ t}$ is the form factor of the diffractively escaping proton, $F_p^2(t) = e^{b_\circ \cdot t}$;
- $g(t)$ is the triple-Pomeron ($I\!PI\!PI\!P$) coupling, which was found experimentally to be independent of $t$ [6];
- $s' \equiv M^2$ is the $I\!P$-$p$ c.m.s. $s$-value, with $M$ the mass of the diffractively excited proton;
- $\xi \approx M^2/s$ is the momentum fraction of the incident proton carried by the Pomeron;
- $s_\circ$ is an energy scale parameter.

In Ref. [5], the unitarity problem arising from the $s^{2\epsilon}$ dependence of the single-diffractive cross section was addressed by interpreting the Pomeron flux factor as the probability of forming a diffractive rapidity gap and *renormalizing* the integrated probability over all phase space in $\xi$ and $t$ to unity if it exceeded unity. Technically, the renormalization was accomplished by dividing the differential diffractive cross section by the flux integral above the $\sqrt{s}$ value of the $\bar{p}p$ collision energy at which the flux became unity.

The renormalization procedure solved an outstanding energy scale problem in diffraction. From Eq. (1), one sees that $\beta^2_{I\!Ppp}(0) \sim s_o^\epsilon$, and therefore in the diffractive cross section given in Eq. (3) the Pomeron flux contains a scale factor $s_o^\epsilon$ while the Pomeron-proton cross section contains a factor $s_o^{\epsilon/2} \cdot g(t)$. Consequently, neither $s_o$ nor $g(t)$ can be independently determined from the measurement of the differential or total diffractive cross sections, but only the product $g(t) \cdot s_o^{\epsilon/2}$. This is of such importance that it deserves being framed:

$$\boxed{f_{I\!P/p}(\xi,t) \sim s_o^\epsilon} \qquad \boxed{\sigma^{I\!Pp}(s',t) \sim s_o^{-\epsilon/2} \cdot g(t)} \quad \Rightarrow \quad \boxed{\sigma_{sd}\text{ determines: } \boldsymbol{g(t) \cdot s_o^{\epsilon/2}}}$$

In Ref. [5], this entanglement was resolved by using $s_o = 1\,\text{GeV}^2$, a value determined from the results displayed in Fig. 1. It was argued that the *knee* in the cross section observed at $\sqrt{s} = 22\,\text{GeV}$ occurs at the energy at which the Pomeron flux integral becomes unity. Since this integral depends on $s$ and $s_o$, determining the $s$-value at which the integral is unity yields $s_o$. The value of $s_o$ was found to be $s_o = 1\,\text{GeV}^2$, and that of the triple-Pomeron coupling $g(t) = 0.69\,\text{mb}^{\frac{1}{2}} = 1.1\ \text{GeV}^{-1}$. It was also mentioned in the paper that the uncertainty in $s_o$ in terms of the uncertainty in the position of the *knee* is $\delta s_o/s_o = -\delta s/s = -4(\delta\sqrt{s}/\sqrt{s})$, and thus a *reasonable* 10% uncertainty in the $\sqrt{s}$-position of the *knee* would result in a 40% uncertainty in the value of $s_o$.

With all the parameters in Eq. (3) experimentally determined, the differential and total single diffractive cross sections at the LHC can be predicted. In Ref. [5], using a linear logarithmic expression $A + B\ln s$ ($s$ in GeV$^2$) in the range $22 < \sqrt{s} < 10\,000$ GeV, the following parameterization was obtained for the total single diffractive cross section:

$$\sigma_{sd}^{p\bar{p}}\Big|_{\xi<0.05} \approx (4.3 + 0.3\ln s)\,\mathrm{mb} \quad (22 < \sqrt{s} < 10\,000\ \mathrm{GeV}). \tag{4}$$

By extrapolating to LHC energies, this formula predicts $\sigma_{sd}^{pp}|_{\xi<0.05} = 10.0\,\mathrm{mb}$ at $\sqrt{s} = 14$ TeV. An uncertainty of $\leq 10\%$ is estimated for the cross section in Eq. (4) given the 5% uncertainty of the CDF measurements and that resulting from the Pomeron trajectory parameters.

**The underlying basis of the renormalization concept** is revealed by a change of variables from $\xi$ to $M^2$ using the relationship $\xi = M^2/s$. This leads to a diffractive cross section:

$$\frac{d^2\sigma_{sd}(s, M^2, t)}{dM^2 dt} = \left[\frac{\sigma_\circ}{16\pi}\sigma_\circ^{I\!Pp}\right] \frac{s^{2\epsilon}}{N(s, s_o)} \frac{e^{bt}}{(M^2)^{1+\epsilon}}, \tag{5}$$

where $b = b_0 + 2\alpha' \ln\frac{s}{M^2}$ is the slope parameter of the $t$-distribution and $N(s, s_o)$ the integrated Pomeron flux. The latter is obtained from a straight-forward integration:

$$\text{definition}:\ f_{I\!P/p}(\xi, t) \Rightarrow N^{-1}(s, s_o)\cdot f_{I\!P/p}(\xi, t),$$

$$N(s, s_o) \equiv \int_{\xi(min)}^{\xi(max)} d\xi \int_{t=0}^{-\infty} dt\, f_{I\!P/p}(\xi, t) \overset{s\to\infty}{\to} \sim s_o^{\epsilon}\cdot s^{2\epsilon}/\ln s. \tag{6}$$

The asymptotic form for $s \to \infty$ is given here to illustrate that division by the integrated flux in Eq. (5) replaces the $\boldsymbol{s^{2\epsilon}}$ term by a $\mathbf{\ln}\,\boldsymbol{s}$ dependence preserving unitarity:

$$\frac{d^2\sigma_{sd}(s, M^2, t)}{dM^2\,dt} \overset{s\to\infty}{\to} \sim \ln s\, \frac{e^{bt}}{(M^2)^{1+\epsilon}}. \tag{7}$$

In view of the above, the renormalization concept can be phenomenologically understood within both multi-Pomeron exchange and QCD inspired models. In either case, the $s^\epsilon$ ($s^{2\epsilon}$) factor in $\sigma_t$ ($\sigma_{sd}$) arises from overlapping rapidity gaps. Renormalization eliminates this type of *double counting* while preserving the ($\xi$, $t$) or ($M^2$, $t$) dependence of the cross section.

Integrating Eq. (7) over $M^2$ and $t$ yields a constant total single diffractive cross section:

$$\sigma_{sd} \overset{s\to\infty}{\to} 2\,\sigma_\circ^{I\!Pp}\ \exp\left[\frac{\epsilon\, b_0}{2\alpha'}\right] = \sigma_{sd}^{\infty} \Rightarrow (16.8 \pm 0.5)\,\mathrm{mb}\ (\text{see text}). \tag{8}$$

In a recent paper [8], it is suggested that since renormalization eliminates the overlaps caused by *wee-parton* exchanges, $\sigma_{sd}^{\infty}$ must be set equal to the value of $\sigma_o$ of $\sigma_t = \sigma_o\, e^{\epsilon}$. The global fit of Ref. [7] yields $\sigma_o = (16.8 \pm 0.5)$ mb, where the uncertainty is obtained from the uncertainty in the value of $\epsilon = 0.104 \pm 0.002$ quoted in the paper using the correlation between the errors in $\sigma_o$ and $\epsilon$ for fixed $\sigma_t$, which results in $\delta\sigma_o = \sigma_o \cdot \ln s \cdot \delta\epsilon$.

Since renormalization converts $s^{2\epsilon} \Rightarrow \ln s$ in the differential diffractive cross section, the $M^2$ distribution would be expected to have no substantial explicit $s$-dependence. This prediction is confirmed by the data, as shown in Fig. 2. The straight line through the data points is not a fit but is shown here to guide the eye. A fit would have to take into account the dependence of the slope parameter $b$ on $\xi$, and this should be done by comparing the data with a Monte Carlo simulation. However, the difference that would be obtained using such a comparison is estimated to be small, and the $M^2$-scaling behaviour exhibited in Fig. 2 does indeed correctly convey the message that renormalization removes the overlaps, which would cause the cross section to follow the disconnected *standard flux* dotted lines shown in the figure for the different collision energies.

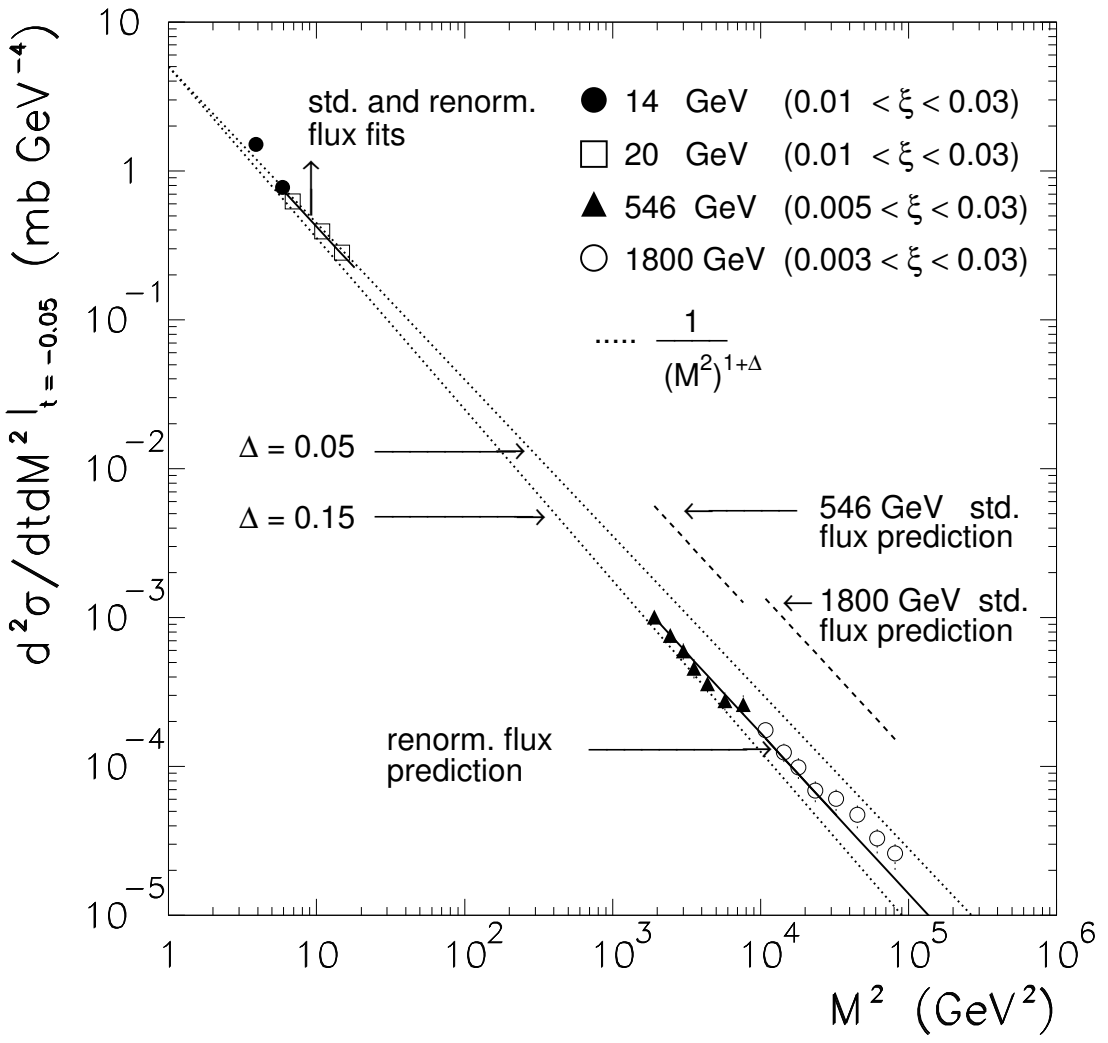

Figure 2: Cross sections $d^2\sigma_{sd}/dM^2dt$ for $p + p(\bar{p}) \to p(\bar{p}) + X$ at $t = -0.05\,(\mathrm{GeV/c})^2$ and $\sqrt{s}$ = 14, 20, 546 and 1800 GeV. Standard (renormalized) flux predictions are shown as dashed (solid) lines (from Ref. [9]).

The renormalization technique used here can be applied to all hadronic diffractive processes, soft and hard alike, which in terms of the final state event topology can be classified into three main categories: forward gap, central gap, and multi-gap diffraction. Moreover, it can be applied to photoproduction and Deep Inelastic Scattering diffractive processes, predicting the factorization breaking observed at the edges of the available phase space, as outlined in the talk on *Factorization Breaking in Diffraction* presented at this conference [10].

# 2 Total Cross Section

## 2.1 The SUPERBALL Model

Theoretical models predicting the total cross section at the LHC must satisfy all unitarity constraints. Available accelerator and cosmic ray data are routinely used to tune the parameters of the models before extrapolating to LHC energies. This process is usually cumbersome, as it involves fitting data which in some cases are not mutually compatible. Using all relevant published data often leads to fits with a $\chi^2$/d.o.f. *pulled* by the *outliers* in the measurements, where *outliers* are data points in clear disagreement with adjacent points from other measurements. Here, we present a model in which these problems are minimized by an inherently unitarized approach based on a saturated Froissart bound above a value of $s = s_F$:

$$\textbf{saturated}\text{ Froissart bound}: \;\; \sigma_t(s > s_F) = \sigma_t(s_F) + \frac{\pi}{s_o} \cdot \ln^2 \frac{s}{s_F}. \tag{9}$$

The saturation occurs in the *wee*-parton exchange governed by the value of the scale parameter $s_o$ that appears in the diffractive cross section in Eq. (3). This parameter is therefore interpreted as the mass-squared of an *object* that is exchanged, and when inserted into the Froissart formula in place of the traditionally used $m_\pi^2$ should saturate the bound above the saturation energy $\sqrt{s}_F = 22$ GeV obtained in Sec. 1 from Fig. 1. As this exchanged *object* resembles a glue-ball, we will refer to this normalization procedure as the **SUPER**GLUE**BALL** or SUPERBALL model.

## 2.2 The Total Cross Section at LHC

Predicting the total cross section at LHC using Eq. (9) would require knowledge of $\sigma(s_F)$ at $\sqrt{s}_F = 22$ GeV. However, the cross section at this energy has substantial Reggeon exchange contributions and also contributions from the interference between the nuclear and Coulomb amplitudes. A complete description would have to take all this into consideration, using the Regge theory amplitudes to describe the Reggeon exchanges, and dispersion relations to obtain the real part of the amplitude from the measured total cross sections up to Tevatron energies. Here, we apply a strategy that bypasses all these complications.

**Strategy:**

- Use the Froissart formula as a *saturated* cross section rather than as a bound above $s_F$:

$$\boxed{\sigma_t(s > s_F) = \sigma_t(s_F) + \frac{\pi}{m^2} \cdot \ln^2 \frac{s}{s_F}}$$

- This formula should be valid above the *knee* in $\sigma_{sd}$ vs. $\sqrt{s}$ at $\sqrt{s}_F = 22$ GeV (Fig. 1) and therefore valid at $\sqrt{s} = 1800$ GeV.

- Use $m^2 = s_o$ in the Froissart formula multiplied by 1/0.389 to convert it to $\text{mb}^{-1}$.

- Note that contributions from Reggeon exchanges at $\sqrt{s} = 1800$ GeV are negligible, as can be verified from the global fit of Ref. [7].

- Obtain the total cross section at the LHC:

$$\boxed{\sigma_t^{\text{LHC}} = \sigma_t^{\text{CDF}} + \frac{\pi}{s_o} \cdot \left( \ln^2 \frac{s^{\text{LHC}}}{s_F} - \ln^2 \frac{s^{\text{CDF}}}{s_F} \right)}$$

For a numerical evaluation of $\sigma^{LHC}$ we use as input the CDF cross section at $\sqrt{s} = 1800$ GeV, $\sigma_t^{\text{CDF}} = (80.03 \pm 2.24)$ mb, the Froissart saturation energy $\sqrt{s}_F = 22$ GeV, and the parameter $s_o$. In Sec. 1, it it was mentioned that a value of $s_o = (1.0 \pm 0.4)\ \text{GeV}^2$ was extracted from the $s$-dependence of the single-diffractive cross section. The extraction of $s_o$ from the data assumed $\epsilon = 0.115 \pm 0.008$, which was the average of the CDF measurements at 540 and 1800 GeV. There is, however, a very strong correlation between the values of $\epsilon$ and $s_o$ through the relationship displayed in Eq. (6). Using a more accurate value of $\epsilon$ extracted in [7], $\epsilon = 0.104 \pm 0.002$, yields $s_o^{\text{CMG}} = 3.7\ \text{GeV}^2$. The resulting prediction for the total cross section at the LHC at $\sqrt{s} = 14\,000$ GeV is:

$$\sigma_{14\,000\,\text{GeV}}^{\text{LHC}} = (80 \pm 3)\,\text{mb} + (29 \pm 12)\,\text{mb} = (109 \pm 12)\,\text{mb}.$$

This result is in good agreement with the value of $\boxed{\sigma_t^{\text{CMG}} = (114 \pm 5)\ \text{mb}}$ obtained by the global fit of Ref. [7] using an *eikonal* approach, where the uncertainty is estimated from that in the value of the parameter $\epsilon$ given in the paper.

The agreement between $\sigma_t^{\text{superball}} = (109 \pm 12)\,\text{mb}$ and $\sigma_t^{\text{CMG}} = (114 \pm 5)\,\text{mb}$ is remarkable, but there are two items to bear in mind: (a) a value of $s_o = 1\,\text{GeV}^2$ was used in the CMG eikonalized evaluation of the cross section, since the result of Ref. [5] was already known by the authors of Ref. [7]; (b) the sensitivity of the present result on the value of $\epsilon$ cannot be overemphasized, and as is the case with the determination of $s_F$, it represents a limiting factor on the accuracy that can be achieved in the prediction of $\sigma^{\text{LHC}}$.

## 3 Summary and Conclusion

The single-diffractive and total *pp* cross sections at the LHC are predicted in a phenomenological approach that obeys all unitarity constraints. The approach is based on the renormalization model of hadronic diffraction, which corrects the double-counting caused by overlapping diffractive rapidity gaps while preserving the dependence of the differential cross section on the fractional momentum loss, $\xi$, and 4-momentum transfer squared, $t$, of the diffracted proton. The renormalization procedure replaces the $s^{2\epsilon}$ dependence of the *differential* diffractive cross section with an $\ln s$ dependence and leads to an asymptotically constant single-diffractive cross section as $s \to \infty$ of $\sigma_{sd}^{\infty} = (16.8 \pm 0.5)\,\text{mb}$.

The total cross section at the LHC is estimated using a *saturated* Froissart bound expression $\sigma_t(s > s_F) = \sigma_t(s_F) + \pi/s_o \cdot \ln^2(s/s_F)$, where the parameters $s_o$ and $s_F$ are experimentally determined from the dependence of the single-diffractive cross section on $\sqrt{s}$. Encoring $\sigma_t$ to the CDF measured value at $\sqrt{s} = 1800\,\text{GeV}$, where Reggeon exchange contributions are negligible, serves to normalize the formula yielding $\sigma_t^{\text{LHC}} = (109 \pm 12)\,\text{mb}$, which is in good agreement with the global fit prediction of $\sigma_t^{\text{CMG}} = (114 \pm 5)\,\text{mb}$ of Ref. [7].

# References

[1] F. Abe et al. (CDF Collaboration), *Measurement of small angle antiproton-proton elastic scattering at* $\sqrt{s} = 546$ *and 1800 GeV*, Phys. Rev. D **50**, 5518-5535 (1994).

[2] F. Abe et al. (CDF Collaboration), *Measurement of the antiproton-proton total cross section at sqrt(s) =546 and 1800 GeV*, Phys. Rev. **50**, 5550-5561 (1994).

[3] F. Abe et al. (CDF Collaboration), *Measurement of* $\bar{p}p$ *single diffraction dissociation at sqrt(s)=546 and 1800 GeV*, Phys. Rev D **50**, 5535 (1994).

[4] K. Goulianos, *Diffractive interactions oh hadrons at high energies*, Phys. Rept. **101**, 169 (1983).

[5] K. Goulianos, *Renormalization of hadronic diffraction and the structure of the pomeron*, Phys. Lett. B **358**, 379 (1995); Erratum: *ib.* **363**, 268 (1995).

[6] R. L. Cool, K. Goulianos, S. L. Segler, H. Sticker, S. N. White, *Diffraction dissociation of* $\pi^{\pm}$, $K^{\pm}$, *and* $p^{\pm}$ *at 100 and 200 GeV/c*, Phys. Rev. Lett. **47**, 701 (1981).

[7] R. J. M. Covolan, J. Montanha and K. Goulianos, *A New determination of the soft pomeron intercept*, Phys. Lett. B **389**, 176 (1996).

[8] K. Goulianos, *Pomeron intercept and slope: the QCD connection*, arxiv:0812.4464v2 [hep-ph] 26 Mar 2009.

[9] K. Goulianos and J. Montanha, *Factorization and scaling in hadronic diffraction*, Phys. Rev. D **59**, 114017 (1999).

[10] K. Goulianos, *Factorization Breaking in Diffraction*, in these Proceedings.

# Amplitudes in the Coulomb Interference Region of pp and p$\bar{\text{p}}$ Scattering

*Anderson Kendi Kohara, Erasmo Ferreira, Takeshi Kodama*

Instituto de Física, Universidade Federal do Rio de Janeiro
C.P. 68528, Rio de Janeiro 21945-970, RJ, Brazil

The determination of the parameters of the pp and p$\bar{\text{p}}$ amplitudes used for the description of scattering in the Coulomb interference region is discussed, with emphasis put on the possibility that the effective slope observed in the differential cross section is formed by different exponential slopes in the real and imaginary amplitudes (called $B_R$ and $B_I$). For this purpose a more general treatment of the Coulomb phase is developed.

The differential cross section data in the range from 19 to 1800 GeV are analysed with four parameters ($\sigma$, $\rho$, $B_I$, $B_R$), and it is observed that unique determination of the parameters cannot be obtained from the available data. Correlations in pairs of the four quantities are investigated, showing ranges leading to the smaller $\chi^2$ values.

## 1 Low |t| Region and Coulomb Phase

Along about 50 years since the beginning of the experimentation of pp and p$\bar{\text{p}}$ scattering at high energies, theoretical models and interpretations are continuously developed, but the experimental data stopped increasing in quantity or quality, restricting the progress in several aspects. The interest in these systems is now renewed [1], due to the data that will come from RHIC and LHC .

The disentanglement of the squared moduli that represent the measured quantities in terms of the imaginary and real amplitudes is not at all an easy phenomenological task. In previous analyses of the pp and p$\bar{\text{p}}$ data, the real and imaginary amplitudes were considered as having the same exponential dependence $\exp(Bt/2)$, where $B$ is the slope of the log plot of $d\sigma/dt$ . This simplifying assumption is not adequate, according to dispersion relations [2] and according to the theorem of A. Martin [3] that says that the position of the zero of the real amplitude is close and approaches $t = 0$ as the energy increases. Both results indicate that the slope of the real amplitude should be larger, and in the present work we investigate the description of the Coulomb interference region allowing for different real and imaginary slopes.

In elastic pp and p$\bar{\text{p}}$ collisions, the combined nuclear and Coulomb amplitude is written

$$F^{C+N}(s,t) = F^C(s,t)e^{i\alpha\Phi(s,t)} + F_R^N(s,t) + i\ F_I^N(s,t)\ , \tag{1}$$

where the Coulomb part $F^C$ and the proton form factor are

$$F^C = (-/+)\ \frac{2\alpha}{|t|}\ F^2_{\text{proton}}\quad ,\quad F_{\text{proton}} = (0.71/(0.71+|t|))^2\ .$$

Our treatment is simplified in the sense that we do not consider different form factors for spin flip amplitudes, assuming that this is fair enough for the purposes of this paper.

The phase $\Phi$ between Coulomb and nuclear amplitudes was first considered by H.A. Bethe and then by West and Yennie [4], followed by different evaluations worked out by several authors [5, 6, 7]. In the present work we extend these investigations in the very low $|t|$ range, considering the possibility of different slopes for the real and imaginary amplitudes.

For small angles we can approximate

$$F^N(s,t) \approx F_R^N(s,0)e^{B_R t/2} + iF_I^N(s,0)e^{B_I t/2} \ . \tag{2}$$

The parameter $\rho$ and the Optical Theorem

$$\rho = \frac{F_R^N(s,0)}{F_I^N(s,0)} \quad , \quad \sigma = 4\pi \ (0.389) \ \text{Im} \ F_I^N(s,0)$$

and the slopes $B_R$, $B_I$ are used to parametrize the differential cross section for small $|t|$. In these expressions, $\sigma$ is in milibarns and the amplitudes $F_R$, $F_I$ are in GeV$^{-2}$ .

For low $|t|$ we have the usual slope $B$ observed in $d\sigma/dt$ data

$$\frac{d\sigma}{dt} = \left|\frac{d\sigma}{dt}\right|_{t=0} e^{Bt} \ , \ \text{with} \ \ B = \frac{\rho^2 B_R + B_I}{1+\rho^2} \ .$$

In our evaluation of $\Phi(s,t)$ we start from the expression obtained by West and Yennie [4]

$$\Phi(s,t) = (-/+)\left[\ln\left(-\frac{t}{s}\right) + \int_{-4p^2}^{0} \frac{dt'}{|t'-t|}\left[1 - \frac{F^N(s,t')}{F^N(s,t)}\right]\right] \ , \tag{3}$$

where the signs $(-/+)$ are applied to the choices pp/p$\bar{\text{p}}$ respectively. The quantity $p$ is the proton momentum in center of mass system, and at high energies $4p^2 \approx s$.

For small $|t|$, assuming that $F^N(s,t')$ keeps the same form for large $|t'|$, we have

$$\frac{F^N(s,t')}{F^N(s,t)} = \frac{c}{c+i} \ e^{B_R(t'-t)/2} + \frac{i}{c+i} \ e^{B_I(t'-t)/2} \ , \ \text{with} \ \ c \ \equiv \ \rho e^{(B_R - B_I)t/2} \ .$$

The integrals that appear in the evaluation are reduced to the form [7]

$$I(B) = \int_{-4p^2}^{0} \frac{dt'}{|t'-t|}\left[1 - e^{B(t'-t)/2}\right] \ , \tag{4}$$

that is solved in terms of exponential integrals [8] as

$$I(B) = E_1\left(\frac{Bs}{2}\right) - E_i\left(-\frac{Bt}{2}\right) + \ln\left(\frac{Bs}{2}\right) + \ln\left(-\frac{Bt}{2}\right) + 2\gamma \ . \tag{5}$$

where we have assumed the high energy simplification $4p^2 + t \to s$.

The real and imaginary parts of the phase are then written

$$\Phi_R(s,t) = (-/+)\left[\ln\left(-\frac{t}{s}\right) + \frac{1}{c^2+1}\left[c^2 I(B_R) + I(B_I)\right]\right] \ ,$$

$$\Phi_I(s,t) = (-/+)\frac{c}{c^2+1}\left[I(B_I) - I(B_R)\right] \ .$$

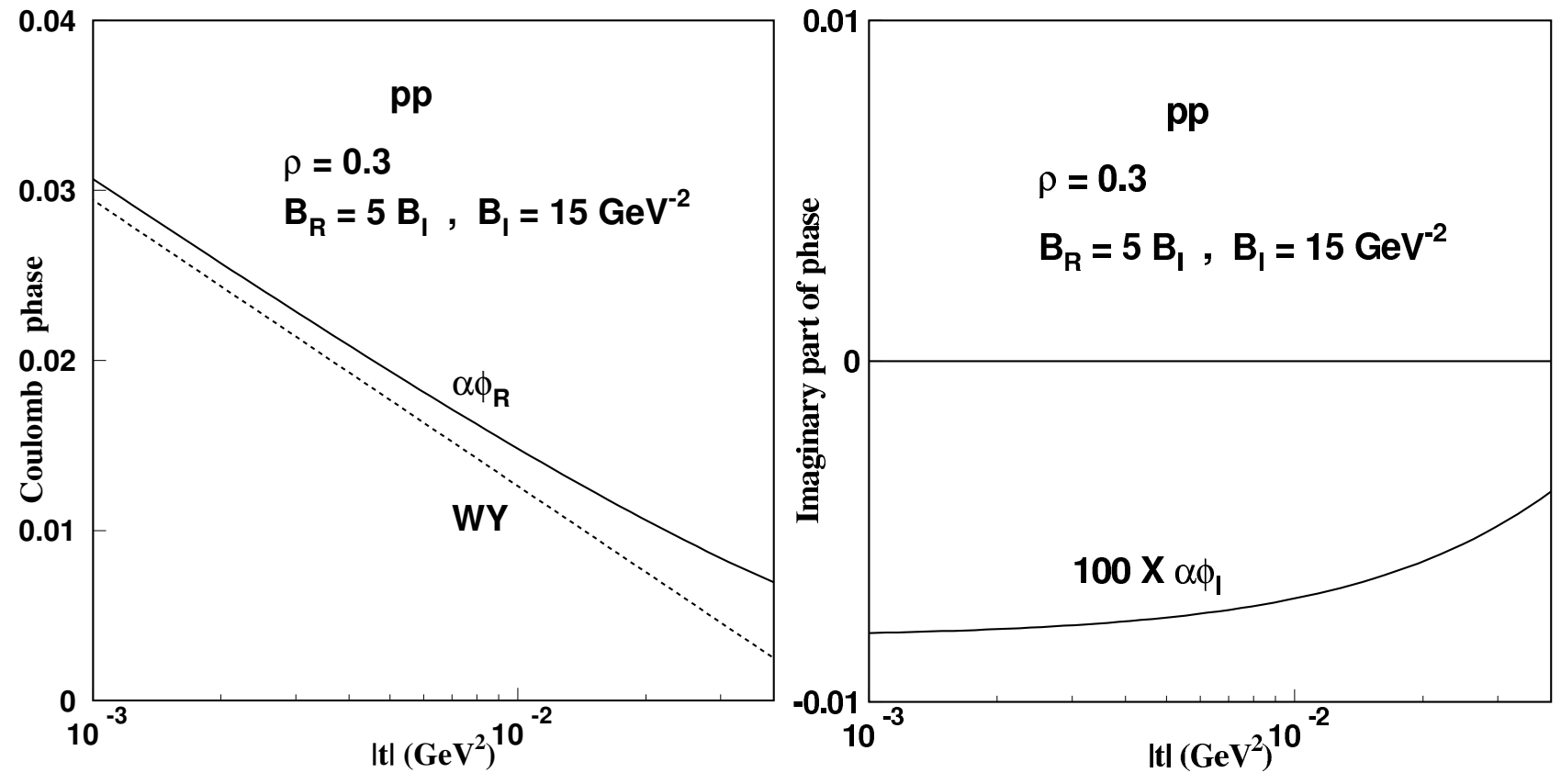

Figure 1: Comparison of the Coulomb phase $\alpha\Phi_R$ from our calculation with the West-Yennie (WY) expression in Eq. (7). The unrealistic large values of parameters $\rho = 0.3$ and $B_R/B_I = 5$ are chosen in order to enhance differences. The calculation is made at $\sqrt{s} = 50\,\mathrm{GeV}$, but the values of $\alpha\Phi_R(s,t)$ do not show explicit dependence on $s$ up to the LHC energy. The plot of $\alpha\Phi_I$ in the RHS shows extremely small values (notice the scale), about 100 times smaller than $\alpha\Phi_R$, so that we can safely put $\alpha\Phi_I$ equal to zero.

In the normalization that we use [9], with $\sigma$ in mb and $t$ in GeV$^2$, the practical expression for $d\sigma/dt$ in terms of the parameters $\sigma$, $\rho$ , $B_I$ and $B_R$ is

$$\frac{d\sigma}{dt} = 0.389\ \pi \left[ \left[ \frac{\rho\ \sigma\ e^{B_R t/2}}{0.389 \times 4\pi} + F^C e^{\alpha\Phi_I}\cos(\alpha\Phi_R)\right]^2 + \left[\frac{\sigma\ e^{B_I t/2}}{0.389\times 4\pi} + F^C e^{\alpha\Phi_I}\sin(\alpha\Phi_R)\right]^2 \right] . \quad (6)$$

The expression for $\alpha\Phi_R$ can be compared with the expression from West and Yennie (WY)

$$\alpha\Phi_{WY} = (-/+)\alpha\left[\gamma + \ln\left(-\frac{Bt}{2}\right)\right] . \quad (7)$$

Figure 1 presents a numerical comparison, where we choose large unrealistic values $\rho = 0.3$, $B_R/B_I = 5$ to enhance possible differences. In the WY formula $B$ is taken as the $d\sigma/dt$ average slope. The plot of $\alpha\Phi_I(s,t)$ in the RHS shows extremely small values so that the imaginary part can be put equal to zero.

# 2 Analysis of Experimental Data

The data published with absolute values for $d\sigma/dt$ depend on evaluation of luminosity, using Coulomb interference forms different from ours, or sometimes take some parameter values (such as the total cross section) from other experiments. To have independent evaluation of the absolute normalization we consider the published data as event rates $dN/dt$, using Eq. (6) with a free normalization factor $a_5$, namely we put the published data as

$$\frac{dN}{dt} = a_5\ \frac{d\sigma}{dt} , \quad (8)$$

Table 1: Results of the analysis at 19.4 - 62.5 GeV. N = number of points fitted, in general with $|t| \leq 0.01\,\text{GeV}^2$; $a_5$ is a normalization factor (see text). We give results for fixed ratios $B_R/B_I$ = 1 and 2. For 23.5 GeV the normalization factor found is 1. $\chi^2 = \sum \chi_i^2/(N-5)$.

| $\sqrt{s}$ | N | $\sigma$(mb) | $\rho$ | $B_I(\text{GeV}^{-2})$ | $B_R$ | $a_5$ | $\chi^2$ |
|---|---|---|---|---|---|---|---|
| 19.4 | 45 | 38.92±0.09 | 0.0064±0.0044 | 11.25±0.84 | $B_I$, $2B_I$ | 1.040±0.004 | 1.015 |
| 19.4 | 45 | 38.92±0.09 | 0.01 (fixed) | 11.25±0.84 | $B_I$, $2B_I$ | 1.042±0.004 | 1.015 |
| 30.6 | 14 | 40.35±0.12 | 0.021±0.003 | 12.00±1.2 | $B_I$ | 0.9878±0.0028 | 0.7301 |
| 30.6 | 14 | 40.34±0.12 | 0.021±0.003 | 12.00±1.2 | $2B_I$ | 0.9877±0.0028 | 0.7301 |
| 44.7 | 26 | 41.36±0.23 | 0.071±0.008 | 12.80±0.25 | $B_I$ | 1.0296±0.0141 | 0.7307 |
| 44.7 | 26 | 41.33±0.23 | 0.072±0.008 | 12.80±0.25 | $2B_I$ | 1.0304±0.0139 | 0.7269 |
| 52.8 | 12 | 42.89±0.08 | 0.063±0.005 | 13.00±0.2 | $B_I$ | 0.9761±0.0030 | 0.0846 |
| 52.8 | 12 | 42.86±0.08 | 0.063±0.005 | 13.00±0.2 | $2B_I$ | 0.9766±0.0030 | 0.0859 |
| 62.5 | 17 | 42.46±0.03 | 0.1043±0.0017 | 13.15±0.2 | $B_I$ | 1.0435±0.0012 | 1.056 |
| 62.5 | 17 | 42.38±0.03 | 0.1064±0.0018 | 13.15±0.2 | $2B_I$ | 1.0465±0.0012 | 1.041 |

with $d\sigma/dt$ written as in Eq. (6), and $a_5$ is determined for each experiment.

We use CERN-Minuit programs (PAW and ROOT) to obtain correlations for the parameters ($\sigma$, $\rho$, $B_R$, $B_I$ and $a_5$), for values of energy where the data seem to have more quality and quantity. As shown in the tables, the values of the physical parameters are sensitive to the freedom given by $a_5$, although this factor is nearly 1. However, the visual quality of fittings are very good in either case (with or without $a_5$), producing superposed curves.

The results obtained depend strongly on the set of N low $|t|$ points selected for the analysis. In close inspection of the data in the range from 19.4 up to 62.5 GeV we observe a subtle knee in the data at about $|t| = 0.01\,\text{GeV}^2$ . The parameter values and $\chi^2$ become unstable when the number of points is larger than N. Examination of the contributions shows that at this position the real nuclear plus the Coulomb parts rapidly become very small compared to the imaginary part, with an exchange of dominating roles.

In the range 19.4 - 62.5 GeV, $\rho$ is very small, the real nuclear amplitude has little influence and the values of $\rho$ and $B_R$ are undetermined, while $\sigma$ and $B_I$ are more stable. Actually we may consider that at these energies $B_I$ is the most reliably determined parameter, as it is less dependent on normalization and on influence of the real amplitude.

## 2.1 Energy Range 19.4 to 62.5 GeV

Our attempts to determine parameters are described below. Some results are given in Table 1, where we stress the role of the normalization factor $a_5$ in the determination of the parameters at the lower energies. At 19.4 GeV the normalization $a_5$ is essential to lead to a positive value of $\rho$. Table 2 presents the results without free normalization factor. Notice the differences in parameter values in comparison with the treatment using $a_5$. To compare $\chi^2$ values, we must take into account that they mean $\sum \chi_i^2/(N-5)$ and $\sum \chi_i^2/(N-4)$ respectively.

Figure 2 shows examples of our fitting solutions, with data at 19.4, 44.7 and 62.5 GeV. At 44.7 and 62.5 GeV there are two lines, with and without factor $a_5$, and all with $B_R = 2B_I$, and following the parameters given in the same table. The lines are superposed, and cannot be distinguished by eye, although $\sigma$ and $\rho$ are different. We thus see the $\chi^2$ criterion cannot determine parameters. It is remarkable that the fitting lines work very well much beyond the fitting limit points (indicated by a vertical dashed lines) given by N; however fitting using a larger number instead of N introduces instability in the determination of the quantities. Thus, finding parameters and producing a line describing well the points are two different questions.

Table 2: Results without free normalization factor. At 23.5 and 30.6 GeV the values of parameters for ratios 1 and 2 are the same in the digits exhibited. $\chi^2 = \sum \chi_i^2/(N-4)$.

| $\sqrt{s}$ | N | $\sigma$(mb) | $\rho$ | $B_I(\mathrm{GeV}^{-2})$ | $B_R$ | $a_5$ | $\chi^2$ |
|---|---|---|---|---|---|---|---|
| 23.5 | 24 | 39.53±0.08 | 0.014±0.003 | 11.20±1.12 | $B_I, 2B_I$ | — | 0.3028 |
| 30.6 | 14 | 40.30±0.10 | 0.034±0.003 | 12.00±1.0 | $B_I, 2B_I$ | — | 0.7160 |
| 44.7 | 26 | 41.81±0.03 | 0.054±0.002 | 12.94±0.41 | $2B_I$ | — | 0.9179 |
| 52.8 | 12 | 42.48±0.08 | 0.077±0.005 | 13.00±0.20 | $B_I$ | — | 0.1422 |
| 52.8 | 12 | 42.46±0.08 | 0.078±0.005 | 13.00±0.20 | $2B_I$ | — | 0.1395 |
| 62.5 | 17 | 43.27±0.03 | 0.086±0.002 | 13.15±0.12 | $B_I$ | — | 1.141 |
| 62.5 | 17 | 43.25±0.03 | 0.086±0.002 | 13.15±0.12 | $2B_I$ | — | 1.146 |

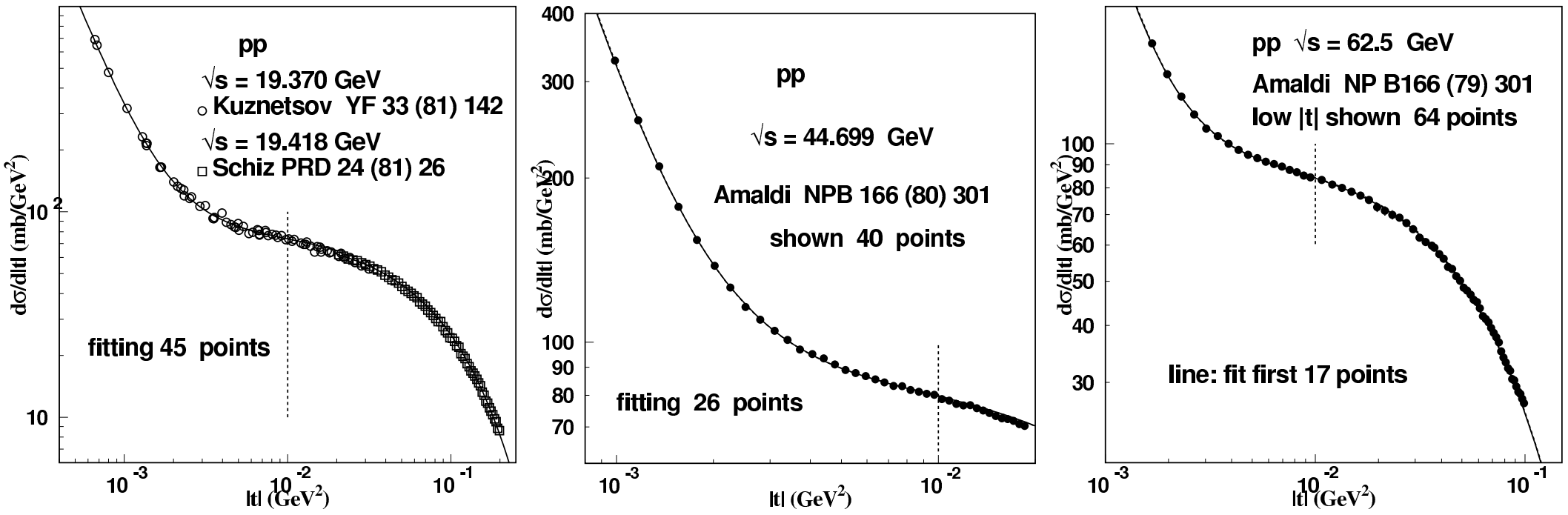

Figure 2: $d\sigma/dt$ and fitting curves at 19.4, 44.7 and 62.5 GeV. At 44.7 and 62.5 GeV there are two superposed curves, solid and dashed, obtained with and without normalization factor $a_5$. At 19.4 GeV the curve drawn up to 0.02 GeV$^2$ includes a factor $a_5 = 1.040$ to describe Kuznetsov's data. The vertical dashed lines show the limit of the N points used in the fitting. The curves follow well the data much beyond these limits.

### 2.2 p$\bar{\mathrm{p}}$ Scattering at 541, 1800 and 1960 GeV

Figure 3 shows the good agreement of $d\sigma/dt$ at low $|t|$, obtained from dN/dt by adjustment of the Coulomb interference, with the data of G. Arnison et al [17] in a $|t|$ range that partially covers the $dN/dt$ event rate, and also the data of Bozzo et al. [18], including high $|t|$ values.

The lowest $|t|$ values reached in measurements of p$\bar{\mathrm{p}}$ elastic scattering at about 540 GeV are reported with event rates [16] only, with 99 points in the interval $0.000875 \le |t| \le 0.11875$ GeV$^2$. We use the Coulomb interference to find the normalization factor connecting event rate and differential cross-section. We fit the 37 points up to $|t| = 0.01$, finding the normalization factor $a_5 = 10.6$ with $\chi^2 = 1.531$ and the parameter values

$$\sigma = (62.93 \pm 0.09)\,\mathrm{mb}\ ,\ \rho = 0.145 \pm 0.004\ ,\ B_I = (15.50 \pm 0.40)\ \mathrm{GeV}^{-2},\ B_R = 2B_I\ .$$

The data at 1800 GeV from two Fermilab experiments are limited by $|t| \ge 0.03$ GeV$^2$ and do not reach the Coulomb interference region. Thus the determination of forward scattering parameters is doubtful. We obtain for E710 (Amos et al.) $\sigma = (72.748 \pm 0.186)\,\text{mb}$, $\rho = 0.14$ (fixed), $B_I = (16.30 \pm 0.04)$ GeV$^{-2}$ ,$B_R = 7B_I$, with $\chi^2 = 0.60$. Figure 4 shows the fittings of the E-710 (Amos, with 51 points) and CDF (Abe, with 26 points).

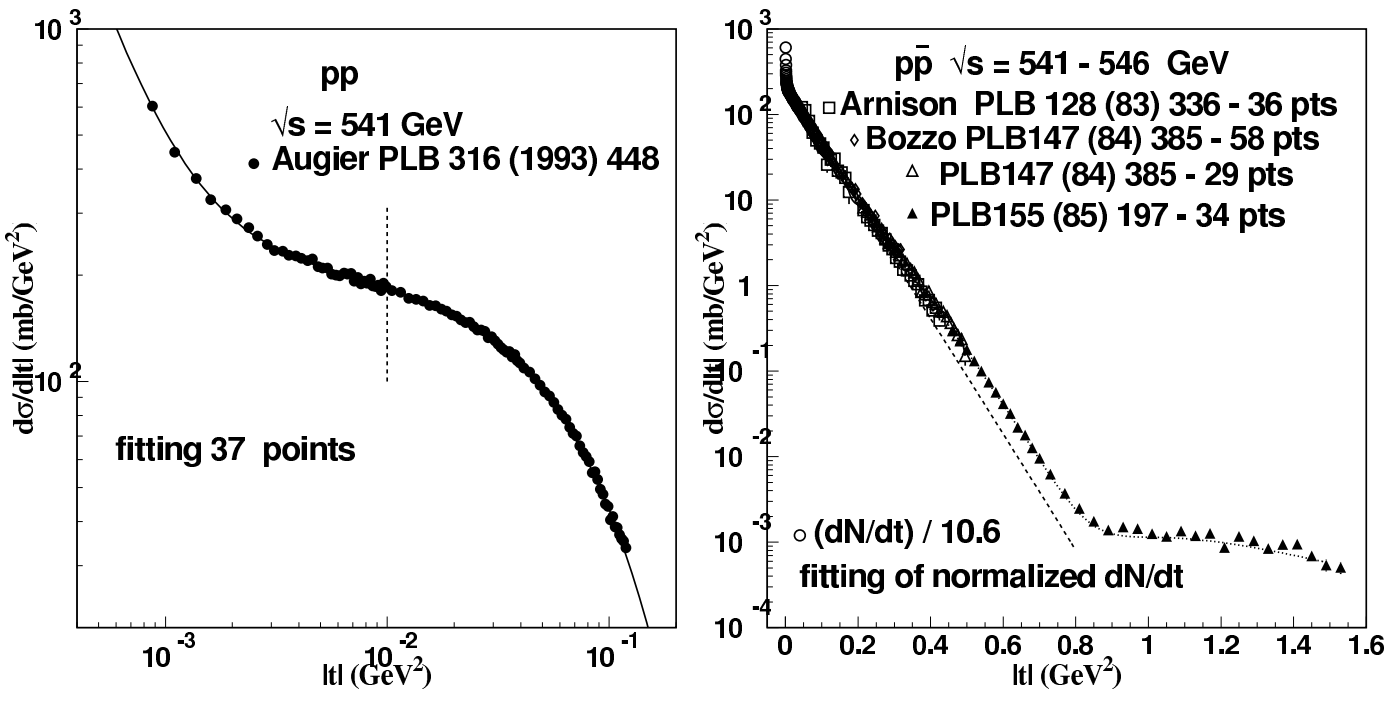

Figure 3: The normalized event rate data at 541 GeV are in good agreement with the $d\sigma/dt$ data [17, 18] at the same energy and higher $|t|$ values.

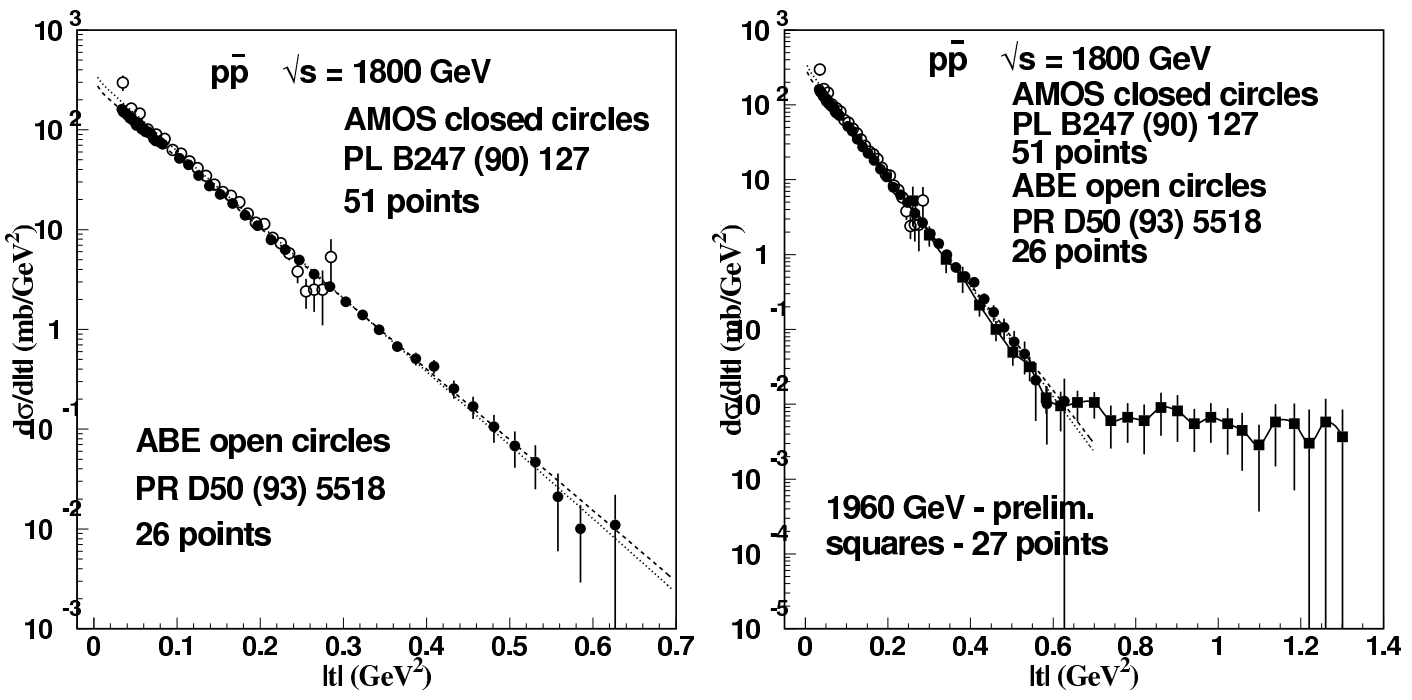

Figure 4: The 1800 GeV data from E710, with 51 points fitted by the dashed line, and from CDF, with 26 points fitted by the dotted line. The RHS plot shows also the preliminary 27 points at 1.96 TeV.

# 3 Remarks and Conclusions

We have performed a detailed analysis of the $d\sigma/dt$ obtained in the ISR/SPS (CERN) and Tevatron (Fermilab) experiments. with freedom of different slopes for the real and imaginary amplitudes, namely $B_R \neq B_I$. The main conclusion is that different values of the slopes, in particular the possibility of $B_R > B_I$, in accordance with the expectations from Martin's theorem [3] and from dispersion relations [2], are perfectly consistent within the present values and errors of experimental data.

We investigate the four physical quantities relevant for the elastic forward processes, namely, $\sigma$, $\rho$, $B_R$ and $B_I$. Studying the behaviour of $\chi^2$ values and the statistically equivalent parameter ranges, we observe that the available data for small $|t|$ at the energies $20 - 2000\,\mathrm{GeV}$ are not sufficient for a good determination of these four parameters, which have ample freedom, with equivalent correlated ranges. The whole analysis shows that the description of the amplitudes requires efficient external inputs, such as dispersion relations.

We stress that this paper does not intend to give the best final values for the scattering parameters. Instead, we show that the analysis of the data leave rather ample possibilities that must be further investigated. We call attention to details that must be taken into account in the acquisition and treatment of future data from RHIC and LHC, expecting to have different situations, with much better statistics and accuracy in the measurements at the Coulomb interference region, together with a systematic energy scan program.

## Acknowledgements

The authors are grateful to support received from the Brazilian agencies CNPq and FAPERJ .

## References

[1] R.Fiore, L.Jenkovszky, R.Orava, E.Predazzi, A.Produkin, O.Selyugin,*Int. J.Mod. Phys. A***24**, 2551 (2009)

[2] E. Ferreira, *Int. J. Mod. Phys. E* **16**, 2893 (2007)

[3] A. Martin , *Phys. Lett. B* **404**, 137 (1997)

[4] G. B. West and D. Yennie , *Ann. of Phys.* **3**, 190 (1958)

[5] O. V. Selyugin , *Phys. Rev. D* **60**, 074028 (1999)

[6] V.A. Petrov, E. Predazzi and A. Prokudin, Eur. Phys. J. C **28**, 525 (2003)

[7] V.Kundrát, M.Lokajicek and I. Vrkoc, *Phys. Lett. B* **656**, 182 (2007);

[8] M. Abramowitz and I. Stegun, *Handbook of Mathematical Functions* , Dover, New York, 1972

[9] P. Gauron, B. Nicolescu and O.V. Selyugin, *Phys. Lett. B* **629**, 83 (2005)

[10] J.R.Cudell, A. Lengyel and E. Martynov, *Phys. Rev. D* **73**, 034008 (2006) [hep-ph/0511073]

[11] U. Amaldi and K. R. Schubert , *Nucl. Phys. B* **166**, 301 (1980)

[12] A. A. Kuznetsov et al. ,*Sov. J. Nucl. Phys.* **33**, 74 (1981), and *Yad. Fiz.* **33**, 142 (1981)

[13] A. Schiz et al , *Phys. Rev. D* **24**, 26 (1981)

[14] N. A. Amos et al., *Nucl. Phys. B* **262**, 689 (1985)

[15] F. Pereira and E. Ferreira, *Phys. Rev. D* **59**, 014008 (1999) ; *Phys. Rev. D* **61**, 077507 (2000) *Int. J. Mod. Phys. E* **16**, 2889 (2007)

[16] C. Augier et al. , CERN UA4/2 Coll. , *Phys. Lett. B* **316**, 448 (1993)

[17] G. Arnison et al. , CERN UA1 Coll. , *Phys. Lett. B* **128**, 336 (1983)

[18] M. Bozzo et al., CERN UA4 Coll., *Phys. Lett. B* **147**, 385 (1984); *Phys. Lett. B* **155**, 197 (1985)

[19] F. Abe , Fermilab CDF Coll., *Phys. Rev. D* **50**, 5518 (1993)

# GPDs of the Nucleons
# and Elastic Scattering at LHC Energies

*O.V. Selyugin*

BLTPH, JINR, Dubna, 141980, Russia

Taking into account the electromagnetic and gravitational form factors, calculated from a new set of $t$-dependent GPDs, a new model including the soft and hard pomerons is built. In the framework of this model the qualitative description of all existing experimental data at $\sqrt{s} \leq 52.8\,\mathrm{GeV}$, including the Coulomb range and large momentum transfers, is obtained with only 4 free parameters. Predictions for LHC energies are made.

## 1 Introduction

The dynamics of strong interactions finds its most complete representation in elastic scattering at small angles. Only in this region of interactions can we measure the basic properties that define the hadron structure: the total cross section, the slope of the diffraction peak and the parameter $\rho(s,t)$. Their values are connected on the one hand with the large-scale structure of hadrons and on the other hand with the first principles which lead to the theorems on the behaviour of the scattering amplitudes at asymptotic energies [1, 2].

The definition of the structure of the high-energy elastic hadron-hadron scattering amplitude at small angles in the new super-high energy range is a fundamental research problem. In this kinematic domain PQCD cannot be directly operative; however, there are numerous results derived in the framework of axiomatic field theory that can guide us both at the theoretical and experimental levels.

The hard pomeron which is obviously present in deeply inelastic processes with the large $\epsilon_2 \approx 0.4$ [3] will lead to a strong decrease of the energy at which the saturation or black-disk regime appears. It is not obvious how the total cross sections will grow with energy, especially in the energy region of the LHC. In the present work I investigate the impact of the hard pomeron on some of the features of elastic proton-proton scattering at LHC energies and small momentum transfer.

The situation is complicated by the possible transition to the saturation regime, as the Black Disk Limit (BDL) will be reached at the LHC [4, 5]. The effect of saturation will be a change in the $t$-dependence of $B$ and $\rho$, which will begin for $\sqrt{s} = 2$ to $6\,\mathrm{TeV}$, and which may drastically change $B(t)$ and $\rho(t)$ at $\sqrt{s} = 14\,\mathrm{TeV}$ [5, 6]. As we are about to see, such a feature can be obtained in very different models.

There are indeed many different models for the description of hadron elastic scattering at small angles [7]. They lead to different predictions for the structure of the scattering amplitude at asymptotic energies, where the diffraction processes can display complicated features [8]. This concerns especially the asymptotic unitarity bound connected with the Black Disk Limit (BDL) [9].

The total helicity amplitudes can be written as $\Phi_i(s,t) = \Phi_i^h(s,t) + \Phi_i^{\rm em}(s,t)e^{\varphi(s,t)}$ , where $\Phi_i^h(s,t)$ comes from the strong interactions, $\Phi_i^{\rm em}(s,t)$ from the electromagnetic interactions and $\varphi(s,t)$ is the interference phase factor between the electromagnetic and strong interactions [10, 11, 12]. For the hadron part the amplitude with spin-flip is neglected, as usual at high energy.

In practice, many different partial waves with $l \to \infty$ must be summed and this leads to the impact parameter representation [13] converting the summation over $l$ into an integration over the impact parameter $b$. In the impact-parameter representation the Born term of the scattering amplitude will be

$$\chi(s,b) \ \sim \ \int d^2q \ e^{i\vec{b}\cdot\vec{q}} \ F_{\rm Born}\left(s,q^2\right) \ , \tag{1}$$

where $t=-q^2$ and dropping the kinematic factor $1/\sqrt{s(s-2m_p^2)}$ and a factor $s$ in front of the scattering amplitude. After unitarisation, the scattering amplitude becomes

$$F(s,t) \sim \int e^{i\vec{b}\vec{q}} \ \Gamma(s,b) \ d^2b \, . \tag{2}$$

The overlap function $\Gamma(s,b)$ can be a matrix, corresponding to the scattering of different spin states. Unitarity of the $S$-matrix, $SS^+ \leq 1$, requires that $\Gamma(s,b) \leq 1$. There can be different unitarisation schemes which map $\chi(s,b)$ to different regions of the unitarity circle [14]. In this work I used the standard eikonal unitarisation scheme which leads to the standard regime of saturation, *i.e.* the BDL [15]:

$$\Gamma(s,b) = 1 - \exp[-i\chi(s,b)]. \tag{3}$$

## 2 Born Amplitude in the Impact-Parameter Representation

In different models one can obtain various pictures of the profile function based on different representations of the hadron structure. In this model I suppose that the elastic hadron scattering amplitude can be divided in two pieces. One is proportional to the electromagnetic form factor. It plays the most important role at small momentum transfer. The other piece is proportional to the matter distribution in the hadron and is dominant at large momentum transfer.

As in the EPSH model [6], I take into account the contributions of the soft and hard pomerons. In this approach the nucleon-nucleon elastic scattering amplitude is proportional to the electromagnetic hadrons form-factors and can be approximated at small $t$ by

$$T(s,t) \ = [ \ k_1 \ (s/s_0)^{\epsilon_1} e^{\alpha_1' \ t \ ln(s/s_0)} \ + \ k_2 \ (s/s_0)^{\epsilon_2} e^{\alpha_2' \ t \ ln(s/s_0)}] \ G_{em}^2(t), \tag{4}$$

where $k_1 = 4.47$ and $k_2 = 0.005$ are the coupling of the "soft" and "hard" pomerons, and $\epsilon_1 = 0.00728$, $\alpha_1' = 0.3$, and $\epsilon_2 = 0.45$, $\alpha_2' = 0.10$ are the intercepts and the slopes of the two pomeron trajectories. The normalization $s_0$ will be dropped below and $s$ contains implicitly the phase factor $\exp(-i\pi/2)$. I shall examine only high-energy nucleon-nucleon scattering with $\sqrt{s} \geq 52.8\,$GeV. So, the contributions of reggeons and odderon will be neglected. This model only includes crossing-symmetric scattering amplitudes. Hence the differential cross sections of the proton-proton and proton-antiproton elastic scattering are equal.

The assumption about the hadron form-factors leads to the amplitude

$$T(s,t)_{Born.} = h_1(F^s_{Born} + F^h_{Born})G^2_{em} + h_2(F^s_{Born} + F^h_{Born})G^2_{grav.}, \tag{5}$$

I suppose a non-linear trajectory for the pomeron and, as a first approximation, assume that the coupling is proportional to the gravitational form factor and that both soft and hard terms in the $F_{Born}(s,t)$ have $\alpha^{'} = 0$ at large $t$.

# 3 Hadron Form Factors

As was mentioned above, all the form factors are obtained from the GPDs of the nucleon. The electromagnetic form factors can be represented as first moments of GPDs

$$F^q_1(t) = \int_0^1 \, dx \; \mathcal{H}^q(x,t); \quad F^q_2(t) = \int_0^1 \, dx \; \mathcal{E}^q(x,t), \tag{6}$$

following from the sum rules [16, 17].

Recently, there were many different proposals for the $t$ dependence of GPDs. We introduced a simple form for this $t$-dependence [18], based on the original Gaussian form corresponding to that of the wave function of the hadron. It satisfies the conditions of non-factorisation, introduced by Radyushkin, and the Burkhardt condition on the power of $(1-x)^n$ in the exponential form of the $t$-dependence. With this simple form we obtained a good description of the proton electromagnetic Sachs form factors. Using the isotopic invariance we obtained good descriptions of the neutron Sachs form factors without changing any parameters.

The Dirac elastic form factor can be written

$$G^2(t) = h_{fa} e^{d_1 \; t} \; + \; h_{fb} e^{d_2 \; t} \; + h_{fc} e^{d_3 \; t}. \tag{7}$$

with $h_{fa} = 0.55$, $h_{fb} = 0.25$, $h_{fa} = 0.20$, and $d_1 = 5.5$, $d_2 = 4.1$, $d_3 = 1.2$. The exponential form of the form factor lets us calculate the hadron scattering amplitude in the impact parameter representation [6].

I shall use this model of GPDs to obtain the gravitational form factor of the nucleon in the impact-parameter representation. This form factor can be obtained from the second momentum of the GPDs. Taking instead of the electromagnetic current $J^\mu$ the energy-momentum tensor $T_{\mu\nu}$ together with a model of quark GPDs, one can obtain the gravitational form factor of fermions

$$\int_{-1}^{1} \, dx \; x[H(x,\Delta^2,\xi) \pm E(x,\Delta^2,\xi)] = A_q(\Delta^2) \pm B_q(\Delta^2). \tag{8}$$

For $\xi = 0$ one has

$$\int_0^1 \, dx \; x[\mathcal{H}(x,t) \pm \mathcal{E}(x,t)] = A_q(t) \pm B_q(t). \tag{9}$$

Calculations in the momentum-transfer representation show that the second moment of the GPDs, corresponding to the gravitational form-factor, can be represented in the dipole form

$$A(t) = L^2/(1-t/L^2)^2. \tag{10}$$

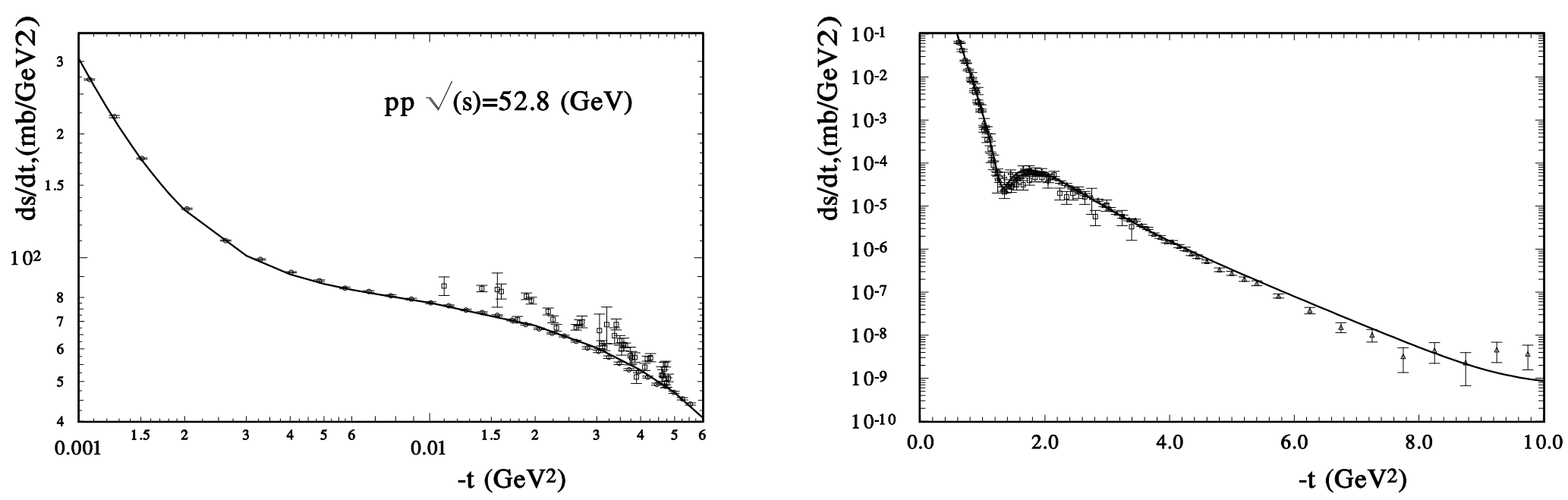

Figure 1: $d\sigma/dt$ at $\sqrt{s} = 52.8\,\mathrm{GeV}$ for $pp$ elastic scattering, at small $|t|$ (left) and at large $|t|$ (right).

with the parameter $L^2 = 1.8\,\mathrm{GeV}^2$. For the scattering amplitude, this leads to

$$A(s,b) = \frac{L^5 b^3}{48} K_3(Lb), \tag{11}$$

where $K_3(Lb)$ is the MacDonald function of the third order. To match both parts of the scattering amplitude, the second part is multiplied by a smooth correction function which depends on the impact parameter

$$\psi(b) = (1 + \sqrt{r_1^2 + b^2}/\sqrt{r_2^2 + b^2}). \tag{12}$$

# 4 Description of the Differential Cross Sections

The model has only four free parameters, which are obtained from a fit to the experimental data:

$$h_1 = 1.09 \pm 0.004; \quad h_2 = 1.57 \pm 0.006; r_1^2 = 1.57 \pm 0.02; \quad r_2 = 5.56 \pm 0.06.$$

I used all the existing experimental data at $\sqrt{s} \geq 52.8\,\mathrm{GeV}$, including the whole Coulomb region and up to the largest momentum transfers experimentally accessible. In the fitting procedure, only statistical errors were taken into account. The systematic errors were used as an additional common normalisation of the experimental data from a given experiment. As a result, one obtains $\sum \chi_i^2/N \simeq 3$. where $N = 924$ is the number of experimental points. Of course, if one sums the systematic and statistical errors, the $\chi^2/N$ decreases, to 2. Note that the parameters are energy-independent. The energy dependence of the scattering amplitude is determined only by the intercepts of the soft and hard pomerons.

In Fig. 1 the differential cross sections for proton-proton elastic scattering at $\sqrt{s} = 52.8\,\mathrm{GeV}$ are presented. At this energy there are experimental data at small (beginning at $-t = 0.0004\,\mathrm{GeV}^2$) and large (up to $-t = 10\,\mathrm{GeV}^2$) momentum transfers. The model reproduces both regions and provides a qualitative description of the dip region at $-t \approx 1.4\,\mathrm{GeV}^2$, for $\sqrt{s} = 53\,\mathrm{GeV}$ and for $\sqrt{s} = 62.1\,\mathrm{GeV}$, although only the low-$t$ and high-$t$ regions have been fitted to.

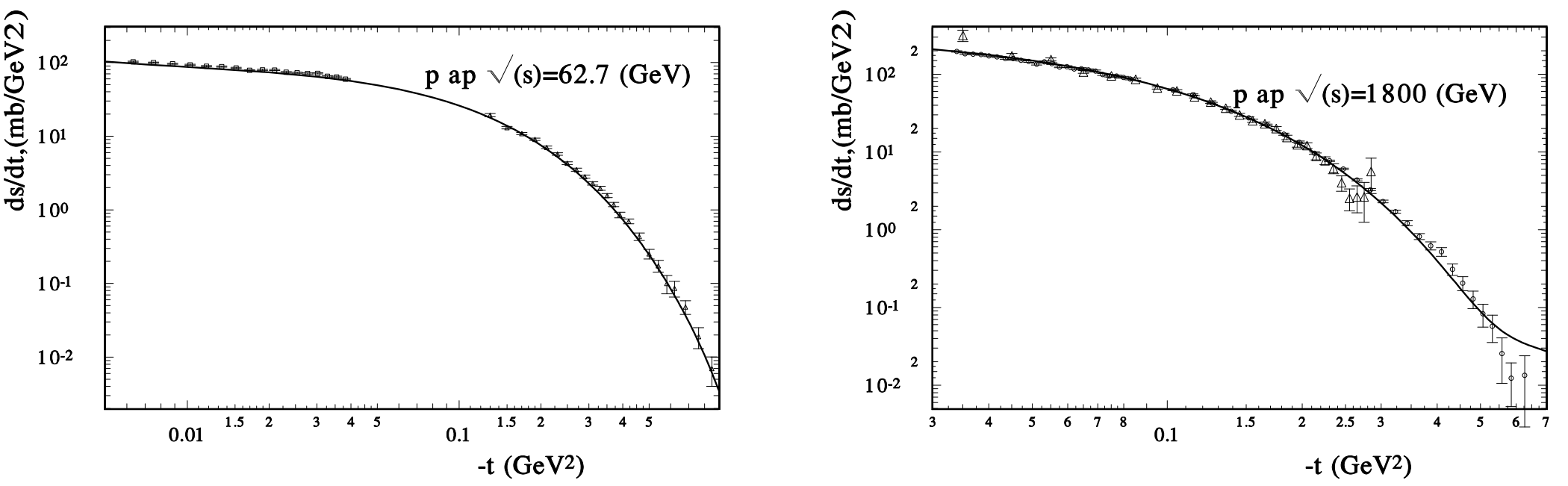

Figure 2: $d\sigma/dt$ for $\bar{p}p$ elastic scattering at small $|t|$, at $\sqrt{s} = 541\,\mathrm{GeV}$ (left) and $\sqrt{s} = 1800\,\mathrm{GeV}$ (right)

Now let us examine the proton-antiproton differential cross sections. In this case at low momentum transfer the Coulomb-hadron interference term plays an important role and has the opposite sign. The model describes the experimental data well. In this case, the first part of the scattering amplitude determines the differential cross sections, and is dominated by the exchange of the soft pomeron.

There was a significant difference between the experimental measurement of $\rho$, the ratio of the real part to the imaginary part of the scattering amplitude, between the UA4 and UA4/2 collaborations at $\sqrt{s} = 541\,\mathrm{GeV}$. As shown in Table 4, the resulting values for $\rho(0)$ appear inconsistent. A more careful analysis [19, 20] shows that there is no contradiction between the measurements of UA4 and UA4/2. Now the present model gives for this energy $\rho(\sqrt{s} = 541\,\mathrm{GeV},\ t = 0) = 0.163$, so, practically the same as in the previous phenomenological analysis.

| $\bar{\rho}$ $(\sqrt{s} = 541\,\mathrm{GeV}, t = 0)$ | | | |
|---|---|---|---|
| experiment | experimental analysis | global analysis | This model |
| UA4 | $0.24 \pm 0.02$ | $0.19 \pm 0.03$ [[19]] | |
| UA4/2 | $0.135 \pm 0.015$ | $0.17 \pm 0.02$ [[20]] | 0.163 |

Table 1: Average values of $\rho$, derived with fixed total cross section (first two columns), and from a global analysis (last two columns).

Now let us examine the data at higher energy, where the contribution of the hard pomeron will be more important. In Fig. 2 the description of the proton-antiproton scattering at $\sqrt{s} = 541\,\mathrm{GeV}$ and at $\sqrt{s} = 1800\,\mathrm{GeV}$ is shown. In this case the Coulomb-hadron interference term leads to a large value of the real part of the scattering amplitude, which is determined by the contribution from the hard pomeron. The good description of the experimental data shows

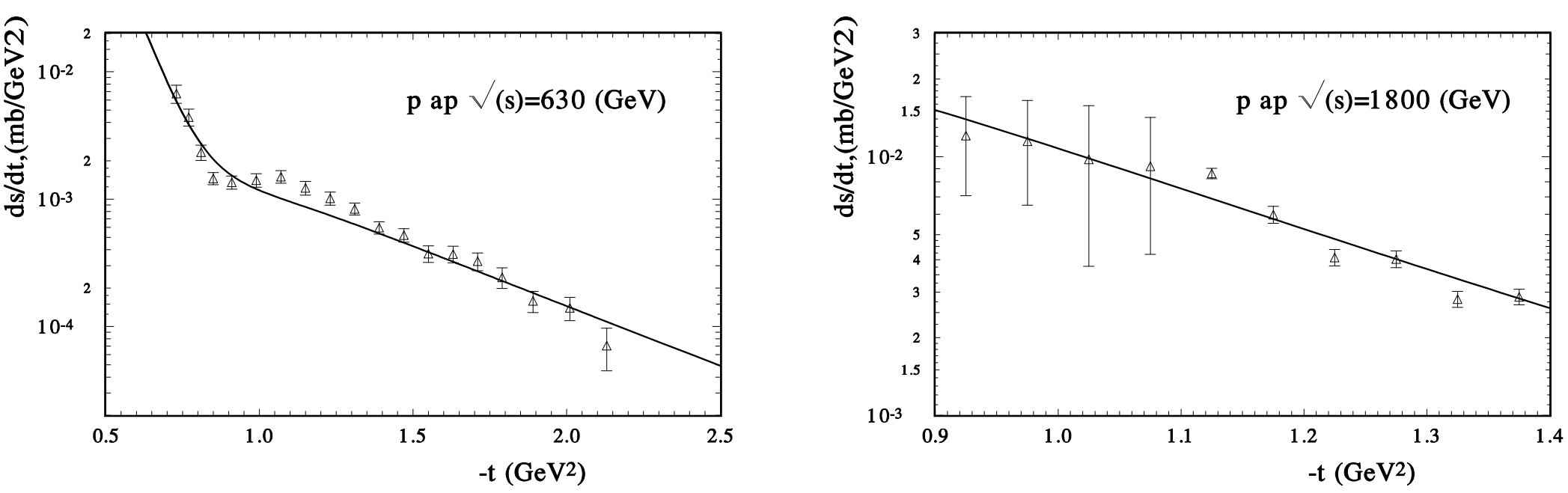

Figure 3: $d\sigma/dt$ for $\bar{p}p$ elastic scattering at large momentum transfer, at $\sqrt{s} = 630\,\mathrm{GeV}$ (left) and $\sqrt{s} = 1800\,\mathrm{GeV}$ (right)

that the parameters of the hard pomeron correspond to the real physical situation.

Fig. 3 shows the description of the experimental data at larger momentum transfers for $\sqrt{s} = 630\,\mathrm{GeV}$ and $\sqrt{s} = 1800\,\mathrm{GeV}$. It is clear that the model leads to a good description of these data. However, one must note that the fine structure of the dip is not reproduced by the model in this case. The model shows only an essential change of the slope in this region.

Saturation of the profile function will surely control the behaviour of $\sigma_{tot}$ at higher energies and will result in a significant decrease of the LHC cross section. For the last LHC energy $\sqrt{s} = 14\,\mathrm{TeV}$ the model predicts $\sigma_{tot} = 146\,\mathrm{mb}$ and $\rho(0) = 0.235$. This result comes from the contribution of the hard pomeron and from the strong saturation from the black disk limit.

## 5 Conclusion

In the presence of a hard Pomeron [21], saturation effects can change the behaviour of the cross sections already at LHC energies. A new model, taking into account the contributions of the soft and hard pomerons and using form factors calculated from the GPDs, successfully describes all the existing experimental data on elastic proton-proton and proton-antiproton scattering at $\sqrt{s} \geq 52.8\,\mathrm{GeV}$, including the Coulomb-hadron interference region, the dip region, and the large-momentum-transfer region. The behaviour of the differential cross section at small $t$ is determined by the electromagnetic form factors, and at large $t$ by the matter distribution (calculated in the model from the second momentum of the GPDs) as was supposed by H.L. Miettinen a long time ago.

The model leads to saturation of the BDL in the TeV region of energy. As a result the parameters of the scattering amplitude $B(s,t)$ and of $\rho(s,t)$ have a complicated dependence on $s$ and $t$ and the scattering amplitude has a non-exponential behaviour at small momentum transfer.

The possibility of a new behaviour of $\rho(s,t)$ and $B(s,t)$ at LHC energies [6] has to be taken into account in the procedure extracting the value of the total cross sections by the standard method [22].

## Acknowledgements

I would like to thank for helpful discussions J.R. Cudell and O.V. Teryaev. I gratefully acknowledge the organisation committee and R. Orava for the financial support to take part in the conference and would like to thank the FRNS and University of Liège where part of this work was done.

## References

[1] A. Martin, F. Cheung: Analytical properties and bounds of the scattering amplitudes, Gordon & Breach, New York, 1970.

[2] S.M. Roy, Phys. Rep. **C 5** (1972) 125 .

[3] A. Donnachie and P. V. Landshoff, Phys. Lett. B **595** (2004) 393 [arXiv:hep-ph/0402081].

[4] O. V. Selyugin and J. R. Cudell, Czech. J. Phys. **54** (2004) B161 [arXiv:hep-ph/0301048].

[5] J. R. Cudell and O. V. Selyugin, Phys. Lett. B **662** (2008) 417 [arXiv:hep-ph/0612046].

[6] J. R. Cudell and O. V. Selyugin, Phys. Lett. B **662** (2008) 417 [arXiv:hep-ph/0612046].

[7] R. Fiore, L. Jenkovszky, R. Orava, E. Predazzi, A. Prokudin, O. Selyugin, Mod.Phys., A24: 2551-2559 (2009)

[8] J. R. Cudell and O. V. Selyugin, Czech. J. Phys. **54** (2004) A441 [arXiv:hep-ph/0309194].

[9] O. V. Selyugin, J. R. Cudell and E. Predazzi, Eur. Phys. J. ST **162** (2008) 37 [arXiv:0712.0621 [hep-ph]].

[10] O.V. Selyugin, Mod. Phys. Lett. A**9** (1994) 1207.

[11] O.V. Selyugin, Mod. Phys. Lett. A**14**, 223 (1999).

[12] O. V. Selyugin, Phys. Rev. D **60** (1999) 074028.

[13] E. Predazzi, Ann. of Phys., **36** (1966) 228, 250.

[14] J. R. Cudell, E. Predazzi and O. V. Selyugin, Phys. Rev. D **79** (2009) 034033 [arXiv:0812.0735 [hep-ph]].

[15] O. V. Selyugin, talk at the International Bogoliubov conference "Problems of theoretical and mathematical physics", August 21-27, 2009, Moscow-Dubna, Russia.

[16] X.D. Ji, Phys. Lett. **78** , (1997) 610; Phys. Rev D **55** (1997) 7114.

[17] Radyushkin, A.V., Phys. Rev. D **56**, 5524 (1997) .

[18] O.V. Selyugin, O.V. Teryaev, Phys. Rev., **D** (2008).

[19] O. Selyugin, Sov. J. Nucl. Phys. **55**, 466 (1992)

[20] O. Selyugin, Phys. Lett. **B333**, 245 (1994)

[21] J. R. Cudell, A. Lengyel, E. Martynov and O. V. Selyugin, Nucl. Phys. A **755** (2005) 587 [arXiv:hep-ph/0501288].

[22] J. R. Cudell and O. V. Selyugin, Phys. Rev. Lett. **102** (2009) 032003 [arXiv:0812.1892 [hep-ph]].

# Bound on Inelastic Total Cross-Sections

*André Martin*

CERN, 1211 Genève 23, Switzerland

We prove that the total inelastic cross-section is bounded by

$$\sigma_{inelastic} < \frac{\pi}{4m_\pi^2}(\ln s)^2 ,$$

i.e. a bound 4 times smaller than the bound obtained by L. Łukaszuk and the author on the total cross-section in 1967. With some extra assumptions, which look reasonable to most people, we show that the bound on the total cross-section itself is divided by 2. For details see [1, 2].

## References

[1] André Martin, *The Froissart bound for inelastic cross sections*, arXiv:0904.3724v4 [hep-ph].

[2] André Martin, *Froissart bound for inelastic cross sections*, Phys. Rev. D **80**, 065013 (2009).

# Measuring Elastic Scattering in the Coulomb Nuclear Interference (CNI) Region in ATLAS – Total Cross Sections and Luminosity

*Matthieu Heller, on behalf of the ATLAS collaboration*

LAL, Orsay, Univ Paris-Sud, CNRS/IN2P3

ATLAS is being complemented with detectors to measure elastic scattering at small angles. This will give a possibility both to measure the total cross section and make an absolute luminosity calibration of the Luminosity monitors of ATLAS. The principle of the measurement, the experimental requirements and the detector performances found in a 2008 test-beam are presented in this article.

## 1 Introduction

The Large Hadron Collider (LHC) will provide proton-proton collisions for the experiments installed at the interaction points (IP). The ATLAS detector is designed to detect high $p_t$ reactions and will focus on the discovery of the Higgs boson and new physics. However ATLAS is now also complemented by detectors in the very forward region to measure small angle elastic scattering. This will give a possibility to measure the total cross section and provide accurate luminosity calibration. The ultimate goal is to reach the Coulomb Nuclear Interference (CNI) region but even without reaching the CNI region precise luminosity calibration can be done. The detectors will be installed in so called "Roman Pots" 240 metres away from the ATLAS interaction point. The scattered particles at the IP will travel along the accelerator lattice and be detected by fiber trackers that are inserted in the Roman Pots. This measurement requires a special optics with very high beta at the interaction point and consequently will be done during dedicated runs.

## 2 Principle of the Measurement

The rate of elastic scattering is linked to the total interaction rate through the optical theorem, which states that the total cross section is directly proportional to the imaginary part of the nuclear forward elastic scattering amplitude extrapolated to zero momentum transfer:

$$\sigma_{tot} \propto \Im\left[F_n(t=0)\right]$$

At very small scattering angle the four-momentum transfer can be written $-t \approx (p\theta)^2$ where $p$ refers to the beam momentum and $\theta$ to the scattering angle at the IP.

Taking into account the optical theorem and the Coulomb term, the rate of elastic scattering at small $t$-values can be written as [1]:

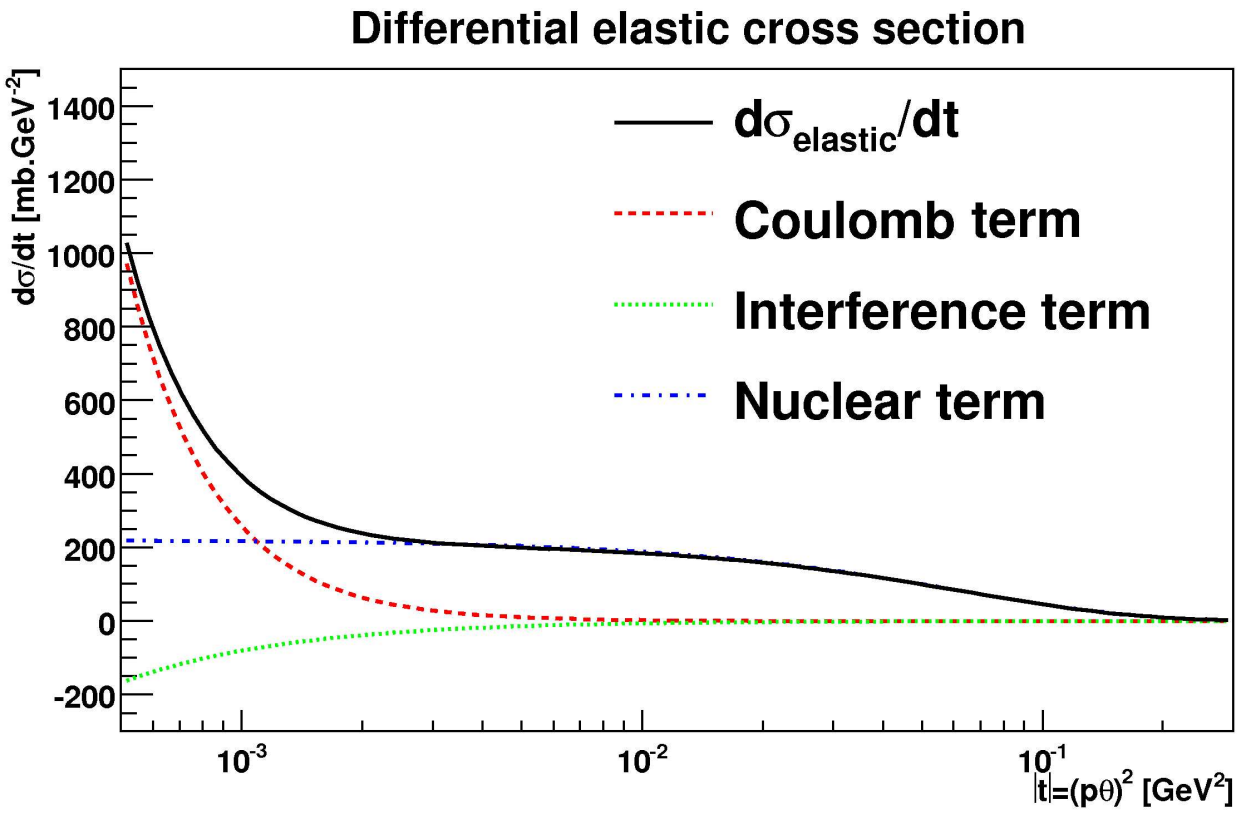

Figure 1: Contribution to the differential elastic cross section of the three components at small $t$ value with $\sigma_{tot}$=100 mb, $\rho$=0.013 and $b$=18 GeV$^{-2}$.

$$\frac{dN}{dt} = L\pi \left| f_C + f_N \right|^2 \approx L\pi \left| \frac{-2\alpha}{|t|} + \frac{\sigma_{tot}}{4\pi} (i + \rho) \, e^{-b|t|/2} \right|^2 \tag{1}$$

where the first term corresponds to the Coulomb and the second to the strong interaction amplitude and $\alpha$ is the fine structure constant, $\sigma_{tot}$ the total proton-proton cross section, $b$ the nuclear slope and $\rho$ the ratio of the real over the imaginary part of the forward elastic scattering amplitude. The formula is oversimplified, and there are also other corrections as the proton electromagnetic form factor and the relative phase between the Coulomb and the strong amplitude that are included in the final analysis. This expression allows to make the distinction between three terms as seen in Fig. 1. Fitting the rate of elastic scattering as a function of $t$ allows to determine $L$, $\sigma_{tot}$, $b$ and $\rho$.

# 3 Experimental Framework

The best absolute accuracy on the luminosity measurement is obtained by approaching as close as possible the Coulomb region where the correlation between the luminosity and the total cross section gets smaller. However at the TeV scale, this means measuring scattering angles of a couple of micro-radians.

At the nominal energy of the LHC (7 TeV) the strong amplitude is equal to the electromagnetic amplitude for $-t = 6\ 10^{-4}$ GeV$^2$. This corresponds to a scattering angle of 3.5 $\mu$rad. A direct measurement of this angle would require to intercept these protons before the first magnetic element. This is in practice impossible because the separation between these protons and the beam core is less than 200 $\mu$m. It was found that the best solution was to use dedicated optics which will separate enough the beam core and the elastically scattered protons at 240 m from the IP (see Fig. 2). In this location Roman Pots (RPs) detectors will be inserted to tag the elastic protons.

The beam conditions required to reach the CNI region are a small emittance and special beam optics. This dedicated beam optics is characterized by a very large $\beta^*$ at the ATLAS IP, zero dispersion and a phase advance of the betatron function between the IP and the detectors of 90 degrees in the vertical plane to perform the parallel-to-point focussing which is sketched in Fig. 3. This has the consequence that particles scattered with the same angle arrive at the same position in the detector. An optics solution fulfilling the above requirements must be used in combination with rather few bunches of low intensity compared to nominal LHC settings. It will produce instantaneous luminosities in the range of $10^{27}$ to $10^{28}$ cm$^{-2}$s$^{-1}$.

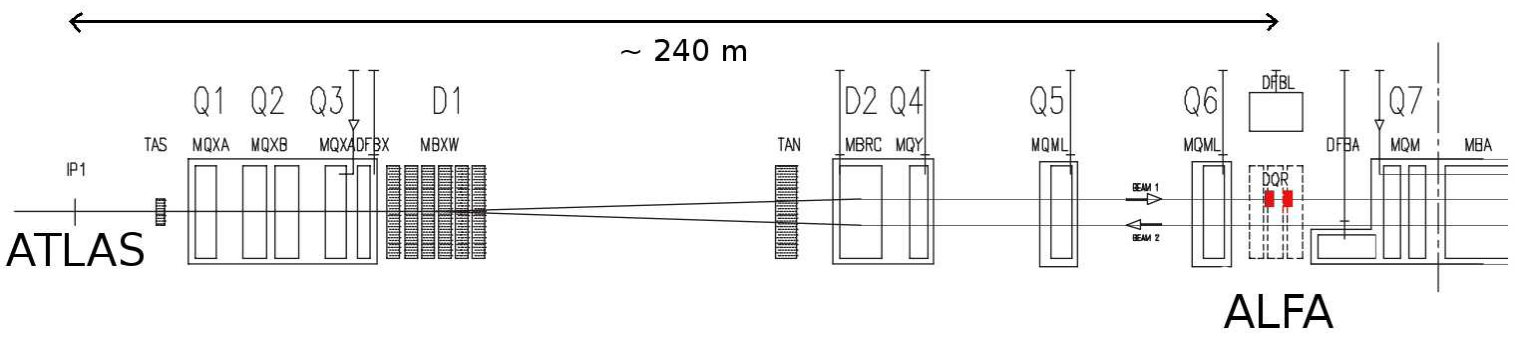

Figure 2: Location of the two stations (in red) on beam 1. Both other stations are on beam 2, on the opposite side with respect to ATLAS.

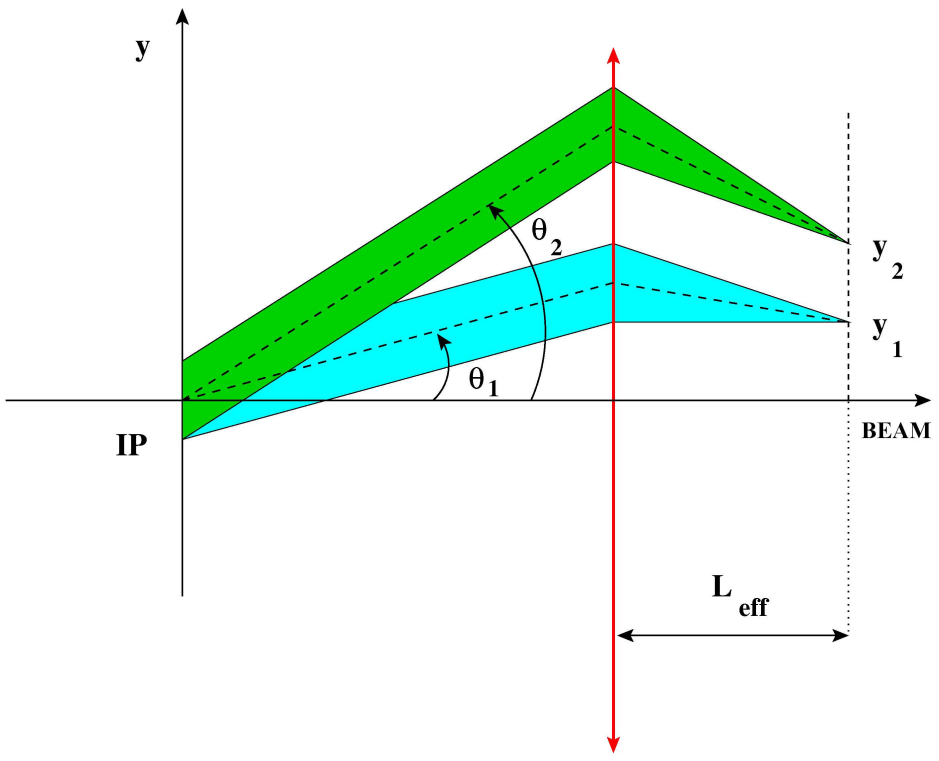

Figure 3: Parallel-to-point focussing where the optics brings all particles scattered with the same angle at the IP to the same location in the detector.

Using the expression of the scattering angle as a function of the position in the detector, the minimum $t$-value that we will be able to detect depends on the particle's momemtum ($p$), the distance to the beam centre expressed in terms of number of $\sigma$ of the beam size ($n_{min}$), the emittance ($\epsilon$), and the value of the betatron function at the IP ($\beta_{IP}$):

$$-t_{min} = p^2 n_{min}^2 \frac{\epsilon}{\beta_{IP}} \tag{2}$$

Assuming that we can approach the beam down to $12\,\sigma$, that we run at 7 TeV and that the normalised emittance is $\epsilon_N = 1\mu$m rad with a $\beta_{IP} = 2625$ m, we obtain $-t_{min} = 3.7\ 10^{-4}$ GeV$^2$ which is equivalent to $\theta_{IP,y} = 3\,\mu$rad.

To intercept these protons, two Roman Pot stations are installed symmetrically in the forward direction on beam 1 and beam 2. Each station is made of an upper and a lower detector. The detectors inserted into the Roman Pots (RPs) are presented in the following section.

# 4 ALFA Detector

The ALFA detector[1] is made of a scintillating fiber tracker inserted in a Roman Pot (see Fig. 4). The tracker is made of 20 layers of 64 squared fibers each organised in a U-V geometry with 90° between two consecutive layers (see Fig. 5).

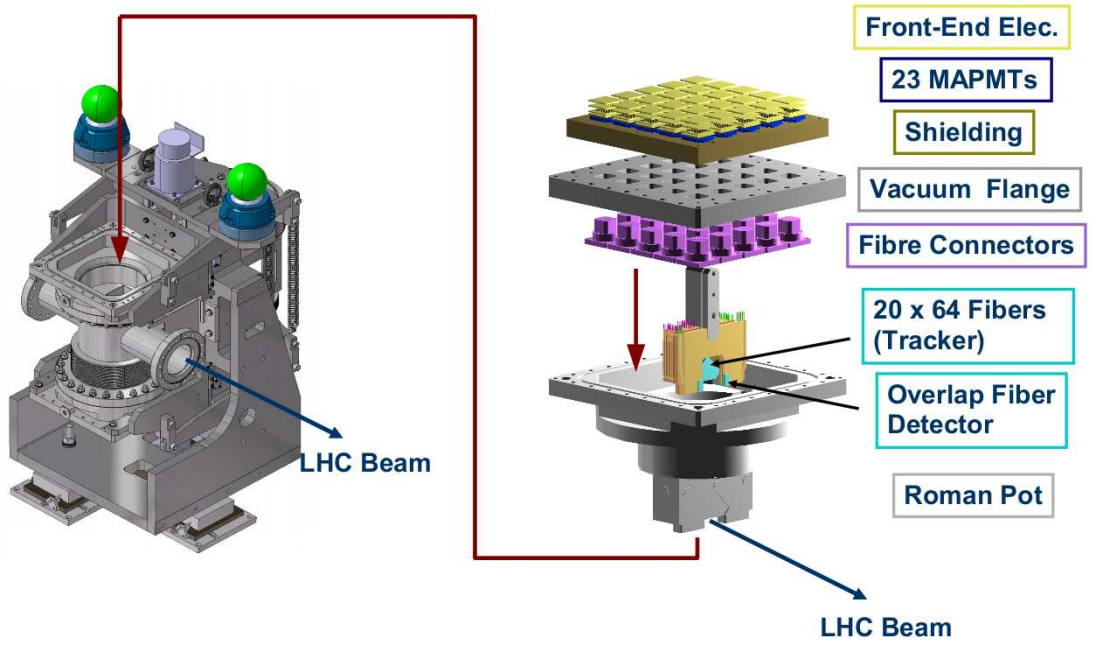

Figure 4: Schematic view of the ALFA detector.

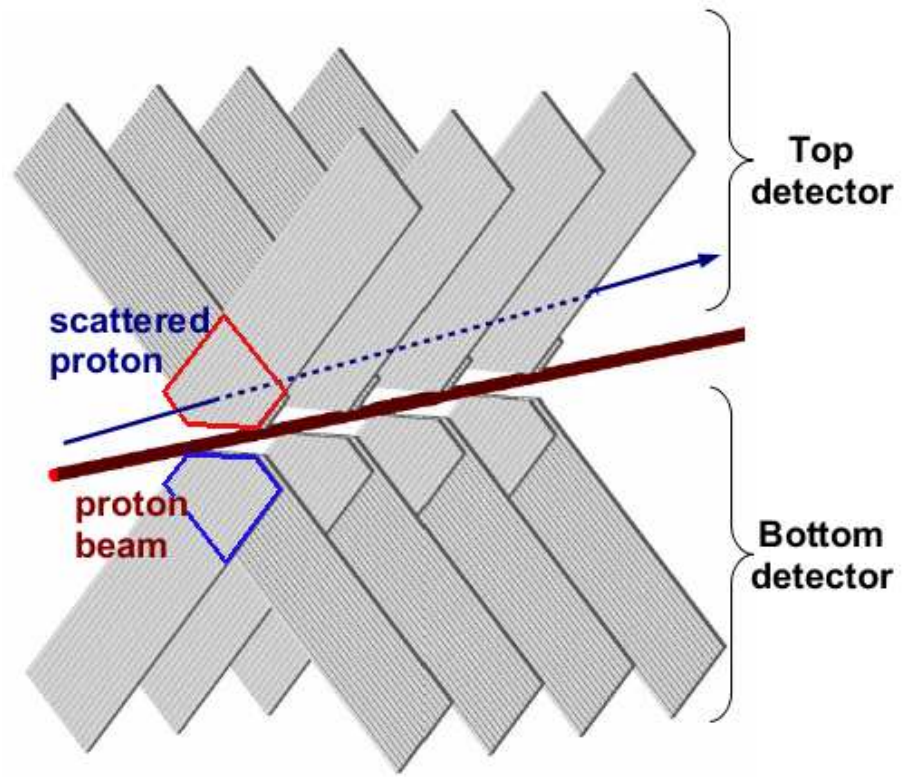

Figure 5: Schematic view of the main tracker. Only four U-V planes out of ten are represented on the picture.

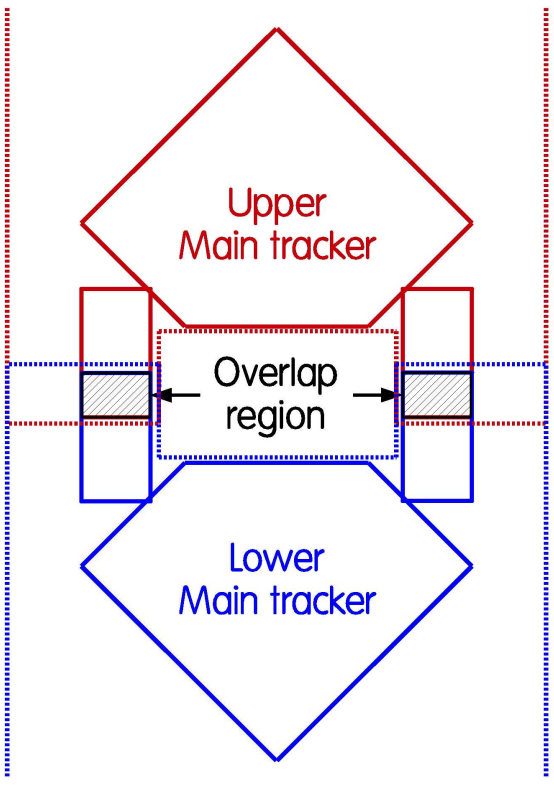

Figure 6: Schematic view of the overlap detectors.

The scintillating fibers are 0.5×0.5 mm$^2$ and a staggering of 10 % is applied between two consecutive planes providing a theoretical resolution close to 15 $\mu$m. However as shown in Section 6, the assembling imperfections worsen this result. Each layer is read out by a Multi Anode PhotoMulTiplier (MAPMT) with 64 channels. The compact front-end electronics called PMF [2], are directly mounted on the back of the MAPMT. The MAROC2 ASIC [2] allows to correct the MAPMT gain non-uniformity and to discriminate the signal. An FPGA ensures the communication with the motherboard fixed on the pot. On top of that, the *overlap detectors* sketched in Fig. 6 are used to measure the relative position between the upper and the lower pot with a precision of 10 $\mu$m. Mechanically fixed to the main tracker, they provide essential information about the $t$-scale that cannot be extracted from the data. This measurement will be done reconstructing the halo particles passing at the same time through the upper and lower overlap detector.

[1] **ALFA** stands for **A**bsolute **L**uminosity **F**or **A**TLAS

# 5 The Simulation

The acceptance of the ALFA detector is a major issue and has to be computed by simulation. Indeed, one has to know as precisely as possible which protons will pass through the sensitive area of the detectors and which will go in between the upper and lower pot or will be lost during the transport from the IP to the RPs. This simulation therefore includes:

- a Monte-Carlo generator (Pythia [3]) to generate the event according to the differential elastic cross section,
- transport software (MadX [4]) to track the protons along the magnetic lattice (see Fig. 7),
- reconstruction software to build the spectrum, correct it and fit it.

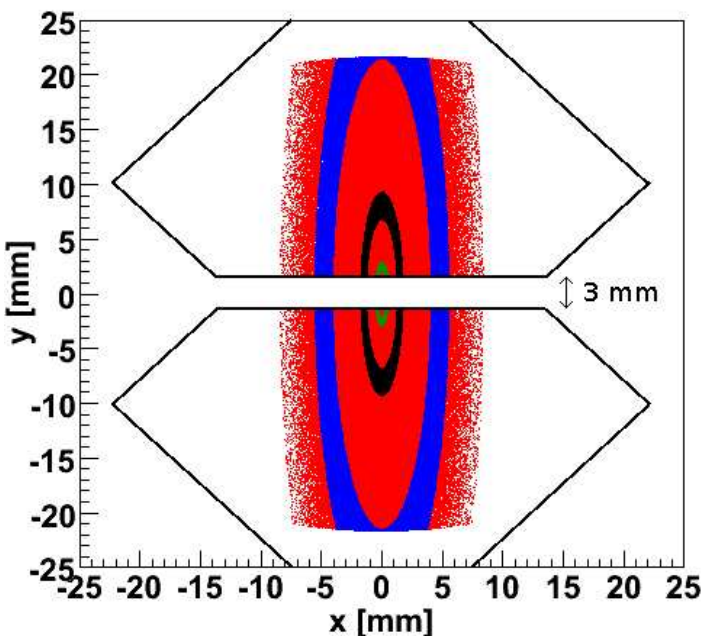

Figure 7: Scattering picture at the RP. The black lines show the diamond shape of the active part of the detector.

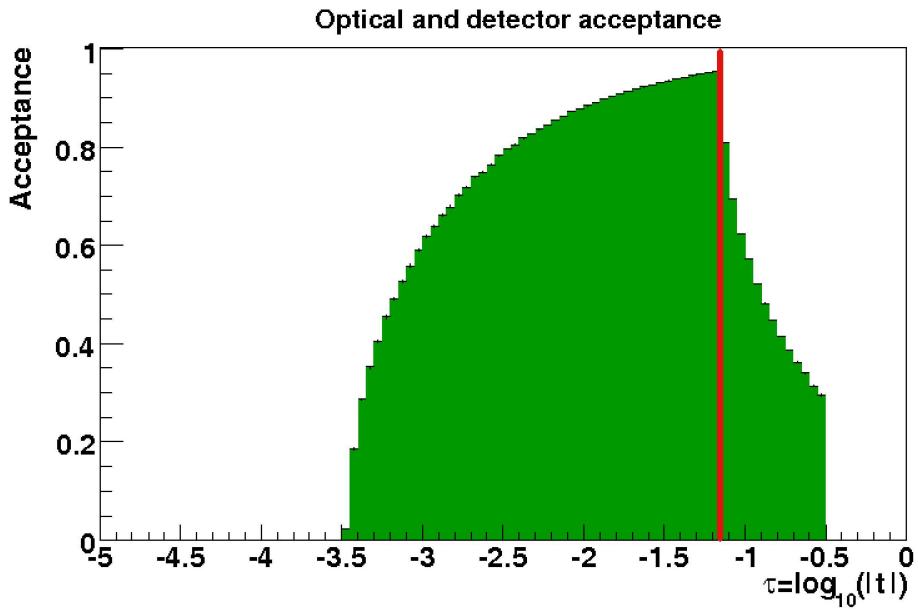

Figure 8: Acceptance obtained with the same parameters as define earlier.

Due to the optics settings, one $t$-value is seen as an ellipse in the detector. In Fig. 7 are sketched (x,y)-maps at the RP for three different $t$-ranges to give an idea of where the scattered protons will hit the detector. The inner ring represents $t \in \left[7\ 10^{-4}; 13\ 10^{-4}\ \text{GeV}^2\right]$, the middle one $t \in \left[7\ 10^{-3}; 13\ 10^{-3}\ \text{GeV}^2\right]$ and the outer one $t \in \left[7\ 10^{-2}; 13\ 10^{-2}\ \text{GeV}^2\right]$. The elastic protons detected allow to reconstruct the spectrum and to determine the acceptance. Fig. 8 shows the acceptance as a function of $t$ for 7 TeV protons with the detectors at 12 $\sigma$ from the beam center. One can observe two distinctive causes of acceptance losses. On the right of the red line, events are mainly lost in the accelerator due to the magnetic element aperture, and on the left, protons are passing between the two pots. The acceptance determination will allow to correct the detected spectrum to recover the underlying physics distribution.

In Table 1 [1, 5] are summarized the contributions to the precision on the luminosity measurement from various sources. In the so-called beam properties part, the major impact will be from the angular divergence as it mainly affects small values of $t$ for which the fit is very sensitive. However the measurement of the optical parameters will also be an important issue. The detector properties will be dominated by the detector alignment with respect to the beam and acceptance corrections. Any error on this measurement will drastically affect the result of the fit. The background is composed of beam-gas events, particles surviving the LHC collimation system and all non-elastic processes such as single diffraction events. Its rejection is done using mainly the back-to-back topology of the elastic event. The simulation has shown that the 2-3 % claimed in [1] on absolute accuracy can be reached.

| | [%] |
|---|---|
| Beam properties | 1.2 |
| Detector properties | 1.4 |
| Backgroud substraction | 1.1 |
| Total uncertainty | 2.1 |
| Statistical error | 1.8 |
| Total | 2.8 |

Table 1: Summary of the different uncertainties obtained on integrated luminosity for $\int L = 3.6\ 10^{32}\ \mathrm{cm}^{-2}$.

## 6 Detector Performances

The first complete tracker (20 planes of 64 fibers each) has been tested in August 2008 on the SPS test-beam line H8 at CERN. However, only half of the detector could be read out. The ALFA detector was coupled to a silicon strip telescope with a resolution of $10\,\mu$m. Using this setup we were able to determine the layer efficiency and the spatial resolution.

The layer efficiency has been computed using the telescope prediction. Indeed the accurate telescope resolution allows to point in each layer to the fiber where the particle went through. Comparing this prediction with the data allows us to determine the layer efficiency which has been found to be $\approx$90 %. Combining layers together this is enough to reach almost 100 % detection efficiency for the entire ALFA tracker.

The reconstruction method consist in projecting all fibers from all planes of the same orientation on a perpendicular plane. If a particle went through the detector, the hit fibers corresponding to the path will overlap on this projection. One can determine then the position of the center of the overlap region and its width. To accept a track in the reconstruction algorithm, a minimum of three fibers for half a detector must define the overlap region i.e. all layers do not have to be efficient, but only 60 % of them. For a full detector this limit is set to a minimum of 6 layers. This condition has a major impact on the resolution: the higher the number of fibers defining the track path is, the greater is the probability to have a small area of overlap providing a better resolution. This is exactly what we see in Fig. 9, which displays the resolution improving with a decreasing overlap region.

The spatial resolution obtained with half a detector is close to $52\,\mu$m in both directions. This results has been compared to a Monte Carlo simulation including layer efficiency and real geometry, and comparable results have been found. For a full detector, where the staggering configuration is optimal, the average overlap region width is lower than $100\,\mu$m. Judging by the Fig. 9 we can expect a resolution around $35\,\mu$m.

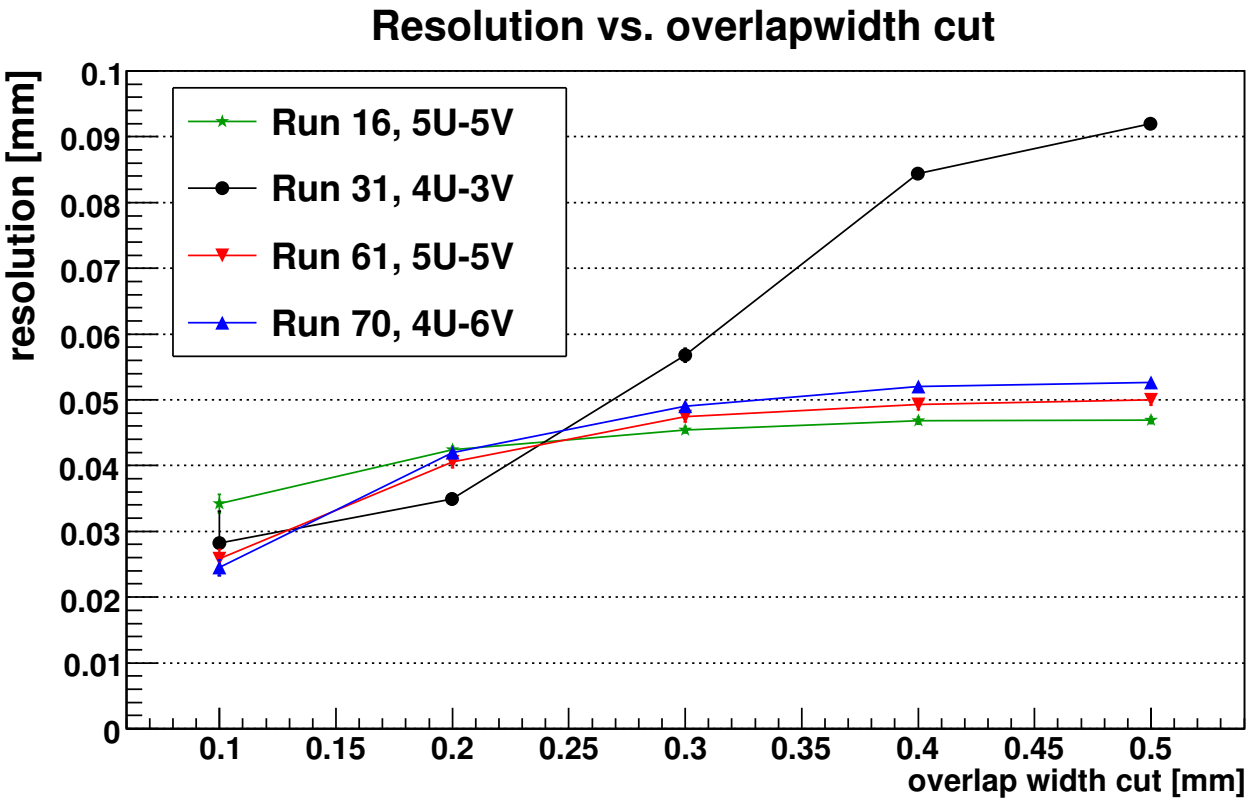

Figure 9: Resolution dependency on the overlap region width, #U and #V standing for the number of U-V layers used.

# 7 Conclusion

The studies of the systematics has shown that 2-3 % absolute accuracy on luminosity can be reached with the ALFA detector. All test-beam studies done so far have validated the technical choices made for the different parts of the ALFA detector. A new test-beam period is planned for October 2009 to definitely validate the electronics and the full tracker.

# References

[1] CERN/LHCC/2008-004, ATLAS TDR 18, 17 January 2008.

[2] *PMF: the front end electronics of the ALFA detector*, Barrillon P., proceedings of 2008 IEEE conference.

[3] Pythia website, http://home.thep.lu.se/ torbjorn/Pythia.html

[4] MAD-X website, http://mad.web.cern.ch/mad/

[5] *Luminosity calibration from elastic scattering*, Stenzel H., CERN, Geneva ,ATL-LUM-PUB-2007-001

# Theoretical Aspects of High Energy Elastic Nucleon Scattering

*Vojtěch Kundrát*[1], *Jan Kašpar*[2,1], *Miloš Lokajíček*[1]

[1]Institute of Physics of the AS CR, v. v. i., 182 21 Prague 8, Czech Republic
[2]CERN, 1211 Genève 23, Switzerland

The eikonal model must be denoted as strongly preferable for the analysis of elastic high-energy hadron collisions. The given approach allows to derive corresponding impact parameter profiles that characterise important physical features of nucleon collisions, e.g., the range of different forces. The contemporary phenomenological analysis of experimental data is, however, not able to determine these profiles unambiguously, i.e., it cannot give the answer whether the elastic hadron collisions are more central or more peripheral than the inelastic ones. However, in the collisions of mass objects (like protons) the peripheral behaviour of elastic collisions should be preferred.

## 1 Coulomb-Hadronic Interference

The first attempt to determine the complete elastic scattering amplitude $F^{C+N}(s,t)$ for high energy collisions of charged nucleons has been made by Bethe [1]:

$$F^{C+N}(s,t) = F^{C}(s,t)e^{i\alpha\Psi(s,t)} + F^{N}(s,t) \tag{1}$$

where $F^{C}(s,t)$ is Coulomb amplitude (known from QED), $F^{N}(s,t)$ - elastic hadronic amplitude, $\alpha\Psi(s,t)$ - real relative phase between Coulomb and hadronic scattering and $\alpha = 1/137.036$ is the fine structure constant. This relative phase has been specified by West and Yennie [2] using the Feynman diagram technique (one photon exchange) as

$$\Psi_{WY}(s,t) = -\ln\frac{-s}{t} - \int_{-4p^2}^{0}\frac{dt'}{|t'-t|}\left[1 - \frac{F^{N}(s,t')}{F^{N}(s,t)}\right]; \tag{2}$$

$p$ representing the value of the incident momentum in the centre-of-momentum system. By their construction (see Eq. (2), the phase $\Psi_{WY}(s,t)$ is to be real.

However, it has been shown in Ref. [3] that this requirement can be fulfilled only provided the phase of the elastic hadronic amplitude $\zeta^{N}(s,t)$ defined in our case as

$$F^{N}(s,t) = i|F^{N}(s,t)|e^{-i\zeta^{N}(s,t)}, \tag{3}$$

is $t$ independent at all kinematically allowed values of $t$. Rigorous proof has been given for $|\zeta^{N}(s,t)| < \pi$, which is fulfilled practically in all standard phenomenological models leading to the central distribution of elastic processes in impact parameter space. It is not yet clear if it holds also for $|\zeta^{N}(s,t)| < 2\pi$ corresponding to peripheral behaviour; see Fig. 1 where both the

types of phase $t$ dependences are represented. The corresponding $t$ dependences of imaginary parts of complex relative phases $\alpha\Psi_{WY}(s,t)$ are shown in Fig. 2.

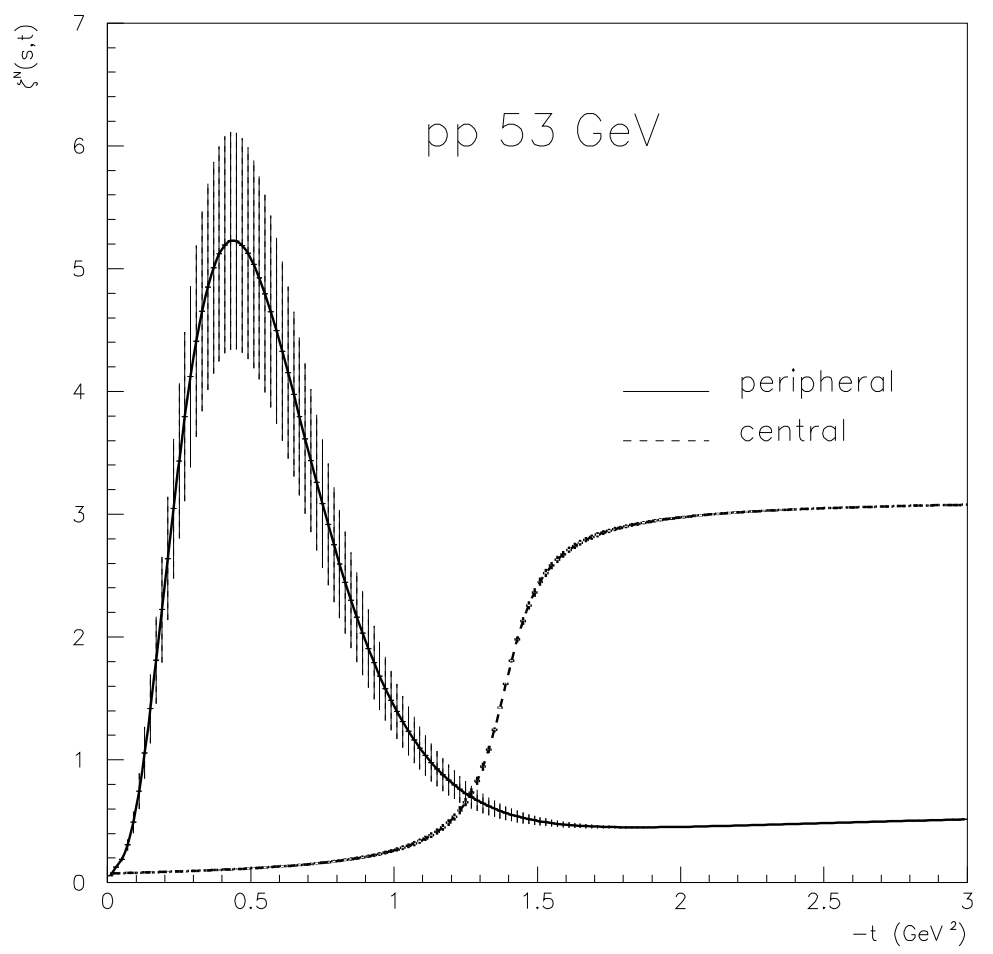

Figure 1: $t$ dependences of the hadronic phases $\zeta^N(s,t)$ leading to the central and peripheral behaviours of elastic *pp* scattering.

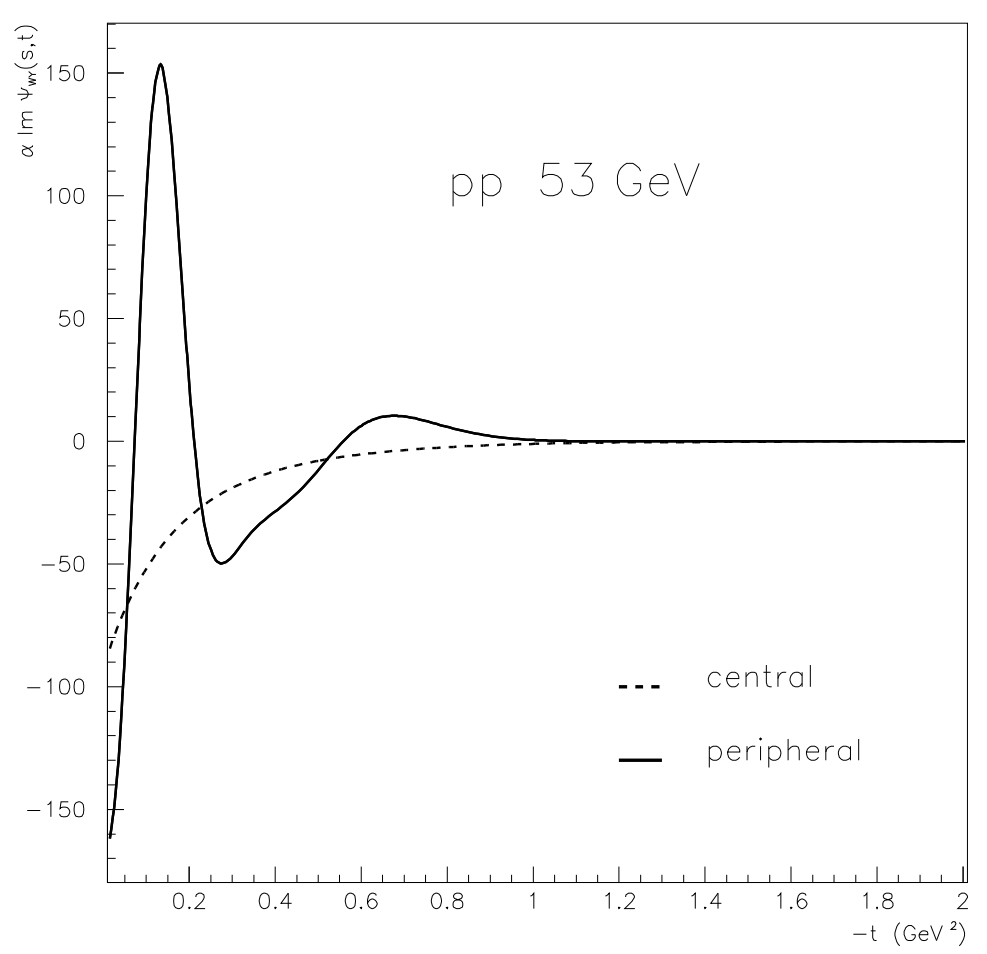

Figure 2: $t$ dependences of imaginary parts of the complex WY phases $\alpha\Psi_{WY}(s,t)$ corresponding to hadronic phases from Fig. 1.

There is, however, no reason why the elastic hadronic phase should be independent of $t$ variable at all measured values of $t$. Thus the other approach different from the West and Yennie formalism and based on the eikonal approach should be preferred - see, e.g., Ref. [4]. In such a case the complete elastic scattering amplitude $F^{C+N}(s,t)$ is related to the complete eikonal $\delta^{C+N}(s,b)$ with the help of Fourier-Bessel (FB) transformation

$$F^{C+N}(s,q^2=-t) = \frac{s}{4\pi i}\int\limits_{\Omega_b} d^2b e^{i\vec{q}.\vec{b}}\left[e^{2i\delta^{C+N}(s,b)}-1\right], \quad (4)$$

where $\Omega_b$ is the two-dimensional Euclidean space of the impact parameter $\vec{b}$. Mathematically consistent use of FB transformation requires, of course, the existence of reverse transformation. However, at finite energies the amplitude $F^{C+N}(s,t)$ is defined in a finite region of $t$ only. Thus the consistent application of formula (4) requires to take into account also the values of elastic amplitude from unphysical region where the elastic hadronic amplitude is not defined; for details see Refs. [5]. This issue has been discussed by Islam [6] who has shown that the problem may be solved in a unique way by continuing analytically the elastic hadronic amplitude $F^N(s,t)$ from physical to unphysical region of $t$ ; see also Ref. [8].

The common influence of both the Coulomb and elastic hadronic scattering then can be described by complete eikonal which is formed by the sum of both the Coulomb $\delta^C(s,b)$ and hadronic $\delta^N(s,b)$ eikonals. Then the complete amplitude (valid at any $s$ and $t$) can be finally written [4] as

$$F^{C+N}(s,t) = \pm\frac{\alpha s}{t}f_1(t)f_2(t) + F^N(s,t)\left[1 \mp i\alpha G(s,t)\right], \quad (5)$$

where

$$G(s,t) = \int\limits_{-4p^2}^{0} dt' \left\{ \ln\left(\frac{t'}{t}\right) \frac{d}{dt'} [f_1(t')f_2(t')] + \frac{1}{2\pi} \left[\frac{F^N(s,t')}{F^N(s,t)} - 1\right] I(t,t') \right\}, \tag{6}$$

and

$$I(t,t') = \int\limits_{0}^{2\pi} d\Phi'' \frac{f_1(t'')f_2(t'')}{t''}, \quad t'' = t + t' + 2\sqrt{tt'} \cos\Phi''. \tag{7}$$

Here the two form factors $f_1(t)$ and $f_2(t)$ reflect the spatial structure of colliding nucleons and should describe it in a sufficiently broad interval of $t$. As the Coulomb amplitude $F^C(s,t)$ is known from QED the complete amplitude is determined practically by the $t$ dependence of the hadronic amplitude $F^N(s,t)$.

# 2 Impact Parameter Picture of Elastic Nucleon Scattering

As it has been already mentioned the mathematically consistent use of FB transformation introducing the impact parameter representation $h_{el}(s,b)$ of elastic scattering amplitude $F^N(s,t)$ requires its definition also in the unphysical region of $t$, i.e., for $t < t_{min} = -s + 4m^2$ ($m$ being the nucleon mass). The function $h_{el}(s,b)$ must be, therefore, subdivided into two parts [5, 6]

$$\begin{aligned} h_{el}(s,b) &= h_1(s,b) + h_2(s,b) = \\ &= \frac{1}{4p\sqrt{s}} \int\limits_{t_{min}}^{0} dt\ F^N(s,t)\ J_0(b\sqrt{-t}) + \frac{1}{4p\sqrt{s}} \int\limits_{-\infty}^{t_{min}} dt\ F^N(s,t)\ J_0(b\sqrt{-t}). \end{aligned} \tag{8}$$

Similar expressions can be obtained also for the impact parameter representation $g_{inel}(s,b)$ of the inelastic overlap function $G_{inel}(s,t)$ introduced in Ref. [7].

The unitarity equation in the impact parameter space can be then written as [5, 6]

$$\Im h_1(s,b) = |h_1(s,b)|^2 + g_1(s,b) + K(s,b) \tag{9}$$

where the correlation function $K(s,b)$ is very small compared to the other functions appearing in Eq. (9) [8].

The total cross section and integrated elastic and inelastic cross sections then equal to

$$\sigma_{tot}(s) = 8\pi\int\limits_0^\infty bdb\ \Im h_1(s,b); \quad \sigma_{el}(s) = 8\pi\int\limits_0^\infty bdb\ |h_1(s,b)|^2; \quad \sigma_{inel}(s) = 8\pi\int\limits_0^\infty bdb\ g_1(s,b). \tag{10}$$

Eqs. (10) are valid provided

$$\int\limits_0^\infty bdb\ \Im h_2(s,b) \ = \ 0, \qquad \int\limits_0^\infty bdb\ g_2(s,b) \ = \ 0. \tag{11}$$

The functions $\Im h_1(s,b)$, $g_1(s,b)$ and $|h_1(s,b)|^2$ represent then the impact parameter profiles; they describe the density of interactions between two colliding nucleons in dependence on their mutual distance. The first two oscillate at grater values of $b$, but their mean squared values characterising the mean ranges of corresponding forces acting between the colliding particles can be established directly from the elastic hadronic amplitude $F^N(s,t)$ [9]. For the mean squared value of elastic impact parameter it has been derived

$$\langle b^2(s)\rangle_{el} = 4\,\frac{\int\limits_{t_{min}}^{0} dt\,|t|\,\left(\frac{d}{dt}|F^N(s,t)|\right)^2}{\int\limits_{t_{min}}^{0} dt\,|F^N(s,t)|^2} + 4\,\frac{\int\limits_{t_{min}}^{0} dt\,|t|\,|F^N(s,t)|^2\left(\frac{d}{dt}\zeta^N(s,t)\right)^2}{\int\limits_{t_{min}}^{0} dt\,|F^N(s,t)|^2}. \tag{12}$$

Similarly the total and inelastic mean squared values equal to

$$\langle b^2(s)\rangle_{tot} = 2B(s,0); \qquad \langle b^2(s)\rangle_{inel} = \frac{\sigma_{tot}(s)}{\sigma_{inel}(s)}\langle b^2(s)\rangle_{tot} - \frac{\sigma_{el}(s)}{\sigma_{inel}(s)}\langle b^2(s)\rangle_{el}. \tag{13}$$

Here the diffractive slope is defined as

$$B(s,t) = \frac{d}{dt}\left[\ln\frac{d\sigma^N}{dt}\right] = \frac{2}{|F^N(s,t)|}\frac{d}{dt}|F^N(s,t)|. \tag{14}$$

# 3 Impact Parameter Profiles for *pp* Scattering at 53 GeV

Basic results concerning the analysis of *pp* elastic scattering data at energy of 53 GeV at the ISR [10] based on the eikonal approach have been published in Ref. [4]. Here we will mention only the results related to the impact parameter profiles.

In the quoted paper we have used the formulas (5)-(7) for the complete elastic scattering amplitude $F^{C+N}(s,t)$ generating the differential cross section

$$\frac{d\sigma_{el}}{dt} = \frac{\pi}{sp^2}|F^{C+N}(s,t)|^2. \tag{15}$$

The elastic hadronic amplitude, i.e., its modulus and phase defined in Eq. (3), has been conveniently parameterised in order to describe the *pp* elastic scattering as central as well as peripheral process. While the $t$ dependence of the modulus can be almost unambiguously determined from the data the phase can be only partially constrained. Both the possible alternatives (central and peripheral) have been presented in Ref. [4]. The $t$ dependences of the corresponding shapes of the hadronic phase $\zeta^N(s,t)$ are shown in Fig. 1.

Once the elastic hadronic amplitude $F^N(s,t)$ has been specified it has been possible to determine corresponding impact parameter profiles together with their statistical errors with the help of FB transformation. And it has been also possible to determine the RMS values of the total, elastic and inelastic profiles with the help of Eqs. (12) and (13) for the central as well as the peripheral pictures of elastic *pp* scattering. Their shapes corresponding to peripheral behaviour are shown in Fig. 3 (for the central case see Ref. [3]); all RMS values are included in Table 1. In the central picture the elastic RMS is much lower than the inelastic one. This result agrees with the result of Miettinen [11]. It means that the protons in 'head-on' collisions should be rather transparent. In the peripheral case it is the second term in Eq. (12) that gives the significant contribution to the elastic RMS value. Some profiles corresponding to the peripheral case exhibit greater oscillations at higher impact parameter values $b$. However, the oscillations can be removed as it will be shown in Ref. [8]. It will be briefly described in the following.

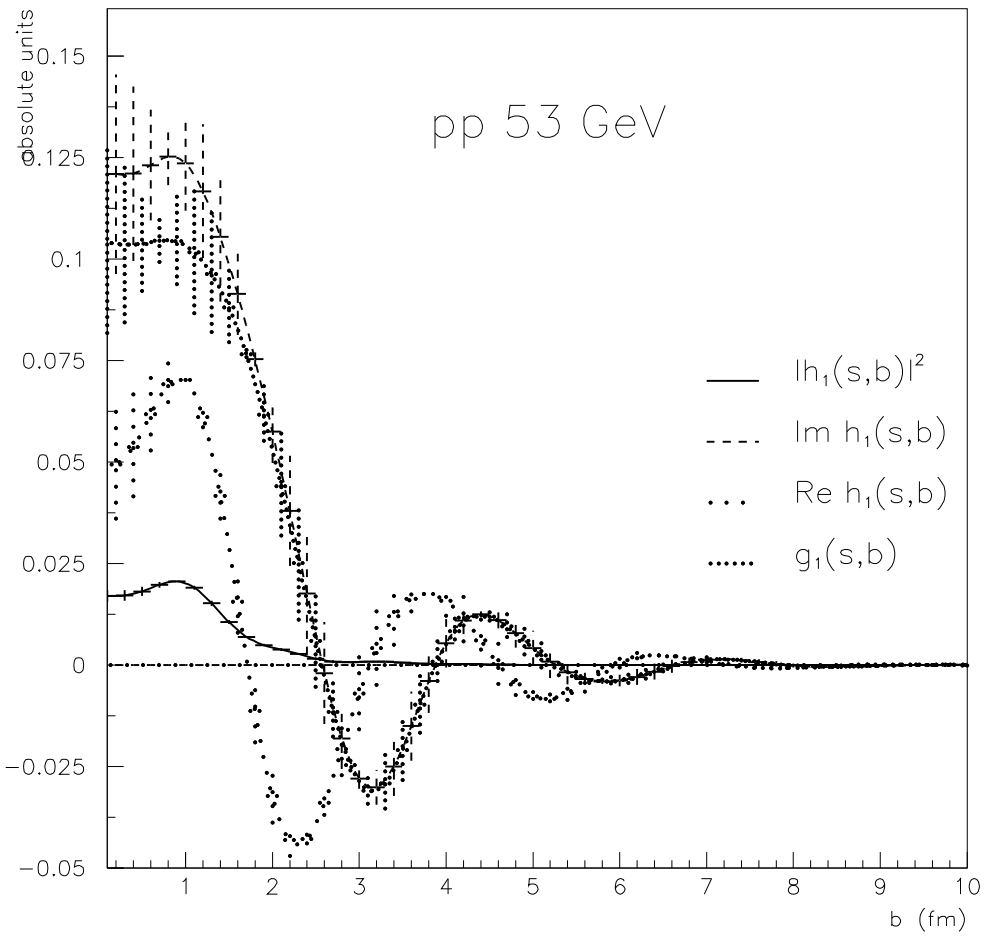

Figure 3: Oscillating peripheral profiles of elastic $pp$ scattering at 53 GeV.

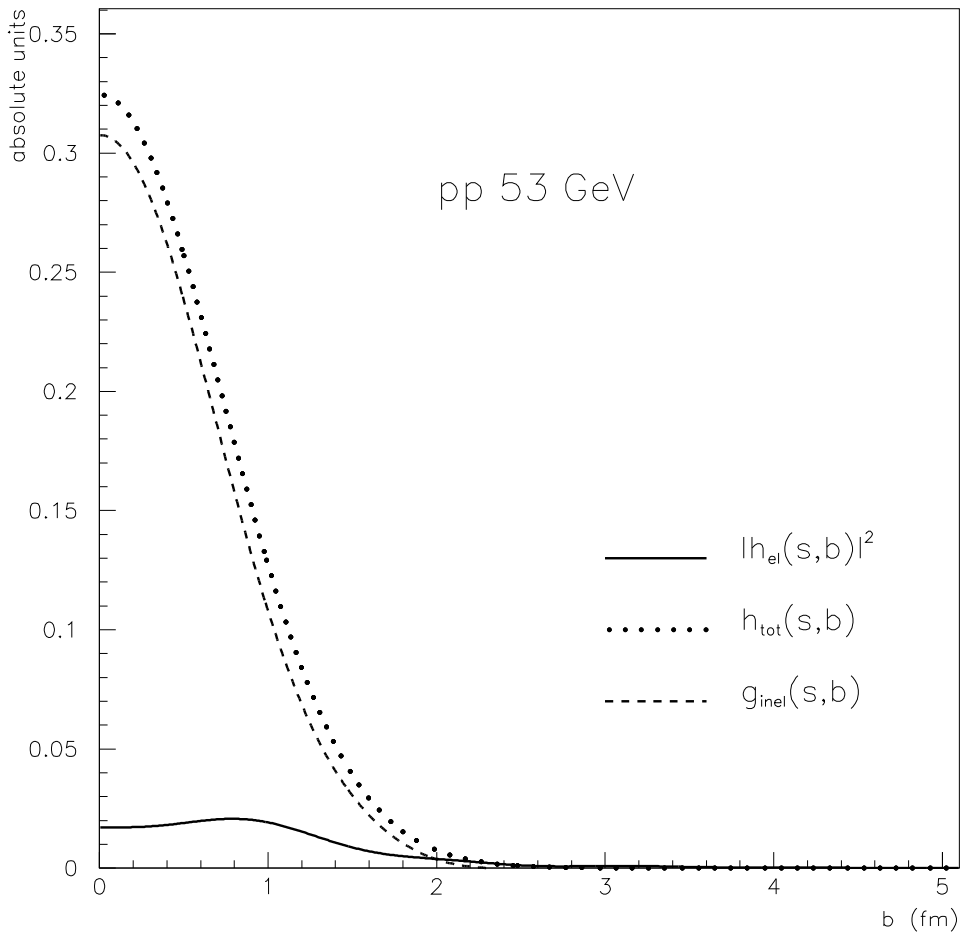

Figure 4: Positive peripheral profiles of elastic $pp$ scattering at 53 GeV.

It is necessary to construct actual (non-negative) $pp$ profiles which correspond to the values of RMS derived with the help of Eqs. (12) and (13). The way how to do it is described in Refs. [8, 9]. It consists in adding an real function $c(s, b)$ to the both sides of the unitarity equation (9). All dynamical characteristics corresponding to elastic hadron scattering will be preserved if the function $c(s, b)$ appearing in Eqs. (17) fulfills some additional conditions.

Unitarity condition (9) may be then expressed as

$$h_{tot}(s,b) = |h_1(s,b)|^2 + g_{inel}(s,b) \tag{16}$$

where

$$h_{tot}(s,b) = \Im h_1(s,b) + c(s,b), \quad g_{inel}(s,b) = g_1(s,b) + K(s,b) + c(s,b) \tag{17}$$

are non-negative functions for all $b$. The function $h_{tot}(s,b)$ must be positively semidefinite (and monotony decreasing function) at all values of $b$. The oscillating function $c(s,b)$ is required to cancel the oscillations from the total and inelastic profiles. The elastic peripheral profile $|h_1(s,b)|^2$ will remain unchanged. And integrated inelastic cross section is preserved if the function $c(s,b)$ fulfills the conditions

$$\int_0^\infty b\, db\, c(s,b) = 0; \qquad \int_0^\infty b^3\, db\, c(s,b) = 0. \tag{18}$$

These two conditions represent the integral conditions limiting the shape of the function $c(s,b)$. According to Islam's approach [6] this function may be identified with the $\Im h_2(s,b)$ for which the conditions (18) are to be required. However, in the standard approach the function $c(s,b)$ can be hardly determined analytically. The best way at the present seems to specify it in a numerical way as it will be illustrated in the following. It may be expected that the total profile entering into modified unitarity condition (16) should be approximately of Gaussian type with the values that may be characterised by integral cross sections and by RMS values shown in Table 1. The elastic profile will remain unchanged. Under these assumptions the total profile shape can be defined (the $s$ dependence being dropped) as $h_{tot}(b) = ae^{-\beta b^2}$. Using formulas 3.461 from Ref. [13] the corresponding integrals needed for calculation of the total cross section and total mean squared value can be analytically determined as

$$\int_0^\infty b\, db\, a\, e^{-\beta b^2} = \frac{a}{2\beta}, \qquad \int_0^\infty b^3\, db\, a\, e^{-\beta b^2} = \frac{a}{2\beta^2} \tag{19}$$

and the values of the constants $a$ and $\beta$ can be determined from experimentally established values. For the peripheral case of elastic $pp$ scattering at energy of 53 GeV their values are: $a = 0.324$ and $\beta = 0.946$. The $b$ dependence of the auxiliary function $c(s, b)$ is then determined with the help of the first equation from (17) where $\Im h_1(s, b)$ is taken from experimental analysis. And the second equation determines then the shape of the inelastic profile.

Table 1: Root-mean-squares of impact parameters for $pp$ collisions at 53 GeV.

| elastic profile | $\sqrt{< b^2 >_{tot}}$ | | $\sqrt{< b^2 >_{el}}$ | | $\sqrt{< b^2 >_{inel}}$ |
|---|---|---|---|---|---|
| | | modulus | phase | sum | |
| | [fm] | [fm] | [fm] | [fm] | [fm] |
| peripheral | 1.033 | 0.676 | 1.671 | 1.803 | 0.772 |
| central | 1.028 | 0.679 | $\sim 0.$ | 0.679 | 1.087 |

Knowing the shapes of the total and inelastic profiles together with the $b$ dependence of the auxiliary function $c(s, b)$ the values of all the integrated cross sections and of all the mean squares can be verified numerically as can be seen from the Table 2. The new values are practically quite comparable with the original ones. Also the values of the integrals of function $c(s, b)$ (see Eqs. (18)) only slightly different from zero are shown in Table 2. The modified profiles are exhibited in Fig. 4. The new total and inelastic profiles are central while the elastic profile remains unchanged and is peripheral.

# 4 Conclusion

Some results concerning elastic nucleon collisions at high energies and based on the validity of optical theorem have been summarised; the eikonal model has been applied to. The approach suitable for the case of finite energies has been presented. It has been shown that elastic processes may be interpreted as peripheral.

Table 2: The values of integrated cross sections and of the total, elastic and inelastic RMS.

| Quantity | | Original values | New values |
|---|---|---|---|
| $\sigma_{tot}$ | [mb] | 42.864 | 42.872 |
| $\sigma_{el}$ | [mb] | 7.479 | 7.479 |
| $\sqrt{< b^2 >_{tot}}$ | [fm] | 1.0 | 1.028 |
| $\sqrt{< b^2 >_{el}}$ | [fm] | 1.803 | 1.803 |
| $\sqrt{< b^2 >_{inel}}$ | [fm] | 0.772 | 0.772 |
| $\int_0^\infty bdb\ c(s,b)$ | [fm$^2$] | - | 0.029 |
| $\int_0^\infty b^3db\ c(s,b)$ | [fm$^4$] | - | 0.097 |

# References

[1] H. Bethe, Ann. Phys. **3** 190 (1958)

[2] G.B. West and D.R. Yennie, Phys. Rev. **172** 1413 (1968)

[3] V. Kundrát, M. Lokajíček and I. Vrkoč, Phys. Lett. **B656** 182 (2007)

[4] V. Kundrát and M. Lokajíček, Z. Phys. **C63**, 619 (1994)

[5] T. Adachi and T. Kotani, Progr. Theor. Phys. Suppl., Extra Number 316 (1965); Progr. Theor. Phys. **35** 463 (1966); **35** 485 (1966); **37-38** 297 (1966); **39** 430 (1968); **39** 785 (1968)

[6] M.M. Islam, Lectures in theoretical Physics, ed. A. O. Barut and W. E. Brittin, Vol. 10B (Gordon and Breach, 1968), p.97; Nucl. Phys. **B104** 511 (1976)

[7] L. van Hove, Nuovo Cim. **27** 2344 (1963)

[8] V. Kundrát, M. Lokajíček and J. Kašpar, “Impact parameter profiles of elastic scattering in hadronic collisions”; to be published

[9] V. Kundrát, M. Lokajíček and D. Krupa, Phys. Lett. **B544** 132 (2002)

[10] K.R. Schubert, Landolt-Bornnstein, Numerical Data and Functional Relationships in Science and Technology, New Series, (Springer, Berlin, 1979), Vol. I/9a

[11] H.G. Miettinen, Proc. Rencontre de Moriond, Meribel les Allues, Vol. 1 (ed. J. Tran Thanh Van, Orsay 1974), Acta Phys. Pol. **B6** 625 (1975)

[12] G. Giacomelli and M. Jacob, Phys. Rep. **55** 1 (1979)

[13] I.S. Gradshteyn, I.M. Ryzhik, Tables of Integrals, Series and Products, Academic Press, San Diego, (1980).

# The Total Cross Section at the LHC: Models and Experimental Consequences

*J.R. Cudell*

IFPA, AGO Dept., Université de Liège, Belgium

I review the predictions of the total cross section for many models, and point out that some of them lead to the conclusion that the standard experimental analysis may lead to systematic errors much larger than expected.

The total cross section is a highly non-perturbative object that we cannot predict from QCD. In fact, we have to rely on theoretical ideas that were developed before QCD, in the context of the analytic $S$ matrix, such as analyticity, or the unitarity of partial waves.

The natural place to discuss these ideas is the complex-$j$ plane, where the singularities of the amplitudes determine their behaviour with $s$. But we do not know these singularities. The simplest ones, which correspond to the exchange of bound states, are simple poles. Accounting for the exchange of meson trajectories is not too hard, as we know their spectrum. But as their contribution falls with $s$, they will not matter at the LHC. One must thus model the pomeron, for which there is little spectroscopic guidance. Again, the simplest idea is to use a simple pole at $j = 1 + \epsilon$. But we know that if there are simple poles, there must be cuts, which correspond to multiple exchanges. We do not know how to calculate these: we only know general properties of the two-pomeron cut. This means that there are many possibilities, such as eikonal models, $U$-matrix unitarisation, extended eikonal/$U$-matrix models, or multi-channel eikonals. These cuts will be needed at the LHC to restore partial-wave unitarity. It is also possible that the pomeron is not an exchange of bound states, so that one should not start from a simple pole, but rather consider multiple poles (double or triple) at $j = 1$, which automatically obey unitarity.

## 1 The COMPETE Fits

The multitude of models has been confronted with data, and the COMPETE collaboration [1] gathered and cleaned the datasets which are now available on the servers of the Particle data group [2], and which gather all the soft data measured at $\sqrt{s} \geq 4$ GeV and $t = 0$, *i.e.* the total cross sections and the $\rho$ parameters when available, for $pp$, $\bar{p}p$, $\pi^{\pm}p$, $K^{\pm}p$, $\gamma p$ and $\gamma\gamma$. These data have some drawbacks. First of all, most cross sections have been measured at ISR energies, and there is a gap from $\sqrt{s} \approx 100$ GeV to $\sqrt{s} \approx 500$ GeV. This gap would have been filled by the pp2pp collaboration at RHIC, but it has unfortunately been stopped. Another problem comes from the highest-energy data, where a $2\sigma$ disagreement exists between E710 and CDF. Finally, the compatibility of the various measurements of $\rho$ at ISR energies is questionable, and not enough information has been published to redo the analyses.

Nevertheless, COMPETE considered several hundreds of possible parametrisations, based on simple, double or triple poles, and kept only those which had a global $\chi^2$/point smaller than 1, for $\sqrt{s} \geq 5$ GeV. From these, one can predict the total cross section at the LHC, and estimate the error due to the extrapolation. Fig. 1 shows the result of this study. The extrapolation of the parametrisations which passed the $\chi^2$ test is given by the outer band. It predicts $84\,\text{mb} \leq \sigma \leq 112\,\text{mb}$ for $\sqrt{s} = 10\,\text{TeV}$, and 90–117 mb for 14 TeV.

The inner band of Fig. 1 shows the best guess, and it is based on a series of statistical indicators explained in [4]. It corresponds to a universal triple-pole ($\log^2 s$) parametrisation. One should note that the multiple-pole fits have some peculiar properties: in the triple-pole case, the contribution of the pomeron *falls* with $s$ for energies smaller than 6 GeV, whereas in the double-pole case, it actually becomes negative below 9.5 GeV. Simple poles seem to be excluded, but they can be considered again if the lower cut-off on the energy is raised to 10 GeV. The multiple-pole parametrisations have been used successfully [5] to reproduce low-$x$ data in deeply inelastic scattering (DIS), showing that universality does not seem to hold in general, as the coefficient of the triple pole depends on $Q^2$. More recently, multiple poles have been used successfully [6] to reproduce the differential elastic cross sections [7].

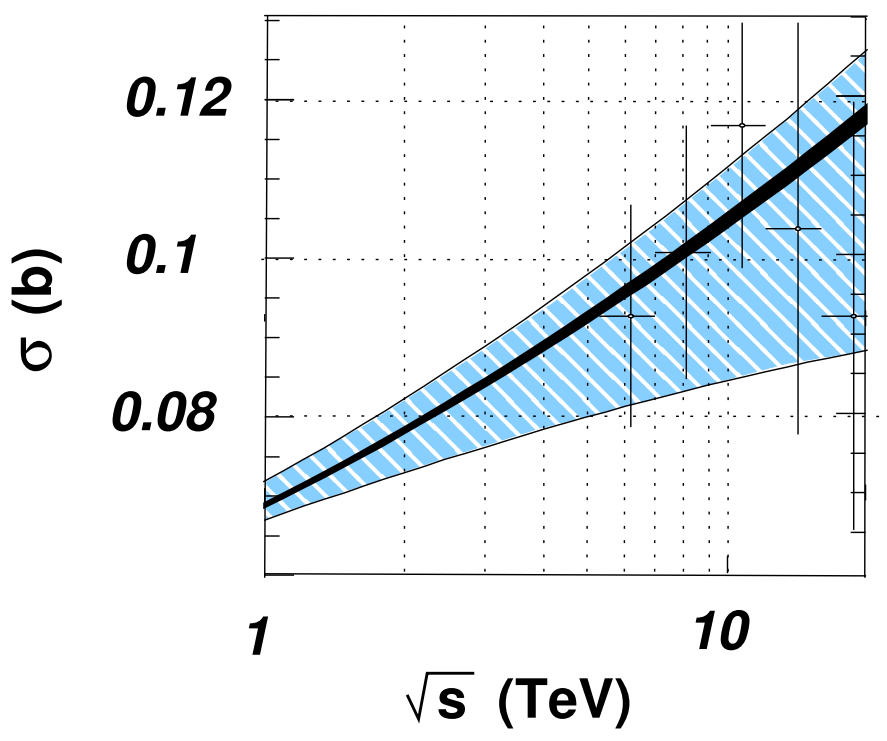

Figure 1: The COMPETE prediction for the total cross section in the LHC energy region. The inner black band represents our best guess with statistical errors only, the outer band all the parametrisations with $\chi^2$/point $\leq 1$.

Hence multiple-pole fits seem to work, but it is hard to imagine how such a structure may emerge from QCD. On the other hand, simple-pole fits have a simple phenomenology, a simple interpretation as the exchange of glueballs, and only a handful of parameters.

## 2 The Simple-Pole Fit: Low Energy

Clearly, one simple-pole pomeron has never been a reasonable choice: in their original soft-pomeron model [8], Donnachie and Landshoff included a two-pomeron cut, which lowers the cross section, and which enables one to go to $\approx 40$ TeV before unitarity is broken. However, such a simple model seems to produce worse fits than multiple poles. Furthermore, HERA data on DIS and on vector-meson production have lead to the introduction of a second singularity [9], taken also as a simple pole, and mimicking the BFKL cut. One can indeed follow the standard transformation to the $j$-plane to convince oneself that hadronic singularities must be present in DIS [10]. This new singularity has a coupling which depends on $Q^2$, or on the vector-meson mass, and, together with the soft pomeron, it enables one to reproduce all HERA measurements [11]. The question then is to know whether such a singularity might be present in soft data.

It actually came as a surprise that, indeed, the introduction of a hard pole brings the simple-pole pomeron fit on equal footing with multiple poles, as far as the $\chi^2$/point is concerned [12]. This is true provided that one stops the fit at $\sqrt{s} \approx 200$ GeV. Beyond that, the hard pole must be unitarised, as its contribution to the total cross section would be much too large. Nevertheless, it is remarkable that a global fit to soft data leads to a hard pomeron with an intercept $\approx 1.4$, perfectly compatible with that obtained by Donnachie and Landshoff in DIS. It must be noted however that the coupling of the hard pomeron is tiny. It contributes at most 7% to the total cross section for $\sqrt{s} < 200$ GeV, as shown in Fig. 2. Previous studies [1] considered only fits to the highest energy, in which case the hard pomeron coupling becomes extremely small. Even if the energy is limited to 200 GeV, the coupling remains of the order of 1% of that of the soft pomeron.

The two-pomeron model was later applied to the differential elastic cross section in [13], where it was shown that the hard pomeron was compatible with data on the first cone ($0.1$ GeV$^2$ $< |t| \leq 0.5$ GeV$^2$), for energies smaller than 100 GeV, although in this case it did not significantly improve the fit. Note also that the form factors of the various exchanges were directly extracted from the data, confirming the presence of a zero in the form factor of the $C = -1$ trajectory, and suggesting one for the $c = +1$ meson trajectory.

One has thus found an alternative to the multiple-pole fits: the two-pomeron model, which can reproduce DIS data, photoproduction data, soft forward data and differential elastic cross section, provided one stays at moderate energy and small $|t|$. As we shall now see, if this model is the correct one, then it may have important consequences at the LHC.

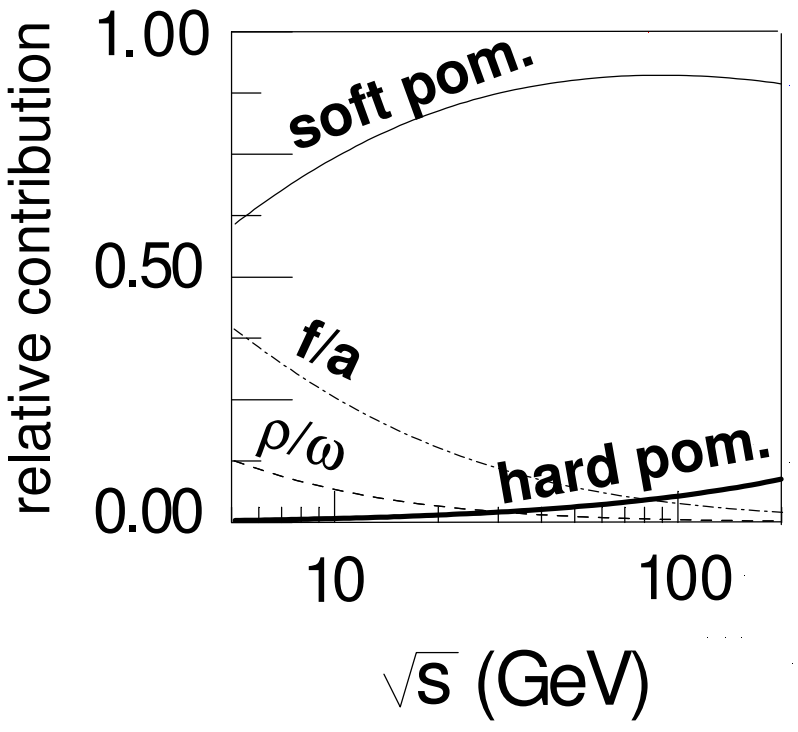

Figure 2: The contribution to the total $pp$ cross section of the two pomerons and the two reggeons, divided by the average of the $pp$ and $\bar{p}p$ total cross sections.

# 3 The Simple-Pole Fit: High Energy

Indeed, to go to higher energies, one must confront the question of unitarisation. At high energy, the partial waves can be obtained by a Fourier transform of the amplitude $\mathcal{A}(s,t)$ to impact-parameter space, as $\ell \approx b\sqrt{s}$. The unitarity constraint can then be written

$$|\mathcal{A}(s,b)|^2 \leq 2Im\left(\mathcal{A}(s,b)\right).$$

It is easy to see that this means that the complex $\mathcal{A}(s,b)$ has to be in a circle of radius 1 centred at $\mathcal{A}(s,b) = i/2$, which we shall call the unitarity circle (Note that another normalisation is frequently used, where the circle has radius 1 and is centred at $i$.)

In our two-pomeron model, the amplitude $\mathcal{A}(s,0)$ leaves the unitarity circle for energies slightly below the Tevatron, and it is quite far out at the LHC, as can be seen in Fig. 3. The problem, of course, is that one does not know exactly how the amplitude will stay in the unitarity circle. It is known that multi-pomeron exchanges, and maybe multi-pomeron vertices, will restore unitarity. But how to implement these ingredients is far from unique [14]. Even the simplest two-pomeron cut is not fully calculable, let alone a full resummation.

Mathematically, unitarisation schemes can be classified into three classes. The first one maps the whole upper-half complex plane onto the unitarity circle. In this class, one finds the eikonal scheme, and the $U$-matrix scheme [16]. Note that the latter is the only one to provide a one-to-one mapping between the half-plane and the circle, but the amplitude also has a new pole at $\mathcal{A}(s,b) = -2i$. The second class is a simple extension of the first one: instead of mapping the upper half-plane to the full unitarity circle, one maps it to a smaller circle, still centred at $\mathcal{A}(s,b) = i/2$. The third class, which we recently proposed [15], maps a given amplitude to the circle, which is far less restrictive than mapping the half plane. In general, these schemes can have very different properties [15]: they can lead to the Pumplin bound ($\sigma_{el} < \sigma_{tot}/2$ [17]) and to shadowing, but it is also possible to have anti-shadowing and $\sigma_{el} \approx \sigma_{tot}$, both for eikonal-like and $U$-matrix-like schemes. I shall not consider these more exotic possibilities here, but one has to bear in mind that they cannot be a priori ruled out.

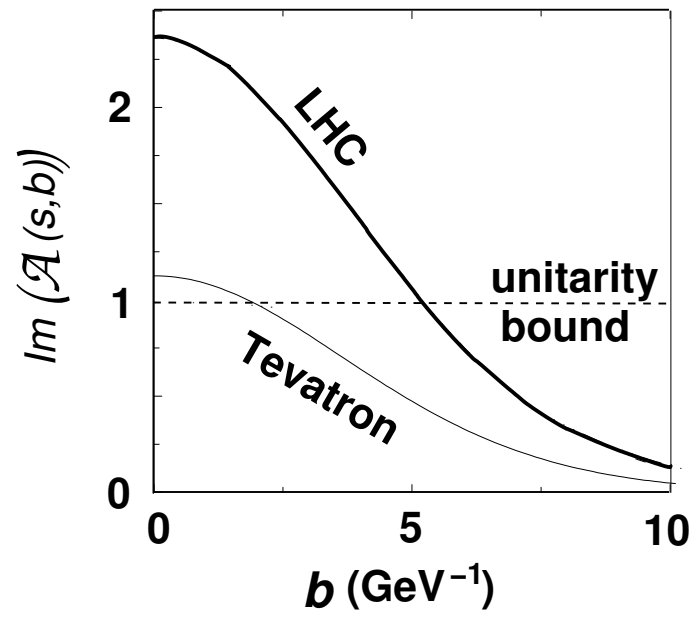

Figure 3: The imaginary part of the amplitude $\mathcal{A}(s,b)$ at the Tevatron and at the LHC, before unitarisation.

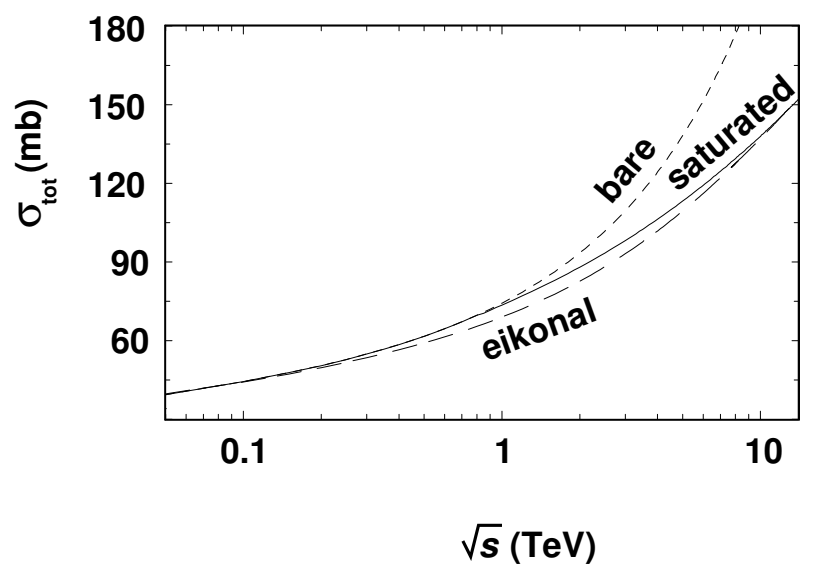

Figure 4: Total cross section in a two-pomeron model without (bare) and with simple unitarisation schemes.

As we do not know what scheme to use, we have decided to be conservative, and to unitarise the two-pomeron model using the simplest and most conservative approach. The minimal scheme assumes that something happens when the amplitude reaches the unitarity bound, and one simply cuts-off the amplitude at 1. We shall call it the saturation scheme. Note that to do that while preserving analytic properties is not very simple, but nevertheless it can be done [18]. The other scheme is the standard one-channel eikonal.

Using either of these schemes produces an amplitude which respects unitarity, and thus can be used to predict the total cross section at the LHC. The first scheme keeps by definition the results of the fit below the Tevatron energy, and thus one can keep the parameters of the low-energy simple-pole fit. The eikonal on the other hand changes slightly the low-energy results, and thus necessitates some re- fitting. The result is shown in Fig. 4. One sees that, in this very simple approach, the total cross section could be as large as 150 mb. It is interesting to note that a similar number has been obtained by Landshoff, in a very different manner [14].

# 4 The Measurement of $\sigma_{\mathrm{tot}}$

The unitarised two-pomeron model has a consequence that has been overlooked so far. As the energy grows, the protons become blacker, and edgier. This in turn changes the diffraction pattern, *i.e.* the exponential slope of the cross section varies quickly with $t$: the differential cross section cannot be approximated any longer by $e^{Bt}$ with $B$ constant. In turn, the real part of the amplitude also develops a strong $t$ dependence, and $\rho(0)$ could be as large as 0.24. We have checked that this is the case for the eikonal scheme, and for the saturation scheme [18], and show the results for the latter in Fig. 5.

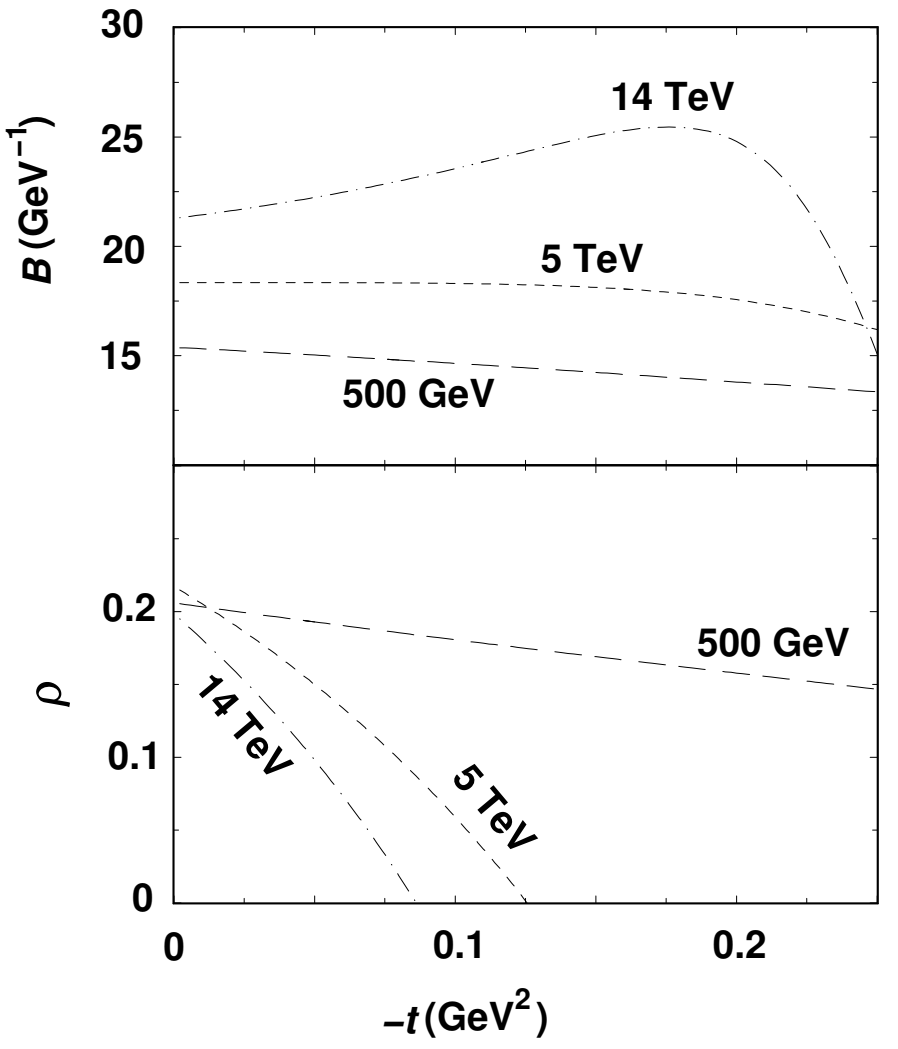

Figure 5: The slope $B$ and the ration of the real to the imaginary parts of the amplitude for $pp$ scattering, for various energies, in the saturation scheme.

A hadronic slope $B$ almost constant with $t$ for $|t| < 0.25$ GeV$^2$ and a small $\rho$ parameter are standard assumptions for the planned measurement of the total cross section [19]. If the two-pomeron model is correct, then these assumptions will be wrong, and we can evaluate the systematic uncertainty that they will generate. So, we have used our unitarised two-pomeron model [20] to simulate data, using bins and errors similar to those of the UA4/2 experiment [21], and including the Coulomb-nucleon interference [22]. We then performed the standard analysis on these simulated data, and extracted a measurement of the total cross section, which we could compare to the input value from our model. The result of this analysis is that the extracted value of the total cross section will systematically overshoot the model value by about 10 mb in a luminosity-dependent method, and about 15 mb in a luminosity-independent one. Hence, an additional study of the $t$ dependence of the slope and of the $\rho$ parameter will be needed before one reaches the 1 % precision level.

# 5 Conclusion

The measurement of the total cross section at the LHC will tell us a lot about the analytic structure of the amplitude, as there is a variety of predictions that span the region from 90 to 230 mb:

- $\sigma_{tot} > 200$ mb: the only unitarisation scheme able to accommodate such a large number is the $U$ matrix [16]. It basically predicts the same inelastic cross section as more standard schemes, but the elastic cross section is much larger, and accounts for the difference.

- 120 mb $< \sigma_{tot} <$ 160 mb: this would be a clear signal for a two-pomeron model, and would also tell us about the unitarisation scheme.

- $\sigma_{tot} \approx 110$ mb: this is the standard prediction not only of the COMPETE fits, but also of many models based on a simple eikonal and only one pomeron pole.

- $\sigma_{tot} < 100$ mb: this would indicate either the validity of double-pole parametrisations, or that of unitarisation schemes in which multiple-pomeron vertices are important [23, 24].

And finally, if one really enters a new regime at the LHC, the study of the $t$ dependence of the elastic cross section, and of the real part of the amplitude, may be of great importance to unravel the underlying dynamics and to improve the experimental measurement of $\sigma_{tot}$.

## Acknowledgements

I acknowledge the contribution of my collaborators of the COMPETE collaboration to the investigations summarised here, and in particular of A. Lengyel, E. Martynov and O.V. Selyugin. I also acknowledge discussions with A. Donnachie, P.V. Landshoff, U. Maor and S. Troshin, and thank Karine Gilson for a careful proofreading.

## References

[1] J. R. Cudell, V. Ezhela, K. Kang, S. Lugovsky and N. Tkachenko, Phys. Rev. D **61** (2000) 034019 [Erratum-ibid. D **63** (2001) 059901] [arXiv:hep-ph/9908218]; J. R. Cudell *et al.*, Phys. Rev. D **65** (2002) 074024 [arXiv:hep-ph/0107219].

[2] See http://pdg.lbl.gov/2009/hadronic-xsections/hadron.html

[3] J. R. Cudell *et al.* [COMPETE Collaboration], Phys. Rev. Lett. **89** (2002) 201801 [arXiv:hep-ph/0206172].

[4] J. R. Cudell *et al.* [COMPETE collaboration], arXiv:hep-ph/0111025.

[5] J. R. Cudell and G. Soyez, Phys. Lett. B **516** (2001) 77 [arXiv:hep-ph/0106307].

[6] E. Martynov and B. Nicolescu, Eur. Phys. J. C **56** (2008) 57 [arXiv:0712.1685 [hep-ph]]; E. Martynov, Phys. Rev. D **76** (2007) 074030 [arXiv:hep-ph/0703248].

[7] A complete dataset for differential elastic cross sections is available at http://www.theo.phys.ulg.ac.be/∼cudell/data/

[8] A. Donnachie and P. V. Landshoff, Nucl. Phys. B **231** (1984) 189; Nucl. Phys. B **267** (1986) 690.

[9] A. Donnachie and P. V. Landshoff, Phys. Lett. B **437** (1998) 408 [arXiv:hep-ph/9806344].

[10] J. R. Cudell, A. Donnachie and P. V. Landshoff, Phys. Lett. B **448** (1999) 281 [Erratum-ibid. B **526** (2002) 413] [arXiv:hep-ph/9901222].

[11] A. Donnachie and P. V. Landshoff, arXiv:0803.0686 [hep-ph]; Phys. Lett. B **550** (2002) 160 [arXiv:hep-ph/0204165]; A. Donnachie and P. V. Landshoff, Phys. Lett. B **518** (2001) 63 [arXiv:hep-ph/0105088].

[12] J. R. Cudell, E. Martynov, O. V. Selyugin and A. Lengyel, Phys. Lett. B **587** (2004) 78 [arXiv:hep-ph/0310198].

[13] J. R. Cudell, A. Lengyel and E. Martynov, Phys. Rev. D **73** (2006) 034008 [arXiv:hep-ph/0511073].

[14] P. V. Landshoff, arXiv:0903.1523 [hep-ph].

[15] J. R. Cudell, E. Predazzi and O. V. Selyugin, Phys. Rev. D **79** (2009) 034033 [arXiv:0812.0735 [hep-ph]].

[16] Sergey Troshin, these proceedings; S. M. Troshin and N. E. Tyurin, Eur. Phys. J. C **21** (2001) 679 [arXiv:hep-ph/0004084]; C. T. Sachrajda and R. Blankenbecler, Phys. Rev. D **12** (1975) 1754.

[17] J. Pumplin, Phys. Rev. D **8** (1973) 2899.

[18] J. R. Cudell and O. V. Selyugin, Phys. Lett. B **662** (2008) 417 [arXiv:hep-ph/0612046].

[19] Jan Kaspar, these proceedings; G. Anelli *et al.* [TOTEM Collaboration], JINST **3** (2008) S08007.

[20] J. R. Cudell and O. V. Selyugin, Phys. Rev. Lett. **102** (2009) 032003 [arXiv:0812.1892 [hep-ph]].

[21] C. Augier *et al.* [UA4/2 Collaboration], Phys. Lett. B **316** (1993) 448.

[22] O. V. Selyugin, Mod. Phys. Lett. A **11** (1996) 2317.

[23] Uri Maor, these proceedings; E. Gotsman, E. Levin, U. Maor and J. S. Miller, Eur. Phys. J. C **57** (2008) 689 [arXiv:0805.2799 [hep-ph]];

[24] M. G. Ryskin, A. D. Martin and V. A. Khoze, Eur. Phys. J. C **60** (2009) 249 [arXiv:0812.2407 [hep-ph]].

# Proton-Proton Elastic Scattering at LHC and Proton Structure

*M. M. Islam*[1], *Jan Kašpar*[2,3], *R. J. Luddy*[1], *A. V. Prokudin*[4]

[1]Department of Physics, University of Connecticut, Storrs, CT 06269 USA
[2]CERN, 1211 Genève 23, Switzerland
[3]Academy of Sciences of the Czech Republic, Prague, Czech Republic
[4]Department of Theoretical Physics, University of Torino and INFN, Torino, Italy

Talk presented by R. J. Luddy

A systematic phenomenological investigation of high energy pp and p̄p elastic scattering has led to a physical picture of the proton. The proton appears to consist of an outer region of qq̄ condensed ground state, an inner shell of topological baryonic charge and a core where valence quarks are confined. It is, therefore, a 'chiral bag' (a low energy model of nucleon structure) enclosed by the qq̄ condensed state – a 'Condensate Enclosed Chiral Bag'. Based on this picture, we predict pp elastic $d\sigma/dt$ at LHC at $\sqrt{s}$= 14 TeV and $|t|$=0 – 10 GeV$^2$ and compare our prediction with those of a number of leading models. Precise measurement of $d\sigma/dt$ by the TOTEM Collaboration will be able to distinguish between these models.

High energy proton-proton (pp) and antiproton-proton (p̄p) elastic scattering measurements have been at the frontier of accelerator research since the early seventies, when pp elastic scattering was measured at the CERN ISR Collider over a wide range of energy and momentum transfer. This was followed by measurement of pp elastic scattering at the Fermilab in a fixed target experiment, then by p̄p elastic scattering measurements at the CERN Super Proton Synchrotron (SPS) Collider and, finally, in the nineties by p̄p elastic scattering measurement at the Fermilab Tevatron. Table 1 chronicles this sustained and dedicated experimental effort by physicists extending over quarter of a century as the centre-of-mass (c.m.) energy increased from the GeV region to the TeV region. With the start-up of the LHC on the horizon, a detailed plan has been underway to measure pp elastic scattering at c.m. energy 14 TeV (seven times the c.m. energy of the Tevatron) over a wide range of momentum transfer ($|t| \simeq 0.003$ – 10.0 GeV$^2$) by the TOTEM Collaboration. On the other hand, the ATLAS Collaboration plans to measure pp elastic scattering at 14 TeV in the very small momentum transfer range ($|t| \simeq 0.0006 - 0.1$ GeV$^2$), where the pp Coulomb amplitude and strong interaction amplitude interfere.

A phenomenological investigation of high energy pp and p̄p elastic scattering was begun by Islam et al. in the late seventies with the goal to quantitatively describe the measured elastic differential cross sections – as the c.m. energy increased and as one proton probed the other at smaller and smaller distances (when the momentum transfer increased). This investigation has now been pursued for three decades and has led to: 1) a physical picture of the proton, and 2) an effective field theory model underlying that picture [1]. The proton appears to have three regions (Fig. 1): an outer region of quark-antiquark (qq̄) condensed ground state, an inner shell of baryonic charge - where the baryonic charge is geometrical or topological in nature (similar to the 'Skyrmion Model' of the nucleon) and a core region of size 0.2 fm - where valence quarks are confined. The part of the proton structure comprised of a topological baryonic charge shell and three valence quarks in a small core has been known as a chiral bag model of the nucleon in low energy studies [2]. What we are finding from high energy elastic scattering then– is that the proton is a 'Condensate Enclosed Chiral Bag'.

| Accelerator | $\sqrt{s}$ | $\|t\|$ | Reference |
|---|---|---|---|
| CERN ISR ( pp ) | 23 - 62 GeV | 0.8 - 10 GeV$^2$ | Nagy et al. Nucl. Phys. B 150, 221 (1979) |
| Fermilab Fixed target ( pp ) | 27.4 GeV | 5.5 - 14.2 GeV$^2$ | Faissler et al. Phys. Rev. D 23, 33 (1981) |
| CERN SPS ( p̄p ) | 546 GeV | 0.03 - 1.55 GeV$^2$ | Bozzo et al. Phys. Lett. B 147, 385 (1984); ibid. 155, 197 (1985) |
| | 630 GeV | 0.7 - 2.2 GeV$^2$ | Bernard et al. Phys. Lett. B 171, 142 (1986) |
| | 541 GeV | 0.00075 - 0.120 GeV$^2$ | Augier et al. Phys. Lett. B 316, 448 (1993) |
| Tevatron ( p̄p ) | 1.8 TeV | 0.03 - 0.63 GeV$^2$ | Amos et al. Phys. Lett. B 247, 127 (1990) Abe et al. Phys. Rev. D 50, 5518 (1994) |
| LHC ( pp ) | 14 TeV | 0.003 - 10 GeV$^2$ | TOTEM Collab. Anelli et al. JINST 3 S08007 (2008) |
| LHC ( pp ) | 14 TeV | 0.0006 - 0.1 GeV$^2$ | ATLAS ALFA: ATLAS Collab. TDR CERN/LHCC/2007 |

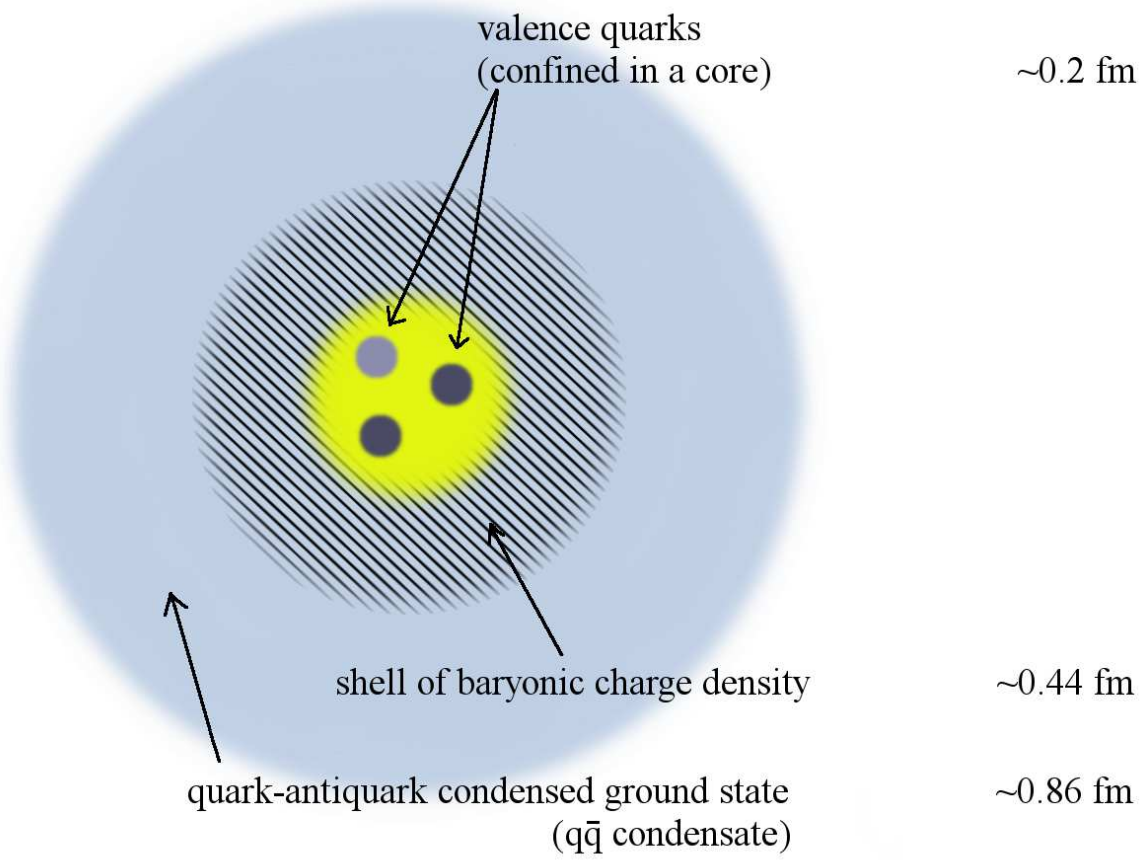

Figure 1: Physical picture of the proton – a condensate enclosed chiral bag.

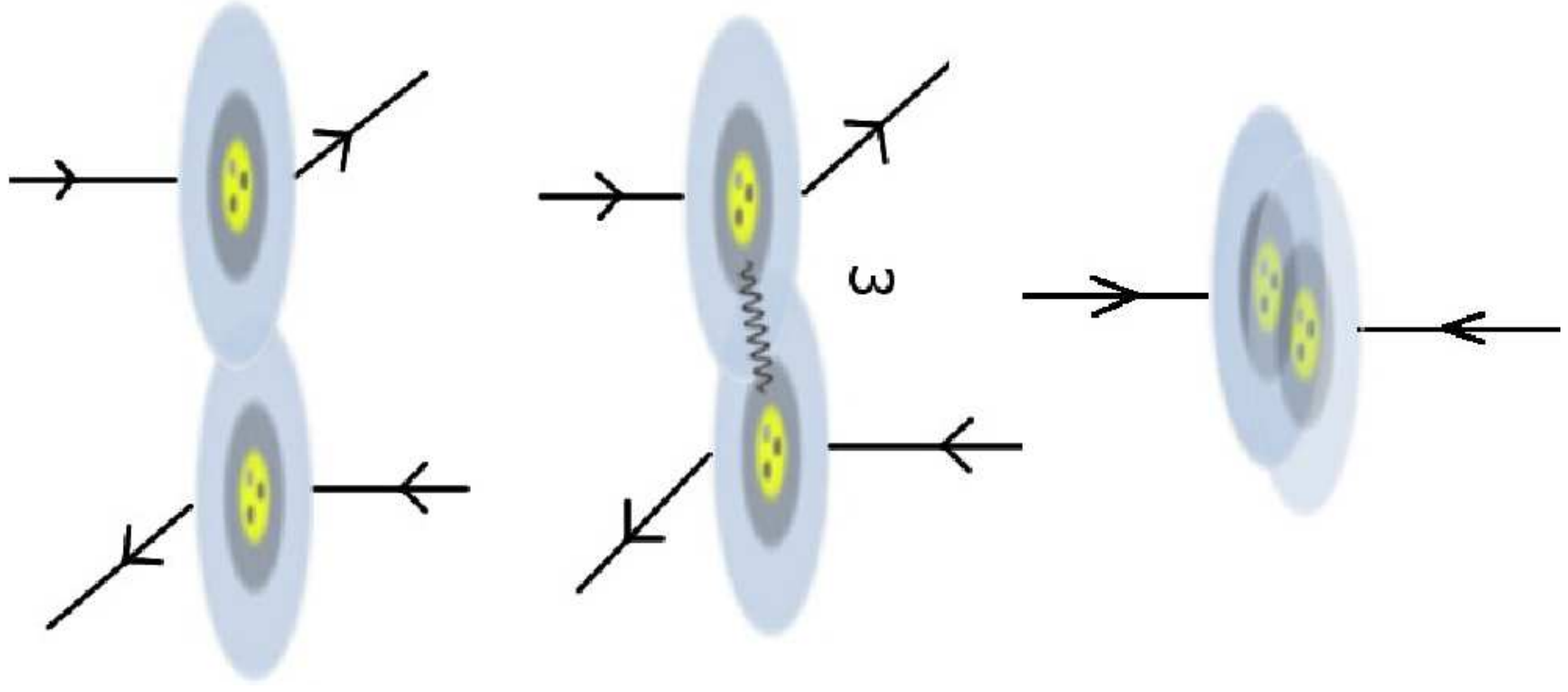

Figure 2: Elastic scattering processes (from left to right): 1) diffraction, 2) $\omega$ exchange, 3) short-distance collision ($b \lesssim 0.1$ fm).

The proton structure shown in Fig. 1 leads to three main processes in elastic scattering (Fig. 2): 1) In the small $|t|$ region, i.e., in the near forward direction, the outer cloud of q$\bar{\text{q}}$ condensate of one proton interacts with that of the other giving rise to diffraction scattering. This process underlies the observed increase of the total cross section with energy and the equality of pp and $\bar{\text{p}}$p total cross sections at high energy. It also leads to diffraction minima, as in optics, which can be seen in our Fig. 7. 2) In the intermediate momentum transfer region ($|t| \simeq 1-4$ GeV$^2$), the topological baryonic charge of one proton probes that of the other via vector meson $\omega$ exchange. This process is analogous to one electric charge probing another via photon exchange. The spin-1 $\omega$ acts like a photon – because of its coupling with the topological baryonic charge. 3) In the large $|t|$ region: $|t| \gtrsim 4$ GeV$^2$, one proton probes the other at transverse distances $b \lesssim 1/q$ ($q = \sqrt{|t|}$), i.e., at transverse distances of the order of 0.1 fm or less. Elastic scattering in this region is viewed in our model as the hard collision of a valence quark from one proton with a valence quark from the other proton as shown in Fig. 3. We refer to elastic scattering with $|t| \gtrsim 4$ GeV$^2$ as deep-elastic scattering.

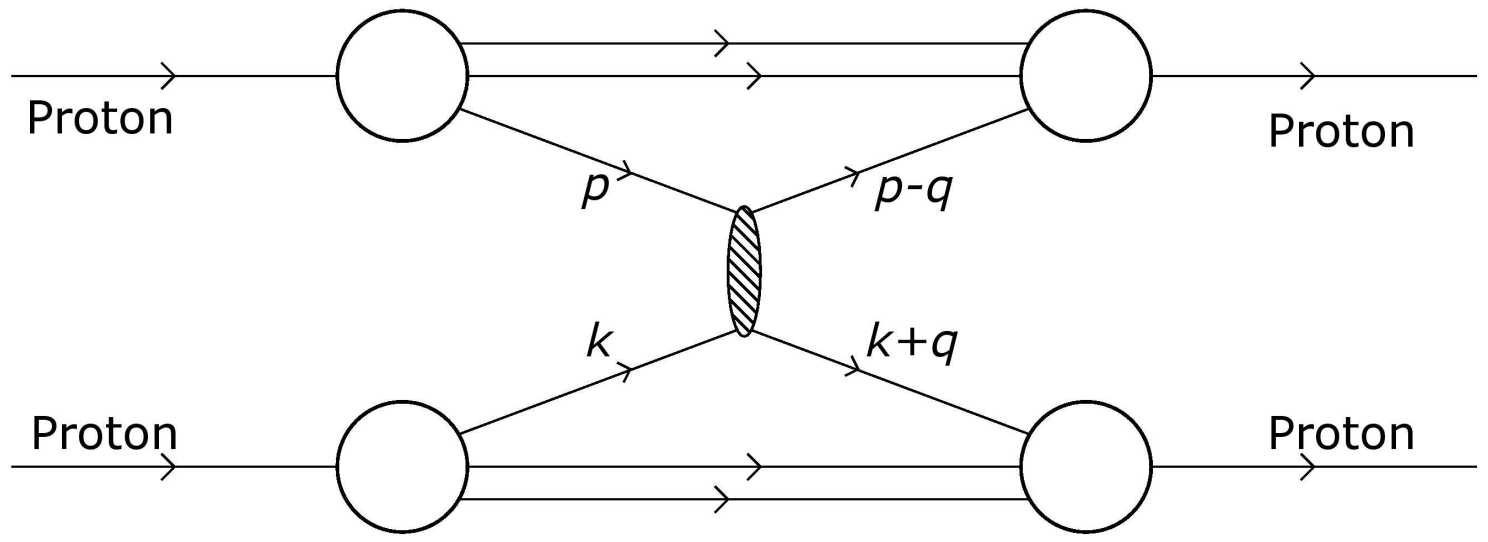

Figure 3: Hard collision of a valence quark of 4–momentum $p$ from one proton with a valence quark of 4–momentum $k$ from the other proton, where the collision carries off the whole momentum transfer $q$.

Let us next see how the above three processes are described in our calculations. Diffraction is described by using the impact parameter representation and a phenominological profile function:

$$T_D(s,t) = i\,p\,W \int_0^\infty b\,\mathrm{d}b\; J_0(b\,q)\Gamma_D(s,b); \tag{1}$$

$q$ is the momentum transfer ($q = \sqrt{|t|}$) and $\Gamma_D(s,b)$ is the diffraction profile function, which is related to the eikonal function $\chi_D(s,b)$: $\Gamma_D(s,b) = 1 - \exp(i\chi_D(s,b))$. We take $\Gamma_D(s,b)$ to be an even Fermi profile function:

$$\Gamma_D(s,b) = g(s)[\frac{1}{1+e^{(b-R)/a)}} + \frac{1}{1+e^{-(b+R)/a}} - 1]. \quad (2)$$

The parameters $R$ and $a$ are energy dependent: $R = R_0 + R_1(\ln s - \frac{i\pi}{2}\ )$, $a = a_0 + a_1(\ln s - \frac{i\pi}{2})$; $g(s)$ is a complex crossing even energy-dependent coupling strength.

The diffraction amplitude we obtain has the following asymptotic properties:

1. $\sigma_{\rm tot}(s) \sim (a_0 + a_1 \ln s)^2$ (Froissart-Martin bound)
2. $\rho(s) \simeq \frac{\pi a_1}{a_0 + a_1 \ln s}$ (derivative dispersion relation)
3. $T_D(s,t) \sim i\ s\ \ln^2 s\ f(|t|\ \ln^2 s)$ (Auberson-Kinoshita-Martin scaling)
4. $T_D^{\bar{\rm p}{\rm p}}(s,t) = T_D^{\rm pp}(s,t)$ (crossing even)

Incidentally, the profile function (2) has been used by Frankfurt et al. to estimate the absorptive effect of soft hadronic interactions (gap survival probability) in the central production of Higgs at LHC [3].

The $\omega$-exchange amplitude in our model is given by

$$T_\omega(s,t)\ =\ \pm\tilde{\gamma}\exp[i\ \hat{\chi}(s,0)]\ s\ \frac{F^2(t)}{m_\omega^2 - t}, \quad (3)$$

where the $\pm$ signs refer to $\bar{\rm p}$p and pp scattering. The factor of $s$ shows that $\omega$ behaves like an elementary spin-1 boson, while the two form factors indicate $\omega$ is probing two baryonic charge distributions – one for each proton. The factor $\exp[i\ \hat{\chi}(s,0)]$ comes from absorption due to the cloud-cloud interaction in $\omega$ exchange (Fig. 2).

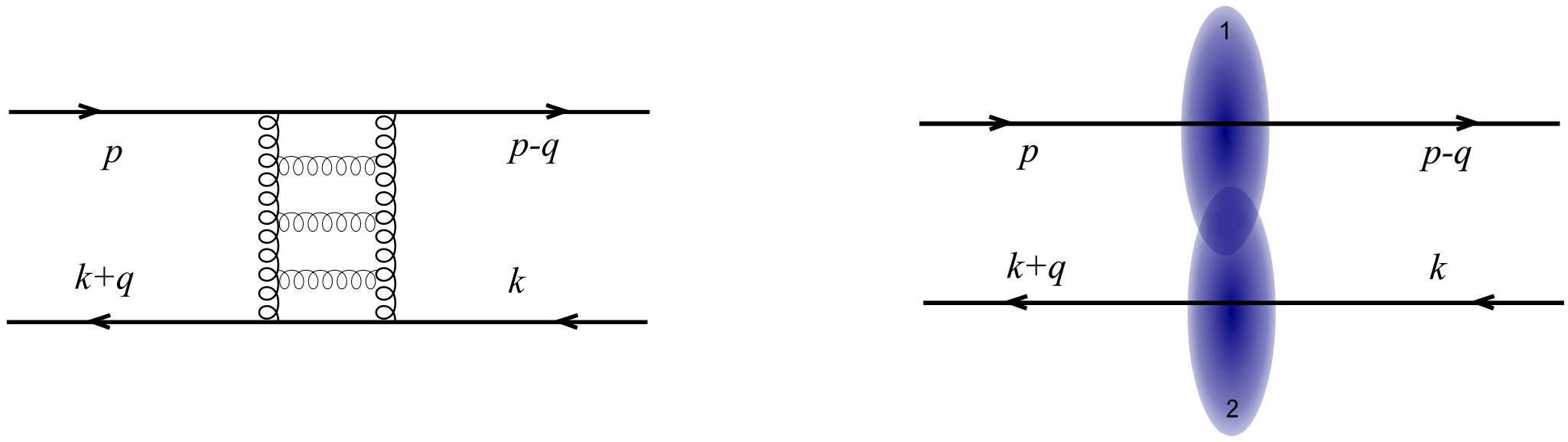

Figure 4: QCD processes for valence quark–quark scattering: (left) exchange of gluons in the form of ladders; (right) low–x gluon cloud of one quark interacting with that of the other.

For deep-elastic qq scattering, we have considered two QCD processes: Fig. 4. The first involves the exchange of reggeized gluon ladders (a BFKL ladder): figure on the left, plus next-to-leading-order corrections. We refer to this process as 'hard pomeron' exchange. Our hard pomeron amplitude is given by [4]

$$\hat{T}(\hat{s},t) = i\ \gamma_{hp}\ \hat{s}\ \left(\hat{s}\ e^{-i\frac{\pi}{2}}\right)^\omega \frac{1}{\left(1+\frac{q^2}{m_h^2}\right)}. \quad (4)$$

The energy dependence of this amplitude goes like $\hat{s}^{1+\omega}$, where $\hat{s}$ is the c.m. energy squared of the two colliding valence quarks. $\hat{s} = (p+k)^2$ and $\omega = 0.13 - 0.18$ [5]. We have used the value $\omega = 0.15$ in our calculations.

The other QCD process we have considered in Fig. 4 is where the low-x gluon cloud of one quark interacts with that of the other: figure on the right. The corresponding amplitude is given by [6]

$$\hat{T}(\hat{s},t) = i\,\gamma_{gg}\;\hat{s}\;\left(\hat{s}\,e^{-i\frac{\pi}{2}}\right)^{\lambda}\,\frac{1}{\left(1+\frac{q^2}{m^2}\right)^{2(\mu+1)}}. \qquad (5)$$

Its energy dependence goes like $\hat{s}^{1+\lambda}$, where $\lambda$ was found to be 0.29 by Golec-Biernat and Wüsthoff (GBW Model) [7]. We used this value in our calculations.

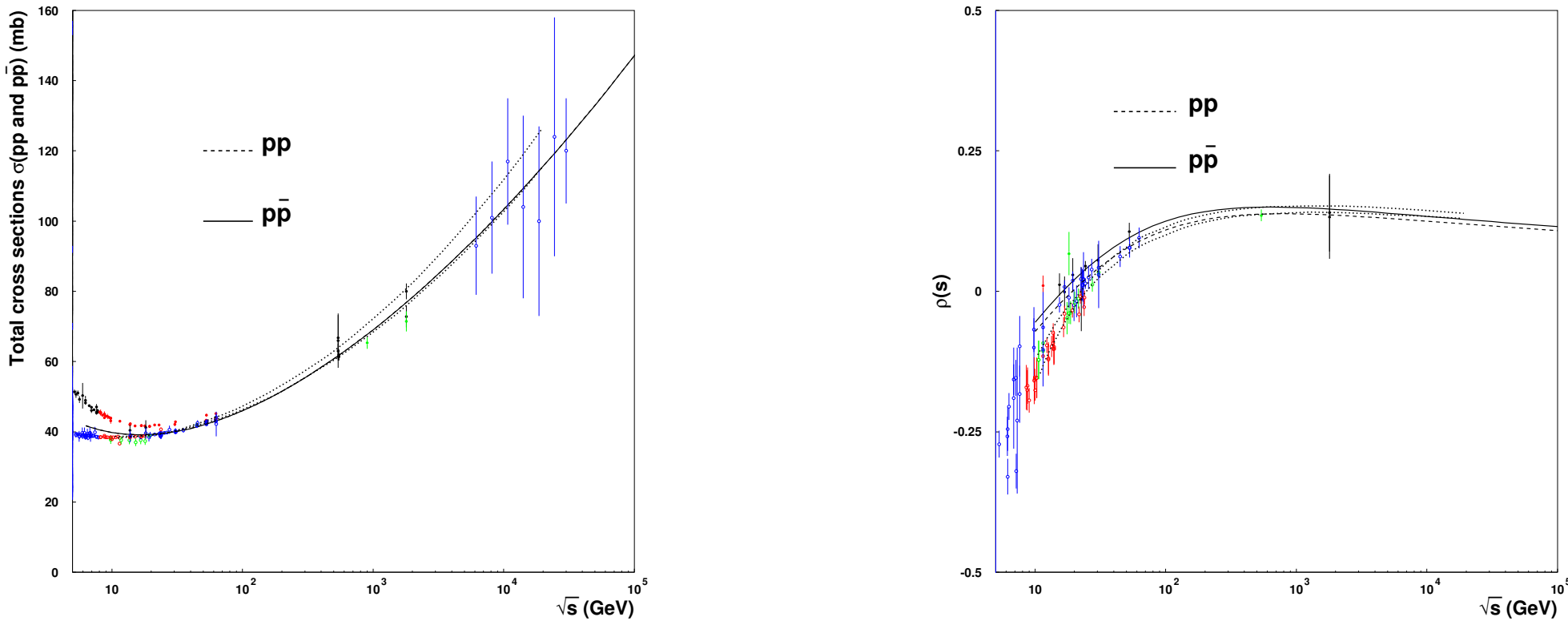

Figure 5: Left figure shows our calculated total cross section $\sigma_{tot}(s)$ as a function of $\sqrt{s}$ (solid curve). Right figure shows our calculated $\rho_{\bar{p}p}$ (solid curve) and $\rho_{pp}$ (dashed curve) as functions of $\sqrt{s}$.

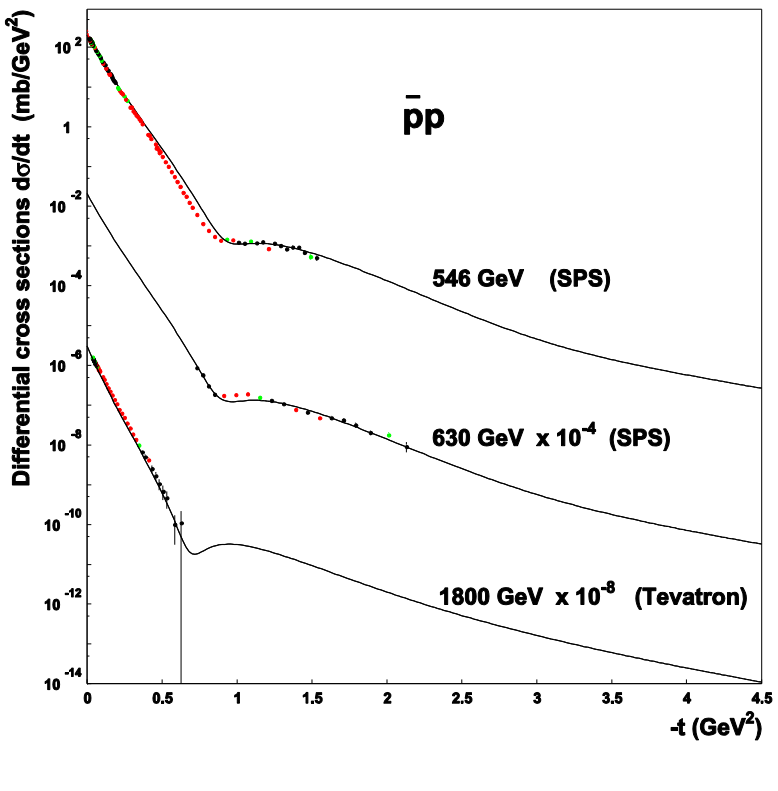

Figure 6: Solid curves show our calculated $d\sigma/dt$ at $\sqrt{s}$ = 546 GeV, 630 GeV and 1.8 TeV.

The parameters in our model are determined by requiring that the model should describe satisfactorily the asymptotic behaviour of $\sigma_{tot}(s)$ and $\rho(s)$ as well as the measured $\bar{p}p$ elastic $d\sigma/dt$ at $\sqrt{s}$= 546 GeV, 630 GeV, and 1.8 TeV. The results of this investigation together with the experimental data are presented in Figs. 5 and 6 [1]. We obtain quite satisfactory descriptions. The dotted curves in Fig. 5 represent the error bands given by Cudell et al. (COMPETE Collaboration) to their best fit [8].

Our predicted pp elastic $d\sigma/dt$ at c.m. energy 14 TeV due to the combined three processes–diffraction, $\omega$ exchange, and valence quark-quark scattering (from hard pomeron exchange) is shown in Fig. 7 [1]. Also shown are separate $d\sigma/dt$'s due to diffraction (dotted curve), $\omega$ exchange (dot-dashed curve), and valence quark-quark scattering (dashed curve). At $\sqrt{s} = 14$ TeV, our calculated values for $\sigma_{tot}$ and $\rho_{\rm pp}$ are 110 mb and 0.120, respectively.

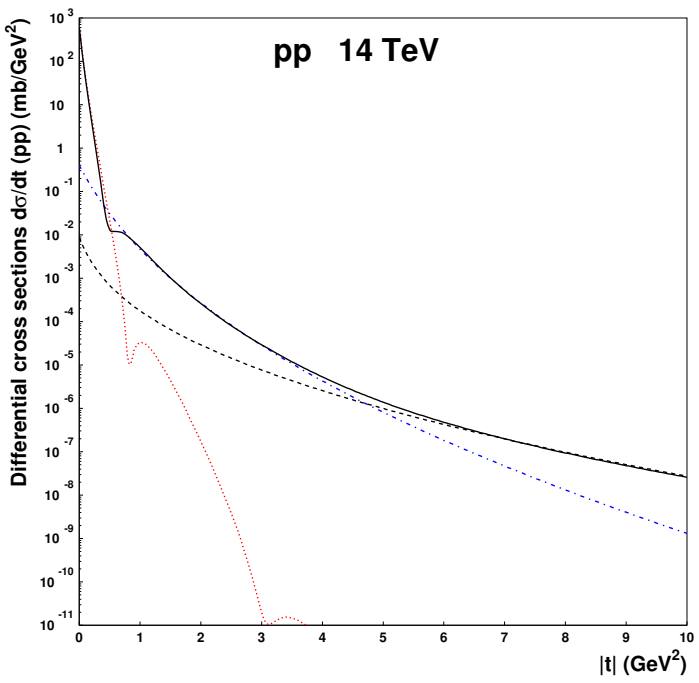

Figure 7: The solid curve shows our predicted $d\sigma/dt$ at LHC at $\sqrt{s} = 14$ TeV due to the combined three processes– diffraction, $\omega$ exchange, and valence quark–quark scattering (from hard pomeron exchange). Also shown are separate $d\sigma/dt$'s due to diffraction (dotted curve), $\omega$ exchange (dot–dashed curve), and valence quark–quark scattering (dashed curve).

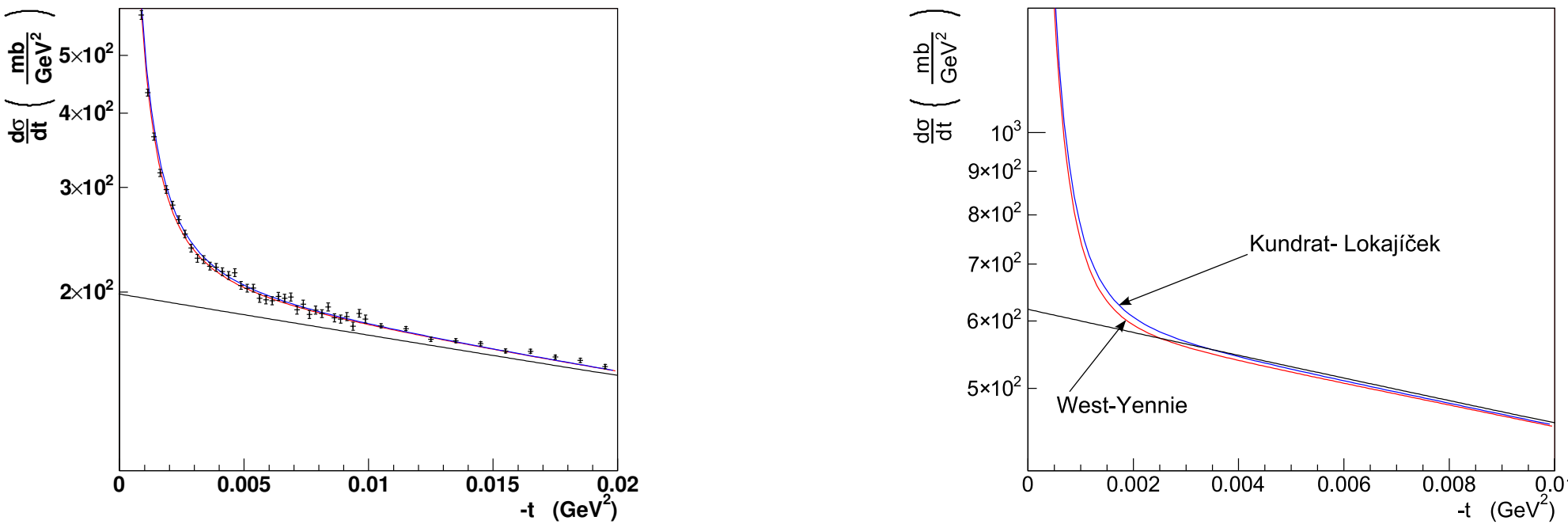

Figure 8: Left figure shows our calculated $d\sigma/dt$ for $\bar{\rm p}$p elastic scattering at $\sqrt{s} = 541$ GeV in the Coulomb–hadronic interference region together with the experimental data[11]. Right figure presents our predicted $d\sigma/dt$ for pp elastic scattering at $\sqrt{s} = 14$ TeV in the Coulomb–hadronic interference region.

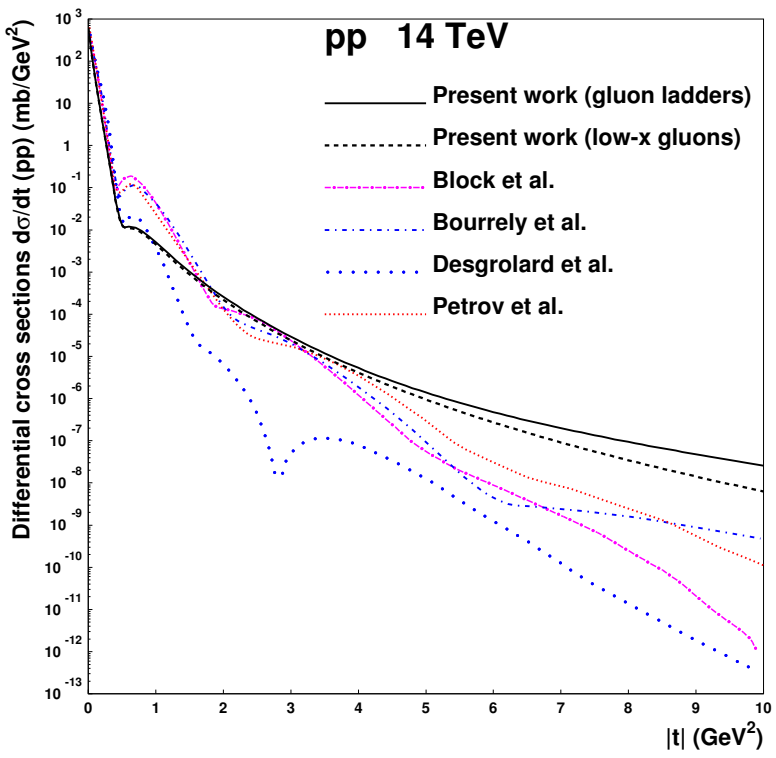

Figure 9: Our predicted $d\sigma/dt$ at LHC at c.m. energy 14 TeV with q–q scattering due to gluon ladders (solid curve) and low–x gluon cloud–cloud interaction (dashed curve). Also shown are predicted $d\sigma/dt$ of four other dynamical models [12, 13, 14, 15].

Fig. 8 (left) shows our calculated $d\sigma/dt$ for $\bar{\rm p}$p elastic scattering at $\sqrt{s} = 541$ GeV [1] in the Coulomb-hadronic interference region using the Kundrát–Lokajíček formulation [9] and the West-Yennie formulation [10]. Experimental data are from Augier et al. [11]. Right figure in Fig. 8 presents our predicted $d\sigma/dt$ for pp elastic scattering at $\sqrt{s} = 14$ TeV [1] in the Coulomb-hadronic interference region using the Kundrát–Lokajíček formulation (upper curve) and the West-Yennie formulation (lower curve). The straight lines in Fig. 8 represent $d\sigma/dt$ due to our hadronic amplitude by itself. The ATLAS ALFA experiment plans to measure pp $d\sigma/dt$ in this region at LHC.

Finally, in Fig. 9, we show our predicted pp elastic $d\sigma/dt$ at $\sqrt{s} = 14$ TeV with quark-quark scattering due to hard pomeron exchange (solid line) and due to low-x gluon cloud-cloud interaction (dashed line) and compare with the predicted $d\sigma/dt$ of four other dynamical models [12, 13, 14, 15]. A distinctive feature of our predicted $d\sigma/dt$ is that the differential cross section falls off smoothly beyond the bump at $|t| \simeq 1$ GeV$^2$. In contrast, the other models predict visible oscillations in $d\sigma/dt$. Furthermore, these models lead to much smaller differential cross sections than ours in the large $|t|$ region: $|t| \gtrsim 5$ GeV$^2$.

If the planned measurement of elastic $d\sigma/dt$ by the TOTEM Collaboration in the momentum transfer range $|t| \simeq 0 - 10$ GeV$^2$ shows quantitative agreement with our predicted $d\sigma/dt$, then the underlying picture of the proton–a 'Condensate Enclosed Chiral Bag' (as shown in Fig. 1) will be supported. The consequent discovery of the structure of the proton at LHC at the beginning of the 21st century will be analogous to the discovery of the structure of the atom from high energy $\alpha$- particle scattering by gold atoms at the beginning of the 20th century.

## Acknowledgements

The authors wish to thank the members of the TOTEM Collaboration for discussions and comments. MMI and RJL would like to thank CERN for Conference support. RJL would also like to thank Northwest Catholic High School for travel support.

## References

[1] M. M. Islam, R. J. Luddy and A. V. Prokudin, *Int. J. Mod. Phys. A***21**, 1 (2006).

[2] A. Hosaka and H. Toki: *Quarks, Baryons and Chiral Symmetry*, World Scientific Publishing Co. (2001).

[3] L. Frankfurt, C. E. Hyde-Wright, M. Strikman and C. Weiss, *Phys. Rev. D* **75**, 054009 (2007).

[4] M. M. Islam, R. J. Luddy and A. V. Prokudin, *Phys. Lett. B* **605**, 115 (2005).

[5] S. Brodsky, V. S. Fadin, V. T. Kim, L. N. Lipatov and G. B. Pivovarov, *JETP Lett.* **70**, 155 (1999).

[6] M. M. Islam, Jan Kašpar and R. J. Luddy, *Mod. Phys. Lett. A* **24**, 485 (2009).

[7] K. Golec-Biernat and M. Wüsthoff, *Phys. Rev. D* **59**, 014017 (1998); *ibid.* **60**, 114023 (1999).

[8] J. R. Cudell *et al.*, COMPETE Collab.,*Phys. Rev. Lett.* **89**, 201801 (2002).

[9] V. Kundrát and M. Lokajíček, *Z. Phys. C* **63**, 619 (1994).

[10] G. B. West and D. R. Yennie, *Phys. Rev.* **172**, 1413 (1968).

[11] C. Augier *et al.*, *Phys. Lett. B* **316**, 448 (1993).

[12] M. M. Block, E. M. Gregores, F. Halzen and G. Pancheri, *Phys. Rev. D* **60**, 054024 (1999).

[13] C. Bourrely, J. Soffer and T. T. Wu, *Eur. Phys. J. C* **28**, 97 (2003).

[14] P. Desgrolard, M. Giffon, E. Martynov and E. Predazzi, *Eur. Phys. J. C* **16**, 499 (2000).

[15] V. Petrov, E. Predazzi and A. V. Prokudin, *Eur. Phys. J. C* **28**, 525 (2003).

# TOTEM Experiment: Elastic and Total Cross Sections

*J. Kašpar on behalf of the TOTEM Collaboration*

CERN, Geneva and Institute of Physics of the AS CR, Prague

The physics programme and the detector apparatus of the TOTEM experiment is presented. Then, the key optics and their goals are summarised. The method to measure the total pp cross section is introduced. One of its essential parts, the extrapolation to $t = 0$, is discussed in detail and extrapolation strategies for $\beta^* = 1535$ and 90 m optics are presented. In particular, an adequate parameterisation and a treatment of Coulomb scattering is proposed. In the last section, a Roman Pot alignment procedure is described.

## 1 Introduction

The TOTEM experiment [1, 2] is dedicated to forward hadronic phenomena. The tree pillars of its physics programme are: an accurate measurement of the total pp cross section, a measurement of elastic scattering in a wide kinematic range and studies of diffractive processes. This paper is focused on the first two, for the latter one we refer to [1, 2, 3].

The programme is touching one of the least explored and understood areas of hadronic physics. This fact can be well demonstrated by Fig. 1. The left plot shows several model predictions for elastic differential cross sections which differ by several orders of magnitude at large $|t|$ (four-momentum transfer squared). The right figure compiles data on the total pp cross section. Due to large uncertainties of cosmic ray experiments and conflicting Tevatron data [4, 5], this data set can hardly favour any of the proposed theoretical descriptions over another. TOTEM shall shed some light onto those open questions by providing precise measurements – see for instance the anticipated error bar for total cross section in Fig. 1.

The challenging programme brings special requirements for the detector apparatus. In particular, *large pseudorapidity coverage* – to detect most fragments from inelastic collisions and excellent *acceptance for outgoing diffractive and elastic protons*. To accomplish this task, TOTEM comprises three subdetectors: the inelastic telescopes T1 and T2 and a system of Roman Pots (RP) for proton detection. For details on instrumentation see [1, 2, 6]. This design results in a unique apparatus with an excellent pseudorapidity coverage, see Fig. 2. The acceptance of the RPs can be further varied by using different optics, as will be discussed in the next section.

## 2 Measurement of the Total Cross Section

The forward protons will be detected by the system of Roman Pots. Their position and acceptance depends on the settings of the accelerator (beam optics) – for details see chapter 6 in [2]. TOTEM plans to exploit the following 3 types of optics.

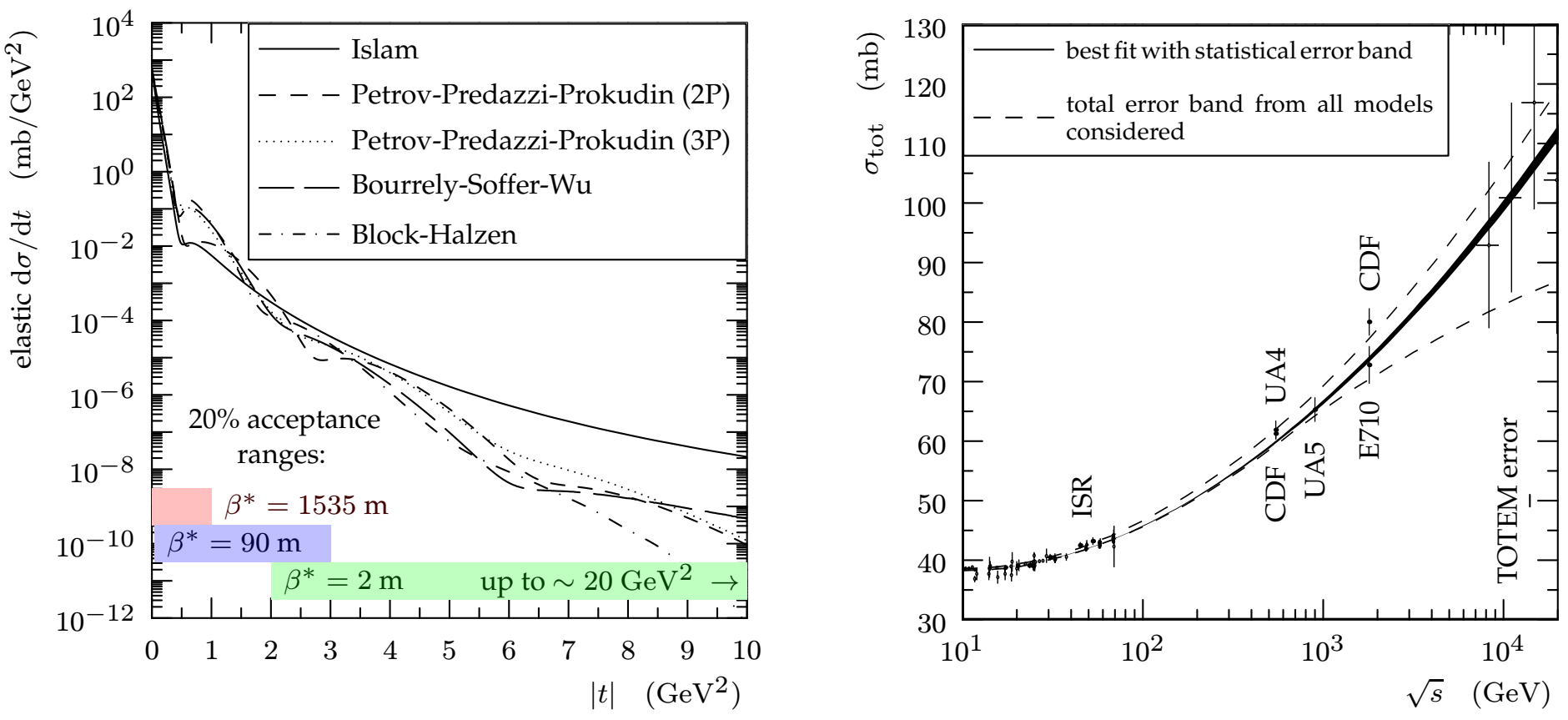

Figure 1: Left: predictions of the elastic differential cross-section at a centre-of-mass energy of 14 TeV by several phenomenological models. Acceptance bands for the main optics (see Sec. 2) are shown at the bottom. Right: a compilation of available data for the total pp cross section with a fit by the COMPETE collaboration [7]. The anticipated ultimate precision (1%) is shown in the bottom right corner.

1. $\beta^* = 1535$ m. This is the ultimate optics for low $|t|$ elastic scattering and precise (1% error) total cross section measurement.

2. $\beta^* = 90$ m is a universal optics allowing for measurement of elastic scattering (medium $|t|$ range), total cross section (5% uncertainty) and also for diffraction studies.

3. $\beta^* = 0.5 \div 3$ m (standard optics) are suited for high $|t|$ elastic scattering and various diffractive measurements.

See Figs. 1 and 2 for a comparison of elastic scattering acceptances for the above optics.

TOTEM intends to measure the total cross section by the *luminosity independent method*. It is based on the Optical Theorem:

$$\sigma_{\rm tot}(s) \propto \Im T^H(s, t=0) \ , \tag{1}$$

relating the total cross section $\sigma_{\rm tot}$ to the hadronic[1] component of the elastic scattering amplitude $T^H(s,t)$. When it is complemented by common definitions for luminosity $\mathcal{L}$ and rates $N$

$$\varrho = \left.\frac{\Re T^H}{\Im T^H}\right|_{t=0}, \qquad \frac{{\rm d}\sigma}{{\rm d}t} \propto |T^H|^2, \qquad {\rm d}N = \mathcal{L}{\rm d}\sigma, \qquad N_{\rm tot} = N_{\rm el} + N_{\rm inel}, \tag{2}$$

one can obtain relations for the total cross section and luminosity:

$$\sigma_{\rm tot} = \frac{1}{1+\varrho^2}\frac{{\rm d}N/{\rm d}t|_{t=0}}{N_{\rm el}+N_{\rm inel}}, \qquad \mathcal{L} = (1+\varrho^2)\,\frac{(N_{\rm el}+N_{\rm inel})^2}{{\rm d}N/{\rm d}t|_{t=0}} \ . \tag{3}$$

[1] There is obviously a second component due to the Coulomb scattering. Their interference is briefly discussed in Sec. 2.

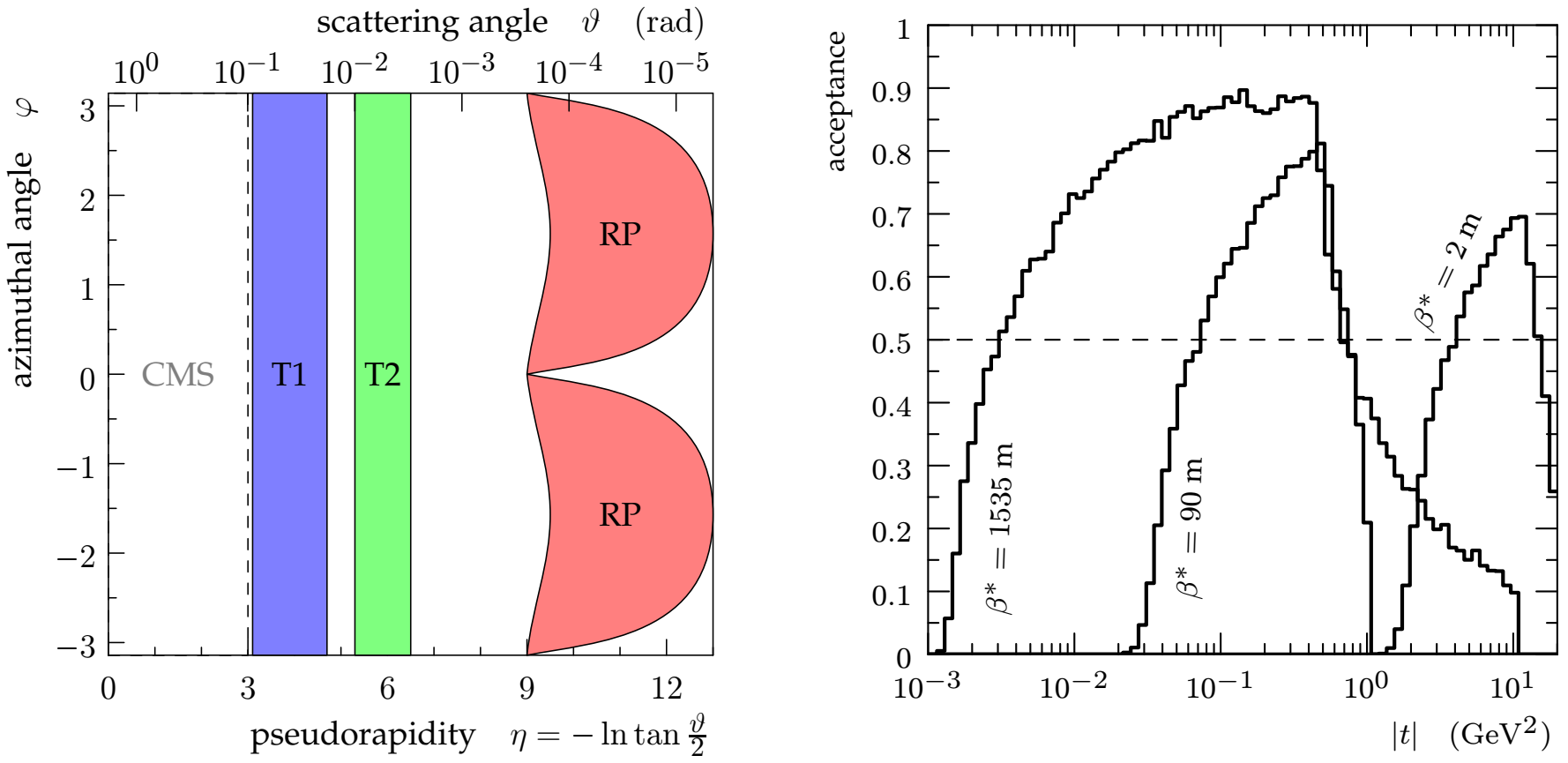

Figure 2: The acceptance of the TOTEM apparatus. Left: The coverage of the three subsystems of TOTEM. The shown acceptance of RPs refers to the $\beta^* = 1535$ m optics. For the other optics, the acceptance is shifted to lower pseudorapidity values, which narrows the gap between RPs and T2. Right: RP acceptances for elastic events for different optics at $\sqrt{s} = 14$ TeV.

Here, $\mathrm{d}N/\mathrm{d}t|_{t=0}$ stands for elastic rate in the Optical Point (i.e. $t = 0$), which is to be obtained by an extrapolation procedure discussed in Sec. 2. $N_{\mathrm{el}}$ is the total elastic rate, it will be measured by RPs and will be adjusted, again, by the extrapolation procedure. $N_{\mathrm{inel}}$ represents the total inelastic rate measured by the telescopes T1 and T2 (more details in Sec. 2.2 in [8]).

The $\varrho$ quantity can only be determined by an analysis of the Coulomb-hadronic interference (see below in Sec. 2) and there is only a small $|t|$ window, where these effects are significant enough. Moreover, for the energy of 14 TeV this region is found around $t = 1 \cdot 10^{-3}$ GeV$^2$ which is on the very edge of TOTEM's acceptance. Therefore TOTEM might not be able to determine the $\rho$ value at the nominal LHC energy, unless allowed to insert the RPs closer than the standard 10 beam-$\sigma$ distance (which would push the acceptance to lower $|t|$). For reduced energies, the prospects are much brighter as the interference region shifts towards higher $|t|$ values. Even if TOTEM was unable to resolve $\rho$, its value could be taken from external predictions (e.g. [7]). Note that expected $\rho$ values are small $\approx 0.14$ and since $\rho$ enters the formulae Eq. (3) only via $1 + \varrho^2$, the influence of any uncertainty is small [2, 8].

## Extrapolation to $t = 0$

The value $\mathrm{d}\sigma/\mathrm{d}t|_0$ is, indeed, not accessible experimentally and thus an extrapolation from a higher $|t|$ region must be applied. A necessary condition for any successful extrapolation is a suitable parameterisation. Looking at Fig. 3, showing several model predictions in a low $|t|$ region, one can observe *almost* exponential decrease of the elastic cross section up to $|t| \leq 0.25$ GeV$^2$. This is further supported by *almost* constant differential slope $B(s,t)$ in the quoted range[2]. The plot (c) hints that the phase of hadronic amplitudes can be described by a polynomial of a low degree. These arguments suggest that the following parameterisation is

[2] The model of Islam et al. is an exception which would be easily recognised (e.g. in large $|t|$ elastic scattering) and a different strategy would be applied.

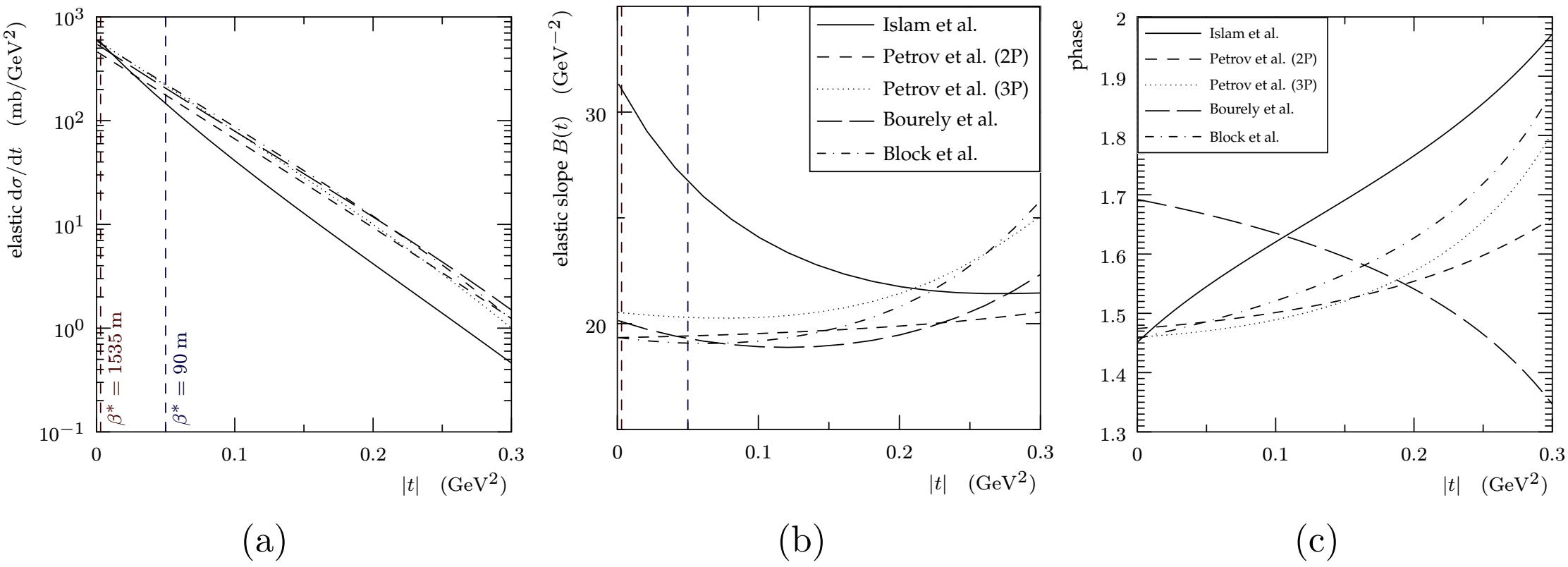

Figure 3: Model predictions for $\sqrt{s} = 14$ TeV in a low $|t|$ region. (a): predictions for the elastic differential cross section. (b): predictions for the elastic slope $B(s,t) = \frac{\mathrm{d}}{\mathrm{d}t} \log \frac{\mathrm{d}\sigma}{\mathrm{d}t}$. (c): predictions for the hadronic phase.

adequate:

$$T^H(s,t) = e^{M(t)} e^{iP(t)}, \quad \frac{\mathrm{d}\sigma}{\mathrm{d}t} = |T^{C+H}(s,t)|^2, \quad \text{with } M, P \text{ polynomials for a fixed } s\ . \tag{4}$$

$T^{C+H}$ stands for the scattering amplitude of the combined Coulomb and hadronic forces and will be discussed below. The questions to be answered are: what is the optimal fit range and what is the optimal degree of the polynomials. It is obvious that if too many free parameters are introduced, they cannot be resolved with confidence. This is mainly a problem for the phase polynomial $P(t)$ since any phase information can only be resolved from a narrow Coulomb interference window, as discussed above. The optimal values shall give good results for most of the models considered; in this way the procedure can be regarded as model-independent.

So far, only the hadronic contribution $T^H$ to the elastic scattering has been discussed. It is clear that Coulomb interaction will play a role and therefore must be taken into account. At the time being, there are two approaches to calculate scattering amplitudes $T^{C+H}$ for the combined interaction: the *traditional* (à la West-Yennie [9]) and the *eikonal* (see e.g. Kundrát-Lokajíček [10]). The traditional approach is based on rather constraining assumptions on the form of the hadronic amplitude, and furthermore it has recently been shown internally inconsistent [11].

As mentioned in Sec. 2, TOTEM plans to measure the total cross section with two optics: $\beta^* = 1535$ m and 90 m. The lowest measurable $|t|$ values differ quite considerably (see Fig. 2 right and vertical marks in Fig. 3) and therefore the extrapolation strategies differ as well.

For the 1535 m optics, the Coulomb interference effects play a role and thus an interference formula must be exploited (the eikonal one has been used in this study). The following configuration has been found optimal: quadratic $B(t)$ and constant phase with upper bound $|t| = 4 \cdot 10^{-2}$ GeV$^2$. Preliminary results are shown in Fig. 4 (a). One can see that most models lie within a band $\pm 0.2\%$ (except for the model of Islam et al. − see footnote 2).

As for what concerns the 90 m optics, the Coulomb effects are negligible and therefore the phase parameterisation becomes irrelevant[3]. On the other hand, the horizontal $t$ component $t_x$ can be resolved with a limited resolution only – see Fig. 4 (b). Since $t = t_x + t_y$, the considerable uncertainties propagate to the full $t$ distribution. A number of solutions might be suggested.

[3]The $T^{C+H}$ coincides with $T^H$ and the phase factor $\exp(iP(t))$ cancels out when differential cross section is calculated according to the Eq. (4).

1. Use the $t$-distribution (i.e. $\mathrm{d}\sigma/\mathrm{d}t$) despite large uncertainties.
2. Using azimuthal symmetry, one can "transform" a $t_y$-distribution in a $t$-distribution:

$$\frac{\mathrm{d}\sigma}{\mathrm{d}t_y} = \frac{\mathrm{d}\sigma}{\mathrm{d}t_x} \quad \Rightarrow \quad \frac{\mathrm{d}\sigma}{\mathrm{d}t}(t) \propto \int_t^0 \mathrm{d}u\, \frac{\mathrm{d}\sigma}{\mathrm{d}t_y}(u)\, \frac{\mathrm{d}\sigma}{\mathrm{d}t_y}(t-u)\ .$$

   However, since low $|t_y|$ information is missing (out of acceptance), an extrapolation step would be needed just for this transformation.
3. "Transform" a $t$-parameterisation in a $t_y$-parameterisation and fit it directly through $t_y$ data:

$$t_y = t\,\sin^2\varphi,\ \text{with } \varphi \text{ uniformly distributed} \quad \Rightarrow \quad \frac{\mathrm{d}\sigma}{\mathrm{d}t_y}(t_y) = \frac{2}{\pi}\int_0^{\pi/2} \frac{\mathrm{d}\varphi}{\sin^2\varphi}\ \frac{\mathrm{d}\sigma}{\mathrm{d}t}\left(\frac{t_y}{\sin^2\varphi}\right)$$

   Considering a parameterisation of type Eq. (4), one can derive an approximate formula:

$$\frac{\mathrm{d}\sigma}{\mathrm{d}t} = e^{a + bt + ct^2 + \dots} \quad \Rightarrow \quad \frac{\mathrm{d}\sigma}{\mathrm{d}t_y}(t_y) \approx \frac{1}{\sqrt{\pi}}\ \frac{e^{a + bt_y + ct_y^2 + \dots}}{\sqrt{|b\,t_y|}}$$

   which can be justified provided the non-linear terms in the exponent ($ct^2, \dots$) do not give an essential contribution – which is the case, see Fig. 3.

Eventually, the third approach has been chosen and a cubic polynomial with an upper bound of $|t| = 0.25\ \mathrm{GeV}^2$ has been found optimal. Preliminary results are plotted in Fig. 4 (c). Most models fall in a band between $-1\%$ and $-3\%$ (Islam's model being again an exception – see footnote 2). The overall offset of $-2\%$ is a consequence of the beam divergence and can be corrected in the data analysis.

# 3 Alignment of Roman Pots

An accurate alignment is of major importance for the TOTEM experiment in order to deliver precise measurements. Among the subdetectors of TOTEM, the alignment of the RPs presents the biggest challenge since they are movable. The importance of alignment is most pronounced at the $\beta^* = 1535$ m optics, where the beam divergence (the dominant smearing effect) is rather low and hence the impact of any misalignment has a large relevance. To give a feeling, a 100 $\mu$m displacement of a vertical RP would lead to angular shift of about 0.4 $\mu$rad (based on an effective length $L_y \approx 270$ m, typical for this optics). This is to be compared to the spread of the beam divergence 0.3 $\mu$rad.

We recall that the system of RPs is composed of two arms, each arm includes two stations, each station comprises two units of 2 vertical and 1 horizontal RP and, finally, each RP contains a package of 10 edge-less silicon detectors. The entire structure is intended to be aligned by the following three steps.

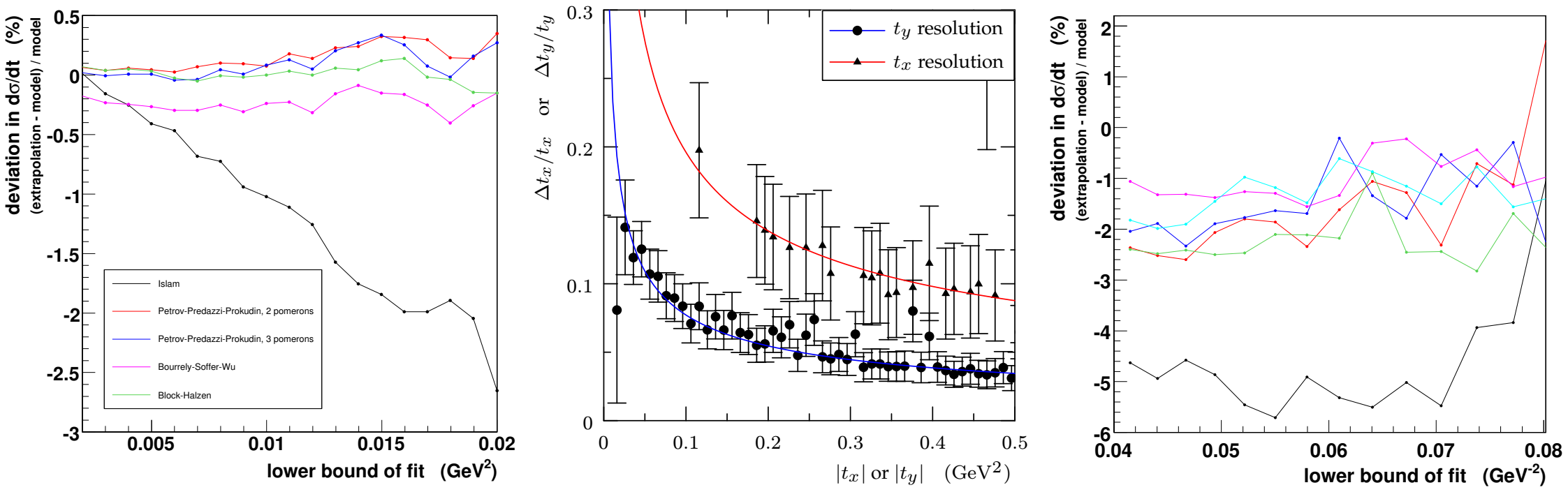

(a) (b) (c)

Figure 4: (a): the extrapolation deviation as a function of fit's lower bound for the $\beta^* = 1535$ m optics. (b): comparison of $t_x$ and $t_y$ resolutions for the $\beta^* = 90$ m optics. (c): the extrapolation deviation as a function of the lower bound of the fit for the 90 m optics.

1. *Internal alignment.* That is an alignment of detectors within one RP with respect to each other. To accomplish this task, a track-based algorithm has been developed. This algorithm is inspired by Millepede [12]; it performs a consistent analysis of track residuals to extract as much alignment information as possible. Any straight tracks traversing the detector package can be used as an input for this method. Thus, test beam, cosmic test data, beam halo tracks etc. can all be used. The procedure has been examined in beam and cosmic tests and its results have been successfully compared to those of a laboratory optical measurement – see Fig. 5 (b).

   As the strip silicon detectors measure one coordinate only, one can not establish more than one component of its (mis)-shift. To resolve the second transverse component of the shift, we are currently investigating an alternative method – the efficiency drop around the sensitive edge may help to pinpoint the position of the edge, hence a second shift component. This is possible as the custom-developed detectors have very small insensitive margin and thus the efficiency drop is well spatially localised.

2. *Station alignment* comprises two aspects – relative alignment of RP detectors within a station and beam position determination. Regarding the first one, the same track-based algorithm as for the internal alignment can be used. The only difference is that one needs tracks passing through several RPs at a time. This is possible thanks to a key design feature – the *overlap* (see Fig. 5 (a)) between vertical and horizontal RPs.

   The alignment with respect to the beam requires usage of physics processes, in particular their hit and angular distributions. Elastic scattering seems the most convenient in this regard due to its full azimuthal symmetry. Then, a horizontal hit profile in vertical RPs and a vertical profile in horizontal detectors shall unveil the beam's position in the horizontal and vertical direction respectively; see an example in Fig. 5 (c). As TOTEM will operate at various optics, the rate of elastic protons will not always be high enough and therefore diffractive processes will be used in addition.

   The Beam Position Monitors (one mounted on each RP unit) can monitor relative beam fluctuations with a precision of several microns. Absolute beam position measurements are, however, exposed to a large uncertainty on the offset and thus need to be cross-calibrated with results of other methods (e.g. the profile method from the previous paragraph).

Another important device is the control of RP motors, which can resolve the position of RP once it has been moved in a working place. After a careful absolute calibration a resolution of about 10 $\mu$m is expected.

3. *Global alignment*, i.e. cross-alignment between stations in both arms of the experiment, will rely on elastic tracks. Elastic protons have exactly (up to smearing effects) opposite directions and thus provide a perfect tool for alignment of the opposite arms. Again, for the third time, the tracks-based algorithm can be exploited.

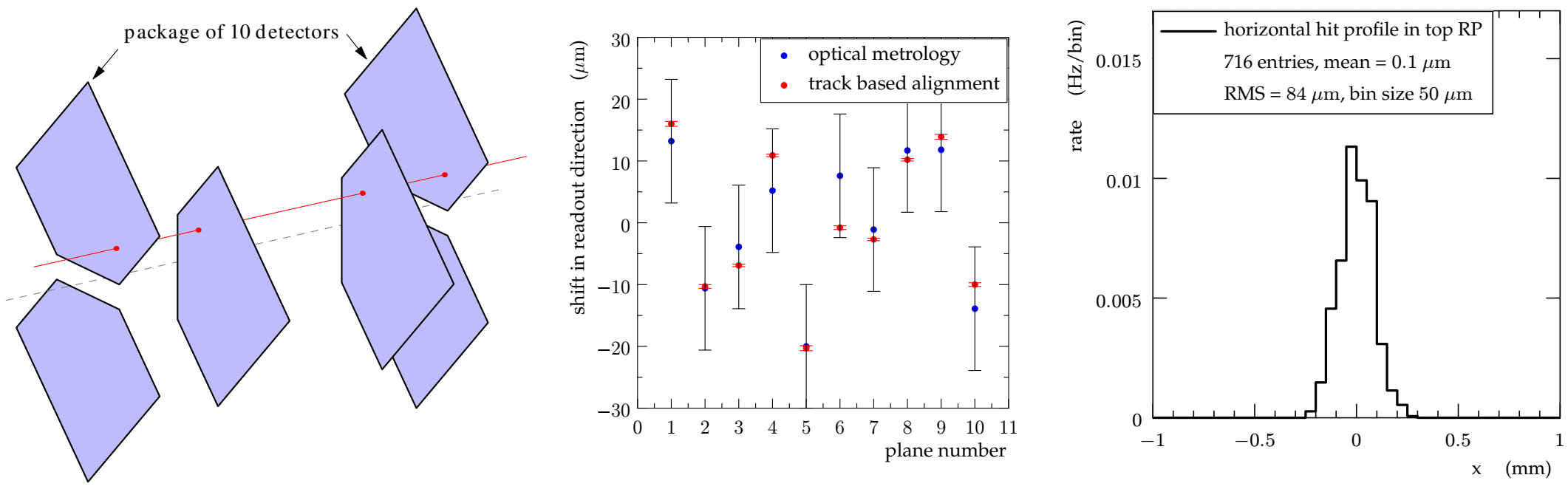

(a) (b) (c)

Figure 5: (a): An illustration of the detector overlap. (b): A comparison of alignment results by the track based algorithm and the optical measurement. (c): An example of a profile to determine the position of the beam – a horizontal profile of elastic hits in a top RP, for $\beta^* = 2$ m optics (the lowest elastic rate).

# References

[1] TOTEM: Technical Design Report, CERN-LHCC-2004-002 (2004); addendum CERN-LHCC-2004-020.

[2] G. Anelli *et al.* [TOTEM Collaboration], JINST **3**, S08007 (2008).

[3] S. Giani [TOTEM Collaboration], *Diffraction at TOTEM*, 13$^{\text{th}}$ International Conference on Elastic and Diffractive Scattering, CERN (2009).

[4] F. Abe *et al.* [CDF Collaboration], Phys. Rev. **D50** (1994) 5550.

[5] N. A. Amos *et al.* [E710 Collaboration], Phys. Rev. Lett. **68** (1992) 2433.

[6] G. Ruggiero [TOTEM Collaboration], *TOTEM*, 13$^{\text{th}}$ International Conference on Elastic and Diffractive Scattering, CERN (2009).

[7] J. R. Cudell *et al.* [COMPETE Collaboration], Phys. Rev. Lett. **89** (2002) 201801.

[8] M. Deile [TOTEM Collaboration], *Total cross-section measurement and diffractive physics with TOTEM*, 12$^{\text{th}}$ International Conference on Elastic and Diffractive Scattering, Hamburg (2007)

[9] G. B. West and D. R. Yennie, Phys. Rev. **172** (1968) 1413.

[10] V. Kundrát and M. Lokajíček, Z. Phys. **C63** (1994) 619.

[11] V. Kundrát, M. Lokajíček and I. Vrkoč, Phys. Lett. **B656** (2007) 182.

[12] V. Blobel, Millepede II documentation at http://www.desy.de/~blobel/Mptwo.pdf .

# Optical Theorem and Elastic Nucleon Scattering

*Miloš V. Lokajíček, Vojtěch Kundrát*

Institute of Physics, AVČR, v.v.i., 18221 Prague

In the theoretical analysis of high-energy elastic nucleon scattering one starts commonly from the description based on the validity of the optical theorem, which allows to derive the value of the total cross section directly from the experimentally measured t-dependence of the elastic differential cross section at the corresponding energy. It may be shown, however, that this theorem has been derived on the basis of one assumption that might be regarded perhaps as acceptable in the case of long-range (e.g. Coulomb) forces but must be denoted as quite unacceptable in the case of finite-range hadron forces. Consequently, the conclusions leading to the increase of the total cross section with energy at higher collision energies must be newly analyzed and reevaluated. It concerns also the value of the beam luminosity derived from elastic data. The necessity of new analysis concerns the derivation of the hadronic t-dependence at very low transverse momenta as the separation of Coulomb scattering may be also strongly influenced. It will be shown in conclusion that all mentioned problems might be solved on the basis of the ontological model proposed quite recently.

## 1 Introduction

It is commonly argued that it is possible to derive also the value of the total cross section directly from the measured t-dependence of the elastic differential cross section when the validity of the optical theorem is taken into account. And the derived values indicate that the total cross section should rise with energy. From these values the beam luminosity is also being derived.

However, we will show that in the derivation of the optical theorem some important assumption has been involved that may be perhaps acceptable for infinite-range Coulomb force, but can be hardly brought to harmony with the finite-range forces between nucleons. It means that neither the total cross section nor the beam luminosity may be derived from the measured differential elastic scattering cross section in the framework of the standard phenomenological theory without adding some assumption or additional property.

The validity of the optical theorem does not correspond to reality and all conclusions based on it should be reevaluated. However, it will be shown in the last additional section that some properties of ontological model might help in solving newly the given problems.

## 2 Optical Theorem and its Derivation

Let us start with the standard derivation approach. We shall follow the approach described in the book of Barone and Predazzi [1]. They have started from Fraunhofer diffraction and from Babinet's principle. The profile function of a hole has been denoted by $\Gamma(b)$ and that of an obstacle by $S(b)$. When the initial state is represented by a plane wave $\psi_{in} = e^{ikz}$, the final state may be expressed in the case of an obstacle as the superposition of the set of individual scattered states (see also [2]):

$$U(x,y.z) = -\frac{ik}{2\pi}U_0\frac{e^{ikr}}{r}\int d^2\mathbf{b}\,S(\mathbf{b})\,e^{-i\mathbf{q}.\mathbf{b}}$$

and similarly for the hole if $S(b)$ is substituted by $\Gamma(b)$. According to the Huygens-Fresnel principle ($S(\mathbf{b})+\Gamma(\mathbf{b})=1$) the same shapes of hole and obstacle combine again to the original plane wave. The whole approach being based on the assumption of small diffraction angles ($\sin\theta \cong \theta$).

The amplitude may be divided in principle into two parts: scattered and unscattered, where the scattered part is represented by the function of transferred momentum $\mathbf{q}$:

$$U_{out}(x,y,z)) \;=\; U_{unsc} \;+\; U_{scatt} = \; U_0\left(\psi_{unsc} + f(\mathbf{q})\frac{e^{ikr}}{r}\;\right)$$

$$\psi_{unsc} = \alpha e^{ikz},\;\; |\alpha| < 1; \qquad \mathbf{q} = \mathbf{k}' - \mathbf{k}, \quad |\mathbf{k}'| = |\mathbf{k}| = k$$

The unscattered part is represented by a plane wave; of course, with some norm less than one. The modulus squared of $U(x,y,z)$ represents the final intensity and $U_0$ - incoming intensity.

The differential elastic cross section is given by the square of $f(\mathbf{q})$, and the total elastic cross section may be expressed in integral form:

$$\frac{d\sigma}{d\Omega} = |f(\mathbf{q})|^2, \qquad \sigma_{el} \cong \frac{1}{k^2}\int |f(\mathbf{q})|^2 d^2\mathbf{q} = \int d^2\mathbf{b}|\Gamma(\mathbf{b}|^2$$

where $\Gamma(\mathbf{b})$ represents the corresponding profile.

And if it is assumed that the unscattered part may be identified with $f(0)$ and if the interaction is normalised to unity one obtains the following expressions for absorption and total cross sections:

$$\sigma_{abs} = \int d^2\mathbf{b}(1 - |S(\mathbf{b})|^2)$$

$$\sigma_{tot} \;=\; \sigma_{el} + \sigma_{abs} = 2\int d^2\mathbf{b}\,Re\Gamma(\mathbf{b})$$

It means that the total cross section should be represented by the imaginary part of scattering amplitude at point $\mathbf{q} = 0$:

$$\sigma_{tot} \;=\; \frac{4\pi}{k}\,Im\,f(\mathbf{q} = 0)$$

However, some additional assumptions have been used that cannot be applied to in the case of finite nucleon force, as it will be shown in the next section.

# 3 Nucleon Force and Optical Theorem

By identifying the unscattered state with one vector of scattered states the function $f(\mathbf{q})$ has been substituted in principle by capital $F(\mathbf{q})$, where one singular point has been added to the previous function:

$$U(x,y,z) \;=\; U_0 F(\mathbf{q})\frac{e^{ikr}}{r}; \qquad F(0) = f(0) + |\psi_{unsc}|\,.$$

However, in all approaches two assumptions concerning the function $F(\mathbf{q})$ have been added; the given function has been taken as continuous, which has meant that the unscattered part has equalled zero, and as decreasing from $\mathbf{q} = 0$ without any actual reason.

Both these assumptions might hold for infinite range Coulomb force, but they can hardly correspond to finite nucleon force where also the whole fully unscattered part lies in the range of measured beam. In any case the optical theorem cannot be applied to in the case of nucleon scattering.

And we must ask: What is actual shape of nucleon scattering amplitude in the case of very small scattering angles? And what are other physical consequences? E.g., how it is with luminosity estimation?

# 4 Elastic Collisions at Very Small Angles

The situation may be demonstrated in Fig. 1 where two pairs of lines are pictured representing the amplitudes of elastic scattering (real parts being neglected). Each pair is represented once by complete (Coulomb and hadronic) amplitude and once by hadronic part. It has been assumed that approximately at $t = -0.02\,\mathrm{GeV}^2$ the contributions of both (Coulomb and hadronic) components are the same.

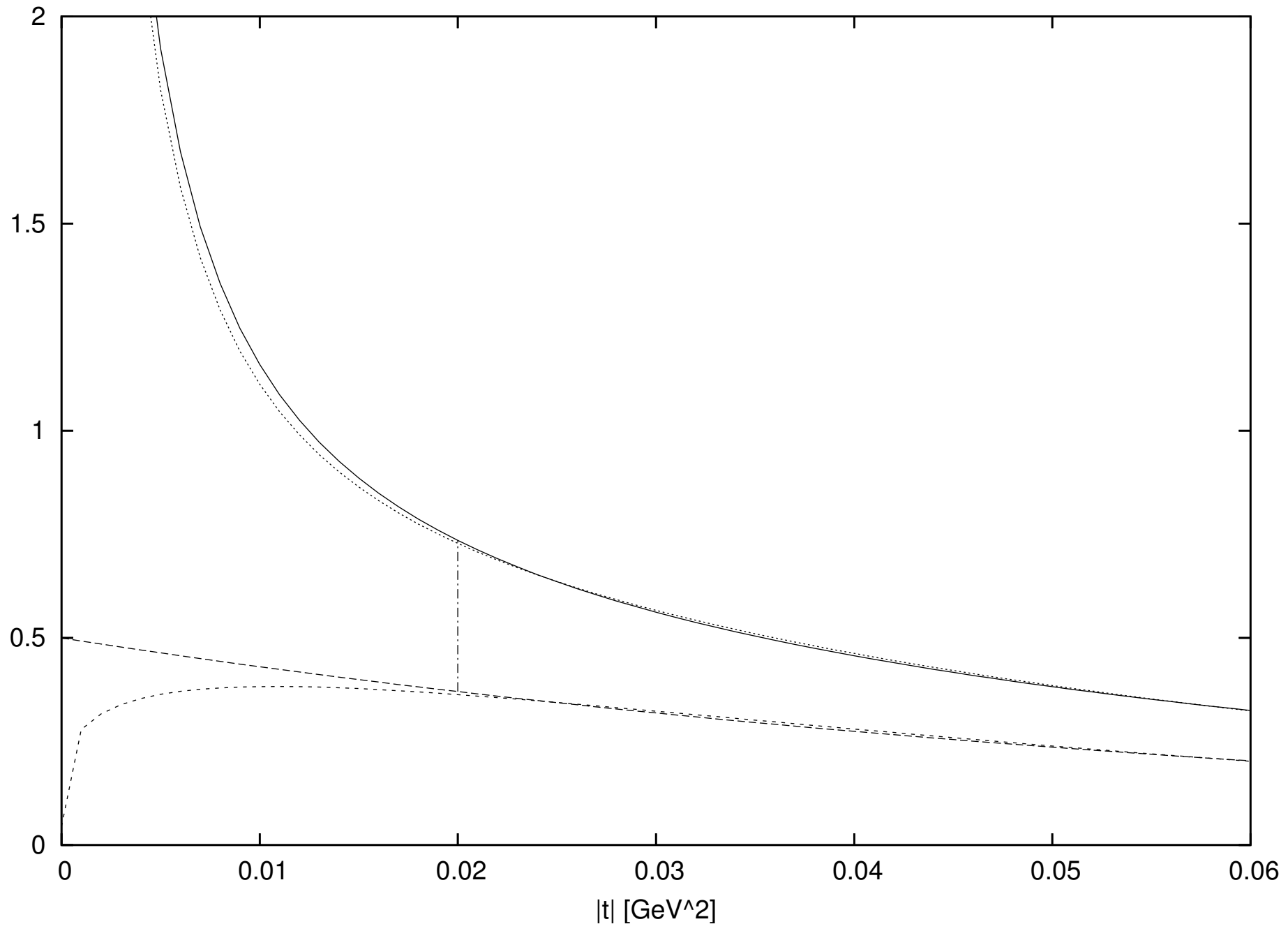

Figure 1: *Different hadronic amplitudes corresponding approximately to elastic data obtained at the energy of 53 GeV.*

The first pair of curves represents the standard approach at ISR energy; the curves corresponding approximately to experimental data obtained at energy of 53 GeV (see, e.g., [3]). The other pair shows then that practically the same complete elastic amplitude may be obtained even if the hadronic part behaves quite differently at very small values of $|t|$; the behaviour corresponding better to the situation at finite-range forces when the profile function $\Gamma(b)$ is equal to zero for all values of $b$ greater than $b_{max}$ ($\Gamma(b > b_{max}) = 0$) and then rises rather strongly with decreasing impact parameter $b$. The validity of the optical theorem cannot be required, of course, in such a case.

It means that the values of hadronic scattering may be strongly overvalued in the region around $t = 0$ in the standard approach. However, it is evident on the other side that in such a case also the value of luminosity $L$ derived from elastic scattering values:

$$\frac{d\sigma_{el}(t)}{dt} = L|F^{C+N}(t)|^2,$$

might be very different, its contemporary values being significantly undervalued.

# 5 Conclusion

It follows from the preceding analysis that the contemporary phenomenological theories of elastic collisions do not allow to establish the total cross sections without some additional experiments or without more detailed analysis based on a model in which elastic and inelastic processes would be mutually correlated on realistic physical grounds. There is also the question how it is with the interpretation of nucleon elastic data for very small values of $\mathbf{q}$. Are they really decreasing from the value at $\mathbf{q} = 0$ or may they exhibit different behaviour being fully hidden under the effect of Coulomb force?

The problem is related very closely to the determination of corresponding total cross section and to establishing luminosity value from elastic scattering data. All these values may be hardly derived from elastic data if one must start from phenomenological models of elastic scattering only.

Important problem must be seen in the fact that the $t$-dependence of elastic differential cross section is the only available experimental set of data from which the mere modulus of complex amplitude function may be established while the phase remains practically undetermined. Quite different impact-parameter characteristics may be then derived according to its choice; see, e.g., [4, 5]. It means that the behaviours ontologically very different may be derived on the basis of standard phenomenological models.

The invalidity of optical theorem brings, therefore, rather inconclusive outlook as to the consequences following from the analysis of experimental elastic data. And one should ask if it is possible to find a procedure how to make use of elastic data in a more effective way. Such an approach seems to follow from applying an ontological model to processes running in particle collisions, which will be shortly described in the last additional section.

# 6 Additional Section: Ontological Model

In this additional section we should like to present a more promising view. We have shown quite recently that the elastic data may be well interpreted on the basis of ontological approach that allows to provide a correlation between elastic and inelastic processes. A nucleon consisting of many constituents, quarks and partons, has been assumed to be a matter object existing in a series of states differing by external dimensions.

The model has been explained and demonstrated with the help of elastic data obtained at energy 53 GeV; see [6]. It has been shown that this data in the interval $t \in (-4., 0.)\,\mathrm{GeV}^2$ may be truly interpreted as the superposition of three collision types between two states that exhibit maximal dimensions. All available elastic data at the given energy (in the interval $t \in (-14., 0.)\,\mathrm{GeV}^2$ may be then well interpreted as the superposition of six collision kinds between three nucleon states of maximal dimensions.

In addition to the already mentioned basic assumption (i.e., nucleon consisting of different internal states) it has been assumed, too, that the probability of each elastic event at any value of impact parameter may be expressed as the product of two probabilities:

$$P_{el}(b) \;=\; P_{tot}(b)\,.\,P_{rat}(b);$$

the first factor representing total collision probability and the other one representing the ratio of elastic processes to all collision processes.

And it may be regarded as quite natural from realistic point of view to add the third assumption that these partial probabilities may be represented by two oppositely monotone functions of impact parameter $b$. And it has been possible to establish both the monotone functions for all involved collision kinds by performing the alternative fit of earlier ISR results.

Here are the corresponding preliminary results. The maximum dimensions of three largest states (involved in the given collision process) should be 1.64, 1.42, 0.88 fm; these states should exist in individual nucleons with following frequencies: $\approx$ 57, 31, 11 %.

Having known the $b$-dependent probabilities of total and elastic collisions for all collision kinds it has been possible to determine also the approximate values of corresponding cross sections:

$$\sigma_{tot} \cong 36\,\mathrm{mb}, \qquad \sigma_{el} = 7.3\,\mathrm{mb}.$$

While the elastic cross section has had practically the same value as standardly introduced the total cross section has differed rather significantly from the value obtained on the basis of the optical theorem validity. This earlier value seems to have been overvalued approximately by 15%. It means, of course, that also the luminosity value taken for ISR collider has been probably undervalued by 15%.

The probability distribution of elastic processes in impact parameter plane have exhibited, of course, clear peripheral behaviour, which has corresponded to similar results obtained on the basis of eikonal approach [3].

# References

[1] V. Barone and E. Predazzi, High-Energy Particle Diffraction; Springer-Verlag (2002).

[2] M. Lokajíček and V. Kundrát, arXiv:0906.3961 [hep-ph] (2009).

[3] V. Kundrát, M. Lokajíček, Zeit. Phys. **C63** 619 (1994).

[4] V. Kundrát, M. Lokajíček and D. Krupa, Phys. Lett. **B544** 132 (2002).

[5] V. Kundrát, M. Lokajíček, Jr. and M. Lokajíček, Czech. J. Phys. **B31** 1334 (1981).

[6] M. Lokajíček and V. Kundrát, arXiv:0909.3199 [hep-ph] (2009).

# Elastic and Diffractive Scattering after AdS/CFT

*Richard Brower*[1], *Marko Djurić*[2], *Chung-I Tan*[2*]

[1]Physics Department, Boston University, Boston, MA 02215, USA.
[2]Physics Department, Brown University, Providence, RI 02912, USA.

At high energies, elastic hadronic cross sections are believed to be dominated by vacuum exchange. In leading order of the $1/N_c$ expansion, this exchange process has been identified as the BFKL *Pomeron* or its strong AdS dual, the closed string graviton [1]. However, the difference of particle anti-particle cross sections is given by a so-called *Odderon* carrying C = -1 quantum numbers identified in weak coupling with odd numbers of exchanged gluons. Here we show that the dual description associates this with the Neveu-Schwartz ($B_{\mu\nu}$) sector of closed string theory. We also discuss the extension of the strong coupling treatment to central diffractive Higgs production at LHC.

## 1 Introduction

The subject of near-forward high energy scattering for hadrons has a long history. The traditional description of high-energy small-angle scattering in QCD has two components — a soft Pomeron Regge pole associated with exchanging tensor glueballs, and a hard BFKL Pomeron at weak coupling. On the basis of gauge/string duality, a coherent treatment of the Pomeron was provided [1]. These results agree with expectations for the BFKL Pomeron at negative $t$, and with the expected glueball spectrum at positive $t$, but provide a framework in which they are unified [2]. Therefore, a firm theoretical foundation for Pomeron in QCD has been established. It is now possible to identify a *dual Pomeron* as a well-defined feature of the curved-space string theory [1].

We focus here on the recent developments based on Maldacena's weak/strong duality, (AdS/CFT), relating Yang-Mills theories to string theories in (deformed) Anti-de-Sitter space [1, 3, 4, 5]. The application of this duality to diffractive scattering and the Pomeron physics represent an important area where a connection with the string-theory-based techniques can be made. Furthermore, it is now possible to extend this treatment to central diffractive production of Higgs at LHC.

In the large 't Hooft coupling limit, Pomeron can be considered as a *Reggeized Massive Graviton*, propagating in a 5-dimensional curved space, the so-called $AdS_5$, where both the IR (soft) Pomeron and the UV (BFKL) Pomeron are dealt in a unified single step. The connection with the stringy aspects in a five-dimensional description is indeed very direct. In gauge theories with string-theoretical dual descriptions, the Pomeron emerges unambiguously. Indeed, Pomeron is directly related to the graviton and its higher spin partners on the leading (five-dimensional) Regge trajectory. In AdS/CFT, confinement is associated with a deformed $AdS_5$ geometry having an effective horizon, e.g., that for a black hole. The solution to this is

* speaker

unknown and represents the major theoretical challenge in model-building. Each model leads to certain unique signature. LHC data can provide guide and direction in this endeavor.

# 2 Pomeron and QCD Parameter-Space

It is useful to take a step back in examining high energy scattering in QCD. From a theoretical stand point it is useful to consider a 3-parameter space varying the number of colors ($N_c$), the 't Hooft coupling ($\lambda = g^2 N_c$), and the virtuality of an external probe $1/Q$, e.g., that of a virtual photon.

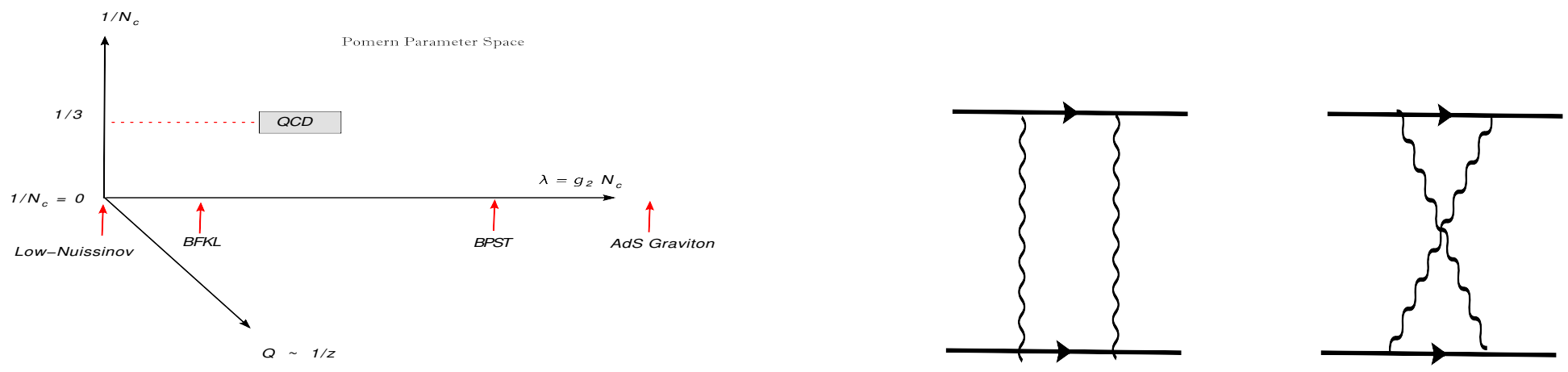

Figure 1: (a) The Pomeron in QCD viewed as a function of colors ($N_c$), the 't Hooft coupling ($\lambda = g^2 N_c$), and the resolution (virtuality: $Q \sim 1/z$) of the probe. (b) Two-gluon exchange as Low-Nussinov Pomeron.

Thus diffractive scattering (or Pomeron exchange) in QCD can now be considered in two steps. First one may consider the leading contribution in the $1/N_c$ expansion holding fixed the 't Hooft coupling $\lambda = g^2 N_c$. For example, instead of $N_c = 3$, the leading term for the Regge limit of 2-to-2 scattering in the this limit is the exchange of a network of gluons with the topology of a cylinder in the 't Hooft topological expansion or, in string language, the exchange of a closed string. This gives rise to what we call the "bare Pomeron" exchange. Taking into account high order terms in the $1/N_c$ leads to two effects: (1) The cylinder diagrams includes closed quark loops, leading to $q\overline{q}$ pairs or multi-hadron production via the optical theorem dominated by low mass pions, kaon etc. (2) The multiple exchange of the bare Pomeron which includes the eikonal corrections (or survival probability) and triple-Pomeron and higher order corrections in a Reggeon calculus. We will focus primarily on the "bare Pomeron" sector and will discuss only briefly higher order effect due to eikonalization.

In weak coupling coupling summations where $\lambda \ll 1$ and $N_c$ large, the leading singularity (prior to full unitarization) is at $j_0 = 1 + (\ln 2/\pi^2)\,\lambda$ where $\lambda = g^2 N_c$ is the 't Hooft coupling. Indeed, in the limit $\lambda \to 0$, this so-called BFKL Pomeron reduces to the Low-Nussinov Pomeron, i.e., two-gluon exchange, as depicted in Fig. 1b. When this description is adequate, hadronic cross sections are expected to rise as a small power $s^{j_0 - 1}$ until unitarity forces compliance with the Froissart bound. However there is an additional probe of the "Pomeron" as a function of virtuality $Q^2$ in off-shell photon scattering.

It is generally acknowledged that diffractive scattering is intrinsically a non-perturbative phenomenon. In the limit where the 't Hooft coupling is large, weak coupling calculations become unreliable. In Ref. [1], it has been shown that the leading singularity in strong coupling in the conformal limit approached $j = 2$. In the language of the AdS/CFT, Pomeron is the graviton pole in the 5-dim $AdS$ space where the $AdS$ radius $r$ serves the 5th dimension. (In what follows, this will be referred to as the strong coupling BPST Pomeron, or simply the BPST Pomeron.)

One of the more interesting developments of the BPST Pomeron is the recognition that the virtuality of an external probe, $1/Q$, can be identified with the $AdS$ radius, $z = 1/r$. Conformal invariance, which allows a simultaneous scale transformations in the transverse size and the probe scale, can now be encoded as the isometry of the transverse Euclidean $AdS_3$.

# 3 Forward Scattering, Gauge/String Duality, and Confinement

For conformally invariant gauge theories, the metric of the dual string theory is a product, $AdS_5 \times W$, $ds^2 = \left(\frac{r^2}{R^2}\right)\eta_{\mu\nu}dx^\mu dx^\nu + \left(\frac{R^2}{r^2}\right)dr^2 + ds^2_W$ , where $0 < r < \infty$. For the dual to $\mathcal{N} = 4$ super-symmetric Yang-Mills theory the AdS radius $R$ is $R^2 \equiv \sqrt{\lambda}\alpha' = (g^2_{\rm YM}N)^{1/2}\alpha'$ , and $W$ is a 5-sphere of this same radius. We will ignore fluctuations over $W$ and also assume that $\lambda \gg 1$, so that the space-time curvature is small on the string scale, and $g^2_{YM} \ll 1$ so that we can use string perturbation theory. (See [1, 3, 4] for more references.)

The fact that 5-dim description enters in high energy collision can be understood as follows. In addition to the usual LC momenta, $p_\pm = p^0 \pm p^z$ (2d), and transverse impact variables, $\vec{b}$ (2d), there is one more "dimension": a "resolution" scale specified by a probe, e.g., $1/Q^2$ of virtual photon in DIS, (see Fig. 2a.) Because of conformal symmetry, these 5 coordinates transform into each others, leaving the system invariant. In the strong coupling limit, conformal symmetry is realized as the $SL(2, C)$ isometries of Euclidean $AdS_3$ subspace of $AdS_5$, where $r$ can be identified with $Q^2$.

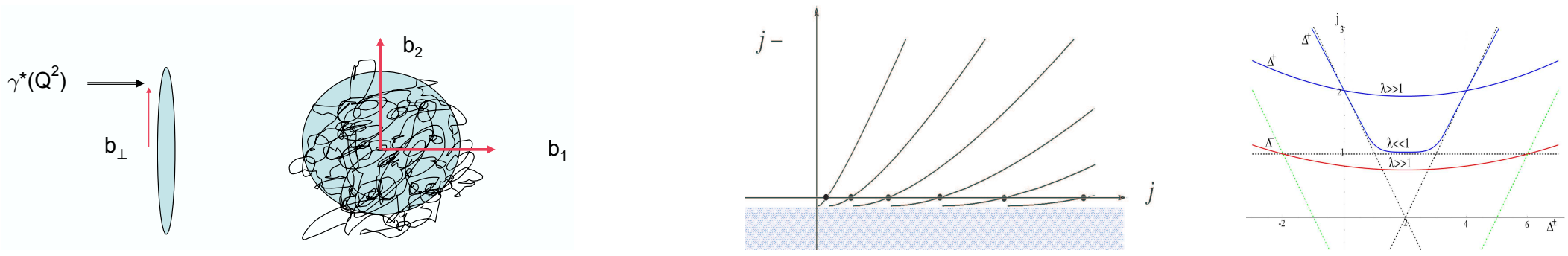

Figure 2: (a) Intuitive picture for $AdS^5$ kinematics. (b) Schematic representation of $J$-plane singularity structure. (c) Schematic form of $\Delta$-$j$ relation for $\lambda \ll 1$ and $\lambda \gg 1$.

One important step in formulating the dual Pomeron involves the demonstration [6] that in exclusive hadron scattering, the dual string theory amplitudes at wide angle, due to the red-shifted local-momenta, $s \to \tilde{s} = (R/r)^2 s$ and $t \to \tilde{t} = (R/r)^2 t$, give the power laws that are expected in a gauge theory. It was also noted that at large $s$ and small $t$ that the classic Regge form of the scattering amplitude should be present in certain kinematic regimes [6]. Equally important is the fact that, with confinement, transverse fluctuations of the metric tensor $G_{MN}$ in $AdS$ acquire a mass and can be identified with a tensor glueball [7, 8]. It was suggested in [8] that, at finite $\lambda$, this will lead to a Pomeron with an intercept below 2. That is, Pomeron can be considered as a *Reggeized Massive Graviton*.

For a conformal theory in the large $N_c$ limit, a dual Pomeron can always be identified with the leading eigenvalue of a Lorentz boost generator $M_{+-}$ of the conformal group [3]. The problem reduces to finding the spectrum of a single $J$-plane Schrödinger operator. One finds that, in the strong coupling limit, conformal symmetry requires that the leading $C = +1$ Regge singularity is a fixed $J$-plane cut, $j_0^{(+)} = 2 - 2/\sqrt{\lambda}$ For ultraviolet-conformal theories with confinement deformation, the spectrum exhibits a set of Regge trajectories at positive $t$, and a leading $J$-plane cut for negative $t$, the cross-over point being model-dependent. (See Fig. 2b.) For theories with logarithmically-running couplings, one finds a discrete spectrum of poles at all $t$, with a set of slowly-varying and closely-spaced poles at negative $t$.

# 4 Conformal Pomeron, Odderon and Analyticity

At high-energy, analyticity and crossing lead to $C = \pm 1$ vacuum exchanges, the *Pomeron* and the *Odderon*. The qualitative picture for Pomeron exchange in weak coupling [9] has been understood for a long time, in leading order expansion in $g^2_{YM}$ and all order sum in $g^2_{YM} log(s/s_0)$. In the conformal limit, both the weak-coupling BFKL Pomeron and Odderons correspond to $J$-plane branch points, e.g., the BFKL Pomeoron is a cut at $j_0^{(+)}$, above $j = 1$. Two leading Odderons have been identified. (See [4, 10] for more references.) Both are branch cuts in the $J$-plane. One has an intercept slightly below 1 [11], and the second has an intercept precisely at 1 [12]. These are summarized in Table 1.

In the strong coupling limit, as we have already mentioned above, conformal symmetry dictates that the leading $C = +1$ Regge singularity is a fixed $J$-plane cut at $j_0^{(+)} = 2 - 2/\sqrt{\lambda} + O(1/\lambda)$. As $\lambda$ increases, the "conformal Pomeron" moves to $j = 2$ from below, approaching the *AdS* graviton. We have recently shown [4] that the strong coupling *conformal odderons* are again fixed cuts in the $J$-plane, with intercepts specified by the AdS mass squared, $m^2_{AdS}$, for Kalb-Ramond fields [13], $j_0^{(-)} = 1 - m^2_{AdS}/2\sqrt{\lambda} + O(1/\lambda)$ . Interestingly, two leading *dual odderons* can be identified, parallel the weak-coupling situation. One solution has $m^2_{AdS,(1)} = 16$. There is also a second solution where $m^2_{AdS,(2)} = 0$. We outline below how these features emerge in *Gauge/String duality.*

| | Weak Coupling | Strong Coupling |
|---|---|---|
| $C = +1$: Pomeron | $j_0^{(+)} = 1 + (\ln 2)\ \lambda/\pi^2 + O(\lambda^2)$ | $j_0^{(+)} = 2 - 2/\sqrt{\lambda} + O(1/\lambda)$ |
| $C = -1$: Odderon | $j_{0,(1)}^{(-)} \simeq 1 - 0.24717\ \lambda/\pi + O(\lambda^2)$<br>$j_{0,(2)}^{(-)} = 1 + O(\lambda^3)$ | $j_{0,(1)}^{(-)} = 1 - 8/\sqrt{\lambda} + O(1/\lambda)$<br>$j_{0,(2)}^{(-)} = 1 + O(1/\lambda)$ |

Table 1: Pomeron and Odderon intercepts at weak and strong coupling.

## 4.1 Flat-Space Expectation for $C = \pm 1$ Sectors

String scattering in 10-d flat-space at high energy leads to a crossing-even and crossing-odd amplitudes, $\mathcal{T}_{10}^{(\pm)}(s,t) \to f^{(\pm)}(\alpha' t)(\alpha' s)^{\alpha_\pm(t)}$ , where $\alpha_+(t) = 2 + \alpha' t/2$ and $\alpha_-(t) = 1 + \alpha' t/2$ respectively. That is, at $t = 0$, a massless state with integral spin is being exchanged, e.g., for $C = +1$, one is exchanging a massless spin-2 particle, the ubiquitous graviton. Of course, the coefficient functions, $f^{(\pm)}(\alpha' t)$, are process-dependent.

Massless modes of a closed string theory can be identified with transverse fluctuations coming from a left-moving and a right-moving level-one oscillators, e.g., states created by applying $a^\dagger_{1,I}\tilde{a}^\dagger_{1,J}$ to the vacuum, i.e., $a^\dagger_{1,I}\tilde{a}^\dagger_{1,J}|0;k^+,k_\perp\rangle$, with $k^2 = 0$. Since a 10-dim closed string theory in the low-energy limit becomes 10-dim gravity; these modes can be identified with fluctuations of the metric $G_{MN}$, the anti-symmetric Kalb-Ramond background $B_{MN}$ [13], and the dilaton, $\phi$, respectively. It is important to note that we will soon focus on $AdS^5$, i.e., one is effectively working at $D = 5$. With $D = 5$, the independent components for $G_{MN}$ and $B_{MN}$ are 5 and 3 respectively, precisely that necessary for having (massive) states with spin 2 and 1 [8]. For oriented strings, it can be shown that the symmetric tensor contributes to $C = +1$ and the anti-symmetric tensor contributes to $C = -1$.

## 4.2 Diffusion in AdS and DGLAP Connection

Let us next introduce diffusion in AdS. We will restrict ourselves to the conformal limit. Regge behavior is intrinsically non-local in the transverse space. For flat-space scattering in 4-dimension, the transverse space is the 2-dimensional impact parameter space, $\vec{b}$. In the Regge limit of $s$ large and $t < 0$, the momentum transfer is transverse. Going to the $\vec{b}$-space, $t \to \nabla^2_b$ , and the flat-space Regge propagator, for both $C = \pm 1$ sectors, is nothing but a diffusion kernel, $\langle\, \vec{b} \mid (\alpha' s)^{\alpha_\pm(0)+\alpha' t \nabla^2_b/2} \mid \vec{b}'\, \rangle$, with $\alpha_+(0) = 2$ and $\alpha_-(0) = 1$ respectively. In moving to a ten-dimensional momentum transfer $\tilde{t}$, we must keep a term coming from the momentum transfer in the six transverse directions. This extra term leads to diffusion in extra-directions, i.e., for $C = +1$,

$$\alpha'\tilde{t} \to \alpha'\Delta_P \equiv \frac{\alpha' R^2}{r^2}\nabla^2_b + \alpha'\Delta_{\perp P} \; .$$

The transverse Laplacian is proportional to $R^{-2}$, so that the added term is indeed of order $\alpha'/R^2 = 1/\sqrt{\lambda}$. To obtain the $C = +1$ Regge exponents we will have to diagonalize the differential operator $\Delta_P$. Using a Mellin transform, $\int_0^\infty d\tilde{s}\; \tilde{s}^{-j-1}$, the Regge propagator can be expressed as

$$\tilde{s}^{2+\alpha'\tilde{t}/2} = \int \frac{dj}{2\pi i}\; \tilde{s}^j\; G^{(+)}(j) = \int \frac{dj}{2\pi i}\; \frac{\tilde{s}^j}{j - 2 - \alpha'\Delta_P/2}$$

where $\Delta_P \simeq \Delta_j$, the tensorial Laplacian. Using a spectral analysis, it leads to a $J$-plane cut at $j_0^{(+)}$.

A similar analysis can next be carried out for the $C = -1$ sector. We simply replace the Regge kernel by

$$\tilde{s}^{1+\alpha'\tilde{t}/2} = \int \frac{dj}{2\pi i}\; \tilde{s}^j\; G^{(-)}(j) = \int \frac{dj}{2\pi i}\; \tilde{s}^j (j - 1 - \alpha'\Delta_O/2)^{-1} \; .$$

The operator $\Delta_O(j)$ can be fixed by examining the EOM at $j = 1$ for the associated supergravity fluctuations responsible for this exchange, i.e., the anti-symmetric Kalb-Ramond fields, $B_{MN}$. One finds two solutions,

$$G^{(-)}(j) = \frac{1}{[j - 1 - (\alpha'/2R^2)(\Box_{Maxwell} - m^2_{AdS,i})]} \; , \; i = 1, 2 \; ,$$

where $\Box_{Maxwell}$ stands for the Maxwell operator. Two allowed values are $m^2_{AdS,1} = 16$ and $m^2_{AdS,2} = 0$. A standard spectral analysis then lead to a branch-cut at $j_0^{(-)}$.

It is also useful to explore the conformal invariance as the isometry of transverse $AdS_3$. Upon taking a two-dimensional Fourier transform with respect to $q_\perp$, where $t = -q_\perp^2$, one finds that $G^{(\pm)}$ can be expressed simply as

$$G^{(\pm)}(z, x^\perp, z', x'^\perp; j) = \frac{1}{4\pi z z'} \frac{e^{(2-\Delta^{(\pm)}(j))\xi}}{\sinh \xi}\ , \tag{1}$$

where $\cosh \xi = 1 + v$, $v = [(x^\perp - x'^\perp)^2 + (z - z')^2]/(2zz')$ the $AdS_3$ chordal distance, and $z = R^2/r$, and $\Delta^{(\pm)}(j) = 2 + \sqrt{2}\ \lambda^{1/4}\sqrt{(j - j_0^{(\pm)})}$ is a $J$-dependent effective $AdS_5$ conformal dimension [1, 3, 4]. The $\Delta - j$ curve for $\Delta^{(\pm)}$ is shown in Fig. 2c.

# 5 Unitarity, Absorption, Saturation and the Eikonal Sum

For simplicity, we will focus here on the $C = +1$ sector, assuming all crossing odd amplitudes vanish. It has been shown in Refs. [3, 5] that, in the strong coupling limit, a 2-to-2 amplitude, $A(s,t)$, in the near-forward limit can be expressed in terms of a "generalized" eikonal representation,

$$A_{2\to 2}(s,t) = \int dz dz' P_{13}(z) P_{24}(z') \int d^2b\ e^{-ib^\perp q_\perp} \widetilde{A}(s, b^\perp, z, z')\ , \tag{2}$$

where $\widetilde{A}(s, b^\perp, z, z') = 2is\left[1 - e^{i\chi(s,b^\perp,z,z')}\right]$ , and $b^\perp = x^\perp - x'^\perp$ due to translational invariance. The probability distributions for left-moving, $P_{13}(z)$, and right moving, $P_{14}(z)$ particles are products of initial (in) and final (out) particle wave functions. The eikonal, $\chi$, can be related to the strong coupling Pomeron kernel [1, 3], and can be expressed as the inverse Mellin transform of $G^{(+)}(j, x^\perp - x'^\perp, z, z')$.

We note the salient feature of eikonal scattering locally in transverse $AdS_3$, and the near-forward field-theoretic amplitude is obtained from a bulk eikonal amplitude after convolution. It is useful to focus our attention on the properties of the bulk eikonal formula $\widetilde{A}(s, b^\perp, z, z')$ itself. For $\chi$ real, it is elastic unitary. On the other hand, when $\chi$ is complex, (with $\mathrm{Im}\chi > 0$), one has inelastic production. Absorption and saturation can now be addressed in this context. It is also important to note that, for Froissart bound, confinement is crucial. Discussion on these and related issues can be found in Ref. [3]. For applications of [1, 3, 4, 5] for DIS, see [14]. For a more proper treatment while taking into account of confinement effects, see [15].

# 6 Diffractive Production of Higgs at LHC

A promising production mechanism for Higgs meson at the LHC involves the forward proton-proton scattering $pp \to pHp$. Because of the exceptional signal to background discrimination, this may even be a discovery channel depending of course on the production cross section. The theoretical estimates generally involve the assumption of perturbative contribution of gluon fusion in the central rapidity region [16]. In most estimates the Pomeron is effectively replaced by two-gluon exchange, e.g., the Low-Nussinov Pomeron. In spite of the plausibility of this approach, there are considerable uncontrolled uncertainties.

We have begun the analysis in strong coupling based on the AdS/CFT correspondence and conformal strong coupling BPST Pomeron [1] This amounts to a generalization of our previous

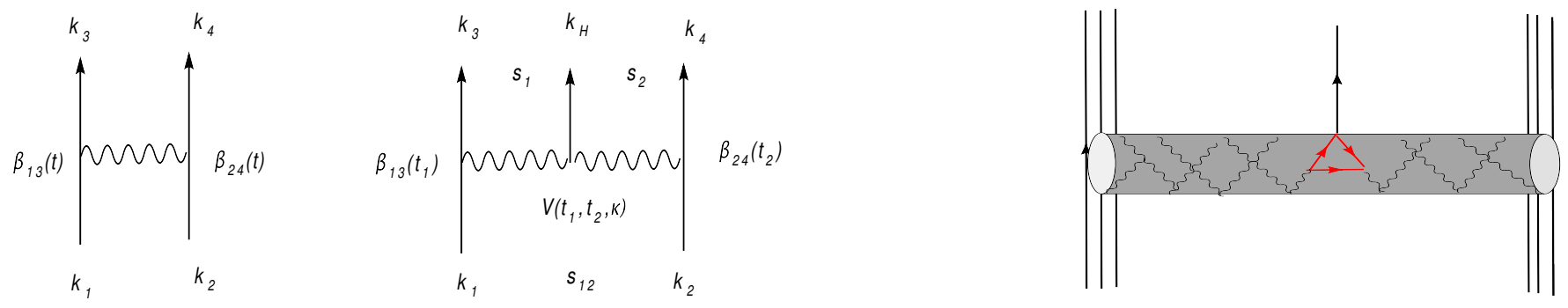

Figure 3: (a) kinematics for single-Regge limit for 2-to-2 amplitudes, (b) Double-Regge kinematics for 2-to-3 amplitudes. (c) Cylinder Diagram for large $N_c$ Higgs Production.

*AdS* for 2-to-2 amplitudes to one for 2-to-3 amplitudes, e.g., from Fig. 3a to Fig. 3b. A more refined analysis for Higgs production involves a careful treatment for that depicted in Fig. 3c. While this also will have its uncertainties, a careful comparison between weak and strong coupling Pomeron should give better bounds on these uncertainties. Ultimately the strong coupling approach calibrated by comparison with experimental numbers for double diffraction heavy quark production, can provide increasingly reliable estimates for Higgs production.

Focusing only on contributions from Pomeron exchange, a flat-space 2-to-2 amplitude in the Regge limit can be expressed as (Fig. 3a)

$$A(s,t) \simeq \beta_{13}(t)\frac{1+e^{-i\alpha(t)}}{\sin \pi\alpha(t)}(\alpha' s)^{\alpha(t)}\beta_{24}(t)\ .$$

For a 5-point amplitude, there are five independent invariants, Fig. 3b. In the kinematic region for diffractive scattering where transverse momenta of all produced particles are limited, $\kappa \equiv s_1 s_2/s$ is fixed, with $\kappa \simeq m_H^2 + q_\perp^2$, in the frame where incoming particles are longitudinal. Using a double $J$-plane representation, in the double-Regge region, a 2-to-3 amplitude can be expressed using a double-$J$-plane representation, as

$$T(s,s_1,s_2,t_1,t_2) \simeq \int_{-i\infty}^{i\infty}\frac{dj_1}{2\pi i}\int_{-i\infty}^{i\infty}\frac{dj_2}{2\pi i}\ \xi(j_1)\ (\alpha' s_1)^{j_1}\ \xi(j_2)\ (\alpha' s_2)^{j_2}\beta_{13}(t_1)$$
$$\cdot\, G_{j_1}(t_1)\mathcal{V}(t_1,t_2,\kappa)G_{j_2}(t_2)\beta_{24}(t_2)$$

where $\xi$ is the signature factor and $\mathcal{V}$ is the Pomeron-Particle-Pomeron coupling.

To move on to *AdS*, we simply need to replace $G_j(t)$ and $\mathcal{V}(t_1,t_2,\kappa)$ by corresponding generalizations. The essential new feature is a new vertex, $\mathcal{V}$, depicted in Fig. 3c, appropriate for a diffractive central Higgs production [17]. From $m_0 \ll m_H \ll m_t$, the Higgs vertex is replaced by a source for $F^a_{\mu\nu}F^a_{\mu\nu}$ at the boundary of AdS ($z \to 0$). The standard AdS/CFT dictionary leads to a bulk to boundary propagator $\Delta(x-x',z)$ for the interior of AdS to this point so that this vertex can be approximated by a factorized product. In a subsequent analysis we will add corrections due to (i) conformal symmetry breaking, (ii) the proton impact factor and (iii) eikonal "survival" probability to obtain phenomenological results for double Pomeron Higgs production at the LHC. These will be reported in future publications [17].

# References

[1] R. C. Brower, J. Polchinski, M. J. Strassler and C. I. Tan, JHEP **0712**, 005 (2007).

[2] E. Levin and C. I. Tan, "Heterotic Pomeron: A Unified Treatment of High-Energy Hadronic Collisions in QCD", "Proc. for International Symposium on Multiparticle Dynamics", Santiago, Spain, July 1992,

and “Workshop on Small-x and Diffractive Physics at Tevatron”, FNAL, Sept. 1992, hep-ph/9302308; S. Bondarenko, E. Levin and C. I. Tan, Nucl. Phys. A **732**, 73 (2004).

[3] R. C. Brower, M. J. Strassler and C. I. Tan, JHEP **0903**, 050 (2009); R. C. Brower, M. J. Strassler and C. I. Tan, JHEP **0903**, 092 (2009).

[4] R. C. Brower, M. Djuric and C. I. Tan, JHEP **0907**, 063 (2009).

[5] L. Cornalba, et al., Nucl. Phys. B **767**, 327 (2007); JHEP **0708**, 019 (2007); **0709**, 037 (2007); **0806**, 048 (2008).

[6] J. Polchinski and M. J. Strassler, Phys. Rev. Lett. **88**, 031601 (2002); R. C. Brower and C. I. Tan, Nucl. Phys. B **662**, 393 (2003).

[7] R. C. Brower, S. D. Mathur and C. I. Tan, Nucl. Phys. B **574**, 219 (2000); R. C. Brower, S. D. Mathur and C. I. Tan, Nucl. Phys. Proc. Suppl. **83**, 923 (2000).

[8] R. C. Brower, S. D. Mathur and C. I. Tan, Nucl. Phys. B **587**, 249 (2000).

[9] L. N. Lipatov, Sov. J. Nucl. Phys. **23**, 338 (1976) [Yad. Fiz. **23**, 642 (1976)].

[10] C. Ewerz, “The odderon: Theoretical status and experimental tests,” arXiv:hep-ph/0511196.

[11] R. A. Janik and J. Wosiek, Phys. Rev. Lett. **82**, 1092 (1999); M. A. Braun, hep-ph/9805394.

[12] J. Bartels, L. N. Lipatov and G. P. Vacca, Phys. Lett. B **477**, 178 (2000) [arXiv:hep-ph/9912423].

[13] M. Kalb and P. Ramond, Phys. Rev. D **9**, 2273 (1974).

[14] E. Levin, J. Miller, B. Z. Kopeliovich and I. Schmidt, JHEP **0902**, 048 (2009); Y. Hatta, E. Iancu and A. H. Mueller, JHEP **0801**, 026 (2008); L. Cornalba and M. S. Costa, Phys. Rev. D **78**, 096010 (2008).

[15] R. C. Brower, M. Djurić, Ina Sarcevic and C. I. Tan, “DIS after AdS/CFT”, (in preparation).

[16] A. D. Martin, M. G. Ryskin and V. A. Khoze, Acta Phys. Polon. B **40**, 1841 (2009); J. R. Cudell, A. Dechambre, O. F. Hernandez and I. P. Ivanov, Eur. Phys. J. C **61**, 369 (2009); E. Gotsman, H. Kowalski, E. Levin, U. Maor and A. Prygarin, Eur. Phys. J. C **47**, 655 (2006).

[17] R. C. Brower, M. Djurić, and C. I. Tan, “Double Diffractive Higgs Production in Strong Coupling”, (in preparation).

# Reflective Elastic Scattering at LHC

*Sergey Troshin, Nikolay Tyurin*

IHEP, Protvino, 142281 Russia

We discuss effects of reflective scattering for hadron and heavy nuclei collisions at the LHC and asymptotic energies. It is shown that the reflective scattering might lead to decreasing matter density with energy beyond the LHC energies. A limiting form of energy dependence of the hadron density is obtained. The unitarity upper bound for the absolute value of the real part of the elastic scattering amplitude and two-particle inelastic binary reaction amplitudes in impact parameter representation is two times less than the corresponding bound for the imaginary part of the elastic scattering amplitude. The former limit restricts a possible odderon contribution.

## 1 Reflective Scattering

A new physical interpretation of unitarity saturation in elastic scattering as a reflective scattering was proposed in [1] proceeding from optical analogy. This interpretation is related to the non-perturbative aspects of strong interactions and follows from the specific property of the unitarity saturation when the elastic $S$-matrix becomes negative and $S(s,b)|_{b=0} \to -1$ at $s \to \infty$. It should be noted that $S(s,b) = 1 + 2if(s,b)$, where $f(s,b)$ is the elastic scattering amplitude in the impact parameter representation.

In particular, we would like to note that the possible values of the elastic $S$ matrix can be negative (in the pure imaginary case). Transition to the reflective scattering mode is naturally reproduced by the $U$-matrix form of elastic unitarisation. The elastic scattering $S$-matrix ($2 \to 2$ scattering matrix element) in the impact parameter representation is written in this unitarisation scheme in the form of a linear fractional transform:

$$S(s,b) = \frac{1 + iU(s,b)}{1 - iU(s,b)}, \qquad (1)$$

where $U(s,b)$ is the generalised reaction matrix, which is considered to be an input dynamical quantity. For simplicity we consider the case of a pure imaginary $U$-matrix and make the replacement $U \to iU$ in (1). The reflective scattering mode ($S(s,b) < 0$) starts to appear at the energy $s_R$, which is determined as a solution of the equation $U(s_R, b = 0) = 1$. At $s > s_R$ the elastic scattering acquires ability for reflection, while the inelastic overlap function $h_{inel}(s,b)$ gets a peripheral impact parameter dependence in the region $s > s_R$. It should be noted that unitarity condition for the elastic scattering amplitude $F(s,t)$, which can be written in the form

$$\mathrm{Im}F(s,t) = H_{el}(s,t) + H_{inel}(s,t), \qquad (2)$$

where $H_{el,inel}(s,t)$ are the corresponding elastic and inelastic overlap functions introduced by Van Hove [2]. The functions $H_{el,inel}(s,t)$ are related to the functions $h_{el,inel}(s,b)$ via the Fourier-Bessel transforms, i.e.

$$H_{el,inel}(s,t) = \frac{s}{\pi^2}\int_0^\infty b\,db\,h_{el,inel}(s,b)J_0(b\sqrt{-t}). \quad (3)$$

The elastic and inelastic cross-sections can be obtained as follows:

$$\sigma_{el,inel}(s) \sim \frac{1}{s}H_{el,inel}(s,t=0). \quad (4)$$

Saturation of unitarity leads to the peripheral dependence of $h_{inel}(s,b)$. It is a manifestation of the self-damping of the inelastic channels at small impact parameters. The function $h_{inel}(s,b)$ reaches its maximum value at $b = R(s)$, note that

$$R(s) \sim \frac{1}{M}\ln s,$$

while an elastic scattering (due to reflection) occurs effectively at smaller values of impact parameter, i.e. $\langle b^2\rangle_{el} < \langle b^2\rangle_{inel}$. At the values of energy $s > s_R$ the equation $U(s,b) = 1$ has a solution in the physical region of impact parameter values, i.e. $S(s,b) = 0$ at $b = R(s)$. Fig. 1 shows the regions where elastic $S$-matrix has positive and negative values. Of course, the reflective scattering exists only in the elastic channel. For example, all inelastic binary reactions have amplitudes $\tilde{f}(s,b)$ in the impact parameter representation, which satisfy the inequality $|\tilde{f}(s,b)| \leq 1/2$, while the elastic scattering amplitude satisfies $|f(s,b)| \leq 1$. It follows from unitarity equation in the impact parameter representation:

$$\mathrm{Im} f(s,b) = h_{el}(s,b) + h_{inel}(s,b);$$

amplitudes $\tilde{f}(s,b)$ contribute to $h_{inel}(s,b)$. The upper bound for the elastic scattering cross-section is four times higher than the upper bound for the inelastic cross-section as it was recently demonstrated in [3]. Scattering dynamics in the elastic channel such as $pp \to pp$ is therefore strikingly different, e.g. from the inelastic binary diffractive process, such as $pp \to pN$, where $N$ is an isobar. The latter reactions should have a peripheral impact parameter profile, which can be related to the dominating contribution of helicity-flip amplitudes. The unitarity limit and black disk limit are the same for the inelastic overlap function, but those limits are different for the elastic overlap function

$$h_{el}(s,b) \equiv |f(s,b)|^2.$$

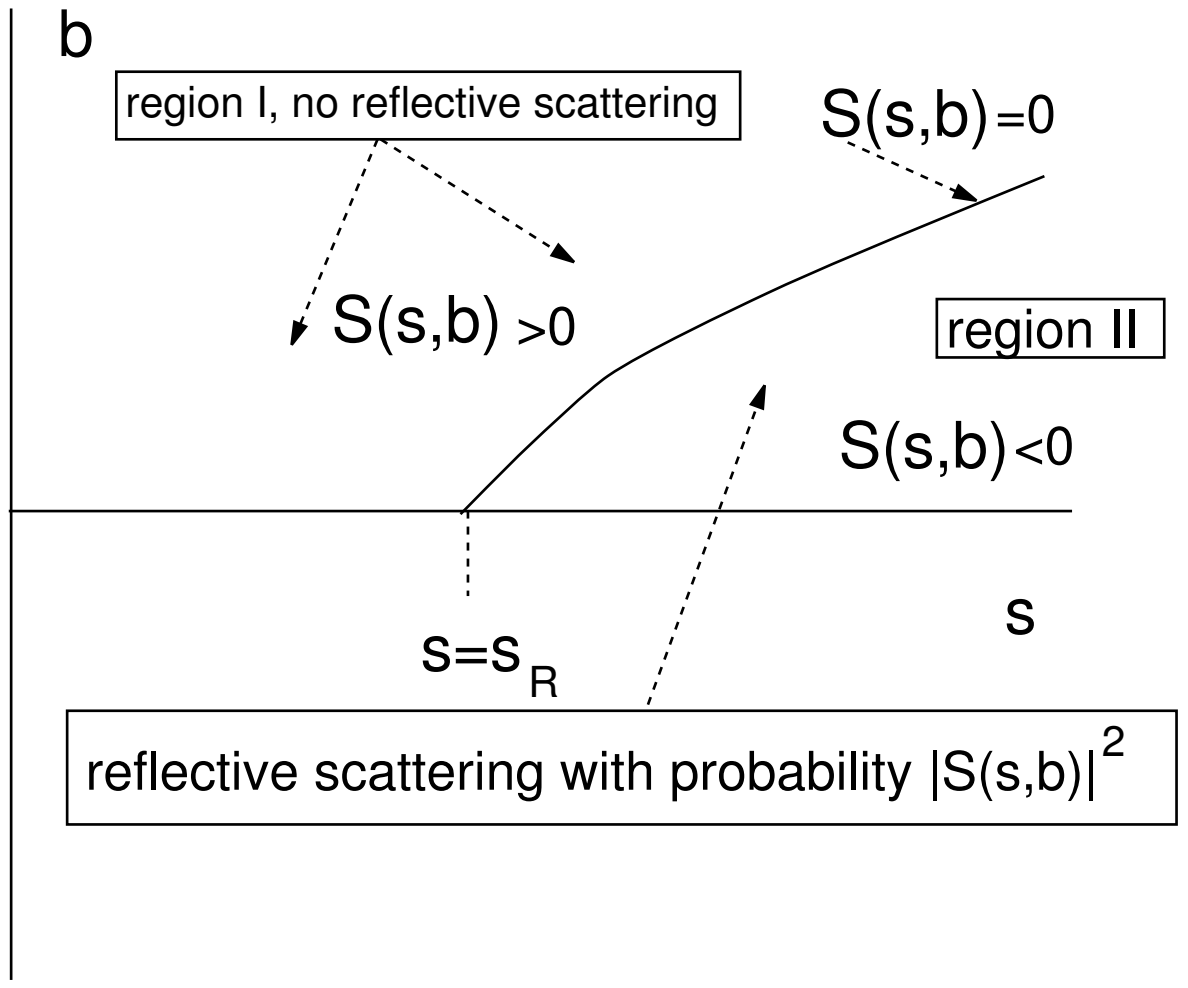

Figure 1: *Regions of positive (absorptive scattering) and negative values (absorptive and reflective scattering) of the function $S(s,b)$ in the $s$ and $b$ plane.*

The unitarity limit for the elastic overlap function is four times higher than the black disk limit. This is an important point for consideration of exclusive limit of inclusive reactions. Saturation of unitarity leads to suppression of the inelastic cross-section, i.e. at fixed impact parameter $(b < R(s))$ $h_{inel}(s,b) \to 0$ at $s \to \infty$ and

$$\sigma_{el}(s) \sim R^2(s), \quad \sigma_{inel}(s) \sim R(s). \tag{5}$$

Thus, $H_{el}(s,t)$ which has the following $t$-dependence

$$H_{el}(s,t) \sim \frac{RJ_1(R\sqrt{-t})}{\sqrt{-t}},$$

dominates over $H_{inel}(s,t)$, which depends on $t$ like

$$H_{inel}(s,t) \sim RJ_0(R\sqrt{-t}),$$

at $-t = 0$, but it is not the case for the scattering in the non-forward directions. In this region these two functions have similar energy dependencies proportional to $R^{1/2}(s)$ at rather large fixed values of $-t$. The mean impact parameter values for elastic and inelastic interactions have also similar energy dependencies

$$\langle b^2 \rangle_{el}(s) \sim R^2(s), \quad \langle b^2 \rangle_{inel}(s) \sim R^2(s), \tag{6}$$

but the value of impact parameter averaged over all interactions (cf. [4])

$$\langle b^2 \rangle_{tot}(s) = \frac{\sigma_{el}(s)}{\sigma_{tot}(s)}\langle b^2 \rangle_{el}(s) + \frac{\sigma_{inel}(s)}{\sigma_{tot}(s)}\langle b^2 \rangle_{inel}(s)$$

acquires the main contribution from elastic scattering according to Eq. (5). Therefore, the inelastic intermediate states will give subleading contribution to the slope of diffraction cone $B(s)$,

$$B(s) \equiv \frac{d}{dt} \ln \left( \frac{d\sigma}{dt} \right) \Bigg|_{t=0},$$

at asymptotic energies. Indeed, since $B(s) \sim \langle b^2 \rangle_{tot}(s)$, it can be written in the form

$$B(s) = B_{el}(s) + B_{inel}(s),$$

where $B_{el}(s) \sim R^2(s)$, while $B_{inel}(s) \sim R(s)$. It should be noted that both terms $B_{el}(s)$ and $B_{inel}(s)$ are proportional to $R^2(s)$ in case of the absorptive scattering.

Under reflective scattering, the behaviour of the function $H_{inel}(s,t)$ is determined by a peripheral impact parameter profile and its $-t$ dependence is different. Meanwhile, the elastic overlap function $H_{el}(s,t)$ has similarities with that function in the case of approach where absorption is only presented. As a result, zeroes and maxima of the functions $H_{el}(s,t)$ and $H_{inel}(s,t)$ will be located at different values of $-t$ and zeroes and maxima of Im$F(s,t)$ will also be located at different position in the cases of absorptive and the reflective scattering. In the case of reflective scattering, dips and maxima will be located in the region of lower values of $-t$. We would like to note that the presence of reflective scattering enhances the large $-t$ region by factor $\sqrt{-t}$ compared to absorptive scattering. Despite that these two mechanisms lead at the asymptotics to the significant differences in the total, elastic and inelastic cross-section dependencies, their predictions for the differential cross-section of elastic scattering are not so much different at small and moderate values of $-t$.

# 2 Intermittent Remark on Unitarity and Real Part of Scattering Amplitude

It is evident that there are serious difficulties in accounting all known dynamical issues and limitations into a particular phenomenological model. But it is equally difficult to expect that the model inconsistent with unitarity (i.e. the one violating probability conservation law) would adequately reflect the dynamics of hadron interaction and provide reliable predictions. To fulfill unitarity condition under a model construction of the elastic amplitudes, it is natural to use unitarisation approaches such as eikonal or $U$-matrix, which consider amplitudes in the impact parameter space. They automatically guarantee that elastic amplitude in the impact parameter representation will obey unitarity condition.

Despite that the full implementation of unitarity is not possible nowadays (cf. e.g. [5]), the amplitude in the impact parameter space should not exceed unity anyway. However, when the amplitude $F(s,t)$ is constructed in the $s$ and $t$ representation, it is a priori not evident that the particular form of this amplitude being transformed into the impact parameter space $f(s,b)$ would satisfy unitarity. This remains to be true, even when the model under consideration leads to the predictions for observables which explicitly agree with axiomatic bounds, e.g. such as well known Froissart-Martin bound for the total cross-sections. Getting agreement with experimental data at finite energies and asymptotic bounds at $s\to\infty$ is not enough since a wide class of functional dependencies can successfully describe experimental data and provide correct asymptotic behaviour. Additional steps to justify that the impact parameter amplitude is *at least* less than unity in the whole region of kinematic variables are necessary.

In the above remarks we supposed that imaginary part of scattering amplitude is a dominating one. Further unitarity restriction exists for models which do not suppose domination of imaginary part of scattering amplitude, such as models with maximal odderon contribution [6].

Indeed, unitarity condition in the impact parameter representation for the elastic scattering amplitude can be rewritten in the form:

$$\mathrm{Im}f(s,b)[1-\mathrm{Im}f(s,b)] = [\mathrm{Re}f(s,b)]^2 + h_{inel}(s,b).$$

Since $0\leq \mathrm{Im}f(s,b)\leq 1$, we obtain that unitarity limits the real part of scattering amplitude in the following way

$$[\mathrm{Re}f(s,b)]^2 \leq 1/4,$$

$$-\frac{1}{2}\sqrt{1-4h_{inel}(s,b)} \leq \mathrm{Re}f(s,b) \leq \frac{1}{2}\sqrt{1-4h_{inel}(s,b)}.$$

The function $\mathrm{Re}f(s,b)$ can be sign changing one in contrast with $\mathrm{Im}f(s,b)$). This limitation, as it was already mentioned, is essential for the models with odderon and is indirectly in favour of the standard procedure of neglecting the real part of scattering amplitude compared to its imaginary part. It also is evident that absolute value of the real part and imaginary part of elastic scattering amplitude $f(s,b)$ cannot reach their maximal values simultaneously, moreover when $\mathrm{Im}f(s,b) \to 1$, saturating unitarity limit at large values of $s$ in the region $b < R(s)$, then $\mathrm{Re}f(s,b) \to 0$ in this kinematic region. It should be noted that this saturation does not suppose that $\mathrm{Re}f(s,b)$ vanish everywhere. It means that $[\mathrm{Re}f(s,b)]^2$ should have a peripheral impact parameter profile. The same conclusion is valid when $\mathrm{Im}f(s,b) \to 1/2$, saturating the black disk limit at large values of $s$ in the region $b < R(s)$, then $\mathrm{Re}f(s,b) \to 0$ because $h_{inel}(s,b) \to 1/4$ in this region. The above difference in the impact parameter profiles would result in the different energy dependencies of $\mathrm{Im}F(s,t=0)$ and $\mathrm{Re}F(s,t=0)$ bringing maximal odderon on the edge of contradiction with unitarity (or black disk) limit saturation. Of course, unitarity or black disk limits saturation itself does not follow from axiomatic field theory, but we would like to note, that it is much more natural to expect that it could be a manifestation of a maximal strength of strong interaction instead of behaviour of the real part of the forward scattering amplitude in the form $\mathrm{Re}F(s,t=0) \sim s\ln^2 s$ as it happens in the models incorporating the maximal odderon regime.

# 3 Reflective Scattering and Deconfinement

Possible existence of the reflective scattering at very high energies implies that confinement becomes stronger and stronger as the collision energy increases and proton collisions resemble more and more collisions of hard spheres. In this section we address one aspect of the broad problem of transition to the deconfined state of matter, namely, we discuss the role of the reflective scattering on the energy dependence of density in the percolation mechanism of the transition to the deconfined state of matter.

The main idea of the percolation mechanism of deconfinement is a formation in the certain volume of a connected hadron cluster due to increasing temperature and/or hadron density [7], i.e. when vacuum as a connected medium disappears, the deconfinement takes place. This process has typical critical dependence on particle density. Thus, it was proposed to use percolation to define the states of matter and consider the disappearance of a large-scale vacuum as the end of hadronic matter existence [7, 8].

The probability of reflective scattering at $b < R(s)$ and $s > s_R$ is determined by the magnitude of $|S(s,b)|^2$; this probability is equal to zero at $s \le s_R$ and $b \ge R(s)$ (region I in Fig. 1). At the energies $s > s_R$ reflective scattering will mimic the presence of a repulsive core in hadron and meson interactions. Presence of the reflective scattering can be accounted for using van der Waals method (cf. [9]). This approach was used originally for description of the fluids behaviour starting from the gas approximation by means of taking into account the nonzero size of molecules. Consider central collision of two identical nuclei having $N$ hadrons in total with centre-of-mass energy $\sqrt{s}$ per nucleon and calculate hadron density $n_R(T,\mu) = N/V$ in the initial state at given temperature $T$ and baryochemical potential $\mu$ in the presence of reflective scattering. The effect of the reflective scattering of hadrons is equivalent to decrease of the volume of the available space which the hadrons are able to occupy in the case when reflective scattering is absent. Thus following to van der Waals method, we must then replace

volume $V$ by

$$V - p_R(s)V_R(s)\frac{N}{2},$$

i.e. we should write

$$n(T,\mu) = \frac{N}{V - p_R(s)V_R(s)\frac{N}{2}},$$

where $n(T,\mu)$ is hadron density without account for reflective scattering and $p_R(s)$ is the averaged over volume $V_R(s)$ probability of reflective scattering:

$$p_R(s) = \frac{1}{V_R(s)}\int_{V_R(s)} |S(s,r)|^2 d^3x.$$

The volume $V_R(s)$ is determined by the radius of the reflective scattering. Here we assume spherical symmetry of hadron interactions, i.e. we replace impact parameter $b$ by $r$ and approximate the volume $V_R(s)$ by $V_R(s) \simeq (4\pi/3)R^3(s)$. Hence, the density $n_R(T,\mu)$ is connected with corresponding density in the approach without reflective scattering $n(T,\mu)$ by the following relation

$$n_R(T,\mu) = \frac{n(T,\mu)}{1+\alpha(s)n(T,\mu)},$$

where $\alpha(s) = p_R(s)V_R(s)/2$. Let us now estimate change of the function $n_R(T,\mu)$ due to the presence of reflective scattering. We can approximate $p_R(s)$ by the value of $|S(s,b=0)|^2$ which tends to unity at $s \to \infty$. It should be noted that the value $\sqrt{s_R} \simeq 2\,\text{TeV}$ [10]. Below this energy there is no reflective scattering, $\alpha(s) = 0$ at $s \le s_R$, and therefore corrections to the hadron density are absent. Those corrections are small when the energy is not too much higher than $s_R$. At $s \ge s_R$ the value of $\alpha(s)$ is positive, and presence of reflective scattering diminishes hadron density. We should expect that this effect would already be noticeable at the LHC energy $\sqrt{s} \simeq 5\,\text{TeV}$ in Pb+Pb collisions. At very high energies $(s \to \infty)$

$$n_R(T,\mu) \sim 1/\alpha(s) \sim M^3/\ln^3 s.$$

This limiting dependence for the hadron density appears due to the presence of the reflective scattering which results in similarity of head-on hadron collisions with scattering of hard spheres. It can be associated with saturation of the Froissart-Martin bound for the total cross-section. It should be noted that this dependence has been obtained under assumption on spherical symmetry of hadron interaction region. Without this assumption, limiting dependence of the hadron density in transverse plane can only be obtained, i.e. transverse plane density of hadrons would have then the following behaviour

$$n_R(T,\mu) \sim M^2/\ln^2 s.$$

To conclude this section, we would like to note that the lower densities of hadron matter are needed for percolation (and transition to the deconfined state) in the presence of reflective scattering. It might be useful to note that the rescattering processes are also affected by the reflective scattering. Reflective scattering would lead to noticeable effects at the LHC energies and beyond and could help in searches of the deconfined state and studies of properties of transition mechanism to this state of matter which might proceed by means of percolation. Thus, it will affect description of initial state dynamics in nuclear interactions at the LHC energies by introducing notion of limiting density of strongly interacting matter at respective energies.

# 4 Conclusion

Thus, at very high energies there would be two different regions of impact parameter distances in particles collisions, namely the outer region (peripheral collisions) where elastic scattering has exclusively a shadow origin and inner region (central collisions) where reflective and absorptive scattering give competing contributions, reflective scattering contribution increases while absorptive scattering contribution decreases at fixed impact parameter. It is not surprising that the model with reflective scattering contribution leads to significantly higher values for total and elastic cross-sections at the LHC energies[1] while it renders the standard values for the inelastic cross-section. In the geometric terms, the generic scattering picture at fixed energy beyond the black disc limit can be described as a scattering off a partially reflective and partially absorptive disk surrounded by the black ring (which becomes grey at larger values of the impact parameter). The evolution with energy is characterised by increasing albedo due to the interrelated increase of reflection and decrease of absorption at small impact parameters. This picture predicts that the scattering amplitude at the LHC energies is beyond the black disk limit at small impact parameters and it provides explanation for the regularities observed in cosmic rays studies, e.g. the existence of the knee in the cosmic rays spectrum. It leads also appearance of limiting density dependent on energy which takes place only at very high energies and has an origin related to unitarity saturation.

# Acknowledgements

We are grateful to J.-R. Cudell, M. Islam, L. Jenkovszky, A. Krisch, U. Maor, A. Martin, E. Martynov, V. Petrov and D. Sivers for the interesting comments and discussions.

# References

[1] S.M. Troshin, N.E. Tyurin, Int. J. Mod. Phys. **A22** 4437 (2007).

[2] L. Van Hove, Nuovo Cimento, **28** 798 (1963).

[3] A. Martin, arXiv: 0904.3724, talk given at this Conference.

[4] Z. Ajduk, Nuovo Cim. **A15**, 390 (1973).

[5] P.V. Landshoff, arXiv: 0903.1523.

[6] R. Avila, P. Gauron, B. Nicolescu, Eur. Phys. J. **C49** 581 (2007);
E. Martynov, B. Nicolescu, Eur. Phys. J. **C56** 57 (2008).

[7] P. Castorina, K. Redlich, H. Satz. Eur. Phys. J. **C59** 67 (2009).

[8] H. Satz, arXiv:0803.1611.

[9] J. Cleymans, K. Redlich, H. Satz, E. Suhonen, Z. Phys. **C33** 151 (1986).

[10] P.M. Nadolsky, S.M. Troshin, N.E. Tyurin, Z. Phys. **C69** 131 (1995).

[1] The value of total cross-section about 150 mb is predicted for *pp*-interactions at the LHC starting up energy $\sqrt{s} = 7\,\text{TeV}$.

# Part III

# Soft Diffraction

# Theoretical Overview on Soft Diffraction.

*Alexei Kaidalov*

Institute of Theoretical and Experimental Physics, B. Cheremushkinskaya 25, 119571 Moscow, Russia

Theoretical approaches to description of diffractive processes at high energies are discussed. It is pointed out that inelastic diffractive processes should be suppressed at small impact parameters. Important role of the pion exchange for analytic structure of the pomeron trajectory is emphasised. Models for large mass diffraction and recent calculations of survival probabilities are reviewed.

## 1 Introduction

Diffractive processes (elastic and inelastic) constitute a substantial part (about 1/2) of the total interaction cross sections of hadrons at high energies. Investigation of these processes provides an important information on mechanisms of high-energy interactions. There are important problems of QCD which can be studied in diffractive processes:
a) Nature of the pomeron in QCD.
b) Role of the s-channel unitarity and multi-pomeron exchanges.
c) Small-x problem and "saturation" of parton densities at $x \to 0$.
d) Violation of Regge and QCD-factorisations in diffractive processes.

Elastic scattering at high energies is a classical example of diffractive process. Absorption of the initial wave due to many inelastic channels leads by unitarity to elastic diffractive scattering.

Inelastic diffractive processes were first considered by E.L. Feinberg and I.Ya. Pomeranchuk [1] and elegant formulation in terms of the eigen-states was given by M.L. Good and W.D. Walker (GW) [2]. In this approach cross section for inelastic diffraction is related to a dispersion of eigen-amplitudes. Thus in a black disk limit inelastic diffraction is absent (exists at the edge of the disk only). This is an s-channel view on diffractive scattering. Note that GW-approach assumes a separation of diffractive and multiparticle states. This is not true in general for production of large mass states, which we will discuss in terms of the t-channel or Regge approach.

In the t-channel approach amplitudes of diffractive processes are described by an exchange of the pomeron, which has vacuum quantum numbers (positive signature and parity and C-parity , isospin I=0) (see for example review [3]. An increase with energy of the total interaction cross sections indicates that an intercept of the pomeron is larger than unity. An exchange by a Regge pole with $\Delta \equiv \alpha_P(0) - 1 > 0$ leads to a violation of the s-channel unitarity and for such "supercritical" pomeron multi-pomeron exchanges in the t-channel are very important. They restore unitarity and make theory consistent with Froissart bound.

## 2 Unitarity Effects in Gribov's Approach

A general method of calculation of multi-pomeron contributions to amplitudes of diffractive processes was formulated by V.N. Gribov [4]. In this approach a contribution of two-pomeron exchange (PP) to the process of elastic scattering can be expressed, using analyticity and unitarity for pomeron-particle scattering amplitudes, as the sum over all intermediate diffractive states in the s-channel.

In the same way amplitudes for nP-exchanges are expressed through all possible diffractive intermediate states. This result is based on general properties of Feynman diagrams and is valid in QCD. It allows to build approximate schemes for calculation of multi-pomeron contributions. The simplest approximation corresponds to an account of elastic intermediate states only. It leads to the eikonal approximation for scattering amplitudes. For elastic scattering amplitudes in the impact parameter space:

$$\mathrm{Im}\, f(s,b) \; = \frac{1}{2}\left(1 \; - \; e^{-\Omega/2}\right), \tag{1}$$

where the eikonal $\Omega = -4i\delta_P(s,b)$ is the Fourier transform of the pomeron pole exchange.

Low mass diffractive states are often approximated by several resonance states. In this case the same method leads to Eq. (1) with $\Omega$ being a matrix, which elements correspond to transitions between different diffractive states. The simplest treatment is a diagonalisation of this matrix. Thus an account of the low mass diffraction in the Gribov's method is equivalent to the Good-Walker [2] approach to inelastic diffraction.

A simplest generalisation of the eikonal model, which takes into account inelastic diffractive intermediate states is so called "quasi-eikonal" model, where the amplitude $f(s,b)$ has the form

$$f(s,b) = \frac{i}{2C}\left[1 - \exp(2iC\delta_P(s,b))\right] \tag{2}$$

The function $\delta_P(s,b) \sim s^{\Delta}$ and it becomes large at very high energies. In this limit the scattering amplitude for elastic scattering $f(s,b) \to i/2$ in the eikonal model (scattering on a black disk) and $f(s,b) \to 1/2C$ in the quasi-eikonal model (scattering on a grey disc). This property of the quasi-eikonal model is closely related to the fact that one of the eigen-states for the diffractive matrix $\hat{\Omega}_P$ has in this model a zero eigen-value.

This is a crude approximation, which takes into account a big difference in the interaction cross section of hadrons with different transverse sizes. Configurations of quarks inside hadrons with small transverse size $r$ have total interaction cross sections $\sim r^2$, because hadrons in QCD interact as colour dipoles. There is a distribution of quarks and gluons inside colliding hadrons with different values of $r$ so one can expect that there will be a slow approach to the black disk limit for elastic scattering amplitude as $s \to \infty$. In this limit the effective radius of interaction increases as $\ln s$. Thus total interaction cross sections for the supercritical Pomeron theory have Froissart type behaviour $\sigma^{(tot)} \sim \ln^2(s)$ as $s \to \infty$.

Unitarisation effects due to multi-pomeron exchanges most strongly influence amplitudes of inelastic diffractive processes. In the eikonal approximation a suppression of cross section for inelastic diffractive process $S^2$ is:

$$S^2 \; = \; \frac{\int \, |\mathcal{M}(s,b)|^2 \, e^{-\Omega(b)} \, d^2b}{\int \, |\mathcal{M}(s,b)|^2 \, d^2b}, \tag{3}$$

This expression is easily generalised to the case of several diffractive channels. The same Eq. (2) is valid for each diagonal state and it is necessary to sum over all diagonal states (with corresponding weights).

The quantity $\Omega$ increases with energy as $s^{\Delta}$ and becomes large at very high energy. According to Eq. (2) cross sections of inelastic diffractive processes becomes negligible at small impact parameter and is concentrated at the edge of interaction region at $b > 1\,\mathrm{fm}$ and the radius of this region increases with energy as $\ln s$. Note that models based on perturbative QCD are not valid in this peripheral region. On the other hand account of $\pi\pi$-cut in the pomeron trajectory play an important role in this region (especially in view of smallness of $\alpha'_P(t=0)$). An imaginary part of the P-trajectory due to $\pi\pi$-cut has the form [5]

$$\mathrm{Im}\,\alpha_P(t) = \frac{b_P(t))(t-4\mu^2)^{\alpha_P(t)+1/2}}{16\pi C(\alpha_P(t))(4s_0)^{\alpha_P(t)}\sqrt{t}}, \tag{4}$$

where $C(j) = \frac{\pi\Gamma(2j+2)}{2^{2j+1}\Gamma(j+1)^2}$, $s_0 = 1\,\mathrm{GeV}^2$, $\mu$ is the pion mass and $B_P(t)$ is the pomeron residue in $\pi\pi$-scattering. Due to smallness of $\mu$ the pion cut has an important influence on the pomeron trajectory [6]. In particular it gives a contribution to the slope of the pomeron trajectory $\delta\alpha'_P(0) = \frac{\sigma^{tot}_{\pi\pi}}{16\pi^3}(\ln(\frac{m}{\mu})-1) \approx 0.05\,\mathrm{GeV}^{-2}$ for $m \approx 1\,\mathrm{GeV}$, which can be considered as a lower bound on the slope of the pomeron. The singularity in the P-trajectory leads also to a substantial curvature in the pomeron trajectory in the small-t region. It is essential for resolving problems with s-channel unitarity in inelastic diffractive processes at super-high energies [3]

The value of $S^2$ is not universal: it depends on behaviour of a matrix element $\mathcal{M}(s,b)$ on impact parameter $b$.

# 3 Interplay of Soft and Hard Diffraction

At very high energies diffractively produced system can contain hard subsystem. For example diffractive production of dijets, W, Z-bosons and heavy quarks. Especially interesting class of hard diffractive processes is exclusive central production of Higgs boson. It allows to study Higgs bosons in a very clean environment and gives a possibility to determine quantum numbers of Higgs. In hard diffraction the subprocess of a heavy state production can be calculated using QCD perturbation theory. The simplest inclusive diffractive process is a diffractive dissociation of a highly virtual photon. In this case the photon interacts with a quark and a study of these processes at HERA gave a possibility to determine distributions of quarks and gluons in the pomeron. These distributions and QCD factorisation can be used to predict cross sections of hard inclusive diffractive processes in hadronic interactions. However both QCD and Regge-factorisations in hard diffractive processes are violated due to multi-pomeron exchanges. They strongly modify predictions based on a single pomeron exchange. This is a manifestation of an interplay of soft and hard diffraction. CDF data [7] show that cross section of diffractive dijet production about an order of magnitude smaller than prediction based on QCD factorisation and partonic distributions extracted from HERA results. Dependence on $\beta$ (distribution of partons in the pomeron) is also substantially different from the predicted one. Calculation of suppression in the two-channel eikonal model [8] allows to reproduce both the observed suppression and $\beta$-dependence.

It is interesting that the same suppression is observed for double gap (double pomeron exchange) events at Tevatron [9]. This observation is in accord with a dominance of eikonal-type rescatterings [10].

The problem of calculation of survival probability and its energy dependence is very important for prediction of cross section of double-pomeron production of Higgs bosons and I shall return to this problem after discussion of influence of large mass diffraction on a magnitude of suppression of hard diffractive processes.

# 4 Large Mass Diffraction and Interaction of Pomerons

So far we have considered the low mass excitations in diffractive intermediate states. A mass of diffractively excited state at large $s$ can be large. The only condition for diffraction dissociation is $M^2 \ll s$. For large masses of excited states $M^2 \approx (1-x)s$ and rapidity gap $\Delta y \approx \ln(s/M^2) \equiv \xi'$. The large mass behaviour of the pomeron-particle amplitudes is described by the triple-pomeron and multi-pomeron diagrams.

The cross section for inclusive single diffraction dissociation in the Regge pole model can be written in the following form

$$\frac{d^2\sigma}{d\xi_2 dt} = \frac{(g_{11}(t))^2}{16\pi} \left|G_p(\xi', t)\right|^2 \sigma_{P2}^{(tot)}(\xi_2, t) \tag{5}$$

where $\xi_2 \equiv \ln(M^2/s_0)$ and $G_p(\xi', t) = \eta(\alpha_p(t)) \exp[(\alpha_p(t)-1)\xi']$ is the pomeron Green function. The quantity $\sigma_{P2}^{(tot)}(\xi_2, t)$ can be considered as the pomeron-particle total interaction cross section [3]. At large $M^2$ this cross section in Regge-model has the same behaviour as usual cross sections

$$\sigma_{P2}^{(tot)}(M^2, t) = \sum_k g_{22}^k(0)\, r_{PP}^{\alpha_k}(t) \left(\frac{M^2}{s_0}\right)^{\alpha_k(0)-1} \tag{6}$$

where the $r_{PP}^{\alpha_k}(t)$ is the triple-reggeon vertex , which describes coupling of two pomerons to reggeon $\alpha_k$.

In this kinematic region $s \gg M^2 \gg m^2$ the inclusive diffractive cross section is described by the triple–Regge diagrams and has the form

$$f^1 = \sum_k G_k(t)(1-x)^{\alpha_k(0)-2\alpha_P(t)} \left(\frac{s}{s_0}\right)^{\alpha_k(0)-1} \tag{7}$$

The pomeron-proton total cross section and triple-Regge vertices $r_{PP}^P, r_{PP}^f$ have been determined from analysis of experimental data on diffractive production of particles in hadronic collisions (see review [3]). Account of multi-pomeron rescattering for triple-reggeon diagram in analysis of large mass diffraction dissociation leads to a substantial change (increase) in values of triple-reggeon couplings (for recent analysis see [12, 13]).

It is clear that for very large masses it is not enough to consider the triple-pomeron contribution only as it violates unitarity for the pomeron-particle scattering amplitude. An important theoretical question is: what is the structure of the vertices for n pomeron to m pomeron transitions. The simplest approximation is to assume an eikonal-type structure for the pomeron-particle amplitudes at large $M^2$:

$$g_{mn} = cg^{m+n} \tag{8}$$

where $c$ and $g$ are some functions of $t$. This behaviour of vertices follows from multi-peripheral model and is natural from the t-channel unitarity point of view. It was used in [11] (KPTM) to sum all diagrams with interactions of pomerons. This model leads to a good description of total, elastic and single diffraction dissociation cross sections ($\sigma_{SD}$) in $pp(p\bar{p})$ interactions [11] with $\Delta \approx 0.2$. It is worth to note that without multi-pomeron effects $\sigma_{SD}$ has too fast increase with energy and exceeds experimentally observed cross section by a factor $\sim 10$.

In recent analysis of single and double diffraction dissociation [13] a simplified version of KPTM was used, where besides eikonal rescatterings between colliding protons eikonalization of each leg of the triple-reggeon diagrams was taken into account. More complicated t-channel iterations, which become important at extremely high energies, were neglected. This model gives a very good description of data on diffractive production from energies of fixed targets up to Tevatron.

An important question is how to apply Abramovsky, Gribov, Kancheli (AGK) cutting rules [14] in presence of multi-pomeron vertices. This is necessary in order to describe processes of multi-particle production in presence of interactions between pomerons. Strictly speaking AGK cutting rules were not proved for $n \to m$ pomeron transitions. For simplest application of AGK rules the problem was considered by S. Ostapchenko [15, 16] and generalisation of the Quark-Gluon Strings Model (QGSM) [17], was formulated in a Monte Carlo version.

In treatment of diagrams with interactions between pomerons it is necessary to take into account that the notion of the pomeron exchange is meaningful for large rapidity gaps only (usual choice $y > y_0$, with $y_o = \ln(10) = 2.3$). Thus a cutoff at small rapidities for each pomeron line should be introduced. It leads to a natural limitation to the number n of the t-channel iterations of pomeron exchanges (or number of gaps) at each initial energy : $n < \ln(s/s_0)/y_0$ with $s_0 = 1\,\mathrm{GeV}^2$. This threshold effect was taken into account in Ref. [11] and should be accounted for in all realistic calculations with pomeron interactions. It plays an important role in calculations of survival probabilities (see below).

The Durham group (KMR) has made recently a new fit of data on cross sections of diffractive processes [18, 19]. All $n \to m$ pomeron transition were taken into account in the framework of a partonic model, which lead to the behaviour $g_{mn} \sim nmg^{n+m}$, which is somewhat different from the one discussed above in the eikonal approximation. Summation of diagrams was performed by numerical solution of a system of highly nonlinear equations for amplitudes. To account for semi-hard and hard interactions three types of pomeron poles were introduced. Formulae for cross sections of different inelastic diffractive processes were obtained using some probabilistic arguments (and not cutting rules as in the standard approach). In this model it is possible to obtain a reasonable description of total cross section for pp-interaction, its elastic cross section in the diffraction cone region and cross sections of single and double diffraction. In the version of the model, which takes into account transverse degrees of freedom [19], for intercepts of pomerons $\Delta = \alpha_P(0) - 1$ values close to 0.3 were obtained.

A different approach was used by the Tel-Aviv group (GLM) [20]. Arguments, based on a small value of the pomeron slope, were used to justify applicability of perturbative QCD (PQCD) for diffractive processes. Motivated by PQCD the authors used the triple-pomeron interaction only with maximal number of pomeron loops. The last assumption may be reasonable for interaction of very small dipoles, but is difficult to justify for interaction of protons. The diagrams without pomeron loops (for example the diagram with a single triple-pomeron vertex) should be taken into account for self-consistency. It is known [21, 22] that the one dimensional approximation, used in calculations of GLM, leads asymptotically to decreasing total cross sections. GLM propose to use their model in a limited energy range. I have emphasised above that inelastic diffractive processes are concentrated at large impact parameters and that nonperturbative effects (for example two-pion cut in the pomeron trajectory) are important in this region. The fit of GLM to total pp-interaction cross section, differential cross section of elastic scattering and SD and DD-integrated cross sections [20] lead to the value $\Delta = 0.33$ for the intercept of the pomeron. Note that the threshold effects, discussed above, have not been taken into account both by GLM and KMR groups

A general feature of models, which take into account interactions between pomerons ("enhanced" diagrams), is a slower increase with energy of total cross sections. For example predictions of both KMR and GLM models for the total cross section of pp interaction at LHC energy are close to 90 mb, which is substantially smaller than in models without these interactions. Same effect exists in the model of Ref. [11], though the corresponding cross section is closer to 100 mb. Values of the pomeron intercept is substantially higher than in the eikonal-type models.

There is an interesting problem of influence of pomeron interactions on survival probabilities for hard processes [8, 23, 24]. The largest difference in KMR and GLM models is in predictions for survival probabilities due to enhanced diagrams. For DPE Higgs production at LHC an account of threshold effects is very important and calculation of KMR in a simplified model, but with account of these effects [25], show that for central Higgs production an extra suppression due to pomeron interactions is insignificant. On the other hand in GLM model a modification of survival probabilities due to enhanced diagrams is very strong: at LHC there is a decrease by a factor $\approx 16$. For DPE processes at Tevatron GLM model predicts a decrease of survival probability by an extra factor 3.5. It is not clear how these factors depend on the mass of the produced hard system. A comparison of CDF data on diffractive dijet production [7] with prediction based on QCD factorisation and survival factor of two channel eikonal model show that extra suppression due to enhanced diagrams does not exceed 50%. Analogous restriction follows from CDF data on DPE dijet production [9, 26].

Thus up to energies of Tevatron interaction between pomerons play a minor role in hard diffractive processes. This is to a large extent related to the phase-space limitations. For soft diffraction enhanced diagrams are important and lead to a change of parameters of the "bare" pomeron in reggeon theory. At LHC effects of enhanced diagrams will be observable in hard diffractive processes. Their influence on survival probabilities can be studied, in particular, in diffractive production of jets (with not too large masses).

My main conclusions are:
unitarity effects due to multi-pomeron interactions are very important for diffractive processes at very high energies. They lead to a violation of both Regge and QCD-factorisations for hard diffraction.

Inelastic diffraction is peripheral in impact parameter space and account of $\pi\pi$-cut in the pomeron trajectory is necessary for proper description of amplitudes.

Experimental investigation of diffractive processes at LHC will give an important information on QCD–dynamics at high energies.

# Acknowledgements

I thank E. Gotsman, V. Khoze, U. Maor and M. Poghosyan for useful discussions. This work was partially supported by the grants RFBR 0602-72041-MNTI, 0602-17012, 0802-00677a and Nsh-4961.2008.2.

# References

[1] E.L. Feinberg, I.Ya. Pomeranchuk, Doklady Akad. Nauk SSSR, **94** 439 (1953); Nuovo Cim. Suppl. **III** 652 (1956).

[2] M.L. Good and W.D. Walker, Phys. Rev. **126** 1857 (1960).

[3] A.B. Kaidalov, Phys. Rep. **50** 157 (1979).

[4] V.N. Gribov, Sov. Phys. JETP **19** 483 (1969).

[5] V.N. Gribov, I.Ya. Pomeranchuk, Nucl. Phys. **38** 516 (1962).

[6] A.A. Anselm, V.N. Gribov, Phys. Lett. **B40** 487 (1972).

[7] CDF Collaboration: T. Affolder et al., Phys. Rev. Lett. **84** 5043 (2000).

[8] A.B. Kaidalov, V.A. Khoze, A.D. Martin and M.G. Ryskin, Eur. Phys. J. **C21** 521 (2001).

[9] CDF Collaboration, T. Affolder et al., Phys.Rev.Lett. **85** 4215 (2000).

[10] A.B. Kaidalov, V.A. Khoze, A.D. Martin and M.G. Ryskin, Phys. Lett. **B559** 235 (2003).

[11] A.B. Kaidalov, L.A. Ponomarev, K.A. Ter-Martirosyan, Sov. J. Nucl. Phys. **44** 486 (1986).

[12] E.G.S. Luna, V.A. Khoze, A.D. Martin and M.G. Ryskin, Eur. Phys. J. **C59** 1 (2009).

[13] M. Poghosyan, Talk at EDS 09 (2009).

[14] V. A. Abramovsky, V. N. Gribov and O. V. Kancheli, Sov. J. Nucl. Phys. **18** 308 (1974).

[15] S. Ostapchenko, Phys. Lett. **B636**; Phys. Rev. **D74**, 014026 (2006).

[16] S. Ostapchenko, Phys. Rev. **D77**, 034009 (2008).

[17] A.B. Kaidalov, Phys. Lett. **116B** 459 (1982); A.B. Kaidalov and K.A. Ter-Martirosyan, Phys. Lett. **117B** 247 (1982); A.B. Kaidalov and K.A. Ter-Martirosyan, Sov. J. Nucl. Phys. **39** 979 (1984); **40** 135 (1984).

[18] M.G. Ryskin, A.D. Martin and V.A. Khoze, Eur. Phys. J. **C54** 199 (2008).

[19] A.D. Martin, M.G. Ryskin and V.A. Khoze, arXiv:0903.2980v2.

[20] E. Gotsman, E. Levin, U. Maor, J.S. Miller, Eur. Phys. J. **C57** 689 (20080.

[21] D. Amati et al., Nucl. Phys. **B112** 107 (1976); D. Amati et al., Nucl. Phys. **B114** 483 (1976).

[22] M.A. Brown and G.P. Vacca, Eur. Phys. J. **C50** 857 (2007).

[23] J. Bartels, S. Bondarenko, K. Kutak, L. Motyka, Phys. Rev. **D73**, 093004 (2006).

[24] L. Frankfurt, C.E. Hyde, M. Strikman, C. Weiss, arXiv:hep-ph/0710.2942;0808.0182.

[25] A.D. Martin, V.A. Khoze and M.G. Ryskin, arXiv:hep-ph/0803.3939.

[26] K. Hatakeyama, Proc. of ISMD 2008, DESY-PROC-2009-01, p.507.

# Description of Soft Diffraction in the Framework of Reggeon Calculus. Predictions for LHC

*Alexei Kaidalov*[1], *Martin Poghosyan*[2]

[1] Institute of Theoretical and Experimental Physics, 117526 Moscow, Russia
[2] Università di Torino and INFN, 10125 Torino, Italy

A model, based on Gribov's Reggeon calculus, is proposed and applied to processes of soft diffraction at high energies. It is shown that by accounting for absorptive corrections for all legs of triple-Regge and loop diagrams a good description of experimental data on inelastic soft diffraction can be obtained. In this paper we give a brief description of the model and of its predictions for LHC energies.

## 1 Introduction

The process of soft single- and double- diffraction dissociation are closely related to small angle elastic scattering in which each of the incoming hadrons may become a system which will then decay into a number of stable final state particles. Regge-pole theory is the main method for description of high-energy soft processes. In this approach (see [1]), the inclusive cross-section of single and double diffraction dissociations is described by triple-Reggeon and loop diagrams, respectively. Triple-Reggeon description is in good agreement with the FNAL and ISR data for soft diffraction dissociation [2]. However, the higher-energy data from SPS and Tevatron do not show the increase of the cross section with energy expected from the simple fits and the contribution of triple-Pomeron vertex (in the elastic scattering amplitude) violates unitarity. A number of different approaches have been proposed in order to be in agreement with the data from higher energy experiments: non-gaussian parameterisation for Reggeon-hadron vertex [3], renormalisation [4] or damping [5] of the Pomeron flux. A more realistic approach suggested in [6] and [7] by the inclusion of initial state elastic scattering corrections to the triple-Reggeon vertices. However, the analysis done in [8] shows that this correction is not enough for restoring the $s$-channel unitarity.

Besides of its own role, the theoretical knowledge of soft diffraction is also important in analysing hard diffraction data, which is an active area of study at HERA and Tevatron and will continue to be interesting at LHC. The knowledge of the interaction between Pomerons (enhanced diagrams) is important for analysing data at very high energies, too. The contribution of these diagrams can be essential in hadron-nucleus and especially in nucleus-nucleus collisions, where the thermalisation and the quark-gluon plasma formation strongly depend on the strength of interactions between Pomerons [9].

In this article we propose to describe data on soft diffraction dissociation in $pp$ and $p\bar{p}$ interactions taking into account all possible non-enhanced absorptive corrections to triple-Regge vertices and loop diagrams. This approach describes available data on high-mass soft diffraction in the energy range from ISR, FNAL to Tevatron. The article is organised as follows: In the next two sections we briefly describe the Regge-pole approach, Gribovs' Reggeon calculus and AGK cutting rules. Our proposed model is presented in Section 4 and its predictions are compared with data in Section 5.

## 2 Single Regge-Pole Approximation

In Regge theory, the simplest singularity in the $j$-plane is a moving pole $\alpha(t)$ in the $t$-channel (see the leftmost graph in Fig. 1) and in the small $t$-region the scattering amplitude, $M(s,t)$, of the process $a + b \to c + d$ can be parameterised as:

$$M(s,t) = \gamma(0)\eta(\alpha(0))(s/s_0)^{\alpha(0)-1}\exp(\lambda(s)t). \tag{1}$$

Here $\eta(\alpha(t))$ is the signature factor, $\gamma(t) \equiv g_{ac}(t)g_{bd}(t)$ is the factorisation residue, and $\lambda(s) = R^2 + \alpha'_R \ln(s/s_0)$. The parameter $R^2$ characterises the $t$-dependence of the product of residue function and of the signature factor. In our notations the normalisation of the scattering amplitude is such that $\sigma_{tot} = 8\pi \operatorname{Im} M(s,0)$ and $d\sigma_{el}/dt = 4\pi|M(s,t)|^2$.
Because the Pomeron's intercept is larger than unity (which is required in order to guarantee the growth of the total cross-section, $\sigma_{tot} \sim s^{\Delta}$ where $\Delta \equiv \alpha_P(0) - 1$), the corresponding cross-section grows as a power function of $s$ and therefore the contribution of the Pomeron-pole in the scattering amplitude violates unitarity. The easiest way to restore the unitarity is to take into account branch points which correspond to multi-Reggeon exchange. The calculation of the multi-Reggeon exchange amplitude is possible in eikonal (or eikonal-like) approximation, where only single particle intermediate states are taken into account.

## 3 Eikonal Approximation and AGK Cutting Rules

Regge poles are not the only singularities in the complex angular momentum plane. Exchange of several Reggeons in $t$-channel leads to moving branch points in the $j$-plane (Fig. 1). A Regge pole exchange can be interpreted as corresponding to single scattering while Regge cuts correspond to multiple scatterings on constituents of hadrons. In case of the 'super-critical' Pomeron ($\Delta > 0$) the contribution of $n$-Pomeron exchange in the scattering amplitude ($M_P^{(n)}(s,0) \sim s^{n\Delta}$) is increasing with the increase of the energy and the entire series of $n$-Pomeron exchange should be summed. On the contrary, the contribution of the branch points concerned with the exchange of several secondary Reggeons decreases very quickly with increasing collision energy and the contribution of such branch points can be neglected with respect to the branch points due to the exchange of one secondary Reggeon and Pomerons that are needed for properly matching low energy data. For instance, in eikonal approximation the amplitude of $n$-Pomeron exchange can be written in the following form [10]:

$$M^{(n)}(s,t) = \frac{(2i)^{n-1}}{n!} \int \prod_{i=1}^{n} \left[ M^{(1)}(s, \mathbf{q}_{i\perp}^2) \frac{d^2 \mathbf{q}_{i\perp}}{2\pi} \right] \delta \left( \mathbf{q}_\perp - \sum_{i=1}^{n} \mathbf{q}_{i\perp} \right) . \qquad (2)$$

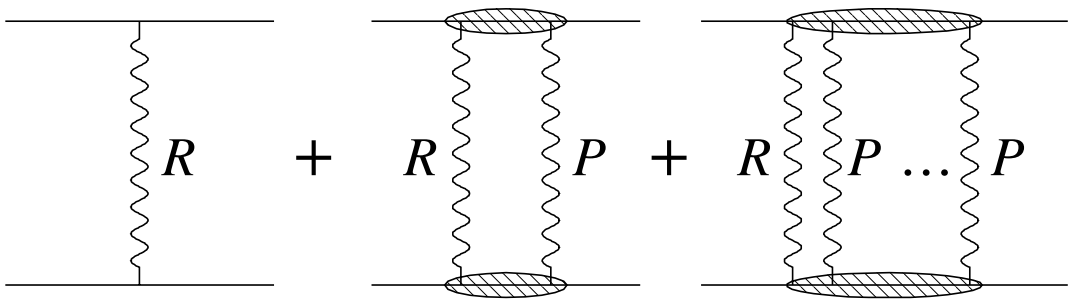

Figure 1: Single pole and $RP^n$ cut contribution in the elastic scattering amplitude. $R$ stands for secondary Reggeon and for Pomeron.

Using the parameterisation (1) for the Regge pole contribution and performing the integration over the transverse momenta of Reggeons in Eq. (2) it can be shown that the account of the multi-Pomeron exchanges results in the unitarisation of the scattering amplitude, which leads to the Froissart behaviour of the total cross section for $s \gg m_N^2$: $\sigma_{tot} \simeq 8\pi\alpha'_P \Delta \ln^2(s/s_0)$.
In the language of Regge poles the multiparticle production processes are related to cut-Reggeon diagrams. Abramovski, Kancheli and Gribov (AGK) proposed rules [11] for calculating the discontinuity of the matrix element that represent the generalisation of the Optical Theorem for the case of multi-Pomeron exchange. The basic results of AGK needed for the following discussion are: a) There is one and only one cut-plane which separates the initial and final states of the scattering. b) Each cut-Pomeron gives an extra factor of $(-2)$ due to the discontinuity of the Pomeron amplitude. c) Each un-cut Pomeron obtains an extra factor of 2 since it can be placed on both sides of the cut-plane.

# 4 The Model

We propose to describe single- and double- diffraction processes by such diagrams where any number of Pomeron exchanges is taken into account together with each $R$ of the triple-Reggeon and loop diagrams and as well as the screening corrections are considered, as shown in Fig. 2. In this figure the solid line accompanied with a dashed line corresponds to one Reggeon (Pomeron or secondary-Reggeon) exchange together with any number Pomeron exchange. The double-dashed lines stand for eikonal screening.

The theory does not give any prediction on the structure of the vertices for $n$ Pomeron to $m$ Pomeron transitions. The simplest approximation is to assume an eikonal-type structure. In this approximation the general approach of constructing elastic scattering amplitude with account of enhanced diagrams has been proposed in [12]. Assuming $\pi$-meson exchange dominance of multi-Pomeron interaction vertices, the authors summed high order enhanced diagrams iterating multi-Pomeron vertices in both, $s$- and $t$- channels. In that article it was demonstrated that the inclusion of these diagrams in most of the cases leads to predictions that are very close to the results of eikonal type models, where a Pomeron with suitably renormalised intercept is used.

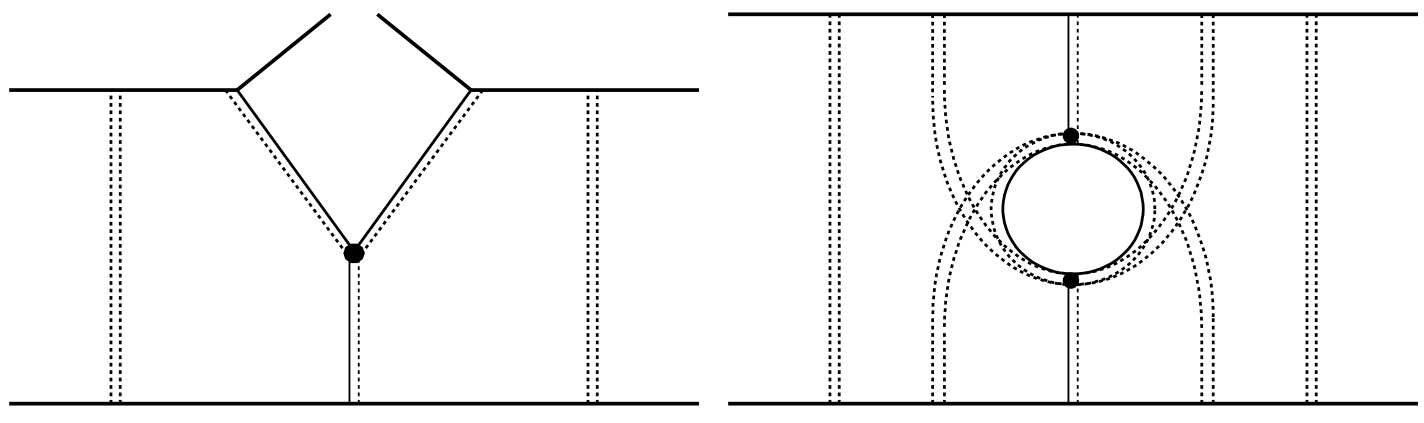

Figure 2: Eikonalised triple-Reggeon and loop diagrams which are proposed to describe single- and double-diffractive processes in hadron-hadron collisions.

For calculating the diagrams shown in Fig. 2 we assume the mentioned $\pi$-meson exchange dominance of multi-Pomeron interaction vertex corresponding to the transition of $n$ Pomerons

to $m$ Pomerons:

$$\lambda^{(n,m)} = r_{3P} g_\pi^{n+m-3} \exp\left(-R_\pi^2 \sum_{i=1}^{n+m} q_i^2\right). \quad (3)$$

As a secondary Regge pole we consider $f$-trajectory. The conservation laws allow us to assume the same pion dominance at the same transition with participation of $f$-trajectory. In these terms according to the AGK cutting rules, the cross-section corresponding to the cut dressed triple-Reggeon graph for the process $a + b \to X + b$ has the form[1]:

$$\frac{d\sigma}{d\zeta} = \frac{1}{2} \sum_{i,j,k=P,R} G_{ijk} \int d\mathbf{b}\, d\mathbf{b_1}\, \Gamma^i_{b\pi}(\zeta_2, \mathbf{b_2}) \Gamma^j_{b\pi}(\zeta_2, \mathbf{b_2}) \Gamma^k_{a\pi}(\zeta, \mathbf{b_1}) \exp\{-2\Omega_{ab}(\xi, \mathbf{b})\} \quad (4)$$

Here we introduce the following notations

$$\zeta = \ln(M_X^2/s_0),\ \zeta_2 = \xi - \zeta,\ \mathbf{b_2} = \mathbf{b} - \mathbf{b_1},\ \Omega_{\alpha\beta}(\zeta, \mathbf{b}) = \frac{g_{\alpha\beta}}{\lambda_{\alpha\beta}} \exp\left\{\Delta\zeta - \frac{\mathbf{b}^2}{4\lambda_{\alpha\beta}}\right\}, \quad (5)$$

$$\Gamma^P_{\alpha\beta}(\zeta, \mathbf{b}) = 1 - e^{-\Omega_{\alpha\beta}(\zeta, \mathbf{b})},\ \Gamma^R_{\alpha\beta}(\zeta, \mathbf{b}) = \frac{g^R_{\alpha\beta}}{\lambda^R_{\alpha\beta}} \exp\left\{(\alpha_R - 1)\zeta - \frac{\mathbf{b}^2}{4\lambda^R_{\alpha\beta}} - \Omega_{\alpha\beta}(\zeta, \mathbf{b})\right\}.$$

$G_{ijk}$ stand for triple-Regge vertices strength. The expression of the cross-section in the $(\zeta, t)$-space is rather long and we do not present it here.

Analogously can be calculated the cross-section corresponding to the cut dressed loop diagram standing for the process $a + b \to X_1 + X_2$, and it has the following form:

$$\frac{d\sigma}{d\zeta_1 d\zeta_2} = \frac{1}{4} \sum_{i,j,k,l=P,R} G_{ijk} G_{lik} \int d\mathbf{b}\, d\mathbf{b_1}\, d\mathbf{b_2}\, \Gamma^i_{\pi a}(\zeta_1, \mathbf{b_1}) \Gamma^l_{\pi b}(\zeta_2, \mathbf{b_2}) \Gamma^j_{\pi\pi}(\zeta_3, \mathbf{b_3}) \Gamma^k_{\pi\pi}(\zeta_3, \mathbf{b_3})$$
$$\times \exp\{-2\Omega_{ab}(\xi, \mathbf{b}) - 2\Omega_{a\pi}(\xi - \zeta_1, \mathbf{b} - \mathbf{b_1}) - 2\Omega_{b\pi}(\xi - \zeta_2, \mathbf{b} - \mathbf{b_2})\} \quad (6)$$

Here in addition to (5) we used the following notations:

$$\zeta_1 = \ln(M_{X_1}^2/s_0),\ \zeta_2 = \ln(M_{X_2}^2/s_0),\ \zeta_3 = \xi - \zeta_1 - \zeta_2,\ \mathbf{b_3} = \mathbf{b} - \mathbf{b_1} - \mathbf{b_2}$$

# 5 Extraction of the Parameters from Experimental Data

Because we do not consider the contribution of the enhanced diagrams in the elastic scattering amplitude it allows us to differentiate data fitting procedure and realise it by two steps. At the first step we fix secondary-Reggeon and Pomeron parameters. The trajectories of secondary-Reggeons are fixed from fit to data on spin vs. mass for corresponding family of mesons, and the following results are found: $\alpha_f(t) = 0.7 + 0.8t$, $\alpha_\omega(t) = 0.4 + 0.9t$, $\alpha_\rho(t) = 0.5 + 0.9t$. Then the residues of secondary-Reggeons and the residues/trajectory of Pomeron are found from fit to data on elastic scattering and total interaction cross-section. At the second step we fix triple-Reggeon interaction vertices' constants from fit to data on high mass soft-diffraction dissociation, using the values of the parameters fixed in the first step as an input.

We take into account $P$-, $f$- and $\omega$- poles in $pp$ and $p\bar{p}$ elastic scattering amplitude. Since we assume pion exchange dominance at the coupling of Reggeons, we fix the parameters of secondary Reggeon- and Pomeron- pion coupling as well. In $\pi^{\pm}p$ elastic scattering amplitude, we take into account $P$-, $f$- and $\rho$- poles. Thus, we assume $M = M_P + M_f \pm M_\omega$ for $pp$ and $p\bar{p}$ collisions and $M = M_P + M_f \pm M_\rho$ for $\pi^+p$ and $\pi^-p$ collisions, respectively. For Pomeron-trajectory we have found the following parameterisation: $\alpha_P(t) = 1.117 \pm 0.252t$, and other parameters are listed in the Tables 1 and 2.

[1] If the transferred momentum is very low and the mass of the diffracted system is high the $\pi$-meson exchange plays an important role. This we take into account based on the OPER model [13].

Table 1: $p$+$p(\bar{p})$ interaction parameters in GeV$^{-2}$ units.

| | |
|---|---|
| $g_N$ | $= 1.366 \pm 0.004$ |
| $R_N^2$ | $= 1.428 \pm 0.006$ |
| $g_N^f$ | $= 2.871 \pm 0.008$ |
| $R_N^{f2}$ | $= 0.918 \pm 0.023$ |
| $g_N^\omega$ | $= 2.241 \pm 0.074$ |
| $R_N^{\omega 2}$ | $= 0.945 \pm 0.026$ |

Table 2: $\pi^{\pm}p$ interaction parameters in GeV$^{-2}$ units.

| | |
|---|---|
| $g_\pi$ | $= 0.85 \pm 0.0004$ |
| $R_\pi^2$ | $= 0.5 \pm 0.002$ |
| $g_{\pi N}^f$ | $= 3.524 \pm 0.001$ |
| $R_{\pi N}^{f2}$ | $= 1. \pm 0.001$ |
| $g_{\pi N}^\rho$ | $= 1.12 \pm 0.017$ |
| $R_{\pi N}^{\rho 2}$ | $= 9.19 \pm 0.837$ |

Table 3: Found values of $G_{ijk}$ in GeV$^{-2}$ units.

| | | |
|---|---|---|
| $G_{PPP}$ | =0.0098 | $\pm$0.0005 |
| $G_{PPR}$ | =0.03 | $\pm$0.004 |
| $G_{RRP}$ | =0.005 | $\pm$0.001 |
| $G_{RRR}$ | =0.05 | $\pm$0.002 |
| $G_{PRP}$ | =0.013 | $\pm$0.001 |
| $G_{PRR}$ | =0.033 | $\pm$0.005 |

Next we fix the triple-Reggeon vertices strengths ($G_{ijk}$) from fit to data on soft single-diffraction dissociation in $pp$ and $p\bar{p}$ interactions. We used the available data on spectra of non-diffracted proton from fixed-target experiments [14] and [15], from ISR [16] and from CDF [4]. Being interested in soft diffraction, we have chosen measurements done for $d^2\sigma/d\zeta dt$ within the diffractive cone (-$t \leq 0.2\,\mathrm{GeV}^2$). The values of $G_{ijk}$ found are reported in Table 3 and the fit result is compared with data in Figs. 3-5.

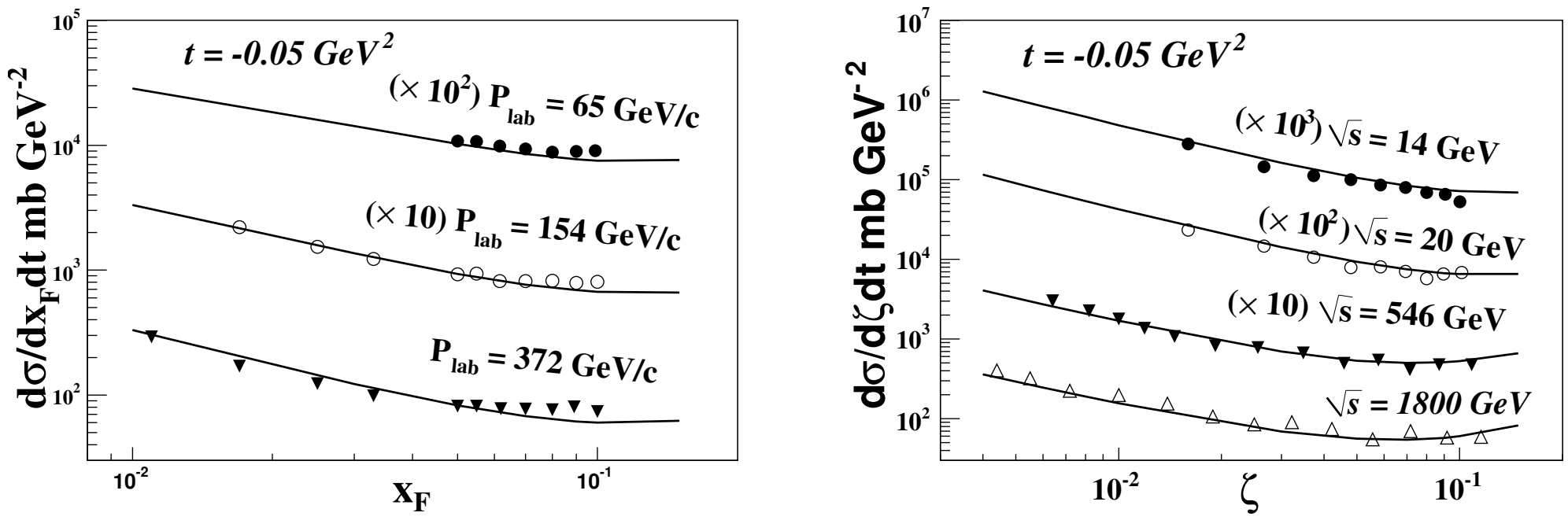

Figure 3: Double differential cross-section $d^2\sigma/d\zeta dt$ for $p(\bar{p}) + p \to p(\bar{p}) + X$ measured at Fermilab at various $\sqrt{s}$ and fixed $t$. The data are taken from [4, 15].

In Fig. 6 we compare predictions of the model on single-diffractive integrated cross-sections with experimental data [17]. In each case the integration is done in accordance with the corresponding measurement as they are indicated in the figure.

In Fig. 7 we compare predictions of the model on double-diffractive integrated cross-sections with experimental data [18, 19]. The data at $\sqrt{s}$ >100 GeV correspond to the cross-section for minimum 3 units of rapidity gap between two produced clusters and are taken from [18]. The rest of data (at $\sqrt{s}$ <100 GeV) are taken from [19] where exclusively and semi-inclusively measured data are reduced to totally inclusive cross-section. The theoretical curve is calculated using Eq. (6) and requiring minimum 3 units of rapidity gap between two diffracted clusters.

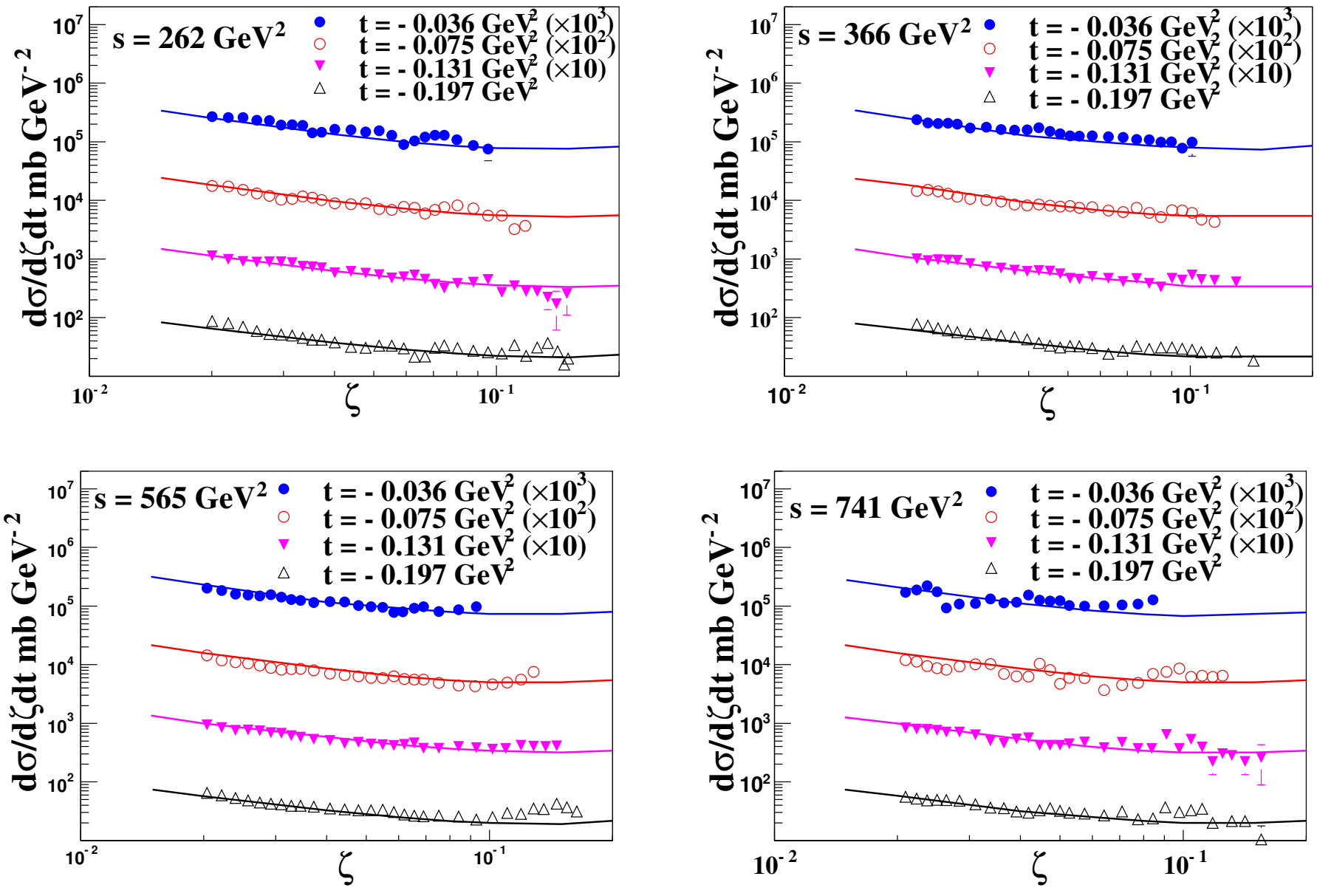

Figure 4: Double differential cross-section $d^2\sigma/d\zeta dt$ for $pp \to pX$ measured at Fermilab at various $\sqrt{s}$ and $t$. The data are taken from [14].

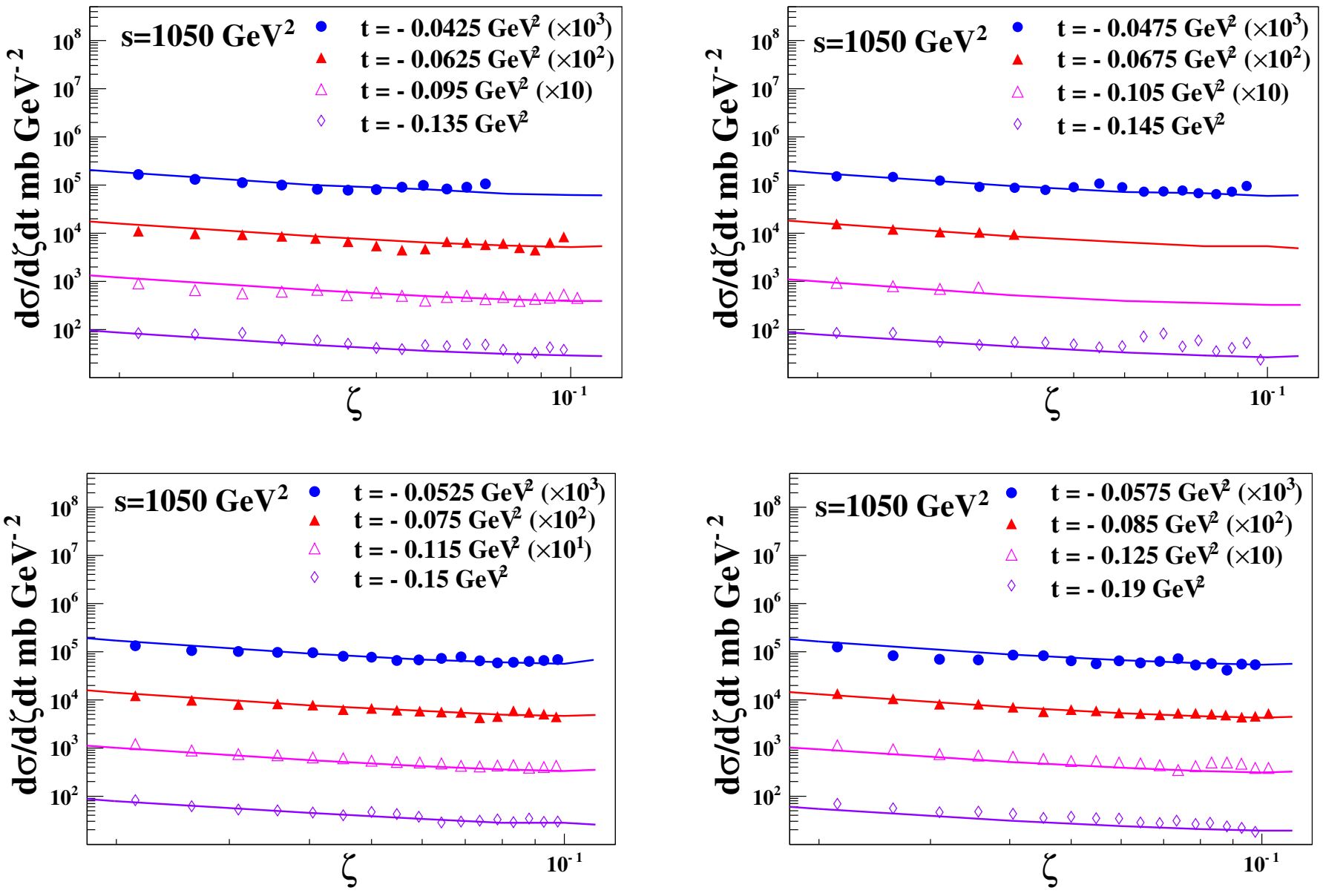

Figure 5: Double differential cross-section $d^2\sigma/d\zeta dt$ for $pp \to pX$ measured at ISR at various $\sqrt{s}$ and $t$. The data are taken from [16].

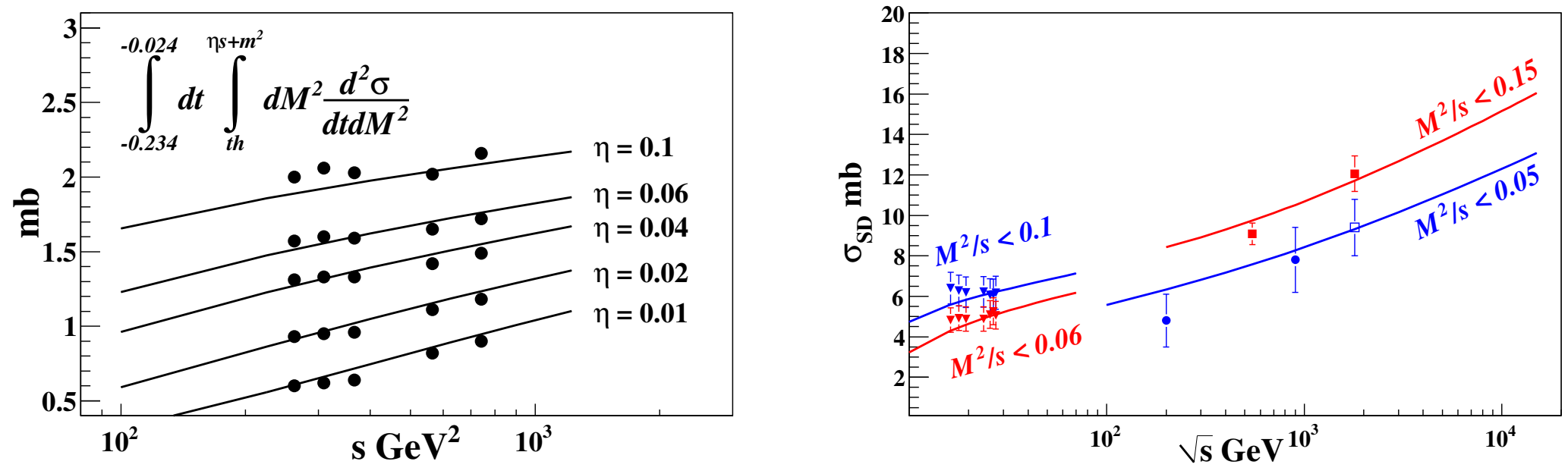

Figure 6: Integrated single-diffractive cross-section as a function of $\sqrt{s}$. The integrations are done in accordance with corresponding measurement as they are indicated in the plots.

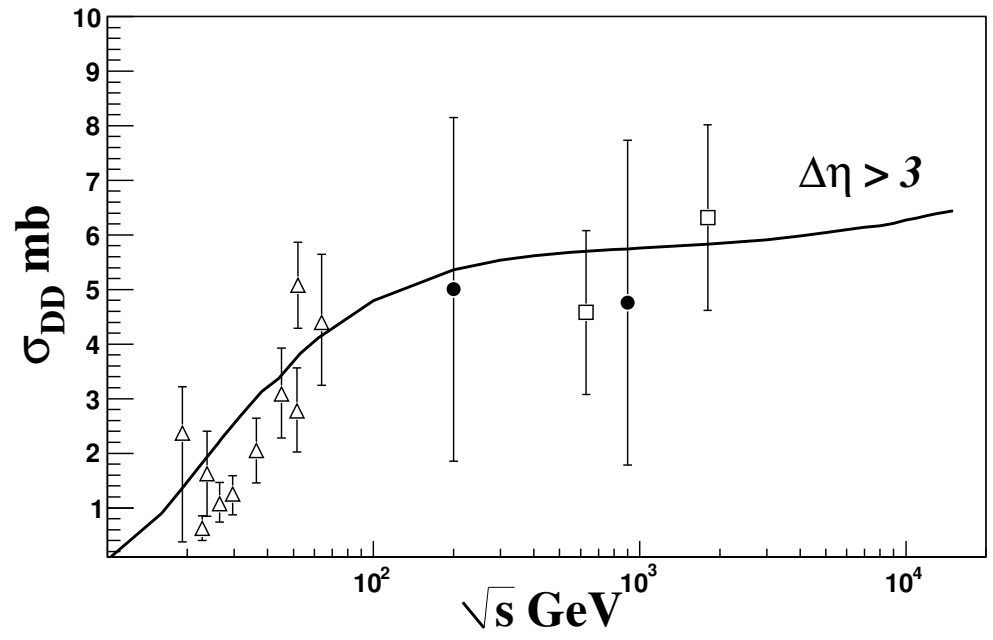

Figure 7: Double-diffractive cross-section as a function of $\sqrt{s}$ . The theoretical curve is calculated requiring minimum 3 units of rapidity gap between two diffracted clusters.

# 6 Summary and Predictions for LHC

In this article we report the results of calculations of all non-enhanced absorptive corrections to triple-Regge vertices and loop diagrams in eikonal approximation using Gribov's Reggeon calculus. Numerically evaluating the model we have found a good description of data on high-mass soft diffraction dissociation in the energy range from ISR, FNAL to Tevatron (from $P_{lab} =$ 65 GeV/c to $\sqrt{s} = 1800$ GeV). It is worth to emphasise that such a detailed description of inclusive diffraction in this broad region of energies is achieved for the first time. In Table 4 we present the predictions of the model on single- and double- diffractions cross-section for different energies of LHC. The single-diffractive cross-section is obtained integrating over masses up to $M^2/s = 0.05$, and the double-diffractive cross-section is obtained requiring minimum 3 units of rapidity gap between two diffracted clusters.

| $\sqrt{s}$ TeV | $\sigma_{SD}$ mb | $\sigma_{DD}$ mb |
|---|---|---|
| 0.9 | 8.2 | 5.7 |
| 7 | 11.6 | 6.1 |
| 10 | 12 | 6.2 |
| 14 | 13 | 6.4 |

Table 4: Predictions for LHC.

## Acknowledgements

We thank A. Grigiryan, J.-P. Revol and K. Safarik for their interest in this work.

## References

[1] A.B. Kaidalov, Phys. Rep. **50** 157 (1979).

[2] A.B. Kaidalov *et al.*, Phys. Lett. **B45** 493 (1973); JETP Lett. **17** 440 (1973).

[3] A. Donnachie and P.V. Landshoff, arXiv:hep-ph/0305246;
R. Fiore *et al.*, Phys.Rev. **D61** 034004 (2000).

[4] K. Goulianos and J. Montanha, Phys. Rev. **D59** 114017 (1999).

[5] E. Erhan and P. Schlein, Phys. Lett. **B427** 389 (1998).

[6] E. Gotsman *et al.*, Phys. Rev. **D49** R4321 (1994).

[7] E.G.S. Luna *et al.*, Eur. Phys. J. **C59** 1 (2009).

[8] E.S. Martynov and B.V. Struminsky, Yad.Fiz. **59** 1817 (1996) [Physics of Atomic Nuclei **59** 1755 (1996)].

[9] A.B. Kaidalov, Nucl. Phys. **A525** 39c (1991).

[10] K.A. Ter-Martirosyan, Preprints ITEP No. 70, 71 (1975); No. 7, 11, 133, 134, 135, 158 (1976).

[11] V.A. Abramovski, O.V. Kanceli and V.N. Gribov, Sov. J. Nucl. Phys. **18** 308 (1974).

[12] A.B. Kaidalov, L.A. Ponomarev and K.A. Ter-Martirosyan, Sov. J. Nucl. Phys. **44** 483 (1986).

[13] K.G. Boreskov *et al.*, Sov. J. Nucl. Phys. **15** 203 (1972); **17** 669 (1973); **19** 565 (1974).

[14] R.D. Schamberger *et al.*, Phys. Rev. **D17** 1268 (1978).

[15] Y. Akimov *et al.*, Phys. Rev. Lett. **39** 1432 (1977).

[16] J.C.M. Armitage *et al.*, Nucl. Phys. **B194** 365 (1982).

[17] R.E. Ansorge *el al.*, Z. Phys. **C33** 175 (1986); G.J. Alner *et al.*, Phys. Rept. **154** 247 (1987); F. Abe *et al.*, Phys. Rev. **D50** 5535 (1994); N. Amos *et al.*, Phys. Lett. **B301** 313 (1993).

[18] T. Affolder *et al.*, Phys. Rev. Lett. **87** 141802-1 (2001).

[19] G. Alberi and G. Goggi, Phys. Rept. **74** 1 (1981).

# Soft Scattering Re–Visited

*Uri Maor*∗

Department of Particle Physics, School of Physics and Astronomy
Raymond and Beverly Sackler Faculty of Exact Science
Tel Aviv University, Tel Aviv 69978, Israel

An updated formulation of soft diffraction, compatible with unitarity is presented. Its consequent soft scattering features are explored. The critical interplay between theory and and data analysis will be discussed.

## 1 Introduction

The present vigorous studies of soft scattering and Pomeron ($I\!\!P$) physics are based on sophisticated utilisation of relatively old theoretical ideas and models, such as Gribov's Reggeon field theory [1], Good and Walker (GW) decomposition of the proton wave function accounting for low mass diffraction [2] and the eikonal approximation [3] which secures the compatibility of the scattering amplitudes with $s$ channel unitarity. Compliance with $t$ channel unitarity is associated with multi Pomeron interactions ($I\!\!P$ enhancement) which are a generalisation of Mueller's triple Pomeron mechanism [4] provided $G_{3I\!\!P}$, the triple Pomeron coupling, is large enough. This mechanism supplements GW diffraction with an additional high mass diffraction.

In this talk I shall discuss the modelling and predictions derived from the above dynamical considerations, stressing the critical interplay between theory and data analysis. The implied gap survival probabilities will be discussed by Gotsman in the following talk. I shall assume a Regge like parametrisation in which the $I\!\!P$ is super critical, i.e. $\alpha_{I\!\!P}(t) = 1 + \Delta_{I\!\!P} + \alpha'_{I\!\!P} t$, where $\Delta_{I\!\!P} > 1$. The above $I\!\!P$ exchange violates $s$-unitarity at high energies. Recall that implementing $s$-unitarity is model dependent. I shall confine myself to eikonal models which have the virtue of simplicity.

In the ISR-Tevatron range $\sigma_{tot}$ and $\sigma_{el}$ are well reproduced by Donnachie-Landshoff (DL) non screened Regge parametrisation with $\Delta_{I\!\!P} = 0.08$ and $\alpha'_{I\!\!P} = 0.25\,\mathrm{GeV}^{-2}$. The energy dependence of the soft diffractive cross sections (notably $\sigma_{sd}$) is much milder, implying that strong screenings initiated by $s$-unitarity must be taken into account. As we shall see, the interplay between theory and data analysis results in strong constraints on both $\Delta_{I\!\!P}$ and $\alpha'_{I\!\!P}$ inputs. This results has profound consequences for the nature of the Pomeron and its QCD foundations, suggesting a unifying interpretation of soft and hard Pomerons.

∗e-mail: maor@post.tau.ac.il

## 2 Good-Walker Eikonal Models

Consider a system of two states, a hadron $\Psi_h$ and a diffractive state $\Psi_D$ which are orthonormal. The GW mechanism stems from the observation that these states do not diagonalise the 2x2 interaction matrix $\mathbf{T}$. Lets introduce two wave functions $\Psi_1$ and $\Psi_2$ which diagonalise $\mathbf{T}$,

$$A_{i,k}^{i',k'} =< \Psi_i\,\Psi_k|\mathbf{T}|\Psi_{i'}\,\Psi_{k'} >= A_{i,k}\,\delta_{i,i'}\,\delta_{k,k'}. \tag{1}$$

i.e. the $A_{i,k}$ amplitudes are constructed from the elastic scattering of $\Psi_i$ and $\Psi_k$. In this representation the observed hadronic states are written

$$\Psi_h = \alpha\,\Psi_1 + \beta\,\Psi_2 \qquad \Psi_D = -\beta\,\Psi_1 + \alpha\,\Psi_2\,, \tag{2}$$

where $\alpha^2 + \beta^2 = 1$. The corresponding unitarity equations are

$$\mathrm{Im}\,A_{i,k}^{S}\,(s,b) = |A_{i,k}^{S}\,(s,b)\,|^2 + G_{i,k}^{in}(s,b), \tag{3}$$

where $G_{i,k}^{in}$ is the summed probability for all non GW inelastic processes induced by an initial $(i,k)$ state. A general solution of Eq. (3) can be written as

$$A_{i,k}^{S}(s,b) = i\left(1 - \exp\left(-\frac{\Omega_{i,k}^{S}(s,b)}{2}\right)\right), \tag{4}$$

$$G_{i,k}^{in}(s,b) = 1 - \exp\left(-\Omega_{i,k}^{S}(s,b)\right), \tag{5}$$

where $\Omega_{i,k}^{S}$ are arbitrary. In the eikonal approximation $\Omega_{i,k}^{S}$ are assumed to be real and determined by the Born (non-screened) input. From Eq. (5) we deduce that the probability that the initial projectiles $(i,k)$ reach the final LRG diffractive interaction unchanged, regardless of the initial state re-scatterings, is given by $P_{i,k}^{S} = \exp\left(-\Omega_{i,k}^{S}(s,b)\right)$. In general, we have to consider four possible $(i,k)$ elastic re-scattering options. For initial $p$-$p$ (or $\bar{p}$-$p$) the two off diagonal amplitudes are equal, $A_{1,2}^{S} = A_{2,1}^{S}$. The corresponding elastic, SD and DD amplitudes are

$$a_{el}(s,b) = i\{\alpha^4 A_{1,1}^{S} + 2\alpha^2\beta^2 A_{1,2}^{S} + \beta^4 \mathcal{A}_{2,2}^{S}\}, \tag{6}$$

$$a_{sd}(s,b) = i\alpha\beta\{-\alpha^2 A_{1,1}^{S} + (\alpha^2 - \beta^2)A_{1,2}^{S} + \beta^2 A_{2,2}^{S}\}, \tag{7}$$

$$a_{dd} = i\alpha^2\beta^2\{A_{1,1}^{S} - 2A_{1,2}^{S} + A_{2,2}^{S}\}. \tag{8}$$

The GW mechanism was originally conceived so as to describe a system of a nucleon plus its diffractive $N^*$ isobars. Obviously, this simplistic approach is not suitable for high energy diffraction where $M_{diff}^2$ is bounded by $0.05s$ [1], implying a continua of diffractive Fock states. Throughout this talk I shall relate to GLMM [5] and KMR [6] models which are conceptually very similar, but differ significantly in both their formalism and data analysis. In the present context, two procedures were devised to overcome the above difficulty: GLMM lump together all GW diffractive states to an effective $|D>$ state, to which we add the non GW $I\!P$ enhanced high mass diffraction. In this approach the GW contribution is very significant and the mass distribution is smooth. KMR and LKMR [7] chose to confine GW diffraction to low

[1] this is an arbitrary bound commonly used.

$M^2_{diff} < 10\,\mathrm{GeV}^2$, to which they add the high mass $I\!P$ enhanced contribution. In this approximation the bulk of the diffractive mass is non GW, and its smoothness at $10\,\mathrm{GeV}^2$ is not secured.

GLMM, KMR and LKMR are multi channel eikonal models in which the initial re-scatterings of the incoming projectiles includes also GW diffractive states.

$$\Omega^S_{i,k}(s,b) = \nu^S_{i,k}(s)\,\Gamma^S_{i,k}(s,b,...)\,, \tag{9}$$

where $\nu^S_{i,k}(s) = g_i g_k(\frac{s}{s_0})^{\Delta_{I\!P}}$ and $\Gamma^S_{i,k}$ are the $b$-space profiles. The profiles are external information in as much as, beside their normalization and asymptotic constraints on their behaviour, they are determined by the data analysis. In GLMM $\Gamma^S_{i,k}$ are given as the $b$-transform of a two $t$-poles expression ($t = -q^2$). Setting $\alpha'_{I\!P}$=0, the profiles are energy independent

$$\frac{1}{(1\,+\,q^2/m_i^2)^2} \times \frac{1}{(1\,+\,q^2/m_k^2)^2} \;\Longrightarrow\; \Gamma^S\left(b; m_i, m_k; \alpha'_{I\!P} = 0\right). \tag{10}$$

GLMM introduce a small energy dependence

$$m_i^2 \;\Longrightarrow\; m_i^2(s) \;\equiv\; \frac{m_i^2}{1\,+\,4m_i^2\alpha'_{I\!P}\ln(s/s_0)}. \tag{11}$$

The above parametrisation is compatible with the requirements of analyticity/crossing symmetry at large $b$, pQCD at large $q^2$ and Regge at small $q^2$. For details see Ref. [5]. KMR and LKMR use a different parametrisation for $\Gamma^S_{i,k}$ which is numerically compatible with GLMM. The 3 groups reproduce $d\sigma_{el}/dt$ well in the forward $t < 0.5\,\mathrm{GeV}^2$ cone.

Consider a model in which diffraction is exclusively GW. This was recently considered by GLMM and LKMR. These, as well as earlier KMR GW models, fit the (different) elastic sectors of their data bases, obtaining output fitted $\Delta_{I\!P} = 0.10 - 0.12$ with $\chi^2/d.o.f. < 1.0$. The above GW models fail to reproduce the diffractive sectors of their data bases. This deficiency is traced to the need to add the enhanced $I\!P$ high mass contributions. This has been done in GLMM and KMR. LKMR model is confined to Muller's $3I\!P$ approximation.

# 3 Multi Pomeron Interactions

Consider a single diffraction channel $p + p \to p + M_{sd}$. Mueller's triple Pomeron mechanism, derived from 3 body unitarity, leads to high SD mass which is non GW. In the leading order

$$M^2_{sd}\frac{d\sigma^{3I\!P}}{dt\,dM^2_{sd}} \;=\; \frac{1}{16\pi^2}g_p^2(t)g_p(0)G_{3I\!P}(t)\left(\frac{s}{M^2_{sd}}\right)^{2\Delta_{I\!P}+2\alpha'_{I\!P}t}\left(\frac{M^2_{sd}}{s_0}\right)^{\Delta_{I\!P}}. \tag{12}$$

The virtue of Eq. (12) is that $\Delta_{I\!P}$ can be determined from either the energy or mass dependences of the SD cross sections. This approximation is valid for $s \gg M^2_{sd} \gg m^2_p$.

CDF analysis suggests [8] a relatively large value of $G_{3I\!P}$. Consequently, we need to consider a very large family of multi Pomeron interactions (enhanced $I\!P$) which are not included in the GW mechanism. As we shall see, this "new" dynamical feature initiates profound differences in the calculated values of soft cross sections and induces additional non GW diffractive gap survival probabilities (soft and hard). Note that these features become significant above the Tevatron energy.

GLMM and KMR treatment of $I\!P$ enhanced interactions stems from Gribov [1] and Kaidalov et al. [9] classical papers on Reggeon calculus. Recall that in this context the soft Pomeron is a simple pole in the J-plane, while the hard (BFKL) Pomeron is a branch cut. KMR model, which is a partonic model, derives directly from these sources. Its summation is confined to semi-enhanced $I\!P$ diagrams (see Fig. 1b). KMR calculations are based on two ad hoc assumptions:
1) The coupling of a multi $I\!P$ point vertex $nI\!P \to mI\!P$ $(n+m>2)$ is $g^n_m = \frac{1}{2} g_N nm\lambda^{n+m-2}$. In this notation $G_{3I\!P} = \lambda g_N$. Note that in Kaidalov et al. $g^n_m = \frac{1}{2} g_N \lambda^{n+m-2}$.
2) Most of LHC non GW diffractive reactions of interest are hard. Given a $3I\!P$ vertex, $G_{3I\!P}$ is unchanged by the interchange of soft and hard Pomerons. This is not self evident. A possible support for the above is obtained from GLMM interpretation of the Pomeron (see below).

As we shall see in the next chapter, the data analysis executed by GLMM and KMR converges to compatible exceedingly small $\alpha'_{I\!P}$ values and high, BFKL like, $\Delta_{I\!P} \simeq 0.30-0.35$. The adjustments of these parameters are correlated. In non screened Regge model $\Delta_{I\!P}$ controls the elastic cross section energy dependence, while $\alpha'_{I\!P}$ controls the energy dependence of the (shrinking) elastic slope. As $\alpha'_{I\!P}$ gets smaller $\Delta_{I\!P}$ becomes larger initiating stronger screening which compensates the reduction of $\alpha'_{I\!P}$, and vice versa. As we saw, the vanishing value of the fitted $\alpha'_{I\!P}$ was induced by the GLMM and KMR b-profiles chosen so as to reproduce the elastic differential cross section. These results have profound implications:

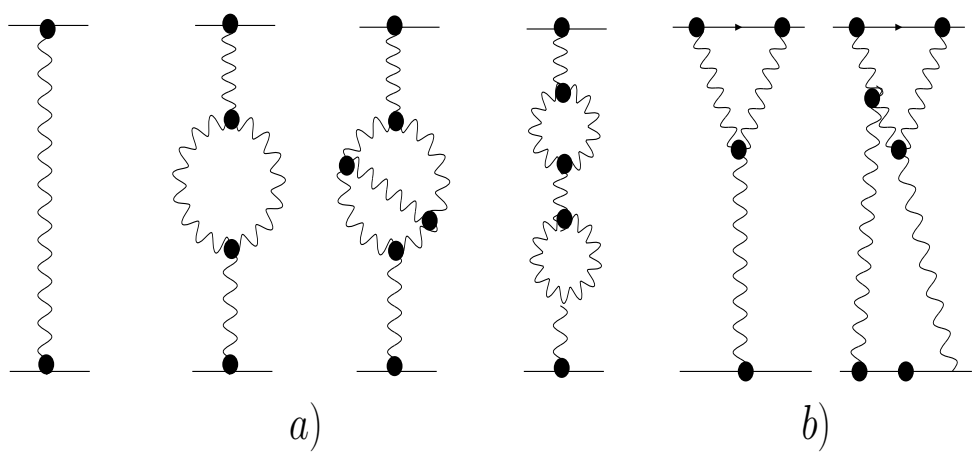

Figure 1: Low order terms of the Pomeron Green's function. a) Enhanced. b) Semi-enhanced.

1) A key observation of GLMM is that the exceedingly small fitted value of $\alpha'_{I\!P}$ implies that the "soft" $I\!P$ is hard enough to be treated perturbatively. Following Gribov we identify the correlation between $\alpha'_{I\!P}$ and $\langle p_t \rangle$, the mean transverse momentum of the partons (actually, colour dipoles) associated with the $I\!P$. $\langle p_t \rangle = 1/\sqrt{\alpha'_{I\!P}}$, from which we deduce that the QCD running coupling constant $\alpha_S \ll 1$. Accordingly, we proceed from Gribov's parton model interpretation to pQCD. GLMM sum over the enhanced diagrams. Technically, we have adopted the MPSI procedure [10] in which $g^n_m$ is reduced to a sequence of triple $I\!P$ vertexes (Fan diagrams). For details see Ref. [5]. This may pose a problem for the calculation of SD cross section for which the lowest order diagram is semi enhanced. To avoid this problem we have added to this calculation a term by term summation of the relevant semi enhanced diagrams.
2) The fitted high value of $\Delta_{I\!P}$ initiates strong screening which results in a renormalisation of the Pomeron exchange amplitudes. As a result $\Delta^{eff}_{I\!P}$ is reduced monotonically with energy. In GLMM calculations $\Delta^{eff}_{I\!P}(2\,\text{TeV}) \simeq 0.070$, $\Delta^{eff}_{I\!P}(14\,\text{TeV}) \simeq 0.045$ and $\Delta^{eff}_{I\!P}(60\,\text{TeV}) \simeq 0.032$. KMR results are compatible with ours, see Table 2. The slow decrease of $\Delta^{eff}_{I\!P}$ raises the question if its value may become negative at high enough energies, larger than $W = 10^5$ GeV which is the bound of validity of both GLMM and KMR. In GLMM we have checked that $\Delta^{eff}_{I\!P}(100\,\text{TeV}) > 0$. I am less clear about KMR. The compatibility between GLMM and KMR is surprising. As noted, GLMM sum over the enhanced diagrams (Fig. 1a) while KMR sum over the semi enhanced diagrams (Fig. 1b). Very intuitively (at the risk of being wrong), it seems that GLMM renormalise the $I\!P$ propagator while KMR renormalise the $I\!P$ vertex. A complete calculation should, obviously, include both contributions.

# 4 The Interplay Between Theory and Data Analysis

The data analysis of interest aims to adjust the theoretical parameters. To this end we construct:
1) A suitable data base adjusted, so as to fix the parameters with a satisfactory resolution.
2) An adjustment procedure, be it a fit (GLMM, LKMR) or tuning by trial and error (KMR).
3) We distinguish between an adjustment of all free parameters which is executed through a reconstruction of the entire data base in one step (GLMM), and a two steps (KMR, LKMR) procedure, in which the first step determines the GW free parameters by adjusting the elastic sector of the data base. These parameters are fixed in the second step in which the rest of the free parameters are determined.
There is a significant difference between the data analysis carried out by GLMM and KMR. This reflects both in the choices of data bases made by the two groups and their adjustment procedures. The starting point of both investigations is the observation that a GW model reproduces the elastic data well, but its reproduction of the diffractive sector is deficient. Both groups claim to achieve an improved reproduction of their over all data base once the contributions of enhanced Pomeron diagrams are included.

GLMM have constructed a global data base so as to simultaneously fit all its free parameters. It includes $\sigma_{tot}$, $\sigma_{el}$, $\sigma_{sd}$, $\sigma_{dd}$ and $B_{el}$ in the ISR-Tevatron range, CDF differential elastic cross sections and SD mass distribution were checked for consistency. The conceptual approach of KMR and LKMR is completely different. Their data base contains only the measured values of $d\sigma_{el}/dt$, which enables to predict $\sigma_{tot}$, and $d\sigma_{sd}/dt\,d(M_{sd}^2/s)$. In my opinion KMR data base is too limited to enable a substantiation of their premises. Specifically:
1) As we saw, the $b$-profiles $\Gamma_{i,k}^S$ control the features of $d\sigma_{el}/dt$ which are only weakly coupled to the proposed dynamics. I have checked 6 models, published over the last 10 years (3KMR+LKMR+2GLMM) with different dynamics, i.e., exclusive GW (GLMM, LKMR), GW+zero order $I\!P$ enhancement (LKMR) and GW+$I\!P$ enhancement (GLMM+2KMR). The output fitted parameters spread over $0.1 \leq \Delta_{I\!P} \leq 0.55$ and $0 \leq \alpha'_{I\!P} \leq 0.066$. All 6 models reproduce, almost identically, the CDF distributions of $d\sigma_{el}/dt$ with $|t| \leq 0.5\,\mathrm{GeV}^2$. The unavoidable conclusion is that a reconstruction of $d\sigma_{el}/dt$ on its own has no resolution power. The only common ingredient to all 6 models is their compatible b-profiles. These profiles constrain $\alpha'_{I\!P}$ to very small values. This is the key observation leading to a pQCD (GLMM) or partonic (KMR) $I\!P$ interpretation.
2) The reconstructions of CDF $d\sigma_{sd}/dt\,d(M_{sd}^2/s)$ by GLMM, LKMR and KMR, are remarkably similar. They support the introduction of high mass multi Pomeron interactions. In my opinion, though, this investigation, in its present state state, is unable to provide a decisive verdict on this issue. GLMM and KMR sum different sectors of the enhanced diagrams. LKMR take into account only the lowest order Mueller's $3I\!P$ diagram. Regardless of these differences, the three groups produce compatible output results which indicate that CDF data, as is, is not sufficient to differentiate between different modes of $I\!P$ enhanced diagram summations.
3) To further clarify the experimental limitations, let us recall that CDF non conventionally define their high mass diffraction bound at $1 - x_L = \frac{M_{sd}^2}{s} \leq 0.15$ (the common bound is 0.05).

KMR and LKMR define a lower bound $M_{sd}^2 > 10\,\mathrm{GeV}^2$ which corresponds to $1 - x_L = 3\cdot 10^{-6}$. Note that CDF SD mass distribution available for analysis corresponds to $1-x_L > 3\cdot 10^{-3}$. This mass distribution covers less than 35% of the expected high diffractive mass spectra. GLMM have a completely different classification in which the GW low mass diffraction reduces monotonically with no arbitrary upper bound and the high mass is defined above $10\,\mathrm{GeV}^2$, identical to KMR. Additional difficulty with the analysis of CDF data with $1 - x_L > 0.03$ is that a large arbitrary background contribution has to be added. It is induced by secondary Regge diagrams such as $I\!PI\!PR$. An added element of ambiguity is that LKMR with a zero order $3I\!P$ calculation is as successful as the high order summations of GLMM and KMR.

| | $\Delta_{I\!P}$ | $\beta$ | $\alpha'_{I\!P}$ $\mathrm{GeV}^{-2}$ | $g_1$ $\mathrm{GeV}^{-1}$ | $g_2$ $\mathrm{GeV}^{-1}$ | $m_1$ GeV | $m_2$ GeV | $\chi^2/d.o.f.$ |
|---|---|---|---|---|---|---|---|---|
| GW | 0.120 | 0.46 | 0.012 | 1.27 | 3.33 | 0.913 | 0.98 | 0.87 |
| GW+$I\!P - enh.$ | 0.335 | 0.34 | 0.010 | 5.82 | 239.6 | 1.54 | 3.06 | 1.00 |

Table 1: Fitted parameters for GLMM GW and GW+$I\!P$-enhanced models.

GLMM fitting procedure aims to reproduce our global data base. A fit with a GW model (no $I\!P$-enh) provides excellent reproduction of our elastic sector while the the reproduction of the diffractive sector is very poor! The repeated fit with a GW+$I\!P$-enh model results with a very good $\chi^2$. The outputs of both models are presented in Table 1. Checking we note that the exceedingly small value of $\alpha'_{I\!P}$ is persistently obtained in both models. The outputs of $\Delta_{I\!P}$ and $g_2$ change drastically once $I\!P$-enh is included. As we shall see this has significant consequences for the approach of $a_{el}(s,b)$ toward the black disc bound. As we have noted, KMR and LKMR tune $g_1$ and $g_2$ through a reproduction of $d\sigma_{el}/dt$ which are frozen in the next phase reproduction of $d\sigma_{sd}/dt\, d(M_{sd}^2/s)$. The values they obtain for $\Delta_{I\!P}$ and $\alpha'_{I\!P}$ are compatible with GLMM.

| | Tevatron | | | LHC | | | W=$10^5$ GeV | | |
|---|---|---|---|---|---|---|---|---|---|
| | GLMM | KMR(07) | KMR(08) | GLMM | KMR(07) | KMR(08) | GLMM | KMR(07) | KMR(08) |
| $\sigma_{tot}$ | 73.3 | 74.0 | 73.7 | 92.1 | 88.0 | 91.7 | 108.0 | 98.0 | 108.0 |
| $\sigma_{el}$ | 16.3 | 16.3 | 16.4 | 20.9 | 20.1 | 21.5 | 24.0 | 22.9 | 26.2 |
| $\sigma_{sd}$ | 9.8 | 10.9 | 13.8 | 11.8 | 13.3 | 19.0 | 14.4 | 15.7 | 24.2 |
| $\sigma_{dd}$ | 5.4 | 7.2 | | 6.1 | 13.4 | | 6.3 | 17.3 | |

Table 2: Comparison of GLMM, KMR(07) and KMR(08) cross sections in mb.

GLMM and KMR high energy Tevatron, LHC and Cosmic Rays predicted cross sections are summarised in Table 2. The elastic and total cross section outputs of the two models are compatible and, above the Tevatron, significantly lower than those obtained in models with no multi-Pomeron contributions. This is a consequence of $\Delta_{I\!P}$ renormalization due to the enhanced $I\!P$ contributions. GLMM and KMR(07) predicted $\sigma_{sd}$ are compatible, where KMR(07) are systematically larger by approximately 10%. KMR(08) predicted $\sigma_{sd}$ are considerably larger than GLMM as well as KMR(07) and are growing at a faster rate. The difference between KMR(07) and GLMM $\sigma_{dd}$ predictions is even more dramatic, where $\sigma_{dd}(KMR(07))/\sigma_{dd}(GLMM) \simeq 3$ at $W = 10^5$ GeV. This very large difference is due to KMR large diffractive high mass predictions. The recent KMR(08) neglects to mention the high diffractive mass sector of double diffraction while showing higher SD cross sections than in KMR(07).

In my opinion the GLMM and KMR compatible predictions of total and elastic cross sections at the LHC and AUGER are of fundamental importance because they are significantly lower than the predicted values based on unitarity models with no $I\!P$ enhancement. These measurements may provide a decisive support for the importance of multi-Pomeron interactions at high enough energies and, consequently, imply that the growth of the total and elastic cross sections with energy is much more moderate than anticipated. This feature reflects in the slow decrease of $\Delta_{I\!P}^{eff}$ shown in Sec 3.

# 5 The Approach Toward the Black Disc Bound

The base amplitudes of GLMM are $A_{i,k}^S$, with $b$ dependences specified in Eq. (10) – Eq. (11). These are the building blocks with which we construct $a_{el}$, $a_{sd}$ and $a_{dd}$ (Eq. (6) – Eq. (8)). The $A_{i,k}^S$ amplitudes are bounded by the black disc unitarity limit of unity. Checking GLMM fitted parameters, presented in Table 1, we observe that $g_1$ and $g_2$, which are comparable in the GW model, significantly change in the GW+$I\!P$ enhanced model where we obtain $g_2 >> g_1$. The implication of of our fitted values of $g_1$ and $g_2$, is that including the diffractive data in our global fit forces a large inequality between the three GW $A_{i,k}^S$ components. $A_{2,2}^S(s, b = 0)$ reaches unity at a very low energy, $A_{1,2}^S(s, b = 0)$ reaches unity at approximately $W = 100\,\mathrm{GeV}$ and $A_{1,1}^S(s, b = 0)$ reaches unity at exceedingly high energies, well above LHC. The observation that one, or even two, of our $A_{i,k}^S(s, b) = 1$ does not imply that the elastic scattering amplitude has reached the unitarity bound at these $(s, b)$ values. $a_{el}(s, b)$ reaches the black disc bound when, and only when, $A_{1,1}^S(s, b) = A_{1,2}^S(s, b) = A_{2,2}^S(s, b) = 1$, independent of $\beta$. The approach of $a_{el}(s, b = 0)$ toward the black bound depends on the rate at which $A_{1,1}^S(s, b)$ increases with energy. Recall that this increase above LHC becomes ever so moderate as a consequence of the renormalization reduction of $\Delta_{I\!P}$. This feature coupled to the smallness of $g_1^2$ implies that $a_{el}(s, b)$ will reach the black bound at energies well above the LHC. Our results are different from the predictions of most available models, notably KMR, in which $a_{el}(s, b = 0)$ reaches unity a few TeV above LHC. Note, though, that GLMM is the only model which includes the diffractive along side the elastic data in its data analysis. All models which predict saturation of $a_{el}(s, b = 0)$ just above LHC have confined their data analysis exclusively to the elastic sector.

A consequence of the input $\Omega_{i,k}^S$ being very large at small $b$, is that $P_{i,k}^S(s, b)$ is exceedingly small at these small $b$ values. As a result, given a diffractive (non screened) input, its output (screened) amplitude is peripheral in $b$. This is a general feature, common to all eikonal models regardless of their b-profiles details. The general behaviour indicated above becomes more extreme at ultra high energies, where $a_{el}$ continues to get darker and expand. Consequently, the inelastic diffractive channels (soft and hard) becomes more and more peripheral and relatively smaller when compared with the elastic channel. Given $(s, b)$ at which $a_{el}(s, b) = 1$, the corresponding diffractive amplitudes, GW and non GW, vanish.

The behaviour of the ratio $R_D = (\sigma_{el} + \sigma_{sd} + \sigma_{dd})/\sigma_{tot}$ conveys information regarding the onset of s-unitarity at very high energies. Assuming diffraction to be exclusively GW, we obtain [11] $R_D \leq 0.5$. Multi $I\!P$ induced diffraction is not included in $R_D$ since it originates from $G_{i,k}^{in}$. Hence its non screened high mass cross section is suppressed by its survival probability which decreases with energy. In GLMM $R_D < 0.5$, decreasing slowly. In KMR(07) $R_D > 0.5$, increasing slowly with energy. The partial information available on KMR(08) suggests that its $R_D$ grows even faster.

# 6 Conclusions

This presentation centred on the phenomenology implied by multi Pomeron dynamics incorporated in soft diffraction and its consequences for soft scattering.

The concept of $I\!P$ enhancement was triggered, at the time, by the assessment that $G_{3I\!P}$ is not too small. Our view of the Pomeron got more focused with the updated data analysis of soft scattering in which we get $\Delta_{I\!P} \simeq 0.30 - 0.35$ and $\alpha'_{I\!P} \simeq 0.01$. The implied KMR Pomeron is hard enough to be treated partonically, in which the traditional classification of the soft $I\!P$ as a simple J-pole and the hard $I\!P$ as a branch cut in the J-plane is maintained. GLMM went further ahead identifying the soft $I\!P$ with the hard $I\!P$. This is, clearly, a fundamental theoretical issue which should be further investigated.

As it stands this dynamics is compatible with the data, but we can not support it, as yet, with a decisive signature. A GLMM and KMR prediction is the expected significant reduction, compared with non screened predictions, of $\sigma_{tot}$ and $\sigma el$ at the LHC. In my opinion this measurement is of a critical value.

Decisive experimental signatures of $I\!P$-enh are expected, essentially, above the Tevatron. Consequently, we should be prudent when evaluating phenomenological models which reproduce the Tevatron data well. This is, obviously, required of a successful model, but is definitely not sufficient.

# Acknowledgements

This research was supported in part by BSF grant # 20004019 and by a grant from Israel Ministry of Science and the Foundation for Basic Research of the Russian Federation.

# References

[1] V.A. Gribov: *Sov. J. Nuclear Phys.* **9** 369 (1969), arXiv:hep-ph/0006158, *Gauge Theories and Quark Confinement* PHASIS, Moscow (2002)

[2] M.L. Good and W.D. Walker: *Phys. Rev.* **120** 1857 (1960).

[3] R.J. Glauber: In *Lectures in Theoretical Physics Vol I*, ed. W.E. Brittin and L.G. Dunham, Interscience Pub., London 1959.

[4] A.H. Mueller: *Phys. Rev.* **D2** 2963 (1970), **D2** 150 (1971).

[5] E. Gotsman, E. Levin, U, Maor and J.S. Miller: *Eur. Phys. J.* **C57** 269 (2008), arXiv: 0901.1540[hep-ph], arXiv: 0903.0247[hep-ph].

[6] M.G. Ryskin, A.D. Martin and V.A. Khoze: *Eur. Phys. J.* **C54** 199 (2008)(GLM07), *Eur. Phys. J.* **C60** 249 (2009), arXive:0811.1481[hep-ph], arXive:0811.4571[hep-ph], arXive:0812.2413[hep-ph](GLM08).

[7] E.G. Luna, V.A. Khoze, A.D. Martin and M.G. Ryskin: *Eur. Phys. J.* **C59** 1 (2009).

[8] CDF Collaboration: *Phys. Rev.* **D50** 5535 (1994); K. Goulianos and J. Montanha: *Phys. Rev.* **D59** 114017 (1999).

[9] A.B. Kaidalov, L.A. Ponomarev and K.A. Ter-Martirosyan: *Sov. J. Nucl. Phys.* **44** 468 (1986).

[10] A. H. Mueller and B. Patel: *Nucl. Phys.* **B425** 471 (1994); A. H. Mueller and G. P. Salam: *Nucl. Phys.* **B475** 293 (1996); G. P. Salam: Nucl. Phys. **B461** 512 (1996); E. Iancu and A. H. Mueller: Nucl. Phys. **A730** 460 (2004), Nucl. Phys. **A730** 494 (2004).

[11] J. Pumplin: *Phys. Rev.* **D8** 2 (1973).

# Aspects of Higgs Production at the LHC

*Errol Gotsman**

Department of Particles Physics, School of Physics and Astronomy
Raymond and Beverly Sackler Faculty of Exact Science
Tel Aviv University, Tel Aviv 69978, Israel

We discuss the main features and predictions of the GLMM model, which is based on a QCD motivated theoretical approach, and successfully describes the experimental data on total, elastic and diffractive cross sections. In addition we calculate the survival probability for a SM Higgs at the LHC, and compare our results with those of the Durham group.

## 1 Introduction

Over the past few years the subject of "soft physics" has reemerged from the shadows, and has aroused the interest of the phenomenological community. This in no small way, due to the realization that the calculation of the probability of detecting a diffractive hard pQCD process e.g. Higgs production at the LHC, also depends on the underlying secondaries which are produced by "soft" rescattering. Central diffractive production e.g. (Higgs boson, 2 jets, 2 $\gamma$'s, $\chi_c$), are accompanied by gaps in rapidity, between the two outgoing projectiles, and the centrally produced particles, which makes their detection easier. The subject of the survival of these rapidity gaps was initiated over twenty years ago [1], and has been refined over the interim period [2, 3].

## 2 Details of our Model

The details on which our two channel (GLM) model is based i.e. the Good-Walker (G-W) mechanism [4] can be found in [2]. See also U. Maor's talk in these proceedings. It is well known that G-W neglects the diffractive production of large mass states (Mueller diagrams [5]), and to successfully describe the diffractive data, these need to be included by adding the relevant triple Pomeron contributions (see l.h. diagrams in Fig. 1). In addition it is also necessary to include diagrams containing Pomeron loops, as shown in the r.h. diagrams in Fig. 1.

To simplify the problem of summing the Pomeron loop diagrams, we assume that at high energies only the triple Pomeron interaction is essential, this conjecture has been proved in perturbative QCD (see Refs. [6, 7]). Mueller [8] has shown that in the leading log $x$ approximation of pQCD for a large number of colours ($N_c \gg 1$), the correct degrees of freedom are colourless dipoles.

*e-mail: gotsman@post.tau.ac.il

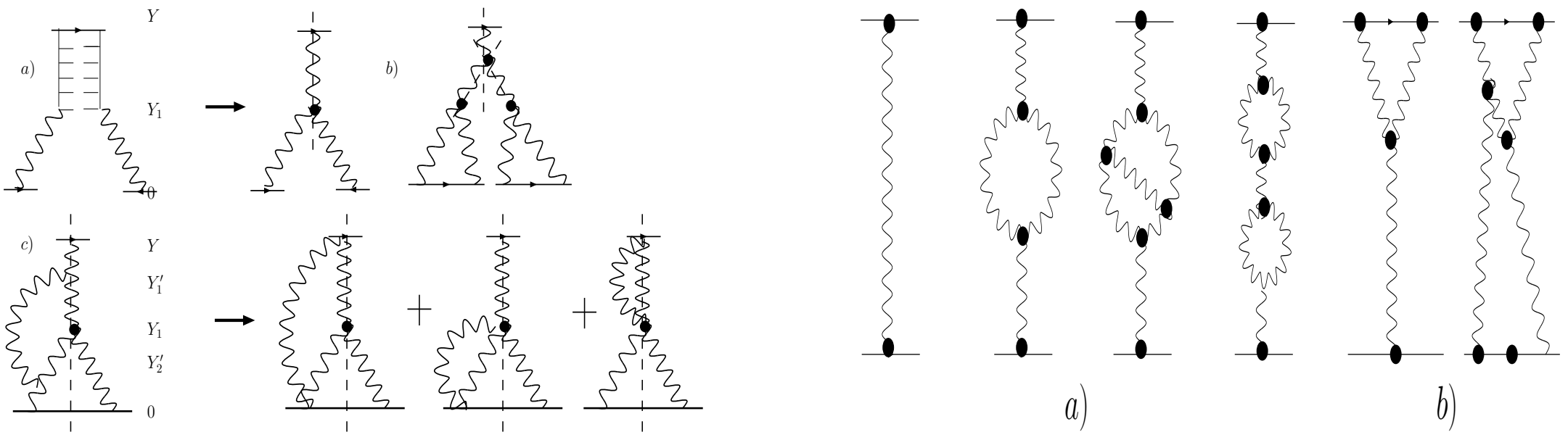

Figure 1: L.h. figure: Examples of Pomeron diagrams not included in G-W mechanism. R.h. figure: Low order terms of the Pomeron Green's function. a) Enhanced. b) Semi-enhanced.

## 2.1 Summing Interacting Pomeron Diagrams

In the leading order approximation of pQCD, only one Pomeron (dipole) splitting into two Pomerons (dipoles), and two Pomerons (dipoles) merging into one Pomeron (dipole) are considered. All other Pomeron vertices do not appear in the leading log $x$ approximation of pQCD. We therefore restrict ourselves to sum only Pomeron diagrams containing triple Pomeron vertices. We add a caveat, that we neglect the $4I\!P$ term (which is needed for s channel unitarity), however, this term is only significant at energies $W > 10^5$ , which is at the limit of the validity of our model.

To make the calculation tractable we further assume that the slope of the Pomeron trajectory $\alpha'_{I\!P} = 0$. The results of our numerical fit to the relevant data, which we will discuss later, ($\alpha'_{I\!P} = 0.01$) lends credence to this assumption.

The theory which includes all the above ingredients can be formulated in terms of a generating function [9, 10]

$$Z(y,\, u) \;=\; \sum_n \; P_n(y)\; u^n, \tag{1}$$

where, $P_n(y)$ is the probability to find $n$-Pomerons (dipoles) at rapidity $y$. At rapidity $y = Y = \ln(s/s_0)$ we can impose an arbitrary initial condition. For example, demanding that there is only one fastest parton (dipole), which is $P_1(y = Y) = 1$, while $P_{n>1}(y = Y) = 0$. In this case we have the following initial condition for the generating function

$$Z(y \,=\, Y) \,=\, u\,. \tag{2}$$

At $u = 1$

$$Z(y,\, u \,=\, 1) \;=\; 1, \tag{3}$$

which follows from the physical meaning of $P_n$ as a probability. The solution, with these two conditions, will give us the sum of enhanced diagrams.

For the function $Z\,(u)$ the following simple equation can be written (see Ref. [2] and references therein)

$$-\frac{\partial\, Z(y,\, u)}{\partial\, \mathcal{Y}} \;=\; -\,\Gamma(1 \to 2)\, u\,(1\,-\,u)\; \frac{\partial\, Z(y,\, u)}{\partial\, u} \;+\; \Gamma(2 \to 1)\, u\,(1\,-\,u)\; \frac{\partial^2\, Z(y,\, u)}{\partial^2\, u}, \tag{4}$$

where, $\Gamma(1 \to 2)$ describes the decay of one Pomeron (dipole) into two Pomerons (dipoles), while $\Gamma(2 \to 1)$ relates to the merging of two Pomerons (dipoles) into one Pomeron (dipole).

Using the functional $Z$, we calculate the scattering amplitude [10, 11], using the following formula:

$$N(Y) \equiv \mathrm{Im}A_{el}(Y) = \sum_{n=1}^{\infty} \frac{(-1)^n}{n!} \left.\frac{\partial^n Z(y, u)}{\partial^n u}\right|_{u=1} \gamma_n(Y = Y_0, b), \tag{5}$$

where, $\gamma_n(Y = Y_0, b)$ is the scattering amplitude of $n$-partons (dipoles) at low energy.

The generating function approach has the advantage that it can be solved analytically (see Ref. [12]), using the MPSI [13] approximation. The exact expression for the Pomeron Green's function is given by

$$G_{I\!P}(Y) = 1 - \exp\left(\frac{1}{T(Y)}\right) \frac{1}{T(Y)} \Gamma\left(0, \frac{1}{T(Y)}\right), \tag{6}$$

where $\Gamma(0, x)$ is the incomplete gamma function, and $T(Y) = \gamma\, e^{\Delta_{I\!P} Y}$. $\gamma$ denotes the amplitude of the two dipole interaction at low energy. The MPSI approximation only takes into account the first term of the expression of the enhanced diagrams, neglecting other terms, as they are suppressed as $e^{-\Delta Y}$. Consequently, this approximation is only reliable in the region $Y \leq min[\frac{1}{\gamma}, \frac{1}{\alpha'_{I\!P} m_i^2}]$.

# 3 Determining the Parameters of the Model and Results of the Fit

The pertinent details of our fit to the experimental data, and our determination of the relevant parameters of the model, needed to describe the soft interactions, are contained in [2]. In this section we only mention the salient features, and results of the fit. Our fit is based on

| | $\Delta_{I\!P}$ | $\beta$ | $\alpha'_{I\!P}$ GeV$^{-2}$ | $g_1$ GeV$^{-1}$ | $g_2$ GeV$^{-1}$ | $m_1$ GeV | $m_2$ GeV | $\chi^2/d.o.f.$ |
|---|---|---|---|---|---|---|---|---|
| GW | 0.120 | 0.46 | 0.012 | 1.27 | 3.33 | 0.913 | 0.98 | 0.87 |
| GW+$I\!P - enh.$ | 0.335 | 0.34 | 0.010 | 5.82 | 239.6 | 1.54 | 3.06 | 1.00 |

Table 1: Fitted parameters for GLMM GW and GW+$I\!P$-enhanced models.

55 experimental data points, which includes the $p$-$p$ and $\bar{p}$-$p$ total cross sections, integrated elastic cross sections, integrated single and double diffraction cross sections, and the forward slope of the elastic cross section in the ISR-Tevatron energy range. The model gives a good reproduction of the data, with a $\chi^2/d.o.f. \approx 1$. In addition to the quantities contained in the data base, we obtain a good description of the CDF [14] differential elastic cross sections and the single diffractive mass distribution at $t = 0.05$ GeV$^2$. An important advantage of our approach, is that the model provides a very good reproduction of the double diffractive (DD) data points. Other attempts to describe the DD data e.g. (see Refs. [15, 3]), were not successful in reproducing the DD experimental results over the whole energy range.

In Table 1 we list the values of the parameters obtained by a least squares fit to the experimental data, both for the G-W formalism (elastic data only), and for the G-W formalism plus enhanced graphs (elastic plus diffractive data). In Table 2 we compare our results with results of the two versions of the model proposed by the Durham group [15, 3]. In Fig. 2 we display the predictions of the GLMM model's values for the various cross sections. Note that $\sigma_{el}$ and $\sigma_{sd}$ have completely different energy dependence, unlike the predictions of [3]. At an energy of 7 TeV the predictions of GLMM are (in mb): $\sigma_{tot}$= 86.0, $\sigma_{el}$= 19.5, $\sigma_{sd}$= 10.7, $\sigma_{dd}$ = 5.9 and the forward slope $B_{el} = 19.4\,\mathrm{GeV}^{-2}$.

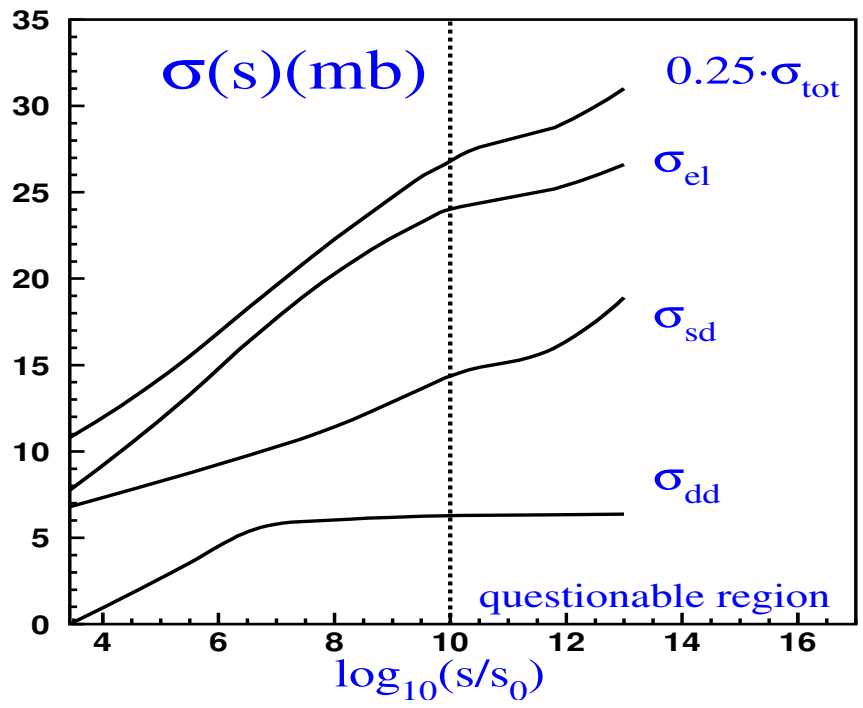

Figure 2: Energy dependence of GLMM cross sections.

| | Tevatron | | | LHC | | | W=$10^5$ GeV | | |
|---|---|---|---|---|---|---|---|---|---|
| | GLMM | KMR(07) | KMR(08) | GLMM | KMR(07) | KMR(08) | GLMM | KMR(07) | KMR(08) |
| $\sigma_{tot}$ | 73.3 | 74.0 | 73.7 | 92.1 | 88.0 | 91.7 | 108.0 | 98.0 | 108.0 |
| $\sigma_{el}$ | 16.3 | 16.3 | 16.4 | 20.9 | 20.1 | 21.5 | 24.0 | 22.9 | 26.2 |
| $\sigma_{sd}$ | 9.8 | 10.9 | 13.8 | 11.8 | 13.3 | 19.0 | 14.4 | 15.7 | 24.2 |
| $\sigma_{dd}$ | 5.4 | 7.2 | | 6.1 | 13.4 | | 6.3 | 17.3 | |
| $\frac{\sigma_{el}+\sigma_{diff}}{\sigma_{tot}}$ | 0.43 | 0.46 | | 0.42 | 0.53 | | 0.41 | 0.57 | |

Table 2: Comparison of GLMM, KMR(07) and KMR(08) cross sections in mb.

# 4 Survival Probability for Central Diffractive Production of the Higgs Boson

A general review of survival probability calculations can be found in [16]. We denote by $\langle |\, S^2 \,| \rangle$ the probability that the Large Rapidity Gap (LRG) survives, and is not filled by secondaries from eikonal and enhanced rescattering effects (see Fig. 3). The expression for the survival probability can be written $\langle |\, S^2_{2ch} \,| \rangle = \frac{N(s)}{D(s)}$ where

$$N(s) = \int d^2 b_1\, d^2 b_2 \left[\sum_{i,k} < p|i >^2 < p|k >^2 \;\; A^i_H(s,b_1)\, A^k_H(s,b_2)(1 - A^{i,k}_S\left((s,(\mathbf{b}_1+\mathbf{b}_2)\right)\right]^2,$$

and

$$D(s) = \int\, d^2 b_1\, d^2 b_2 \left[\sum_{i,k} < p|i >^2 < p|k >^2 \;\; A^i_H(s,b_1)\, A^k_H(s,b_2)\right]^2$$

$A_S(b,s)$ denotes the "soft" strong amplitude of our model [2]. While for the "hard" amplitude $A_H(b,s)$, we assume an input Gaussian $b$ dependence. i.e. $A^H_{i,k} = A_H(s)\,\Gamma^H_{i,k}(b)$ with $\Gamma^H_{i,k}(b) = \frac{1}{\pi (R^H_{i,k})^2}\, e^{-\frac{2\,b^2}{(R^H_{i,k})^2}}$. The "hard" radii are constants determined from HERA data on elastic and inelastic $J/\Psi$ production. We introduce two hard b-profiles $A^{pp}_H(b) = \frac{V_{p\to p}}{2\pi B^H_{el}} \exp\left(-\frac{b^2}{2\,B^H_{el}}\right)$, and $A^{pdif}_H(b) = \frac{V_{p\to dif}}{2\pi B^H_{in}} \exp\left(-\frac{b^2}{2B^H_{in}}\right)$. The values $B^H_{el}$=5.0 GeV$^{-2}$ and $B^H_{in}$=1 GeV$^{-2}$ have been taken from ZEUS data. The value $B^H_{el}$= (3.6) GeV$^{-2}$ was used in [2], this has now been changed in light of the latest measurements of the "hard" slope, by the H1 group. This is in contrast to KMR treatment [15] where they assume: $A^{pp}_H(b) = A^{pdif}_H(b) \propto \exp\left(-\frac{b^2}{2B^H}\right)$ with $B^H_{el} = B^H_{inel} = 4$ or 5.5 GeV$^{-2}$ The sensitivity of our results to the parameters of the "hard" amplitude are shown in Fig. 4 (left), note that for $B^H_{in}$ =1 GeV$^{-2}$, changing the value of $B^H_{el}$ from 3.6 to 5.0 GeV$^{-2}$, increases $\langle\mid S^2_{2ch}\mid\rangle$ by $\approx$ 70 % . Our results for $\langle\mid S^2\mid\rangle$ = $\langle\mid S^2_{2ch}\mid\rangle$ x $\langle\mid S^2_{enh}\mid\rangle$ is given by the full line in Fig. 4 (right), it decreases with increasing energy, due to the behaviour of $\langle\mid S^2_{enh}\mid\rangle$.

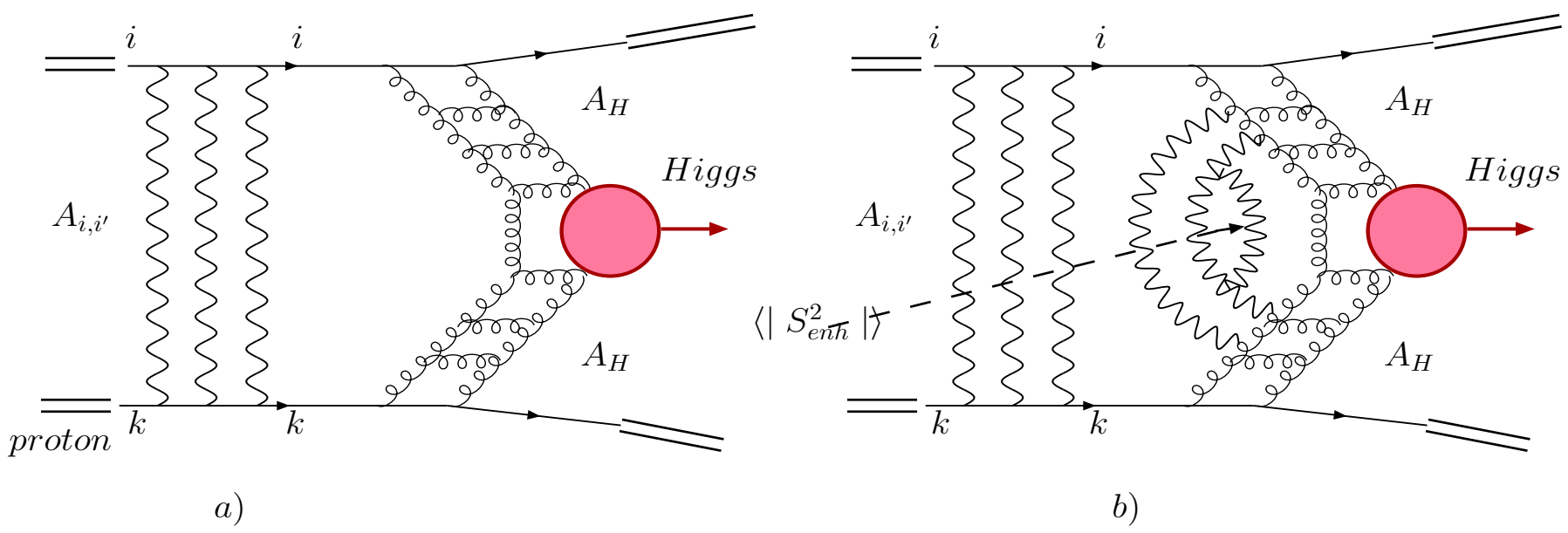

Figure 3: a) the survival probability in the G-W mechanism, b) illustrates the origin of the additional factor $\langle\mid S^2_{enh}\mid\rangle$.

Our results and the Durham group's results for Survival Probability are given in Table 3. At an energy of 7 TeV we predict a value $\langle\mid S^2\mid\rangle \approx 0.6\%$. We have also succeeded in summing the semi-enhanced contribution (see r.h. side of Fig. 1) to the Survival Probability, and find that it is almost energy independent, and has a value $\approx$ 100% at Tevatron and LHC energies [17].

| | Tevatron | | | LHC (14 TeV) | | | W=$10^5$ GeV | | |
|---|---|---|---|---|---|---|---|---|---|
| | GLMM | KMR(07) | KMR(08) | GLMM | KMR(07) | KMR(08) | GLMM | KMR(07) | KMR(08) |
| $S^2_{2ch}(\%)$ | v 5.3 | 2.7-4.8 | | 3.9 | 1.2-3.2 | | 3.2 | 0.9-2.5 | |
| $S^2_{enh}(\%)$ | 28.5 | 100 | | 6.3 | 100 | 33.3 | 3.3 | 100 | |
| $S^2(\%)$ | 1.51 | 2.7-4.8 | | 0.24 | 1.2-3.2 | 1.5 | 0.11 | 0.9-2.5 | |

Table 3: Comparison of results obtained for Survival Probability in Tel Aviv and Durham models

# 5 Discussion and Conclusions

We present a model for soft interactions having two components: (i) G-W mechanism for elastic and low mass diffractive scattering, and (ii) Pomeron enhanced contributions for high mass diffractive production. In addition we find from our fit that the slope of the Pomeron $\alpha'_{I\!P} \approx 0.01$. This is consistent with what one expects in pQCD, since for a BFKL Pomeron $\alpha'_{I\!P} \propto 1/Q_s^2 \to 0$ as $s \to \infty$. Having $\alpha'_{I\!P} \to 0$, provides a necessary condition that links strong (soft) interactions with the hard interactions described by pQCD. A key hypothesis in our model is that the soft processes are not "soft", but originate from short distances. We have only one Pomeron. There is no requirement for a "soft" and "hard" Pomeron. This is in accord with Hera data for $F_2$, which is smooth throughout the transition region [18].

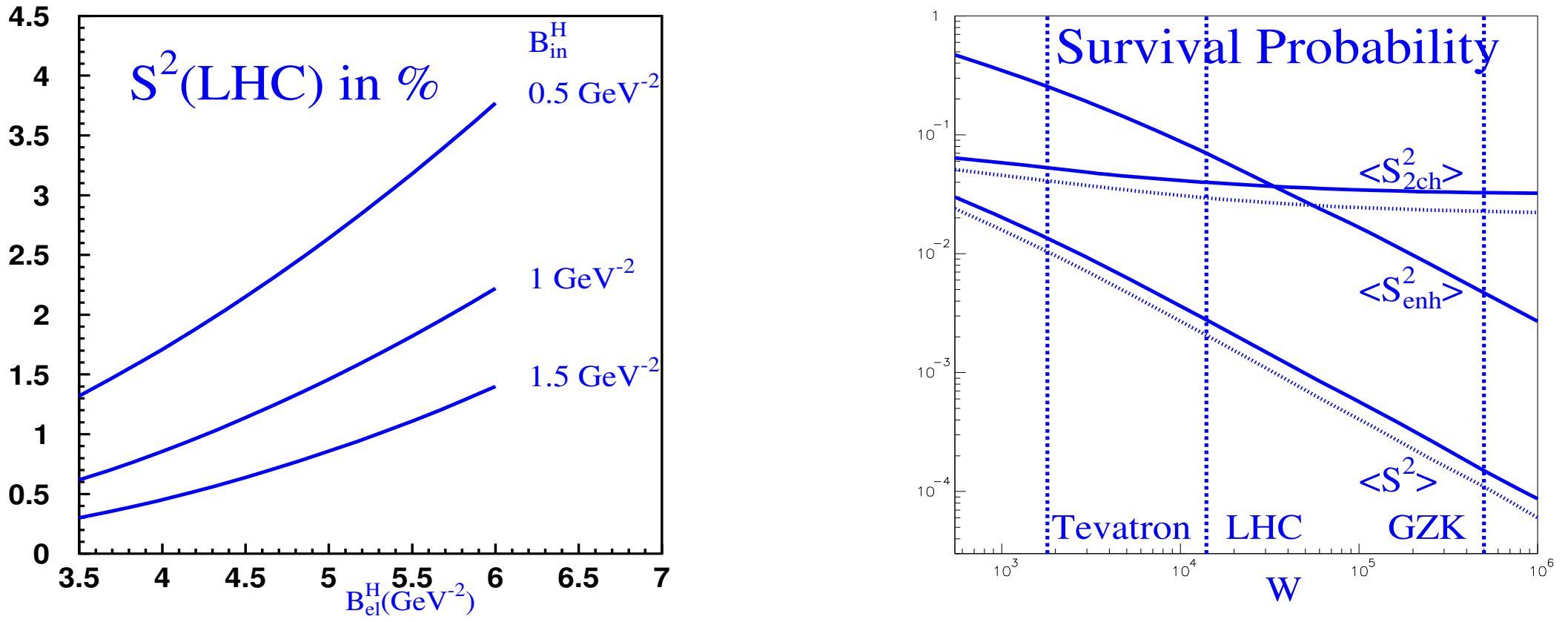

Figure 4: L.h. figure: The dependence of $S^2$ at the LHC on $B^H_{el}$ and $B^H_{in}$. R.h. figure: Energy dependence of $S^2$ for centrally produced Higgs. The full (dashed) line is for $B^H_{el} = 5.0$ (3.6) $\mathrm{GeV}^{-2}$.

To illustrate our achievements and problems, we compare our approach with the work of the Durham group [15, 3]. The main difference in the underlying philosophy of the two groups is that, the Durham approach is based on the parton model where there is only a short range rapidity interaction between partons, while we, due to exchange of gluons in QCD, have a long range rapidity interaction. Both approaches consider $\alpha'_{I\!P}$ as being small. In both programs the Pomeron interaction was taken into account. The difference between the two approaches is that KMR made an *ad hoc* "reasonable" assumption, that the multi-Pomeron vertices have the following form, for the transition of $n$ Pomerons to $m$ Pomerons

$$g^n_m = n\,m\,\lambda^{n+m-2}\,g_N/2 = n\,m\,\lambda^{n+m-3}\,g_{3I\!P}/2. \qquad (7)$$

No theoretical arguments or theoretical models were offered in support of this assumption, which certainly contradicts the pQCD approach [6, 7]. In spite of these differences, the values obtained for the $\sigma_{tot}$ and $\sigma_{el}$ are in surprisingly close agreement (see Table 2). It is only in the latest version of the Durham model [3], which includes three components of the Pomeron, with different transverse momenta of the partons in each component (to mimic BFKL diffusion in $k_t$), that there are fairly large discrepancies in the diffractive sector i.e. $\sigma_{sd}$ and $\sigma_{dd}$. KMR [3] find that at higher energies $\sigma_{sd}$ and $\sigma_{el}$ have comparable values and similar energy dependence, this is not so in our description [2] (see Fig. 2). We note that the results presented by Poghosyan at this conference for $\sigma_{sd}$ and $\sigma_{dd}$ [19] agree both in magnitude and energy dependence with those obtained in the GLMM model. There is also disagreement in the result for the calculation of the Survival Probability. In [3], the Survival Probability is now multiplied by a "renormalizing" factor $(\langle p_t^2 \rangle B)^2$ and referred to as $\langle S^2_{eff} \rangle$. The result for LHC energy is $\langle S^2_{eff} \rangle = 0.015^{+0.01}_{-0.005}$. It is not clear whether there is in addition a factor of $\langle S^2_{enh} \rangle \approx 1/3$, that needs to be incorporated. If affirmative, then the discrepancy between our result and that of the Durham group for $\langle S^2 \rangle$ at Tevatron energies is small, but the discrepancy becomes larger for the LHC energy range, as we predict that $\langle S^2_{enh} \rangle$ decreases as the rapidity between the projectiles increases, while Durham claim little (if any) energy dependence for $\langle S^2_{enh} \rangle$.

## Acknowledgements

This research was supported in part by BSF grant # 20004019 and by a grant from Israel Ministry of Science and the Foundation for Basic Research of the Russian Federation.

## References

[1] Y.L. Dokshitzer, S.I. Troian and V.A. Khoze, Sov. J. Nucl. Phys. **46** 712 (1987); Y.L. Dokshitzer, V.A. Khoze and T. Sjöstrand, Phys. Lett. **B274** 116 (1992); J.D. Bjorken, Intr. J. Mod. Phys. **A7** 4189 (1992); Phys. Rev. **D47** 101 (1993); E. Gotsman, E. Levin and U. Maor, Phys. Lett. **B309** 199 (1993).

[2] E. Gotsman, E. Levin, U, Maor and J.S. Miller, Eur. Phys. J. **C57** 269 (2008).

[3] M.G. Ryskin, A.D. Martin and V.A. Khoze, Eur. Phys. J. **C60** 249 (2009).

[4] M.L. Good and W.D. Walker, Phys. Rev. **120** 1857 (1960).

[5] A.H. Mueller, Phys. Rev. **D2** 2963 (1970).

[6] J. Bartels, M. Braun and G.P. Vacca, Eur. Phys. J. **C40** 419 (2205); J. Bartels and C. Ewerz, JHEP **9909** 026 (1999); J. Bartels and M. Wusthoff, Z. Phys. **C6** 157 (1995); A.H. Mueller and B. Patel, Nucl. Phys. **B425** 471 (1994); J. Bartels, Z. Phys. **C60** 471 (1993).

[7] M.A. Braun: Phys. Lett. **B632** 297 (2006), Eur. Phys. J. **C16** 337 (2000); Phys. Lett. **B483** 115 (2000); Eur. Phys. J. **C33** 113 (2004); M.A. Braun and G.P. Vacca, Eur. Phys. J. **C6** 147 (1999).

[8] A.H. Mueller, Nucl. Phys. **B415** 373 (1994); **B437** 107 (1995).

[9] A.H. Mueller, Nucl. Phys. **B415** 107 (1995).

[10] E. Levin and M. Lubinsky, Nucl. Phys. **A763** 172 (2005); Phys. Lett. **B607** 131 (2005); Nucl. Phys. **A730** 191 (2004).

[11] Y.V. Kovchegov, Phys. Rev. **D60** 034008 (1999).

[12] M. Kozlov and E. Levin, Nucl. Phys **A779** 142 (2006).

[13] A.H. Mueller and B. Patel, Nucl. Phys. **B425** 471 (1994); A.H. Mueller and G.P. Salam, Nucl. Phys. **B475** 293 (1996); G.P. Salam, Nucl. Phys. **B461** 512 (1996); E. Iancu and A.H. Mueller, Nucl. Phys. **A730** 460 (2004).

[14] CDF Collaboration, Phys. Rev. **D50**, 5535 (1994).

[15] M.G. Ryskin, A.D. Martin and V.A. Khoze, Eur. Phys. J. **C54** 199 (2008).

[16] E. Gotsman, E. Levin, U. Maor, E. Naftali, and A. Prygarin, in :HERA and the LHC- A Workshop on the Implications of HERA for LHC Physics: Proceedings Part A, p. 221 (2005). arXiv:hep-ph-0601012

[17] E. Gotsman, E. Levin and U. Maor (to be published).

[18] H1 Collaboration, arXiv:hep-ex/0904.0929 (2009)

[19] A.B. Kaidalov and M.G. Poghosyan, arXiv:hep-ph/0909.5156

# One-Particle Inclusive Distribution in the Unitarized Pomeron Models

*A. Alkin*$^{(1)}$, *E. Martynov*$^{(1)}$, *E. Romanets*$^{(2)}$, *V. Pauk*$^{(2)}$

[1] Bogolyubov Institute for Theoretical Physics, *Metrologichna 14b, Kiev, UA-03680, Ukraine*
[2] Taras Shevchenko Kiev National University, *Volodimirska 60, Kiev, UA-03101, Ukraine.*

It is shown that starting from the pomeron contribution with intercept $\alpha_P(0) > 1$ one can obtain in a quasi-eikonal approach inclusive cross section which is similar to contribution of triple pole (at $t = 0$) pomeron. Generalizing this analogy we consider tripole and dipole pomeron contributions to inclusive cross section. They lead to $\langle n\rangle \propto \ln^3 s$ (tripole) or $\langle n\rangle \propto \ln^2 s$ (dipole) and describe well the data on charged hadron distributions in $\bar{p}p$. Predictions of one particle $p_t$ and rapidity distributions for LHC energies are given.

The model of simple pole pomeron with intercept $\alpha_P(0) = 1 + \varepsilon, \quad \varepsilon > 0$ [1] gives a simple and compact parametrization for many high-energy soft processes (elastic and deep inelastic scattering, diffraction and others), it describes well many experimental data at high energies.

On the other side at $s \to \infty$ contribution of such pomeron violates unitarity explicitly. The model leads to total cross section of hadron interaction $\sigma_t(s) \propto (s/s_0)^\varepsilon$ ($s_0 = 1$ GeV$^2$) in contradiction to the Froissart bound $\sigma_t \leq (\pi/m_\pi^2)\ln^2(s/s_0)$.

Thus the model is only a phenomenological tool and must be improved in order to restore unitarity. There are a few known ways to avoid at least a rough violation of unitarity bound. The most simple method for that is to sum multipomeron diagrams of Fig. 1.

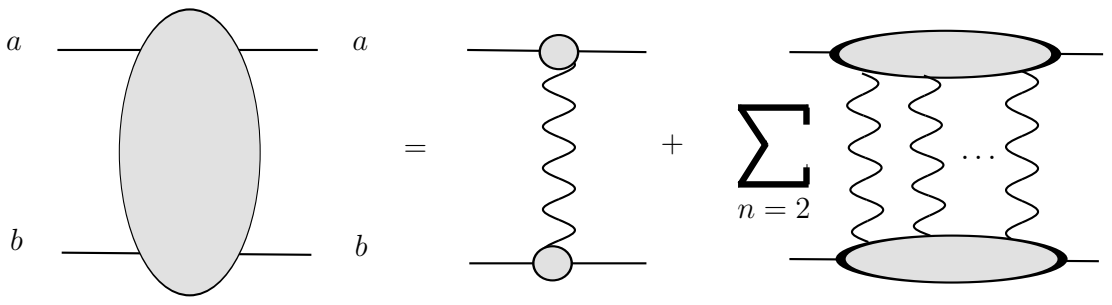

Figure 1: Multipomeron contributions to elastic scattering amplitude.

Starting with one pomeron exchange written in the form

$$a(s,t) = \eta_P(t)\tilde{g}_{ab}(t)\left(\frac{s}{s_0}\right)^{\alpha_P(t)} = -g_a(t)g_b(t)\left(-i\frac{s}{s_0}\right)^{\alpha_P(t)},$$

where $s_0 = 1\text{GeV}^2$ and $\eta_P(t) = \frac{1+\exp(-i\pi\alpha_P(t))}{-\sin(\pi\alpha_P(t))}$, then going to impact parameter representation

$$h(s,b) = \frac{1}{8\pi s}\int\limits_0^\infty dq\, q\, J_0(qb)\, a(s,-q^2)$$

one can obtain under some simplifying assumption (see below)

$$H(s,b) = -\frac{1}{2i}\sum_{n=1}^{\infty}\frac{G_a(n)G_b(n)}{n!}[-2ih(s,b)]^n.$$

The amplitude $H(s,b)$ takes this form if we assume that two-hadrons-$n$-pomeron amplitude is proportional to the product of two-hadron-pomeron vertices (a pole approximation for intermediate states) as shown in Fig. 2.

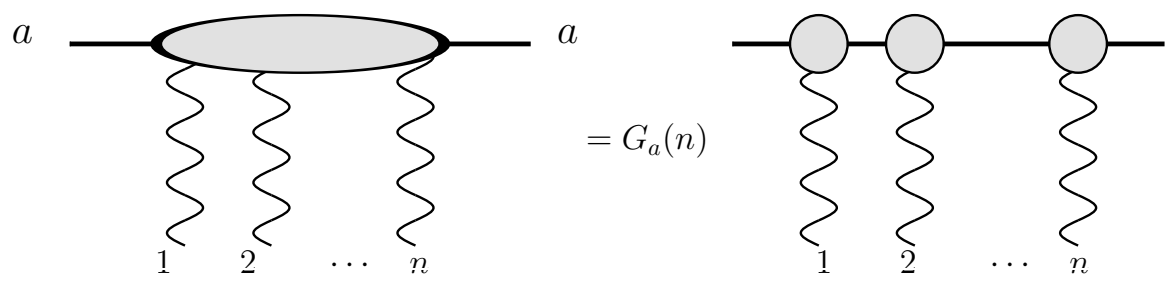

Figure 2: Amplitude of interaction of two hadrons with $n$ pomerons in a pole approximation but with phenomenological factor $G_a(n)$.

Moreover, assuming either $G(n) = C^n$ or $G(n) = C^n\sqrt{n!}$ we obtain two well known schemes of pomeron unitarization: quasi-eikonal [2] or quasi-$U$-matrix models [3, 4].

$$H(s,b) = \begin{cases} \frac{1}{2iC_aC_b}\left(1 - e^{-2iC_aC_bh(s,b)}\right), & \text{if} \quad G_{a,b}(n) = C^n_{a,b} \\ \frac{h(s,b)}{1+2iC_aC_bh(s,b)}, & \text{if} \quad G_{a,b}(n) = C^n_{a,b}\sqrt{n!} \end{cases}$$

If $\alpha_P(t) = 1 + \varepsilon + \alpha'_P\, t$ and $g_{a,b}(t) = \exp(B_{a,b}\, t)$ one can find that at $s \to \infty$ in both models

$$\sigma_t^{ab}(s) \approx 8\pi\varepsilon R^2(s)\ln(s/s_0) \approx 8\pi\varepsilon\alpha'_P\ln^2(s/s_0),$$

where $R^2(s) = B_a + B_b + \alpha'_P\ln(s/s_0)$.

This result gives a ground for another method of constructing amplitude. One can consider just from the beginning more complicated singularities of partial amplitudes than usual simple angular momentum poles. Because the factorization of residues is valid not only for simple $j$-poles but also for any isolated $j$-singularity [5] one can consider, for instance, double pole (dipole pomeron) [6] or triple pole (at $t \neq 0$ because of analyticity it must be a pair of hard branch points collided to a triple pole at $t = 0$) instead of simple pole. In these models $\sigma_t(s) \propto \ln(s/s_0)$ (dipole) or $\sigma_t(s) \propto \ln^2(s/s_0)$ (tripole) at $s \to \infty$. Both models lead to a very good description of the hadron total cross sections as well as of the differential elastic cross sections, deep inelastic scattering and vector meson photoproduction.

Therefore it is interesting to see how many-particle processes are described in these unitarized pomeron models. We consider here one particle distribution in rapidity and pseudorapidity.

# 1 Multipomeron Exchanges in the Model with $\alpha_{\mathbf{P}}(0) > 1$

Due to the generalized Optical Theorem the differential cross section of one particle inclusive production ($a + b \to c + X$) in the central kinematic region where

$$\begin{array}{lcll} s & = & (p_a+p_b)^2 \to \infty, & \\ t & = & (p_a-p_c)^2, & |t|\to\infty, \\ u & = & (p_b-p_c)^2, & |u|\to\infty, \\ M^2 & = & (p_a+p_b-p_c)^2, & s/M^2\to 1, \\ tu/s & = & m_t^2 = m_c^2+p_l^2 & \end{array}$$

is related with the diagram of Fig. 3.

More exactly, at large energy and for the simple pomeron pole with $\alpha_P(0)-1=\varepsilon$

$$E\frac{d^3\sigma}{d^3p} = E\frac{d^3\sigma}{dp_l d^2p_t} = 8\pi\,\mathrm{Disc}_{M^2}\mathcal{M}(a+b+\bar{c}\to a+b+\bar{c}) = g_a(0)\left(\frac{|t|}{s_0}\right)^{\varepsilon} v_c(p_t^2)\left(\frac{|u|}{s_0}\right)^{\varepsilon} g_b(0).$$

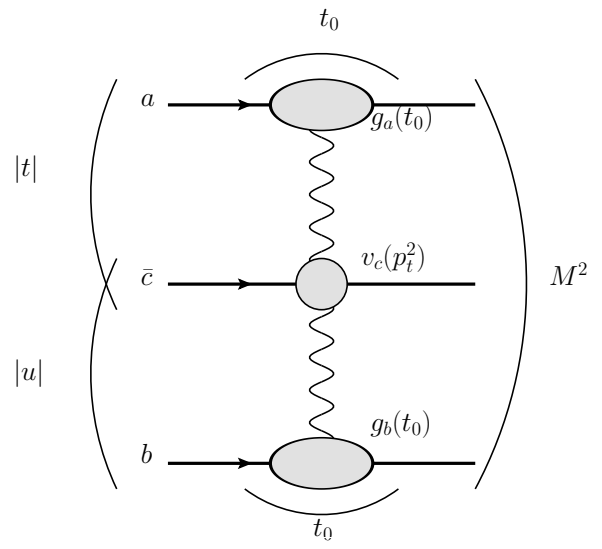

Figure 3: Pomeron contribution to inclusive production in the central region.

where $E, p_l, \vec{p}_t$ are energy and momenta of the inclusive hadron $c$, $g_{a,b}(t_0=0)$ are the coupling vertices $aPa, bPa$, $s_0 = 1\mathrm{GeV}^2$.

It is more convenient for what follows to use another set of variables, $(p_t, y)$ or $(p_t, \eta)$

$$y=\frac{1}{2}\ln\frac{E+p_l}{E-p_l}, \quad |y|\le y_0 = 1/2\ln(s/m_t^2), \qquad \eta=-\ln(\tan\vartheta/2),$$

where $\vartheta$ is the scattering angle of hadron $c$ in the centre-of-mass system. With the rapidity variable the cross section in the one pole approximation is read as

$$E\frac{d^3\sigma}{d^3p} = g_a(0)e^{\varepsilon(y_0-y)}v_c(p_t^2)e^{\varepsilon(y_0+y)}g_b(0) = g_a(0)v_c(p_t^2)g_b(0)e^{2\varepsilon y_0}. \tag{1}$$

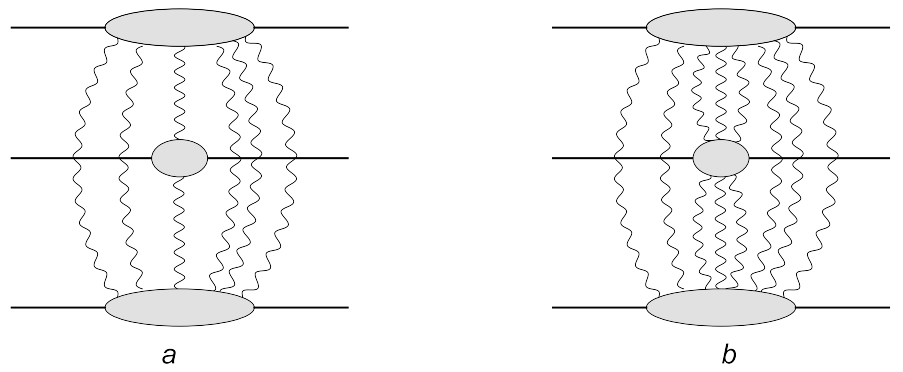

Figure 4: Multipomeron exchange diagrams for one particle inclusive production, ($a$) - diagrams calculated in [8], ($b$) - diagrams calculated in [10, 11].

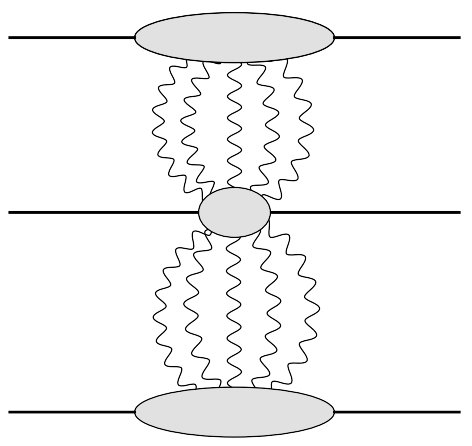

Figure 5: The dominating contribution to central inclusive production at $s\to\infty$.

In Ref. [8] the contribution to inclusive cross section of the diagrams given in Fig. 4a can be calculated. It was shown that due to Abramovsky-Gribov-Kancheli rules [9] only input diagram with one pomeron exchange contributes, rest sum of diagrams vanishes. Thus the cross section again is given by Eq. (1).

Making use of the sum rule

$$\int\frac{d^3p}{E}E\frac{d^3\sigma(ab\to cX)}{d^3p} = \langle n_c\rangle\sigma_t^{ab}(s) \tag{2}$$

one can calculate the mean multiplicity of hadrons as well as taking into account that $\sigma_t \approx \sigma_0 \ln^2(s/s_0)$ to find at $s \to \infty$

$$\frac{dn_c}{dy} = \frac{1}{\sigma_t(s)} 8\pi g_a(0)\, \tilde{V}_c g_b(0)\, (s/s_0)^{\varepsilon} \propto \frac{(s/s_0)^{\varepsilon}}{\ln^2(s/s_0)},$$

where $\tilde{V}_c$ is the integral of $v_c(p_t^2)$ over $p_t^2$. Thus, integrating over $y$ we obtain a power growth for mean multiplicity, $\langle n_c \rangle \propto (s/s_0)^{\varepsilon}/\ln(s/s_0)$. Let us notice that $dn_c/dy$ does not depend on $y$, however it is not supported by the experimental data.

More diagrams must be added to calculate the inclusive cross section under interest. These diagrams are shown in Fig. 4b and were calculated in [10, 11]. Likewise the case of diagrams Fig. 4a the sum of all contributions with the reggeons between hadrons $a$ and $b$ vanishes.

As result at $s \to \infty$ the inclusive cross section in the central region is dominated by the contribution of the diagram in Fig. 5 and can be written in a general form as

$$E\frac{d^3\sigma}{d^3p} = g_a(0)\mathcal{F}(y_0 - y)v_c(p_t^2)\mathcal{F}(y_0 + y)g_b(0), \quad \text{where} \quad \mathcal{F}(y_0 \pm y) = (y_0 \pm y)^2.$$

It is necessary to note that this result exactly coincides with those which can be obtained if we assume from the beginning that pomeron at $t = 0$ is a triple $j$-pole.

This fact and similar ones valid for the elastic amplitude allow us to make more general assumptions and consider the diagram of Fig. 3 with pomerons of arbitrary hardness at $t = 0$.

If the pomeron contribution to the partial amplitude (of elastic scattering) at $t = 0$ is proportional to $1/(j-1)^{\nu+1}$ then

$$\mathcal{F}(y_0 \pm y) = (y_0 \pm y)^{\nu} \qquad \text{and} \qquad \frac{dn}{dy} \propto (y_0 - y)^{\nu}(y_0 + y)^{\nu}.$$

We would like to remark that such a behaviour of $dn/dy$ (at $\nu > 0$) is in a qualitative agreement with high energy experimental data, which show a rise $dn/dy$ at $y_0$ and a parabolic form. Taking into account that such pomeron leads to $\sigma_t(s) \propto \ln^{\nu}(s/s_0)$ one can find

$$\frac{dn}{dy}(y = 0) \propto \ln^{\nu}(s/s_0) \quad \text{and} \quad \langle n \rangle \propto \ln^{1+\nu}(s/s_0).$$

It is known that excellent description of mean hadron multiplicity is achieved within a logarithmic energy dependence with $\nu = 2$ or $\nu = 3$. All mentioned properties of unitarized pomeron models concerning one particle inclusive distribution are rather attractive, but they should be checked out quantitatively with the data. We do that in the next section.

# 2 Comparison of the Unitarized Pomeron Models with the Data

**Experimental data.** Our aim is not the detailed description of all data, we would like to demonstrate only a possibility of the considered models to reproduce the main features of the high energy data. Evidently, at lower energy we need to add more Regge contributions increasing the number of the fitting parameters. To avoid extra number of contributions and parameters we consider the data on $E\,d^3\sigma/d^3p$ at $\sqrt{s}$ = 200, 540, 630, 900, 1800 GeV (240 points) and on $dn/d\eta$ normalized to $\sigma_{in}$ (48 points) [12].

Even for the high energies chosen there is a nontrivial dependence of cross sections on $p_t$ (Fig. 6), their slope is changing with energy. The dependence on $p_t$ in the pomeron contribution is coming only from the vertex function $v_c(p_n^2)$, therefore one has to conclude that the slope effect can be explained in the model only due to sub-asymptotic contributions. Besides this, an exponential increasing $E\,d^3\sigma/d^3p$ at small transverse momenta $p_t < 1\,\text{GeV}$ is changed for a power-like behaviour at higher $p_t$ (larger than 1 GeV).

Another set of data, namely, $dn/d\eta$ is more interesting for our aim. It can be obtained from $Ed^3\sigma/d^3p$ by integration over $p_t$ and with a replacing $y$ for $\eta$. To perform the integration one has to know the vertex functions $v_c(p_t^2)$ which are not determined within any Regge model, we parameterize it in a some form to reproduce existing experimental data. The explicit form of $p_t$-dependence is not crucial for models under interest. It plays only subsidiary role in obtaining $d\sigma/d\eta$ or $dn/d\eta$. The dependence of the differential cross section on $y$ is more important for a verification of our models.

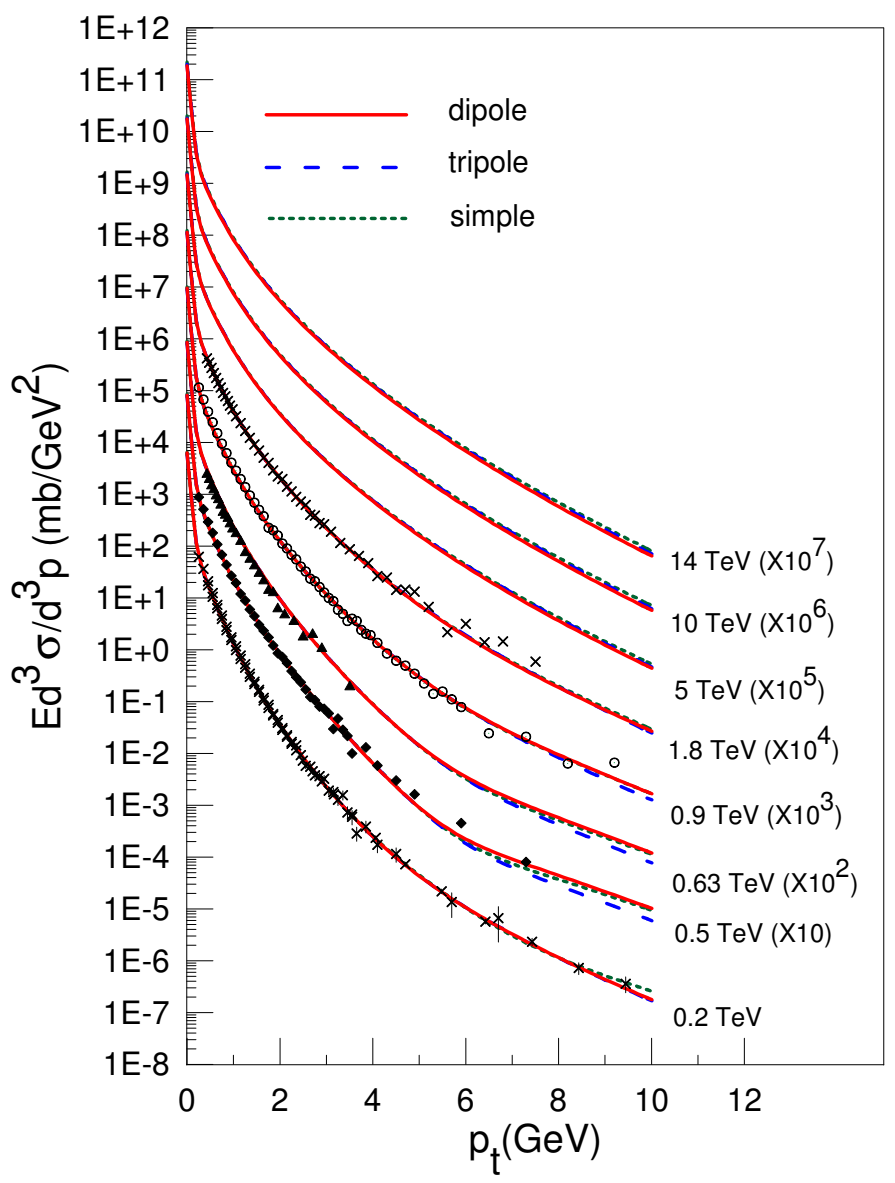

Figure 6: $p_t$-dependence of inclusive cross sections at high energies.

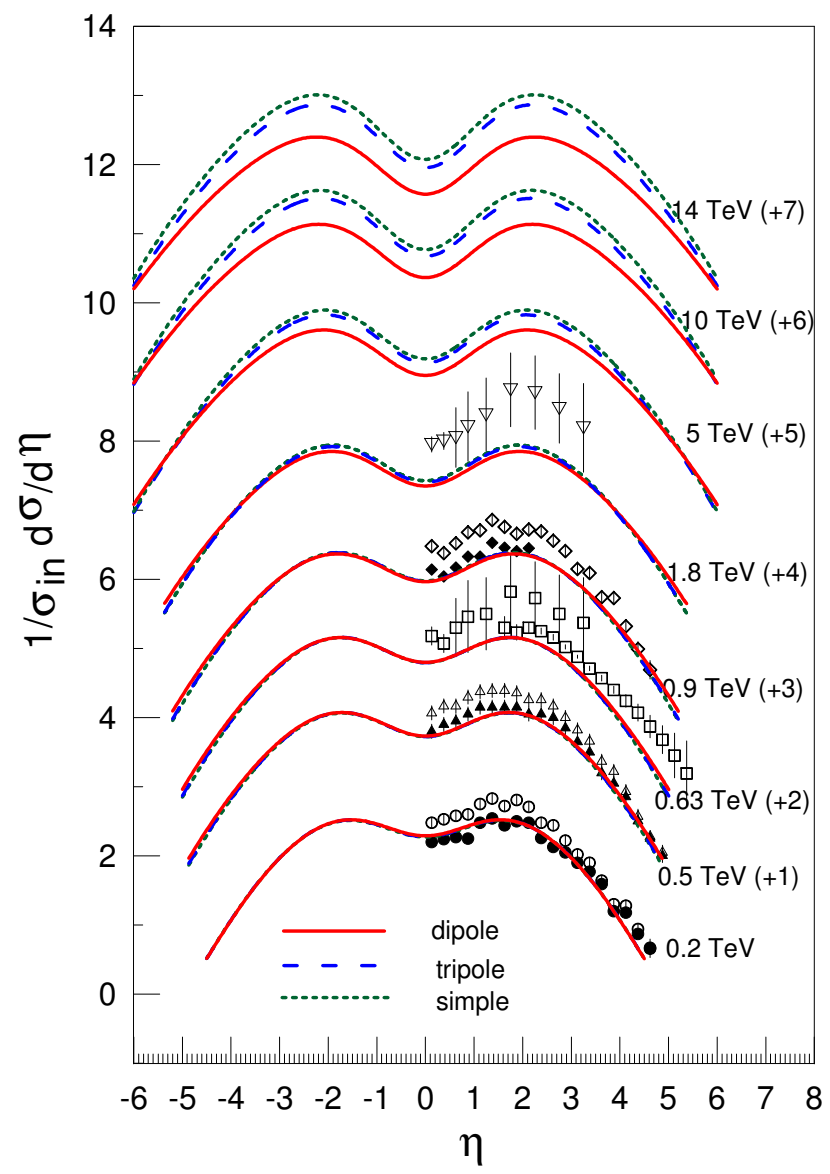

Figure 7: Density of the produced hadrons.

**Pomeron models.** At $\sqrt{s} \le 200\,\text{GeV}$ in the dipole pomeron model we take into account in the diagram in Fig. 3 the dipole (d) and simple (p) poles, both with $\alpha(0) = 1$, and $f$-reggeon for the upper and lower parts of the diagram. In the simple pole model instead of the dipole we have considered simple pole with $\alpha(0) = 1 + \varepsilon$. In the tripole pomeron case a triple pole has to be added. However, to avoid too many parameters we consider here a simplified model for tripole pomeron. The general form of the cross section for all considered models is the following:

$$
\begin{aligned}
E\frac{d^3\sigma}{d^3p} = \;& g_{11}v_{11}(p_t)z_u z_d + g_{12}v_{12}(p_t)(z_u + z_d) + g_{22}v_{22}(p_t)+\\
& g_{13}v_{13}(p_t)[z_u e^{\varepsilon_f(y_0+y)} + z_d e^{\varepsilon_f(y_0-y)}]+\\
& g_{23}v_{23}(p_t)[e^{\varepsilon_f(y_0-y)} + e^{\varepsilon_f(y_0+y)}] + g_{33}v_{33}(p_t)e^{2\varepsilon_f y_0}
\end{aligned}
$$

where $g_{ab}$ with $a, b = 1, 2, 3$ are constants, $\varepsilon_f = \alpha_f(0) - 1$, $v_{ab}(p_t)$ are vertex functions

$$v_{ab}(p_t) = (1 + p_t^2/p_0^2)^{-\mu_{ab}}[e^{-Bp_t} + c(1 + p_t^2/p_1^2)^{-\mu}].$$

For the dipole pomeron model

$$z_u = (y_0 - y), \qquad z_d = (y_0 + y),$$

for the tripole pomeron model

$$z_u = \beta(y_0 - y)^2 + (y_0 - y), \qquad z_d = \beta(y_0 + y)^2 + (y_0 + y),$$

and for the simple pomeron model

$$z_u = e^{\varepsilon(y_0-y)}, \quad z_d = e^{\varepsilon(y_0+y)}.$$

**The data fit.** The description of the data is demonstrated in Figs. 6 and 7. Data are taken from [12]. The red solid line – dipole pomeron model, blue long dashed line – tripole pomeron model, green doted line – simple pomeron pole with $\alpha(0) > 1$. Predictions for three LHC energies are also shown. In Fig. 7 the solid symbols correspond to the data normalized to $\sigma_{in}$, open symbols correspond to data normalized to $\sigma_{NSD}$ (not used in the fit procedure). Parameters of the models as well as $\chi^2$ will be given in a more complete paper [13].

One can see that the theoretical curves in the three models are very close to each other, at least for energies where data exist. The fit gives $\chi^2$/n.d.f. = 3.01 (dipole), 2.98 (tripole) and 2.96 (simple pole). It is not a surprise because the parameters of the tripole and simple models in fact mimic at not very high energy the dipole pomeron model. In the tripole model parameter $\beta$ is equal to 0.03, thus the terms containing $(y - y_0)^2$ are not important at the achieved energy, $\sqrt{s} \leq 1800\,\mathrm{GeV}$. A similar situation occurs in the simple pomeron model where a strong cancellation among the $S$- and $P$-terms occurs.

$$g_{ss}(s/s_0)^{\varepsilon} + g_{pp} \approx g_{ss} + g_{pp} + g_{ss}\varepsilon\ln(s/s_0)$$

However, a difference between the models' predictions is increasing with energy. It can be seen clearly in Figs. 8 and 9 which demonstrate the behaviour of $dn(\eta = 0)/d\eta$, $\langle n\rangle$ in energy.

# 3 Conclusion

We have shown that the high energy experimental data on one-particle inclusive distribution can be described well in the models of unitarized pomeron, which do not violate unitarity restrictions. They predict a small difference for differential cross sections, and mean multiplicities at LHC energies, giving $dn/d\eta(y = 0) \propto \ln^4 s(\ln^3 s)$ and $\langle n\rangle \propto \ln^3 s(\ln^2 s)$ for the dipole (tripole) pomeron model, correspondingly.

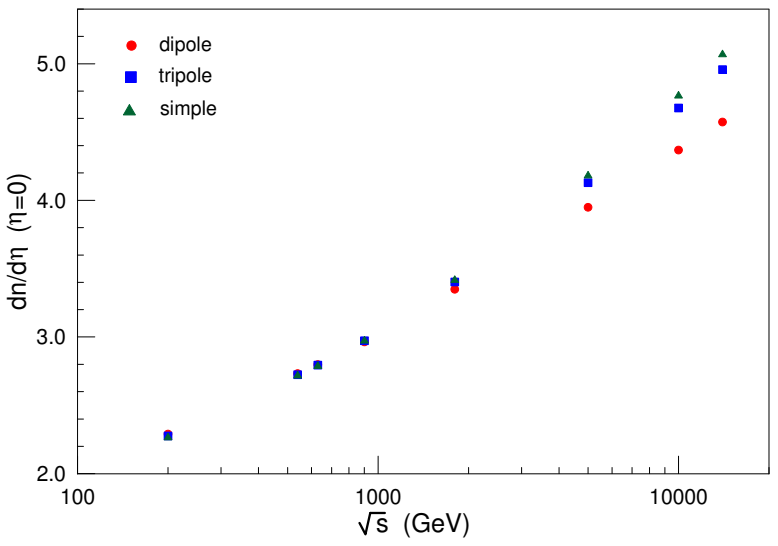

Figure 8: Density of the produced hadrons at $\eta = 0$ as function of energy.

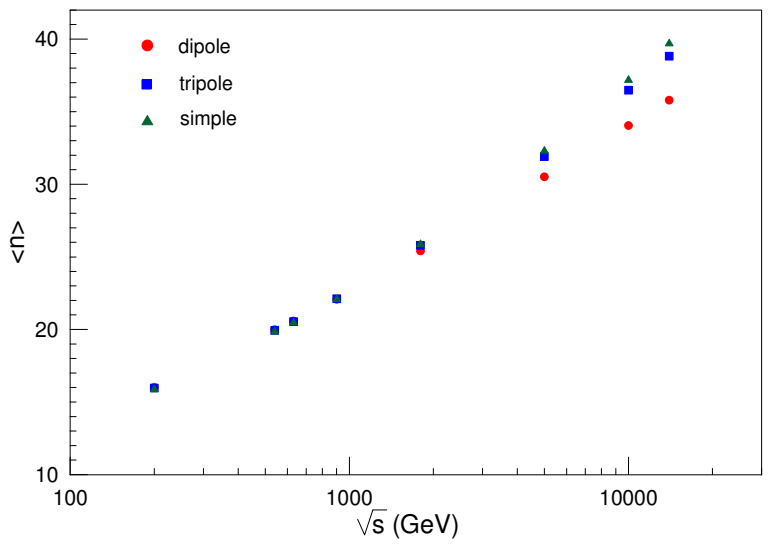

Figure 9: Mean multiplicity of the produced hadrons as function of energy, calculated in the interval $-3.5 \leq \eta \leq 3.5$.

# Acknowledgement

We thank M. Poghosyan for helpful discussions.

# References

[1] S. Donnachie, G. Dosch, P. Landshoff, O. Nachtmann, Pomeron Physics and QCD, Cambridge University Press, 2002.

[2] K.A. Ter-Martirosyan, Sov. ZhETF Pisma **15**, 519(1972).
Discussion of an eikonal approximation in the Regge theory as well as references to the original papers on this subject can be found in P.D.B. Collins, An introduction to Regge theory & high energy physics, Cambridge University press (Cambridge) 1977.

[3] S.M. Troshin and N.E. Tyurin, Phys. Part. Nucl. **35**, 555 (2004).

[4] N.I. Glushko, N.A. Kobylinsky. E.S. Martynov, V.P. Shelest, Yad. Fiz. (Sov. Nucl. Phys.) **38**, 180 (1983).

[5] J.R. Cudell, E. Martynov, G. Soyez, Nucl. Phys. **B682** (2004) 391.

[6] E. Martynov, Phys. Rev. **D76**, 074030, (2007) and references therein.

[7] M.S. Dubovikov, K.A. Ter-Martirosyan, Nucl.Phys. **B124**, 163 (1977). See also [6] and refs. therein.

[8] K.A. Ter-Martirosyan, Preprint of the ITEP, ITEP-135, 1976, Moscow (in Russian).

[9] V. A. Abramovsky, V. N. Gribov, and O. V. Kancheli, Sov. J. N ucl. Phys. **18**, (1974) 308.

[10] E.S. Martynov, Preprint of Bogolyubov Institute for Theoretical physics, ITP-85-26-E, 1985, Kiev.

[11] A.K. Likhoded, O.P. Yushchenko, Int. Journ. Modern Physics A, 6 (1991) 913.

[12] F. Abe, Phys. Rev. Lett. **61**, 1819 (1988); F. Abe et al. [CDF Collaboration], Phys. Rev. **D41**, 2330 (1990); G.J. Alner et al. (UA5 Collaboration), Phys. Rept. **154**, 247 (1987); J. Adams et. al., Phys. Rev. Lett. **91**, 172302 (2003); C. Albajar et al., Nucl. Phys. **B335**, 261 (1990); G.J. Alner et al. (UA5 Collaboration), ZP **C33**, 1 (1986); R. Harr et al., Phys. Lett. **B401**, 176 (1997).

[13] A.Alkin, E.Martynov, E.Romanets, V.Pauk, to be published.

# Factorization Breaking in Diffraction

*Konstantin Goulianos*

The Rockefeller University, 1230 York Avenue, New York, NY 10065-9965, USA

Factorization breaking in diffraction has been experimentally observed in soft and hard $pp$ and $\bar{p}p$ processes, as well as in photoproduction and in low $Q^2$ deep inelastic scattering. In this paper, relevant experimental results are presented and phenomenologically connected through a *common thread* provided by the renormalization model of hadronic diffraction.

## 1 Introduction

Factorization breaking in diffraction has been a topic of interest in high energy physics since the observation of a breakdown of factorization in diffractive dijet production in $\bar{p}p$ collisions at $\sqrt{s} = 630$ GeV by the UA8 collaboration published in 1992 [1]. A suppression of a factor $\sim 4$ was reported relative to theoretical expectations based on parton densities extracted from diffractive deep inelastic scattering (DDIS) at HERA. This result was later confirmed by the CDF collaboration [2], where a suppression of $\mathcal{O}(10)$ was found at $\sqrt{s} = 1800$ GeV. Equally important is a 1994 CDF result of a breakdown of factorization in soft diffraction: the total $\bar{p}p$ diffractive cross section at $\sqrt{s} = 540$ GeV [$\sqrt{s} = 1800$ GeV] was found to be suppressed by a factor of $\sim 4$ [factor of $\mathcal{O}(10)$] relative to Regge theory expectations [3].

The similarity of the suppression between soft and hard processes is in contrast with diffractive photon dissociation results [4] and DDIS, where only a $\sim 30\%$ suppression is seen in $\gamma p$ but no suppresion was seen in high-$Q^2$ DDIS. Recently, HERA experiments reported factorization breaking in $\gamma p$ and $\gamma^* p$ processes, including vector meson production and dijet production (see HERA talks in these proceedings). The breakdown generally occurs at low $Q^2$ with a magnitude dependant on scale, such as the mass of the vector meson or the dijet mass.

We review relevant experimental data from the Tevatron and from HERA, and offer a phenomenological interpretation based on renormalizing the *rapidity gap probability* to unity, which effectively removes overlapping rapidity gaps generally appearing in other models as multi-Pomeron exchanges (see [5]). The renormalization model (RENORM) is briefly discussed in Sect. 4. By removing potential contributions from overlapping rapidity gaps, RENORM leads to a scaling behaviour in single-diffraction and an asymptotically constant total cross section, $\sigma_t^{SD} \stackrel{s\to\infty}{\to}$ constant [6].

The paper is organized in five sections:

1. Introduction
2. $pp$ and $\bar{p}p$ results
3. $\gamma p$ and $\gamma^* p$ results
4. RENORM: *the common thread*
5. Summary and conclusions

# 2 $pp$ and $\bar{p}p$ Results

Figure 1 shows the soft and hard diffractive $\bar{p}p$ processes studied at CDF.

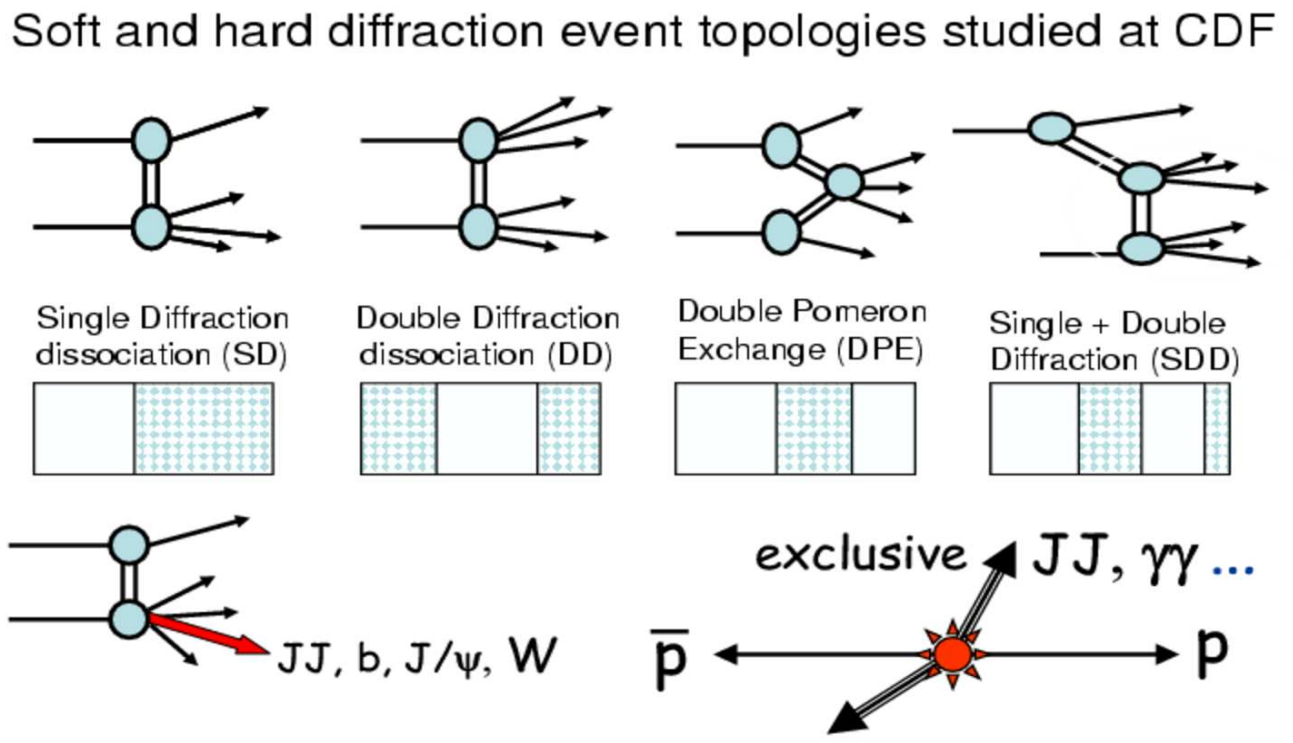

Figure 1: Event topologies of processes studied in $\bar{p}p$ collisions at CDF.

## 2.1 Soft Single-Diffraction

The first result on factorization breaking was the discovery that the total single-diffractive cross section did not exhibit the $s^{2\epsilon}$ dependence expected by Regge factorization but was suppressed by a factor of $\mathcal{O}(10)$ at $\sqrt{s} = 1800$ GeV, as shown in Fig. 2 (left). In contrast, $d^2\sigma_t^{SD}/dt\,dM^2|_{t=0.05}$ which was expected to vary as $s^{2\epsilon}$ was found to have no explicit $s$-dependence – see Fig. 2 (right). This $\boldsymbol{M^2}$**-scaling** behaviour leads to an asymptotically constant $\sigma_t^{SD}$ as $s \to \infty$ and forms the basis of the RENORM model, which is used in predicting the ratio of the intercept of the Pomeron trajectory to its slope [7] and the total cross section at the LHC [6].

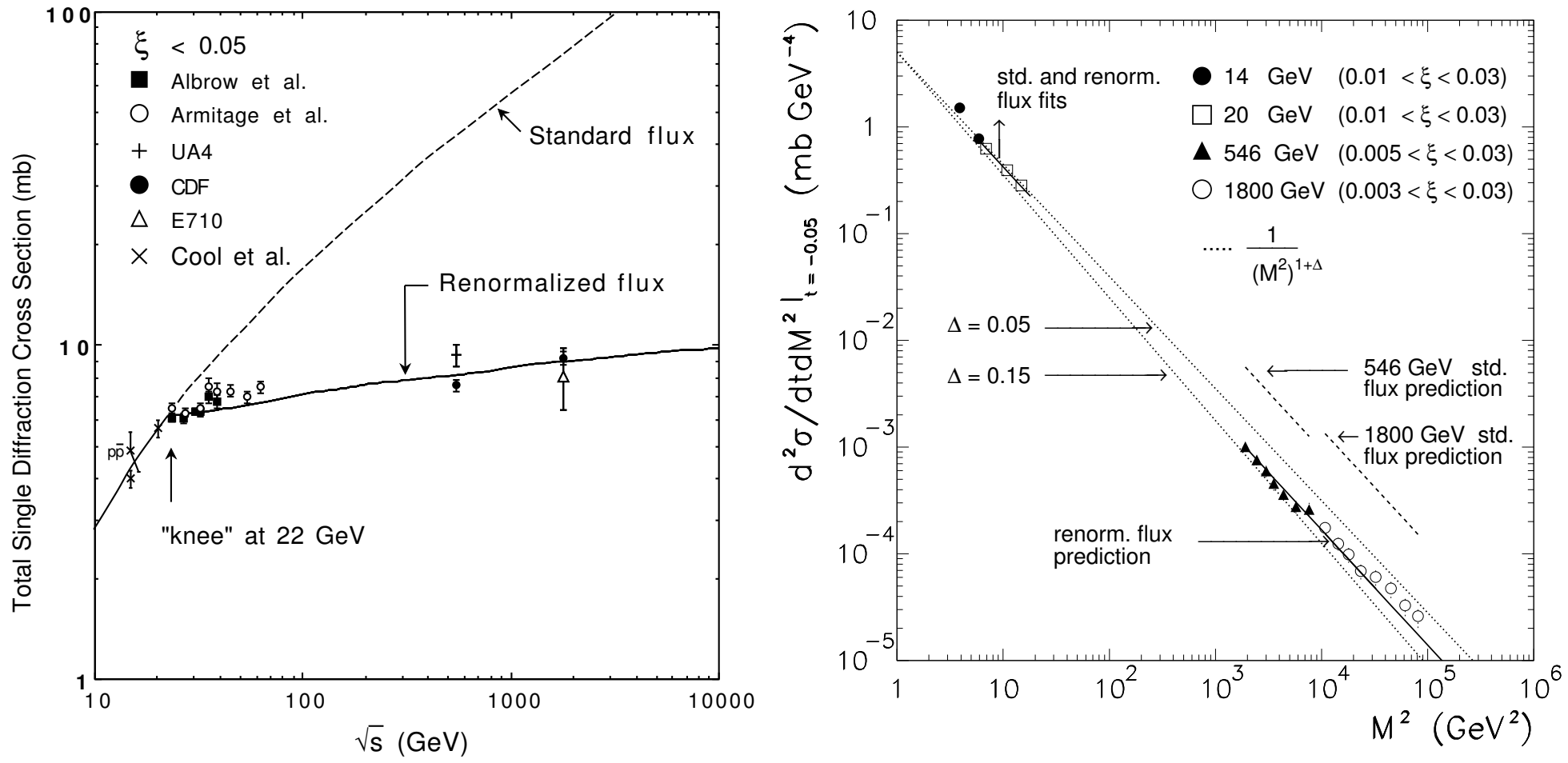

Figure 2: *(left)* $\sigma_t^{SD}$ vs. $\sqrt{s}$; *(right)* $d^2\sigma_t^{SD}/dt\,dM^2|_{t=0.05}$ compared with Regge predictions.

## 2.2 Soft Double and Multi-Gap Diffraction

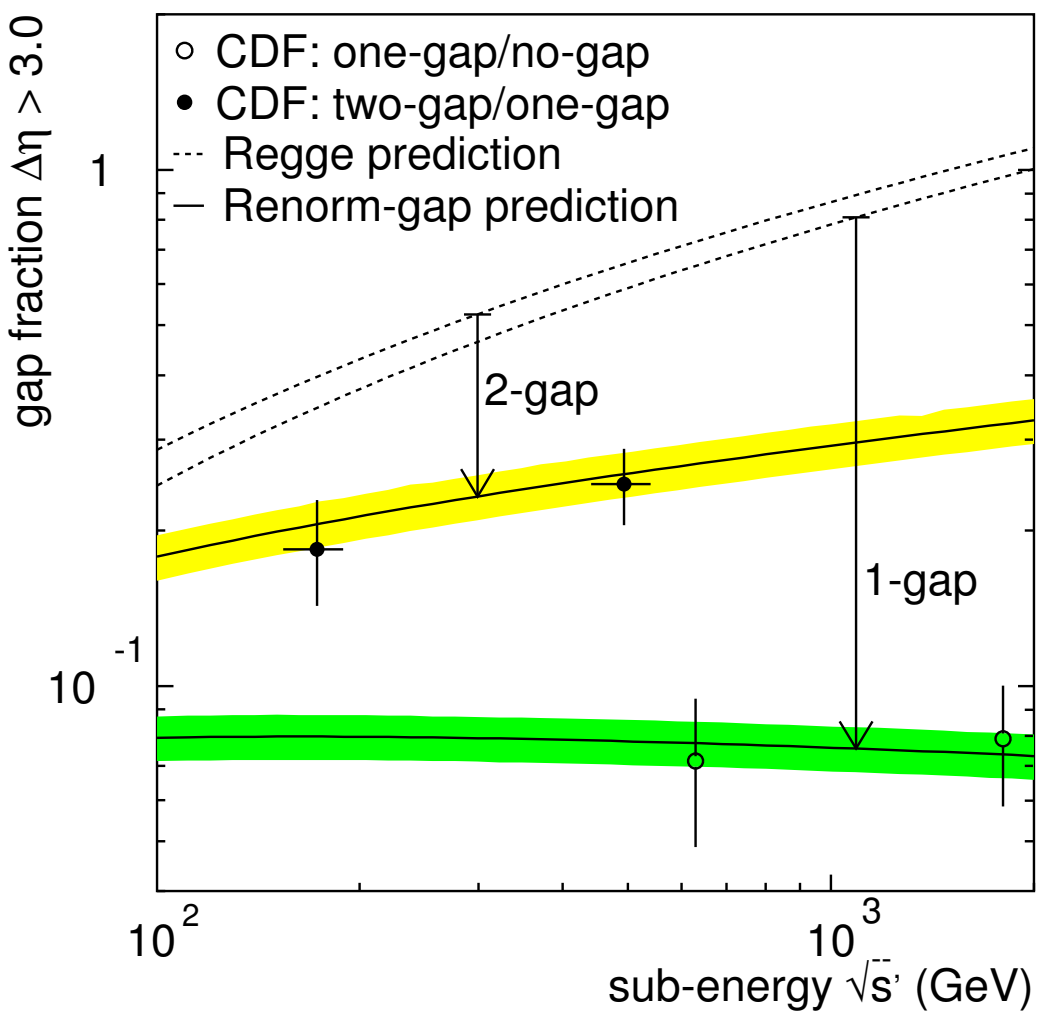

Figure 3: Ratios of two-gap (SDD) to one-gap (SD) rates (solid) and one-gap DD to no-gap (total cross section) vs. $\sqrt{s}_{I\!P-p}$ and $\sqrt{s}_{\bar{p}p}$, respectively.

An important input to deciphering the mechanism of factorization breaking in diffraction is provided by the study of processes with multiple diffractive rapidity gaps. Two such processes were studied by CDF, DPE and SDD (see Fig 1). The $\eta$-range available at the Tevatron is not large enough to observe multi-gap events with more than two rapidity gaps, but the lessons learnt from two-gap diffraction studies can be used to pave the way to multi-gap diffraction studies at the LHC.

Classified by the number of rapidity gaps in an event, the following soft diffraction processes were studied at CDF:

- 0-gap: total cross section,
- 1-gap: SD and DD, and
- 2-gap: DPE and SDD.

It was found that while factorization breaking of the same magnitude is observed in the 1-gap to no-gap ratios, the 2-gap to 1-gap ratios are much less suppressed.

## 2.3 Hard Diffraction

As shown in Fig. 1, CDF has obtained results for several single diffractive hard processes involving $JJ$, $b$-quark, $J/\psi$ and $W$ production (and also $Z$ production in Run II). Two types of results have been extracted from the data: diffractive fractions (ratios of diffractive to total production rates) and diffractive structure functions. The general features of the Run I results are summarized below.

- *Diffractive fractions:* at the same collision energy, all measured diffractive fractions are approximately equal; at $\sqrt{s} = 1800$ GeV the fractions are $\approx 1\%$; differences among the measured fractions can be attributed to kinematics.
- *Diffractive structure functions:* the most precise structure functions were extracted from dijet production in SD [2] and in DPE [11]; results are shown in Fig. 4.

The following conclusions were drawn:

(a) *factorization breaking:* a factorization breaking of $\mathcal{O}(10)$ relative to expectations from diffractive parton densities extracted from DDIS at HERA was found, which is similar to that observed in soft diffraction relative to Regge expectations.

(b) *restoration of factorization:* the 2-gap to 1-gap ratio is not as strongly suppressed, just as in soft diffraction.

In addition to the results obtained in Run I, there are also several results obtained in Run II at $\sqrt{s} = 1960$ GeV. The factorization breakdown in the diffractive structure function from SD dijets was confirmed, but there are other results that show the relationship between the diffractive and non-diffractive structure functions and point to a saturation of the rapidity gap probability as the main controlling factor of the factorization breakdown.

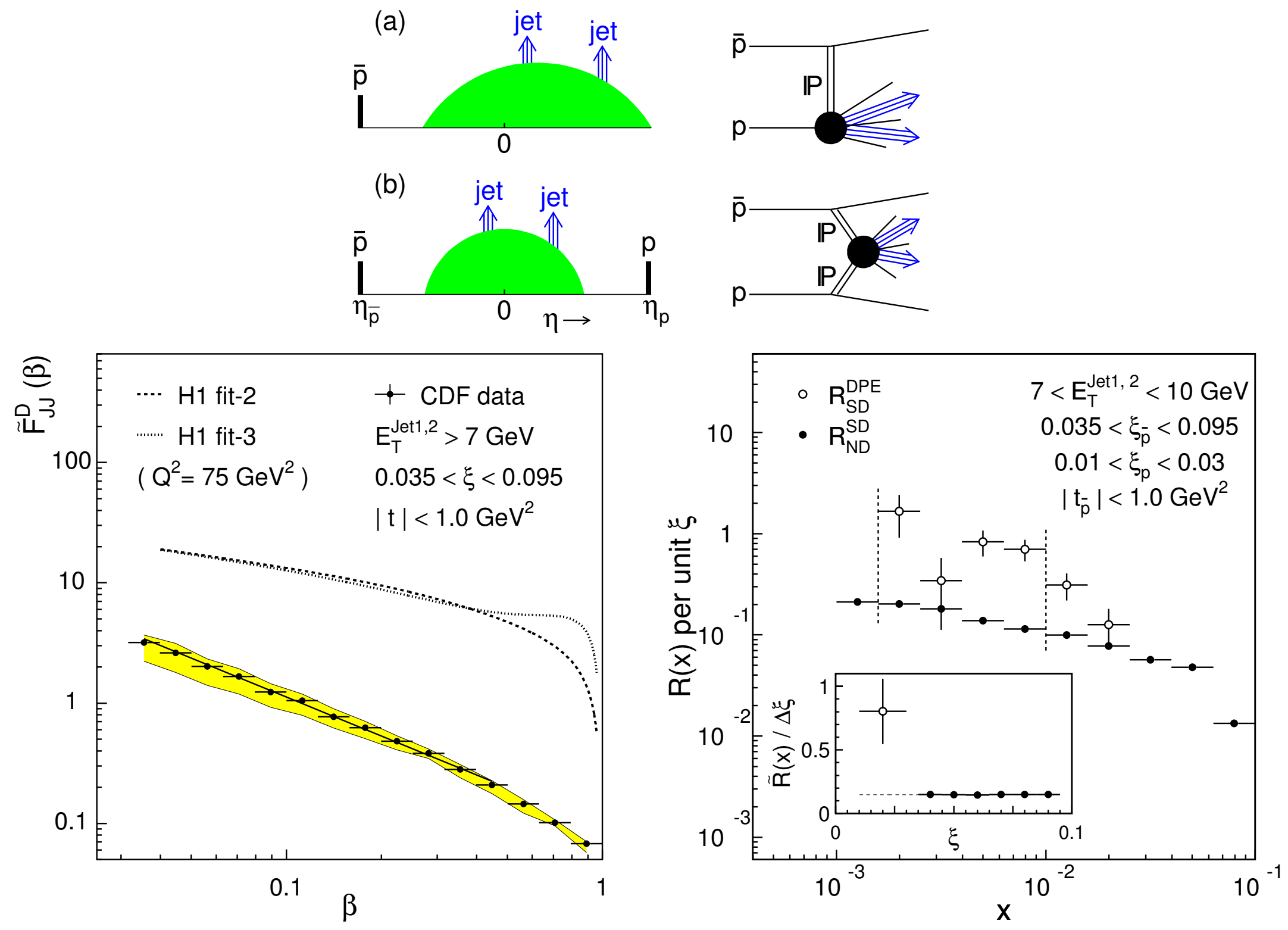

Figure 4: Dijet production in (a) SD and (b) DPD; *(left)* $F_{JJ}^{D}(\beta)$ vs. $\beta$; *(right)* ratios of DPE to SD and SD to ND rates per unit $\xi$ vs. $x$-Bjorken.

The following Run II results from diffractive events (SD) triggered by the Roman Pot Spectrometer (RPS) and non-diffractive ones (ND) triggered by a dijet event with a calorimeter tower above 5 GeV (Jet5 sample) illustrate the scale *independence* of the suppression factor in dijet production:

- Dijet $E_T^* = (E_T^{jet1} + E_T^{jet2})/2$ distributions,
- $x$-Bjorken distributions, and
- $t$-distributions.

These results are presented in Figs. 5 and 6.

→ Figure 5 shows the $E_T^*$ distribution for SD and ND events. The two distributions are practically identical.

→ Figure 6 (left) shows the SD to ND ratio as a function of Bjorken-x for different $Q^2$ values. In the range of $10^2$-$10^4$ in $Q^2$, within which $E_T^*$ varies by a factor of 100, this ratio varies by less that a factor of two.

→ Figure 6 (right) displays the slope of the diffractive $t$ distribution over the $Q^2$ range of the RPS triggered data normalized to the value from inclusive RPS triggered data which are dominated by soft diffraction. As seen, there is no scale dependence in the slope of the $t$ distribution in the range $\sim 1\,\mathrm{GeV}^2 < Q^2 < 10^4\,\mathrm{GeV}^2$.

The above results suggest that SD interactions have the same QCD origin as non-diffractive ones, *i.e.* originate from the *proton* low-x parton densities. The suppression in rate relative to theoretical expectations is due to the colour constraint imposed by the requirement of exchanging another parton that forms a colour-singlet with vacuum quantum numbers, commonly referred to as *Pomeron*. This picture is reinforced by the CDF finding that the final state event topologies, namely pseudorapidity and $ET$ distributions, are very similar for SD and ND events when compared at the same $I\!P - p$ collision energy $\sqrt{s}'$ as for $\bar{p}p$ collisions at $\sqrt{s}$. This is further discussed in Sect. 3.

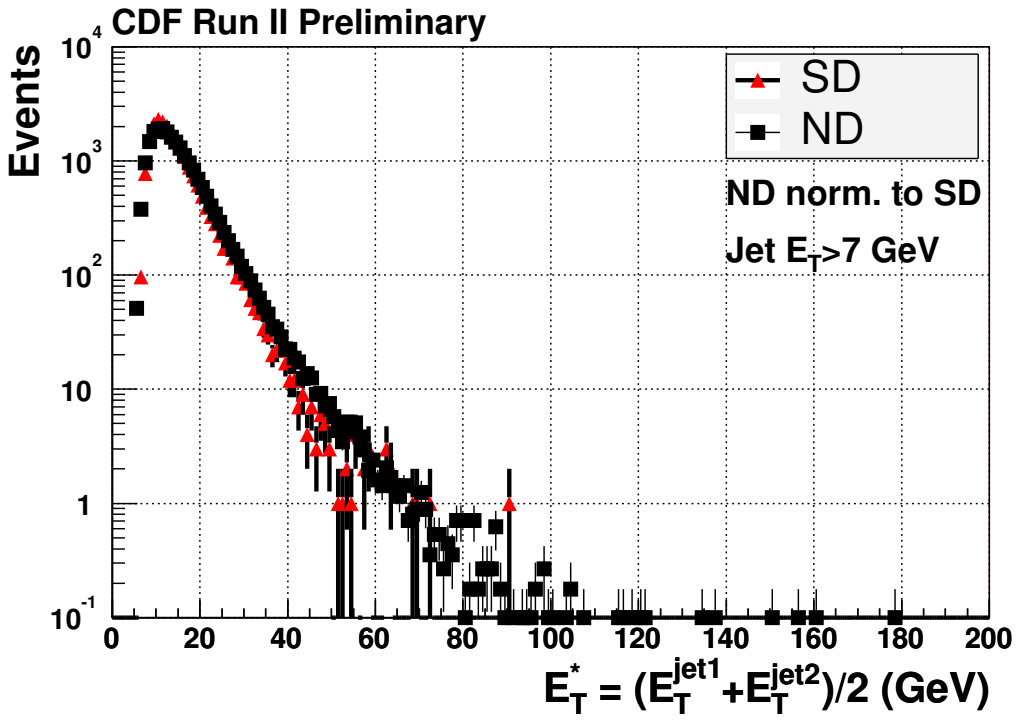

Figure 5: Mean dijet transverse energy distribution for SD and ND events.

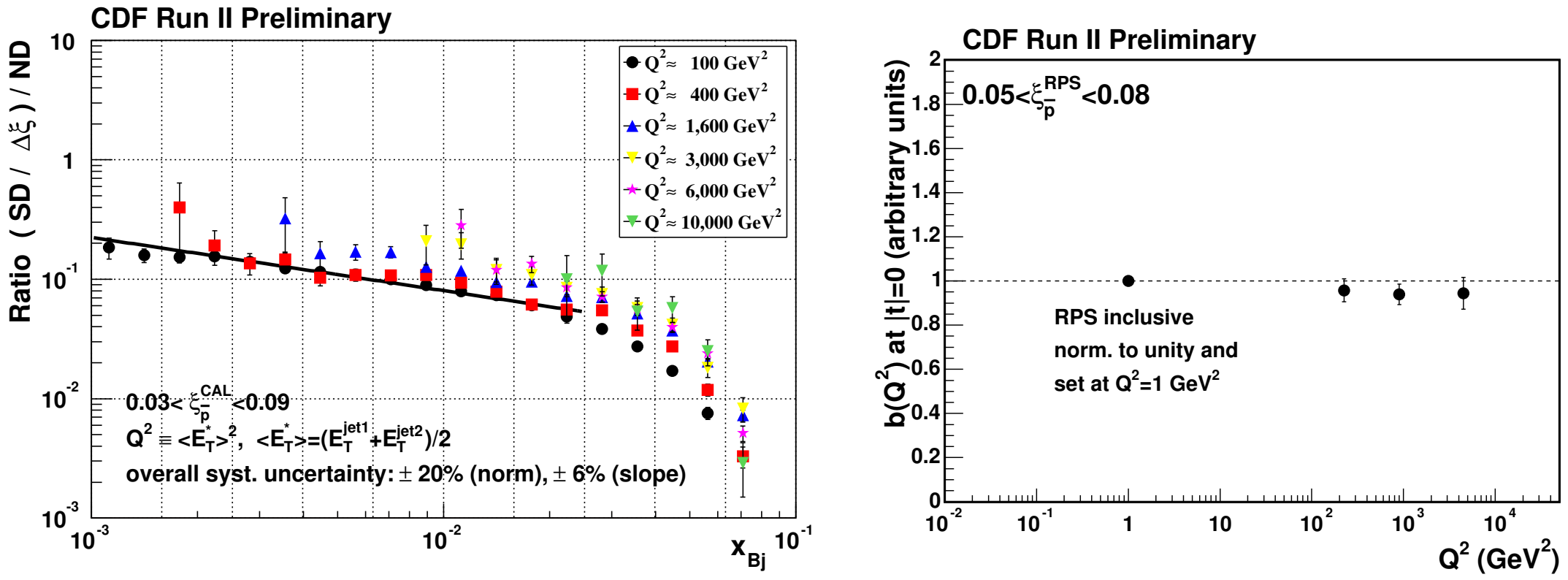

Figure 6: (left) Ratio of diffractive to non-diffractive dijet event rates as a function of $x_{\mathrm{Bj}}$ (momentum fraction of parton in the antiproton) for different $Q^2$ values; (right) the slope parameter of the $t$ distribution $b_1$ near $t = 0$ vs. $Q^2$ (normalized to RPS inclusive data sample)

# 3 $\gamma p$ and $\gamma^* p$ Results

Diffractive photoproduction and DDIS results have been presented at this conference (see [12, 13] and references therein). Below, we present selected results pertaining to factorization breaking, and in the next section we relate the magnitude of the observed effect to that found in $\bar{p}p$ collisions at the Tevatron.

The processes we discuss are vector meson production in $\gamma p$ and $\gamma^* p$ and dijet photopro-duction. Of particular interest is the dependence of the factorization breakdown effect on scale, such as the vector meson mass and the jet $E_T$. Since no scale dependence is observed at the Tevatron, the observation of such dependence at HERA could provide clues for the source of the mechanism of the breakdown.

**a) Diffractive vector meson production:**

1. *$W^{\Delta}$-dependence on $M_{VM}$:* $\Delta$ increases (*) with $M_{VM}$.
2. $b$-slope *of $t$-distribution:* $b$ increases (*) with $M_{VM}$.

**b) Diffractive dijet production:**

1. *direct and resolved processes:* violation observed in both components.
2. *$E_T^{jet}$-dependence:* violation increases with $E_T^{jet}$.

(*) The effect could be a suppression at low $M_{VM}$ in (1) or with decreasing $E_T^{jet}$ in (2).

In all cases, the maximum factorization breaking effect observed is up to $\sim 50\%$.

# 4 RENORM: *the Common Thread*

The renormalization model for hadronic diffraction was introduced in [14] and was later extended to a model of renormalizing the *gap probability* to include DD and multi-gap diffractive processes. RENORM is inspired by the Regge description of diffraction, in which the differential cross section factorizes into two parts, one depending on the pseudorapidity space in which particles are produced and the other on the space occupied by rapidity gaps. This second part is interpreted as the rapidity gap probability and should saturate when it reaches unity.

The collision energy at which saturation occurs can be *read off* from Fig. 2 as $\sqrt{s} = 22$ GeV, which corresponds to a rapidity span of $\ln s = 2 \times \ln 22 = 6.2$ units. For any process where the rapidity span in which particles can be produced exceeds 6.2 units, saturation will occur expressed as a suppression of the cross section. The magnitude of the effect can be determined from Fig. 2 as the ratio of the values of the cross section represented by the *renormalized flux* (solid) to *standard flux* (dashed) curves. All CDF results presented here are all in agreement with RENORM predictions (see listed references). In each case, care was taken to asses the rapidity span available for particle production.

Renormalization can equally well be applied to $\gamma p$ and $\gamma^* p$ collisions at HERA. The HERA data were taken at a c.m.s. $ep$ collision energy of 320 GeV, which corresponds to a rapidity span of 11.5 units. In DDIS and in processes with a hard scale in the final state, rapidity space occupied by this scale becomes unavailable for particle production reducing the probability of overlaps and thereby the suppression factor.

In soft diffraction, the entire rapidity span is available for particle production, and therefore from Fig. 2 a suppression of a factor $\sim 3$ would be expected at this energy, in agreement with the data.

In vector meson production, rapidity space occupied by the vector meson mass ($\ln M_{VM}^2$) and by the $|t|$ scale of the recoil proton ($\ln |t|$) must be subtracted from the value of 11.5 before evaluating the suppression factor. Therefore, one would expect the suppression to increase as the $M_{VM}$ and $|t|$ decrease. This is precisely what is observed in the data: as the $\ln Q^2$ decreases, including contributions from any hard scale present in the final state, the suppression increases.

In diffractive dijet photoproduction, hard scales are introduced my $E_T^{jet}$ and $|t|$. For the data samples studied, a suppression factor of $\sim 2$ would be expected, both for the **direct** and **resolved** components.

# 5 Summary and Conclusions

Results from the Tevatron on factorization breaking in soft and hard diffraction in $\bar{p}p$ collisions obtained by the CDF collaboration have been presented, including single-gap and multi-gap processes. Factorization breaking in diffractive vector meson and dijet production at HERA has also been discussed and compared with the Tevatron results. The renormalization model *RENORM*, which handles double-counting caused by overlapping rapidity gaps was offered as a *common-thread* to explain under the same principle both the Tevatron and HERA results.

# References

[1] UA8 Collaboration, A. Brandt *et al.*, Phys. Lett. **B 297**, 417 (1992); **B 421**, 395 (1998).

[2] T. Affolder *et al.* (CDF Collaboration), "Diffractive Dijets with a Leading Antiproton in $\bar{p}p$ Collisions at $\sqrt{s} = 1800\,\text{GeV}$", Phys. Rev. Lett. **84**, 5083 (2000).

[3] F. Abe *et al.* (CDF Collaboration), "Measurement of $\bar{p}p$ Single Diffraction Dissociation at $\sqrt{s} = 546$ and 1800 GeV", Phys. Rev D **50**, 5535 (1994).

[4] T.J. Chapin et al., "Diffraction Dissociation of Photons on Hydrogen", Phys. Rev. D **31**, 17 (1985).

[5] K. Goulianos, "Hadronic diffraction: where do we stand"? in *Les Rencontres de Physique de la Vallée d'Aoste: Results and Perspectives in Particle Physics, La Thuile, Italy, February 27 - March 6, 2004*, Frascati Physics Series, Special 34 Issue, edited by Mario Greco; arXiv:hep-ph/0407035v1.

[6] K. Goulianos, "Diffractive and Total Cross Sections at LHC", in these proceedings.

[7] K. Goulianos, "Pomeron intercept and slope: the QCD connection", arxiv:0812.4464v2 [hep-ph] 26 Mar 2009.

[8] T. Affolder et al. (CDF Collaboration), "Double Diffraction Dissociation at the Fermilab Tevatron Collider", Phys. Rev. Lett. **87**, 141802-(1-6) (2001).

[9] D. Acosta et al. (CDF Collaboration), "Inclusive Double-Pomeron Exchange at the Fermilab Tevatron pbar-p Collider", Phys. Rev. Lett. **93**, 141601-(1-7) (2004).

[10] D. Acosta et al. (CDF Collaboration), "Central pseudorapidity gaps in events with a leading antiproton at the Fermilab Tevatron $\bar{p}p$ Collider", Phys. Rev. Lett. **91**, 011802-(1-6) (2003).

[11] Affolder, T. *et al.* (CDF Collaboration), "Dijet Production by Double Pomeron Exchange at the Fermilab Tevatron", Phys. Rev. Lett. **85**, 4215 (2000).

[12] A. Proskuryakov, "Inclusive hard diffraction at HERA", in these proceedings.

[13] P. Marage, Université Libre de Bruxelles, "Exclusive hard diffraction at HERA", in these proceedings.

[14] K. Goulianos, "Renormalization of Hadronic Diffraction and the Structure of the Pomeron", Phys. Lett. B **358**, 379 (1995); Erratum-*ib.* **363**, 268 (1995).

# Soft Interaction Processes at HERA: Leading Baryon Production, Multi-Parton Interactions

*Armen Bunyatyan*

DESY, Notkestraße 85, 22607 Hamburg, Germany

(for the H1 and ZEUS Collaborations)

Experimental results from HERA on the production of leading protons and neutrons in *ep* collisions are reviewed. The measurements are well described by the models which include virtual meson exchange. Assuming the validity of pion exchange model, the leading neutron data are used to constrain the pion structure function. A refinement of the simple factorisation picture is provided by baryon absorption, occurring through rescattering. Exchange models accounting for absorption describe the $Q^2$ evolution of the data.

Furthermore, the influence of underlying event and multi-parton interactions on the charged particle multiplicity in the photoproduction of jets is investigated.

## 1 Introduction

The QCD hardness scale for secondary particles production in semi-inclusive deep inelastic scattering (DIS) gradually decreases from $Q^2$, the photon virtuality which determines the hard scale in the virtual photon (current) fragmentation region, to a soft, hadronic, scale in the proton fragmentation region. Production of leading baryons (LB) in a process with a hard scale provides a probe of the relationship between the QCD of quarks and gluons and the strong interaction of hadrons. The LBs are produced with small transverse momentum $p_T$, ensuring the presence of a soft process with its related long-range correlations.

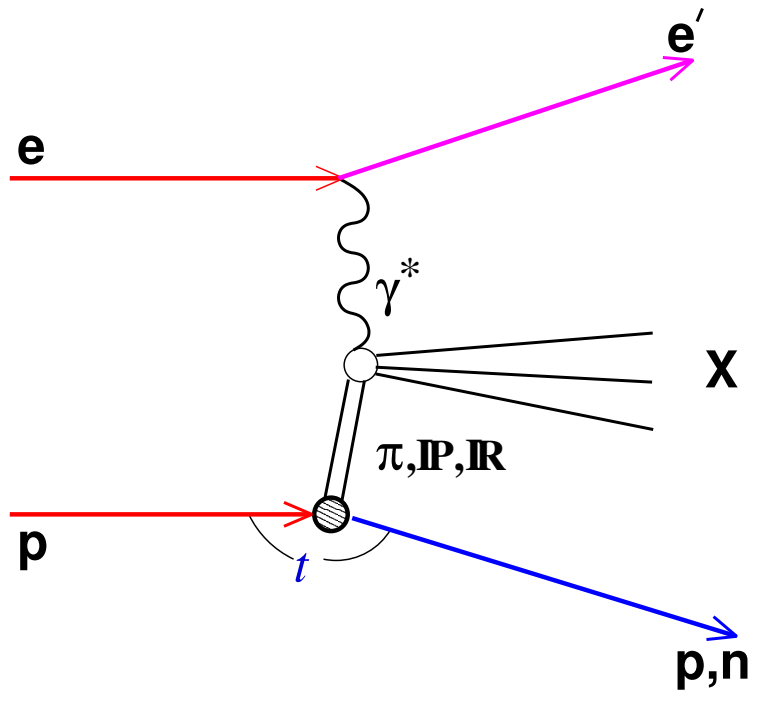

Figure 1: Leading baryon production via the colour singlet exchange processes.

The observation of events with neutrons or protons carrying a large fraction $x_L$ of the incident proton beam energy in electron-proton scattering at HERA [1–6] has led to renewed interest in the QCD evolution and factorisation properties of proton fragmentation to LBs in DIS [7–18]. Although a fraction of these LBs may result from the hadronisation of the proton remnant, the $t$-channel exchange of colour singlet virtual particles is expected to contribute significantly [7, 8, 19, 20]. In this picture, the proton fluctuates into a virtual meson-baryon state; the virtual photon subsequently interacts with a parton from the pion, leaving a fast forward baryon in the final state (Figure 1). The production of leading neutron (LN) in the virtual exchange model occurs through the exchange of isovector states, and $\pi^+$ exchange is expected to dominate. For leading proton (LP) production, isoscalar exchanges also contribute, including diffractive Pomeron mediated interactions. In the simple exchange picture, the cross section is factorised and LB production is largely independent of the variables describing the photon vertex (vertex factorisation). For example, if pion exchange dominates LN production, the cross section can be written as $d\sigma_{\gamma^* p \to nx} = f_{\pi/p}(x_L, t) \times d\sigma_{\gamma^* \pi \to X}$. Here $f_{\pi/p}$ is the flux of virtual pions in the proton, a factor constrained from low energy hadronic data. Such a reaction can thus be used to probe the structure function $F_2^\pi$ of the exchanged pion.

The H1 and ZEUS experiments measured leading baryons in DIS and photoproduction events. Leading protons were measured with position sensitive detectors placed along the proton beam downstream of the interaction point. Leading neutrons were measured with lead-scintillator forward calorimeters at the zero-degree point after the proton beam was bent vertically; magnet apertures limited neutron detection to scattering angles less than 0.75 mrad.

## 2 Leading Baryon Production Cross Sections and Models

Figure 2 shows the cross sections of LP and LN production in DIS [4, 6] normalised to the inclusive DIS cross section as function of $x_L$. For LPs, the rate is approximately flat up to the diffractive peak, where it increases by a factor of about six. For LNs, the cross section rises from the lowest $x_L$ due to the increase in $p_T^2$ space, reaches a maximum near $x_L = 0.7$, and falls to zero at $x_L = 1$. The right side of Figure 2 shows a comparison of normalised LP and LN cross sections restricted to the same $p_T^2$ range of $p_T^2 < 0.04\,\mathrm{GeV}^2$. In the range $0.32 < x_L < 0.92$, there are approximately twice as many protons as neutrons. In a particle exchange model, the exchange of isovector particles would result in half as many protons as neutrons. Thus, exchange of isoscalars must be invoked to account for the observed proton rate.

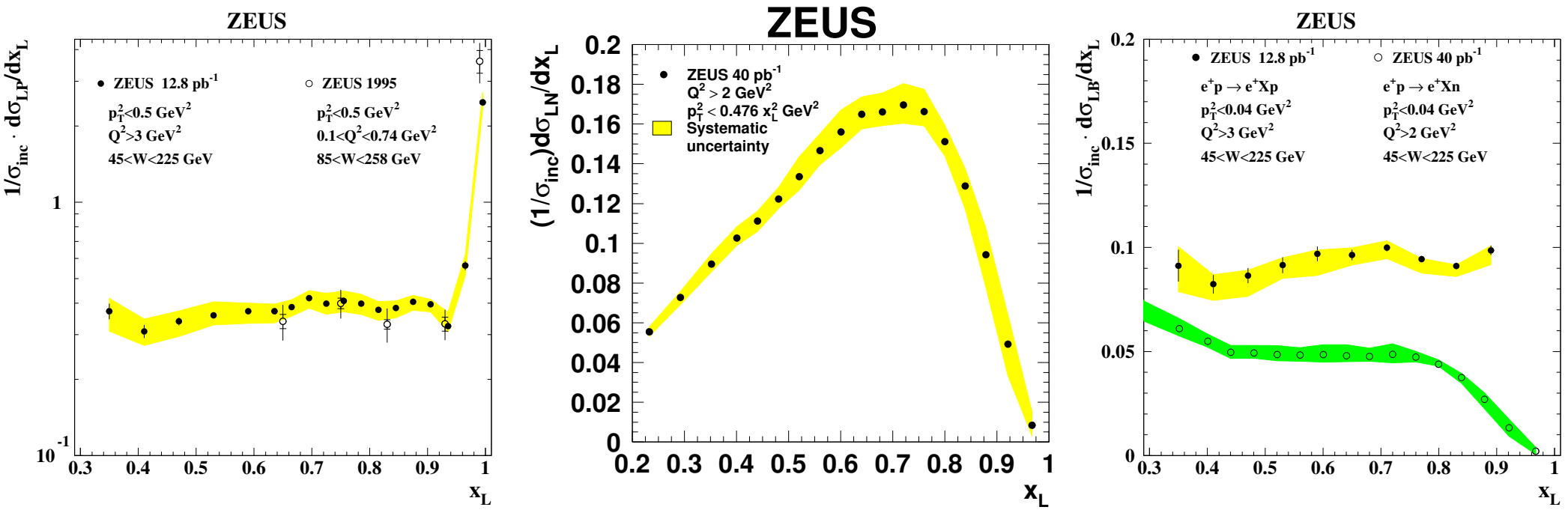

Figure 2: The distributions $(1/\sigma_{inc})d\sigma_{LB}/dx_L$ for protons (left) and neutrons (central) in DIS. The right side figure shows the comparison of leading proton and neutron $x_L$ spectra to restricted range in $p_T^2 < 0.04\,\mathrm{GeV}^2$.

Figure 3 shows the double differential cross-section for LN and LP production in DIS as a function of $p_T^2$ in bins of $x_L$, normalised to the inclusive DIS cross section. In each $x_L$ bin the data are well described by an exponential distribution $a(x_L) \cdot e^{-b(x_L)p_T^2}$. The LN $x_L$ cross section, the intercepts $a$ and slopes $b$ are compared in Figure 4 to several Monte Carlo (MC) models [21, 22]. None of the models incorporating only standard fragmentation predicts the observed LN yield [4, 5]. The mixture of the standard fragmentation and $\pi$-exchange models gives a better description of the shape of the $x_L$ distribution, and also predicts the rise of $b$ with $x_L$, although with too high values.

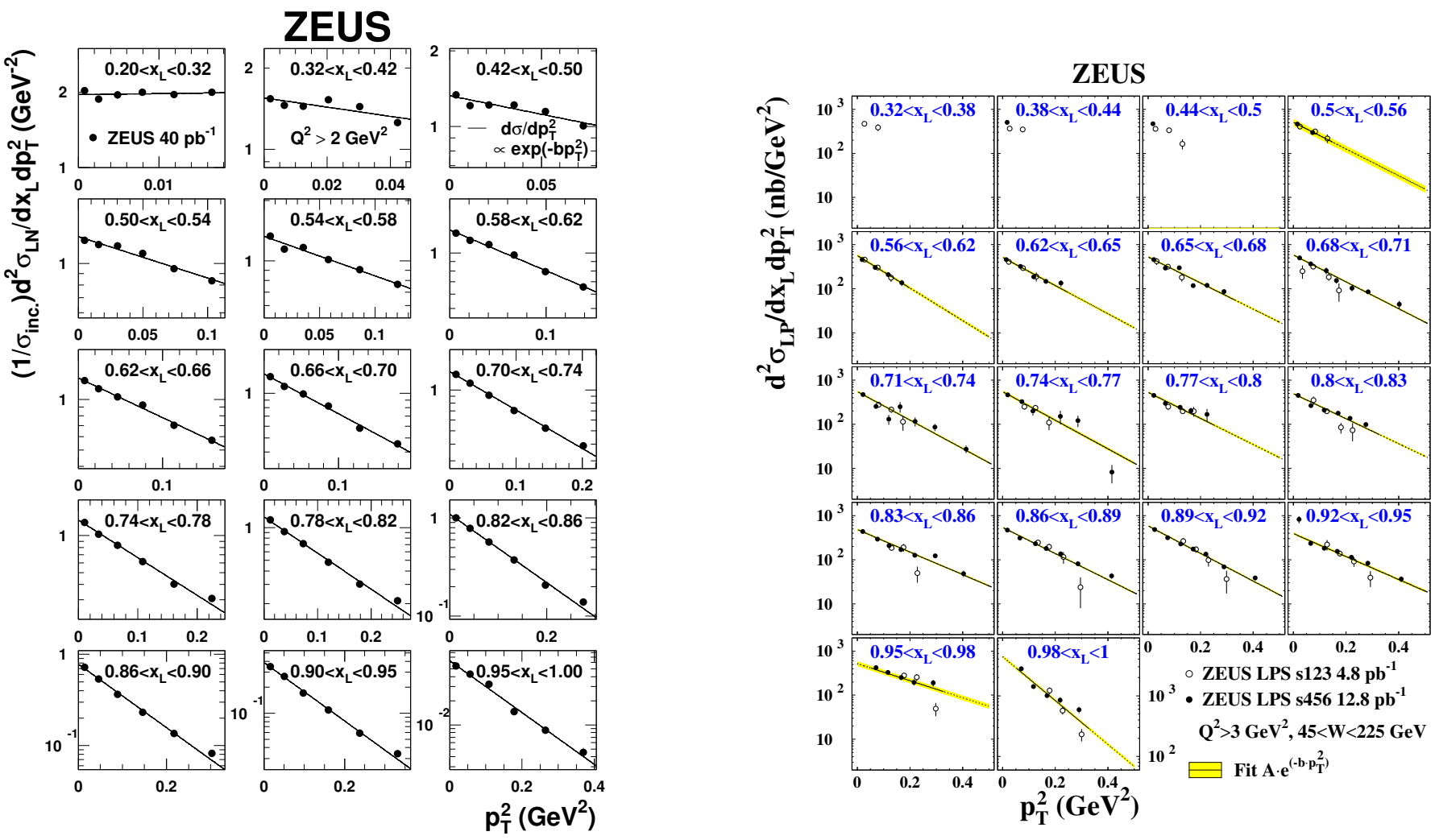

Figure 3: LN and LP cross-sections in DIS as a function of $p_T^2$ in bins of $x_L$, normalised to the inclusive DIS cross section. Lines are the results of fits to an exponential function.

A similar failure to describe the data is observed for LP production in DIS [6]. The left side of Figure 5 shows a comparison of the $x_L$ distribution and the $p_T^2$ exponential slope $b$ to the predictions of MC models. None of them can reproduce either the flat dependence of the cross section versus $x_L$ below the diffractive peak at $x_L = 1$ or the magnitude and dependence of $b$ on $x_L$. The same data are compared on the right side of Figure 5 to a Regge-based model [10] incorporating the isovector and isoscalar exchanges, and including the Pomeron for diffraction. A good description of the $x_L$ distribution and the slopes is obtained by adding a substantial contribution of isoscalar Reggeon exchanges, which turn out to be the dominant processes below the diffractive peak.

# 3 Vertex Factorisation and Absorptive Effects

A refinement of the simple factorisation picture is provided by baryon absorption, which can occur through rescattering [16–18]. In a geometrical picture [17], if the size of the meson-baryon system is small compared to the size of the photon, the baryon may also scatter on the photon and migrate to lower $x_L$ or higher $p_T$, thus escaping detection. This results in a relative depletion of observed forward baryons. Since the size of the photon is inversely related to the photon virtuality $Q^2$, more absorption is expected in photoproduction ($Q^2 \sim 0$) than in DIS. Also, since the size of the meson-baryon system is inversely proportional to the baryon $p_T$, rescattering results in a depletion of high $p_T$ baryons in photoproduction relative to DIS.

To investigate the $Q^2$ dependence of LN production, the $x_L$ distributions for photoproduction and for DIS in three bins of increasing $Q^2$ are shown on the left side of Figure 6. The yield of LNs decreases monotonically with decreasing $Q^2$. This is in qualitative agreement with the expectations of an increase of absorption as $Q^2$ decreases. A similar $Q^2$ dependence of the

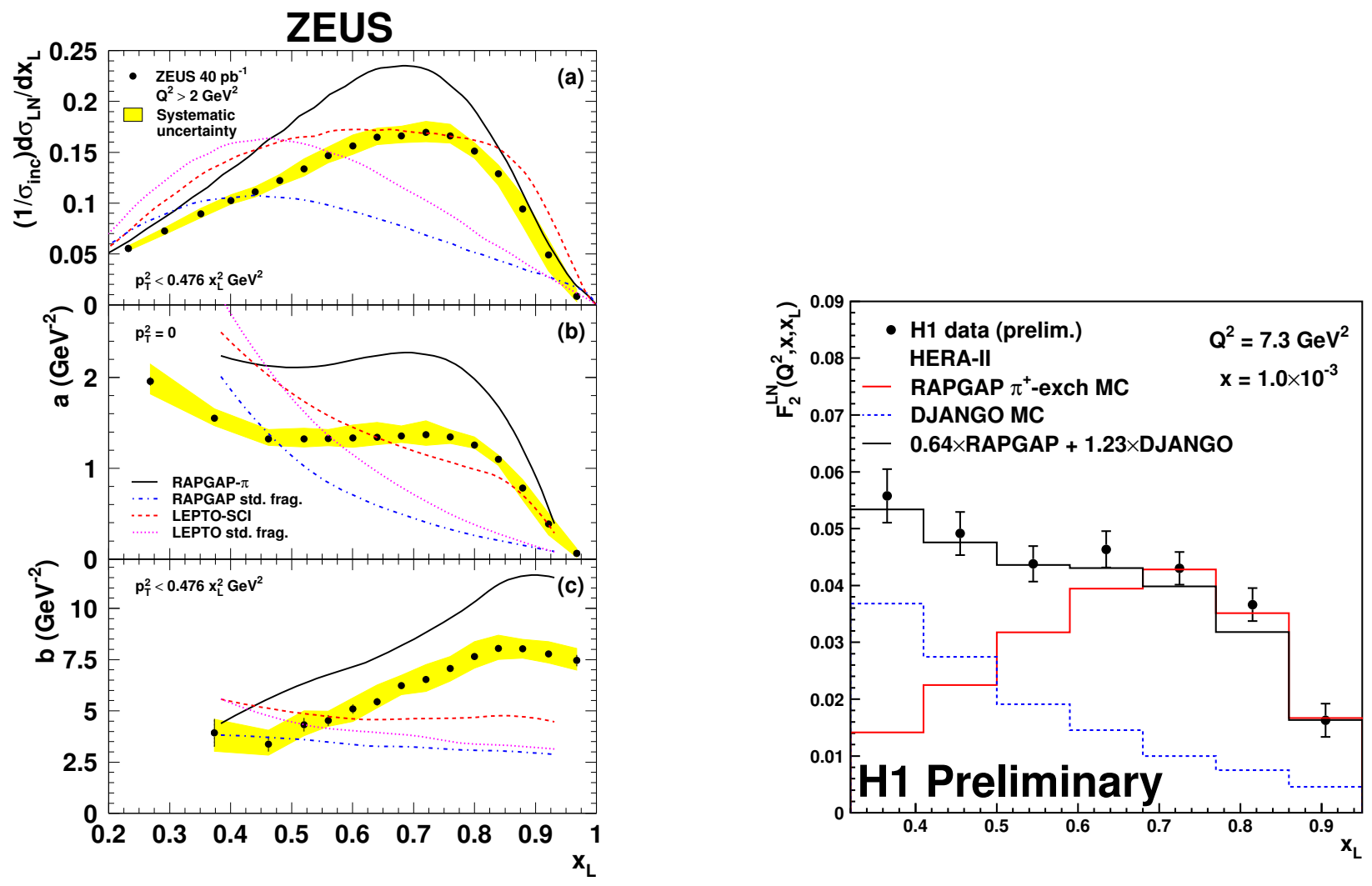

Figure 4: Left: Normalised LN cross section, intercept $a$ and exponential slope $b$ as a function of $x_L$, compared to Monte Carlo models. Right: LN $x_L$ distribution with an optimised mixture of exchange and fragmentation models.

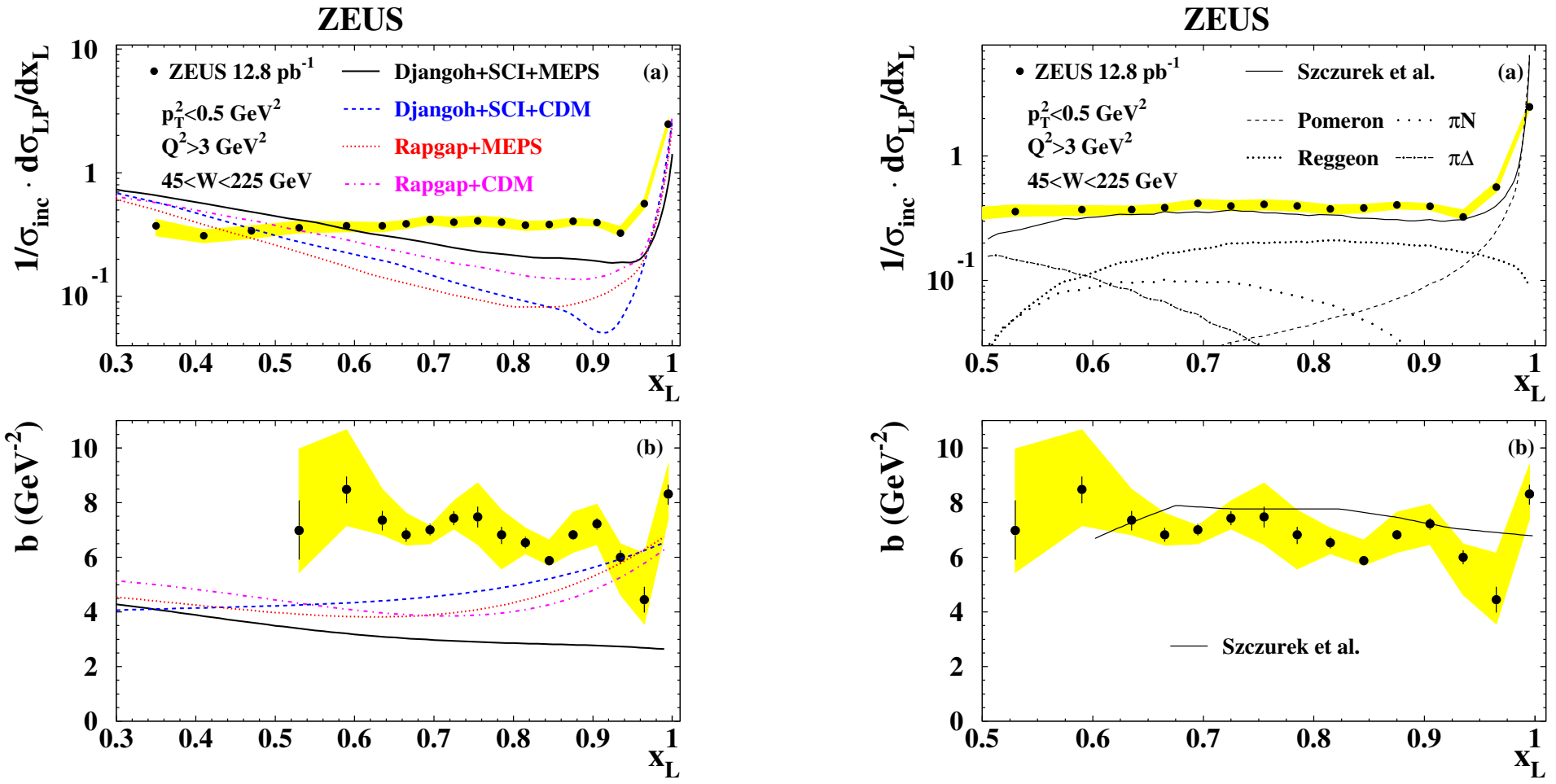

Figure 5: Normalised LP cross section and exponential slope $b$ of LPs in DIS as a function of $x_L$, compared to Monte Carlo models (left) and a Regge-based model [10] (right).

yieldis also observed in the LP data [4]

A calculation of LN production through pion exchange with neutron absorption, based on multi-pomeron exchanges, has become available [18]. It also accounts for the migration of the neutrons in $x_L$ and $p_T$ after rescattering and includes secondary exchanges of $\rho$ and $a_2$ mesons. The prediction of this model for the $x_L$ neutron distribution in photoproduction is shown on the right side of Figure 6. A fair description of both the shape and the magnitude of the distribution is observed. However, the model with pion exchange predicts a higher value of the slope $b$. Extending the model to include also secondary exchanges, a better description of the observed slopes is obtained still while maintaining a fair description of the $x_L$ distribution.

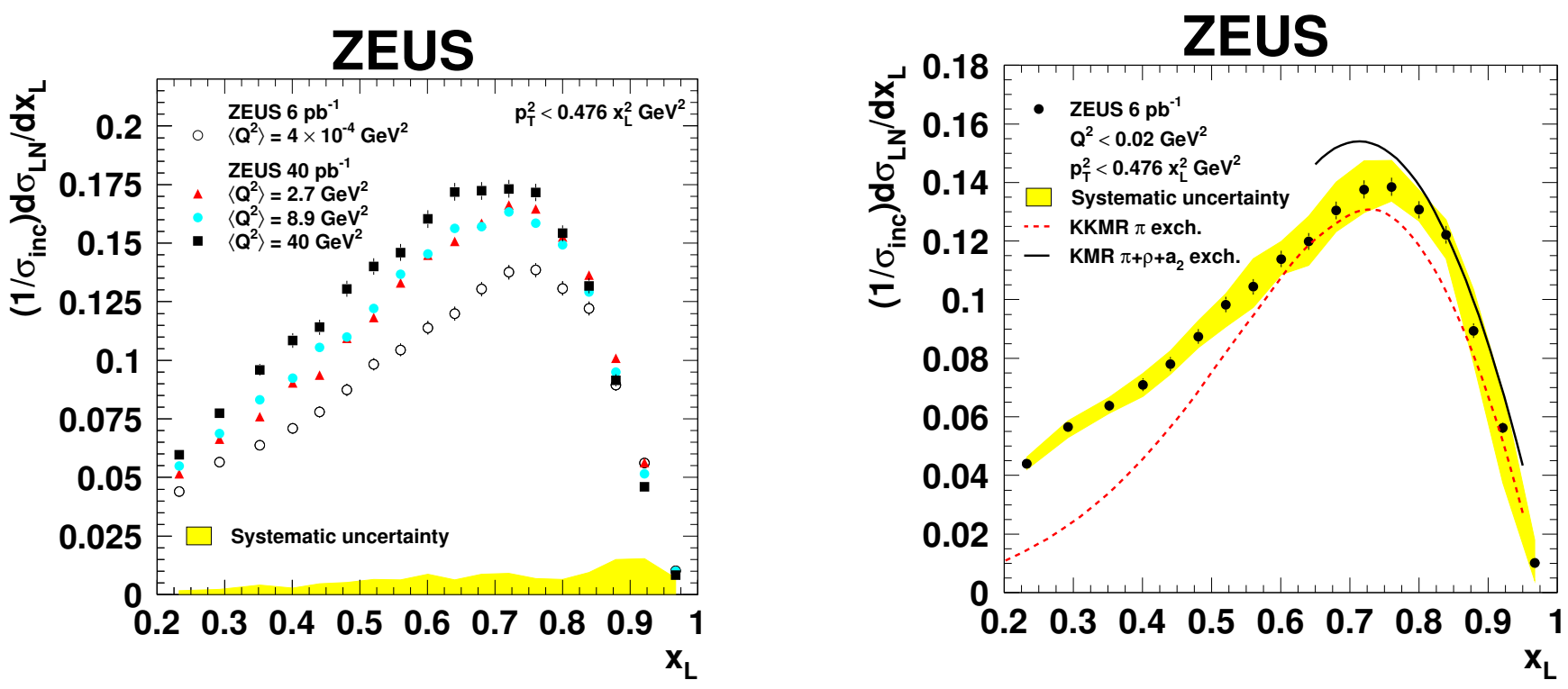

Figure 6: Left: normalised LN cross section as a function of $x_L$ for photoproduction and for three bins of $Q^2$ in DIS. Right: LN $x_L$ distribution for photoproduction compared to exchange models including absorption effects.

# 4 Leading Neutron Production and the Pion Structure Function

Analogous to the inclusive proton structure function $F_2(x, Q^2)$, the semi-inclusive LN structure function $F_2^{LN(3)}(x, Q^2, x_L)$ is defined as

$$\frac{d^3\sigma}{dQ^2 dx\, dx_L} = \frac{4\pi\alpha^2}{xQ^4}[1 - y + \frac{y^2}{2}] \cdot F_2^{LN(3)}(x, Q^2, x_L).$$

Figure 7 shows the ratios $F_2^{LN(3)}/F_2$ in bins of $x$ and $x_L$ as a function of $Q^2$ [5], where $F_2^{LN}$ values are measured from LN production in DIS, and $F_2$ is obtained from the H1 PDF 2000 parameterisations [23]. At fixed $x_L$, the ratios are almost flat in all $(x, Q^2)$ bins, suggesting the validity of vertex factorisation, i.e. independence of the photon and proton vertices.

Assuming that the pion exchange dominates the LN production at high $x_L$, $F_2^{LN}$ can be presented as a product of a pion flux factor $f_{\pi/p}(x_L, t)$ and the pion structure function $F_2^{\pi}(\beta, Q^2)$, where $\beta = x/(1 - x_L)$ is the fraction of the pion momentum carried by the struck parton. The parton distributions in the pion have been previously constrained from Drell-Yan processes and direct photon production in pion-nucleon collisions, and are limited to high $\beta$ ($\gtrsim 0.1$) values. This measurement of $F_2^{LN(3)}$ allows to test these parameterisations at lower $\beta$. The pion structure function can be estimated as $F_2^{\pi} = F_2^{LN}/\Gamma_\pi$, where $\Gamma_\pi$ is the integrated over $t$ pion flux. The value of $\Gamma_\pi$ depends on pion flux parameterisation: for the parameterisation from [7], $\Gamma_\pi = 0.131$. The right side of Figure 7 shows $F_2^{LN}/\Gamma_\pi$ as a function of $\beta$ for fixed values of $Q^2$. The data are compared to the two parameterisations of the pion structure function [24, 25] as well as to the H1 PDF 2000 [23] parameterisation of the proton structure function scaled by the factor 2/3, according to the naive expectation based on the number of valence quarks in the pion and the proton respectively. The measured data show a steep rise with decreasing $\beta$, in accordance with $F_2^{\pi}$ parameterisations, but are slightly below the expectations, suggesting that additional phenomena, like absorption, may play a role. Also, the theoretical uncertainties on the pion flux factor need to be carefully considered before any conclusion can be drawn.

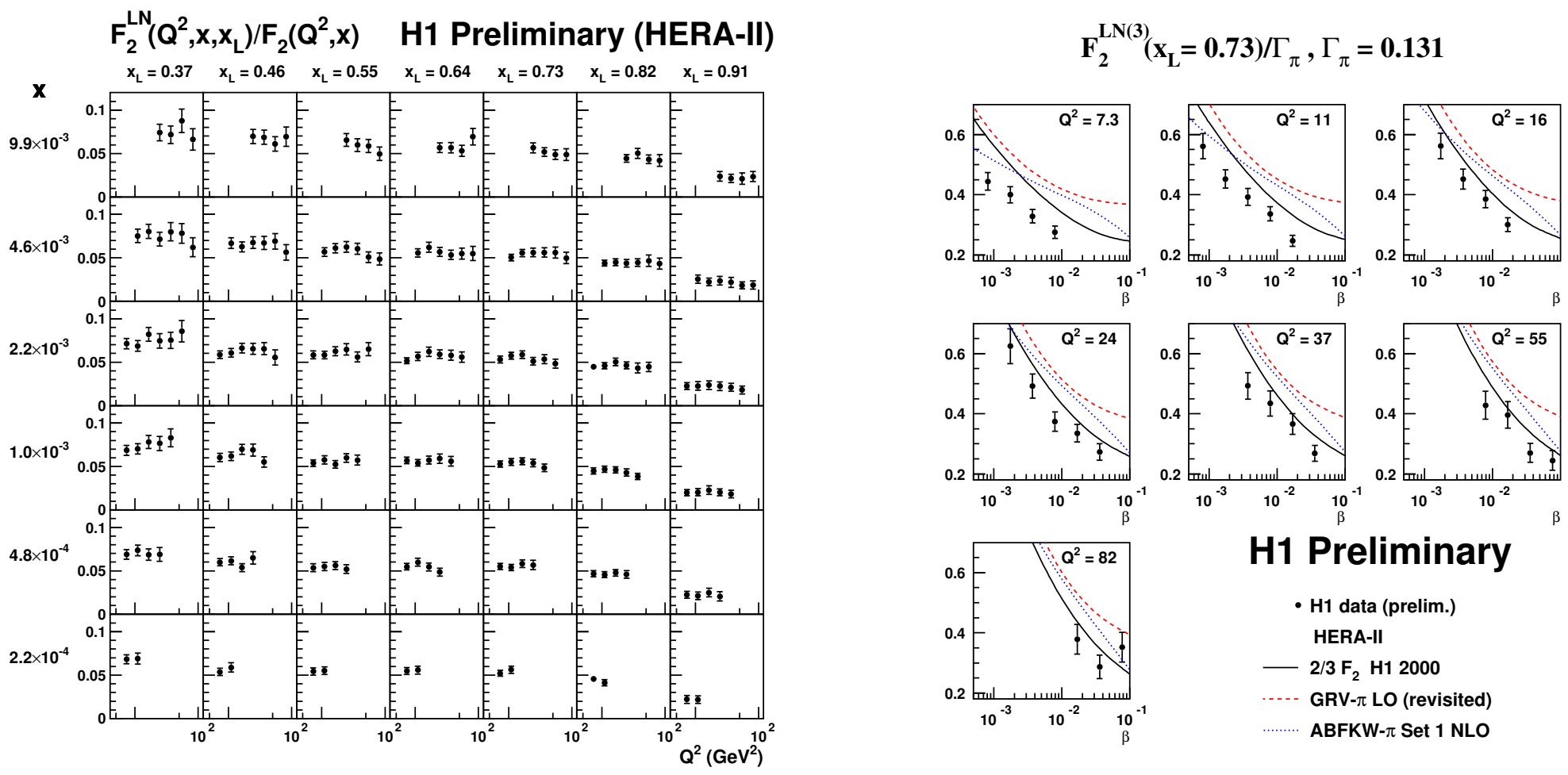

Figure 7: Left: ratio of the semi-inclusive LN to inclusive structure functions as a function of $Q^2$ in bins of $x$ and $x_L$. Right: The semi-inclusive LN structure function $F_2^{LN(3)}$ divided by the integrated pion flux plotted as function of $\beta = x/(1 - x_L)$ in bins of $Q^2$. The curves are the proton structure function scaled by 2/3 and two parameterisations of the pion structure function.

# 5 Study of Multiple Interactions in Photoproduction

The phenomena of absorption and rescattering discussed above are closely related to the multi-parton interactions (MI) which play an important role in hadronic interactions. The MI take place when the density of partons in the colliding beams is large enough that more than one interaction happens within one collision. MI have been required to describe the transverse momentum and particle multiplicity distributions in a region transverse to the jets at the TeVatron. In quasi-real photoproduction ($Q^2 \sim 0$) the photon has a point-like as well as a hadronic (resolved) component. Measurements in photoproduction at HERA have the advantage that the transition from a point-like photon towards a resolved photon can be studied in detail as function of the variable $x_\gamma$. MI are expected within the model of [26, 27] for resolved photons ($x_\gamma < 1$) but not for the point-like photons which have $x_\gamma = 1$.

In the photoproduction of dijets at HERA the effects of MI and underlying event can be studied in a fashion similar to the studies done at the TeVatron [28]. The underlying event is defined as everything in addition to the dijet production and includes MI as well as the contributions coming from higher order QCD radiation and hadronisation. The average track multiplicity as a function of the difference in azimuthal angle $\Delta\phi = \phi_{jet} - \phi_{track}$ is shown in Figure 8 for the two $x_\gamma$ regions: a resolved photon enriched region with $x_\gamma < 0.7$ and a point-like photon enriched region with $x_\gamma > 0.7$ [29]. The contributions from leading, subleading and third jets are clearly visible. The data are compared to predictions of PYTHIA MC [27]. For large $x_\gamma$ the effect of MI is very small, while at $x_\gamma < 0.7$ the inclusion of MI contributes as a pedestal to the track multiplicity and improves the description of the data.

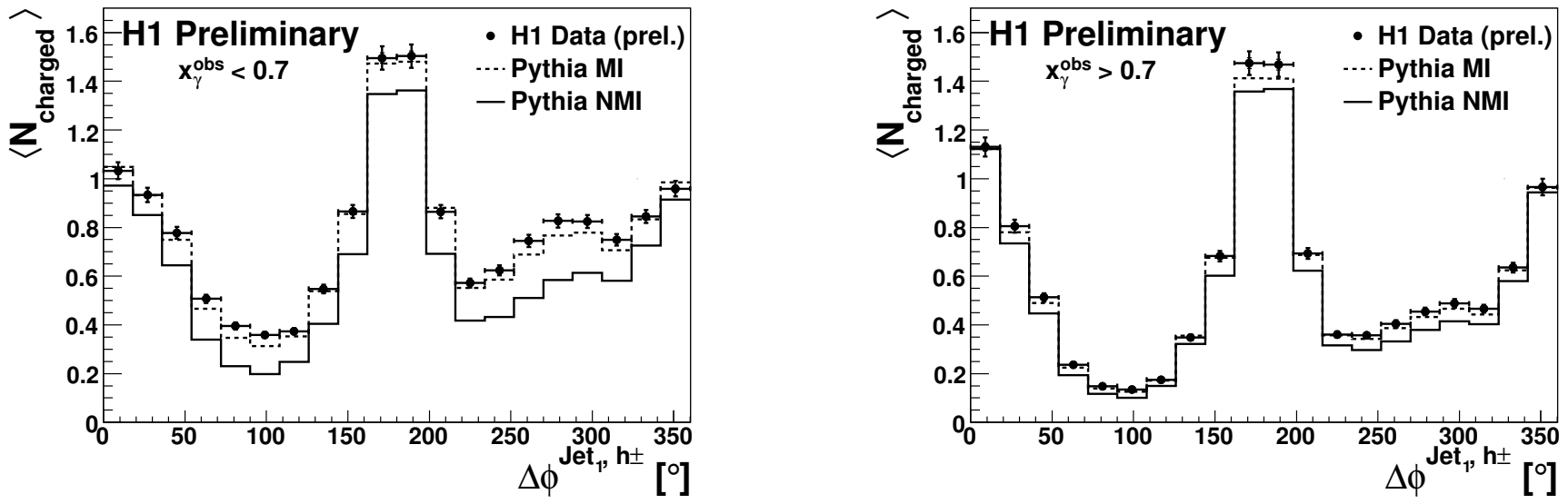

Figure 8: Charged particle multiplicity as a function of the difference in azimuthal angle $\Delta\phi = \phi_{jet} - \phi_{track}$ for $x_\gamma < 0.7$ (left) and for $x_\gamma > 0.7$ (right). Data is compared to PYTHIA MC predictions with and without multi-parton interactions.

In Figure 9 the average charge particle multiplicity is shown for the transverse regions of leading and subleading jets (*toward* and *away* regions). In general, the average track multiplicity rises with increasing $P_T^{Jet}$ from around 4-5 particles at $P_T^{Jet} \sim 5\,\text{GeV}$ to around 7 particles at higher $P_T^{Jet}$ depending on $x_\gamma$. In the region of $x_\gamma > 0.7$ the measurements are reasonably well described with a simulation containing only one hard interaction together with parton showers and hadronisation, whereas at low $x_\gamma$ good agreement is achieved only if MI are included in the simulation. The simulation including MI gives also a reasonable description of the average charged particle multiplicity in the transverse region over the full phase space region.

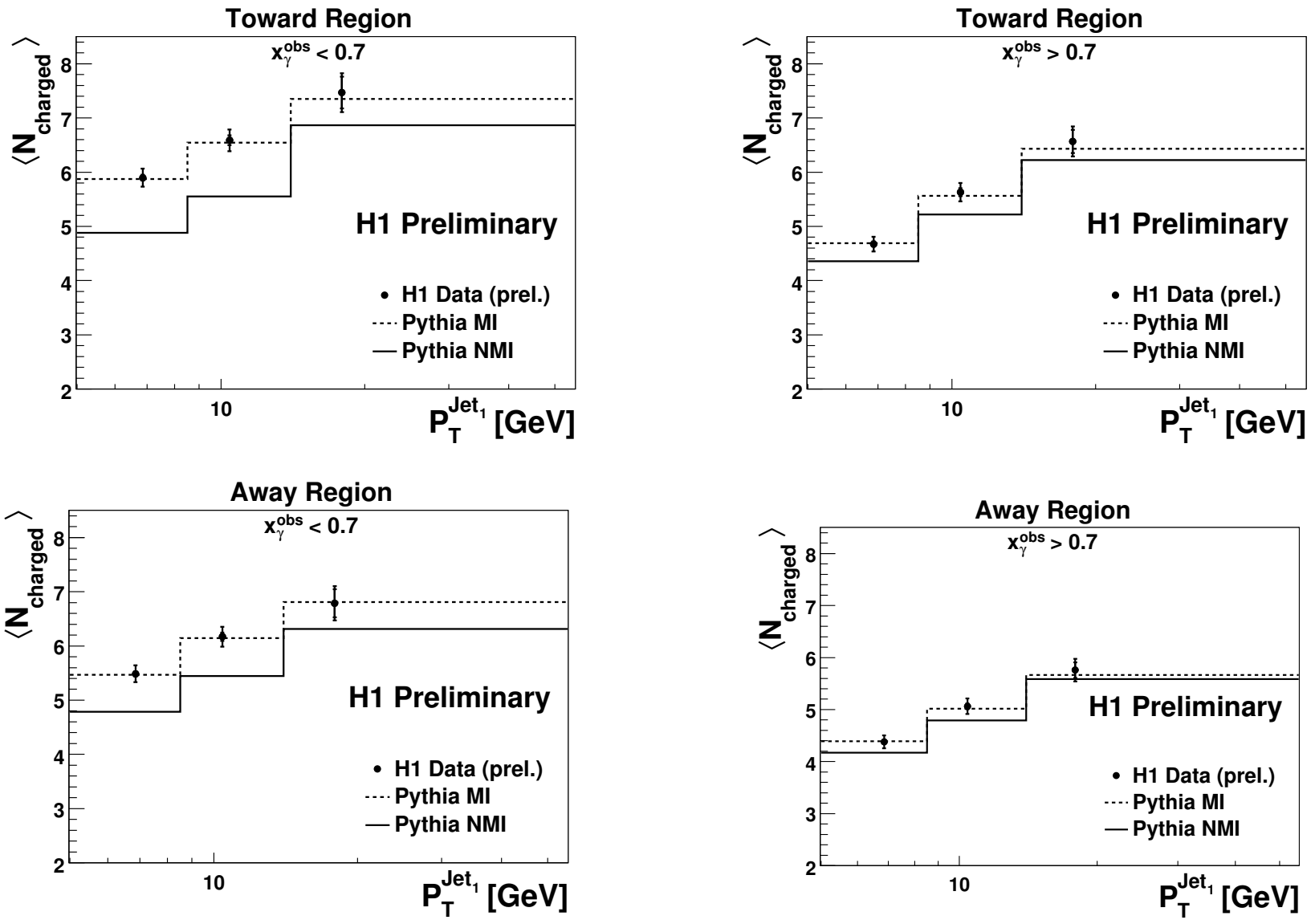

Figure 9: Charged particle multiplicity for $x_\gamma < 0.7$ (left) and for $x_\gamma > 0.7$ (right). The data are compared to PYTHIA MC predictions with and without multi-parton interactions.

# 6 Conclusions

The presence of a leading baryon in the final state provides information on the relationship between the soft and hard aspects of the strong interaction. The production of leading baryons has been studied as a function of several kinematic variables. There is a clear evidence that both contributions from fragmentation processes and from the exchange of colour-neutral particles such as isoscalars are required to describe the data. Thus the data show sensitivity to fragmentation models. With the assumption that pion exchange dominates leading neutron production the leading neutron data is sensitive to the pion structure function.

The charged particle multiplicity is studied in the photoproduction of jets. The measurements is described by PYTHIA simulation which includes multi-parton interactions.

# References

[1] C. Adloff *et al.* [H1 Collaboration], Eur. Phys. J. **C6** (1999) 587, [hep-ex/9811013].

[2] S. Chekanov *et al.* [ZEUS Collaboration], Nucl. Phys. **B637** (2002) 3, [hep-ex/0205076].

[3] A. Aktas *et al.* [H1 Collaboration], Eur. Phys. J. **C41** (2005) 273, [hep-ex/0501074].

[4] S. Chekanov *et al.* [ZEUS Collaboration], Nucl. Phys. **B776** (2007) 1, [hep-ex/0702028].

[5] H1 Collaboration,"Leading Neutron production in DIS at HERA", H1prelim-08-111.

[6] S. Chekanov *et al.* [ZEUS Collaboration], JHEP **0906** (2009) 074, [arXiv:0812.2416 [hep-ex]].

[7] H. Holtmann *et al.*, Phys. Lett. **B338** (1994) 363.

[8] B. Kopeliovich, B. Povh and I. Potashnikova, Z. Phys. **C73** (1996) 125, [hep-ph/9601291].

[9] M. Przybycień, A. Szczurek and G. Ingelman, Z. Phys. **C74** (1997) 509, [hep-ph/9606294].

[10] A. Szczurek, N. N. Nikolaev and J. Speth, Phys. Lett. **B428** (1998) 383, [hep-ph/9712261].

[11] L. Frankfurt, W. Koepf and M. Strikman, Phys. Lett. **B405** (1997) 367.

[12] M. Grazzini, L. Trentadue, and G. Veneziano, Nucl. Phys. **B519** (1998) 394.

[13] J.C. Collins, Phys. Rev. **D57** (1998) 3051; Erratum - ibid, **D61** (2000) 019902.

[14] J. Benecke et al., Phys. Rev. **188** (1969) 2159.

[15] T.T. Chou and C.N. Yang, Phys. Rev. **D50** (1994) 590.

[16] N. N. Nikolaev, J. Speth and B. G. Zakharov, [hep-ph/9708290].

[17] U. D'Alesio and H. J. Pirner, Eur. Phys. J. **A7** (2000) 109, [hep-ph/9806321].

[18] V. A. Khoze, A. D. Martin and M. G. Ryskin, Eur. Phys. J. **C48** (2006) 797, [hep-ph/0606213].

[19] J. D. Sullivan, Phys. Rev. **D5** (1972) 1732.

[20] M. Bishari, Phys. Lett. **B38** (1972) 510.

[21] K. Charchula, G. A. Schuler and H. Spiesberger, Comput. Phys. Commun. **81** (1994) 381, DJANGOH 1.4.

[22] H. Jung, Comp. Phys. Commun. **86** (1995) 147, RAPGAP 3.1.

[23] C. Adloff *et al.* [H1 Collaboration], Eur. Phys. J. **C21** (2001) 33, [hep-ex/0012053].

[24] M. Glück, E. Reya and I. Schienbein, Eur. Phys. J. **C10** (1999) 313, [hep-ph/9903288].

[25] P. Aurenche *et al.*, Phys. Lett. **B233** (1989) 517.

[26] T. Sjostrand and M. van Zijl, Phys. Rev. **D36** (1987) 2019.

[27] T. Sjostrand, S. Mrenna and P. Skands, "PYTHIA 6.4 Physics and Manual," JHEP **0605** (2006) 026, [hep-ph/0603175].

[28] A. A. Affolder *et al.* [CDF Collaboration], Phys. Rev. **D65** (2002) 092002.

[29] H1 Collaboration, "Study of Multiple Interactions In photoproduction at HERA", H1prelim-08-036.

# Part IV

# Hard Diffraction

# Analytic Properties of DPE Amplitudes *or* Collinear Factorisation for Central Exclusive Production

*O. V. Teryaev*[1]

[1]Bogoliubov Laboratory of Theoretical Physics, JINR, Dubna 141980, Russia

Analytic and crossing properties of amplitudes of the central exclusive production (CEP) are considered using the formalism of collinear Generalised Parton Distributions (GPDs). The analytic continuation from unphysical region is considered which leads to the finite expression. The natural interpretation of the emerging cuts corresponds to double spectral density in overlapping channel due to the instability of produced particle and inapplicability of Steinmann relations. The relations of CEP amplitudes to the exclusive decay rates are discussed. The direct calculation in physical region results in violation of factorisation similar to the discussed recently for pion transition and electromagnetic form-factors. The similarity between Feynman mechanism for form-factor and Durham model is pointed out.

## 1 Introduction

Currently the standard QCD mechanism for the diffractive production of heavy central system is provided by the model of Kaidalov, Khoze, Martin and Ryskin (Durham group, KKMR) developed for Higgs production at the LHC (see Refs. [1, 2, 3]). It is nevertheless interesting whether collinear QCD factorisation may be also applied here. This opportunity was discussed earlier [4] at the Blois workshop of 2005 (which marked the 20th anniversary of these meetings).

## 2 Collinear QCD Factorisation and "Standard" Durham Model

The basic idea is the generation of new hard processes by the "substitution" of Distribution Amplitudes (DA's) by Generalised Parton Distributions (GPD's), which may be considered as a generalised crossing: here not only the momenta of initial and final particles may be interchanged, but their number increased.

The first stage is just the pion form-factor, and the hard meson electroproduction may be considered as a substitution of one of DA's by GPD's. The next stage would be the substitution by another GPD of the remaining DA, so that one gets the amplitude for Central Exclusive Drell-Yan (CEDY) process $p_1 p_2 \to p_1' p_2' Q$ (the Abelian counterpart of Higgs production, requiring to change quark GPDs to gluon ones). The explicit calculation of the cross-section in the *physical* region, however, results in the violation of factorisation. However, now this possibility does not seem so dangerous as four years ago. Indeed, the BABAR data (discussed at this meeting by S. Eidelman) may imply the violation of factorisation [5] for pion transition and electromagnetic form-factors. Note that the Feynman mechanism for the pion electromagnetic form-factor, which is supposed to dominate in the case of violation of collinear factorisation, directly correspond to the KKMR approach, as in both cases the parton with almost zero longitudinal momentum fraction (soft quark or screening gluon with $x' \ll x$) appears. Still,

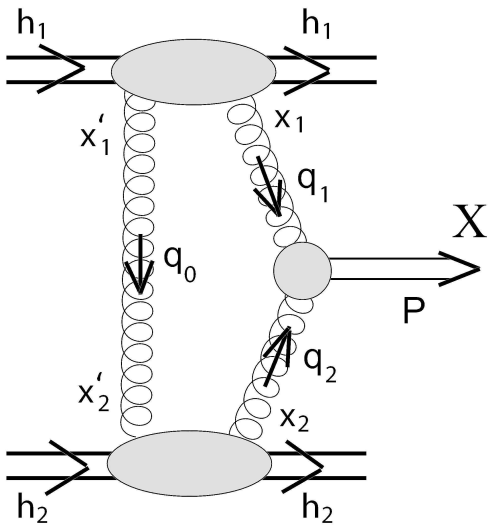

Figure 1: The KKMR mechanism of central exclusive production is similar to Feynman mechanism for form-factors: screening gluon with soft momentum $q_0$ corresponds to soft quark for form-factor

the direct fit of the collinear factorisation expression regularised by the finite width of gluon propagator may be of interest.

At the same time, the consideration of the unphysical region $|\xi_{1,2}| > 1$, where $\xi_{1,2} = s_{2,1}/s$ and $s_i = (p_i' + Q)^2$, results in the factorised amplitude

$$\mathcal{H}(\xi) = \int_{-1}^{1} dx\, dy \frac{H(x,\xi_1)}{x-\xi_1} \frac{H(y,\xi_2)}{y-\xi_2}. \tag{1}$$

Let us stress that the dimension parameter appearing in CEP amplitude in front of this expression is just the dilepton (or Higgs) mass squared, in complete similarity to momentum transfer $Q^2$ for pion form-factor. One may [4] recast it in the form of (double and single) subtracted spectral representations. However, the analytic continuation to the physical region is now more subtle, as the cuts in $s$ and $s_i$ provide the different signs for the $\imath\varepsilon$ addition to $\xi_{1,2}$. The symmetric contribution of the combinations of the cuts in $s, s_1$ and $s, s_2$ would lead to the pure real amplitude.

At the same time, the double cut in $s_1, s_2$ is not forbidden by Steinmann relations, as the produced particle (virtual photon or Higgs) is unstable[1]. It is therefore quite natural to perform analytic continuation in a symmetric way: $\xi_{1,2} \to \xi_{1,2} + \imath\varepsilon$. In this case both imaginary and real parts of CEP amplitudes are controlled.

Moreover, the crossing from GPDs to Generalised Distribution Amplitudes (GDA), describing the hard production of hadron pairs, relates the CEP amplitudes to that of exclusive decay of Higgs to two $p\bar{p}$ pairs. For the latter amplitude the region $|\xi_{1,2}| > 1$ (being unphysical for CEP) is a physical one and (1) is applicable. Let us stress, that it is the factorised expression which make this continuation possible, connecting the processes with the very different invariant masses of final state.

# 3 Conclusions

The possible violation of collinear QCD factorisation in the simplest lepton-pion processes makes the applicability of the (regularised) collinear factorisation to CEP amplitudes less dramatic. At the same time, the analytic continuation form unphysical region may be performed by relating the cuts in scaling variables $\xi_{1,2}$ to the cuts in overlapping variables which are not forbidden by Steinmann relations. This also allows to obtain a crossing relations between CEP and exclusive decay amplitudes. The similarity between various exclusive amplitudes appearing when non-perturbative inputs are changes provides a new insight for standard KKMR mechanism relating it to Feynman mechanism for pion form-factor.

[1] I am indebted to V.S. Fadin and L.N. Lipatov for the illuminating discussion of this problem.

## Acknowledgements

Useful discussions with V.S. Fadin, A.B. Kaidalov, V.A. Khoze and L.N. Lipatov are gratefully acknowledged. This study was partially supported by the Russian Foundation for Fundamental Research, grants No. 09-02-01149, 09-02-00732-, 08-02-00896-.

## References

[1] V.A. Khoze, A.D. Martin and M.G. Ryskin, Phys. Lett. B **401**, 330 (1997);
V.A. Khoze, A.D. Martin and M.G. Ryskin, Eur. Phys. J. C **23**, 311 (2002).

[2] A.B. Kaidalov, V.A. Khoze, A.D. Martin and M.G. Ryskin, Eur. Phys. J. C **33**, 261 (2004).

[3] A.B. Kaidalov, V.A. Khoze, A.D. Martin and M.G. Ryskin, Eur. Phys. J. C **31**, 387 (2003) [arXiv:hep-ph/0307064].

[4] O. V. Teryaev, arXiv:hep-ph/0510031.

[5] A. V. Radyushkin, arXiv:0906.0323 [hep-ph].

# Low-$x$ Gluon Distribution from Discrete BFKL Pomerons

*D.A. Ross*[1], *J. Ellis*[2], *H. Kowalski*[3], *L. Lipatov*[4], *G. Watt*[5]

[1]School of Physics & Astronomy, Southampton SO17 1BJ, U.K.
[2]CERN, 1211 Genève 23, Switzerland
[3]DESY, Notketraße 85, 22607 Hamburg, Germany
[4] St. Petersburg Institute of Nuclear Physics, Gatchina, Russia
[5] Institute for particle Physics Phenomenology, Durham DH1 3LE, UK

Using a modification of the BFKL equation which generates discrete Regge pole solutions, we obtain a good fit to the low-$x$ deep-inelastic data from HERA as well as an integrated gluon distribution which is everywhere positive.

In a recent paper [1], we obtained a good fit to the HERA deep-inelastic data at low-$x$ using a discretized version of the BFKL pomeron [2], which is in line with the Regge picture of diffractive events (and hence deep-inelastic events at low-$x$) whereby the amplitude is dominated by an isolated Regge pole (the "pomeron").

The purely perturbative BFKL equation predicts a cut rather than a pole. However, in 1986, Lipatov [3] suggested the following modifications to the BFKL equation:

1. Accounting for the running of the coupling as a function of the transverse momentum, $\mathbf{k}$, of the exchanged gluons, which spans a large range as one moves away from the top or bottom of the "ladder".

2. Assuming that the non-perturbative (infrared) sector of QCD imposes a fixed phase, $\eta$, on the oscillatory eigenfunctions of the BFKL kernel at some low value, $k_0$, of gluon transverse momentum.

This leads to a discrete set of eigenfunctions, $f_i(\mathbf{k})$ with discrete eigenvalues, $\omega_i$, which can be interpreted as isolated Regge poles., i.e. the scattering of a gluon with transverse momentum $\mathbf{k}$ off some target with CM energy $\sqrt{s}$, has an amplitude which can be written in the form

$$\mathcal{A}(\mathbf{k}, s) \;=\; \sum_i a_i f_i(\mathbf{k}) s^{\omega_i}$$

The eigenfunctions have an oscillating behaviour with a decreasing frequency up to a value of transverse momentum $\mathbf{k}_{crit}$, above which they decay exponentially with $\ln(k)$

A very good fit was obtained using only the first four such eigenfunctions. The only unknown quantity is the proton impact factor $\Phi_p(\mathbf{k})$, which encodes the coupling of the proton to the gluon-scattering amplitude. Since the eigenfunctions form a complete orthonormal set, this impact factor can be expanded in the form

$$\Phi_p(\mathbf{k}) \;=\; \sum_i b_i f_i(\mathbf{k}),$$

where the first four coefficients, $b_i$ were fit to data.

Unfortunately, when we tried to reproduce the full impact factor from this fit, we obtained an un-integrated gluon density $\tilde{g}(x, k^2)$, which becomes negative over a sufficient range that the (integrated) gluon density

$$g(x, Q^2) \equiv \int^{Q^2} \tilde{g}(x, k^2) dk^2,$$

is also negative.

We therefore sought a solution in which the impact factor has a "sensible" form such as

$$\Phi_p(k) = Ak^2 e^{-bk^2}, \quad \text{or} \quad \frac{A}{(k^2 + \mu^2)^\alpha}.$$

This suggested that taking only the first four eigenfunctions was insufficient. Indeed, if we take $n_0$ eigenfunctions , where the first eigenvalue is $\omega_1$ and the last is $\omega_{n_0}$ then we expect an error of order $x^{(\omega_1 - \omega_{n_0})}$. This is 30% for $x \sim 10^{-2}$ and 17% for $x \sim 10^{-3}$.

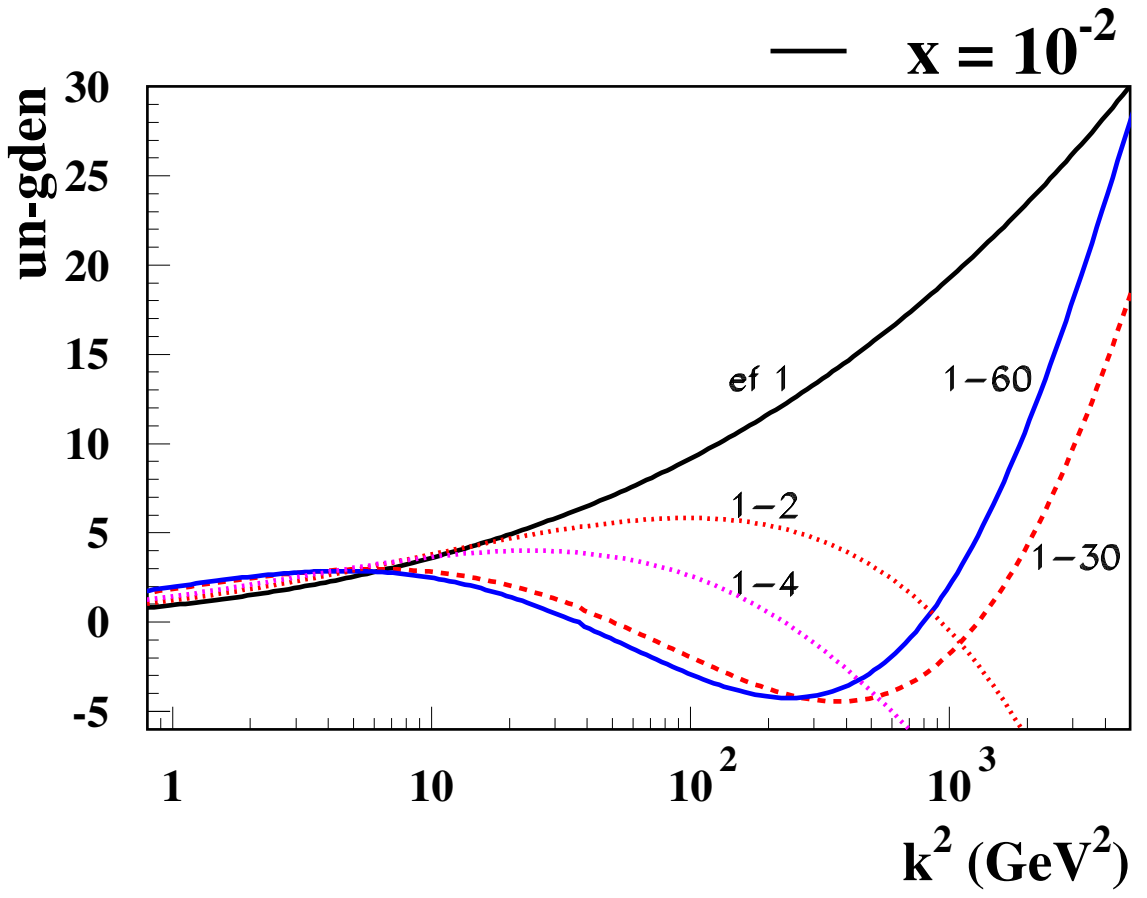

Figure 1: The effect on the un-integrated gluon density from increasing the number of eigenfunctions included.

We are now able to construct many more eigenfunctions, but we find that, although the un-integrated gluon density becomes positive at relatively large **k** when 30 eigenfunctions are used, the range of negative values still generates a physically unacceptable negative gluon density and that a further increase in the number of eigenfunctions taken only marginally improves this situation, as can be seen in Fig. 1.

We now understand why this is the case. A detailed explanation will appear in a forthcoming publication [4]. Within the context of a fixed phase for the oscillations at low transverse momentum, an adjacent eigenfunction has a larger $k_{crit}$ by approximately one half wavelength, whereas the frequency of the oscillations in the relevant range of $k$ is almost identical. This is demonstrated in Fig. 2.

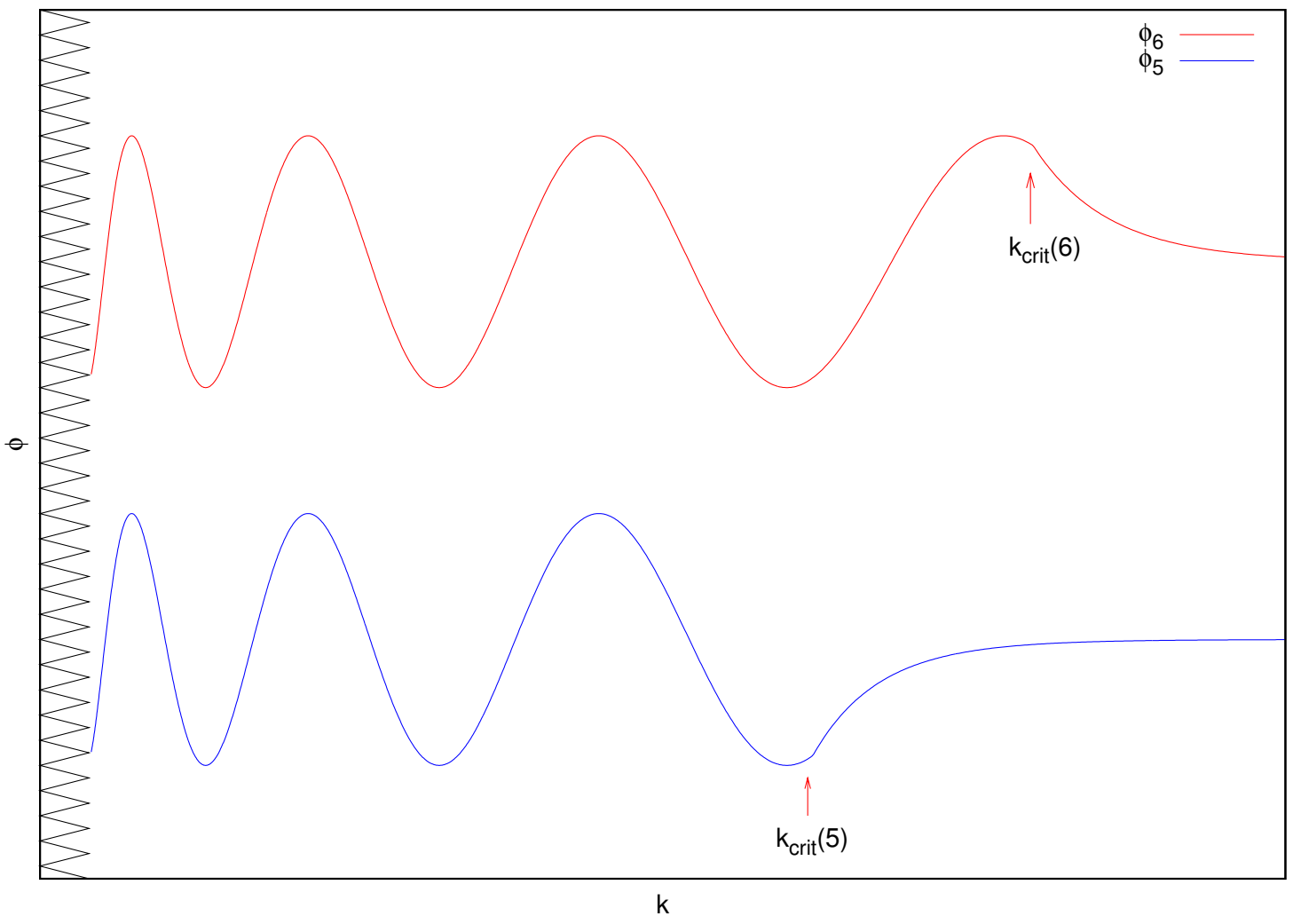

Figure 2: Sketch of eigenfunctions numbers 5 and 6.

In Fig. 3 we plot the initial frequencies (i.e. the frequencies for $k \ll k_{crit}$) for the first 30 eigenfunctions and note that they accumulate at a value $\nu_{max} \sim 0.7$. This means that we can only expect to expand an impact factor as a sum of these eigenfunctions provided the impact factor has non-negligible support up to a value of transverse momentum $k_{max}$ where

$$\nu_{max} \ln\left(\frac{k_{max}}{k_0}\right) \gg \pi.$$

The minimum value of $k_0$ that can be taken without encountering serious perturbative instabilities is $k_0 \sim 0.3\,\mathrm{GeV}$, which leads to a $k_{max}$ far larger than the expected value for a proton impact factor which should be $\mathcal{O}(\Lambda_{QCD})$.

Put another way, this means that an impact factor with support for $k \leq \Lambda_{QCD}$ is *not* compatible with a fixed phase at $k_0 \sim 0.3\,\mathrm{GeV}$. This in turn implies that the second assumption of [3] needs to be revisited.

At leading order, we can write the BFKL equation with running coupling in the form

$$\int \mathcal{K}_0(\mathbf{k},\mathbf{k}')f_i(\mathbf{k}')d^2\mathbf{k}' \;=\; \frac{\omega_i}{\overline{\alpha}_s(k^2)}f_i(\mathbf{k}).$$

In the infrared limit, as $\overline{\alpha}_s$ increases, the RHS goes to zero and it was therefore argued in Ref. [3] that the infrared limit of the eigenfunctions, $f_i(\mathbf{k})$, would be independent of $\omega$ and hence possess a universal phase.

In practice, however, with an infrared cutoff $k_0 \sim 0.3\,\mathrm{GeV}$, the ratio $\omega_i/\overline{\alpha}_s(k^2)$ is *not* negligibly small and so we might expect the infrared phase, $\eta$, to have a dependence on $\omega$.

Within perturbation theory (recalling that the eigenvalues $\omega$ can be expanded in powers of $\overline{\alpha}_s$ starting at first order), such a dependence is expected to be analytic, so that one would

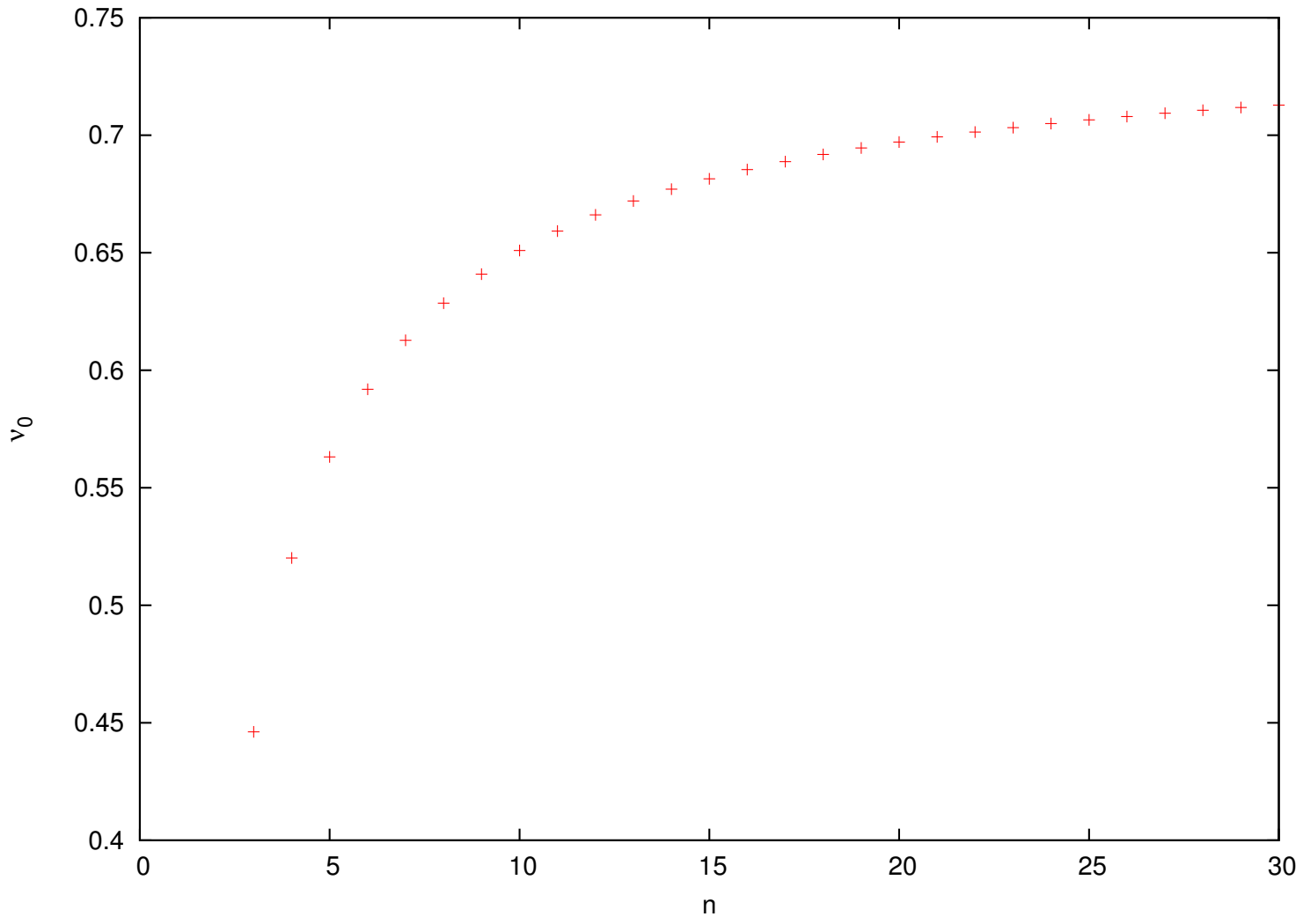

Figure 3: The oscillation frequencies for gluon transverse momentum $k \ll k_{crit}$ for the first 30 eigenfunctions.

expect an improved fit (with small $\omega$) with a phase $\eta(\omega)$ of the form

$$\eta(\omega) \;=\; \eta_0 + \eta'\omega$$

Unfortunately, we were neither able to obtain a satisfactory fit using this ansatz for $\eta$, nor to rectify the problem of a negative gluon density. The best fit has a $\chi^2$/DoF of 3.0.

However, since we are probing the non-perturbative behaviour of QCD, we are entitled to drop the requirement that $\eta$ should be an analytic function of $\omega$ and try, for example, an $\omega$ dependence of the form

$$\eta(\omega) \;=\; \eta_0 + \eta'\sqrt{\omega}\,.$$

We found that this can generate a gluon distribution which is positive everywhere, as shown in Fig. 4, and produce a fit to HERA data with a $\chi^2$/DoF of 1.1.

In Fig. 5 we show the best fits for the linear (dotted line) and non-linear (solid line) fits to the Zeus low-$x$ data, using an impact factor of the form

$$\Phi_p(\mathbf{k}) \;=\; Ak^2 e^{-bk^2}$$

The three free parameters used (apart from the overall normalization - which serves as a fourth parameter) are

| | Linear | Non-linear |
|---|---|---|
| $b$ [GeV$^{-2}$] | 2.0 | 2.0 |
| $\eta_0$ | $-0.74\pi$ | $-0.74\pi$ |
| $\eta'$ | $2.8\pi$ | $1.4\pi$ |

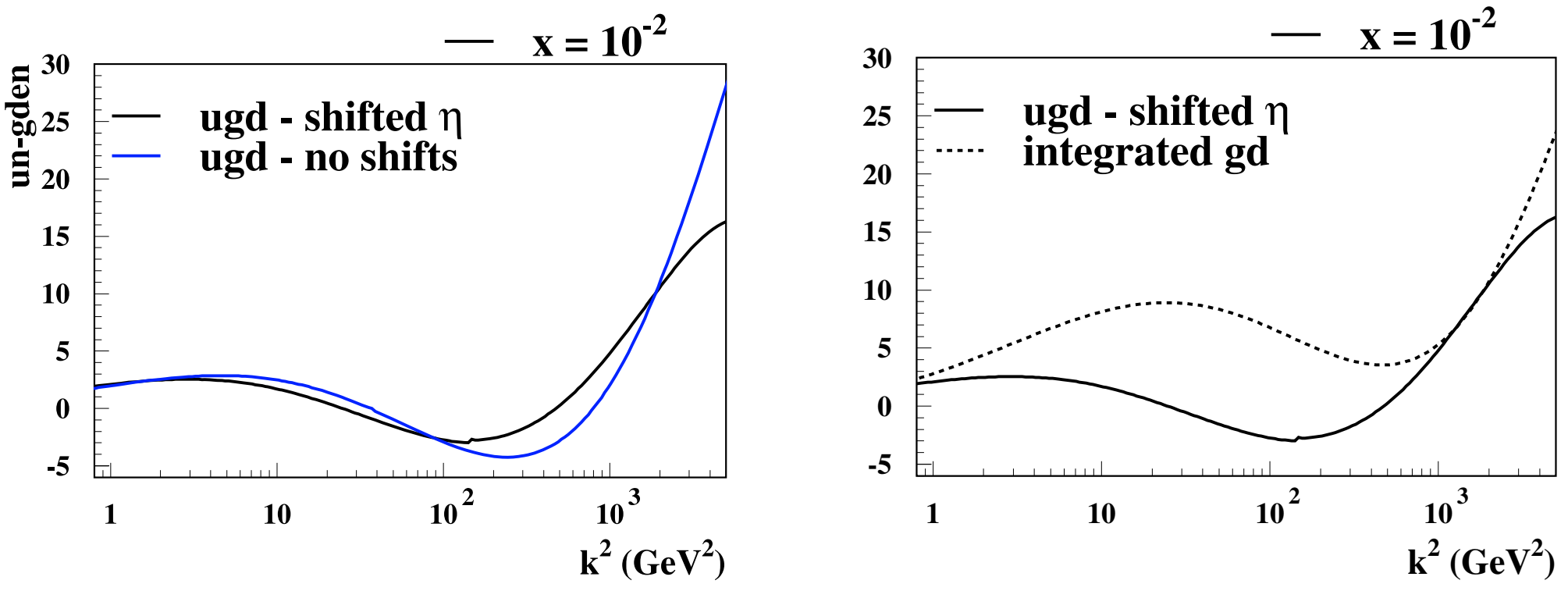

Figure 4: The left-hand graph shows the un-integrated gluon density with a fixed infrared phase and with a non-linear $\omega$-dependent phase. The right-hand graph shows the un-integrated (solid line) with the non-linear $\omega$-dependent infrared phase, and the corresponding (integrated) gluon density (dotted line).

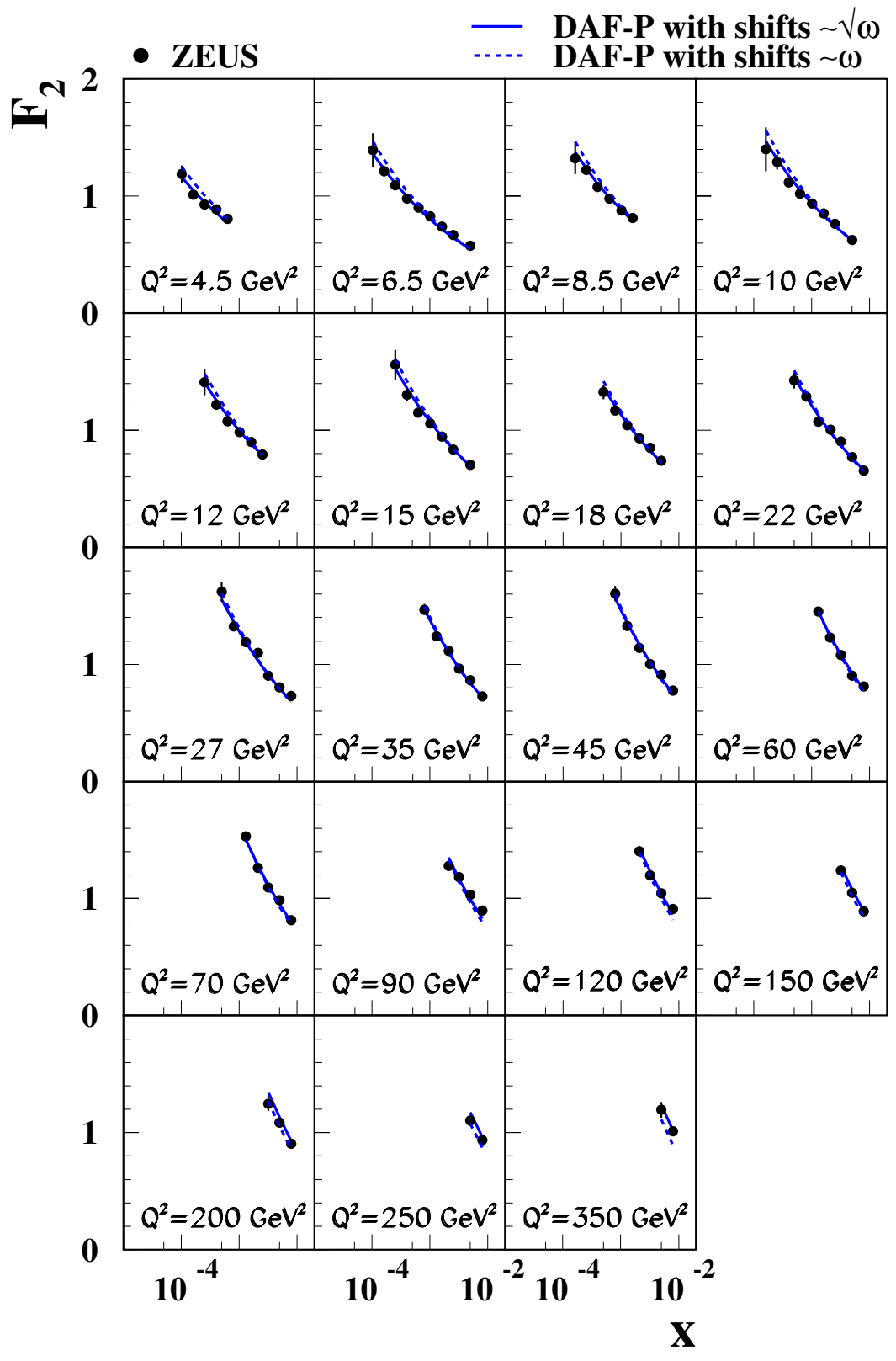

Figure 5: Fit to Zeus [5] low-$x$ data with linear $\omega$-dependence (dotted line) and non-linear $\omega$-dependence (solid line)

We therefore have a low-$x$ gluon density which is everywhere positive, fits the HERA data, and is consistent with the modified BFKL equation provided one allows a non-analytic dependence of the infrared phase, $\eta$, on the eigenvalue, $\omega$, thereby reflecting the non-perturbative nature of the infrared effects.

This gluon density can now be tested by applying it to the prediction of cross-sections (such as jet production) at LHC which are dominated by the low-$x$ gluon distribution.

## References

[1] J. Ellis, H. Kowalski, and D.A. Ross Phys. Lett. **B668** 51 (2008).

[2] Y.Y. Balitski and L.N. Lipatov, Sov. J. Nucl. Phys.**28** 82 (1978);
V.S. Fadin, E.A. Kuraev, and L.N. Lipatov, Sov. Phys. JETP **44** 443 (1976)

[3] L.N. Lipatov, Sov. Phys. JETP **63** 904 (1986)

[4] J. Ellis, H. Kowalski, L.N. Lipatov, D.A. Ross and G. Watt, in preparation.

[5] J. Breitweg et. al (Zeus collaboration), Phys. Lett. **B487** 53 (2000);
S. Chekanov et. al. (Zeus collaboration), Eur. Phys. J. **C21** 443 (2001)

# Soft Gluon Resummation for Gaps between Jets

*Simone Marzani, Jeffrey Forshaw and James Keates*

School of Physics & Astronomy, University of Manchester,
Oxford Road, Manchester, M13 9PL, United Kingdom

We investigate soft gluon resummation for the gaps-between-jets cross-section. After reviewing the theoretical framework that enables one to sum logarithms of the hard scale over the veto scale to all orders in perturbation theory, we then present a study of the phenomenological impact of Coulomb gluon contributions and super-leading logarithms on the gaps between jets cross-section at the LHC.

## 1 Jet Vetoing: Gaps between Jets

We study dijet production with a veto on the emission of a third jet in the inter-jet rapidity region, $Y$, harder than $Q_0$. We shall refer generically to the "gaps between jets" process, although the veto scale is chosen to be large, $Q_0 = 20$ GeV, so that we can rely on perturbation theory. Thus a "gap" is simply a region of limited hadronic activity.

Because of the magnitude of the dijet cross section, gaps between jets will be studied with early LHC data, hence computation of a reliable theoretical prediction has become an urgent task. It is an interesting and at the same time challenging calculation because this observable is sensitive to a remarkably diverse range of QCD phenomena. For instance, the limit of large rapidity separation corresponds to the limit of high partonic centre of mass energy and BFKL effects are expected to become important [1]. On the other hand one can study the limit of emptier gaps, becoming more sensitive to wide-angle soft gluon radiation. Furthermore, if one wants to investigate both of these limits simultaneously, then the non-forward BFKL equation enters the game [2]. In the following we limit ourselves to only wide-angle soft emissions only.

Accurate studies of these effects are important also in relation to other processes, in particular the production of a Higgs boson in association with two jets. This process can occur via gluon-gluon fusion and weak-boson fusion (WBF). QCD radiation in the inter-jet region is clearly different in the two cases and, in order to enhance the WBF channel, one can veto emission between the jets [3,4]. This situation is very closely related to gaps between jets since the Higgs carries no colour charge, and QCD soft logarithms can be resummed using the same technique [5].

## 2 Soft Gluons in Gaps between Jets

Given a hard scattering process, we can study how it is modified by the addition of soft radiation. If the observable is inclusive enough, then we have no effects because soft contributions cancel when real and virtual corrections are added together, as a result of the Bloch-Nordsieck theorem. However, if we restrict the real radiation to a corner of the phase space, as happens for the gap cross-section, we encounter a miscancellation and are left with a logarithm of the ratio of the hard scale and veto scale, $Q/Q_0$. The resummation of wide-angle soft radiation in the gaps between jets process was originally performed assuming that the real–virtual cancellation is perfect outside the gap, so that one needs only to consider virtual gluon corrections integrated over momenta for which real emissions are forbidden, i.e. over the "in gap" region of rapidity and with $k_T$ above the veto scale $Q_0$ [6–8]. We shall refer to these contributions as global logarithms. The resummed squared matrix element can be written as:

$$|\mathcal{M}|^2 = \frac{1}{V_c}\langle m_0|e^{-\xi\mathbf{\Gamma}^\dagger} e^{-\xi\mathbf{\Gamma}}|m_0\rangle\,, \quad \xi = \frac{2}{\pi}\int_{Q_0}^{Q}\frac{dk_T}{k_T}\alpha_s(k_T)\,, \tag{1}$$

where $V_c$ is an averaging factor for initial state colour. The vector $|m_0\rangle$ represents the Born amplitude and the operator $\mathbf{\Gamma}$ is the soft anomalous dimension:

$$\mathbf{\Gamma} = \frac{1}{2}Y\mathbf{t}_t^2 + i\pi\mathbf{t}_a\cdot\mathbf{t}_b + \frac{1}{4}\rho_{\text{jet}}(Y,|\Delta y|)(\mathbf{t}_c^2+\mathbf{t}_d^2)\,, \tag{2}$$

where $\mathbf{t}_i$ is the colour charge of parton $i$ and the function $\rho_{\text{jet}}(Y,\Delta y)$ is related to the jet definition. The operator $\mathbf{t}_t^2$ represents the colour exchanged in the $t$-channel:

$$\mathbf{t}_t^2 = (\mathbf{t}_a+\mathbf{t}_c)^2 = \mathbf{t}_a^2+\mathbf{t}_c^2+2\,\mathbf{t}_a\cdot\mathbf{t}_c\,. \tag{3}$$

The imaginary part of Eq. (2) is due to Coulomb gluon exchange. These contributions play an important role in the proof of QCD factorisation and they are also responsible for super-leading logarithms [9, 10]. We notice that for processes with less than four coloured particles, such as deep-inelastic scattering or Drell-Yan processes, the imaginary part of the anomalous dimension does not contribute to the cross-section. For instance, if we consider three coloured particles, then colour conservation implies that $\mathbf{t}_a+\mathbf{t}_b+\mathbf{t}_c=0$, and consequently

$$i\pi\,\mathbf{t}_a\cdot\mathbf{t}_b = \frac{i\pi}{2}\left(\mathbf{t}_c^2-\mathbf{t}_a^2-\mathbf{t}_b^2\right)\,, \tag{4}$$

which contributes as a pure phase. Coulomb gluons do play a role in dijet production, but they are not implemented in angular-ordered parton showers. We shall evaluate the impact of these contributions on the cross-section in the next section.

It was later realised [11] that the above procedure is not enough to capture the full leading logarithmic behaviour. Real gluons emitted outside of the gap are forbidden to re-emit back into the gap and this gives rise to a new tower of logarithms, formally as important as the primary emission corrections, known now as non-global logarithms. The leading logarithmic accuracy is therefore achieved by considering all $2\to n$ processes, i.e. $n-2$ out-of-gap gluons, dressed with "in-gap" virtual corrections, and not only the virtual corrections to the $2\to 2$ scattering amplitudes. The colour structure quickly becomes intractable and, to date, calculations have been performed only in the large $N_c$ limit [11–13].

A different approach was taken in [9, 10], where the specific case of only one gluon emitted outside the gap, dressed to all orders with virtual gluons but keeping the full $N_c$ structure, was considered. That calculation had a very surprising outcome, namely the discovery of a new class of "super-leading" logarithms (SLL), formally more important than the "leading" single logarithms. Their origin can be traced to a failure of the DGLAP "plus-prescription", when the out-of-gap gluon becomes collinear to one of the incoming partons. Real and virtual contributions do not cancel as one would expect and one is left with an extra logarithm. This miscancellation first appears at the fourth order relative to the Born cross-section and it is caused by the imaginary part of loop integrals, induced by Coulomb gluons. The presence of SLL has been also confirmed by a fixed order calculation in [15]; in this approach SLL have been computed at $\mathcal{O}(\alpha_s^5)$ relative to Born, i.e. going beyond the one out-of-gap gluon approximation. The SLL contributions originating from one gluon outside the gap have been recently resummed to all orders [14]. The result takes the form:

$$|\mathcal{M}_1^{\mathrm{SLL}}|^2 = -\frac{2}{\pi}\int_{Q_0}^{Q}\frac{dk_T}{k_T}\int_{\mathrm{out}} dy\ \left(\Omega_R^{\mathrm{coll}} + \Omega_V^{\mathrm{coll}}\right)\,, \tag{5}$$

where $\Omega_{R(V)}^{\mathrm{coll}}$ is the resummed real (virtual) contribution in the limit where the out-of-gap gluon becomes collinear to one of the incoming partons.

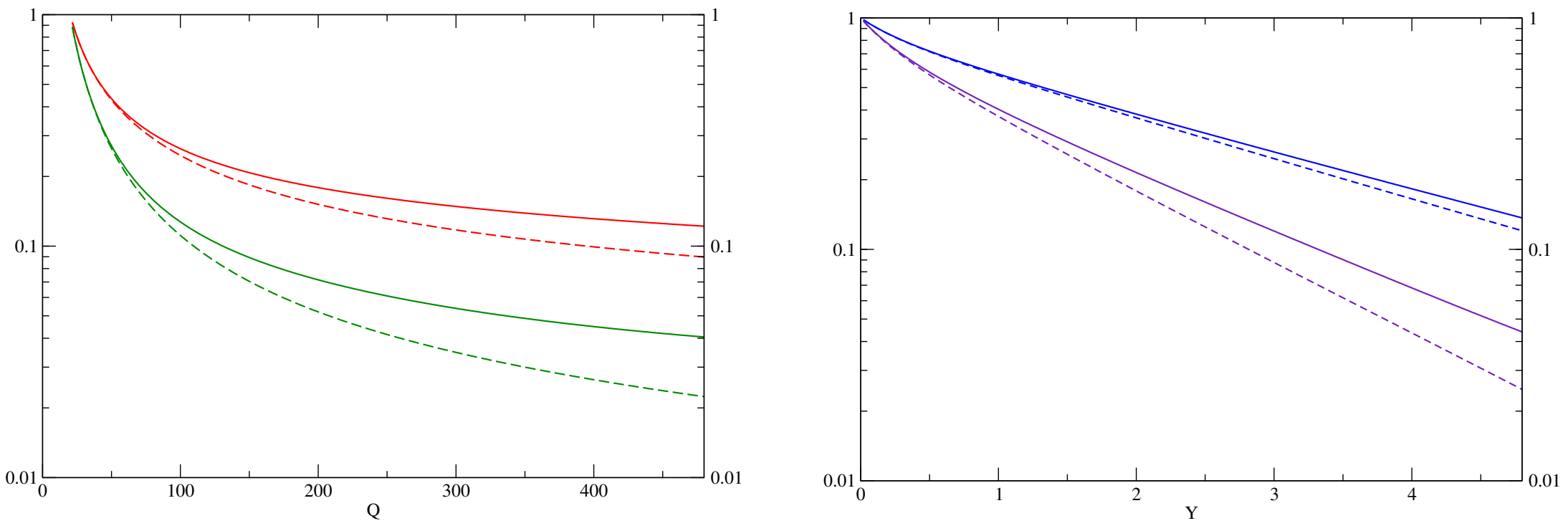

Figure 1: On the left we plot the gap fraction for $Y = 3$ (upper red curves) and $Y = 5$ (lower green curves) as a function of $Q$ and on the right as a function of $Y$, for $Q = 100$ GeV (upper blue curves) and $Q = 500$ GeV (lower violet curves). The solid lines are the full resummation of global logarithms, while the dashed ones are obtained by omitting the $i\pi$ terms in the anomalous dimension. The veto scale is $Q_0 = 20$ GeV and the jet radius $R = 0.4$.

# 3 LHC Phenomenology

In this section we perform two different studies. Firstly we consider the resummation of global logarithms and we study the importance of Coulomb gluon contributions, comparing the resummed results to the ones obtained with a parton shower approach. We then turn our attention to super-leading logarithms and we evaluate their phenomenological relevance. In both studies we consider $\sqrt{S} = 14$ TeV, $Q_0 = 20$ GeV, jet radius $R = 0.4$ and we use the MSTW 2008 LO parton distributions [16].

## 3.1 Comparison to Parton Shower

Soft logarithmic corrections are implemented in HERWIG++ via angular ordering of successive emissions. Such an approach cannot capture the contributions coming from the imaginary part of the loop integrals. We evaluate the importance of these contributions in Fig. 1. On the left we plot the gap cross-section, normalised to the Born cross-section (i.e. the gap fraction), as a function of $Q$ at two different values of $Y$ and, on the right, as a function of $Y$ at two different values of $Q$. The solid lines represent the results of the resummation of global logarithms; the dashed lines are obtained by omitting the $i\pi$ terms in the soft anomalous dimension, Eq. 2. As a consequence, the gap fraction is reduced by 7% at $Q = 100$ GeV and $Y = 3$ and by as much as 50% at $Q = 500$ GeV and $Y = 5$. Large corrections from this source herald the breakdown of the parton shower approach. In Fig. 2 we compare the gap cross-section obtained after resummation to that obtained using Herwig++ [17–20] after parton showering ($Q$ is taken to be the mean $p_T$ of the two leading jets). The broad agreement is encouraging and indicates that effects such as energy conservation, which is included in the Monte Carlo, are not too disruptive to the resummed calculation. Nevertheless, the histogram ought to be compared to the dotted curve rather than the solid one, because Herwig++ does not include the Coulomb gluon contributions. In the bottom part of the figure the ratio of the resummation prediction over the Monte Carlo is plotted. The resummation approach and the parton shower differ in several aspects: some non-global logarithms are included in the Monte Carlo and the shower is performed in the large $N_c$ limit. Of course the resummation would benefit from matching to the NLO calculation and this should be done before comparing to data.

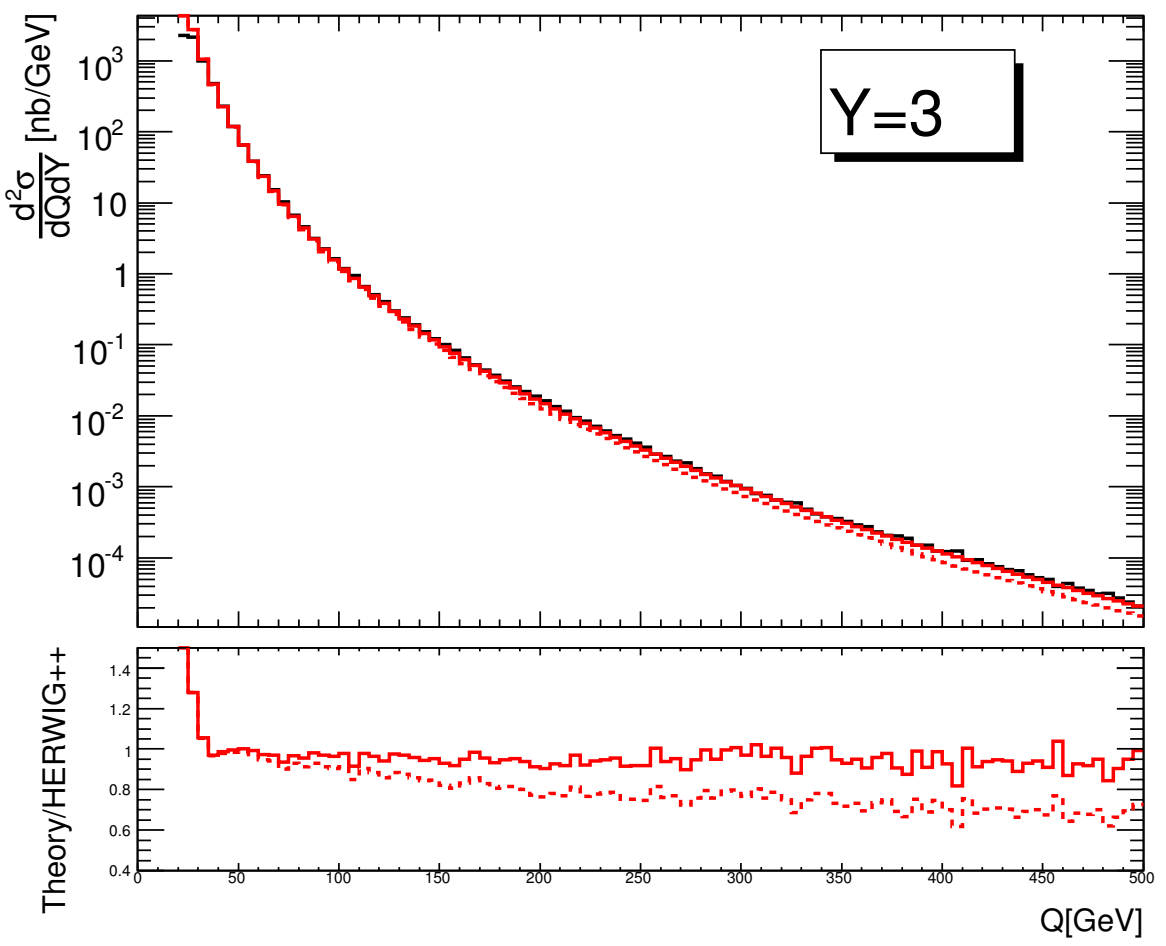

Figure 2: The gap cross-section obtained using Herwig++ (black histogram) is compared to the one from resummation (red curves). As before the solid line is the full result, while the dashed line is obtained by omitting the Coulomb gluon contributions. At the bottom we plot the ratio between the results obtained from the resummation and the one from Herwig++.

## 3.2 Super-Leading Logarithms

In the following we study the phenomenology of super-leading logarithms. In [15], the coefficients of the SLL have been computed order-by-order in perturbation theory up to $\mathcal{O}\left(\alpha_s^5\right)$, with respect to the Born cross-section. At this order in the perturbative expansion one must consider the contributions coming from one and two gluons outside the rapidity gap. We start by computing the impact of the super-leading contributions on the hadronic cross-section order-by-order in perturbation theory. To this end we define

$$K = \frac{\sigma^{(0)} + \sigma^{\mathrm{SLL}}}{\sigma^{(0)}}, \tag{6}$$

where $\sigma^{(0)}$ contains the resummed global logarithms and $\sigma^{\mathrm{SLL}}$ is the super-logarithmic contributions. We plot these $K$-factors in Fig. 3 as a function of $Q$, for $Y = 3$ (left) and for $Y = 5$ (right). The curves on the left (right) are obtained by considering one out-of-gap gluon at $\mathcal{O}(\alpha_s^4)$ (dotted) and $\mathcal{O}(\alpha_s^5)$ (dashed). The two-gluons outside the gap contribution is included in the dash-dotted curves. We also plot these $K$-factors for two values of $Q$ as a function of the rapidity separation $Y$ in Fig. 4. The plot on the left is for $Q = 100$ GeV, while the one on the right is for $Q = 500$ GeV. The different line styles are as in Fig. 3. From these plots it is clear that, for jets with rapidity separation $Y \geq 3$ and transverse momentum bigger than 200 GeV, the inclusion of fixed order SLL contributions leads to sizeable but unstable $K$-factors: higher order contributions are important and their resummation is necessary. Currently the resummation of SLL contributions has been carried out only for the case of one gluon outside the gap. Figs. 3 and 4 offer some hope that the impact of the two-or-more gluons outside the gap contribution may be modest, since the difference between the $K$-factors which include the two-out-of-gap gluon contributions (dash-dotted curves) and the ones that do not (dashed curves) is not large. The resummed results are added (as the solid lines) to the plots in Fig. 3 and Fig. 4. Generally the effects of the SLL are modest, reaching as much as 15% only for jets with $Q > 500$ GeV and rapidity separations $Y > 5$. Remember that we have fixed the value of the veto scale $Q_0 = 20$ GeV and that the impact will be more pronounced if the veto scale is lowered.

# 4 Conclusions and Outlook

We have discussed two phenomenological studies concerning the gaps between jets cross section at the LHC. In particular there are significant contributions arising from the exchange of Coulomb gluons, especially at large $Q/Q_0$ and/or large $Y$, which are not implemented in the parton shower Monte Carlos. However before comparing to data, there is a need to improve the resummed results by matching to the fixed order calculation. Matching to NLO is work in progress. These observations will have an impact on jet vetoing in Higgs-plus-two-jet studies at the LHC.

We have also studied the super-leading logarithms that occur because gluon emissions that are collinear to one of the incoming hard partons are forbidden from radiating back into the vetoed region. Even if their phenomenological relevance is generally modest, they deserve further study because they are deeply connected to the fundamental ideas behind QCD factorization.

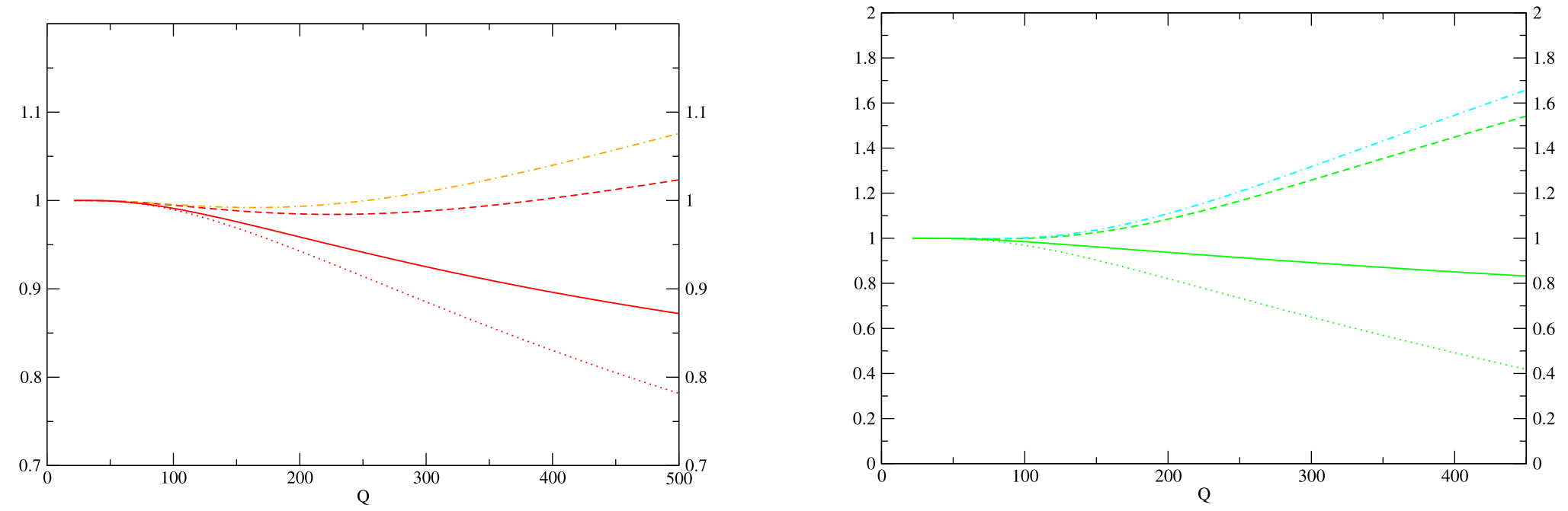

Figure 3: Plots of the $K$-factors as a function of $Q$, for $Y = 3$ on the left and for $Y = 5$ on the right. The curves are obtained by considering the one out-of-gap gluon cross-section at $\mathcal{O}(\alpha_s^4)$ (dotted) and $\mathcal{O}(\alpha_s^5)$ (dashed). The two-gluons outside of the gap contribution is included in the dash-dotted curves. The solid line corresponds to the resummation of the one out-of-gap gluon contributions.

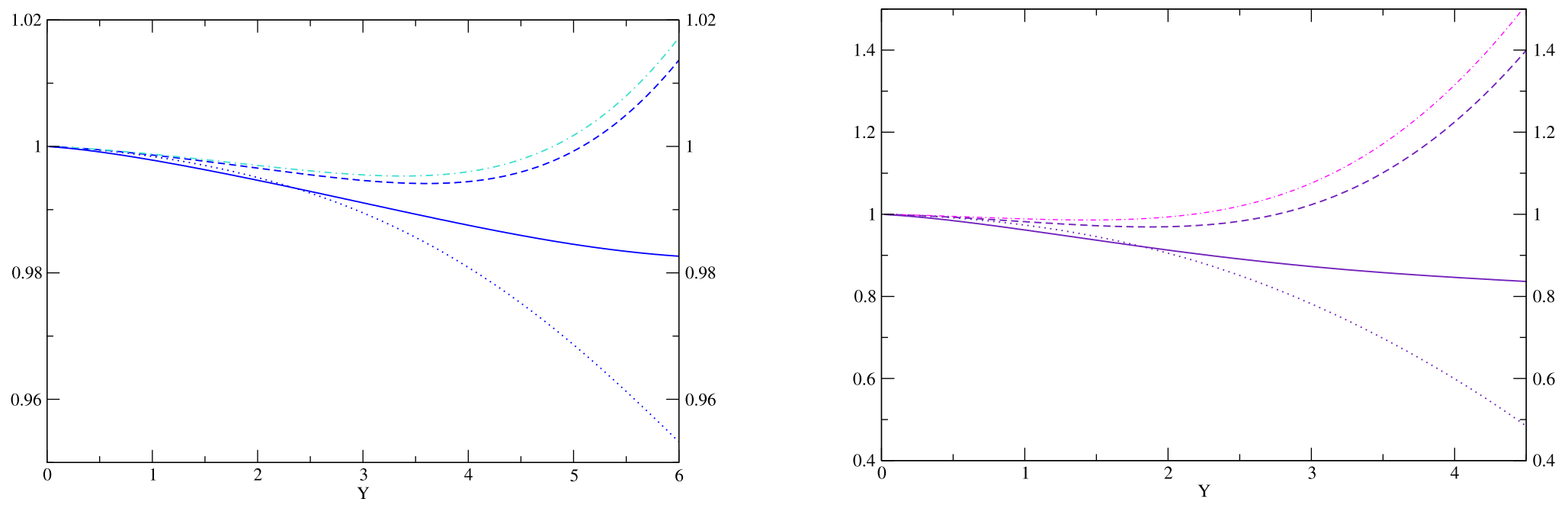

Figure 4: Plots of the $K$-factors as a function of $Y$, for $Q = 100$ GeV on the left and for $Q = 500$ GeV on the right. The curves are obtained by considering the one out-of-gap gluon cross-section at $\mathcal{O}(\alpha_s^4)$ (dotted) and $\mathcal{O}(\alpha_s^5)$ (dashed). The two-gluons outside the gap contribution is included in the dash-dotted curves. The solid line corresponds to the resummation of the one out-of-gap gluon contributions. Notice that for small values of $Y$, the resummed curves dip below the $\mathcal{O}\left(\alpha_s^4\right)$ curves. This behaviour is completely explained by the running of the coupling. In the fixed oder case we have $\alpha_s = \alpha_s(Q)$, while in the resummed case the running coupling is evaluated at $\alpha_s = \alpha_s(k_T)$, i.e. it sits inside the transverse momentum integration.

# References

[1] A. H. Mueller and H. Navelet, Nucl. Phys. B **282**, 727 (1987).

[2] A. H. Mueller and W. K. Tang, Phys. Lett. B **284** (1992) 123.

[3] V. D. Barger, R. J. N. Phillips and D. Zeppenfeld, Phys. Lett. B **346** (1995) 106 [arXiv:hep-ph/9412276].

[4] N. Kauer, T. Plehn, D. L. Rainwater and D. Zeppenfeld, Phys. Lett. B **503** (2001) 113 [arXiv:hep-ph/0012351].

[5] J. R. Forshaw and M. Sjödahl, JHEP **0709** (2007) 119 [arXiv:0705.1504 [hep-ph]].

[6] N. Kidonakis, G. Oderda and G. Sterman, Nucl. Phys. B **531** (1998) 365 [arXiv:hep-ph/9803241].

[7] G. Oderda and G. Sterman, Phys. Rev. Lett. **81** (1998) 3591 [arXiv:hep-ph/9806530].

[8] G. Oderda, Phys. Rev. D **61** (2000) 014004 [arXiv:hep-ph/9903240].

[9] J. R. Forshaw, A. Kyrieleis and M. H. Seymour, JHEP **0608** (2006) 059 [arXiv:hep-ph/0604094].

[10] J. R. Forshaw, A. Kyrieleis and M. H. Seymour, JHEP **0809** (2008) 128 [arXiv:0808.1269 [hep-ph]].

[11] M. Dasgupta and G. P. Salam, Phys. Lett. B **512** (2001) 323 [arXiv:hep-ph/0104277].

[12] R. B. Appleby and M. H. Seymour, JHEP **0212** (2002) 063 [arXiv:hep-ph/0211426].

[13] A. Banfi, G. Marchesini and G. Smye, JHEP **0208** (2002) 006 [arXiv:hep-ph/0206076].

[14] J. Forshaw, J. Keates and S. Marzani, arXiv:0905.1350 [hep-ph].

[15] J. Keates and M. H. Seymour, JHEP **0904** (2009) 040 [arXiv:0902.0477 [hep-ph]].

[16] A. D. Martin, W. J. Stirling, R. S. Thorne and G. Watt, arXiv:0901.0002 [hep-ph].

[17] L. Lönnblad, Comput. Phys. Commun. **118** (1999) 213.

[18] M. Bahr *et al.*, *Herwig++ Physics and Manual*, arXiv:0803.0883 [hep-ph].

[19] R. Kleiss et al, CERN 89-08, vol. 3, pp 129-131.

[20] S. Gieseke, P. Stephens and B. Webber,JHEP **0312** (2003) 045.

# CDF Results on Diffraction

*Christina Mesropian*[1]

[1] *on behalf of the CDF collaboration*
The Rockefeller University
1230 York Avenue, New York, NY, USA

We report results on diffraction obtained by the CDF collaboration in $p\bar{p}$ collisions at the Fermilab Tevatron collider at $\sqrt{s} = 1.96$ TeV. The single diffractive dissociation processes such as diffractive dijet, W and Z productions are discussed. We also present results on double diffractive dissociation for central gap production. The first experimental observation of exclusive dijets is presented.

## 1 Introduction

Diffractive reactions, which constitute a substantial fraction of the total cross section in hadron-hadron scattering, can be described in terms of the *pomeron* exchange, hypothetical object with the quantum numbers of the vacuum. The experimental signatures of the diffraction consist in particular kinematic configurations of the final states: presence of non-exponentially suppressed large rapidity gaps and/or presence of the intact leading particles. The diffractive processes became an important tool in understanding many interesting aspects of QCD such as low-$x$ structure of the proton, behaviour of QCD in the high density regime. Recently, a lot of attention was drawn to the possibility of discovering diffractively produced Higgs boson at the Large Hadron Collider (LHC). CDF collaboration contributed extensively [1]-[9] to significant progress in understanding diffraction by studying wide variety of diffractive processes at three different centre-of-mass energies: 630 GeV, 1800 GeV, Run I of Tevatron, and 1960 GeV - Run II. Some of the important results include the observation of the QCD factorisation breakdown in hard single diffractive processes, discovery of large rapidity gaps between two jets, study of diffractive structure function in double pomeron exchange dijet events.

## 2 CDF II Detector

The identification of diffractive events requires either tagging of the leading particle or observation of a rapidity gap, thus the forward detectors are very important for the implementation of a diffractive program. The schematic layout of CDF II detector is presented in Fig. 1. The Forward Detectors include the Roman Pot fiber tracker Spectrometer (RPS) used to select events with a leading $\bar{p}$, the Beam Shower Counters (BSCs), covering the pseudorapidity range $5.5 <\mid \eta \mid< 7.5$, detecting particles travelling in either direction from the interaction point along the beam-pipe, and the Miniplug Calorimeters (MP) [10] measuring energy and lateral position of electromagnetic and hadronic showers in the pseudorapidity region of $3.5 <\mid \eta \mid< 5.1$.

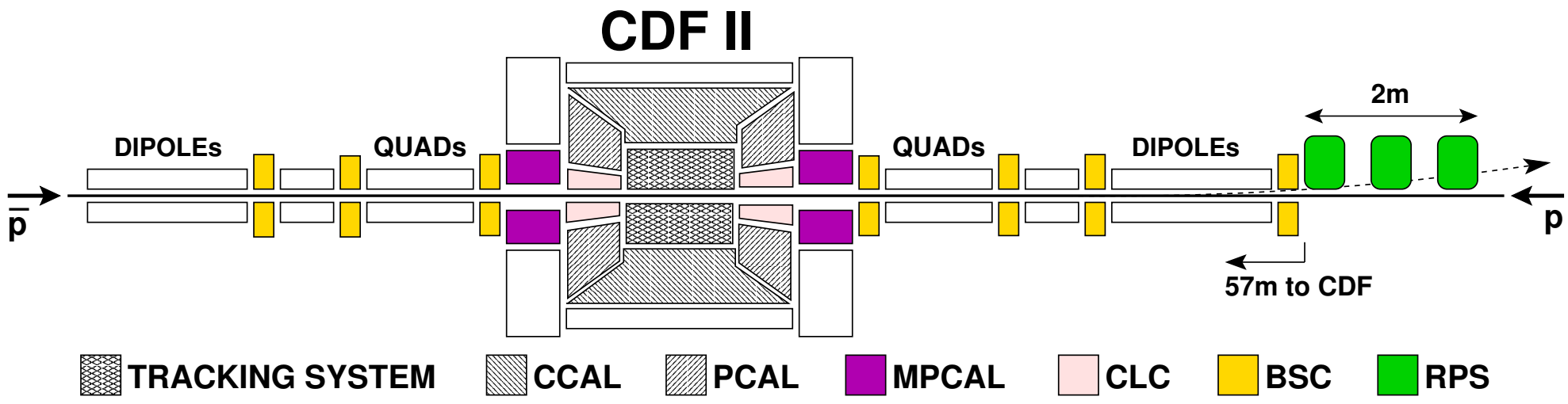

Figure 1: Schematic drawing of the CDF II detector.

# 3 Single Diffractive Dissociation

The signature of single diffractive (SD) dissociation at the Tevatron is either forward rapidity gap along the direction of one of the initial particles, or a presence of leading particle, antiproton, with $\xi < 0.1$. The process $\bar{p}p \to \bar{p}X$, which can be described by assuming that a pomeron is emitted by the incident antiproton and undergoes a hard scattering with the proton, is an ideal reaction to study the partonic content of the pomeron, and the diffractive structure function.

## 3.1 Diffractive Dijet Production

One of the single diffractive processes studied during Run I was the diffractive dijet production [6]. We compared two samples of dijet events, diffractive, triggered by the presence of intact antiproton detected in the RPS, and non-diffractive (ND). By taking the ratio of SD dijet rates to ND, which in a good approximation is the ratio of the diffractive to the known proton structure function, the diffractive structure function can be extracted. Our studies of the diffractive structure function were continued in Run II. One of the major challenges in diffractive studies during Run II is the presence of multiple $p\bar{p}$ interactions, which by overlapping with the diffractive event spoil signature characteristics of those, such as rapidity gaps. The rejection of *overlaps* was done by reconstructing $\xi$ from the calorimeter towers, and considering only $\xi^{cal} < 0.1$, thus rejecting overlap events which have high $\xi^{cal}$ values as a result of having more energy than just the diffractive interactions.

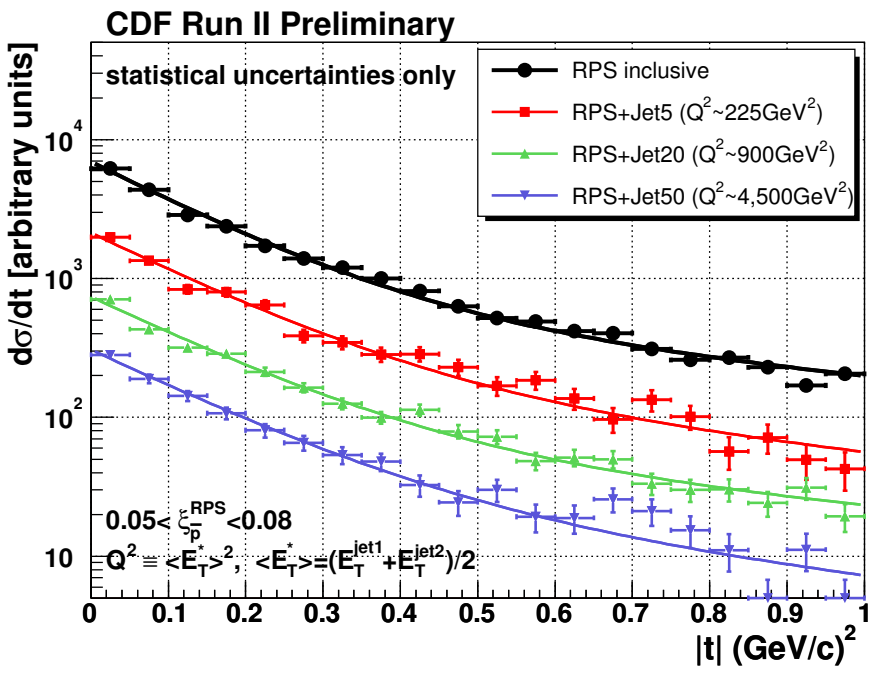

Figure 2: $t$ distributions in soft and hard SD events for different $Q^2$ ranges.

We extended our results from Run I by studying $310\,\mathrm{pb}^{-1}$ SD data sample to examine the $Q^2$ dependence of the structure function up to $10^4\,\mathrm{GeV}^2$, where $Q^2$ is defined as an average value of mean dijet $E_T$. No appreciable $Q^2$ dependence was observed. We also studied the $t$ distribution in single diffractive dijet events up to $Q^2 \sim 4500\,\mathrm{GeV}^2$, and no $Q^2$ dependence of the shape of the $t$ distribution was observed, see Fig. 2.

## 3.2 Diffractive W, Z Production

The diffractive W/Z production is an important process for probing the quark content of the pomeron, since to leading order, the W/Z is produced through a quark, while the gluon associated production is suppressed by a factor of $\alpha_S$ and can be identified by an additional jet.

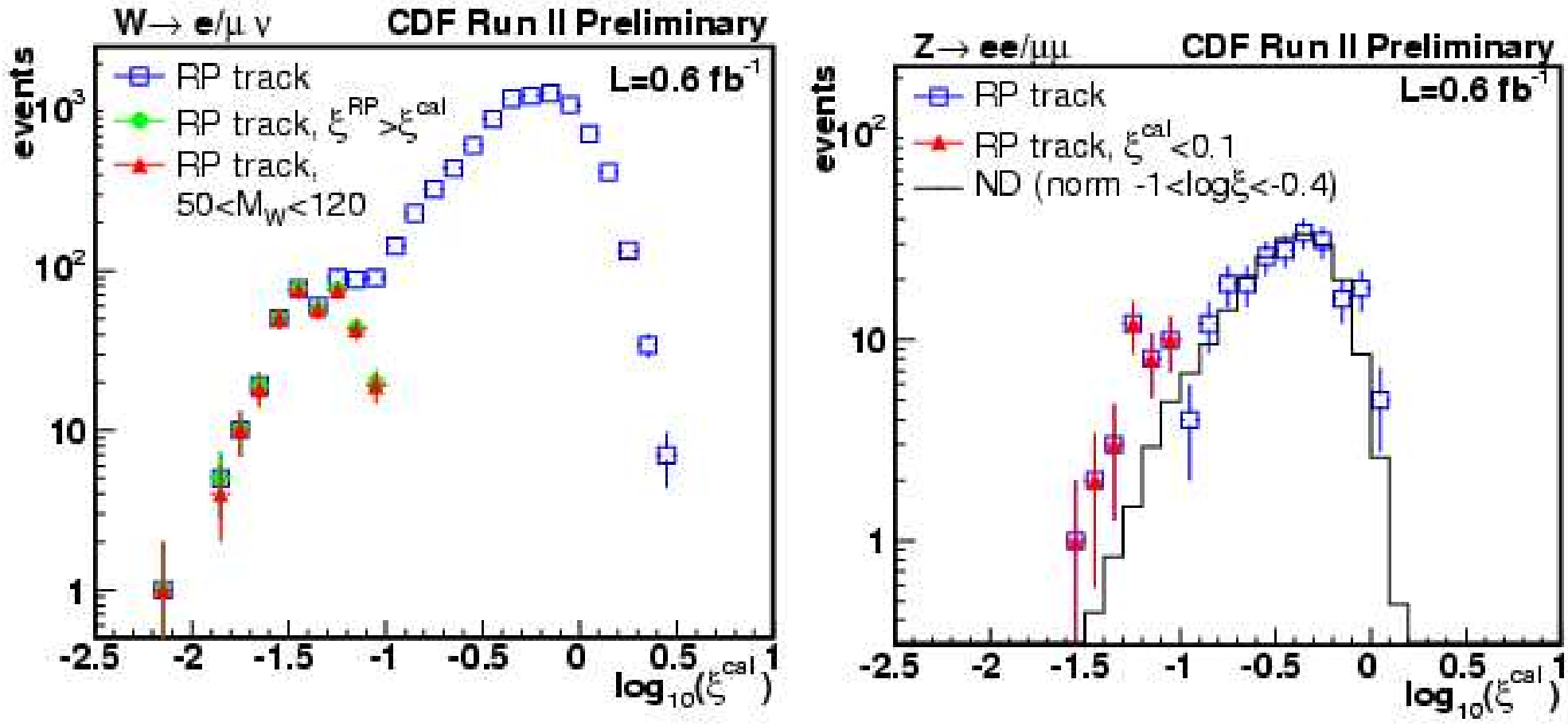

Figure 3: Calorimeter $\xi$ distribution in W (left) and Z (right) events with a RPS track.

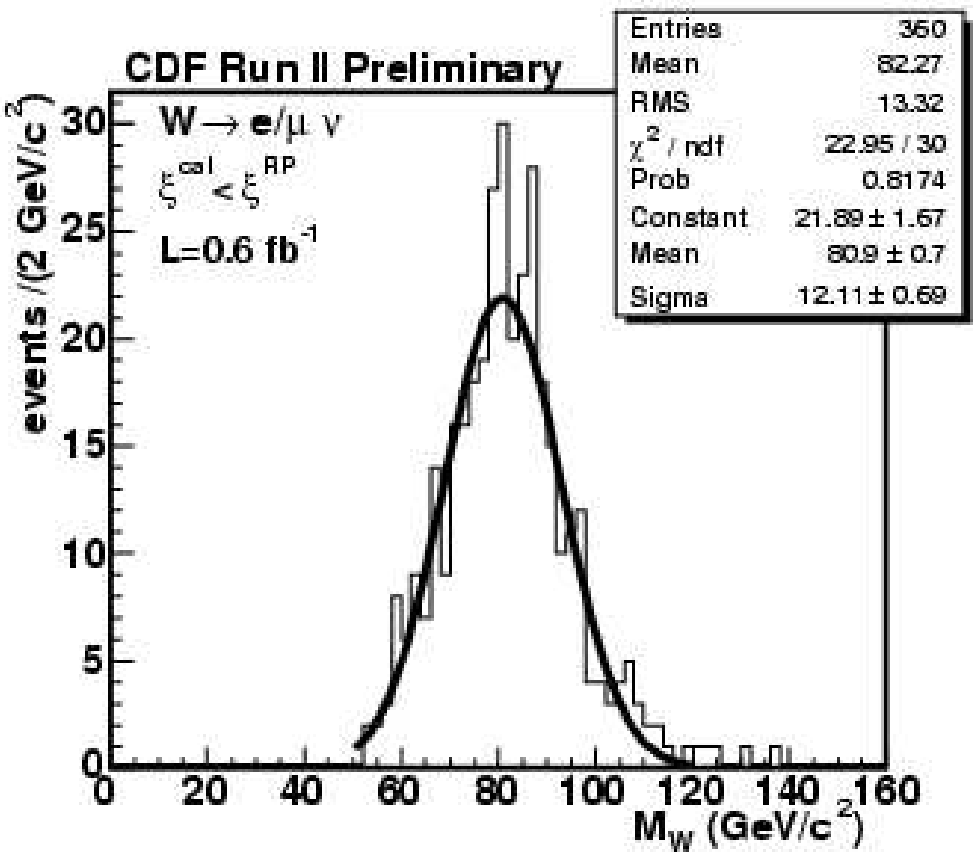

Figure 4: Reconstructed W mass in diffractive W candidate events.

In Run II, we developed a method that completely reconstructs W kinematics. The events are selected by utilising "intact leading antiproton" signature of the diffractive event. The RP spectrometer allows very precise $\xi$ measurement, eliminating the problem of *gap survival probability.* The presence of a hit in the RPS trigger counters and a RPS reconstructed track with $0.03 < \xi < 0.1$ and $|t| < 1\,\mathrm{GeV}^2$ is required. The novel feature of the analysis, the determination of the full kinematics of the $W \to l\nu$ decay, is made possible by obtaining the neutrino ${E_T}^{\nu}$ from the missing $E_T$, $\not{E}_T$, and $\eta_\nu$ from the formula $\xi^{RPS} - \xi^{cal} = \frac{\not{E}_T}{\sqrt{s}} e^{-\eta_\nu}$, where $\xi^{RPS}$ is true $\xi$ measured in RPS and $\xi^{cal} = \sum_{i(towers)} (E_T^i/\sqrt{s}) exp(-\eta^i)$. Since we expect $\xi^{cal} < \xi^{RPS}$, we impose this requirement to remove overlap events to ensure that Ws were produced diffractively. Figures 3a,b show $\xi^{cal}$ distributions for W and Z events with a RPS track. Figure 4 shows reconstructed W mass in diffractive W candidates. The requirements for W/Z selection are following: ${E_T}^l > 25\,\mathrm{GeV}$, $40 < M_T^W < 120\,\mathrm{GeV}$, $66 < M^Z < 116\,\mathrm{GeV}$, and $z-$vertex coordinate $z_{vtx} < 60\,\mathrm{cm}$. The fractions of diffractive W and Z events are measured to be $[0.97 \pm 0.05(stat.) \pm 0.11(syst.)]\%$ and $[0.85 \pm 0.20(stat.) \pm 0.11(syst.)]\%$ for the kinematic range $0.03 < \xi < 0.10$ and $|t| < 1\,\mathrm{GeV}^2$. The measured diffractive W fraction is consistent with the Run I CDF result when corrected for the $\xi$ and $t$ range.

# 4 Double Diffractive Dissociation

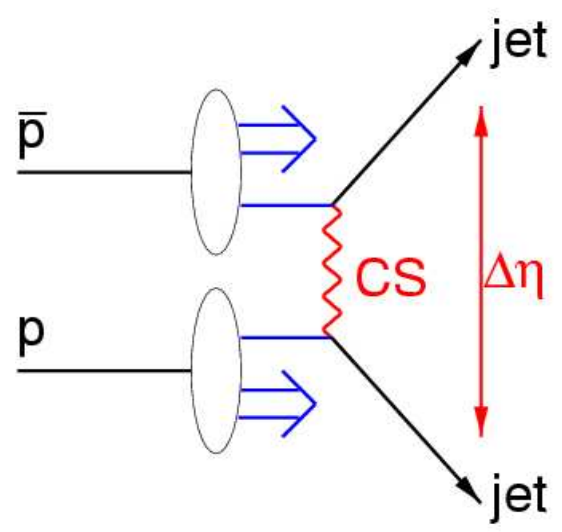

Figure 5: Schematic diagram of an event with a rapidity gap between jets produced in $p\bar{p}$ collisions.

Double diffractive (DD) dissociation is the process in which two colliding hadrons dissociate into clusters of particles (jets in case of hard DD dissociation) producing events with a large non-exponentially suppressed central pseudo-rapidity gap, see Fig. 5. Events with pseudorapidity gaps are presumed to be due to the exchange across the gap of a colour singlet (CS) object, *pomeron*, with vacuum quantum numbers. The dependence of the gap fraction on the width and on the position of the centre of the gap, $\Delta\eta^{gap}$ and $\eta_c{}^{gap}$, and the jet characteristics of the event are of great interest, as they allow testing different theoretical models. Measurements of DD have been performed for hard and soft diffraction processes [1, 4, 5, 8] for $p\bar{p}$ collisions at $\sqrt{s} = 630$ and 1800 GeV by the CDF collaboration. The extended rapidity coverage provided by the MP calorimeters ($3.5 < |\eta| < 5.1$) makes CDF II a powerful detector for hard DD studies, as it provides up to 8 units of $\eta$ between jets for jets in the MPs. Using this data sample, we select events with two jets in the MPs with $E_T > 2$ GeV, $3.5 < |\eta^{jet}| < 5.1$, and $\eta_1\eta_2 < 0$. The $E_T$ cut is designed to maximise the jet signal while minimising the effect of energy deposited by single particles and misidentified jets. We use this data sample to study the dependence of the gap fraction on the width of the gap in "hard" and "soft" DD production. For this aspect of our analysis we use a definition of rapidity gap similar to that of our Run I study [8], where the rapidity gap variable, $\Delta\eta$, is defined as $\Delta\eta = \eta_{max} - \eta_{min}$, where $\eta_{max}$ is the pseudorapidity of the particle (tower) closest to $\eta = 0$ in the proton direction and $\eta_{min}$ is the pseudorapidity of the particle closest to the $\bar{p}$ direction. For events with gaps which overlap $\eta = 0$ these are effectively the edges of the rapidity gap. The data sample mentioned above is used to study "hard" diffraction production, where a hard structure (jet) is present in the event. "Soft" diffractive production is analyzed by examining low luminosity data collected with a min-bias trigger. Fig. 6 shows a comparison of the gap fraction, as a function of $\Delta\eta$, between "hard" and "soft" DD production when a rapidity gap is required within $-1.1 < \eta < 1.1$.

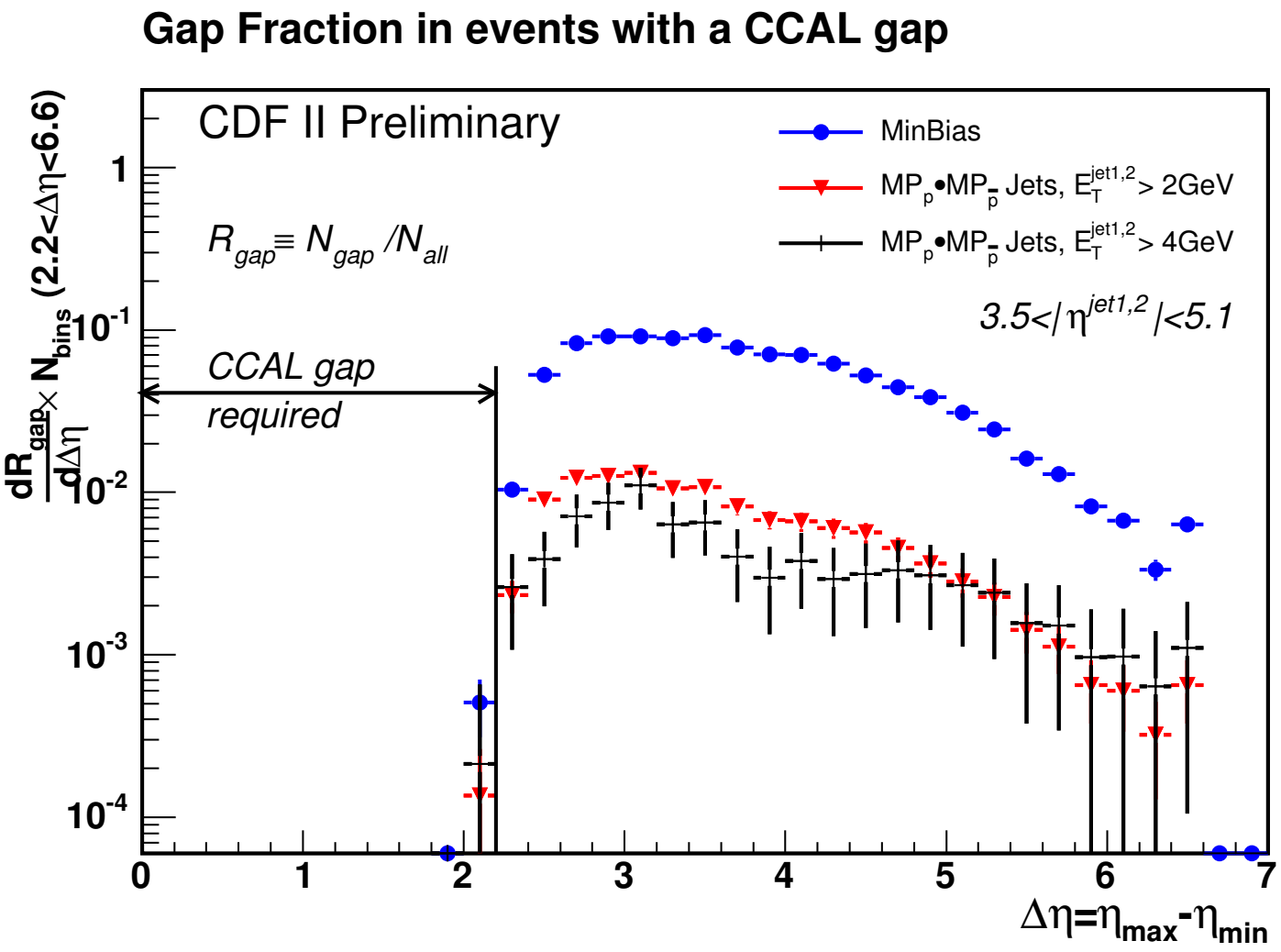

Figure 6: The distribution of the gap fraction $R_{gap} = N_{gap}/N_{all}$ vs. $\Delta\eta = \eta_{max} - \eta_{min}$ for min-bias and MiniPlug jet events of $E_T^{jet1,2} > 2$ GeV and $E_T^{jet1,2} > 4$ GeV.

This comparison is relatively free of systematic uncertainties, as detector and beam related effects cancel out. The distributions are similar in shape, demonstrating that the gap fraction decreases with increasing $\Delta\eta$ for both "hard" and "soft" DD productions.

# 5 Exclusive Production

One of a very interesting topics of study at CDF is the central exclusive production. In leading order QCD such exclusive processes can occur through exchange of a colour-singlet two gluon system between nucleons, leaving large rapidity gaps in forward regions. One of the gluons participates in a hard interaction and additional screening gluon is exchanged to cancel the colour of the interacting gluons, allowing the leading hadrons to stay intact. This is also a special case of dijet/diphoton production in Double Pomeron Exchange (DPE), $p + \bar{p} \to p + X + \bar{p}$. Several different heavy systems $X$ can be considered, but the main motivation for these studies is to use the process $p+\bar{p} \to p+H+\bar{p}$, see Fig. 7(a), to discover Higgs boson at the LHC. Although the cross section for the exclusive Higgs production is too small to be observed at the Tevatron, several processes mediated by the same mechanism but with the higher production rates can be studied to check theoretical predictions. Fig. 7(b-d) shows processes corresponding to studies in exclusive physics at the Tevatron: exclusive dijet, $\chi_c$ meson, and diphoton productions.The last two final states are discussed by M. Albrow and J. Pinfold in these proceedings.

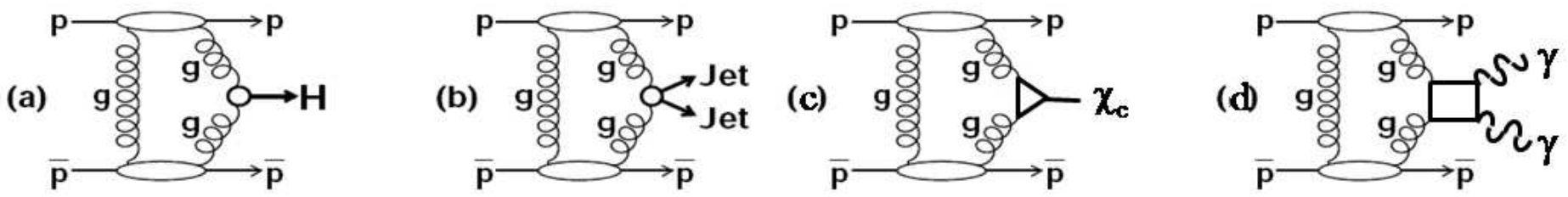

Figure 7: Diagrams of exclusive production of (a) Higgs boson, (b) dijet, (c) $\chi_c$, and (d) a photon-photon pair.

## 5.1 Exclusive Dijet Production

The exclusive dijet production was first studied by CDF in Run I data and the limit of $\sigma_{excl} < 3.7\,\text{nb}$ (95% CL) was placed [11]. The data sample of $313\,\text{pb}^{-1}$ for the exclusive dijet production was collected in Run II with the dedicated trigger requiring a BSC gap on the proton side in the addition to the requirement for the leading anti-proton in the RPS and at least one calorimeter tower with $E_T > 5\,\text{GeV}$. The events in data sample also passed the offline requirement of additional gap in MP on the proton-side. The exclusive signal is extracted using the dijet mass fraction method: from the energies and momenta of the jets in an event, the ratio $R_{jj} \equiv M_{jj}/M_X$ of the dijet mass $M_{jj}$ to the total mass $M_X$ of the final state (excluding the $p$ and $\bar{p}$) is formed and used to discriminate between the signal of exclusive dijets, expected to appear at $R_{jj} = 1$, and the background of inclusive DPE dijets, expected to have a continuous distribution concentrated at lower $R_{jj}$ values. Because of smearing effects in the measurement of ${E_T}^{jet}$ and jet and gluon radiation from the jets, the exclusive dijet peak is broadened and shifts to lower $R_{jj}$ values. The exclusive signal is obtained by a fit of the $R_{jj}$ distribution to expected signal and background shapes generated by Monte Carlo (MC) simulations. The data clearly show an excess at high $R_{jj}$, see Fig. 8 (left) over the non-DPE background events and inclusive DPE predictions obtained from POMWIG MC. The shape of excess is well described by exclusive dijet MC based on two models (ExHuME [12], DPEMC [13]); however, the measured cross section [14], see Fig. 8 (right), disfavours DPEMC. Predictions [15] are found to be consistent with data within its factor of 3 uncertainty. The results could also be checked with an event sample of heavy quark flavor dijets, for which exclusive production is expected to be suppressed in LO QCD by the $J_z = 0$ selection rule of the hard scattered digluon system, where $J_z$ is the projection of the total angular momentum of the system along the beam direction. Fig. 9 shows comparison between exclusive dijet results obtained with MC-based method and the data based suppression of the exclusive heavy flavor to inclusive dijet production rates. The results are consistent with each other.

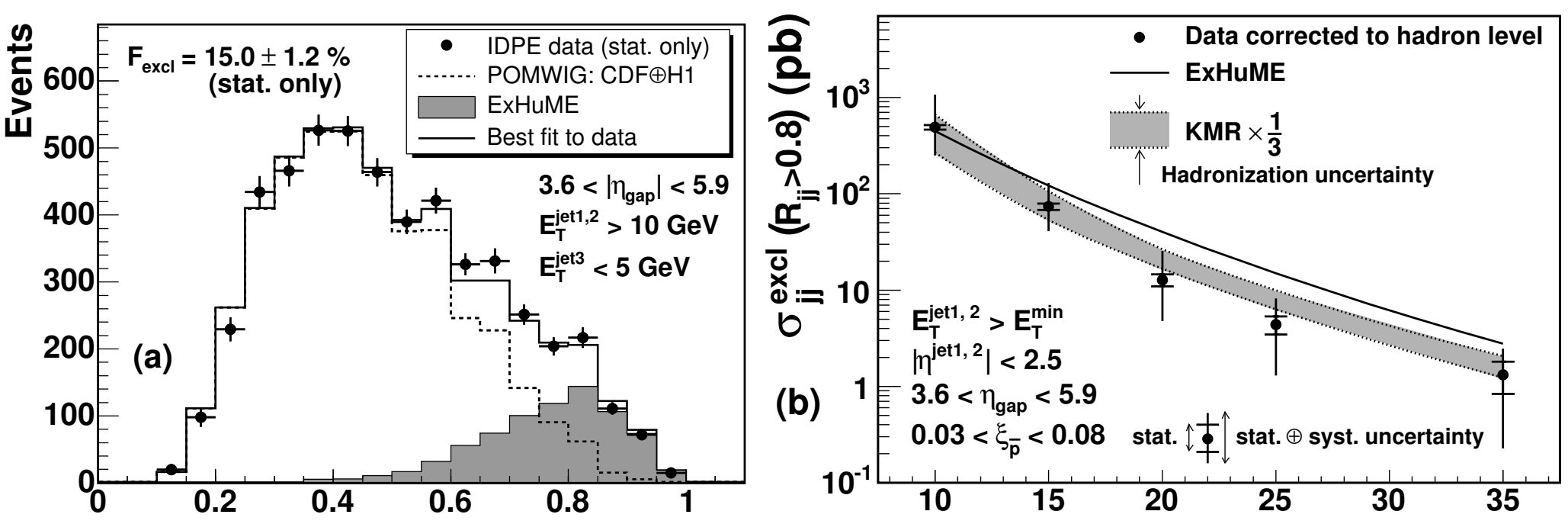

Figure 8: Dijet mass fraction in data (points), POMWIG MC and exclusive dijet (shaded histogram) MC events (left); Exclusive dijet cross sections for events with two jets and with $R_{jj} > 0.8$ (right).

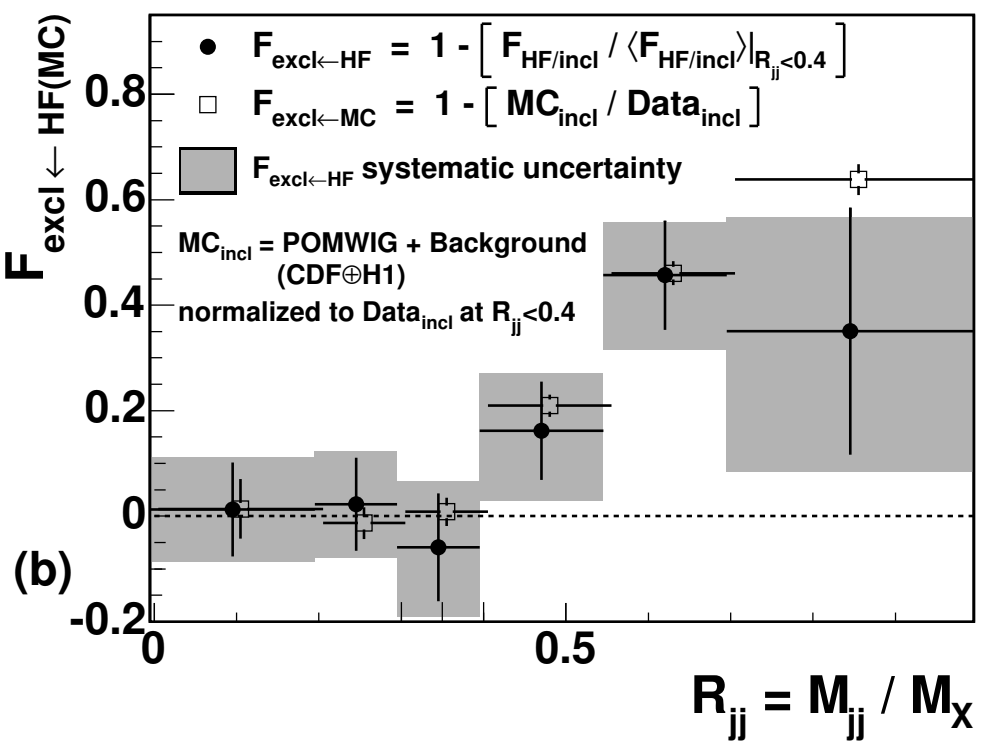

Figure 9: Measured ration $F_{HF}$ (filled circles) of heavy flavor jets to all inclusive jets and the exclusive dijet ratio $F_{excl}$ (open squares) obtained by comparison between inclusive dijet events and POMWIG MC.

## 6 Conclusions

The CDF collaboration continues the long-standing extensive program of diffractive studies. Our Run II results include measurements of diffractive structure function and $t$ distribution from single diffractive dijet production, the measurement of diffractive W/Z production using Roman Pot Spectrometers, and study of double diffractive processes in events with a central rapidity gap. In addition, the first observation of exclusive dijet production by CDF provides an important calibration for the predictions of the exclusive Higgs production at the LHC.

## Acknowledgements

I would like to thank the organisers of 13th International Conference on Elastic & Diffractive Scattering for warm hospitality and for an exciting conference.

## References

[1] F. Abe *et al.,* Phys.Rev.Lett. **74**, 855 (1995).

[2] F. Abe *et al.,* Phys.Rev.Lett. **78**, 2698 (1997).

[3] F. Abe *et al.,* Phys.Rev.Lett. **79**, 2636 (1997).

[4] F. Abe *et al.,* Phys.Rev.Lett. **80**, 1156 (1998).

[5] F. Abe *et al.,* Phys.Rev.Lett. **81**, 5278 (1998).

[6] T. Affolder *et al.,* Phys.Rev.Lett. **84**, 5043 (2000).

[7] T. Affolder *et al.,* Phys.Rev.Lett. **85**, 4215 (2000).

[8] T. Affolder *et al.,* Phys.Rev.Lett. **87**, 141802-1 (2001).

[9] T. Affolder *et al.,* Phys.Rev.Lett. **87**, 241802-1 (2001).

[10] K. Goulianos *et al.,* Nucl.Instrum.Methods A **518**, 24 (2004).

[11] T. Affolder *et al.,* Phys.Rev.Lett. **85**, 4215 (2000).

[12] J. Monk and A. Pilkington, Comput.Phys.Commun. **175**, 232 (2006).

[13] M. Boonekamp and T. Kuss, Comput.Phys. Commun. **167**, 217 (2005).

[14] T. Aaltonen *et al.,* Phys.Rev. D **77** 052004 (2008).

[15] V.A. Khoze *et al.,* Eur. Phys.J. C **14**, 525 (2000); V.A. Khoze *et al.,* arXiv:0705.2314.

[16] A.D. Martin *et al.*, Eur. Phys. J. **C39** 155 (2005).

# Inclusive Hard Diffraction at HERA

*Alexander Proskuryakov*

Moscow State University, Leninskie gory, Moscow 119991, Russia

(On behalf of the H1 and ZEUS Collaborations)

Recent data from the H1 and ZEUS experiments on hard inclusive diffraction are discussed. Results of QCD analyses of the diffractive deep-inelastic scattering processes are reported. Predictions based on the extracted parton densities are compared to diffractive dijet measurements.

## 1 Introduction

One of the most interesting results obtained at HERA is the observation that a significant amount of events in deep-inelastic scattering (DIS) is of diffractive nature. Diffractive processes are usually described in Regge theory [1] as the exchange of a trajectory with the vacuum quantum numbers, the Pomeron trajectory. However, the observation of jets in diffractive $\bar{p}p$ scattering [2] opened the possibility to study diffraction in the framework of quantum chromodynamics (QCD). Diffractive reactions in DIS provide additional information on the structure of diffraction offering an opportunity to study the transition from soft to hard diffraction. The concept of diffractive parton distribution functions (DPDFs) plays an important role in the study of diffractive reactions in DIS and is essential input to calculations of hard diffractive processes at the LHC.

This paper briefly reports recent results on hard inclusive diffraction from the H1 and ZEUS experiments.

## 2 Kinematics and Cross Sections

Fig. 1 shows a schematic diagram for the diffractive process $ep \to eXp$. The data used for the measurements reported in the paper were collected at HERA $ep$ collider, when HERA collided positron (electron) of 27.5 GeV with protons of 920 GeV. The kinematics of the diffractive reaction $ep \to eXp$ is described by the following variables:

- $Q^2 = -(p_e - p_{e'})^2$, the negative four-momentum squared of the virtual photon, where $p_e(p_{e'})$ is the four-momentum the incident (scattered) electron;
- $W^2 = (p_\gamma + p_p)^2$, the squared centre-of-mass energy of the photon-proton system, where $p_\gamma$ and $p_p$ are the four-momenta of the virtual photon and the incident proton respectively;
- $M_X^2 = (p_\gamma + p_p - p_{p'})^2$, the squared mass of the system X, where $p_{p'}$ is the four-momentum of the scattered proton;
- $t = (p_p - p_{p'})^2$, the squared four-momentum transfer at the proton vertex;
- y=$(p_\gamma \cdot p_p)/(p_e \cdot p_p)$, the fraction of the electron energy transferred to the proton in the proton rest frame.

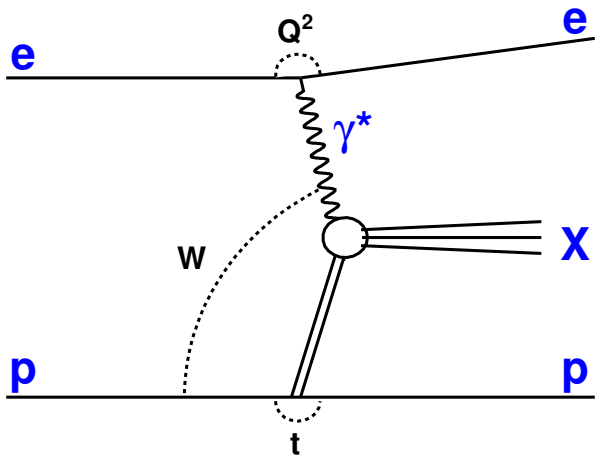

Figure 1: Schematic diagram for the reaction $ep \to eXp$.

The two dimensionless variables $x_{IP}$ and $\beta$ often used instead of $W$ and $M_X$ are defined as $x_{IP} = (p_p - p_{p'}) \cdot p_\gamma / p_p \cdot p_\gamma$ and $\beta = Q^2/2(p_p - p_{p'}) \cdot p_\gamma$.

The cross section for the diffractive process $ep \to eXp$ can be expressed in terms of the diffractive reduced cross section $\sigma_r^{D(4)}(\beta, Q^2, x_{ip}, t)$:

$$\frac{d\sigma^{ep}}{dx_{IP}d\beta dQ^2dt} = \frac{4\pi\alpha^2}{\beta Q^4}(1 - y + \frac{y^2}{2})\sigma_r^{D(4)},$$

where $\sigma_r^{D(4)}$ depends on the diffractive structure functions $F_2^{D(4)}$ and $F_L^{D(4)}$ as

$$\sigma_r^{D(4)} = F_2^{D(4)} - \frac{y^2}{2(1 - y + y^2/2)}F_L^{D(4)}.$$

The diffractive structure function $F_2^{D(3)}(\beta, Q^2, x_{IP})$ and the reduced cross section $\sigma_r^{D(3)}(\beta, Q^2, x_{IP})$ are obtained by integrating $F_2^{D(4)}(\beta, Q^2, x_{IP}, t)$ and $\sigma_r^{D(4)}(\beta, Q^2, x_{IP}, t)$ over $t$.

# 3 Selection of Diffractive Events

At HERA, diffractive events were selected either by the detection of the final state proton [3, 4] or on the basis of a large rapidity gap between the system X and the outgoing proton [4, 5]. The diffractive contribution was also identified by the $M_X$ method [6] based on the shape of the mass distribution of the system X.

In the methods based on the large rapidity gap (LRG) or $M_X$ selections the measured cross section includes a contribution from events of the type $ep \to eXN$, in which the proton dissociates into a low mass state N. The contribution from these proton-dissociative events is estimated from a Monte Carlo simulation.

The method based on the final state proton detection does not have background from proton-dissociative events. It allows a direct measurement of the variable $t$ and gives access to higher values of $x_{IP}$. However, the statistical precision is limited by the acceptance of the proton taggers.

Within the normalisation uncertainties the results from the different methods agree reasonably well [4].

# 4 Results

## 4.1 $t$ Dependence

The differential cross section $d\sigma/dt$ for the diffractive reaction $ep \rightarrow eXp$ is well approximated by the exponential function $e^{-b|t|}$. The differential cross section in the kinematic range $0.0002 < x_{IP} < 0.01$ and $0.01 < x_{IP} < 0.1$ is presented in Fig. 2. The value of the slope parameter $b$ obtained from the fit to the data is $b = 6 - 7$ GeV$^{-2}$ [3, 4].

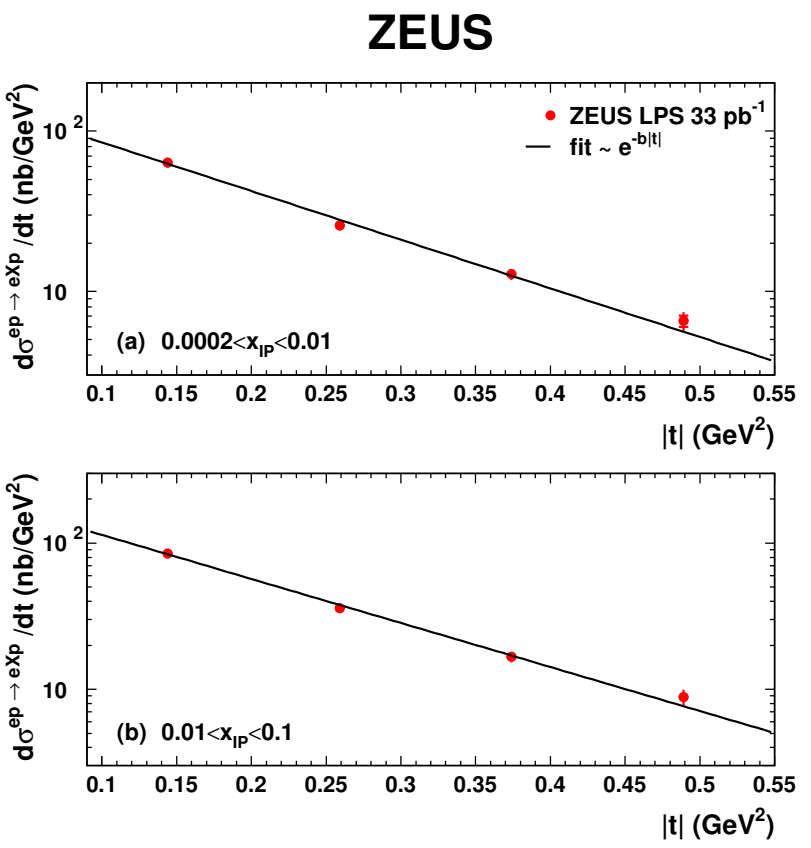

Figure 2: The differential cross section $d\sigma/dt$ for for (a) $0.0002 < x_{IP} < 0.01$ and (b) $0.01 < x_{IP} < 0.1$.

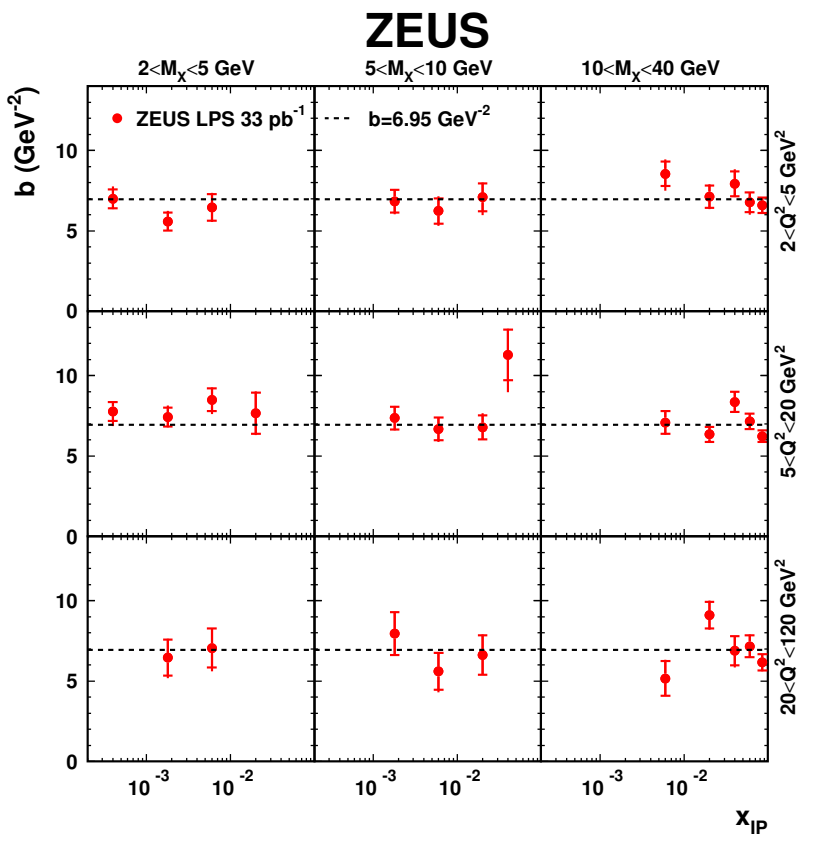

Figure 3: The value of the exponential-slope parameter $b$ of the differential cross section $d\sigma^{ep\rightarrow eXp}/dt \propto e^{-b|t|}$ as a function of $x_{IP}$ in bins of $Q^2$ and $M_X$.

The values of the slope parameter $b$ in different bins of $x_{IP}$, $Q^2$ and $M_X$ are shown in Fig. 3. The shape of the $t$ distribution does not depend on $x_{IP}$, $Q^2$ and $M_X$.

## 4.2 $x_{IP}$ Dependence

In the framework of Regge phenomenology the $x_{IP}$ dependence of the diffractive structure functions is related to the parameters of the Pomeron trajectory parameterised as $\alpha_{IP}(t) = \alpha_{IP}(0) + \alpha'_{IP}t$. The Pomeron intercept, $\alpha_{IP}(0)$, in soft hadronic interactions is $1.096^{+0.012}_{-0.09}$ [7]. However, the same parameter is significantly larger in the diffractive production of heavy vector mesons [9]. The slope of the Pomeron trajectory, $\alpha'_{IP}$ was found to be 0.25 GeV$^{-2}$ [8]. The parameters of the Pomeron trajectory can also be determined from the measurements of the diffractive reaction $ep \rightarrow eXp$.

Fig. 4 shows the reduced cross section $\sigma_r^{D(4)}$ in two $t$ bins, $0.09 < |t| < 0.19$ GeV$^2$ and $0.19 < |t| < 0.55$ GeV$^2$. The data presented in the figure were fitted to the form

$$F_2^{D(4)} = f_{IP}(x_{IP}, t) \cdot F_2^{IP}(\beta, Q^2) + n_R \cdot f_R(x_{IP}, t) \cdot F_2^R(\beta, Q^2)$$

where $n_R$ is a normalisation term. It was assumed that $F_2^{D(4)} = \sigma_r^{D(4)}$ and the fit was limited to $y < 0.5$ to reduce the influence of $F_L^D$. The Pomeron and Reggeon fluxes were parameterised

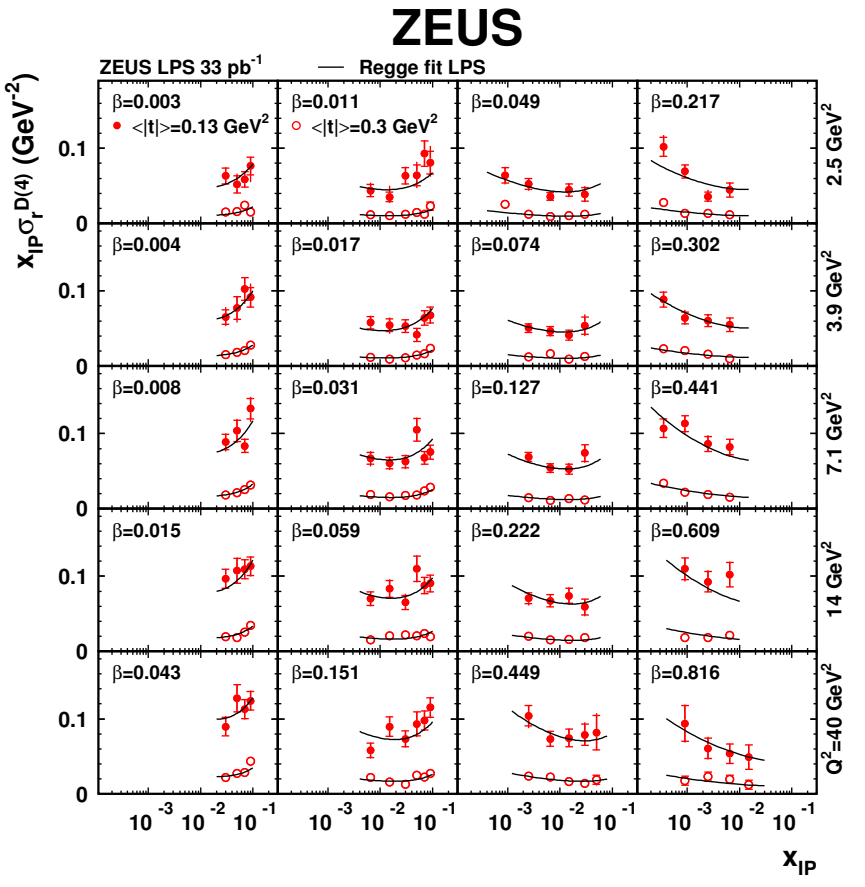

Figure 4: The reduced diffractive cross section multiplied by $x_{IP}$, $x_{IP}\sigma_r^{D(4)}$, in two $t$ bins as a function of $x_{IP}$ for different values of $Q^2$ and $\beta$.

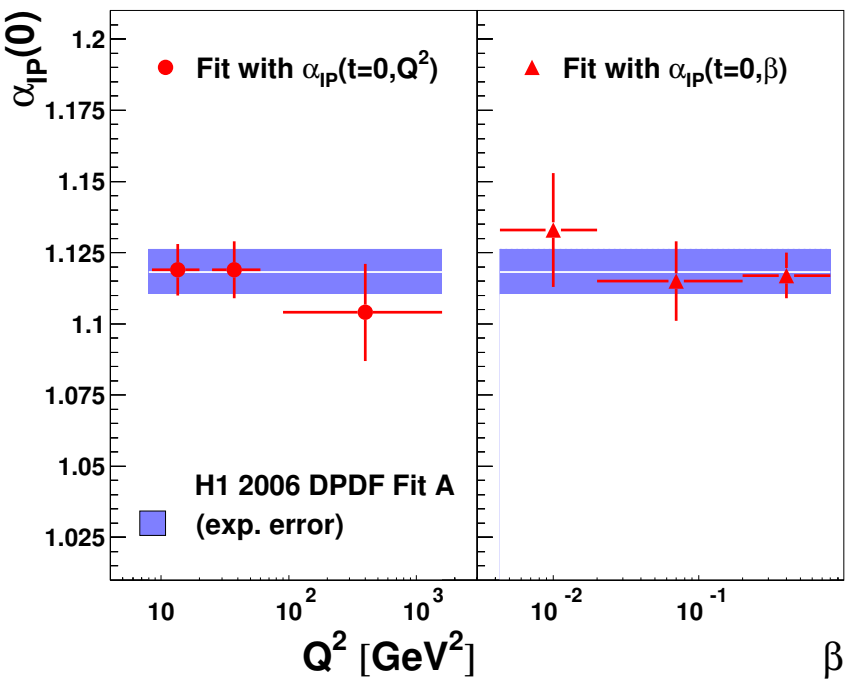

Figure 5: The Pomeron intercept $\alpha_{IP}(0)$ as a function of $Q^2$ and $\beta$.

as [1]

$$f_{IP,R}(x_{IP},t) = \frac{e^{B_{IP,R}t}}{x_{IP}^{2\alpha_{IP,R}(t)-1}},$$

with linear trajectories $\alpha_{IP,R}(t) = \alpha_{IP,R}(0) + \alpha'_{IP,R}t$. The Reggeon structure function, $F_2^R$, was taken to be equal to the pion structure function as parameterised by GRV [10, 11, 12]. The resulting Pomeron intercept is $\alpha_{IP}(0) = 1.11 \pm 0.02(\text{stat.})^{+0.01}_{-0.02}(\text{syst.}) \pm 0.02(\text{model})$. The slope of the Pomeron trajectory is $\alpha'_{IP} = -0.01 \pm 0.06(\text{stat.})^{+0.04}_{-0.08}(\text{syst.}) \pm 0.04(\text{model})$ GeV$^{-2}$.

Similar fits were performed to the LRG data [4, 5]. Fig. 5 shows the values of $\alpha_{IP}(0)$ as a function of $Q^2$ and $\beta$. The present data do not exhibit a significant dependence on $Q^2$ and $\beta$.

## 4.3 QCD Analysis of Diffractive Data

A QCD analysis of diffractive data is based on the QCD factorisation theorem [13, 14, 15, 16]. The theorem allows to express the diffractive structure functions $F_{2/L}^{D(3}$ as a convolution of coefficient functions and DPDFs:

$$F_{2/L}^{D(3)}(\beta, Q^2, x_{IP}) = \sum_i \int_\beta^1 \frac{dz}{z} C_{2/L,i}(\frac{\beta}{z}) f_i^D(z, x_{IP}, Q^2),$$

where the sum runs over partons of type $i$ and $z$ is the momentum fraction of the parton, entering the hard subprocess. The coefficient functions $C_{2/L,i}$ are the same as in inclusive DIS. The DPDFs $f_i^D(z, x_{IP}, Q^2)$ are densities of partons of type $i$ with fractional momentum $zx_{IP}$ in a proton.

The $x_{IP}$ dependence of the DPDFs is parameterised as a sum of two contributions, separately factorisable into a term depending on $x_{IP}$ and a term depending on $z$ and $Q^2$,

$$f_i^D(z, x_{IP}, Q^2) = f_{IP}(x_{IP}) f_i^{IP}(z, Q^2) + f_R(x_{IP}) f_i^R(z, Q^2).$$

This assumption of proton vertex factorisation is a good approximation for the data used in this analysis [4, 5]. The flux factors $f_{IP}$ and $f_R$ describing the $x_{IP}$ dependence of the Pomeron and Reggeon contributions were parameterised using Regge theory. The DPDFs were parameterised at the starting scale $Q_0^2$ in term of quark singlet and gluon distributions as $zf(z, Q_0^2) = Az^B(1-z)^C$. The parameters describing the quarks and gluons at the starting scale were fitted to the data using NLO DGLAP [17, 18, 19] evolution to all values of $Q^2$.

The DPDFs have been extracted in several different analyses [5, 20, 21]. It was shown that the gluon distribution obtained in the fits to the inclusive data only has a large uncertainty at high $z$.

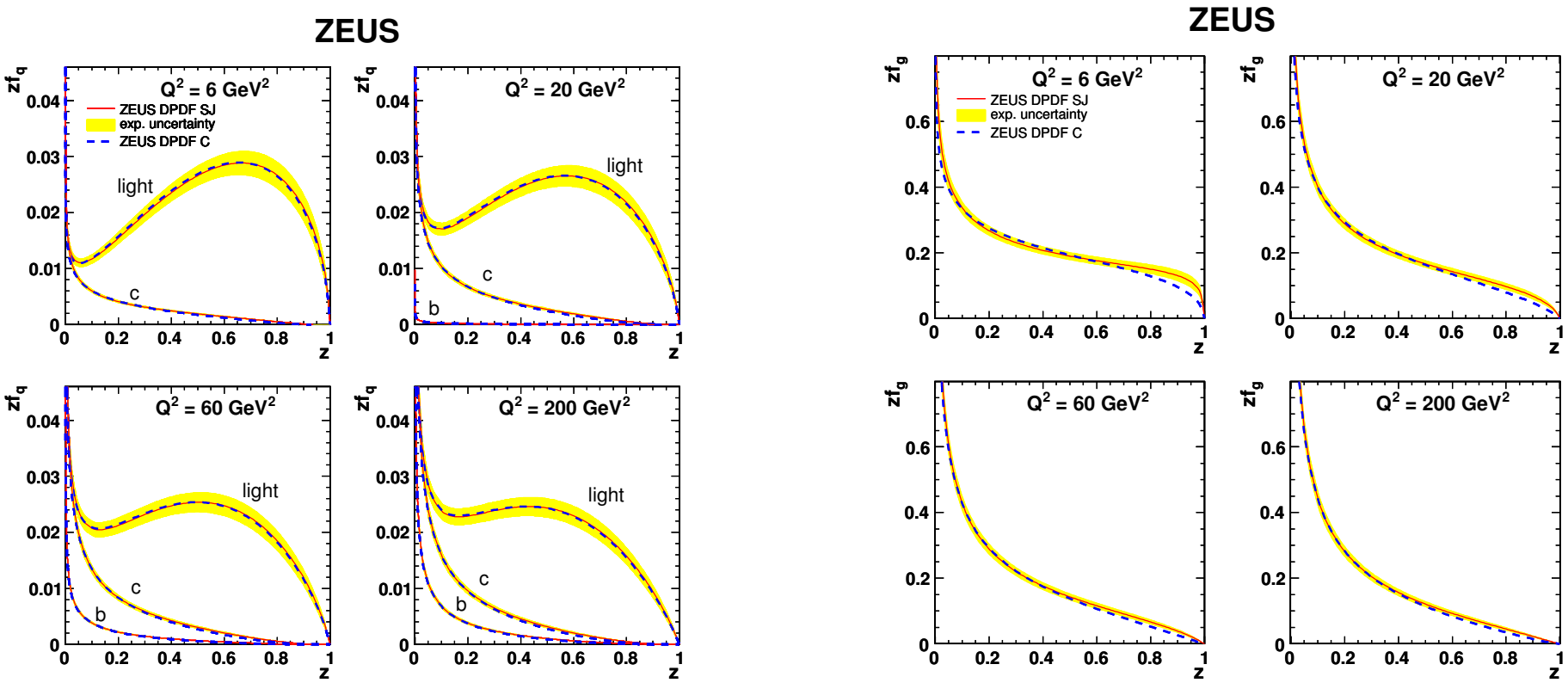

Figure 6: The quark(left plot) and gluon(right plot) distributions obtained from fit to the ZEUS inclusive and dijet data for four different values of $Q^2$. The shaded error bands show the experimental uncertainty.

The inclusion of the diffractive DIS dijet cross sections in the fit provides an additional constraint on the gluon, allowing the determination of the gluon density with better accuracy [21, 22]. The quark and gluon densities from the fit to the ZEUS inclusive and dijet data are shown with their experimental uncertainties for $Q^2 = 6, 20, 60$ and 200 GeV$^2$ in Fig. 6. The fraction of the momentum of the diffractive exchange carried by the gluons is $(60-70)\%$ [21, 22].

## 4.4 Diffractive Dijet Photoproduction

The QCD factorisation theorem does not hold in hadron-hadron collisions. The same can be expected in the dijet photoproduction process where the photon can behave as a hadron-like particle at low $x_\gamma$, the fraction of the photon energy invested in producing the dijet system.

Fig. 7 shows the dijet photoproduction cross section as a function of $x_\gamma$ together with the NLO QCD predictions based on several DPDFs and photon PDFs parameterisations.

The results from the ZEUS experiment are consistent with the NLO QCD predictions. The H1 experiment observes a suppression of NLO QCD predictions by factor 0.5 at all values of $x_\gamma$.

The disagreement can be explained by the different kinematic regions used in both experiments. The ZEUS analysis selected jets with higher transverse energies. The new measurements of the dijet photoproduction cross section performed by the H1 experiment [23] indicates that the suppression factor can depend on the transverse energy of the selected jets.

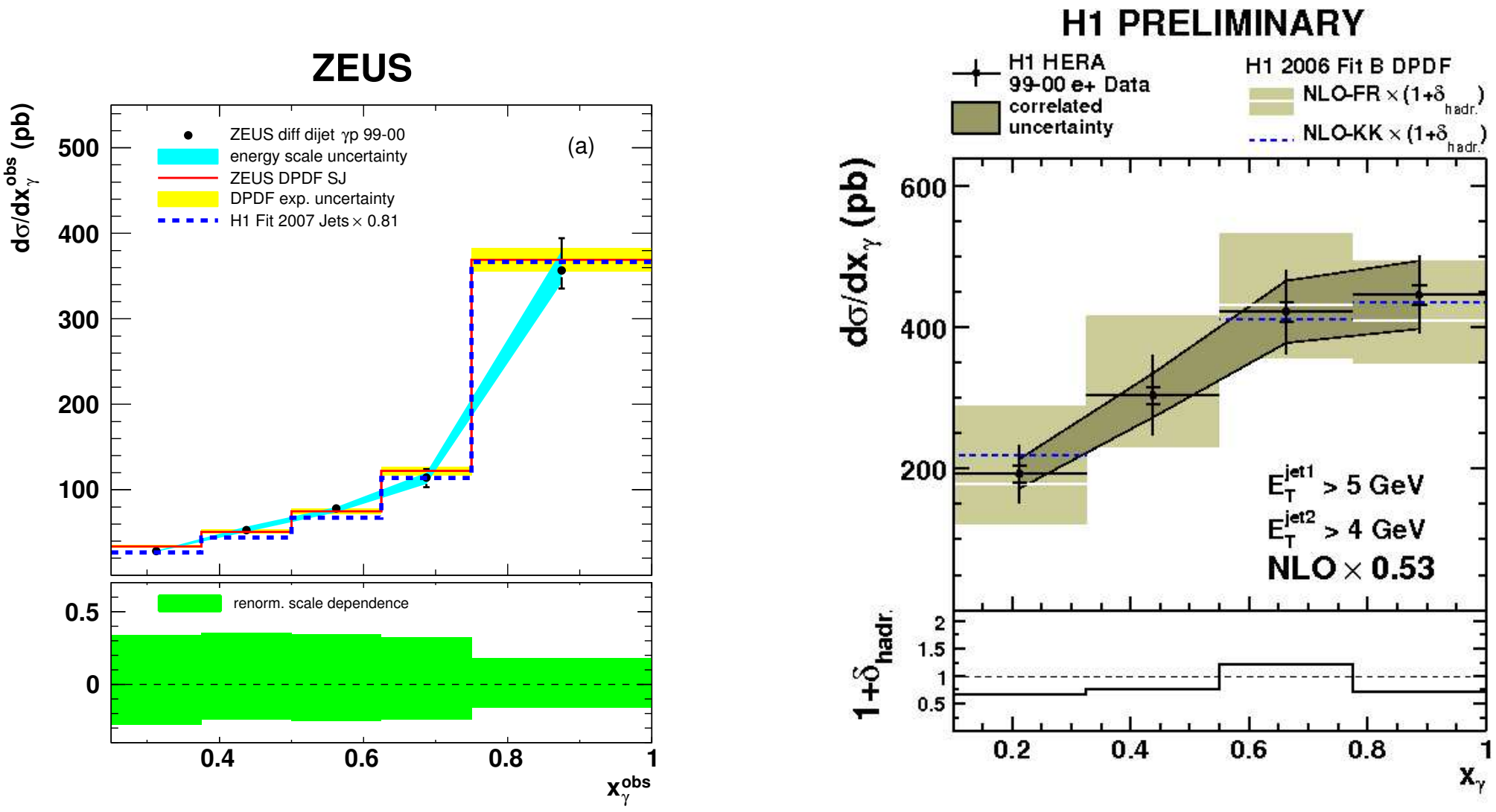

Figure 7: The differential cross section for the diffractive photoproduction of dijets as a function of $x_\gamma$ as measured by ZEUS (left plot) and H1 (right plot). NLO QCD predictions for several DPDFs parameterisations are also shown (scaled by a factor of 0.53 on the right plot). The shaded bands show the uncertainty resulting from the variation of the renormalisation scale.

### 4.5 Longitudinal Diffractive Structure Function

The first measurement of the longitudinal structure function $F_L^D$ [24] performed by the H1 experiment is presented in Fig. 8. The analysis is based on the data samples with different proton beam energies of 460, 575 and 920 GeV. The results are compatible with the predictions from the QCD fits to the previous inclusive data.

## 5 Conclusions

Recent results from HERA provided a lot on new information on diffraction. The cross sections of the diffractive reaction $ep \to eXp$ have been measured in a very wide range of $Q^2$, $W$ and $M_X$. Different methods used to select diffractive processes give consistent results. Diffractive data have been used in QCD analyses and the diffractive parton density functions have been determined. The extracted diffractive parton density functions describes the diffractive charm and dijet production. They are an important input for calculation of the hard diffractive processes at the LHC.

## Acknowledgements

It is a pleasure to thank the organisers for this stimulating and enjoyable workshop.

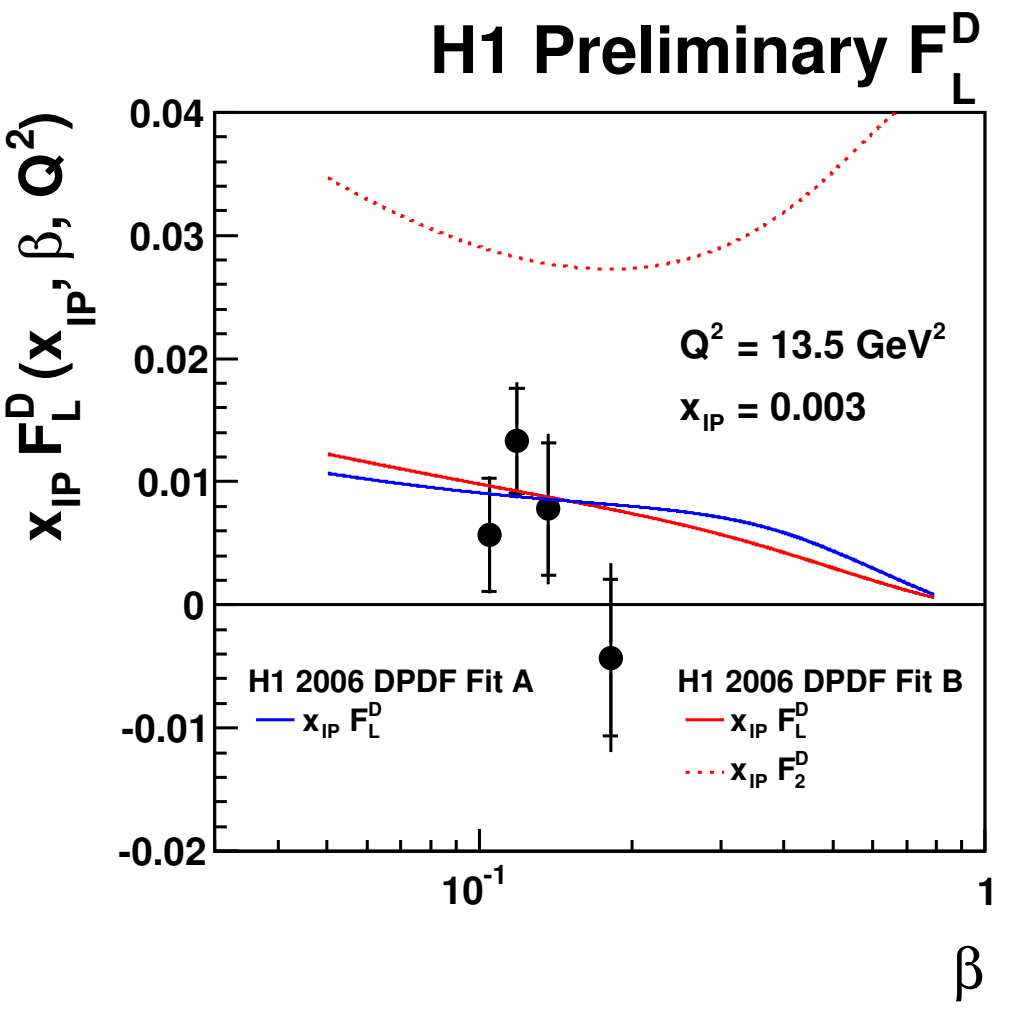

Figure 8: The diffractive longitudinal structure function $F_L^D$ multiplied by $x_{IP}$ as a function of $\beta$. The predictions from NLO QCD fits are also shown.

# References

[1] P.D.B. Collins, An Introduction to Regge Theory and High Energy Physics, Cambridge University Press, Cambridge (1977). Cambridge University Press, Cambridge.

[2] UA8 Coll., A. Brandt et al., Phys. Lett. **B297** 417 (1992).

[3] H1 Coll., A. Aktas et al., Eur. Phys. J. **C48** 749 (2006).

[4] ZEUS Coll., S. Chekanov et al., Nucl. Phys.**B816** 1 (2009).

[5] H1 Coll., A. Aktas et al., Eur. Phys. J. **C48** 715 (2006).

[6] ZEUS Coll., S. Chekanov et al., Nucl. Phys.**B800** 1 (2008).

[7] J.-R. Cudell, K. Kang and S.K. Kim, Phys.Lett. **B395** 311 (1997).

[8] V. Barone and E. Pedazzi, High-Energy Particle Diffraction,Springer Verlag, Heidelberg (2002).

[9] H. Abramowicz, Int. J. Mod. Phys. **A15 S1b** 495 (2000).

[10] M. Glück, E. Reya and A. Vogt, Z. Phys. **C53** 127 (1992)

[11] M. Glück, E. Reya and A. Vogt, Z. Phys. **C53** 651 (1992)

[12] M. Glück, E. Reya and A. Vogt, Z. Phys. **C67** 433 (1995)

[13] J.C. Collins, Phys. Rev. **D57** 3051 (1998).

[14] Erratum, ibid **D61** 019902 (2000).

[15] L. Trentadue and G. Veneziano, Phys. Lett **D323** 201 (1994).

[16] A. Berera and D.E. Soper, Phys. Rev. **D53** 6162 (1996).

[17] V.N. Gribov and L.N. Lipatov, Sov. J. Nucl, Phys. **15** 438 (1972).

[18] Yu.L. Dokshitzer, Sov. Phy. JETP **46** 641 (1977).

[19] G. Altarelli and G. Parisi, Nucl. Phys. **B126** 298 (1977).

[20] A.D. Martin, M.G. Ryskin and G. Watt, Phys. Lett. **B644** 311 (2007).

[21] H1 Coll., A. Aktas et al., JHEP **0710:042** (2007).

[22] ZEUS Coll., to be published.

[23] K. Cerny, Dijets in diffractive photoproduction, in XVI International Workshop on Deep Inelastic Scattering, DIS 2008. April 2008, London

[24] H1 Coll., to be published.

# Exclusive Hard Diffraction at HERA (DVCS and Vector Mesons)

*Pierre Marage*

Université Libre de Bruxelles, Boulevard du Triomphe, B-1050 Bruxelles, Belgium
*On behalf of the H1 and ZEUS Collaborations*

Recent results obtained at HERA on deeply virtual Compton scattering and exclusive vector meson production are reviewed, with the emphasis on the transition from soft to hard diffraction and on spin dynamics.

## 1 Introduction

Since the beginning of HERA data taking, a large number of studies have been performed of deeply virtual Compton scattering (DVCS) and of vector meson (VM) production. The exclusive final states include real photons [1, 2, 3], light ($\rho$ [4, 5, 6, 7, 8, 9], $\omega$ [10, 11] and $\phi$ [12, 13, 7]) and heavy VMs ($J/\psi$ [14, 15, 16, 17, 18], $\psi(2s)$ [19] and $\Upsilon$ [20, 21]). Cross sections are expressed in terms of $\gamma^* p$ scattering.

In the presence of a hard scale, these processes provide unique information on the mechanisms of diffraction, in particular on the transition from soft to hard diffraction and on spin dynamics. A hard scale is provided by the VM mass $M_V$, by the negative square of the photon four-momentum, $Q^2$ (with $Q^2 \simeq 0$ for photoproduction and $1.5 \leq Q^2 \leq 90$ GeV$^2$ for electroproduction), or by the square of the four-momentum transfer at the proton vertex, $t$.

DVCS and VM production with small $|t|$ values ($|t| \leq 0.5$ GeV$^2$ for elastic scattering and $|t| \leq 2.5$ GeV$^2$ for proton dissociation) are interpreted in terms of two complementary QCD approaches. Following a collinear factorisation theorem, the DVCS process, the electroproduction of light VMs by longitudinally polarised photons, and the production of heavy VMs can be described by the convolution of the hard process with generalised parton distributions in the proton (GPDs). High energy DVCS and VM production can also be described through the factorisation of virtual photon fluctuation into a $q\bar{q}$ colour dipole, diffractive dipole–proton scattering, and $q\bar{q}$ recombination into the final state photon or VM. The interaction scale $\mu$ is given by the characteristic transverse size of the dipole, with $\mu^2 \simeq (Q^2 + M_V^2)/4$ for light VM electroproduction by longitudinal photons and for heavy VM production, whereas this value may be significantly reduced for light VM electroproduction by transversely polarised photons, because of end-point contributions in the photon wave function. For DVCS, LO contributions are present, which suggests that the relevant scale may be $Q^2$ rather than $Q^2/4$ as for VMs.

Several models, based on either approach, have been proposed. They differ in particular in the way the VM wave function is taken into account, in the parameterisation of the parton distributions and of the dipole–proton scattering, and in the extension to non-zero $|t|$ values of the scattering amplitude.

High $|t|$ photoproduction, with $2 \leq |t| \leq 30$ GeV$^2$, of real photons [3], $\rho$ and $\phi$ mesons [8, 9],

and $J/\psi$ mesons [17, 18] offer specific testing grounds for the BFKL evolution.

Comparisons of models with the data are discussed in particular in [1, 2, 3, 6, 7, 9, 17, 18, 22].

# 2 Kinematic Dependences

**DVCS** The kinematic dependences of DVCS production, presented in Fig. 1, are well described by models using either GPDs or a dipole approach [1, 2].

The interference of the DVCS and Bethe-Heitler processes gives access, through the measurement of beam charge asymmetry, to the ratio $\rho$ of the real to imaginary parts of the DVCS amplitude. The measurement $\rho = 0.20 \pm 0.05 \pm 0.08$ [2] is in agreement with the value $\rho = 0.25 \pm 0.03 \pm 0.05$ obtained from a dispersion relation using the $W$ dependence of the cross section.

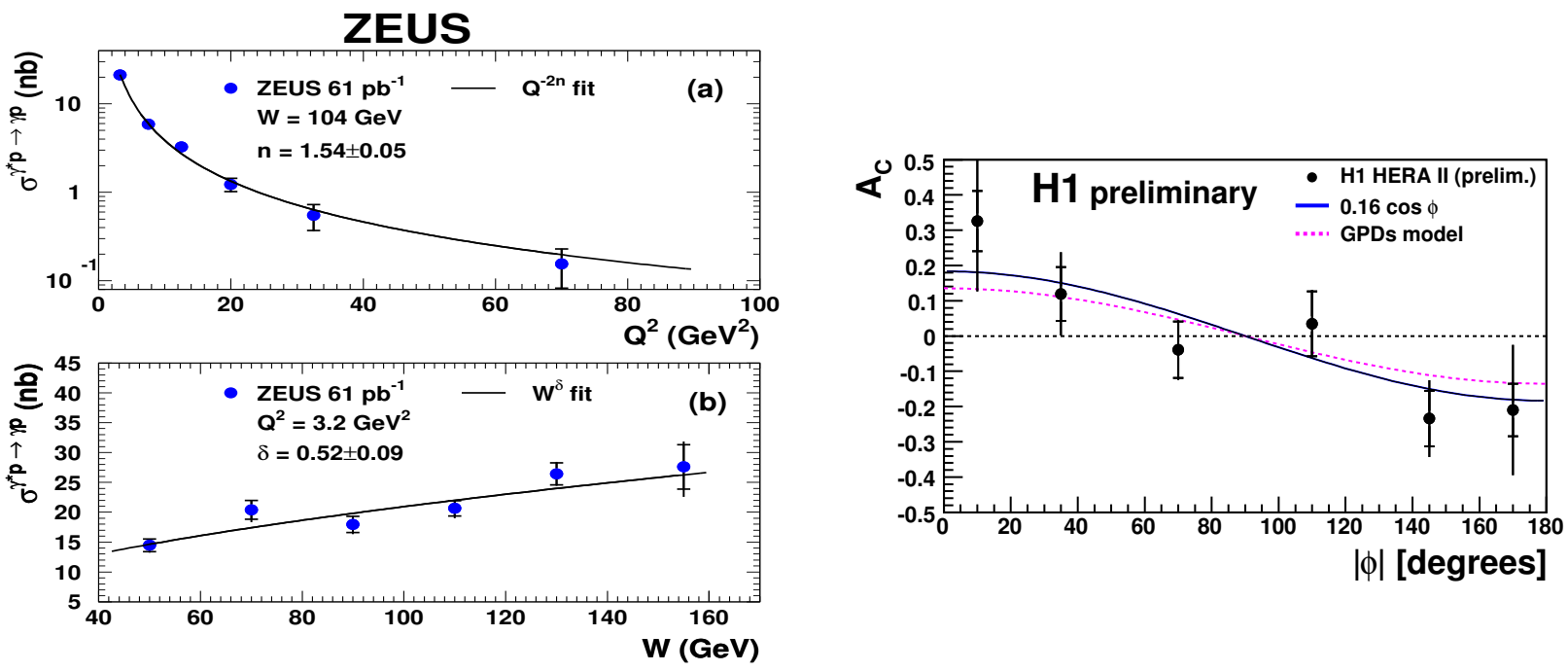

Figure 1: (left) $Q^2$ and $W$ dependences of DVCS production, with simple fit parameterisations [1]; (right) beam charge asymmetry, $\cos\phi$ fit and predictions of a GPD model [2].

**$Q^2$ dependence of light VM production** The cross sections for elastic and proton dissociative $\rho$ and $\phi$ electroproduction have been measured with high precision [6, 7, 13]. The $Q^2$ dependence, shown for $\rho$ mesons in Fig. 2, is reasonably described by several models, using either the GPD or the dipole approach.

Although the production cross sections for light and heavy VMs differ by several orders of magnitude at $Q^2 \simeq 0$, it is striking that the ratios are nearly constant when they are studied as a function of the scaling variable $(Q^2 + M_V^2)/4$, with values close to unity when scaled according to the quark charge content of the VMs, $\rho : \omega : \phi : J/\psi = 9 : 1 : 2 : 8$. This confirms the relevance of the dipole size to the cross sections, even though the agreement with SU(4) universality is not perfect, indicating that wave function effects may need to be taken into account.

The ratio of the production cross sections with proton dissociative and elastic scattering at $|t| = 0$ is found to be independent of $Q^2$. Consistent values around 0.160 are measured for $\rho$ and $\phi$ production with dissociative mass $M_Y < 5$ GeV [7]. This observation supports the independence of the hard and soft vertex contributions to the scattering amplitudes, known as proton vertex or "Regge" factorisation.

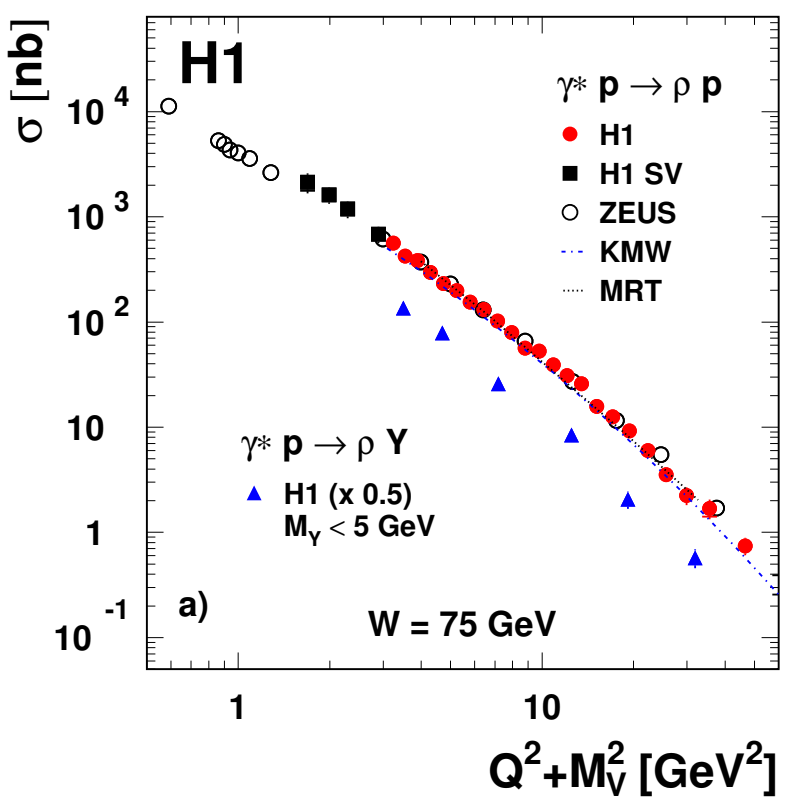

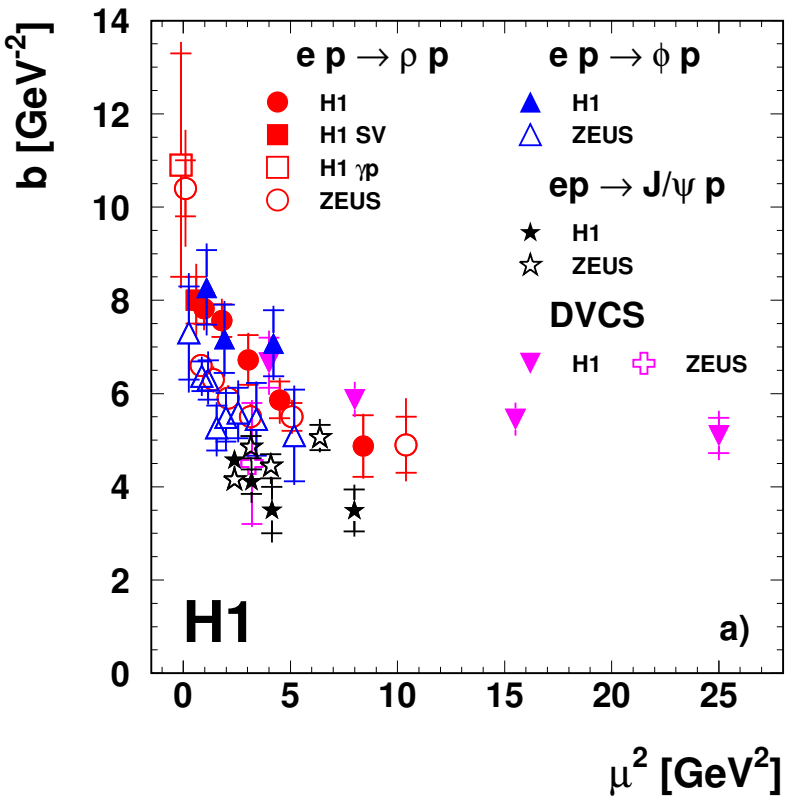

Figure 2: $Q^2$ dependence of elastic and proton dissociative electroproduction cross sections of $\rho$ mesons, and model predictions [7].

Figure 3: Elastic $b$ slopes, as a function of $\mu^2 = (Q^2+M_V^2)/4$ for VM production and $\mu^2 = Q^2$ for DVCS [7].

**$t$ slopes** Exponentially falling $|t|$ distributions, with $\mathrm{d}\sigma/\mathrm{dt} \propto \mathrm{e}^{-\mathrm{b}|\mathrm{t}|}$, are measured for DVCS, light and heavy VM production, in both the elastic and the proton dissociative channels. In an optical model approach, the slope $b$ is given by the convolution of the transverse sizes of the $q\bar{q}$ dipole, of the diffractively scattered system (which vanishes for proton dissociation) and of the exchange (a contribution which is expected to be small). As shown in Fig. 3, the elastic slopes for light VMs strongly decrease with increasing $Q^2$. They reach values of the order of 5 GeV$^{-2}$, similar to those measured in $J/\psi$ production, for values of the scaling variable $(Q^2+M_V^2)/4 \gtrsim 5$ GeV$^2$. This evolution reflects the shrinkage with $Q^2$ of the light quark colour dipole. A similar evolution is observed for DVCS as a function of the variable $Q^2$. The proton dissociative slopes similarly decrease with the increasing scale, down to values around 1.5 GeV$^{-2}$ for $\rho$ production and DVCS, and values slightly below 1 GeV$^{-2}$ for $J/\psi$ production.

The difference between the elastic and proton dissociative slopes, $b_{el} - b_{p.diss.}$, provides another test of proton vertex factorisation. A value of $3.5 \pm 0.1$ GeV$^{-2}$ is measured for $J/\psi$ [14, 16], with a similar value for DVCS [2]. The difference is higher, around 5.5 GeV$^{-2}$, for $\rho$ and $\phi$ mesons, with however an indication of a decrease toward the $J/\psi$ value with increasing $(Q^2+M_V^2)/4$ [7].

**Energy dependence and effective Regge trajectory** The energy dependence of DVCS and VM production is well described by a power law, $\mathrm{d}\sigma/\mathrm{dW} \propto \mathrm{W}^{\delta}$.

Figure 4 (left) shows that the energy dependence is significantly stronger for heavy quark photoproduction, with $\delta \sim 0.8 - 1.2$, than for (soft) hadron–hadron interactions and light VM photoproduction, with $\delta \sim 0.2$. This is explained by the fact that the photoproduction of VMs formed of heavy quarks is a hard process, characterised by small transverse dipoles which probe the low-$x$ gluon density in the proton at a scale where it is quickly increasing with $1/x$.

For light VM production, the $W$ dependence is hardening with $Q^2$, with values of $\delta$ similar to the $J/\psi$ values for $(Q^2+M_V^2)/4 \gtrsim 5$ GeV$^2$. This feature is explained by the shrinkage of the colour dipoles at large $Q^2$ values.

In a Regge inspired parameterisation, the energy dependence of the cross section and its correlation with $t$ are given by $\delta(t) = 4\ (\alpha_{I\!P}(t) - 1)$, with $\alpha_{I\!P}(t) = \alpha_{I\!P}(0) + \alpha' \cdot\ t$, where $\alpha'$ describes the shrinking of the diffractive peak with energy. The hard behaviour of $J/\psi$ production and the hardening with $(Q^2+M_V^2)/4$ of the energy dependence of light VM production for $t = 0$ are shown in Fig. 4 (centre), where values of 1.08 or 1.11 for $\alpha_{I\!P}(0)$ are typical of soft hadron-hadron interactions. As shown in Fig. 4 (right), the slope of the effective trajectory for VM production, including $\rho$ photoproduction [5], is smaller than the value 0.25 GeV$^{-2}$, typical for hadronic interactions. For DVCS $\alpha' = 0.03 \pm 0.09 \pm 0.11$ GeV$^{-2}$ [2]; for $J/\psi$ photoproduction at high $|t|$, combining H1 [17] and ZEUS [18] measurements, $\alpha' = -0.02 \pm 0.01 \pm 0.01$ GeV$^{-2}$.

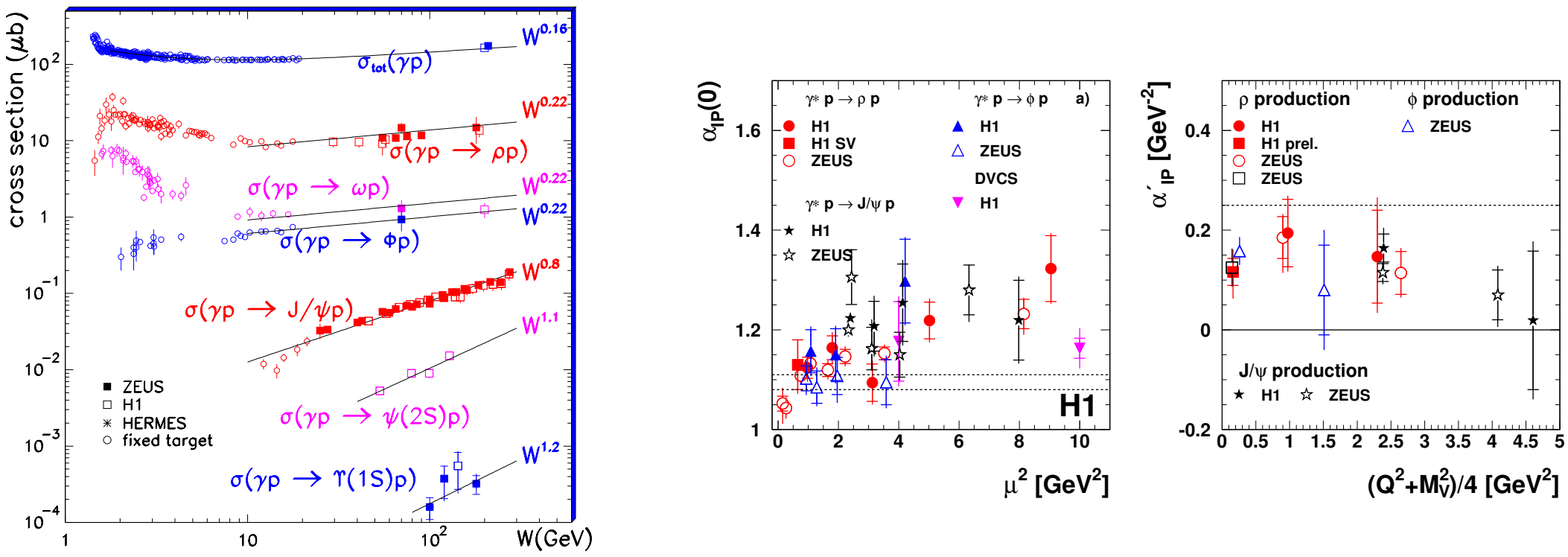

Figure 4: (left) $W$ dependence of VM photoproduction [22]; measurement of the intercept $\alpha_{I\!P}(0)$ (centre) and of the slope $\alpha'$ (right) of the effective Regge trajectory, as a function of the scale $\mu^2 = (Q^2+M_V^2)/4$ for VM production and $\mu^2 = Q^2$ for DVCS [7].

**Remarks on the interaction scales** The energy dependence of the total $ep$ cross section at fixed values of $Q^2$ can be parameterised as $F_2 \propto x^{-\lambda}$, with values of $\lambda$ increasing with $Q^2$, a feature which is attributed to the increase with $Q^2$ of the parton density at small $x$. The prediction that for VM production $\delta = 2\lambda$, when taken at the same scale, can thus provide information on the relevant effective scale for the reaction. The present results clearly indicate that the variable $(Q^2+M_V^2)/4$ is a better candidate than $Q^2$ for such a unified scale, but high precision measurements of the energy dependence of $\rho$ and $J/\psi$ electroproduction remain necessary to settle the scale issue [22].

For the DVCS process, where both LO and NLO (dipole-type) diagrams contribute, the present high energy data seem to favour an effective scale $\approx Q^2$ rather than $\approx Q^2/4$ in order to ensure diffraction universality, but here also more precise data are required.

## 3 Spin Dynamics

The VM production and decay angular distributions allow the measurement of fifteen spin density matrix elements, which are bilinear combinations of helicity amplitudes. Under natural parity exchange, five $T_{\lambda_V \lambda_\gamma}$ amplitudes are independent: two $s$-channel helicity conserving (SCHC) amplitudes ($T_{00}$ and $T_{11}$), two single helicity flip amplitudes ($T_{01}$ and $T_{10}$) and one double flip amplitude ($T_{-11}$).

The $Q^2$ dependence of the matrix elements for $\rho$ and $\phi$ production indicates that the five elements which contain products of the SCHC amplitudes are non-zero, whereas those formed with the helicity violating amplitudes are generally consistent with 0. A notable exception is the element $r_{00}^5$, which involves the product of the dominant $T_{00}$ SCHC amplitude with $T_{01}$, which describes the transition from a transversely polarised photon to a longitudinal $\rho$ meson [6, 7].

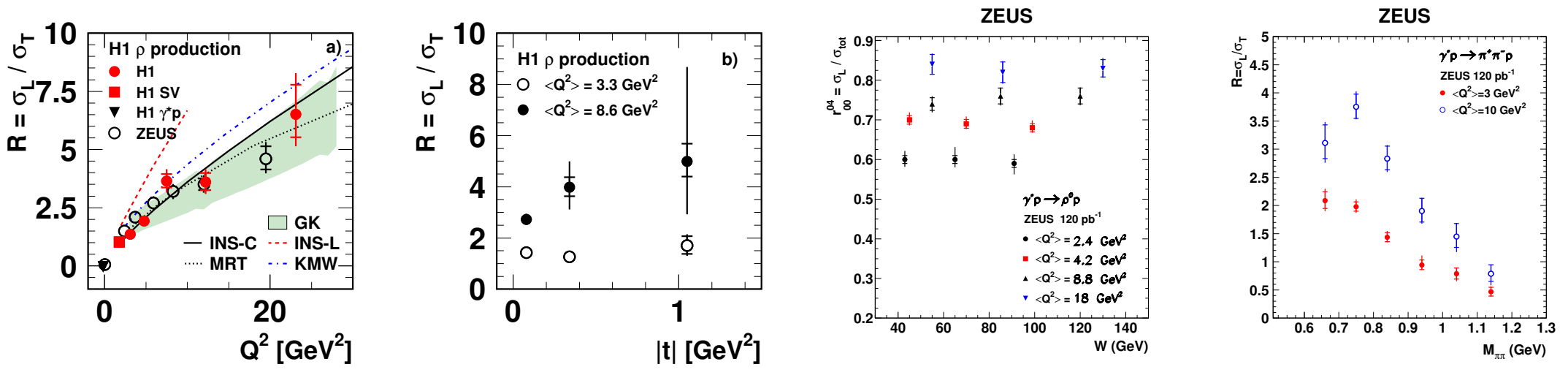

Figure 5: Measurement of $R = \sigma_L/\sigma_T$, as a function of $Q^2$ and $|t|$ [7], $W$, and the invariant mass, $M_{\pi\pi}$, of the two decay pions [6], for $\rho$ electroproduction.

The ratio $R = \sigma_L/\sigma_T$ of the longitudinal to transverse cross sections for $\rho$ production is shown in Fig. 5 as a function of $Q^2$, $W$, $t$, and the invariant mass of the two decay pions, for several domains in $Q^2$.

A strong increase of $R$ with $Q^2$ is observed, which is tamed at large $Q^2$. These features are relatively well described by GPD and dipole models. The $Q^2$ dependence of $R$ for $\rho$, $\phi$ and $J/\psi$ production follows a universal trend when plotted as a function of $Q^2/M_V^2$ [7]. With the present data, no $W$ dependence is observed, but it should be stressed that the lever arms in $W$ for fixed $Q^2$ values are relatively limited.

No $|t|$ dependence is observed for $R$ by ZEUS with $|t| \leq 1$ GeV$^2$ [6], whereas an increase of $R$ with $|t|$ is observed by H1 for $Q^2 > 5$ GeV$^2$, $|t| \leq 3$ GeV$^2$ [7]. This increase can be translated into a measurement of the difference between the longitudinal and transverse $t$ slopes, through the relation $R(t) = \sigma_L(t)/\sigma_T(t) \propto e^{-(b_L - b_T)|t|}$. A slight indication ($1.5\sigma$) is thus found for a negative value of $b_L - b_T$ ($-0.65 \pm 0.14^{+0.41}_{-0.51}$), suggesting that the average transverse size of dipoles for transverse amplitudes is larger than for longitudinal amplitudes

The strong dependence of $R$ with the dipion mass, observed by both experiments [6, 7], cannot be attributed solely to the interference of resonant $\rho$ and non-resonant $\pi\pi$ production, and indicates that the spin dynamics of $\rho$ production depends of the effective $q\bar{q}$ mass.

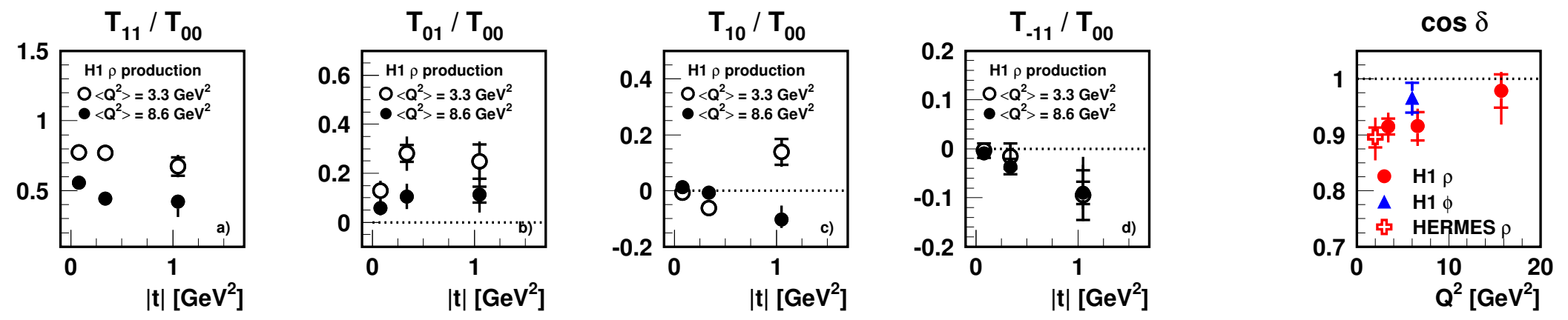

Figure 6: (a-d) Helicity amplitude ratios, as a function of $t$; (right plot) phase difference between the two SCHC amplitudes, $T_{00}$ and $T_{11}$ [7].

Helicity amplitude ratios are measured, under the approximation that they are in phase, through fits to the 15 matrix elements. The four ratios to the dominant $T_{00}$ amplitude are presented in Fig. 6 as a function of $t$, for two domains in $Q^2$. At large $Q^2$, a $t$ dependence compatible with the expected $\sqrt{|t|}$ law is observed for both single helicity flip amplitudes. A significant double-flip amplitude $T_{-11}$ is observed, which may be related to gluon polarisation in the proton. The $t$ dependence of $T_{11}/T_{00}$ at large $Q^2$, a $3\sigma$ effect, is related to the $t$ dependence of $R$ and supports the indication of a difference between the transverse sizes of dipoles in transversely and longitudinally polarised photons.

A small non-zero phase difference between the two SCHC amplitudes, which decreases with increasing $Q^2$, is visible in Fig. 6. Through dispersion relations, this non-zero value is suggestive of different $W$ dependences of the longitudinal and transverse amplitudes.

# 4 Large $|t|$ VM Production

In exclusive real photon and VM production at high energy and large $|t|$, a hard scale is present at both ends of the gluon ladder which extends over a large rapidity range, between the struck parton in the proton (mostly gluons at small $x$) and the quark or antiquark from the photon fluctuation. These processes thus offer a unique testing ground for the BFKL evolution, since no strong $k_T$ ordering along the ladder is expected. This is at variance with high $Q^2$ VM production at low $|t|$, where a large scale is present at the photon end of the ladder and a small scale at the proton end, a configuration which is described by the DGLAP evolution. For real photon and $J/\psi$ production, there is little uncertainty related to the wave functions.

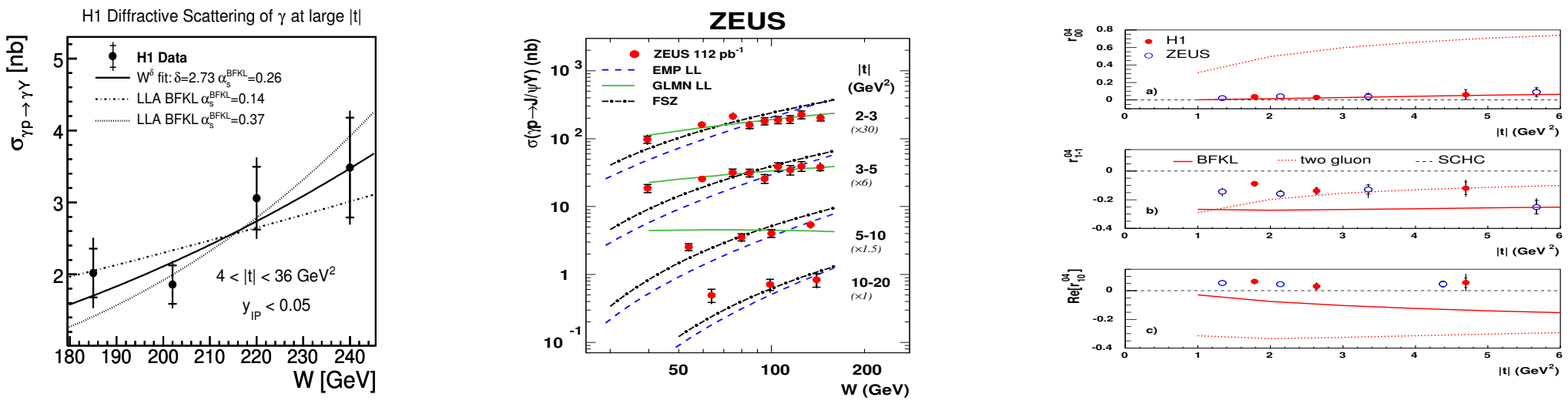

Figure 7: Large $|t|$ photoproduction measurements: $W$ dependence of real photon [3] (left) and $J/\psi$ [18] production (centre); spin density matrix elements for $\rho$ production [9].

A specific QCD prediction for large $|t|$ production is the power-law dependence of the $|t|$ distribution, at variance with the exponential dependence for $|t| \lesssim$ a few GeV$^2$. The $t$ dependences for $|t| \geq 4$ GeV$^2$ of $\gamma$ and $J/\psi$ production are indeed well described by power laws with exponents $n = 2.60 \pm 0.19^{+0.03}_{-0.08}$ [3], and $n = 3.0 \pm 0.1$ [18], respectively.

Figures 7 (left) and (centre) present the $W$ evolutions of high $|t|$ real photon and $J/\psi$ production, respectively. A strong $W$ dependence is observed, compatible with calculations based on the BFKL approach, whereas the DGLAP evolution (valid for $|t| \leq m_{\psi}^2$), predicts a significantly weaker dependence.

The spin density matrix elements for $J/\psi$ production are in agreement with SCHC [17, 18], whereas substantial helicity flip contributions are observed in Fig. 7 (right) for $\rho$ production with $1.5 \lesssim |t| \lesssim 10$ GeV$^2$ [8, 9], which can be understood in a BFKL approach with a chiral-odd component of the photon wave function.

## Acknowledgements

It is a pleasure to thank numerous colleagues from the ZEUS and H1 collaborations as well as theorists for enlightening discussions, and the workshop organisers for the lively discussions and the pleasant atmosphere of the meeting.

## References

[1] S. Chekanov *et al.* [ZEUS Collab.], Phys. Lett. **B573** 46 (2003) [hep-ex/0305028]; *idem*, JHEP 0905 108 (2009) [arXiv:0812.2517].

[2] F.D. Aaron *et al.*, [H1 Collab.], Phys. Lett. **B659** 796 (2008) [arXiv:0709.4114]; *idem*, *Deeply Virtual Compton Scattering and related Beam Charge Asymmetry in ep Collisions at HERA*, DESY-09-109 (2009) [arXiv:0907.5289].

[3] F.D. Aaron *et al.* [H1 Collab.], Phys. Lett. **B672** 219 (2009) [arXiv:0810.3096].

[4] J. Breitweg *et al.* [ZEUS Collab.], Eur. Phys. J. **C14** 213 (2000) [hep-ex/9910038]; *idem*, Eur. Phys. J. **C2** 247 (1998) [hep-ex/9712020]; M. Derrick *et al.* [ZEUS Collab.], Z. Phys. **C73** 253 (1997).

[5] S. Aid *et al.* [H1 Collab.], Nucl. Phys. **B463** 3 (1996) [hep-ex/9601004]; B. List [H1 Collab.], *Extraction of the Pomeron Trajectory from a Global Fit to Exclusive $\rho^0$ Meson Photoproduction Data*, Proc. of the XVII Int. Workshop on DIS and Related Subjects (DIS 2009), Madrid (2009) [arXiv:0906.4945].

[6] S. Chekanov *et al.* [ZEUS Collab.], PMC Phys. **A1** 6 (2007) [hep-ex/0708.1478].

[7] F.D. Aaron *et al.* [H1 Collab.], *Diffractive electroproduction of $\rho$ and $\phi$ mesons*, DESY-09-093 (2009).

[8] S. Chekanov *et al.* [ZEUS Collab.], Eur. Phys. J. **C26** 389 (2003) [hep-ex/0205081].

[9] A. Aktas *et al.* [H1 Collab.], Phys. Lett. **B638** 422 (2006) [hep-ex/0603038].

[10] M. Derrick *et al.* [ZEUS Collab.], Z. Phys. **C73** 73 (1996) [hep-ex/9608010].

[11] J. Breitweg *et al.* [ZEUS Collab.], Phys. Lett. **B487** 273 (2000) [hep-ex/0006013].

[12] M. Derrick *et al.* [ZEUS Collab.], Phys. Lett. **B377** 259 (1996) [hep-ex/9601009].

[13] S. Chekanov *et al.* [ZEUS Collab.], Nucl. Phys. **B718** 3 (2005) [hep-ex/0504010].

[14] S. Chekanov *et al.* [ZEUS Collab.], Eur. Phys. J. **C24** 345 (2002) [hep-ex/0201043].

[15] S. Chekanov *et al.* [ZEUS Collab.], Nucl. Phys. **B695** 3 (2004) [hep-ex/0404008].

[16] A. Aktas *et al.* [H1 Collab.], Eur. Phys. J. **C46** 585 (2006) [hep-ex/0510016].

[17] A. Aktas *et al.* [H1 Collab.], Phys. Lett. **B568** 205 (2003) [hep-ex/0306013].

[18] [ZEUS Collab.], *Diffractive photoproduction of $J/\psi$ mesons with large momentum transfer at HERA*, DESY-09-137 (2009).

[19] C. Adloff *et al.* [H1 Collab.], Phys. Lett. **B541** 251 (2002) [hep-ex/0205107].

[20] C. Adloff *et al.* [H1 Collab.], Phys. Lett. **B483** 23 (2000) [hep-ex/0003020].

[21] S. Chekanov *et al.* [ZEUS Collab.], Phys. Lett. **B680** 4 (2009); doi:10.1016/j.physletb.2009.07.066.

[22] A. Levy [ZEUS Collab.], *Electroproduction of vector mesons*, Proc. of the XVII Int. Workshop on DIS and Related Subjects (DIS 2009), Madrid (2009) [arXiv:0907.2178].

# Latest DVCS Results from HERMES

*Sergey Yaschenko for the HERMES collaboration*

DESY, Platanenallee 6, D-15738 Zeuthen, Germany

Hard exclusive leptoproduction of real photons on nucleons and nuclei, Deeply Virtual Compton Scattering (DVCS), is one of the theoretically cleanest ways to access Generalized Parton Distributions (GPDs). During 1996–2006 the HERMES experiment at DESY, Hamburg, collected data on DVCS utilizing the HERA polarized electron or positron beams with energies of 27.6 GeV, and longitudinally and transversely polarized or unpolarized gas targets (H, D or heavier nuclei). The azimuthal asymmetries measured in the DVCS process allow access to the imaginary and/or real part of certain combinations of GPDs. An overview of the latest HERMES results on DVCS is presented.

## 1 Introduction

The theoretical framework of Generalized Parton Distributions (GPDs) includes Parton Distribution Functions and Form-Factors as limiting cases and moments of GPDs, respectively, and can provide a three-dimensional representation of the structure of hadrons at the partonic level. GPDs depend on three kinematic variables: the squared four-momentum transfer $t$ to the nucleon and $x$ and $\xi$, which represent respectively the average and half the difference of the longitudinal momentum fractions carried by the probed parton in initial and final states. For the proton, there are four twist-2 GPDs per quark flavor: $H_q$, $E_q$, $\widetilde{H}_q$ and $\widetilde{E}_q$. Among other hard exclusive processes, Deeply Virtual Compton Scattering is one of the theoretically cleanest ways to access GPDs. DVCS is experimentally indistinguishable from the electromagnetic Bethe-Heitler (BH) process because they share the same final state. The real photon is radiated (Fig. 1) from the struck quark in DVCS, or from the initial or scattered lepton in BH.

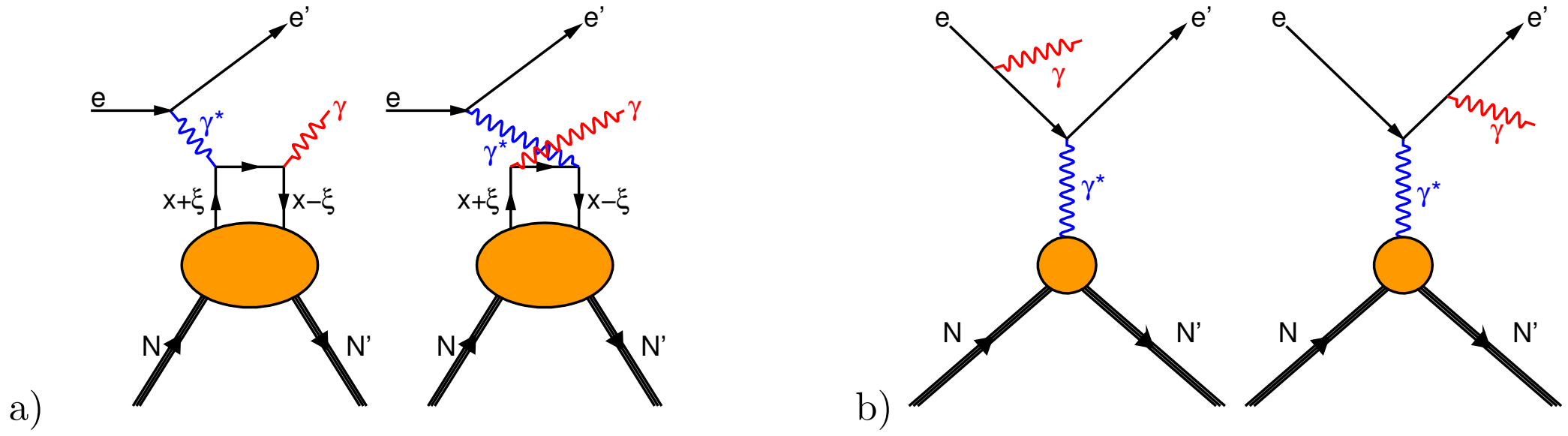

Figure 1: Leading order diagrams for the deeply virtual Compton scattering (a) and the Bethe–Heitler (b) processes.

The cross section of the exclusive photoproduction process can be written as [1]

$$\frac{d\sigma}{dx_B dQ^2 dt d\phi} = \frac{\alpha^3 x_B y}{16\pi^2 Q^2 e^3} \frac{2\pi y}{Q^2} \frac{|T_{DVCS}|^2 + |T_{BH}|^2 + I}{\sqrt{1 + 4x_B^2 M_N^2/Q^2}}, \tag{1}$$

where $T_{DVCS}$ ($T_{BH}$) is the DVCS (BH) amplitude, $I$ is the interference term, $x_B$ is the Bjorken scaling variable and $-Q^2$ is the squared four-momentum transferred by the virtual photon. The amplitude of the BH process can be precisely calculated from measured elastic form factors of the nucleon. The BH process dominates at HERMES kinematics. However, the kinematic dependences of the cross section terms generate a set of azimuthal asymmetries which depend on the azimuthal angle $\phi$ between the real-photon production plane and the lepton scattering plane.

# 2 Azimuthal Asymmetries in DVCS

The cross section for a longitudinally polarized lepton beam scattered off an unpolarized proton target $\sigma_{LU}$ can be related to the unpolarized cross section $\sigma_{UU}$ by

$$\sigma_{LU}(\phi; P_B, C_B) = \sigma_{UU}(\phi) \cdot \left[1 + P_B A_{LU}^{DVCS}(\phi) + C_B P_B A_{LU}^{I}(\phi) + C_B A_C(\phi)\right], \tag{2}$$

where $A_{LU}^I$ ($A_{LU}^{DVCS}$) is the charge (in)dependent beam-helicity asymmetry (BSA) and $A_C$ is the beam charge asymmetry (BCA), $C_B(P_B)$ denotes the beam charge (polarization). In the analysis effective asymmetry amplitudes are extracted, which include $\phi$ dependencies from the BH propagators and the unpolarized cross section. Each asymmetry can be expanded in a Fourier series in $\phi$ as

$$A_{LU}^I(\phi) = \sum_{n=1}^{2} A_{\mathrm{LU,I}}^{\sin(n\phi)} \sin(n\phi) + \sum_{n=0}^{1} A_{\mathrm{LU,I}}^{\cos(n\phi)} \cos(n\phi), \tag{3}$$

$$A_{LU}^{DVCS}(\phi) = \sum_{n=1}^{2} A_{\mathrm{LU,DVCS}}^{\sin(n\phi)} \sin(n\phi) + \sum_{n=0}^{1} A_{\mathrm{LU,DVCS}}^{\cos(n\phi)} \cos(n\phi), \tag{4}$$

$$A_C(\phi) = \sum_{n=0}^{3} A_{\mathrm{C}}^{\cos(n\phi)} \cos(n\phi) + A_{\mathrm{C}}^{\sin\phi} \sin\phi. \tag{5}$$

By combining the data taken with different beam charges and helicities, the amplitudes were fitted simultaneously using a Maximum Likelihood method described in detail in Ref. [2]. In the case of unpolarized beam and transversely polarized target, the transverse target-spin azimuthal asymmetry (TTSA) can be measured, which in addition to $\phi$ also depends on the angle $\phi_S$ between the lepton scattering plane and the direction of the target polarization vector.

# 3 Experiment and Data Analysis

The HERMES experiment [3] utilized longitudinally polarized 27.6 GeV electron/positron beams of the HERA storage ring at DESY together with longitudinally and transversely polarized or unpolarized gas targets (H, D or heavier nuclei). Exclusive events were selected requiring the detection of exactly one scattered lepton and of exactly one photon. In addition, as the recoiling proton has not been detected, the missing mass was required to match the proton mass within the resolution of the spectrometer, which defines the "exclusive region". Without recoil proton detection it is not possible to separate the elastic DVCS/BH events from the "associated" process, where the nucleon in the final state is excited to a resonant state. Within the exclusive region, its contribution is estimated from a Monte Carlo simulation to be about 12%, which is taken as part of the signal. The main background contribution of about 3% is originating from semi-inclusive $\pi^0$ production and is corrected for. The contribution from exclusive $\pi^0$ production is estimated to be less than 0.5%. The systematic uncertainties are obtained from a Monte Carlo simulation estimating the effects of limited acceptance, smearing, finite bin width and alignment of the detectors with respect to the beam. Other sources are background contributions and a shift of the position of the exclusive missing mass peak between the data taken with different beam charges.

# 4 Results

In Figures 2 and 3 results obtained with the hydrogen target are shown [4]. The first four rows of Fig. 2 represent different cosine amplitudes of the BCA, whereas the last row displays the fractional contributions of the associated BH process. In the first column the integrated result is shown, in the other columns the amplitudes are binned in $-t$, $x_B$ or $Q^2$. The error bars show the statistical uncertainties and the bands the systematic uncertainties. The magnitudes of the first two cosine moments $A_{\mathrm{C}}^{\cos 0\phi}$ and $A_{\mathrm{C}}^{\cos\phi}$ increase with increasing $-t$, with opposite signs, in agreement with theoretical expectations. In HERMES kinematics, both relate to the real part of the GPD $H$, but the constant term is suppressed relative to the first moment. The second cosine moment appears in twist-3 approximation and is found to be compatible with zero like the third cosine moment, which is related to gluonic GPDs. The first sine moment $A_{\mathrm{LU,I}}^{\sin\phi}$ is large and negative in the covered kinematics (see Fig. 3). This amplitude relates to the imaginary part of the GPD $H$. Also shown in the figures are GPD model calculations based on the framework of double distributions [6]. The model includes a Regge-inspired $t$-ansatz and a factorized $t$-ansatz. The BCA amplitudes favour the double-distribution model with a Regge-inspired $t$-dependence, if the D-term is neglected. Both model calculations fail to describe the data except for small $-t$. The charge-independent BSA moments are found to be compatible with zero. Results obtained with a deuterium target (see Fig. 4) are found to be consistent in most kinematic regions [5]. The proton and deuteron results for the amplitude $\mathrm{A}_{\mathrm{LU,I}}^{\sin(2\phi)}$ integrated over the acceptance differ by 2.5 sigma for the total experimental uncertainties. This possible discrepancy is most evident at large $-t$ or large $x_B$ (or $Q^2$) and has no obvious explanation.

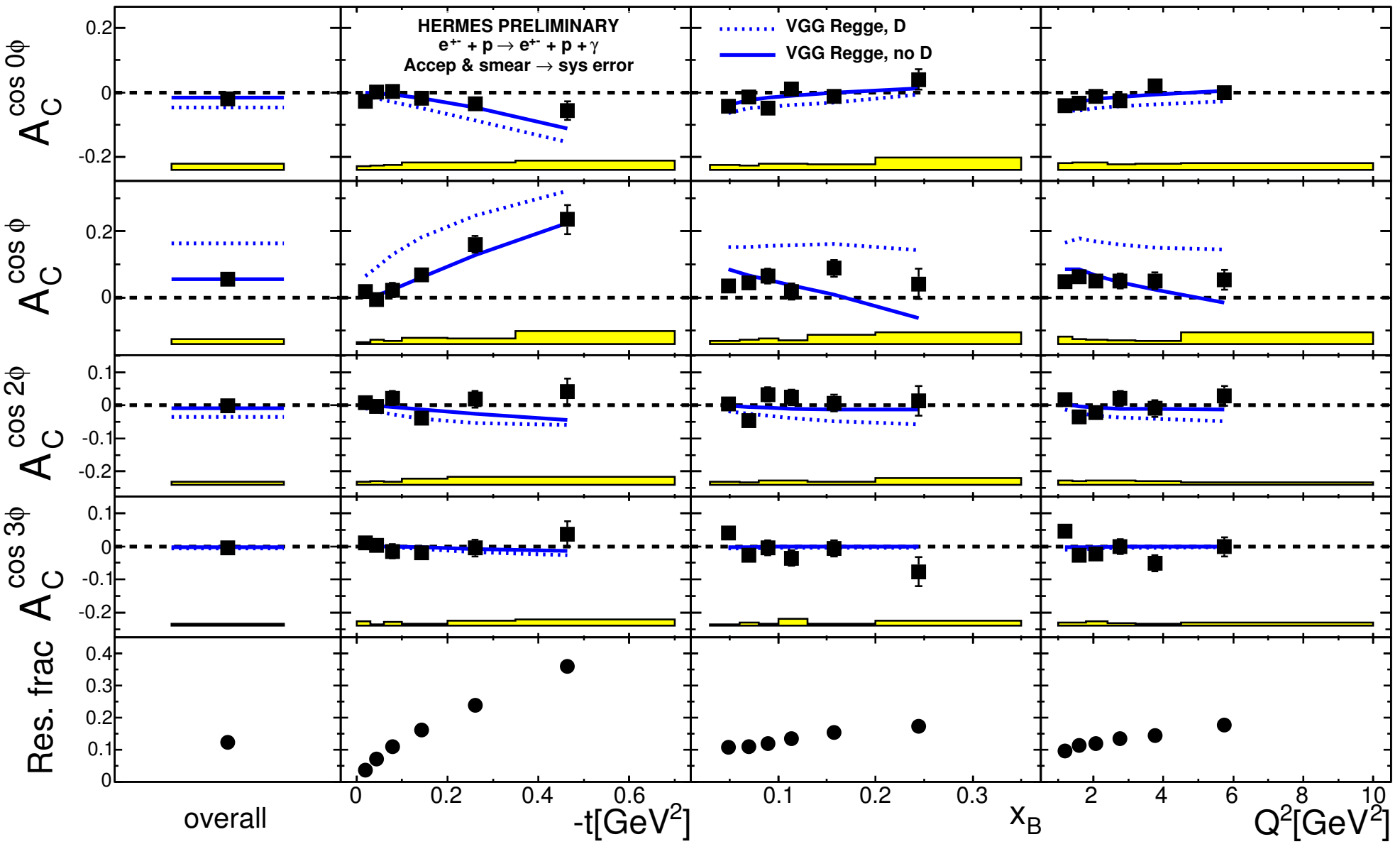

Figure 2: The amplitudes of the beam charge asymmetry extracted from hydrogen data (black bullets) [4]. The error bars (bands) represent the statistical (systematic) uncertainties. The curves are predictions of a double-distribution GPD model [6].

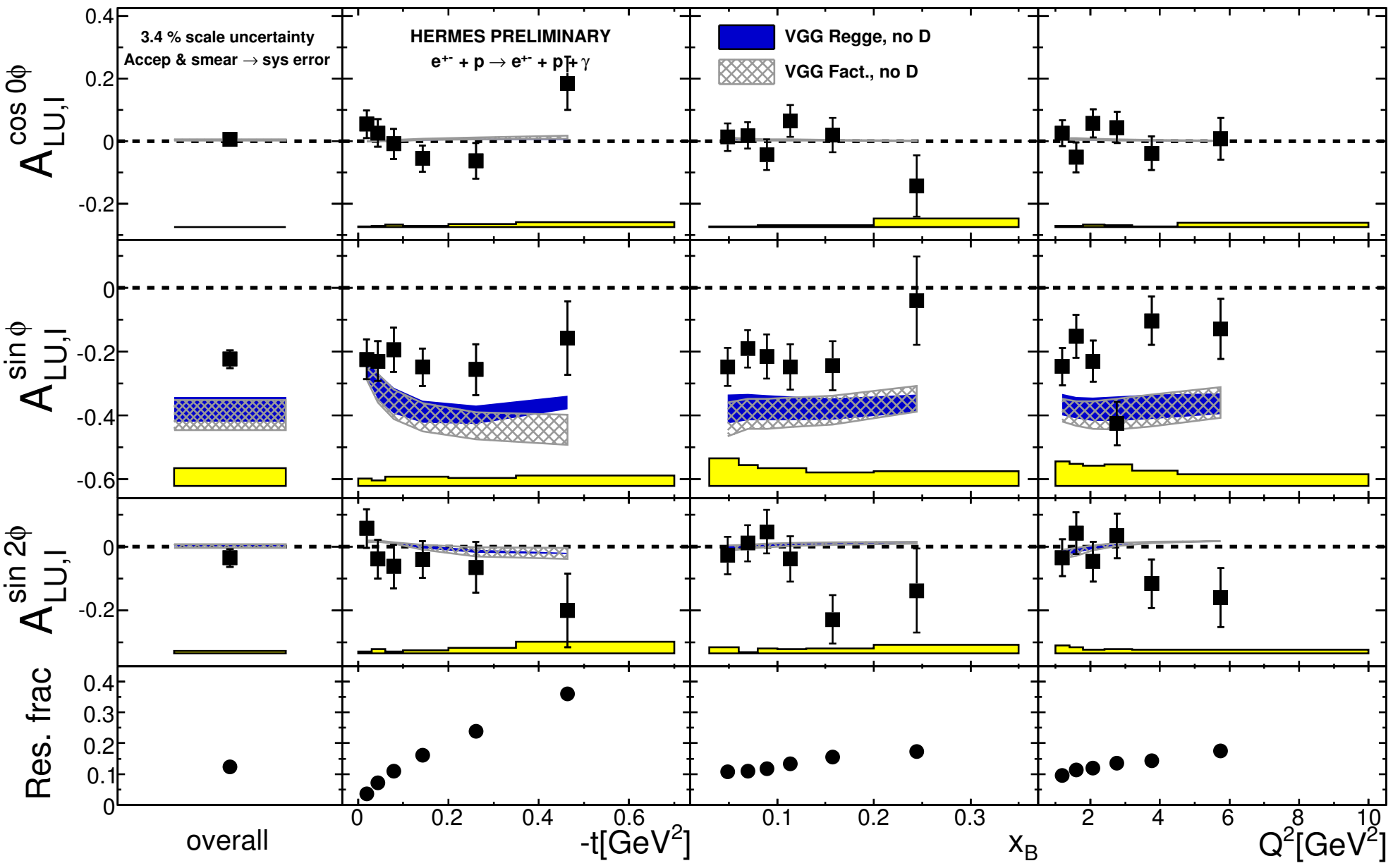

Figure 3: The amplitudes of the beam-helicity asymmetry from the interference term on the unpolarized hydrogen target [4]. The error bars (bands) represent the statistical (systematic) uncertainties. The curves are predictions of a double-distribution GPD model [6].

For data taken with the transversely polarized target, the beam charge asymmetry $A_{\mathrm{C}}(\phi)$ and the TTSAs $A_{\mathrm{UT}}^{\mathrm{DVCS}}(\phi)$ and $A_{\mathrm{UT}}^{\mathrm{I}}(\phi)$ from DVCS and interference term, respectively, have been extracted simultaneously. By comparing GPD model calculations [6] with the measured BCA and TTSA, a model-dependent constraint on the total angular momenta carried by up- and down- quarks of the nucleon is obtained as $J_u + J_d/2.8 = 0.49 \pm 0.17\,(\mathrm{exp_{tot}})$ [2]. However, the double-distribution model of [6] cannot explain all existing DVCS data.

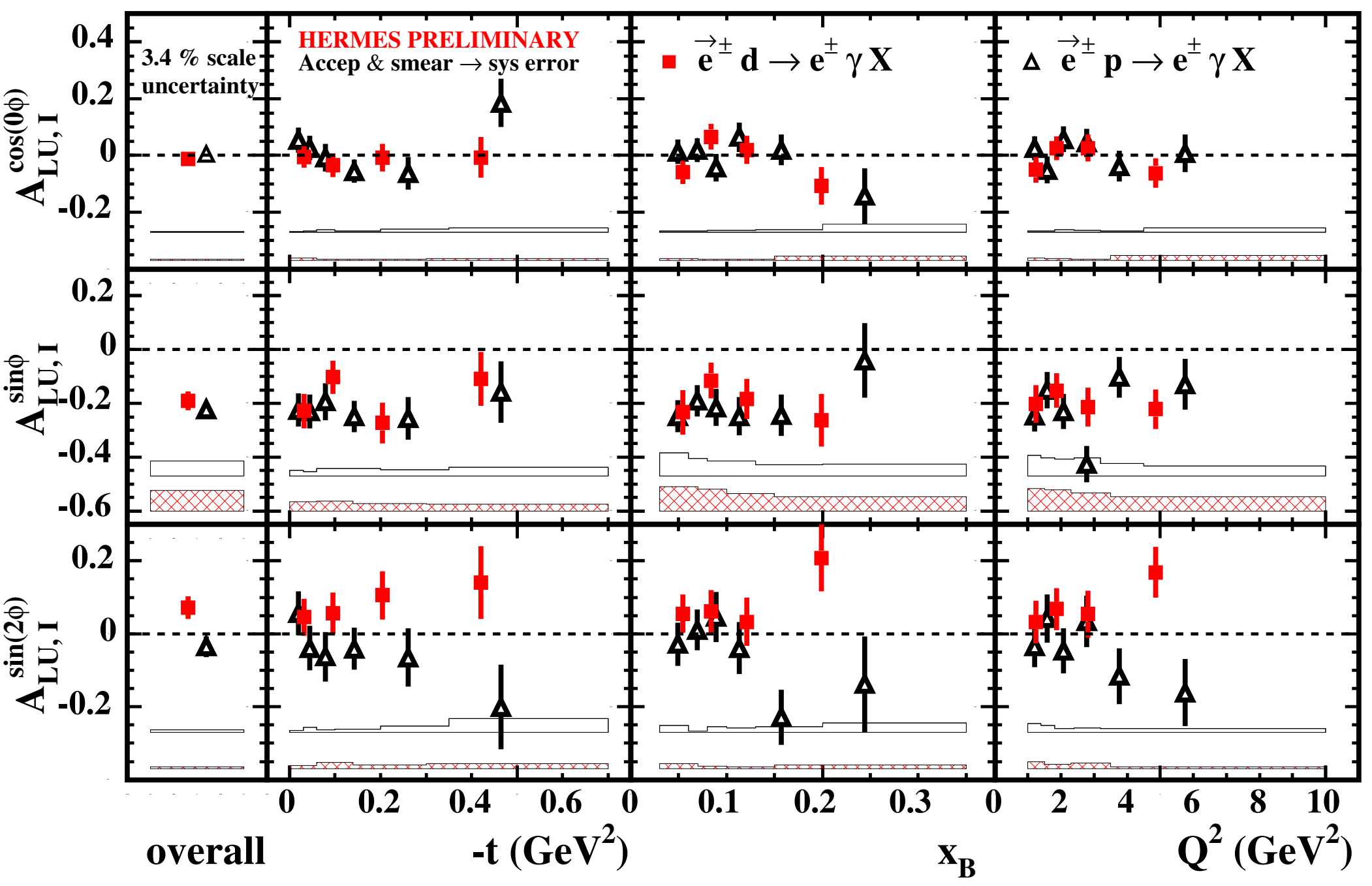

Figure 4: The amplitudes of the beam-helicity asymmetry that are sensitive to the interference term, extracted from deuteron data (squares) and from proton data (triangles) [5]. The error bars (bands) represent the statistical (systematic) uncertainties.

The nuclear-mass dependence of beam-helicity azimuthal asymmetries has been measured for targets ranging from hydrogen to xenon [7]. For hydrogen, krypton and xenon, data were taken with both beam charges. For both DVCS and BH, coherent scattering occurs at small values of $-t$ and rapidly diminishes with increasing $|t|$. Coherent and incoherent-enriched samples are selected according to a $-t$ threshold that is chosen to vary with the target such that for each sample approximately the same kinematic conditions are obtained for all target types. The nuclear-mass dependence of the beam-charge and beam-helicity azimuthal asymmetries is presented separately for the coherent and incoherent-enriched samples in Figs. 5 and 6. The cos$\phi$ amplitude of the beam-charge asymmetry is consistent with zero for the coherent-enriched samples for all three targets, while it is about 0.1 for the incoherent-enriched samples. The sin$\phi$ amplitude of the beam-helicity asymmetry shown in Fig. 6 has values of about $-0.2$ for both the coherent and incoherent-enriched samples. No nuclear-mass dependence of the beam-charge and beam-helicity asymmetries is observed within experimental uncertainties. This is in agreement with models that approximate nuclear GPDs by nucleon GPDs neglecting bound state effects. The data do not support the enhancement of nuclear asymmetries compared to the free proton asymmetries for coherent scattering on spin-0 and spin-1/2 nuclei as anticipated by various models [8, 9, 10]. They also contradict the predicted strong A dependence of the beam-charge asymmetry resulting from a contribution of meson exchange between nucleons to the scattering amplitude [10].

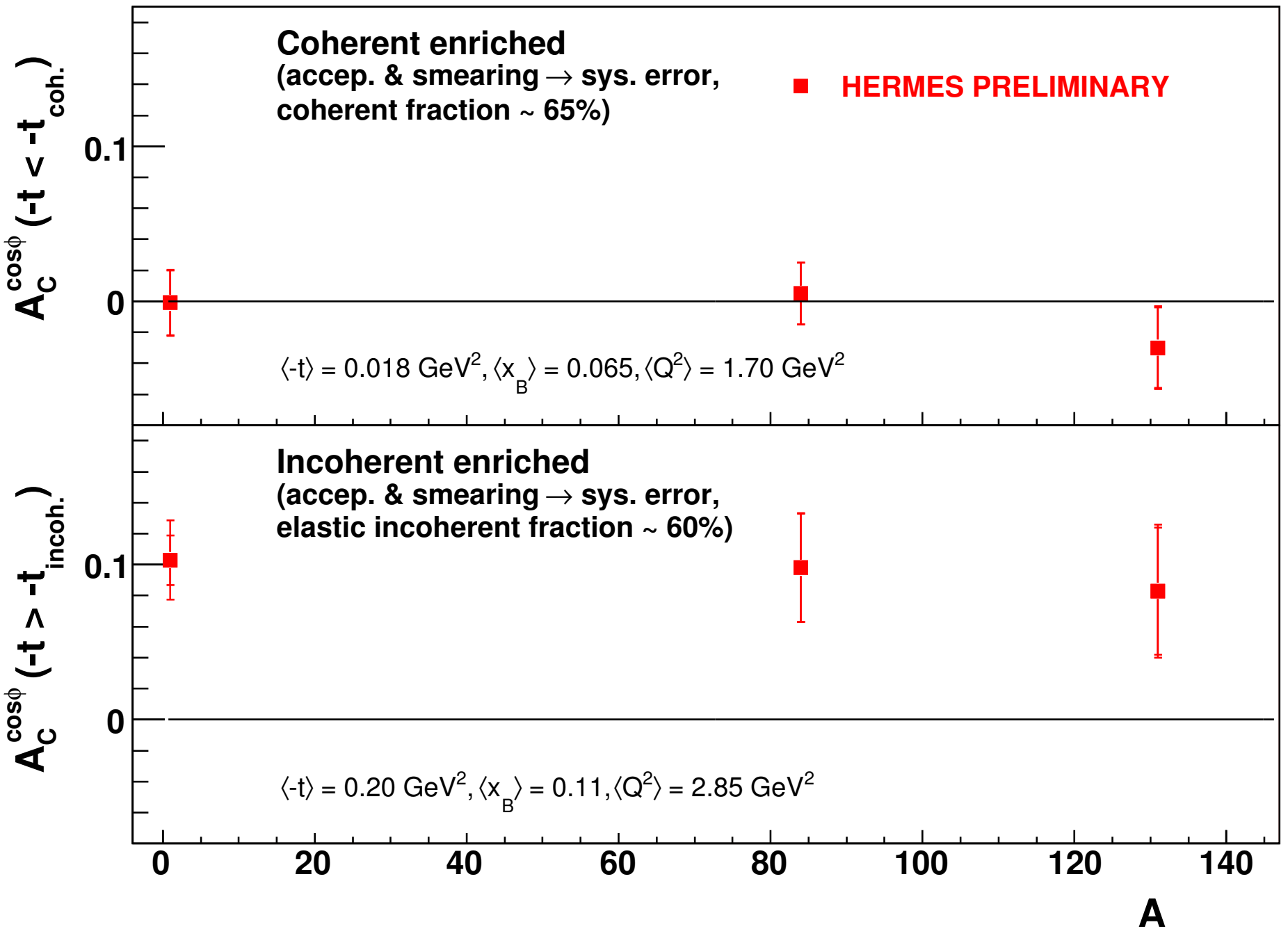

Figure 5: Nuclear-mass dependence of the cos$\phi$ amplitude of the beam-charge asymmetry for the coherent-enriched (upper panels) and incoherent-enriched (lower panels) data samples. The inner (full) errors bar represent the statistical (total) uncertainties.

## 5 The HERMES Recoil Detector

In the results presented above, DVCS events were selected without detection of recoiling particles using missing-mass method. In order to ensure exclusivity and reduce the background from the associated BH process, a recoil detector was installed at HERMES in winter of 2005–2006. The recoil detector consists of a silicon strip detector, a scintillating fiber tracker and a photon detector located in a 1 T solenoidal magnetic field. The main task of the photon detector is to detect photons from $\pi^0$ decay and thereby suppress the contribution from associated $\Delta^+$ production. The recoil detector was in operation during the high luminosity run of HERA using unpolarized hydrogen and deuterium targets from 2006 to 2007. A rich data set was collected in these two years and is being analyzed.

## 6 Conclusion

HERMES has made significant measurements of cosine moments of beam-charge asymmetry and sine moments of charge dependent beam-helicity asymmetry in DVCS from hydrogen and deuterium targets. The statistical precision of the data allows to put constraints on theoretical calculations. The unknown contribution from the associated process can be understood from data taken with the recoil detector. A model-dependent constraint on the total angular mo-menta carried by up and down quarks of the nucleon is obtained by comparing GPD model calculations with the measured BCA and TTSA. Beam-charge and beam-helicity asymmetries have been measured for targets ranging from hydrogen to xenon. No nuclear-mass dependence of the asymmetry amplitudes is observed within experimental uncertainties. The obtained results provide constraints on nuclear GPD models.

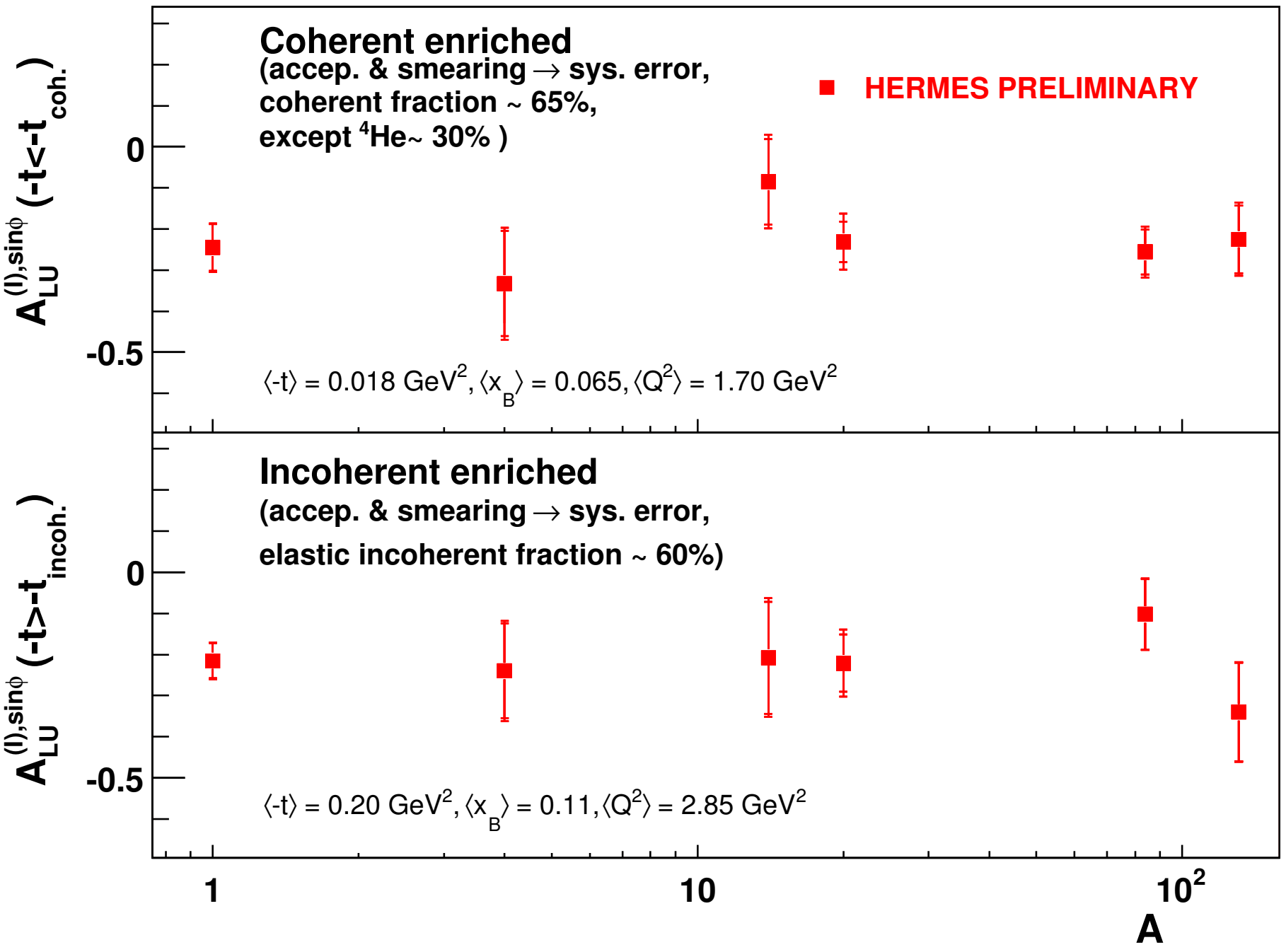

Figure 6: Nuclear-mass dependence of the sin$\phi$ amplitude of the beam-helicity asymmetry for the coherent-enriched (upper panels) and incoherent-enriched (lower panels) data samples. The inner (full) errors bar represent the statistical (total) uncertainties.

# References

[1] A. V. Belitsky, D. Müller and A. Kirchner, Nucl. Phys. **B629** 323 (2002).

[2] A. Airapetian *et al.* [HERMES Collaboration], JHEP **06** 066 (2008).

[3] K. Ackerstaff *et al.* [HERMES Collaboration], Nucl. Instrum. Meth. **A417** 230 (1998).

[4] D. Zeiler on behalf of the HERMES Collaboration, Proc. of DIS08, London, England.

[5] H. Marukyan on behalf of the HERMES Collaboration, AIP Conf. Proc. **1149** 619 (2009).

[6] M. Vanderhaeghen, P. A. M. Guichon, and M. Guidal, Phys. Rev. **D60** 094017 (1999).

[7] H. Ye on behalf of the HERMES Collaboration, Proc. of DIS08, London, England.

[8] A. Kirchner, D. Müller, Eur. Phys. J. **C32** 347 (2003).

[9] V. Guzey, M. I. Strikman, Phys. Rev. **C68** 015204 (2003).

[10] V. Guzey, M. Siddikov, J. Phys. **G32** 251 (2006).

# Low $x$ and Diffractive Physics at a Large Hadron Electron Collider

*Paul Newman*

School of Physics & Astronomy, University of Birmingham, B15 2TT, UK

The prospect of a high energy $ep$ and $eA$ collider (an LHeC) involving an LHC hadron beam and a new electron accelerator is discussed. The low $x$ physics possibilities of such a facility are explored in particular.

## 1 The LHeC Project

Energy-frontier physics will be dominated for the foreseeable future by the proton and heavy ion beams of the LHC, whose unprecedented energy and intensity herald a new era in the field. In the context of a CERN-ECFA-NuPECC commissioned workshop [1, 2], the fledgling LHeC project is investigating whether these hadron beams could be exploited as part of a new high performance electron-proton ($ep$) and electron-ion ($eA$) 'Large Hadron electron Collider' [3–6]. Through its unique sensitivity to the lepton-quark vertex, this could be complementary to the LHC $pp$, $pA$ and $AA$ programmes and to a possible pure lepton future collider in revealing physics at the TeV energy scale. The large achievable luminosities in particular set the LHeC aside from previously evaluated possible future high energy $ep$ colliders [7]. Work is ongoing to assess the physics potential of an LHeC, as well as its accelerator, interaction region and detector requirements and the impact on the existing LHC programme.

Two basic configurations are under study [8]. A new electron ring based on slim (20 cm $\times$ 10 cm) dipole magnet elements, carried on top of the LHC proton ring, yields the largest luminosities. With 50 GeV beam electrons at 50 MW power consumption, an electron ring could deliver $5 \times 10^{33}$ cm$^{-2}$s$^{-1}$, a factor of 100 beyond the highest luminosity achieved at HERA. An alternative solution is a linear electron accelerator arriving tangentially. A luminosity of $5 \times 10^{32}$ cm$^{-2}$s$^{-1}$ could be achievable for 50 GeV electron energy and 50 MW power, assuming an LHC luminosity upgrade. Linac energies up to 150 GeV are under consideration, which might be possible if energy recovery techniques may be applied. The possibility also exists of replacing the protons with, in the first instance, lead ions. Other heavy ion species and deuterons are also under consideration.

## 2 Overview of the LHeC Physics Programme

An overview of LHeC physics can be found for example in [9]. The accessible kinematic plane for $ep$ collisions assuming a 7 TeV proton and a 140 GeV electron beam is compared with previous experiments in Figure 1a. The coverage is extended compared with HERA towards low Bjorken $x$ at fixed $Q^2$ or towards high $Q^2$ at fixed $x$ by the ratio of squared centre of mass energies $s_{\text{LHeC}}/s_{\text{HERA}} \sim 20$. With sufficient luminosity to overcome the basic $1/Q^4$ cross section dependence, squared 4-momentum transfers $Q^2 \sim 10^6$ GeV$^2$ are accessible, probing distance scales below $10^{-19}$ m. As well as sensitivity to new physics [10], an LHeC would allow a full flavour decomposition of the parton distribution functions (PDFs) of the proton and would clarify many issues at the highest $x$ [8]. It would permit measurements of the strong coupling constant and the light quark electroweak couplings to unprecedented precision [11]. As discussed in more detail in Sections 4.1 and 4.3, the ultra-high parton density region $x \lesssim 10^{-4}$ will be accessed for the first time at sufficiently large $Q^2$ for perturbative QCD techniques to be applied. When the LHC runs with heavy ions, the LHeC becomes the first ever $eA$ colliding beam machine (Section 4.2).

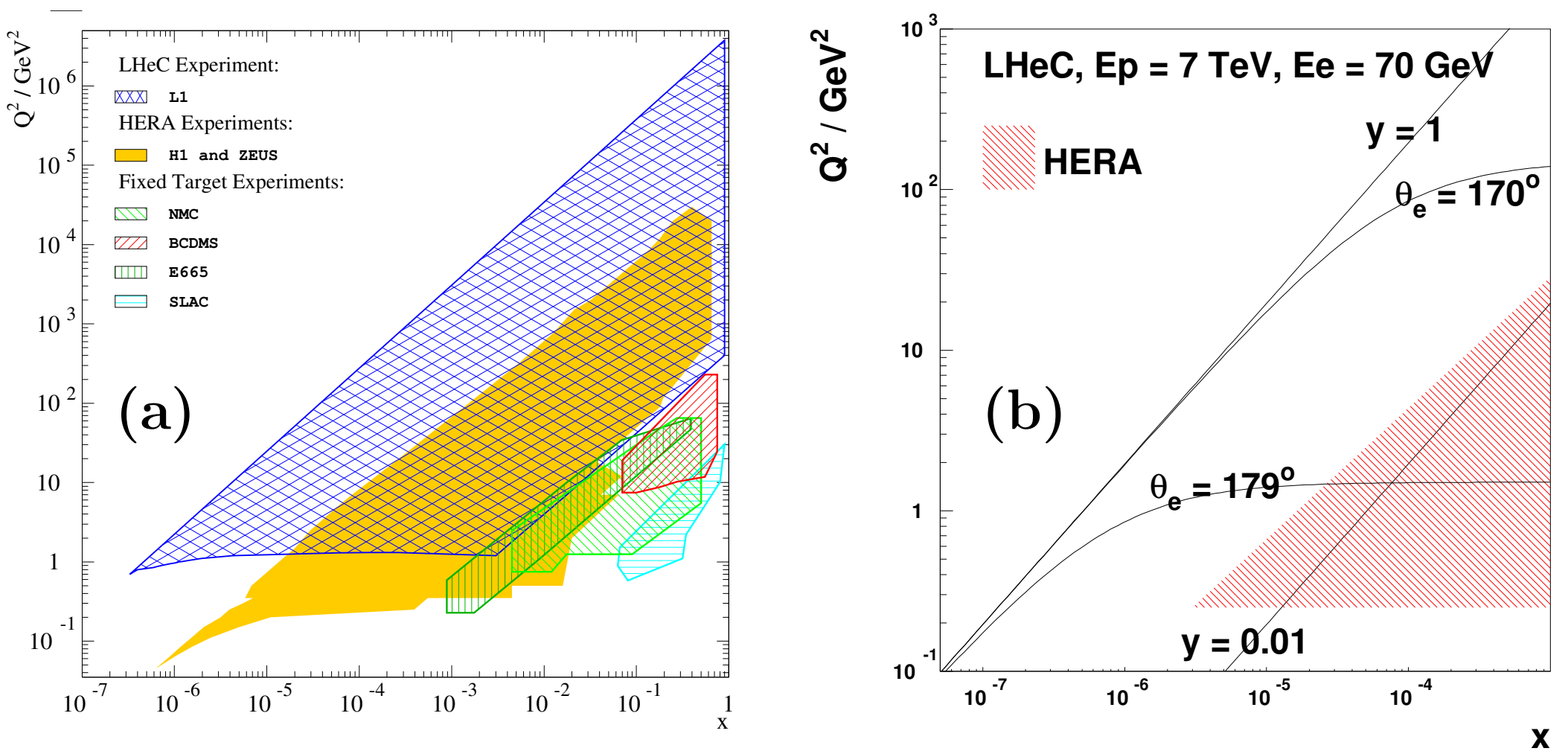

Figure 1: (a) Kinematic plane for $ep$ collisions, showing the coverage of fixed target experiments, HERA and an LHeC. (b) Zoomed view of the low $x$ corner of the kinematic plane, showing the acceptances for two different cuts on electron scattering angle $\theta_e$ at the LHeC.

Accessing the full available phase space brings challenges in the detector and interaction region design [12], as illustrated for the example of the scattered electron kinematics with a 70 GeV beam in Figure 1b. If the electron detection acceptance extends to scatterings through a 1° angle ($\theta_e = 179°$), full coverage of the region $Q^2 > 1$ GeV$^2$ is obtained, reaching below $x = 10^{-6}$. In contrast, with detector components restricted to $\theta_e < 170°$, there is little acceptance for $Q^2 < 100$ GeV$^2$ or $x < 10^{-4}$. Optimising the luminosity by including beam focusing elements close to the interaction region [3], similar to those installed for the HERA-II upgrade, must therefore be evaluated against the corresponding loss of small angle detector acceptance. In order to obtain good sensitivity to both high cross section low-$x$ physics and rare high transverse momentum processes, a two stage programme may be necessary.

# 3 Low $x$ Physics and Electron-Hadron Scattering

At sufficiently large $Q^2$ in the low $x$ region, the 'asymptotically free' quarks of DIS meet a high background density of partons, and various novel effects are predicted. Ultimately, unitarity constraints become important and a 'black body' limit is approached [13], in which the cross section reaches the geometrical bound given by the transverse proton size. This limit is characterised by new effects such as $Q^2$ dependences which differ fundamentally from the usual logarithmic variations and diffractive cross sections approaching 50% of the total [14]. Applying the black body bound to the inelastic cross section for the interaction of a colour dipole, formed from a $\gamma^* \to q\bar{q}$ splitting, leads to an approximate constraint on the gluon density $xg(x, Q^2) < Q^2/\alpha_s$ [15], comparable to expectations for the gluon at the lowest LHeC $x$ values. "Parton saturation" effects are therefore expected in the low $x$ region at the LHeC.

Although no conclusive saturation signals have been observed in parton density fits to existing HERA data, hints have been obtained by fitting the data to dipole models [16–20], which are applicable at very low $Q^2$ values, beyond the range in which quarks and gluons can be considered to be good degrees of freedom. The typical conclusion [19] is that HERA data in the perturbative regime do not exhibit any evidence for saturation. However, when data in the $Q^2 < 1$ GeV$^2$ region are included, only models which include saturation effects are successful. Similar conclusions have been reached by studying the change in fit quality in the NNPDF NLO QCD PDF fit framework as low $x$ and $Q^2$ data are progressively omitted [21].

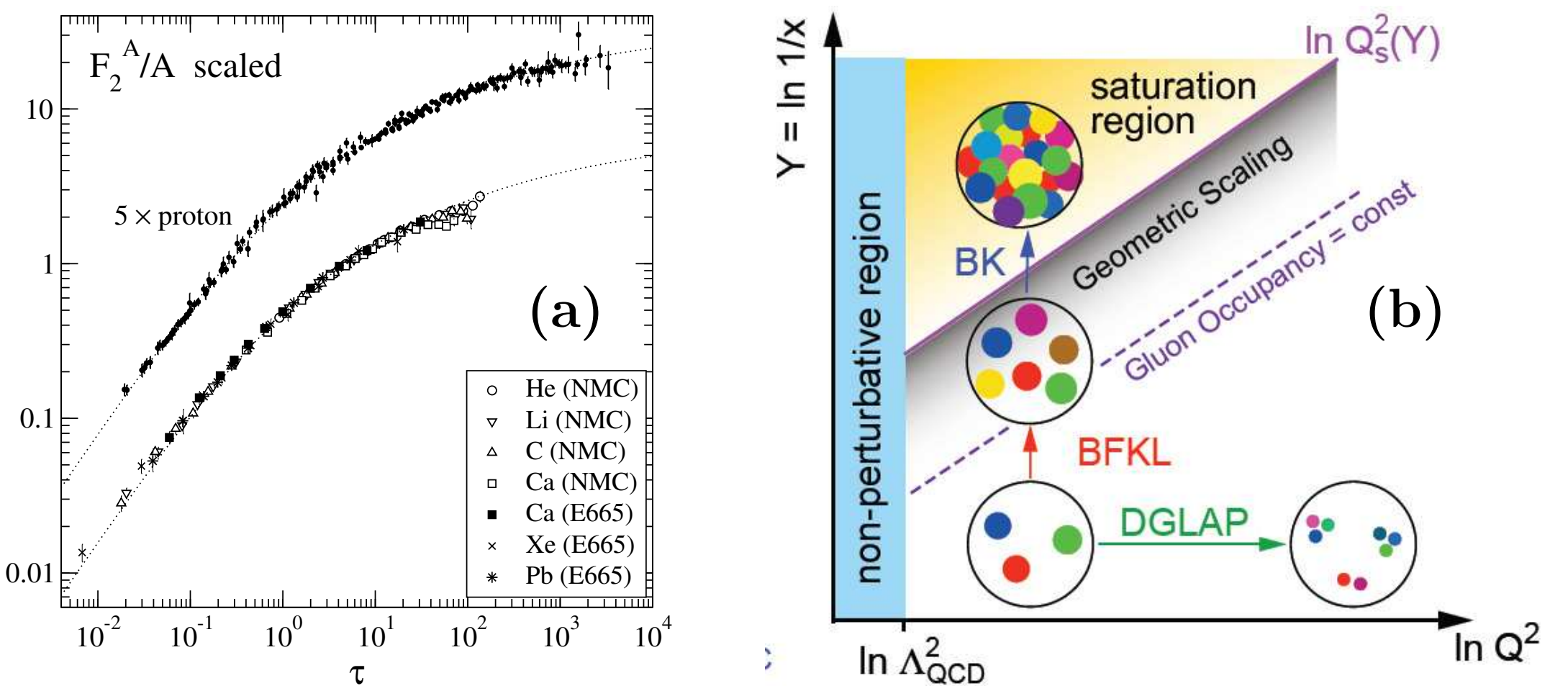

Figure 2: (a) Geometric scaling plot for protons and for nuclei (see text for details). (b) Illustration of the DIS kinematic plane, showing the transition to the saturation region.

The 'geometric scaling' [22] feature of the data reveals that to good approximation the low $x$ cross section is a function a single variable $\tau = Q^2/Q_s^2(x)$, where $Q_s^2 = Q_0^2\, x^{-\lambda}$ is an $x$ dependent 'saturation scale'. This parameterisation works well for scattering from both protons and heavy ions, as shown in Figure 2a [23]. An interpretation of this feature is that the cross section is invariant along lines of constant 'gluon occupancy' or 'blackness'. As illustrated in Figure 2b, such lines are diagonals in the $\ln 1/x$ v $\ln Q^2$ kinematic plane, due to two competing effects in the growth of the blackness: increasing parton densities as $x$ decreases and dilution of the system as $Q^2$ grows and the resolution improves. When viewed in detail, there is a change in behaviour in the geometric scaling plot, Figure 2a, near $\tau = 1$, which has been interpreted as a transition to the saturation region shown in Figure 2b. However, data with $\tau < 1$ exist only at very low, non-perturbative, $Q^2$ values to date, precluding a partonic interpretation.

Whether or not the low $Q^2$ HERA saturation signal is confirmed, a central aim of the LHeC programme is to observe how unitarisation impacts on the proton structure. Understanding the mechanisms involved in terms of parton dynamics, for example the gluon recombination process $gg \to g$ [24], should be possible in the low $x$ and moderate $Q^2$ region at the LHeC.

# 4 Simulated LHeC Low $x$ Performance

This section describes some first explorations of low $x$ physics possibilities with an LHeC. It is by no means exhaustive. Among the important topics which are under study, but are not covered here are forward jets and their relation to parton cascade dynamics and Deeply Virtual Compton Scattering (DVCS). More details on these and other topics may be found at [1, 2].

## 4.1 Inclusive $ep$ Cross Sections

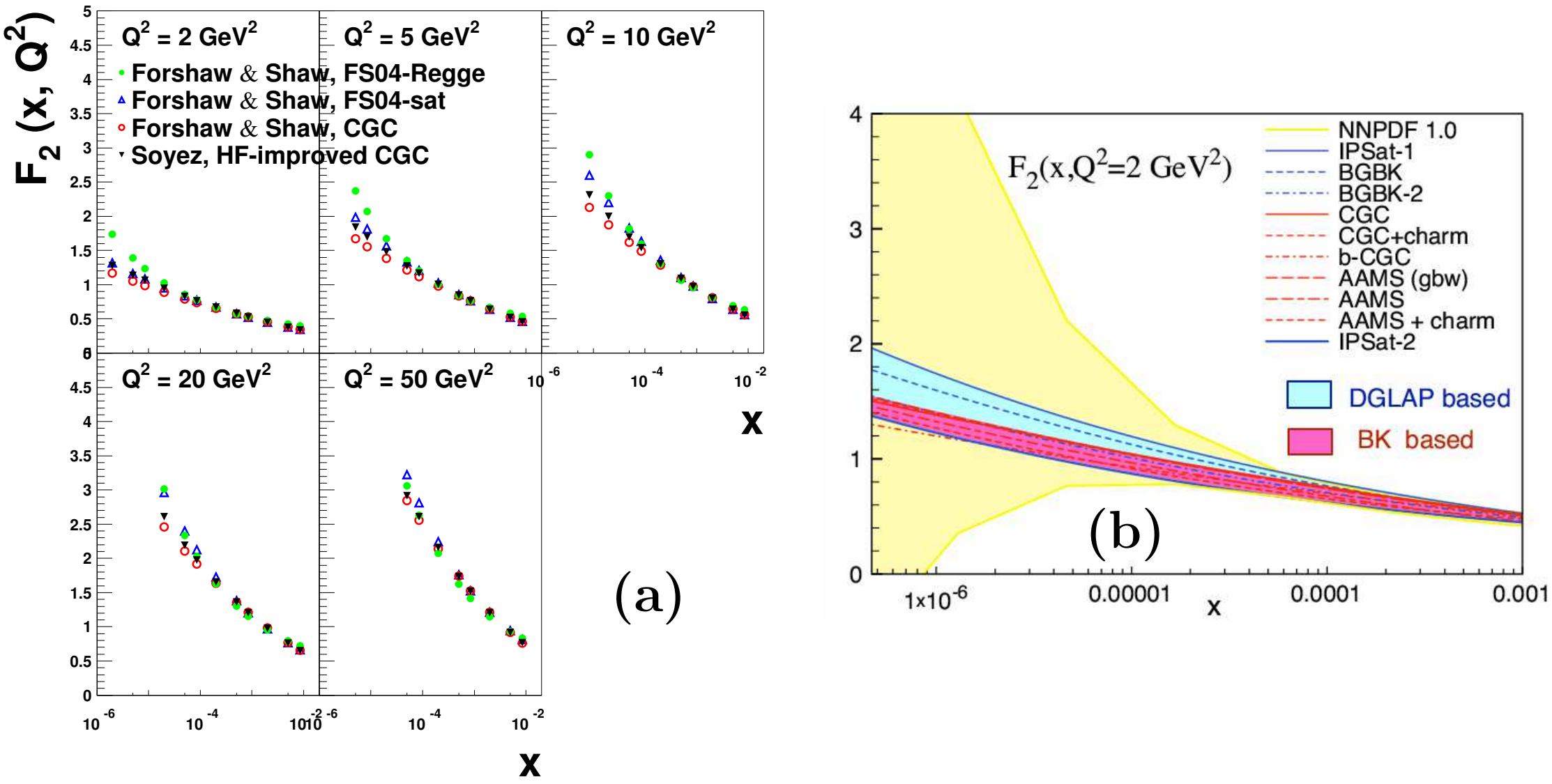

Figure 3: Extrapolations into the LHeC low $x$ region of a variety of models of the inclusive structure function $F_2$. (a) Simulated data points in four dipole models at several $Q^2$ values. (b) A wide range of dipole model predictions at $Q^2 = 2$ GeV$^2$, also compared with the extrapolation uncertainty band from the NNPDF QCD fit to current data.

Figure 3a shows extrapolations of four dipole models constrained by fits to HERA data to predict the structure function $F_2(x, Q^2)$ in the LHeC kinematic range, which is shown in the form of simulated measurements [25]. At the lowest $x$ and $Q^2$, there is a clear distinction between the 'FS04-Regge' model [19], which does not include saturation, and all others [19, 20], which include saturation effects as estimated from low $Q^2$ HERA data. However, any such sensitivity is lost by around $Q^2 = 50$ GeV$^2$, emphasising the importance of low angle scattered electron acceptance.

Figure 3b [26] shows a wider selection of dipole models, all of which include unitarisation effects, at a low scale, $Q^2 = 2$ GeV$^2$. The predictions have been grouped into two classes, according to whether the low $x$ saturation is generated from eikonalisation of two gluon exchange within a DGLAP framework or from the non-linear BK equation [27] or Colour-Glass

Condensate [28] approach. It is interesting to note that the range of variation among these dipole models with QCD-based input is substantially smaller than the full range which is formally allowed by extrapolating the reasonable approach to parameterisation uncertainties in the NNPDF PDF fit [29]. The expected experimental precision (Figure 3a) is certainly good enough to distinguish between many of the different models.

Whilst such extrapolations of dipole fits to HERA data give encouraging indications, the unequivocal establishment of parton saturation at the LHeC is likely to be challenging. Two studies using very different approaches to PDF fitting are in progress [30, 31]. They both subject LHeC pseudo-data derived from saturating dipole models to NLO DGLAP fits, to determine whether saturation effects could be masked, for example by the flexibility in the parton parameterisations. It is not yet clear whether a breakdown of pure DGLAP dynamics may be visible with $F_2$ data alone. If not, the two ongoing analyses agree that the addition of $F_L$ data as a second observable in the fits would prove conclusive.

## 4.2 Inclusive $eA$ Cross Sections

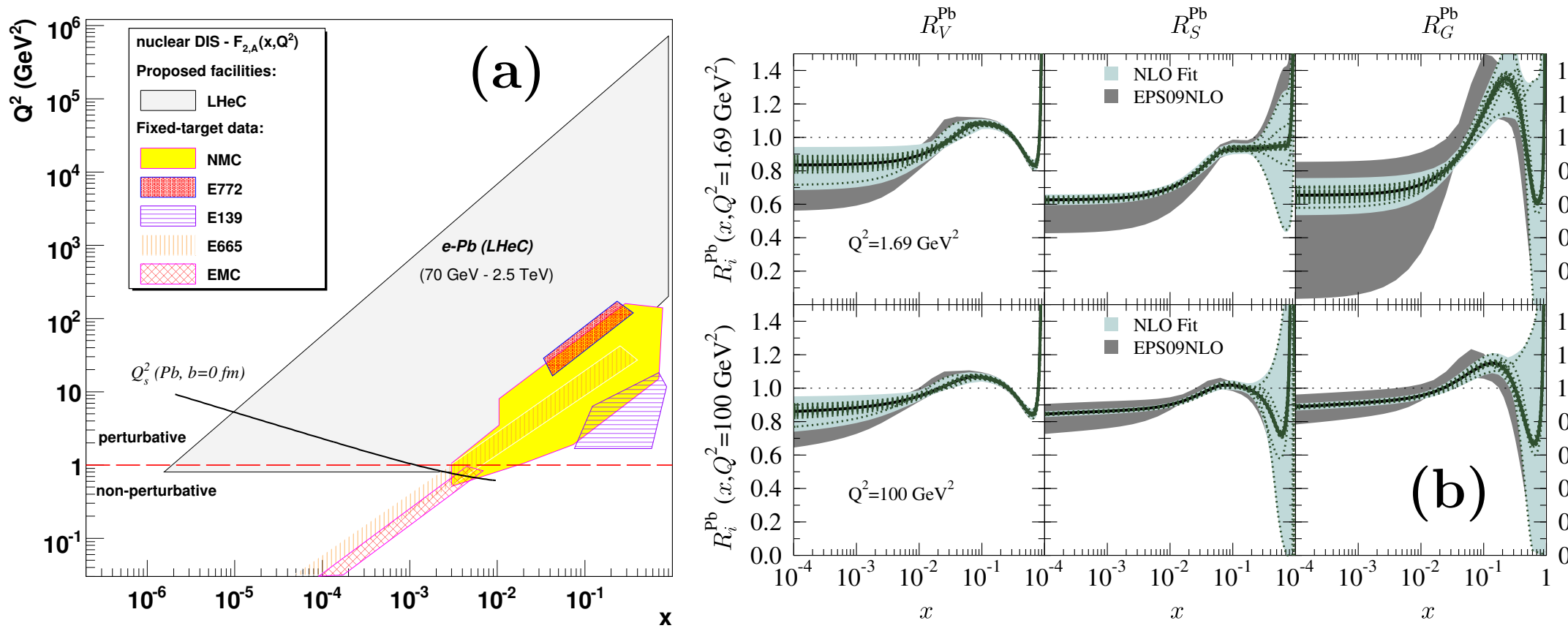

Figure 4: (a) Kinematic plane for $eA$ collisions, showing existing fixed target coverage and the potential LHeC range. (b) Nuclear ratios (see text) as extracted in the framework of the EPS09 nuclear PDF fits. Uncertainty bands are shown with ("NLO Fit") and without ("EPS09NLO") the addition of the LHeC pseudo-data.

Since $eA$ collisions have previously been achieved only in fixed target experiments, the parton distributions of nuclei are completely unknown at low $x$. Indeed, as illustrated in Figure 4a [32], the LHeC offers an extension in the kinematic range by around four orders of magnitude towards lower $x$ at fixed $Q^2$ or towards higher $Q^2$ at fixed $x$. The LHeC is thus unique in its sensitivity to the initial state of heavy ion ($AA$) collisions in the LHC energy range.

The small $x$ nuclear gluon density $g_A$ at central impact parameters is enhanced relative to that ($g_N$) in a nucleon by a factor $(g_A/\pi R_A^2)/(g_N/\pi R_N^2) \simeq A^{1/3} g_A/A g_N \simeq A^{1/3}$ [14]. Scattering from nuclei thus offers enhanced sensitivity to unitarisation phenomena compared with $ep$ collisions, if such effects can be unfolded from nuclear shadowing corrections due to the coherent scattering of the lepton from more than one nucleon. Figure 4a includes an estimate of the critical saturation line for electron-lead collisions. There is a substantial low $x$ region within the LHeC acceptance below this line. The prospects of unfolding and understanding saturation effects when $ep$ and $eA$ data are considered together are very strong.

The influence of simulated LHeC data on fits to nuclear PDFs has been evaluated in the framework of the EPS09 NLO QCD analysis of existing nuclear data [33]. Figure 4b [34] illustrates this in the form of the nuclear ratio, (e.g. $R^A(x,Q^2) = F_2^A(x,Q^2)/AF_2^p(x,Q^2)$ for the total quark contribution), for the specific case of lead ($A = 207$, $A^{1/3} \sim 6$). If only existing fixed target $eA$ and $pA$ Drell-Yan / leading pion data are included, the uncertainties on the valence quark ($R_V$), sea quark ($R_S$) and gluon ($R_G$) ratios are all large, the gluon at low $x$ and $Q^2$ being particularly problematic. Adding LHeC data resolves the low $x$ region in a manner which is sensitive to saturation effects [34]. More detail on the potential synergies between $eA$, $pA$ and $AA$ scattering can be found in [35].

## 4.3 Diffraction

Non-inclusive observables promise to enhance the LHeC sensitivity to non-linear evolution and saturation phenomena. Diffractive channels are promising, due to the underlying exchange of a pair of gluons. The cleanest processes experimentally are Deeply-Virtual Compton Scattering (DVCS, $ep \to e\gamma p$) and exclusive vector meson production ($ep \to eVp$), which have both played a major role at HERA [36]. Simulations of LHeC elastic vector meson photoproduction ($Q^2 \sim 0$) have yielded encouraging results, especially for the $J/\psi$, as illustrated in Figure 5a [37, 38]. With acceptance for the muon decay products extending to within 1° of the beampipe,[1] invariant photon-proton masses $W$ of well beyond 1 TeV are accessible, extending substantially beyond HERA coverage and clearly distinguishing between models in which saturation effects are present and where they are absent [18]. Similar studies of elastic $J/\psi$ photoproduction in LHeC $eA$ collisions have been proposed as a direct means of extracting the nuclear gluon density [39].

First studies [37] have been made of LHeC possibilities with the inclusive diffractive DIS process, $ep \to eXp$. Similarly to fully inclusive DIS, fractional struck quark momenta relative to the diffractive exchange, $\beta = x/x_{I\!P}$, a factor of around 20 lower than at HERA are accessible at the LHeC. Large improvements in diffractive parton densities (DPDFs) [40] are possible from NLO DGLAP fits to diffractive structure function, dijet and heavy flavour data. The extended phase space towards large $Q^2$ at fixed $x$ increases the lever-arm for extracting the diffractive gluon density and opens the possibility of significant weak gauge boson exchange, which would allow a quark flavour decomposition for the first time. Figure 5b shows a comparison between HERA and the LHeC in terms of the invariant masses $M_X$ which could be produced in diffractive processes with $x_{I\!P} < 0.05$ (RAPGAP Monte Carlo model [41]). Diffractive masses up to several hundred GeV are accessible, such that diffractive final states involving beauty quarks and $W$ and $Z$ bosons, or even exotic states with $1^-$ quantum numbers, could be produced.

Leading twist diffraction has been related [13, 42] to the leading twist component of the nuclear shadowing phenomenon. Measuring diffractive DIS together with nuclear structure functions (Section 4.2) in the LHeC range therefore tests the unified picture of complex strong interactions and leads to a detailed understanding of the shadowing mechanism, possibly essential in interpreting saturation signatures in $eA$ interactions.

[1]This is likely to be achievable, even if tracking and calorimetry extend only to within 10° of the beampipe.

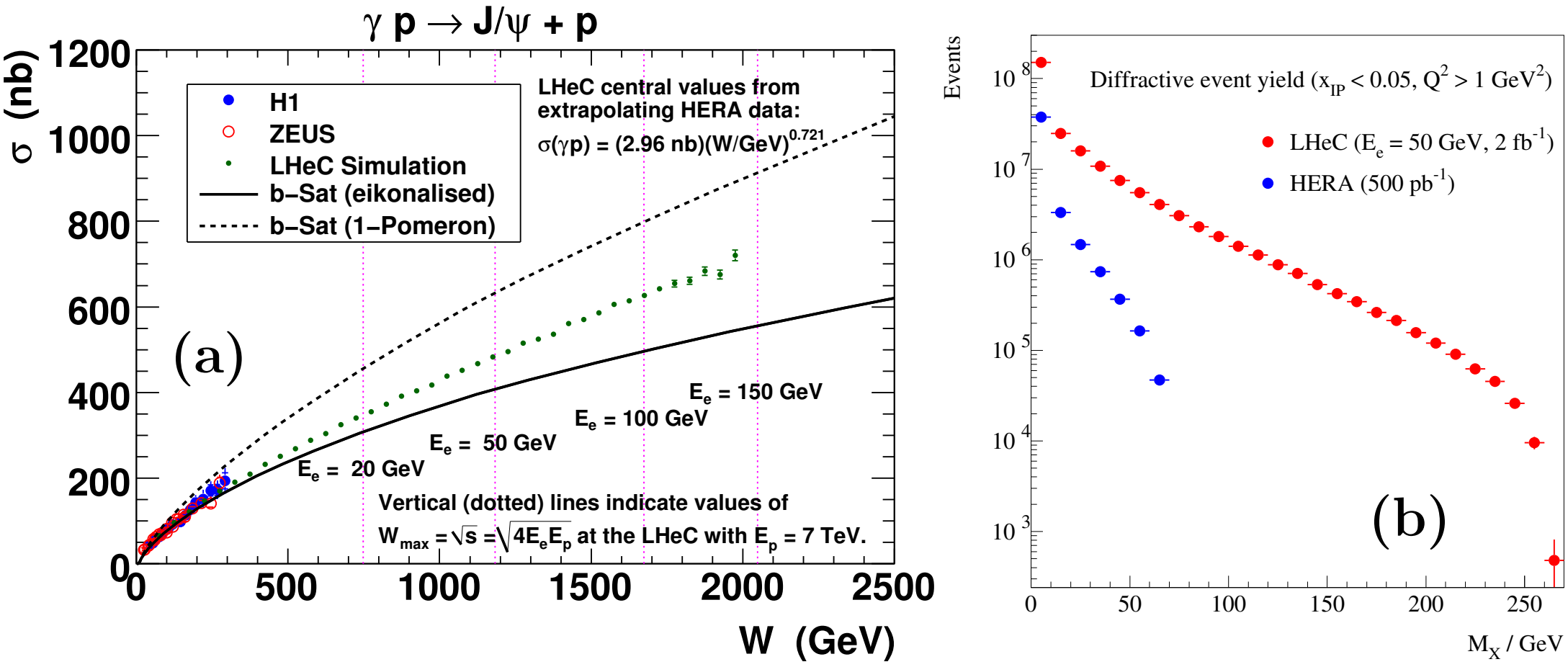

Figure 5: (a) An LHeC simulation of elastic $J/\psi$ photoproduction cross section data for a 150 GeV electron beam, compared with HERA data and dipole model predictions with ("b-Sat - eikonalised") and without ("b-Sat - 1 pom") non-linear effects. (b) Comparison between expected LHeC and HERA diffractive mass ($M_X$) distributions.

## 5 Summary

An investigation of the possible exploitation of the LHC proton beams for $ep$ physics is well underway in the framework of the LHeC project. If realised, an LHeC facility would become an integral part of the quest to fully understand the new Terascale physics which will emerge as the LHC era unfolds. Integral to this, the prospects for understanding the influence of unitarity constraints on low $x$ physics in terms of new parton dynamics are particularly promising. Further evaluations of the full physics potential and the various detector, interaction region and accelerator lay-out options are ongoing. Frequent updates on the progress towards a Conceptual Design Report can be found at [4].

## Acknowledgements

Thanks to all colleagues who have contributed to the evaluation of the LHeC low $x$ programme to date, in particular, my co-convenors of the working group on physics at high parton densities, Nestor Armesto, Brian Cole and Anna Stasto.

## References

[1] 1st ECFA-CERN LHeC Workshop: `http://indico.cern.ch/conferenceDisplay.py?confId=31463`.

[2] 2nd ECFA-CERN-NuPECC LHeC Workshop: `http://indico.cern.ch/conferenceDisplay.py?confId=59304`.

[3] J. Dainton et al., JINST **1**, P10001 (2006).

[4] The LHeC project web page: `http://cern.ch/lhec`.

[5] M. Klein and P. Newman, CERN Courier, April 2009, 22.

[6] M. Klein, plenary presentation to ECFA, November 2009, available from [4].

[7] J. Feltesse, R. Rückl and A. Verdier, presentations at LHC workshop, Aachen, 1990, CERN 90-10 (1990); *'The THERA Book: ep Scattering at $\sqrt{s} \sim 1$ TeV'*, (DESY 01-123F vol 4); *'QCD Explorer Based on LHC and CLIC-I'*, D. Schulte and F. Zimmermann, presentation at EPAC'04 (Lucerne), CERN-AB-2004-079 (2004).

[8] M. Klein, *'The Large Hadron Electron Collider Project'*, proc. of DIS09 [hep-ex/0908.2877].

[9] P. Newman, *'Deep Inelastic Scattering at the TeV Energy Scale and the LHeC Project'*, proc. of the 2008 Ringberg workshop on New Trends in HERA Physics, Nucl. Phys. **B191** (Proc. Suppl.) 307 (2009) [hep-ex/0902.2292].

[10] E. Perez, *'New Physics Summary Talk'*, in [2].

[11] O.Behnke, *'QCD and Electroweak Physics Summary Talk'*, in [2].

[12] P. Newman, *'LHeC Low x Detector Requirements'*, in [2].

[13] V. Gribov, Sov. Phys. JETP **30**, 709 (1970).

[14] L. Frankfurt et al., *'Electron-Nucleus Collisions at THERA'*, in THERA Book (see [7]), [hep-ph/0104252].

[15] L. Frankfurt, W. Koepf and M. Strikman, Phys. Rev. **D54**, 3194 (1996); E. Gotsman et al., J. Phys. **G27**, 2297 (2001).

[16] K. Golec-Biernat, M. Wüsthoff, Phys. Rev. **D59**, 014017 (1999);

[17] E. Iancu, K. Itakura and S. Munier, Phys. Lett. **B590**, 199 (2004);

[18] H. Kowalski and D. Teaney, Phys. Rev. D **68** (2003) 114005; H. Kowalski, L. Motyka and G. Watt, Phys. Rev. **D74**, 074016 (2006).

[19] J. Forshaw and G. Shaw, JHEP **0412**, 052 (2004).

[20] G. Soyez, Phys. Lett. **B655**, 32 (2007).

[21] F. Caola, 'Deviations from NLO DGLAP in HERA Data', in [2].

[22] A. Stasto, K. Golec-Biernat and J. Kwiecinski, Phys. Rev. Lett. **86** (2001) 596.

[23] A. Freund, K. Rummukainen, H. Weigert and A. Schafer, Phys. Rev. Lett. **90** (2003) 222002.

[24] V. Gribov, E. Levin, G. Ryskin, Phys. Rept. **100**, 1 (1983).

[25] M. Klein, *'Parton Distributions at the LHeC'*, proc. of DIS07, Munich (2007).

[26] J. Albacete, N. Armesto, J. Milhano and C. Salgado, Phys. Rev. D **80** (2009) 034031; J. Albacete, *'Predictions for Low x inclusive LHeC data'*, in [2].

[27] I. Balitsky, Nucl. Phys. B **463** (1996) 99; Y. Kovchegov, Phys. Rev. D **60** (1999) 034008.

[28] E. Iancu, A. Leonidov and L. McLerran, Nucl. Phys. A **692** (2001) 583.

[29] R. Ball *et al.*, Nucl. Phys. B **809** (2009) 1 [Erratum-ibid. B **816** (2009) 293].

[30] P Newman, *'Physics at High Parton Densities'*, in [1].

[31] J. Rojo-Chacon, *'Lox x Physics at LHeC with NNPDFs'*, in [1]; A. Guffanti, *'NNPDF fits of LHeC Pseudodata'*, in [2].

[32] D. d'Enterria, hep-ex/0706.4182; D. d'Enterria, private communication.

[33] K. Eskola, H. Paukkunen and C. Salgado, JHEP **0904** (2009) 065.

[34] K. Eskola, *'Nuclear Parton Distributions'*, in [2]; H.Paukkunen, private communication.

[35] T. Ullrich, *'The Symbiotic Relationship between Heavy Ion Physics and DIS'*, in [2].

[36] A. Bruni, X. Janssen and P. Marage, *'Exclusive Vector Meson Production and Deeply Virtual Compton Scattering at HERA'*, procedings of the HERA-LHC workshops, 2006-8.

[37] P. Newman, *'LHeC Diffraction and Vector Meson Pseudo-Data'*, in [2].

[38] G. Watt, private communication.

[39] A. Caldwell, H. Kowalski, hep-ex/0909.1254.

[40] H1 Coll., Eur. Phys. J. **C48**, 715 (2006).

[41] H. Jung, Comput. Phys. Commun. **86**, 147 (1995).

[42] L. Frankfurt and M. Strikman, Eur. Phys. J **A5**, 293 (1999).

# The $J/\psi$ Way to Nuclear Structure

*A. Caldwell*[1], *H. Kowalski*[2]
*talk given by H. Kowalski*

[1]Max Planck Institute for Physics, München
[2]DESY, Notkestraße 85, 22607 Hamburg, Germany

We propose to investigate the properties of nuclear matter by measuring the elastic scattering of $J/\psi$ on nuclei with high precision. The $J/\psi$ mesons are produced from the photons emitted in high energy electron-proton or electron-nucleus scattering in the low-x region. The measurement could be performed at the future ENC, EIC or LHeC facilities.

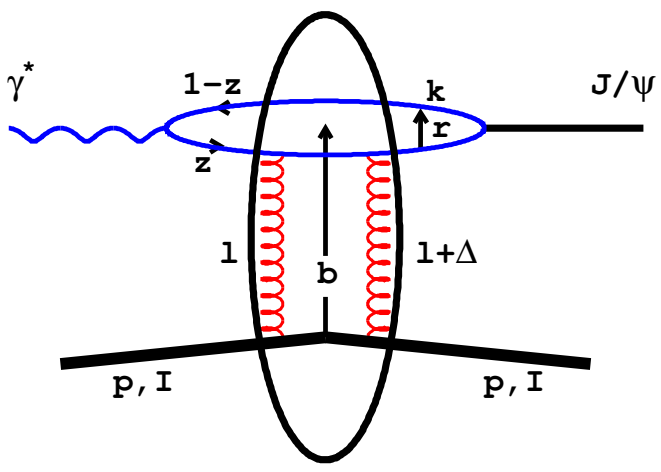

Figure 1: Elastic scattering of a $J/\psi$ meson on a proton or ion.

In recent years HERA has shown that deep inelastic electron-proton reactions at high energy can be described by the scattering of small quark-antiquark dipoles. In this so called low-$x$ region, the incoming target proton frequently remains intact. This happens because, in leading order QCD, the $q\bar{q}$ pair interacts elastically with the nucleon by exchanging *two* gluons with transverse momenta, $\vec{l}$ and $\vec{l}+\vec{\Delta}$. The transverse momenta of the gluons are large, however their difference, $\vec{\Delta}$, can be very small so that the net momentum transfer to the nucleon is small. The reaction can therefore leave a proton or a nuclear target in its ground state or in a slightly excited state.

Of particular interest is the charmed dipole elastic or quasi-elastic scattering on nuclei with a subsequent transformation into the $J/\psi$ vector meson. In this reaction the momentum transfer can be well measured because it is equal to the difference between the transverse momenta of the incoming virtual photon and the final meson. The advantage of $J/\psi$ scattering is its high cross section and smallness of the dipole. In addition, the momenta of the decay products of $J/\psi \rightarrow \mu^+\mu^-$ or $e^+e^-$ can be very precisely measured. The smallness of the dipole in low-$x$ reactions assures that the interaction is mediated by gluon exchange only. Thus, the deflection of the $J/\psi$ measures directly the intensity and the spatial distribution of the nuclear fields.

The measurement of $J/\psi$ scattering on nuclei could become an important source of information on nuclear structure and high density QCD. The interaction of a dipole with a nucleus can be viewed as a sum of dipole scatterings on the nucleons forming the nucleus. The size of the charmed dipole in elastic $J/\psi$ scattering is around 0.15 fm; i.e., it is much smaller than the nucleon radius. It is therefore possible that dipoles interact with smaller objects than nucleons; e.g., with constituent quarks or hot spots. Taking the conventional point of view and assuming that the nucleus is built out of nucleons and that dipoles scatter on the ensemble of nucleons the dipole model predicts the nuclear cross sections shown in Figure 2.

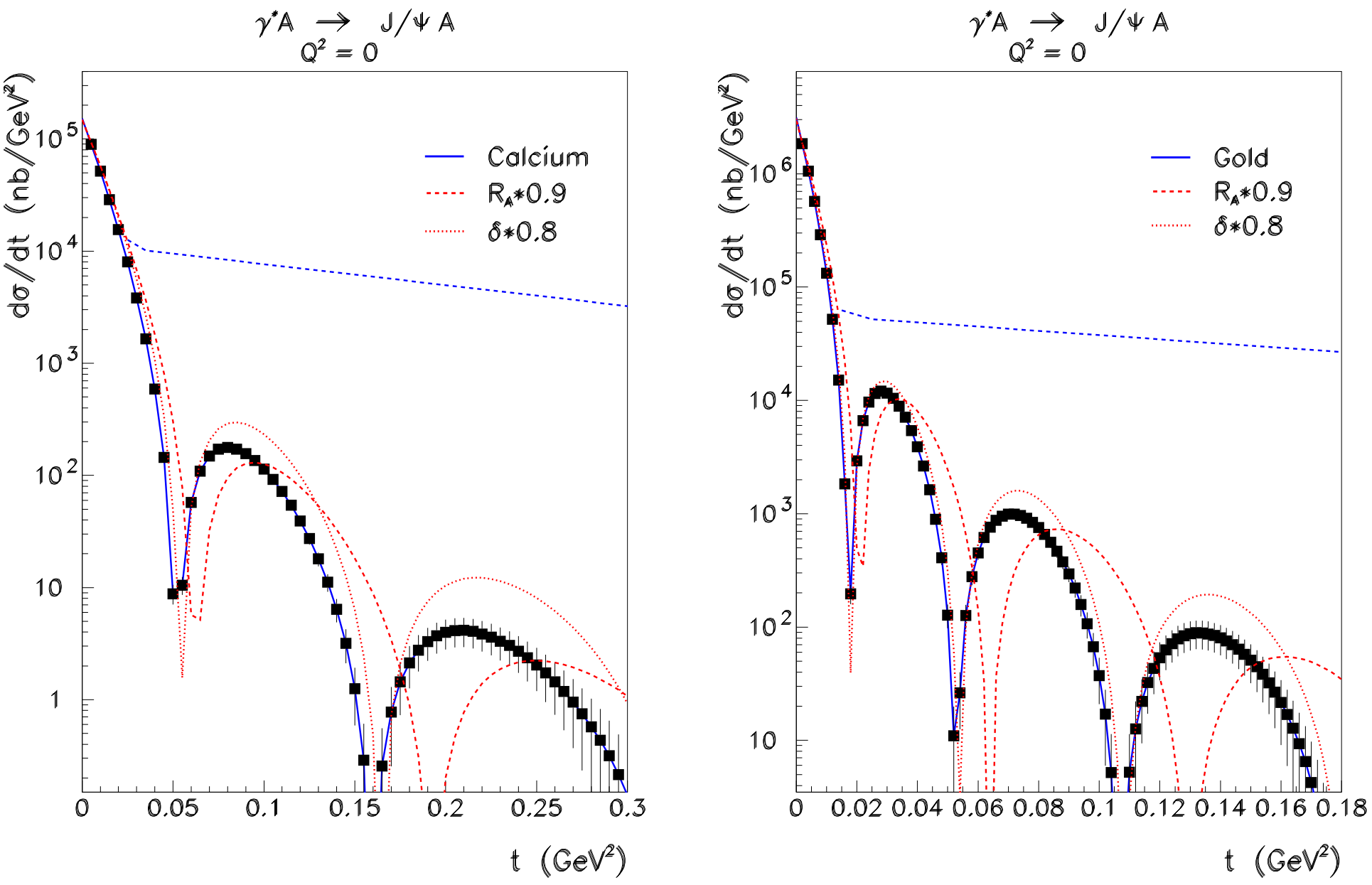

Figure 2: The prediction of the dipole model for the $t$ distribution of coherent $J/\psi$ photoproduction on nuclei assuming that the single nucleon distribution can be identified with the Woods-Saxon distribution. The statistical errors of the simulated measurements are based on the assumed collected sample of $10^6$ events. The upper dashed line shows the sum of the coherent and incoherent process in case of no correlations.

Figure 2 shows the coherent and incoherent cross sections for scattering on nuclei. In the coherent process the nucleus remains in its ground state. In the incoherent process the nucleus gets excited and frequently breaks into nucleons or nucleonic fragments. Experimentally we expect to be able to distinguish cases where the nucleus remains intact and cases where the nucleus breaks up. In the nuclear breakup process, there are several free neutrons and protons in the final state, as well as other fragments. The number of free nucleons could depend on the value of the momentum transfer. The free nucleons and fragments have high momenta and different charge-to-mass ratios than the nuclear beam, and should therefore be measurable in specialized detectors. However, we do not have a one-to-one correspondence between an intact nucleus and a coherent scattering process since incoherent processes can lead to an intact nucleus in an excited state. The excited states of the nucleus without breakup can be, at least partly, identified and well measured.

The measurement of $t$-distributions together with the measurement of nuclear debris could become a source of invaluable information about the nuclear equation of state. Thus, the $J/\psi$ is an ideal probe to investigate the inner structure of gluonic fields which keep the nuclei together [1].

# References

[1] A. Caldwell and H. Kowalski, arXiv:0909.1254, DESY 09-145, MPP-2009-163

# Part V

# Central Exclusive Production

# Exclusive High Mass Di-Leptons in CDF

*Michael Albrow*

Fermi National Accelerator Laboratory, Batavia, IL60510, USA

In the Collider Detector at Fermilab, CDF, we have measured central exclusive production, $p + \bar{p} \to p + X + \bar{p}$, where $X$ is a pair of leptons or photons and nothing else. In this talk I focus on central masses $M(X) > 8$ GeV/c$^2$. We measured QED production $\gamma\gamma \to e^+e^-, \mu^+\mu^-$ up to $M(X) = 75$ GeV/c$^2$, and candidates for photoproduction of Upsilons, $\gamma I\!\!P \to Y(1S), Y(2S), Y(3S)$. I report a search for exclusive photoproduction of $Z$-bosons, and the status of searches for exclusive two-photons: $p+\bar{p} \to p+\gamma\gamma+\bar{p}$. These measurements constrain the cross section $\sigma(p+p \to p+H+p)$ at the LHC.

## 1 Introduction

By central exclusive production CEP (also called central exclusive diffraction, CED) at the Fermilab Tevatron we mean reactions $p + \bar{p} \to p + X + \bar{p}$, where $X$ is a simple system fully measured, and "+" are large rapidity gaps ($\Delta y \gtrsim 5$ units) with *no* particles. In the Collider Detector at Fermilab, CDF [1], we cannot detect the forward $p$ or, except for some large $M(X)$ events, the $\bar{p}$. However we installed scintillation counters (beam shower counters, BSC) along the beam pipe, which detected showers from particles out to pseudorapidity $|\eta| = 7.4$. Requiring them to be empty selects events in which the $p$ and $\bar{p}$ did not fragment and went down the beam pipe.

In the Standard Model the only significant coherent $t$-channel exchanges over such large rapidity gaps are colour singlets with charge $Q = 0$ and spin $J$, or effective spin $\alpha(t = 0)$, $\geq 1$. These are the photon, $\gamma$, pomeron $I\!\!P$ ($C = +1$) and odderon $O$ ($C = -1$). $Z$-boson exchange would be allowed, but the proton would inevitably break up. The odderon has not yet been convincingly observed; in our observation of exclusive $J/\psi$ and $\psi(2S)$ reported in this meeting by Pinfold [2, 3] we placed a new limit. In this talk I will report on CDF measurements of exclusive lepton pairs, $p+\bar{p} \to p+l^+l^- +\bar{p}$ above the charmonium region, see Fig. 1. This includes non-resonant (QED) $\gamma\gamma \to e^+e^-, \mu^+\mu^-$, and photoproduction: $\gamma I\!\!P \to \Upsilon, Z$. We can also search for exclusive $I\!\!P I\!\!P \to \chi_b \to \Upsilon + \gamma$, but the cross section is very small and its observation probably requires no additional collisions (no pile-up). This process, like exclusive $\chi_c$, is a good test of exclusive Higgs boson production, as the QCD part of the Feynman diagrams is identical. While our CDF studies provide good tests of QCD with large rapidity gaps, and hard pomeron behaviour, they are also precursors to processes with $M(X) \gtrsim 100\,\mathrm{GeV}/c^2$ at the LHC [4], where $X$ can be $h, H, W^+W^-, \tilde{l}^+\tilde{l}^-$ etc., including any exotic particles that couple to gluons or photons and have the right quantum numbers. If states such as Higgs bosons are seen this way at the LHC, their mass, width, spin, C-parity and coupling $\Gamma(Xgg)$ can be determined in unique ways. Even a pair of nearby states, e.g. MSSM $h(140) \to b\bar{b}$ and $H(150) \to b\bar{b}$ can be resolved, which is impossible by other means.

## 2 CDF Detectors

The central detector of CDF [1] has layers of silicon tracking and drift chambers surrounded by a time-of-flight scintillation counter barrel, in a solenoidal field. This is surrounded by electromagnetic and hadronic calorimeters, muon scintillators and tracking chambers. The forward region, $\theta < 3°$, has, on each side, a 48-channel Cherenkov luminosity counter hodoscope, a "miniplug" calorimeter, a set of beam shower counters, BSC, and on the outgoing $\bar{p}$ side, scintillating fiber trackers in Roman pots. The BSC were very important as rapidity gap detectors in no-pile-up events, and for triggering on exclusive events. They are relatively simple scintillation counters around the beam pipes covering $5.5 < |\eta| < 7.4$. Only BSC-1 sees primary particles, and it has two radiation lengths in front to convert photons; the others see showers created in the beam pipe. We have proposed them for CMS [5]; all LHC experiments should have them! Another recommendation to all experiments is to record zero-bias, or bunch-crossing, triggers routinely, e.g. at 1 Hz. These were essential in our CDF exclusive studies. We divided those events into two classes: [a] = probably no interaction, e.g. no tracks, and [b] at least one interaction, with tracks from the beam line. Then for each subdetector, e.g. BSC-1 which had 8 PMTs, we plotted the "hottest" PMT as Log(max ADC counts in BSC-1) for events in classes [a] and [b]. One can also plot the sum $\sum_i(ADC_i)$. Repeating for all subdetectors, one finds cuts that select events with all the CDF detectors empty, except for the state $X$.

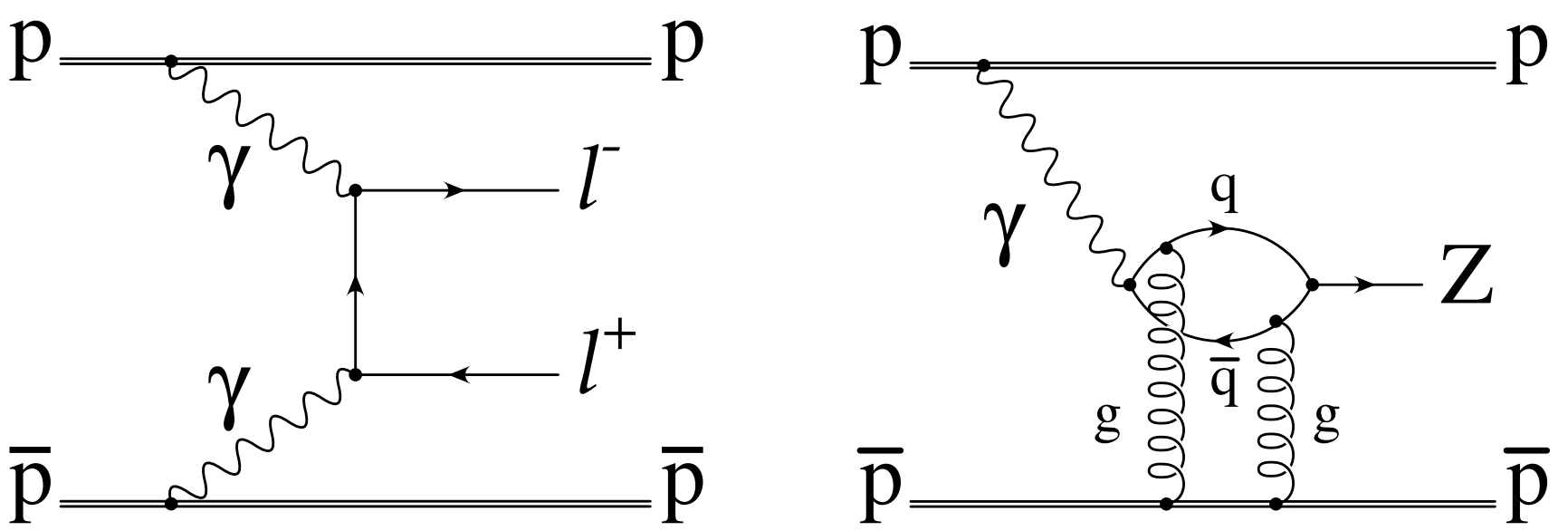

(a) Two-photon production of lepton pairs

(b) Z photoproduction (or $\Upsilon$ photoproduction if $q = b$).

Figure 1: Exclusive two-photon production of lepton pairs, and exclusive photoproduction.

## 3 CDF Exclusive Physics Programme

In March 2001 some of us proposed [6] to add very forward proton tracking detectors to CDF to look for exclusive Higgs boson production. At that time some theorists (cited in Ref. [6]) had suggested that the cross section could be as high as 10-100 fb, and a signal could be seen using the missing mass technique [7]. The Durham group prediction [8] was a factor $\gtrsim 10^2$ lower, at 0.06 fb (for M(H) = 120 GeV), impossibly low for the Tevatron. Incidentally in Ref. [6] we suggested for the first time that exclusive $\gamma\gamma$ production is a good test of the theory, and that troublesome pile-up can be highly suppressed with "fast timing Cherenkov counters". Given the very large theoretical uncertainty on $\sigma(H)$, the proposal could not proceed at that time, but we embarked on a program to measure related exclusive processes with cross sections accessible even in the Durham model. We have now measured $p + X + \bar{p}$ final states with $X = \gamma\gamma, \chi_c$ and dijets $JJ$, all of which have very similar QCD diagrams and issues, and all three are consistent with the Durham group predictions (which have a factor "a few" uncertainty). One can now be certain that it is possible to produce a Higgs boson (if it exists) at the LHC with no other particles, and reasonably confident that their prediction $\sigma(SMH(120)) \sim 1$ - 10 fb at $\sqrt{s} = 14$ TeV is not unrealistic. The cross section can be much higher in some Beyond SM scenarios.

The exclusive processes measured in CDF have different strengths and weaknesses. $X = \gamma\gamma$ is the cleanest as the photons, like the Higgs, have no strong interactions. Looking at the Feynman diagrams through a "QCD-only filter" they are identical; simply the $q$-loop (mostly $u$ and $c$) becomes a $t$-loop and $\gamma\gamma$ is replaced by $H$. But the cross section is very small. The Durham prediction [9] is 36 fb for $|\eta(\gamma)| < 1$ and $E_T(\gamma) > 5$ GeV. This corresponds to $0.8^{+1.6}_{-0.5}$ events in the CDF search [10]; 3 candidates were found, of which 2 were "perfect" with two single narrow electromagnetic showers, while the third had some characteristics of $\pi^0\pi^0$ (broader showers). Note that the two golden candidates were distilled from $10^{12}$ inelastic collisions; such things can be done! $X = \chi_c$ is also clean and has a much bigger cross section: $\frac{d\sigma}{dy}(y = 0) = 76 \pm 10(stat) \pm 10(syst)$ nb (CDF [2, 3]), compared with 90 nb (with a large uncertainty) predicted by Durham [11]. The weakness is that the charm mass is small so the process is not very perturbative, the $\chi_c$ is colourless but it is still a hadron with final state interactions, and resolving $\chi_{c0}(1P), \chi_{c1}(1P)$, and $\chi_{c2}(1P)$ is difficult. The $\chi_b$ is more perturbative than the $\chi_c$, so the theory is under better control, but the cross section is expected to be only about 1/500th that of the $\chi_c$ and the decay modes are not well known. For the $\chi_{b0}$ the Particle Data Group gives only $\chi_{b0}(1P) \to \gamma\Upsilon(1S) < 6\%$, with no other modes well known, so it will not be possible to give a cross section soon. (Some hadronic decays with poorly known branching fractions are impossible to trigger on in a normal collider environment.) Finally CDF also measured [12] exclusive di-jets, produced by the process $gg \to gg$ with a colour-cancelling gluon exchange. The cross section is quite large, $\approx 100$ pb for two jets with $E_T > 15$ GeV and $|\eta| < 2.5$, with the dijet having $> 80\%$ of the total central mass, i.e. $R_{JJ} > 0.8$. In this region there is an excess of dijets compared with inclusive dijet expectations. The data, out to Jet $E_T^{min} = 35$ GeV, agree within a factor $\sim 3$ with the Durham prediction [13], but are more than an order of magnitude lower than the DPEMC prediction [14].

The main particle states that can be produced exclusively in double pomeron exchange, $D\mathbb{P}E$, are $\sigma(600), f_0(980), \chi_c, \chi_b$ and $H$. The first two were observed at the ISR [15] and, less cleanly, at the SPS (fixed target), and the $\chi_c$ is now established [2, 3]. The other accessible states are $\gamma\gamma$ and $JJ$. It will be very hard to get a good ($\lesssim 25\%$) measurement of the $\chi_b$, but $H$ appears to be in reach at the LHC, provided the two protons can be measured. (Double proton tagging for $p + \chi_b + p$ at high enough luminosity to get some events is probably not feasible, due to its low mass.)

In CDF we observed $\gamma\gamma \to e^+e^-$ collisions with $E_T(e) > 5$ GeV for the first time in hadron-hadron collisions. We published [16] 16 events (background $\sim 1.9$) with $M(e^+e^-) > 10$ GeV/c$^2$ in excellent agreement with the LPAIR Monte Carlo. The highest mass event was at 38 GeV/c$^2$, and all pairs are very back-to-back with $(180° - \Delta\phi) < 2.4°$. We followed that with a search [17] for exclusive $Z$, in the process finding eight $e^+e^-$ or $\mu^+\mu^-$ events with $M(l^+l^-)$ from 40 to 75 GeV/c$^2$. Again "QED rules", and the lepton pairs all have $(180° - \Delta\phi) < 0.75°$. This mass reach is as high as (or higher than) $e^+e^-$ at LEP (which has no strongly interacting background) and $ep$ at HERA. This demonstrates that exclusive dileptons can be extracted from the huge backgrounds in hadron-hadron collisions, which is good news as they provide an excellent (probably the best) calibration of the momentum scale and resolution of the high precision proton spectrometers being planned [4] for ATLAS and CMS. One does not need to see both protons to calibrate the spectrometers, as each one is very well known (ultimately limited by the incoming beam momentum spread, $\frac{dp}{p} \sim 10^{-4}$). The use of exclusive QED dileptons, with precisely known cross section, to measure the machine luminosity (integrated over a period, perhaps days), has been suggested. Unfortunately the precision is likely to be limited by unseen proton dissociation, and knowledge of efficiencies, acceptance and non-exclusive background. (One cannot require no pile-up, as the result would then depend on $\sigma_{inel}$ which is *a priori* unknown).

Exclusive $Z$ production is allowed in the Standard Model: a radiated virtual photon fluctuates to a $q\bar{q}$ pair, which scatters by hard pomeron exchange on the other proton, followed by $q\bar{q} \to Z$, as in exclusive vector meson photoproduction, see Fig. 1. However the SM cross section is much too small at the Tevatron: 0.3 fb [18] or 0.21 fb [19]. At the LHC (14 TeV) the predictions are 13 fb [18] and 69 fb [19], which may make an observation possible. A signal at the Tevatron, or a significantly higher cross section at the LHC, would be evidence for new particles with strong and electroweak couplings. White's theory [20] of the supercritical pomeron predicts colour-sextet quarks coupling strongly to the pomeron and to the $W$ and $Z$, and he expects a much enhanced cross section, but without a quantitative prediction. We used a sample of $3.17 \times 10^5$ lepton pairs with $M(l^+l^-) > 40$ GeV/c$^2$, of which $1.83 \times 10^5$ were in the $Z$ peak. We required exclusivity over the full range $-7.4 < |\eta| < +7.4$, finding 8 events, agreeing with QED expectations. Fig. 2 shows the mass and azimuthal difference distributions of these events. All events were very back-to-back in the transverse plane, with $(\pi - \Delta\phi) < 0.75°$. One event with $M(\mu^+\mu^-) = 66$ GeV/c$^2$ had a $\bar{p}$ track in the Roman pots; for the others the $\bar{p}$ was out of their acceptance or they were not operational. None of the eight exclusive events were $Z$-candidates, and a limit was placed: $\sigma_{excl}(Z) < 0.96$ pb at 95% C.L.[1] A nice check of the exclusivity analysis comes from $W \to l^{\pm}\nu$ events, which cannot be exclusive, but otherwise are very similar to $Z$, and which are more abundant. It may be possible to improve this limit using a factor 2-3 more data and including pile-up events, using the requirements of no associated tracks on the $l^+l^-$ vertex, small $(\pi - \Delta\phi)$ and $p_T(l^+l^-)$. We are testing this method on the Upsilon region, with $8 < M(l^+l^-) < 40$ GeV/c$^2$. The QED is a good control, and the photoproduced $Y$ states have cross sections that are within reach, although not very well known. (The HERA data [21] do not resolve clearly the $Y(1S), Y(2S)$ and $Y(3S)$ states and have quite large uncertainties.) Predictions for $\frac{d\sigma}{dy}(Y(1S), y = 0)$ are [18, 22, 23, 24] are around $5 - 14$ pb. Applying the branching fraction to $\mu^+\mu^-$ or $e^+e^-$, $B = 0.025$, would give a few hundred events in 2 fb$^{-1}$ ($\times$ the acceptance and efficiency). HERA has provided a nice compilation [25] of exclusive cross sections for vector mesons from $\rho$ to $Y(1S)$ vs. $W = \sqrt{s(\gamma p)}$. In CDF with $y = 0$ we have $W(J/\psi) \sim 80$ GeV and $W(Y(1)) \sim 136$ GeV. At HERA the ratio of these cross sections is $\sim 300$. We are studying this region both in $\mu^+\mu^-$ and $e^+e^-$ events.

For the dimuons we used a trigger with two muons with $p_T(\mu) > 4$ GeV/c and $|\eta| < 0.6$. The inclusive $M(\mu^+\mu^-)$ spectrum shows the states $Y(1S), Y(2S)$ and $Y(3S)$ as well-separated peaks, see e.g. Ref. [26]; the mass resolution is only $\sigma(M) \approx 50$ MeV/c$^2$, less than the mass differences. Separation is important in order to measure exclusive $\chi_b \to \Upsilon + \gamma$, which feed differently the three states. Inclusively the ratio $\Upsilon(1S)$:continuum is about 8:1. Requiring

[1] Intriguingly, an event with $M(e^+e^-) = 92$ GeV/c$^2$, and $(\pi - \Delta\phi) = 1.25°$ (larger than expected for QED) was rejected; it failed the exclusivity requirement only in the BSC counters on one arm. It has the characteristics of a photoproduced $Z$ but with a proton dissociation. However with only one event no claim can be made.

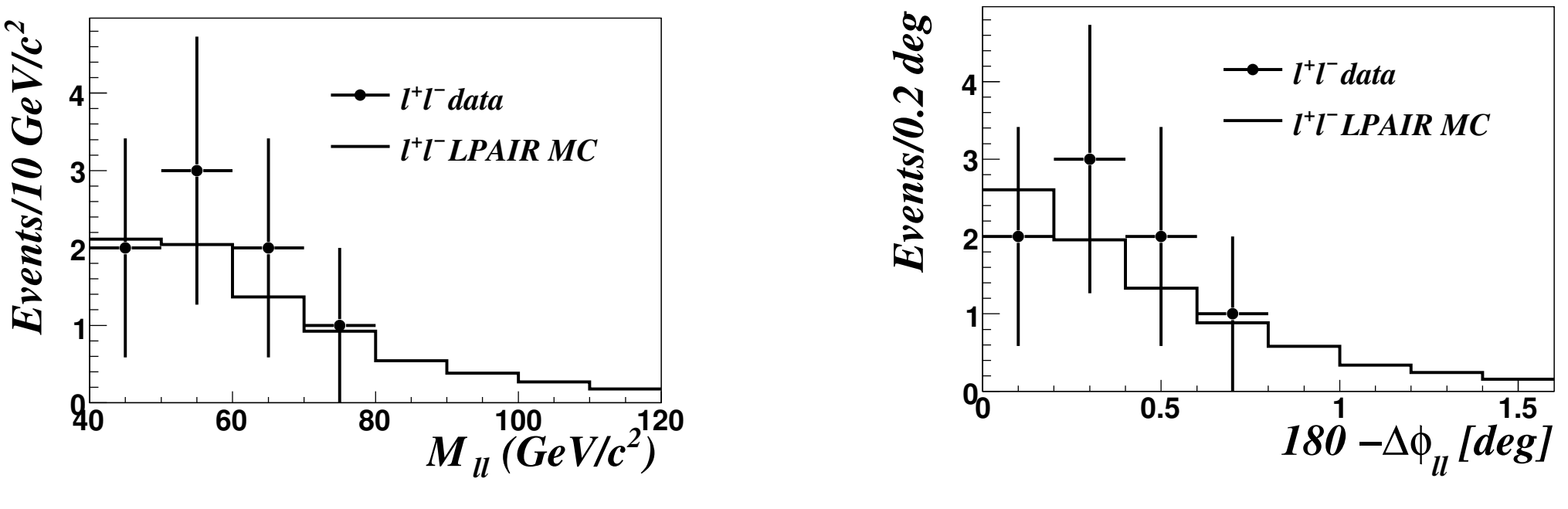

(a) Mass spectrum of exclusive lepton pairs.

(b) $180° - \Delta\phi$ between the leptons.

Figure 2: Mass distribution of exclusive lepton pairs and azimuthal angle difference (from 180°), data and normalised LPAIR prediction.

no other tracks on the $\mu^+\mu^-$ vertex, $\pi - \Delta\phi < 0.1$ rad and $p_T(\mu^+\mu^-) < 1.5$ GeV/c retains both $\Upsilon$ and continuum events, but we cannot yet claim that these events are exclusive. These cuts should be efficient for the QED events and for most of the $Y$'s; the issue is non-exclusive backgrounds in the presence of other interactions. We are now studying the continuum to see if it is exactly as expected for QED, as a control, and can then give $Y$-photoproduction cross sections, depending on unknown backgrounds from $\chi_b \to Y + \gamma$. Unfortunately neither the $\chi_b$ production rates nor their radiative decays are known. The $p_T(Y)$ distribution is broader for $\chi_b$-daughters, which may help, but one would like to reconstruct the photons, which is probably not possible with pile-up. The same issues will confront us at the LHC.

For the $Y \to e^+e^-$ channel we chose to veto pile-up, and used a trigger requiring two electromagnetic showers with $E_T > 2$ GeV and $|\eta| < 2$m, and forward gaps (BSC-1 empty). Our prime motivation for this trigger is to search for additional exclusive $\gamma\gamma$ candidates, and hopefully to make a definitive observation. Compared with our earlier search, we lowered the trigger threshold from 4 GeV to 2 GeV, took more data, will expand the $\eta$-coverage and use better background ($\pi^0$) rejection. Analysis is underway. The same trigger collects QED $e^+e^-$ pairs with $M(e^+e^-) > 8$ GeV/c$^2$, and $Y \to e^+e^-$ decays. The QED data is a good control of our exclusivity cuts, and the $Y$ candidates can be compared with our $Y \to \mu^+\mu^-$ candidates. As these events are without pile-up, we may have some $Y + \gamma$ candidates attributable to $\chi_b$ (but not many are expected). The decay photon tends to be soft (namely 391(442) MeV in the $Y$-frame) for the $\chi_{b0(b2)}$, and is poorly measured. Other exclusive final states are probably not useful, mainly because of trigger limitations.

Can exclusive $Y$ photoproduction be seen with a forward proton tag at 420 m at the LHC? We have $(1 - x_F)_{1,2} = \frac{1}{\sqrt{s}} \sum_{leptons} p_T e^{+(-)\eta}$ ($1 - x_F$ is the fractional momentum loss of the proton), and the most likely kinematics are $p_{T,1} = p_{T,2} \approx 4.75$ GeV/c, $\eta_1 \approx \eta_2$. To have one proton with $1 - x_F$ as large as (say) 0.01 we need a muon pair with $\eta \sim 2.5$, at the limit of the muon coverage in CMS. A more serious problem with using $p + Y(\to \mu^+\mu^-) + p$ for proton calibration is that many $\Upsilon$s, perhaps even most, will be decay products of $\chi_b$ states and the proton momenta are then not known. For these reasons the QED $\gamma\gamma \to \mu^+\mu^-$ with $M(\mu^+\mu^-) \gtrsim 10$ GeV/c$^2$ will probably be the calibration channel. Electron pairs are less favourable due to final state radiation and bremsstrahlung. With $|\eta_{max}| = 2$ and $M_{min} =$ 20 GeV/c$^2$ the cross section is 1.6 pb. One proton will usually have much too small $\xi$, but fortunately both protons are known and one can be used to calibrate the spectrometers.

In conclusion, five years ago the predictions for exclusive Higgs production had more than two orders of magnitude spread. Since then in CDF we have measured three related processes, exclusive $JJ, \gamma\gamma$ and $\chi_c$, all consistent with the predictions of the Durham group within the quoted factor of "a few". Especially the $\chi_c$ observation means that exclusive Higgs production must happen, if there is a Higgs boson. Our observation of exclusive photoproduced vector mesons demonstrates that exclusive $Z$ photoproduction must be possible, albeit with a small cross section (in the Standard Model) at the LHC. Our observations of exclusive $\gamma\gamma \to e^+e^-, \mu^+\mu^-$ (with a forward proton detected) are encouraging for forward spectrometer calibrations, and mean that $\gamma\gamma \to W^+W^-$ (and $\tilde{l}^+\tilde{l}^-$ if sleptons exist) could be measured. This bodes well for a rich physics program with high precision forward spectrometers at the LHC.

## Acknowledgements

I acknowledge support from the U.S. Dept. of Energy through Fermi National Accelerator Laboratory.

## References

[1] D. Acosta et al. (CDF Collaboration), Phys. Rev. D **71**, 032001 (2005) and references therein.

[2] T. Aaltonen et al. (CDF Collaboration), *Observation of exclusive charmonium and $\gamma\gamma \to \mu^+\mu^-$ in $p\bar{p}$ collisions at $\sqrt{s}$ = 1.96 TeV*, Phys. Rev. Lett. **102**, (2009) 242001.

[3] J. Pinfold, *Exclusive charmonium production in CDF*, talk at this conference.

[4] M.G. Albrow et al. (FP420 R&D Collaboration), *Higgs and new physics with forward protons at the LHC*, arXiv:0806.0302[hep-ex], to be published in J.Inst.

[5] M.G. Albrow et al., *Forward Shower Counters (FSC) in CMS, a proposal*, CMS Internal (Dec. 2007); M.G. Albrow et al., *Forward physics with rapidity gaps at the LHC*, arXiv:0811.0120, submitted to J. Inst.

[6] M.G. Albrow et al., *A search for the Higgs boson using very forward tracking detectors with CDF* (2001), hep-ex/0511057.

[7] M.G. Albrow and A. Rostovtsev, *Searching for the Higgs boson at hadron colliders using the missing mass method*, hep-ph/0009336.

[8] V.A. Khoze, A.D. Martin and M.G. Ryskin, *Can the Higgs be seen in rapidity gap events at the Tevatron or the LHC?*, Eur.Phys J **C14**, 525 (2000).

[9] V.A. Khoze, A.D. Martin, M.G. Ryskin and W.J. Stirling, *Diffractive $\gamma\gamma$ production at hadron colliders*, Eur. Phys. J. **C 38**, 475 (2005).

[10] A. Abulencia et al. (CDF Collaboration), *Search for exclusive $\gamma\gamma$ production in hadron-hadron collisions*, Phys. Rev. Lett. **99**, (2007) 242002.

[11] V.A. Khoze, A.D. Martin, M.G. Ryskin and W.J. Stirling, *Double diffractive $\chi$ meson production at hadron colliders*, Eur. Phys. J. **C 35**, 211 (2004).

[12] T. Aaltonen et al.(CDF Collaboration), *Observation of exclusive dijet production at the Fermilab Tevatron $\bar{p}p$ collider*, Phys. Rev. D **77**, 052004 (2008).

[13] J. Monk and A. Pilkington (ExHuME), Comput. Phys. Commun. **175**, 232 (2005).

[14] M. Boonekamp and T. Kucs (DPEMC), Comput. Phys. Commun. **167**, 217 (2005).

[15] T. Åkesson et al. (AFS Collaboration), *A search for glueballs and a study of double pomeron exchange at the CERN ISR*, Nucl. Phys. B **264**, 154 (1986).

[16] A. Abulencia et al. (CDF Collaboration), *Observation of exclusive electron-positron production in hadron-hadron collisions*, Phys. Rev. Lett. **98**, (2007) 112001.

[17] T. Aaltonen et al. (CDF Collaboration), *Search for exclusive Z-boson production and observation of high-mass $p\bar{p} \to p\gamma\gamma\bar{p} \to pl^+l^-\bar{p}$ events in $p\bar{p}$ collisions at $\sqrt{s}$ = 1.96 TeV*, Phys. Rev. Lett. **102**, (2009) 222002.

[18] L. Motyka and G. Watt, *Exclusive photoproduction at the Fermilab Tevatron and CERN LHC within the dipole picture*, Phys. Rev. D **78**, 014023 (2008).

[19] V.P. Goncalves and M.V.T. Machado, *Diffractive photoproduction of Z bosons in coherent interactions at the CERN LHC*, Eur. Phys. J. **C 56**, 33 (2008); Erratum-ibid. **C 61**:351 (2009).

[20] A.R. White, *Physics of a sextet quark sector*, Phys. Rev. D **72**, 036007 (2005).

[21] ZEUS Collaboration, S. Chekanov et al., *Exclusive photoproduction of $\Upsilon$ mesons at HERA*, arXiv:0903.4205 (2009).

[22] S. Klein and J. Nystrand, *Photoproduction of $J/\psi$ and $\Upsilon$ in $pp$ and $\bar{p}p$ collisions*, hep-ph/0310223.

[23] A. Bzdak, L. Motyka, L. Szymanowski and J.R. Cudell, *Exclusive $J/\psi$ and $\Upsilon$ hadroproduction and the QCD odderon*, Phys. Rev. D **75**, 094023 (2007).

[24] A. Rybarska, W. Schafer, and A. Szczurek, *Exclusive photoproduction of $\Upsilon$: from HERA to Tevatron*,Int. J. Mod. Phys. **A 24**, 557 (2009); Phys. Lett. **B 668**, 126 (2008).

[25] See e.g. M. Klein and R. Yoshida, *Collider physics at HERA*, Prog.Part.Nucl.Phys. **61**, 343 (2008); arXiv:0805.3334.

[26] D. Acosta et al. (CDF Collaboration), *$\Upsilon$ production and polarization in $p\bar{p}$ collisions at $\sqrt{s}$ = 1.8 TeV*, Phys. Rev. Lett. **88**, 161802 (2002).

# Exclusive Charmonium Production at CDF

*James L. Pinfold*[1]

[1]Physics Department, University of Alberta, Edmonton,
Alberta T6G 0V1, Canada

We have observed the reactions $p\bar{p} \to pX\bar{p}$, with X being a centrally produced $J/\psi$, $\psi(2S)$, or $\chi_{c0}$, and $\gamma\gamma \to \mu^+\mu^-$ in $p\bar{p}$ collisions at $\sqrt{s} = 1.96\,\text{TeV}$. The required event topology consists of two oppositely charged central muons, and either no other particles or one additional photon detected. Exclusive vector meson production is as expected for elastic photoproduction, $\gamma p \to J/\psi(\psi(2S))p$, observed here for the first time in hadron-hadron collisions. We also observe exclusive $\chi_{c0} \to J/\psi + \gamma$ decays. The cross sections $\frac{d\sigma}{dy}|_{y=0}$ for $J/\psi$, $\psi(2S)$, and $\chi_{c0}$ are (3.92 ± 0.25(stat) ± 0.52(syst)) nb, (0.53 ± 0.09(stat) ± 0.10(syst)) nb, and (76 ± 10(stat) ± 10(syst)) nb, respectively, and the continuum is consistent with QED.

## 1 Introduction

In central exclusive production processes, $p+\bar{p} \to p+X+\bar{p}$ the colliding hadrons emerge intact with small transverse momenta, and the produced state X is in the central region, with small rapidity $|y|$, and is fully measured. If regions of rapidity exceeding about 5 units are devoid of particles, only photon and Pomeron [1], *IP*, exchanges are significant, where *IP* consists mostly of two gluons in a colour singlet state with charge parity $C = 1$ state [2].

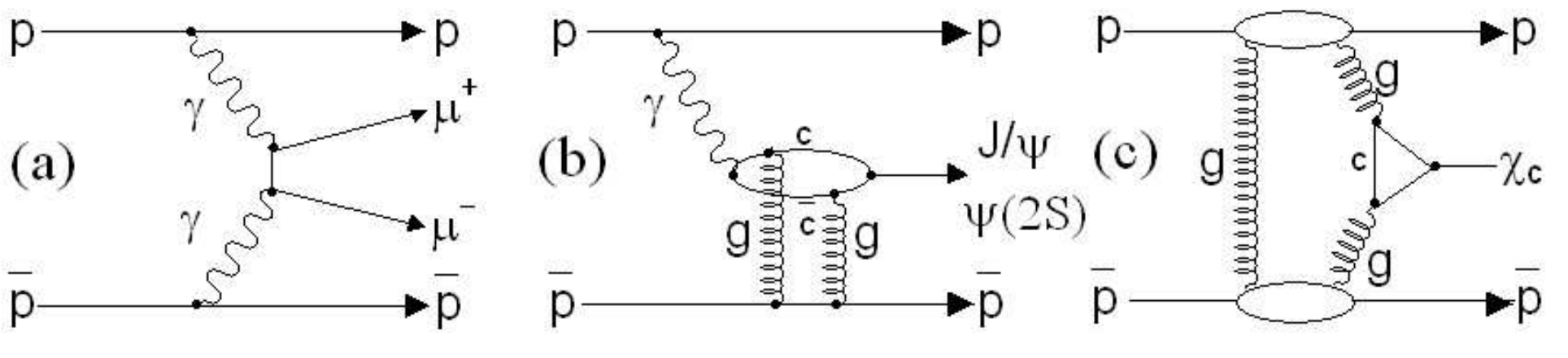

Figure 1: Feynman diagrams for (a) $\gamma\gamma \to \mu^+\mu^-$, (b) $\gamma IP \to J/\psi(\psi(2S))$, and (c) $IPIP \to \chi_c$, with the 2-gluon exchange forming a Pomeron.

In these proceedings we report measurements of exclusive dimuon production, $X \to \mu^+\mu^-$, with 3.0 GeV/c$^2$ $\leq M \leq$ 4.0 GeV/c$^2$, directly [QED, Figure 1(a)], or from photoproduced $J/\psi$ or $\psi(2S)$ [Figure 1(b)] decay, and $\chi_{c0} \to J/\psi + \gamma \to \mu^+\mu^-\gamma$ [Figure 1(c)]. Lower masses were excluded by muon range, and higher masses by trigger rate limitations. Exclusive photoproduction of vector mesons has been measured in ep collisions at HERA [3], but not previously observed in hadron-hadron collisions. The theoretical uncertainty on the QED cross section is $< 0.3\%$; this process is distinct from Drell-Yan production, which is negligible in this regime.

## 2 The CDF Detector

We used $p\bar{p}$ collision data at $\sqrt{s} = 1.96\,\text{TeV}$ corresponding to an integrated luminosity L = 1.48 $\text{fb}^{-1}$ delivered to the CDF-II detector. This is a general purpose detector described in more detail elsewhere [4]. Surrounding the collision region is a tracking system consisting of silicon microstrip detectors and a cylindrical drift chamber in a 1.4 Tesla solenoidal field. The tracking system has 100% efficiency for reconstructing isolated tracks with $p_T \geq 1\,\text{GeV/c}$ and pseudorapidity $|\eta| < 0.6$.

A barrel of 216 time-of-flight counters outside the cylindrical drift chamber is surrounded by calorimeters with separate electromagnetic (EM) and hadronic sections covering the range $|\eta| < 3.6$. Drift chambers outside the calorimeters were used to measure muons with $|\eta| < 0.6$ [5]. The regions $3.6 < |\eta| < 5.2$ are covered by lead-liquid scintillator calorimeters [6]. Gas Cherenkov counters covering $3.7 < |\eta| < 4.7$ determined the luminosity with a 6% uncertainty [7]. We did not have detectors able to measure the forward $p$ and $\bar{p}$, but beam shower scintillation counters (BSC1- BSC3), located along the beam pipe, can detect products of $p(\bar{p})$ fragmentation over the range $|\eta| < 7.4$.

### The Event Selection

The level 1 trigger required at least one muon track with $p_T > 1.4\,\text{GeV/c}$ and no signal in BSC1 ($5.4 \leq |\eta| \leq 5.9$), and a higher level trigger required a second track with opposite charge. The offline event selection closely followed that described in [8] where we observed exclusive $e^+e^-$ production. We required two oppositely charged muon tracks, each with $p_T > 1.4\,\text{GeV/c}$ and $|\eta| < 0.6$, accompanied by either (a) no other particles in the event or (b) only one additional EM shower with $E_T^{EM} > 80\,\text{MeV}$ and $|\eta| < 2.1$. Condition (a) defines an exclusive dimuon event. The exclusivity efficiency $\varepsilon_{exc}$ is the probability that the exclusive requirement is not spoiled by another inelastic interaction in the same bunch crossing, or by noise in a detector element. This efficiency was measured as the fraction of bunch crossing triggers that pass the exclusivity requirement (a). We found $\varepsilon_{exc} = 0.093$ with negligible uncertainty. The product $\varepsilon_{exc}L = L_{eff} = (139 \pm 8)\,\text{pb}^{-1}$ is the effective luminosity for single interactions.

After these selections, cosmic rays were the main background. They were all rejected, with no significant loss of real events, by timing requirements in the time-of-flight counters and by requiring the three-dimensional opening angle between the muon tracks to be $\Delta\theta_{3D} < 3.0\,\text{rad}$. Within a Fiducial Kinematic Region (FKR) [$|\eta(\mu)| < 0.6$ and $3.0\,\text{GeV}/c^2 \leq M_{\mu\mu} \leq 4.0\text{GeV}/c^2$], there are 402 events with no EM shower. The $M_{\mu\mu}$ spectrum is shown in Figure 2. The $J/\psi$ and $\psi(2S)$ are prominent, together with a continuum. The spectrum is well fitted by two Gaussians with expected masses and widths (dominated by the resolution) and a continuum whose shape is given by the product of the QED spectrum ($\gamma\gamma \to \mu^+\mu^-$), acceptance, and efficiency, as shown in Figure 2(inset).

Backgrounds to exclusive $\mu^+\mu^-$ events are (a) proton fragmentation, if the products are not detected in the forward detectors, (b) for the $J/\psi$ , $\chi_{c0}$ events with a photon that did not give an EM shower above 80 MeV, and (c) events with some other particle not detected. The probability of a $p$ or $\bar{p}$ fragmenting at the $p\gamma p(p^*)$ vertex was calculated with the LPAIR Monte

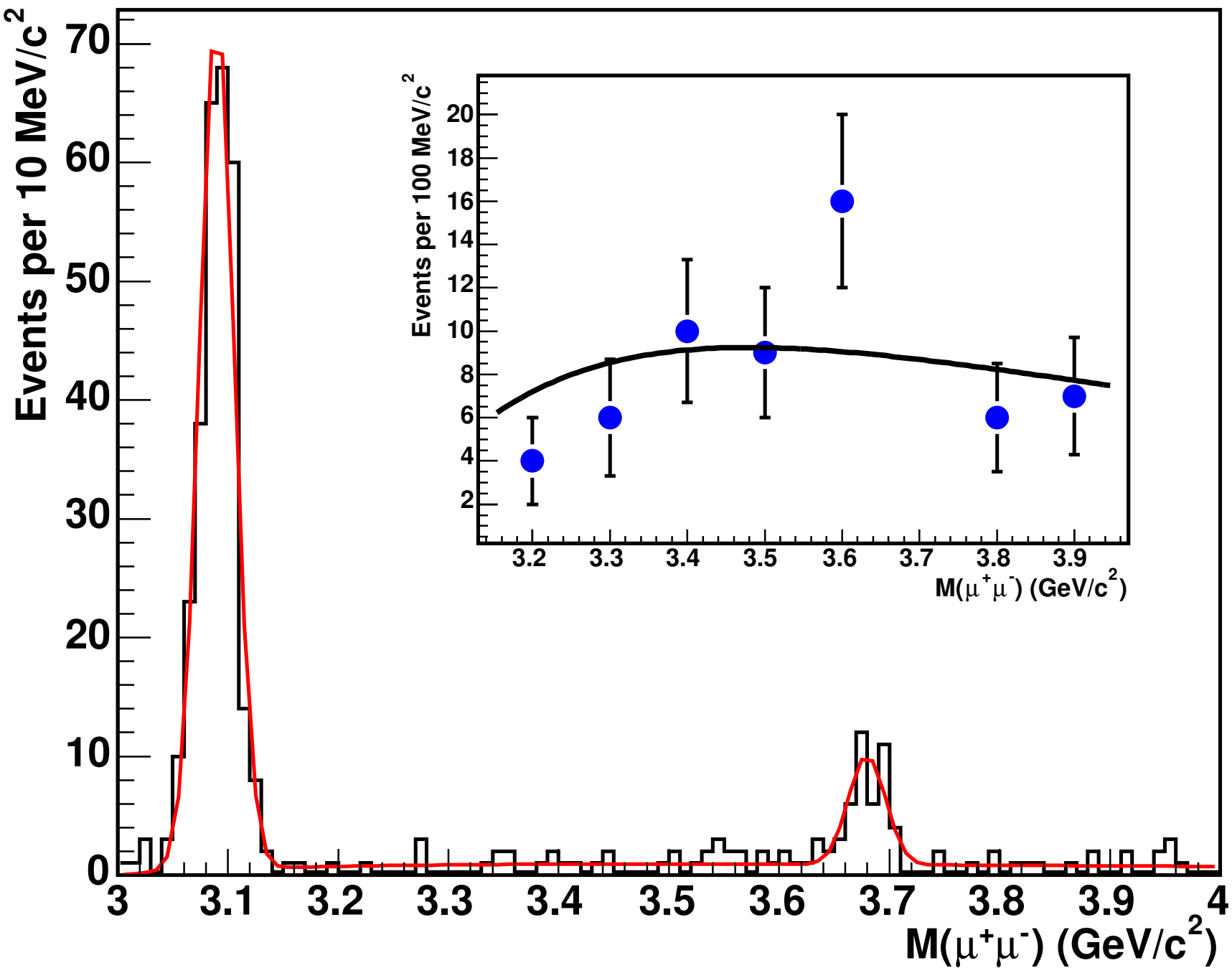

Figure 2: Mass distribution of 402 exclusive events, with no EM shower (histogram, together with a fit to two Gaussians for the $J/\psi$ and $\psi(2S)$, and a QED continuum. All three shapes are predetermined, with only the normalisations floating. Inset - Data above the $J/\psi$ and excluding $3.65 \leq M_{\mu^+\mu^-} \leq 3.75\,\mathrm{GeV}/c^2$ [$\psi(2S)$] with the fit to the QED spectrum times acceptance (statistical uncertainties only).

Carlo (MC) simulation [9] to be $0.17\pm0.02$ (syst), and the probability that all the fragmentation products have $|\eta| < 7.4$ to be $0.14 \pm 0.02$ (syst). If a proton fragments, the decay products may not be detected through BSC inefficiency, estimated from data to be $0.08 \pm 0.01$. The fragmentation probability at the $pIPp(p^*)$ vertex was taken from the ratio of single diffractive fragmentation to elastic scattering at the Tevatron [10] to be $0.24 \pm 0.05$.

We compared the kinematics of the muons, e.g. $p_T(\mu^+\mu^-)$ and $\Delta\phi_{\mu\mu}$, with simulations for the $J/\psi$, $\psi(2S)$ [11], and QED [9] with $3.2 \leq M_{\mu\mu} \leq 3.6$ and $3.8 \leq M_{\mu\mu} \leq 4.0\,\mathrm{GeV/c}^2$ to exclude the $J\psi$ and $\psi(2S)$. The distributions agree well with the simulations; the few events that are outside expectations are taken to be nonexclusive background. As expected, $\langle p_T \rangle$ is smaller for the QED process, and the data agree well with STARLIGHT [11], apart from two events with $p_T > 0.8\,\mathrm{GeV/c}$ where no events are expected. Comparing data with LPAIR we estimate that the nonexclusive background is $9 \pm 5\%$ of the observed (QED) events. The $\psi(2S)$data are well fitted by the STARLIGHT photoproduction simulation [11]. The distribution of $p_T(J/\psi)$ is well fitted by STARLIGHT, apart from five events with $p_T$ >1.4 GeV/c. These could be due to nonexclusive background, some $\chi_{c0}$ radiative decays with an undetected photon, or an Odderon component.

To measure $\chi_{c0}$ production we required one EM shower $E_T^{EM} > 80$ MeV in addition to two muons. There are 65 events in the $J/\psi$ peak and eight continuum events; these are likely to be $\gamma\gamma \to \mu^+\mu^-$ with a bremsstrahlung. We interpret the 65 events as $\chi_{c0} \to J/\psi + \gamma$ production and decay. The distribution of the mass formed from the $J/\psi$ and the EM shower energy, while broad, has a mean value equal to the $\chi_{c0}$ mass. The $E_T^{EM}$ spectrum is well fitted by an empirical function which extrapolates to only 3.6 ± 1.3(syst) $\chi_{c0}$ candidates with showers below 80 MeV. The $p_T(J/\psi)$ and $\Delta\phi_{\mu\mu}$ distributions for the events with an$E_T^{EM}$ signal are consistent with all these $J/\psi$ being from $\chi_{c0}$ decay, as simulated by the CHICMC Monte Carlo [12]. Additional photon inefficiency comes from conversion in material, 7 ± 2%, and dead regions of the calorimeter, 5.0 ± 2.5%, giving a total inefficiency 17 ± 4%, which gives a background to exclusive $J\psi$ of 4.0 ±1.6% (all errors systematic).

Figure 2 (inset) shows the subset of the data above 3.15 GeV/c$^2$ (to exclude the $J/\psi$), excluding the bin 3.65-3.75 GeV/c$^2$ which contains the $\psi(2S)$. The curve shows the product of the QED spectrum and acceptance × efficiency, A$\varepsilon$, with only the normalization floating, from the 3-component fit to the full spectrum. The continuum data agrees with the QED expectation. The integral from 3 GeV/c$^2$ to 4 GeV/c$^2$ 2 is 77 ± 9(stat) events, and after correcting for backgrounds and efficiencies, the measured cross section for QED events with $|\eta|(\mu^\pm) < 0.6$ and $3.0 \leq M_{\mu\mu} \leq$ GeV/c$^2$ is $\sigma = 2.7 \pm 0.3(stat) \pm 0.4$ pb, in agreement with the QED prediction 2.18 ± 0.01 pb [9].

For the prompt $J/\psi$ and $\psi(2S)$ cross sections, we took the number of events from the Gaussian fits, subtracted backgrounds, and corrected for A$\varepsilon$ to obtain $BR.\sigma_{FKR}$ for both muons in the fiducial kinematic region. To obtain $\frac{d\sigma}{dy}|_{y=0}$ from $\sigma_{FKR}$ we used the STARLIGHT MC program, which gives the ratio of these two cross sections for each resonance, and divided by the branching fractions BR. We found $\frac{d\sigma}{dy}_{y=0}(J/\psi) = 3.93 \pm 0.25(\text{stat}) \pm 0.52(\text{syst})$ nb. This agrees with the predictions $2.76^{+0.6}_{-0.2}$nb [11] and 3.4 ± 0.4 nb [13] among others [14, 15]. We found $\frac{d\sigma}{dy}_{y=0}(\psi(2S)) = 0.53 \pm 0.09(\text{stat}) \pm 0.10(\text{syst})$ nb compared with a prediction [11] $0.46^{+0.11}_{-0.04}$. The ratio $\text{R} = \frac{\psi(2S)}{J/\psi} = 0.14 \pm 0.05$ is in agreement with the HERA value [3] R = 0.166 ± 0.012 at similar $\sqrt{(\gamma p)}$.

After correcting the 65 $\chi_{c0}$ candidates for backgrounds and efficiencies, and applying the branching fraction $BR(\chi_{c0} \to J/\psi + \gamma) = 0.128 \pm 0.0011$ [16], we found $\frac{d\sigma}{dy}_{y=0}(\chi_{c0}) = 76 \pm 10(\text{stat}) \pm 10(\text{syst})$ nb. Reference [17] predicted $\frac{d\sigma}{dy}_{y=0}(\chi_{c0}) = 130$ nb; however, the Particle Data Group (PDG) value [16] of the $\chi_c$ width has been reduced by a factor 1.45 correcting this prediction to 90 nb. Yuan [18] predicted 160 nb (again the factor 1.45 should be applied) and Bzdak [19] 45 nb.

In conclusion we have observed, for the first time in hadron-hadron collisions, exclusive photoproduction of $J/\psi$ and $\psi(2S)$, exclusive double Pomeron production of $\chi_{c0}$, and the QED process $\gamma\gamma \to \mu^+\mu^-$. The photoproduction process has previously been studied in *ep* collisions at HERA, with similar kinematics ($\sqrt{s(\gamma p)}$ ~100 GeV), and t the cross sections are in agreement. Our observation of exclusive $\chi_{c0}$ production implies that exclusive Higgs boson production should occur at the LHC [20] and imposes constraints on the $p + \bar{p} \to p + H + \bar{p}$ cross section.

# References

[1] See, e.g., J. R. Forshaw and D. A. Ross, Quantum Chromodynamics and the Pomeron (Cambridge University Press, Cambridge, England, 1997); S. Donnachie et al., Pomeron Physics and QCD (Cambridge University Press, Cambridge, England, 2002).

[2] A. Schafer, L. Mankiewicz, and O. Nachtmann, Phys. Lett. B 272, 419 (1991); V. A. Khoze, A. D. Martin, and M. G. Ryskin, Eur. Phys. J. C 24, 459 (2002); C. Ewerz, arXiv:hep-ph/0306137.

[3] See, e.g., H. Jung, Acta Phys. Pol. Supp. 1, 531 (2008), and references therein.

[4] D. Acosta *et al.* (CDF Collaboration), Phys. Rev. D 71, 032001 (2005) and references therein.

[5] G. Ascoli *et al.*, Nucl. Instrum. Methods Phys. Res., Sect. A 268, 33 (1988).

[6] M. Gallinaro *et al.*, IEEE Trans. Nucl. Sci. 52, 879 (2005).

[7] D. Acosta *et al.*, Nucl. Instrum. Methods Phys. Res., Sect. A 494, 57 (2002).

[8] A. Abulencia *et al.* (CDF Collaboration), Phys. Rev. Lett. 98, 112001 (2007).

[9] J. A. M. Vermaseren, Nucl. Phys. B 229, 347 (1983); S. P. Baranov *et al.*, in Proc. Physics at HERA (DESY, Hamburg, 1991), p. 1478.

[10] F. Abe *et al.* (CDF Collaboration), Phys. Rev. D 50, 5518 (1994); 50, 5535 (1994).

[11] S. Klein and J. Nystrand, Phys. Rev. Lett. 92, 142003 (2004); (private communication).

[12] W. J. Stirling (private communication).

[13] L. Motyka and G. Watt, Phys. Rev. D 78, 014023 (2008).

[14] W. Schafer and A. Szczurek, Phys. Rev. D 76, 094014 (2007).

[15] V. P. Goncalves and M.V. T. Machado, Eur. Phys. J. C 40, 519 (2005).

[16] C. Amsler *et al.*, Phys. Lett. B 667, 1 (2008).

[17] V. A. Khoze, A. D. Martin, M. G. Ryskin, and W. J. Stirling, Eur. Phys. J. C 35, 211 (2004); V. A. Khoze, A. D. Martin, and M. G. Ryskin, Eur. Phys. J. C 19, 477 (2001); 20, 599(E) (2001); (private communication).

[18] F. Yuan, Phys. Lett. B 510, 155 (2001).

[19] A. Bzdak, Phys. Lett. B 619, 288 (2005).

[20] M. G. Albrow et al., arXiv:0806.0302 [J. Inst. (to be published)].

# Central Exclusive $\chi_c$ Production

*Oleg V. Teryaev*[1], *Roman S. Pasechnik*[1] *and Antoni Szczurek*[2]

[1]Bogoliubov Laboratory of Theoretical Physics, JINR, Dubna 141980, Russia
[2]Institute of Nuclear Physics PAN, PL-31-342 Cracow, Poland and
University of Rzeszów, PL-35-959 Rzeszów, Poland

The amplitudes of the central exclusive production of $\chi_c$ mesons are calculated using different unintegrated gluon distribution functions (UGDFs). The procedure of generalisation of UGDF for non-forward case by saturation of positivity constraints is suggested. We compare exclusive production of all charmonium states $\chi_c(0^+)$, $\chi_c(1^+)$ and $\chi_c(2^+)$ including branching fraction for radiative $J/\Psi + \gamma$ decay channel. Kinematical enhancement of the maximal helicity amplitudes is shown.

## 1 QCD Factorisation and Durham Model

It is well known that the exclusive diffractive Higgs production provides a very convenient tool for Higgs searches at hadron colliders due a very clean environment unlike the inclusive production [1].

The QCD mechanism for the diffractive production of heavy central system has been proposed recently by Kaidalov, Khoze, Martin and Ryskin (Durham group, KKMR) for Higgs production at the LHC (see Refs. [1, 2, 3]). The QCD factorisation implies the separation of the amplitude of the exclusive $pp \to pXp$ process to the hard subprocess amplitude describing the fusion of two off-shell gluons into a heavy system $g^*g^* \to X$, and the soft hadronic parts containing information about emission of the relatively soft gluons from the proton lines (see Fig. 1). In the framework of $k_\perp$-factorisation approach these soft parts are written in terms of so-called off-diagonal unintegrated gluon distributions (UGDFs) and cannot be calculated perturbatively. The QCD factorisation is rigorously justified in the limit of very large factorisation scale being the transverse mass of the central system $M_\perp$.

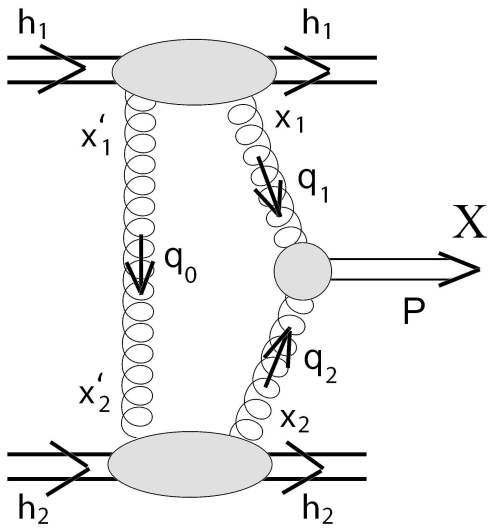

Figure 1: The QCD mechanism of diffractive production of the heavy central system $X$.

In order to check the underlying production mechanism it is worth to replace Higgs boson by a lighter (but still heavy enough to provide the QCD factorisation) meson which is easier to measure. In this respect the exclusive production of heavy quarkonia is under special interest from both experimental and theoretical points of view [4]. Testing the KKMR approach against various data on exclusive meson production at high energies is a good probe of nonperturbative dynamics of partons described by UGDFs.

Recently, the signal from the diffractive $\chi_c(0^+, 1^+, 2^+)$ charmonium production in the radiative $J/\Psi + \gamma$ decay channel has been measured by the CDF Collaboration [5]: $d\sigma/dy|_{y=0}(pp \to pp(J/\psi + \gamma)) \simeq (0.97 \pm 0.26)\,\mathrm{nb}$. In the very forward limit the contributions from $\chi_c(1^+, 2^+)$ vanish due to the $J_z = 0$ selection rule (see [6] and references therein); however, for general kinematics this might not be true. In particular, it was shown in Ref. [9] that the axial-vector $\chi_c(1^+)$ production, due a relatively large branching fraction of its radiative decay, may not be negligible and gives a noticeable contribution to the total signal measured by the CDF Collaboration. As shown below, the same holds also for the tensor $\chi_c(2^+)$ meson contribution.

The production of the axial-vector $\chi_c(1^+)$ meson has an additional suppression w.r.t. $\chi_c(0^+)$ and $\chi_c(2^+)$ in the limit of on-shell fusing gluons due to the Landau-Yang theorem [9]. Such an extra suppression may lead to the dominance of the $\chi_c(2^+)$ contribution in the radiative decay channel. Off-shell effects play a significant role also for the scalar $\chi_c(0^+)$ production reducing the total cross section by a factor of 2 – 5 depending on UGDFs [10].

According to the KKMR approach the amplitude of the exclusive double diffractive colour singlet production $pp \to pp\chi_{cJ}$ is [11, 10]

$$\mathcal{M}_{J,\lambda} = const \cdot \delta_{c_1 c_2} \Im \int d^2 q_{0,t} V_{J,\lambda}^{c_1 c_2}(q_1, q_2, P) \frac{f_{g,1}^{\mathrm{off}}(x_1, x_1', q_{0,t}^2, q_{1,t}^2, t_1) f_{g,2}^{\mathrm{off}}(x_2, x_2', q_{0,t}^2, q_{2,t}^2, t_2)}{q_{0,t}^2\, q_{1,t}^2\, q_{2,t}^2}\,, \quad (1)$$

where $t_{1,2}$ are the momentum transfers along the proton lines, $q_0$ is the momentum of the screening gluon, $q_{1,2}$ are the momenta of fusing gluons, and $f_{g,i}^{\mathrm{off}}(x_i, x_i', q_{0,t}^2, q_{i,t}^2, t_i)$ are the off-diagonal UGDFs.

Traditional (asymmetric) form of the off-diagonal UGDFs is taken in the limit of very small $x' \ll x_{1,2}$ as proportional to conventional diagonal unintegrated density in analogy to collinear off-diagonal gluon distributions (with factorised $t$-dependence), i.e.

$$\begin{aligned} f_{g,1}^{\mathrm{off}} &= R_g\, f_g^{(1)}(x_1, {Q_{1,t}^{\mathrm{eff}}}^2, \mu^2) \cdot F_N(t_1)\,, \\ f_{g,2}^{\mathrm{off}} &= R_g\, f_g^{(2)}(x_2, {Q_{2,t}^{\mathrm{eff}}}^2, \mu^2) \cdot F_N(t_2), \quad \mu^2 = \frac{M_\perp^2}{2} \end{aligned} \quad (2)$$

with a nearly constant prefactor $R_g \simeq 1.4$, ${Q_{1/2,t}^{\mathrm{eff}}}^2 = \min(q_{0,t}^2, q_{1/2,t}^2)$ are the effective gluon transverse momenta, as adopted in Ref. [1, 11], $F_N(t)$ is the form factor of the proton vertex, which can be parameterised as $F_N(t) = \exp(b_0 t)$ with $b_0 = 2\,\mathrm{GeV}^{-2}$ [12], or by the isoscalar nucleon form factor $F_1(t)$ as we have done in Ref. [10].

Our results in Ref. [10] showed up the strong sensitivity of the KKMR numerical results on the definition of the effective gluon transverse momenta $Q_{1/2,t}^{\mathrm{eff}}$ and the factorisation scales $\mu_{1,2}$. This behaviour is explained by the fact that for heavy $q\bar{q}$ production the great part of the diffractive amplitude (1) comes from nonperturbatively small $q_{0,t} < 1\,\mathrm{GeV}$. It means that the total diffractive process is dominated by very soft screening gluon exchanges with no hard scale. So, the perturbatively motivated KMR UGDFs [1] based on the Sudakov suppression and conventional parton densities are not completely justified in the soft part of the gluon ladder [10], whereas for fusing gluons it can be still reliable due to a large scale $\mu$ there.

In principle, factor $R_g$ in Eq. (2) should be a function of $x'$ and $x_1$ or $x_2$. In this case the off-diagonal UGDFs do not depend on $x'$ and $q_{0,t}^2$ (or $q_{1/2,t}^2$), and their evolution is reduced to diagonal UGDFs evolution corresponding to one "effective" gluon. In general, factor $R_g$ can depend on UGDF and reflects complicated and still not well known dynamics at small $x$ region.

## 2 Skewed UGPDs and Positivity Constraints

In order to test this small $x$ dynamics and estimate the theoretical uncertainties related with introducing of one "effective" gluon instead of two gluons in Eq. (2), in Refs. [10, 13] we have suggested more general symmetrical prescription for the off-diagonal UGDFs. Actually, it is possible to calculate the off-diagonal UGDFs in terms of their diagonal counterparts as follows

$$\begin{aligned} f_{g,1}^{\text{off}} &= \sqrt{f_g^{(1)}(x_1', q_{0,t}^2, \mu_0^2) \cdot f_g^{(1)}(x_1, q_{1,t}^2, \mu^2)} \cdot F_N(t_1)\,, \\ f_{g,2}^{\text{off}} &= \sqrt{f_g^{(2)}(x_2', q_{0,t}^2, \mu_0^2) \cdot f_g^{(2)}(x_2, q_{2,t}^2, \mu^2)} \cdot F_N(t_2)\,, \end{aligned} \tag{3}$$

where

$$x_1' = x_2', \quad \mu_0^2 = q_{0,t}^2, \quad \mu^2 = \frac{M_\perp^2}{2}.$$

This form of skewed two-gluon UGDFs (3) is inspired by the positivity constraints for the collinear Generalised Parton Distributions [14], and may be considered as a saturation of the Cauchy-Schwarz inequality for the density matrix [15]. One may doubt what is the reason of such a saturation. Usually it happens when the dimension of the linear space where inequality is studied is small. Physically this corresponds to the small number of intermediate on-shell states in the imaginary part of the amplitudes, which is likely to happen at large rather than small $x$. However, the decreasing of the contribution of intermediate states with their invariant mass growing may effectively reduce the dimensionality of space relevant for saturation of positivity constraints.

It allows us to incorporate the actual dependence of the off-diagonal UGDFs on longitudinal momentum fraction of the soft screening gluon $x'$ and its transverse momentum $q_{0,t}^2$ in explicitly symmetric way.

However, trying to incorporate the actual dependence of UGDFs on (small but nevertheless finite) $x'$ we immediately encounter the problem. The kinematics of the double diffractive process $pp \to pXp$ does not give any precise expression for $x'$ in terms of the phase space integration variables. From the QCD mechanism under consideration one can only expect the general inequality $x' \ll x_{1,2}$ and upper bound $x' \lesssim q_{0,t}/\sqrt{s}$ since the only scale appearing in the left part of the gluon ladder is the transverse momentum of the soft screening gluon $q_{0,t}$.

To explore the sensitivity of the final results on the values of $x'$, staying in the framework of traditional KKMR approach, one can introduce naively $x' = \xi \cdot q_{0,t}/\sqrt{s}$ with an auxiliary parameter $\xi$ [9]. In our earlier papers [10, 13] we considered the limited case of maximal $x'$ (with $\xi = 1$). However, our recent results incorporating tensor $\chi_c(2^+)$ contribution [16] showed that the experimental CDF data demand smaller $x'$ (softer gluon), i.e. $\xi < 1$. We will analyze this issue in greater details in the next section.

The hard vertex function $V_{J,\lambda}^{c_1c_2}(q_1, q_2, P)$ describes the coupling of two virtual gluons to $\chi_{cJ}$ mesons and appears also in the studies of their inclusive production [7, 8]. It can be found by using the next-to-leading-logarithmic-approximation (NLLA) BFKL $g^*g^*(q\bar{q})$-vertex in quasi-multi-Regge kinematics (QMRK) and projecting it out to the colour singlet bound state $\chi_{cJ}$ employing the pNRQCD technique (for scalar and axial-vector case, see Refs. [10, 9]). We do not take into account the NLO QCD corrections here, $K_{NLO} = 1$, until otherwise is mentioned.

# 3 Results

Results for the total cross section of diffractive $\chi_c(0^+, 1^+, 2^+)$ meson production at Tevatron energy $W = 1960\,\mathrm{GeV}$ are shown in Table 1. As have been pointed out in Ref. [6, 17] the absorptive corrections are quite sensitive to the meson spin-parity. This was studied before in the context of scalar and pseudoscalar Higgs production in Ref. [2]. In the last column we show the results for the expected observable signal in $J/\psi + \gamma$ channel summed over all $\chi_c$ spin states

$$\frac{d\sigma_{obs}}{dy}\bigg|_{y=0} = K_{NLO}^2 \sum_{\chi} \langle S_{\rm eff}^2 \rangle \frac{d\sigma_{J/\psi\gamma}}{dy} \tag{4}$$

where we adopt the following effective gap survival factors (for $\langle p_t \rangle \simeq 0.5\,\mathrm{GeV}$), calculated for different spins in Ref. [6, 17]: $\langle S_{\rm eff}^2(\chi_c(0^+))\rangle \simeq 0.02$, $\langle S_{\rm eff}^2(\chi_c(1^+))\rangle \simeq 0.05$ and $\langle S_{\rm eff}^2(\chi_c(2^+))\rangle \simeq 0.05$. The NLO corrections factor in the $g^*g^* \to q\bar{q}$ vertex is assumed to be the same for all $\chi_c$ states $K_{NLO} = 1.5$.

We see from the Table that the calculated signal is below the CDF data for off-diagonal UGDFs calculated as in Eq. (3) with $\xi = 1$. This provides an argument that $x'$ should be smaller than used, i.e. $\xi < 1$. Indeed, for Kutak-Stasto UGDF [19] and $\xi = 0.1$ we get the value for $d\sigma_{obs}/dy(y = 0) \simeq 0.8$, which is within the CDF error bars.

Table 1: Differential cross section $d\sigma_{\chi_c}/dy(y = 0)$ (in nb) of the exclusive diffractive production of $\chi_c(0^+, 1^+, 2^+)$ mesons and their partial and total signal in radiative $J/\psi + \gamma$ decay channel $d\sigma_{J/\psi\gamma}/dy(y = 0)$ at Tevatron for different UGDFs, $t$-dependent form factors $F_N(t)$ and values of auxiliary parameter $\xi$ controlling the characteristic $x'$ values.

| UGDF | $\xi$ | $\chi_c(0^+)$ | | $\chi_c(1^+)$ | | $\chi_c(2^+)$ | | ratio | signal |
|---|---|---|---|---|---|---|---|---|---|
| | | $\frac{d\sigma_{\chi_c}}{dy}$ | $\frac{d\sigma_{J/\psi\gamma}}{dy}$ | $\frac{d\sigma_{\chi_c}}{dy}$ | $\frac{d\sigma_{J/\psi\gamma}}{dy}$ | $\frac{d\sigma_{\chi_c}}{dy}$ | $\frac{d\sigma_{J/\psi\gamma}}{dy}$ | $\frac{\chi_c(2^+)\to J/\psi}{\chi_c(0^+)\to J/\psi}$ | $\frac{d\sigma_{obs}}{dy}$ |
| GBW [18], (3) | 1.0 | 48.4 | 0.55 | 0.8 | 0.27 | 15.6 | 3.03 | 5.5 | 0.40 |
| | 0.1 | 35.4 | 0.40 | 1.3 | 0.44 | 8.6 | 1.67 | 4.2 | 0.26 |
| KS [19], (3), | 1.0 | 72.5 | 0.83 | 0.5 | 0.17 | 7.9 | 1.53 | 1.8 | 0.23 |
| $F_1(t)$ | 0.1 | 260 | 2.96 | 1.6 | 0.55 | 27 | 5.24 | 1.8 | 0.78 |
| KS [19], (3), $\exp(b_0 t)$ | 0.1 | 238 | 2.71 | 1.2 | 0.41 | 20.3 | 3.94 | 1.45 | 0.61 |
| KMR [1], (2) $R_g = 1.0$ | – | 216 | 2.5 | 1.4 | 0.5 | 13.5 | 2.6 | 1.04 | 0.46 |

The relative contributions of different charmonium states in $J/\psi + \gamma$ channel are found to

be:

$$\sigma(0^+ \to J/\psi + \gamma) : \sigma(1^+ \to J/\psi + \gamma) : \sigma(2^+ \to J/\psi + \gamma) = \begin{cases} 1 : 0.71 : 4.64, & \text{KL} \\ 1 : 1.94 : 13.47, & \text{GBW} \\ 1 : 0.49 : 4.85, & \text{KS} \\ 1 : 0.55 : 2.81, & \text{KMR} \end{cases}$$

We see that the contribution of the tensor $\chi_c(2^+)$ meson dominates over $\chi_c(0^+, 1^+)$ for all UGDFs. As a normalization we took the contribution of $\chi_c(0^+)$ meson. Its production rate for KMR UGDF is calculated as $K^2_{NLO} \cdot \langle S^2_{\text{eff}}(\chi_c(0^+))\rangle \cdot R^4_g \cdot \sigma(\chi_c(0^+))/dy(y=0) \simeq 37$ nb, which is very close to original KMR result 35 nb [17]. At the same time, the discrepancy with KMR results for higher spin mesons remains to be investigated.

## Acknowledgements

Useful discussions and helpful correspondence with Mike Albrow, Sergey Baranov, Włodek Guryn, Alexei Kaidalov, Valery Khoze, Alan Martin, Francesco Murgia, Mikhail Ryskin and Wolfgang Schäfer are gratefully acknowledged. This study was partially supported by the Polish grant of MNiSW N N202 249235, the Russian Foundation for Fundamental Research, grants No. 07-02-91557 and No. 09-02-01149.

## References

[1] V.A. Khoze, A.D. Martin and M.G. Ryskin, Phys. Lett. B **401**, 330 (1997); Eur. Phys. J. C **14**, 525 (2000); Eur. Phys. J. C **23**, 311 (2002).

[2] A.B. Kaidalov, V.A. Khoze, A.D. Martin and M.G. Ryskin, Eur. Phys. J. C **33**, 261 (2004).

[3] A.B. Kaidalov, V.A. Khoze, A.D. Martin and M.G. Ryskin, Eur. Phys. J. C **31**, 387 (2003) [arXiv:hep-ph/0307064].

[4] M. G. Ryskin, A. D. Martin, V. A. Khoze and A. G. Shuvaev, J. Phys. G **36**, 093001 (2009) [arXiv:0907.1374 [hep-ph]]; J. L. Pinfold, Int. J. Mod. Phys. A **24**, 351 (2009); C. Royon, arXiv:0904.2217 [hep-ph].

[5] T. Aaltonen *et al.* [CDF Collaboration], Phys. Rev. Lett. **102**, 242001 (2009) [arXiv:0902.1271].

[6] A. D. Martin, M. G. Ryskin and V. A. Khoze, Acta Phys. Polon. B **40**, 1841 (2009) [arXiv:0903.2980 [hep-ph]].

[7] P. Hagler, R. Kirschner, A. Schafer, L. Szymanowski and O. V. Teryaev, Phys. Rev. Lett. **86**, 1446 (2001) [arXiv:hep-ph/0004263].

[8] Ph. Hagler, R. Kirschner, A. Schafer, L. Szymanowski and O. V. Teryaev, Phys. Rev. D **63**, 077501 (2001) [arXiv:hep-ph/0008316].

[9] R. S. Pasechnik, A. Szczurek and O. V. Teryaev, Phys. Lett. B **680**, 62 (2009) [arXiv:0901.4187 [hep-ph]].

[10] R. S. Pasechnik, A. Szczurek and O. V. Teryaev, Phys. Rev. D **78**, 014007 (2008) [arXiv:0709.0857 [hep-ph]].

[11] V. A. Khoze, A. D. Martin, M. G. Ryskin and W. J. Stirling, Eur. Phys. J. C **35**, 211 (2004) [arXiv:hep-ph/0403218]; L. A. Harland-Lang, V. A. Khoze, M. G. Ryskin and W. J. Stirling, arXiv:0909.4748, accepted for publication in Eur. Phys. J. C.

[12] V. A. Khoze, A. D. Martin and M. G. Ryskin, Eur. Phys. J. C **18**, 167 (2000) [arXiv:hep-ph/0007359].

[13] A. Szczurek, R. S. Pasechnik and O. V. Teryaev, Phys. Rev. D **75**, 054021 (2007) [arXiv:hep-ph/0608302].

[14] B. Pire, J. Soffer and O. Teryaev, Eur. Phys. J. C **8**, 103 (1999) [arXiv:hep-ph/9804284].

[15] X. Artru, M. Elchikh, J. M. Richard, J. Soffer and O. V. Teryaev, Phys. Rept. **470**, 1 (2009) [arXiv:0802.0164 [hep-ph]].

[16] R. S. Pasechnik, A. Szczurek and O. V. Teryaev, arXiv:0909.4498 [hep-ph].

[17] V. A. Khoze, Talk given at 13th International Conference on Elastic and Diffractive Scattering (13th "Blois Workshop"), CERN, 29th June - 3rd July 2009.

[18] K. Golec-Biernat and M. Wüsthoff, Phys. Rev. D **60**, 114023-1 (1999).

[19] K. Kutak and A.M. Staśto, Eur. Phys. J. **C41**, 341 (2005).

# Central Exclusive Production: Vector Mesons, Dijets, Higgs Boson

*J.R. Cudell*

IFPA, AGO Dept., Université de Liège, Belgium

I review the situation of theoretical predictions of central exclusive production, and show that the CDF dijet data can be used to constrain the prediction of central exclusive Higgs boson production. I also show that central exclusive production might be used as a discovery tool for the odderon.

## 1 Development of Central Exclusive Production and Data

Central exclusive production has been studied for a long time as it is a potential discovery channel for new physics coupled to quarks and gluons. The original idea [1, 2, 3] concerned the production of a light Higgs boson, which would predominantly decay into bottom quarks, and thus be extremely hard to observe in inelastic channels. Over the years, calculations of exclusive production have progressed through the implementation of several crucial features. The first attempt to embed Higgs boson production into a pomeron [1] used non-perturbative gluons, and the calculation was later translated into a perturbative one in [4], at the price of introducing an unknown proton form factor. The possibility of protons breaking could then be modelled, but only in an eikonal framework [3]. This was later generalised [5, 6] for any amplitude, provided that the production is at much smaller distance than the rescatterings. Finally, large perturbative corrections at the production vertex – the so-called Sudakov form factor – were identified in [7].

All these ingredients may be sufficient to estimate the cross section for the production of Higgs bosons and other heavy systems at the LHC. The best known model which incorporates all the above ingredients is that of the Durham group [8]. It successfully predicted the order of magnitude of the cross sections later measured by CDF, for dijets [9], diphotons [10] and $\chi_c$ [11]. Indeed, disagreement among theorists was finally settled, as CDF did observe exclusive production of high-mass systems, going up to 130 GeV, and hence one believes that all the ingredients of the Durham model are indeed necessary.

However, several of these elements can be improved, and the general feeling that the uncertainties are of the order of a factor 3 must be reassessed. Hence the first goal of this contribution is to summarise the findings of [12] concerning exclusive production.

The second purpose is to examine central exclusive production not as a means of producing new physics, but rather as genuine new physics in its own right. Indeed, central exclusive production of vector mesons may be used as a discovery channel for the odderon. The general structure of the calculation [13] is similar to that in the pomeron case, and backgrounds due to photon exchange do not seem prohibitive. But, as I shall explain, one is also limited here by the presence of large uncertainties in the theory. Here again, data from CDF are becoming available [11], and may help reduce these uncertainties.

# 2 Skeleton of an Exclusive Calculation

One must insist first on the fact that the calculations are very inspired by perturbation theory. However, as we shall see later, a large part of the amplitude lies in the soft region, so that one cannot *d*erive the steps of the calculation, but one hopes that the nonperturbative region is not too different from the perturbative one, at least at high $s$.

The first step [4] is to model pomeron or odderon exchange à la Low-Nussinov, *i.e.* to consider the smallest number of gluons that need to be exchange between quarks to produce the final state via colour-singlet exchange, as shown in Fig. 1. In the pomeron case, one uses cutting rules to calculate the imaginary part of the amplitude, which one assumes to be dominant. In the odderon and photon cases, the calculation is more involved as the odderon-photon and odderon-pomeron amplitudes have different phases. Apart from colour factors, these amplitudes can be calculated either directly or using the BFKL vertices. They are not yet physical, as quark-quark scattering via singlet exchange is infrared divergent.

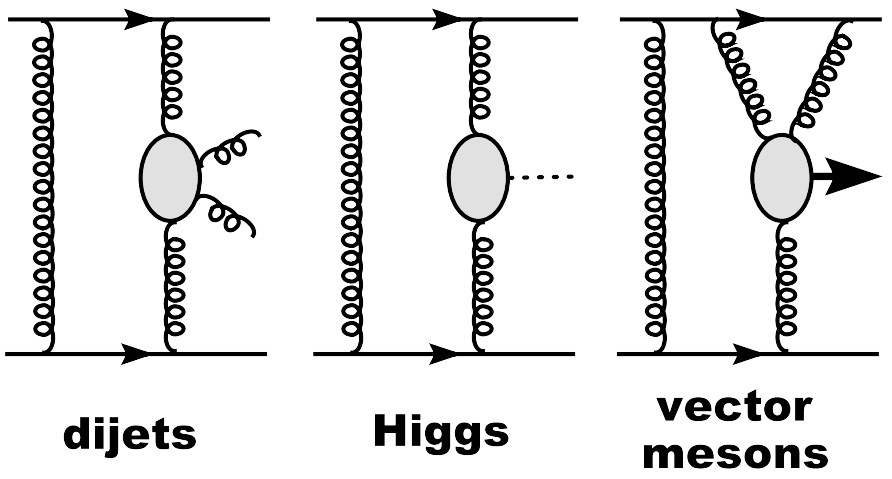

Figure 1: Some of the lowest-order diagrams for the three processes considered here.

To get a finite answer, one needs to consider scattering of colour-singlet objects rather than colour charges, as shown in Fig. 2. In the terminology of BFKL or of Cheng and Wu, this is called the *i*mpact factor, which takes into account the fact that the exchanged gluons can be connected to other quarks or gluons, and leads to convergent integrals in the infrared region. The problem here is that we do not know in general what these objects are. One possibility is to model them via light-cone wave functions [14, 15] but the latter are unknown, too, so that only general properties can be derived. This is the best one can do for odderon (3-gluon) exchange. For 2-gluon exchange, one can do slightly better [16] by forcing the parametrisation to agree with (skewed) off-diagonal structure functions when the gluons are hard. One nevertheless has to take into account the contribution of soft gluons, and ensure that the impact factor goes to zero when one of the gluons goes on shell. The Durham group neglects both of these constraints, and considers a parametrisation which is correct only for hard gluons.

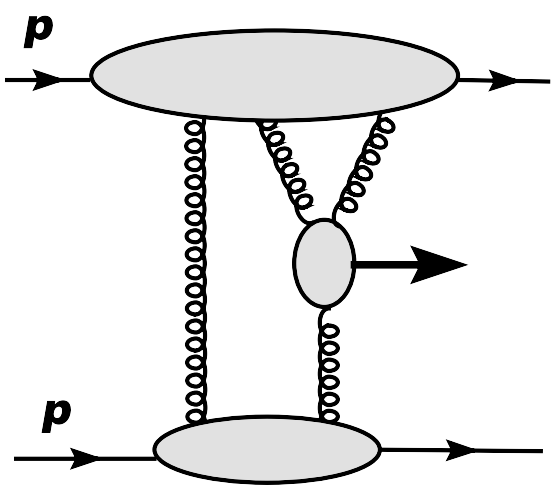

Figure 2: The two impact factors entering the vector-meson calculation.

If one produces a high-mass system, one needs to worry about large virtual corrections at the vertex. Indeed, if the produced system has a scale $M$, and is linked to gluons of virtuality $\mu$, one knows that there are large Sudakov double logarithms $\log^2(M/\mu)$. The trick to evaluate them is based on the infrared finiteness of inclusive corrections. One knows that, if the gluons go on-shell, then the virtual corrections will cancel the infrared divergences of bremsstrahlung. So the logarithms can be calculated by considering the bremsstrahlung diagrams. The double-logs are under theoretical control, and can be resummed [17]. The situation with single logarithms $\log(M/\mu)$ is more complicated. Some of them can be resummed, and some others cannot, depending on the process. Finally, all this holds if constant terms are small. We found [12] that, for $M \approx 20$ GeV (*i.e.* the first dijet points), this is not the case.

In the Higgs boson case, the upper scale $M$ is given by the Higgs boson mass, and the single logs were evaluated in [18] and lead to angular ordering, together with a determination of the lower scale $\mu$. In the dijet case, other diagrams lead to single logarithms, which cannot be resummed. The extra single logs are fortunately small [19], so that the general structure of the Sudakov form factor is similar to that in the Higgs boson case. However, the gluons-to-jets vertex changes if the jets have sufficient transverse energy, as an extra propagator then enters the loop integrals. This modifies the power of the logarithms in the answer, and hence standard Sudakov techniques apply only if one chooses the transverse energy as an upper scale. In the vector-meson case, the situation is much more complicated, but fortunately the scales involved are small, and the logs cannot be very large.

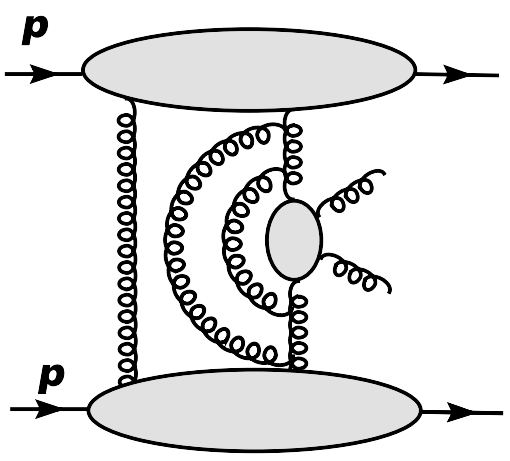

Figure 3: The large Sudakov vertex corrections in the dijet case.

As a final generic ingredient, one has to take into account that factorisation does not hold when one goes from $\gamma p$ to $\bar{p}p$. Hence the impact factors, derived from structure functions, have corrections due to screening. Nobody really knows how to implement these, as knowing them would amount to being able to unitarise pomeron exchange. Many estimates agree within a factor 3 [20], but they are all based on eikonal or multi-channel eikonal schemes. These screening corrections should be folded with the one-pomeron exchange amplitude [5, 6], but we shall simply treat them as an effective factor – the "gap survival probability".

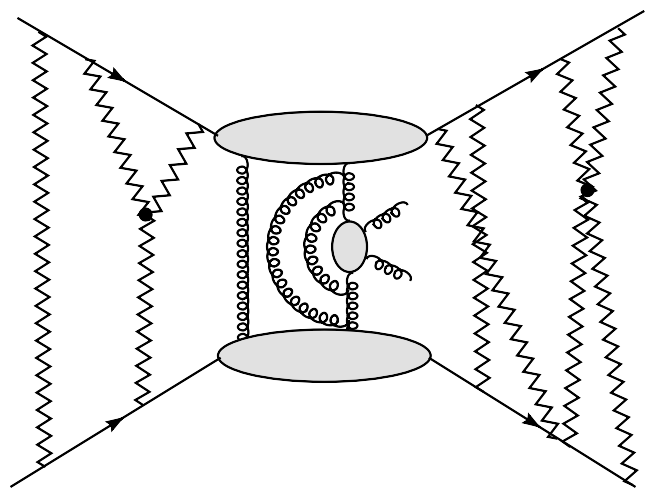
Figure 4: Screening corrections.

Finally, process-specific corrections still have to be performed. In the jet case, some of the particles coming from the partons are missed by the jet-finding algorithm, so that the jet transverse energy is smaller than the parton one. As the cross section falls fast with energy this brings in a rather large correction.

All the above corrections go in the same direction, decreasing the cross section by a factor of the order of 600, as shown in the second column of Table 1. This is the well-known problem of exclusive calculations: although the lowest order is calculable, there are huge corrections coming from nonperturbative or higher-order effects, which overwhelm the lowest order.

# 3 Properties of the Amplitudes

First of all, as is well-known, it is possible to reproduce almost exactly the dijet data measured by CDF. Figure 5 shows one of the possible curves, for specific choices of the various correction factors outlined above. But this is one of the many possible choices. As we we are about to see, all the corrections have large uncertainties (*factors*), so that a change in one can be compensated by a change in the other.

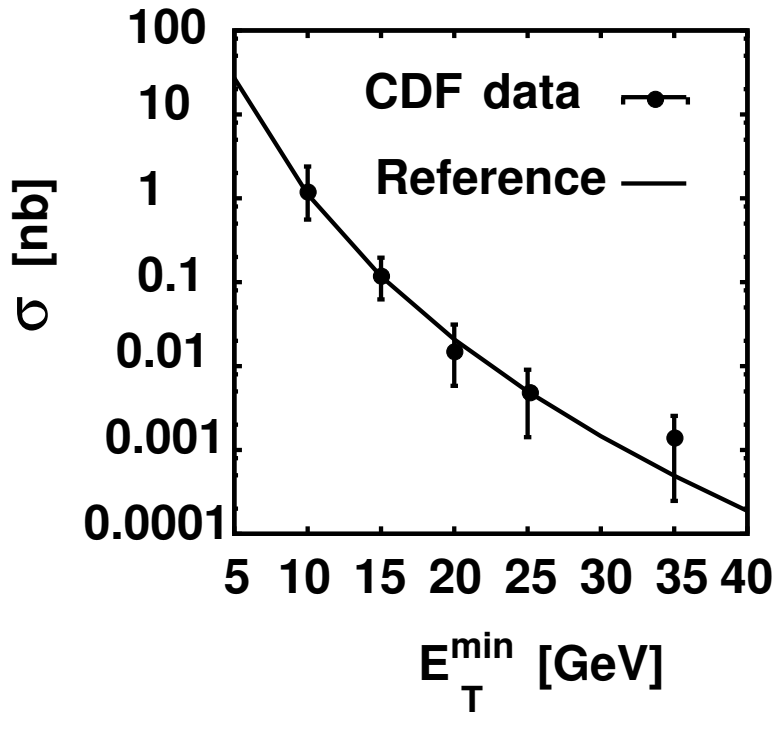

Figure 5: Dijet data and a possible curve.

The second point is indeed that all these corrections have rather large uncertainties. Modest changes of scales in the Sudakov form factor, slightly different parametrisations of impact factors, or modifications in the unitarisation scheme to calculate the splash-out all lead to appreciable differences, as shown in the last three columns of Table 1.

The final property is the most worrisome one. It is well-known that two-gluon or three-gluon exchange between protons has a strong infrared contribution in elastic scattering: although the cross sections are finite, the typical gluon off-shellness is of the order of 600 MeV, and comes directly from the size of the proton, which is included in the impact factors. One might hope that, in the case of exclusive production of heavy objects, the situation would be different, and it has been claimed that the Sudakov form factor would shift the calculation to the perturbative region. This is the case for the Durham model, but it may not be correct. Indeed, one cannot allow highly off-shell partons to come out of the proton without paying a price. In our case, this comes from the impact factor (omitted in [8]), that suppresses highly off-shell partons most of the time. So the shift due to the Sudakov form factor is mostly compensated by the impact factor. If one produces a 100 GeV object, more than half of the cross section comes from a region where one of the gluons has an off-shellness smaller than 1 GeV. Hence, the core of the calculation has a strong non-perturbative component. Note that folding with the gap survival probability (instead of taking it as a constant factor) will increase the long-distance contribution as the gap survival probability is larger at values of the impact parameter.

| | Dijets | | Higgs boson | Vector mesons |
|---|---|---|---|---|
| | $\sigma(E_T > 10$ GeV) [nb] | uncertainty factor | | |
| Impact factor | 600 | 3 | 3 | $> 3$ |
| Sudakov form factor | 25 | 20 | 7 | 1 |
| Gap survival | 3 | 3 | 3 | 3 |
| Slash-out | 1 | 2 | – | – |

Table 1: The second column gives the value of the Tevatron dijet cross section after various corrections are included, for $E_T^{min} = 10\,\mathrm{GeV}$. The next three columns show the factor of uncertainty (maximum / minimum) of the correction.

# 4 Results

## 4.1 Dijets

As we have seen, it is possible to reproduce the CDF data. Conversely, these give a very useful constraint to reduce the theoretical uncertainties, especially as they extend to the mass region of a standard Higgs boson. Hence, we can consider a set of curves going through the error bars of the CDF points, and see how they extrapolate to the LHC. We apply cuts typical of FP420 (proton fractional momentum loss between 0.2% and 2%, jet rapidity less than 1 and mass of the dijet system greater than 50 GeV), to obtain the cross section shown in Fig. 6. The outer band corresponds to the theoretical curves going through the CDF dijet error bars, and the inner band shows the intrinsic extrapolation errors: all the curves making up that band are identical at the Tevatron (and the same as the curve of Fig. 5), but spread when extrapolated to the LHC. We see that the cross section is large enough for a measurement of the dijet cross section in the early LHC. This would further help reduce the ambiguities in the theory.

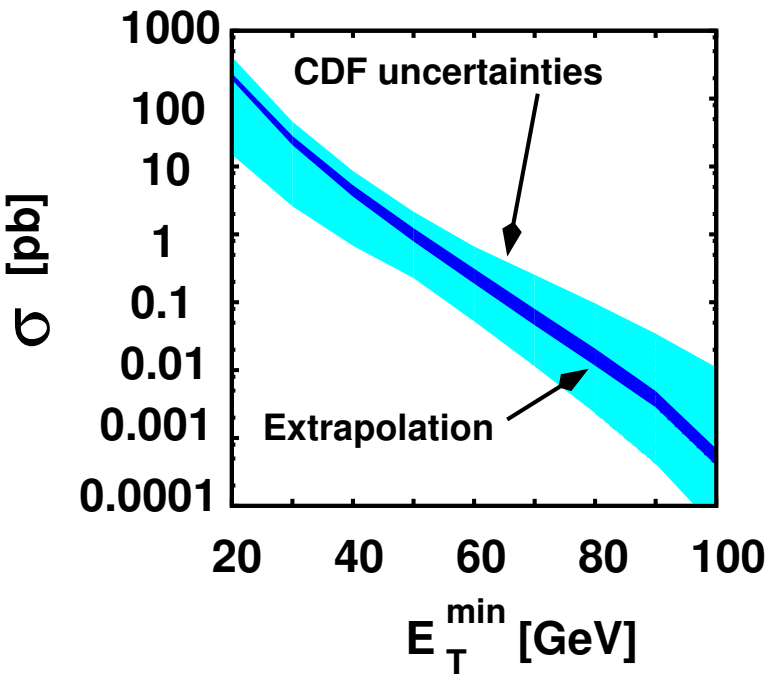

Figure 6: The dijet cross section for FP420 cuts, for $\sqrt{s} = 14$ TeV.

## 4.2 Higgs Boson

As above, we can keep the sets of parameters which reproduce the CDF dijet data, and see what they give for the Higgs boson. Although the set of diagrams is not identical [12], it turns out that the dominant ones are, so that the results can be directly translated from the dijets to the Higgs boson. The cross sections predicted for CDF, for a standard Higgs boson heavier than 110 GeV, are always smaller that 0.03 fb, and hence of little interest. At the LHC, the cross section will then be at most 8 fb for a Higgs-boson mass of 100 GeV, and will drop to at most 1 fb for a Higgs-boson mass of 145 GeV [21], again for cuts typical of FP420 (proton fractional momentum loss between 0.2% and 2%, and Higgs-boson rapidity less than 1 ).

## 4.3 Vector Mesons

The problem here is the background. In the dijet and Higgs boson cases, the background is negligible. Unfortunately, it is possible to produce a vector meson either via odderon exchange or via photon exchange, as shown in Fig. 7. Both cross sections are of the same order of magnitude, and given that we do not know the impact factor of the odderon, it is hard to be more precise. We show in Table 2 the possible ratios of odderon to photon cross sections, for various vector mesons , and for the Tevatron or the LHC.

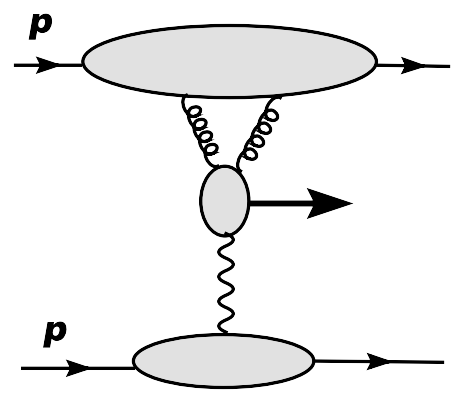

Figure 7: The analog of Fig. 2 for photon exchange.

The various uncertainties in the calculation lead to a range of values for the odderon and for the photon cross section. These uncertainties are somewhat lower in the ratio of these cross sections. We see that the best place to look for the odderon may be the Tevatron, although the best channel ($\Upsilon$ production) is unfortunately the hardest one experimentally. It may be worth pointing out that, due to an interference between pomeron-odderon and odderon-pomeron exchanges, the odderon-pomeron cross section for forward production of vector mesons is close to maximum in $\bar{p}p$ collisions, whereas it vanishes in the $pp$ case, and has its maximum around 600 MeV. CDF has published [11] an upper limit $d\sigma/dy|_{y=0} < 2.3$ nb for the odderon cross section, corresponding to a ratio of 90% for the cross sections, thus getting close to the detection level for the odderon.

| ratios of $d\sigma/dy\|_{y=0}$ | $J/\psi$ odderon / photon | $\Upsilon$ odderon / photon |
|---|---|---|
| Tevatron | 26–56 % | 80 – 170 % |
| LHC | 6 – 15 % | 15 -38 % |

Table 2: Ratios of the pomeron-odderon and pomeron-photon cross sections for exclusive $J/\psi$ and $\Upsilon$ production in $pp$ and $p\bar{p}$ collisions.

In order to enhance the signal, the most obvious way would be to concentrate on high-$|t|$ data, as photon exchange fall much faster with $|t|$ than odderon exchange. For instance, cutting the momentum transfer to the proton or the antiproton to be greater than 500 MeV would enhance the odderon signal by a factor 10 [22]. The other way would be to cut on the vector-meson transverse energy. A cut $p_T > 1$ GeV enhances the odderon signal by a factor 4 [22].

# 5 Conclusion

We have seen that central exclusive production can be reproduced by models which include a number of corrections to the naïve estimates based on lowest-order calculations. These corrections can be approximated (in the case of Sudakov form factors), fitted (in the case of impact factors) or guessed (in the case of gap survival probabilities), and are thus prone to large uncertainties. These uncertainties can be somewhat reduced using the recent CDF data on dijet or vector-meson production.

Further reduction of these uncertainties would be possible if the dijet cross section is measured at the LHC, especially as the extrapolation of the gap survival probability to higher energies is far from certain. At present, one can state that the production cross section for a standard Higgs of 120 GeV should be between 0.3 and 2 fb.

As for the discovery of the odderon, it seems that there is a chance to disentangle it from the photon exchange background, and that CDF is getting close to the level of statistics needed to detect it.

# Acknowledgements

I acknowledge the contribution of my collaborators A. Bzdak, A. Dechambre, O.F. Hernández, I.P. Ivanov, L. Motyka, L. Szymanowski to the investigations summarised here. I also thank P.V. Landshoff, M. Albrow, K. Goulianos, V. Khoze, A. Martin and M. Ryskin for sharing their insights with me, and Karine Gilson for a careful proofreading.

# References

[1] A. Bialas and P. V. Landshoff, Phys. Lett. B **256** (1991) 540.

[2] A. Schafer, O. Nachtmann and R. Schopf, Phys. Lett. B **249** (1990) 331.

[3] J. D. Bjorken, Phys. Rev. D **47** (1993) 101.

[4] J. R. Cudell and O. F. Hernandez, Nucl. Phys. B **471**, 471 (1996) [arXiv:hep-ph/9511252].

[5] L. Frankfurt, C. E. Hyde, M. Strikman and C. Weiss, Phys. Rev. D **75** (2007) 054009 [arXiv:hep-ph/0608271].

[6] S. M. Troshin and N. E. Tyurin, Eur. Phys. J. C **39** (2005) 435 [arXiv:hep-ph/0403021].

[7] A. Berera and J. C. Collins, Nucl. Phys. B **474** (1996) 183 [arXiv:hep-ph/9509258].

[8] V. A. Khoze, A. D. Martin and M. G. Ryskin, Eur. Phys. J. C **14** (2000) 525 [arXiv:hep-ph/0002072].

[9] T. Aaltonen *et al.* [CDF Collaboration], Phys. Rev. D **77** (2008) 052004 [arXiv:0712.0604 [hep-ex]].

[10] T. Aaltonen *et al.* [CDF Collaboration], Phys. Rev. Lett. **99** (2007) 242002 [arXiv:0707.2374 [hep-ex]].

[11] T. Aaltonen *et al.* [CDF Collaboration], Phys. Rev. Lett. **102** (2009) 242001 [arXiv:0902.1271 [hep-ex]].

[12] J. R. Cudell, A. Dechambre, O. F. Hernandez and I. P. Ivanov, Eur. Phys. J. C **61**, 369 (2009) [arXiv:0807.0600 [hep-ph]].

[13] A. Bzdak, L. Motyka, L. Szymanowski and J. R. Cudell, Phys. Rev. D **75**, 094023 (2007) [arXiv:hep-ph/0702134].

[14] J. R. Cudell and B. U. Nguyen, Nucl. Phys. B **420** (1994) 669 [arXiv:hep-ph/9310298].

[15] M. Fukugita and J. Kwiecinski, Phys. Lett. B **83** (1979) 119.

[16] I. P. Ivanov and N. N. Nikolaev, Phys. Rev. D **65** (2002) 054004 [arXiv:hep-ph/0004206]; I. P. Ivanov, N. N. Nikolaev and A. A. Savin, Phys. Part. Nucl. **37** (2006) 1 [arXiv:hep-ph/0501034].

[17] Y. L. Dokshitzer, D. Diakonov and S. I. Troian, Phys. Rept. **58** (1980) 269.

[18] A. B. Kaidalov, V. A. Khoze, A. D. Martin and M. G. Ryskin, Eur. Phys. J. C **33** (2004) 261 [arXiv:hep-ph/0311023].

[19] A. Dechambre and I.P. Ivanov, in preparation.

[20] See *e.g.* A. Achilli, R. Hegde, R. M. Godbole, A. Grau, G. Pancheri and Y. Srivastava, Phys. Lett. B **659** (2008) 137 [arXiv:0708.3626 [hep-ph]].

[21] J.R. Cudell, A. Dechambre, O.F. Hernández and I.P. Ivanov, in preparation.

[22] L. Motyka, arXiv:0808.2216 [hep-ph], in the proceedings of 16th International Workshop on Deep Inelastic Scattering and Related Subjects (DIS 2008), London, England, 7-11 Apr 2008, p. 73.

# BSM Higgs Studies at the LHC in the Forward Proton Mode

*S. Heinemeyer*[1], *V.A. Khoze*[2], *M.G. Ryskin*[3], *M. Taševský*[4] *and G. Weiglein*[2]

[1]Instituto de Física de Cantabria (CSIC), Santander, Spain,
[2]IPPP, Department of Physics, Durham University, Durham DH1 3LE, U.K.,
[3]Petersburg Nuclear Physics Institute, Gatchina, St. Petersburg, 188300, Russia,
[4]Institute of Physics of the ASCR, Na Slovance 2, 18221 Prague 8, Czech Republic

The prospects for central exclusive diffractive (CED) production of BSM Higgs bosons at the LHC are reviewed. This comprises the production of MSSM and 4th generation Higgs bosons. The sensitivity of the searches in the forward proton mode for the Higgs bosons as well as the possibility of a coupling structure determination are briefly discussed.

## 1 Introduction

In recent years, there has been a growing interest in the possibility to complement the standard LHC physics menu by installing near-beam proton detectors in the LHC tunnel. Projects to install the proton detectors at 220 m and 420 m from the interaction points are now under review inside ATLAS and CMS [1–4]. The combined detection of both outgoing protons and the centrally produced system gives access to a rich program of studies of QCD, electroweak and BSM physics, see for instance [3, 5]. Importantly, these measurements will provide valuable information on the Higgs sector of MSSM and other popular BSM scenarios, see [6–11].

As it is well known, many models of new physics require an extended Higgs sector. The most popular extension of the SM is the MSSM, where the Higgs sector consists of five physical states. At lowest order the MSSM Higgs sector is $\mathcal{CP}$-conserving, containing two $\mathcal{CP}$-even bosons, the lighter $h$ and the heavier $H$, a $\mathcal{CP}$-odd boson, $A$, and the charged bosons $H^{\pm}$. It can be specified in terms of the gauge couplings, the ratio of the two vacuum expectation values, $\tan\beta \equiv v_2/v_1$, and the mass of the $A$ boson, $M_A$. The Higgs phenomenology in the MSSM is strongly affected by higher-order corrections (see [12] for reviews).

Another very simple example of physics beyond the SM is a model which extends the SM by a fourth generation of heavy fermions (SM4), see, for instance, [13]. Here the masses of the 4th generation quarks and leptons are assumed to be (much) heavier than the mass of the top-quark. In this case, the effective coupling of the Higgs boson to two gluons is three times larger than in the SM, and all branching ratios change correspondingly.

Proving that a detected new state is, indeed, a Higgs boson and distinguishing the Higgs boson(s) of the SM, the SM4 or the MSSM from the states of other theories will be far from trivial. In particular, it will be of utmost importance to determine the spin and $\mathcal{CP}$ properties of a new state and to measure precisely its mass, width and couplings.

The CED processes are of the form $pp \to p \oplus H \oplus p$, where the $\oplus$ signs denote large rapidity gaps on either side of the centrally produced state. If the outgoing protons remain intact and scatter through small angles then, to a very good approximation, the primary digluon system obeys a $J_z = 0$, $\mathcal{CP}$-even selection rule [14]. Here $J_z$ is the projection of the total angular momentum along the proton beam. This permits a clean determination of the quantum numbers of the observed resonance which will be dominantly produced in a $0^+$ state. Furthermore, because the process is exclusive, the proton energy losses are directly related to the central mass, allowing a potentially excellent mass resolution, irrespective of the decay channel. The CED processes allow in principle all the main Higgs decay modes, $b\bar{b}$, $WW$ and $\tau\tau$, to be observed in the same production channel. In particular, a unique possibility opens up to study the Higgs Yukawa coupling to bottom quarks, which, as it is well known, may be difficult to access in other search channels at the LHC. Here it should be kept in mind that access to the bottom Yukawa coupling will be crucial as an input also for the determination of Higgs couplings to other particles [15, 16].

Within the MSSM, CED production is even more appealing than in the SM. The lightest MSSM Higgs boson coupling to $b\bar{b}$ and $\tau\tau$ can be strongly enhanced for large values of $\tan\beta$ and relatively small $M_A$. On the other hand, for larger values of $M_A$ the branching ratio of $H \to b\bar{b}$ is much larger than for a SM Higgs of the same mass. As a consequence, CED $H \to b\bar{b}$ production can be studied in the MSSM up to much higher masses than in the SM case.

Here we briefly review the analysis of [7] where a detailed study of the CED MSSM Higgs production was performed (see also Refs. [6, 8, 17] for other CED studies in the MSSM). This is updated by taking into account recent theoretical developments in background evaluation [18,19] and using an improved version [20] of the code `FeynHiggs` [21] employed for the cross section and decay width calculations. The regions excluded by LEP and Tevatron Higgs searches are evaluated with `HiggsBounds` [22]. These improvements are applied for the CED production of MSSM Higgs bosons [7] in the $M_h^{\max}$ benchmark scenario (defined in [23]), and in the SM4.

# 2 Signal and Background Rates and Experimental Aspects

The Higgs signal and background cross sections can be approximated by the simple formulae given in [6, 7]. For CED production of the MSSM $h, H$-bosons the cross section $\sigma^{\rm excl}$ is

$$\sigma^{\rm excl}\,{\rm BR}^{\rm MSSM} = 3\,{\rm fb}\left(\frac{136}{16+M}\right)^{3.3}\left(\frac{120}{M}\right)^{3}\frac{\Gamma(h/H\to gg)}{0.25\ {\rm MeV}}\,{\rm BR}^{\rm MSSM}, \tag{1}$$

where the gluonic width $\Gamma(h/H \to gg)$ and the branching ratios for the various MSSM channels, ${\rm BR}^{\rm MSSM}$, are calculated with *FeynHiggs*`2.6.2` [20]. The mass $M$ (in GeV) denotes either $M_h$ or $M_H$. The normalization is fixed at $M = 120$ GeV, where $\sigma^{\rm excl} = 3$ fb for $\Gamma(H^{\rm SM} \to gg) = 0.25$ MeV. In Ref. [6, 7] the uncertainty in the prediction for the CED cross sections was estimated to be below a factor of $\sim 2.5$. According to [1, 7, 18, 24], the overall background to the $0^+$ Higgs signal in the $b\bar{b}$ mode can be approximated by

$$ {\rm d}\sigma^B/{\rm d}M \approx 0.5\,{\rm fb/GeV}\left[A(120/M)^6 + 1/2\,C(120/M)^8\right]\ , \tag{2}$$

with $A = 0.92$ and $C = C_{\rm NLO} = 0.48 - 0.12 \times (\ln(M/120))$. The expression (2) holds for a mass window $\Delta M = 4 - 5$ GeV and summarizes several types of backgrounds: the prolific $gg^{PP} \to gg$ subprocess can mimic $b\bar{b}$ production due to the misidentification of the gluons as $b$ jets; an admixture of $|J_z| = 2$ production; the radiative $gg^{PP} \to b\bar{b}g$ background; due to the non-zero $b$-quark mass there is also a contribution to the $J_z = 0$ cross section of order $m_b^2/E_T^2$. The first term in the square brackets corresponds to the first three background sources [7], evaluated for $P_{g/b} = 1.3\%$, where $P_{g/b}$ is the probability to misidentify a gluon as a $b$-jet for a $b$-tagging efficiency of 60%. The second term describes the background associated with bottom-mass terms in the Born amplitude. The NLO correction suppresses this contribution by a factor of about 2, or more for larger masses [18].

The main experimental challenge of running at high luminosity, $10^{34}\,\mathrm{cm}^{-2}\,\mathrm{s}^{-1}$, is the effect of pile-up, which can generate fake signal events within the acceptances of the proton detectors as a result of the coincidence of two or more separate interactions in the same bunch crossing, see [2, 3, 7, 8] for details. Fortunately, as established in [8], the pile-up can be brought under control by using time-of-flight vertexing and cuts on the number of charged tracks. Also in the analysis of [7] the event selections and cuts were imposed such as to maximally reduce the pile-up background. Based on the anticipated improvements for a reduction of the overlap backgrounds down to a tolerable level, in the numerical studies in [2, 7] and in the new results below the pile-up effects were not included.

At nominal LHC optics, proton taggers positioned at a distance ±420 m from the interaction points of ATLAS and CMS will allow a coverage of the proton fractional momentum loss $\xi$ in the range 0.002–0.02, with an acceptance of around 30% for a centrally produced system with a mass around 120 GeV. A combination with the foreseen proton detectors at ±220 m [4, 25] would enlarge the $\xi$ range up to 0.2. This would be especially beneficial because of the increasing acceptance for higher masses [7]. The main selection criteria for $h, H \to b\bar{b}$ are either two $b$-tagged jets or two jets with at least one $b$-hadron decaying into a muon. Details on the corresponding selection cuts and triggers for $b\bar{b}$, $WW$ and $\tau\tau$ channels can be found in [2, 7, 26]. Following [7] we consider four luminosity scenarios: "60 fb$^{-1}$" and "600 fb$^{-1}$" refer to running at low and high instantaneous luminosity, respectively, using conservative assumptions for the signal rates and the experimental sensitivities; possible improvements of both theory and experiment could lead to the scenarios where the event rates are higher by a factor of 2, denoted as "60 fb$^{-1}$ eff×2" and "600 fb$^{-1}$ eff×2".

# 3 Updated Sensitivities for CED Production of the $\mathcal{CP}$-Even MSSM Higgs Bosons

Below we extend the analysis of the CED production of $H \to b\bar{b}$ and $H \to \tau\tau$ carried out in [7] and consider the $M_h^{\max}$ benchmark scenario of [23]. The improvements consist of the incorporation of the one-loop corrections to the mass-suppressed background [18] and in employing an updated version of *FeynHiggs* [20, 21] for the cross section and decay width calculations. Furthermore we now also display the limits in the $M_A$–$\tan\beta$ planes obtained from Higgs-boson searches at the Tevatron. For the latter we employed the new code `HiggsBounds`, see [22] (where also the list of CDF and D0 references for the incorporated exclusion limits can be found).

The two plots in Fig. 1 exemplify our new results for the case of $h$ production in the $M_h^{\max}$ scenario [23]. They display the contours of $3\sigma$ statistical significance (left) and $5\sigma$ discovery (right) in the $h \to b\bar{b}$ channel. The left-hand plot shows that while the allowed region at high $\tan\beta$ and low $M_A$ can be probed also with lower integrated luminosity, in the "600 fb$^{-1}$ eff×2" scenario the coverage at the $3\sigma$ level extends over nearly the whole $M_A$–$\tan\beta$ plane, with the exception of a window around $M_A \approx 130 - 140$ GeV (which widens up for small values of $\tan\beta$). The coverage includes the case of a light SM-like Higgs, which corresponds to the region of large $M_A$. It should be kept in mind that besides giving an access to the bottom Yukawa coupling, which is a crucial input for determining all other Higgs couplings [15], the forward proton mode would provide valuable information on the Higgs $\mathcal{CP}$ quantum numbers and allow a precise Higgs mass measurement and maybe even a direct determination of its width.

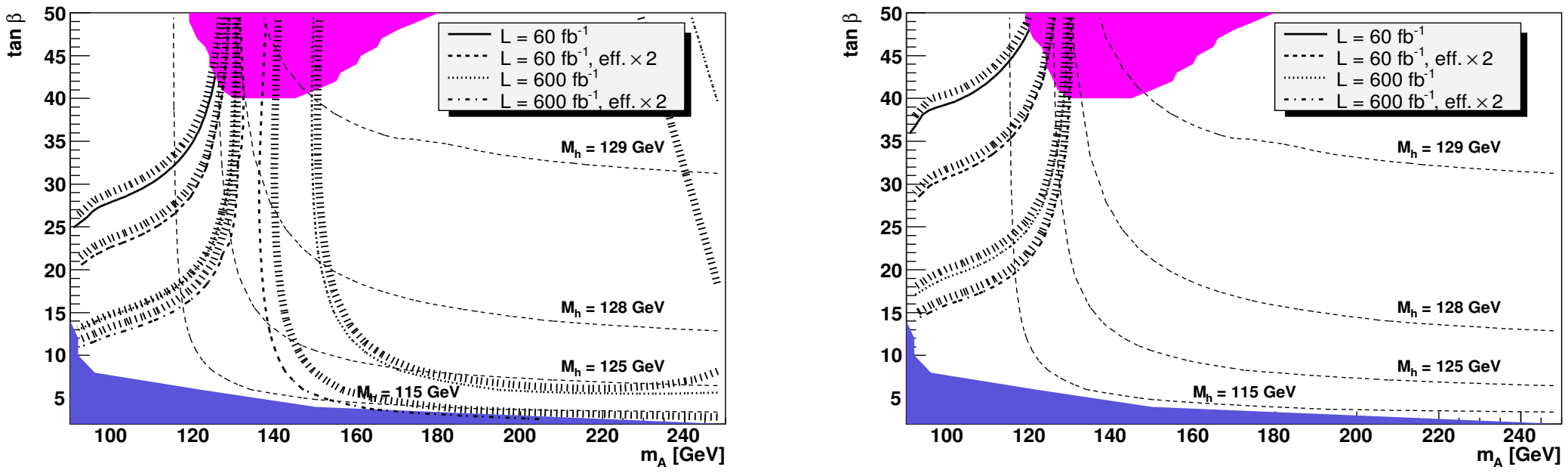

Figure 1: Contours of $3\sigma$ statistical significance (left) and $5\sigma$ discovery (right) contours for the $h \to b\bar{b}$ channel in the $M_h^{\max}$ benchmark scenario with $\mu = +200$ GeV. The results were calculated using Eqs. (1) and (2) for $A = 0.92$ and $C = C_{\rm NLO}$ for effective luminosities of "60 fb$^{-1}$", "60 fb$^{-1}$ eff×2", "600 fb$^{-1}$" and "600 fb$^{-1}$ eff×2". The values of $M_h$ are shown by the contour lines. The medium dark shaded (blue) regions correspond to the LEP exclusion bounds, while the Tevatron limits are shown by the dark shaded (purple) regions.

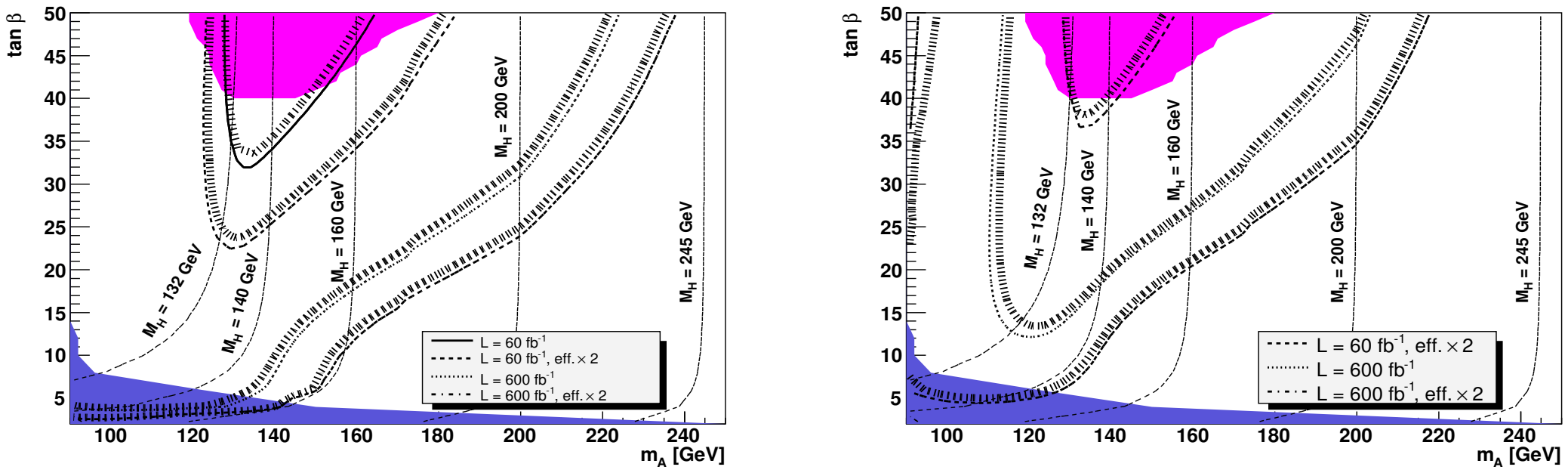

Figure 2: Contours of $3\sigma$ statistical significance (left) and $5\sigma$ discovery (right) contours for the $H \to b\bar{b}$ channel, see Fig. 1.

The properties of the heavier boson $H$ differ very significantly from the ones of a SM Higgs with the same mass in the region where $M_H \gtrsim 150$ GeV. While for a SM Higgs the BR($H \to b\bar{b}$) is strongly suppressed, the decay into bottom quarks is the dominant mode for the MSSM Higgs boson $H$. The $3\sigma$ significance and $5\sigma$ discovery contours in the $M_A$–$\tan\beta$ plane are displayed in the left and right plot of Fig. 2, respectively. While the area covered in the "60 fb$^{-1}$" scenario is to a large extent already ruled out by Tevatron Higgs searches [22], in the "600 fb$^{-1}$ eff×2" scenario the reach for the heavier Higgs at the $3\,\sigma$ level goes beyond $M_H \approx 235$ GeV in the large $\tan\beta$ region. At the $5\sigma$ level the reach is slightly reduced, but still extends beyond $M_H \approx 200$ GeV. Thus, CED production of the $H$ with the subsequent decay to $b\bar{b}$ provides a unique opportunity for accessing its bottom Yukawa coupling in a mass range where for a SM Higgs boson the $b\bar{b}$ decay rate would be negligibly small. In the "600 fb$^{-1}$ eff×2" scenario the discovery of a heavy $\mathcal{CP}$-even Higgs with $M_H \approx 140$ GeV will be possible for all allowed values of $\tan\beta$.

Concerning the determination of the spin and the CP properties of Higgs bosons the standard methods rely to a large extent on the coupling of a relatively heavy SM-like Higgs to two gauge bosons. The first channel that should be mentioned here is $H \to ZZ \to 4l$. This channel provides detailed information about spin and $\mathcal{CP}$-properties if it is open [27].

Within a SM-like set-up it was analyzed how the tensor structure of the coupling of the Higgs boson to weak gauge bosons can be determined at the LHC [28–30]. A study exploiting the difference in the azimuthal angles of the two tagging jets in weak vector boson fusion has shown that for $M_{H^{\rm SM}} = 160$ GeV the decay mode into a pair of $W$-bosons (which is maximal at $M_{H^{\rm SM}} = 160$ GeV) allows the discrimination between the two extreme scenarios of a pure $\mathcal{CP}$-even (as in the SM) and a pure $\mathcal{CP}$-odd tensor structure at a level of 4.5 to $5.3\,\sigma$ using only about 10 fb$^{-1}$(assuming the production rate is that of the SM, which is currently probed by the Tevatron [31].) A discriminating power of two standard deviations at $M_{H^{\rm SM}} = 120$ GeV in the tau lepton decay mode requires an integrated luminosity of 30 fb$^{-1}$ [30].

For $M_H \approx M_A \gtrsim 2M_W$ the lightest MSSM Higgs boson couples to gauge bosons with about SM strength, but its mass is bounded from above by $M_h \lesssim 135$ GeV [21], i.e. the light Higgs stays below the threshold above which the decay to $WW^{(}(*)$ or $ZZ^{(*)}$ can be exploited. On the other hand, the heavy MSSM Higgs bosons, $H$ and $A$, decouple from the gauge bosons. Consequently, the analysis for $M_{H^{\rm SM}} = 160$ GeV cannot be taken over to the MSSM. This shows the importance of channels to determine spin and $\mathcal{CP}$-properties of the Higgs bosons without relying on (tree-level) couplings of the Higgs bosons to gauge bosons. CED Higgs production can yield crucial information in this context [5–7]. The $M_{H^{\rm SM}} = 120$ GeV analysis, on the other hand, can in principle be applied to the SUSY case. However, the coupling of the SUSY Higgs bosons to tau leptons, in this case does not exhibit a (sufficiently) strong enhancement as compared to the SM case, i.e. no improvement over the $2\sigma$ effect within the SM can be expected. The same would be true in any other model of new physics with a light SM-like Higgs and heavy Higgses that decouple from the gauge bosons.

# 4 Sensitivity to Higgs Bosons in the SM4

A very simple example of physics beyond the SM is a model, "SM4", which extends the SM by a fourth generation of heavy fermions, see, for instance, [13]. In particular, the masses of the 4th generation quarks and leptons are assumed to be (much) heavier than the mass of the top-quark. In this case, the effective coupling of the Higgs boson to two gluons is three times larger than in the SM. No other coupling, relevant to LEP and Tevatron searches, changes significantly. Essentially, only the partial decay width $\Gamma(H \to gg)$ changes by a factor of 9 and, with it, the total Higgs width and therefore all the decay branching ratios [32]. The new total decay width and the relevant decay branching ratios can be evaluated as,

$$\begin{aligned}
\Gamma_{\rm SM}(H \to gg) &= {\rm BR}_{\rm SM}(H \to gg)\,\Gamma^{\rm SM}_{\rm tot}(H)\,,\\
\Gamma_{\rm SM4}(H \to gg) &= 9\,\Gamma_{\rm SM}(H \to gg)\,,\\
\Gamma^{\rm SM4}_{\rm tot}(H) &= \Gamma^{\rm SM}_{\rm tot}(H) - \Gamma_{\rm SM}(H \to gg) + \Gamma_{\rm SM4}(H \to gg)\,.
\end{aligned}$$

In Fig. 3 we show the bounds on $M_{H^{\rm SM4}}$ from LEP and Tevatron searches (taken from [22], where also an extensive list of experimental references can be found.) Shown is the experimen-tally excluded cross section divided by the cross section in the SM and the SM4, respectively. The SM4 (SM) is given by the dashed (solid) line. In the red/light grey part the LEP exclusion provides the strongest bounds, while for the blue/dark grey part the Tevatron yields stronger limits. On can see that the exclusion bounds on $M_{H^{\rm SM4}}$ are much stronger than on $M_{H^{\rm SM}}$, and only a window of 112 GeV $\lesssim M_{H^{\rm SM4}} \lesssim$ 145 GeV is still allowed. At larger masses (not shown) $M_{H^{\rm SM4}} \gtrsim$ 220 GeV also remains unexcluded. Consequently, we can focus our studies on the still allowed regions.

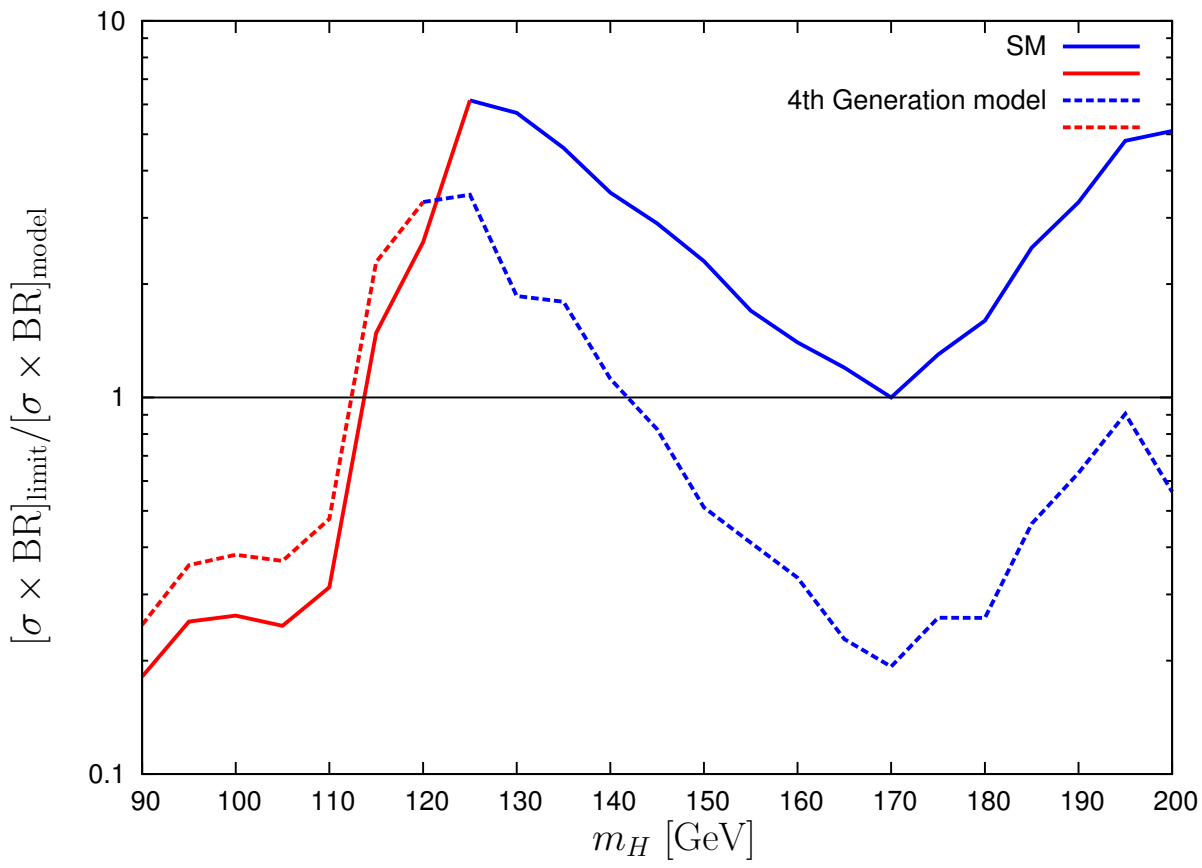

Figure 3: Cross section ratio (see text) for the most sensitive channel as a function of the Higgs mass $m_H$: 4th Generation Model versus SM. The colors indicate whether the most sensitive search channel is from LEP (lighter grey) or the Tevatron (darker grey), taken from [22].

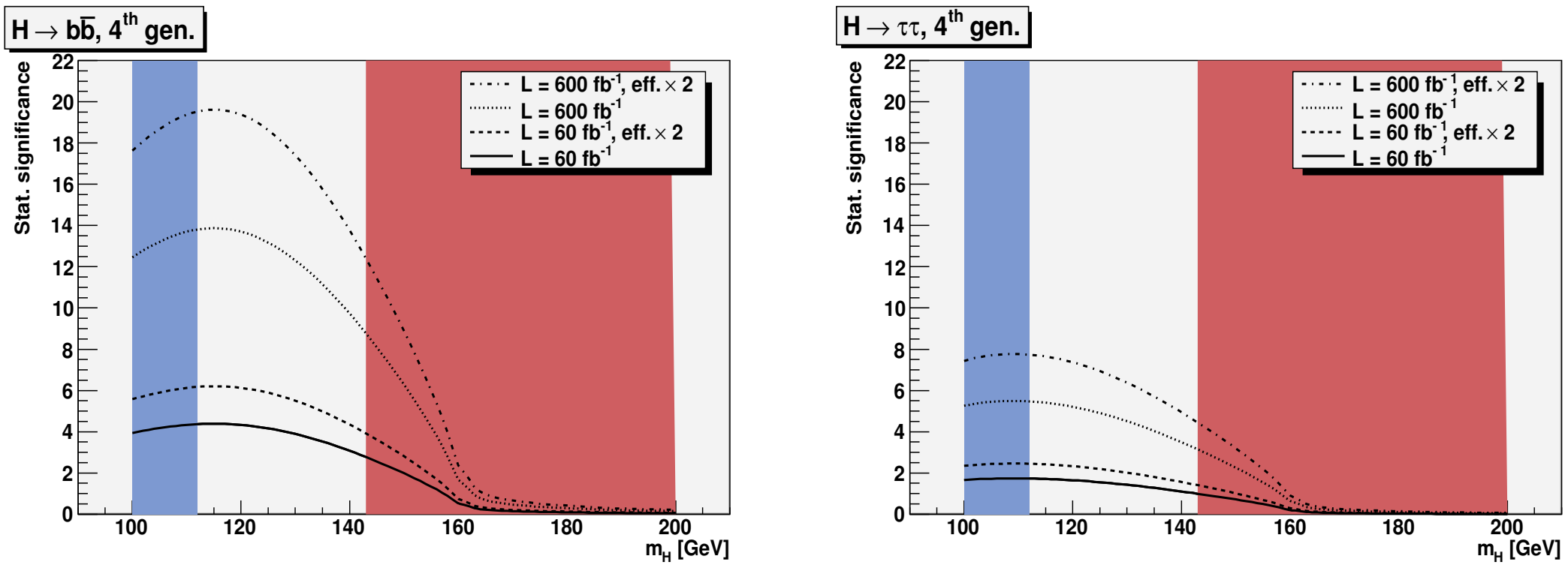

Figure 4: Significances reachable in the SM4 in the $H \to b\bar{b}$ (left) and $H \to \tau^+\tau^-$ (right) channel for effective luminosities of "60 fb$^{-1}$", "60 fb$^{-1}$ eff$\times$2", "600 fb$^{-1}$" and "600 fb$^{-1}$ eff$\times$2". The regions excluded by LEP appear as blue/light grey for low values of $M_{H^{\rm SM4}}$ and excluded by the Tevatron as red/dark grey for larger values of $M_{H^{\rm SM4}}$.

As for the MSSM we have evaluated the significances that can be obtained in the channels $H \to b\bar{b}$ and $H \to \tau^+\tau^-$. The results are shown in Fig. 4 as a function of $M_{H^{\rm SM4}}$ for the four luminosity scenarios. The regions excluded by LEP appear as blue/light grey for low values of $M_{H^{\rm SM4}}$ and regions excluded by the Tevatron appears as red/dark grey for larger values of $M_{H^{\rm SM4}}$. The $b\bar{b}$ channel (left plot) shows that even at rather low luminosity the remaining window of 112 GeV $\lesssim M_{H^{\rm SM4}} \lesssim$ 145 GeV can be covered by CED Higgs production. Due to the smallness of BR($H^{\rm SM4} \to b\bar{b}$) at $M_{H^{\rm SM4}} \gtrsim$ 160 GeV, however, the CED channel becomes irrelevant for the still allowed high values of $M_{H^{\rm SM4}}$. The $\tau^+\tau^-$ channel (right plot) has not enough sensitivity at low luminosity, but might become feasible at high LHC luminosity. At masses $M_{H^{\rm SM4}} \gtrsim 220$ GeV it might be possible to exploit the decay $H \to WW, ZZ$, but no analysis has been performed up to now.

# References

[1] A. De Roeck et al., *Eur. Phys. J.* **C 25**, 391 (2002).

[2] CERN/LHCC 2006-039/G-124, CMS Note 2007/002, TOTEM Note 06-5.

[3] M. Albrow et al. [FP420 R&D Collaboration], arXiv:0806.0302 [hep-ex].

[4] The AFP project in ATLAS, Letter of Intent.

[5] V.A. Khoze, A.D. Martin and M.G. Ryskin, *Eur. Phys. J.* **C 23**, 311 (2002).

[6] A.B. Kaidalov et al., *Eur. Phys. J.* **C 33**, 261 (2004).

[7] S. Heinemeyer, V.A. Khoze, M.G. Ryskin, M. Tasevsky and G. Weiglein, *Eur. Phys. J.* **C 53**, 231 (2008) [arXiv:0708.3052 [hep-ph]].

[8] B. Cox, F. Loebinger and A. Pilkington, *JHEP* **0710**, 090 (2007) [arXiv:0709.3035 [hep-ph]].

[9] J. R. Forshaw et al., *JHEP* **0804**, 090 (2008) [arXiv:0712.3510 [hep-ph]].

[10] S. Heinemeyer et al., arXiv:0811.4571 [hep-ph].

[11] M. Chaichian, P. Hoyer, K. Huitu, V. A. Khoze and A. D. Pilkington, arXiv:0901.3746 [hep-ph].

[12] S. Heinemeyer, *Int. J. Mod. Phys.* **A 21** 2659 (2006) [arXiv:hep-ph/0407244]; A. Djouadi, *Phys. Rept.* **459** (2008) 1 [arXiv:hep-ph/0503173].

[13] P. H. Frampton, P. Q. Hung and M. Sher, *Phys. Rept.* **330** (2000) 263 [arXiv:hep-ph/9903387].

[14] V.A. Khoze, A.D. Martin and M. Ryskin, *Eur. Phys. J.* **C 19** 477 (2001) [Errat.-ibid. **C 20** 599 (2001)] [arXiv:hep-ph/0011393].

[15] M. Dührssen, S. Heinemeyer, H. Logan, D. Rainwater, G. Weiglein and D. Zeppenfeld, *Phys. Rev.* **D 70** (2004) 113009 [arXiv:hep-ph/0406323].

[16] R. Lafaye, T. Plehn, M. Rauch, D. Zerwas and M. Dührssen, arXiv:0904.3866 [hep-ph].

[17] V.A. Khoze, A.D. Martin, M. Ryskin, *Eur. Phys. J.* **C 34** 327 (2004) [arXiv:hep-ph/0401078]; J. Ellis, J. Lee, A. Pilaftsis, *Phys. Rev.* **D 70** 075010 (2004) [arXiv:hep-ph/0404167] *Phys. Rev.* **D 71** 075007 (2005) [arXiv:hep-ph/0502251]; M. Boonekamp et al., *Phys. Rev.* **D 73** 115011 (2006) [arXiv:hep-ph/0506275].

[18] A. G. Shuvaev et al., *Eur. Phys. J.* **C 56**, 467 (2008) [arXiv:0806.1447 [hep-ph]].

[19] V. A. Khoze, M. G. Ryskin and A. Martin, arXiv:0907.0966 [hep-ph].

[20] See: `www.feynhiggs.de` .

[21] S. Heinemeyer, W. Hollik and G. Weiglein, *Comp. Phys. Commun.* **124** (2000) 76 [arXiv:hep-ph/9812320]; *Eur. Phys. J.* **C 9** (1999) 343 [arXiv:hep-ph/9812472]; G. Degrassi et al., *Eur. Phys. J.* **C 28** (2003) 133 [arXiv:hep-ph/0212020]; M. Frank et al., *JHEP* **0702** (2007) 047 [arXiv:hep-ph/0611326].

[22] P. Bechtle, O. Brein, S. Heinemeyer, G. Weiglein and K. E. Williams, to appear in *Comput. Phys. Commun.*, arXiv:0811.4169 [hep-ph]; arXiv:0909.4664 [hep-ph]. The code can be obtained from `www.ippp.dur.ac.uk/HiggsBounds`

[23] M. Carena, S. Heinemeyer, C. Wagner and G. Weiglein, *Eur. Phys. J.* **C 26** 601 (2003) [arXiv:hep-ph/0202167]; *Eur. Phys. J.* **C 45** 797 (2006) [arXiv:hep-ph/0511023].

[24] V.A. Khoze, M. Ryskin and W.J. Stirling, *Eur. Phys. J.* **C 48** 797 (2006) [arXiv:hep-ph/0607134].

[25] V. Berardi et al. [TOTEM Collab.], TDR, CERN-LHCC-2004-002, TOTEM-TDR-001, January 2004.

[26] B. E. Cox et al., *Eur. Phys. J.* **C 45** (2006) 401 [arXiv:hep-ph/0505240].

[27] V. Buescher and K. Jakobs, *Int. J. Mod. Phys.* **A 20** (2005) 2523 [arXiv:hep-ph/0504099].

[28] T. Plehn, D. L. Rainwater and D. Zeppenfeld, *Phys. Rev. Lett.* **88** (2002) 051801 [arXiv:hep-ph/0105325].

[29] V. Hankele, G. Klamke, D. Zeppenfeld and T. Figy, *Phys. Rev.* **D 74** (2006) 095001 [arXiv:hep-ph/0609075].

[30] C. Ruwiedel, N. Wermes and M. Schumacher, *Eur. Phys. J.* **C 51** (2007) 385.

[31] [CDF Collaboration and D0 Collaboration], arXiv:0903.4001 [hep-ex].

[32] G. Kribs, T. Plehn, M. Spannowsky and T. Tait, *Phys. Rev.* **D 76** (2007) 075016 [arXiv:0706.3718 [hep-ph]].

# Exclusive Higgs Production in a Triplet Scenario

*M. Chaichian*[1], *P. Hoyer*[1], *K. Huitu*[1], *V.A. Khoze*[2,3], *A.D. Pilkington*[3]

[1]Department of Physics, University of Helsinki, and Helsinki Institute of Physics, P.O. Box 64, FIN-00014 University of Helsinki, Finland

[2]Institute for Particle Physics Phenomenology, University of Durham, Durham, DH1 3LE, UK

[3]School of Physics & Astronomy, University of Manchester, Manchester M13 9PL, UK

In this talk, we discuss searching for the neutral Higgs boson of a triplet model in central exclusive production at the Large Hadron Collider. In a detailed Monte Carlo analysis, it is found that for appropriate values of the model parameters, an excellent Higgs mass measurement is possible, and that distinguishing the triplet model Higgs boson from the Higgs boson of the Standard Model is possible.

## 1 Introduction

It is well known that in the Standard Model (SM), there is one Higgs doublet responsible for the electroweak symmetry breaking, and consequently there is one physical Higgs boson in the model. The one physical Higgs boson can be considered as the minimal choice, since in addition for the mechanism providing masses for all the particles in the SM, it also takes care of the unitarity of the SM.

However, in most extensions of the SM more Higgs representations occur. In supersymmetric models one has necessarily at least two doublets. Singlets occur in many extensions of the SM. One motivation for including a singlet in a supersymmetric model is to include in a natural way the dimensionful coupling of the minimal supersymmetric standard model, the so-called $\mu$-parameter. In left-right symmetric models, triplets are added to generate a small mass for the neutrinos. Although the new scalars do not always take part in the electroweak symmetry breaking, they affect the properties of the Higgs boson through mixing.

Models with an extended Higgs sector typically contain charged scalars. A large number of studies have previously investigated the possibility of studying the doubly or singly charged components of higher representations. However, the charged scalars may be considerably heavier than the light neutral bosons. Therefore, it would be instructive to study the properties of the light neutral Higgs particles in order to reveal the manifestation of new representations [1].

Higgs triplets are an especially attractive possibility [2]. A tiny neutrino mass may indicate that the mass is being generated by the seesaw mechanism containing the coupling of neutrinos to the triplet. In addition, composite Higgs models contain several multiplets, including the triplet ones. Triplets also occur in the little Higgs models.

Determining that a new detected state is indeed a Higgs boson and distinguishing it from the Higgs boson of the SM will be far from trivial. This task will require a comprehensive programme of precision Higgs measurements. In particular, it will be of utmost importance to determine the spin and $CP$ properties of a new state and to measure precisely its mass, width and couplings.

Based on [3], we discuss here searching for the lightest neutral Higgs boson $H_1^0$ of a model containing triplets, and at the same time identifying the representation of the found $H_1^0$. For this it was found in [3] that the central exclusive production (CEP) mechanism (see, for example, [4]) is very beneficial, if forward proton detectors are installed at ATLAS and/or CMS (see [5]).

# 2 Models with General Higgs Representations

We start with the Standard Model gauge group $\mathrm{SU}(2)_L \times \mathrm{U}(1)_Y$ for the electroweak sector. The masses of the gauge bosons are then obtained from the kinetic part of Lagrangian,

$$L_{kin} = \sum_k (D^\mu \phi_k)^* (D_\mu \phi_k) + \frac{1}{2} \sum_i (D^\mu \xi_i)^T (D_\mu \xi_i), \tag{1}$$

where $\phi_k$ are complex representations and $\xi_i$ are real ones. The covariant derivative is written as $D_\mu = \partial_\mu + igW_\mu^a T^a + \frac{Y}{2} g' B_\mu$, where $T^a$ is the generator of SU(2) in the appropriate representation (with $\mathrm{Tr}(T^a T^b) = \frac{1}{2}\delta^{ab}$) and $Y$ is the U(1) hypercharge. Here $W^a$ and $B$ are the SU(2) and U(1) gauge bosons respectively, and the mixing angle $\theta_W$ of the $Z$ boson and photon is obtained by diagonalising the neutral sector. The $W$ and $Z$ boson masses are given by

$$m_Z^2 = (g^2 + g'^2) \sum_i T_{3i}^2 v_i^2, \quad m_W^2 = g^2 \sum_i T_{3i}^2 v_i^2, \tag{2}$$

where $T_{3i}$ is the isospin third component and $v_i$ is the VEV of particle $i$. It is clear from Eq. (2) that the doublet VEV decreases when several representations obtain non-vanishing VEVs. Furthermore, since the left-handed fermions are in doublets, the charged fermions can only get their masses through the Higgs doublet representation, $m_f = y_f v_{doublet}$, and the fermion Yukawa coupling, $y_f$, must increase to produce the fermion masses. This, for example, leads to an enhancement in the production cross section for Higgs production via gluon fusion, where the dominant contribution is due to the top quark loop. A further enhancement is present in the branching ratio to fermion anti-fermion pairs. The possibility arises, therefore, of observing a very different prediction to that of the Standard Model.

The higher Higgs representations are severely restricted by the electroweak $\rho$-parameter. The $\rho$-parameter in the Standard Model is defined by the ratio of the gauge boson masses,

$$\rho = \frac{m_W^2}{m_Z^2 \cos^2 \theta_W}, \tag{3}$$

which at tree level is exactly unity in the Standard Model. In a model with several scalar representations, whose neutral component develops a VEV, the $\rho$-parameter is given at tree level by [6]

$$\rho = \frac{\sum_i r_i \left( T_i(T_i+1) - T_{3i}^2) v_i^2 \right)}{\sum_i 2 T_{3i}^2 v_i^2}. \tag{4}$$

Here $T_i$ is the weak isospin and $r_i = 1/2(1)$ for real (complex) representations. Finally, the $\rho$-parameter is experimentally constrained to be [7] (quoted errors are at $2\sigma$),

$$\rho - 1 = 0.0002 \begin{array}{c} +0.0024 \\ -0.0009 \end{array}. \tag{5}$$

## 3 Higgs Bosons in a Triplet Model with $\rho = 1$

In order to fulfill the experimental constraint on the $\rho$-parameter in Eqn. (5), the triplet VEV has to be small. Using Eqs. (4) and (5), one finds that the upper limit for the triplet VEV is a few GeV. An alternative method to satisfy the experimental constraint at tree-level is to have representations which add up to $\rho = 1$.

We consider the model studied in [8] in which additional representations are chosen in such a way that the tree-level value of $\rho$ remains unity. The $\rho$-parameter is fixed to one by choosing one complex scalar doublet ($\phi_{Y=1}$) and two triplets, one real ($\xi_{Y=0}$) and one complex ($\chi_{Y=2}$). These can be written as

$$\phi = \begin{pmatrix} \phi^{0*} & \phi^+ \\ \phi^- & \phi^0 \end{pmatrix}, \quad \chi = \begin{pmatrix} \chi^0 & \xi^+ & \chi^{++} \\ \chi^- & \xi^0 & \chi^+ \\ \chi^{--} & \xi^- & \chi^{0*} \end{pmatrix}. \tag{6}$$

The VEVs of the neutral components of the Higgs fields are denoted by $\langle\chi^0\rangle = \langle\xi^0\rangle = b$ and $\langle\phi^0\rangle = a/\sqrt{2}$. For doublet-triplet mixing, the standard notation is employed:

$$c_H \equiv \frac{a}{\sqrt{a^2+8b^2}}, \quad s_H \equiv \frac{\sqrt{8}b}{\sqrt{a^2+8b^2}}, \quad v^2 \equiv a^2 + 8b^2. \tag{7}$$

As we are interested in this model mainly to illustrate the possibility of studying a neutral triplet Higgs sector, the tree-level results of this triplet model are sufficient for demonstrating the phenomenology of the higher representations. In this case, the neutral doublet and triplet do not mix and the neutral mass eigenstates are

$$H_1^0 = \phi^{0r}, \quad H_1^{0'} = \frac{1}{\sqrt{3}}(\sqrt{2}\chi^{0r} + \xi^0), \quad H_3^0 = c_H\chi^{0i} + s_H\phi^{0i}, \quad H_5^0 = \frac{1}{\sqrt{3}}(\sqrt{2}\xi^0 - \chi^{0r}), \tag{8}$$

where $\chi^0 = (\chi^{0r} + i\chi^{0i})/\sqrt{2}$. The mass of $H_1^0$ can be written as $m^2_{H_1^0} = 8c_H^2\lambda_1 v^2$, where $\lambda_1$ is the coupling between four doublets in the potential. Here we will assume $H_1^0$ is the lightest scalar, which can be the case if either $c_H$ or $\lambda_1$ is small.

The couplings of this lightest neutral scalar to the fermions and the gauge bosons are

$$H_1^0 q\bar{q} : -\frac{gm_q}{2m_W c_H}, \quad H_1^0 W^+W^- : gm_W c_H, \quad H_1^0 ZZ : \frac{g}{\cos^2\theta_W} m_W c_H. \tag{9}$$

It is clear that, at tree-level, the coupling of the $H_1^0$ to fermions is always enhanced by the factor of $1/c_H$. Importantly, the gauge boson couplings to $H_1^0$ are suppressed by a factor $c_H$ with respect to the SM and the role of vector boson fusion mechanism for $H_1^0$ production is reduced if $c_H$ is small.

The mass limits for $H_1^0$ can be deduced from the LEP results. If we assume that the number of $b$-quark pairs gives the Higgs boson mass limit, it must be heavier than 73 GeV (40 GeV) for $c_H = 0.5$ ($c_H = 0.2$). Unitarity further constrains most masses, requiring them to be less than of the order of 1 TeV. The Yukawa couplings are constrained by perturbativity, which limits the $H_1^0$ coupling to top,

$$\frac{gm_{top}}{2m_W c_H} < \sqrt{4\pi}. \tag{10}$$

From this it follows that $c_H > 0.2$, which is the most stringent tree-level limit for $c_H$.

When calculating the branching ratios, it is necessary to consider also the loop induced decays of the Higgs bosons to gluons and photons. Taking these into account, the branching ratios of $H_1^0$ are presented in Fig. 1 for $m_{H_1^0} = 120$ GeV and 150 GeV.

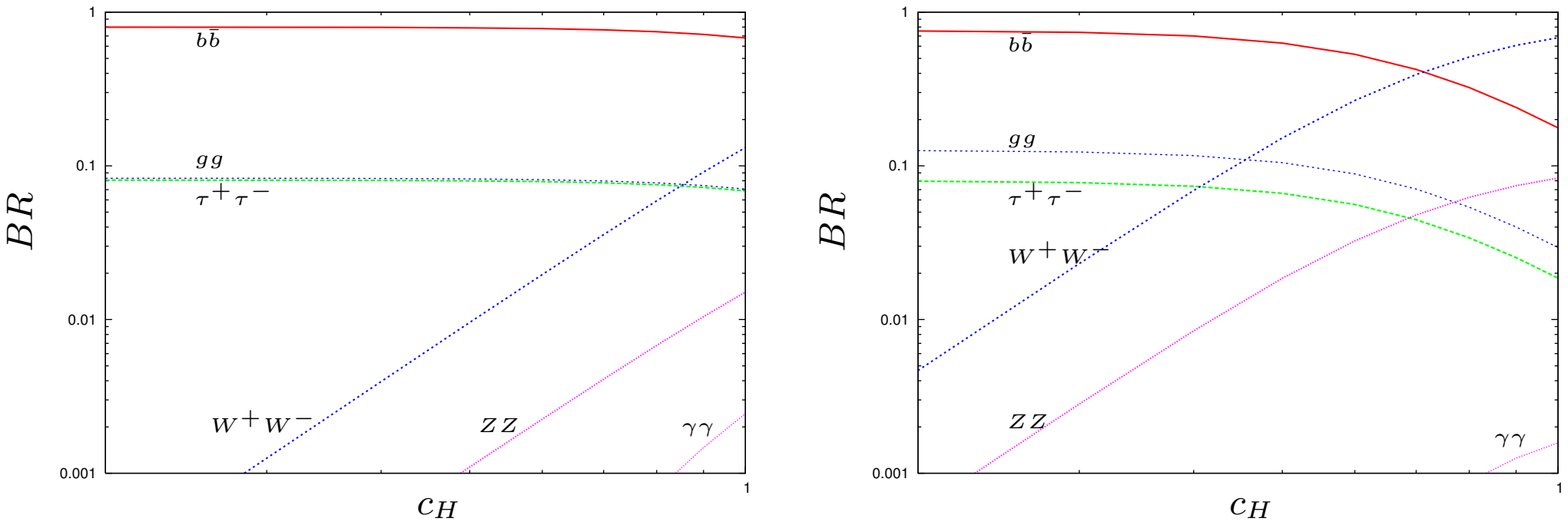

Figure 1: Branching ratios of $H_1^0$ to the Standard Model particles for $m_H = 120$ GeV (left) and $m_H = 150$ GeV (right).

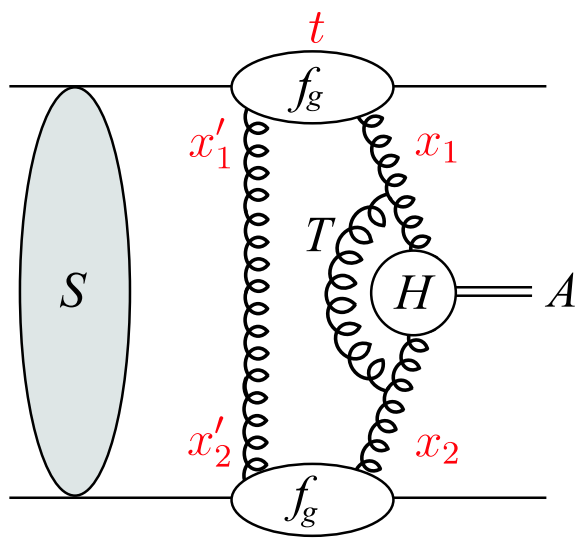

Figure 2: A symbolic diagram for the central exclusive production of a Higgs boson $H$.

# 4 Central Exclusive Diffractive Production of the Triplet Higgs Boson

The central exclusive production (CEP) of a Higgs boson is defined as $pp \to p \oplus H \oplus p$, where the $\oplus$ denote the presence of large rapidity gaps between the outgoing protons and the decay products of the central system. Schematically the process is shown in Fig. (2). It has been suggested in recent years that CEP offers a unique complimentary measurement to the conventional Higgs search channels. Firstly, if the outgoing protons scatter through small angles then, to a very good approximation, the primary active di-gluon system obeys a $J_z = 0$, $CP$-even selection rule [9]. The observation of the Higgs boson in the CEP channel therefore determines the Higgs quantum numbers to be $J^{PC} = 0^{++}$. Secondly, because the process is exclusive, all of the energy/momentum lost by the protons during the interaction goes into the production of the central system. Measuring the outgoing proton allows the central mass to be measured to just a few GeV, regardless of the decay products of the central system. A mass measurement of this type will require new forward proton detectors to be installed at ATLAS and/or CMS.

For a Standard Model Higgs boson, central exclusive diffraction could allow the main decay channels ($b\bar{b}$, $WW$ and $\tau\tau$) to be observed in the same production channel, which provides the opportunity to study the Higgs coupling to $b$-quarks. This may be very difficult to access in other search channels at the LHC, despite the fact that $H \to b\bar{b}$ is by far the dominant decay mode for a light SM Higgs boson. In [3], we propose that CEP is beneficial if higher representations of the Higgs sector are realised, in particular, in searches for the Higgs triplets discussed in Section 3. The CEP of the MSSM Higgs bosons is discussed, for instance, in Ref. [10]. The theoretical formalism [11] for central exclusive production contains distinct parts, as illustrated in Fig. 2.

| $\sigma_{H\to b\bar{b}}$ (fb) | $m_H = 120$ GeV | $m_H = 150$ GeV |
|---|---|---|
| $c_H = 0.2$ | 113.5 | 55.2 |
| $c_H = 0.5$ | 18.0 | 7.4 |
| $c_H = 0.8$ | 6.6 | 1.5 |

Table 1: Generator level cross sections, $\sigma_{H\to b\bar{b}}$, for central exclusive Higgs boson production for $m_H = 120, 150$ GeV and $c_H = 0.2, 0.5, 0.8$.

# 5 Simulation of Higgs Production in the Triplet Model

We restrict our discussion and analysis to the ATLAS interaction point (IP), and note that a similar result would be obtained at CMS. There are three important aspects of forward proton tagging at the LHC that need to be considered for the purposes of this analysis; the acceptance and resolution of the proposed forward proton detectors and the ability of the detectors to measure the time-of-flight of each proton from the interaction point.

The central exclusive signal and background events are simulated with full parton showering and hadronisation effects using the ExHuME v1.3.4 event generator [12]. ExHuME contains a direct implementation of the KMR calculation [11, 4] of central exclusive diffraction given in Sec. 4. The CTEQ6M [13] parton distribution functions are used to calculate the generalised gluon distributions, $f_g$. The generator level cross sections for central exclusive $H \to b\bar{b}$ production in the triplet model are presented in Table 1 for $m_H = 120, 150$ GeV and $c_H = 0.2, 0.5, 0.8$.

The backgrounds to $H \to b\bar{b}$ can be broken down into three broad categories; central exclusive, double pomeron exchange and overlap. See the discussion on backgrounds in [3]. To enhance the signal, we follow the experimental method used in a previous study of $H \to b\bar{b}$ in the SM and the MSSM [14], which imposes a number of exclusivity cuts. These include a cut on di-jet mass fraction, comparison of rapidity of the system found from different input, back-to-back structure of di-jets, and number of charged tracks in signal and background.

A major experimental challenge for central exclusive jet analyses is developing a trigger strategy to retain enough events. As discussed in [3], several different triggering strategies are needed.

We estimate the significance of observing a neutral Higgs boson in the triplet model for the mentioned parameter choices. As the overlap background is luminosity dependent we must specify how the data was collected. For example, we examine the significance for an integrated luminosity of 60 fb$^{-1}$, which corresponds to between three and four years of data acquisition given a peak luminosity of $2\times10^{33}$ cm$^{-2}$ s$^{-1}$. We also present in [3] results for 300 fb$^{-1}$ of data, which corresponds to between three and four years of data acquisition given a peak luminosity of $10^{34}$ cm$^{-2}$ s$^{-1}$.

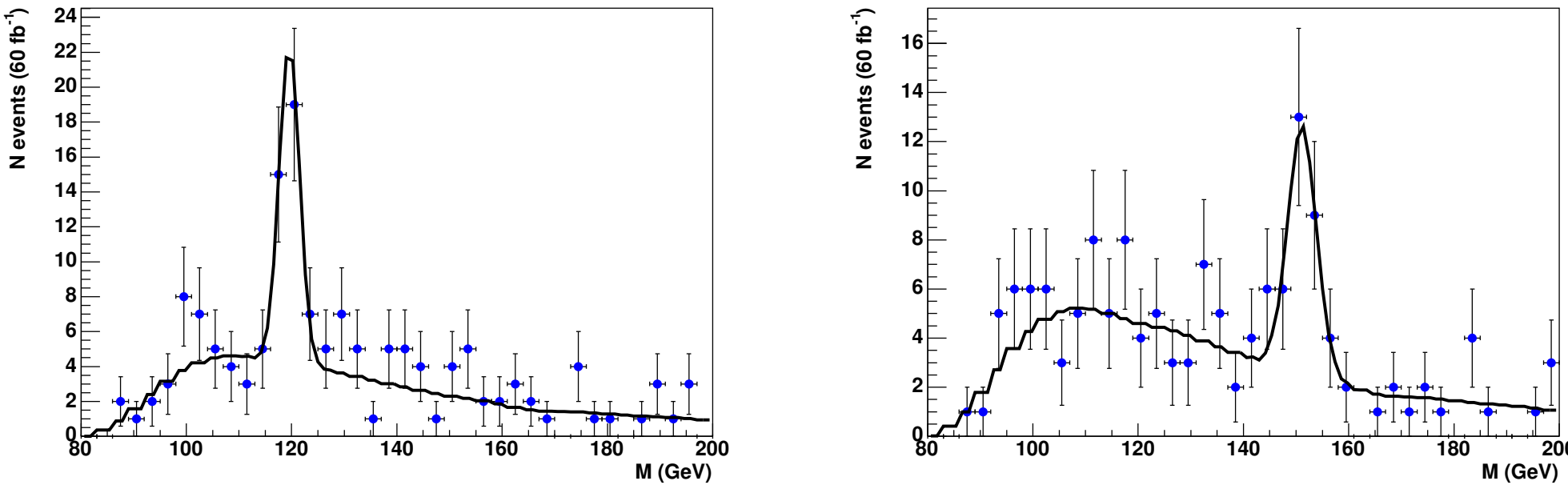

Figure 3: Expected mass distributions given 60 $\text{fb}^{-1}$ of data, collected at $2\times10^{33}$ $\text{cm}^{-2}$ $\text{s}^{-1}$ using a JR25+MU6 trigger, for the following parameter choices: (left) $m_H = 120$ GeV and $c_H = 0.5$, significance is $4.5\sigma$. (right) $m_H = 150$ GeV and $c_H = 0.5$, significance is $3.9\sigma$.

# 6 Conclusions

Searches for the manifestation of the extended Higgs sector at the LHC may allow new insight in the nature of electroweak symmetry breaking. The central exclusive production mechanism would provide a very powerful tool to complement the standard strategies at the LHC for studying Higgs particles. Here we focus on the production of the neutral Higgs boson of the triplet model in the forward proton mode. We assume a model with the tree-level value of the electroweak $\rho$-parameter consistent with experiment, $\rho = 1$. Although this model is used as a benchmark model for the triplets, our results are more general. An extra contribution from other representations enhances the doublet Yukawa couplings resulting in a different experimental signature to that of the SM. We show that a factor of two enhancement of the fermion couplings due to the higher representations implies a significant difference to the doublet case. Let us emphasise that in the case of the current model, all the fermion couplings to the Higgs boson, which is responsible for the fermion masses, increase. This is in contrast with, for instance, the MSSM, where couplings of up-type and down-type fermions change from the Standard Model differently, due to the fact that there are only doublets in the model. It is a common feature of higher Higgs representations that the doublet couplings are enhanced, which thus indicates that higher representations are involved.

We present a detailed Monte Carlo analysis of the central exclusive production of a triplet model Higgs boson for a number of parameter choices. For $c_H \leq 0.5$, we have shown that a light $H_1^0$ Higgs boson (of mass 120-150 GeV) can be observed with a $4\sigma$ (or better) significance if a fixed rate trigger is used. The expected error in the Higgs mass measurement using forward proton detectors is small. Regardless of the parameter choice, the mass measurement can always be made to better than 2 GeV if a fixed rate single jet trigger is used to retain events in which both protons are tagged at 420 m from the IP.

# Acknowledgements

KH gratefully acknowledges support from the Academy of Finland (Project No. 115032). This work was funded in the UK by the STFC.

# References

[1] A. Kundu and B. Mukhopadhyaya, Int. J. Mod. Phys. A **11** (1996) 5221 [arXiv:hep-ph/9507305]; G. L. Kane, S. F. King and L. T. Wang, Phys. Rev. D **64** (2001) 095013 [arXiv:hep-ph/0010312]; C. P. Burgess, J. Matias and M. Pospelov, Int. J. Mod. Phys. A **17** (2002) 1841 [arXiv:hep-ph/9912459]; D. Choudhury, A. Datta and K. Huitu, Nucl. Phys. B **673** (2003) 385 [arXiv:hep-ph/0302141].

[2] for a recent review see E. Accomando *et al.*, arXiv:hep-ph/0608079, see 'Higgs triplets' by John F. Gunion and Chris Hays, p. 497.

[3] M. Chaichian, P. Hoyer, K. Huitu, V.A. Khoze, A.D. Pilkington, JHEP **05** (2009) 011 [arXiv:0901.3746].

[4] V.A. Khoze, A.D. Martin, M.G. Ryskin, Eur. Phys. J. C23 (2002) 311 [arXiv:hep-ph/0111078].

[5] M. G. Albrow *et al.* [FP420 R&D Collaboration]], arXiv:0806.0302 [hep-ex].

[6] J. F. Gunion, H. E. Haber, G. L. Kane and S. Dawson, "The Higgs Hunter's Guide," Westview Press; J. F. Gunion, H. E. Haber, G. L. Kane and S. Dawson, arXiv:hep-ph/9302272.

[7] J. Erler and P. Langacker, in W. M. Yao *et al.* [Particle Data Group], J. Phys. G **33** (2006) 1.

[8] H. Georgi and M. Machacek, Nucl. Phys. B **262** (1985) 463; M. S. Chanowitz and M. Golden, Phys. Lett. B **165** (1985) 105; J. F. Gunion, R. Vega and J. Wudka, Phys. Rev. D **42** (1990) 1673; J. F. Gunion, R. Vega and J. Wudka, Phys. Rev. D **43** (1991) 2322.

[9] V.A. Khoze, A.D. Martin and M. Ryskin, *Eur. Phys. J.* **C 19** (2001) 477 [Erratum-ibid. **C 20** (2001) 599], hep-ph/0011393.

[10] S. Heinemeyer, V. A. Khoze, M. G. Ryskin, W. J. Stirling, M. Tasevsky and G. Weiglein, LHC," Eur. Phys. J. C **53**, 231 (2008) [arXiv:0708.3052 [hep-ph]]; S. Heinemeyer, V. A. Khoze, M. G. Ryskin, M. Tasevsky and G. Weiglein, arXiv:0909.4665.

[11] V.A. Khoze, A.D. Martin, M.G. Ryskin, Eur. Phys. J. C **14** (2000) 525 [arXiv:hep-ph/0002072]; V.A. Khoze, A.D. Martin and M.G. Ryskin, Eur. Phys. J. C **18** (2000) 167 [arXiv:hep-ph/0007359]; M.G. Ryskin, A.D. Martin and V.A. Khoze, *Eur. Phys. J.* **C54**, 199 (2008); E. G. S. Luna, V. A. Khoze, A. D. Martin and M. G. Ryskin, Eur. Phys. J. C **59** (2009) 1 [arXiv:0807.4115 [hep-ph]]; M. G. Ryskin, A. D. Martin and V. A. Khoze, Eur. Phys. J. C **60** (2009) 249 [arXiv:0812.2407 [hep-ph]]; V. A. Khoze, A. D. Martin and M. G. Ryskin, Eur. Phys. J. C **55**, 363 (2008) [arXiv:0802.0177 [hep-ph]].

[12] J. Monk and A. Pilkington, Comput. Phys. Commun. **175**, 232 (2006) [arXiv:hep-ph/0502077].

[13] J. Pumplin, D. R. Stump, J. Huston, H. L. Lai, P. Nadolsky and W. K. Tung, JHEP **0207** (2002) 012 [arXiv:hep-ph/0201195].

[14] B. E. Cox, F. K. Loebinger and A. D. Pilkington, JHEP **0710** (2007) 090 [arXiv:0709.3035 [hep-ph]].

# Measuring Central Exclusive Processes at LHC

*Marek Taševský*

Institute of Physics, Academy of Sciences of the Czech republic, Na Slovance 2, 182 21 Prague, Czech republic

Diffractive physics program for experiments at the Large Hadron Collider is discussed with emphasis on measurements of central exclusive processes. At low luminosities, a L1 trigger based on requiring rapidity gaps can be used, while at high luminosities, the use of proton taggers proposed to be placed at 220 m and 420 m from the interaction point is foreseen.

# 1 Introduction

In this contribution, we focus on measurements of exclusive processes that could be performed at the multi-purpose LHC detectors, ATLAS and CMS. The central exclusive production (CEP) of new particles has received a great deal of attention in recent years (see [1] and references therein). The process is defined as $pp \to p \oplus \phi \oplus p$ and all of the energy lost by the protons during the interaction (a few percent) is used in the production of the central system, $\phi$. The final state therefore consists of a centrally produced hard subprocess, two very forward protons and no other activity. The '$\oplus$' sign denotes the regions devoid of activity, often called rapidity gaps. In Double Pomeron Exchange (DPE), the central system contains remnants from the diffractive exchange in addition to the hard subprocess.

# 2 Low Luminosity Running

At low luminosity, the diffractive processes can be detected using rapidity gaps. A possible L1 trigger would be based on a requirement of a rapidity gap on one or both sides from the IP and an activity in the central detector with energy over a certain threshold. The gap may span the region from the forward calorimeters of the ATLAS detector [2] or CMS detector [3] over the luminosity detectors (LUCID [4] in ATLAS or TOTEM [5] in CMS) up to ZDC detectors in ATLAS [6] or in CMS [3]. Measurements which would be straightforward and hence suitable for analyses of the very early LHC data are ratios of the kind of X+gaps/X(incl.), where X may be W, Z, dijet, heavy quark and dilepton, and X(incl.) means measuring X without requiring rapidity gaps. Measurements of ratios are convenient since many sources of systematic uncertainties are cancelled, particularly that of the luminosity at the early phase and among other, they also serve as valuable checks of different components of the formalism used to predict the CEP cross section by the KMR group [7, 8]. The soft survival probability, $S^2$, can be studied in electroweak processes, such as W+gaps or Z+ gaps. $S^2$ is defined as a probability that additional soft secondaries will not populate the gaps and it explains the factorisation breaking observed at hadron colliders when diffractive parton density functions (dPDF) obtained in Single Diffraction (SD) at HERA were applied in measurements of SD by CDF [9]. The generalised gluon distribution, $f_g$, can be probed in exclusive $\Upsilon$ production proceeding via either a photon or an odderon exchange. The higher-order QCD effects, especially Sudakov-like factors and also a possible role of the enhanced absorptive corrections can be studied in exclusive two- or three-jet events. When the proton tagging becomes available, the t-dependence of the elastic, SD and DPE cross sections can be obtained and hence effect of individual components of the pile-up background can be evaluated.

### 2.1 Dijet Production in DPE and CEP

Without proton tagging, the dijet production in DPE and CEP can be measured by requiring two central jets and rapidity gaps on both sides of the IP in forward calorimeter, LUCID/TOTEM and ZDC. In DPE, the rapidity gaps may be spoilt by particles from the pomeron remnants and although the cross section is about two orders of magnitude larger than the CEP cross section at the same dijet mass, the CEP cross section will dominate if the forward calorimeter is required to be devoid of activity. The measurement of the dijet production in CEP at 7–14 TeV may be compared with a similar measurement made at Tevatron from which models used to describe the data may be constrained.

### 2.2 Gaps between Jets

By selecting events with two jets each in opposite side of forward calorimeter and a rapidity gap in the central detector, ATLAS and CMS can improve an existing measurement of this type by D0 [10] at centre-of-mass (c.m.s.) energy of 1.8 TeV. Different colour singlet exchange models can be tested by comparing data with predictions for the gap fractions as functions of rapidity between the jets.

## 3 High Luminosity Running

A great attention is recently devoted to the possibility of complementing the standard LHC physics menu by adding forward proton detectors (FPD) to the ATLAS and CMS detectors. They would detect a great part of the energy flow that escapes undetected by the main detectors.

### 3.1 SM and BSM Higgs Boson Production

The forward proton tagging will provide an exceptionally clean environment to search for new phenomena at the LHC and to identify their nature. Of particular interest in this context is the CEP which gives access to the generalised (or skewed) PDFs. The CEP of a SM (or MSSM) Higgs boson is attractive for two reasons: firstly, if the outgoing protons remain intact and scatter through small angles then, to a good approximation, the central system $\phi$ must be produced in a $J_z = 0$, CP even state, therefore allowing a clean determination of the quantum numbers of any observed resonance. Here $J_z$ is the projection of the total angular momentum along the proton beam axis. Secondly, from precise measurements of proton momentum losses, $\xi_1$ and $\xi_2$, the mass of the central system can be measured much more precisely than from the dijet mass measured in the calorimeters, by the so-called missing mass method, $M^2 = \xi_1 \xi_2 s$, which is independent of the decay mode. The simplest decay mode from an experimental perspective is the WW decay mode, in which one (or both) of the W bosons decay leptonically. With standard single and double lepton trigger thresholds at ATLAS (or CMS), approximately 6 events are expected for Higgs boson mass around 160 GeV with luminosity of 30 fb$^{-1}$ [11]. In the $b\bar{b}$ decay mode, the quantum number selection rules in CEP strongly suppress the QCD b-jet background, nevertheless severe requirements necessary to get rid of the pile-up background make the event yield rather modest [12, 13]. Full details of the calculation of the background to this channel are described in [14, 15].

In certain regions of the MSSM parameter space the cross section for the CEP of Higgs bosons is significantly enhanced and hence making the $b\bar{b}$ decay mode attractive [15, 16]. In Fig. 1 an example mass spectrum is shown for MSSM Higgs boson candidates of mass of 120 GeV decaying into $b\bar{b}$ for $\tan\beta = 40$ (corresponding to the final cross section of about 18 fb) after 3 years of data taking at luminosity of $2 \cdot 10^{33}$ cm$^{-2}$s$^{-1}$ or 3 years at $10^{34}$ cm$^{-2}$s$^{-1}$. At the low luminosity, the pile-up background can be completely eliminated and the statistical significance is around $3.5\,\sigma$. At the highest luminosity, fast timing detectors are necessary to reduce the pile-up background - significance of $5\,\sigma$ is achieved with time resolution of 2 ps (see Section 4.4).

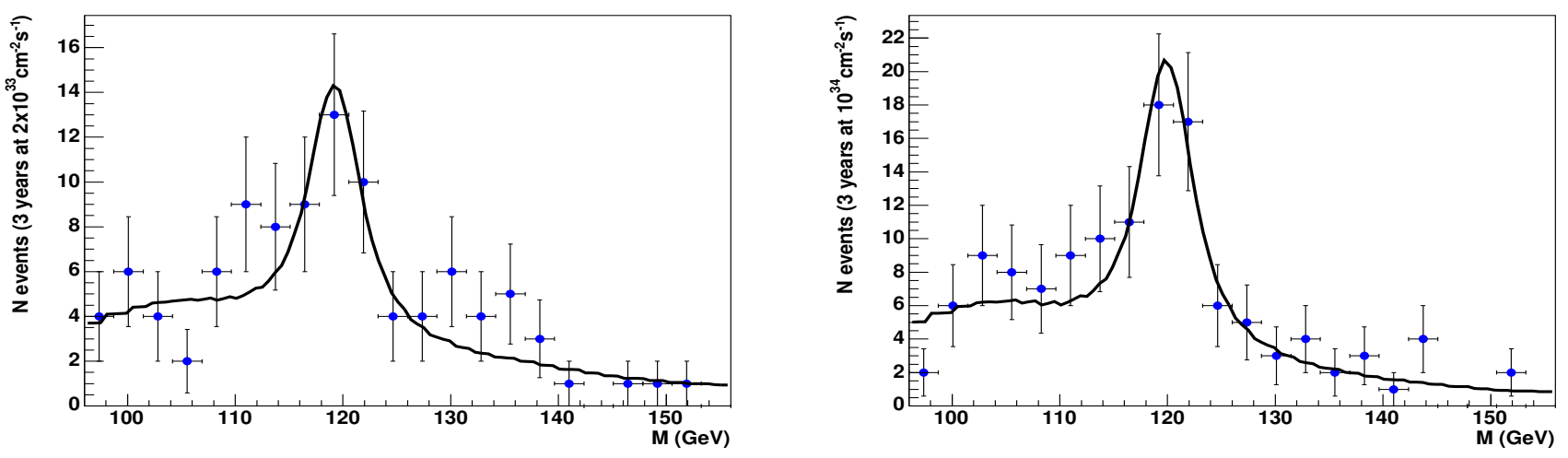

Figure 1: A typical mass fit of the $H \to b\bar{b}$ signal and its backgrounds for 3 years of data taking with ATLAS and the 420+420 detector configuration (a) at $2 \cdot 10^{33}$cm$^{-2}$s$^{-1}$ (60 fb$^{-1}$). The significance of the fit is $3.5\,\sigma$. (b) at $10^{34}$cm$^{-2}$s$^{-1}$ after removing the pile-up background completely. The significance is $5\,\sigma$. Figures from [16].

## 3.2 Central Exclusive Jet Production

The tagging of both protons in FPDs will enable a measurement of the proton transverse momenta and azimuthal angles which allows us to study the opacity of the incoming protons, and more generally, to test the dynamics of the soft survival probability by studying the correlations between the outgoing protons [7, 17]. This can be carried out with the CEP of dijets as the cross section is large. Thanks to the $J_z = 0$ selection rule which is applicable to all CEP processes, quark jet production is suppressed and the CEP can then be recognised as reduced ratio of b-jets to all jets when compared to other production processes.

## 3.3 Diffraction and QCD

The SD and DPE processes serve to provide an information about the low-x structure of the proton and the dPDFs. Inclusive jet and heavy quark production are mainly sensitive to the gluon component of the dPDFs, while vector boson production is sensitive to quarks. The kinematic region covered expands that explored at HERA and Tevatron, with values of $\beta$ (the fractional momentum of the struck parton in the diffractive exchange ) as low as $10^{-4}$ and of $Q^2$ up to $10^4$ GeV$^2$. More information about the SD dijet and W boson production can be found elsewhere in these proceedings (V. Juranek and W. Carvalho)

## 3.4 Photon-Photon Physics

As the LHC beams act also as a source of high-energy photons a rich program of photon-photon and photon-proton physics can be pursued. Photon-induced processes have been extensively studied at LEP and HERA. However at LHC, these processes can be investigated in an unexplored region of the phase-space. The final state topology is similar to CEP, i.e. a central system, X, separated on each side by large rapidity gap from a very forward proton detected in the FPD. Different average proton transverse momenta make it possible to separate between diffractive and photon-induced events.

The W- and Z-pair production (as a tool to study anomalous triple and quartic gauge couplings) is discussed elsewhere in these proceedings (O. Kepka). The SUSY particle production is described by K. Piotrzkowski in these proceedings and in [18].

### 3.4.1 Lepton Pair Production

Two-photon exclusive production of muon pairs has a well known QED cross section, including very small hadronic corrections [19]. Very recently, such event candidates have been observed by the CDF [20] and their cross sections found in a good agreement with theory. After applying simple selection criteria such as $p_T > 10$ GeV, $|\eta| < 2.5$ and requiring one forward proton tag, the cross section is 1.3 pb [1, 21]. This corresponds to approximately 50 muon pairs detected in a 12 hour run at a mean luminosity of $10^{33}\text{cm}^{-2}\text{s}^{-1}$. The large event rate coupled with a small theoretical uncertainty makes this process a perfect candidate for the absolute LHC luminosity calibration [22] and also of the FPD system at 420 m [1]. The $e^+e^-$ production can also be studied, although the trigger thresholds will be larger and hence the final event rate reduced.

## 3.5 Photoproduction

The high luminosity and the high c.m.s. energies available for photoproduction processes at the LHC allows us to study electroweak interactions and to search Beyond the Standard Model up to the TeV scale [21].

### 3.5.1 Associated WH Production

As shown in [21], the cross section for the associated WH production $(pp \to (\gamma p \to WHq') \to pWHq'Y$ after applying selection criteria and considering five different final states is 0.17 fb at $\mathrm{m}_H = 115$ GeV and 0.29 fb at $\mathrm{m}_H = 170$ GeV. The most promising channel seems to be the $jjl^{\pm}l^{\pm}$ at $\mathrm{m}_H = 170$ GeV where the signal to irreducible background ratio is 0.22 fb/0.28 fb, so luminosity of 100 $\text{fb}^{-1}$ might reveal the $HWW$ gauge coupling.

### 3.5.2 Single Top Quark and Anomalous Top Quark Production

Photoproduction of single top quark $(pp \to (\gamma p \to Wt) \to pWtY)$ is dominated by t-channel amplitudes in association with a W boson which all are proportional to the CKM matrix element $|V_{tb}|$. The ratio of associated Wt production cross section to the sum of all top production cross sections is 5% for parton-parton interactions, while it is 50% in photoproduction. In [21] two topologies were studied, namely $lbjj$ and $llb$. The signal cross section after selection cuts of about 44 fb with a signal to irreducible background ratio of 0.6 suggest that this mechanism and hence $|V_{tb}|$ may be easily measurable even with luminosity of 1 $\text{fb}^{-1}$.

At LHC the exclusive single top quark photoproduction can only occur via flavour changing neutral current processes which are not present at tree level of SM but appear in many extensions of SM such as two Higgs-doublet models or R-parity violating supersymmetry. The final state of this $pp \to (\gamma p \to t) \to ptY$ process is composed of a b-jet and a W boson. In [21] the leptonic $lb$ topology was studied and only photoproduction $\gamma p \to W + jet$ background considered. With an integrated luminosity of 1 fb$^{-1}$ the expected limits for anomalous couplings $k_{tu\gamma}$ and $k_{tc\gamma}$ at 95% CL are greatly improved with respect to existing best estimates.

### 3.5.3 Photoproduction of Jets

In photoproduction of jets, the fraction of the photon, $x_\gamma$, and proton, $x_p$, four-momentum carried by a parton involved in binary hard scattering is calculated using the energies and angles of jets in the central detector and of the protons in the FPDs. The direct photon processes are characterised by $x_\gamma \sim 1$ and resolved photon processes by $x_\gamma < 1$. The H1 and ZEUS collaborations have constrained the region $x_p, x_\gamma > 0.1$ for diffractive photoproduction. At LHC we expect to reach values of $x_p$ and $x_\gamma$ of an order of magnitude lower than at HERA. Furthermore, the diffractive photoproduction of dijet systems at the LHC promises to shed light on the issue of the QCD factorisation breaking recently reported in the same process by the H1 experiment [23].

### 3.5.4 Exclusive $\Upsilon$ Production

Exclusive $\Upsilon$ photoproduction, $\gamma p \to \Upsilon p$ can be studied using FPDs [8], although only one proton can be tagged due to the low mass of the $\Upsilon$. The cross section is expected to be approximately 1.25 pb for the decay channel $\Upsilon \to \mu^+\mu^-$ and is sensitive to the same skewed unintegrated gluon densities of the proton as the CEP of Higgs boson. Measuring this process thus helps to constrain the $f_g$ as the soft survival factor is expected to be close to 1. The $\gamma p \to \Upsilon p$ process can also occur via odderon exchange and this channel could be the first evidence for the odderon's existence.

# 4 Future Forward Proton Upgrades at the LHC

The forward detectors and possible upgrades at ATLAS and CMS have been described elsewhere by A. Zoccoli in these proceedings.

## 4.1 FP420

The FP420 R&D collaboration [1], with members from ATLAS, CMS and LHC studied the possibility of installing high precision tracking and timing detectors at 420 m from the IP. Detection of the protons will be achieved by two 3D silicon detector stations at each end of the FP420 region. This novel technique provides high radiation-resistive detectors close to the beam with an insensitive area as small as 5 $\mu$m and with a resolution of about 15 $\mu$m. The tracking and timing detectors will be attached to a movable beam pipe. As the beam pipes in the 420 m region are contained in an interconnecting cryostat and the sensitive detectors are best operated at room temperature, a new connection cryostat has been designed using a modified Arc Termination Module at each end.

## 4.2 Coverage of the Region of 220–240 m

Both ATLAS and CMS collaborations work on equipping the region of 220 m (ATLAS) or 240 m (CMS) by FPDs. The proposed equipment would be very similar to that at 420 m; differences are mainly in no need to change the cryostat and in an addition of a detector to be used for L1 trigger. In ATLAS, the combined effort to install FPDs at 220 and 420 m led in the AFP project (ATLAS Forward Proton) [24].

## 4.3 Acceptance and Resolution

With the position resolution of 15 $\mu$m we expect a mass resolution of the order of 1–2% for the 420+420 and about 3% for the 420+220 configurations over a mass range of 120–200 GeV. For given dipole apertures and collimator settings and a thin window of 200 $\mu$m, the expected $\xi$ range is 0.002–0.02 for 420+420 and 0.01–0.15 for the 420+220 configuration.

The low-$\xi$ (and therefore low mass) acceptance depends critically on the distance of approach of the active area of the sensitive detectors from the beam. The final distance of approach will depend on the beam conditions, machine-induced backgrounds and collimator positions, and the RF impact of the detector on the beams. At 420 m the nominal operating position is assumed to be between 5 and 7.5 mm, at 220 m it is between 2.0 and 2.5 mm. For masses above about 120 GeV, the 220 m detector adds to the acceptance with increasing importance as the central mass increases. The differences between ATLAS and CMS acceptances for the 420+420 as well as 420+220 configurations (see Fig. 2) are due to a different crossing angle which is in the vertical plane for IP1 (ATLAS) and horizontal plane at IP5 (CMS).

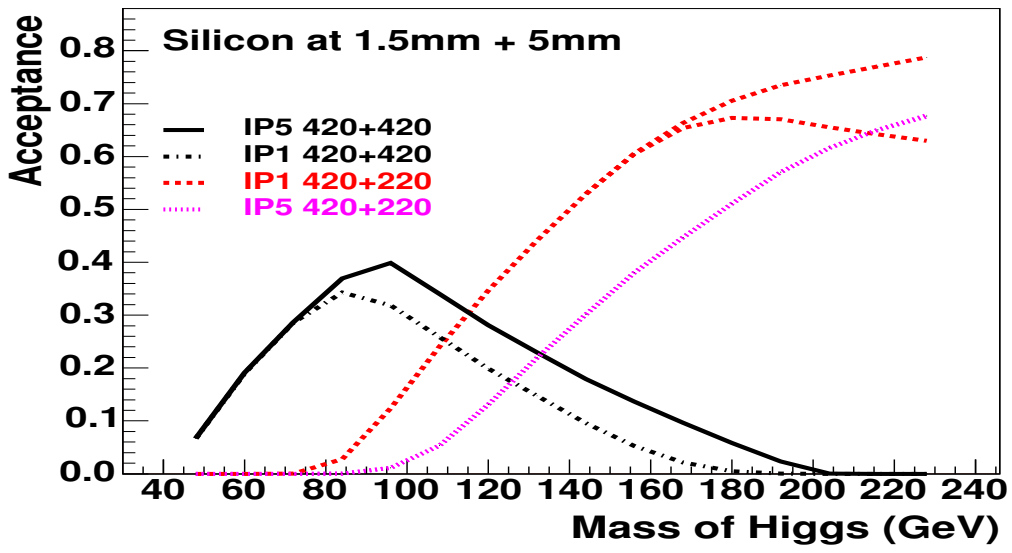

Figure 2: Mass acceptance for the 420+420 and 420+220 detector configurations (5 mm from the beam for 420 m and 1.5 mm for 220 m detectors) for IP1 and IP5. From [1].

## 4.4 Timing Detectors

The most prominent background to many diffractive physics analyses comes from an overlap of two soft SD events from pile-up and one ND event produced at a hard scale. Fast timing detectors with an expected sub-10 ps time resolution corresponding to a vertex resolution of better than 2.1 mm should be able to assign a vertex to the proton detected in the FPD and to reject about 97% of cases that appear to be CEP events but where the protons in reality originated from coincidences with pile-up events. Presently two detector options are studied, namely Quartz and Gas Cerenkov which may be read out with a Constant Fraction Discriminator allowing the time resolution to be significantly improved compared to usual electronics.

### 4.5 Trigger

Due to a limited L1 trigger latency, detectors at 420 m are far away from the central detectors to be included in the L1 trigger in normal running conditions, while detectors at 220 m can in principle be included. The trigger strategy depends on the mass of the diffractively produced object [12]. Demanding standard L1 triggers such as those for high mass H→WW/ZZ or high-$p_T$ dijet trigger would result in an acceptable output rate which may be further reduced by requiring the double proton tag at 220 m.

Triggering on low mass objects is more difficult but in principle feasible as documented in [25, 26] where diffractive L1 triggers for a case of $H \to b\bar{b}$ at $m_H$=120 GeV have been proposed. If the FPD trigger at 220 m is capable of triggering only on hits in the inner 4 mm part and if the L1 calorimeter is capable of defining exclusivity criteria using $E_T, \eta$ and $\phi$, then the final output rate is well below a 2 kHz limit at $\mathcal{L} = 2 \cdot 10^{33}\mathrm{cm}^{-2}\mathrm{s}^{-1}$ and slightly above this limit at $\mathcal{L} = 10^{34}\mathrm{cm}^{-2}\mathrm{s}^{-1}$. Other reductions are under study [26].

## Acknowledgements

Supported by the project AV0-Z10100502 of the Academy of Sciences of the Czech republic and project LC527 of the Ministry of Education of the Czech republic.

## References

[1] FP420 R&D Collab., J. Inst.: 2009-JINST-4-T10001, arXiv:0806.0302 [hep-ex].
[2] ATLAS Collab., J. Inst. 0803: S08003 (2008).
[3] CMS Collab., J. Inst. 0803: S08004 (2008).
[4] ATLAS Collab., CERN-LHCC-2004-010 (2004); A. Zoccoli, these proceedings.
[5] TOTEM Collab., CERN-LHCC-2004-002 (2004); S. Giani, these proceedings.
[6] ATLAS Collab., CERN-LHCC-2007-001 (2007).
[7] V. A. Khoze, A. D. Martin and M. G. Ryskin, Eur. Phys. J. **C 23** (2002) 311.
[8] V. A. Khoze, A. D. Martin and M. G. Ryskin, arXiv:0802.0177 [hep-ph].
[9] CDF Collab., Phys. Rev. Lett. **84** (2000) 5043.
[10] D0 Collab., Phys. Lett. **B440** (1998) 189.
[11] B. E. Cox *et al.*, Eur. Phys. J. **C45** (2006) 401.
[12] M. Albrow *et al.*, CERN-LHCC-2006-039 (2006).
[13] A. Brandt, V. Juránek, A. Pal and M. Taševský, Atlas note in preparation.
[14] V. Khoze *et al.*, Eur.Phys.J.**C25** (2002) 392; A.G. Shuvaev *et al.*, Eur.Phys.J.**C56** (2008) 467.
[15] S. Heinemeyer *et al.*, Eur.Phys.J.**C53** (2008) 231.
[16] B.E. Cox, F.K. Loebinger, A.D. Pilkington, JHEP 0710:090 (2007).
[17] CDF Collab., arXiv:0712.0604 [hep-ex].
[18] K. Piotrzkowski and N. Schul, arXiv:0806.1097 [hep-ph].
[19] V. A. Khoze, A. D. Martin, R. Orava and M. G. Ryskin, Eur. Phys. J. **C 19** (2001) 313.
[20] CDF Collab., Phys. Rev. Lett. **98** (2007) 112001; M. Albrow, these proceedings.
[21] J. Favereau de Jeneret *et al.*, CP3-09-37 (August 2009), arXiv:0908.2020 [hep-ph].
[22] K. Piotrzkowski, Phys. Rev. **D63** (2001) 071502.
[23] H1 Collab., Eur. Phys. J. **C 51** (2007) 549.
[24] ATLAS Collab., Letter of Intent of the AFP project.
[25] M. Grothe *et al.*, CERN-CMS-NOTE-2006-054.
[26] M. Campanelli *et al.*, ATLAS-COM-DAQ-2009-062.

# Part VI

# Diffraction at the LHC

# Diffraction at CMS

*Wagner de Paula Carvalho*[1] , for The CMS Collaboration

[1]Universidade do Estado do Rio de Janeiro, Rua Sao Francisco Xavier 524, 20559-900 Rio de Janeiro, Brazil

A summary of studies on diffraction with CMS detector and of planned measurements with early CMS data is presented.

# 1 Introduction

A long term program in Forward Physics is envisaged to be carried out with CMS [1]. Inclusive single diffraction (SD) and double pomeron exchange (DPE) at low luminosities, diffraction in the presence of a hard scale (jets, heavy quarks, vector bosons) at moderate luminosities and central exclusive production at the highest luminosities are some of the topics to be pursued.

LHC is expected to deliver a few hundred $\text{pb}^{-1}$ over its first running period, scheduled to start by the end of this year. With such amount of data collected under low instantaneous luminosity conditions, a variety of studies will become accessible: observation of hard diffractively produced W bosons and di-jets; assessment of the rapidity gap survival probability at LHC energies; probing of the diffractive parton distribution functions (PDF); observation of exclusively photoproduced $\Upsilon$ and study of its production dynamics.

# 2 Forward Detectors at CMS

Although there are plans to add proton tagging detectors to CMS, in the near future all diffractive analyses will have to rely on the rapidity gap signature and the coverage provided by the most forward CMS subsystems: the Hadronic Forward (HF), CASTOR and Zero Degree (ZDC) calorimeters.

Located at 11.2m from the interaction point (IP), at both sides of CMS, the HF is a steel/quartz fibre calorimeter covering the pseudorapidity range $3.0 < |\eta| < 5.0$. It is $\eta - \phi$ segmented, amounting to 900 towers of typical size $0.175 \times 0.175$.

CASTOR (Centauro And STrange Object Research) is a tungsten/quartz plates calorimeter, placed 14.3 m away from the IP. It is longitudinally and azimuthally segmented, but has no segmentation in $\eta$. The azimuthal segmentation defines 16 sectors. For the LHC start, there will have only one CASTOR, covering the pseudorapidity region $-6.6 < \eta < -5.2$.

ZDC is also a tungsten/quartz calorimeter, located 140m away from the IP at both sides of CMS. It will measure very forward photons and neutrons at $|\eta| > 8.1$.

Further details about these subsystems or CMS apparatus can be found elsewhere [2].

# 3 Studies in Preparation for Data

In this section we present prospective studies based on Monte Carlo simulations for diffractive processes at CMS. These studies were performed in preparation for the LHC start up. A scenario with centre-of-mass energy of 14 TeV and no pile-up was assumed.

## 3.1 Single Diffractive Production of W Bosons and Di-Jets

The single diffractive (SD) production of W bosons and of di-jets are both hard diffractive processes, characterised by the presence of a hard scale and a large rapidity gap (LRG) in the final state. These processes are sensitive to the diffractive structure function of the proton; the W production is mainly sensitive to its quark content and the di-jet is sensitive to its gluon content.

Both analyses, described in details in references [3, 4], used samples produced under similar conditions and the same methodology.

### 3.1.1 Monte Carlo Simulation

In order to simulate the SD process, the POMWIG generator [5], v2.0 beta, was used. For the diffractive PDF and Pomeron flux, the NLO H1 2006 fit B [6] was used, while for the proton PDF, the CTEQ6 [7] parametrisation was adopted. A rapidity gap survival probability ($\langle|S^2|\rangle$) of 0.05 was assumed [8]. The non-diffractive background was simulated by PYTHIA [9] for W production and by MADGRAPH [10] for the di-jet production. All samples were subject to full detector simulation, trigger emulation and reconstruction. The CASTOR information, however, is treated at generator level as this subsystem was not included in the full simulation chain.

### 3.1.2 Event Selection

W or di-jet candidates are selected by applying standard trigger and offline requirements. For $W \to \mu\nu$ selection, the same criteria of reference [11] was used. Basically, to be accepted an event was required to have one muon candidate in $|\eta| < 2.0$ and with $p_T > 25$ GeV. It should also have a transverse mass $M_T > 50$ GeV. Additional muon isolation cuts and cuts to reject contributions from top quark were applied. Events with more than one muon candidate with $p_T > 20$ GeV were rejected. For di-jets, at the trigger level events were selected by requiring at least 2 jets with average uncorrected transverse energy greater than 30 GeV. Jets were reconstructed with the SiSCone5 [12] algorithm and jet-energy scale corrections were applied. Finally, at least two jets with $E_T > 55$ GeV were required.

### 3.1.3 Gap Side Definition and Central Track Multiplicity

On average, SD events have less particles and energy deposited in the side that contains the scattered proton, when compared to non-diffractive events. This becomes clear from the generated energy-weighted $\eta$ distribution for stable particles (excluding neutrinos) in diffractive and non-diffractive $W \to \mu\nu$ events, shown in Figure 1 (also shown in this figure is the $\eta$ range covered by HF and CASTOR). In order to select diffractive candidates, a *gap side* was defined as the side with the lowest energy sum in HF. When applied to simulated SD samples, this definition wrongly selected the gap side (side of the scattered proton) $\sim 30\%$ of times for $W \to \mu\nu$ events and $\sim 10\%$ of times for di-jet events. One additional cut was applied to the di-jet candidates, exploiting the anti correlation between gap side and jets side: if the gap is found to be at positive rapidity, the two leading jets are required to be in the range $-4 < \eta_{jet} < 1$, otherwise, if the gap is at negative rapidity, the two leading jets are then required to be in the range $-1 < \eta_{jet} < 4$. Finally, a cut on the maximum $\eta$ separation between the two leading jets was applied: $|\Delta\eta_{jets}| < 3$.

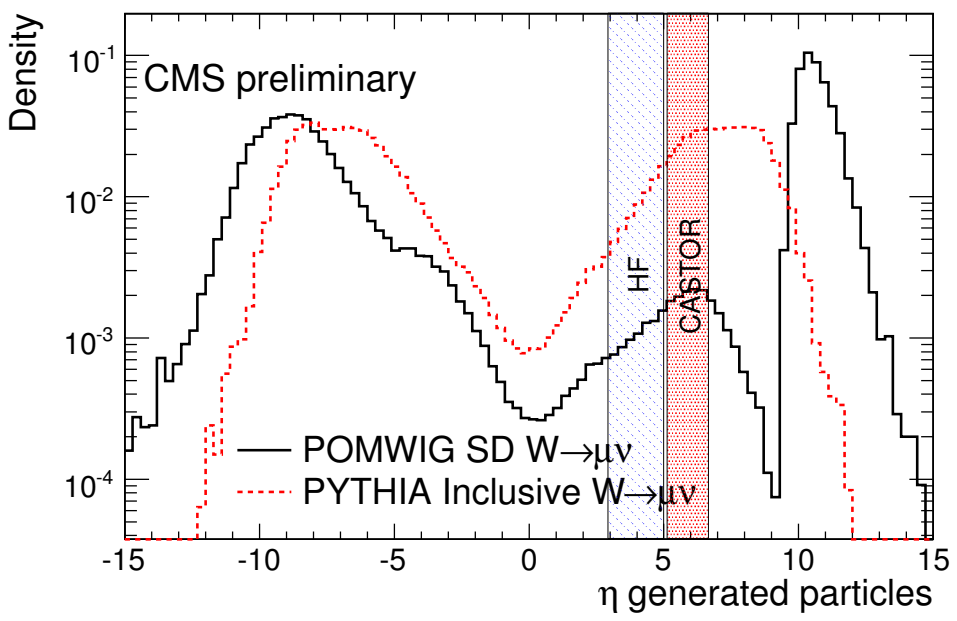

Figure 1: Generated energy-weighted $\eta$ distribution for stable particles (excluding neutrinos) in diffractive and non-diffractive $W \to \mu\nu$ events. The peak at $\eta >\sim 10$ for the diffractive sample corresponds to the scattered proton.

The track multiplicity in the central region can also be used for discriminating diffractive and non-diffractive events. Figure 2 shows the multiplicity distribution for tracks with $p_T > 900\,\mathrm{MeV}$. Diffractive events have a distribution that peaks at zero, contrary to non-diffractive events, and this feature was exploited for introducing a multiplicity cut for tracks with $|\eta_{tracks}| < 2$. Three values were used for these studies: $N_{tracks} \leq 1$, $N_{tracks} \leq 5$ and no cut at all.

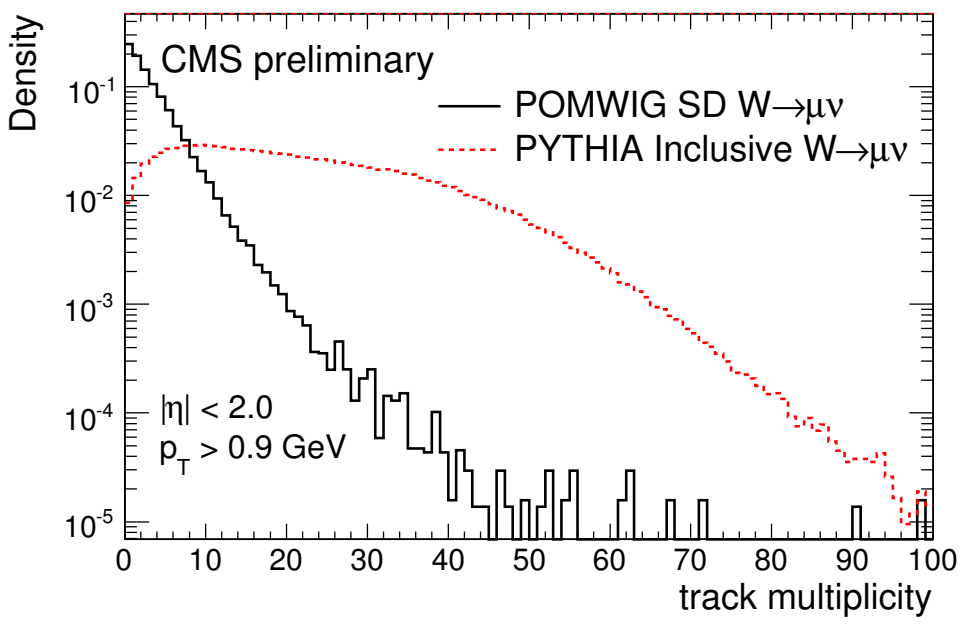

Figure 2: Central tracker multiplicity distributions for diffractively and non-diffractively produced $W \to \mu\nu$ events, excluding the track from the $\mu$ candidate.

### 3.1.4 HF and CASTOR Multiplicity Distributions

For the events passing the selection criteria, two-dimensional (2D) distributions of the activity in the forward calorimeters (HF and CASTOR) were obtained and used to assess the possibility of observing SD in early data, in a similar way to analyses carried out at the Tevatron and at HERA. For HF, activity was quantified by the number of towers with deposited energy above threshold. For CASTOR, the number of $\phi$ sectors hit by hadrons with energy above 10 GeV was used to this purpose. As there was no detector simulation/reconstruction implemented for this subsystem, the generated particle information was used in turn.

Two possible experimental scenarios were considered: *1)* no forward detectors beyond HF and *2)* additional $\eta$ coverage provided by CASTOR.

In the first scenario, HF towers are grouped in two slices as a function of their $\eta$ coordinate: "low-$\eta$ slice" for $2.9 < |\eta| < 4.0$ and "forward slice" for $4.0 < |\eta| < 5.2$. Figure 3 shows the 2D tower multiplicity for $W \to \mu\nu$ events with $N_{tracks} \leq 5$. Top plots present the gap side multiplicity distributions for SD events with the generated gap in the positive (left) and negative (right) sides. They show a clear peak at the zero multiplicity bin. Conversely, the bottom left distribution for non-diffractive events shows no enhancement at the zero bin. The bottom right plot shows the sum of the two distributions, which are normalised to the same integrated luminosity of $100 \text{pb}^{-1}$. This is the kind of distribution expected from data. An excess due to diffractive signal is clearly visible at the zero multiplicity bin. This excess becomes even more significant as the central tracks multiplicity cut becomes stricter.

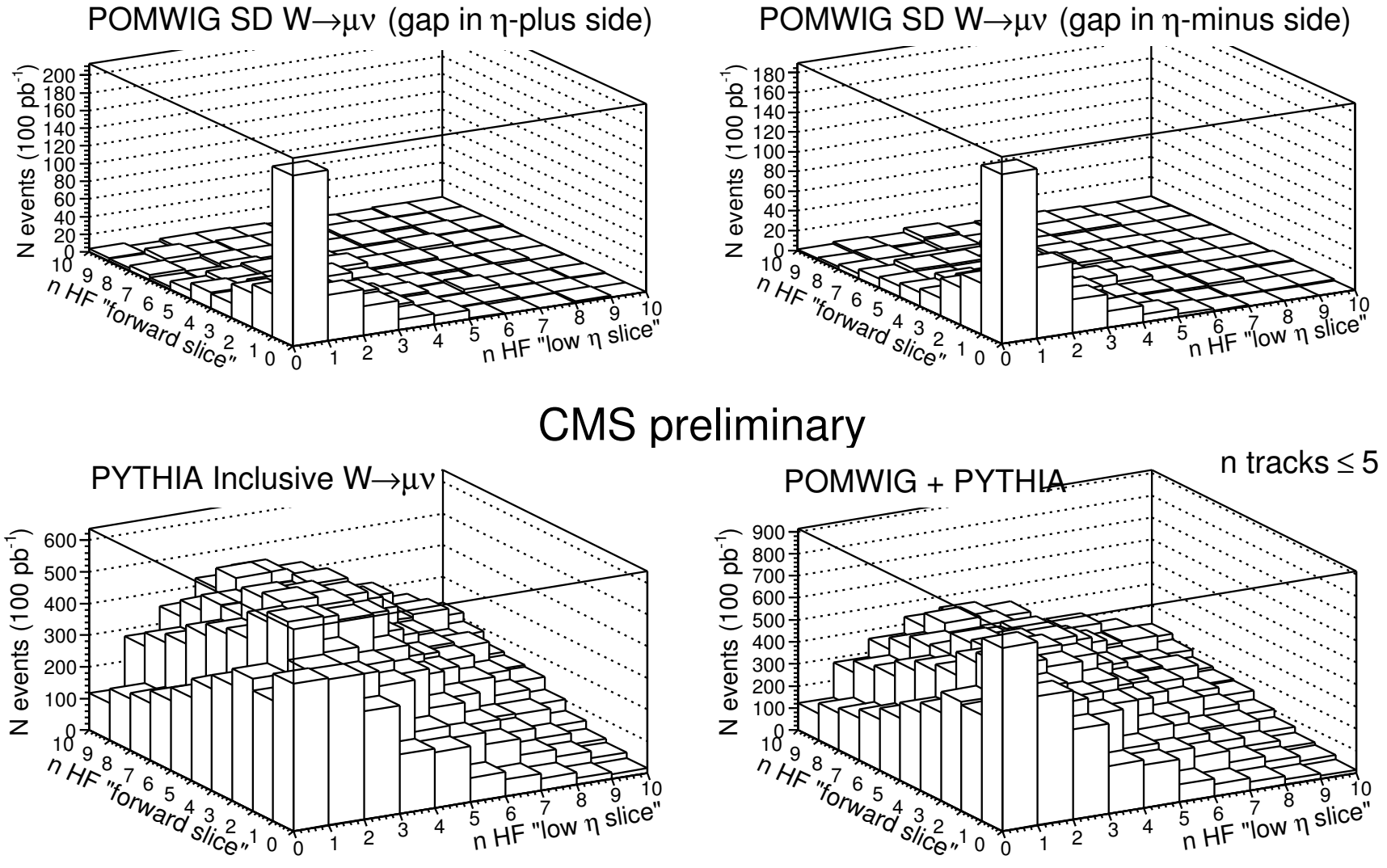

Figure 3: Tower multiplicity distributions in HF.

Similar distributions can be obtained when CASTOR is taken into consideration. Figure 4 shows the HF tower vs. CASTOR $\phi$ sector multiplicity distributions for $W \to \mu\nu$ events with $N_{tracks} \leq 5$. Because CASTOR will be available only in the negative side for the first running period, only events with gap in that side (as defined in Section 3.1.3 were considered). The top left plot shows those events for which the gap has been wrongly determined. The other plots are qualitatively analogous to those in Figure 3. With the extra coverage provided by CASTOR, the signal to background ratio ($S/B$) got greatly improved. For $N_{tracks} \leq 5$, the ratio is $\mathcal{O}(1)$ for HF-only and $\mathcal{O}(10)$ for HF-CASTOR.

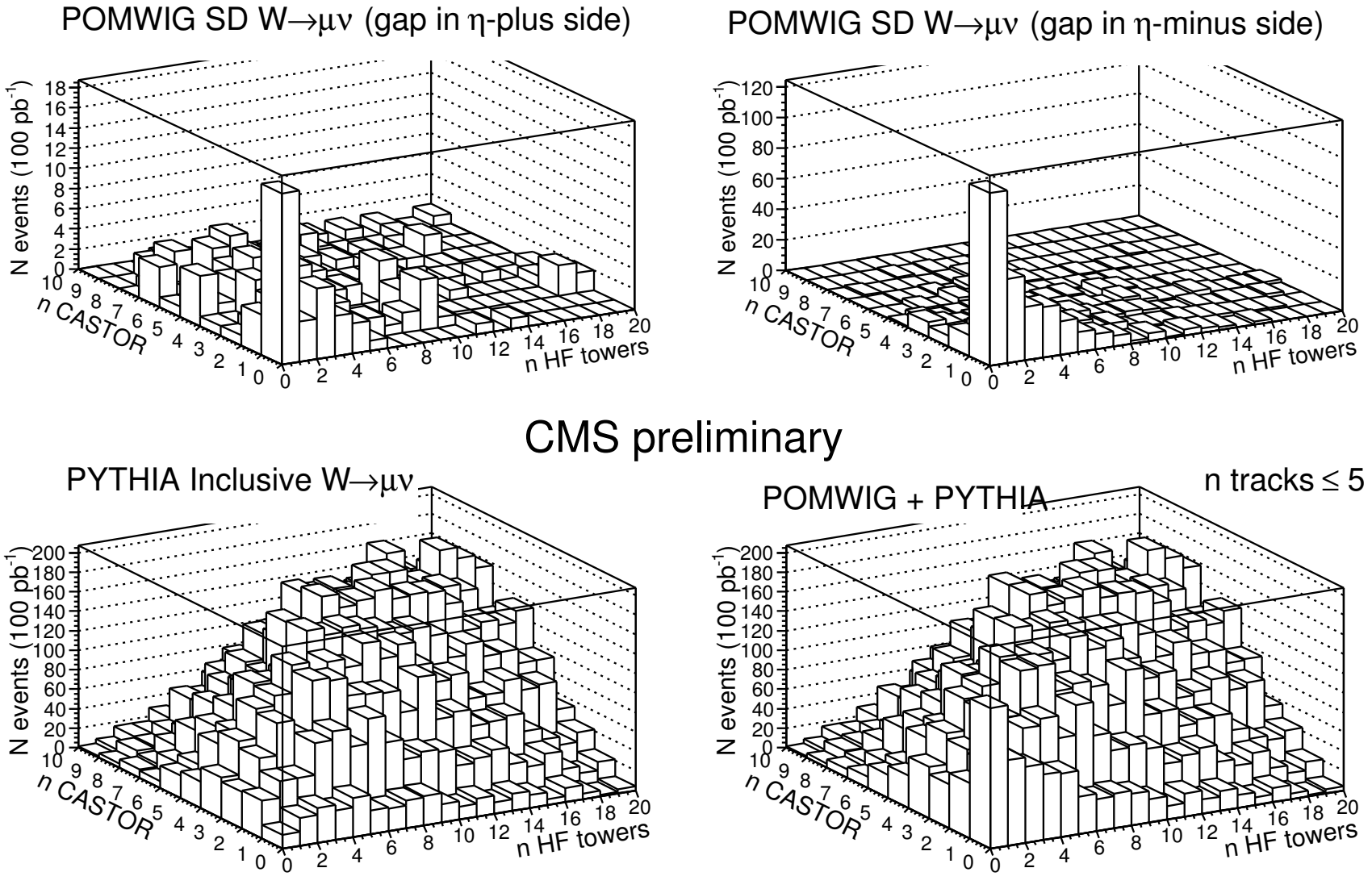

Figure 4: Multiplicity distributions for HF and CASTOR.

The same exercise was carried out for the SD di-jets. In the very same way, the addition of CASTOR results in a improvement by more than one order of magnitude of $S/B$.

### 3.1.5 Establishing Diffractive Signal in Data

The presence of a diffractive signal in data can be demonstrated without relying on MC. This can be achieved by varying the diffractive selection criteria and showing that the diffractive peak at zero multiplicity varies in a predictable way. Table 1 illustrates for SD di-jet production how $S/B$ improves by tightening the central tracker multiplicity cut, even in the less favourable HF-only scenario. It must be pointed out that such behaviour is the opposite of what would be expected, were the excess due to a statistical fluctuation.

| $N_{tracks}^{max}$ | $\infty$ | 5 | 1 |
|---|---|---|---|
| $S/B$ | 0.6 | 0.9 | 1.3 |

Table 1: Evolution of S/B as a function of $N_{tracks}^{max}$, for SD di-jet production.

### 3.1.6 Feasibility Studies and Sensitivity to $\langle|S^2|\rangle$

One goal of these studies was determining the feasibility of observing SD production of di-jets and W, through its semileptonic decay $W \to \mu\nu$, with a limited sample of early CMS data. Under the simulated conditions, signals of $\mathcal{O}(400)$ events are expected in the SD di-jet channel with the first 10pb$^{-1}$ and of $\mathcal{O}(100)$ events in the SD $W \to \mu\nu$ channel with the first 100pb$^{-1}$, if CASTOR is available.

Another possible result that can be achieved is exclude extreme values of the rapidity gap survival probability at LHC energies. Values as low as 0.004 and as high as 0.23 have been proposed [13]. Simulations show that these values would lead to marginally observable signals in the first case, only detectable by profiting of CASTOR extended $\eta$ coverage, and to very prominent signals in the second case, easily detectable by HF alone. Measured event yields could eventually exclude these extreme values.

## 3.2 Exclusive $\Upsilon$ Photoproduction

The exclusive production of $\Upsilon \to \mu^+\mu^-$ through $\gamma p$ interaction, represented in Figure 5, is another diffractive process accessible with early CMS data. As for SD di-jets and W analyses, a scenario with centre-of-mass energy of 14 TeV and no pile-up was assumed in this study [14].

### 3.2.1 Monte Carlo Simulation and Event Selection

The signal samples were generated with STARLIGHT [15] adopting the following predicted values of cross-section times branching fraction for the first three $\Upsilon$ resonances (1S, 2S, 3S): 39.0 pb, 13.0 pb, and 10.0 pb. LPAIR [16] was used to simulate the inelastic two-photon events, while PYTHIA was used for all other backgrounds (Drell-Yan, quarkonium decays, heavy-flavor jets). All samples were subject to full detector simulation, trigger emulation and reconstruction.

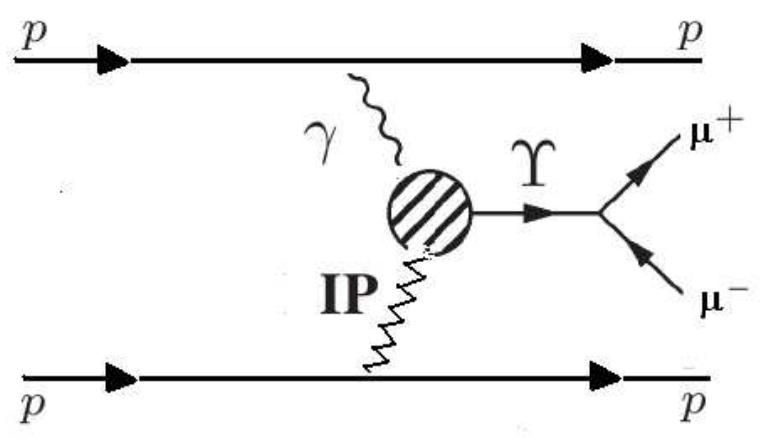

Figure 5: Feynman diagram for $\gamma p \to \Upsilon p \to \mu^+\mu^- p$.

At the trigger level, dimuon candidates were selected by requiring two muons with $p_T > 3$ GeV (high trigger thresholds essentially kill the corresponding dielectron channel). Major backgrounds were suppressed by cutting on the muon pair kinematics and on the additional detector activity. Dimuons were required to be well balanced in transverse momentum, satisfying the condition $|\Delta p_T(\mu^+\mu^-)| < 2.0$ GeV, and nearly back-to-back in the azimuthal angle, $|\Delta\phi(\mu^+\mu^-)| > 2.9$. A calorimeter exclusivity condition was applied by requiring less than 5 extra towers above noise threshold to be present in the event. Finally, no extra charged track, beyond the 2 muon candidates, was allowed in the event.

### 3.2.2 Dimuon Spectrum and $\Upsilon$ Yield

After all the selection and exclusivity conditions are applied, the dominant remaining background comes from inelastic photon-exchange events, in which the proton remnants lay outside HF coverage. However, a significant fraction of these events could be detected by CASTOR and ZDC detectors. Based on the generator level information and considering a configuration with CASTOR at only one side, it was estimated that approximately 2/3 of the remaining inelastic background could be rejected by vetoing on CASTOR and ZDC.

Figure 6 shows clearly visible signals for the first three $\Upsilon$ resonances with a simulated integrated luminosity of 100pb$^{-1}$. With such event yield, studies of the $\Upsilon$ production dynamics might even be possible with early CMS data.

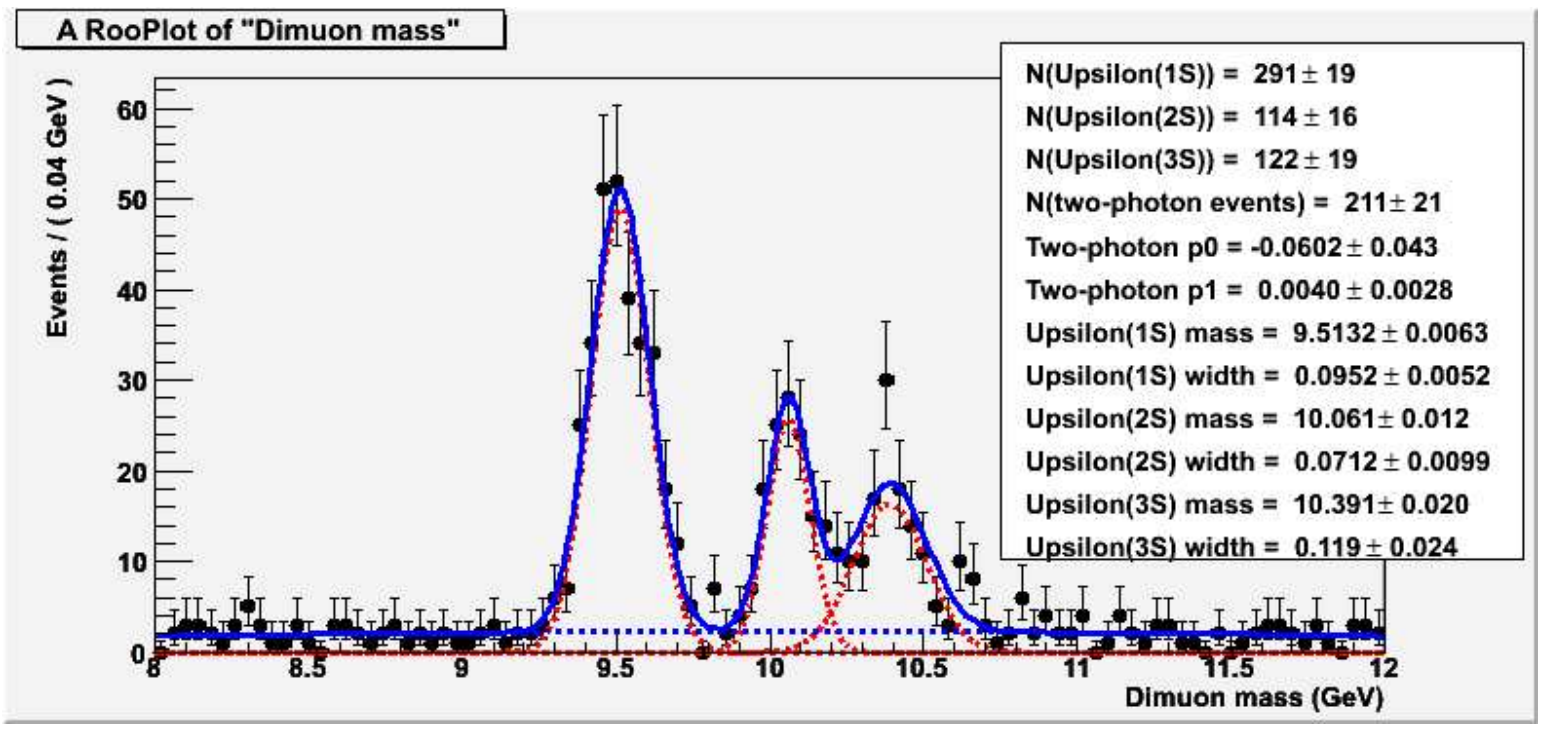

Figure 6: Dimuon invariant mass in the $\Upsilon$ resonances (1S, 2S, 3S) region.

# 4 Summary

CMS is ready to study hard diffractive processes with the LHC early data, using the large rapidity gap and exclusivity techniques. Monte Carlo studies, assuming low instantaneous luminosities and no pile-up, have shown that with the first $10\,\text{pb}^{-1}$ of data it may be possible to observe $\mathcal{O}(300)$ single diffractively produced di-jets for $\langle|S^2|\rangle = 0.05$. Significant deviations from this expected event yield might allow to put constraints on $\langle|S^2|\rangle$ values. When $100\,\text{pb}^{-1}$ of data becomes available, then it should be possible to observe $\mathcal{O}(100)$ single diffractively produced $W \to \mu\nu$ events, again assuming $\langle|S^2|\rangle = 0.05$. At this point, it may also be possible to observe clear signals of $\Upsilon$ resonances photoproduction and even study some aspects of their production dynamics.

# References

[1] The CMS Collaboration, J. Phys. **G34** 995 (2007).

[2] CMS Collaboration, JINST **0803** S08004 (2008).

[3] CMS Collaboration, CMS PAS DIF-07-002 (2007).

[4] CMS Collaboration, CMS PAS FWD-08-002 (2008).

[5] B. E. Cox and J. R. Forshaw, Comput. Phys. Commun. **144** 104 (2002).

[6] H1 Collaboration, A. Aktas et al., Eur. Phys. J. **C48** 715 (2006).

[7] J. Pumplin, D. R. Stump, J. Huston, H. L. Lai, P. Nadolsky and W. K. Tung, JHEP **0207** 012 (2002).

[8] V. A. Khoze, A. D. Martin and M. G. Ryskin, Phys. Lett. **B643** 93 (2006).

[9] T. Sjostrand, S. Mrenna and P. Skands, JHEP **0605** 026 (2006).

[10] J. Alwall et al., JHEP **0709** 028 (2007).

[11] CMS Collaboration, CMS PAS EWK-07-002 (2007).

[12] G. P. Salam and G. Soyez, JHEP **0705** 086 (2007).

[13] J. S. Miller, Eur. Phys. J. **C56** 39 (2008).

[14] CMS Collaboration, CMS PAS DIF-07-001 (2007).

[15] S. R. Klein and J. Nystrand, Phys. Rev. Lett. **92** 142003 (2004);
J. Nystrand, Nucl. Phys. **A752** 470 (2005).

[16] J. A. M. Vermaseren, Nucl. Phys. **B229** 347 (1983).

# Diffraction at TOTEM

*G. Antchev*[1], *P. Aspell*[1], *V. Avati*[1,9], *M.G. Bagliesi*[5], *V. Berardi*[4], *M. Berretti*[5], *M. Besta*[1], *M. Bozzo*[2], *E. Brücken*[6], *A. Buzzo*[2], *F. Cafagna*[4], *M. Calicchio*[4], *M.G. Catanesi*[4], *R. Cecchi*[5], *M.A. Ciocci*[5], *P. Dadel*[1], *M. Deile*[1], *E. Dimovasili*[1,9], *K. Eggert*[9], *V. Eremin*[11], *F. Ferro*[2], *A. Fiergolski*[1], *F. Garcia*[6], *S. Giani*[1*], *V. Greco*[5], *L. Grzanka*[1], *J. Heino*[6], *T. Hildén*[6], *J. Kašpar*[1,7], *J. Kopal*[1,7], *V. Kundrát*[7], *K. Kurvinen*[6], *S. Lami*[5], *G. Latino*[5], *R. Lauhakangas*[6], *R. Leszko*[1], *E. Lippmaa*[8], *M. Lokajíček*[7], *M. Lo Vetere*[2], *F. Lucas Rodriguez*[1], *M. Macrí*[2], *G. Magazzù*[5], *M. Meucci*[5], *S. Minutoli*[2], *H. Niewiadomski*[1,9†], *G. Notarnicola*[4], *E. Oliveri*[5], *F. Oljemark*[6], *R. Orava*[6], *M. Oriunno*[1], *K. Österberg*[6,], *E. Pedreschi*[5], *J. Petäjäjärvi*[6], *J. Prochazka*[7], *M. Quinto*[4], *E. Radermacher*[1], *E. Radicioni*[4], *F. Ravotti*[1], *G. Rella*[4], *E. Robutti*[2], *L. Ropelewski*[1], *M. Rostkowski*[1], *G. Ruggiero*[1], *A. Rummel*[8], *H. Saarikko*[6], *G. Sanguinetti*[5], *A. Santroni*[2], *A. Scribano*[5], *G. Sette*[2], *W. Snoeys*[1], *F. Spinella*[5], *A. Ster*[10], *C. Taylor*[3], *A. Trummal*[8], *N. Turini*[5], *J. Whitmore*[9], *J. Wu*[1], *M. Zalewski*[1]

[1]CERN, Genève, Switzerland,
[2]Università di Genova and Sezione INFN, Genova, Italy,
[3]Case Western Reserve University, Dept. of Physics, Cleveland, OH, USA,
[4]INFN Sezione di Bari and Politecnico di Bari, Bari, Italy,
[5]INFN Sezione di Pisa and Università di Siena, Italy,
[6]Helsinki Institute of Physics and Department of Physics, University of Helsinki, Finland,
[7]Institute of Physics of the Academy of Sciences of the Czech Republic, Praha, Czech Republic,
[8]National Institute of Chemical Physics and Biophysics NICPB, Tallinn, Estonia.
[9]Penn State University, Dept. of Physics, University Park, PA, USA.
[10]MTA KFKI RMKI, Budapest, Hungary.
[11]Ioffe Physico-Technical Institute, Polytechnicheskaya Str. 26, 194021 St-Petersburg, Russian Federation.

The primary objective of the TOTEM experiment at the LHC is the measurement of the total proton-proton cross section with the luminosity-independent method and the study of elastic proton-proton cross-section over a wide $|t|$-range. In addition TOTEM also performs a comprehensive study of diffraction, spanning from cross-section measurements of individual diffractive processes to the analysis of their event topologies. Hard diffraction will be studied in collaboration with CMS taking advantage of the large common rapidity coverage for charged and neutral particle detection and the large variety of trigger possibilities even at large luminosities. TOTEM will take data under all LHC beam conditions including standard high luminosity runs to maximise its physics reach. This contribution describes the main features of the TOTEM diffractive physics programme including measurements to be made in the early LHC runs.

* corresponding author: Simone Giani (simone.giani@cern.ch)
† corresponding author: Hubert Niewiadomski (hubert.niewiadomski@cern.ch)

# 1 Introduction

The TOTEM experiment [1, 2] will measure the total proton-proton cross-section with the luminosity-independent method based on the Optical Theorem, which requires a detailed study of the elastic scattering cross-section down to a squared four-momentum transfer of $|t| \sim 10^{-3}\,\mathrm{GeV}^2$ and the measurement of the total inelastic rate. Furthermore, TOTEM's physics programme aims at a deeper understanding of the proton structure by studying elastic scattering with large momentum transfers, and via the diffractive processes — partly in cooperation with CMS [3], located at the same interaction point, IP5.

To perform these measurements, TOTEM requires a good acceptance for particles produced at very small angles with respect to the beam. TOTEM's coverage in the pseudorapidity range of $3.1 \leq |\eta| \leq 6.5$ on both sides of the interaction point is accomplished by two telescopes for inelastically produced particles. This is complemented by detectors in special movable beam-pipe insertions — so-called Roman Pots — placed at 147 and 220 m from the interaction point, designed to track leading protons at a few mm from the beam centre.

For the luminosity-independent total cross-section measurement, TOTEM has to reach the lowest possible $|t|$ values in elastic $pp$ scattering. Elastically scattered protons close to the beam can be detected downstream on either side of the IP if the displacement at the detector location is large enough and if the beam divergence at the IP is small compared to the scattering angle. To achieve these conditions special LHC optics with high beta value at the IP ($\beta^*$) is required. According to Liouville's theorem [4, 5], the larger the $\beta^*$, which determines the beam size, the smaller the beam divergence ($\sim 1/\sqrt{\beta^*}$). Two optics are proposed: an ultimate one with $\beta^* = 1535\,\mathrm{m}$ and another one, likely foreseen for 2010, with $\beta^* = 90\,\mathrm{m}$. The latter uses the standard injection optics ($\beta^* = 11\,\mathrm{m}$) and beam conditions typical for early LHC running: zero degree crossing-angle and consequently at most 156 bunches together with a low number of protons per bunch.

# 2 Diffraction in TOTEM

Diffractive final states will comprise almost 50 % of all final states at the LHC. Their study will shed light on the proton structure and will help in understanding the transition between the non-perturbative regime of low-$t$ elastic scattering and that of hard diffraction, where rapidity gaps and forward protons co-exist with large $\mathbf{p}_T$ final states.

A general definition of hadronic diffractive process states that diffraction is a reaction in which no quantum numbers are exchanged between the colliding particles [6]. The traditional theoretical framework for diffraction — Regge theory — explains diffractive reactions at high energies in terms of Pomeron exchange. Many consider the concept of Pomeron misleading and thus current research aims at understanding this exchange in terms of QCD.

It is difficult to determine experimentally if the outgoing systems have the same quantum numbers as the incoming particles. Therefore, the operational definition is introduced, which classifies a process to be diffractive on the basis of the non-exponentially suppressed rapidity gap $\Delta\eta$ in the final state: $\mathrm{d}N/\mathrm{d}\Delta\eta \sim \mathrm{const}$, while the non-diffractive events are exponentially suppressed: $\mathrm{d}N/\mathrm{d}\Delta\eta \sim \mathrm{e}^{-\Delta\eta}$.

Given its unique coverage for charged particles at high rapidities, TOTEM is ideal for studies of forward phenomena. Since energy flow and particle multiplicity of inelastic events peak in the forward region, the large rapidity coverage and proton detection on both sides allow the study of a wide range of processes in inelastic and diffractive interactions.

Diffractive scattering comprises Single Diffraction (SD), Double Diffraction (DD), Double Pomeron Exchange (DPE) (Central Diffraction), and higher order multi-Pomeron processes. Together with elastic scattering these processes represent about 50 % of the total cross-section. Many details of these processes with close ties to proton structure and low-energy QCD are still poorly understood.

Majority of diffractive events ($\sim 35\,\%$) (Figure 1) exhibit intact protons in the final state,

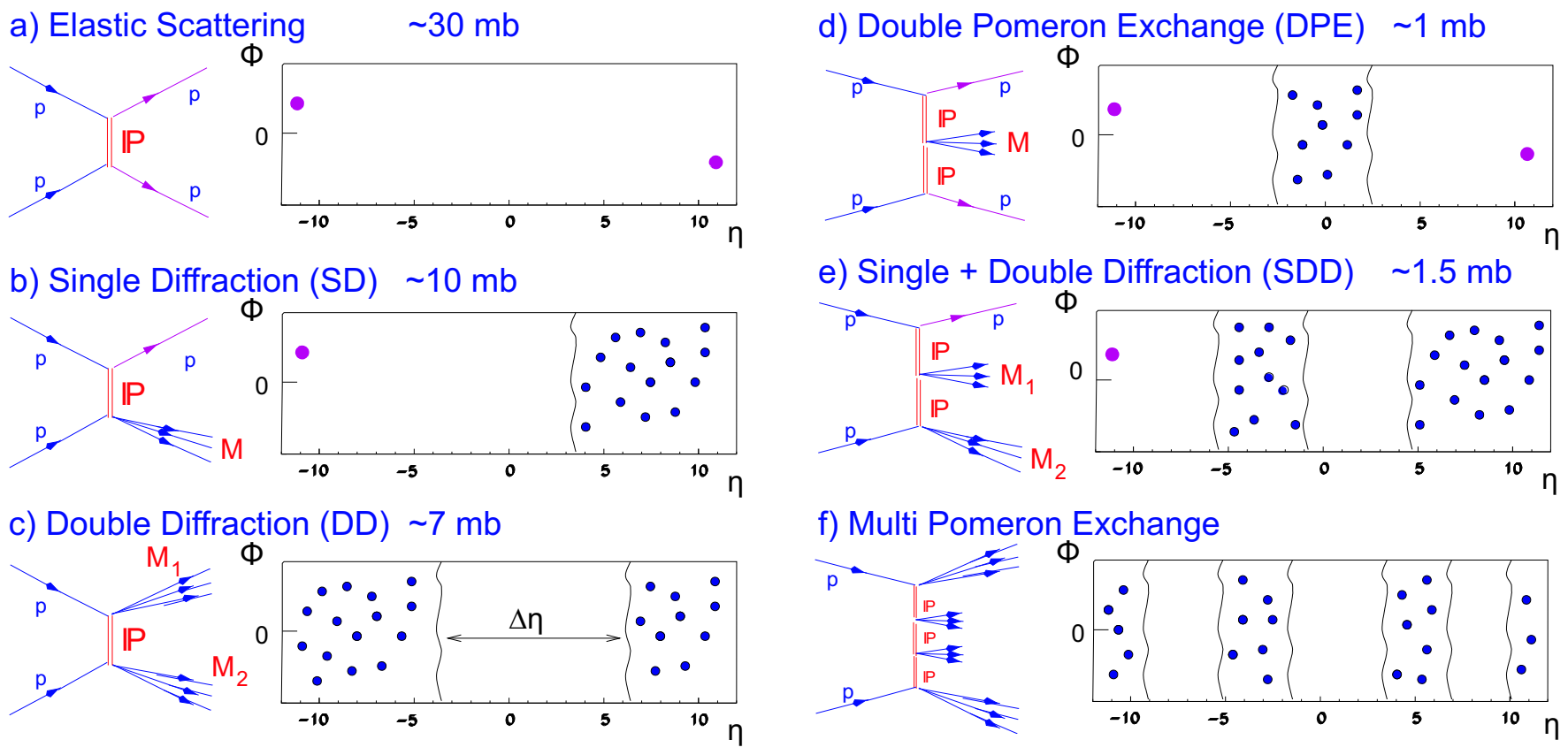

Figure 1: Different classes of diffractive processes and their cross-sections as estimated for the LHC at $\sqrt{s} = 14\,\mathrm{TeV}$.

characterised by their $t$ and by their $\xi$. For large $\beta^*$ (see Figure 2, right) most of these protons can be detected in the RP detectors.

## 3 LHC Optics

The transport of protons in the LHC lattice can be expressed by two optical functions $L_{x,y}$ (effective length) and $v_{x,y}$ (magnification). According to the general form of the transport matrix (see [5]), their values, at a distance $s$ from the IP, depend upon the betatron function $\beta(s)$ and the phase advance $\Delta\mu(s)$. Since $\beta(s)$ is at extremum in the IP, the functions $v(s)$ and $L(s)$ can be expressed as $v(s) = \sqrt{\frac{\beta(s)}{\beta^*}}\cos\Delta\mu(s)$ and $L(s) = \sqrt{\beta(s)\beta^*}\sin\Delta\mu(s)$ with $\Delta\mu(s) = \int_0^s \frac{1}{\beta(s')}ds'$.

The transverse displacement $(x(s), y(s))$ of a proton at a distance $s$ from the IP is related to its transverse origin $(x^*, y^*)$ and its momentum vector (expressed by the horizontal and vertical scattering angles $\Theta_x^*$ and $\Theta_y^*$, and by $\xi = \Delta p/p$) at the IP via the above optical functions and the horizontal dispersion $D_x(s)$ of the machine:

$$\begin{aligned} y(s) &= v_y(s)\cdot y^* + L_y(s)\cdot\Theta_y^* \\ x(s) &= v_x(s)\cdot x^* + L_x(s)\cdot\Theta_x^* + \xi\cdot D_x(s). \end{aligned} \tag{1}$$

As a consequence of the high $\beta^*$, the beam size at the IP is large ($\sigma_{beam}^* \sim \sqrt{\beta^*}$), which reduces the luminosity for such a running scenario. To eliminate the dependence on the transverse position of the proton at the collision point, the magnification has to be chosen close to zero (parallel-to-point focussing, $\Delta\mu = \pi/2$ ). At the same time, a large effective length ensures a sizeable displacement of scattered protons from the beam centre.

Having in mind the above optimisation for the position of the RP220 station, two scenarios have been studied. Their optical functions are compared in Table 1, together with standard

| $\beta^*$ [m] | $L_x$ [m] | $L_y$ [m] | $v_x$ | $v_y$ | $D_x$ [m] | $x_{\mathrm{RP}}$ [mm] | $y_{\mathrm{RP}}$ [mm] | $\sigma(x^*)$ [$\mu$m] | $\sigma(\Theta^*_x)$ [$\mu$rad] |
|---|---|---|---|---|---|---|---|---|---|
| 0.5 | 1.5 | 18 | -3.9 | -3.8 | -0.080 | 1.26 | 6.0 | 16.6 | 30 |
| 2 | 0.49 | 18 | -3.5 | -4.0 | -0.086 | 1.6 | 3.7 | 32 | 16 |
| 90 | 0 | 262 | -1.9 | 0.0 | -0.041 | 4.5 | 6.8 | 212 | 2.3 |
| 1535 | 100 | 270 | 0 | 0 | -0.05 | 0.8 | 1.3 | 450 | 0.3 |

Table 1: Optics parameters at IP5 and at the RP220 station for the TOTEM running scenarios. $L_x$, $L_y$, $v_x$, $v_y$ and $D_x$ are the parameters of Equation 1 at RP220, $x_{\mathrm{RP}}$ and $y_{\mathrm{RP}}$ are the distances of the horizontal and vertical Roman Pots of the RP220 station from the beam centre, respectively, $\sigma(x^*)$ is the beam size at IP5 and $\sigma(\Theta^*_x)$ is the beam divergence at IP5.

low-$\beta^*$ running scenarios. For $\beta^* = 1535\,\mathrm{m}$, the parallel-to-point focussing is achieved in both projections whereas for $\beta^* = 90\,\mathrm{m}$ only in the vertical one. In both cases, the large $L_y$ pushes the protons vertically into the acceptance of the RP detectors.

Both optics also offer the possibility of detecting diffractive protons almost independent of their momentum loss. To be able to measure the momentum loss $\xi$ with an acceptable resolution, $L_x$ has to vanish to eliminate the dependence on the horizontal scattering angle $\Theta^*_x$ (cf. Equation (1)). This condition can only be achieved with the $\beta^* = 90\,\mathrm{m}$ optics (Table 1).

# 4 TOTEM Running Scenarios

The versatile physics programme of TOTEM requires different running scenarios that have to be adapted to the LHC commissioning and operation in the first years. A flexible trigger can be provided by the T1 and T2 telescopes and the Roman Pot detectors. TOTEM will take data under all optics conditions, adjusting the trigger schemes to the luminosity. The DAQ will allow trigger rates up to a few kHz without involving a higher level trigger. The high-$\beta^*$ runs (Table 2) with 156 bunches, zero degree crossing-angle and maximum luminosity between $10^{29}$ and $10^{30}\,\mathrm{cm}^{-2}\mathrm{s}^{-1}$, will concentrate on low-$|t|$ elastic scattering, total cross-section, minimum bias physics and soft diffraction. A large fraction of forward protons will be detected even at the lowest $\xi$ values. Low-$\beta^*$ runs (Table 2) with more bunches and higher luminosity ($10^{32}$ – $10^{34}\,\mathrm{cm}^{-2}\mathrm{s}^{-1}$) will be used for large-$|t|$ elastic scattering and diffractive studies with $-\xi > 0.02$. Hard diffractive events come within reach. In addition, early low $\beta^*$ runs will provide first opportunities for measurements of soft diffraction at LHC energies and for studies of forward charged multiplicity.

# 5 Diffractive Proton Acceptance and Resolution

The acceptance and the reconstruction resolution are defined by the optics parameters via Equation 1. In addition, the beam divergence limits the $\Theta^*_{x,y}$-reconstruction resolution and thus the $t$-resolution, while the beam energy uncertainty $\sigma(p)/p$ limits the precision of longitudinal momentum loss reconstruction.

| $\beta^*$ [m] | $k$ | $N/10^{11}$ | $\mathcal{L}$ [cm$^{-2}$s$^{-1}$] | $\|t\|$-range [GeV$^2$] @ $\xi = 0$ | $\xi$-range |
|---|---|---|---|---|---|
| 1540 | 43 – 156 | 0.6 – 1.15 | $10^{28} - 2 \cdot 10^{29}$ | 0.002 – 1.5 | $< 0.2$ |
| 90 | 156 | 0.1 – 1.15 | $2 \cdot 10^{28} - 3 \cdot 10^{30}$ | 0.03 – 10 | $< 0.2$ |
| 11 | 43 – 2808 | 0.1 – 1.15 | $\sim 10^{30} - 5 \cdot 10^{32}$ | 0.6 – 8 | 0.02 – 0.2 |
| 0.5 – 3 | 43 – 2808 | 0.1 – 1.15 | $\sim 10^{30} - 10^{34}$ | 2 – 10 | 0.02 – 0.2 |

Table 2: Running scenarios at different LHC optics ($k$ is a number of bunches, $N$ — number of particles per bunch, and $\mathcal{L}$ — estimated luminosity). The $|t|$ ranges for elastically scattered protons correspond to the $\geq 50\%$ combined RP147 and RP220 acceptance.

The acceptance and resolution of diffractive protons reconstructed with the Roman Pot devices, located at 220 m from IP5, are summarised in Tables 3–5.

In case of the low-$\beta^*$ optics (Table 3), which is characterised by short effective length

| $\beta^*$ [m] | variable | acceptance | resolution |
|---|---|---|---|
| 2 | $\xi = \Delta p/p$ | $0.02 < -\xi < 0.25$ | $\sigma(\xi) \approx (1 - 6) \times 10^{-3}$ |
| | DPE $M$ [GeV] | $250 < M < 3000$ | $\sigma(M)/M = 1 - 5\%$ |
| | $t$ [GeV$^2$] | complete $t$-range for $-\xi > 0.02$ | $\sigma(t) \approx 0.3\sqrt{t}$ |
| | | $2 < -t_y < 10\,\mathrm{GeV}^2$ @ $\xi = 0$ | |
| | $\phi$ [rad] | $0 \leq \phi < 2\pi$ | $0.15 < \sigma(\phi) < 0.5$ for $1 > -t > 0.05$ GeV$^2$ |

Table 3: Roman Pot acceptance and reconstruction summary for $\beta^*$=2 m optics at $\sqrt{s} = 14$ TeV.

(see Table 1), the protons are primarily accepted in the Roman Pots due to their fractional momentum loss $\xi$, independently of $t$. Because of the machine dispersion, they are shifted towards the horizontal Roman Pot. In this way all the $t$-range is accepted for $-\xi > 0.02$, as is demonstrated in Figure 2, left. The figure presents one of the early running scenarios characterised by reduced beam energy to 5 TeV.

On the contrary, in case of high-$\beta^*$ optics (Tables 4 and 5), the protons are detected because of their four momentum transfer squared $t$, since the effective lengths are high (except $L_x$ for $\beta^* = 90$ m). This results in full $\xi$-range acceptance for $-t > 0.02$ GeV and $-t > 0.003$ GeV for $\beta^* = 90$ m and 1535 m, respectively, as is illustrated by Figure 2, right. On average $\sim 50\,\%$ and $\sim 90\,\%$ of diffractive protons are accepted by RP detectors for $\beta^* = 90$ m and 1535 m, respectively.

The high-$\beta^*$ scenarios, due to high effective lengths and low beam angular divergence, are characterised by good $t$- and $\phi$-reconstruction capabilities and by acceptance of low $|t|$ values. The resolution in $t$ changes depending upon the proton azimuth angle $\phi$, its fractional momentum loss $\xi$ (machine chromaticity) and the azimuth angle of proton. In case of $\beta^* =$ 1535 m optics, for horizontal protons $\sigma(t_x) = (0.04 - 0.4)\sqrt{-t_x}$ while for vertical ones $\sigma(t_y) = (0.002 - 0.02)\sqrt{-t_y}$. The $t$-reconstruction precision in horizontal direction is generally lower than in the vertical one because of shorter effective horizontal lengths and due to the fact that both the horizontal component of the scattering angle and the momentum loss displace the proton in the horizontal direction. As a result, the reconstruction of the horizontal scattering angle and of the momentum loss both depend upon the horizontal displacement, which limits the resolution. The reconstruction of the azimuth angle $\phi$ is possible for all accepted $t$-values, however its resolution greatly improves for higher four momentum transfers.

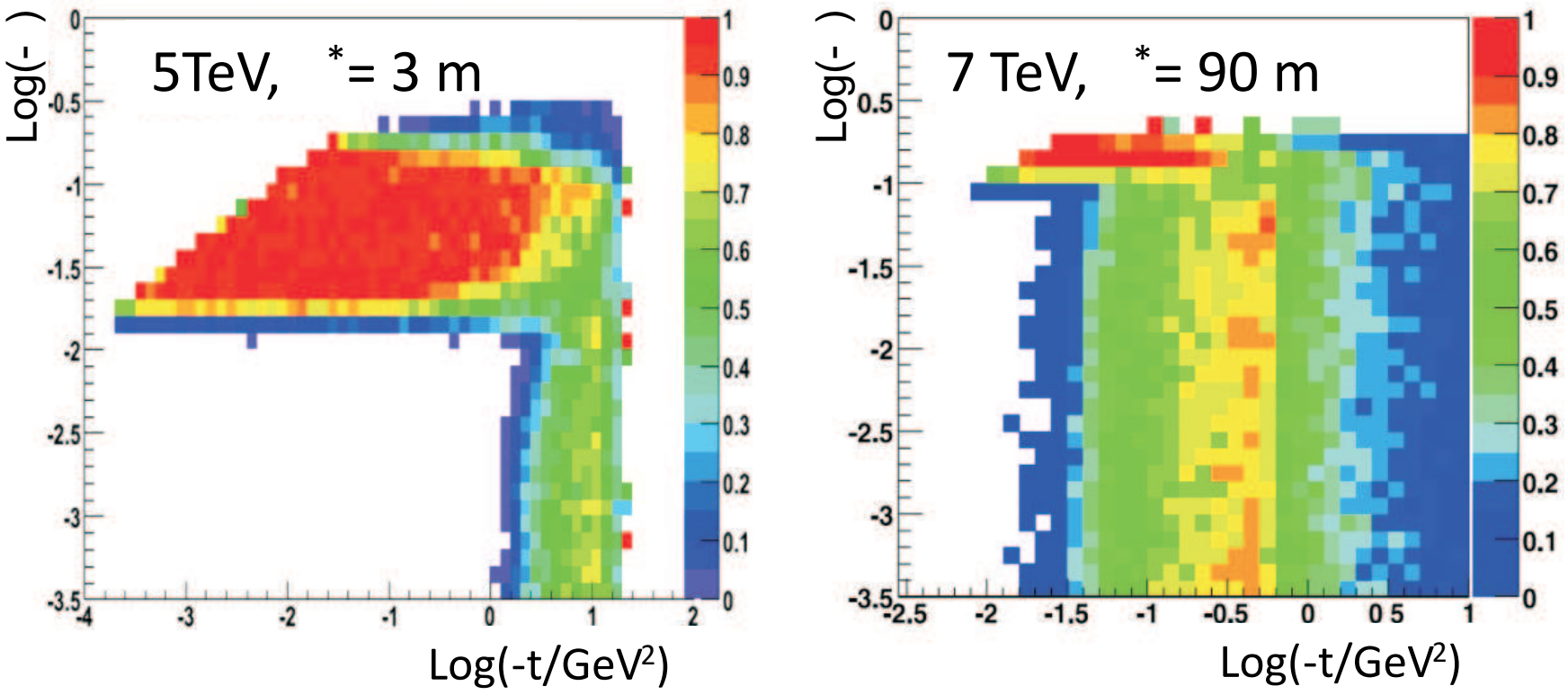

Figure 2: Acceptance in $t$ and $\xi$ of diffractively scattered protons at the RP220 station for the $\beta^* = 3\,\mathrm{m}$ optics at $\sqrt{s} = 10\,\mathrm{TeV}$ (left) and for the $\beta^* = 90\,\mathrm{m}$ optics at $\sqrt{s} = 14\,\mathrm{TeV}$ (right).

| $\beta^*$ [m] | variable | acceptance | resolution |
|---|---|---|---|
| 90 | $\xi = \Delta p/p$ | complete $\xi$-range | $\sigma(\xi) \approx 0.006$<br>$\sigma(\xi) \approx 0.0015$ (CMS vtx.) |
| | DPE $M$ [GeV] | complete $M$-range | $\sigma(M)/M < 1/2$ for $M > 150$ |
| | | | $\sigma(M)/M < 1/2$ for $M \gtrsim 40$ (CMS vtx.) |
| | $t$ [GeV$^2$] | $0.02 < -t < 10\,\mathrm{GeV}^2$ | $\sigma(t_x) \approx (0.3 - 0.4)\sqrt{-t_x}$<br>$\sigma(t_y) \approx 0.04\sqrt{-t_y}$ |
| | $\phi$ [rad] | $0 \le \phi < 2\pi$ | $0.02 < \sigma(\phi) < 1$<br>for $1 > -t > 0.01$ GeV$^2$ |

Table 4: Roman Pot acceptance and reconstruction summary for $\beta^* = 90\,\mathrm{m}$ optics at $\sqrt{s} = 14\,\mathrm{TeV}$. 'CMS vtx.' denotes the reconstruction scenario when primary vertex position, reconstructed in the CMS Tracker, is used in TOTEM proton reconstruction.

| $\beta^*$ [m] | variable | acceptance | resolution |
|---|---|---|---|
| 1535 | $\xi = \Delta p/p$ | complete $\xi$-range | $\sigma(\xi) \approx 0.01$ (RP220) |
| | | | $0.002 < \sigma(\xi) < 0.006$ (RP220 + RP150) |
| | DPE $M$ [GeV] | complete $M$-range | $0.05 < \sigma(M)/M < 0.6$ for $M > 300\,\mathrm{GeV}$ |
| | $t$ [GeV$^2$] | $0.003 < -t < 1.5$ | $\sigma(t_x) \approx (0.04 - 0.4)\sqrt{-t_x}$<br>$\sigma(t_y) \approx (0.002 - 0.02)\sqrt{-t_y}$ |
| | $\phi$ [rad] | $0 \le \phi < 2\pi$ | $0.002 < \sigma(\phi) < 0.02$ for $-t = 1$ GeV$^2$;<br>$0.01 < \sigma(\phi) < 0.1$ for $-t = 0.05$ GeV$^2$; |

Table 5: Roman Pot acceptance and reconstruction summary for $\beta^*$=1535 m optics at $\sqrt{s} = 14\,\mathrm{TeV}$.

In case of the low-$\beta^*$ optics, the $t$-reconstruction is primarily limited by the beam divergence and is practically independent from the azimuth angle $\phi$. The $\phi$-reconstruction is possible only for $-t > 0.18$ GeV$^2$ and $-t > 0.05$ GeV$^2$, for $\beta^* = 0.5$ m and $\beta^* = 2$ m, respectively.

The cases of $\beta^* = 0.5$, 2 and 90 m are characterised by good $\xi$ resolution of $\sigma(\xi) = 0.001$ – 0.006.

# 6 TOTEM Early Measurements

TOTEM will be operable from the LHC start and plans to profit from the early LHC beams. The first collisions are planned at the injection energy $\sqrt{s} = 900$ GeV and $\beta^* = 11$ m. For this scenario, due to large beam sizes and aperture limits, TOTEM RP acceptance is very limited [7]. Therefore, this scenario will be primarily used for experiment commissioning, alignment and calibration.

However, the LHC collisions planned at reduced beam energy of a few TeV and low-$\beta^*$ [8] will allow for detection of SD and DPE events in T1, T2 and RP detectors (see Figure 1). In addition to the commissioning and alignment activities, TOTEM will conduct the preliminary measurements of the differential SD and DPE cross-sections $d\sigma_{\mathrm{SD,DPE}}/dM$ in the mass range of $1.4 < M < 4.2$ TeV and $0.2 < M < 1.8$ TeV, respectively, with mass reconstruction precision $\sigma(M)/M = 2$–$4\,\%$. In addition, T1 and T2 telescopes will study the topology of diffractive events. In particular, they will make an attempt to measure the rapidity gaps $\Delta\eta$ of diffractive systems which can be further compared to the reconstructed proton momentum loss $\xi = \Delta p/p$ via the relation $\Delta\eta \approx -\ln|\xi|$. Also the $pp$ elastic scattering differential cross-section $d\sigma/dt$ for $2 < -t < 10$ GeV will be measured in this scenario with a precision $\sigma(t)/t \approx 0.3\sqrt{|t|}$.

TOTEM will also contribute to the multiplicity studies of forward particles, which are essential not only for minimum bias MC simulations but also for cosmic rays' physics. The energy and mass of ultra-high-energy cosmic rays are obtained with the help of Monte Carlo codes which describe the shower development (dominated by forward and soft QCD interactions) in the upper atmosphere. Various high-energy hadronic interaction models differ of factors up to three, with significant inconsistencies in the forward region ($|\eta| < 5$). The measurements of forward events' topology, carried out by T1 and T2 telescopes, should help to tune the MC codes.

# 7 Later Diffractive Studies

Due to large rapidity coverage of T1 and T2 detectors ($3.1 \le |\eta| \le 6.5$), TOTEM alone can study diffraction extensively by means of rapidity gaps and forward proton detection. In addition, the physics programme can be further extended by cooperation between TOTEM and CMS, which will provide the largest acceptance detector ever built at a hadron collider [3]. This will create a unique opportunity to measure the rapidity gaps in the range of $|\eta| < 6.5$ together with efficient proton reconstruction. In addition, the trigger, based on forward protons, on central system topology and multiplicity, and on calorimetry information, will provide clean signatures for specific processes and efficient background rejection. This will allow for a versatile diffractive physics programme which will cover soft-diffraction, studied at high-$\beta^*$ optics and low luminosity runs (see Table 2), semi-hard and hard diffractive processes, such as jet-jet production, characterised by low cross-sections, which require higher luminosities and thus low-$\beta^*$ optics [3].

The CMS TOTEM cooperation not only extends the available acceptance but it also improves the reconstruction. In particular, the CMS Tracker can provide the primary vertex position with a resolution of $\sim 30\,\mu$m, which highly reduces the uncertainty of longitudinal proton momentum reconstruction for the $\beta^* = 90$ m case, as can be seen in Table 4. The calorimetry system of CMS, in addition to a $\mathbf{p}_T$-trigger, yields an additional means of diffractive system mass determination [9, 10] based on measurement of transverse energy $E_{T_i}$ of centrally produced objects at rapidities $\eta_i$: $\xi_{1,2} = \sum_i E_T^i \mathrm{e}^{-\eta_i}/\sqrt{s}$. The prospects for diffractive physics are discussed in detail in [3].

# 8 Conclusions

The TOTEM physics program aims at deeper understanding of proton structure by measuring the total and elastic *pp* cross sections and by studying a comprehensive menu of diffractive processes. TOTEM will run under all LHC beam conditions to maximise the coverage of the studied processes. Special high-$\beta^*$ runs are needed for the total *pp* cross section measurement with the luminosity-independent method and for soft-diffraction with large forward proton acceptances. Early low-$\beta^*$ runs will provide first opportunities for measurements of soft diffraction in central and single diffractive events, as well as studies of the forward charged multiplicity in inelastic *pp* events. Finally, hard diffraction as well as many forward physics subjects will be studied in collaboration with CMS taking advantage of the unprecedented rapidity coverage for charged and neutral particles.

# References

[1] The TOTEM Collaboration, *Technical Design Report*, CERN-LHCC-2004-002; addendum CERN-LHCC-2004-020.

[2] The TOTEM Collaboration, *The TOTEM Experiment at the LHC*, 2008 JINST 3 S08007, doi: 10.1088/1748-0221/3/08/S08007.

[3] The CMS and TOTEM diffractive and forward working group, *Prospects for Diffractive and Forward Physics at the LHC*, CERN/LHCC 2006-039/G-124, 2006.

[4] E. Wilson, *An Introduction to Particle Accelerators*, Oxford University Press, 2006.

[5] H. Niewiadomski, *Reconstruction of Protons in the TOTEM Roman Pot Detectors at the LHC*, PhD dissertation, Manchester, 2008.

[6] V. Barone, E. Predazzi, *High-Energy Particle Diffraction*, Springer 2002.

[7] H. Niewiadomski, *Acceptance of Protons in the TOTEM RP Detectors for LHC V6.500 Optics of Injection Energy ($E = 450$ GeV, $\beta^* = 11$ m)*, TOTEM-NOTE-2009-001, 2009.

[8] H. Niewiadomski, *Acceptance of Protons in the TOTEM RP Detectors for LHC V6.500 Early Collision Optics ($E = 5$ TeV, $\beta^* = 3$ m)*, TOTEM-NOTE-2009-002, 2009.

[9] D. Acosta et al., CDF Collaboration, Phys. Rev. Lett. 93 (2004) 141601.

[10] V. M. Abazov et al., D0 Collaboration, *Hard single diffraction in $p\bar{p}$ collisions at $\sqrt{s} = 630$ GeV and* 1800 GeV, Phys. Lett. B531 (2002) 52.

# ATLAS Plans on Soft and Hard Diffraction at the Early LHC

*Vojtěch Juránek*[1]
*on behalf of the ATLAS collaboration*

[1]Institute of Physics, Academy of Sciences, Na Slovance 2, 182 21 Prague, Czech Republic

A brief review of ATLAS forward detector system is presented. Short review of forward physics measurements that are planned to be done with first ATLAS data are introduced. This includes several topics, mainly QCD processes, but also photon physics.

## 1 Forward Physics with First ATLAS Data

ATLAS first data will be taken at very low luminosity in very clean environment, where almost no additional $p-p$ scattering will be present. It expected that first data will be taken up in late 2009 or in beginning of 2010 at centre-of-mass energy of 7 TeV which will be increased to 10 TeV later on. Peak luminosity during this first data taking will be from $\mathcal{L} = 5 \cdot 10^{31}$ cm$^{-2}$s$^{-1}$ to $\mathcal{L} = 2 \cdot 10^{32}$ cm$^{-2}$s$^{-1}$. At these luminosities the number of additional $p-p$ interactions will be small. Bunch spacing of 75 ns (expected during the first data taking) can lead up to 1.8 interactions per bunch crossing, which is still quite clean environment. Total integrated luminosity taken up during the first 100 days will be about 100 pb$^{-1}$, and about 200 pb$^{-1}$ during the next 100 day of operation. With this amount of data a lot of forward physics measurements can be done. There are analysis that can be done with 10 pb$^{-1}$ of data or less.

## 2 ATLAS Forward Detectors

ATLAS [1] is general purpose detector at the LHC designed to measure the broadest possible range of signals [2]. The main parts of the ATLAS detector are Inner Detector, electromagnetic and hadronic calorimeters, muon spectrometer and forward detectors. The Inner Detector covers pseudo-rapidity range up to $|\eta| < 2.5$, central electromagnetic and hadronic calorimeters up to $|\eta| < 3.2$ and forward calorimeters up to $|\eta| < 4.9$. The $\eta$ coverage of muon spectrometer is up to $|\eta| < 2.7$.

To have good coverage in forward regions, there are MBTS and three additional smaller detectors: LUCID, ZDC and ALFA. Nearest forward detector is LUCID, which is $\pm 17$ m from the interaction point. ZDC is situated at $\pm 140$ m from interaction point and ALFA at $\pm 240$ m.

### 2.1 MBTS

MBTS stands for Minimum Bias Trigger Scintillators. MBTS is placed between inner detector and end-cap cryostats. It covers pseudo-rapidity in range $2.09 < |\eta| < 3.84$. MBTS consist of 16 wedges on each side, there are two rings in pseudo-rapidity and each ring has 8 azimuthal wedges. MBTS will be not active during whole lifetime of ATLAS. Because of heavy radiation, it is expected that MBTS will become increasingly inefficient after a few months of higher luminosity operation.

## 2.2 LUCID

LUCID (Luminosity measurement using Cerenkov Integrating Detector) is the main ATLAS on-line monitor for relative luminosity measurement. This measurement is done by detecting particles coming from inelastic scattering of the protons. One of the aims of LUCID is to reduce the uncertainty of relative luminosity measurement below 5%. LUCID is designed to count individual charged particles, because instantaneous luminosity measurement is based on the fact that the average number of interaction in bunch crossing is proportional to the number of particles detected in the detector. LUCID has also very good time resolution to be able to measure each individual bunch crossing. More over it has to be highly resistant to radiation.

The LUCID detectors are installed in the end-cap regions of the ATLAS detector. It consists of 20 aluminium tubes that surround the beam pipe in the distance of about 10 cm from the beam-line. The tubes point towards the interaction point. LUCID has good acceptance for minimum bias events and will be used for triggering of these events.

## 2.3 ZDC

ZDC (Zero Degree Calorimeter) is designed for measurements of forward neutrons in heavy-ion collisions. It is placed at 140 m on both sides of interaction point in Target Absorber Neutral (TAN) between the tubes at the point where LHC beam-pipe splits into two separate tubes. ZDC consists of 4 modules, one electromagnetic and three hadronic. It covers pseudo-rapidity $|\eta| > 8.3$. In the very early data taking period, the ZDC electromagnetic module will not be installed and its position is taken by the LHCf experiment [3].

ZDC can be also used for diffractive measurements during the start-up low luminosity runs (below $10^{33}$ cm$^{-2}$s$^{-1}$). The time resolution of ZDC is about 100 ps, so by requiring activity on both sides of ZDC, it can be also used for vertex determination with accuracy of about 3 cm in $z$ direction. This can be used for excluding events like beam halo and beam gas. The only disadvantage for forward physics measurements is that ZDC can detect only neutral particles $(n, \gamma)$. As soon as the luminosity of $10^{33}$ cm$^{-2}$s$^{-1}$ will be reached, the ZDC modules will be removed for $p - p$ scattering and reinstalled only for heavy ions runs (in order to minimise radiation damage of ZDC).

## 2.4 ALFA

ALFA (Absolute Luminosity For ATLAS) is dedicated to measurements of absolute luminosity using Roman Pot detectors. This will be done by measurements of $p - p$ elastic scattering. Protons scatter through very small angles in this reaction, so the detector has to be placed far from the interaction point and very close to the beam. Moreover for this measurement special runs (high $\beta^*$ optics and reduced beam emittance) are required. ALFA roman pots will be able to come as close to the beam as 1 mm. There will be two roman pots on each side. The distance between these two pots is 4 m. ALFA is not yet installed and will not be used during 2009/2010.

### 2.5 Forward Detectors Used in Early Data

The majority of the physics studies that are presented in the following sections make use of the ATLAS central detectors and then propose to use the forward detectors as either a trigger or to measure the properties of the diffractive events in early ATLAS data. The MBTS is used to trigger both minimum bias and diffractive events and can be used to impose a rapidity gap (i.e. particle veto) in the regions $2.1 < |\eta| < 3.8$. In addition to MBTS, LUCID and the ZDC will be used to trigger events. In principle, both LUCID and the ZDC can also be used to define rapidity gaps in the very forward region in order to select central exclusive events (Sec. 5) and single diffractive (Sec. 6). This possibility is currently being investigated using the ATLAS simulation.

## 3 Soft Single Diffraction

Soft single diffraction is a low $t$-process ($t$ being standard Mandelstam variable), where colour singlet is exchanged in $t$ channel between two protons and one of the protons breaks up into a dissociative system. A similar process is soft double diffraction, but in this process both protons breaks up into dissociative system. As the exchanged object is colour singlet, there is a large rapidity gap between intact proton and dissociative system (or between dissociative systems in case of double diffraction). Expected cross sections at LHC are about 12 mb in case of soft single diffraction and about 7 mb in case of soft double diffraction.

Soft single diffraction can be measured by ALFA by tagging the outgoing proton, but it can be measured also with the more central detectors by imposing a pseudo-rapidity gap in the forward regions. With such high cross section, only about two weeks of data taking at lowest luminosity ($10^{31}$ cm$^{-2}$s$^{-1}$) is required by this analysis (it is expected to collect sample of about one million events in two weeks at luminosity of $10^{31}$ cm$^{-2}$s$^{-1}$). The measurement of soft single diffraction at ATLAS will be focused on the diffractive mass distribution, $M_X$, and the fractional energy loss of proton, $\xi = \frac{M_X}{\sqrt{s}}$. ATLAS will cover several orders in magnitude in $\xi$. MBTS, LUCID and ZDC will be used as triggers.

When installed, ALFA will be able to measure directly proton momentum loss $\xi$ during the special runs with high-$\beta^*$ optics and luminosity of $10^{27}$ cm$^{-2}$s$^{-1}$.

## 4 BFLK Jet Evolution and Colour Singlet Exchange

ATLAS has potential to measure a lot of interesting observables inspired by BFKL. A process of interest is QCD $2 \to 2$ scattering mediated by $t$-channel exchange. It is interesting to study jet evolution and compare the results with DQLAP and BFKL predictions. In particular, interesting process is $t$-channel colour singlet exchange, as possible candidate for colour singlet exchange is BFKL pomeron [4].

BFKL predicts different shower evolution to DGLAP. In BFKL gluon splitting is not ordered in $E_T$, which can lead to final state with jets in central region with similar $E_T$ to those in forward regions. BFKL also predicts larger decorrelation than DGLAP in $\Delta\Phi (= \pi - \phi_{jet1} + \phi_{jet2})$ between the jets, see e.g. [5, 6]. This azimuthal decorrelation increases[1] with pseudo-rapidity separation of jets, $\Delta\eta$.

[1]This effect is dominant in leading log approximation. In NLL the decorrelation decrease and becomes similar to the DGLAP predictions [7]

Another interesting observable is so called gap fraction - a fraction of events containing no or small radiation in the centre of the detector, i.g. with large central rapidity gap. This observable has been previously measured at HERA [8, 9] and Tevatron [10, 11]. However, at the Tevatron the gap fraction was not observed as, for example, the centre-of-mass energy was too small. The fraction of events with suppressed activity in central region should rise with rapidity gap between jets. The separation of jets in pseudo-rapidity increases with centre-of-mass energy as $\Delta\eta \simeq \ln \frac{x_1 x_2 s}{Q_1 Q_2}$, where $Q_i \approx E_{T,i}$. This means that situation at LHC will be improved with comparison with Tevatron and it is expected that a sufficiently large statistics sample of events with gaps can be collected. The signature of the process will be two jets in forward calorimeter, one jet in each forward calorimeter, and gap in central calorimeter. The 2FJ18 trigger will be used to trigger the events - trigger on two jets in FCAL with $E_T$ >18 GeV.

# 5 Central Exclusive Di-Jet Production

Central exclusive di-jet production (CEP) is defined as $pp \to p \oplus jj \oplus p$, where all the proton energy lost during the interaction is used in production of di-jet system, $jj$. Symbol $\oplus$ denotes rapidity gap. CEP processes were measured by CDF and are in good agreement with theoretical predictions [12]. However, there are still large uncertainties in theoretical calculations. The main aim of this measurement is to measure cross section as function of $E_T$ to constraint the uncertainty of theoretical models (namely KMR model [13], by factor of 2-3) and understand CEP di-jet production as a background for other interesting processes like CEP production of Higgs boson. Possible strategy will be measurement of di-jet mass fraction

$$R_{jj} = \frac{M_{jj}}{M_{calo}} \simeq \sqrt{\frac{x_1 x_2}{\xi_1 \xi_2}}, \tag{1}$$

where $M_{jj}$ is invariant mass of di-jet system, $M_{calo}$ is the mass of all energy deposit in the calorimeter and

$$x_{1,2} = \frac{1}{\sqrt{s}} \sum_{jets} E_T^i \exp(\pm\eta_i) \tag{2}$$

$$\xi_{1,2} = \frac{1}{\sqrt{s}} \sum_{clus} E_T^i \exp(\pm\eta_i). \tag{3}$$

$R_{jj}$ should be near unity for CEP di-jet, while $R_{jj} \ll 1$ for inclusive and diffactive events.

The trigger strategy is to require a jet with $E_T > 10$ GeV and a veto on at least one side of MBTS trigger (L1_J10_MV). The trigger is expected to be run unprescaled at luminosity around $10^{31}$ cm$^{-2}$s$^{-1}$. It is expected to collect a few hundred exclusive di-jet events in 10 pb$^{-1}$ of data.

# 6 Diffractive Di-Jet Production

Diffractive di-jet production is a process where one proton (single diffraction, SD) or both protons (double pomeron exchange, DPE) remains intact and remnant from exchanged particles (pomerons) are present. This means that not all energy of exchanged objects is used to create central object as in case of exclusive production and also rapidity gaps are smaller because of presence of pomeron remnants. These types of processes were measured at HERA [14, 15] and Tevatron [16, 17], where factorisation breaking were observed in comparison with HERA results.

The aim of this analysis is to measure ratio of di-jet production in hard single diffraction (SD) to non-diffractive di-jet production and ratio of double pomeron exchange (DPE) di-jet production to SD di-jet production. From these measurements, conclusions about gap survival probability at LHC energies can be made and theoretical uncertainties can be reduced. Also diffractive structure functions can be determined from relation

$$\frac{\sigma(SD_{jj})}{\sigma(ND_{jj})} = \frac{F_{jj}^D(x)}{F_{jj}(x)}. \tag{4}$$

POMWIG Monte Carlo [18] predicts SD di-jet production cross section of about 3 $\mu$b for LHC energies with gap survival probability equal to 0.075, $E_T^{jet} > 17$ GeV and $\xi < 0.1$. The cross section is quite large, but unfortunately the triggering SD events is quite problematic at ATLAS. Low $E_T$ jet trigger (J_18) is heavily prescaled (the considered prescale is 6000) and forward detectors are not very helpful - MBTS veto passes only very small amount of signal (only events with very small $\xi$, where rapidity gap is very large). A LUCID veto passes almost all signal, but also a lot of background (non-diffractive di-jets, about 70%), which again require heavy prescaling. ZDC could be very helpful as most of the events which pass veto in ZDC have rapidity gaps in the forward regions. Trigger design for diffractive di-jets is still in progress.

# 7 Photon Physics

Another interesting topics is photon induced forward physics processes, especially photon induced CEP WW, CEP $\Upsilon$ production and CEP di-lepton production. CEP WW is interesting process itself, but it attracts attention also for another reason: using this process allows studying anomalous triple WW$\gamma$ and quartic WW$\gamma\gamma$ gauge coupling. It has be shown [19], that LHC is more sensitive to quartic anomalous coupling than to anomalous triple gauge coupling. It is expected that about 100 pb$^{-1}$ amount of data will be sufficient to observe CEP WW and significantly improve conclusion about WW$\gamma\gamma$ anomalous coupling. Details about measurements of this process can be found in [19] and [20].

# References

[1] The ATLAS Collaboration, The ATLAS Experiment at the CERN Large Hadron Collider, J. of Instr., 3(08):S08003 (2008)

[2] The ATLAS Collaboration, Expected Performance of the ATLAS Experiment: Detector, Trigger and Physics. CERN, Geneva (2008), CERN-OPEN-2008-020

[3] The LHCf Collaboration, The LHCf detector at the CERN Large Hadron Collider, J. of Instr., 3(08):S08006 (2008)

[4] A. Mueller, W. Tang, Phys. Lett. B 284, 123 (1992)

[5] C. Marquet, C. Royon, Phys.Rev.D79:034028

[6] A. Sabio Vera, S. Schwennsen, Nucl. Phys. B 776, 170 (2007)

[7] F. Chevallier, O. Kepka, C. Marquet, C. Royon, Phys.Rev.D79:094019 (2009)

[8] C. Adolf et. al. (H1 Collaboration), Eur. Phys. J. C 24, 517 (2002)

[9] M. Derrick el al. (ZEUS Collaboration), Phys. Lett. B 369, 55 (1996)

[10] B. Abbot et. al. (DO Collaboration), Phys. Lett. B 440, 189 (1998)

[11] F. Abe et. al. (CDF Collaboration), Phys. Rev. Lett. 81, 5278 (1998)

[12] T. Aaltonen et al. (CDF Collaboration), Phys.Rev.D77, 052004 (2008)

[13] V.A. Khoze, A.D. Martin and M.G. Ryskin, Eur. Phys. J. C 23, 311 (2002)

[14] C. Adloff, et al. (H1 Collaboration), Z. Phys. C 76:613 (1997)

[15] M. Derrick et. al. (ZEUS Collaboration), Phys. Lett. B 356:129 (1995)

[16] T. Affolder et al. (CDF Collaboration), Phys. Rev. Lett. 84:5043 (2000)

[17] T. Affolder et al. (CDF Collaboration), Phys. Rev. Lett. 88:151802 (2002)

[18] B.E. Cox and J.R. Forshaw, Comput.Phys.Commun.144, 104 (2002)

[19] O. Kepka, C. Royon, Phys. Rev. D78:073005 (2008)

[20] O. Kepka, these proceedings

# Diffraction at ALICE

*Rainer Schicker*[1],

[1]Physikalisches Institut, Philosophenweg 12, 69120 Heidelberg, Germany

The ALICE detector at the Large Hadron Collider (LHC) consists of a central barrel, a muon spectrometer, zero degree calorimeters and additional detectors which are used for trigger purposes and for event classification. The performance of the ALICE central barrel is discussed. The trigger strategy for diffractive events is outlined. The physics potential of studying diffractive processes at the LHC is presented by investigating observables of the Pomeron and Odderon. Possible signatures of gluon saturation in diffractive events are discussed.

## 1 The ALICE Experiment

The ALICE experiment is presently being built and commissioned at the Large Hadron Collider (LHC) [1, 2]. The ALICE experiment consists of a central barrel covering the pseudorapidity range $-0.9 < \eta < 0.9$ and a muon spectrometer in the range $-4.0 < \eta < -2.4$. Additional detectors for trigger purposes and for event classification exist such that the range $-4.0 < \eta < 5.0$ is covered. The ALICE physics program foresees data taking in pp and PbPb collisions at luminosities up to $\mathcal{L} = \bigtriangledown \times 10^{30}\text{cm}^{-2}\text{s}^{-1}$ and $\mathcal{L} = 10^{27}\text{cm}^{-2}\text{s}^{-1}$, respectively. An asymmetric system pPb will be measured at a luminosity of $\mathcal{L} = 10^{29}\text{cm}^{-2}\text{s}^{-1}$.

## 2 The ALICE Central Barrel

The detectors in the ALICE central barrel track and identify hadrons, electrons and photons in the pseudorapidity range $-0.9 < \eta < 0.9$. The magnetic field strength of 0.5 T allows the measurement of tracks from very low transverse momenta of about $100\,\text{MeV}/c$ to fairly high values of about 100 GeV/c. The tracking detectors are designed to reconstruct secondary vertices from decays of hyperons, D and B mesons. The granularity of the central barrel detectors is chosen such that particle tracking and identification can be achieved in a high multiplicity environment of up to 8000 particles per unit of rapidity. The main detector systems for these tasks are the Inner Tracking System (ITS), the Time Projection Chamber (TPC), the Transition Radiation Detector (TRD) and the Time of Flight array (TOF). These systems cover the full azimuthal angle within the pseudorapidity range $-0.9 < \eta < 0.9$. Additional detectors with partial coverage of the central barrel are a PHOton Spectrometer (PHOS), an electromagnetic calorimeter (EMCAL) and a High-Momentum Particle Identification Detector (HMPID).

### 2.1 The Central Barrel Performance

The ITS, TPC and TRD detectors are the main tracking detectors in the central barrel. With the information from these detectors, pions with momenta as low as 100 MeV/c can be tracked. The combined transverse momentum resolution from the ITS, TPC and TRD detector is expected to be about 3% at a transverse momentum of $p_T = 100\,\mathrm{GeV}/c$.

Particle identification is achieved in the central barrel by different methods. The specific energy loss is measured by the TPC, the TRD and the strip and drift detectors of the ITS.

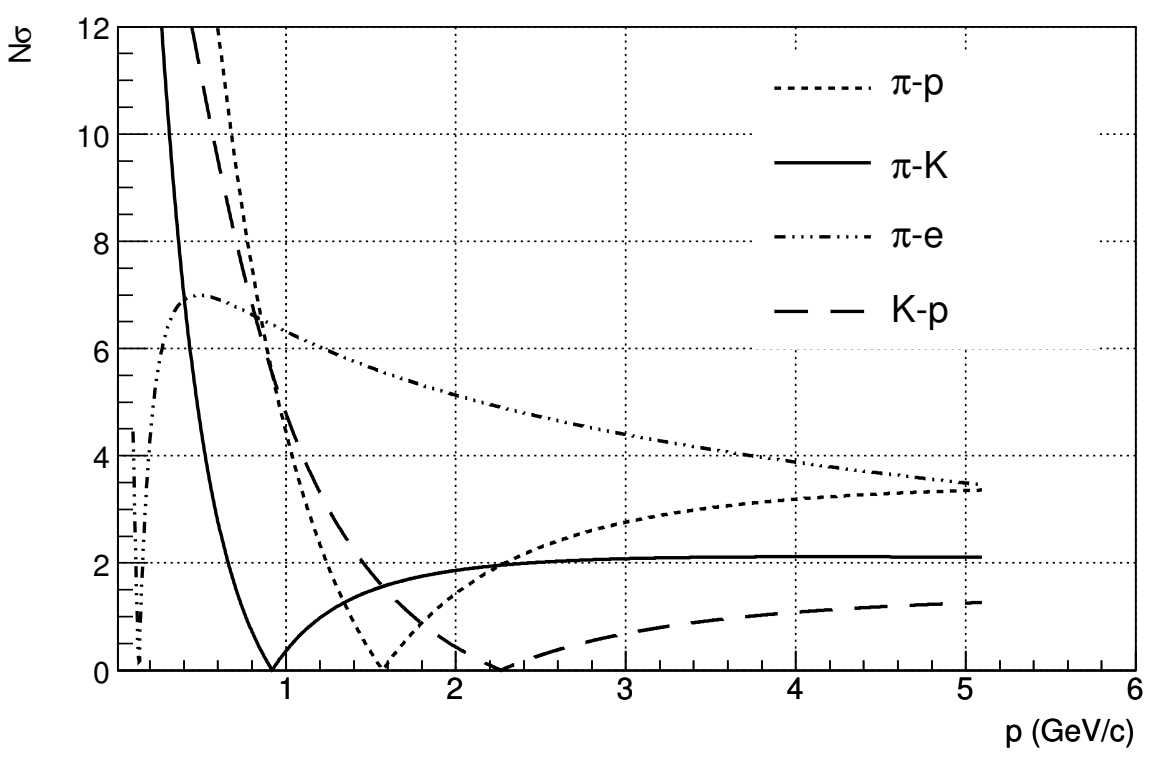

Figure 1: Particle identification by dE/dx measurement.

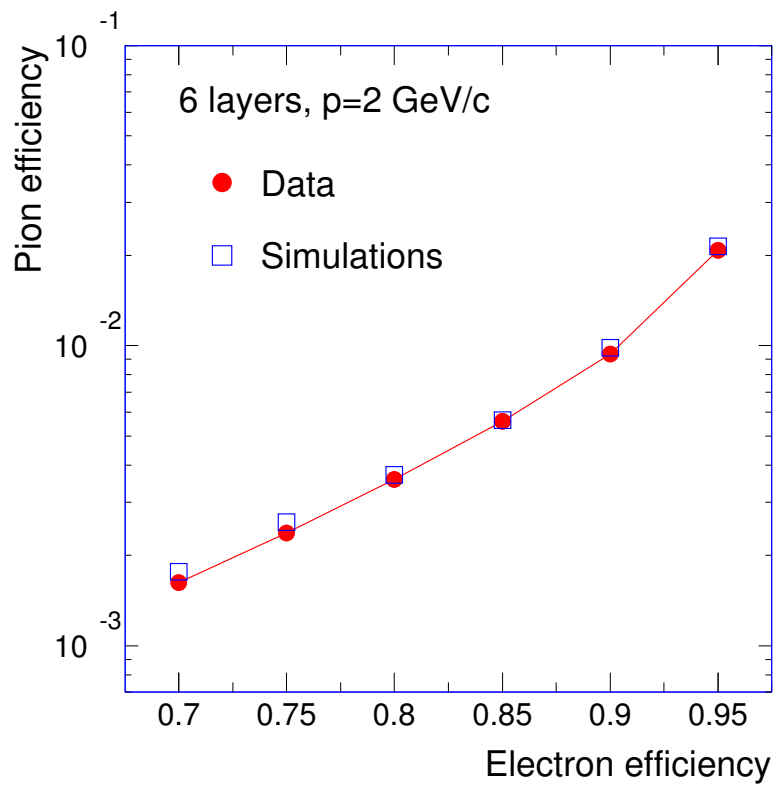

Figure 2: Electron-pion separation in the TRD.

Fig. 1 shows the separation power of the TPC as a function of momentum. The separation of pairs of different particle species is shown in units of the resolution of the $dE/dx$ measurement.

The electron-pion separation at high momenta is significantly improved by the information of the TRD system. Fig. 2 shows the pion efficiency in the TRD as a function of the electron efficiency. Here, pion efficiency indicates that a pion is misidentified as an electron. The expected TRD performance for a full stack of six layers is shown by the squares and compared to test beam data represented by the circles.

## 3 The ALICE Zero Degree Neutron Calorimeter

The Zero Degree Neutron Calorimeters (ZDC) are placed on both sides of the interaction point at a distance of 116 m [3]. The ZDC information can be used to select different diffractive topologies. Events of the type $pp \rightarrow ppX$ do not deposit energy in these calorimeters, single diffractive dissociation events $pp \rightarrow pp^*X$ will have energy in one of the calorimeters whereas double diffractive dissociation events $pp \rightarrow p^*p^*X$ will have energy deposited in both calorimeters. Here, $p^*$ denotes a diffractively excited state of the proton and $X$ represents a centrally produced diffractive state from which the diffractive L0 trigger is derived as described below.

## 4 The ALICE Diffractive Gap Trigger

Additional detectors for event classification and trigger purposes are located on both sides of the ALICE central barrel. First, two arrays of scintillator detectors V0A and V0C cover a pseudorapidity interval of about two units on either side of the central barrel with a fourfold segmentation of half a unit. The azimuthal coverage is divided into eight segments of 45 degrees hence each array is composed of 32 individual counters. Second, a Forward Multiplicity Detector (FMD) is located on both sides of the central barrel. The pseudorapidity coverage of this detector is $-3.4 < \eta < -1.7$ and $1.7 < \eta < 5.1$, respectively.

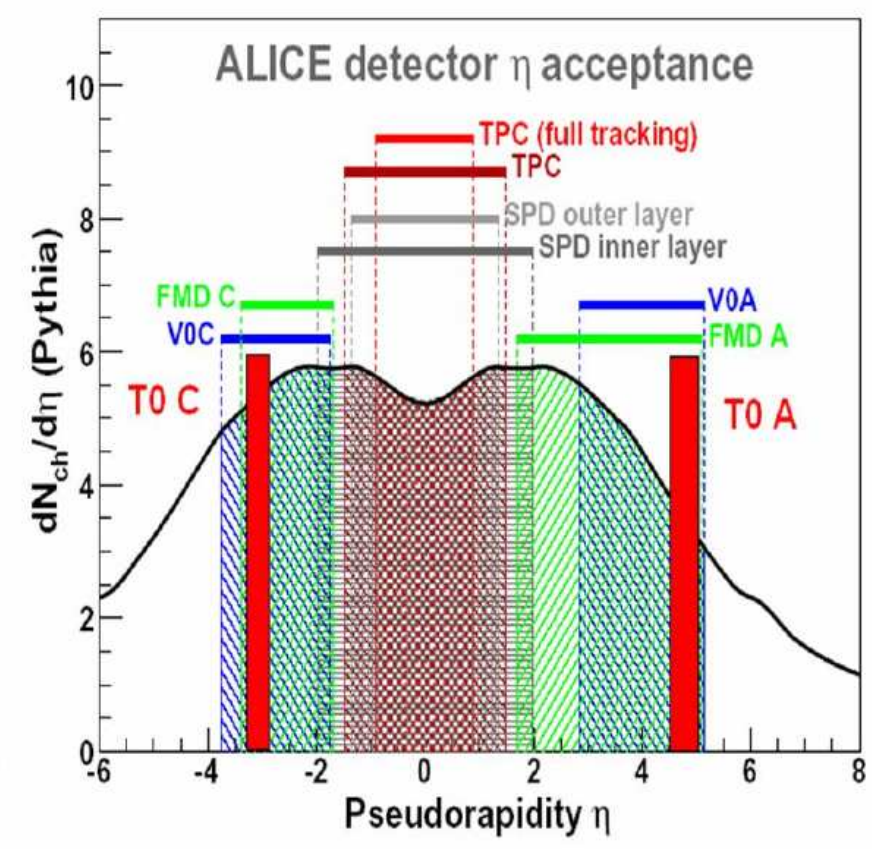

Figure 3: Pseudorapidity coverage of trigger detectors and of detectors in central barrel.

Fig. 3 shows the pseudorapidity coverage of the detector systems described above. The geometry of the ALICE central barrel in conjunction with the additional detectors V0A,V0C and FMD is well suited for the definition of a rapidity gap trigger. The ALICE trigger system is designed as a multi-level scheme with L0, L1 and L2 levels and a High-Level Trigger (HLT). The L0 signal can be taken from the TOF detector or from the pixel detector of the ITS [4].

The HLT has access to the information of all the detectors shown in Fig. 3 and will hence be able to select events with rapidity gaps in the range $-4 < \eta < -1$ and $1 < \eta < 5$. These gaps extend over seven units of pseudorapidity and are hence expected to suppress minimum bias inelastic events by many orders of magnitude.

# 5 ALICE Diffractive Physics

The tracking capabilities at very low transverse momenta in conjunction with the excellent particle identification make ALICE an unique facility at LHC to pursue a long term physics program of diffractive physics. The low luminosity of ALICE as compared to the other LHC experiments restricts the ALICE physics program to reactions with cross section at a level of a few nb per unit of rapidity.

Fig. 4 shows the transverse momentum acceptance of the four main LHC experiments. Not shown in this figure is the acceptance of the TOTEM experiment which has a physics program of measurements of total cross section, elastic scattering and soft diffraction [5]. The acceptance of the TOTEM telescopes is in the range of $3.1 < |\eta| < 4.7$ and $5.3 < |\eta| < 6.5$. The combined data taking of TOTEM and CMS represents the largest rapidity interval covered at the LHC. The CMS transverse momentum acceptance of about $1\,\mathrm{GeV}/c$ shown in Fig. 4 represents a nominal value. The CMS analysis framework foresees the reconstruction of a few selected data samples to values as low as 0.2 GeV/$c$ [6].

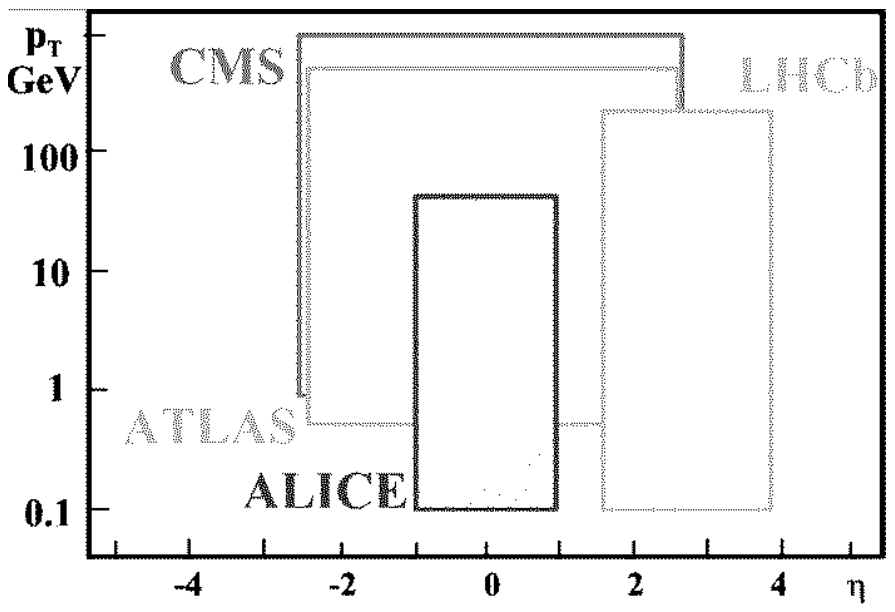

Figure 4: Rapidity and transverse momentum acceptance of the LHC experiments.

# 6 Signatures of the Pomeron

The geometry of the ALICE experiment is suited for measuring a centrally produced diffractive state with a rapidity gap on either side. Such a topology can result, among other, from double Pomeron exchange with subsequent hadronization of the central state. It is expected that the secondaries from Pomeron-Pomeron fusion events show markedly different characteristics as compared to secondaries from inelastic minimum bias events.

First, it is expected that the production cross section of glueball states in Pomeron fusion is larger as compared to inelastic minimum bias events. It will therefore be interesting to study the resonances produced in the central region when two rapidity gaps are required [7].

Second, the slope $\alpha'$ of the Pomeron trajectory is rather small: An analysis of proton-proton, proton-antiproton and photon-proton cross sections and of the proton structure function $F_2$ derives $\alpha' \sim 0.25\,\text{GeV}^{-2}$ whereas a value $\alpha' \sim 0.1\,\text{GeV}^{-2}$ is found in $J/\Psi$ photoproduction at HERA [8]. These values of $\alpha'$ in conjunction with the small t-slope ($< 1\,\text{GeV}^{-2}$) of the triple Pomeron vertex indicate that the mean transverse momentum $k_t$ in the Pomeron wave function is relatively large $\alpha' \sim 1/k_t^2$, most probably $k_t > 1\,\text{GeV}$. The transverse momenta of secondaries produced in Pomeron-Pomeron interactions are of the order of this $k_t$. Thus the mean transverse momenta of secondaries produced in Pomeron-Pomeron fusion is expected to be larger as compared to inelastic minimum bias events.

Third, the large $k_t$ described above corresponds to a large effective temperature. A suppression of strange quark production is not expected. Hence the K/$\pi$ ratio is expected to be enhanced in Pomeron-Pomeron fusion as compared to inelastic minimum bias events [9]. Similarly, the $\eta/\pi$ and $\eta'/\pi$ ratios are expected to be enhanced due to the hidden strangeness content and due to the gluon components in the Fock states of $\eta, \eta'$.

# 7 Signatures of the Odderon

The Odderon was first postulated in 1973 and is represented by color-singlet exchange with negative C-parity [10]. Finding experimental evidence of the Odderon has, however, turned out to be extremely difficult [11].

## 7.1 Signatures of Odderon Cross Sections

Signatures of Odderon exchanges can be looked for in exclusive reactions where the Odderon (besides the photon) is the only possible exchange. Diffractively produced C-odd states such as vector mesons $\phi, J/\psi, \Upsilon$ can result from photon-Pomeron or Odderon-Pomeron exchange. Any excess beyond the photon contribution would be indication of an Odderon exchange.

Fig. 5 and Fig. 6 show the Feynman diagram for vector meson production by Pomeron-Odderon fusion with the protons staying intact and with breakup, respectively. The two different reaction channels can be identified by the information of the ZDC. To each of these two diagrams exists a diagram in which the Odderon is replaced by a photon.

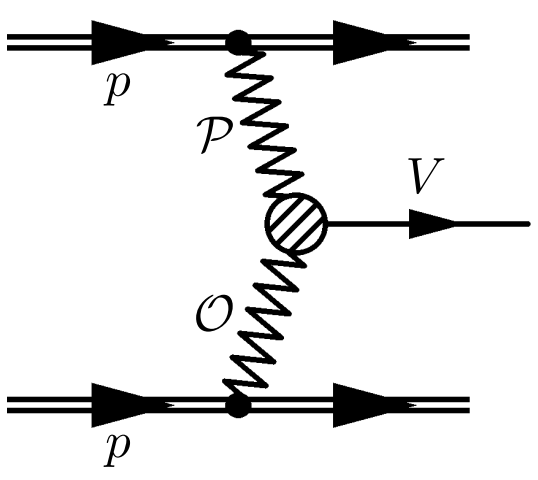

Figure 5: Vector Meson production by Odderon-Pomeron fusion without proton breakup.

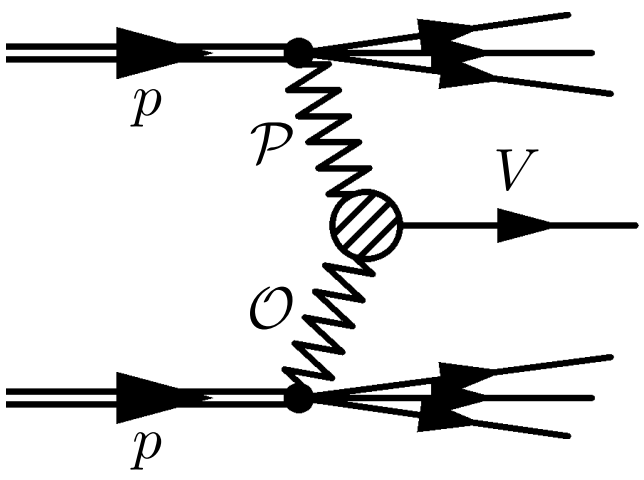

Figure 6: Vector Meson production by Odderon-Pomeron fusion with proton breakup.

Cross sections for diffractively produced $J/\psi$ in pp collisions at LHC energies were first estimated by Schäfer et al. [12]. Calculations by Bzdak et al result in t-integrated photon and Odderon contributions of $\frac{d\sigma}{dy}\,|_{y=0} \sim 15\,\text{nb}$ and $\frac{d\sigma}{dy}\,|_{y=0} \sim 1\,\text{nb}$, respectively [13].

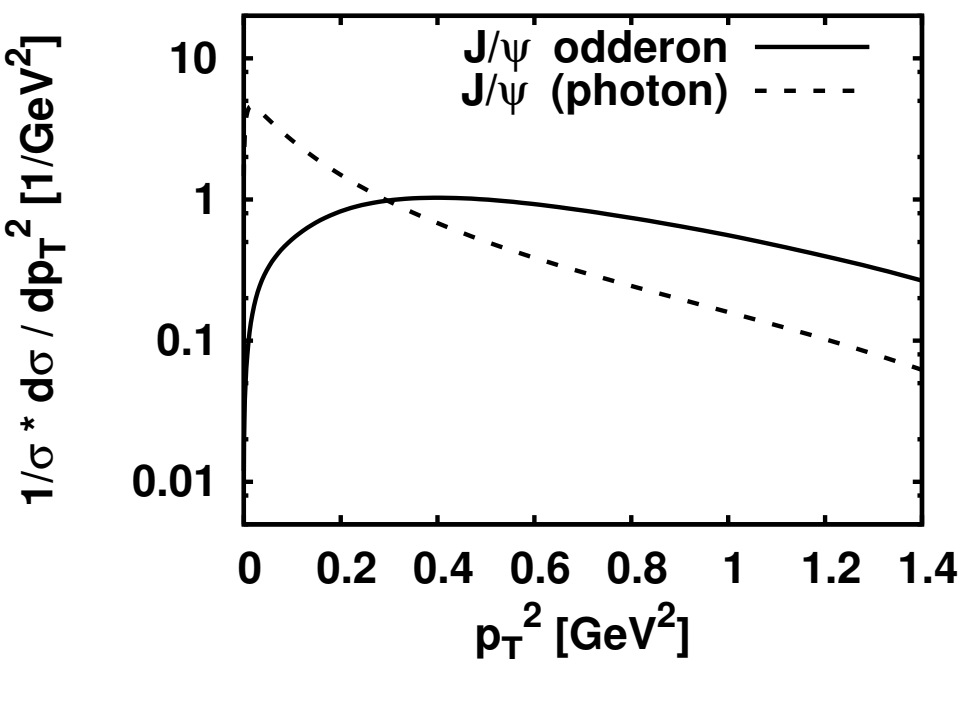

Figure 7: The J/$\psi$ transverse momentum distribution for the photon and Odderon contributions.

The calculated cross sections for the Odderon and photon contribution carry large uncertainties, the upper and lower limit vary by about an order of magnitude. This cross section is, however, at a level where in $10^6\,$s of data taking the $J/\psi$ can be measured in its $e^+e^-$ decay channel at a level of 4% statistical uncertainty. Due to the different t-dependence, the two contributions result in different $p_T$ distributions of the $J/\psi$. The photon and Odderon contributions are shown in Fig. 7 by the dotted and solid lines, respectively [14]. A careful analysis of the $J/\psi$ transverse momentum might therefore allow to disentangle the Odderon contribution.

### 7.2 Signatures of Odderon Interference Effects

If the diffractively produced final state is not an eigenstate of C-parity, then interference effects between photon-Pomeron and photon-Odderon amplitudes can be analysed.

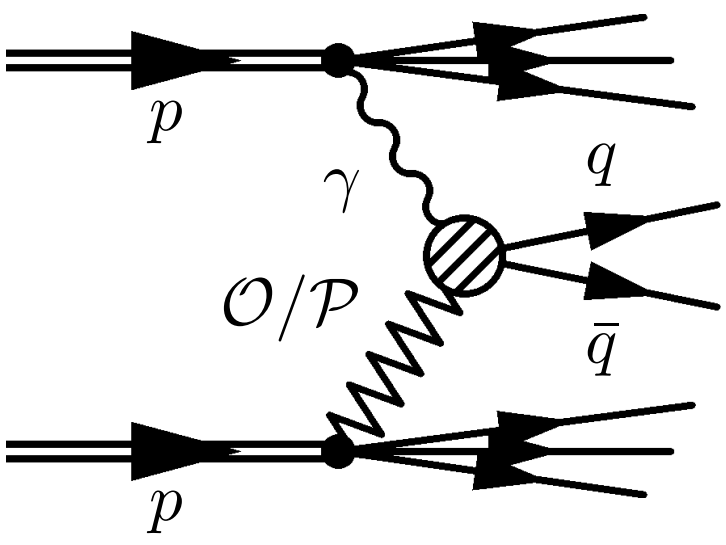

Figure 8: photon-Pomeron and photon-Odderon amplitudes.

Fig. 8 shows the photon-Pomeron and the photon-Odderon amplitudes for $q\bar{q}$ production.

A study of open charm diffractive photoproduction estimates the asymmetry in fractional energy to be on the order of 15% [15]. The forward-backward charge asymmetry in diffractive production of pion pairs is calculated to be on the order of 10% for pair masses in the range $1\,\mathrm{GeV}/c^2 < m_{\pi^+\pi^-} < 1.3\,\mathrm{GeV}/c^2$ [16, 17].

# 8 Photoproduction of heavy quarks

Diffractive reactions involve scattering on small-x gluons in the proton. The number density of gluons at given $x$ increases with $Q^2$, as described by the DGLAP evolution. Here, $Q^2$ and $x$ denote the kinematic parameters used in deep inelastice ep scattering. The transverse gluon density at a given $Q^2$ increases with decreasing $x$ as described by the BFKL evolution equation. At some density, gluons will overlap and hence reinteract. In this regime, the gluon density saturates and the linear DGLAP and BFKL equation reach their range of applicability. A saturation scale $Q_s(x)$ is defined which represents the breakdown of the linear regime. Nonlinear effects become visible for $Q < Q_s(x)$.

Diffractive heavy quark photoproduction represents an interesting probe to look for gluon saturation effects at LHC. The inclusive cross section for $Q\bar{Q}$ photoproduction can be calculated within the dipole formalism. In this approach, the photon fluctuates into a $Q\bar{Q}$ excitation which interacts with the proton as a color dipole. The dipole cross section $\sigma(x, r)$ depends on $x$ as well as on the transverse distance $r$ of the $Q\bar{Q}$ pair. A study of inclusive heavy quark photoproduction in pp collisions at LHC energy has been carried out [18]. These studies arrive at differential cross sections for open charm photoproduction of $\frac{d\sigma}{dy}\mid_{y=0} \sim 1.3\,\mu$b within the collinear pQCD approach as compared to $\frac{d\sigma}{dy}\mid_{y=0} \sim 0.4\,\mu$b within the colour glass condensate (CGC). The cross sections are such that open charm photoproduction seems measurable with good statistical significance. The corresponding numbers for the cross section for bottom photoproduction are $\frac{d\sigma}{dy}\mid_{y=0} \sim 20\,$nb and 10 nb, respectively.

Diffractive photoproduction is characterized by two rapidity gaps in the final state. In the dipole formalism described above, the two gluons of the color dipole interaction are in color singlet state. Diffractive heavy quark photoproduction cross sections in pp, pPb and PbPb collisions at LHC have been studied [19]. The cross sections for diffractive charm photoproduction are $\frac{d\sigma}{dy}\mid_{y=0} \sim 6\,$nb in pp, $\frac{d\sigma}{dy}\mid_{y=0} \sim 9\,\mu$b in pPb and $\frac{d\sigma}{dy}\mid_{y=0} \sim 11\,$mb in PbPb collisions. The corresponding numbers for diffractive bottom photoproduction are $\frac{d\sigma}{dy}\mid_{y=0} \sim 0.014\,$nb in pp, $\frac{d\sigma}{dy}\mid_{y=0} \sim 0.016\,\mu$b in pPb and $\frac{d\sigma}{dy}\mid_{y=0} \sim 0.02\,$mb in PbPb collisions.

Heavy quarks with two rapidity gaps in the final state can, however, also be produced by central exclusive production, i.e. two Pomeron fusion. The two production mechanisms have a different t-dependence. A careful analysis of the transverse momentum $p_T$ of the $Q\bar{Q}$ pair might therefore allow to disentangle the two contributions.

# Acknowledgements

I thank Otto Nachtmann and Carlo Ewerz for illuminating discussions and Leszek Motyka for preparing and communicating Figure 7.

This work is supported in part by German BMBF under project 06HD197D and by WP8 of the hadron physics program of the $7^{th}$ EU program period.

# References

[1] F. Carminati et al, ALICE Collaboration, 2004, J.Phys. G: Nucl. Part. Phys. **30** 1517
[2] B. Alessandro et al, ALICE Collaboration, 2006, J.Phys. G: Nucl. Part. Phys. **32** 1295
[3] R. Arnaldi et al, Nucl. Instr. and Meth. A **564** (2006) 235
[4] The ALICE collaboration, K. Aamodt et al, The ALICE experiment at the CERN LHC, (2008) JINST_3_S08002
[5] S. Giani, Diffraction at TOTEM, these proceedings
[6] D. d'Enterria et al, Addendum CMS technical design report, J. Phys. G34:2307-2455, 2007
[7] F. Close, A. Kirk, G. Schuler, Phys.Lett. B **477** (2000) 13
[8] A. Donnachie, P. Landshoff, Phys.Lett. B**595** (2004) 393
[9] T. Åkesson et al, Nucl. Phys. B **264** (1986) 154
[10] L. Lukaszuk, B. Nicolescu, Lett. Nuovo Cim. **8** (1973) 406
[11] C. Ewerz, Proceedings XII Rencontres de Blois (2005) 377
[12] A. Schäfer, L. Mankiewicz, O. Nachtmann, Phys.Lett. B **272** (1991) 419
[13] A. Bzdak, L. Motyka, L. Szymanowski, J.R. Cudell, Phys.Rev. D **75** (2007) 094023
[14] L. Motyka, private communication
[15] S.J. Brodsky, J. Rathsman, C. Merino, Phys.Lett. B **461** (1999) 114
[16] P. Hägler, B. Pire, L. Szymanowski, O.V. Teryaev, Phys.Lett. B **535** (2002) 117
[17] I.F. Ginzburg, I.P. Ivanov, N.N. Nikolaev, Eur.Phys.J. C**5** (2003) 02
[18] V.P. Goncalves, M.V. Machado, Phys.Rev.D **71** (2005) 014025
[19] V.P. Goncalves, M.V. Machado, Phys.Rev.D **75** (2007) 031502

# Part VII

# Cosmic Ray Physics and Related Studies at LHC

# Cosmic Rays and Extensive Air Showers

*Todor Stanev*[1]

[1]Bartol Research Institute, Department of Physics and Astronomy, University of Delaware, Newark, DE 19716, U.S.A.

We begin with a brief introduction of the cosmic ray energy spectrum and its main features. At energies higher than $10^5$ GeV cosmic rays are detected by the showers they initiate in the atmosphere. We continues with a brief description of the energy spectrum and composition derived from air shower data.

## 1 Introduction

Cosmic rays are defined as charged nuclei of non-solar origin, i.e. particles accelerated in the Galaxy or, at higher energy, in extragalactic astrophysical objects. The energy spectrum of the cosmic rays is a smooth power law spectrum $F(E) = const \times E^{-\alpha}$ with only two identifiable features. The first one is the cosmic ray *knee* at about $3\times10^6$ GeV where the spectrum steepens from $\alpha = 2.7$ to 3.1 and the other one is the *ankle*, at about $3\times10^9$ GeV. where the spectrum becomes again flatter.

The common wisdom is that cosmic rays below the knee are accelerated at galactic supernova remnants. Gamma-rays of energy up to 10 TeV have been observed from sources in the vicinity of well known supernova remnants which is an indication that these are indeed sources of cosmic rays acceleration. The acceleration spectra of cosmic rays are considerably flatter (with smaller $\alpha$) than those of the cosmic rays at Earth. This is believed to to be a propagation effect in the Galaxy, where the lower energy cosmic rays are contained for longer time. At energies above the cosmic ray knee we have no idea about the cosmic ray sources, except that the highest energy particles are certainly of extragalactic origin. It is remarkable that some astrophysical objects can accelerate particles to three orders of magnitude higher than the LHC equivalent

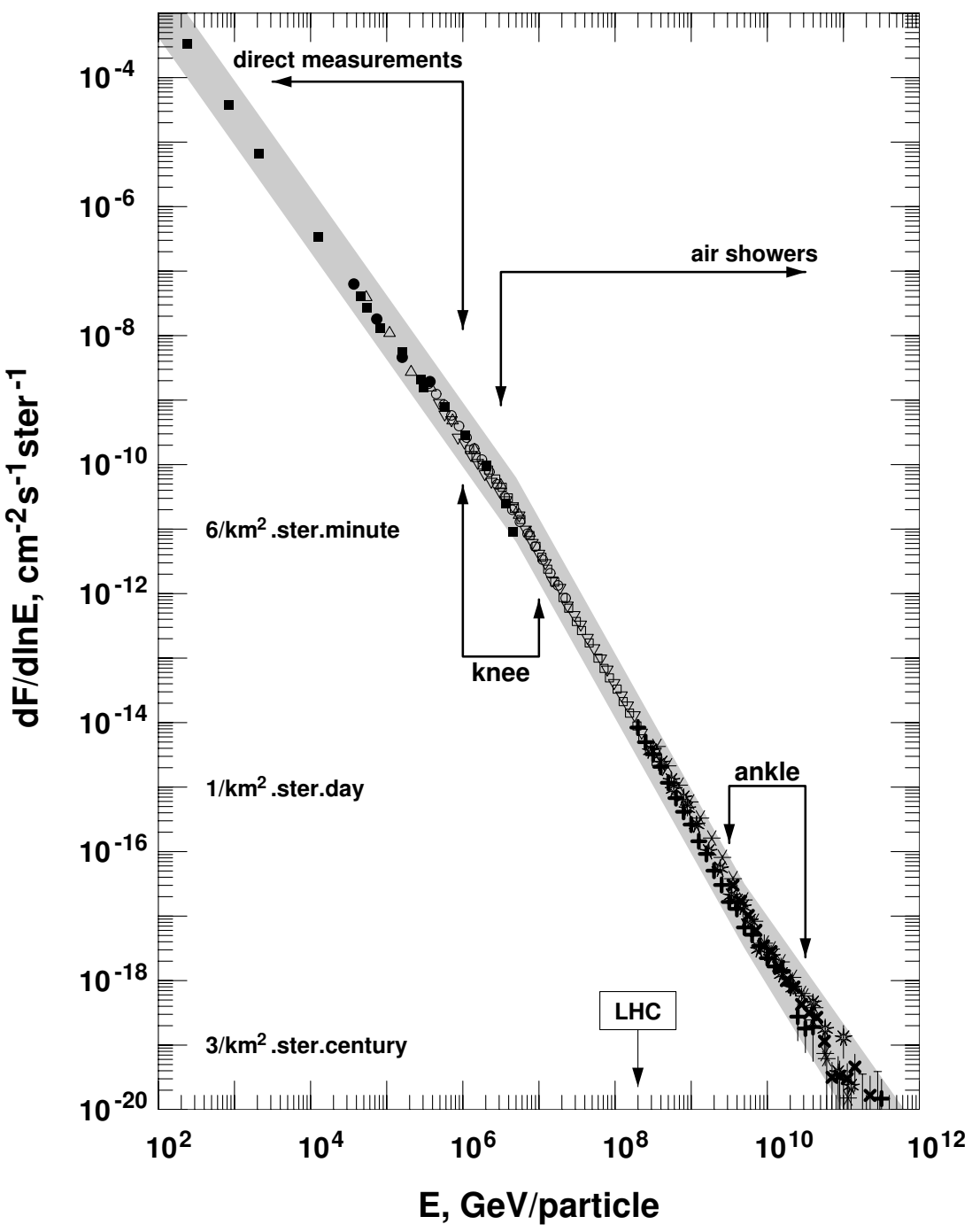

Figure 1: Cosmic rays energy spectrum above 100 GeV.

Lab energy.

The cosmic ray spectrum is shown in Fig. 1. The figure indicates the energy range where the cosmic ray spectrum is measured directly by balloon and satellite experiments. When the energy starts to exceed significantly 1,000 GeV the cosmic ray flux is too small and the cosmic rays are measured by the showers they generate in the atmosphere.

There are different types of air shower detectors:

- Air shower arrays consist of particle detectors that are spaced at different distances from each other depending on the energy range of the detectors. If the design is for detection of $10^6$ GeV air showers the distance between detectors is several tens of meters. In the Auger southern observatory, which aims at shower energy exceeding $10^9$ GeV the distance between detectors is 1,500 m. The shower arrays trigger when several detectors fire in coincidence. The reconstruction of the primary energy depends heavily of the hadronic interaction model that is used by the detector Monte Carlo simulation.

- Air Cherenkov detectors detect the Cherenkov light emitted by the shower charged particles (mainly electrons and positrons) in the atmosphere. Most of the light comes when the shower is at maximum.

- Fluorescent detectors detect the fluorescent light from the Nitrogen atoms in the atmosphere that are excited by the ionisation of the shower charged particles. Unlike the Cherenkov light the fluorescent light is isotropic. High energy showers can be observed from as far as 40 km away. Fluorescent detectors integrate over the shower longitudinal development to estimate the primary particle energy after adding the *invisible* energy, contained in high energy particles and neutrinos.

Different observational methods are now combined as in the case of the southern Auger observatory and the new Telescope Array detector.

# 2 Rough Estimates of the Shower Parameters

As mentioned earlier, shower Monte Carlo calculations are used for calculations of the efficiency of the detectors and estimations of its effective area. The main features of the air shower development can be understood on the basis of the *toy model* of the shower development created by Heitler [1]. Heitler assumed that the shower consists of one type of particles. At each interaction length $\lambda$ two new particles are created each one of them carrying 1/2 of the energy. This continues until the particle energy is less than the critical energy $E_c$ under which particles do not interact. The maximum number of particles in the cascade is then $N_{max} = E_0/E_c$. The depth of maximum is proportional to the logarithm of the ratio of the primary and the critical energies $E_0/E_c$: $X_{max} = \lambda \log_2(E_0/E_c)$ .

Hadronic cascades are much more complicated but one can still use Heitler's approach to derive approximate expressions for some shower parameters. Assuming that the air shower development depends only on the first cosmic ray interaction, one can estimate the depth of the shower maximum in the atmosphere as [2]

$$X_{max} \;=\; X_0 \,\ln\left[\frac{2(1-K_{el})E_0}{(\langle m\rangle/3)\varepsilon_0}\right] \;+\; \lambda_N(E_0)\ , \qquad (1)$$

and the number of electrons at $X_{max}$ as

$$N_e^{max} = \frac{1}{2}\frac{\langle m\rangle}{3}\frac{(1-K_{el})E_0}{\varepsilon_0}\ , \qquad (2)$$

where $m$ is the effective meson multiplicity and the 1/3 factor accounts for the multiplicity of neutral mesons. $K_{el}$ is the elasticity coefficient of the first interaction (roughly 1/2) and $\epsilon_0$ is the critical energy of the electrons in air (81 MeV). Replacing the primary energy $E_0$ with $E_0$/A (the mass of a nucleus) one can derive the expressions for showers initiated by primary nuclei. The conclusions are that $X_{max}$ in such showers is smaller (showers develop higher in the atmosphere): $X^A_{max} = X^p_{max} - X_0 \ln A$ and the muon/electron ratio in showers initiated by nuclei is higher by $A^{1-\beta}$ ($\beta = 0.85$) than in proton showers. These two parameters are most often used in studies of the cosmic ray chemical composition.

After this short introduction it is important to remember that cosmic ray shower experiments are observations, rather than experiments in the accelerator experiment sense. We have no idea of the energy and type of the primary particle or of the first interaction point in the atmosphere. We have to measure as many shower parameters as possible, compare them to Monte Carlo calculations, and derive the energy and composition of the primary particles. This not easy because of the large inherent fluctuations in the shower development. Figure 2 shows the shower longitudinal profiles of ten simulated proton showers of primary energy $10^5$ GeV and their average.

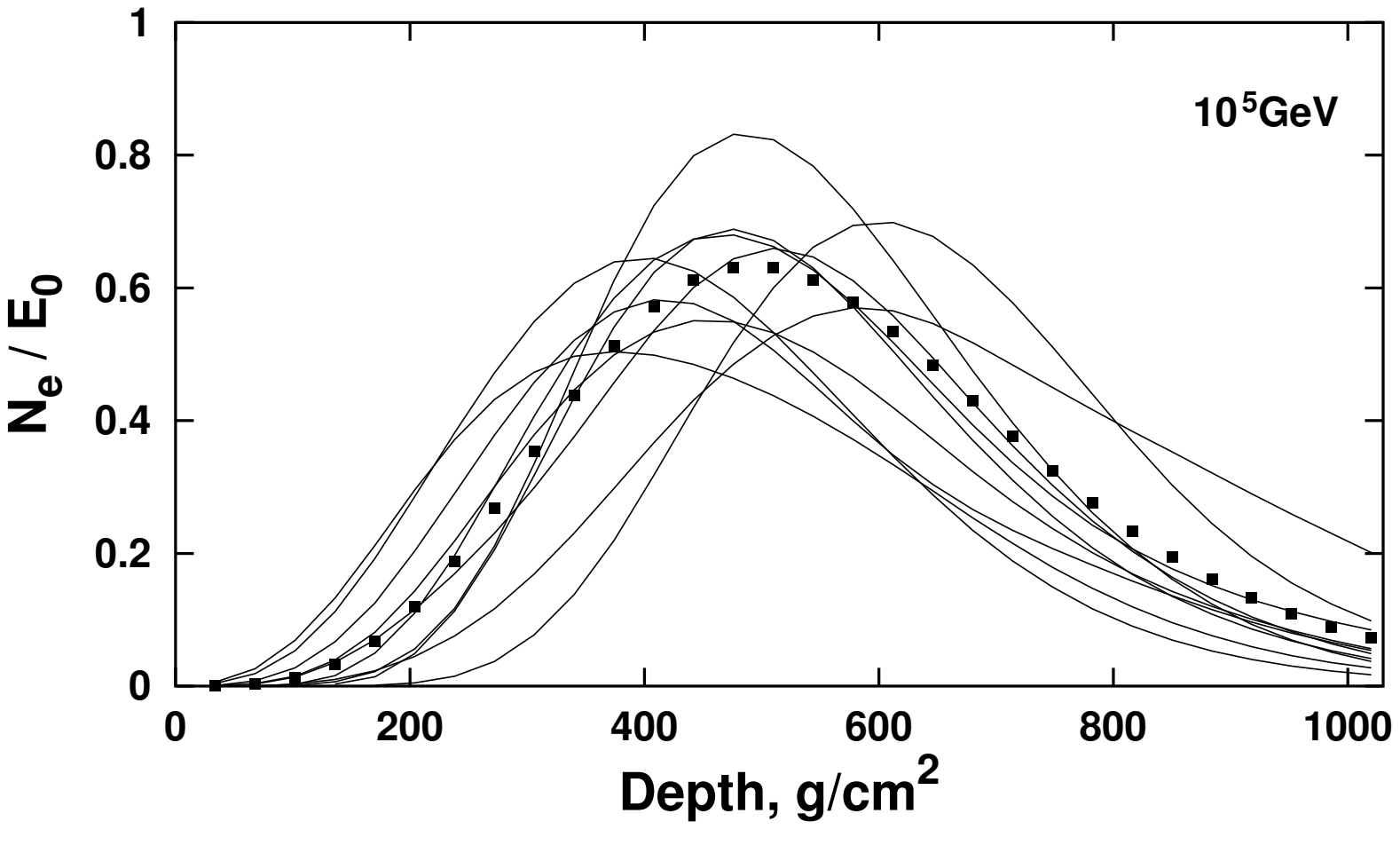

Figure 2: Shower profiles of ten simulated proton showers. The average shower profiles is shown with the points.

For this reason the reconstruction of individual showers is quite uncertain and we have to work with large statistical samples in the investigation of the cosmic ray energy spectrum and composition.

# 3 Air Shower Reconstruction and Analysis

The reconstruction of the air shower parameters starts with the determination of the position of the air shower core and the arrival direction of the air showers. Both this parameters require the knowledge of the lateral distribution of the air shower particles and the curvature of the air shower front.

The lateral distribution is strongly influenced by the type of the air shower detectors. Thin scintillator counters, for example, practically do not detect the gamma-rays in the air shower and count the shower electrons and muons the same way. Thick detectors, like the water Cherenkov tanks of Auger [3] and the frozen tanks of IceTop [4] convert the majority of the gamma-rays in electron-positron pairs. Muons generate much higher signals in such detectors.

Although the shower moves through the atmosphere with the speed of light its shape is not a plane. Particles deflected at large angles during the shower development have higher path lengths and arrive at the observation level slightly later and with more fluctuations. For these reasons the shower front is relatively thin in time close to the shower core (several nanoseconds) and considerably thicker at large distances from the core.

When all these effects are accounted for in an air shower array one uses them in a iterative procedure to calculate the exact position of the shower core from the lateral distribution of the shower particles and its arrival direction from the timing in the different detectors. The error in the shower core position is several meters and the direction has an error of about $1^o$.

The next step is the determination of the shower energy and primary particle mass. The shower energy is strongly correlated with the shower particles density (or signal strength) at certain distance from the core. The distance used is estimated for each detector as the position where the signal strength of showers from different nuclei fluctuates the least. It strongly depends on the distance between detectors and is usually slightly smaller than this distance.

The composition analysis is more complicated since it needs either the detection of the shower muons with underground (or well shielded) detectors or observation of the shower maximum with optical detectors. The showers with the smallest $e/\mu$ ratio (or deeper $X_{max}$ are from primary photons and the showers on the opposite side of the average are from heavy nuclei. The actual analysis procedure is different for each group but these basic principles are always used.

# 4 Experimental Data on the Cosmic Ray Energy Spectrum

Figure 1, which contains seventeen orders of magnitude in flux, gives the reader the impression that all measurements of the cosmic ray energy spectrum agree with each other. The truth is different, the air shower measurements (as well as the direct ones) do not agree very well with each other. Figure 3 demonstrates the current situation.

The cosmic rays spectrum shown in Fig. 3 is multiplied by $E^{2.7}$ to emphasise the features in the energy spectrum, which are otherwise almost invisible. Such a presentation, however, significantly amplifies the differences between different experiments. A difference of 25% (typical systematic uncertainty) in the energy estimate of two experiments leads to a visible difference of 1.5 in the presented spectrum plus a shift in the energy scale. The biggest differences in Fig. 3 are in two energy ranges: just before the cosmic ray knee, and at energies higher than $10^9$ GeV. In the lower energy range most of the errors should be systematic as for many experiments this is either the beginning or the end of their sensitivity range. For these reasons the exact position of the cosmic ray knee is not determined.

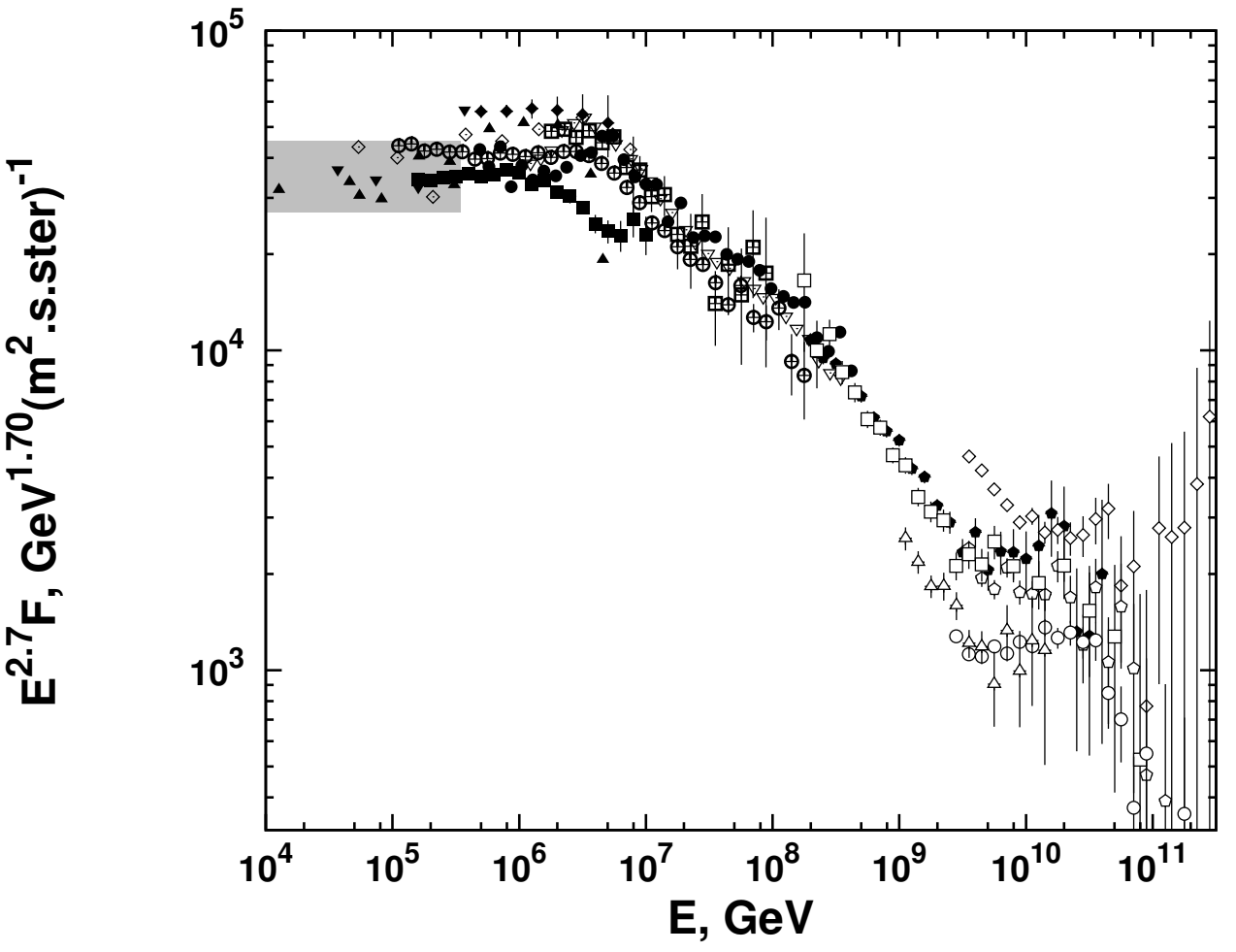

Figure 3: Cosmic ray energy spectrum measured by different air shower experiments [6]. The only set of direct measurements is shown with filled squares.

A highly respected analysis of the cosmic ray knee region comes from the Kascade group [5]. The shower array Kascade consists of 252 electron and muon detectors covering an area of 200×200 m$^2$ with a 320 m$^2$ calorimeter in the middle. The primary energy estimate is a combination of weighted electron and muon numbers in the shower. The basis of the analysis is a two dimensional $N_e$ and $N_\mu$ distribution of all detected showers above the threshold $N_e$ and $N_\mu$ values. Since showers generated by different primary nuclei this distribution was used to generate the energy spectra of the different components, and thus the energy dependence of the cosmic ray chemical composition. The collaboration paid much attention to the results using two different hadronic interaction models for the analysis.

If the interaction model QGSjet01 [7] were used the knee of the all nuclei cosmic ray spectrum is at $4\times10^6$ GeV and the two spectral indices before and after the knee are 2.7 and 3.1. In the case of SIBYLL 2.1 [8] the knee is at $5.7\times10^6$ GeV and the spectral indices are the same. The energy spectra of the individual five components (H, He, C, Si, Fe) show a domination of the heavy nuclei at energy above $2\times10^7$ GeV while before the knee the spectra of He and H have the highest fluxes. It was not possible to derive exactly the *knees* of the individual components. The calculations with the Sibyll interaction model show a slightly heavier nuclear composition.

At much higher energy (above $10^{10}$ GeV) there are very different measurements of the cosmic ray flux. The highest values were obtained by the AGASA experiment [9], which published eleven events with energy exceeding $10^{11}$ GeV. More recently these events were re-analyzed with the standard hadronic interaction models and the energy estimate was decreased. The highest statistics measurements are those of the High Resolution Fly's Eye (HiRes) and especially the Auger Southern Observatory.

Both HiRes and Auger observed the GZK [10] cutoff of the cosmic ray spectrum [11, 12] that is caused by the interactions of the ultrahigh energy cosmic rays (UHECR) with the mi-crowave background (MBR). The UHECR energy is so high that photoproduction interactions are possible in the MBR for cosmic rays of energy above $3\times10^{10}$ GeV. The energy loss length of $10^{11}$ GeV protons is of order 100 Mpc ($3\times10^{24}$ m) and decreases to about 15 Mpc at higher energy. Heavy nuclei lose energy in photodisintegration, where they lose one or two nucleons per interaction. Since the required CMS energy is much smaller the interaction starts at much smaller energy per nucleon, $E^0/A$. UHE gamma-rays lose energy even faster than protons. These distances define the GZK horizon within which we look for the sources of UHECR.

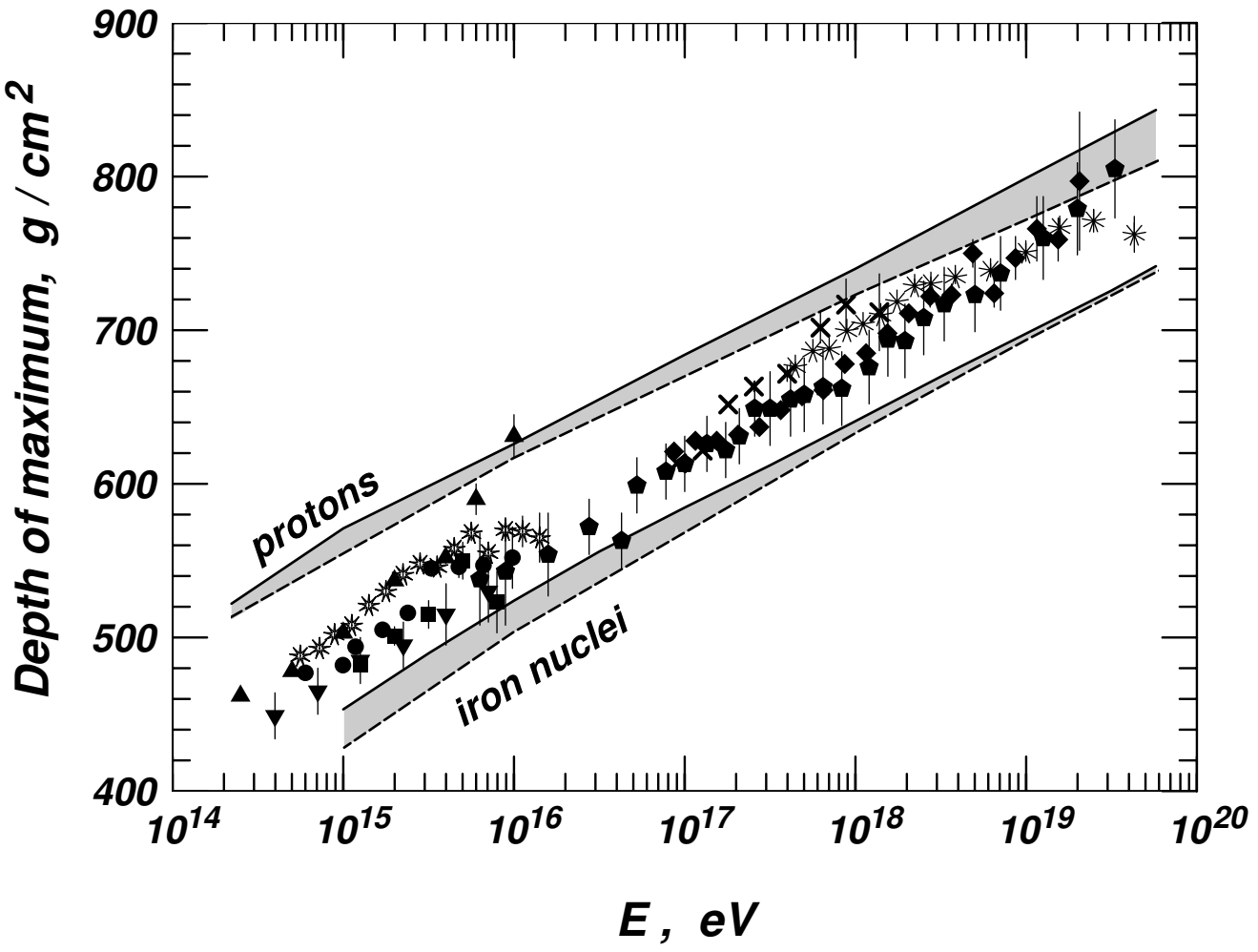

Figure 4: Depth of maximum measurements as a function of the shower energy. The shaded areas are the predictions of different models for protons and iron nuclei.

The chemical composition of UHECR is not well established yet. At these energies the composition is measured by the depth of shower maximum ($X_{max}$). HiRes measures $X_{max}$ consistent with pure proton composition of UHECR, while Auger sees $X_{max}$ distributions indicating medium heavy nuclei such as Oxygen or Carbon as shown in Fig. 4. Our previous expectations were that extragalactic cosmic rays would mostly consist of protons with a 10% admixture of He nuclei in accordance with the average composition of extragalactic matter.

If the cosmic ray composition measured by Auger is confirmed we will have to look for the sources of the ultrahigh energy cosmic rays among the astrophysical objects where there are enough heavy nuclei to become accelerated. On the other hand heavy nuclei have a maximum acceleration energy higher by the charge $Z$ from that of protons.

# Acknowledgements

This short review comes from work performed in collaboration with P.L. Biermann, R. Engel, T.K. Gaisser, A. Reimer, D. Seckel and others. Partial support for my work comes from the US Department of Energy contract DE-FG02-91ER40690.

# References

[1] W. Heitler, Quantum Theory of Radiation (Oxford University Press, 1944)

[2] J. Matthews, Astropart. Phys., **22**, 387 (2005)

[3] See *http://auger.org*

[4] X. Bai for the IceCube collaboration, Nucl. Phys. B (Proc. Suppl) **175**, 415 (2008)

[5] T. Antoni et al, Astropart. Phys. 24:1 (2005)

[6] T.K. Gaisser and T. Stanev, Cosmic Rays, in Review of Particle Physics. Particle Data Group (C. Amsler et al.), Phys.Lett.B**667**:1 (2008)

[7] N.N. Kalmykov, S.S. Ostapchenko & A.I. Pavlov, Nucl. Phys. B (Proc. Suppl.), **52B**, 17 (1997)

[8] R. Engel et al., Proc. 26th Int. Cosmic Ray Conf. (Salt Lake City), **1**, 415 (1999)

[9] M. Takeda et al., Phys. Rev. Lett., **81**, 1163

[10] K. Greisen, Phys. Rev. Lett., **16**, 748 (1966); G.T. Zatsepin & V.A. Kuzmin, JETP Lett. **4** 78 (1966)

[11] R. Abbasi et al. (HiRes Collaboration), Phys. Rev. Lett. **100**:101101 (2008)

[12] J. Abraham et al. (Auger Collaboration), Phys. Rev. Lett. **101**:061101 (2008)

# Current Concepts in Theory and Modelling of High Energy Hadronic Interactions

*K. Werner*[1], *T. Pierog*[2]
[1]SUBATECH, Univ. of Nantes – IN2P3/CNRS– EMN, Nantes, France
[2]Forschungszentrum Karlsruhe, Inst. f. Kernphysik, Karlsruhe, Germany

We discuss some key observations of cosmic ray experiments. It will be shown that prediction from air shower simulation using different hadronic interaction models differ by large amounts. We try to understand this by investigating the theoretical concepts behind these models.

The interpretation of the results of air shower experiments depends heavily on simulations. Whereas the electromagnetic part of an air shower (the so-called electromagnetic cascade) is well under control, the hadronic part is not accessible from first principles, and is therefore treated via phenomenological hadronic interaction models.

We will focus in this article on two "key" observables of air showers: the number of muons and the number of electrons at the observation level (defined by the geographical location of the detector array of the corresponding experiment).

It has been a longstanding problem in air shower physics that the number of muons obtained from air shower simulations has always been too low compared to the measurements. Any attempt to modify the hadronic interaction models in order to get more muons created other problems. In 2006, non of the existent models (QGSJET [1], SIBYLL [2]) could consistently describe all cosmic ray air shower data.

Starting to use EPOS [3, 4, 5] as interaction model, it was found that one gets significantly more muons, without changing observables like $X_{\max}$ too much, see [6]. As an example, we show in Fig. 1 the muon density at a fixed distance from the core, as measured by the MIA collaboration [7], compared to simulations based on QGSJET and EPOS. Significantly more muons are produced in the EPOS simulations. Similar results have been obtained more recently from the AUGER collaboration [8].

Why are there more muons produced in EPOS ? Because EPOS produces more baryons! In Fig. 2, we plot the antiproton over pion ratio in p+Air collisions for EPOS, QGSJET, and SIBYLL, as a function of the energy. Knowing that the pion rate in the three models is similar, we can see that the antiproton production increases much more in EPOS compared to the other models. This fact is also observed for other baryons.

The particular role of the baryons concerning muon production is easily understood. The main property of the baryon, in this context, is the fact that it is not a $\pi^0$. The latter particle decays immediately into two photons, its energy is given to the corresponding electromagnetic cascade, no muons can be produced in the following. On the contrary, a baryon can still interact, producing charged pions, which then decay into muons. Also, baryons have a softer pion spectrum than pions in the next generation, leading to less energy lost in the electromagnetic channel in case of $\pi^0$ production in the next collision with air. So although baryons are not the most abundant particles in the cascade, their role is very important concerning the muon rate.

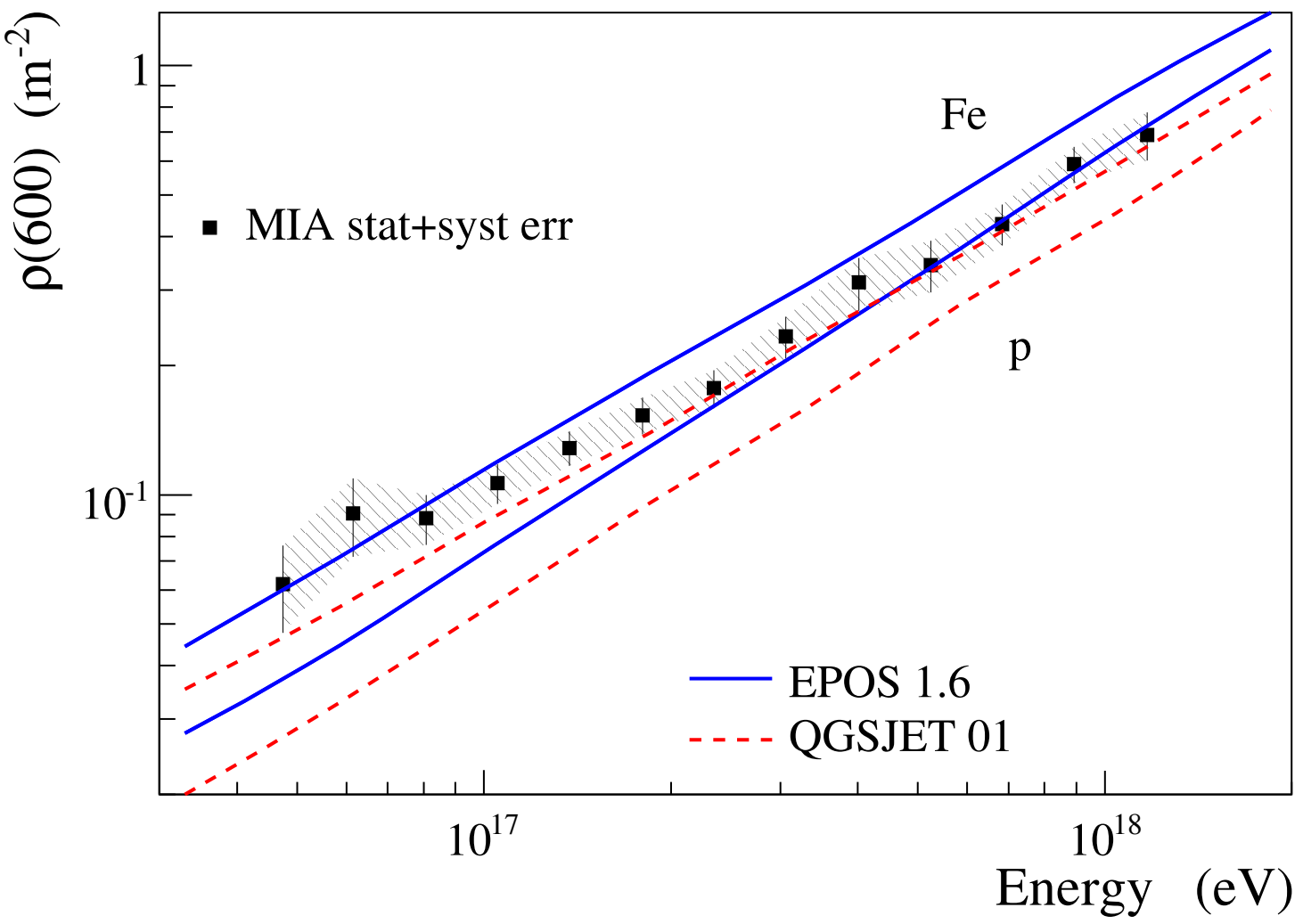

Figure 1: The muon density at a fixed distance from the core, as measured by the MIA collaboration [7], compared to simulations based on QGSJET and EPOS.

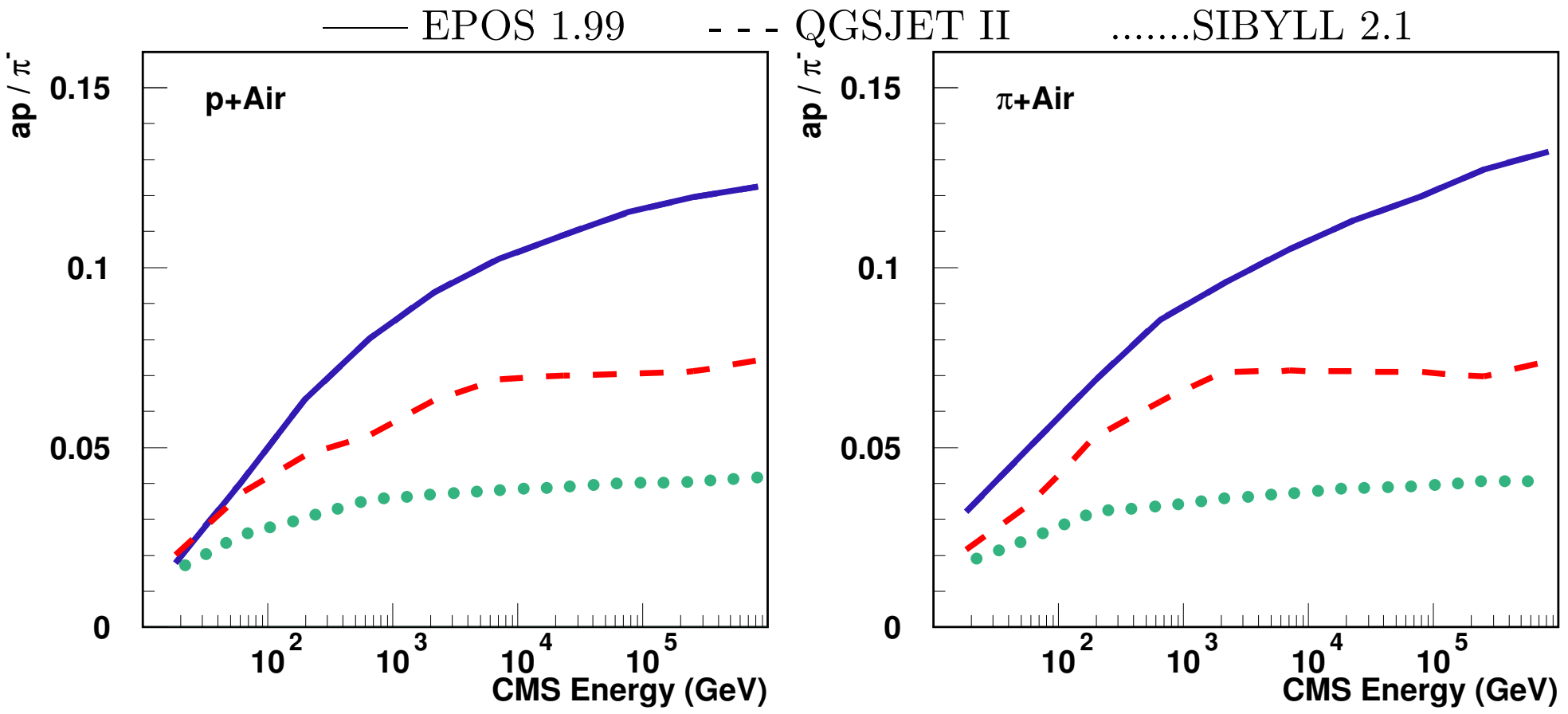

Figure 2: The antiproton over pion ratio in p+Air collisions for EPOS, QGSJET, and SIBYLL, as a function of the energy.

EPOS has been designed (and optimised) to understand ALL types of hadrons by carefully studying baryon production in accelerator experiments, without thinking about CR applications. In Fig. 3(upper panel), we plot the yields of different kinds of baryons in proton-proton collisions at 158 GeV, from EPOS calculations, compared to data from SPS/NA49 [9]. An enor-mous amount of pp ($p\bar{p}$) data has been considered, at SPS, ISR, RHIC, TEVATRON, also $\pi$p, pA and $\pi A$ collisions. As another example, we show in Fig. 3 (lower panel) the antiproton over pion ratio in pp scattering at 1800 GeV (left), and proton production in pion carbon scattering at 100 GeV (right).

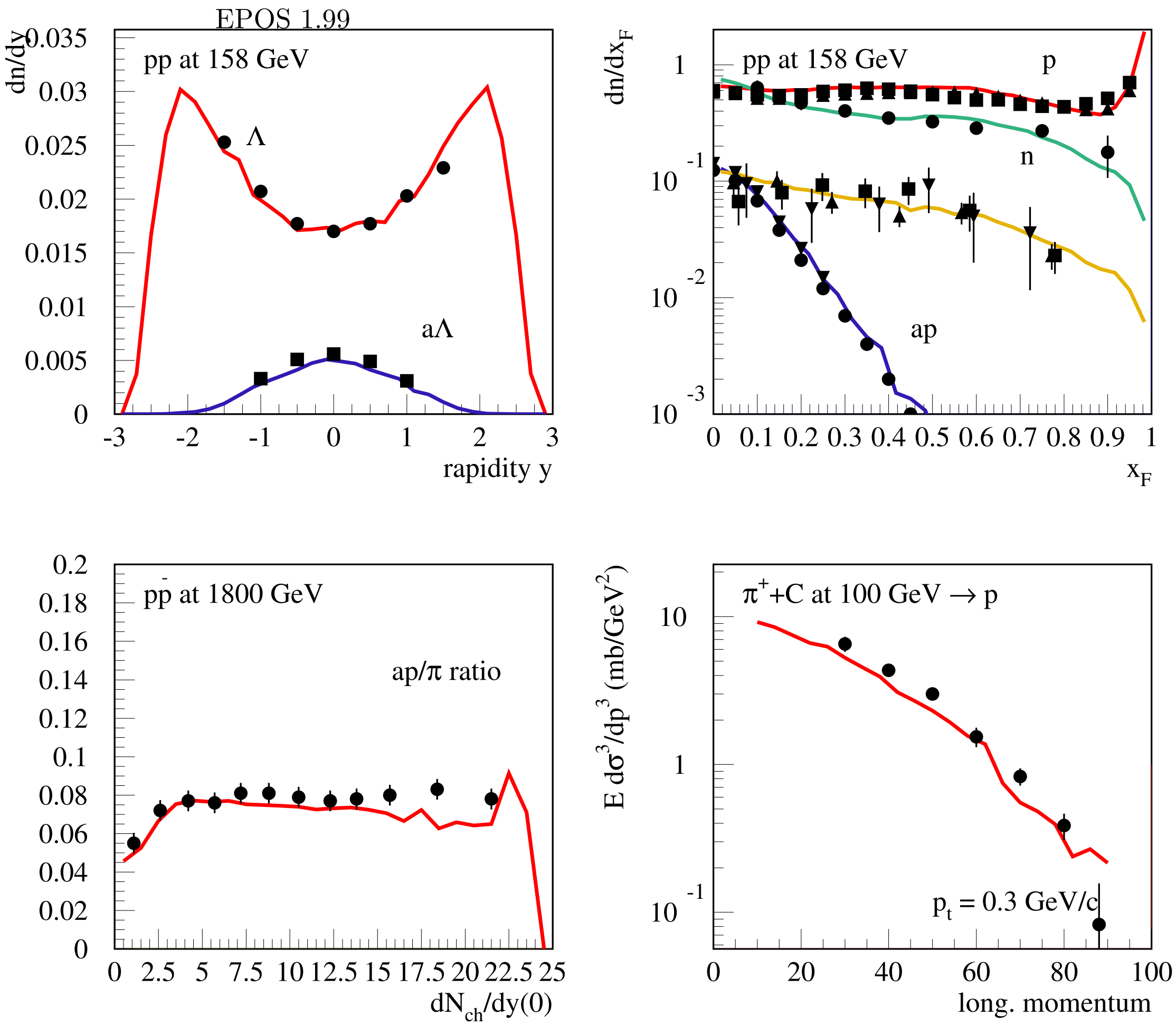

Figure 3: Upper panel: The yields of different kinds of baryons in proton-proton collisions at 158 GeV, compared to data from SPS/NA49 [9]. Lower panel: Antiproton over pion ratio as a function of the multiplicity density, in pp scattering at 1800 GeV (left), and proton production in pion carbon scattering at 100 GeV (right). We compare EPOS calculations with data [10, 11].

If we compare EPOS to QGSJET and SIBYLL, we find similar results concerning pions, but big differences concerning baryons, see Fig. 4, where we show pion production (left) and proton production (right) in pion carbon scattering at 100 GeV. We compare calculations from different models with data [11]. Clearly visible the large difference between EPOS and the other models, in case of protons. Whereas EPOS is close to the data, the other models are lower by as much as a factor of 2-3.

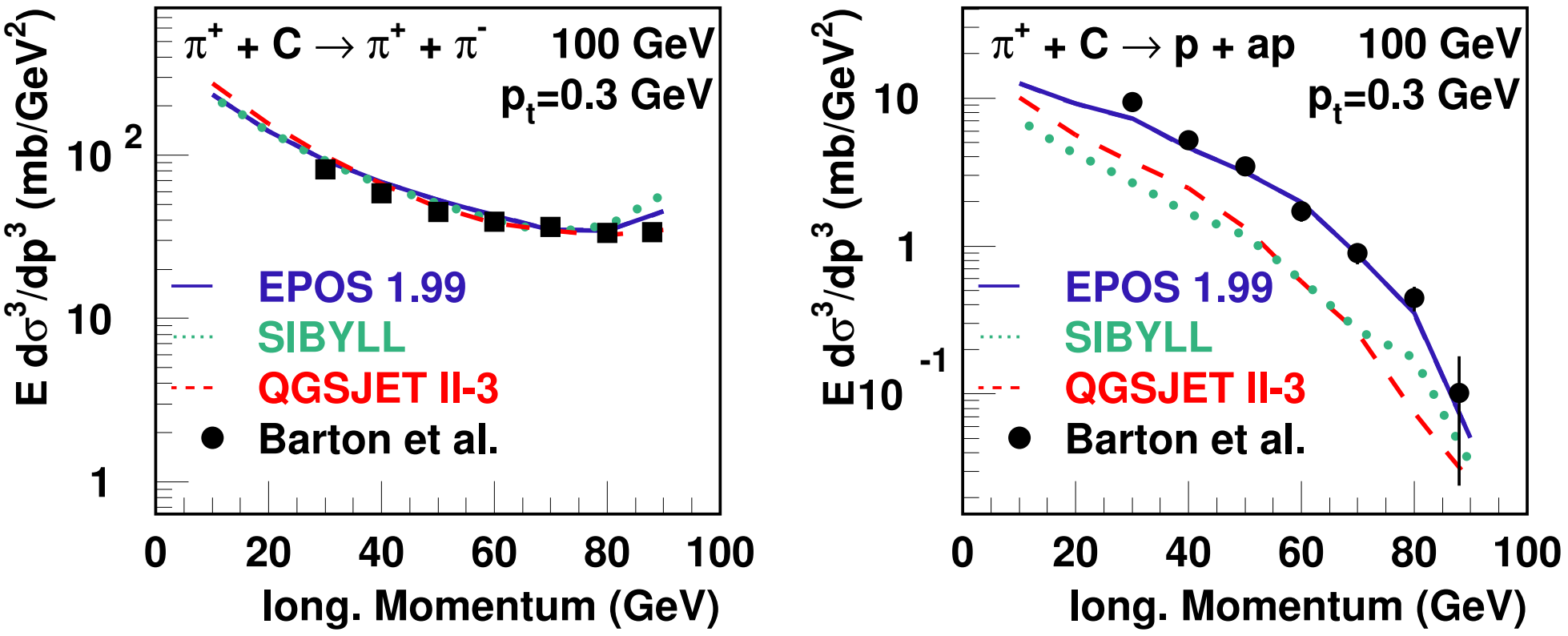

Figure 4: Pion production (left) and proton production (right) in pion carbon scattering at 100 GeV.

Having increased the muon number without affecting too much the electrons leads, however, to some contradictions. The problems comes from KASCADE data [12], where the number of muons is correlated with the number of electrons. Here, QGSJET and SIBYLL seem to work, so increasing the muons and not the electrons will give a wrong electron-muon correlation. The solution is related to a completely different subject: non-linear effects (already considered for particle production) should also be taken into account for cross section calculations (which has not been done in earlier EPOS versions). Introducing non-linear effects as discussed in [1] also for cross section calculations, we obtain the results as shown in Fig. 5. Both cross sections from EPOS calculations are below the results from the other models. There is also a trend in the data towards lower values, in more recent measurement compared to older data. Using our new results (with lower cross sections compared to other models), we get more electrons at ground, since the shower gets deeper into the atmosphere. We seem to be in agreement with the KASCADE muon-electron correlations, but having both more electrons and more muons compared to the other models. Studies are under way to make precise comparisons between the new EPOS and KASCADE.

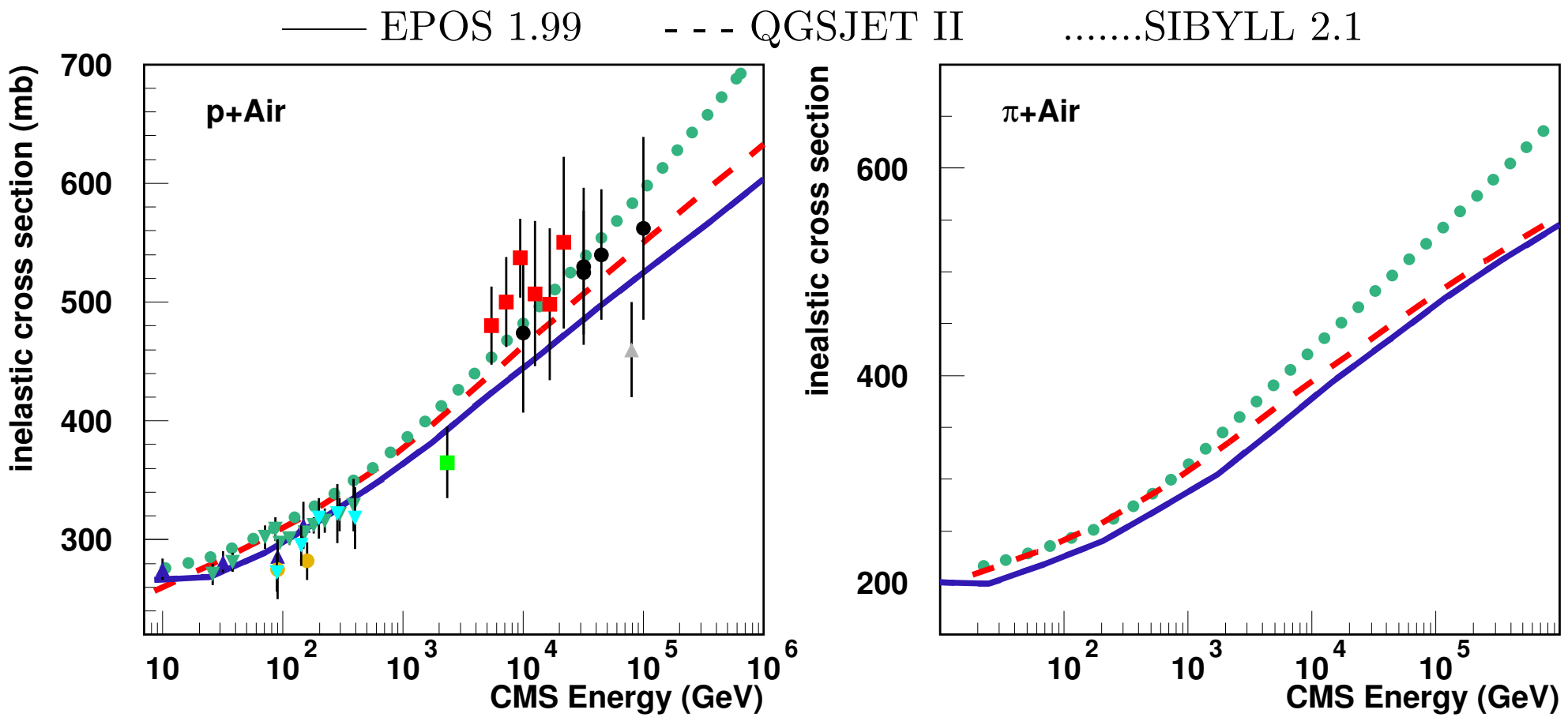

Figure 5: The inelastic cross section in p-Air and $\pi$-Air collisions, for different models.

Having a smaller inelastic cross section compared to earlier calculations (and other models) has also an impact on Xmax: it will be bigger, see Fig. 6.

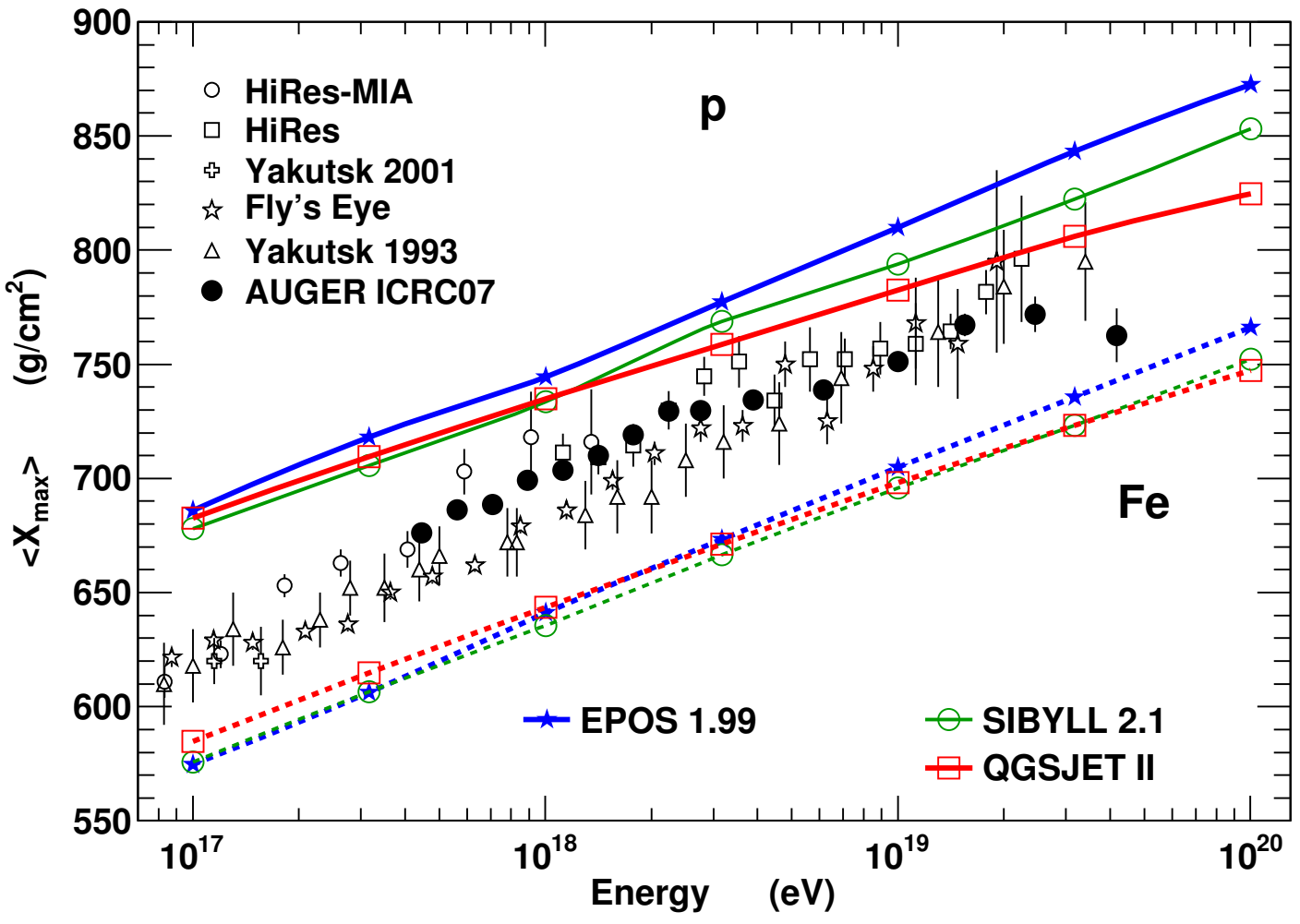

Figure 6: Xmax from different models compared to data.

In the following, we will discuss very briefly the physics of the interaction models.

EPOS and QGSJET are multiple scattering model in the spirit of the Gribov-Regge approach, see Fig. 7 (left). Here, one does not mean simply multiple hard scatterings, the elementary processes corresponds to complete parton ladders, which means hard scatterings plus initial state radiation. In this case, this elementary process carries an important fraction of the available energy. This is why in EPOS one treats very carefully the question of energy sharing in the multiple scattering process. Particle production comes from remnants and string decay. In SIBYLL, one distinguishes between a primary interaction leading to two $q - qq$ string, and subsequent scatterings of the type $g + g$, leading to $\bar{q} - q$ strings after splitting of the gluons, see Fig. 7 (right).

All there models treat in some way so-called non-linear effects due to high parton densities. In EPOS one first parameterises the numerically obtained results for an elementary interaction (more precisely: the imaginary part of the corresponding amplitude in $b-$space) as $\alpha(x^+)^\beta(x^-)^\beta$, which is then changed into $\alpha(x^+)^{\beta+\epsilon_P}(x^-)^{\beta+\epsilon_T}$ , where $\epsilon_P$, $\epsilon_T$ mimic the effect of rescattering of ladder partons (or Pomeron-Pomeron interactions), see Fig. 8 (left), and the corresponding screening effects. Here, $x^+, x^-$ are the light cone momentum fractions of the first ladder partons. The exponents $\epsilon_P$, $\epsilon_T$ depend on $\log s$ and the number of participating nucleons in case of pA or AA scattering. So high density effects are treated in an effective fashion, but energy is perfectly conserved (the only model which does so). In QGSJET, Pomeron-Pomeron interactions are taken into account to all orders, see Fig. 8 (right), but in this case energy conservation for multi-Pomeron diagrams is no longer imposed. In SIBYLL, an energy dependent saturation scale is introduced, which serves as a $p_t$ cutoff. Energy conservation is not imposed either.

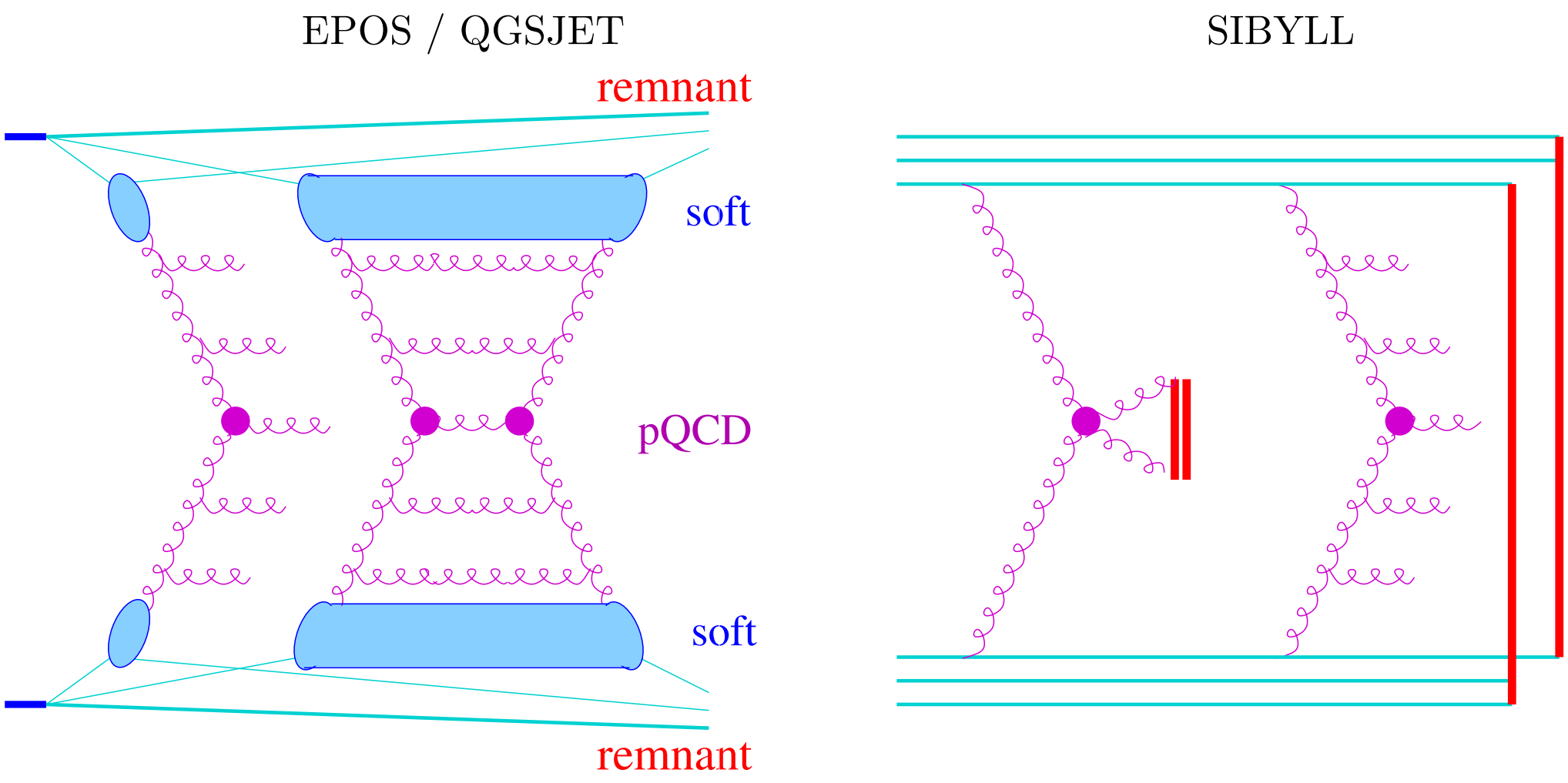

Figure 7: Multiple scattering diagram in EPOS, QGSJET (left) and SIBYLL (right).

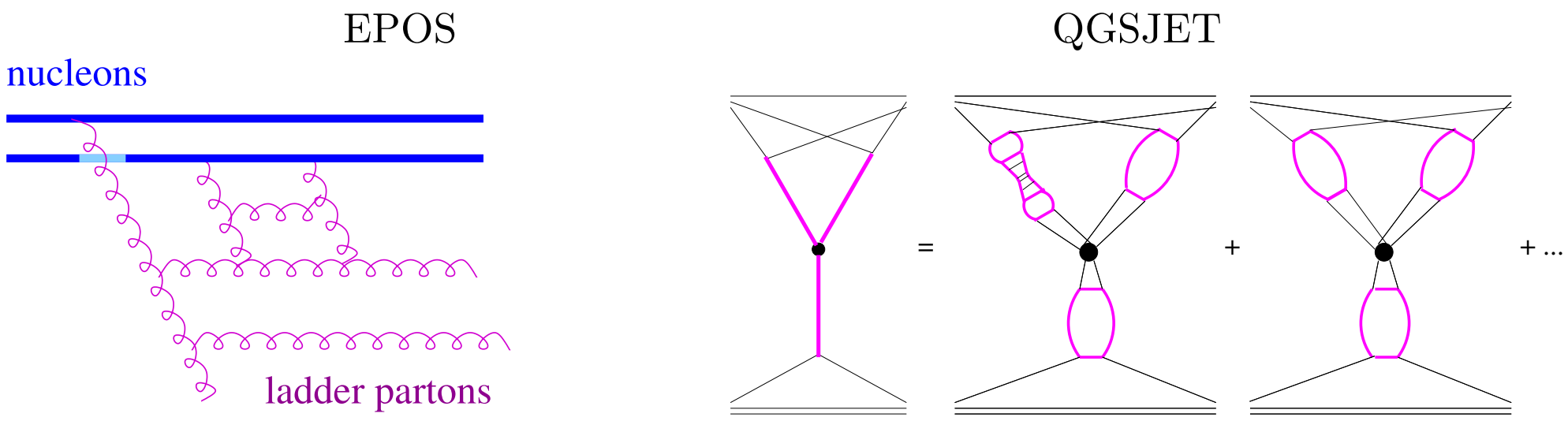

Figure 8: Non-linear effects: effective treatment of rescattering of ladder partons (Pomeron-Pomeron interactions) in EPOS (left); explicit treatment of triple Pomeron graphs (and higher orders) in QGSJET (right).

Finally, based on the experience with heavy ion collisions, EPOS treats high density proton-proton events collectively, via a three-dimensional hydrodynamical evolution of a quark-gluon plasma / hadron gas, with subsequent freeze out.

To summarise: air shower simulations with EPOS provide more muons, due to more baryon production, compared to QGSJET and SIBYLL, which is due to more baryon production in the former model. Despite the large differences in their predictions, the basic theoretical concepts of the three models are similar (multiple scattering of Gribov-Regge type, strings, non-linear effects). But the practical implementation is quite different.

# References

[1] N.N. Kalmykov, S.S. Ostapchenko, A.I. Pavlov, Nucl. Phys. B (Proc. Suppl.) 52, 17 (1997); S. Ostapchenko, Phys. Lett. B 636, 40 (2006); S. Ostapchenko, Phys. Rev. D74, 014026 (2006)

[2] R. Engel et al., Proc. 26th ICRC, Salt Lake City, 415 (1999)

[3] H. J. Drescher, M. Hladik, S. Ostapchenko, T. Pierog and K. Werner, Phys. Rept. 350, 93, 2001

[4] Klaus Werner, Fu-Ming Liu, and Tanguy Pierog. Phys. Rev. C 74, 044902 (2006), arXiv: hep-ph/0506232

[5] K. Werner, Phys. Rev. Lett. 98, 152301 (2007)

[6] T. Pierog, Klaus Werner, Phys. Rev. Lett. 101, 171101 (2008). arXive: astro-ph/0611311

[7] T. Abu-Zayyadet al., Astrophys. J. 557, 686 (2001); T. Abu-Zayyad et al., Phys. Rev. Lett. 84, 4276 (2000)

[8] A. Astellina [PAO Collaboration], Proceeding of the 31st ICRC, Lodz, Poland (2009)

[9] T. Anticic et al. [NA49 Collaboration], arXiv:0904.2708 [hep-ex]; T. Anticic et al., Fizika B10, 269 (2001)

[10] D. E. Acosta et al. [CDF Collaboration], Phys. Rev. D65 (2002) 072005

[11] D. S. Barton et al., Phys. Rev. D 27, 2580 (1983)

[12] K. Apel [KASCADE-Grande Collaboration], J. Phys. G36, 035201 (2009)

# Test of Hadronic Interaction Models with Air Shower Data

*Jörg R. Hörandel*

Radboud University Nijmegen, P.O. Box 9010, 6500 GL Nijmegen, The Netherlands; http://particle.astro.kun.nl

The description of high-energy hadronic interactions plays an important role in the (astrophysical) interpretation of air shower data. The parameter space important for the development of air showers (energy and kinematic range) extends well beyond today's accelerator capabilities. Therefore, accurate measurements of air showers are used to constrain modern models to describe high-energy hadronic interactions. The results obtained are complementary to information gained at accelerators and add to our understanding of high-energy hadronic interactions.

## 1 Introduction

The understanding and modelling of extensive air showers (particle cascades in the atmosphere) brings together the particle physics and astroparticle physics communities. To strengthen the connections and the scientific exchange between those communities is very fruitful for both sides and yields complementary information on the understanding of high-energy hadronic interactions.

When high-energy cosmic rays impinge onto the atmosphere they initiate cascades of secondary particles – the extensive air showers. Observations of air showers are used to indirectly infer the properties of cosmic rays at energies exceeding $10^{14}$ eV. The interpretation of air shower data faces a twofold challenge: the (exact) mass composition of cosmic rays is not known at those energies and, additionally, the properties of high-energy interactions taking place in air showers are partly unknown. Direct measurements of cosmic rays (fully ionised atomic nuclei) at energies below $10^{14}$ eV indicate that they are mostly composed of elements from hydrogen (protons) up to iron [1, 2]. The abundance of heavier elements is significantly smaller. Hence, in the following, we assume that cosmic rays comprise elements from hydrogen to iron.

We will focus on results from the KASCADE experiment [3], one of the most advanced air shower detectors in the energy range around $10^{15}$ eV. It has a unique set-up which allows to measure simultaneously the electromagnetic, muonic, and hadronic shower components. This is in particular valuable to test the consistency of hadronic interaction models. For about a decade [4, 5] systematic checks of interaction models have been performed with air shower data, and the most stringent constraints on interaction models, derived from air shower data, have been obtained with KASCADE measurements.

KASCADE consists of several detector systems. A $200 \times 200$ m$^2$ array of 252 detector stations, equipped with scintillation counters, measures the electromagnetic and, below a lead/iron shielding, the muonic parts of air showers. An iron sampling calorimeter of $16 \times 20$ m$^2$ area detects hadronic particles [6]. It has been calibrated with a test beam at the SPS at CERN up to 350 GeV particle energy [7]. For a detailed description of the reconstruction algorithms see Ref. [8].

## 2 Quantitative Tests

The principal idea of the consistency tests of hadronic interaction models is to simulate air showers initiated by protons and iron nuclei as the two extremes of possible primary particles. The shower simulations were performed using CORSIKA [9], applying different embedded hadronic interaction models. In order to determine the signals in the individual detectors, all secondary particles at ground level are passed through a detector simulation program using the GEANT package [10]. [1] The predicted observables at ground level, such as e.g. the number of electrons, muons, and hadrons or the energy of the hadrons are then compared to the measurements. If the measured values are inside the predicted interval for proton and iron induced showers, the particular interaction model used for the simulations is compatible with the data. On the other hand, if the measured values are outside the proton-iron interval, there is a clear hint for an incompatibility between the model under investigation and the measurements.

Hadronic interactions at low energies ($E_h < 80$ and 200 GeV, respectively) were modelled using the **GHEISHA** [14] and **FLUKA** [15, 16] codes. Both models are found to describe the data equally well [11]. High-energy interactions were treated with different models as discussed below, several models have been systematically tested over the last decade.

First quantitative tests [4, 17, 5] established **QGSJET** 98 [18] as the most compatible code. Similar conclusions have been drawn for the successor code QGSJET 01 [11]. The next version of the code, QGSJET-II-2, has been investigated recently [19]. The analyses yield inconsistencies, in particular for hadron-electron correlations. An example is shown in Fig. 1. For a given interval of the number of electrons the frequency of the maximum hadron energy registered in each shower is plotted. It can be recognised that for small hadron energies, the measured values are outside the range (proton-iron) predicted by QGSJET-II-2. Studies of the latest version, QGSJET-II-3, are in progress.

Predictions of **SIBYLL** 1.6 [20] were not compatible with air shower data, in particular there were strong inconsistencies for hadron-muon correlations [4]. These findings stimulated the development of SIBYLL 2.1 [21]. This model proved to be very successful, the predictions of this code are fully compatible with KASCADE air shower data [22, 23, 11].

Analyses of the predictions of the **DPMJET** model yield significant problems in particular for hadron-muon correlations for the version DPMJET 2.5 [24], while the newer version DPMJET 2.55 is found to be compatible with air shower data [11].

Investigations of the **VENUS** [25] model revealed some inconsistencies in hadron-electron correlations [5]. The predictions of **NEXUS** 2 [26] were found to be incompatible with the KASCADE data, in particular, when hadron-electron correlations have been investigated [11].

Recently, predictions of the interaction model **EPOS** 1.61 [27, 28, 29] have been compared to KASCADE air shower data [12]. This model is a recent development, historically emerging from the VENUS and NEXUS codes. The analysis indicates that EPOS 1.61 delivers not enough hadronic energy to the observation level and the energy per hadron seems to be too small. This is illustrated in Fig. 2: the predicted hadronic energy sum, relative to the measured values is plotted as function of the number of reconstructed hadrons. In this representation the measured values are at the zero line. Shown are results for two interaction models and two primary particle species. The values for protons and iron nuclei for QGSJET 01 are above and below zero, respectively, as expected. However, for EPOS the predictions for both primary particle types are significantly below zero – a strong hint that the predictions are not compatible with the data. Most likely, the incompatibility of the EPOS predictions with the KASCADE measurements is caused by too high inelastic cross sections for hadronic interactions implemented in the EPOS code. These findings stimulated the development of a new version EPOS 1.9 [30]. Corresponding investigations with this new version are under way.

[1] For details on the event selection and reconstruction, see Ref. [11, 12, 13].

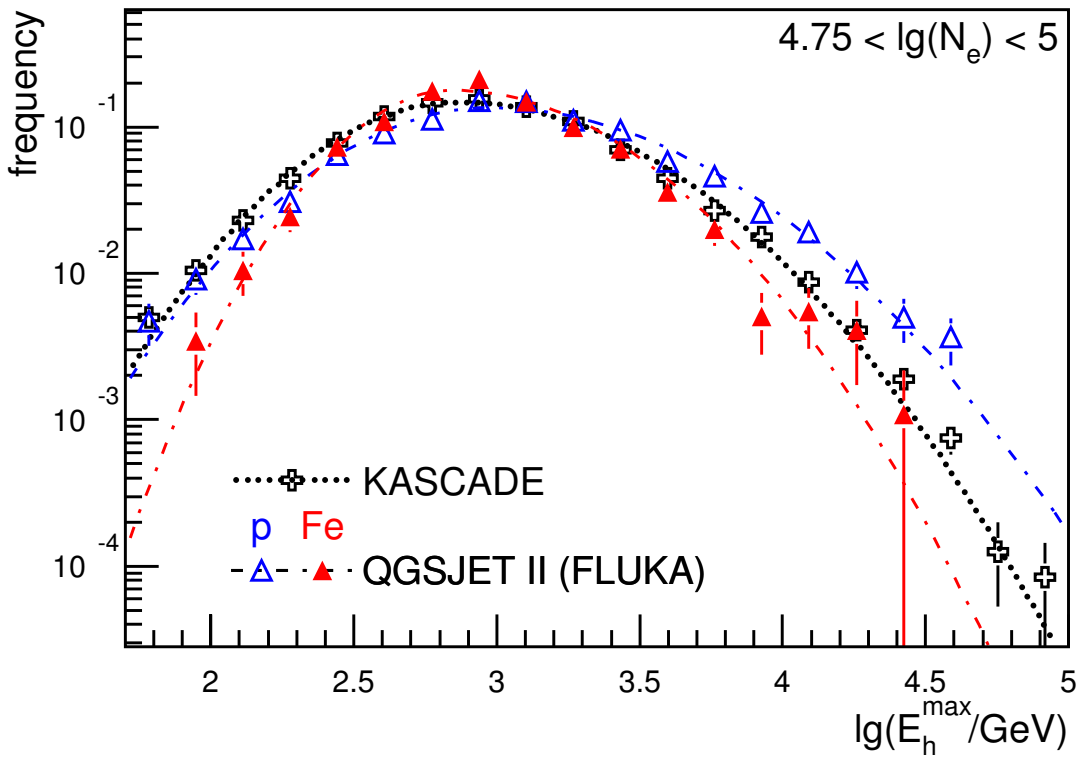

Figure 1: Energy of the most energetic hadron reconstructed at observation level. Predictions of QGSJET II are compared to measured values [19].

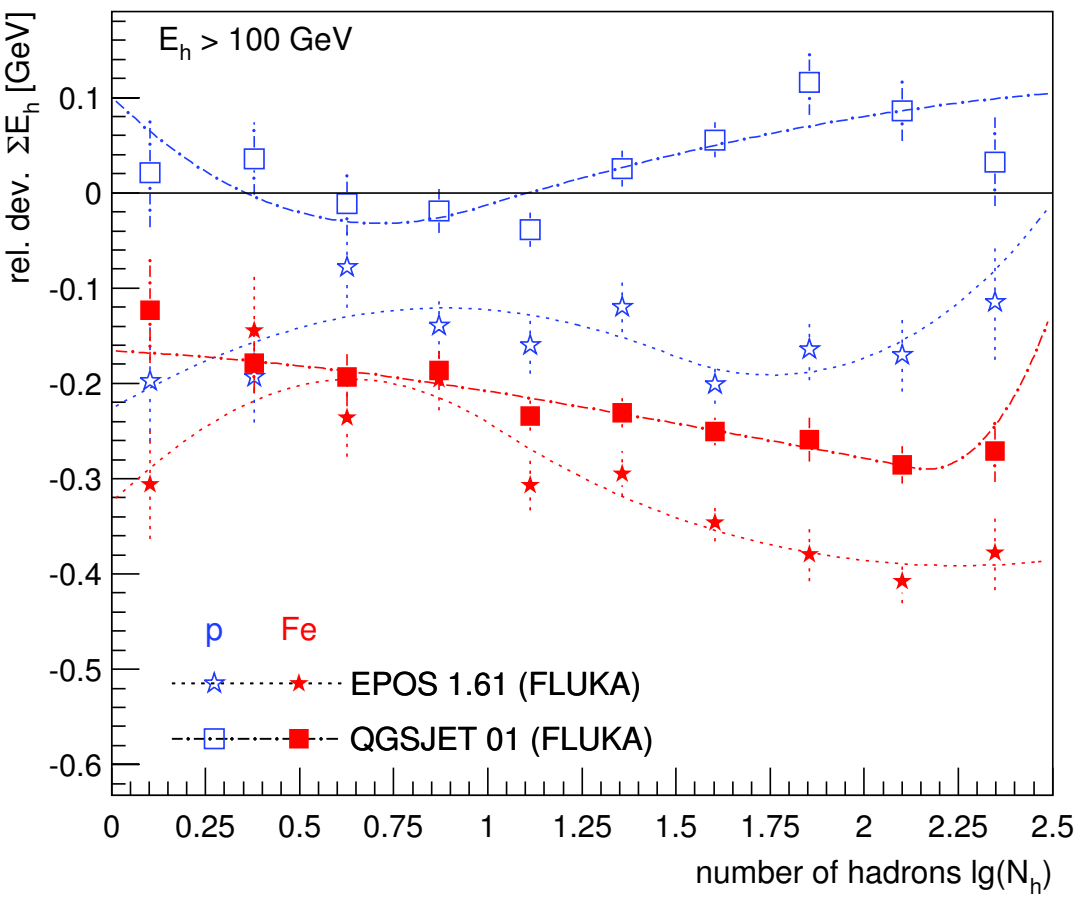

Figure 2: Relative hadronic energy sum $(\sum E_h^{sim} - \sum E_h^{meas})/\sum E_h^{meas}$ as a function of the reconstructed number of hadrons for two interaction models and two primary particle species [12].

Presently, the most compatible predictions are obtained from the models QGSJET 01 and SIBYLL 2.1.

For a more detailed test of the interaction models one has to assume a mass composition in the simulations to compare a single simulation curve (instead of a proton-iron range) with the measured distribution. This can be done consistently by taking a mass composition derived from other observables using the same combination of low-energy and high-energy models [23]. Energy spectra for elemental groups in cosmic rays have been obtained by applying an unfolding procedure to the measured two-dimensional electron and muon number spectra [31]. This composition of cosmic rays has been used as input for the air shower simulations, and the predicted observables at ground level have been compared to the measurements. The investigations reveal that the deviations between the model predictions and the measurements are of the order of 15% [23]. This number illustrates the present accuracy of the quantitative description of the development of air showers.

At much higher energies (exceeding $10^{18}$ eV) investigations of hadronic interactions are under way with data from the Pierre Auger Observatory. At present, all models investigated exhibit problems in predicting the correct number of muons in air showers [32, 33, 34, 35].

# 3 Uncertainties of Accelerator Measurements Extrapolated to Air Shower Observables

Studies have been carried out to evaluate the effect of uncertainties in the description of individual interactions on the development of air showers. An example is the variation of the inelastic proton-proton cross section and the elasticity of the interactions within the error bounds given by accelerator measurements [36]. For the studies parameters in the hadronic interaction model QGSJET 01 have been modified. For illustration, we use here the models 3 and 3a from Ref. [36].[2] With respect to the original QGSJET 01 code the inelastic hadronic cross sections have been lowered, e.g. the proton-air cross section at $10^6$ GeV is reduced by 5% from 385 mb to 364 mb and the elasticity has been increased by about 12%.

A lower cross section implies a longer mean free path for the hadrons in the atmosphere and thus a reduction of the number of interactions. A larger elasticity means that more energy is transferred to the leading particle. Both changes applied result in showers which penetrate deeper into the atmosphere. For example, the average depth of the shower maximum for protons at 100 PeV is shifted by 24 g/cm$^2$ due to the lower cross section and by 10 g/cm$^2$ due to the higher elasticity [36].

The shift of the shower maximum also affects the number of particles registered at ground level. Since the maximum moves closer to the observation level one expects an increase of the number of particles. However, reducing the number of interactions due to a lower cross section also reduces the possibility to produce secondary particles and an increase of the elasticity implies at the same time that less energy is available for multi-particle production. This means that we are faced with two competing processes influencing the number of particles observed.

Simulations reveal that an increase of the elasticity enhances the particle numbers for all species observed (electrons, muons, and hadrons). An increase is registered for both, primary protons and iron nuclei. This means the effect of deeper penetrating cascades seems to dominate. As an example, the increase of the number of muons when increasing the elasticity is illustrated in Fig. 3 [37]. Shown are the relative changes in the number of muons for model 3a relative to model 3 ($\delta N_\mu = (N_\mu^{3a} - N_\mu^3)/N_\mu^3$) for primary protons and iron induced showers as function of primary energy.

The increase of the number of muons $N_\mu$ as function of primary energy $E_0$ has been estimated using a Heitler model to be $N_\mu = (E_0/\xi_c^\pi)^\beta$, where $\xi_c^\pi \approx 20\ GeV$ is the critical energy for pions at which the probability for an interaction and decay are about equal [38]. The exponent $\beta$ depends on the elasticity of the interaction as $\beta \approx 1 - 0.14(1-\kappa)$. Using the energy dependence of $\kappa$ for the two modifications of QGSJET [36] and introducing an energy dependent $\beta$, an increase of the number of muons as function of energy is expected as indicated by the line in Fig. 3. The general trend of the simple estimate is reflected by the detailed simulations, but the absolute values are about 5% larger for the simple estimate as compared to the full simulation. This illustrates the sensitivity of air shower observables to properties of hadronic interactions. Another example is discussed in the following.

Recent investigations [13] revealed that the attenuation length $\lambda_E$, defined as $\Sigma E_H = E_0 \exp(-X/\lambda_E)$ is very sensitive to the inelastic hadronic cross sections. $E_0$ is the energy of the primary particle and $\sum E_H$ the hadronic energy sum registered at ground level. Thus, $\lambda$ is a measure for the hadronic energy transported to ground. A detailed inspection of the attenuation lengths obtained for showers induced by light and heavy elements indicates that the cross sections in the hadronic interaction model QGSJET 01 may be too large and the elasticity may be too small. A modification with altered parameters (model 3a as discussed above) improves the situation.

[2] Model 3 and 3a refer to nomenclature used in Ref. [36]. For model 3 the cross section has been lowered, and, in addition for model 3a the elasticity has been modified.

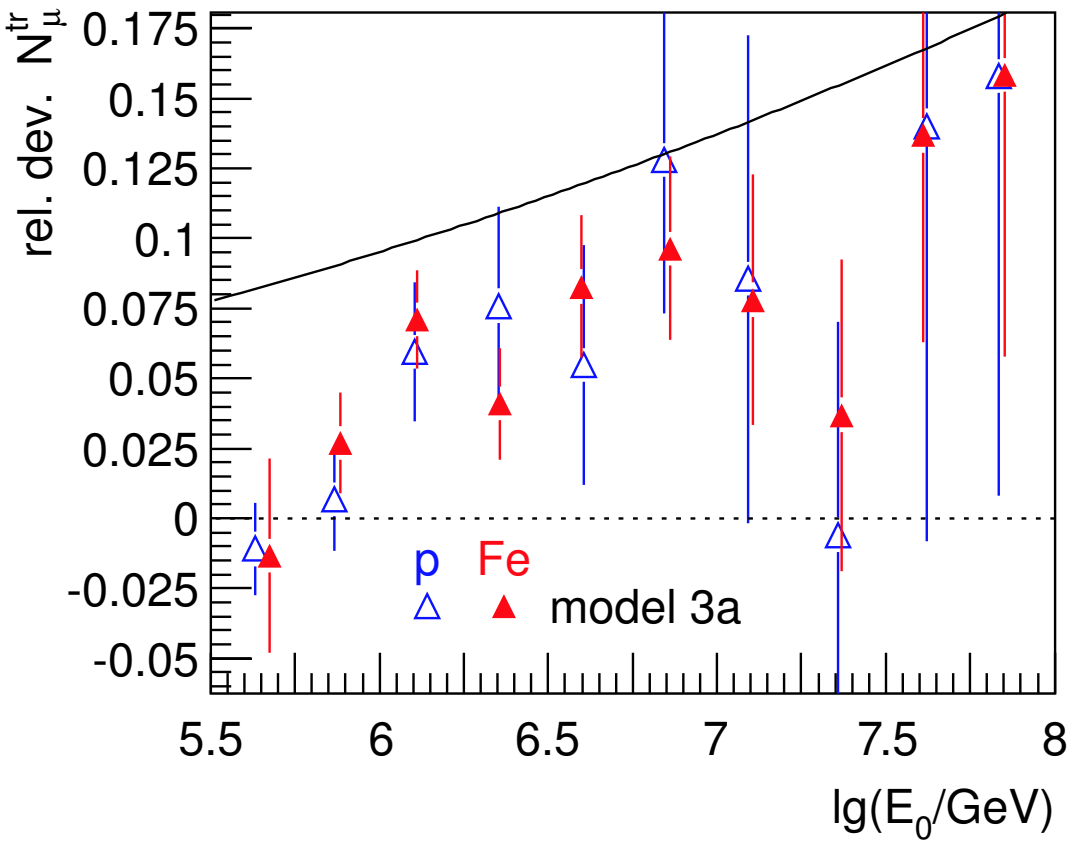

Figure 3: Relative deviation of the number of muons in model 3a relative to model 3, i.e. the change of the number of muons related to an increase of the elasticity as function of primary energy. The line indicates an estimate according to a simple Heitler model [37].

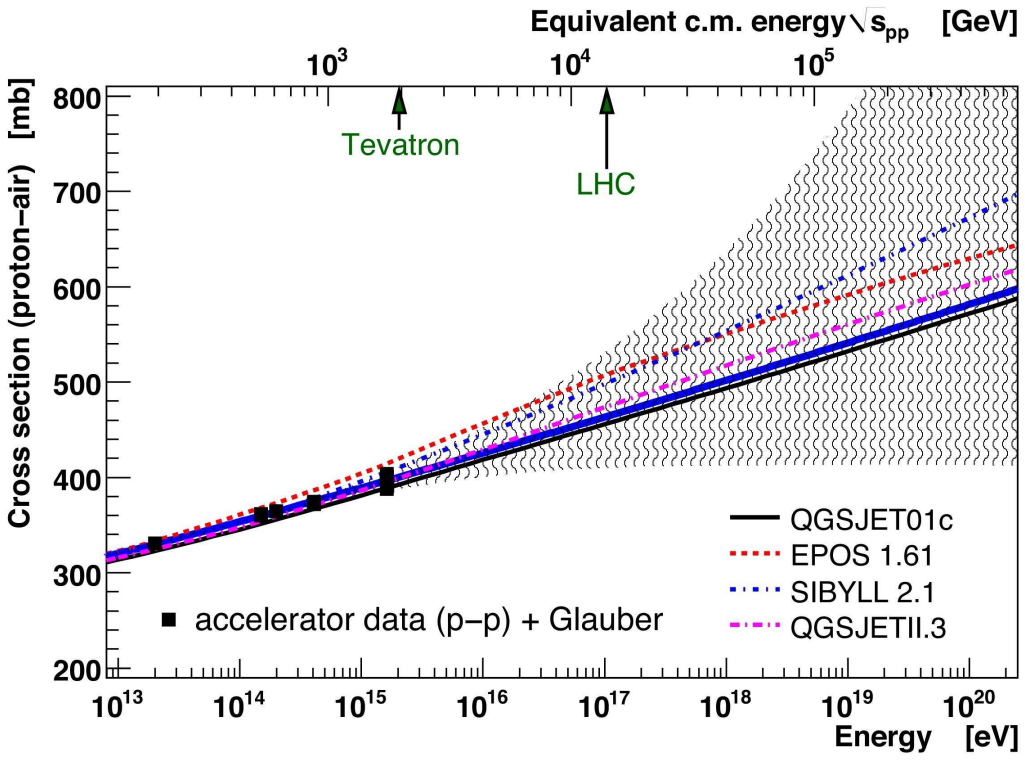

Figure 4: Uncertainties of the extrapolation of the proton-air cross section from accelerator to cosmic-ray energies [39].

Also the influence of the transverse momentum $p_\perp$ in hadronic interactions on (hadronic) air shower observables has been analyzed by the KASCADE group [40]. It turned out that the geometrical distributions of the most energetic hadrons at ground level are sensitive to this parameter. The maximum geometrical distance $d_{max}$ between the four highest-energy hadrons in each shower are sensitive to $p_\perp$. Altering $p_\perp$ in air shower simulations results in different $d_{max}$ distributions.

# 4 Accelerator Data Needed for Cosmic-Ray Physics

Complementary to the investigation of air showers more information about hadronic interactions is needed from accelerator experiments to fully understand cosmic rays, as discussed in the following.

**Air shower measurements** In high-energy interactions most energy is escaping the interaction region in the forward direction, i.e. at large pseudorapidity values $\eta$. For example, the energy flow at the LHC at $E_{cm} = 14$ TeV, corresponding to $E_{lab} \approx 2E_{cm}^2/m_p = 10^{17}$ eV, peaks at pseudorapidity values around 7 to 10. The forward region with values $|\eta| > 4$ is of great importance for air shower experiments.

Of particular interest are the total (inelastic) cross sections, the elasticity/inelasticity of the interactions, as well as the production cross sections of secondary particles and their parameter distributions, like multiplicity, transverse momentum, energy, and pseudorapidity. As projectiles protons and pions are of interest to study the elementary interactions but also beams of heavier nuclei (such as C, N, O, or Fe, being dominant in the cosmic-ray composition) are desirable. Targets are preferably air constituents, i.e. nitrogen, oxygen, (and carbon). In particular, at the LHC the study of p-p and p-N interactions is of great importance.

The uncertainties introduced in the proton-air cross section by extrapolating from accelerator data to highest energies is illustrated in Fig. 4 [39]. It is obvious that LHC data will drastically reduce the uncertainties in the regime of the highest-energy cosmic rays.

**Direct measurements** Further input from accelerator experiments is also required for the interpretation of data from balloon-borne cosmic-ray detectors, delivering unique information about the propagation of cosmic rays in our galaxy. The systematic uncertainties of measurements of the boron-to-carbon ratio are presently dominated by uncertainties in the production cross section of boron in the residual atmosphere above the detector [41]. Boron is produced through spallation of the relatively abundant elements of the CNO group in the atmosphere.[3] Thus, the production cross sections of boron for protons and CNO nuclei impinging on nitrogen targets are of great interest at energies significantly exceeding 100 GeV/n.

# 5 Outlook

We are looking forward to new data from the LHC in the next few years. They will improve current models used for air shower simulations. Complementary data – at energies much higher than at the LHC – from the Pierre Auger Observatory will yield further information on high-energy hadronic interactions. A close cooperation between the high-energy physics and astroparticle physics communities will help to improve our understanding of elementary processes in nature.

# Acknowledgements

It was a pleasure for me to serve together with Gianni Navarra as convener for the cosmic-ray session at the EDS 09 meeting. I thoroughly regret that Gianni Navarra passed away in August 2009 — we miss him and will always remember him.

I would like to thank the organisers of EDS 09 for the possibility to strengthen the connection between the accelerator and cosmic-ray communities. In particular, I acknowledge the great efforts of Mario Deile to organise a perfect meeting.

I'm grateful to my colleagues from the KASCADE-Grande experiment for fruitful discussions.

# References

[1] J.R. Hörandel. *Adv. Space Res.*, 41:442, 2008.

[2] J. Blümer, R. Engel, and J.R. Hörandel. *Prog. Part. Nucl. Phys.*, 63:293, 2009.

[3] T. Antoni et al. *Nucl. Instr. & Meth. A*, 513:490, 2003.

[3]The detectors float typically below a residual atmosphere of about $3 - 5$ g/cm$^2$.

[4] J. R. Hörandel et al. *Nucl. Phys. Proc. Suppl.*, 75A:228–233, 1999.

[5] T. Antoni et al. *J. Phys. G: Nucl. Part. Phys.*, 25:2161, 1999.

[6] J. Engler et al. *Nucl. Instr. & Meth. A*, 427:528, 1999.

[7] S. Plewnia et al. *Nucl. Instr. & Meth. A*, 566:422, 2006.

[8] T. Antoni et al. *Astropart. Phys.*, 14:245, 2001.

[9] D. Heck et al. Report FZKA 6019, Forschungszentrum Karlsruhe, 1998.

[10] Geant 3.21 detector description and simulation tool. CERN Program Library Long Writeup W5013, CERN, 1993.

[11] W.D. Apel et al. *J. Phys. G: Nucl. Part. Phys.*, 34:2581, 2007.

[12] W.D. Apel et al. *J. Phys. G: Nucl. Part. Phys.*, 36:035201, 2009.

[13] W.D. Apel et al. *Phys. Rev. D*, 80:022002, 2009.

[14] H. Fesefeldt. Report PITHA-85/02, RWTH Aachen, 1985.

[15] A. Fasso et al. CERN-2005-10, INFN/TC-05/11, SLAC-R-773, 2005.

[16] A. Fasso et al. arXiv:hep-ph/0306267, 2003.

[17] J.R. Hörandel. *Proc. 26th Int. Cosmic Ray Conf., Salt Lake City*, 1:131, 1999.

[18] N.N. Kalmykov et al. *Nucl. Phys. B (Proc. Suppl.)*, 52B:17, 1997.

[19] J.R. Hörandel et al. *Proc. 31th Int. Cosmic Ray Conf., Lodz*, # 227, 2009.

[20] J. Engel et al. *Phys. Rev. D*, 46:5013, 1992.

[21] R. Engel et al. *Proc. 26th Int. Cosmic Ray Conf., Salt Lake City*, 1:415, 1999.

[22] J. Milke et al. *Acta Physica Polonica B*, 35:341, 2004.

[23] J. Milke et al. *Proc. 29th Int. Cosmic Ray Conf., Pune*, 6:125, 2005.

[24] J. Ranft. *Phys. Rev. D*, 51:64, 1995.

[25] K. Werner. *Phys. Rep.*, 232:87, 1993.

[26] H.J. Drescher et al. *Phys. Rep.*, 350:93, 2001.

[27] K. Werner, F.M. Liu, and T. Pierog. *Phys. Rev. C*, 74:044902, 2006.

[28] T. Pierog et al. *Proc. 30th Int. Cosmic Ray Conf., Merida*, 4:629, 2008.

[29] T. Pierog and K. Werner. arXiv:astro-ph 0611311, 2006.

[30] T. Pierog et al. *Proc. 31th Int. Cosmic Ray Conf., Lodz*, # 428, 2009.

[31] T. Antoni et al. *Astropart. Phys.*, 24:1, 2005.

[32] R. Engel et al. *Proc. 30th Int. Cosmic Ray Conf., Merida*, 4:385, 2008.

[33] F. Schmidt et al. arXiv:0902.4613, 2009.

[34] A. Castellina et al. *Proc. 31th Int. Cosmic Ray Conf., Lodz*, # 33, 2009.

[35] R. Ulrich et al. arXiv:0906.4691, 2009.

[36] J.R. Hörandel. *J. Phys. G: Nucl. Part. Phys.*, 29:2439, 2003.

[37] J.R. Hörandel et al. *Proc. 29th Int. Cosmic Ray Conf., Pune*, 6:121, 2005.

[38] J. Matthews. *Astropart. Phys.*, 22:387, 2005.

[39] R. Ulrich et al. arXiv 0906.3075, 2009.

[40] T. Antoni et al. *Phys. Rev. D*, 71:072002, 2005.

[41] H.S. Ahn et al. *Astropart. Phys.*, 30:133, 2008.

# Forward Experiments at LHC

*Alessia Tricomi*[1]

[1]University of Catania and INFN Catania, Catania, Italy

Observations of Cosmic Rays over a wide energy range provide useful information to understand high energy phenomena in the Universe. Large experiments for the detection of secondary particles produced in the interaction of primary Cosmic Rays are providing valuable inputs and progress in the field. However, the uncertainty caused from the poor knowledge of the interaction between very high energy primary cosmic ray and the Earth's atmosphere prevents the deduction of astrophysical parameters from the observational data. The Large Hadron Collider (LHC) provides the best opportunity for calibrating the hadron interaction models in the most interesting energy range, between $10^{15}$ and $10^{17}$ eV. To constrain the models used in the extensive air shower simulations the measurements of very forward particles are mandatory. Among the LHC experiments, measurements expected by TOTEM, ZDCs and LHCf will give crucial forward particle data for cosmic ray studies. In this paper, the impact of LHC forward experiments for Cosmic Ray Physics is discussed.

## 1 Introduction

The capability to measure the energy spectrum, the arrival direction and the chemical composition of the cosmic rays reaching Earth's atmosphere is of fundamental importance to understand the origin of High Energy Cosmic Rays and the high energy phenomena happening in our Universe. Despite in the last decades a huge step forward in the understanding of High and Ultra High Energy Cosmic Rays (UHECR) has been achieved thanks to large Extensive Air Shower experiments (EAS), still the origin and nature of cosmic rays with energies between $10^{15}$ eV and the Greisen-Zatsepin-Kuzmin (GZK) cutoff at about $10^{20}$ eV, remains a central open question in high-energy astrophysics. In fact, due to the low observed flux of high energy primary cosmic ray (Fig. 1, left), no direct measurement is possible and primary cosmic ray properties can be inferred only by indirect (yet complementary) measurements of shower particles produced in the interaction with the atmosphere which acts as a "calorimeter" medium. The first method relies on measuring the fluorescence light emitted by air molecules excited by the cascade of secondaries. The second one relies on the use of either water Čerenkov tanks or scintillators to sample the shower at ground. However, the interpretation of EAS data in terms of primary cosmic ray properties is not straightforward since it is strongly affected by the knowledge of the nuclear interactions in the Earth's atmosphere and results are not always in agreement between different experiments. This is true, for instance, for the determination of the primary energy spectrum in the UHE region, in particular the existence of events above the so called GZK cut-off, and the chemical composition of cosmic rays. Contradictory results have been reported for the existence of events over the GZK cut-off. Indeed, evidence of such UHECRs, above the cut-off, have been reported by the AGASA experiment [1], while the results of the HiRes [2] experiment and, more recently, the ones of the Pierre Auger Collaboration [3] are consistent with the existence of the cut-off. A key point which raises observing the cosmic ray energy spectra (Fig. 1) is the importance of the energy scale calibration between different experiments. As can be seen from Fig. 1, a systematic difference from the previous measurements is present also for the Auger results. It has been noted that with an energy rescale of AGASA, HiRes and Auger results most of the discrepancies disappear [4, 5, 6].

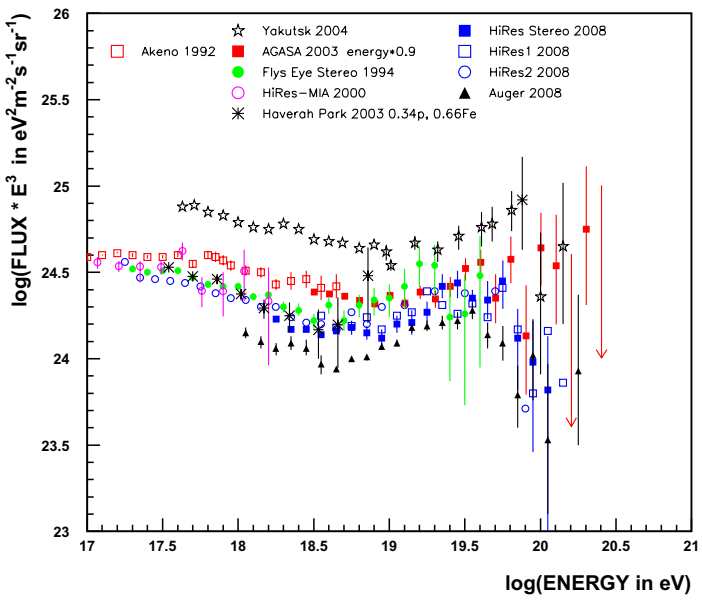

Figure 1: Energy spectrum of cosmic rays at the highest energies. A comparison of the results from main EAS experiments is shown. A clear discrepacy between AGASA, HiRes and Auger can be seen in the region above $10^{20}$ eV. From Ref. [5].

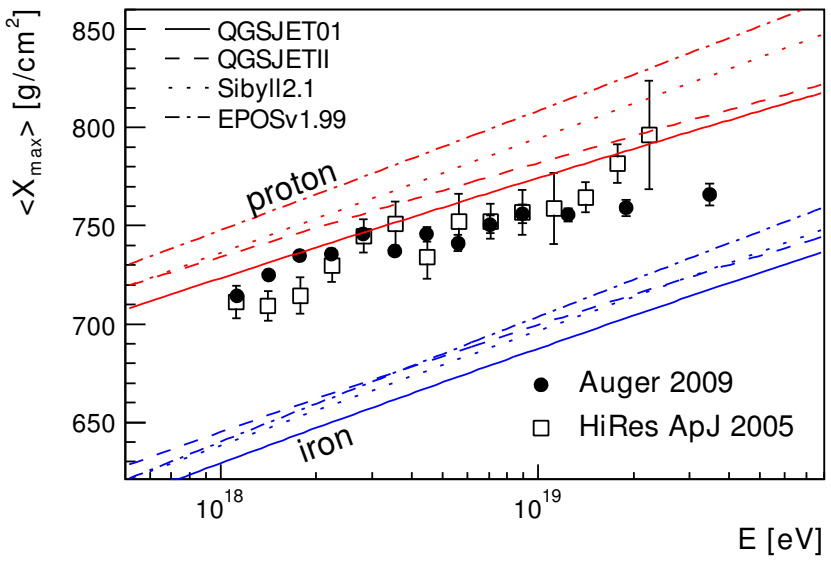

Figure 2: The position of the shower maximum $X_{MAX}$ is shown as a function of the primary cosmic ray energy. The various lines correspond to predictions made by different Monte Carlo models. Plot from Ref. [7].

Similar considerations hold for the interpretation of cosmic ray chemical composition. Cosmic rays are not purely protons but they contain also heavy nuclei. Nuclear cascade showers initiated by the disintegration of heavy nuclei develop more rapidly in comparison with the showers initiated by protons. The position of the shower maximum, $X_{MAX}$, cleary depends on the composition of the cosmic rays.

Recently, Auger results pose another puzzle for the highest energy extra-galactic cosmic rays. Auger results on the correlation between the arrival direction of the cosmic rays and the direction of AGNs seem to indicate that the highest energy cosmic rays are protons [8] even if a subsequent analysis seems to indicate less correlation than what was claimed in the first paper [9]. On the other hand, florescence based measurements of air shower elongation seem to indicate that the composition of the highest energy cosmic-rays favour a mixed composition [10]. Fig. 2 shows a comparison of the most recent results on the $X_{MAX}$ distribution as obtained by HiRes and Pierre Auger collaborations. Superimposed are the distribution expected by different Monte Carlo models [7].

Because the deduction of primary cosmic ray composition from the elongation parameter has a strong model dependence, especially in the highest energy region, the reduction of the uncertainty in the hadronic interaction models is important for solving this puzzle as for the correct interpretation of the primary spectrum. Accelerator experiments validating the interaction model chosen are hence mandatory. As a matter of fact air shower development is dominated by the forward products of the interaction between the primary particle and the atmosphere. The bulk of the primary particle production is dominated by forward and soft QCD interactions, usually modeled in Regge-Gribov-based [11] approaches with parameters constrained by the existing collider data ($E_{lab} \preceq 10^{15}$ eV). The only available data on the pro-duction cross-section of neutral pions emitted in the very forward region have been obtained more than twenty years ago by the UA7 Collaboration [12] at CERN. They measured the photon distribution within an emission angle of as little as 1.8 milli-radians from the beam axis and up to an energy of 630 GeV in the center of mass system, corresponding to a laboratory frame energy of $2 \times 10^{14}$ eV, well below the knee region. When extrapolated to energies around the GZK-cutoff, the current MCs predict energy and multiplicity flows differing by factors as large as three, with significant inconsistencies especially in the forward region. The LHC accelerator, thanks to its unprecedented energy of 14 TeV in the center of mass system, corresponding to $10^{17}$ eV in the laboratory reference frame, offers a unique opportunity to measure both neutral and charged particles emitted in the very forward region. It should be stressed that with LHC not only the energy frontier will be boosted but also, thanks to the complementarity of the different detectors installed, the capability to cover almost the full range of pseudorapidity will be reached (Fig. 3). Measurement of forward particle production in p-p, p-Pb, and Pb-Pb collisions at LHC energies will thus provide strong constraints on nuclear interaction models and allow for a better understanding of high energy cosmic ray properties.

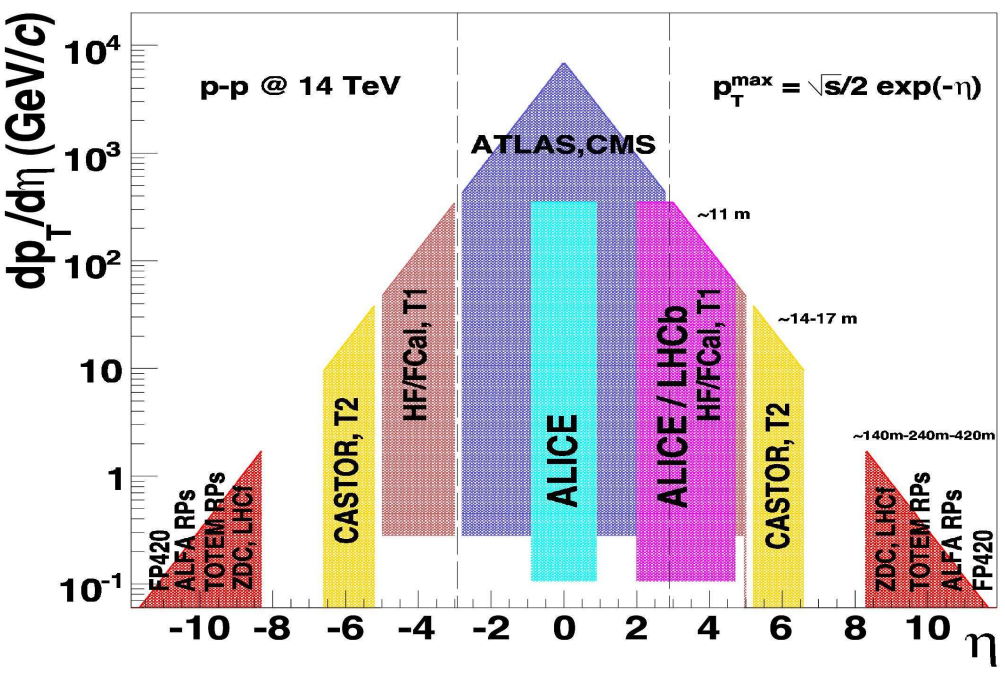

Figure 3: Acceptance in the $p_T - \eta$ plane of the current (and proposed) LHC experiments.

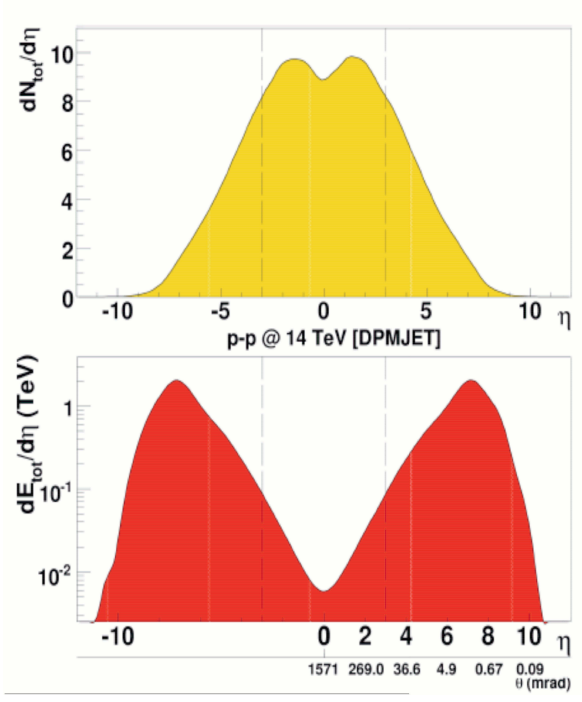

Figure 4: Total multiplicity (top) and energy flow (bottom) in 14 TeV p-p collisions at LHC.

## 2 Forward Physics at LHC

At LHC center of mass energy, secondary particles emitted in the forward direction (which are the bulk of the air showers) carry most of the energy (Fig. 4). Three fundamental parameters for air shower studies can be measured by accelerator experiments:

- Total/inelastic cross sections;
- elasticity/inelasticity;
- particle energy spectra.

The determination of these three parameters are mandatory to determine the longitudinal and lateral spread of air showers. Measurements near zero degree collision angle are also needed to determine the total cross section based on the optical theorem. In order to extend the Physics program of general purpose detectors at LHC additional dedicated detectors for measuring the very forward particles produced in the collisions have been designed, as part of the major experiments or as independent experiments. The techniques used for measurement of the forward emitted particles can be classified into three different categories:

- detectors that surround the beam pipe in the forward region. In this case very forward particles remaining in the beam pipe cannot be covered;

- detectors inside the beam pipe that can be moved close to the beam. This is generally called the Roman Pot method after the Italian experiment that first employed this technique. This is the ideal method for measuring charged particles close to zero degree collision angle and is the standard method used to determine the total cross section via the optical theorem;

- detectors beyond the dipole separating the colliding beams and at zero degree collision angle (Zero Degree Calorimeters or ZDCs). This is the ideal method for measuring forward neutral particles since charged particles are swept out by the beam separation dipole and zero degree collision angle is accessible.

LHC experiments feature an unprecedented rapidity coverage thanks to dedicated forward detectors which complement the mid-rapidity coverage (Fig. 3). Both ATLAS [13] and CMS [14] has been instrumented also in the forward part. Forward calorimetry is available at $\pm 11$ m (the FCal and HF hadronic calorimeters), at $\pm 14$ m (CMS CASTOR sampling calorimeter) [15], and at $\pm 140$ m (the Zero-Degree-Calorimeters, ZDCs) [16, 17] from Interaction Points (IPs). In addition, ATLAS has Roman Pots (RPs) at $\pm$ 220 and 240 m, and both ATLAS and CMS are planning to install a new proton-tagger system at 420 m (FP420) from each IPs. Also ALICE [18] and LHCb [19] are equipped with forward detectors: both have muon spectrometers covering the region $2 \preceq |\eta| \preceq 5$, not covered by ATLAS or CMS thus complementing their information. In addition, two independent forward experiments TOTEM [20] and LHCf [21] have been installed at LHC. TOTEM shares IP5 with CMS and it consists of two types of trackers (T1 and T2 telescopes) which surround the beam pipe covering the region $3.1 < |\eta| < 4.7$ and $5.2 < |\eta| < 6.5$ respectively, plus Roman Pots installed at $\pm 147$ and $\pm 220$ m covering the very forward ($\eta \sim 10$ ) elastically scattered particles that are near the outgoing beams. Combining these measurements, TOTEM can determine the total cross section with a precision of $\pm 1$ mb. Last but not least, the LHCf experiment is a fully dedicated astroparticle experiment at LHC installed in the same region of the ATLAS ZDCs, $\pm 140$ m away from IP1. The detector consists of two sampling and imaging calorimeters made by 16 layers of plastic scintillators interleaved by tungsten layers as converter. Additionally, a set of four X-Y position sensitive layers, made by 1 mm$^2$ scintillating fibers in one calorimeter and silicon micro-strip layers in the other, provide incident shower positions, in order to obtain the transverse momentum of the incident primary and to correct for the effect of leakage from the edges of the calorimeters. LHCf is a kind of ZDC, but designed with a very different concept from the ATLAS, CMS, and ALICE ZDCs. Like the ZDCs LHCf also measures the neutral particles emitted at and near zero degree collision angle. The single particle energy and position resolutions of LHCf have been optimized for discriminating between the hadron interaction models used in cosmic-ray studies.

# 3 Monte Carlo Model Discrimination

While the knowledge of nuclear interaction models is mandatory to infer primary cosmic ray properties from EAS experiments, unfortunately the bulk of particle production in high-energy hadronic collisions can still not be calculated from first-principles QCD. Monte Carlo models frequently used to simulate cosmic ray cascades, like DPMJET [22], QGSJET 01 and II [23], SYBILL [24] and EPOS [25], can be regarded as phenomenological "QCD-inspired" models. They are indeed based on general principles such as unitarity and analyticity often combined with perturbative QCD predictions for high-$p_T$ processes to obtain an almost complete description of the final states. Soft processes are described within Gribovs Reggeon theory [11] and hadrons are produced mainly in the fragmentation of color strings. A detailed description of the difference between these models can be found in Ref. [26].

LHC data will play a fundamental role to calibrate all these models up to an energy of $10^{17}$ eV thanks to the complementarity of the forward detectors described in previous Section. In particular, data collected by TOTEM/ALFA and by LHCf/ZDC will be mandatory to achieve this goal.

The energy spectra of single $\gamma$-rays and neutrons expected to be measured by the LHCf experiment and calculated using the different hadron interaction models are shown in Fig. 5. In this calculation, a 1000 sec of the 14 TeV LHCf operation at a luminosity of $10^{29}\,\mathrm{cm}^{-2}\mathrm{s}^{-1}$ is assumed which will be achieved at the beginning of the LHC commissioning phase. 5% and 30% energy resolution respectively for $\gamma$ and neutron and statistical errors are included in the calculation. As can be noted from Fig. 5 clear discrimination between the calculated spectra for the various models is possibile, especially for neutron spectra even if the energy resolution is significantly worse than for $\gamma$ reconstruction. The measurement of forward neutron energy spectrum by LHCf/ZDC is of particular importance since it is related to the elasticity parameter discussed in Sect. 2 and can provide useful information to characterize the event in heavy ion collisions.

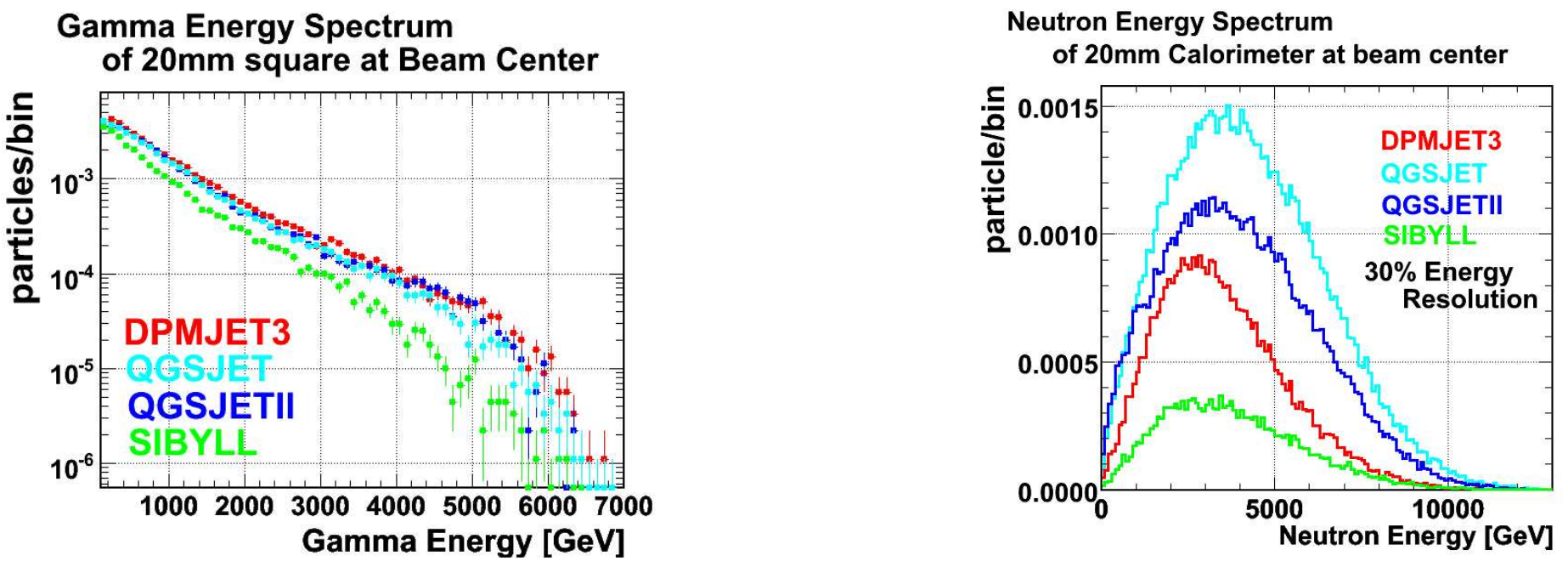

Figure 5: Expected energy spectrum for $\gamma$s and neutrons according to different interaction models at 7+7 TeV center of mass energy. For $\gamma$ a 5% energy resolution has been taken into account while for neutrons a 30% energy resolution has been included in the calculated spectrum.

The calibration of Monte Carlo codes asks for a precision measurement of the energy scale. For this reason LHCf relies on a very precise reconstruction of the $\pi^0$ mass, by reconstructing the 2 $\gamma$ from $\pi^0$ decays each impinging one of the two towers of the calorimeter.

MC simulations for $1.04 \times 10^7$ and $1.17 \times 10^7$ p-p collisions, each corresponding to about 20 minutes operation during the LHC beam commissioning with 43 bunches and $10^{29}\,\mathrm{cm^{-2}s^{-1}}$ of luminosity, were carried out with the DPMJET3.03 model for two different detector position (one in which the detector is in the nominal position and one in which the detector is 10 mm down), respectively. Details of the analysis can be found in [27]). As can be seen in Fig. 6, the reconstructed $\pi^0$ energy spectrum well reproduced the original production spectrum. A good discrimination between different Monte Carlo models is hence feasible.

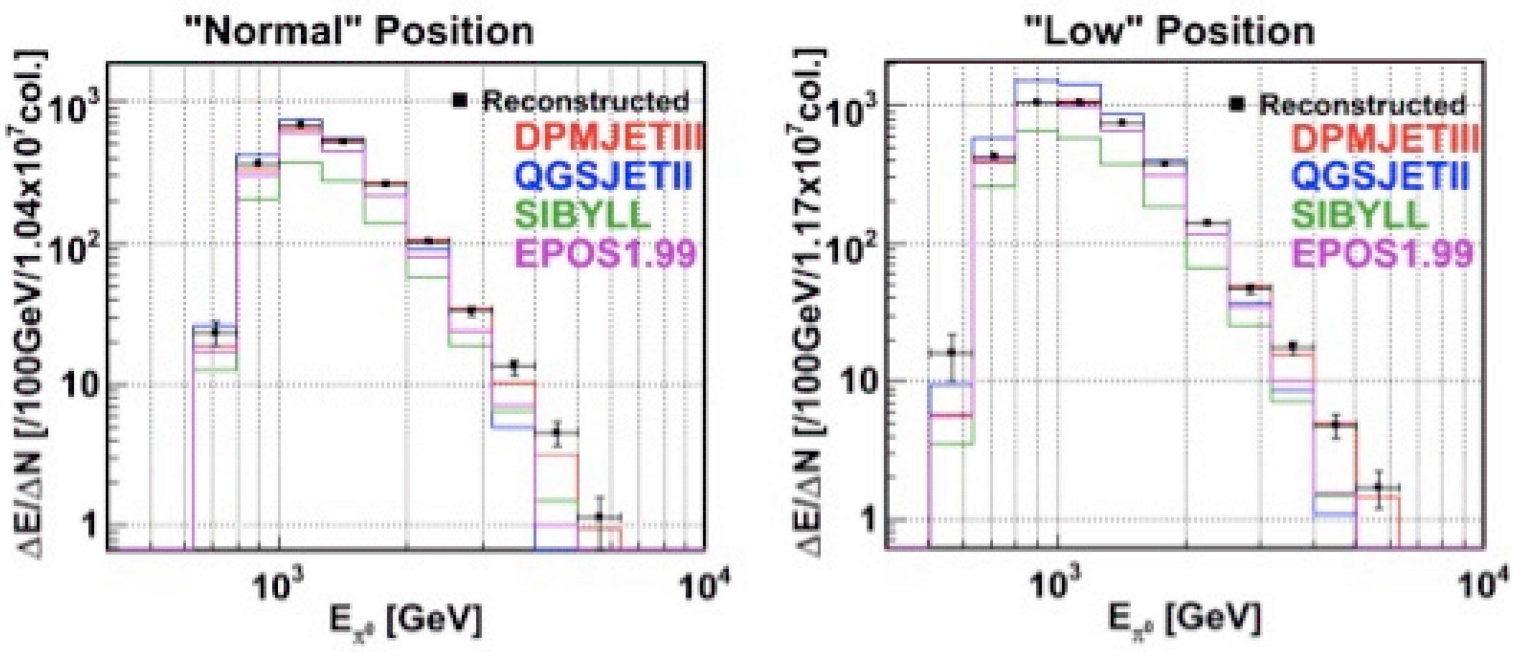

Figure 6: Reconstructed $\pi^0$ energy spectrum compared with expectations from different MC models for two LHCf detector positions. Events correspond to about 20 minutes at $10^{29}\,\mathrm{cm^{-2}s^{-1}}$ luminosity for 14 TeV p-p collisions. The reconstructed spectrum well reproduce the production spectrum. The main systematic error is due to the uncertainty on the absolute energy scale of the calorimeter.

In addition to detecting $\pi^0 \to \gamma\gamma$ and forward going neutrons, the LHCf/ZDC can detect and reconstruct $\eta \to \gamma\gamma$ (Fig. 7 left), $\Lambda, \Delta \to n\pi^0$, $\Sigma \to \Lambda\gamma$ and $K_S^0 \to \pi^0\pi^0$ (Fig. 7 right), and measure their production cross section, energy spectrum, and angle within the detector acceptance thus ensuring redundant tools for energy scale calibration.

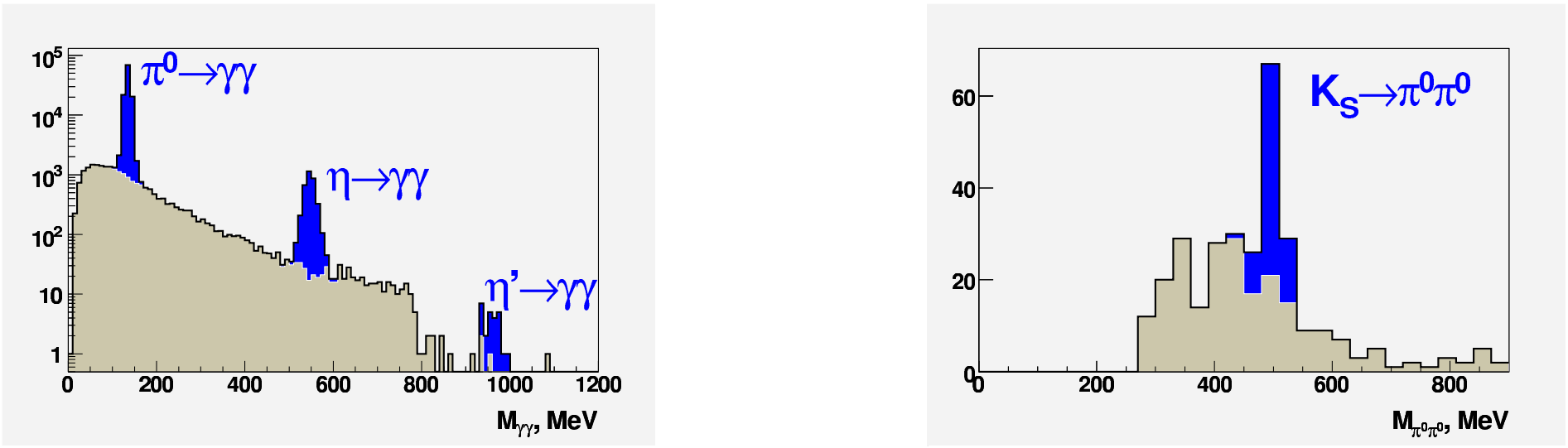

Figure 7: Simulated $\gamma\gamma$ (left) and $\pi^0\pi^0$ (right) mass spectra. Reconstructed $\pi^0$, $\eta$ and $\eta$' peaks are clearly visible (left) above $\gamma\gamma$ background as well as $K_S^0$ peak (right) above $\pi^0\pi^0$ background. Events corresponds to $10^6$ p-p collisions at 14 TeV generated with Pythia 6.3. Plots from Ref. [16].

Figures 8 and 9 from Ref. [28] show for p-p collisions at 14 TeV and p+Pb collisions at 8.8 TeV, respectively, the inclusive multiplicity and energy flows predicted by the models for all pseudo-rapidities, as well as the energy deposit in the acceptances covered by the CASTOR/T2 ($5.2 < |\eta| < 6.6$) and ZDC/LHCf ($|\eta| > 8.1$ for neutrals) detectors. In some cases, differences as large as 60% are observed between prediction of different models.

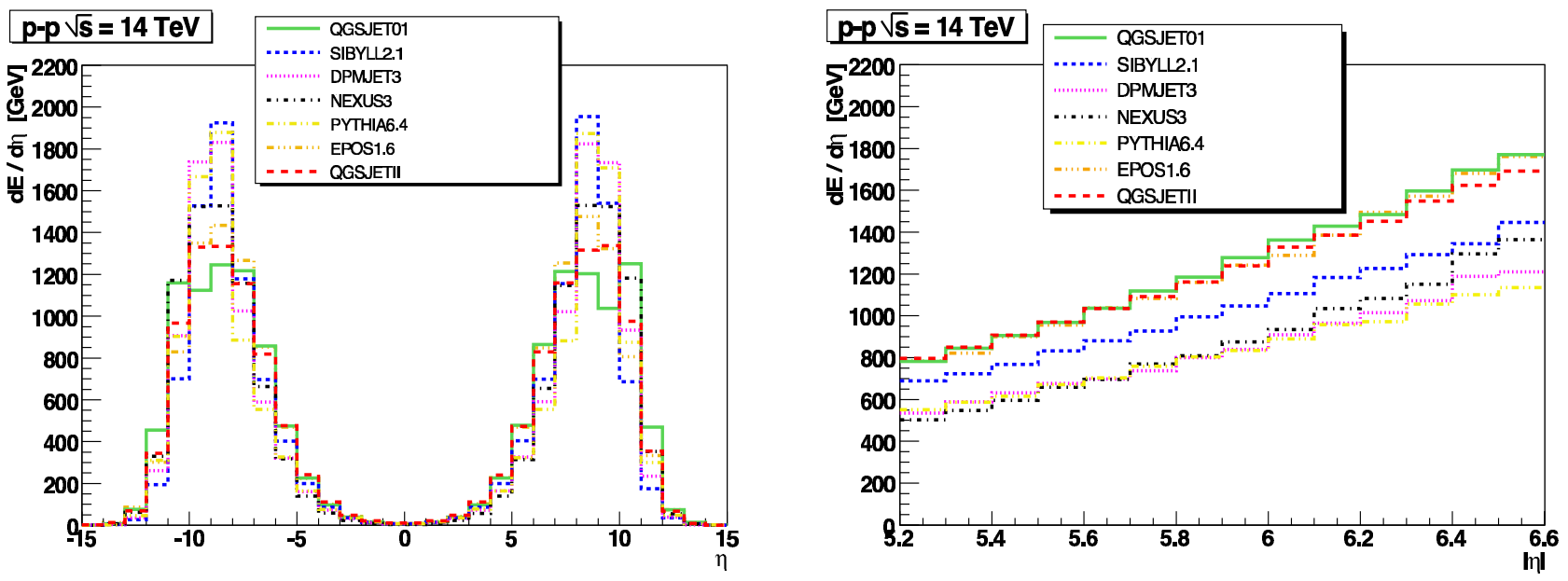

Figure 8: Energy flow in the whole pseudorapidity range (left) and in the CASTOR/T2 acceptance ($5.2 < |\eta| < 6.6$) (right) as predicted by different Monte Carlo models used in cosmic ray Physics for p-p collision at 14 TeV.

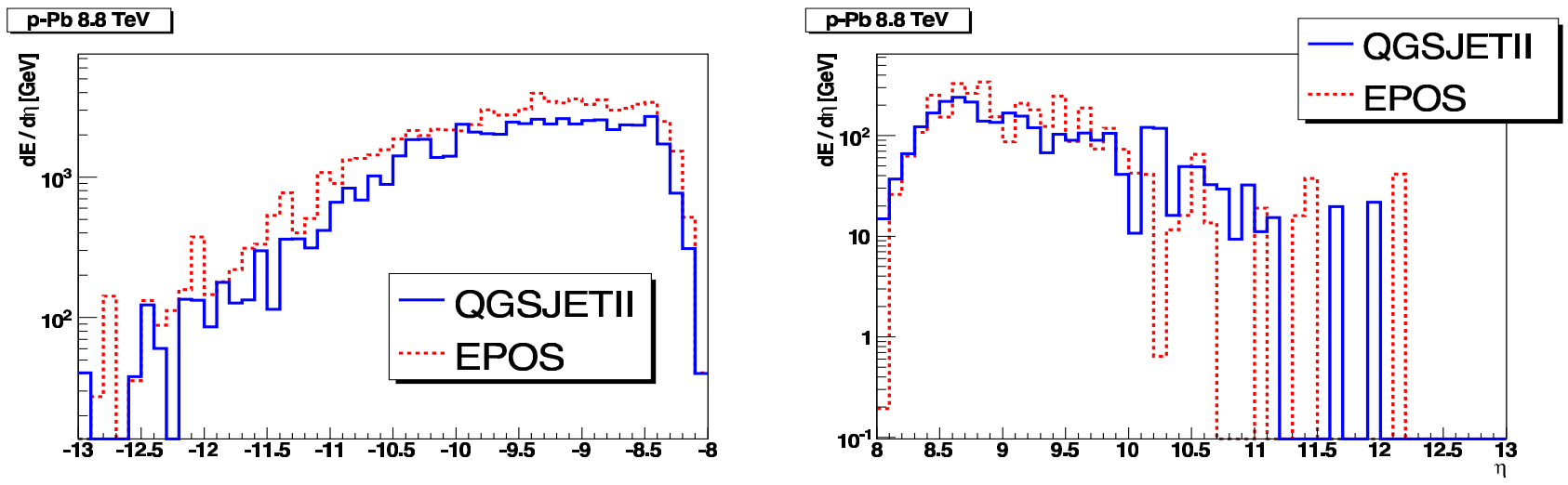

Figure 9: Neutral energy density in LHCf/ZDC acceptance for p+Pb collisions at 8.8 TeV as predicted by different Monte Carlo models used in cosmic ray Physics.

# 4 Conclusions

The new generation of HECR experiments provide valuable data to understand high-energy phenomena and, at the same time, pose interesting questions. To solve many of these puzzles the reduction of the uncertainty due to the hadron interaction model used in simulation is mandatory. Data collected at LHC will provide a fundamental instrument to calibrate Monte Carlo codes in the most critical energy range of the CR spectrum, between the knee and the GZK cut-off. In particular, LHC experiments feature an unprecedented rapidity coverage thanks to dedicated forward detectors which will provide the most significant data to discriminate the currently used models or set constraints for the future models.

# Acknowledgements

I wish to thank all the people who helped me to collect useful material for this paper, especially O. Adriani, K. Eggert, D. D'Enterria, P. Grafström, M. Grothe and S. White.

# References

[1] M. Takeda et al. Extension of the cosmic-ray energy spectrum beyond the predicted Greisen-Zatsepin-Kuzmin cutoff. *Phys. Rev. Lett.*, 81:1163–1166, 1998, astro-ph/9807193.

[2] R. Abbasi et al. Observation of the GZK cutoff by the HiRes experiment. *Phys. Rev. Lett.*, 100:101101, 2008, astro-ph/0703099.

[3] J. Abraham et al. Observation of the suppression of the flux of cosmic rays above $4 \times 10^{19}$eV. *Phys. Rev. Lett.*, 101:061101, 2008, astro-ph/0806.4302.

[4] V. Berezinsky. Transition from galactic to extragalactic cosmic rays. 2007, astro-ph/0710.2750.

[5] M. Nagano. Search for the end of the energy spectrum of primary cosmic rays. *New J. Phys.*, 11:065012, 2009.

[6] J. Bluemer, R. Engel, and J. R. Hoerandel. Cosmic Rays from the Knee to the Highest Energies. *Prog. Part. Nucl. Phys.*, 63:293–338, 2009, astro-ph.HE/0904.0725.

[7] M. Unger. Study of the Cosmic Ray Composition with the Pierre Auger Observatory. *Talk at SOCoR, Trondheim 2009*, 2009.

[8] J. Abraham et al. Correlation of the highest energy cosmic rays with nearby extragalactic objects. *Science*, 318:938–943, 2007, astro-ph/0711.2256.

[9] J. Abraham et al. Correlation of the highest-energy cosmic rays with the positions of nearby active galactic nuclei. *Astropart. Phys.*, 29:188–204, 2008, astro-ph/0712.2843.

[10] J. Abraham et al. Studies of Cosmic Ray Composition and Air Shower Structure with the Pierre Auger Observatory. 2009, astro-ph/0906.2319.

[11] V. N. Gribov. A REGGEON DIAGRAM TECHNIQUE. *Sov. Phys. JETP*, 26:414–422, 1968.

[12] E. Pare et al. Inclusive Production of $\pi^0 s$ and Feynman Scalin Test in the Fragmentation Region at the S $\bar{p}p$ S Collider. *Phys. Lett.*, B242:531–535, 1990.

[13] G. Aad et al. The ATLAS Experiment at the CERN Large Hadron Collider. *JINST*, 3:S08003, 2008.

[14] R. Adolphi et al. The CMS experiment at the CERN LHC. *JINST*, 0803:S08004, 2008.

[15] S. Basegmez et al. Performance studies of the full length prototype for the CASTOR forward calorimeter of the CMS experiment. CERN-CMS-CR-2008-090.

[16] Zero degree calorimeters for ATLAS. CERN-LHCC-2007-01.

[17] O. A. Grachov et al. Performance of the combined zero degree calorimeter for CMS. *J. Phys. Conf. Ser.*, 160:012059, 2009, nucl-ex/0807.0785.

[18] ALICE technical design report on forward detectors: FMD, T0 and V0. CERN-LHCC-2004-025.

[19] A. Augusto Alves et al. The LHCb Detector at the LHC. *JINST*, 3:S08005, 2008.

[20] G. Anelli et al. The TOTEM experiment at the CERN Large Hadron Collider. *JINST*, 3:S08007, 2008.

[21] O. Adriani et al. The LHCf detector at the CERN Large Hadron Collider. *JINST*, 3:S08006, 2008.

[22] Stefan Roesler, Ralph Engel, and Johannes Ranft. The Monte Carlo event generator DPMJET-III. 2000, hep-ph/0012252.

[23] Sergey Ostapchenko. Status of QGSJET. *AIP Conf. Proc.*, 928:118–125, 2007, hep-ph/0706.3784.

[24] R. S. Fletcher, T. K. Gaisser, Paolo Lipari, and Todor Stanev. SIBYLL: An Event generator for simulation of high-energy cosmic ray cascades. *Phys. Rev.*, D50:5710–5731, 1994.

[25] Klaus Werner. The hadronic interaction model EPOS. *Nucl. Phys. Proc. Suppl.*, 175-176:81–87, 2008.

[26] Klaus Werner. *These Proceedings*, 2009.

[27] Hiroaki Menjo. Simulation Study for the performance of the LHCf experiment. *Proc. of 31st International Cosmic Ray Conference, Łódź, Poland.*

[28] David d'Enterria, Ralph Engel, Thomas McCauley, and Tanguy Pierog. Cosmic-ray Monte Carlo predictions for forward particle production in p-p, p-Pb, and Pb-Pb collisions at the LHC. 2008, astro-ph/0806.0944.

# HERA, LHC and Cosmic Rays

*Armen Bunyatyan*

DESY, Notkestraße 85, 22607 Hamburg, Germany

At high energy, cosmic rays can only be studied by measuring the extensive air showers which they produce in the atmosphere of the Earth. The development of extensive air showers strongly relies on the physics of forward region of hadronic interactions. Measurement of forward particle production at HERA and LHC colliders constrain the physics used in hadronic interaction models and allow for more reliable determinations of the cosmic ray energy and composition.

## 1 Introduction

The origin and nature of cosmic rays (CRs) with energies between $10^{15}$ eV and the Greisen-Zatsepin-Kuzmin (GZK) cutoff at about $10^{20}$ eV remains a central open question in high-energy astrophysics (more details in the contributions at this conference [1–3]). At these energies the flux of CRs is so low that it cannot be measured directly using particle detectors. Therefore all CR measurements of higher energy are based on analysing the secondary particle showers, called extensive air showers (EAS), which they produce in the atmosphere of the Earth. To interpret the characteristics of EAS in terms of primary particle type and energy, detailed modelling of the various interaction and decay processes of the shower particles is needed (for details see [4, 5]). In particular, the elemental composition of the CR flux reconstructed form air shower data depends very much on the assumptions on hadronic multiparticle production. Knowing the CR composition is essential for understanding the phenomena such as the *knee* and *ankle*, the changes of the power-law index of CR flux distribution at about $3 \times 10^{15}$ eV and $3 \times 10^{18}$eV (see Figure 1), and for confirming or ruling out models proposed for the sources of ultra-high energy CRs.

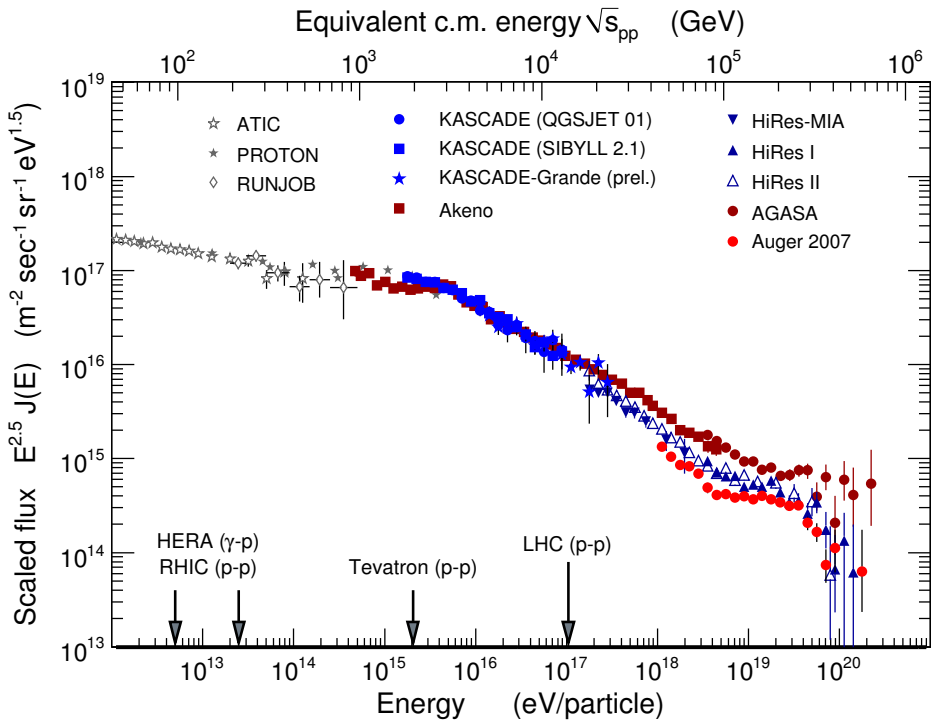

Figure 1: The flux of cosmic rays in the energy range from $10^{12}$ eV. In addition, the equivalent energies of colliders, referring to proton-proton collisions, are indicated by arrows.

Here, we discuss the relation between hadronic multiparticle production and EAS observables and the constraints given by accelerator data. Due to the huge difference between the energy ranges accessible at colliders and in the CR experiments it is very difficult to make direct comparison of their measurements. Nevertheless, it is possible to relate particle production pro-cesses in CR interactions to those studied in collider experiments. The current understanding of interaction processes is realised in the Monte Carlo (MC) event generators, which describe the interactions of the primary in the upper atmosphere. Such event generators allow us to study hadron production at colliders as in CR interactions [6]. The event generators combine theoretical predictions with phenomenological models and parameterisations and have to be tuned by comparing their predictions to measurements at accelerators. Considering the underlying theory and models entering MC programs, almost all data measured at colliders are relevant for understanding of very high energy cosmic ray interactions [7], e.g. limits on physics beyond the Standard Model, parton densities, low-$x$ dynamics and saturation, reliability and range of applicability of perturbative QCD, transition between soft and hard physics, heavy flavour production, etc. In particular, the lack of experimental data on forward hadron production is one of the main source of model uncertainties, as the bulk of the primary particle production is dominated by forward and soft QCD interactions [8]. When extrapolated to energies around the GZK-cutoff, the current MCs predict energy and multiplicity flows differing by factors as large as three, with significant inconsistencies in the forward region. Thus, the modelling of CR interactions strongly depends on the input from accelerator experiments. The measurements of leading proton and neutron distributions from HERA experiments are the highest energy data available at present ($E_{lab} \approx 5 \times 10^{13}$ eV). These data provide important input for CR model tuning. The measurement of forward particle production in $pp$, $pA$ and $AA$ collisions at LHC ($E_{lab} \approx 10^{17}$ eV) will provide further strong constraints on these models and allow for more reliable extrapolations of the CR energy and composition around the GZK cut-off.

## 2 Forward Baryons at HERA

In $ep$ scattering at HERA, a significant fraction of events contains a low-transverse momentum baryon carrying a large fraction $x_L$ of the incoming proton energy. Although a fraction of these *leading baryons* may result from the hadronisation of the proton remnant, the $t$-channel exchange of colour singlet virtual particles is expected to contribute significantly [9–12]. In this picture, the proton fluctuates into a virtual meson-baryon state; the virtual photon subsequently interacts with a parton from the pion, leaving a fast forward baryon in the final state (Figure 2). The production of leading neutron in the virtual exchange model occurs through the exchange of isovector states, and $\pi^+$ exchange is expected to dominate. For leading proton production, isoscalar exchanges also contribute, including diffractive Pomeron mediated interactions (more details in the contribution at this conference [13]).

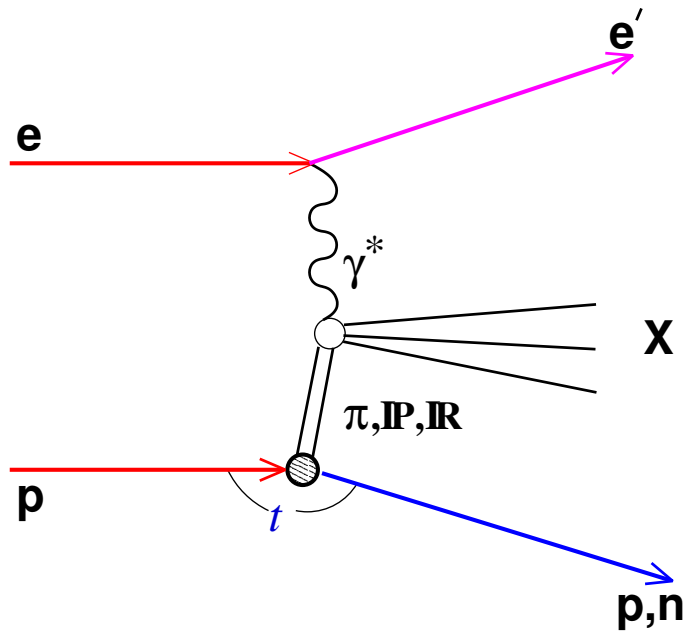

Figure 2: Leading baryon production $ep \to eXN$ via the colour singlet exchange processes.

To measure the forward baryons the H1 and the ZEUS experiments have been equipped with dedicated detectors. Forward protons were measured with position sensitive detectors (Roman Pots) placed along the proton beam downstream of the interaction point. Leading neutrons were measured with lead-scintillator forward calorimeters (FNC) at the zero-degree point; magnet apertures limited neutron detection to scattering angles less than 0.75 mrad.

The cross section of leading proton production in DIS normalised to the inclusive DIS cross section ($1/\sigma_{tot} \cdot d\sigma_{LP}/dx_L$) as function of $x_L$ is shown in Figure 3 as well as the exponential slope $b$ of squared transverse momentum ($p_T^2$) distribution [14]. The rate of leading protons is approximately flat up to the diffractive peak, where it increases by a factor of about six. In the left side of Figure 3 the distributions are compared to the predictions of MC models DJANGO and RAPGAP [15, 16]. which are based on standard fragmentation. These models don't reproduce either the flat dependence of the cross section versus $x_L$ below the diffractive peak at $x_L \sim 1$ or the magnitude and dependence of $b$ on $x_L$. The same data are compared to a Regge-based model [17] incorporating the isovector and isoscalar exchanges, and including the Pomeron for diffraction. A good description of the $x_L$ distribution and the slopes is obtained by adding a substantial contribution of isoscalar Reggeon exchanges, which turn out to be the dominant processes below the diffractive peak.

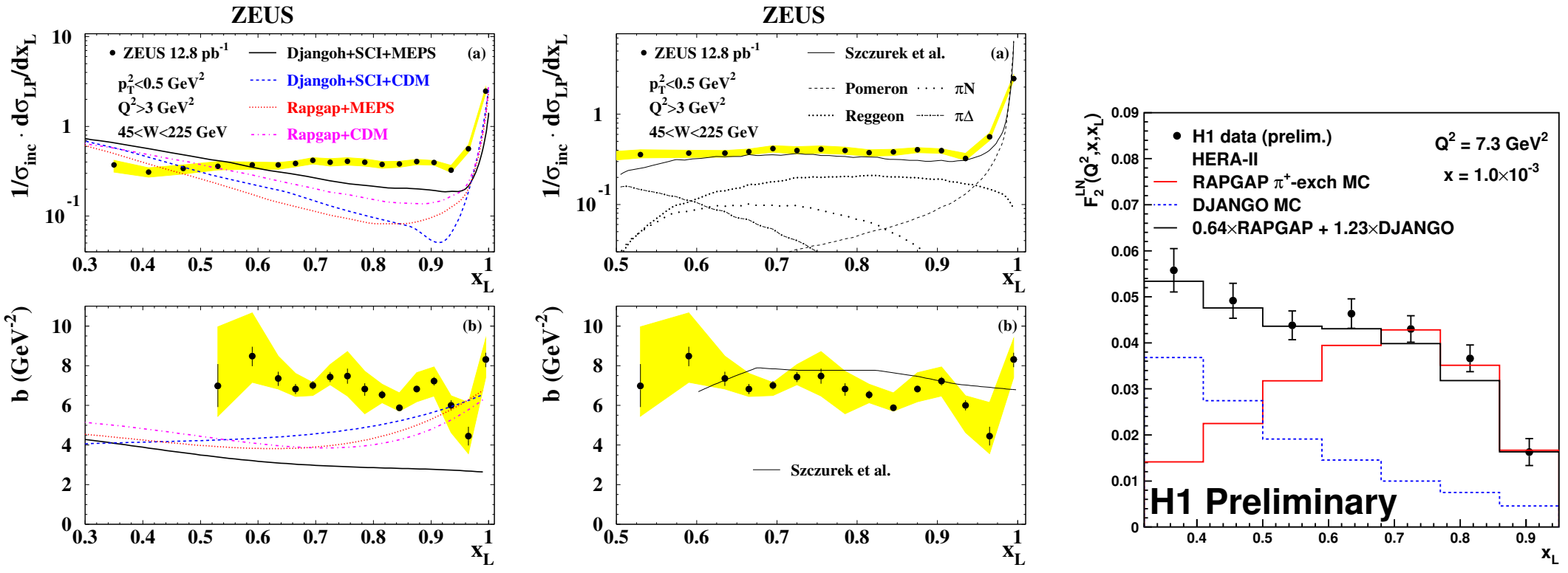

Figure 3: Normalised LP cross section and exponential slope $b$ of LPs in DIS as function of $x_L$, compared to Monte Carlo models (left) and a Regge-based model [17] (centre). (Right) leading neutron $x_L$ distribution compared with Monte Carlo simulation which represents an optimised mixture of exchange and fragmentation models.

The leading neutron $x_L$ cross section is compared in right side Figure 3 to the prediction of RAPGAP MC model, which here simulates the neutron production via $\pi^+$ exchange and the DJANGO MC for inclusive DIS [18]. The DJANGO model which incorporates only standard fragmentation can not describe the observed LN yield, while the mixture of the standard fragmentation and $\pi$-exchange models gives a better description of the shape of the $x_L$ distribution.

# 3 Forward Particles at HERA and Cosmic Rays

The measurements of forward particles at HERA may provide valuable information for the physics of ultra-high energy CRs. The models used for the CR analyses can make predictions for HERA kinematics, which can be compared to the experimental measurements. Below the comparisons are presented with the several models of hadronic interaction commonly used to simulate air showers: EPOS 1.6 [19, 20], QGSJET 01 [21], QGSJET II [22–24], and SIBYLL 2.1 [25–27].

Comparison of the leading proton and the leading neutron spectra measured at HERA with the predictions of the CR models are shown in Figures 4 and 5 [28]. As expected, the HERA measurements are sensitive to the differences between the models and can be used for the tuning of model parameters.

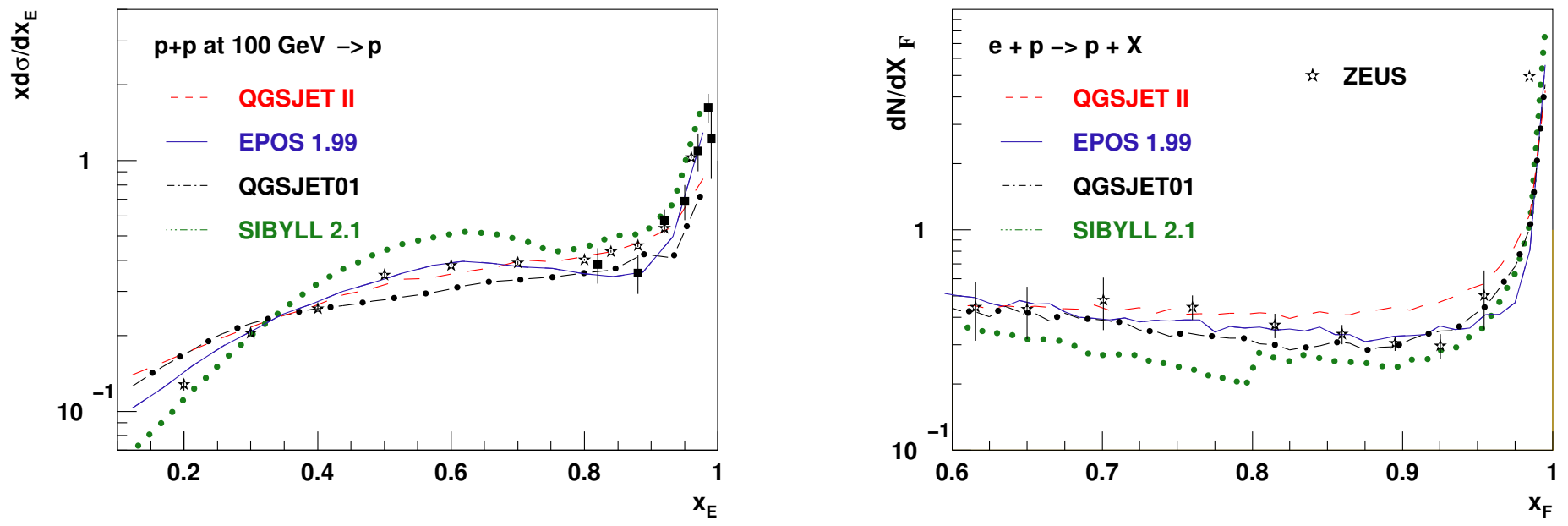

Figure 4: Comparison of the leading proton spectra measured at HERA with the predictions of cosmic ray interaction models [28].

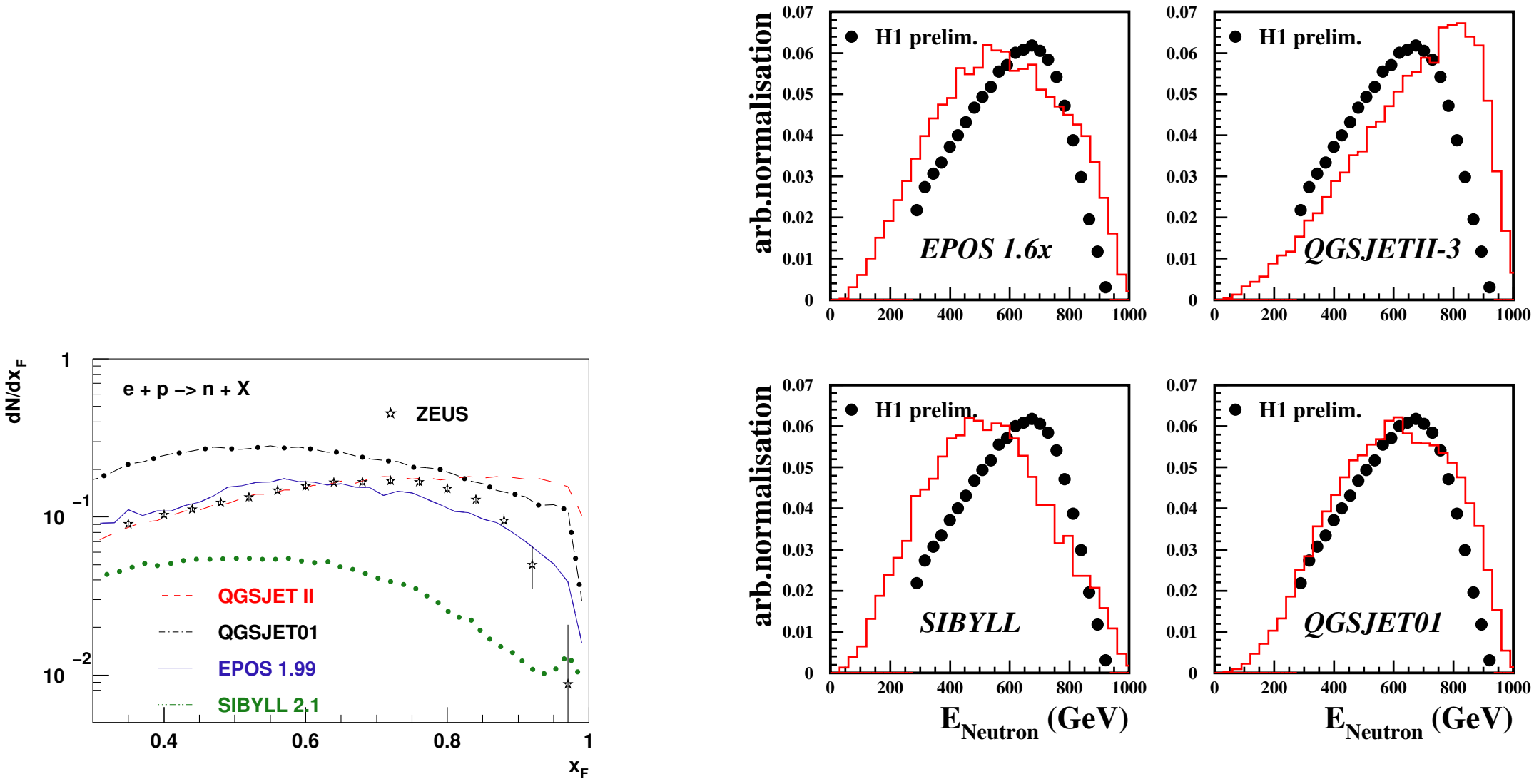

Figure 5: Comparison of leading neutron production energy distribution at HERA with the predictions of the cosmic ray interaction models [28].

# 4 LHC and Cosmic Rays

The coming energy frontier for hadron collisions will be reached by the LHC collider. The LHC will open up a phase space for particle production in an unprecedented range spanning about 20 units (see left side panel of Figure 6). As a general feature, particle production in hadronic collisions is peaked at central rapidities, whereas most of the energy is emitted at very low angles. The sub-detectors of the two large experiments ATLAS and CMS (ATLAS

LUCID and CMS CASTOR calorimeters, the Zero-Degree-Calorimeters, ATLAS Roman Pots) and two independent experiments LHCf and TOTEM are capable of measuring very forward particles. Coverage of each experiment in pseudo-rapidity is also indicated in Figure 6 by arrows. Because each experiment has different capability (charged or neutral particle measurement, hadron or electromagnetic calorimeter, calorimeter or tracker, infinite or finite pseudo-rapidity coverage, aperture, position/energy resolutions), they provide complementary data for total understanding of the very forward particles.

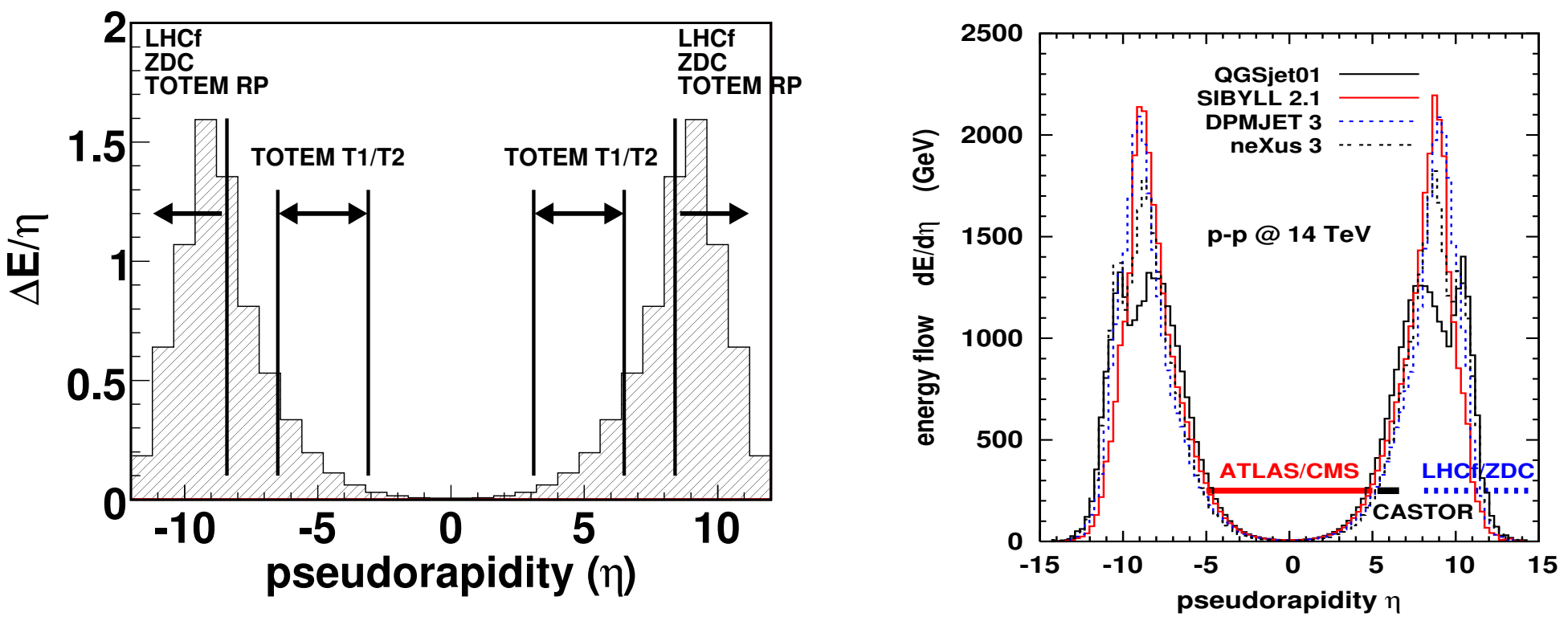

Figure 6: Left:Approximate $p_T$-$\eta$ coverage of LHC detectors. Right:Pseudo-rapidity energy distribution for $pp$ at the LHC predicted by four MC models commonly used in ultra-high energy cosmic rays physics.

The right side of Figure 6 compares the predictions of QGSJET [21], DPMJET [29], NEXUS [30] and EPOS [19,20] for the energy flow ($dE/d\eta$) in $pp$ collisions at $\sqrt{s} = 14\,TeV$. In the range covered by detectors like CASTOR or TOTEM (around $|\eta| \approx 6$) and ZDC or LHCf (beyond $|\eta| \approx 8$, for neutrals), the model predictions differ by up to 60%.

The LHCf experiment [31] placed at 140 m from the interaction point is dedicated for the very forward neutral particle measurements for the efficient cosmic ray model tuning. LHCf will take the data in the early stage of the LHC commissioning. Figure 7 shows the comparison of models for the neutral particle distributions (neutrons, $\pi^0$) [28]. The figure demonstrates the potential of LHCf experiment to distinguish the models.

A good test of the fundamental properties of the hadronic interaction models is provided by the multiplicity distribution of charged particles. Figure 8 shows the MC model predictions for the different multiplicity distributions at LHC [28]. The discrepancies at LHC energy range can be larger than a factor of two in the shape of distributions. The charged multiplicity distribution will be one of the first measurements of the LHC experiments and provide reliable constrain for CR interaction models.

Measurement of forward particle production in $pp$, $pA$, and $AA$ collisions will thus provide strong constraints on these models and allow for more reliable determinations of the CR energy and composition at the highest energies.

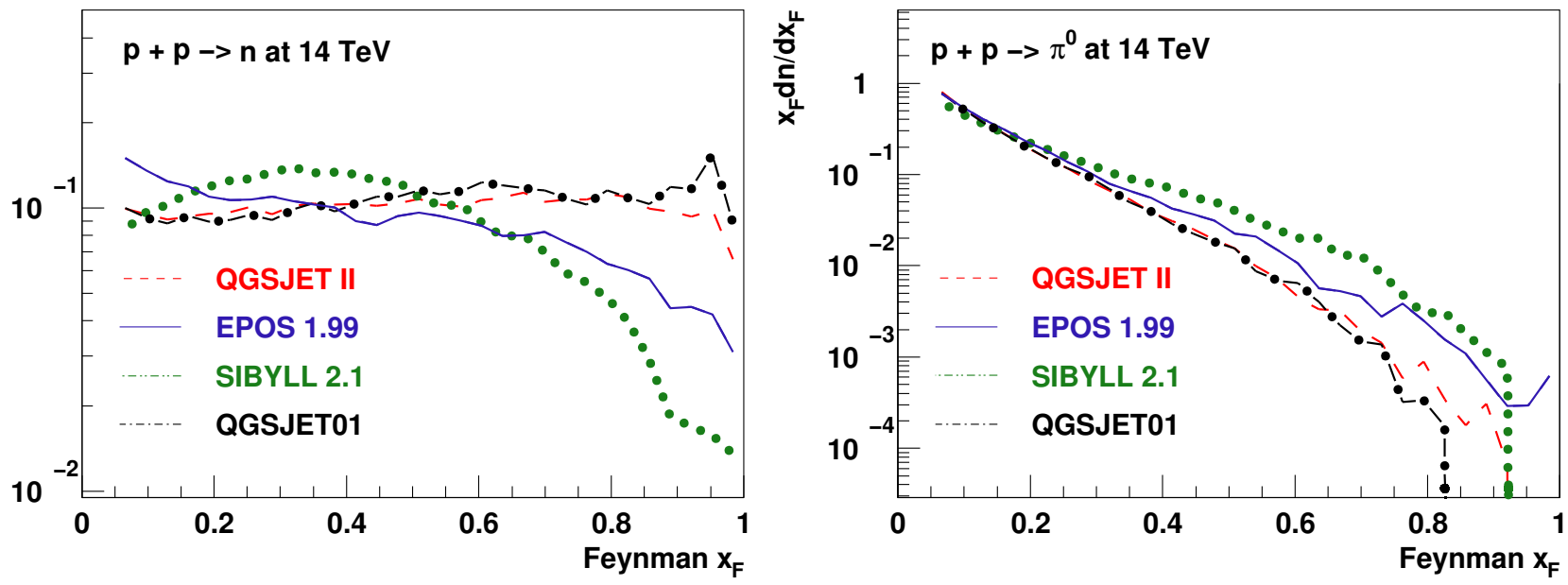

Figure 7: Feynman-$x_F$ distribution of forward neutrons and $\pi^0$ from $pp$ collision at 14 TeV [28].

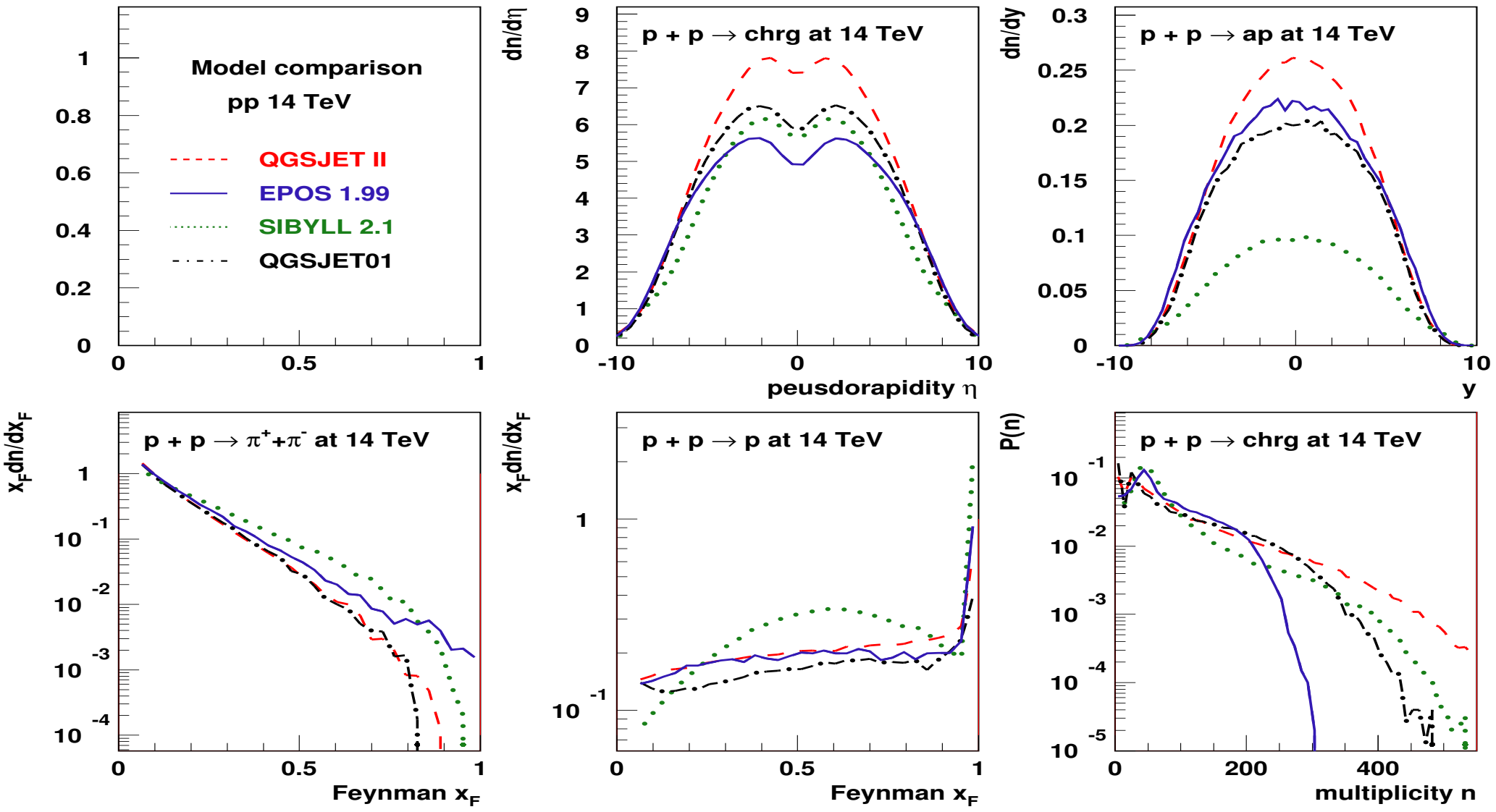

Figure 8: Multiplicity distribution of $pp$ collision at 14 TeV energy calculated with EPOS 1.99, QGSJET II, QGSJET 01 and SIBYLL 2.1 Monte Carlo models [28].

# 5 Conclusions

The energy and mass of the primary ultra-high energy cosmic rays are obtained with the help of Monte Carlo models of hadronic interaction. These models strongly depend on the experimental measurements at collider experiments, in particular in the forward region.

The HERA experiments provide a wealth of measurements of leading baryon production in $ep$ interactions. These measurements give an important input for an improved theoretical understanding of the proton fragmentation mechanism and help to reduce the uncertainties in the model predictions for cosmic ray showers. Forward measurements at LHC in $pp$, $pA$ and $AA$ collisions will provide further strong constraints to calibrate and tune these models and make more reliable predictions for the cosmic rays energy and composition.

## Acknowledgements

I would like to thank the organisers for this interesting, stimulating and enjoyable conference.

## References

[1] T. Stanev, *"Cosmic Rays and Extensive Air Showers"*, these proceedings.

[2] K. Werner, *"Current concepts in theory and modelling of high energy hadronic interactions"*, these proceedings.

[3] J. Hörandel, *"Test of hadronic interaction models with EAS data"*, these proceedings.

[4] J. Knapp *el al., Astropart. Phys.* **19** (2003) 77, [astro-ph/0206414].

[5] L. Anchordoqui *et al., Ann. Phys.* **314** (2004) 145, [hep-ph/0407020].

[6] D. d'Enterria, R. Engel, Th. McCauley and T. Pierog, [arXiv:0806.0944[astro-ph]].

[7] R. Engel, *Nucl. Phys. Proc. Suppl.* **75A** (1999) 62, [astro-ph/9811225].

[8] R. Engel, *Nucl. Phys. Proc. Suppl.* **151** (2006) 437, [astro-ph/0504358].

[9] J. D. Sullivan, *Phys. Rev.* **D5** (1972) 1732.

[10] M. Bishari, *Phys. Lett.* **B38** (1972) 510.

[11] H. Holtmann et al., *Phys. Lett.* **B338** (1994) 363.

[12] B. Kopeliovich, B. Povh and I. Potashnikova, *Z. Phys.* **C73** (1996) 125, [hep-ph/9601291].

[13] A. Bunyatyan, *"Soft Interaction Processes at HERA: Leading Baryon Production, Multiple Interactions"*, these proceedings.

[14] S. Chekanov et al., *JHEP* **06** (2009) 074, [arXiv:0812.2416[hep-ex]].

[15] K. Charchula, G. A. Schuler and H. Spiesberger, *Comput. Phys. Commun.* **81** (1994) 381.

[16] H. Jung, *Comp. Phys. Commun.* **86** (1995) 147.

[17] A. Szczurek, N.N. Nikolaev and J. Speth, *Phys. Lett.* **B428** (1998) 383, [hep-ph/9712261].

[18] H1 Collaboration, *"Leading Neutron Production in DIS at HERA"*, H1prelim-08-111.

[19] H. J. Drescher *et al., Phys. Rept.* **350** (2001) 93, [hep-ph/0007198].

[20] K. Werner, F.-M. Liu and T. Pierog. *Phys. Rev.* **C74** (2006) 044902, [hep-ph/0506232].

[21] N. N. Kalmykov, S. S. Ostapchenko and A. I. Pavlov, *Nucl. Phys. Proc. Suppl.* **52B** (1997) 17.

[22] S. Ostapchenko, *Nucl. Phys. Proc. Suppl.* **151** (2006) 143, [hep-ph/0412332].

[23] S. Ostapchenko, *Phys. Rev.* **D74** (2006) 014026, [hep-ph/0505259].

[24] S. Ostapchenko, *Phys. Lett.* **B636** (2006) 40, [hep-ph/0602139].

[25] R. S. Fletcher, T. K. Gaisser, P. Lipari and T. Stanev, *Phys. Rev.* **D50** (1994) 5710.

[26] J. Engel, T. K. Gaisser, T. Stanev and P. Lipari, *Phys. Rev.* **D46** (1992) 5013.

[27] R. Engel, T. K. Gaisser, T. Stanev and P. Lipari, *"Air shower calculations with the new version of SIBYLL"*, prepared for 26th International Cosmic Ray Conference (ICRC 99), Salt Lake City, Aug 1999.

[28] Model predictions provided by T. Pierog.

[29] J. Ranft, *"DPMJET version II.5: Sampling of hadron hadron, hadron nucleus and nucleus nucleus interactions at accelerator and cosmic ray energies according to the two-component dual parton model: Code manual"*, [hep-ph/9911232].

[30] K. Werner, F. M. Liu, S. Ostapchenko and T. Pierog, *Heavy Ion Phys.* **21** (2004) 279, [hep-ph/0306151].

[31] O. Adriani et al., *"Technical proposal for the CERN LHCf experiment: Measurement of photons and neutral pions in the very forward region of LHC"*, CERN-LHCC-2005-032.

# Unusual Event Alignment Topologies in Cosmic Rays and Expectation for the LHC

*A. De Roeck*[1], *I.P. Lokhtin*[2], *A.K. Managadze*[2], *L.I. Sarycheva*[2], *A.M. Snigirev* [2]
[1]CERN, 1211 Geneva 23, Switzerland
[2]M.V.Lomonosov Moscow State University, D.V.Skobeltsyn Institute of Nuclear Physics, 119991, Vorobievy Gory, Moscow, Russia

Based on the observation of the so called alignment phenomenon in cosmic ray emulsion experiments, namely a strong collinearity of shower cores related to coplanar scattering of secondary particles in the interaction, events with an unusual topology in the mid-forward rapidity region are expected to be produced at the LHC.

## 1 Introduction

The intricate phenomenon of coplanarity of the most energetic cores of $\gamma$-ray-hadron secondary particles (families) has been observed since a long time ago in mountain-based [1, 2] and stratospheric [3] x-ray-emulsion chamber experiments. So far no simple satisfactory explanation for these cosmic ray observations as been given, in spite of numerous attempts (see, e.g. [2, 4] and references therein). Among these explanations, a jet-like mechanism [5] looks very attractive and can give a natural explanation of the alignment of three spots, i.e. the particles resulting from the energy deposits of the secondaries in showering material, along a straight line, resulting from momentum conservation in a simple parton scattering picture. The relation between the observed alignment of spots in the x-ray film in cosmic ray emulsion experiments and the characteristics of events dominated by jets at very high energies, was tested in our earlier work [6, 7]. Based on these studies we now report on predictions from the alignment phenomenon for the CERN Large Hadron Collider (LHC).

## 2 Problem under Consideration

For clarity, let us recall that in the Pamir experiment [1, 2] families with a total energy of the $\gamma$-quanta above a given threshold, and with the requirement of at least one hadron (identified by the travel length in the material), were selected and analyzed. The alignment effect becomes clearly apparent for event with $\sum E_\gamma > 0.5$ PeV (which corresponds to interaction energies with a CMS energy $\sqrt{s} \gtrsim 4$ TeV). The families are produced dominantly by an incident proton with an energy $\gtrsim 10^4$ TeV interacting at a height $h$ of several hundred meters to several kilometres in the atmosphere above the detector [1, 2]. The collision products are observed as spots within a radial distance $r_{\max}$ up to several centimetres in the emulsion, where the spot separation $r_{\min}$ is of the order of 1 mm.

Our analysis [6, 7] shows that a jet-like mechanism can, in principle, provide an explanation of the results of these emulsion experiments. For this explanation to work it is necessary that particles from both hard jets (with rapidities close to zero in the centre-of-mass system) hit the detection region as a result of the large Lorentz factor from the transformation from the centre-of-mass system to the laboratory one. This is possible when the combination of $h$, $\sqrt{s}$ and $r_{\max}$ meets the following condition:

$$2hm_p/\sqrt{s} \lesssim kr_{\max}, \tag{1}$$

where $m_p$ is the proton mass. A value of $k \sim 1/2$ is needed in order to have particles with adjacent positive and negative rapidities in the centre-of-mass system to hit the detection region. At a height of $h = 1000\,\text{m}$ (which is a standard height used in emulsion experiment estimations) and $r_{\max} = 15$ mm, the condition in (1) is fulfilled for an energy energy $\sqrt{s} \gtrsim 270$ TeV , i.e. much higher than the LHC high energy range for heavy ions and protons $\sqrt{s} \simeq 5.5 \div 14$ TeV and the threshold interaction energies after which the alignment appears $\sqrt{s_{\text{eff}}} \simeq 4$ TeV [1, 4], corresponding to the alignment phenomenon. Eq. (1) can be fulfilled and at the LHC energy (14 TeV) also, but at the considerably less height $h \lesssim 50$ m which is different from the traditional emulsion experiment assumption of about 1 km.

On the other hand if particles from the central rapidity region and the jet-like mechanism are insufficient to describe the observed alignment, and there is another *still unknown* mechanism of its appearance at the energy $\sqrt{s} \sim 5.5 \div 14$ TeV and the accepted height $h \sim 1000\,\text{m}$, then in any case some sort of alignment should arise at the LHC too in the mid-forward rapidity region (following from the laboratory acceptance criterion for, e.g., pp collisions) [6, 7]:

$$r_{\min} < r_i \Longrightarrow \eta_i < \eta_{\max} = \ln(r_0/r_{\min}) \simeq 4.95, \tag{2}$$

$$r_i < r_{\max} \Longrightarrow \eta_i > \eta_{\min} = \ln(r_0/r_{\max}) \simeq 2.25, \tag{3}$$

where

$$r_0 = 2h/e^{\eta_o} = 2hm_p/\sqrt{s}, \tag{4}$$

$\eta_0 = 9.55$ is the rapidity of centre-of-mass system in the laboratory reference frame, $\eta_i$ is the particle rapidity in the centre-of-mass system, $r_i$ is the radial particle spacing in the $x$-ray film. In that case at the LHC a *strong azimuthal anisotropy of energy flux will be observed, namely almost all of the energy will be deposited along a radial direction* for all events with the total energy deposition in the rapidity interval (2) and (3) above a threshold. Note that at present there are no models or theories which give such azimuthal anisotropy which can explain the experimentally observed alignment phenomenon at $\sqrt{s} \gtrsim \sqrt{s_{\text{eff}}} \simeq 4\text{TeV}$ and $h \sim 1000\,\text{m}$ [1, 4].

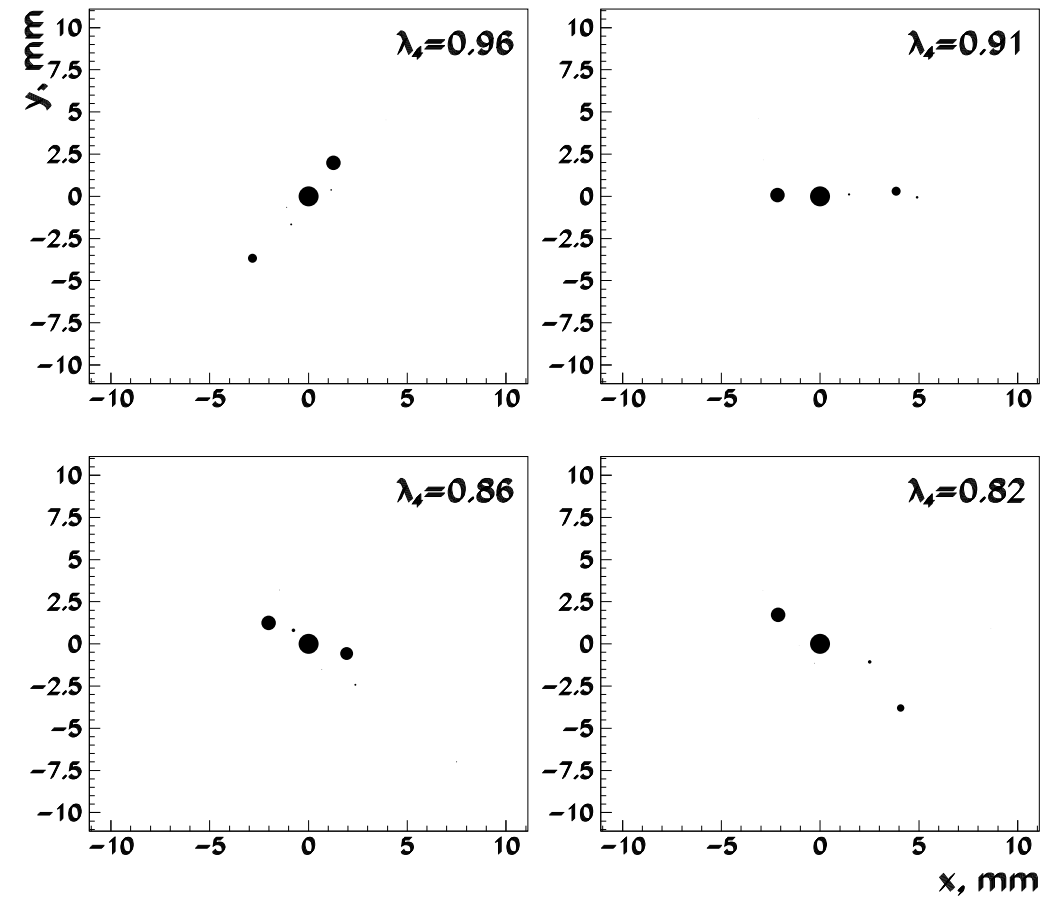

Figure 1: Samples of core distributions for simulated events with $E_{\Sigma}^{\text{thr}} = 10$ PeV and $\lambda_4 > 0.8$. The size of spots is proportional to their energy (except for the central spot which is not to scale).

# 3 Topology of Alignment Events

Here we would like to draw the attention to the unusual topology of such alignment events in the centre-of-mass system, in order to design a selection at the LHC. As was mentioned above, hard enough i.e. high $p_T$ jets ($p_T^{\rm jet} \gtrsim 3$ TeV), centrally produced in the centre-of-mass system, can imitate the appropriate topology of events with the large degree of alignment $P_N$ in the laboratory system at a small enough height $h \sim 50$ of the primary interaction. The spatial distribution of the most energetic clusters in the transverse ($xy$)-plane for a few generated events along with the corresponding values of $\lambda_N$ are presented in Figs. 1 and 2 (from the work [6]) to give the reader a feeling for the topology of alignment events in the laboratory reference frame close to experimentally observed ones [1, 2].

The alignment parameter $\lambda_N$, for $N$ spots, is conventionally defined as [2]:

$$\lambda_N = \frac{\sum_{i\neq j\neq k}^{N} \cos(2\phi_{ijk})}{N(N-1)(N-2)}, \tag{5}$$

Here $\phi_{ijk}$ is the angle between two vectors $(\mathbf{r_k} - \mathbf{r_j})$ and $(\mathbf{r_k} - \mathbf{r_i})$ (for the central spot $\mathbf{r} = \mathbf{0}$). This parameter characterises the location of $N$ points along a straight line and varies from $-1/(N-1)$ to 1. For instance, in the case of a symmetrical, close to the most probable random configuration of three points in a plane (the equilateral triangle) $\lambda_3 = -0.5$. The case of perfect alignment corresponds to $\lambda_N = 1$ , when all points lie exactly along a straight line, while for an isotropic distribution $\lambda_N < 0$. The degree of alignment $P_N$ is defined as a fraction of events with $\lambda_N > 0.8$ [2], for events with a number of points larger or equal to $N$.

The threshold on the total energy of all $(N-1)$ selected clusters $E_\Sigma^{\rm thr} \sim E_{\rm lab}/2$ (without taking into account the energy deposition in the central cluster around $r = 0$), $\sum_{l=1}^{N-1} E_l > E_\Sigma^{\rm thr}$, was introduced to select the events with hard jets. To be specific we consider a collision of two protons and fix a primary energy in the laboratory system $E_{\rm lab} \simeq 9.8 \times 10^4$ TeV, that is equivalent to $\sqrt{s} \simeq 14$ TeV — the maximum energy at LHC – (the rapidity shift from laboratory system to CMS is then $\eta_0 \simeq 9.55$ after a transformation from the centre-of-mass system to the laboratory one). To simulate $pp$ collision a these energies we use the Monte Carlo generator PYTHIA [8], which is expected to give a fair description of –be it low multiplicity– jet events in hadron-hadron interactions and has been tuned using the available experimental data.

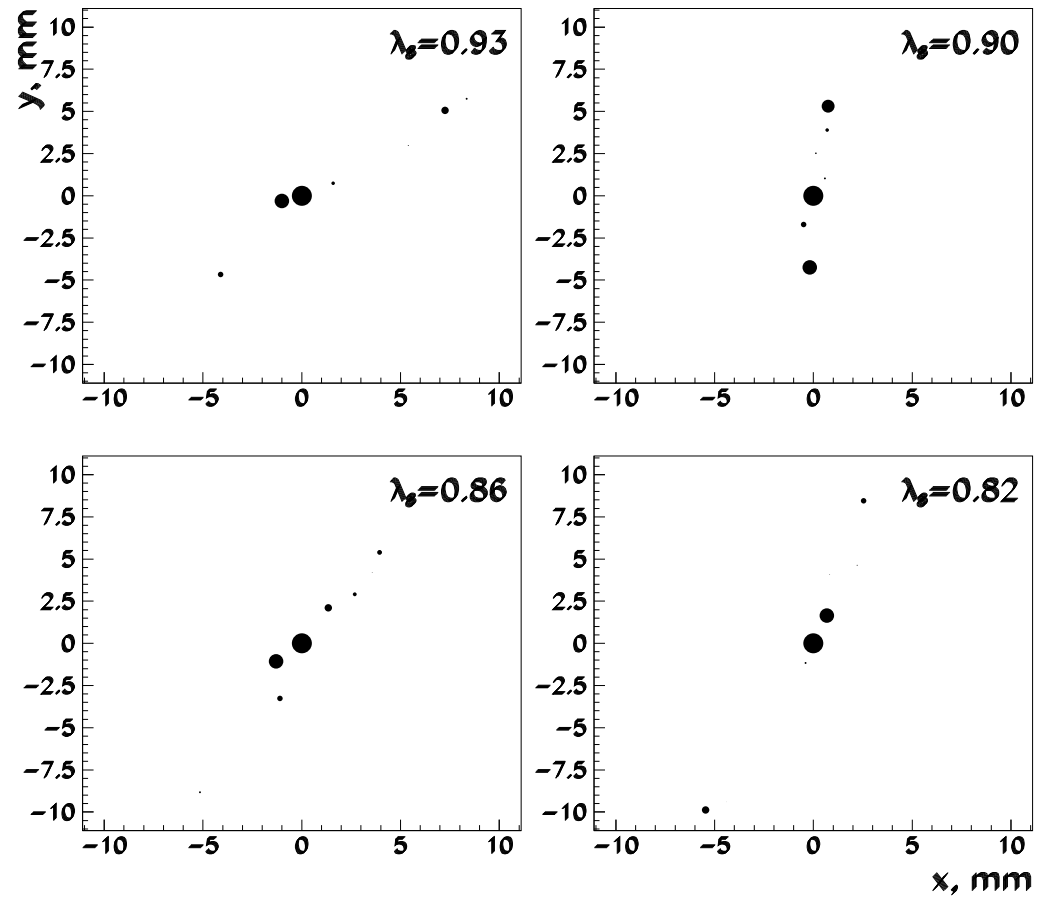

Figure 2: Samples of core distributions for PYTHIA simulated events with $E_\Sigma^{\rm thr} = 10$ PeV and $\lambda_8 > 0.8$. The size of spots is proportional to their energy (except for the central spot which is not to scale).

In the centre-of-mass system the alignment events with jets with sufficiently high $p_T$ have two pronounced concentrations of energy deposition in $\eta \times \phi$-space (rapidity×azimuthal angle) with the azimuthal separation close to $\pi$. The typical structure (topology) of energy deposition for such events is presented in Fig. 3. As it was shown [6, 7], for $p_T^{\text{jet}} \gtrsim 3$ TeV, particles from these hard jets together with particles flying close to the $z$-axis (within the transverse radius $\lesssim 1$ mm in the laboratory reference frame) result in a degree of alignment $P_N$ comparable with the experimentally observed one (Fig. 4, from [6]).

We can introduce a quantitative measure of energy deposition allowing the alignment events to be selected in the centre-of-mass system, namely:

$$v_2^{p_T} = \frac{1}{N_{\text{event}}} \sum_{\text{event}} \frac{\sum_i p_{Ti}^2 \cos 2(\phi_i - \phi_{\text{axis}})}{\sum_i p_{Ti}^2}, \tag{6}$$

where $\phi_i$ is the azimuthal angle for the $i$th particle with the transverse momentum $p_{Ti}$. The sum runs over all particles under consideration. For two jet events

$$\phi_{\text{axis}} = (\phi_{\text{jet1}} + \phi_{\text{jet2}} - \pi)/2, \tag{7}$$

where $\phi_{\text{jet1}}$ and $\phi_{\text{jet2}}$ are the azimuthal angles for the first and second jets respectively. Experimentally they can be defined as the directions of leading particles. Without a weight factor $p_{Ti}$ this definition (6)

$$v_2 = \frac{1}{N_{\text{event}}} \sum_{\text{event}} \frac{\sum_i \cos 2(\phi_i - \phi_{\text{axis}})}{\sum_i} \tag{8}$$

coincides with the definition of the elliptic flow coefficient relative to the azimuthal angle of the reaction plane (instead of $\phi_{\text{axis}}$) used in a standard flow analysis [10, 11].

Figure 5 shows the elliptic anisotropy coefficients $v_2^{p_T}$ and $v_2$ as functions of jet hardness $p_{T\text{hard}}^{\text{min}}$ (the minimum transverse momentum of hard parton-parton subscattering, a parameter of PYTHIA) for jets from the central rapidity region. Thus events with a high degree of alignment, e.g. the upper dashed curve in Fig. 4, can be also characterised by a relatively large value of the topological parameters of the energy anisotropy in the centre-of-mass system:

$$\begin{aligned} & v_2^{p_T} \gtrsim v_2^{p_T}(p_{T\text{hard}}^{\text{min}} = 3\text{TeV}) = 0.98 \\ \text{or} \quad & v_2 \gtrsim v_2(p_{T\text{hard}}^{\text{min}} = 3\text{TeV}) = 0.6 \end{aligned} \tag{9}$$

as it follows from Fig. 5 and from our previous study [6, 7]. One should note that the alignment parameter $\lambda_N$ is defined in the $r \times \phi$-space (radial distance×azimuthal angle) of the laboratory reference frame while the elliptic anisotropy coefficients $v_2^{p_T}$ and $v_2$ are defined in the $p_T \times \phi$-space (transverse

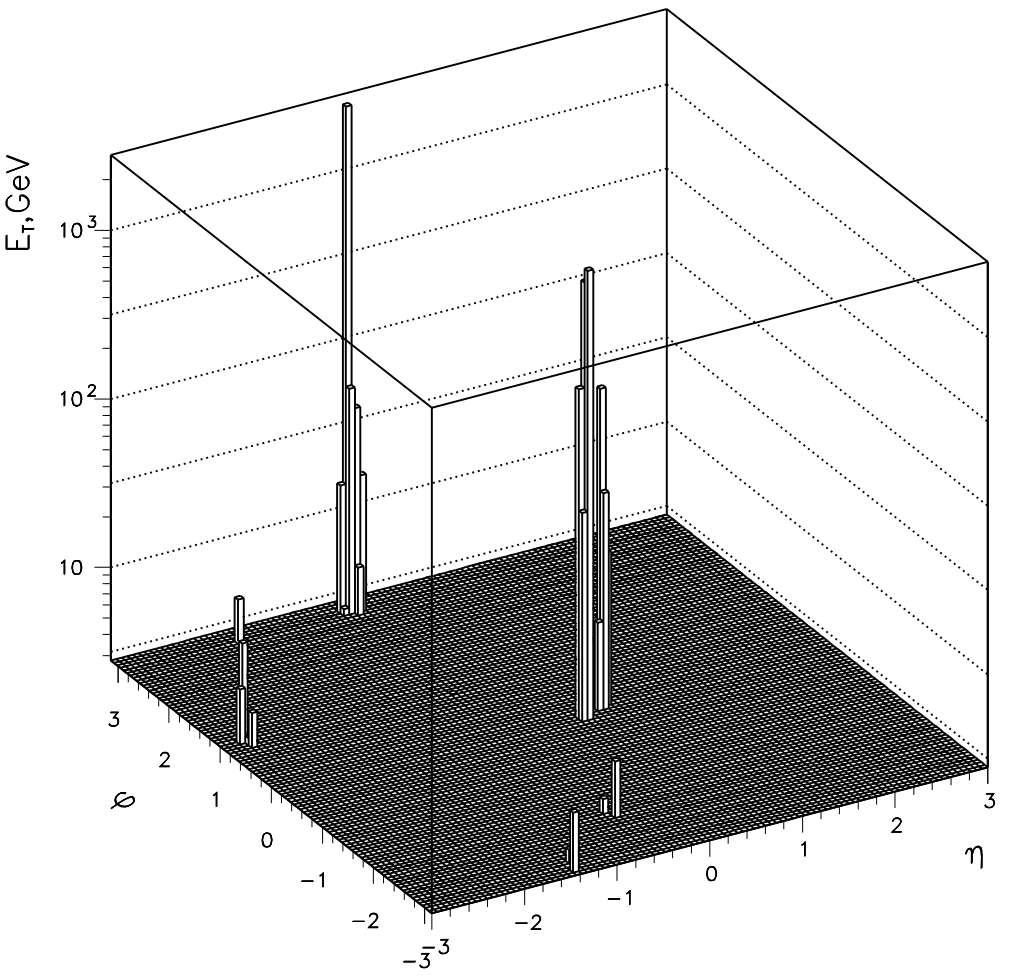

Figure 3: The typical energy deposition in the centre-of-mass system for alignment events in the laboratory frame.

momentum×azimuthal angle) of the centre-of-mass system. $\lambda_N$ characterises the location of $N$ points just along the straight line while $v_2^{p_T}$ and $v_2$ characterise the narrowness of energy thrust which is necessary because of the Lorentz invariance of the azimuthal angle (but not yet sufficient) to allow for the observation of a large degree of alignment $P_N$. Therefore we conclude that a relatively large value of the elliptic anisotropy coefficients $v_2^{p_T}$ and $v_2$ can result in a large value of the alignment parameter $\lambda_N$. We prefer also to consider $v_2^{p_T}$ and $v_2$ as a quantitative measure of alignment events in the centre-of-mass system instead of some analog of $\lambda_N$ in the $p_T \times \phi$-space because the radial particle spacing $r_i$ in the $x$-ray film is practically independent of the value of the transverse momentum for ultrarelativistic particles ($p_{Ti} \gg m_i$) and is mainly determined by the particle rapidity $\eta_i$ [6, 7] (see also Eqs. (2) and (3)).

# 4 Discussion and Conclusions

Let us discuss what follows from the experimental observation of alignment phenomenon and our proposal to describe it with a jet-like mechanism. We found that only jets with high enough $p_T$ can imitate the appropriate topological characteristics of alignment events. These hard jets can be selected in a "natural" physical way, namely, by the introduction of the threshold on the energy deposition in the detection region [6, 7]. We are not interested in the exact relation between the threshold energy and the jet hardness but note that the introduction of a threshold on the energy deposition in the detection region results in an appearance of the azimuthal anisotropy (preferred direction) of energy deposition. This azimuthal anisotropy (the elliptic anisotropy coefficients $v_2^{p_T}$ and $v_2$) becomes stronger with increasing threshold value. One should note that in the Pamir experiment the events with a threshold on the total energy deposition (and by taking into account the energy deposition in the central cluster around $r = 0$ unlike the procedure used in this paper) were selected and analyzed. The Pamir Collaboration consequently concludes the presence of threshold on *the energy of primary interaction* for the onset of the alignment phe-nomenon, and estimates this threshold energy to be $\sqrt{s} \gtrsim \sqrt{s_{\rm eff}} \simeq 4\,{\rm TeV}$. While we conclude that at the *fixed* energy of primary interaction the threshold on the total energy of all $(N-1)$ selected clusters (without taking into account the energy deposition in the central cluster around $r = 0$) must exist *also* to allow for the observation of a large degree of alignment, since for alignment events the energy of all these $(N-1)$ selected clusters is comparable with the energy in the central cluster around $r = 0$. Our studies with jets strongly supports this conclusion. The central cluster is a special one and should be treated separately to compare the colliding beams experiment results with those of fixed target ones more consistently. For instance, all ultrarelativistic particles ($p_{Ti} \gg m_i$) from the rapidity interval $4.95 \lesssim \eta_i \lesssim 9.55$ in the centre-of-mass system form one central cluster around $r = 0$ with the size $\sim 1$ mm in the laboratory reference frame at the accepted height $h \sim 1000$ m and the LHC energy due to the strong Lorentz compression. Therefore the energy of central cluster is not taken into account in our threshold analysis.

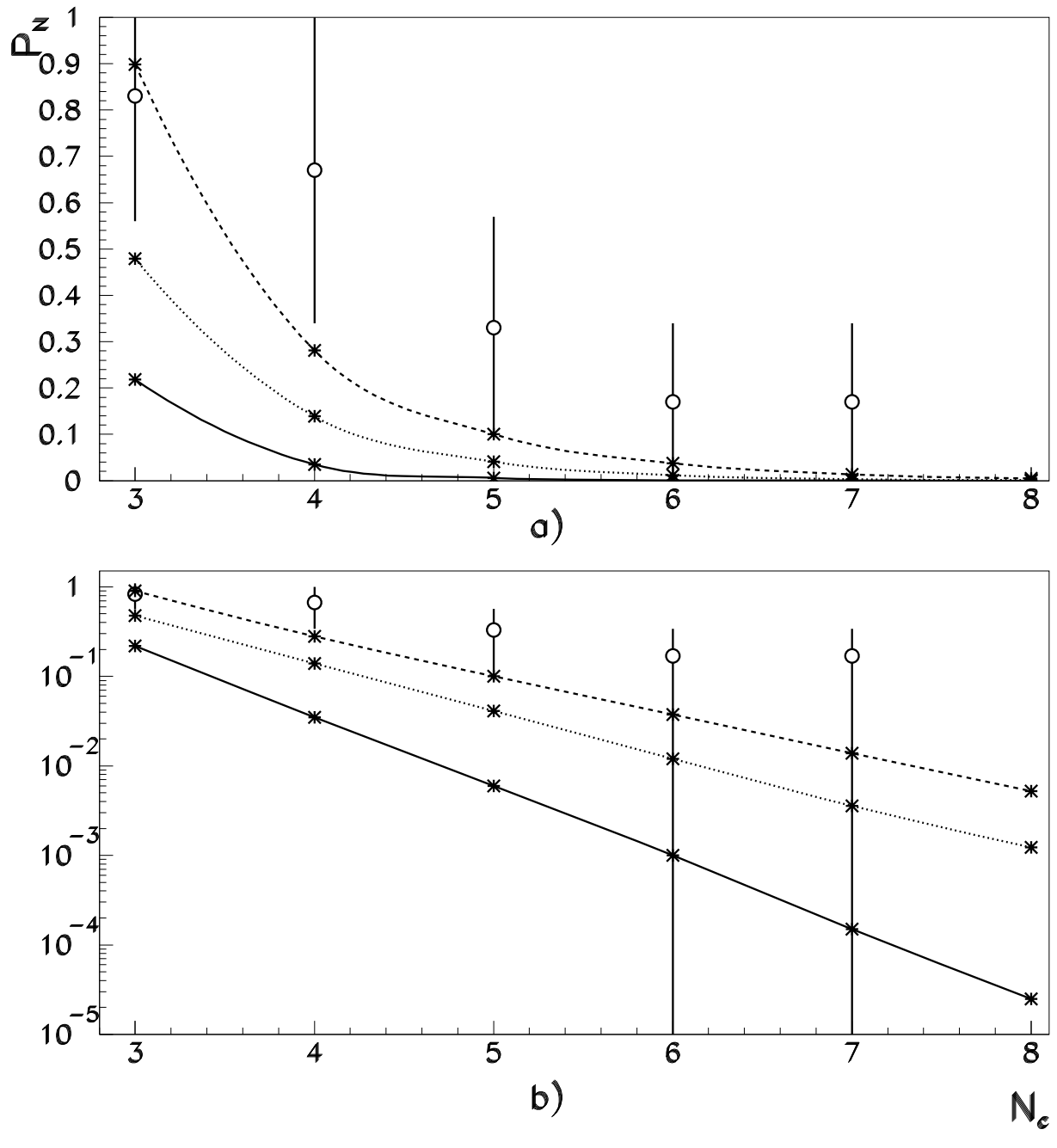

Figure 4: The alignment degree $P_N$ as a function of cluster number $N = N_c$ at $h = 50$ m and $\sqrt{s} = 14$ TeV in linear (a) and logarithmic (b) scales. The solid curve is the result (coincident with one at $h = 1000$ m) without restriction on the minimum value of process hardness $p_{T\text{hard}}^{\min}$, the dotted curve — at $p_{T\text{hard}}^{\min} = 300$ GeV, the dashed curve — at $p_{T\text{hard}}^{\min} = 3$ TeV. Points (○) with errors are experimental data from [9].

At the accepted height $h \sim 1000$ m the particles from the restricted rapidity interval between (2) and (3) in the centre-of-mass system form all remaining $(N-1)$ energetic selected clusters in the laboratory reference frame. Note that the absolute rapidity interval can be shifted corresponding to the height: it is necessary only that the difference $(\eta_{\rm max} - \eta_{\rm min})$ is equal to $\simeq 2.7$ in accordance with the variation of the radial distance by a factor of $\sim 15$ $(r_{\rm max}/r_{\rm min} = 15)$ due to the relationship $r_i \simeq r_0/e^{\eta_i}$ (independently of $r_0(h)$). A strong azimuthal anisotropy (the large elliptic anisotropy coefficients $v_2^{p_T}$ and $v_2$) must be experimentally observed in this restricted rapidity interval (2) and (3) in the centre-of-mass system for the energy deposition above threshold, if the alignment phenomenon exists.

In the mid-forward rapidity region discussed for this analysis, jets can also lead to events with the required values for the thrust of energy deposition in the $\eta \times \phi$-space but with the large dispersion of the azimuthal separation relative to $\pi$ and not enough large values of $v_2^{p_T}$ and $v_2$. The reason is simple: jets are not hard enough to provide for the strong momentum correlations (memory) with the primary scattering plane due to the kinematic restriction [12]

$$e^{\eta^{\rm jet}} p_T^{\rm jet} \lesssim \sqrt{s}/2. \qquad (10)$$

For instance if $\eta^{\rm jet} = 4$ then $p_T^{\rm jet}({\rm max}) \simeq 130\,{\rm GeV}$, $v_2^{p_T}({\rm max}) \simeq 0.83$ and $v_2({\rm max}) \simeq 0.38$ as one can estimate from Fig. 5. Such a maximal azimuthal anisotropy is small in comparison with needed one (9) for the observation of a large degree of alignment.

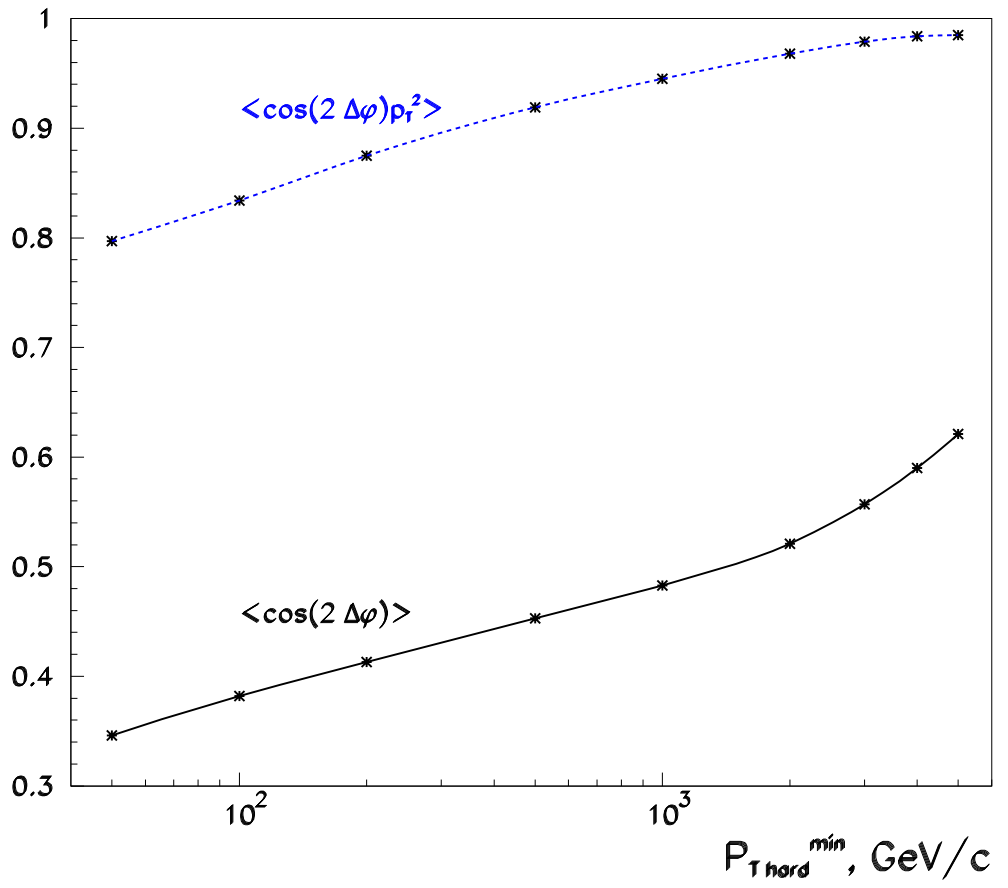

Figure 5: The elliptic anisotropy coefficients $v_2^{p_T}$ and $v_2$ as functions of jet hardness $p_{T{\rm hard}}^{\rm min}$ for jets from central rapidity region $|\eta| < 2$.

Thus basing on our dealing with jets and the existence of the alignment phenomenon we predict anomalously large values (on the level of our estimation (9)) of the elliptic anisotropy coefficients $v_2^{p_T}$ and $v_2$ in comparison with its "background" values from jets in the mid-forward rapidity interval (2) and (3) beginning from some threshold on the energy deposition in this rapidity region. The order of value of this transverse threshold energy can be estimated again from the jet background: $\sim 100$ GeV and higher since we do not know any mechanism of a large degree of alignment in this case.

The suggested investigation of the azimuthal anisotropy of the energy deposition in dependence on the threshold energy both in pp and in heavy ion collisions (to differentiate between hadronic and nuclear interaction effects) at the LHC can clarify the origin of the alignment, give the new restrictions on the values of height and energy, and possibly discover new still unknown physics.

## Acknowledgements

It is pleasure to thank A.I. Demianov, S.V. Molodtsov, R.A. Mukhamedshin, L.G. Sveshnikova and G.T. Zatsepin for discussions. A.M.S is specially thankful to the organisers of EDS'09 for the warm welcome and hospitality. This work is supported by Russian Foundation for Basic Research (grants No 08-02-91001 and No 08-02-92496), Grants of President of Russian Federation (No 107.2008.2 and No 1456.2008.2) and Dynasty Foundation.

## References

[1] Pamir Collaboration, A. Borisov *et. al.*, in Proceedings of 4th International Symposium on Very High Energy Cosmic Ray Interactions, Beijing, edited by D. Linkai, 4 (1986);
Pamir Collaboration, in Proceedings of the 21st International Cosmic Ray Conference, Adelaide, Australia (1989), edited by R.J.Protheroe (University of Adelaide, Australia), 227 (1990);
S.A. Slavatinsky, in Proceedings of the 5th International Symposium on Very High Energy Cosmic Ray Interactions, Lodz, Poland (1988), edited by M. Giler (University of Lodz, Lodz, Poland), 90 (1989).

[2] V.V. Kopenkin, A.K. Managadze, I.V. Rakobolskaya and T.M. Roganova, Phys. Rev. D **52**, 2766 (1995).

[3] A.V. Apanasenko *et. al.*, in Proceedings of 15th International Cosmic Ray Conference, Plovdiv, **7** 220 (1977);
A.K. Managadze *et. al.*, in Proceedings of 27th International Cosmic Ray Conference, Hamburg, **1** 1426 (2001).

[4] R.A. Mukhamedshin, J. High Energy Phys. IHEP**05**, 049 (2005).

[5] F. Halzen and D.A. Morris, Phys. Rev. D **42**, 1435 (1990).

[6] I.P. Lokhtin, A.K. Managadze, L.I. Sarycheva and A.M. Snigirev, Eur. Phys. J **C 44**, 51 (2005).

[7] I.P. Lokhtin, A.K. Managadze, L.I. Sarycheva and A.M. Snigirev, Phys. Atom. Nucl. **69**, 113 (2006).

[8] T. Sjostrand, Comp. Phys. Com. **135**, 238 (2001).

[9] V.V. Kopenkin, A.K. Managadze, I.V. Rakobolskaya and T.M. Roganova, Izv. Rus. Akad. Nauk. Ser. Fiz. **58**, 13 (1994).

[10] S. Voloshin and Y. Zang, Z. Phys. C **70**, 665 (1996).

[11] A.M. Poskanzer and S.A. Voloshin, Phys. Rev. C **58**, 1671 (1998).

[12] Yu.L. Dokshitzer, D.I. Dyakonov and S.I. Troyan, Phys. Rep. **58**, 269 (1980).

# Part VIII

# Photon-induced Physics

# Ultra-Peripheral Collisions in PHENIX

*Zaida Conesa del Valle, for the PHENIX Collaboration*

Laboratoire Leprince-Ringuet, Ecole Polytechnique - CNRS/IN2P3, Palaiseau, France

Ultra-peripheral nuclei collisions provide means to study photon-induced interactions in a nuclear environment. We discuss the PHENIX collaboration results on $J/\psi$ and $e^+e^-$ photoproduction in Au+Au ultra-peripheral collisions at $\sqrt{s_{NN}} = 200$ GeV [1, 2]. Their production cross-section and transverse momentum spectra are presented. The results are compared and found to be consistent with various theoretical calculations.

## 1 Motivation

High energy particle collisions are the experimental tools to improve our understanding of the elementary particles and their interactions [1]. Photon-induced reactions are an interesting approach to study strong and electromagnetic interactions, complementarily to $e^+e^-$, $ep$ (DIS), $pp$ and $p\bar{p}$ collisions. Traditionally, $\gamma\gamma$ and $\gamma p$ interactions have been studied in fixed target experiments with electron beams (e.g. PEP, LEP), in $e^+e^-$ colliders, and with the HERA electron-proton collider. Recently, the photon fluxes generated at proton or nuclei collisions have attained a large enough luminosity to open the possibility to study $\gamma\gamma$, $\gamma n$ (photon-nucleon) and $\gamma A$ (photon-nuclear) interactions as well. See Fig. 1 (left) for a comparison of the equivalent photon luminosities at various colliders [3]. These studies become experimentally possible in the ultra-peripheral collisions (UPC). UPC are protons or nuclei reactions where the impact parameter is larger than twice the nuclear radius such as there is no strong interaction. The main advantages of hadron vs. lepton colliders are the larger photon luminosity (the photon spectrum is proportional to $Z^2$, the two-photon luminosity goes as $Z^4$) and the opportunity to probe strong electromagnetic fields (coupling $\propto Z\sqrt{\alpha}$ instead of $\propto \sqrt{\alpha}$).

The ultra-peripheral Au+Au collisions at $\sqrt{s_{NN}} = 200$ GeV are possible thanks to the BNL RHIC collider and give access to photon-nucleon and two-photon interactions at a maximum centre-of-mass energy of $W_{\gamma n}^{max} \sim 34$ GeV and $W_{\gamma\gamma}^{max} \sim 6$ GeV respectively [1]. These energies allow the study of photoproduction of dileptons and vector mesons [3]. The STAR collaboration has measured $\rho^0$ photoproduction at RHIC [4], and this proceeding concentrates on the first measurement ever of $J/\psi \to e^+e^-$ and high-mass $e^+e^-$ in UPC heavy-ion collisions with the PHENIX experiment [1, 2]. Exclusive $J/\psi$ photoproduction can proceed via Pomeron-exchange (two-gluon picture) either through coherent photon-nuclear ($\gamma A \to J/\psi$) or incoherent photon-nucleon ($\gamma n \to J/\psi$) reactions. The $J/\psi$ photoproduction cross section is then related to the gluon nuclear distribution functions, $G_A(x, Q^2)$. This allows one to constrain them in the small Bjorken-$x$ region, $x = m_{J/\psi}^2/W_{\gamma A}^2 \cdot e^{\pm y}$ which for $|y| \leq 0.35$ gives $x \sim 10^{-2}$, a relatively unexplored region (see Fig. 1 right [5]), and to eventually probe quarkonia propagation in normal nuclear matter (the so called nuclear absorption). On the other hand, exclusive dilepton ($e^+e^-$) pair production can occur through a pure electromagnetic process ($\gamma\gamma \to e^+e^-$) and tests QED on a strongly interacting regime.

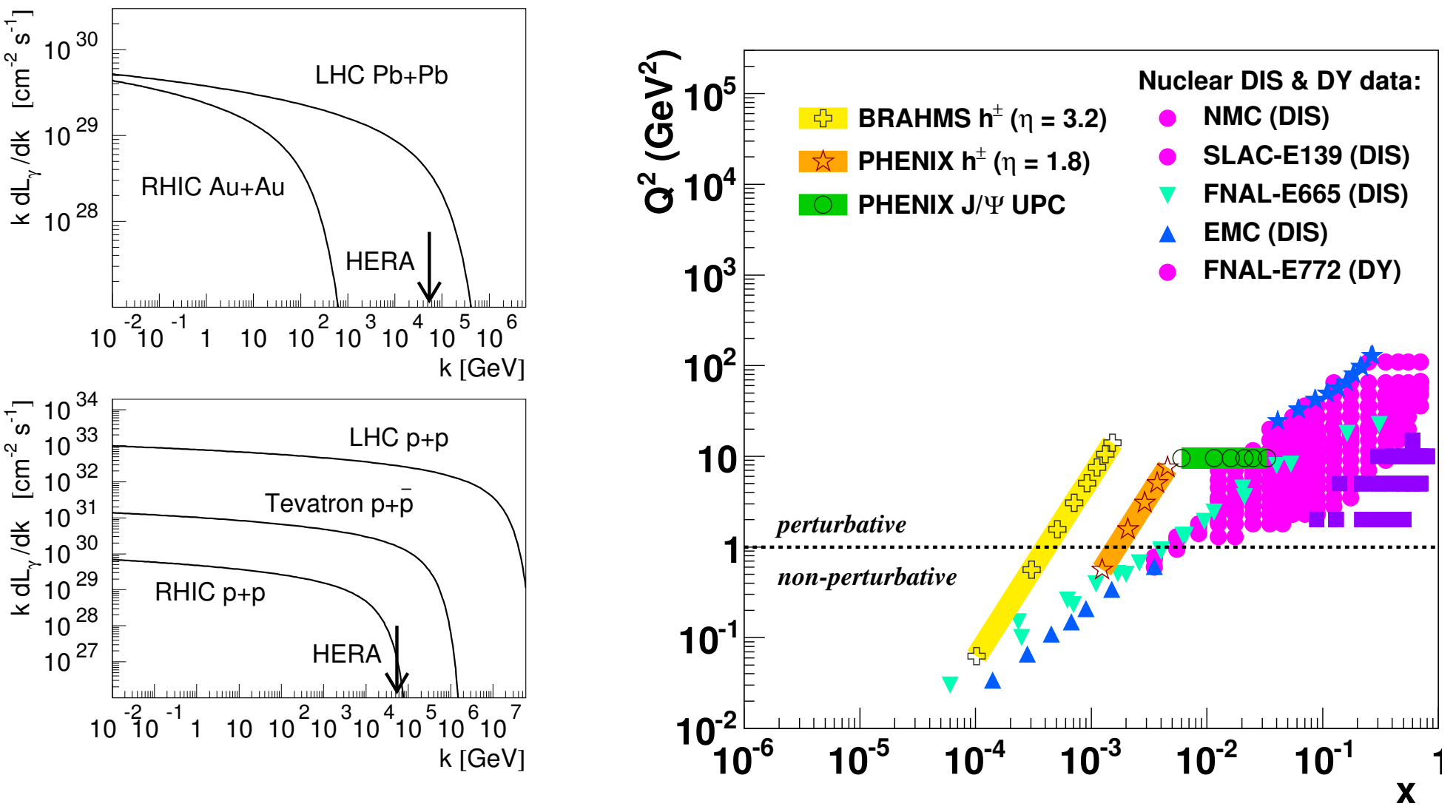

Figure 1: Left: Equivalent photon luminosity in heavy ion (top) and proton (bottom) collisions [3]. The arrow represents the maximum photon energy in HERA *ep* collisions. Right: Kinematic $Q^2$ versus Bjorken-$x$ map of the regions explored to probe the nuclear PDFs [5].

## 2 The Measurement

The measurement of the ultra-peripheral probes of the interaction is experimentally challenging. Here we concentrate on the $J/\psi \to e^+e^-$ measurement in UPC Au+Au collisions at $\sqrt{s_{NN}} =$ 200 GeV. Therefore, we examine the apparatus capabilities to detect and identify electrons and to trigger on UPC.

### 2.1 Experimental Setup

The PHENIX experiment was designed to study the most central heavy-ion collisions at the BNL RHIC collider [2]. The apparatus is formed by four spectrometers (north, south, east, west) and a few global detectors. The forward arms (north and south spectrometers, $1.2 < |\eta| < 2.2$) detect and identify muons, whereas the central arms (east and west spectrometers, $|\eta| < 0.35$) measure electrons, photons and hadrons. Figure 2 presents a scheme of the experimental setup as it was on year 2004 [2]. Minimum bias events are triggered with the help of the Beam-Beam Counters (BBC, at $3.1 < |\eta| < 3.9$) and Zero Degree Calorimeters (ZDC, located at $\pm 18$ m from the interaction point and covering $|\theta| < 2$ mrad). Electron tracking and identification is possible thanks to (from inner to outer): the multi-layer drift chambers (DC), the multi-wire proportional chambers (PC), the Ring-Imaging-Cherenkov detectors (RICH) and the electromagnetic calorimeters (EMCal) with two different technologies: lead-scintillator sandwich (PbSc), and lead-glass Cherenkov (PbGl) calorimeters.

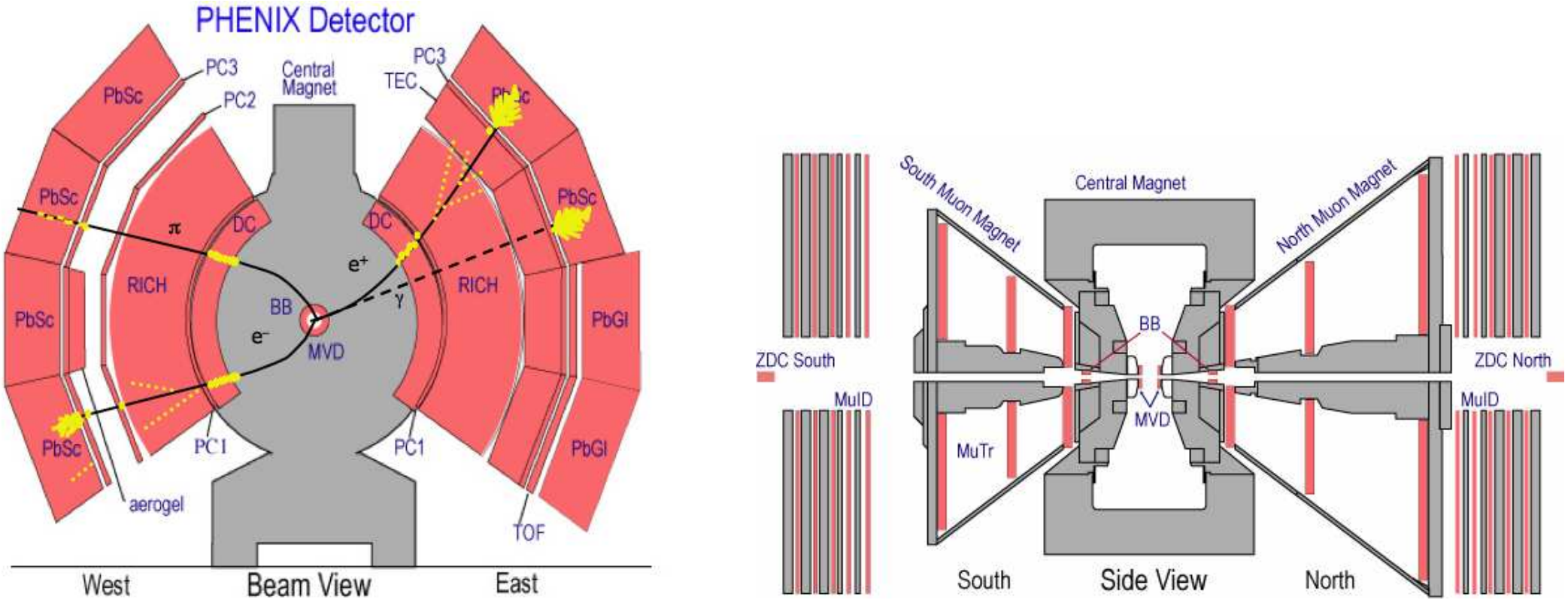

Figure 2: Section of the PHENIX central (left) and forward (right) arms as they were on year 2004 [2]. For illustration, passage of electrons, positrons, photons and pions through the central detectors are depicted.

## 2.2 Trigger Strategy

The strong electromagnetic fields associated with the ultra-relativistic heavy-ions lead to a large probability to exchange additional soft photons which can excite the interacting nuclei. The Giant-Dipole-Resonance (GDR) mechanism is the dominant excitation mode, which most probably decays by emitting neutrons at forward rapidities. The probability of $J/\psi$ coherent production ($\gamma A \to J/\psi$) in coincidence with Au Coulomb excitation leading to the emission of at least one neutron at forward rapidity is of $55 \pm 6\%$, whereas it is of $\approx 100\%$ for $J/\psi$ incoherent production ($\gamma n \to J/\psi$) [6]. This characteristic has been exploited for tagging the UPC events, and, as the probability of exchanging one or several photons factorise, it does not introduce any bias on the determination of the $J/\psi$ photoproduction cross-section. Therefore, the chosen trigger configuration was to:

1. detect these neutrons in one or both ZDC,

2. impose a veto on coincident signals on both BBCs, which selects exclusive-like events with at least one large rapidity gap and helps to reject peripheral or beam-gas interactions,

3. use an EMCal trigger to consider events with one or more electrons of energy $> 1$ GeV on the final state.

The efficiency of this trigger setup for $e^+e^-$ pairs of $m_{e^+e^-} > 2$ GeV/c$^2$ is $90 \pm 10\%$ (see details in reference [1]).

## 2.3 Data Analysis

The total equivalent sampled luminosity by PHENIX during the 2004 RHIC Au+Au run at $\sqrt{s_{NN}} = 200$ GeV is of $\mathcal{L}_{integrated} = 141 \pm 12\ \mu\text{b}^{-1}$. The subsample of events analyzed correspond to those centred in the detector fiducial area (vertex in $\pm 30$ cm) and with only two charged particles (electrons) in the final state. This is a restrictive criteria to identify exclusive processes characterised by a few isolated particles. Electron identification criteria are enumerated in reference [1].

A total of 28 $e^+e^-$ pairs and zero like-sign pairs $(e^+e^+ + e^-e^-)$ with $m_{ee} > 2$ GeV/c$^2$ have been measured (Fig. 3 left). The continuum and $J/\psi$ components are isolated by fitting the invariant mass spectra with a continuum (exponential) plus a $J/\psi$ (Gaussian) shape:

$$\frac{dN}{dm_{e^+e^-}} = \frac{A}{\sqrt{2\pi}\cdot\sigma}\cdot e^{-(m_{e^+e^-}-\mu)^2/(2\sigma^2)} + C\cdot e^{B\,m_{e^+e^-}}\,,$$

where $A$, $\sigma$ & $\mu$ are the yield, width and mean position of the $J/\psi$ peak, and $C$ & $B$ are the continuum normalization and slope respectively. Despite the poor statistics, the trend is compatible with STARLIGHT [6] MonteCarlo simulations and a full reconstruction with our experimental setup. The choice of the fit functional form is justified by these simulations. The number of $J/\psi$ is of $N[J/\psi] = 9.9 \pm 4.1(\text{stat}) \pm 1.0(\text{syst})$, and of continuum $e^+e^-$ counts $N[e^+e^-] = 13.7 \pm 3.7(\text{stat}) \pm 1.0(\text{syst})$ for $m_{ee} \in [2.0, 2.8]$ GeV/c$^2$. The quoted systematical uncertainties consider variations of the fit function; we also test a continuum power law form and study the dependence on the slope parameters. The statistical and systematical uncertainties on the $e^+e^-$ continuum contribution to the invariant mass distribution are represented by dashed-lines in Fig. 3 left. After correcting for the limited experimental acceptance and by the inefficiencies, we compute the doubly differential dielectron photoproduction cross-sections per invariant mass range (see Tab. 1), and the $J/\psi$ differential photoproduction cross section at mid-rapidity ($|\eta| < 0.35$), which is:

$$\frac{d\sigma_{J/\psi+Xn}}{dy} = 76 \pm 31(\text{stat}) \pm 15(\text{sys})\;\;\mu\text{b}\,.$$

Remark that all these cross-sections calculations correspond to particle production in coincidence with forward neutron emission (labelled as "$+Xn$"). In addition, the pairs transverse momentum ($p_T$) distribution is shown in Fig. 3 (right) as a function of their invariant mass.

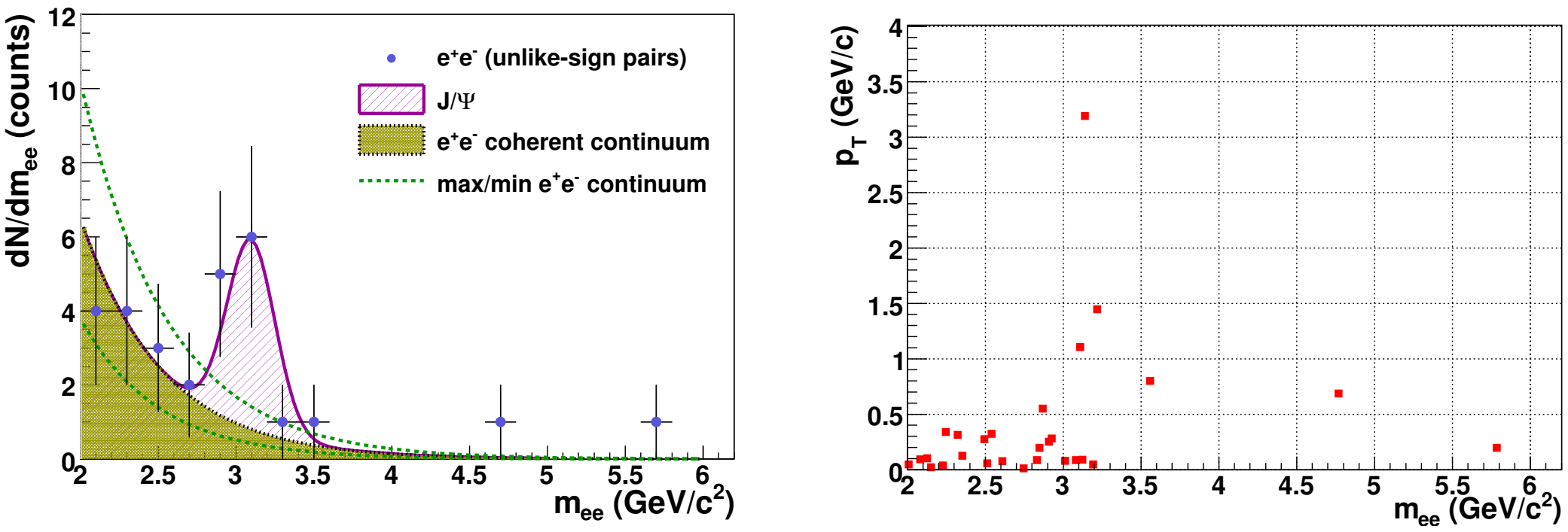

Figure 3: $e^+e^-$ invariant mass distribution as measured in Au+Au UPC at $\sqrt{s_{NN}} = 200$ GeV with PHENIX. The left plot depicts its decomposition into continuum and $J/\psi$. The right figure presents the invariant mass versus pair $p_T$ correlation.

| $m_{e^+e^-}$ [GeV/c$^2$] | $d^2\sigma/dm_{e^+e^-}dy\|_{y=0}$ [$\mu$b/(GeV/c$^2$)] DATA | STARLIGHT |
|---|---|---|
| [2.0, 2.8] | $86 \pm 23$ (stat) $\pm$ 16 (syst) | 90 |
| [2.0, 2.3] | $129 \pm 47$ (stat) $\pm$ 28 (syst) | 138 |
| [2.3, 2.8] | $60 \pm 24$ (stat) $\pm$ 14 (syst) | 61 |

Table 1: Measured $e^+e^-$ continuum photoproduction cross sections at mid-rapidity in UPC Au+Au collisions (accompanied with forward neutron emission) at $\sqrt{s_{NN}} = 200$ GeV. For comparison, the STARLIGHT predictions are also quoted [6].

# 3 Results Interpretation

The dielectron pairs (of $m_{e^+e^-} \in [2.0, 2.8]$ GeV/c$^2$) low transverse momentum (see Fig. 3 right), and their doubly differential photoproduction cross section (see Tab. 1), are in good agreement with the STARLIGHT calculations [6] for coherent ($\gamma\gamma \to e^+e^-$) photoproduction accompanied with forward neutron emission ($e^+e^- + Xn$). Consequently, the measurements are in agreement with LO QED theoretical calculations (STARLIGHT) even in this strongly interacting regime. Further conclusions are limited by, on the one hand, the poor statistics (the measurement uncertainties), and, on the other hand, by the deficit of other theoretical calculations in the kinematic regime of interest, and the lack of knowledge of the higher order QED contributions [7, 8].
The $J/\psi$ $p_T$ distribution (Fig. 3 right) suggests the presence of both coherent ($\gamma A$) and incoherent ($\gamma n$) contributions, consistent with low and intermediate to high $p_T$ respectively, and in agreement with the theoretical expectations [9]. The $J/\psi + Xn$ photoproduction cross-section is compared in Fig. 4 and Fig. 5 to various theoretical calculations. Note that the predictions for both coherent and incoherent contributions are drawn separately (Fig. 4 left) and, whenever possible, also summed up (Fig. 4 right). Besides, Fig. 5 illustrates the sensitivity of the coherent $J/\psi$ production as computed by [12] to different shadowing schemes. The predictions of references [9, 11, 12] have been scaled down according to [6] to account for the cross-section reduction when requiring forward neutron emission. The measurement is thus consistent with different model calculations even though the current precision precludes yet any detailed conclusion on the basic ingredients: shadowing and nuclear absorption.
Finally, one can also compare to HERA $ep$ data of $J/\psi$ photoproduction. A rough comparison is possible dividing the production cross-sections by the equivalent (theoretical) photon spectrum ($dN/d\omega$) [6]:

$$\sigma_{\gamma A \to J/\psi A} = \frac{d\sigma_{AA\to J/\psi AA}}{dy} \cdot \frac{1}{2\,dN/d\omega} \; ;$$
$$\sigma_{\gamma A \to J/\psi A} \approx A^{\alpha}\, \sigma_{\gamma p \to J/\psi p} \; .$$

The equivalent photon spectrum at midrapidity for $\langle W_{\gamma p}\rangle = 24$ GeV is $2\,dN/d\omega = 6.7$ (10.5) for coherent (incoherent) $J/\psi$ production [6]. At HERA, $\sigma_{\gamma p\to J/\psi p} = 30.5 \pm 2.7$ nb at $\langle W_{\gamma p}\rangle =$ 24 GeV [13]. If we assume, for the sake of simplicity, that the coherent and incoherent contributions to $J/\psi$ are of 50%-50%, we obtain that the scaling with the number of nucleons ($A$) is: $\alpha_{coh} = 1.01 \pm 0.07$ for the coherent and $\alpha_{incoh} = 0.92 \pm 0.08$ for the incoherent components respectively. This is consistent with the naive expectation of a scaling with the number of colliding nucleons of the photonuclear cross-section for hard probes.

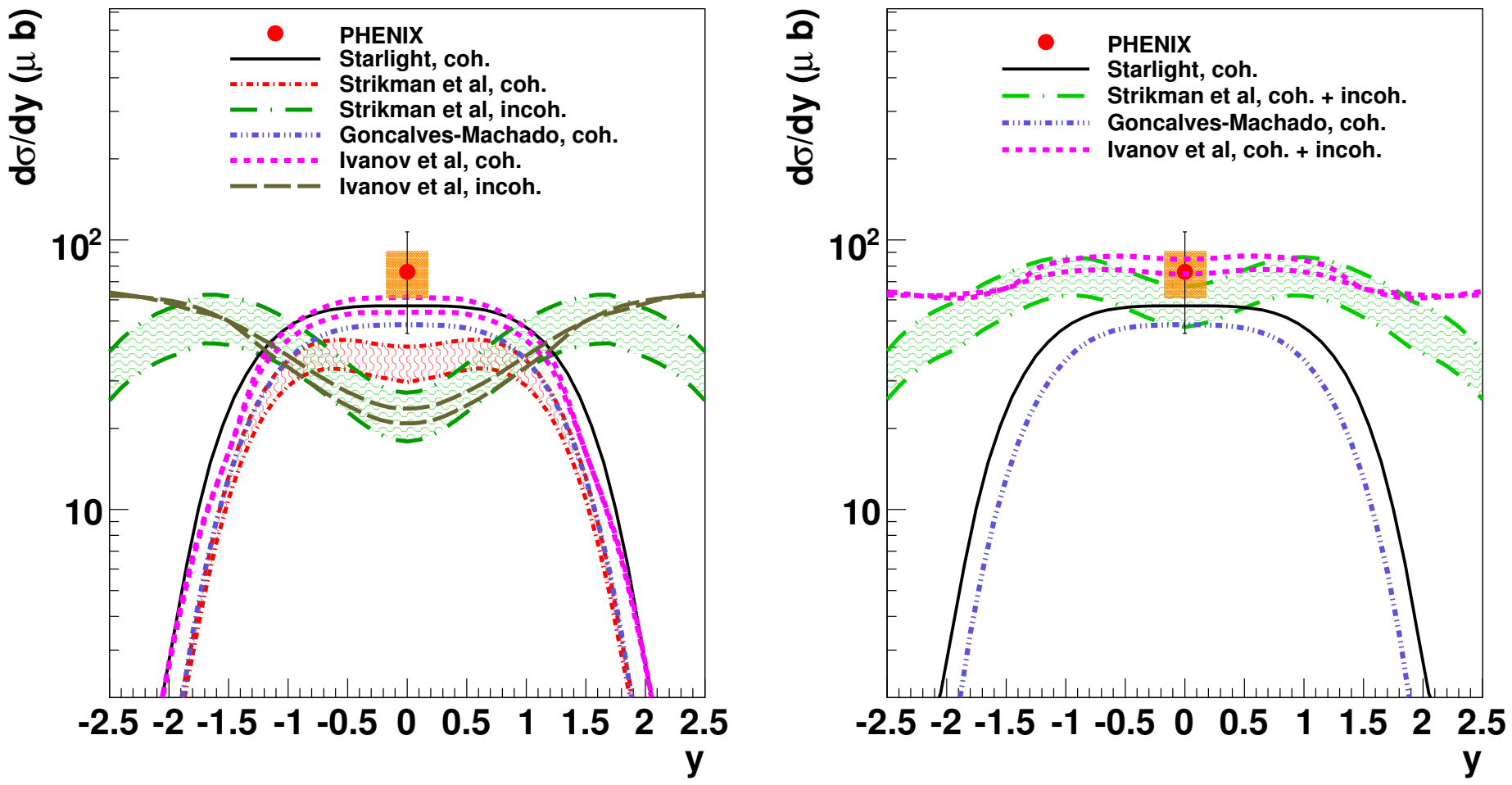

Figure 4: Comparison of the measured $J/\psi + Xn$ photoproduction cross section with various theoretical calculations. Some of them include both coherent and incoherent components (labelled as "coh. + incoh.") : Strikman *et al* [9], Ivanov *et al* [11]. The rest only compute the coherent contribution: STARLIGHT [6], Gonçalves *et al* [10].

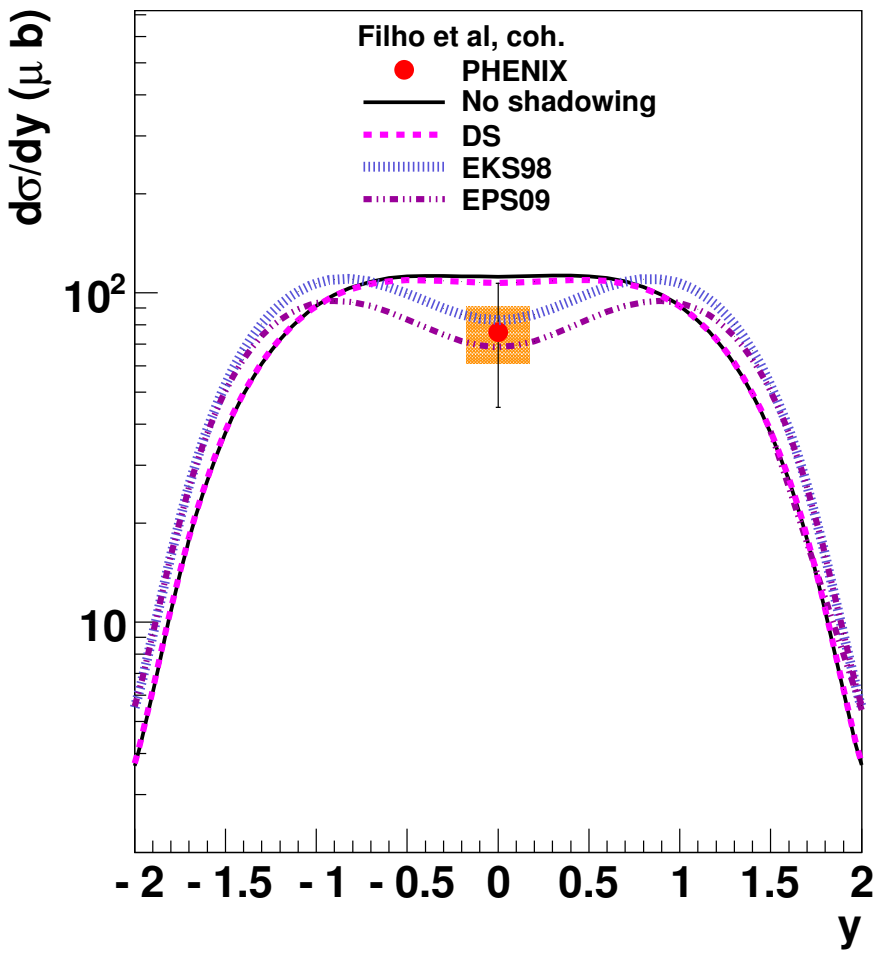

Figure 5: Comparison of the measured $J/\psi + Xn$ photoproduction cross section with the theoretical calculations of the coherent contribution from Filho *et al* [12] with respect to different shadowing parameterisations.

# 4 Summary

The PHENIX experiment has proven its versatility by presenting the first measurement ever of high-mass $e^+e^-$ and $J/\psi$ photoproduction in ultra-peripheral heavy-ion reactions [1] in Au+Au collisions at $\sqrt{s_{NN}} = 200$ GeV (accompanied by Au Coulomb nuclear breakup). Dielectron photoproduction cross sections are in agreement with theoretical STARLIGHT LO QED calculations for coherent two-photon production, $\gamma\gamma \to e^+e^-$. $J/\psi$ photoproduction cross section and its $p_T$ pattern are consistent with the expectations [6, 9, 10, 11, 12] and favour the possibility of both coherent ($\gamma A \to J/\psi$) and incoherent ($\gamma n \to J/\psi$) contributions. Even though the poor statistics prevents one from ruling out any hypothesis, it shows itself to be a promising tool to learn about $J/\psi$ production processes and probe the gluon nuclear PDFs at low-$x$. Future analysis at RHIC with higher luminosities and the imminent LHC collisions will provide more discriminating tools.

# References

[1] PHENIX Collaboration, A. Afanasiev et al, *Phys. Lett.* **B679**, 321 (2009) , and references therein.

[2] PHENIX Collaboration, K. Adcox et al, *Nucl. Instrum. Meth.* **A449** (2003) 469-479.

[3] J. Nystrand, *Nucl. Phys.* **A787** 29 (2007); *Nucl. Phys.* **A752** 470-479 (2005).

[4] STAR Collaboration, B.I. Abelev et al, *Phys. Rev.* **C77** (2008) 034910; C. Adler et al, *Phys. Rev. Lett.* **89** (2002) 272302.

[5] D. d'Enterria, *Nucl. Phys. Proc. Suppl.* 184:158-162 (2008), *arXiv:* 0711.1123 (2007).

[6] A.J. Baltz, S.R. Klein and J. Nystrand, *Phys. Rev. Lett.* **89** 012301 (2002); S.R. Klein and J. Nystrand, *Phys. Rev.* **C60** 014903 (1999); A.J. Baltz, Y. Gorbunov, S.R. Klein, and J. Nystrand, *Phys. Rev.* **C80** 044902 (2009).

[7] A.J. Baltz, *Phys. Rev. Lett.* **100** 062302 (2008).

[8] U.D. Jentschura and V. G. Serbo, *arXiv:0908.3853*, to appear in Eur. Phys. J C.

[9] M. Strikman, M. Tverskoy, M. Zhalov, *Phys. Lett.* **B626** (2005) 72-79.

[10] V.P. Gonçalves and M.V.T. Machado, *arXiv:*0706.2810 (2007); *J. Phys.* **G32**:295-308 (2006).

[11] Y.P. Ivanov, B.Z. Kopeliovich, I. Schmidt, *arXiv:*0706.1532 (2007).

[12] A.L. Ayala Filho, V.P. Gonçalves and M.T. Griep, *Phys. Rev.* **C78**, 044904 (2008).

[13] ZEUS Collaboration, S. Chekanov et al, *Eur. Phys. J. C* **24** (2002) 345;
H1 Collaboration, A. Aktas et al, *Eur. Phys. J. C* **46** (2006) 585.

# Photoproduction in Ultra-Peripheral Relativistic Heavy Ion Collisions at STAR

*Boris Grube*[1] for the STAR Collaboration

[1]Excellence Cluster Universe, Technische Universität München, Garching, Germany

In ultra-peripheral relativistic heavy ion collisions the beam ions scatter at impact parameters larger than the sum of their radii, so that they interact via long range electromagnetic forces. Due to the Lorentz-boost of the beam particles, the exchanged virtual photons have high energies and can induce the photoproduction of vector-mesons. We present recent results of the STAR experiment at RHIC on $\rho^0(770)$ production in Au-Au ultra-peripheral collisions at various energies. STAR has also observed the photoproduction of $\pi^+\pi^-\pi^+\pi^-$, which is related to the still poorly known excited states of the $\rho^0$.

## 1 Introduction

The electromagnetic field of a nucleus which is moving at relativistic velocities can be approximated by a flux of quasi-real virtual photons using the Weizsäcker-Williams approach [1]. The number of photons scales with the atomic charge $Z$ squared, so that fast moving heavy nuclei create intense photon fluxes. Relativistic heavy ions may thus be used as photon sources or targets.

In Ultra-Peripheral relativistic heavy ion Collisions (UPCs) the long-range electromagnetic interactions are separated from the otherwise indistinguishable hadronic interactions by requiring impact parameters $b$ larger than the sum of the nuclear radii $R_A$ of the beam ions. Due to the large Lorentz-boosts of the beam particles, it is possible to study photonuclear reactions as well as photon-photon interactions at high energies in UPCs [2].

The photoproduction of vector mesons is a typical process in UPCs. A virtual photon, radiated by the "emitter" nucleus, fluctuates into a $q\bar{q}$ pair, which scatters elastically off the "target" nucleus and emerges as a real vector meson (cf. Fig. 1a). At high energies the scattering can be described in terms of soft Pomeron exchange. The cross section is strongly enhanced at low transverse momenta $p_T \lesssim 2\hbar/R_A$ of the produced meson, because the $q\bar{q}$ pair couples coherently to the entire nucleus. For these coherent processes the cross section depends on the nuclear form factor $F(t)$, where $t$ is the squared four-momentum transfer to the target nucleus. For larger $p_T$ the $q\bar{q}$ pairs couple to the individual nucleons within the target nucleus resulting in a smaller cross section which scales approximately with the mass number $A$ modulo corrections for the nuclear absorption of the meson.

Because of the intense photon flux in the case of heavy beam ions, the photoproduction of vector mesons may be accompanied by Coulomb excitation of the beam particles. The excited ions decay mostly via neutron emission [3] which is a distinctive event signature that can be utilised in the trigger decision. In lowest order the vector meson photoproduction accompanied by mutual nuclear dissociation of the beam ions is a three-photon process. One

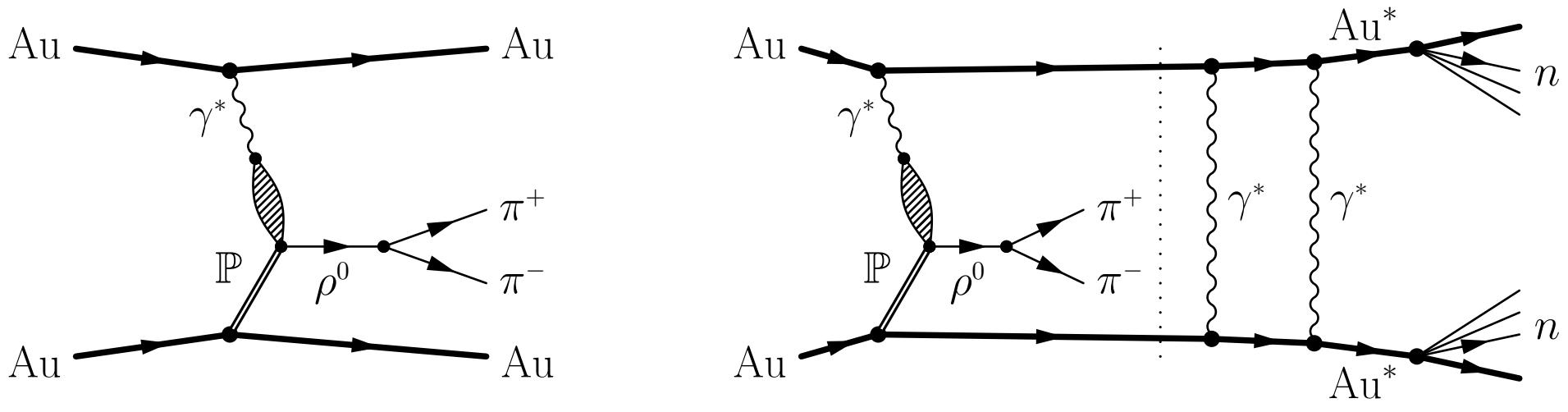

**a)** **b)**

Figure 1: Schematic view of the photonuclear production of a $\rho^0(770)$ meson in an ultra-peripheral Au-Au collision and its subsequent decay into two charged pions. The meson is produced in the fusion processes of a virtual photon $\gamma^*$ and a Pomeron $\mathbb{P}$. a) shows the exclusive reaction, b) the one with mutual Coulomb excitation of the beam ions and following neutron emission.

photon produces the vector meson and two additional photons excite the nuclei (see Fig. 1b). In good approximation all three photon exchanges are independent so that the cross section can be factorised [3]:

$$\sigma_{V,\,xn\,xn} = \int \mathrm{d}^2 b\,[1 - P_{\mathrm{had}}(b)] \cdot P_V(b) \cdot P_{xn,1}(b) \cdot P_{xn,2}(b),$$

where $P_{\mathrm{had}}(b)$ is the probability for hadronic interaction, $P_V(b)$ the probability to produce a vector meson $V$, and $P_{xn,i}(b)$ the probability that nucleus $i$ emits $x$ neutrons. Compared to exclusive photonuclear vector meson production, reactions with mutual Coulomb excitation have smaller median impact parameters.

In this paper we present recent results from the STAR experiment at the Relativistic Heavy Ion Collider (RHIC). The Solenoidal Tracker At RHIC (STAR) uses a large cylindrical Time Projection Chamber (TPC) [4] with 2 m radius and 4.2 m length, operated in a 0.5 T solenoidal magnetic field to reconstruct charged tracks. For tracks with pseudorapidity $|\eta| < 1.2$ and transverse momentum $p_T > 100$ MeV/$c$ the tracking efficiency is better than 85 %. The UPC trigger is based on two detector systems: The two Zero Degree Calorimeters (ZDCs) [5] which sit at $\pm 18$ m from the interaction point and measure neutral particles emitted in very forward direction. They have an acceptance close to unity for the neutrons originating from nuclear dissociation of the beam ions. The second trigger detector system used to select UPC events is the Central Trigger Barrel (CTB) [6]. It is an array of 240 plastic scintillator slats that surrounds the TPC and provides information about the charged-particle multiplicity.

Two basic types of trigger algorithms are used: The “topology” trigger requires a low overall charged-particle multiplicity and subdivides the CTB into four azimuthal quadrants. Events with coincident hits in the left and right quadrants are recorded thereby selecting roughly back-to-back pion pairs. The top and bottom quadrants are used to veto cosmic rays which otherwise could be reconstructed as unlike-sign particle pairs with zero transverse momentum and rapidity. Since there is no requirement on the energy deposit in the ZDCs, the “topology” data mainly contain exclusively produced vector mesons. In contrast to this the “minimum bias” trigger selects UPC events, where both beam ions dissociated by requiring coincident energy deposits in the ZDCs in addition to a low total charged-particle multiplicity in the

CTB.

In the offline analysis two- and four-prong events are selected by requiring two and four charged tracks, respectively, in the TPC to have zero net charge and to form a common (primary) vertex. All tracks are assumed to be pions. In order to suppress backgrounds from beam-gas interactions, peripheral hadronic interactions, and pile-up events in addition a low overall charged-track multiplicity is required. Backgrounds from pile-up events, beam-gas interactions, and — in the case of the two-prong sample — cosmic rays are reduced by selecting events with their primary vertex in a region close to the interaction diamond. Cosmic ray backgrounds in the two-prong "topology" sample are suppressed further by excluding events with rapidities $y_\rho \approx 0$. Due to the ZDC requirement in the "minimum bias" trigger, the cosmic ray background is nearly completely removed. Finally, coherent events are selected by requiring small transverse momenta of $p_T < 150$ MeV/$c$ for the produced vector mesons.

## 2 Coherent Photoproduction of $\rho^0(770)$

There are at least three models that describe the production of $\rho^0(770)$ mesons in ultra-peripheral collisions: The model of Klein and Nystrand (KN) [7] uses the Vector Dominance Model (VDM) for the virtual photon and a classical mechanical approach for the scattering on the target nucleus, based on data from $\gamma p \to \rho^0(770)\, p$ experiments. The Frankfurt, Strikman, and Zhalov (FSZ) model [8] employs a generalised VDM to describe the virtual photon and a QCD Gribov-Glauber approach for the scattering. The model of Gonçalves and Machado (GM) [9] takes into account nuclear effects and parton saturation phenomena by using a QCD colour dipole approach.

The coherent cross section of $\rho^0(770)$ production accompanied by mutual nuclear dissociation of the beam ions $\sigma^{\text{coh}}_{\rho,xnxn}$ is measured using "minimum bias" data. The $\rho^0$ yield is estimated by fitting the invariant mass peak of the acceptance corrected $m_{\pi^+\pi^-}$ distribution and extrapolating the result from the experimentally accessible rapidity range of $|y_\rho| < 1$ to the full solid angle using the KN model [7]. In Au-Au collisions at $\sqrt{s_{NN}} = 200$ GeV the cross section was measured to be $\sigma^{\text{coh}}_{\rho,xnxn} = 31.9 \pm 1.5_{\text{stat.}} \pm 4.5_{\text{syst.}}$mb; at $\sqrt{s_{NN}} = 130$ GeV the value is $28.3 \pm 2.0_{\text{stat.}} \pm 6.3_{\text{syst.}}$mb [10].

Since the efficiency of the "topology" trigger is not well known, the total cross section $\sigma^{\text{coh}}_{\rho,\text{tot}}$ is estimated by applying coherent cross section ratios for different nuclear excitation states, which are extracted from the "topology" data, to the cross section values for mutual excitation. This way the total coherent $\rho^0$ production cross section is measured to be $530 \pm 19_{\text{stat.}} \pm 57_{\text{syst.}}$mb at $\sqrt{s_{NN}} = 200$ GeV and $460 \pm 220_{\text{stat.}} \pm 110_{\text{syst.}}$mb at $\sqrt{s_{NN}} = 130$ GeV [10]. In Fig. 2a the total cross sections and the cross sections with mutual nuclear dissociation are compared to the KN model predictions.

Figure 2b shows the measured total coherent $\rho^0$ production cross section as a function of rapidity and compares to the various model predictions. Due to the limited experimentally accessible rapidity range of $|y_\rho| < 1$, it is not possible to discriminate the models based on the shape of their rapidity distribution. Considering the cross section values, the KN model agrees best with the data.

In $\rho^0$ production an interesting interference phenomenon is caused by the fact that the $\rho^0$ is produced close ($\mathcal{O}(1$ fm$)$) to the target nucleus and that the emitter and the target nucleus are indistinguishable. Because the impact parameter is larger than the sum of the nuclear radii of the projectiles, the system essentially acts like a two-slit interferometer with slit separation $|\vec{b}|$. Either nucleus A emits a virtual photon which scatters off nucleus B or vice versa. The two indistinguishable processes are related by parity transformation and, since the $\rho^0$ has negative intrinsic parity, the amplitudes have to be subtracted [11]:

$$\sigma(\vec{p}_T, \vec{b}, y_\rho) = \left| A(p_T, b, y_\rho) - A(p_T, b, -y_\rho)\, e^{i\vec{p}_T \cdot \vec{b}} \right|^2$$

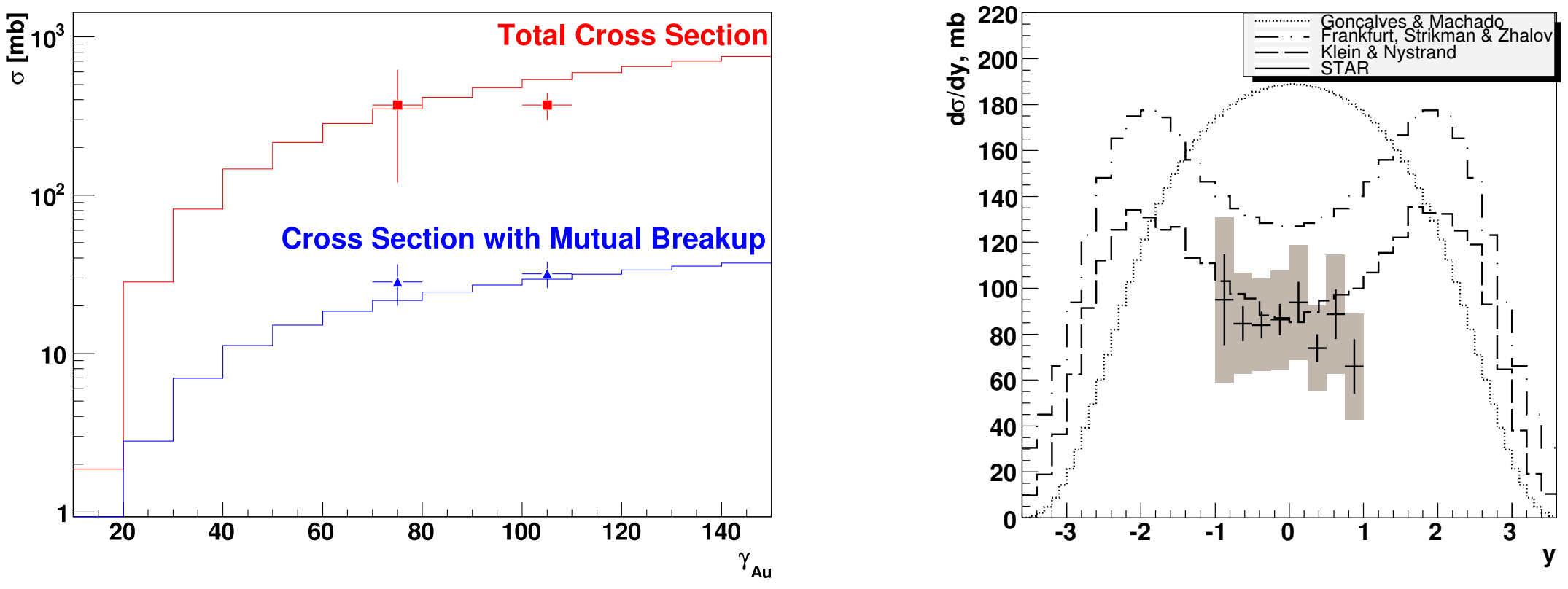

a) b)

Figure 2: a) Energy dependence of the total coherent cross section (red) and the one with mutual nuclear dissociation (blue) in comparison to the KN model predictions (continuous histogram). b) Comparison of the measured total cross section for coherent $\rho^0$ production with theoretical predictions [10]. The vertical line at each point shows the statistical error. The shaded area displays the sum of statistical and systematic errors. The dashed line represents the KN [7], the dash-dotted line the FSZ [8], and the dotted one the GM model [9].

At mid-rapidity $A(p_T, b, y_\rho) \approx A(p_T, b, -y_\rho)$ so that the above equation simplifies to

$$\sigma(\vec{p}_T, \vec{b}, 0) = 2\,|A(p_T, b, 0)|^2 \left[1 - \cos(\vec{p}_T \cdot \vec{b})\right]$$

The interference is destructive for transverse momenta $p_T \lesssim \hbar/\langle b \rangle$. Figure 3 shows the $t(\approx p_T^2)$ distribution, which is roughly exponential at larger $t$, but has a significant downturn for $t < 0.0015\ (\mathrm{GeV}/c)^2$, consistent with the Monte-Carlo simulation that includes the interference effect.

The flight path $\beta\gamma c\tau$ of the produced $\rho^0$ is much smaller than the impact parameter so that the $\rho^0$ decays at two well-separated points in space-time. This means that the amplitudes overlap and interfere only *after* the decay and that the interference must involve the $\pi^+\pi^-$ final state. Interference is only possible, if the final state wave function is entangled, nonlocal, and not factorisable into individual $\pi^\pm$ wave functions.

The strength of the interference is extracted from the data by fitting the $t$ distribution with the function

$$\frac{\mathrm{d}N}{\mathrm{d}t} = a\, e^{-kt} \left[1 + c\,(R(t) - 1)\right]$$

where $k$ is the slope parameter and $c$ the spectral modification parameter that measures the interference. A value of $c = 0$ would correspond to no interference, a value of $c = 1$ to the interference predicted by the KN model [7, 11]. The deviation of the $t$ distribution from the exponential shape due to the interference effect is parameterised by the function $R(t)$ which is determined from the ratio of the simulated $t$ spectrum with and without interference. The measured spectral modification parameter of $87 \pm 5_{\text{stat.}} \pm 8_{\text{syst.}}\%$ shows that the interference is significant [12], which means that the $\pi^+\pi^-$ final state wave function retains amplitudes for all possible $\rho^0$ decays, long after the decay occurred. The system is thus an example of the Einstein-Podolsky-Rosen paradox [16] with continuous variables momentum and position.

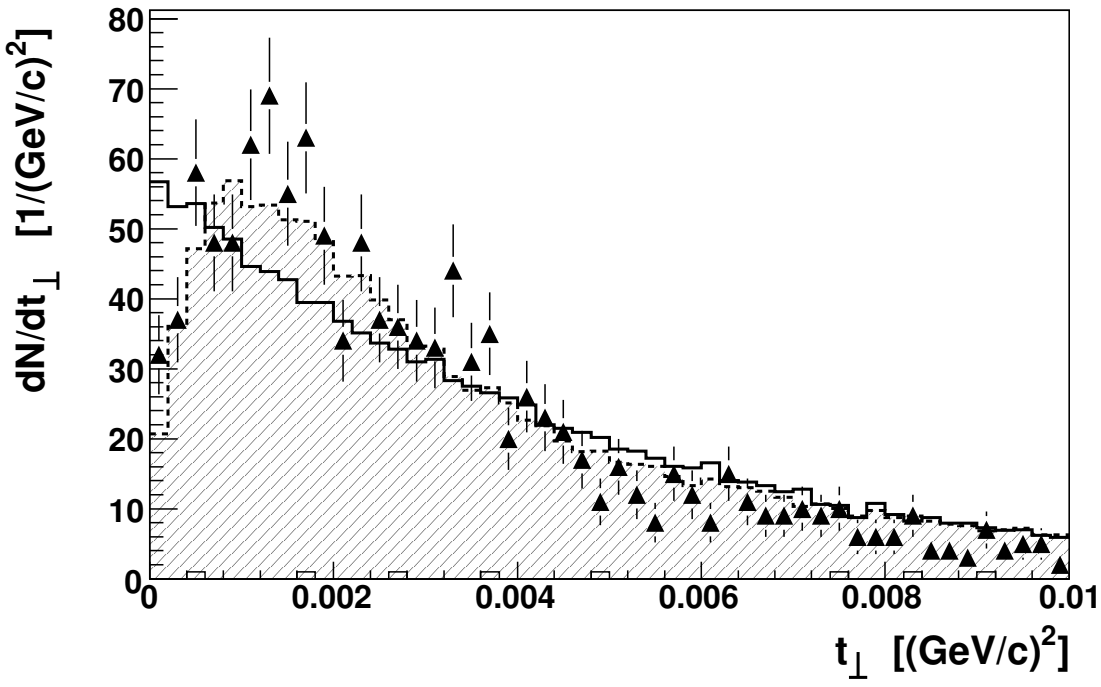

Figure 3: Uncorrected $t$ spectrum of $\rho^0$ in the rapidity range $|y_\rho| < 0.5$ for the "minimum bias" data [12]. The points represent the data. The dashed (filled) histogram is a simulation that includes the interference effect, whereas the solid histogram is a simulation without interference.

# 3 Coherent Photoproduction of $\pi^+\pi^-\pi^+\pi^-$ Final States

Coherent $\pi^+\pi^-\pi^+\pi^-$ production in ultra-peripheral collisions accompanied by mutual nuclear dissociation of the beam ions was measured in Au-Au collisions at $\sqrt{s_{NN}} = 200$ GeV using the "minimum bias" data. The transverse momentum spectrum of the neutral four-prongs exhibits an enhancement at low $p_T$ characteristic for coherent production (cf. Fig. 4a). The data show a broad peak in the $\pi^+\pi^-\pi^+\pi^-$ invariant mass distribution (see Fig. 4b), similar to what was seen in earlier fixed-target photoproduction experiments [13, 14]. This peak is usually attributed to the excited $\rho^0$ states $\rho(1450)$ and $\rho(1700)$. However, the exact nature of these states is still controversial.

The $\pi^+\pi^-\pi^+\pi^-$ invariant mass distribution was fitted with a $S$-wave Breit-Wigner modified by a phenomenological Ross-Stodolsky factor [15]:

$$f(m) = A \cdot \left(\frac{m_0}{m}\right)^n \cdot \frac{m_0^2\Gamma_0^2}{(m_0^2 - m^2)^2 + m_0^2\Gamma_0^2} + f_{\text{BG}}(m) \tag{1}$$

The non-interfering background $f_{\text{BG}}$ was parameterised by a second order polynomial which was extracted from the invariant mass distribution of +2 or −2 charged four-prongs. Taking into account the experimental acceptance, the fit yields a resonance mass of $1540 \pm 40$ MeV/$c^2$ and a width of $570 \pm 60$ MeV. The Ross-Stodolsky exponent has a value of $n = 2.4 \pm 0.7$, however, mass and width depend strongly on the value of $n$.

Using the acceptance-corrected $\pi^+\pi^-\pi^+\pi^-$ yield from the above fit and the respective $\rho^0(770)$ yield from the $\pi^+\pi^-$ invariant mass distribution the cross section ratio $\sigma^{\text{coh}}_{4\pi,xnxn}/\sigma^{\text{coh}}_{\rho,xnxn}$ is estimated to be $13.4 \pm 0.8$ %, where again the KN model [7] was used to extrapolate from the experimentally accessible rapidity region $|y| < 1$ to the full solid angle. Using the measured coherent $\rho^0$ production cross section $\sigma^{\text{coh}}_{\rho,xn,xn}$ the $\pi^+\pi^-\pi^+\pi^-$ production cross section is $4.3 \pm 0.3$ mb.

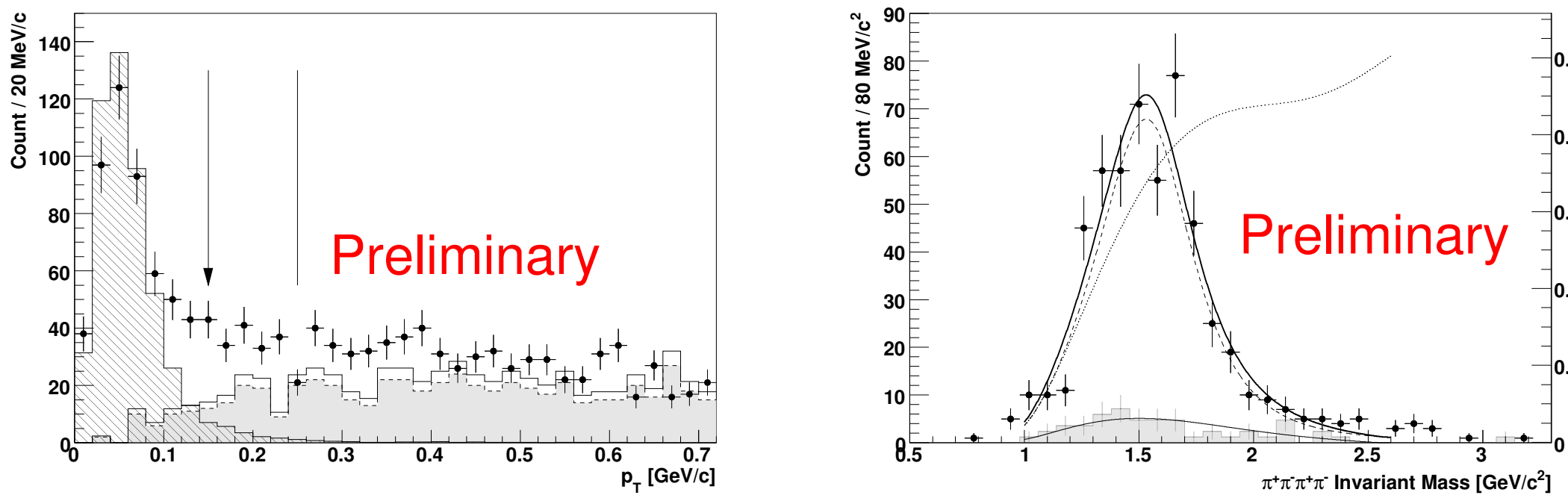

a) b)

Figure 4: a) $\pi^+\pi^-\pi^+\pi^-$ transverse momentum distribution: At low transverse momenta the four-prong couples coherently to the entire nucleus leading to a strong enhancement of the cross section. The hatched filled histogram shows the expected distribution from simulation. The background for the coherent part is estimated from $+2$ or $-2$ charged four-prong combinations by normalising their $p_T$ distribution (gray filled histogram) to that of the neutral four-prongs in the region of $p_T > 250$ MeV/$c$ (vertical line) yielding the unfilled histogram. b) Invariant mass distribution of coherently produced $\pi^+\pi^-\pi^+\pi^-$: The points represent the data, the gray filled histogram is the background estimated from charged four-prongs. The thick black line shows the fit of the modified $S$-wave Breit-Wigner of Eq. (1) on top of a second order polynomial background (thin black line) taking into account the detector acceptance in the region $|y| < 1$ (rising dotted line). The dashed curve represents the signal curve without background.

Figure 5a shows that $\pi^+\pi^-\pi^+\pi^-$ events mainly consist of a low mass $\pi^+\pi^-$ pair accompanied by a $\rho^0(770)$. This motivated the Monte-Carlo decay model $\rho' \to \rho^0(770)\ f_0(600)$ which is used to estimate the acceptance corrections. As can be seen in Fig. 5a this model reproduces the data well.

In photoproduction on carbon targets the $\rho'$ was seen not only in the $\pi^+\pi^-\pi^+\pi^-$ decay mode, but also in $\pi^+\pi^-$ final states [14]. Figure 5b shows the high mass region of the measured $m_{\pi^+\pi^-}$ spectrum. In order to suppress backgrounds, in particular cosmic rays, tighter cuts are applied. The data do not show any significant enhancement around 1540 MeV/$c^2$.

# 4 Summary

STAR has measured photonuclear production of $\rho^0(770)$ in ultra-peripheral relativistic heavy ion collisions. The measured cross sections agree with model predictions. STAR also measured for the first time the interference effect in $\rho^0$ production which indicates that the decoherence induced by the $\rho^0$ decay is small and that the $\pi^+\pi^-$ final state wave function is entangled and nonlocal. In addition STAR has observed coherent photoproduction of $\pi^+\pi^-\pi^+\pi^-$ final states in UPCs. The $\pi^+\pi^-\pi^+\pi^-$ invariant mass spectrum exhibits a broad peak around 1540 MeV/$c^2$ and no corresponding enhancement is seen in the $m_{\pi^+\pi^-}$ distribution. The coherent $\pi^+\pi^-\pi^+\pi^-$ production cross section is $13.4 \pm 0.8$ % of that of the $\rho^0(770)$ meson.

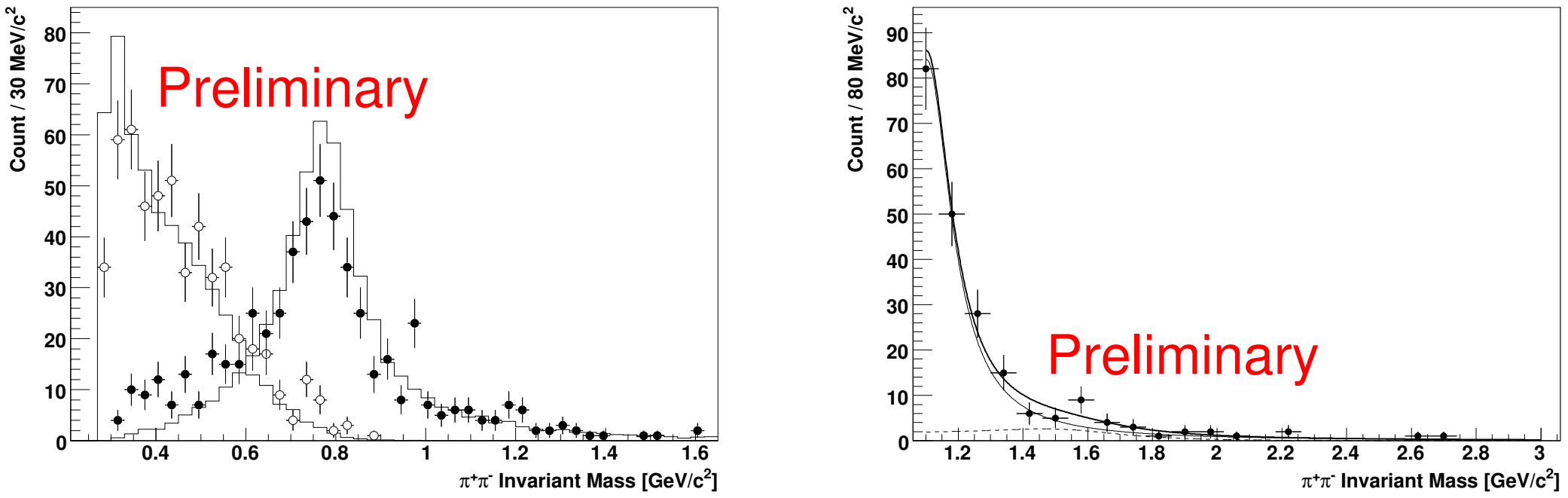

a) b)

Figure 5: a) Invariant Mass distribution of two-pion subsystems: The open circles show the measured invariant mass spectrum of the lightest $\pi^+\pi^-$ pair in the event. The filled circles represent the invariant mass distribution of the $\pi^+\pi^-$ that is recoiling against the lightest pair. The spectrum exhibits a clear peak in the $\rho^0(770)$ region. The solid line histograms show the prediction from simulation assuming the relative $S$-wave decay $\rho' \to \rho^0(770)\ f_0(600)$. b) High mass region of the $m_{\pi^+\pi^-}$ spectrum with tighter cuts applied in order to suppress background: The points represent the data. No significant enhancement is seen in the region around 1540 MeV/$c^2$ where the $\pi^+\pi^-\pi^+\pi^-$ invariant mass spectrum exhibits a peak.

# References

[1] C. F. von Weizsäcker, Z. Phys. **88**, 612 (1934); E. J. Williams, Phys. Rev. **45**, 729 (1934).

[2] G. Baur, K. Hencken, D. Trautmann, S. Sadovsky, and Yu. Kharlov, Phys. Rept. **364**, 359 (2002); F. Krauss, M. Greiner, and G. Soff, Prog. Part. Nucl. Phys. **39**, 503 (1997); C. A. Bertulani, S. R. Klein, and J. Nystrand, Ann. Rev. Nucl. Part. Sci. **55**, 271 (2005); A. J. Baltz *et al.*, Phys. Rept. **458**, 1 (2008).

[3] A. J. Baltz, S. R. Klein, and J. Nystrand, Phys. Rev. Lett. **89**, 012301 (2002); G. Baur, K. Hencken, A. Aste, D. Trautmann, and S. R. Klein, Nucl. Phys. **A729**, 787 (2003).

[4] M. Anderson *et al.*, Nucl. Intrum. Methods **A499**, 659 (2003); **A499**, 679 (2003).

[5] C. Adler *et al.*, Nucl. Intrum. Methods **A470**, 488 (2001).

[6] F. S. Bieser *et al.*, Nucl. Intrum. Methods **A499**, 766 (2003).

[7] S. Klein and J. Nystrand, Phys. Rev. C **60**, 014903 (1999).

[8] M. Frankfurt, M. Strikman and M. Zhalov, Phys. Lett. **B537**, 51 (2002); Phys. Rev. C **67**, 034901 (2003).

[9] V. P. Gonçalves and M. V. T. Machado, Eur. Phys. J. **C40**, 519 (2005).

[10] C. Adler *et al.*, Phys. Rev. Lett. **89**, 272302 (2002); B. I. Abelev *et al.*, Phys. Rev. C **77**, 34910 (2008).

[11] S. R. Klein and J. Nystrand, Phys. Rev. Lett. **84**, 2330 (2000); Phys. Lett. A308, 323 (2003)

[12] B. I. Abelev *et al.*, Phys. Rev. Lett. **102**, 112301 (2009).

[13] H. H. Bingham *et al.*, Phys. Lett. **41B**, 635 (1972)); P. Schacht, I. Derado, D. C. Fries, J. Park, and D. Yount, Nucl. Phys. **B81**, 205 (1974); G. Alexander, O. Benary, J. Gandsman, D. Lissauer, A. Levy, Y. Oren and L. M. Rosenstein, Phys. Lett. **57B**, 487 (1975); D. P. Barber *et al.*, Z. Phys. **C4**, 169 (1980); D. Aston *et al.*, Nucl. Phys. **B189**, 15 (1981); M. Atkinson *et al.*, Z. Phys. **C26**, 499 (1985).

[14] M. S. Atiya *et al.*, Phys. Rev. Lett. **43**, 1691 (1979).

[15] M. Ross and L. Stodolsky, Phys. Rev. **149**, 1172 (1966).

[16] A. Einstein, B. Podolsky, and N. Rosen, Phys. Rev. **47**, 777 (1935)

# Two-Photon Interactions at Belle and BaBar

*Simon Eidelman*

Budker Institute of Nuclear Physics, Lavrentyev Ave. 11,
Novosibirsk, 630090, Russia
(for the Belle and BaBar Collaborations)

Results on two-photon physics obtained in experiments at the $B$ factories are discussed. BaBar used single-tag $\gamma\gamma$ collisions to measure the transition form factor of the $\pi^0$ meson. Belle studied no-tag $\gamma\gamma$ collisions to measure cross sections of exclusive production of two baryons and two mesons. Experimental results are confronted with QCD predictions.

# 1 Introduction

Two experiments at B factories (BaBar at SLAC and Belle at KEK) collected huge integrated luminosities: about 560 fb$^{-1}$ at BaBar and 950 fb$^{-1}$ at Belle. In addition to copiously produced $B$ meson pairs, this statistics gives access to studying two-photon physics including processes with small cross sections.

It is worth mentioning some special features of two-photon collisions:

- it is a clean source of hadrons with positive $C$-parity;
- peculiar kinematics: the final $e^{\pm}$ fly in the same direction as the initial $e^{\pm}$ and lose little energy; the products of $\gamma\gamma$ have small transverse momentum;
- the cross section grows as $\ln^3 E_{\rm CM}$;
- different types of experiments are possible: no-tag – both $e^{\pm}$ undetected, single-tag – one $e^{\pm}$ detected, double-tag – both $e^{\pm}$ detected;
- it is an excellent laboratory for QCD tests in $\gamma\gamma$ production of hadrons.

# 2 $\pi^0$ Transition Form Factor

BaBar used 442 fb$^{-1}$ collected at 10.54 and 10.58 GeV to study the $\pi^0$ transition form factor in the single-tag mode, i.e. when one of the photons is almost real while the second is strongly off-shell with a momentum transfer $q^2 \equiv -Q^2$, $4 < Q^2 < 40$ GeV$^2$ [1]. This is serious progress compared to the previous experiments in which CELLO studied the momentum range from 0.7 to 2.2 GeV$^2$ [2] and CLEO from 1.6 to 8 GeV$^2$ [3]. The distribution of the invariant mass of two photons shows a clear peak from the $\pi^0$, Fig. 1.

About 13200 events of $\gamma\gamma^* \to \pi^0$ were selected at BaBar compared to 127 at CELLO and 1219 at CLEO. The main background comes from virtual Compton scattering, $e^+e^- \to e^+e^-\gamma$, with one final $e^{\pm}$ at small angles, while the other $e^{\mp}$ and $\gamma$ scatter at large angles. The major peaking background – $e^+e^- \to e^+e^-\pi^0\pi^0$, $\sim$ 1600 events detected.

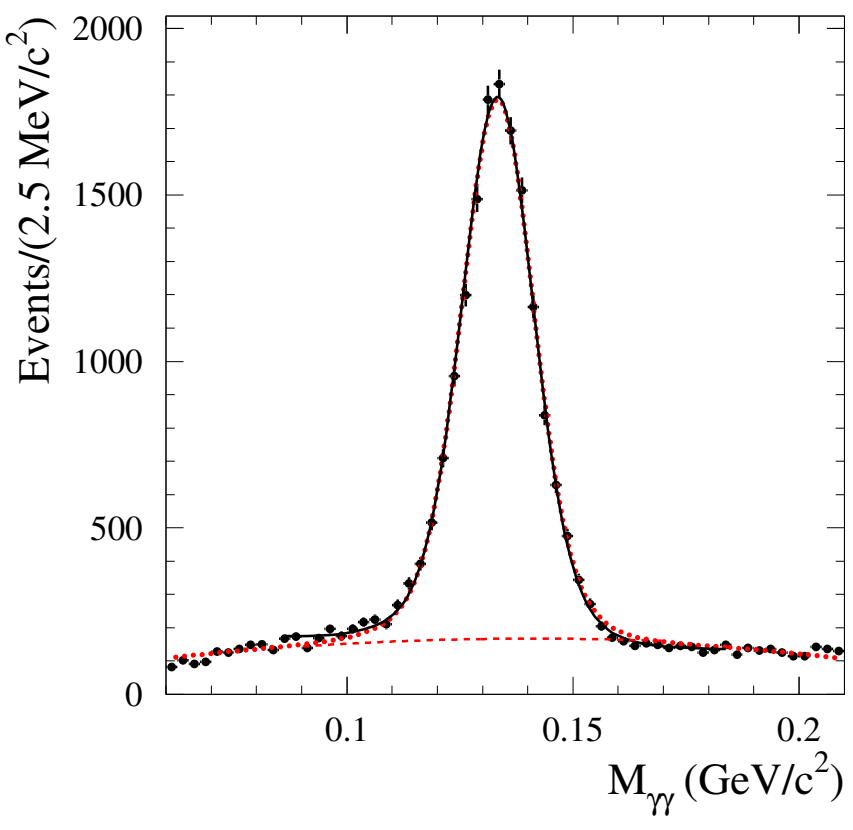

Figure 1: Invariant mass of two photons.

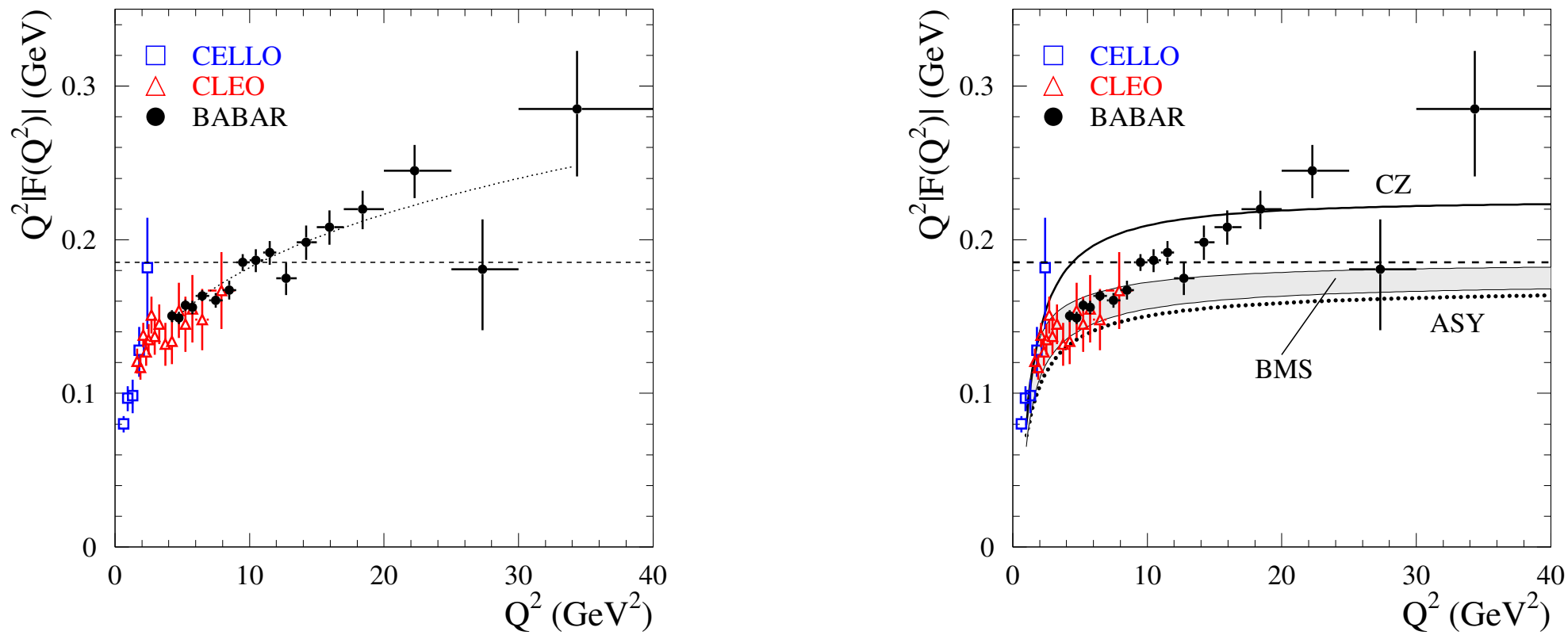

Figure 2: $Q^2$ dependence of the $\pi^0$ transition form factor.

To describe the $Q^2$ dependence, they fit the form factor using the function $Q^2|F(Q^2)| = A\left(\frac{Q^2}{10\ \mathrm{GeV}^2}\right)^\beta$ and obtain $A = 0.182 \pm 0.002$ GeV, $\beta = 0.25 \pm 0.02$. The effective $Q^2$ dependence of the form factor ($\sim 1/Q^{3/2}$) differs significantly from the leading-order pQCD prediction ($\sim 1/Q^2$) [4], demonstrating the importance of higher-order pQCD and power corrections in the $Q^2$ region under study.. The horizontal dashed line in Fig. 2, left, indicates the asymptotic limit $Q^2|F(Q^2)| = \sqrt{2}f_\pi \approx 0.185$ GeV for $Q^2 \to \infty$. The measured form factor exceeds the limit for $Q^2 > 10$ GeV$^2$ contradicting most models for the pion wave function $\phi_\pi$, which give form factors approaching this limit from below. Fig. 2, right, shows some theoretical predictions obtained using the light-cone sum rules [5] at NLO pQCD with twist-4 for three types of $\phi_\pi$: that of Chernyak-Zhitnitsky (CZ) [6], the asymptotic (ASY) [7] and the one derived from QCD sum rules with non-local condensates (BMS) [8]. For all three $\phi_\pi$ the $Q^2$ dependence is almost

flat for $Q^2 > 10$ GeV$^2$, whereas the data show significant growth between 8 and 20 GeV$^2$. This indicates that the approximation mentioned above is not adequate for $Q^2$ less than $\sim 15$ GeV$^2$. In the $Q^2$ range from 20 to 40 GeV$^2$, where uncertainties due to higher-order pQCD and power corrections are expected to be smaller, the BaBar data lie above the asymptotic limit and are consistent with the CZ model.

Several papers appeared after the BaBar result: in Ref. [9] it is shown that the form factor growth above 10 GeV$^2$ can not be explained in terms of NNLO higher-order perturbative corrections while in Refs. [10, 11] it is argued that the $Q^2$ dependence observed by BaBar can be explained with the flat pion wave function.

# 3 Results from Belle

For the exclusive pair production $\gamma\gamma \to h_1 h_2$ in the leading order (quark-counting rule) $\frac{d\sigma}{dt} \propto \frac{f(\cos\theta^*)}{s^{n-2}}$, where $s = W_{\gamma\gamma}^2 = W^2$ and $n$ is the number of "elementary" fields [12].

Scaling behaviour is expected in the QCD asymptotic regime ($s \to \infty$): $\sigma \propto 1/s^3$ for mesons and $\sigma \propto 1/s^5$ for baryons. The handbag model predicts that at intermediate energies amplitudes are dominated by soft non-perturbative terms [13].

Belle studied various two-body final states $\gamma\gamma \to p\bar{p}, \pi^+\pi^-$, $K^+K^-$, $K_S^0 K_S^0$, $\pi^0\pi^0$, $\eta\pi^0$ at $W$ up to 4 GeV [14, 15, 16, 17, 18]. These studies allow various QCD tests to be performed of which we'll discuss the energy dependence of the cross section.

The $\gamma\gamma \to p\bar{p}$ cross section was measured for $W$ between 2.025 and 4.0 GeV with an integrated luminosity of 89 fb$^{-1}$ [14]. If they fit the data with a power law $\sigma \propto W^{-n}$ with $n$ floating (Fig. 3, left), they obtain $n = 15.1^{+0.8}_{-1.1}$ at $2.5 < W < 2.9$ GeV and $n = 12.4^{+2.4}_{-2.3}$ at $3.2 < W < 4.0$ GeV. In Fig. 3, right, we show the results of the fits with $n$ fixed at 10 and 15. Although for both ranges a good fit can be obtained at $n$=15, a smaller power, $n$=10, describes the data above 3.2 GeV reasonably well. This may imply that lower power terms become dominant at higher energies, which is an indication for the transition to asymptotics.

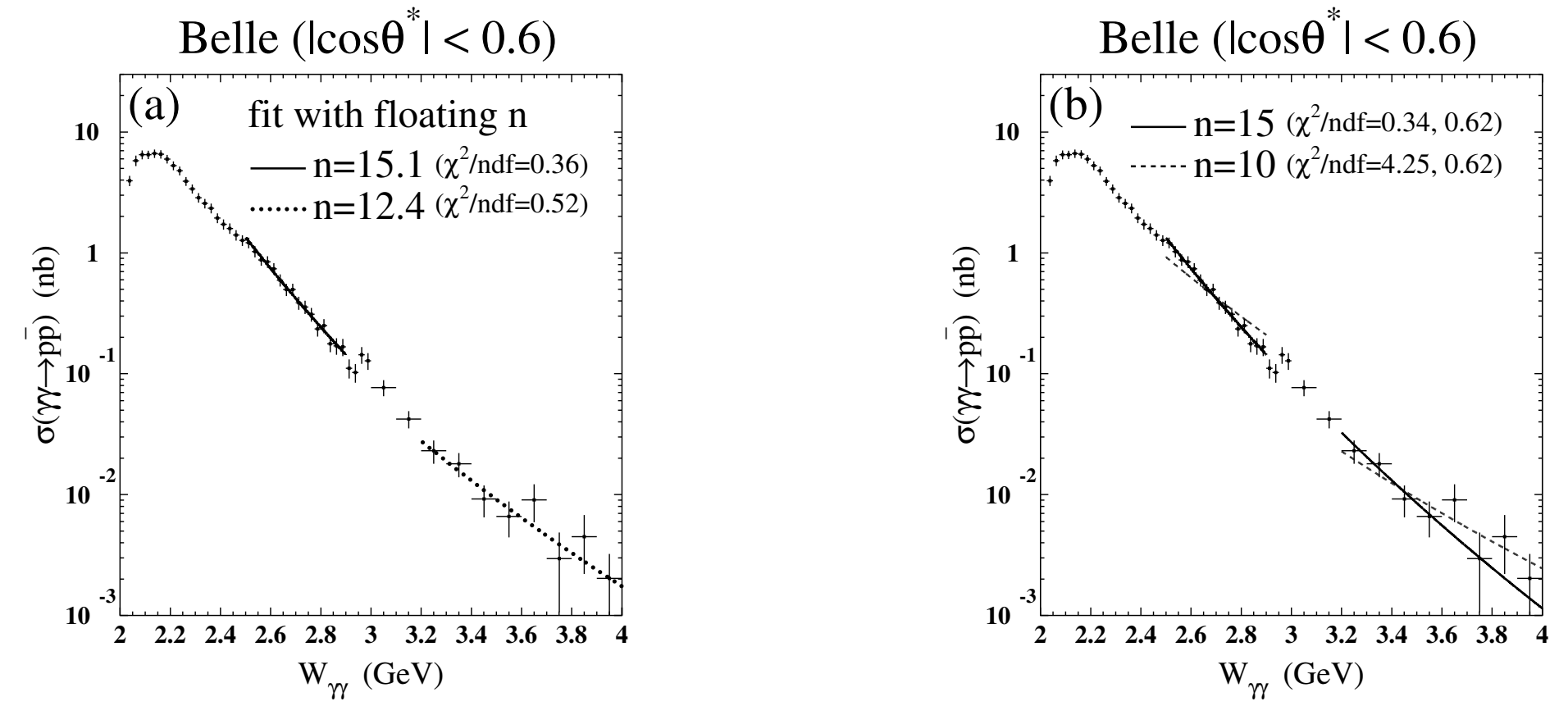

Figure 3: $W$ dependence for the process $\gamma\gamma \to p\bar{p}$.

The $\gamma\gamma \to \pi^+\pi^-$, $K^+K^-$ cross sections were measured for $W$ between 2.4 and 4.1 GeV with an integrated luminosity of 87.7 fb$^{-1}$ [15]. Fig. 4 shows the observed cross sections for $\gamma\gamma \to \pi^+\pi^-$ (left) and $\gamma\gamma \to K^+K^-$ (right) and compares them to the ALEPH measurement [19]. Above 3 GeV ALEPH data as well as much more precise data from Belle (more than 6000 events for each of the processes) agree with $\sigma \propto 1/W^6$. Direct fits of the Belle data to $\sigma \propto W^n$ for $W$ between 3.0 and 4.1 GeV give somewhat steeper dependence $n = -7.9 \pm 0.4 \pm 1.5$ for $\pi^+\pi^-$ and $n = -7.3 \pm 0.3 \pm 1.5$ for $K^+K^-$, but still not contradicting to the $W^{-6}$ dependence.

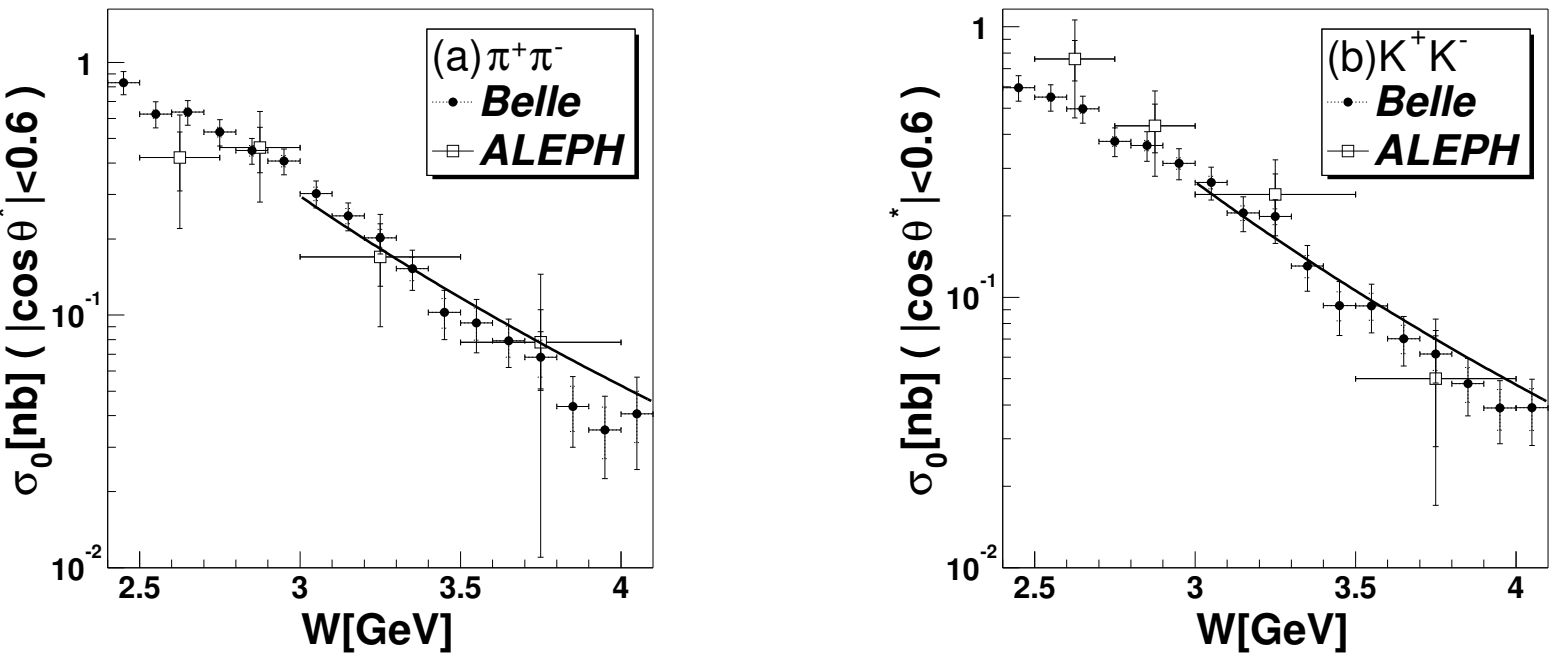

Figure 4: $W$ dependence for the processes $\gamma\gamma \to \pi^+\pi^-$ (left) and $\gamma\gamma \to K^+K^-$ (right).

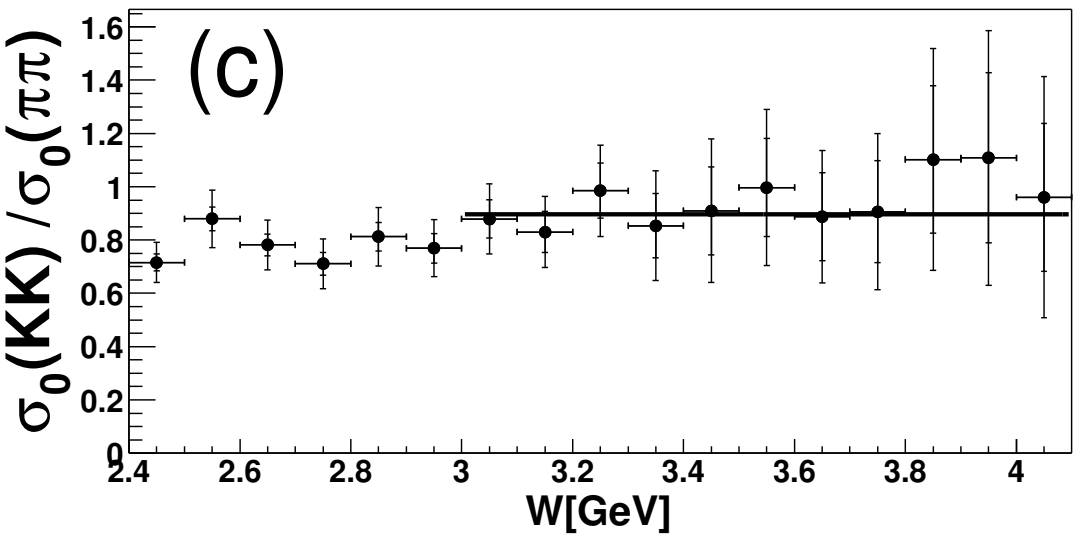

Figure 5: The ratio of the cross sections $\gamma\gamma \to \pi^+\pi^-$ and $\gamma\gamma \to K^+K^-$.

Fig. 5 shows the ratio of the cross sections $\sigma(\gamma\gamma \to K^+K^-)/\sigma(\gamma\gamma \to \pi^+\pi^-)$ as a function of $W$. The ratio is energy independent above 3.0 GeV in accordance with the QCD prediction. The obtained value of the ratio is $0.89 \pm 0.04 \pm 0.15$ consistent with 1.08 predicted in Ref. [20] and significantly lower than 2.23 following from Ref. [12]. The value predicted in [20] is based on consistent consideration of SU(3) breaking effects using different wave functions for pions and kaons derived from the QCD sum rules whereas in [12] the same wave functions are used so that the ratio behaves as the fourth power of the ratio of the kaon and pion decay constants.

Belle has also measured for the first time the cross section of $\gamma\gamma \to K_S^0 K_S^0$ cross sections for $W$ from 2.4 to 4.0 GeV using a data sample of 397.6 fb$^{-1}$ [16]. Fig. 6, left, shows the observed cross section. The fit to the data gives a $W^{-n}$ dependence with $n = 10.5 \pm 0.6 \pm 0.5$ and suggests that the values of $W$ are not yet large enough to neglect power corrections not taken into account in Refs. [12, 20]. The ratio $\sigma_0(K_S^0 K_S^0)/\sigma_0(K^+K^-)$ shown in Fig. 6 decreases from $\sim 0.13$ to $\sim 0.01$. Such energy dependence is inconsistent with the prediction of Ref. [13] that the ratio should be $\approx 2/25$ in the SU(3) symmetry limit.

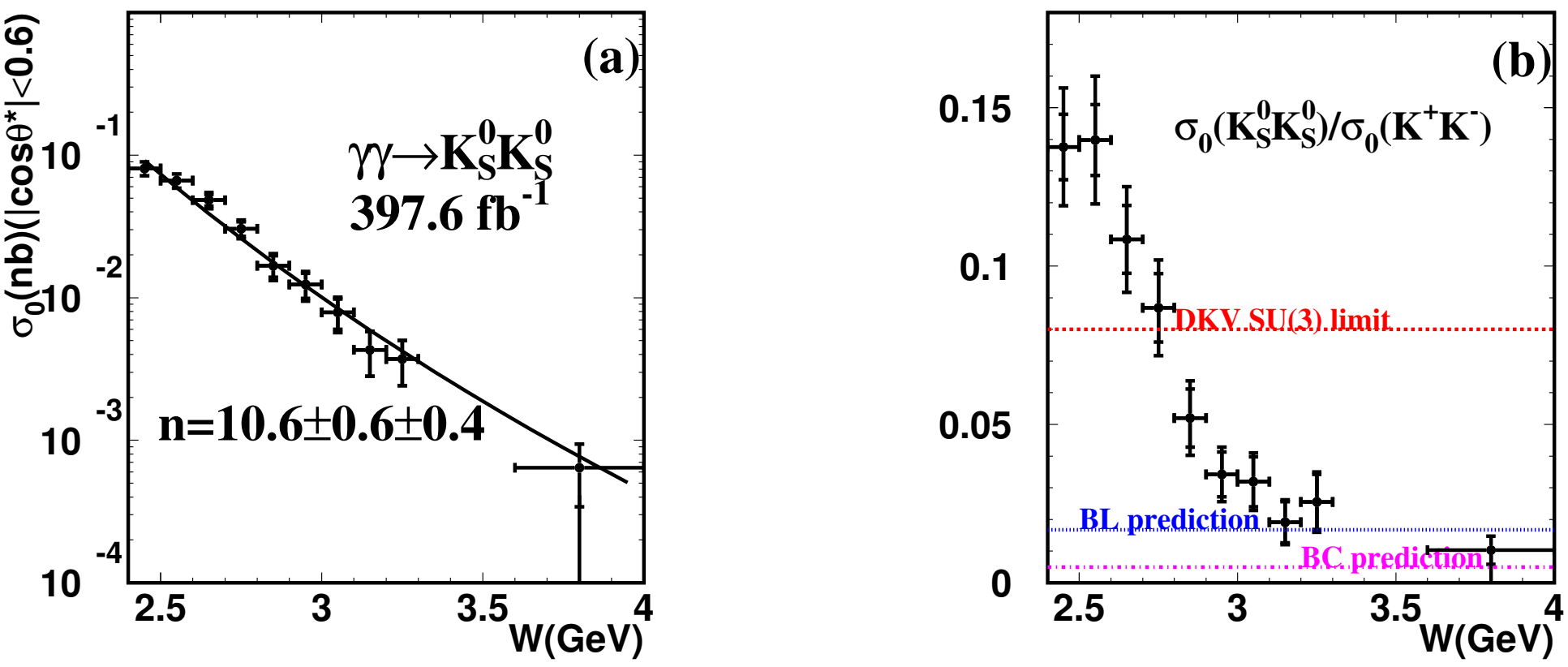

Figure 6: The cross section of $\gamma\gamma \to K_S^0 K_S^0$ and its ratio to the cross section of $\gamma\gamma \to K^+K^-$.

Finally, Belle used a data sample of 223 fb$^{-1}$ to measure the cross sections of $\gamma\gamma \to \pi^0\pi^0$ for $W$ from 0.6 to 4.1 GeV [17] and of $\gamma\gamma \to \eta\pi^0$ for $W$ from 0.84 to 4.0 GeV [18].

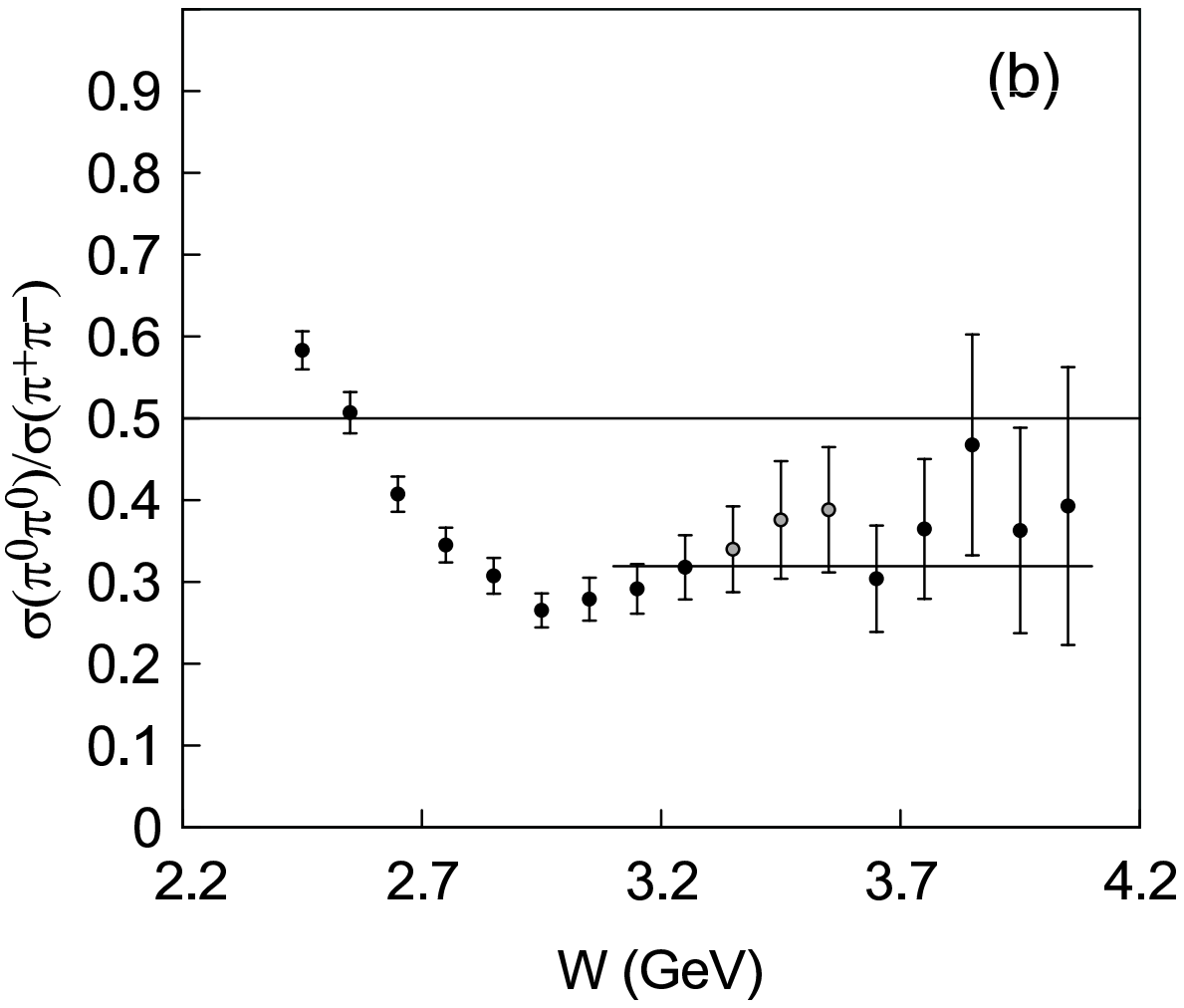

Figure 7: The ratio of the cross sections of $\gamma\gamma \to \pi^0\pi^0$ and $\gamma\gamma \to \pi^+\pi^-$.

Fig. 7 shows the ratio of the cross sections of $\gamma\gamma \to \pi^0\pi^0$ and $\gamma\gamma \to \pi^+\pi^-$. The ratio is falling at low energies, but above 3.1 GeV is almost constant with an average of $0.32 \pm 0.03 \pm 0.05$ that is significantly larger than the leading-order QCD prediction [12, 20] and lower than 0.5 suggested by isospin invariance [13].

For $\gamma\gamma \to \eta\pi^0$, a fit with $W^{-n}$ gives $n = 10.5 \pm 1.2 \pm 0.5$ compatible with $K_S^0 K_S^0$, but higher than for $\pi^0\pi^0$. A fit of the ratio of the $\gamma\gamma \to \eta\pi^0$ and $\gamma\gamma \to \pi^0\pi^0$ cross sections gives $0.48 \pm 0.05 \pm 0.04$ with 0.46 predicted in QCD.

We summarise all results on the $W$ dependence in Table 1.

| Mode | $n$ | $\int L\,dt$, fb$^{-1}$ | $W$ range, GeV | $\lvert\cos\theta^*\rvert$ range |
|---|---|---|---|---|
| $\pi^+\pi^-$ | $7.9 \pm 0.4 \pm 1.5$ | 87.7 | [3.0,4.1] | $<$ 0.6 |
| $K^+K^-$ | $7.3 \pm 0.3 \pm 1.5$ | 87.7 | [3.0,4.1] | $<$ 0.6 |
| $K^0_S K^0_S$ | $10.5 \pm 0.6 \pm 0.5$ | 397.6 | [2.4,3.3],[3.6,4.0] | $<$ 0.6 |
| $\pi^0\pi^0$ | $6.9 \pm 0.6 \pm 0.7$ | 223 | [3.1,3.3],[3.6,4.1] | $<$ 0.6 |
| $\pi^0\pi^0$ | $8.0 \pm 0.5 \pm 0.4$ | 223 | [3.1,3.3],[3.6,4.1] | $<$ 0.8 |
| $\eta\pi^0$ | $10.5 \pm 1.2 \pm 0.5$ | 223 | [3.1,4.1] | $<$ 0.8 |
| $p\bar{p}$ | $15.1^{+0.8}_{-1.1}$ | 89 | [2.5,2.9] | $<$ 0.6 |
| | $12.4^{+2.4}_{-2.3}$ | 89 | [3.2,4.0] | $<$ 0.6 |

Table 1: $W$ dependence of the cross sections of various processes

# 4 Conclusions

- Huge integrated luminosity collected at the $B$ factories has already resulted in high-statistics studies of some rare phenomena
- BaBar measured the $\gamma\gamma^* \to \pi^0$ transition form factor from 4 to 40 GeV$^2$; below 15 GeV$^2$ the NLO pQCD with twist-4 is inadequate, above 20 GeV$^2$ the data lie above the asymptotic limit; the $\eta_c$ form factor will appear soon; the $\eta$, $\eta'$ form factors are under study. These results can be important for models of form factors in the light-by-light contribution to the muon anomaly.
- Belle performed tests of QCD at $3 < W < 4$ GeV with
  $\gamma\gamma \to p\bar{p}$, $\pi^+\pi^-$, $K^+K^-$, $K^0_S K^0_S$, $\pi^0\pi^0$, $\eta\pi^0$;
  for $\sigma(W) \sim W^{-n}$ $n$ follows pQCD
- There were also many interesting studies of hadronic resonances: $f_0$'s in $\pi^+\pi^-$, $\pi^0\pi^0$, $a_0$'s in $\eta\pi^0$, $f_2$'s in $K^+K^-$ at Belle; $\eta_c$ and $\eta_c(2S)$ at BaBar and Belle, $\chi_{c2}(2P)$ was discovered at Belle in $\gamma\gamma \to D\bar{D}$
- High-statistics $\gamma\gamma$ production has good potential for discovering new states, measuring transition form factors and $\mathcal{B}$'s, testing QCD predictions

# Acknowledgements

I'm grateful to M. Deile for excellent organisation of the Conference. Thanks are also due to V. Chernyak, V. Druzhinin, K. Trabelsi and S. Uehara for useful discussions.
This work was supported in part by grants RFBR 07-02-00816, 08-02-13516, 08-02-91969 and 09-02-01143 and PST.CLG.980342.

# References

[1] B. Aubert et al. (BaBar Collab.), Phys. Rev. D **80** 052002 (2009).

[2] H.J. Behrend et al. (CELLO Collab.), Z. Phys. C **49** 401 (1991).

[3] J. Gronberg et al. (CLEO Collab.), Phys. Rev. D **57** 33 (1998).

[4] G.P. Lepage and S.J. Brodsky, Phys. Rev. D **22** 2157 (1980).

[5] A.P. Bakulev, S.V. Mikhailov and N.G. Stefanis, Phys. Rev. D **67** 074012 (2003).

[6] V.L. Chernyak and A.R. Zhitnitsky, Nucl. Phys. B **201** 492 (1982), Erratum-ibid B **214** 547 (1983).

[7] G.P. Lepage and S.J. Brodsky, Phys. Lett. B **87** 359 (1979).

[8] A.P. Bakulev, S.V. Mikhailov and N.G. Stefanis, Phys. Lett. B **508** 279 (2001), Erratum-ibid, B **590** 309 (2004).

[9] S.V. Mikhailov and N.G. Stefanis, Nucl. Phys. B **821** 291 (2009).

[10] A.V. Radyushkin, arXiv:hep-ph/0906.0323 (2009).

[11] M.V. Polyakov, arXiv:hep-ph/0906.0538 (2009).

[12] S. Brodsky, P. Lepage, Phys. Rev. D **24** 1808 (1981).

[13] M. Diehl, P. Kroll, C. Vogt, Phys. Lett. B **532** 99 (2002).

[14] C.C. Kuo et al. (Belle Collab.), Phys. Lett. B **621** 41 (2005).

[15] H. Nakazawa et al. (Belle Collab.), Phys. Lett. B **615** 39 (2005).

[16] W.T. Chen et al. (Belle Collab.), Phys. Lett. B **651** 15 (2005).

[17] S. Uehara et al. (Belle Collab.), Phys. Rev D **79** 052009 (2009).

[18] S. Uehara et al. (Belle Collab.), Phys. Rev. D **80** 032001 (2009).

[19] A. Heister et al. (ALEPH Collab.), Phys. Lett. B **569** 140 (2003).

[20] M. Benayoun, V.L. Chernyak, Nucl. Phys. B **329** 285 (1990).

# Photon Physics at CMS

*Jonathan Hollar*[1]

[1]Lawrence Livermore National Laboratory, 3000 East Ave., Livermore, California, USA

The Large Hadron Collider will allow studies of $\gamma\gamma$ and photoproduction interactions at energies significantly higher than previous experiments, in both $pp$ and heavy ion collisions. In this article, studies of the feasibility of measuring $\gamma\gamma \to \ell^+\ell^-$, $\gamma p \to \Upsilon p \to \ell^+\ell^- p$, and $\gamma A \to \Upsilon A \to \ell^+\ell^- A$ processes in early LHC data with the CMS detector are presented.

# 1 Introduction

Exclusive dilepton production in $pp$ collisions at CMS can occur through the processes $\gamma\gamma \to \ell^+\ell^-$ and $\gamma p \to \Upsilon p \to \ell^+\ell^- p$. Due to its precisely known cross-section the QED $\gamma\gamma \to \ell^+\ell^-$ process is a potential candidate for absolute luminosity measurements at the LHC, if the experimental systematics can be controlled. The cross-section for the $\gamma p \to \Upsilon p \to \ell^+\ell^- p$ process is expected to be related to the generalised gluon density of the proton as $\sigma \sim [g(x)]^2$. The increasing gluon density at low values of the fractional momentum $x$ leads to a predicted dependence of $\sigma \sim W^{1.7}$ in leading order perturbative QCD, where $W$ is the $\gamma p$ centre-of-mass energy. This results in large predicted cross-sections at the LHC, which will probe values of $\langle W \rangle$ significantly higher than previous measurements in $ep$ collisions [10, 9]. At higher luminosities, photon interactions will reach sufficient energies to probe a wide range of physics beyond the Standard Model, such as Supersymmetric $\gamma\gamma \to \tilde{\ell}^+\tilde{\ell}^-$ production [2, 3, 4, 5], anomalous gauge boson couplings in $\gamma\gamma \to W^+W^-$ [5, 6], and other exotic models [7, 8]. In addition, the $pp \to p\ell^+\ell^- p$ sample will serve as an alignment sample for proposed forward proton spectrometers [12].

In heavy ion collisions, the effective luminosity of photon interactions is enhanced by the large electromagnetic field of the nucleus: by a factor $Z^4$ in $\gamma\gamma$ interactions, and a factor $Z^2$ in photoproduction. This allows studies of higher-order QED effects in a strongly-interacting regime ($\alpha Z \sim 0.6$), and studies of nuclear PDF's in vector-meson photoproduction. The $\gamma\gamma \to \ell^+\ell^-$ and $\gamma A \to J/\psi A^{(*)} \to \ell^+\ell^- A^{(*)}$ processes have been observed at RHIC [13]; the higher energies of the LHC will also allow studies of $\gamma A \to \Upsilon A^{(*)} \to \ell^+\ell^- A^{(*)}$.

# 2 Exclusive $\gamma\gamma \to \ell^+\ell^-$ and $\gamma p \to \Upsilon p \to \ell^+\ell^- p$ in $pp$ Collisions

A Monte Carlo study of the prospects for measuring $\gamma\gamma \to \ell^+\ell^-$ and $\gamma p \to \Upsilon p \to \ell^+\ell^- p$ at CMS with 100 pb$^{-1}$ of integrated luminosity at $\sqrt{s} = 14$ TeV has been performed. At low instantaneous luminosities the rate of pileup (multiple interactions in the same bunch crossing) is expected to be small, meaning the signal can be distinguished by the presence of two leptons and no other activity above the noise thresholds. Zero pileup is assumed throughout the current study.

Signal events are produced using the LPAIR [11] (for $\gamma\gamma \to \ell^+\ell^-$) and STARLIGHT [14] (for $\gamma p \to \Upsilon p \to \ell^+\ell^- p$) generators. The largest background after all selection criteria are applied are inelastic photon-exchange events, in which the proton remnants escape undetected. The other sources of dilepton backgrounds relevant to CMS include Drell-Yan production, quarkonium decays, and heavy-flavor semileptonic decays. The Pythia generator is used to generate all of these samples. Both signal and background samples are passed through a full detector simulation, trigger emulation, and reconstruction.

# 3 Event Selection

## 3.1 Trigger and Lepton Selection

As the signal consists primarily of very low $p_T$ leptons, the lowest possible trigger thresholds are required to select a large sample of events. In the dimuon channel, the standard CMS dimuon trigger for a luminosity of $10^{32}\,\mathrm{cm}^{-2}\mathrm{s}^{-1}$ is assumed, requiring two muons with $p_T > 3$ GeV. In the dielectron channel, the threshold is lowered using a dedicated trigger. In the Level 1 (hardware) trigger, exactly two EM candidates and no additional jets with $p_T > 10$ GeV are required. In the High Level Trigger (HLT), the EM candidates are required to be matched to charged tracks, and to be back-to-back in $\phi$ and balanced in the transverse energy $E_T$.

Starting from the triggered sample of dileptons, we require that the offline reconstruction find exactly two same flavor opposite-sign dileptons in the event. Further selections on the acoplanarity ($|\Delta\phi(\ell^+\ell^-)|$) and transverse momentum balance ($|\Delta p_T(\ell^+\ell^-)|$) are applied, exploiting the constrained kinematics of the signal. In the $\mu^+\mu^-$ channel, we require $|\Delta\phi(\mu^+\mu^-)| > 2.9$ and $|\Delta p_T(\mu^+\mu^-)| < 2.0$ GeV. In the $e^+e^-$ channel, we require $|\Delta\phi(e^+e^-)| < 2.7$ and $|\Delta E_T(e^+e^-)| < 5.0$ GeV.

## 3.2 Exclusivity Selection

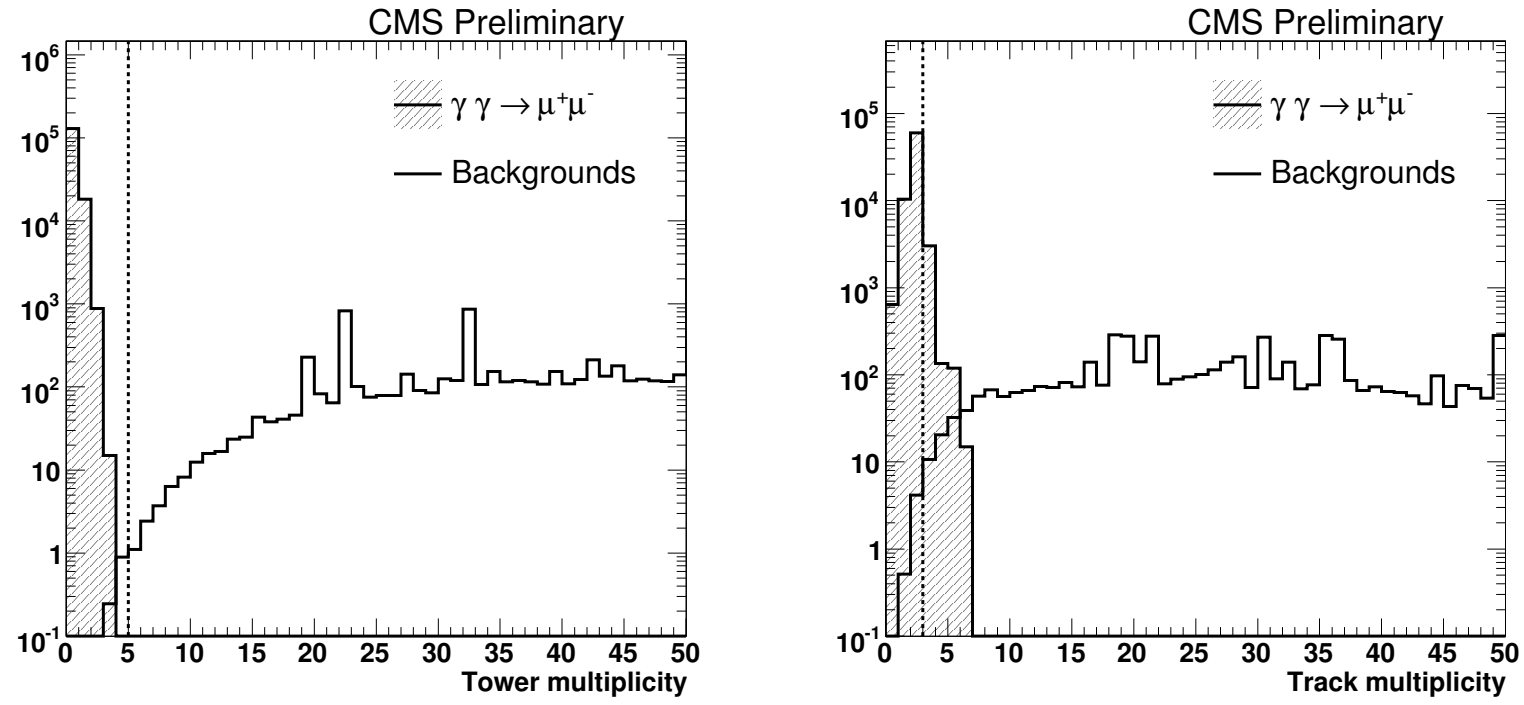

Figure 1: Multiplicity of extra calorimeter towers (left) and charged track multiplicity (right) for signal (shaded histogram) and background (open histogram), with arbitrary normalization.

In the no-pileup startup scenario assumed here, the signal is distinguished by having no calorimeter activity that is not associated with the leptons, and no charged tracks in addition to the two signal leptons. A common exclusivity selection is applied for the dimuon and dielectron channels.

The calorimeter exclusivity requirement is implemented by requiring there be no more than 5 "extra" calorimeter towers with $E > 5$ GeV, where extra towers are defined as those separated from either of the lepton candidates by $\Delta R > 0.3$ in the $\eta - \phi$ plane. The resulting distribution for the $\gamma\gamma \to \mu^+\mu^-$ signal and the sum of backgrounds are shown in Figure 1 (left).

The track exclusivity requirement provides additional background suppression in the $|\eta| < 2.5$ region covered by the tracker. The track multiplicity distributions for signal and background are shown in Figure 1 (right). We require the track multiplicity be $N(tracks) < 3$.

### 3.2.1 Forward Detector Vetos

Since inelastic photon-exchange events will have a high efficiency for passing the selection described in the previous sections, a veto on activity in the forward detectors is useful in suppressing these backgrounds. The *ZDC* (Zero Degree Calorimeter) detector covers the region $|\eta| > 8.2$, and will detect high energy neutrons and photons. The*CASTOR* detector provides a nearly continuous extension of coverage from the hadronic calorimeter to the region $5.2 < |\eta| < 6.6$.

Since a full simulation and reconstruction is not yet available for these detectors, we estimate their ability to reject inelastic backgrounds based on their generator level acceptance. We assume a startup scenario in which a *CASTOR* detector is available on one side of CMS only, and ZDC detectors are available on both sides. With these assumptions, approximately 2/3 of the remaining inelastic background can be vetoed using the combination of *CASTOR* and *ZDC*.

# 4 Results

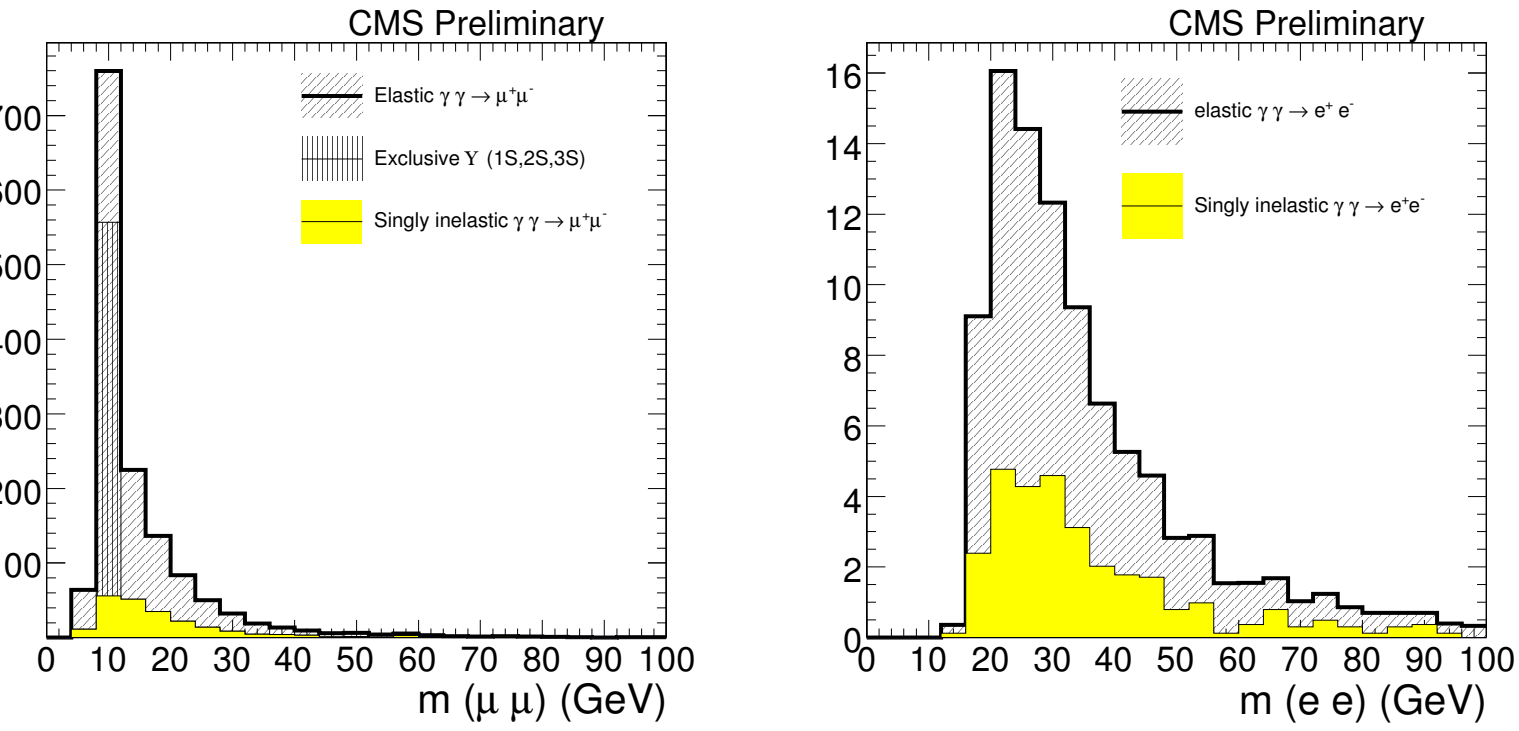

Figure 2: $m(\mu^+\mu^-)$ (left) and $m(e^+e^-)$ (right) for events passing all selection criteria. The elastic two-photon (open histogram) $\Upsilon$ photoproduction (shaded histogram), and singly inelastic background (solid histogram) are shown.

The dilepton invariant mass distributions after application of all trigger and selection criteria are shown in Figure 2. The expected elastic $\gamma\gamma \to \mu^+\mu^-$ signal yields in $100\,\mathrm{pb}^{-1}$ are:

$$N_{elastic}(\gamma\gamma \to \mu^+\mu^-) = 709 \pm 27,$$

where the statistical error is $\sqrt{N_{elastic}}$. The expected contribution from singly inelastic two-photon events, assuming the ZDC and CASTOR vetoes described previously, is:

$$N_{inelastic}(\gamma\gamma \to \mu^+\mu^-) = 223 \pm 15 \pm 42(\mathrm{model}),$$

where the first uncertainty is taken as $\sqrt{N_{inelastic}}$, and the second corresponds to a 19% model-dependent uncertainty on the inelastic cross-section. Without the ZDC and Castor vetoes, the singly inelastic contribution would be significantly larger:

$$N_{inelastic}(\gamma\gamma \to \mu^+\mu^-) = 636 \pm 25 \pm 121(\mathrm{model}),$$

where the errors are $\sqrt{N_{inelastic}}$ and model-dependence.

In the $\gamma\gamma \to e^+e^-$ channel, the expected yields are significantly smaller due to the higher trigger threshold and lower efficiency for reconstructing low $E_T$ electrons. After all trigger and selection criteria are applied the expected elastic signal yields in $100\,\mathrm{pb}^{-1}$ are:

$$N_{elastic}(\gamma\gamma \to e^+e^-) = 67 \pm 8,$$

where the error is $\sqrt{N_{elastic}}$. The expected contribution from singly inelastic two-photon events, assuming the ZDC and CASTOR vetoes, is:

$$N_{inelastic}(\gamma\gamma \to e^+e^-) = 31 \pm 6 \pm 6(\mathrm{model}),$$

where the first uncertainty is $\sqrt{N_{inelastic}}$, the second corresponds to a 19% model-dependent uncertainty on the inelastic cross-section. Without the ZDC and CASTOR vetoes, the singly inelastic contribution would be:

$$N_{inelastic}(\gamma\gamma \to e^+e^-) = 82 \pm 9 \pm 15(\mathrm{model}),$$

where the errors are $\sqrt{N_{inelastic}}$ and model-dependence.

# 5 Applications

## 5.1 Luminosity Studies

The small theoretical uncertainty on the $\gamma\gamma \to \ell^+\ell^-$ cross-section makes it a candidate for use as an absolute luminosity normalization sample. Due the contribution of the theoretically less clean inelastic events, the elastic signal cannot be extracted on an event-by-event basis. However, the $\Delta\phi$ and $\Delta p_T$ distributions provide a means of statistically separating the two contributions. This is shown in Figure 3 for the sample of $\gamma\gamma \to \mu^+\mu^-$ events passing all selection requirements.

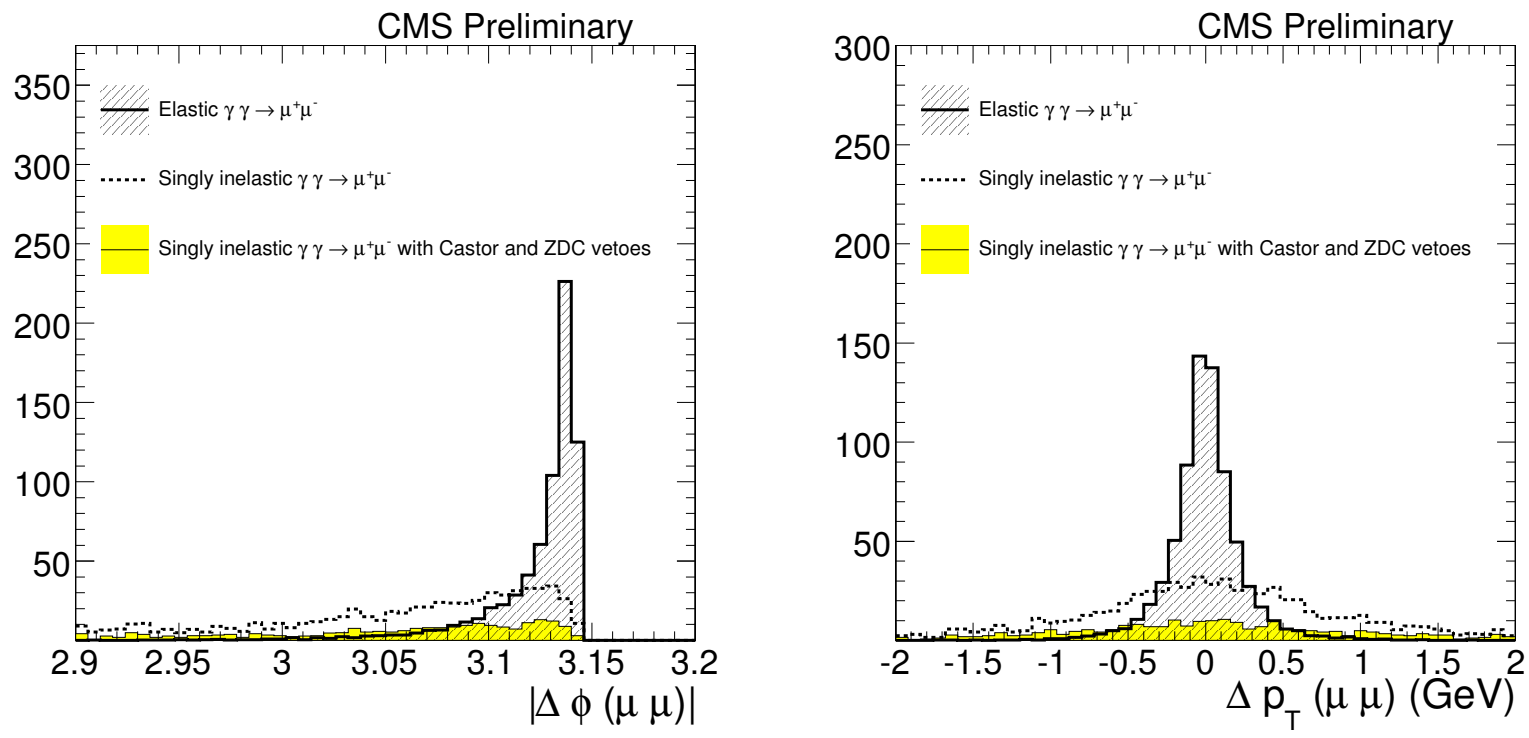

Figure 3: Distributions of $|\Delta\phi(\mu^+\mu^-)|$ (left) and $|\Delta p_T(\mu^+\mu^-)|$ (right) for $\gamma\gamma \to \mu^+\mu^-$ events passing all selection requirements. The elastic signal is denoted by the open histogram, the inelastic background is shown with no $CASTOR/ZDC$ vetos (dashed line), and with the vetos described in the text (solid histogram).

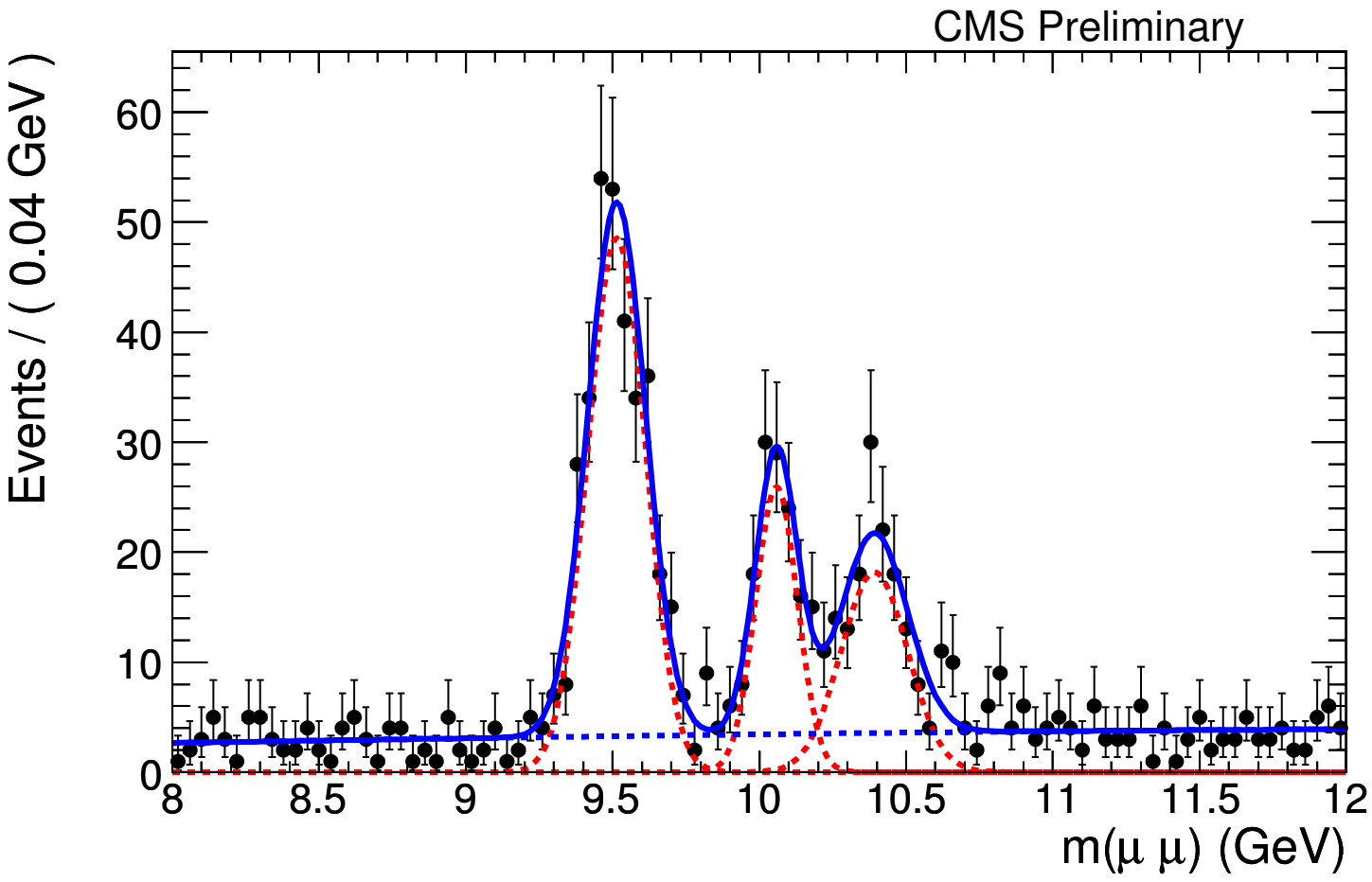

Figure 4: Dimuon invariant mass in the range $8 < m(\mu^+\mu^-) < 12$ GeV. The lines show the result of a fit, where the dashed line is the $\Upsilon$ component, the dotted line is the two-photon continuum, and the solid line is the sum of the two.

## 5.2 $\Upsilon$ Physics

The $\Upsilon$ photoproduction signal can be extracted by performing a fit to the dimuon invariant mass distribution in the range $8 < m < 12$ GeV. The $1S$, $2S$, and $3S$ $\Upsilon$ resonances are fitted to single Gaussians, while the sum of elastic and inelastic $\gamma\gamma \to \mu^+\mu^-$ contributions are fitted to a second order polynomial (Figure 4). For an assumed integrated luminosity of $100\,\text{pb}^{-1}$ and the cross-section predicted from $STARLIGHT$, the three $\Upsilon$ resonances are clearly visible above the $\gamma\gamma$ continuum. The $\gamma p$ centre-of-mass energy for the selected events is $\langle W\rangle_{visible} = 537$ GeV.

With hundreds of events, further studies of $\Upsilon$ production, such as measuring $t$ or $\eta$ distributions, may be performed. While $t$ cannot be measured directly, the $p_T^2$ of the $\Upsilon$ is expected to provide a good approximation. Preliminary studies with STARLIGHT signal samples confirm this, with the measured slope of the reconstructed $p_T^2$ distribution $b = (3.50 \pm 0.06)\,\text{GeV}^{-2}$ in agreement with the generator-level value of $b = (4.03 \pm 0.04)\,\text{GeV}^{-2}$.

# 6 Heavy Ion Interactions

In heavy ion collisions, the effective cross-section is enhanced over the $pp$ case by a factor of $Z^4$ for $\gamma\gamma$ interactions ($Z^2$ for $\gamma A$ interactions), resulting in large event yields for relatively low integrated luminosities. A full simulation study has been performed using events generated with the STARLIGHT model, for an assumed Pb-Pb run of $0.5\,\text{nb}^{-1}$ collected at $\sqrt{s} = 5.5$ TeV. The trigger selection requires identification of a $\mu^+\mu^-$ or $e^+e^-$ pair of any energy, no significant activity within the forward hadronic calorimeter (covering $3 < |\eta| < 5$), and a neutron from the Coulomb breakup of the the ion detected in the ZDC.

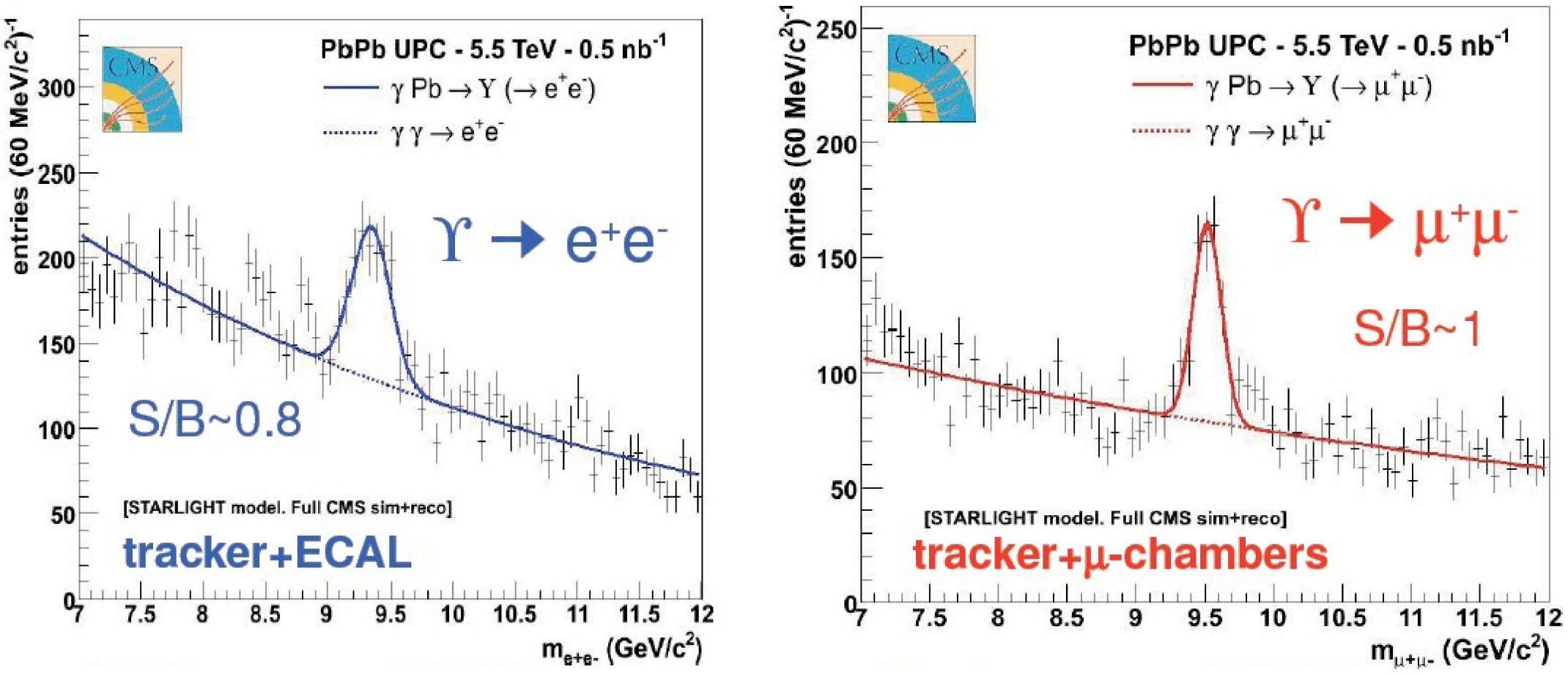

Figure 5: Dielectron (left) and dimuon (right) invariant masses in the range $7 < m(\mu^+\mu^-) < 12$ GeV. The lines show the result of a fit to the $\Upsilon(1S)$ signal and continuum background.

The residual non-photon exchange background is subtracted using a sample of like-sign dileptons. After the trigger selection and background subtraction are performed, a sample of $\sim 180$ $\Upsilon(1S) \to \mu^+\mu^-$ and $\sim 220$ $\Upsilon(1S) \to e^+e^-$ events are expected in $0.5\,\text{nb}^{-1}$. As in the $pp$ case, the good resolution in the dimuon channel should allow resolution of the $\Upsilon(2S)$ and $\Upsilon(3S)$ resonances (not included in the simulation used here). Systematic uncertainties on the Upsilon yields are estimated to be $\sim 10\%$, based on variations in the shape of the continuum background.

## 7 Conclusions

With $100\,\mathrm{pb}^{-1}$ of integrated luminosity at $\sqrt{s} = 14$ TeV, a large sample of $\gamma\gamma \rightarrow \mu^+\mu^-$ and $\gamma p \rightarrow \Upsilon p \rightarrow \mu^+\mu^- p$ events can be triggered and reconstructed in the CMS experiment, using a common selection for both samples. A smaller sample of $\gamma\gamma \rightarrow e^+e^-$ events will also be collected. With minimal pileup these events can be distinguished using exclusivity requirements, and the inelastic backgrounds reduced using forward detector vetos. The $\Upsilon$ sample will allow measurements of cross-sections and production dynamics at significantly higher energies than previous experiments. The $\gamma\gamma \rightarrow \ell^+\ell^-$ sample will serve as a calibration sample for studies of luminosity and lepton reconstruction. In Heavy Ion interactions, a significant signal for $\gamma A \rightarrow \Upsilon A^{(*)} \rightarrow \ell^+\ell^- A^{(*)}$ may be observed in both the $\mu^+\mu^-$ and $e^+e^-$ channels in a nominal $0.5\,\mathrm{nb}^{-1}$ Pb-Pb run at $\sqrt{s} = 5.5\,\mathrm{TeV}$.

## References

[1] CERN-LHCC-2006-01 CMS Collaboration

[2] Phys.Lett.B 328 (1994) 369-373, J. Ohnemus, T.F. Walsh, and P.M. Zerwas

[3] Phys.Rev.D50 (1994) 2335-2338 M. Drees, R.M. Godbole, M. Nowakowski, and S.D. Rindani

[4] Phys.Rev.D53 (1996) 2371-2379, G. Bhattacharya, Pat Kalyniak, and K.A. Peterson

[5] arXiv:0908.2020, J. de Favereau de Jeneret, et al.

[6] Phys.Rev.D78 (2008) 073005, O. Kepka, and C. Royon

[7] Phys.Rev.Lett.102 (2009) 222002 T. Aaltonen et al.

[8] Phys.Rev.Lett.81 (1998) 524-529 B. Abbott et al.

[9] Phys. Lett.B (2000) 483 C. Adloff, et al.

[10] arXiv:0903.4205 ZEUS Collaboration

[11] Nucl.Phys.B 229 (1983) 347, J.A.M. Vermaseren

[12] arXiv:0806.0302, M. Albrow, et al.

[13] Phys.Lett.B 679 (2009) 321-329, S. Afanasiev et al.

[14] hep-ph/0311164J. Nystrand and S. Klein

[15] hep-ph/0412096J. Nystrand and S. Klein

[16] J. Phys.G 23 (1997) 2069-2080 Angelis, A. L. S. and Panagiotou, A. D.

[17] AIP Conf. Proc. 867 (2006) 258-265 Grachov, Oleg A., et al.

# Anomalous Quartic $WW\gamma\gamma$ and $ZZ\gamma\gamma$ Couplings in Two-Photon Processes at the LHC

*E. Chapon*[1], *O. Kepka*[1,2,3], *C. Royon*[1]

[1] IRFU/Service de physique des particules, CEA/Saclay, 91191 Gif-sur-Yvette cedex, France
[2] IPNP, Faculty of Mathematics and Physics, Charles University, Prague
[3] Center for Particle Physics, Institute of Physics, Academy of Science, Prague

Two-photon production of $WW/ZZ$ pairs is used to calculate sensitivities to anomalous quartic couplings between photon and massive electroweak bosons at the LHC. We show how the signal can be selected from the diffractive and two-photon backgrounds and that the current sensitivities can be improved by almost two orders of magnitude with early data. Using the full LHC luminosity and the forward detectors installed at 220 m and 420 m from the CMS or ATLAS detectors, sensitivities on anomalous couplings of the order of $10^{-6}\,\mathrm{GeV}^{-2}$ can be achieved.

## 1 Two-Photon Diboson Production

In the Standard Model (SM), the interaction between the electroweak bosons is dictated by the underlying non-abelian symmetry of the SM Lagrangian. Although the triple gauge boson couplings are measured quite precisely, the constraints of the quartic couplings come from LEP experiments and are less stringent as they required three bosons to be detected. In this work we focus on deriving sensitivities to anomalous quartic vertices involving two photons and a pair of $W$ or $Z$ bosons in exclusive interactions induced by the exchange of two photons at the LHC. Only the fully-leptonic decays of the electroweak bosons are considered. The analysis follows the first investigation [1], but with a broader set of backgrounds considered and also calculated for two different luminosity scenarios.

In two-photon interactions, two almost real photons are emitted from high energetic proton beams, interact with each other and yield various final states such as the pair of leptons, jets, and electroweak bosons, see Figure 1 (left). Since protons leave the interaction intact, and are scattered at small angles, there is no proton remnant, and only the object created from two photons populates the central detector. The fractional momentum loss of the proton is $\xi \equiv 1 - |\mathbf{p}|/|\mathbf{p}_b|$, where $\mathbf{p}_b$ is the beam momentum and $\mathbf{p}$ is the scattered proton momentum. The process is exclusive and can be selected either by tagging the outgoing forward protons or by requiring the detection of the created system from two photons in the central detector and nothing else. The two-photon production is calculated in the framework of the Equivalent Photon Approximation (EPA) which is in detail described in [2].

The two-photon production of $W$ pairs has a total cross section of 95.6 fb (using the value of the fine-structure constant $\alpha_{\mathrm{QED}} = 1/137$ at the scale $Q^2 = 0\,\mathrm{GeV}^2$) whereas the production of $Z$ pairs is forbidden in the SM in the leading order since neither the photon nor the $Z$ boson carry the electric or weak charges.

The sensitivities to potential Beyond Standard Model (BSM) physics is investigated in terms of four anomalous parameters $a_0^W/\Lambda^2$, $a_0^Z/\Lambda^2, a_C^W/\Lambda^2$, $a_C^Z/\Lambda^2$ using effective anomalous Lagrangians conserving C− and P−parities separately which are added to the SM [3]. They read

$$\mathcal{L}_6^0 = \frac{-e^2}{8}\frac{a_0^W}{\Lambda^2}F_{\mu\nu}F^{\mu\nu}W^{+\alpha}W^-_\alpha - \frac{e^2}{16\cos^2\theta_W}\frac{a_0^Z}{\Lambda^2}F_{\mu\nu}F^{\mu\nu}Z^\alpha Z_\alpha,$$

$$\mathcal{L}_6^C = \frac{-e^2}{16}\frac{a_C^W}{\Lambda^2}F_{\mu\alpha}F^{\mu\beta}(W^{+\alpha}W^-_\beta + W^{-\alpha}W^+_\beta) - \frac{e^2}{16\cos^2\theta_W}\frac{a_C^Z}{\Lambda^2}F_{\mu\alpha}F^{\mu\beta}Z^\alpha Z_\beta,$$

where $F_{\mu\nu} = \partial_\mu A_\nu - \partial_\nu A_\mu$ is the electromagnetic stress-energy tensor and $\theta_W$ is the Weinberg angle. $\Lambda$ is a typical scale of new physics (e.g. the mass of new particle) whose exact value is, however, not important since sensitivities to entities including this scale are derived, $a_0^W/\Lambda^2$ for instance. The additional Lagrangians are the lowest order operators involving two photons which can be constructed having a correct Lorentz structure and obeying custodial symmetry protecting the value of $\rho = M_W/(M_Z\cos\theta_W)$ to be close to the experimentally measured value 1. They have a dimension six in terms of energy. Notice that in $\mathcal{L}_6^0$ Lorentz indices are decoupled and the operator can be interpreted as a low energy exchange of a massive scalar field. In fact, since anomalous couplings of the order of $10^{-6}\,\mathrm{GeV}^{-2}$ emerge from 1-loop corrections in Higgs-less theories, the measurement proposed in this paper can give us an important information about the symmetry breaking mechanism and mass generation.

Adding these new operators, the total cross section is greatly enhanced as depicted in Figure 1 (right). In the SM, the $\gamma\gamma \to WW$ cross section is constant in the high energy limit due to the cancellation between the involved diagrams ($s-$, $t-$channel, direct four-boson diagrams). Adding new quartic term, the cancellation does not hold any more and the cross section growth as a function of the centre-of-mass energy is driven by diagrams in which the dibosons in the final state are longitudinally polarised. This eventually leads to the violation of unitarity in the process. The main problem is that the unitarity is violated already at the LHC energies which makes the interpretation of new terms as the effective description of new physics impossible. The usual approach in this case is to supplement the anomalous parameters with form factors which weaken the effect of the coupling at high centre-of-mass energy and mimic the exchange of a particle of mass which is beyond the reach of the accelerator. Following the study [1] we use the following parameterisation of the coupling form factor $a \to a/\left[1 + (W_{\gamma\gamma}/2\,\text{TeV})^2\right]^2$, where $W_{\gamma\gamma} = \sqrt{s\xi_1\xi_2}$ is the invariant mass of two photons and 2 TeV is the scale beyond which the effect of the coupling is suppressed. More detailed discussion on the form factors in our analysis will be given in an up-coming publication.

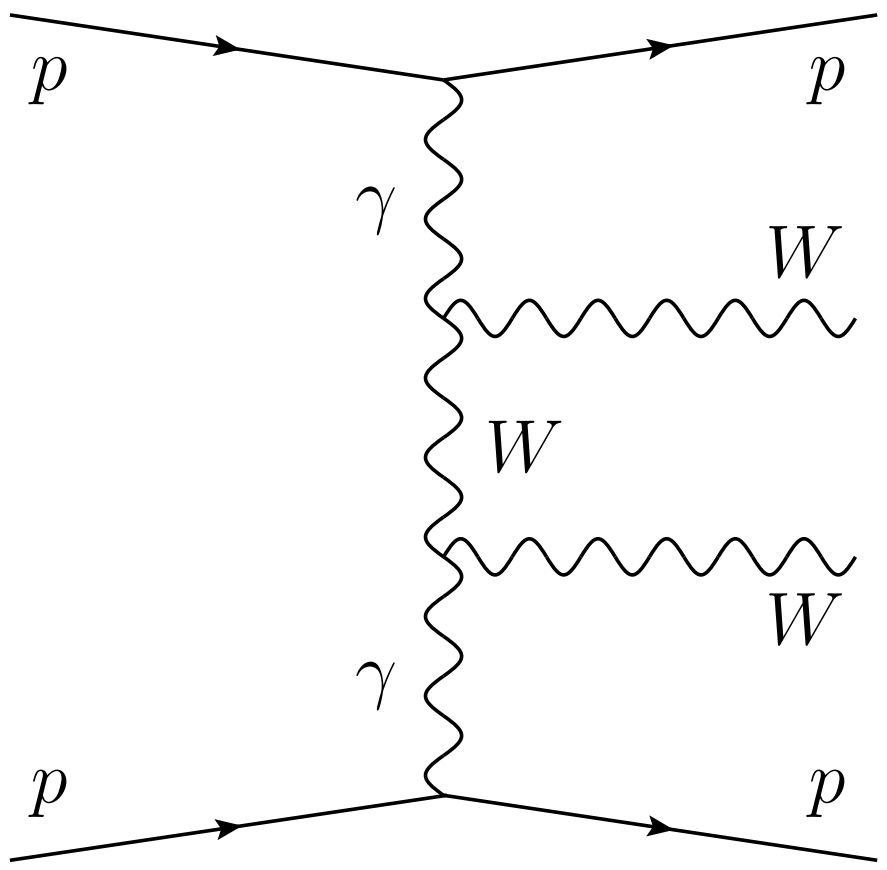

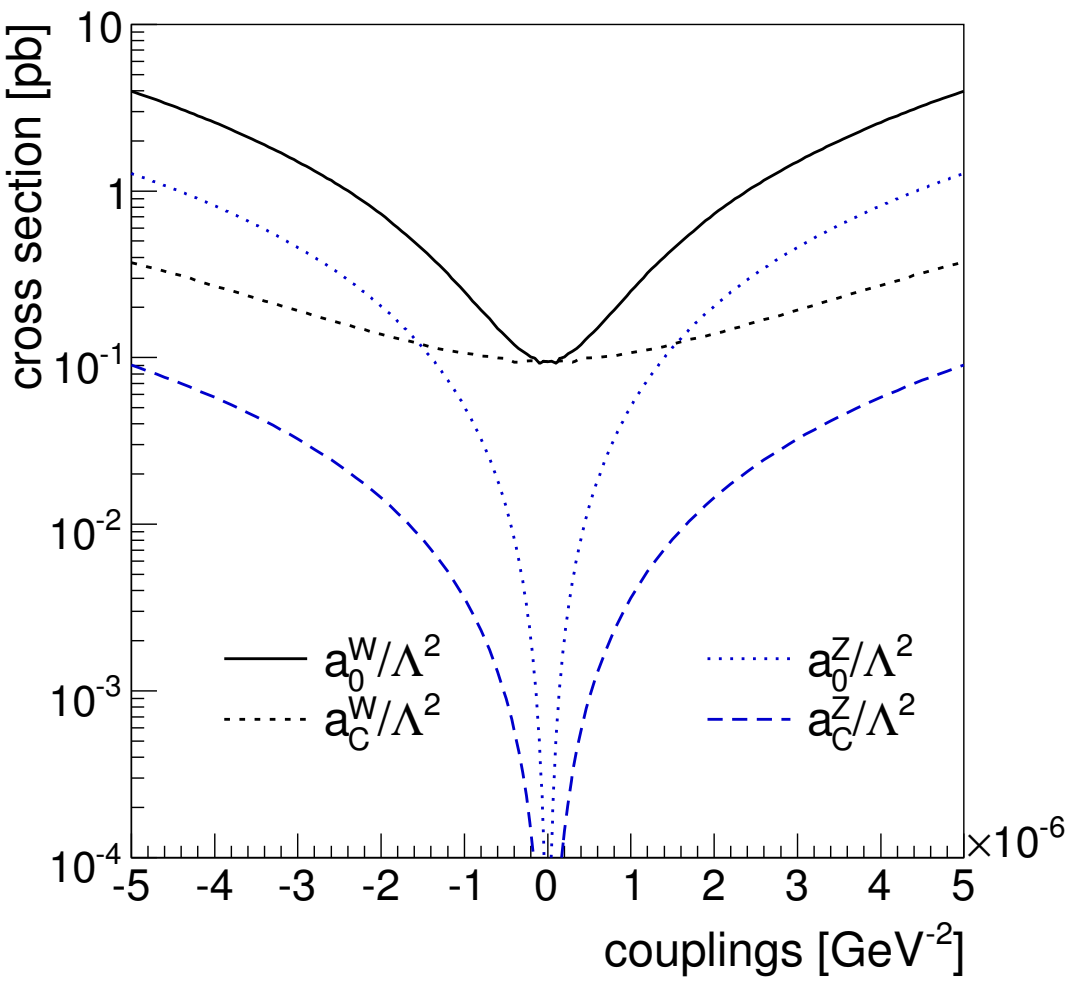

Figure 1: Leading order Feynman diagram of two-photon $WW$ production (left) and enhancement of the total $W$ and $Z$ pair production cross sections in two-photon events from their SM values 95.6 fb and 0 fb, respectively (right). No survival probability which should be 0.9 for two-photon events is applied.

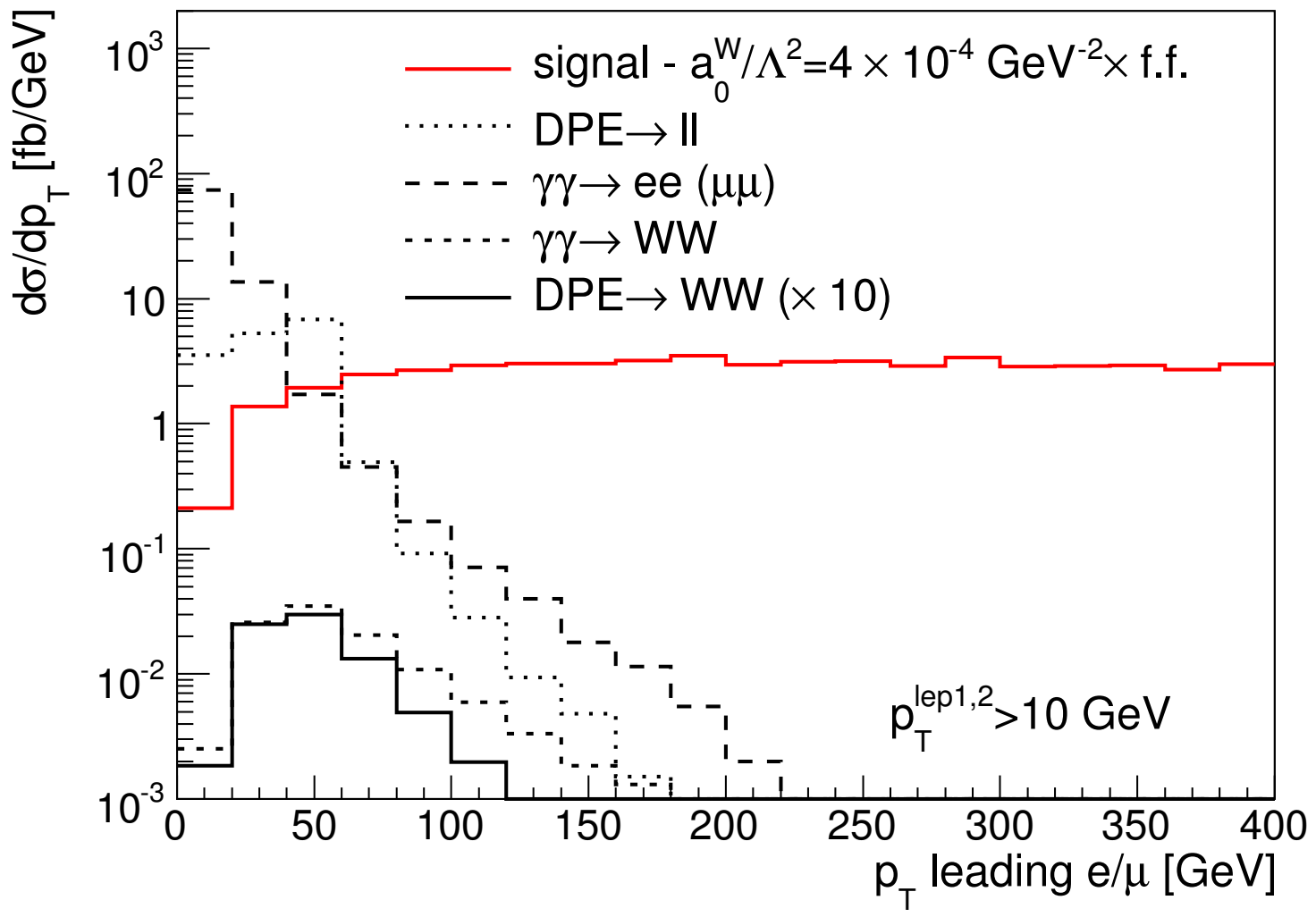

Figure 2: $p_T$-distribution of the leading electron/muon for signal and background processes. As expected, the signal appears at high $p_T$. The largest background are two-photon $ee(\mu\mu)$ production and DPE→ $l\bar{l}$.

Since the effect of anomalous parameters on the cross section is large even when form factors are taken into account, the study is divided into two steps. First we present an analysis using the central detector only and limited collected luminosity of 10 (100) pb$^{-1}$ at a reduced centre-of-mass LHC energy $\sqrt{s} = 10$ TeV. To achieve the ultimate sensitivity, the high LHC luminosity 30 (200) fb$^{-1}$ at $\sqrt{s} = 14$ TeV is used in which the two-photon events are selected using very ATLAS Forward Proton (AFP) detectors to be installed at 220 m and 420 m on both sides around the interaction point at few millimetres from the beam. These detectors were recently recognised by the ATLAS management as a potential upgrade of the ATLAS detector and the ATLAS community showed an interest in the diffractive program. The analysis using the CMS detector and the corresponding forward detectors would be similar.

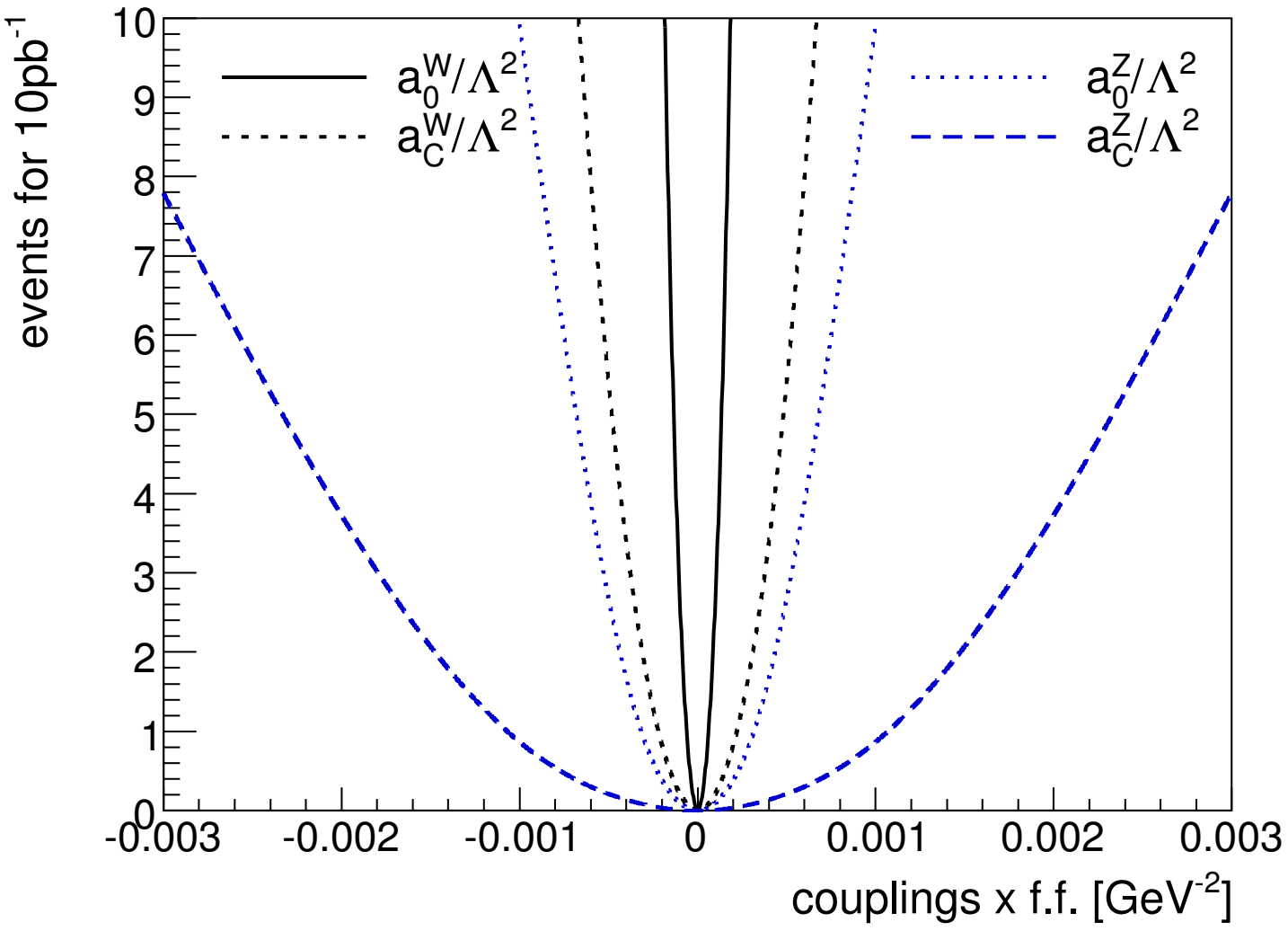

Figure 3: Expected number of signal events due to anomalous quartic couplings for $10\,\mathrm{pb}^{-1}$ after considered cuts (see text).

# 2 Background

The two-photon $WW/ZZ$ events in which both bosons decay leptonically were studied with respect to the following background: 1) non-diffractive $WW/ZZ$ production which has a large energy flow in the forward region and large number of tracks due to the break-up of the collided protons 2) $\gamma\gamma \to l\bar{l}$ - two-photon dilepton production in which leptons fly exactly back-to-back in the transverse plane 3) DPE$\to$ $l\bar{l}$ - dilepton production through double Pomeron exchange (DPE) in which the partonic structure of the hard diffractive Pomeron is probed in Drell-Yan like process and higher number of tracks is expected due to the Pomeron remnants 4) DPE$\to$ $WW$ - diboson production through double Pomeron exchange, the same production as in the precedent case.

Two-photon or diffractive processes with intact protons were generated using the `FPMC` generator [4] and simulated using `ATLFAST++`, fast standalone simulation of ATLAS inside `ROOT`. The matrix elements with anomalous couplings are calculated using the `CompHEP` program [5] interfaced with `FPMC`. The implemented survival probability factors are 0.9 for the QED two-photon processes and 0.03 for the Pomeron exchanges [6].

The comparison between the anomalous signal and the considered backgrounds with intact protons is shown in Figure 2. The signal is dominant at high transverse lepton momenta and also at large missing mass $W_{\gamma\gamma}$ which sets the analysis strategy.

# 3 Anomalous Quartic Couplings at Low Luminosity

Let us describe our selection criteria for the low luminosity scenario $10\,(100)\,\mathrm{pb}^{-1}$ in which only one interaction per bunch crossing is present. Such amount of data is expected to be collected during the first months of LHC running. In order to reject non-diffractive and DPE backgrounds with large number of particles due to proton and Pomeron remnants, we require an exclusivity cut $n_{\mathrm{tracks}} \leq 2$. To further suppress the background from two-photon dilepton events, we request large transverse lepton ($e$, $\mu$) momentum $p_T^{lep1} > 160\,\mathrm{GeV}$, $p_T^{lep2} > 10\,\mathrm{GeV}$ and large missing transverse energy $\not{E}_T > 20\,\mathrm{GeV}$ which is a natural characteristics of $W$ events decaying to leptons and neutrinos.

| Couplings | OPAL limits $[\mathrm{GeV}^{-2}]$ | Sensitivity @ $\mathcal{L} = 10\ (100)\,\mathrm{pb}^{-1}$ 95% CL $[\mathrm{GeV}^{-2}]$ | Sensitivity @ $\mathcal{L} = 30\ (200)\ \mathrm{fb}^{-1}$ 95% CL $[\mathrm{GeV}^{-2}]$ |
|---|---|---|---|
| $a_0^W/\Lambda^2$ | [-0.020, 0.020] | $1.0{\times}10^{-4}$ $(3.3 \times 10^{-5})$ | $2.6 \times 10^{-6}$ $(1.4 \times 10^{-6})$ |
| $a_C^W/\Lambda^2$ | [-0.052, 0.037] | $3.5 \times 10^{-4}$ $(1.1 \times 10^{-4})$ | $9.4 \times 10^{-6}$ $(5.2 \times 10^{-6})$ |
| $a_0^Z/\Lambda^2$ | [-0.007, 0.023] | $5.2 \times 10^{-4}$ $(1.7 \times 10^{-4})$ | $6.4 \times 10^{-6}$ $(3.7 \times 10^{-6})$ |
| $a_C^Z/\Lambda^2$ | [-0.029, 0.029] | $1.8 \times 10^{-3}$ $(5.9 \times 10^{-4})$ | $35 \times 10^{-6}$ $(14 \times 10^{-6})$ |

Table 1: 95% confidence limits on the anomalous quartic couplings using $10\,(100)\,\mathrm{pb}^{-1}$ of early data (third column) and using $30\,(200)\,\mathrm{fb}^{-1}$ with forward detectors (last column). Coupling form factors are used as described in the text. The OPAL limits [7] can be improved up to four orders of magnitude.

The $ZZ$ signal is background free because two leptons of the same charge are created when both $Z$s decay leptonically. The requirement which was used to select the $ZZ$ signal was either to have $\geq 2$ leptons of the same charge, or $\geq 3$ leptons. Leptons are required to have a transverse momentum $p_T^{lep1}$ >160 GeV and $p_T^{lep2}$ >25 GeV. In addition, no jet can be seen in the event. Such requirements are sufficient to reject all non-diffractive, two-photon or DPE exchange background.

The expected number of signal events for $10\,\mathrm{pb}^{-1}$ after the mentioned cuts is shown in Figure 3. The sensitivities to anomalous couplings using low luminosity are shown in Table 1 (third column). They are obtained by varying one of the anomalous parameters while keeping the others to their zero SM values. The sensitivities can be improved by two orders of magnitude with respected to the limits from the OPAL Collaboration [7] with a small amount of data.

# 4 Anomalous Quartic Couplings at High Luminosity

To obtain sensitivities to anomalous couplings of the order of $10^{-6}\,\mathrm{GeV}^{-2}$ which would allow to test several Higgs-less theories, the full LHC collected luminosity must be used. During the nominal LHC runs about 32 interactions will be present at the peaked luminosity and the number of interactions per bunch crossing will not drop below 13 during one store (the total cross section $\sigma_{\mathrm{tot}} = 100\,\mathrm{mb}$ is assumed). In this case, protons have to be tagged in forward detectors to select the two-photon or diffractive events. Moreover, fast timing detectors will be used to reject the overlap background in which one inelastic event in the central detectors is overlaid with two soft diffractive events giving proton hits in the forward detectors.

In our analysis we select events with fractional momentum loss inside the generic acceptance of the proposed AFP detectors $0.0015 < \xi < 0.15$. To select $WW$ signal we apply the $\not{E}_T > 20\,\mathrm{GeV}$ cut. The corresponding missing mass spectrum $W_{\gamma\gamma}$ for the $WW$ anomalous signal and all backgrounds is shown in Figure 4. To further suppress the contribution of DPE and two-photon dilepton signal, we require the invariant lepton mass to be far from the $Z$ pole $M_{ll} \notin \langle 80, 100\rangle$ GeV, dileptons not back-to-back $\Delta\phi < 3.13$ and large missing mass $W_{\gamma\gamma} > 800\,\mathrm{GeV}$, $p_T^{lep1} > 160\,\mathrm{GeV}$, $p_T^{lep2} > 10\,\mathrm{GeV}$. After these cuts the total considered background is reduced to $\approx 1$ event for $\mathcal{L} = 30\,\mathrm{fb}^{-1}$.

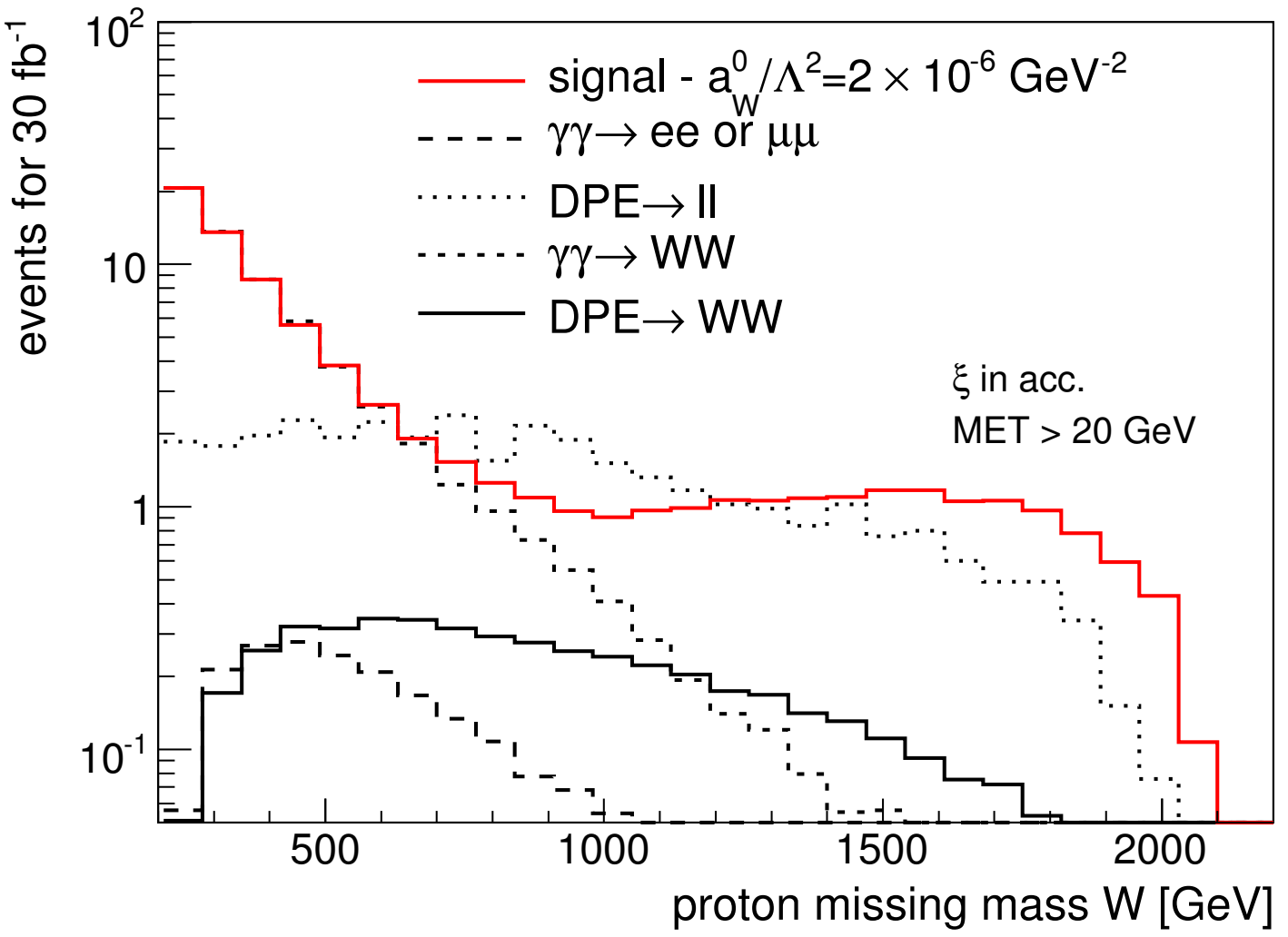

Figure 4: Proton missing mass in the AFP acceptance $0.0015 < \xi < 0.15$ and after $\not{E}_T > 20\,\mathrm{GeV}$ for signal and all backgrounds for $30\,\mathrm{fb}^{-1}$.

The selection of the $ZZ$ signal is similar to the low luminosity analysis. The complete set of used cuts is ($n_{lep} \geq 2$, 2 of same charge) or $n_{lep} \geq 3, 0.0015 < \xi < 0.15, p_T^{lep1} > 160\,\mathrm{GeV}, p_T^{lep2} > 25\,\mathrm{GeV}, n_{jet} = 0$. The overlaid background is assumed to be negligible with the use of timing detectors which detect the arrival time of the protons and allow to distinguish whether protons in the forward detectors come from the same vertex as the tracks in the central detector.

The calculated sensitivities using integrated luminosity $\mathcal{L} = 30\,(200)\,\mathrm{fb}^{-1}$ are shown in Table 1. The improvement with respect to OPAL limits is almost four orders of magnitude. The charged and neutral terms in the anomalous Lagrangians can partially cancel each other. This effect is seen in two dimensional discovery limits in Figure 5. They are calculated by varying either the pair $a_0^W/\Lambda^2$, $a_C^W/\Lambda^2$ or the pair $a_0^Z/\Lambda^2$, $a_C^Z/\Lambda^2$ at the same time.

# 5 Conclusion

Two-photon interactions at the LHC are exclusive processes which make an important part of the forward physics program. Since heavy mass objects can be created, these events will be used to test the SM in a new ways. The anomalous couplings between photons and pairs of electroweak bosons of the order of $10^{-6}\,\mathrm{GeV}^{-2}$ can be probed which might be used to distinguish between some of the Higgs-less models.

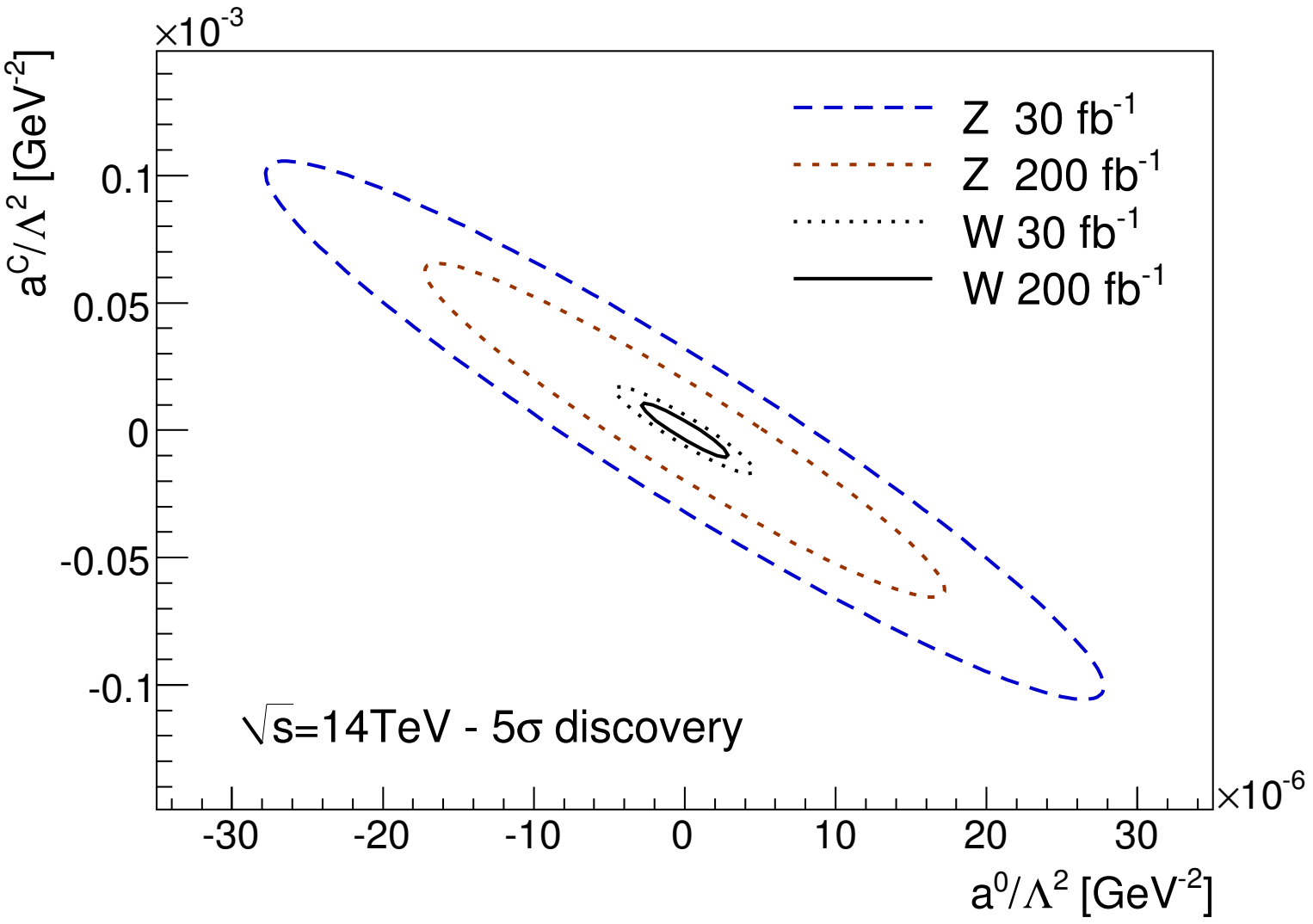

Figure 5: Two dimensional discovery limits using high luminosity and forward detectors. The significance is calculated as $S/\sqrt{B+1}$ where $S$ is the enhancement of the cross section due to anomalous couplings and $B$ is the total background including SM two-photon production. Sensitivities are shown for $30\,\mathrm{fb}^{-1}$ and $200\,\mathrm{fb}^{-1}$ of collected luminosity.

# References

[1] T. Pierzchala and K. Piotrzkowski, Nucl. Phys. Proc. Suppl. **179-180** (2008) 257 [arXiv:0807.1121 [hep-ph]].

[2] O. Kepka and C. Royon, Phys. Rev. D **78** (2008) 073005 [arXiv:0808.0322 [hep-ph]].

[3] G. Belanger and F. Boudjema, Phys. Lett. B **288**, 201 (1992).

[4] M. Boonekamp, V. Juránek, O. Kepka, C. Royon, Forward Physics Monte Carlo, Proceedings of the Workshop of the Implications of HERA for LHC physics; arXiv:0903.3861 [hep-ph]; `http://cern.ch/fpmc`.

[5] E. Boos *et al.* [CompHEP Collaboration], Nucl. Instrum. Meth. A **534** (2004) 250 [arXiv:hep-ph/0403113].

[6] V. A. Khoze, A. D. Martin and M. G. Ryskin, Eur. Phys. J. C **23** (2002) 311.

[7] G. Abbiendi *et al.* [OPAL Collaboration], Phys. Rev. D **70** (2004) 032005 [arXiv:hep-ex/0402021].

# Part IX

# Heavy-Ion Physics

# QCD and Heavy Ions: RHIC Overview

*Raphael Granier de Cassagnac*

Laboratoire Leprince-Ringuet, École polytechnique, CNRS-IN2P3, Palaiseau, France

Nowadays, the most violent heavy ion collisions available to experimental study occur at the Relativistic Heavy Ion Collider (RHIC) of the Brookhaven National Laboratory. There, gold ions collide at $\sqrt{s_{NN}} = 200$ GeV. The early and most striking RHIC results were summarised in 2005 by its four experiments, BRAHMS, PHENIX, PHOBOS and STAR, in their so-called *white papers* [1, 2, 3, 4] that will be largely referenced thereafter. Beyond and after this, a wealth of data has been collected and analysed, providing additional information about the properties of the matter created at RHIC. It is categorically impossible to give a comprehensive review of these results in a 20 minutes talk or a 7 pages report. Here, I have made a selection of some of the most striking or intriguing signatures: jet quenching in Section 2, quarkonia suppressions in Section 3 and thermal photons in Section 4. A slightly longer and older version of this review can be found in [5]. Some updates are given here, as well as emphasis on new probes recently made available.

## 1 Multiplicities and Densities

The first obvious *things* that come out of heavy ion collisions are *a lot* of particles. The number of charged particles was measured for various collision energies and centralities by the four RHIC experiments, and in particular by the dedicated PHOBOS collaboration over a broad range of 10.8 units of pseudorapidity [6]. At midrapidity, the number of charged particle reaches $dN_{ch}/d\eta|_{\eta=0} \simeq 670$ in the most violent Au+Au collisions and they sum up to about 6000 particles (of any charge) over the full rapidity range. These huge numbers were in fact *lower* than expected from various simple models, extrapolating lower energy results (for more details and complete references, see Section 2.1 in Ref. [3]). This moderation of the produced particles is an indication that the gluon density in the initial state starts to saturate, or similarly to be shadowed. In other words, low momentum gluons from neighbour nucleons overlap and recombine. In short, these results show that **the (initial) matter is gluon saturated**.

The second obvious manifestation of the collision violence is the transverse (i.e. unboosted by the initial parton longitudinal momenta) energy liberated. Measuring it allows one to estimate the energy density $\varepsilon$ of the medium after a given time $\tau_0$, through the Bjorken formula [7]: $\varepsilon = dE_T/dy|_{y=0}/\tau_0 A_T$, where $A_T$ is the transverse area of the collision. The four RHIC experiments measure consistent values of $dE_T/dy|_{y=0}$ that correspond to an energy density of at least 5 GeV/fm$^3$ at $\tau_0 = 1$ fm/$c$, and for the most central collisions. The time to be considered is certainly lower than 1 fm/$c$ (see Section 2 of Ref. [2]) and thus the *lower* energy density estimate is much higher than the threshold for the transition to a quark gluon plasma, as predicted by QCD on the lattice [8]: $\varepsilon_c \sim 1$ GeV/fm$^3$.

This tells us that **the matter should be deconfined**, i.e. made of free quarks and gluons. The following sections review some of the measurements that indicate that it is indeed the case.

# 2 Jet Quenching

## 2.1 High Transverse Momentum Suppression

Figure 1 is an illustration of the first and most striking QGP signature seen at RHIC, namely the quenching of jets [9, 10]. Displayed is, for various particles, the nuclear modification factor $R_{AA}$ defined as the yield of particles seen in A+A collisions, normalised by the same yield from p+p collisions scaled by the average number of binary collisions $\langle N_{coll} \rangle$ corresponding to the considered centrality: $R_{AA} = dN_{AA}/\langle N_{coll} \rangle dN_{pp}$. Hard processes (high $p_T$ particles in particular) are expected to respect such a scaling ($R_{AA} = 1$). This is indeed the case of the direct photon[1] [13] (grey squares), while the corresponding $\pi^0$ (blue circles) and $\eta$ (red triangles) are suppressed by a factor of five at large $p_T$. This is understood as an energy loss of the scattered partons going through a very dense matter, and producing softened jets and leading (high $p_T$) particles. This medium is so dense that it cannot be made of individual hadrons, but rather of quarks and gluons. In [14], PHENIX has released data on $\pi^0$ modification factors up to 20 GeV/$c$, and performed a quantitative estimate of the constraints on theoretical models. As an example, gluon densities of $dN_g/dy = 1400^{+270}_{-150}$ are needed to produce such a strong quenching in the model depicted in [15].

High $p_T$ suppressions are seen for various particles with various $p_T$ reaches and by the four experiments [1, 2, 3, 4]. It gets stronger for more central collisions. Checking that *normal* nuclear matter cannot induce what is seen in heavy ion collisions is a crucial test for *any* QGP signature and property. It is usually done through p+A like collisions. And indeed, high $p_T$ suppression is not observed in d+Au collisions (in particular for neutral pions [16] to be compared to the ones in Figure 1) where a moderate enhancement is even seen as a function of $p_T$, probably due to multiple scattering of the incoming partons providing additional transverse momentum (the so-called Cronin effect).

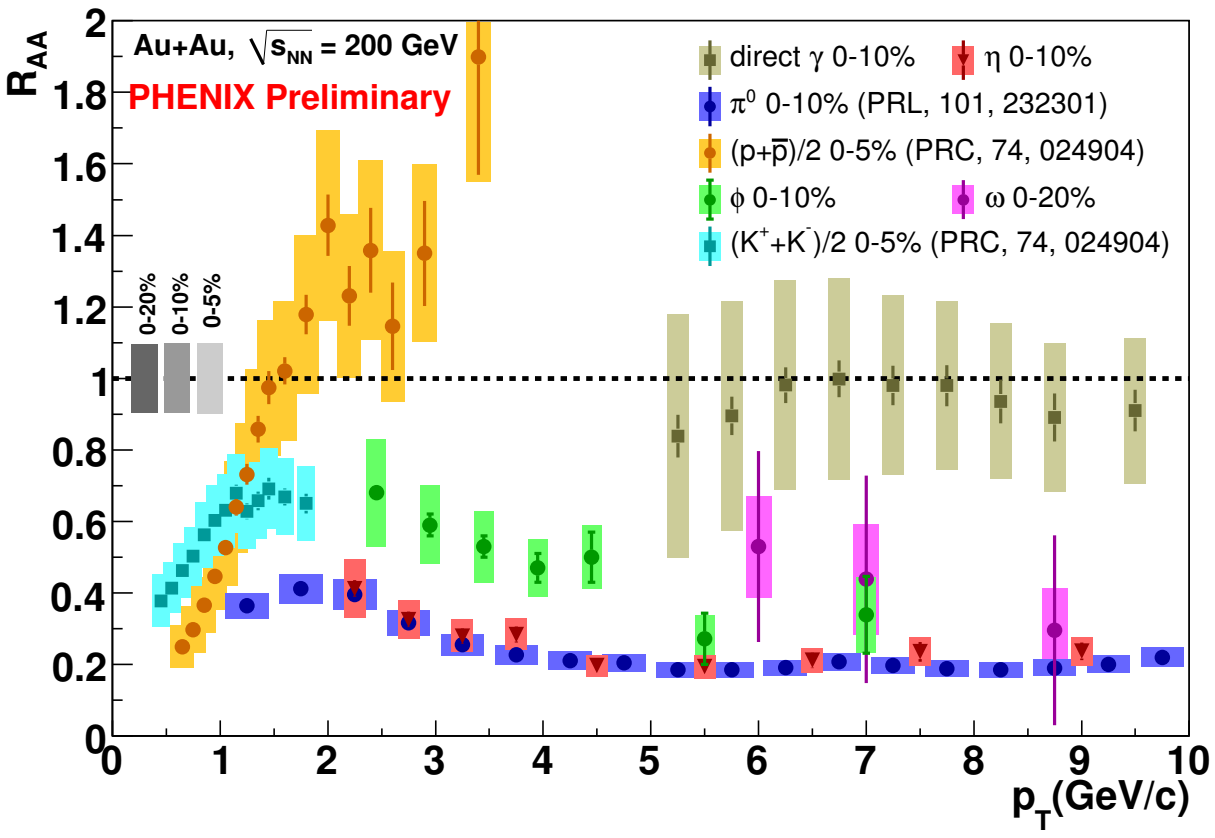

Figure 1: Nuclear modification factors for photons, $\pi^0$, $\eta$, protons, $\phi$, $\omega$ and kaons for central collisions, from the PHENIX experiment.

In any case, the quenching of high $p_T$ particles shows that **the matter they traverse is dense**.

[1] PHENIX has released preliminary photons up to 18 GeV/$c$ [11], which start to deviate below unity. As discussed for instance in [12], this can be explained by several phenomena (nucleus to proton isospin difference, EMC effect, or quark energy loss prior to photon emission) which have nothing to do with a QGP.

## 2.2 Azimuthal Correlations

Another way to look at jets is to consider back to back high transverse momentum hadron correlations. Figure 2 shows the measurements of such correlations for various collision types performed by the STAR experiment and reported in Section 4.2 of reference [4].

Displayed are the azimuthal distributions of hadrons around a *trigger* particle of high enough $p_T$ to reflect the main direction of jets (4 GeV/$c$ for the trigger particle and 2 GeV/$c$ for the others in this example). In p+p collisions (black histogram), one clearly sees particles belonging to both the narrower same ($\Delta\phi = 0$) and broader opposite ($\Delta\phi = \pi$) jets, while in central Au+Au collisions (blue stars) the away-side jet disappears [17]. This is also attributed to jet quenching, the away-side jet being absorbed by the dense matter produced at RHIC. As for the high $p_T$ suppression we saw in the previous section, this effect is not observed in d+Au collisions (red circles) in which away-side hadrons are clearly distinguishable [18].

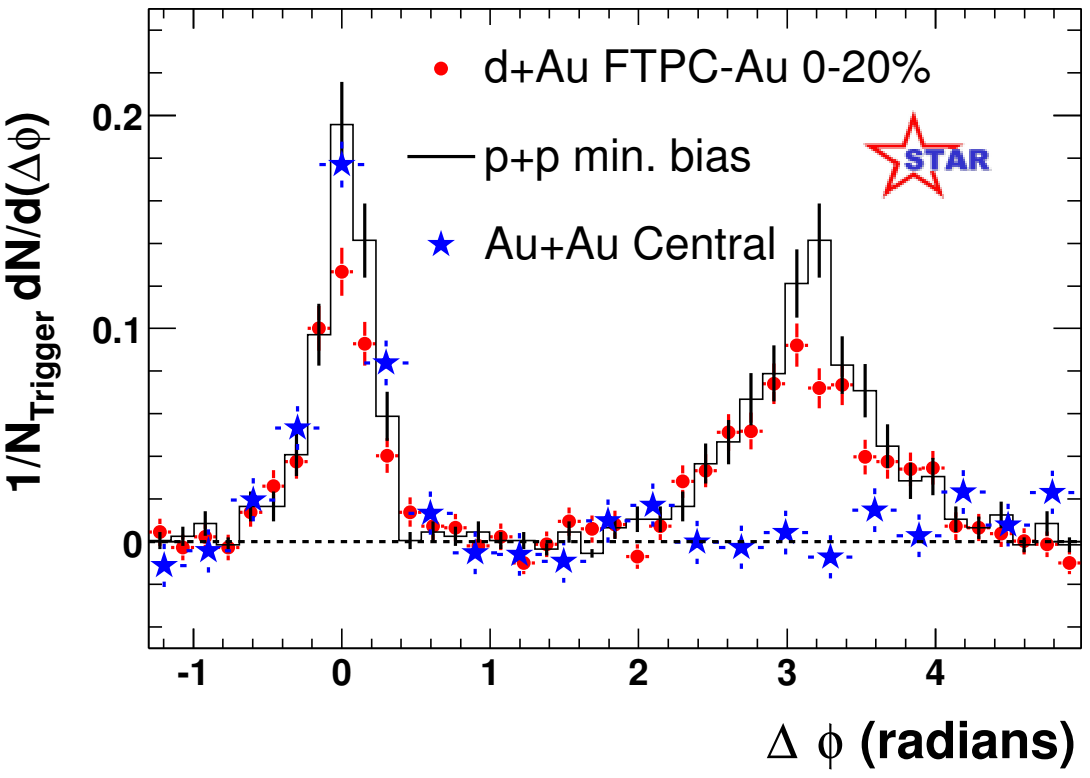

Figure 2: Dihadron azimuthal correlations in p+p, d+Au and Au+Au central collisions, from the STAR experiment.

Jet-induced hadron production has been further and extensively investigated at RHIC and various effects corroborate the jet quenching hypothesis, among which:

- In Au+Au collisions, the away-side disappearance grows with centrality. In fact, the most peripheral collisions exhibit a very similar away-side pattern as in p+p and d+Au collisions.

- The jets emitted in the reaction plane are less suppressed than in the perpendicular direction, where they have more matter to traverse [19, 20]. In fact, the high $p_T$ (near-side) particles we see in central Au+Au collisions are likely to come from the periphery, the *corona*, of the collision.

- By lowering the $p_T$ requirements (down to $\sim$1 GeV/$c$), one can find back the away-side jets [21].

- These weakened away-side jets are depleted at $\Delta\phi = \pi$ and exhibit two displaced maxima around $\Delta\phi = \pi \pm 1.1$ radians [22, 23]. This camel-back or conical-like shape provides insight in the quenched parton interactions with the medium. Various scenarios are proposed, such as radiative loss [24], Čerenkov-like or Mach-cone emissions [25]. The later allows one to compute an average speed of sound in the medium of $c_S \sim 0.45$.

- Analyses of three particles correlations also exhibit the conical pattern [26].

- The near-side jet exhibits a *ridge* along pseudorapidity (thus perpendicular to the azimuthal structure) that suggests the jets are indeed flowing with the expanding matter [21, 23, 27].

In brief, these high $p_T$ dihadron correlation studies show that **the matter is opaque** to jets to a first approximation, and clearly modifying their remaining structure.

### 2.3 New Tools

In addition to all the above, new tools were recently made available, thanks to the statistics accumulation at RHIC:

- The correlation of a jet (or leading hadron) with a high energy photon helps calibrating the jet, since the photon, essentially unmodified by the medium, should balance its initial transverse momentum. Both PHENIX [28] and STAR [29] have seen away side jets and released preliminary analyses of the so called $I_{AA}$ (or $I_{CP}$), which is the jet particle yield per photon seen in central A+A collisions with respect to p+p (or peripheral A+A). Though limited by statistics, $I_{AA}$ exhibit similar suppressions as $R_{AA}$.
- Another long awaited tool was the full reconstruction of jets in a heavy ion collision environment. Reconstructed jets have been shown recently by the STAR [30] and PHENIX [31] experiments, in Au+Au and Cu+Cu collisions respectively. The STAR preliminary result exhibits jet broadening with $R_{AA}$ going from close to unity to much lower values ($R_{AA} < 0.1$) when varying the radius of the jet reconstruction cone (from $R = 0.4$ to 0.2).

Both these novel methods should allow, in the near future, to derive medium-modified fragmentation functions.

## 3 Quarkonia Suppression

The bulk (low $p_T$) charm production scales to first order with the number of binary collisions ($R_{AA} \sim 1$), as shown in [32]. This forms a good baseline for the study of bound states made of charm-anticharm quarks, the more stable among them being the $J/\psi$ particle. In fact, charmonia were predicted to melt in the QGP, due to Debye screening of the colour charge in the medium [33]. Furthermore, $J/\psi$ suppression was indeed observed at lower energy ($\sqrt{s_{NN}} =$ 17.3 GeV) by the NA50 experiment [34] and is the main signature that led CERN to claim for the discovery of QGP. It was thus an awaited measurement at RHIC energies. Figure 3 shows $J/\psi$ nuclear modification factors as measured by the PHENIX experiment [35], for both mid (red circles, $|y| < 0.35$) and forward rapidity (blue squares, $1.2 < |y| < 2.2$), as a function of centrality (given by the number of participants $N_{part}$). These results brought two surprises:

- First, the midrapidity result is surprisingly similar to the one observed by the NA50 experiment which also lies close to midrapidity (black crosses, $0 < y < 1$). There is no fundamental reason for this to happen since the energy density for a given $N_{part}$ is higher at RHIC and should further melt quarkonia.
- Even more surprising is the fact that, at forward rapidity, $J/\psi$ are further suppressed (by $\sim 40\%$), while any density induced suppression scenario, such as the Debye screening mentioned above would predict the opposite trend.

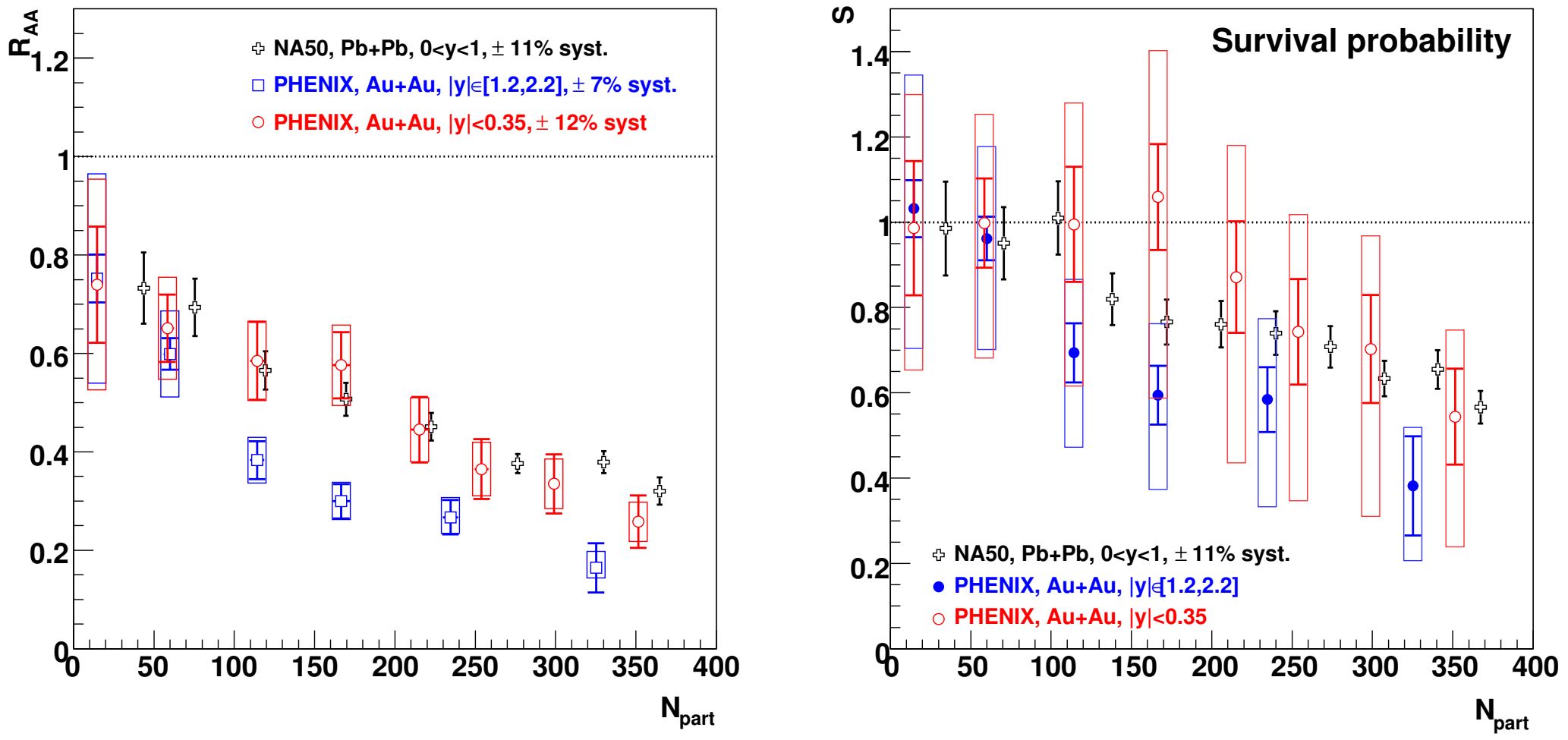

Figure 3: $J/\psi$ suppression measured by the PHENIX and NA50 experiments, as a function of centrality, given by the number of participants. Left: nuclear modification factor. Right: $J/\psi$ survival probabilities after normal nuclear effects subtraction.

But one needs to be careful in interpreting these results since $J/\psi$ are known to be suppressed by regular nuclear matter as it is seen in p+A or d+A collisions [34, 36]. In order to compare the two regimes, one thus first needs to subtract these *normal* nuclear matter effects. At RHIC, they are poorly constrained by a relatively low statistics d+Au data set. Several methods, summarised in [36], can nevertheless be used to estimate them. The most data-driven one, inspired by [37] is used to obtain the right part of Figure 3. The very large error bar displayed as a box is essentially reflecting the large normal suppression uncertainties. It illustrates that the two surprises mentioned above may be caused by normal effects: anomalous suppression could be different at SPS and RHIC, and similar at forward and rapidity at RHIC. More RHIC d+Au data were very recently released [38] that will help to reduce the normal suppression uncertainty. However, we clearly see that $J/\psi$ are suppressed beyond normal nuclear effects, both at SPS and RHIC (especially at forward rapidity).

An alternate scenario was (prematurely) proposed to explain the RHIC rapidity difference. $J/\psi$ could indeed be recreated in the plasma by recombination of independent charm and anticharm quarks (a large variety of recombination or coalescence models exists, see references in [39]). This beautiful idea of *reconfinement*, and thus of *deconfinement*, unfortunately does not provide very quantitative predictions of the nuclear modification factors (recombination models suffering from the lack of input charm quark distributions). Other observables ($p_T$ dependence, elliptic flow, feed-down contributions...) start to be available[2] but so far, they do not allow any firm conclusion.

However, even if the details of the mechanisms responsible for the exact $J/\psi$ yield at RHIC are not known, we do not need them to reckon that $J/\psi$ seem to melt beyond normal nuclear effects, at least in the most central collisions. This is a sign that **the matter is deconfining**. It is to be noted that the era of $\Upsilon$ studies ($b\bar{b}$ bound states) was recently opened and should provide new insights in quarkonia suppression. At present, preliminary result gives $R_{AA} < 0.64$ with a confidence level of 90% for minimum-bias upsilon-mass dielectrons [40], while $R_{dA} = 0.98 \pm 0.32 \pm 0.28$ [41]. These do not yet allow one to derive strong conclusions.

[2]For a comprehensive review on the subject, see [39].

# 4 Thermal Radiation

Any equilibrated and hot system should emit thermal radiation. We saw in Figure 1 that photons are unmodified by the medium and the nuclear modification factor is compatible with unity. This holds for large $p_T$ (typically larger than 4 GeV/$c$), but lower $p_T$ photons exhibit an *enhancement*.

In Figure 4, the bottom curves and points show the p+p photon spectrum (stars) from PHENIX compared to NLO pQCD calculation. The upper spectra are from various centrality selections of Au+Au collisions. The dashed lines are derived from the p+p collisions and scaled up by the number of collisions. The lowest $p_T$ photons (obtained through the "internal conversion" method [42]) clearly exhibit an enhancement. Various hydrodynamical models (for a review, see [43]) fairly reproduce the data assuming early (typically at a time of the order of 0.15 to 0.6 fm/$c$) temperature of 300 to 600 MeV, well above the critical temperature of $T_c \simeq 190$ MeV predicted by lattice QCD [8] as the phase transition boundary to a quark-gluon plasma.

We thus see thermal photons that demonstrate that **the matter is hot**.

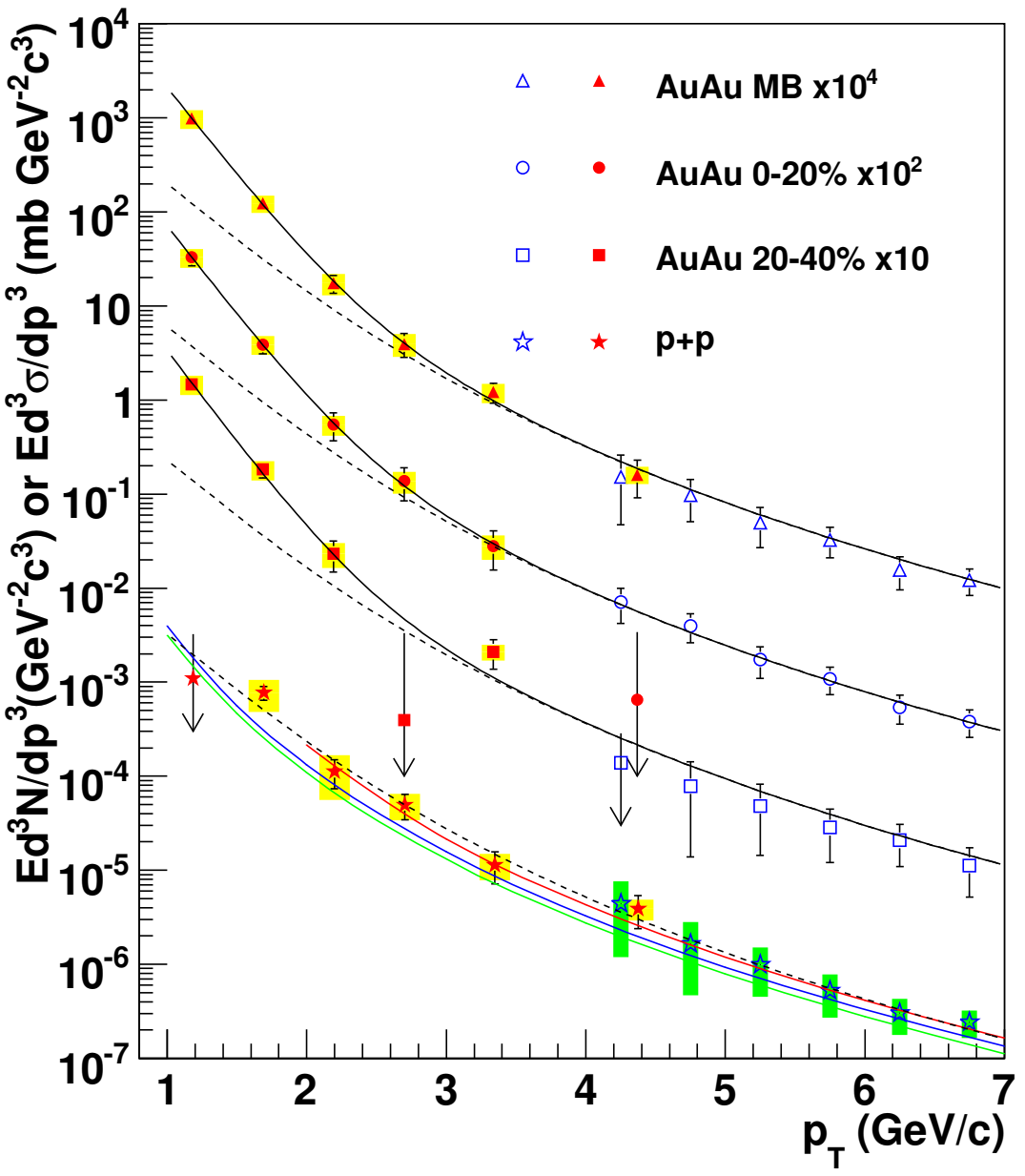

Figure 4: Thermal + perturbative QCD fits to the photon yield in Au+Au collisions, as seen by the PHENIX experiment. The lower points are from p+p collisions and are matched to perturbative QCD only.

# 5 Conclusions

Even if we have not (yet) observed any sharp change in the behaviour of the Au+Au observables related to the predicted phase transition, nor numbered the degrees of freedom of the system, it is clear that the matter produced at RHIC behaves very differently than ordinary hadronic matter. Indeed, we saw that the matter is gluon saturated, dense, opaque, deconfining and hot. Other observables [5] show that it is also strongly interacting and liquid-like, as well as of partonic nature. It is thus very likely to be formed by deconfined quarks and gluons.

# References

[1] I. Arsene et al. Nucl. Phys. A757 (2005) 1, nucl-ex/0410020.
[2] K. Adcox et al. Nucl. Phys. A757 (2005) 184, nucl-ex/0410003.
[3] B. B. Back et al. Nucl. Phys. A757 (2005) 28, nucl-ex/0410022.
[4] J. Adams et al. Nucl. Phys. A757 (2005) 102, nucl-ex/0501009.
[5] R. Granier de Cassagnac. Int. J. Mod. Phys. A22 (2008) 6043, arXiv:0707.0328.
[6] B. B. Back et al. Phys. Rev. Lett. 91 (2003) 052303, nucl-ex/0210015.
[7] J.D. Bjorken. Phys. Rev. D27 (1983) 140.
[8] F. Karsch. Lect. Notes Phys. 583 (2002) 209.
[9] A. Adare et al. Phys. Rev. Lett. 101 (2008) 232301, arXiv:0801.4020.
[10] S. S. Adler et al. Phys. Rev. C75 (2007) 024909, nucl-ex/0611006.
[11] T. Isobe. J. Phys. G34 (2007) S1015, nucl-ex/0701040.
[12] F. Arleo. JHEP 09 (2006) 015, hep-ph/0601075.
[13] S. S. Adler et al. Phys. Rev. Lett. 94 (2005) 232301, nucl-ex/0503003.
[14] A. Adare et al. Phys. Rev. C77 (2008) 064907, 0801.1665.
[15] I. Vitev and M. Gyulassy. Phys. Rev. Lett. 89 (2002) 252301, hep-ph/0209161.
[16] S. S. Adler et al. Phys. Rev. Lett. 91 (2003) 072303, nucl-ex/0306021.
[17] C. Adler et al. Phys. Rev. Lett. 90 (2003) 082302, nucl-ex/0210033.
[18] J. Adams et al. Phys. Rev. Lett. 91 (2003) 072304, nucl-ex/0306024.
[19] J. Adams et al. Phys. Rev. Lett. 93 (2004) 252301, nucl-ex/0407007.
[20] S. Afanasiev et al. accepted by Phys. Rev. C., arXiv:0903.4886.
[21] J. Adams et al. Phys. Rev. Lett. 95 (2005) 152301, nucl-ex/0501016.
[22] S. S. Adler et al. Phys. Rev. Lett. 97 (2006) 052301, nucl-ex/0507004.
[23] A. Adare et al. Phys. Rev. C78 (2008) 014901, arXiv:0801.4545.
[24] A. D. Polosa and C. A. Salgado. Phys. Rev. C75 (2007) 041901, hep-ph/0607295.
[25] J. Ruppert and B. Muller. Phys. Lett. B618 (2005) 123, hep-ph/0503158.
[26] B. I. Abelev et al. Phys. Rev. Lett. 102 (2009) 052302, arXiv:0805.0622.
[27] J. Adams et al. Phys. Rev. C73 (2006) 064907, nucl-ex/0411003.
[28] A. Adare et al. Phys. Rev. C80 (2009) 024908, arXiv:0903.3399.
[29] A. M. Hamed (for the STAR collaboration). arXiv:0907.4523.
[30] M. Ploskon (for the STAR collaboration). arXiv:0908.1799.
[31] Y.-S. Lai (for the PHENIX collaboration). arXiv:0907.4725.
[32] A. Adare et al. Phys. Rev. Lett. 98 (2007) 172301, nucl-ex/0611018.
[33] T. Matsui and H. Satz. Phys. Lett. B178 (1986) 416.
[34] B. Alessandro et al. Eur. Phys. J. C39 (2005) 335, hep-ex/0412036.
[35] A. Adare et al. Phys. Rev. Lett. 98 (2007) 232201, nucl-ex/0611020.
[36] A. Adare et al. Phys. Rev. C77 (2008) 024912, Erratum-ibid. C79 (2009) 059901, arXiv:0711.3917.
[37] R. Granier de Cassagnac. J. Phys. G34 (2007) S955, hep-ph/0701222.
[38] C.-L. da Silva (for the PHENIX collaboration). arXiv:0907.4696.
[39] R. Granier de Cassagnac. J. Phys. G35 (2008) 104023, arXiv:0806.0046.
[40] E.-T. Atomssa (for the PHENIX collaboration). arXiv:0907.4787.
[41] H. Liu (for the STAR collaboration). arXiv:0907.4538.
[42] A. Adare et al. submitted to Phys. Rev. Lett., arXiv:0804.4168.
[43] D. d'Enterria and D. Peressounko. Eur. Phys. J. C46 (2006) 451, nucl-th/0503054.

# Progress in Jet Reconstruction in Heavy Ion Collisions

*Juan Rojo*
INFN, Sezione di Milano, Via Celoria 16, I - 20133, Milano (Italy)

We review recent developments related to jet clustering algorithms and jet reconstruction, with particular emphasis on their implications in heavy ion collisions. These developments include fast implementations of sequential recombination algorithms, new IRC safe algorithms, quantitative determination of jet areas and quality measures for jet finding, among many others. We also show how jet reconstruction provides a useful tool to probe the characteristics of the hot and dense medium created in heavy ion collisions, which allows to distinguish between different models of parton-medium interaction.

## 1 Recent Developments in Jet Algorithms

With the upcoming start-up of the proton-proton and heavy ion programs at the LHC, jet reconstruction techniques have been the subject of intense research in the recent years [1, 2, 3, 4]. In this contribution we briefly review this progress with special emphasis on their implications in heavy ion collisions.

An important development has been the fast implementation of the $k_T$ [5] and Cambridge/Aachen [6, 7] jet algorithms. Prior to 2005, existing implementations scaled as $N^3$, with $N$ the number of particles to be clustered, thus making it unpractical for high multiplicity proton-proton collisions, and even more in heavy ion collisions (HIC). Thanks to computational geometry methods, the performance of these algorithms was made to scale as $N \ln N$ [8]. These fast implementations are available through the `FastJet` package [9], together with area-based subtraction methods and plugins to external jet finders.

Another important achievement has been the formulation of a practical (scaling as $N^2 \ln N$) infrared and collinear (IRC) safe cone algorithm, SISCone [10]. Unlike all other commonly used cone algorithms, SISCone is IRC safe to all orders in perturbation theory by construction. This property allows one to compare any perturbative computation with experimental data, which for IRC unsafe algorithms is impossible beyond some fixed order [10]. The phenomenological implications of SISCone when compared with the (IRC unsafe) commonly used MidPoint cone algorithm range from few percent differences in the inclusive jet spectrum, somewhat larger in the presence of realistic Underlying Event (UE), up to 50% differences for more exclusive observables, like the tails of jet-mass spectra in multi-jet events.

There has been historically some confusion about the concept of the *size* of a jet, specially since the naive jet area is ambiguous beyond LO. The situation was recently clarified by the introduction of quantitative definitions of jet areas based on the *catchment* properties of hard jets with respect to very soft particles, called *ghosts* in [11]. On top of their theoretical importance, jet areas have important applications related to the subtraction of soft backgrounds coming from the UE or from Pile-Up (PU), both in proton-proton and in heavy-ion collisions, as discussed in [12].

Another recently developed IRC safe jet algorithm is the anti-$k_t$ algorithm [13]. This algorithm is related to $k_T$ and Cam/Aa by its distance measure, $d_{ij} \equiv \min(k_{ti}^{2p}, k_{tj}^{2p})\Delta R_{ij}^2/R^2$, with $p = -1$ ($p = 1$ corresponds to $k_T$ and $p = 0$ to Cam/Aa). The anti-$k_T$ algorithm has the property of being soft-resilient, that is, due to its distance soft particles are always clustered with hard particles first. This property leads to rather regular jet areas, which become perfectly circular in the limit in which all hard particles are separated in the $(y, \phi)$ plane by at least a distance $R$ [13]. Another important advantage of the anti-$k_t$ algorithm is that it has a very small back-reaction, that is, the presence of a soft background has reduced effects on which hard particles are clustered into a given jet. This is a particularly relevant advantage of the anti-$k_t$ algorithm for jet reconstruction in very dense environments like heavy ion collisions.

## 2 Performance of Jet Algorithms

A recurring question in jet studies is "what is the best jet definition for a given specific analysis under certain experimental conditions"? Most existing techniques either use as a reference unphysical Monte Carlo partons (an ambiguous concept beyond LO) and/or assume some shape for the measured kinematic distributions. To overcome these disadvantages, a new strategy to quantify the performance of jet definitions in kinematic reconstruction tasks has been recently introduced [14], which was designed to make use exclusively of physical observables. Related analytical studies which address the same question were discussed in Ref. [15].

In Ref. [14] two quality measures respecting the above requirements are proposed, and applied to the kinematic reconstruction of invariant mass distributions in dijet events in simulated LHC proton-proton collisions for a wide range of energies. These quality measures can in turn be mapped into an effective luminosity ratio, defined as

$$\rho_{\mathcal{L}}(\mathrm{JD}_2/\mathrm{JD}_1) \equiv \frac{\mathcal{L}(\text{needed with JD}_2)}{\mathcal{L}(\text{needed with JD}_1)} = \left[\frac{\Sigma\,(\mathrm{JD}_1)}{\Sigma\,(\mathrm{JD}_2)}\right]^2 . \qquad (1)$$

Given a certain signal significance $\Sigma$ with jet definition $\mathrm{JD}_2$, $\rho_{\mathcal{L}}(\mathrm{JD}_2/\mathrm{JD}_1)$ indicates the factor more luminosity needed to obtain the same significance as with jet definition $\mathrm{JD}_1$.

The results of [14] over a large range of jet definitions,[1] summarised in Fig.. 1, indicate that for gluon jets, and in general for TeV scales, there are significant benefits to be had from using larger radii that those commonly used, up to $R \gtrsim 1$. In general, SISCone and C/A-filt (Cam/Aa supplemented with a filtering procedure [17]) show the best performance. These conclusions are robust in the presence of high-luminosity PU, when subtracted with the jet area technique [12]. The same techniques could be applied to heavy ion collisions to determine how in such case the optimal jet definition is affected by the overwhelming underlying event present.

## 3 Jet Reconstruction in Heavy Ion Collisions

While QCD jets are ubiquitous in pp collisions, until recently [18, 19] no real jet reconstruction had been obtained in the much more challenging environment of HIC. Indeed, usually in HIC one refers to the leading particle of the event as a *jet*. However, reconstructing full QCD jets provides a much more precise window to the properties of the hot and dense medium created in the collision than just leading particles.

[1] There results can also be accessed through an interactive web tool [16] which allows the user to compare the jet finding quality for a wide range of parameters (jet algorithm, $R$, value of PU, ...).

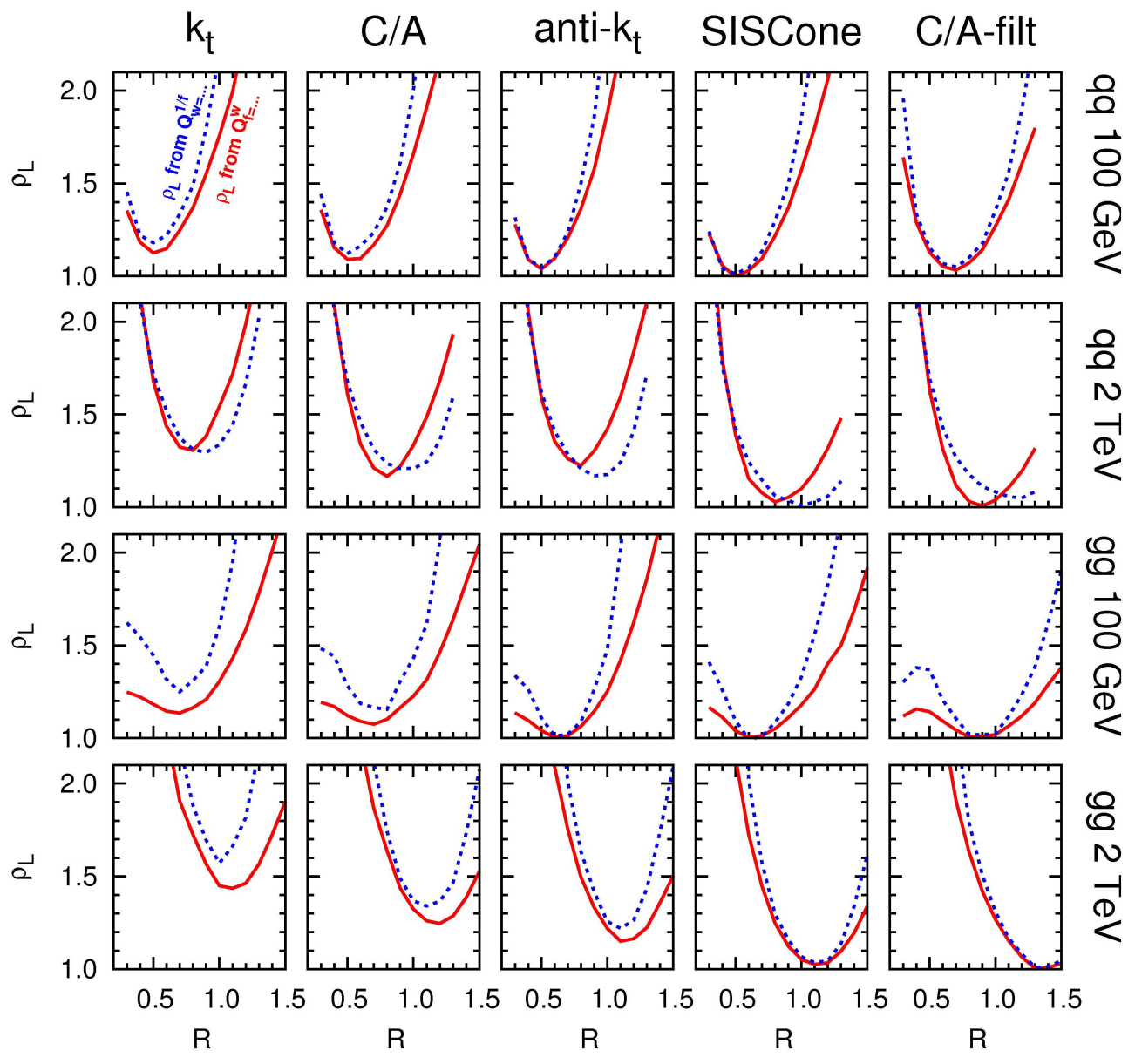

Figure 1: The effective luminosity ratio, Eq. 1, for quark and gluon jets at 100 GeV and 2 TeV, from Ref. [14].

The difficulty in reconstructing jets in HIC stems from the huge backgrounds, which need to be subtracted in order to compare with baseline results, like proton-proton or proton-ion collisions at the same energy. There are various techniques to subtract such large backgrounds. In Ref. [12] it was shown how the area method introduced above could efficiently subtract large UE backgrounds in HIC for LHC conditions with good accuracy, see Fig. 2.

It is therefore important for precision measurements to control the accuracy of the subtraction procedure intrinsic to jet reconstruction in HIC, as well as to understand the differences between the performances of different jet algorithms. Ongoing studies [20] suggest that one of the dominant sources of systematic error in HIC jet reconstruction is back-reaction [11], therefore anti-$k_t$ is potentially interesting in this situation due to its small back-reaction [13]. In Ref. [20] also the use of local ranges for the determination of the background level $\rho$ as a technique to reduced the effects of point-to-point background fluctuations is investigated.

Related to progress in jet reconstruction algorithms, an important development in the recent years has been the development of several exclusive Monte Carlo event generators for heavy ion collisions which account for the interaction of partons propagating within the hot and dense medium created in the collision [21, 22, 23, 24]. These event generators, which assume different models for the parton-medium interaction [25], can be used together with modern jet reconstruction techniques in order to determine, under realistic experimental conditions, which jet-related observables are at the same time more robust and more sensitive to the different

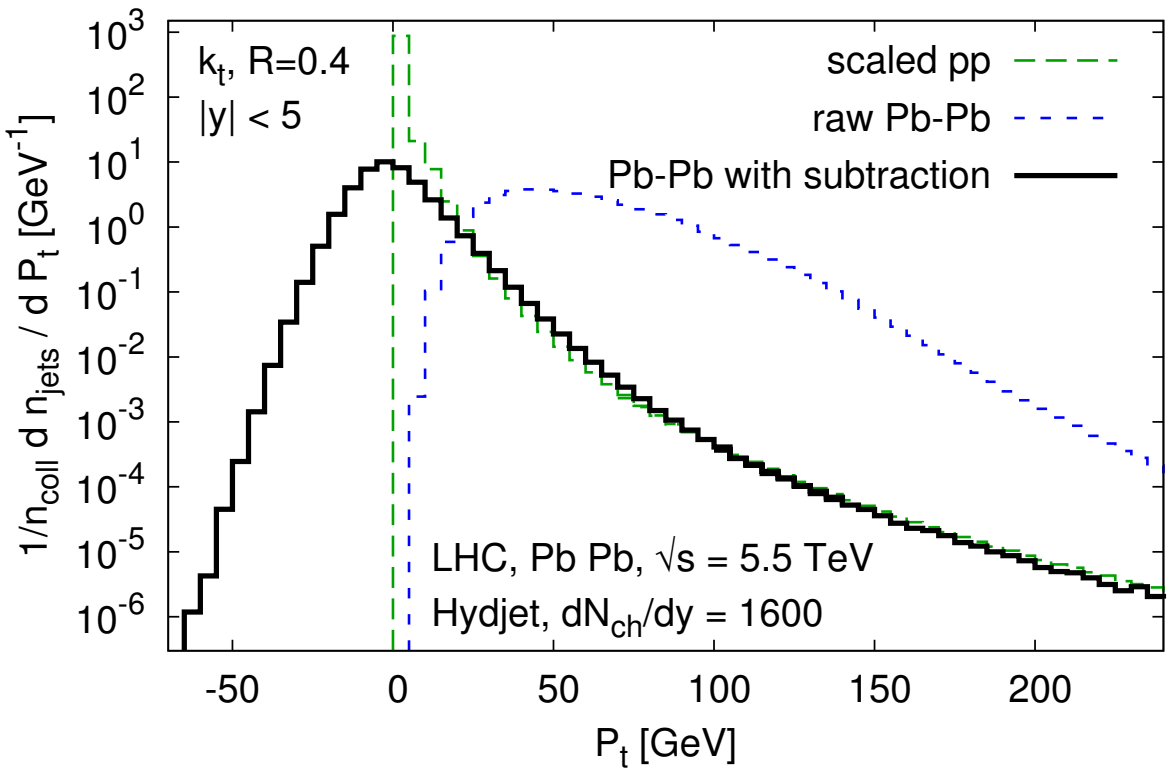

Figure 2: The simulated inclusive jet spectrum at the LHC with the $k_T$ algorithm, including subtraction, from Ref. [12].

scenarios for the hot and dense medium dynamics.

As an illustration of how these Monte Carlo programs with medium effects can be coupled to jet reconstruction techniques and be used to determine medium properties in HIC, in Fig. 3 we show preliminary results for two observables which are sensitive to medium effects: the dijet azimuthal correlations and the jet shape. Jet shapes are defined analogously to [21]. To obtain these results, hard events with $p_T \geq 100$ GeV are generated and then propagated through a model of the hot medium by means of the Q-PYTHIA Monte Carlo [21], for different values of the medium parameters. The resulting hadronic event is embedded into a minimum bias PbPb event generated with the PSM Monte Carlo [26] for different scenarios of central multiplicity at the LHC. Jet reconstruction is performed with different algorithms of the FastJet package, and UE subtraction is performed with the jet area method. In the particular examples of Fig. 3 the jet radius is chosen to be $R = 0.5$.

For the two examples of Fig. 3, we show the proton-proton baseline results with and without parton-medium interactions, whose strength is characterised by the parameter $\hat{q}L$ [21], with $L$ the medium length. The medium effects are clearly visible for the two observables, inducing a broadening of the jet shape and a decorrelation of the dijet azimuthal spectrum. These curves, with no HIC UE, are labelled as *'No PbPb'*. Then we also show the corresponding results when the pp event is embedded into the PbPb event, in which case the UE event has been subtracted with the area method. These curves are labelled as *'PbPb, subtraction'*. We observe how after the background subtraction the baseline proton-proton results are reasonably recovered, in both cases with ($\hat{q}L \neq 0$) and without ($\hat{q}L = 0$) medium effects.

These preliminary results indicate that medium-sensitive jet related observables can be accurately reconstructed even in the presence of large backgrounds, and are thus useful probes of the details of the parton-medium interactions. More work however is required to quantify the accuracy with which the hot and dense medium created in HIC, and the values of the parameters which characterise it, can be studied by reconstructed jets and related observables.

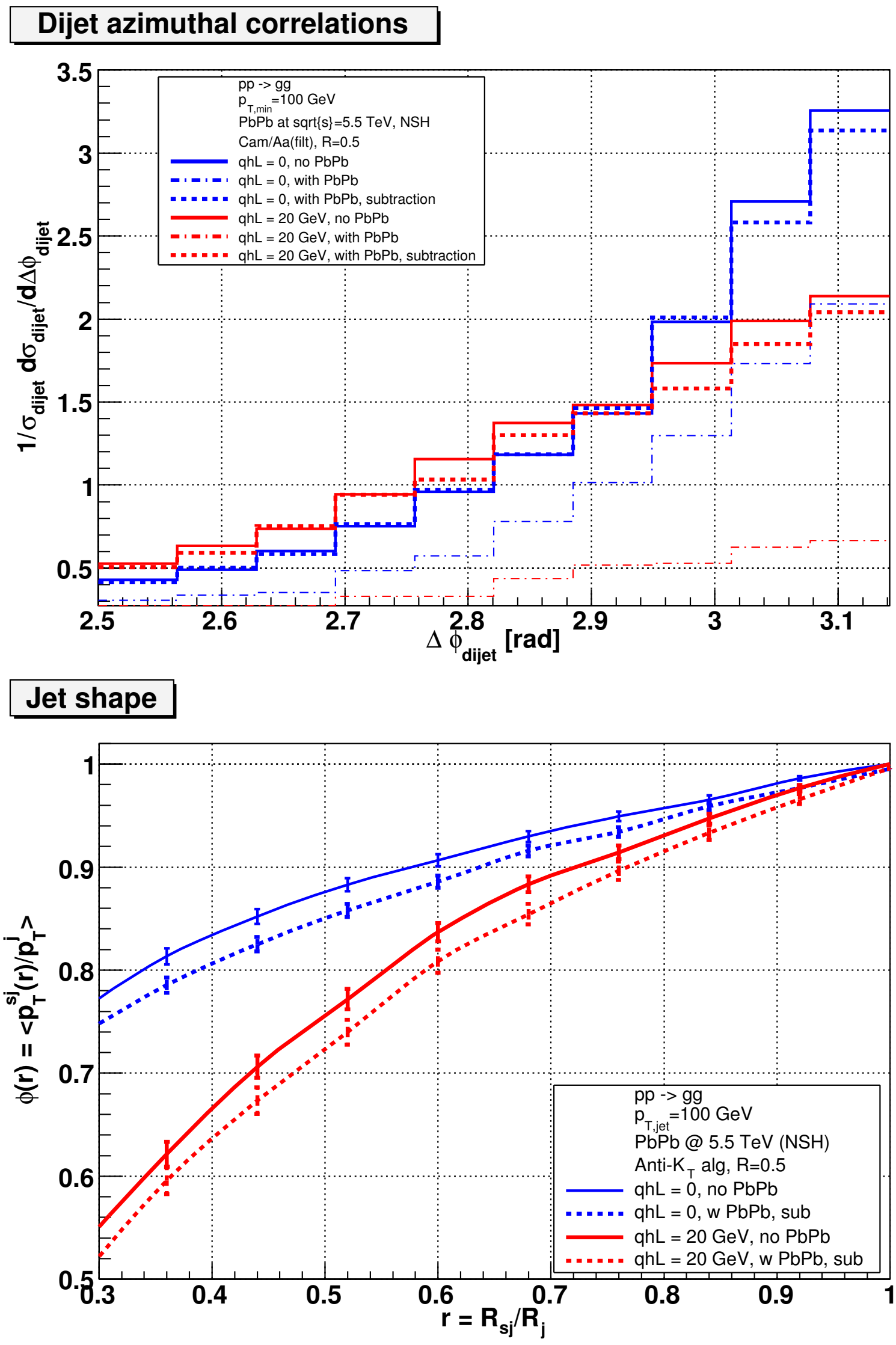

Figure 3: Preliminary results for full jet reconstruction, including background subtraction, of jet-related observables under realistic experimental conditions at the LHC: dijet azimuthal correlations (upper plot) and jet shapes (lower plot). See text for details.

# Acknowledgements

The author wants to acknowledge M. Cacciari, G. Salam and G. Soyez for collaboration in this research, as well as N. Armesto and C. Salgado for assistance with Q-PYTHIA.

# References

[1] S. D. Ellis, J. Huston, K. Hatakeyama, P. Loch, and M. Tonnesmann. Jets in Hadron-Hadron Collisions. *Prog. Part. Nucl. Phys.*, 60:484–551, 2008.

[2] C. Buttar et al. Standard Model Handles and Candles Working Group: Tools and Jets Summary Report. 2008.

[3] Gavin P. Salam. Towards Jetography. 2009.

[4] Matteo Cacciari. Recent Progress in Jet Algorithms and Their Impact in Underlying Event Studies. 2009.

[5] S. Catani, Yuri L. Dokshitzer, M. H. Seymour, and B. R. Webber. Longitudinally invariant $k_t$ clustering algorithms for hadron hadron collisions. *Nucl. Phys.*, B406:187–224, 1993.

[6] S. Catani, Yuri L. Dokshitzer, M. Olsson, G. Turnock, and B. R. Webber. New clustering algorithm for multi - jet cross-sections in $e^+e^-$ annihilation. *Phys. Lett.*, B269:432–438, 1991.

[7] M. Wobisch and T. Wengler. Hadronization corrections to jet cross sections in deep- inelastic scattering. 1998.

[8] Matteo Cacciari and Gavin P. Salam. Dispelling the $N^3$ myth for the $k_t$ jet-finder. *Phys. Lett.*, B641:57–61, 2006.

[9] Matteo Cacciari, Gavin P. Salam, and Gregory Soyez. 2005-2009.

[10] Gavin P. Salam and Gregory Soyez. A practical seedless infrared-safe cone jet algorithm. *JHEP*, 05:086, 2007.

[11] Matteo Cacciari, Gavin P. Salam, and Gregory Soyez. The Catchment Area of Jets. *JHEP*, 04:005, 2008.

[12] Matteo Cacciari and Gavin P. Salam. Pileup subtraction using jet areas. *Phys. Lett.*, B659:119–126, 2008.

[13] Matteo Cacciari, Gavin P. Salam, and Gregory Soyez. The anti-$k_t$ jet clustering algorithm. *JHEP*, 04:063, 2008.

[14] Matteo Cacciari, Juan Rojo, Gavin P. Salam, and Gregory Soyez. Quantifying the performance of jet definitions for kinematic reconstruction at the LHC. *JHEP*, 12:032, 2008.

[15] Mrinal Dasgupta, Lorenzo Magnea, and Gavin P. Salam. Non-perturbative QCD effects in jets at hadron colliders. *JHEP*, 02:055, 2008.

[16] Matteo Cacciari, Juan Rojo, Gavin P. Salam, and Gregory Soyez. 2008.

[17] Jonathan M. Butterworth, Adam R. Davison, Mathieu Rubin, and Gavin P. Salam. Jet substructure as a new Higgs search channel at the LHC. *Phys. Rev. Lett.*, 100:242001, 2008.

[18] Yue-Shi Lai. Probing medium-induced energy loss with direct jet reconstruction in p+p and Cu+Cu collisions at PHENIX. 2009.

[19] Sevil Salur. Full Jet Reconstruction in Heavy Ion Collisions. 2009.

[20] Matteo Cacciari, Juan Rojo, Gavin P. Salam, and Gregory Soyez. Quantifying the performance of jet definitions and subtraction in HIC at the LHC. in preparation.

[21] Nestor Armesto, Leticia Cunqueiro, and Carlos A. Salgado. Q-PYTHIA: a medium-modified implementation of final state radiation. 2009.

[22] Korinna Zapp, Gunnar Ingelman, Johan Rathsman, Johanna Stachel, and Urs Achim Wiedemann. A Monte Carlo Model for 'Jet Quenching'. *Eur. Phys. J.*, C60:617–632, 2009.

[23] I. P. Lokhtin et al. Heavy ion event generator HYDJET++ (HYDrodynamics plus JETs). *Comput. Phys. Commun.*, 180:779–799, 2009.

[24] Nestor Armesto, Gennaro Corcella, Leticia Cunqueiro, and Carlos A. Salgado. Angular-ordered parton showers with medium-modified splitting functions. 2009.

[25] Carlos A. Salgado. Lectures on high-energy heavy-ion collisions at the LHC. 2009.

[26] N. S. Amelin, N. Armesto, C. Pajares, and D. Sousa. Monte Carlo model for nuclear collisions from SPS to LHC energies. *Eur. Phys. J.*, C22:149–163, 2001.

# Spin Correlations in the $\Lambda\Lambda$ and $\Lambda\bar{\Lambda}$ Systems Generated in Relativistic Heavy-Ion Collisions

*V.L. Lyuboshitz, V.V. Lyuboshitz* †

Joint Institute for Nuclear Research, 141980, Dubna, Moscow Region, Russia
† E-mail: Valery.Lyuboshitz@jinr.ru

Spin correlations for the $\Lambda\Lambda$ and $\Lambda\bar{\Lambda}$ pairs, generated in relativistic heavy ion collisions, and related angular correlations at the joint registration of hadronic decays of two hyperons with non-conservation of space parity are analyzed. Within the conventional model of one-particle sources, correlations vanish at enough large relative momenta. However, under these conditions, in the case of two non-identical particles ($\Lambda\bar{\Lambda}$) a noticeable role is played by two-particle annihilation (two-quark, two-gluon) sources, which lead to the difference of the correlation tensor from zero. In particular, such a situation may arise when the system passes through the "mixed phase".

## 1 General Structure of the Spin Density Matrix of the Pairs $\Lambda\Lambda$ and $\Lambda\bar{\Lambda}$

Spin correlations for $\Lambda\Lambda$ and $\Lambda\bar{\Lambda}$ pairs, generated in relativistic heavy ion collisions, and respective angular correlations at joint registration of hadronic decays of two hyperons, in which space parity is not conserved, give important information on the character of multiple processes.

The spin density matrix of the $\Lambda\Lambda$ and $\Lambda\bar{\Lambda}$ pairs, just as the spin density matrix of two spin-1/2 particles in general, can be presented in the following form [1–3]:

$$\hat{\rho}^{(1,2)} = \frac{1}{4}\left[\hat{I}^{(1)} \otimes \hat{I}^{(2)} + (\hat{\boldsymbol{\sigma}}^{(1)}\mathbf{P}_1) \otimes \hat{I}^{(2)} + \hat{I}^{(1)} \otimes (\hat{\boldsymbol{\sigma}}^{(2)}\mathbf{P}_2) + \sum_{i=1}^{3}\sum_{k=1}^{3} T_{ik}\hat{\sigma}_i^{(1)} \otimes \hat{\sigma}_k^{(2)}\right]; \qquad (1)$$

in doing so, $tr_{(1,2)}\hat{\rho}^{(1,2)} = 1$.

Here $\hat{I}$ is the two-row unit matrix, $\boldsymbol{\sigma} = (\hat{\sigma}_x,\, \hat{\sigma}_y,\, \hat{\sigma}_z)$ is the vector Pauli operator $(x, y, z \to 1, 2, 3)$, $\mathbf{P}_1$ and $\mathbf{P}_2$ are the polarisation vectors of first and second particle ($\mathbf{P}_1 = \langle\hat{\boldsymbol{\sigma}}^{(1)}\rangle$, $\mathbf{P}_2 = \langle\hat{\boldsymbol{\sigma}}^{(2)}\rangle$), $T_{ik} = \langle\hat{\sigma}_i^{(1)} \otimes \hat{\sigma}_k^{(2)}\rangle$ are the correlation tensor components. In the general case $T_{ik} \neq P_{1i}\, P_{2k}$. The tensor with components $C_{ik} = T_{ik} - P_{1i}P_{2k}$ describes the spin correlations of two particles.

## 2 Spin Correlations and Angular Correlations at Joint Registration of Decays of Two $\Lambda$ Particles into the Channel $\Lambda \to p + \pi^-$

It is essential that any decay of an unstable particle may serve as an analyser of its spin state.

The normalised angular distribution at the decay $\Lambda \to p + \pi^-$ takes the form:

$$\frac{d\,w(\mathbf{n})}{d\,\Omega_{\mathbf{n}}} = \frac{1}{4\,\pi}(1 + \alpha_\Lambda \mathbf{P}_\Lambda\,\mathbf{n}) \tag{2}$$

Here $\mathbf{P}_\Lambda$ is the polarisation vector of the $\Lambda$ particle, $\mathbf{n}$ is the unit vector along the direction of proton momentum in the rest frame of the $\Lambda$ particle, $\alpha_\Lambda$ is the coefficient of $P$-odd angular asymmetry ($\alpha_\Lambda = 0.642$). The decay $\Lambda \to p+\pi^-$ selects the projections of spin of the $\Lambda$ particle onto the direction of proton momentum; the analysing power equals $\boldsymbol{\xi} = \alpha_\Lambda \mathbf{n}$.

Now let us consider the double angular distribution of flight directions for protons formed in the decays of two $\Lambda$ particles into the channel $\Lambda \to p + \pi^-$, normalised by unity (the analysing powers are $\boldsymbol{\xi}_1 = \alpha_\Lambda \mathbf{n}_1$, $\boldsymbol{\xi}_2 = \alpha_\Lambda \mathbf{n}_2$). It is described by the following formula [2,3]:

$$\frac{d^2\,w(\mathbf{n}_1,\mathbf{n}_2)}{d\,\Omega_{\mathbf{n}_1}\,d\,\Omega_{\mathbf{n}_2}} = \frac{1}{16\,\pi^2}\left[1 + \alpha_\Lambda \mathbf{P}_1\,\mathbf{n}_1 + \alpha_\Lambda \mathbf{P}_2\,\mathbf{n}_2 + \alpha_\Lambda^2 \sum_{i=1}^{3}\sum_{k=1}^{3} T_{ik} n_{1i} n_{2k}\right], \tag{3}$$

where $\mathbf{P}_1$ and $\mathbf{P}_2$ are polarisation vectors of the first and second $\Lambda$ particle, $T_{ik}$ are the correlation tensor components, $\mathbf{n}_1$ and $\mathbf{n}_2$ are unit vectors in the respective rest frames of the first and second $\Lambda$ particle, defined in the common (unified) coordinate axes of the c.m. frame of the pair $(i, k = \{1, 2, 3\} = \{x, y, z\})$.

The polarisation parameters can be determined from the angular distribution of decay products by the method of moments [2,3] – as a result of averaging combinations of trigonometric functions of angles of proton flight over the double angular distribution.

The angular correlation, integrated over all angles except the angle between the vectors $\mathbf{n}_1$ and $\mathbf{n}_2$ and described by the formula [2 – 5]

$$d\,w(\cos\theta) = \frac{1}{2}\left(1 + \frac{1}{3}\,\alpha_\Lambda^2 T\,\cos\theta\right)\sin\theta\,d\theta = \frac{1}{2}\left[1 - \alpha_\Lambda^2\left(W_s - \frac{W_t}{3}\right)\cos\theta\right]\sin\theta\,d\theta, \tag{4}$$

is determined only by the "trace" of the correlation tensor $T = W_t - 3W_s$ ($W_s$ and $W_t$ are relative fractions of the singlet state and triplet states, respectively), and it does not depend on the polarisation vectors (single-particle states may be unpolarised).

## 3 Correlations at the Joint Registration of the Decays $\Lambda \to p + \pi^-$ and $\bar{\Lambda} \to \bar{p} + \pi^+$

Due to CP invariance, the coefficients of $P$-odd angular asymmetry for the decays $\Lambda \to p + \pi^-$ and $\bar{\Lambda} \to \bar{p} + \pi^+$ have equal absolute values and opposite signs: $\alpha_{\bar{\Lambda}} = -\alpha_\Lambda = -0.642$. The double angular distribution for this case is as follows [2,3]:

$$\frac{d^2\,w(\mathbf{n}_1,\mathbf{n}_2)}{d\,\Omega_{\mathbf{n}_1}\,d\,\Omega_{\mathbf{n}_2}} = \frac{1}{16\,\pi^2}\left[1 + \alpha_\Lambda \mathbf{P}_\Lambda\,\mathbf{n}_1 - \alpha_\Lambda \mathbf{P}_{\bar{\Lambda}}\,\mathbf{n}_2 - \alpha_\Lambda^2 \sum_{i=1}^{3}\sum_{k=1}^{3} T_{ik} n_{1i} n_{2k}\right], \tag{5}$$

(here $-\alpha_\Lambda = +\alpha_{\bar{\Lambda}}$ and $-\alpha_\Lambda^2 = +\alpha_\Lambda\alpha_{\bar{\Lambda}}$).

Thus, the angular correlation between the proton and antiproton momenta in the rest frames of the $\Lambda$ and $\bar{\Lambda}$ particles is described by the expression:

$$d\,w(\cos\theta) = \frac{1}{2}\left(1 - \frac{1}{3}\,\alpha_\Lambda^2 T\,\cos\theta\right)\sin\theta\,d\theta = \frac{1}{2}\left[1 + \alpha_\Lambda^2\left(W_s - \frac{W_t}{3}\right)\cos\theta\right]\sin\theta\,d\theta, \qquad (6)$$

where $\theta$ is the angle between the proton and antiproton momenta.

# 4 Spin Correlations at the Generation of $\Lambda\,\Lambda$ Pairs in Multiple Processes

Further we will use the model of one-particle sources [6], which is the most adequate one in the case of collisions of relativistic ions.

Two $\Lambda$ particles are identical particles. Spin and angular correlations in this case arise due to the Fermi statistics and final-state interaction.

Indeed, it is easy to see that the Fermi-statistics effect leads not only to the momentum–energy $\Lambda\,\Lambda$-correlations at small relative momenta (correlation femtoscopy), but to the spin correlations as well.

The following relation holds, in consequence of the symmetrisation or industrialisation of the total wave function of any identical particles with nonzero spin (bosons as well as fermions) [7] :

$$(-1)^{S+L} = 1. \qquad (7)$$

Here, $S$ is the total spin and $L$ is the orbital momentum in the c.m. frame of the pair. At the momentum difference $q = p_1 - p_2 \to 0$ the states with nonzero orbital momenta "die out", and only states with $L = 0$ and even total spin $S$ survive.

Since the $\Lambda$-particle spin is equal to 1/2, at $q \to 0$ the $\Lambda\,\Lambda$ pair is generated only in the singlet state with $S = 0$.

Meantime, at the 4-momentum difference $q \neq 0$ there are also triplet states generated together with the singlet state.

Within the conventional model of one-particle sources emitting unpolarised particles, the triplet states with spin projections +1, 0, and −1 are produced with equal probabilities. If correlations are neglected, the singlet state is generated with the same probability – the relative "weights" are $\widetilde{W_t} = 3/4$, $\widetilde{W_s} = 1/4$.

When taking into account the Fermi statistics and $s$-wave final-state interaction, which is essential at close momenta (at orbital momenta $L \neq 0$ the contribution of final-state interaction is suppressed), the fractions of triplet states and the singlet state are renormalised.

We will perform here the analysis of spin $\Lambda\Lambda$ correlations in the c.m. frame of the $\Lambda\Lambda$ pair. In the c.m. frame, we have: $q = \{0, 2\mathbf{k}\}$, where $q$ is the difference of 4-momenta of the $\Lambda$ particles, $\mathbf{k}$ is the momentum of one of the particles. In doing so, the momentum $\mathbf{k}$ is connected with the relative momentum $\mathbf{q}$ in the laboratory frame by the Lorentz transformation [8] (we use the unit system with $\hbar = c = 1$):

$$\mathbf{k} = \frac{1}{2}\left[\mathbf{q} + (\gamma - 1)\frac{(\mathbf{qv})\mathbf{v}}{|\mathbf{v}|^2} - \gamma\mathbf{v}\,q_0\right]; \qquad (8)$$

here, $\mathbf{v} = (\mathbf{p}_1 + \mathbf{p}_2)/(\varepsilon_1 + \varepsilon_2)$ is the velocity of the $\Lambda\Lambda$ pair in the laboratory frame, $\gamma = (1 - v^2)^{-1/2}$ is the Lorentz factor, $\mathbf{q} = \mathbf{p}_1 - \mathbf{p}_2$ and $q_0 = \varepsilon_1 - \varepsilon_2$ are the laboratory relative momentum and energy, respectively.

The Lorentz transformations of 4-coordinates are given by the expressions:

$$\mathbf{r}^* = \mathbf{r} + (\gamma - 1)\,\frac{(\mathbf{rv})\mathbf{v}}{|\mathbf{v}|^2} - \gamma \mathbf{v}\,t, \qquad t^* = \gamma(t - \mathbf{vr}), \tag{9}$$

where $\mathbf{r} = \mathbf{r}_1 - \mathbf{r}_2$ and $t = t_1 - t_2$ are the differences of coordinates and times for one-particle sources in the laboratory frame.

The interference term connected with identity (quantum statistics) is determined by the expression:

$$\langle \cos 2\mathbf{kr}^* \rangle = \int W_{\mathbf{v}}(\mathbf{r}^*) \cos(2\mathbf{kr}^*) d^3\mathbf{r}^*, \tag{10}$$

where

$$W_{\mathbf{v}}(\mathbf{r}^*) = \int W(x)\, d\,t^* = \int W(\mathbf{r}^*, t^*)\, d\,t^*$$

is the distribution of coordinate difference between two sources in the c.m. frame of the $\Lambda\Lambda$ pair.

Meantime, the contribution of $s$-wave final-state interaction is expressed as follows (at the sizes of the generation region in the c.m. frame, exceeding the effective radius of interaction of two $\Lambda$ particles):

$$B_{\text{int}}(q) = B^{(\Lambda\Lambda)}(\mathbf{k}, \mathbf{v}) = \int W_{\mathbf{v}}(\mathbf{r}^*)\, b(\mathbf{k}, \mathbf{r}^*) d^3\mathbf{r}^*, \tag{11}$$

where the function $b(\mathbf{k}, \mathbf{r}^*)$ has the structure [2, 8, 9]:

$$b(\mathbf{k}, \mathbf{r}^*) = |f^{(\Lambda\Lambda)}(k)|^2\, \frac{1}{(r^*)^2} + 2\,\text{Re}\left( f^{(\Lambda\Lambda)}(k)\, \frac{e^{ikr^*} \cos \mathbf{kr}^*}{r^*} \right) - 2\,\pi\, |f^{(\Lambda\Lambda)}(k)|^2\, d_0^{(\Lambda\Lambda)} \delta^3(\mathbf{r}^*). \tag{12}$$

Here, $k = |\mathbf{k}|$, $r^* = |\mathbf{r}^*|$, $f^{(\Lambda\Lambda)}(k)$ is the amplitude of low-energy $\Lambda\Lambda$ scattering. In the framework of the effective radius theory [7, 10]:

$$f^{(\Lambda\Lambda)}(k) = a_0^{(\Lambda\Lambda)} \left( 1 + \frac{1}{2}\, d_0^{(\Lambda\Lambda)}\, a_0^{(\Lambda\Lambda)}\, k^2 - i\, k\, a_0^{(\Lambda\Lambda)} \right)^{-1}, \tag{13}$$

where, by definition, $(-a_0^{(\Lambda\Lambda)})$ is the length of $s$-wave scattering and

$$d_0^{(\Lambda\Lambda)} = \frac{1}{k}\, \frac{d}{d\,k} \left( \text{Re}\, \frac{1}{f^{(\Lambda\Lambda)}(k)} \right)$$

is the effective radius.

The integral (11), with expression (12) inside, approximately takes into account the difference of the true wave function of two interacting $\Lambda$ particles with the momenta $\mathbf{k}$ and $(-\mathbf{k})$ at small distances from the asymptotic wave function of continuous spectrum [9, 11].

Information about the parameters of $\Lambda\Lambda$ scattering is contained in the works studying double hypernuclei and pair correlations in the reactions with formation of two $\Lambda$ particles (see, for example, [12–14]). Analysis of the experimental data leads to the conclusion that the length of $\Lambda\Lambda$ scattering is comparable by magnitude ($\approx (-20)$ fm) with the length of neutron–neutron scattering [14].

# 5 Spin Correlations at the Generation of $\Lambda\bar{\Lambda}$ Pairs in Multiple Processes

In the framework of the model of independent one-particle sources, spin correlations in the $\Lambda\bar{\Lambda}$ system arise only on account of the difference between the interaction in the final triplet state ($S = 1$) and the interaction in the final singlet state. At small relative momenta, the $s$-wave interaction plays the dominant role as before, but, contrary to the case of identical particles ($\Lambda\Lambda$), in the case of non-identical particles ($\Lambda\bar{\Lambda}$) the total spin may take both the values $S = 1$ and $S = 0$ at the orbital momentum $L = 0$. In doing so, the interference effect, connected with quantum statistics, is absent.

If the sources emit unpolarised particles, then, in the case under consideration, the correlation function describing momentum-energy correlations has the following structure (in the c.m. frame of the $\Lambda\bar{\Lambda}$ pair):

$$R(\mathbf{k},\mathbf{v}) = 1 + \frac{3}{4}\,B_t^{(\Lambda\bar{\Lambda})}(\mathbf{k},\mathbf{v}) + \frac{1}{4}\,B_s^{(\Lambda\bar{\Lambda})}(\mathbf{k},\mathbf{v}). \tag{14}$$

The components of the correlation tensor for the $\Lambda\bar{\Lambda}$ pair are as follows:

$$T_{ik} = \frac{B_t^{(\Lambda\bar{\Lambda})}(\mathbf{k},\mathbf{v}) - B_s^{(\Lambda\bar{\Lambda})}(\mathbf{k},\mathbf{v})}{4 + 3\,B_t^{(\Lambda\bar{\Lambda})}(\mathbf{k},\mathbf{v}) + B_s^{(\Lambda\bar{\Lambda})}(\mathbf{k},\mathbf{v})}\,\delta_{ik}; \tag{15}$$

here the contributions of final-state triplet and singlet $\Lambda\bar{\Lambda}$ interaction are determined by the expression (analogously to Eqs. (11), (12) for the $\Lambda\Lambda$ interaction [2, 9], with the replacement $\cos\mathbf{kr}^* \to e^{i\,\mathbf{kr}^*}$ in Eq. (12) owing to the non-identity of the particles $\Lambda$ and $\bar{\Lambda}$ [8]):

$$B_{s(t)}^{(\Lambda\bar{\Lambda})}(\mathbf{k},\mathbf{v}) = \left|f_{s(t)}^{(\Lambda\bar{\Lambda})}(k)\right|^2 \left\langle \frac{1}{(r^*)^2} \right\rangle + 2\,\mathrm{Re}\left( f_{s(t)}^{(\Lambda\bar{\Lambda})}(k) \left\langle \frac{e^{ikr^*} e^{i\,\mathbf{kr}^*}}{r^*} \right\rangle \right) -$$

$$-\frac{2\pi}{k}\left|f_{s(t)}^{(\Lambda\bar{\Lambda})}(k)\right|^2 \frac{d}{dk}\left( \mathrm{Re}\,\frac{1}{f_{s(t)}^{(\Lambda\bar{\Lambda})}(k)} \right) W_{\mathbf{v}}(0), \tag{16}$$

where $f_{s(t)}^{(\Lambda\bar{\Lambda})}(k)$ is the amplitude of the $s$-wave low-energy singlet (triplet) $\Lambda\bar{\Lambda}$ scattering.

At sufficiently large values of $k$, one should expect that [9]:

$$B_s^{(\Lambda\bar{\Lambda})}(\mathbf{k},\mathbf{v}) = 0, \qquad B_t^{(\Lambda\bar{\Lambda})}(\mathbf{k},\mathbf{v}) = 0.$$

In this case the angular correlations in the decays $\Lambda \to p + \pi^-$, $\bar{\Lambda} \to \bar{p} + \pi^+$, connected with the final-state interaction, are absent :

$$T_{ik} = 0, \qquad T = 0.$$

# 6 Angular Correlations in the Decays $\Lambda \to p + \pi^-$ and $\bar{\Lambda} \to \bar{p} + \pi^+$ and the "Mixed Phase"

Thus, at sufficiently large relative momenta (for $k \gg m_\pi$) one should expect that the angular correlations in the decays $\Lambda \to p + \pi^-$ and $\bar{\Lambda} \to \bar{p} + \pi^+$, connected with the interaction of the $\Lambda$ and $\bar{\Lambda}$ hyperons in the final state (i.e. with one-particle sources) are absent. But, if at the considered energy the dynamical trajectory of the system passes through the so-called "mixed phase", then the two-particle sources, consisting of the free quark and antiquark, start playing a noticeable role. For example, the process $s\,\bar{s} \to \Lambda\bar{\Lambda}$ may be discussed.

In this process, the charge parity of the pairs $s\bar{s}$ and $\Lambda\bar{\Lambda}$ is equal to $C = (-1)^{L+S}$, where $L$ is the orbital momentum and $S$ is the total spin of the fermion and antifermion. Meantime, the $CP$ parity of the fermion–antifermion pair is $CP = (-1)^{S+1}$.

In the case of one-gluon exchange, $CP = 1$, and then $S = 1$, i.e. the $\Lambda\bar{\Lambda}$ pair is generated in the triplet state; in doing so, the "trace" of the correlation tensor $T = 1$.

Even if the frames of one-gluon exchange are overstepped, the quarks $s$ and $\bar{s}$, being ultra-relativistic, interact in the triplet state ($S = 1$). In so doing, the primary $CP$ parity $CP = 1$, and, due to the $CP$ parity conservation, the $\Lambda\bar{\Lambda}$ pair is also produced in the triplet state. Let us denote the contribution of two-quark sources by $x$. Then at large relative momenta $T = x > 0$.

Apart from the two-quark sources, there are also two-gluon sources being able to play a comparable role. Analogously with the annihilation process $\gamma\gamma \to e^+e^-$, in this case the "trace" of the correlation tensor is described by the formula (the process $g\,g \to \Lambda\bar{\Lambda}$ is implied):

$$T = 1 - \frac{4\,(1-\beta^2)}{1 + 2\,\beta^2\,\sin^2\theta - \beta^4 - \beta^4\,\sin^4\theta}, \tag{17}$$

where $\beta$ is the velocity of $\Lambda$ (and $\bar{\Lambda}$) in the c.m. frame of the $\Lambda\bar{\Lambda}$ pair, $\theta$ is the angle between the momenta of one of the gluons and $\Lambda$ in the c.m. frame (see [15]). At small $\beta$ ($\beta \ll 1$) the $\Lambda\bar{\Lambda}$ pair is produced in the singlet state (total spin $S = 0$, $T = -3$), whereas at $\beta \approx 1$ – in the triplet state ($S = 1$, $T = 1$). Let us remark that at ultra-relativistic velocities $\beta$ (i.e. at extremely large relative momenta of $\Lambda$ and $\bar{\Lambda}$) both the two-quark and two-gluon mechanisms lead to the triplet state of the $\Lambda\bar{\Lambda}$ pair ($T = 1$).

In the general case, the appearance of angular correlations in the decays $\Lambda \to p + \pi^-$ and $\bar{\Lambda} \to \bar{p} + \pi^+$ with the *nonzero* values of the "trace" of the correlation tensor $T$ at large relative momenta of the $\Lambda$ and $\bar{\Lambda}$ particles may testify to the passage of the system through the "mixed phase".

# 7 Summary

So, it is advisable to investigate the spin correlations of $\Lambda\Lambda$ and $\Lambda\bar{\Lambda}$ pairs produced in relativistic heavy ion collisions.

The spin correlations are studied by the method of angular correlations – method of moments.

The spin correlations, as well as the momentum-energy ones, make it possible to determine the space-time characteristics of the generation region and, besides, the parameters of low-energy scattering of $\Lambda$ on $\Lambda$ and $\Lambda$ on $\bar{\Lambda}$. They should be investigated jointly with the momentum-energy correlations.

# References

[1] V.L. Lyuboshitz and M.I. Podgoretsky, Yad. Fiz. **60** 45 (1997)

[2] V.L. Lyuboshitz, Proceedings of XXXIV PNPI Winter School . Physics of Atomic Nuclei and Elementary Particles, St-Petersburg (2000), p.402.

[3] R. Lednicky and V.L. Lyuboshitz, Phys. Lett. **B508** 146 (2001).

[4] G. Alexander and H.J. Lipkin, Phys. Lett. **B352** 162 (1995) .

[5] R. Lednicky, V.V. Lyuboshitz and V.L. Lyuboshitz, Yad. Fiz. **66** 1007 (2003)

[6] M.I. Podgoretsky, Fiz. Elem. Chast. At. Yadra **20** 628 (1989)

[7] L.D. Landau and E.M. Lifshitz, *Quantum Mechanics. Nonrelativistic Theory* (in Russian) (Nauka, Moscow, 1989); §§ 62, 133 .

[8] V.L. Lyuboshitz and V.V. Lyuboshitz, in: *Proceedings of XXXVII and XXXVIII Winter Schools of the Petersburg Institute of Nuclear Physics. Physics of Atomic Nuclei and Elementary Particles*, St.-Petersburg (2004), p.390 .

[9] R. Lednicky and V.L. Lyuboshitz, Yad. Fiz. **35** 1316 (1982)

[10] H.A. Bethe and P. Morrison, *Elementary Nuclear Theory* (New York, 1956); § 10 .

[11] V.L. Lyuboshitz, Yad. Fiz. **41** 820 (1985) [ Sov. J. Nucl. Phys. **41** 529 (1985)].

[12] S. Iwao, M. Chako, I. Kazanava, Progr. Theor. Phys. **48** 1412 (1972).

[13] A.O. Onichi *et al.*, Nucl. Phys. **A670** 2970 (2000).

[14] R. Afnan, Nucl. Phys. **A639** 550 (1998).

[15] H. McMaster, Rev. Mod. Phys. **33** 8 (1961)

# Part X

# Forward Physics and Low-$x$ QCD

# HERA Results on Small x and Forward Jets

Lev Khein

Skobeltsyn Institute of Nuclear Physics, Moscow State University
MSU, GSP-2, Vor.Gory, Moscow 119992, Russia

Results on measurements at small x at HERA performed by H1 and ZEUS collaborations are reviewed. Parton dynamics, i.e. DGLAP vs BFKL evolution, is studied with use of data on inclusive forward jet and forward jet accompanied by a dijet system via comparison with fixed order QCD calculations and leading-order parton-shower Monte Carlo models. Signatures of saturation at HERA are discussed.

## 1 Introduction

The conventional approximation of perturbative QCD, collinear factorisation and associated with it DGLAP equations, have successfully described the main volume of the data on hard processes. Since in this approach terms $(\log 1/x)^n$ are either omitted (Monte Carlo with LO matrix element and leading logarithm parton showers) or summed only up to first order (collinear NLO), at small $x_{\rm Bj}$ it is expected to become invalid. An alternative approach, resumming $\log 1/x$ terms and realised in the BFKL equations, should substitute DGLAP at small $x_{\rm Bj}$. HERA has reached as low $x_{\rm Bj}$ as $n \cdot 10^{-5}$. A question to HERA appears if these $x_{\rm Bj}$ are low enough to reveal deficiency of DGLAP (collinear factorisation)and applicability of BFKL ($k_T$ factorisation).

The key difference in the two approaches immediately influencing hadronic final states is the ordering of partons in the transverse momentum, $p_T$. The DGLAP approach features strongly ordered $p_T$ chains of partons while the BFKL approach features non-ordered $p_T$ chains. This difference could be best distinguished by measuring the forward, i.e. close in rapidity to the proton, jet cross-sections. At small $x_{\rm Bj}$ this provides large space for the BFKL evolution over x. If to suppress the DGLAP evolution over virtuality by requirement the forward jet $p_t$ to be close to the $p_t$ of the central hard jets the forward jet cross-section from BFKL should largely overshoot that from DGLAP.

The second expected important feature of small x regime is saturation. Since the BFKL structure function (and DGLAP as well) steeply rises with decrease of $x_{\rm Bj}$, unitarity should be violated at some very low x. To preserve unitarity saturation thus is inevitable. The second question to HERA appears if it can provide distinct indications of saturation.

Understanding the small $x_{\rm Bj}$ regime is of particular relevance since at the LHC many important processes, such as the production of electroweak gauge bosons or Higgs particles, involve collisions of partons with a small fraction of the proton momentum which could be lower than $x_{\rm Bj}$ of HERA by more than one order of magnitude.

## 2 Calculations

DGLAP is represented by the next-to-leading codes DISENT and NLOJET++ and the Monte Carlo code RAPGAP (RG-DIR), in which leading order matrix element is supplemented by the leading logarithm parton showers. Alternative, the BFKL-like branch of calculations is represented by the ARIADNE MC, which is an an implementation of the Colour Dipole Model (CDM), where gluons form a cascade of independently radiating dipoles, thus providing non-ordered in the transverse momentum of the partons, i.e. BFKL-like, evolution, and the CASCADE MC, which is an implementation of the CCFM equations converging to BFKL at small $x_{\rm Bj}$. Only gluon evolution is considered in CASCADE and two sets of unintegrated parton (gluon) density function (uPDF), J2003 set-1 and set-2 are probed. Another way to break strict DGLAP-specific $p_t$ ordering is to add to the DGLAP evolution starting from the proton oppositely directed DGLAP evolution from the photon side, i.e. include the resolved component of the photon. This is also implemented in the RAPGAP code (RG-DIR+RES).

## 3 Forward Jets

The neutral current DIS HERA data were obtained with electrons or positrons of $E_e = 27.5$ GeV energy colliding with protons of energy $E_p = 920$ GeV. Jets were reconstructed with the inclusive $k_T$-algorithm. Forward jets were studied inclusively and within trijets, i.e. when the forward jet was supplemented by a dijet. Events with the forward jet were selected as follows.

| | H1 | ZEUS |
|---|---|---|
| $Q^2(GeV^2)$ | 5-85 | 20-100 |
| y | 0.1-0.7 | 0.04-0.7 |
| $x_{\rm Bj}$ | $10^{-4} - 4 \cdot 10^{-3}$ | $4 \cdot 10^{-4} - 5 \cdot 10^{-3}$ |
| $p_{T,jet}(GeV)$ | 3.5 | 5 |
| $\eta$ | $1.74 - 2.79$ | $2 - 4.3$ |
| $x_{jet}$ | $> 0.035$ | $> 0.036$ |
| ${p_{T,jet}}^2/Q^2$ | $0.5 - 5.0$ | $0.5 - 2.0$ |

The two last requirements were imposed to enhance the BFKL evolution and to suppress the DGLAP evolution: the jet should have a large fraction of the proton momentum $x^{\rm jet} = p_Z^{\rm jet}/p > 0.036$ and the transverse energy of the order of the virtuality of the photon $0.5 < (E_T^{\rm jet})^2/Q^2 < 2$. For forward jet+dijet study the $x^{\rm jet}$ cut was omitted.

The measured by H1 collaboration [1] single differential forward jet cross sections are compared with LO ($\alpha_S$) and NLO ($\alpha_S^2$) (a) calculations and with QCD Monte Carlo models (b and c) in Fig. 1. The shaded band around the data points shows the error from the uncertainties in the calorimetric energy scales. The hatched band around the NLO calculations illustrates the theoretical uncertainties in the calculations. The dashed line in (a) shows the LO contribution.

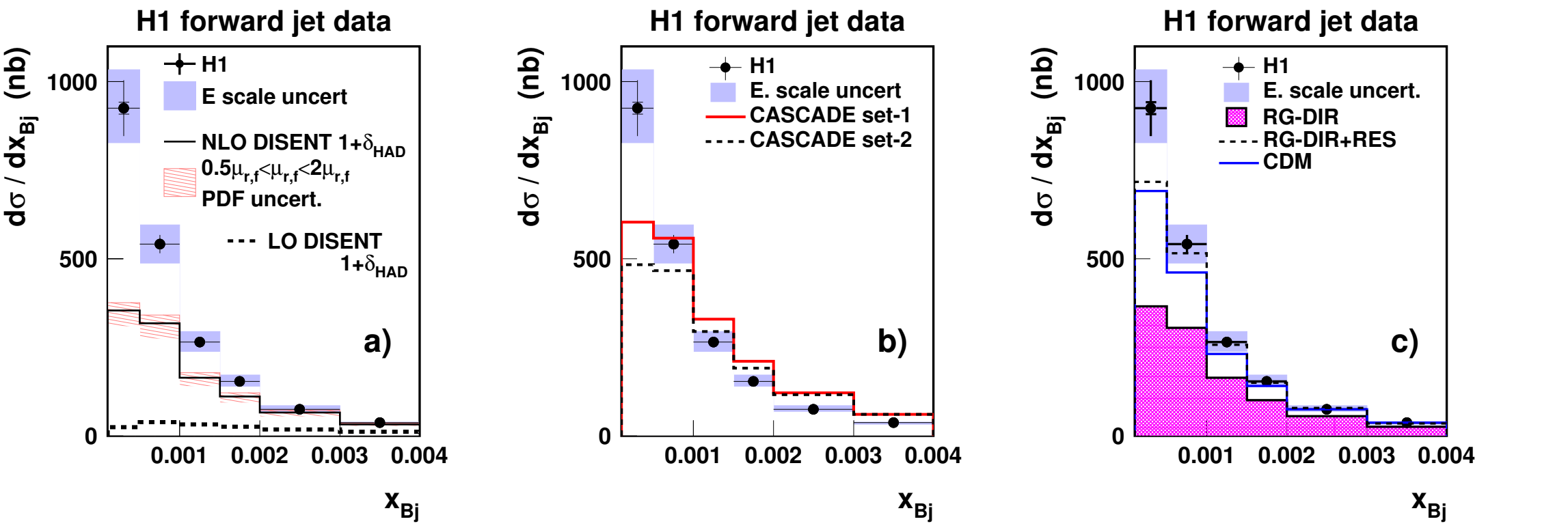

Figure 1: Inclusive forward jet cross section as a function of $x_{\rm Bj}$. H1 data are compared with LO ($\alpha_S$) and NLO ($\alpha_S^2$) (a) calculations and with QCD Monte Carlo models (b and c). The shaded band around the data points shows the error from the uncertainties in the calorimetric energy scales. The hatched band around the NLO calculations illustrates the theoretical uncertainties in the calculations. The dashed line in (a) shows the LO contribution.

A comparison of CASCADE with ZEUS data [2] in distributions over different quantities is presented in Fig. 2. Neither of two uPDF sets provides overall satisfactory agreement with the measurements. This points to need to further gluon PDF adjusting and/or inclusion of quarks into the evolution.

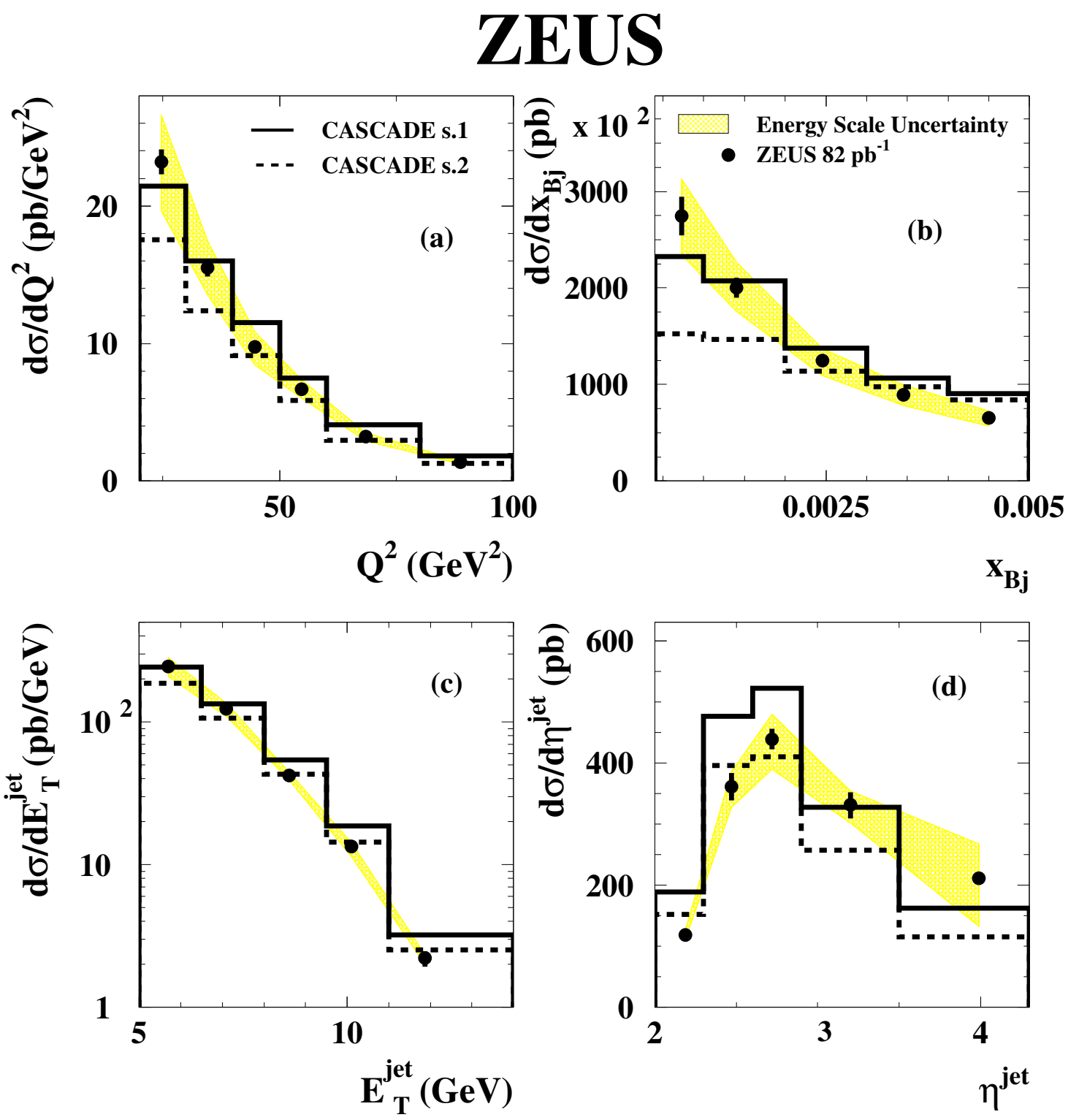

Figure 2: Inclusive forward jet cross sections. ZEUS data are compared with CASCADE with two sets of uPDF.

For trijets with a forward jet study, the three jets were ordered in pseudorapidity such that $\eta^{\mathrm{jet}_1} < \eta^{\mathrm{jet}_2} < \eta^{\mathrm{fjet}}$. Cross sections were plotted as functions of two pseudorapidity separations $\Delta\eta_1 = \eta^{\mathrm{jet}_2} - \eta^{\mathrm{jet}_1}$ and $\Delta\eta_2 = \eta^{\mathrm{fjet}} - \eta^{\mathrm{jet}_2}$. Additionally, the cross section as a function of $\Delta\eta_2$ was plotted for two intervals of $\Delta\eta_1$, $\Delta\eta_1 < 1$ and $\Delta\eta_1 > 1$.

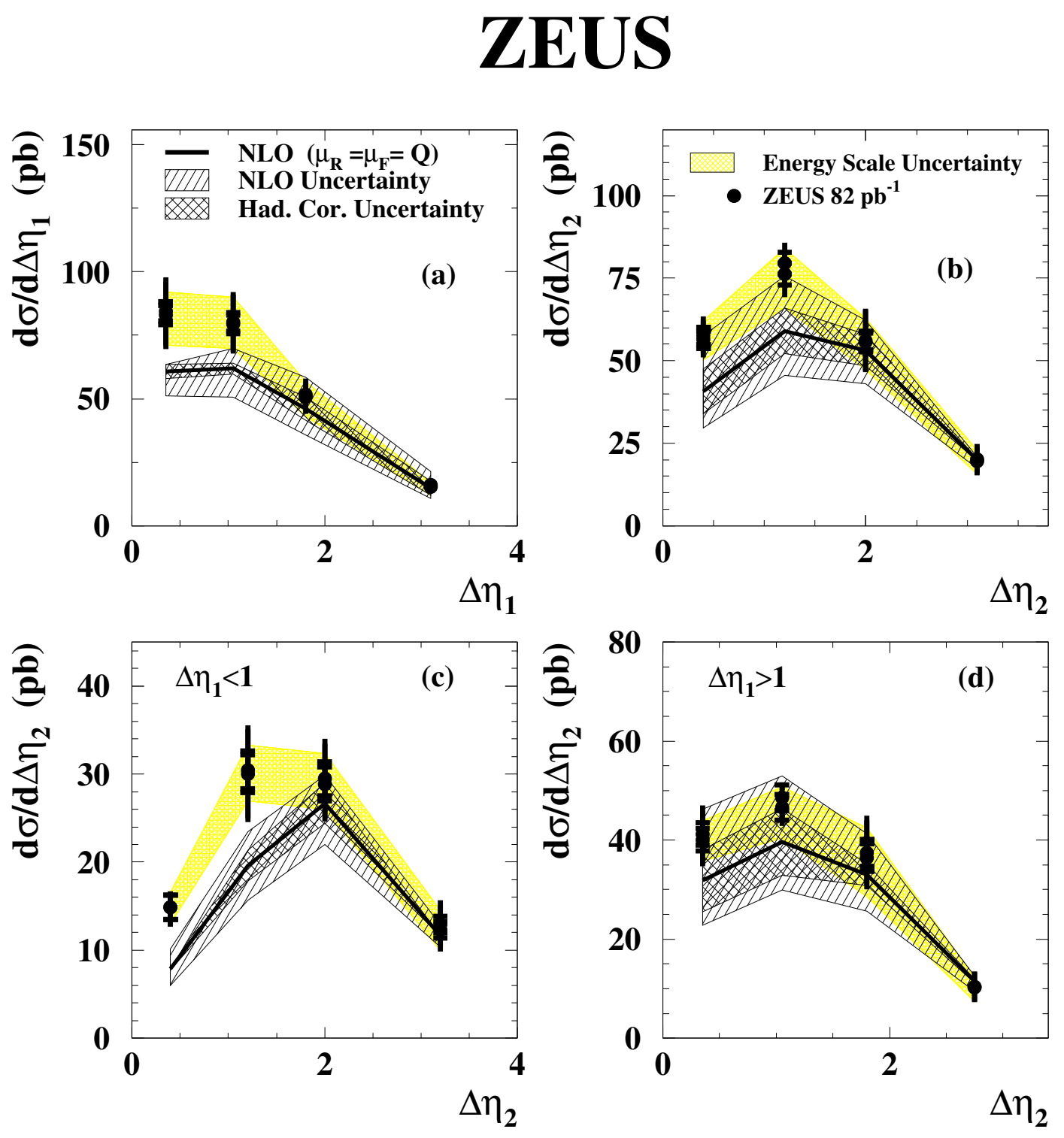

Figure 3: The forward jet+dijet cross sections. ZEUS data are compared with the NLO QCD calculations (solid line). The hatched area shows the theoretical uncertainties. The shaded area shows the cumulative uncertainty of the CAL and FPC energy scales.

Fig. 3 shows the comparison of the ZEUS data [2] with the calculations of NLOJET++. The calculations agree with the data at large $\Delta\eta_2$, while do not describe the data at small $\Delta\eta_2$, especially when $\Delta\eta_1$ is small. The large $\Delta\eta_2$ kinematics at low $x_{\mathrm{Bj}}$ favours dijets originating from the photon-gluon fusion, with an additional gluon responsible for the forward jet. This case is well treated by NLOJET++. The small $\Delta\eta_1$ and $\Delta\eta_2$ region corresponds to the event configuration in which all the three jets tend to go forward, away from the hard interaction. This configuration favours multigluon emission, which is lacking in NLOJET++.

Fig. 4 presents comparison of the H1 data [1] with the strict DGLAP MC (RAPGAP RG-DIR), DGLAP with resolved photon (RAPGAP RG-DIR+RES) and the CDM represented by the ARIADNE MC. The colour dipole model gives good agreement in all cases, whereas the DGLAP models give cross sections that are too low except when both $\Delta\eta_1$ and $\Delta\eta_2$ are large. In the latter topology all models and the NLO calculation agree with the data since little space is left for additional (to partons producing hard jets and the forward jet) gluon radiation. The dijet+forward jet sample differentiates CDM and the DGLAP-resolved model, in contrast to the more inclusive samples where CDM and RG-DIR+RES give the same predictions.

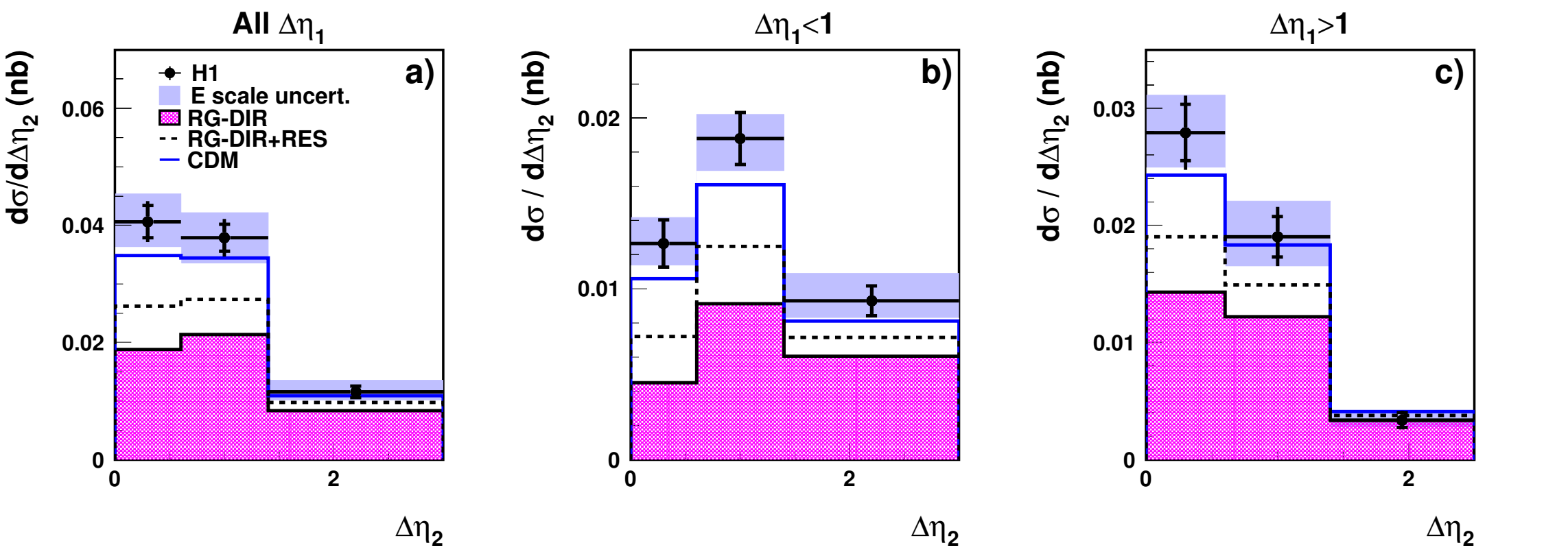

Figure 4: The forward jet+dijet cross sections. H1 data are compared with RAPGAP DIR, RAPGAP DIR+RES and CDM (Ariadne).

A comparison of the Cascade MC with the ZEUS data is presented in Fig. 5. Like that for the inclusive forward jets neither of two uPDF sets provides a satisfactory description of the data.

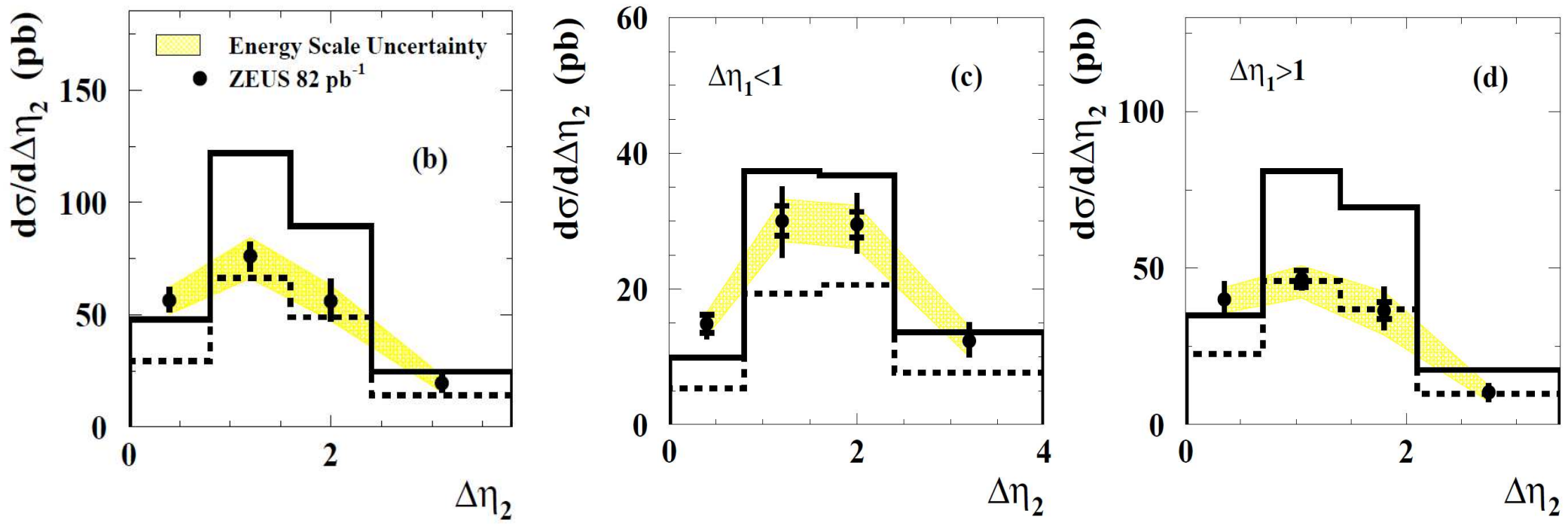

Figure 5: The forward jet+dijet cross sections. ZEUS data are compared with the Cascade with two sets of uPDF.

# 4 Saturation

There are two experimental facts that are considered as main arguments for relevance of saturation to HERA. The first one is geometric scaling. It quite naturally follows from the dipole with saturation models [3] and means that cross-sections are functions of $\tau = Q^2/Q_S^2(x)$ rather than $x$ and $Q^2$ separately, where $Q_S^2(x)$ has a meaning of the saturation scale. Both inclusive and diffractive data of HERA exhibit the geometric scaling, see Fig. 6. A problem is that the geometric scaling can also be derived from the linear BFKL and even DGLAP equations.

Another fact in favour of saturation at HERA is the flatness of dependence on x (Fig. 7 left, ZEUS data [4]) or W (Fig. 7 right, H1 data [5]) of the ratio of the diffractive to the inclusive cross-sections. This behaviour is incompatible with a simple Regge approach, where both the diffractive and inclusive cross sections are driven by the exchange of the same pomeron trajectory, whereas again is successfully described in the colour dipole with saturation model [3]. It is worth to be noted that the flatness is seen at non-small x as well (see Fig. 7 right), i.e. outside the saturation domain.

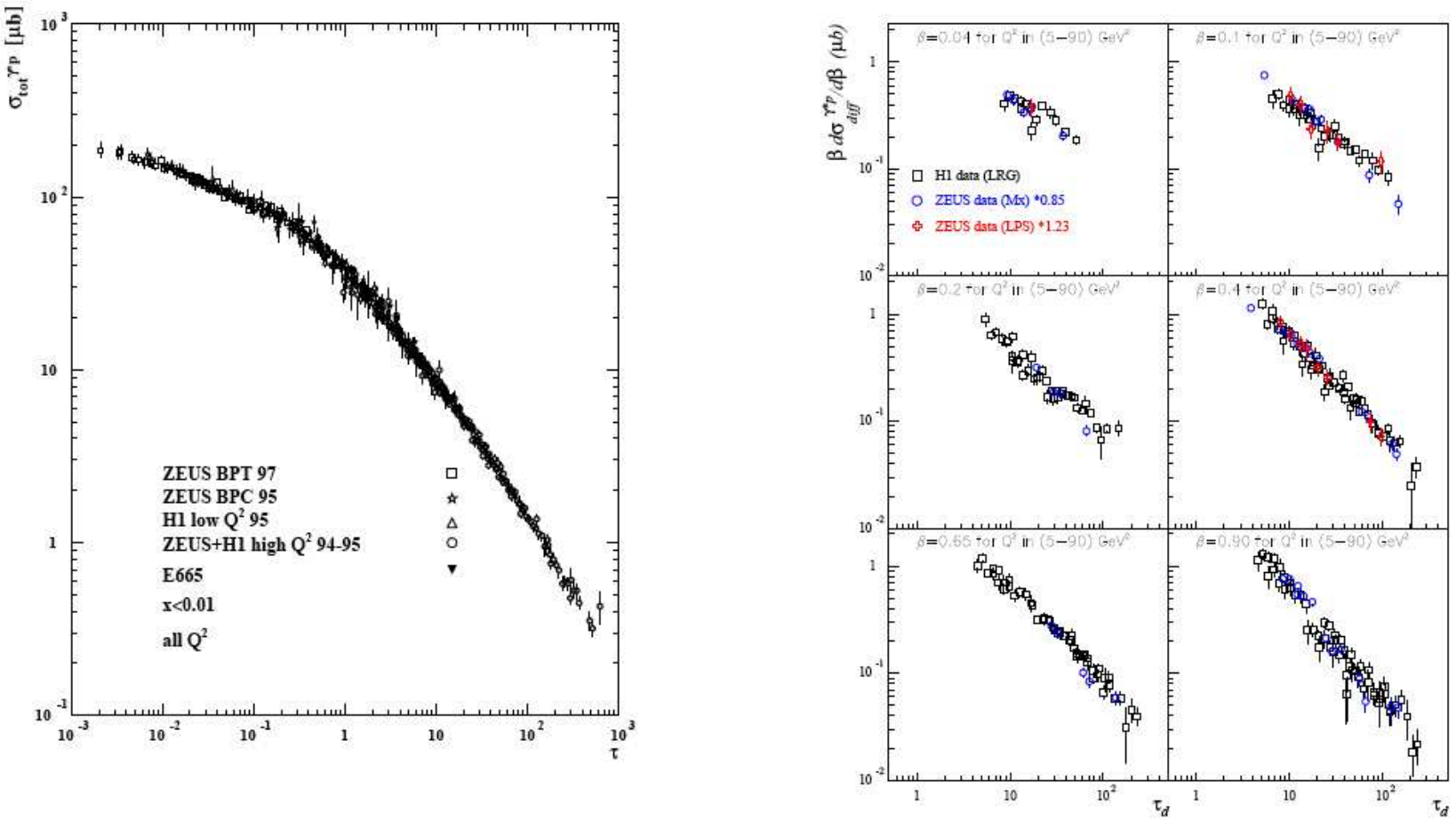

Figure 6: The total cross section as a function of $\tau = Q^2/Q_S^2(x)$ for $x_{\rm Bj} < 0.01$ (left) and the diffractive cross section as a function of $\tau_d = Q^2/Q_S^2(x_P)$ in bins of $\beta$ for $x_P < 0.01$

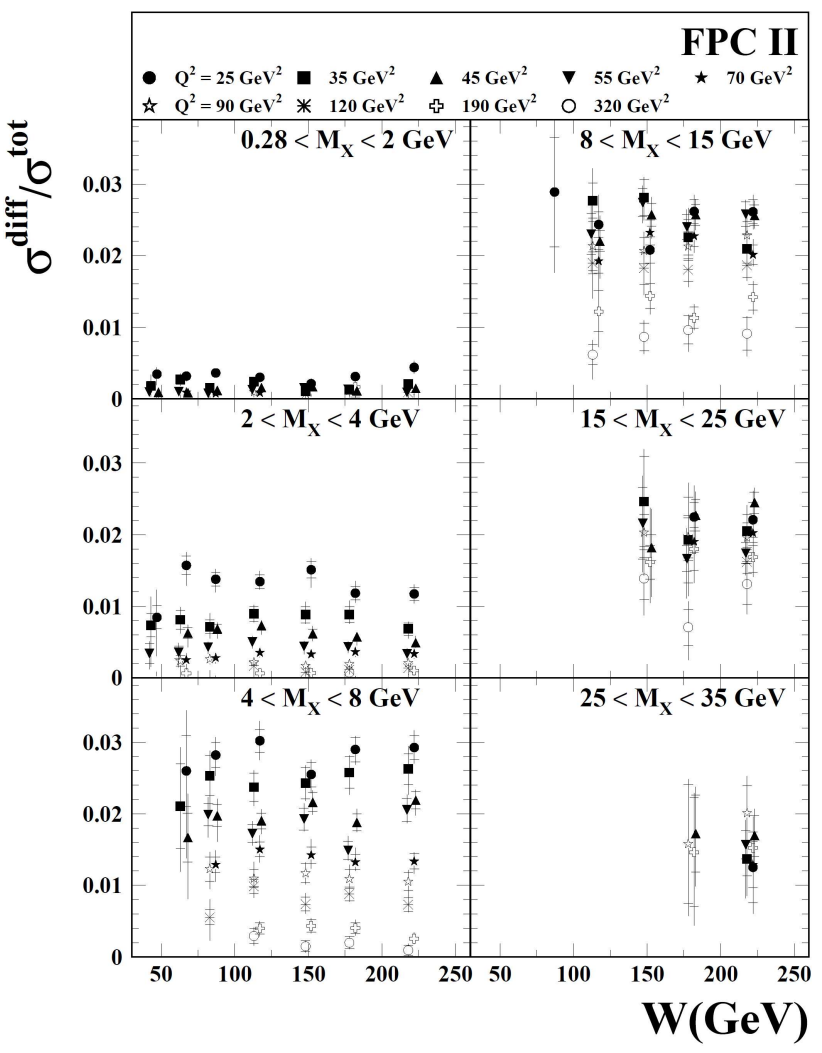

Figure 7: The ratio of the diffractive cross section, integrated over the $M_x$ intervals indicated to the total cross section measured by ZEUS as a function of $W$ for the $Q^2$ values indicated.

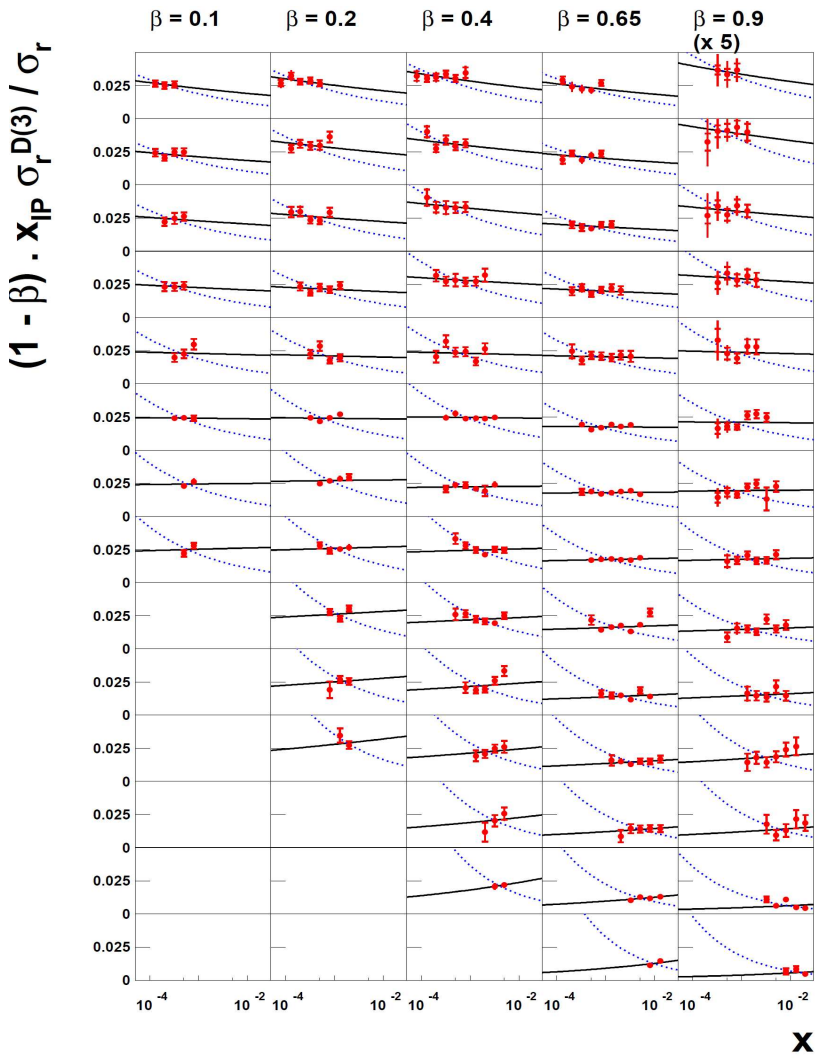

Figure 8: The ratio of the diffractive to the inclusive reduced cross section measured by H1 multiplied by $(1-\beta)\cdot x_P$. The data are compared with models motivated by the Regge phenomenology for which the diffractive and inclusive effective pomeron trajectories are the same ($\alpha_P = \alpha_P^{incl}$) and different ($\alpha_P \neq \alpha_P^{incl}$).

# 5 Conclusion

Forward jets at HERA small $x_{\mathrm{Bj}}$ manage to reveal failure of leading log DGLAP and deficiencies of NLO DGLAP. Breaking of $k_T$ ordering by inclusion of the resolved photon component improves description but fails at the smallest $x_{\mathrm{Bj}}$ in the forward jet+dijet case. LO CCFM based MC, CASCADE, cannot fully describe data on forward jets, other sets of uPDF are to be probed and/or inclusion of quarks in the evolution, and/or accounting for saturation are to be considered. Only CDM (ARIADNE MC), featured by BFKL-like non-ordered in $k_T$ parton cascade, is capable of successful description of the whole volume of the data on forward jets. A problem could to be, nevertheless, that largely being based on phenomenology ARIADNE is too free in tuning.

HERA unlikely can provide clear indications of saturation. Effects which are considered as indications of saturation, geometrical scaling and constant ratio of diffractive to total cross-sections vs x, could not be considered to be definitive enough. However, the success of the dipole with saturation models, with impressing easiness describing the otherwise difficult for treatment small x data, hints to relevance of saturation at HERA.

# References

[1] A. Aktas *et al.*, Eur. Phys. J. **C46** 27 (2006) [arXiv:hep-ex/0502029].

[2] S. Chekanov *et al.*, Phys. Lett. **B632** 13 (2006) [arXiv:hep-ex/0508055].

[3] K. Golec-Biernat and M. Wusthoff, Phys. Rev. **D59** 014017 (1999) [arXiv:hep-ph/9807513]; K. Golec-Biernat and M. Wusthoff, Phys. Rev. **D60** 114023 (1999) [arXiv:hep-ph/9903358]; J. Bartels, K. Golec-Biernat and H. Kowalski, Phys. Rev. **D66** 014001 (2002) [arXiv:hep-ph/0203258].

[4] S. Chekanov *et al.*, Nucl. Phys. **B800** 1 (2008) [arXiv:hep-ex/0802.3017].

[5] A. Aktas *et al.*, Eur. Phys. J. **C48** 715 (2006) [arXiv:hep-ex/0606004].

# Physics with forward jets in ATLAS, CMS and LHCb

*David d'Enterria*[1]

[1]ICC-UB & ICREA, Univ. de Barcelona, 08028 Barcelona, Catalonia

The capabilities of the ATLAS, CMS and LHCb detectors to reconstruct jets at forward rapidities ($|\eta| > 3$) in *p-p* collisions at the CERN Large Hadron Collider are reviewed. The QCD and Higgs physics motivations for such measurements are summarised. Details are given on studies that provide information on the parton structure and evolution at small values of fractional momenta in the proton.

## 1 Introduction

The ATLAS, CMS and LHCb experiments[1] feature detection capabilities at forward rapidities ($|\eta| > 3$, see Fig. 1 left) which allows them to reconstruct jets in a kinematic range of interest for various Higgs and QCD physics studies in *p-p* collisions at TeV energies.

On the one hand, having the possibility to reconstruct jets beyond $|\eta| \approx 3$ in ATLAS and CMS is crucial to signal Higgs boson production in vector-boson-fusion (VBF) processes, $qq \xrightarrow{VV} qqH$ (with $V = W, Z$), where the valence quarks $q$ from each proton fragment into jets in the forward and backward hemispheres [2]. The presence of such low-angle jets is instrumental to significantly reduce the QCD backgrounds in various VBF Higgs discovery channels at the LHC, particularly for low $H$ masses [3, 4]. In the case of LHCb, given the excellent secondary vertex capabilities of the detector, forward jet studies have focused on the reconstruction of $b$-jets aiming at the $H \to b\bar{b}$ decay channel in Higgs production associated with vector bosons (about one third of the cross section falls within the LHCb acceptance) [5].

On the other hand, forward jet production is in its own right an interesting perturbative QCD (pQCD) process whose study yields important information on the underlying parton structure and its dynamical evolution in the proton. In particular, it provides valuable information on the gluon density $xG(x, Q^2)$ in a regime of low momentum fraction, $x = p_{parton}/p_{hadron} < 10^{-2}$, where standard deep-inelastic *e-p* data can only indirectly constrain its value [6], and where its evolution is expected to be affected by non-linear QCD dynamics [7]. Indeed, in *p-p* collisions, the *minimum* parton momentum fractions probed in each proton in a $2 \to 2$ process with a jet of momentum $p_T$ produced at pseudo-rapidity $\eta$ are

$$x_2^{min} = \frac{x_T\, e^{-\eta}}{2 - x_T\, e^{\eta}} \;, \quad \text{and} \quad x_1^{min} = \frac{x_2\, x_T\, e^{\eta}}{2x_2 - x_T\, e^{-\eta}} \;, \text{ where } \quad x_T = 2p_T/\sqrt{s}\,, \qquad (1)$$

[1]ALICE has jet reconstruction capabilities at central [1] but not forward rapidities.

i.e. $x_2^{min}$ decreases by a factor of ~10 every 2 units of rapidity. The extra $e^{\eta}$ lever-arm motivates the interest of *forward* jet production measurements to study the PDFs at small values of $x$. From Eq. (1), it follows that the measurement at the LHC of jets with transverse momentum $p_T$ = 20 GeV/c at rapidities $\eta \approx 5$ allows one to probe $x$ values as low as $x_2 \approx 10^{-5}$ in partonic collisions with highly asymmetric longitudinal momenta in the initial-state (Fig. 1, right).

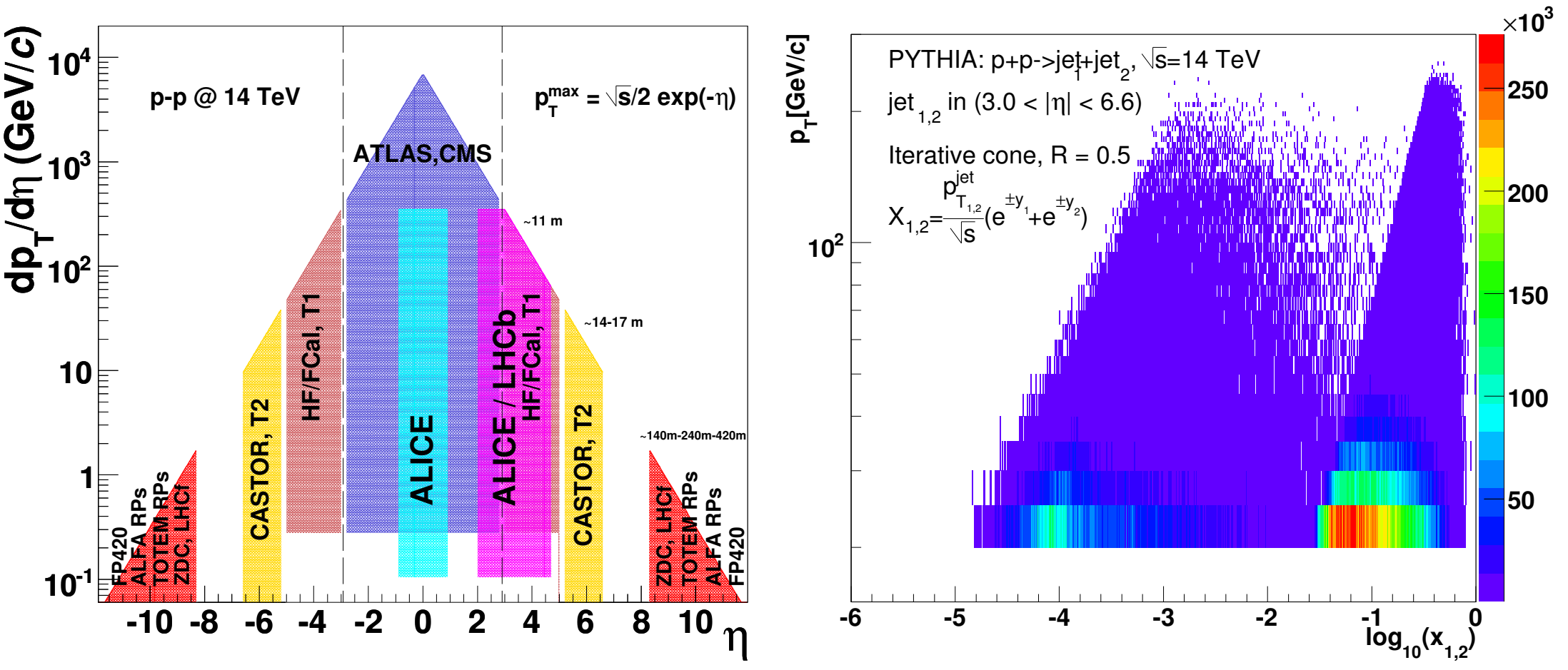

Figure 1: Left: Acceptance of the LHC detectors in the $p_T$–$\eta$ plane ('forward' detectors are beyond the dashed vertical line) [8]. Right: Log($x_{1,2}$) distribution of two partons producing at least one jet above $p_T$ = 20 GeV/c at forward rapidities in *p*-*p* collisions at $\sqrt{s}$ = 14 TeV [9].

In this contribution we summarise first the forward jet reconstruction capabilities of the three LHC experiments (Section 2) and, then, in Sections 3 and 4 we present simulation studies of two CMS forward-jet measurements [9]:

1. single inclusive jet cross section at moderate transverse momenta ($p_T \approx$ 20 – 120 GeV/c),

2. azimuthal (de)correlations of "Mueller-Navelet" [10] dijet events, characterised by jets with similar $p_T$ separated by a large rapidity interval ($\Delta\eta \approx 6 - 10$), which are sensitive, respectively, to the small-$x_2$ (and high-$x_1$) proton PDFs, as well as to low-$x$ QCD evolution of the BFKL [11], CCFM [12] and/or saturation [7] types.

## 2 Experimental Performances

In ATLAS and CMS, jets can be reconstructed calorimetrically at forward rapidities in the FCal [13] and HF [14] calorimeters[2] (3< $|\eta|$ <5), by means of standard jet algorithms of the cone or sequential-clustering types [16]. The jet radii are often chosen relatively small (e.g. $\mathcal{R}$ = 0.5 for the cone and $D$ = 0.4 for the $k_T$ algorithms) so as to minimise the effects of hadronic activity inside the jet due to the underlying event and beam-remnants. Figure 2

[2] In addition, in CMS one can further extend jet reconstruction up to $|\eta| \approx 6.6$ with the CASTOR detector [15].

(left) shows the energy resolution for forward jets reconstructed in CMS with three different algorithms (iterative cone, SISCone and $k_T$) [9]. The obtained $p_T$ relative resolutions are of $\mathcal{O}(20\%)$ at 20 GeV/c decreasing to 10% above 100 GeV/c. Similar results are obtained for ATLAS [17]. We note that though the forward calorimeters have a coarser granularity than the barrel and endcap ones, the energy resolution is *better* in the forward direction than at central rapidities because (i) the *total* energy of the jet is boosted at forward rapidities, and (ii) the forward jets are more collimated and, thus, the ratio of jet-size/detector-granularity is more favourable. The position ($\eta$, $\phi$) resolutions (not shown here) for forward jets are also very good: $\sigma_{\phi,\eta} \approx 0.045$ at $p_T = 20$ GeV/c, improving to $\sigma_{\phi,\eta} \approx 0.02$ above 100 GeV/c [9]. Good $\phi$-$\eta$ resolutions are important when it comes to detailed studies of the azimuthal decorrelation as a function of the pseudorapidity separation in events with forward-backward dijets (see Section 4). Figure 2 (right) shows the efficiency and purity of forward jets reconstructed with the seeded cone finder ($\mathcal{R} = 0.4$) in the ATLAS FCal calorimeter [17]. Above ~35 GeV/c, the efficiency saturates at around 95% with a purity below 4%.

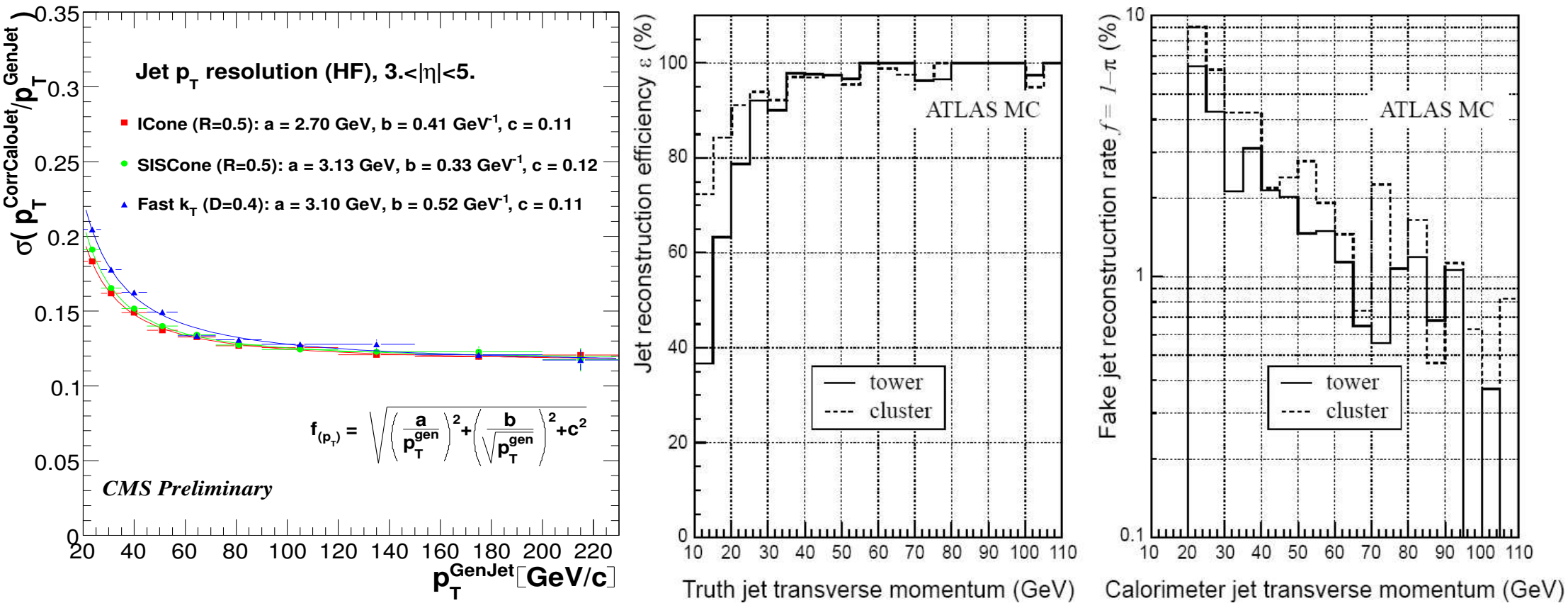

Figure 2: Left: Forward jet relative $p_T$ resolutions for various jet algorithms in the CMS HF calorimeter [9]. Right: Forward jet reconstruction efficiency as function of the true jet $p_T$, and fake reconstruction rate versus the reconstructed jet $p_T$ in the ATLAS FCal calorimeter [17].

In LHCb, jet reconstruction has focused on $b$-jets given the excellent vertexing capabilities of the detector. The physics motivation is so far centered on the measurement of the $H \to b\bar{b}$ channel for intermediate-mass Higgs production associated with vector bosons (30% of such a signal falls within the LHCb acceptance) [5]. Both seeded-cone- and $k_T$-algorithms have been tested including information from the calorimeters and the tracking devices in a "particle flow" type of approach. A neural-network is trained to identify $b$-jets and optimise the jet energy reconstruction. The main current limitation of the measure is the saturation of the calorimeters (designed originally mostly for single particle triggering/measurements at low and moderate $p_T$'s) for jets with total energy beyond $E_{tot} \approx 1.5$ TeV (i.e. $p_T \approx 20 - 150$ GeV/c for $\eta \approx 3 - 5$) given the very strong $\cosh(\eta)$ energy boost at large rapidities.

## 3 Inclusive Forward Jet Spectrum: Low-$x$ PDFs

Figure 3 (left) shows the forward jet $p_T$ spectrum generated with PYTHIA and reconstructed in the CMS HF calorimeter with the SISCone finder [9]. The spectrum is compared to fastNLO jet predictions [18] with the MRST03 and CTEQ6.1M PDFs. The right plot shows the percent differences between the reconstructed spectrum and the two theoretical predictions. The single jet spectra obtained for different PDFs are similar at high $p_T$, while differences as large as $\mathcal{O}(60\%)$ appear below ~60 GeV/c. The error bars include the statistical (a total integrated luminosity of 1 pb$^{-1}$ is assumed) and the energy-resolution smearing errors. The thin violet band around zero is the PDF uncertainty from the CTEQ6.1M set alone. The main source of systematic uncertainty is due to the calibration of the jet energy-scale (JES). Assuming a conservative $5-10\%$ JES error, one finds propagated uncertainties of the order $30-40\%$ in the jet yields at $p_T = 35-60$ GeV/c (yellow band) which are similar to the theoretical uncertainty associated to the PDF choice. If the JES can be improved at the 5% level or below, and the PDF uncertainties are indeed as large as the differences between MRST03 and CTEQ6M, a forward jet measurement could help constrain the underlying PDF in global-fit analyses.

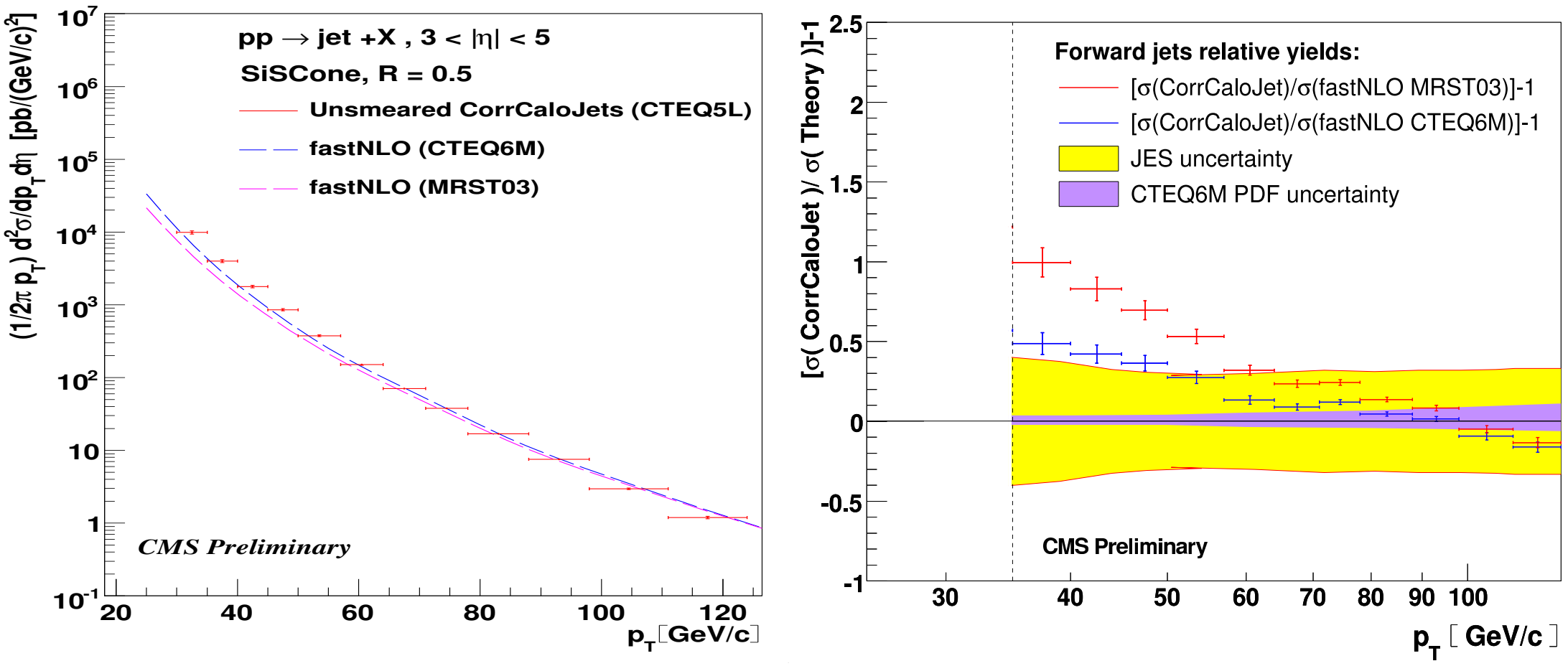

Figure 3: Left: Reconstructed forward jet spectrum (only stat errors shown) in $p$-$p$ at 14 TeV compared to fastNLO predictions with MRST03 and CTEQ6.1M PDFs. Right: Percent differences between the forward jet $p_T$ spectrum and the two fastNLO predictions. The yellow band shows the propagated yield uncertainty for a $5-10\%$ jet-energy scale (JES) error.

## 4 Forward-Backward Dijet Correlations: Low-$x$ QCD

The interest in forward jet measurements goes beyond the *single* inclusive cross sections: the production of *dijets* with similar $p_T$ but separated by large rapidities, the so-called "Mueller-Navelet jets" [10], is a particularly sensitive measure of non-DGLAP QCD evolutions. The large rapidity interval between the jets (e.g. up to $\Delta\eta \approx 12$ in the extremes of CMS forward calorimeters) enhances large logarithms of the type $\Delta\eta \sim log(s/p_{T,1}p_{T,2})$ which can be appropriately resummed within the BFKL [19], CCFM [12] and/or saturation [20, 21] frameworks. One of the phenomenological implications of this type of dynamics is an enhanced radiation between the two jets which results in a larger azimuthal decorrelation for increasing $\Delta\eta$ separations compared to collinear pQCD approaches. CMS [9] has carried out an analysis with PYTHIA [22] and HERWIG [23] selecting events with forward jets (ICone, $\mathcal{R} = 0.5$) which satisfy the following Mueller-Navelet (MN) type cuts:

- $p_{T,i} > 35$ GeV/c (good parton-jet matching and good jet trigger efficiencies in HF)
- $|p_{T,1} - p_{T,2}| < 5$ GeV/c (similar $p_T$ to minimise DGLAP evolution)
- $3 < |\eta_{1,2}| < 5$ (both jets in HF)
- $\eta_1 \cdot \eta_2 < 0$ (each jet in a different HF, i.e. their separation is $\Delta\eta \gtrsim 6$)

The data passing the MN-cuts are divided into 4 equidistant pseudorapidity bins with separations $\Delta\eta$=6.5, 7.5, 8.5 and 9.5 and the dijet cross section computed as $d^2\sigma/d\eta dQ = N_{jets}/(\Delta\eta\Delta Q \int \mathcal{L}\mathrm{d}t)$, where $Q = p_{T,1} \approx p_{T,2}$ and $N$ is the observed number of jets in the $\Delta\eta, \Delta Q$ bin. For 1 pb$^{-1}$, one expects a few 1000s (100s) MN jets with separations $\Delta\eta > 6$ (9). Figure 4, left, shows the expected PYTHIA yields passing the MN cuts for $\Delta\eta \approx 7.5$. The obtained dijet sample appears large enough to carry out detailed studies of the $\Delta\eta$ dependence of the yields, and look e.g. for a possible "geometric scaling" behaviour in the Mueller-Navelet yields [21].

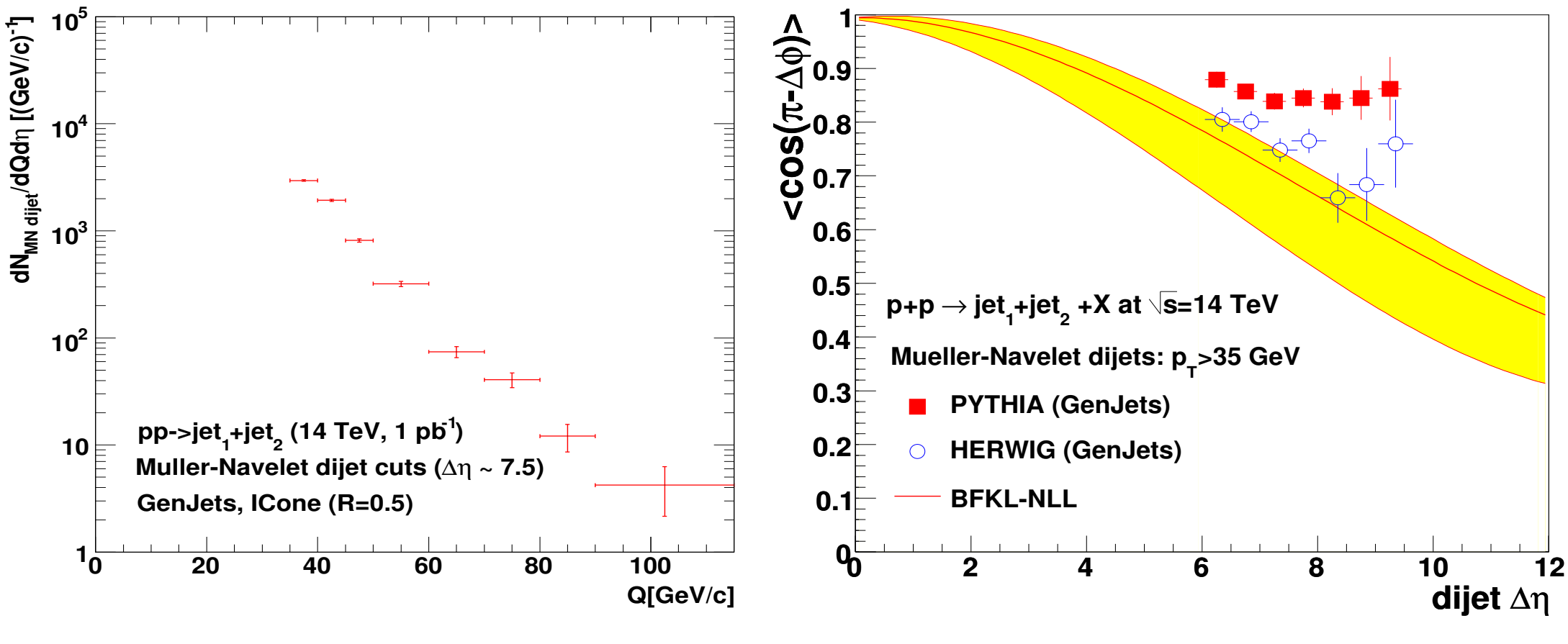

Figure 4: CMS study of dijet events passing the Mueller-Navelet cuts (see text) [9]. Left: Expected PYTHIA yields (1 pb$^{-1}$) for $\Delta\eta \approx 7.5$. Right: Average of $cos(\pi - \Delta\phi)$ versus $\Delta\eta$ in PYTHIA-, HERWIG-generated events compared to BFKL (yellow band) [19] analytical estimates.

An enhanced azimuthal decorrelation for increasing rapidity separation, measured e.g. by the average value (over events) of the cosine of the $\Delta\phi$ difference between the MN jets $\langle cos\,(\pi - \Delta\phi)\rangle$ versus the $\Delta\eta$ between them, is the classical "smoking-gun" of BFKL radiation [19, 20]. One expects $\langle cos(\pi - \Delta\phi)\rangle = 1$ (0) for perfect (de)correlation between the two jets. The results are shown in Fig. 4 (right) for the two highest-$p_T$ jets in the event passing the MN cuts. Only the dominant (statistical) errors are presented. At the Monte Carlo truth level (not shown here), the originating partons in PYTHIA or HERWIG are almost exactly back-to-back for all $\Delta\eta$ in each such jet-pair events. Yet, at the *generator-level*, the $\langle cos(\pi - \Delta\phi)\rangle$ decorrelation increases to 15% (25%) for PYTHIA (HERWIG), $\langle cos(\pi - \Delta\phi)\rangle \approx 0.85$ (0.75), due to parton showering and hadronization effects. Yet, the forward dijet decorrelation observed in both MCs is smaller (and less steep as a function of $\Delta\eta$) than found in BFKL approaches (yellow band) [19, 20].

# 5 Summary

We have summarised the forward jet reconstruction capabilities of the ATLAS, CMS and LHCb experiments. The measurement of forward jets opens up the possibility to carry out interesting studies in the Higgs (tagging of vector-boson-fusion or Higgs-to-$b\bar{b}$ in associated $WZ$ production) and QCD (low-$x$ parton densities and dynamics) sectors of the Standard Model.

# Acknowledgements

The author thanks Salim Cerci for his valuable collaboration with the CMS forward jets analysis, as well as Marco Musy for discussions on LHCb jet reconstruction. Support by the 7th EU Framework Programme (contract FP7-ERG-2008-235071) is acknowledged.

# References

[1] A. Morsch [ALICE Collaboration], AIP Conf. Proc. **1026**, 72 (2008).

[2] A. Djouadi, Phys. Rept. **457** (2008) 1.

[3] A. de Roeck (ed.) *et al.* [CMS Collaboration], J. Phys. G **34**, 995 (2007).

[4] S. Asai *et al.* [ATLAS Collaboration], Eur. Phys. J. C **32S2**, 19 (2004).

[5] A. Bay and C. Potterat [LHCb Collaboration], LHCb Note 2009-023, LHCb Note 2009-024.

[6] D. d'Enterria, Eur. Phys. J. A **31**, 816 (2007).

[7] See e.g. F. Gelis, T. Lappi and R. Venugopalan, *Int. J. Mod. Phys. E* **16**, 2595 (2007).

[8] D. d'Enterria, Proceeds DIS 2007, pp. 1141; arXiv:0708.0551 [hep-ex].

[9] S. Cerci and D. d'Enterria [CMS Collaboration], AIP Conf. Proc. **1105** (2009) 28; CMS PAS FWD-08-001 (CMS-AN/2008-060).

[10] A. H. Mueller and H. Navelet, *Nucl. Phys. B* **282**, 727 (1987).

[11] L.N. Lipatov, *Sov. J. Nucl. Phys.* **23**, 338 (1976); E.A. Kuraev, L.N. Lipatov and V.S. Fadin, *Zh. Eksp. Teor. Fiz* **72**, 3 (1977); Ya.Ya. Balitsky, L.N. Lipatov, *Sov. J. Nucl. Phys.* **28**, 822 (1978).

[12] H. Jung and G. P. Salam, Eur. Phys. J. C **19** (2001) 351; M. Deak, F. Hautmann, H. Jung and K. Kutak, JHEP **0909** (2009) 121.

[13] A. Artamonov *et al.*, JINST **3** (2008) P02010.

[14] A. S. Ayan *et al.*, *J. Phys. G* **30**, N33 (2004).

[15] X. Aslanoglou *et al.*, *Eur. Phys. J. C* **52**, 495 (2007).

[16] C. Buttar *et al.*, arXiv:0803.0678 [hep-ph].

[17] G. Aad *et al.* [The ATLAS Collaboration], arXiv:0901.0512.

[18] T. Kluge, K. Rabbertz and M. Wobisch, arXiv:hep-ph/0609285; and K. Rabbertz, private comm.

[19] A. Sabio Vera, F. Schwennsen, *Nucl. Phys. B* **776**, 170 (2007); and private communication.

[20] C. Marquet and C. Royon, *Nucl. Phys. B* **739**, 131 (2006); and arXiv:0704.3409 [hep-ph].

[21] E. Iancu, M. S. Kugeratski and D. N. Triantafyllopoulos, Nucl. Phys. A **808**, 95 (2008).

[22] T. Sjostrand, S. Mrenna and P. Skands, *J. High Energy Phys.* **0605** (2006) 026.

[23] G. Marchesini *et al.*, *Comput. Phys. Commun.* **67**, 465 (1992).

# CCFM Evolution with Unitarity Bound

*Emil Avsar*

Penn State University, University Park, PA 16802, USA

We perform a detailed study of the CCFM evolution, complemented with an absorptive boundary which mimics the non-linear dynamics of parton saturation. We present results for the evolution of the unintegrated gluon density in the presence of saturation and extract the energy dependence of the emerging saturation momentum. We show that CCFM and BFKL evolution lead to almost identical predictions after including the effects of gluon saturation and of the running of the coupling.

## 1 Introduction

In a series of studies, [1, 2], we have proposed a method for effectively implementing saturation and unitarity within a generic linear evolution equation for the unintegrated gluon distribution, so like the BFKL [3] and the CCFM [4] equations. In these proceedings we shall summarise some of the results from [2] and we shall discuss prospects for the future and open questions. For details of the topics to be presented here we refer to the cited articles.

At HERA, for not too low $Q^2$ values, schemes based on the NLO DGLAP evolution have been quite successful at describing data. However, the situation at lower $Q^2$ is not so clear. A very recent study [5] indeed points to some deviations from NLO DGLAP predictions in HERA data, for smaller $x$ values. One should, however, realise that deviations from the standard collinear approach may be seen even at relatively large momenta $Q$, well above $Q_s$. Interesting observables are in particular forward jets, or Mueller-Navelet jets. At LHC, due to the high collisions energy, even events with large $Q^2$ can contain effects of small $x$ evolution. Also interesting is the description of angular (de)correlations in multi-jet events.

The search for non-linear effects in data has so far mostly concentrated on inclusive observables, like $F_2$ and $F_L$, which are generally poor in discriminating between various models (though the use of both of them simultaneously gives stronger constraints). One has moreover mostly used simple dipole models, which contain few fit parameters, but also lack in dynamics since for example energy dependence is fitted by hand. Obviously, it would be a great advantage to be able to study more exclusive observables, as this would make the various predictions from the different dynamics harder to mask via fit parameters or input distributions. Secondly, it is desirable to have an approach which includes non-linear effects and at the same time contains more dynamical information on the underlying evolution. Also, it is by now rather well-known that kinematical effects such as energy conservation are significant. As such effects are hard to include in the analytical formalism, it is an advantage to have a numerical procedure where they can be included together with the small-$x$ evolution and the non-linear effects related to parton saturation.

Our strategy is to enforce an absorptive boundary condition at low transverse momenta which prevents the gluon phase–space occupation numbers to grow beyond their physical values at saturation. The method is the extension of a strategy originally introduced in relation with analytic studies of the BFKL evolution in the presence of saturation [6, 7]. Importantly, in [1] we demonstrated the power of this method, and its efficiency by comparing the numerical solutions to the BFKL equation with the absorptive boundary condition against those of the non-linear Balitsky–Kovchegov (BK) equation [8, 9]. We have shown that the absorptive boundary method successfully reproduces the results of the BK equation for both fixed and running coupling, and for *all* the energies, not only the asymptotic ones. Therefore, the simplicity and efficiency of this method makes it a very compelling tool for phenomenological studies at LHC. In addition, the great advantage is that it can directly be implemented in present Monte Carlo event generators, such as CASCADE [10] and LDCMC [11]. Thus one does not need to write new and complicated codes but can simply modify existing ones.

Of course, being such a simple method, our approach has some drawbacks. One is that one is limited studying the region above $Q_s$ where the saturation effects are harder to identify. Below $Q_s$ one would need to take into the full non-linear dynamics, which also includes complex many-body correlations and cannot be treated with standard pdfs. Even though saturation certainly plays a greater role in the small momenta region, covering the underlying event at LHC, the analysis of it will be extremely complex and difficult to study. Besides, for such low momenta, it may be difficult to separate saturation physics from the genuinely non–perturbative physics. By staying in the higher $Q$ region we can study saturation and small $x$ effects in the more clean environment of perturbative QCD.

# 2 Short Overview of CCFM

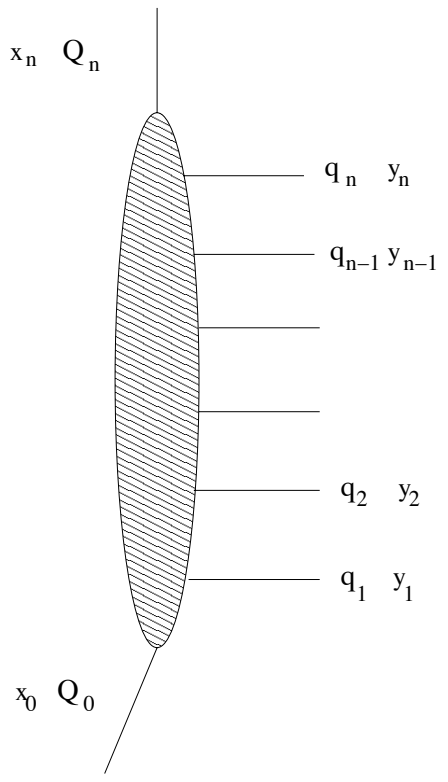

Figure 1: *Kinematics of the gluon radiation. The last t-channel propagator is denoted by $Q_n$, while the real s-channel gluons are represented by the horizontal lines.*

In Figure 1 we define the kinematics of the evolution. The figure represents a gluon ladder as produced by CCFM where by $q_i$ we denote the transverse momenta of the real, $s$-channel, gluons, and we define $Q_i = -\sum_k^i q_k$ where we have assumed the incoming virtual, $t$-channel gluon to be collinear with the proton (not shown in the figure). The important aspects of CCFM can be summarised as follows.

Only those diagrams are kept which give the leading factors of $\alpha_s \ln s$ where $s$ is the cms energy. This implies a strong ordering in the energy of the real gluons:

$$x = x_n \ll y_n \ll y_{n-1} \ll \cdots \ll y_2 \ll y_1 \approx 1\,, \tag{1}$$

where by $y_i$ we denote the energy fraction of the $i$ th emitted real gluon. The real gluons are distributed with the standard logarithmic weights $\bar{\alpha}_s dq_i^2/q_i^2 \cdot dy_i/y_i$. In BFKL the $t$–channel propagators have precisely momenta $Q_i$ since there the energy ordering agrees with the order of emission. In CCFM, however, the real gluon emissions are instead ordered according to their angles $\xi_i \equiv q_i^2/(y_i^2 E^2)$ ($E$ is the energy of the proton). However, it turns out that also in CCFM can one approximately identify the $Q_i$ as the momenta of the $t$–channel propagators. This is crucial in rewriting CCFM in the form which is commonly used.

In addition there are in CCFM two types of virtual corrections which both exponentiate, $S^2_{eik}$ and $S^2_{ne}$. They are related to corrections involving the eikonal and non-eikonal components of the three-gluon vertex. As shown in [2] one has at each vertex $k$

$$S^2_{ne}(k) \cdot S^2_{eik}(k) = \Delta^{BFKL}_{ne}(k). \tag{2}$$

where $\Delta^{BFKL}_{ne}(k)$ is the so-called "non–Sudakov" form factor in BFKL.

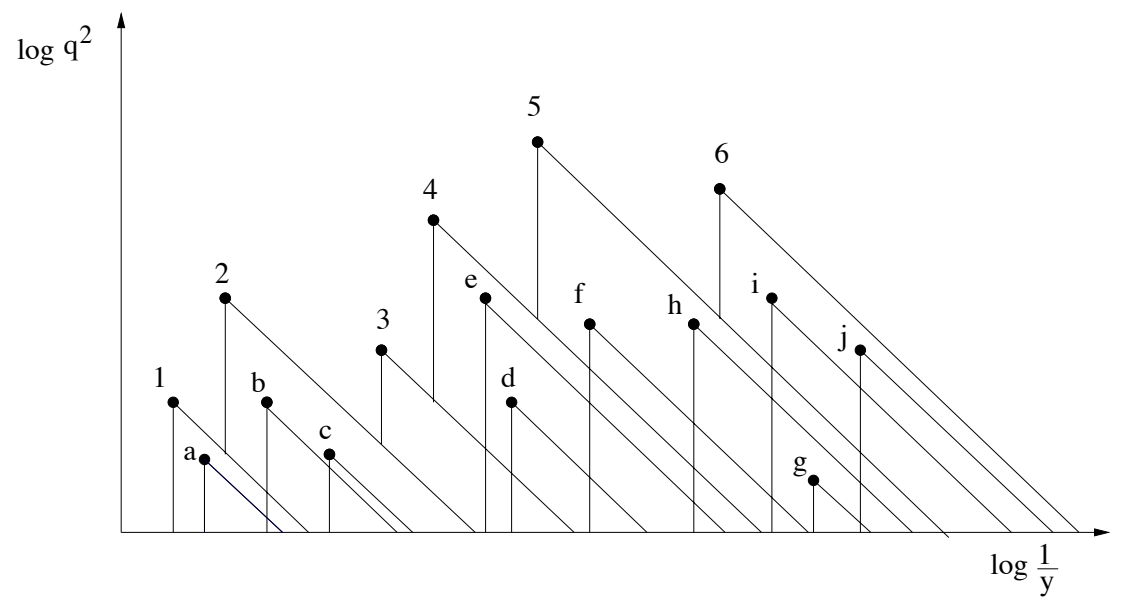

Figure 2: *Dividing the CCFM radiation into two subsets: hard emissions (enumerated emissions) and soft emissions (indexed by small letters) All real emissions are represented by fat dots. The vertical axis denote the transverse momenta while the horizontal axis the inverse energy fraction. The diagonal lines indicate the angles of the emissions. One sees that soft emissions also have small transverse momenta.*

A very important feature of the CCFM cascade is the identification of so called soft and hard emissions in the ladder. The hard emissions are those real emissions which are ordered in *both* energy and angle, while the soft emissions are such that they are always followed in the angular ordering by a gluon with higher energy. If the ladder consisted only of hard emissions, the $t$–channel propagators in CCFM would have momenta exactly equal to $Q_i$ as then the angular and the energy orderings coincide. However, the soft emissions which break the exact correspondence are soft not only in energy but also in momenta, see Figure 2. This implies that they, to a first approximation, do not change the momenta of the virtual propagators. Therefore one can indeed identify the $Q_i$ with the propagator momenta, even in the angular ordered cascade. This argument shows that there is an implicit approximation in the CCFM formalism — the fact that soft gluon emissions are assumed not to change the virtual transverse momenta. The reason for classifying the emissions as hard and soft is because it allows one to rewrite the gluon distribution in a much simplified form.

Within these approximations, the CCFM evolution equation for the gluon density can be rewritten in three different levels. The first one is the most exclusive one including all[1] soft and hard emissions from the gluon ladder. It is given by

$$\mathcal{A}(x,k,\bar{p}) = \bar{\alpha}_s \int_x^1 dz \int \frac{\mathrm{d}^2 p}{\pi p^2}\, \theta(\bar{p}-zp)\Delta_s(\bar{p},zp) \left( \frac{\Delta_{ns}(k,z,p)}{z} + \frac{1}{1-z} \right) \times \mathcal{A}\left(\frac{x}{z}, |k+(1-z)p|, p\right). \tag{3}$$

Here we are using rescaled momenta within the integrand: $\bar{p} = \bar{q}/(1-x)$ and $p = q/(1-z)$. The third argument of the gluon distribution indicates the maximally allowed emission angle, $\bar{\xi} = \bar{q}^2/(x^2E^2)$. This is a feature specific to CCFM and follows from the coherence property of the emissions. The $1/z$ piece comes from the hard emissions while $1/(1-z)$ comes from the soft emissions. Now, the soft emissions can be exactly compensated by the Sudakov form factors $\Delta_s$ so that this expression is equivalent to the more inclusive expression

$$\mathcal{A}(x,k,\bar{p}) = \bar{\alpha}_s \int_x^1 \frac{\mathrm{d}z}{z} \int \frac{\mathrm{d}^2 p}{\pi p^2}\, \theta(\bar{p}-zp)\Delta_{ns}(k,z,p)\mathcal{A}\left(\frac{x}{z}, |k+(1-z)p|, p\right) \tag{4}$$

which includes only the hard emissions explicitly. It does not make sense to say one of the equations is preferred to the other, within the approximations of CCFM they are equivalent. Thus one could base an event generator on (4) to construct the gluon distribution but then the soft emissions must be later included a final state emissions to get the correct final state. Nevertheless, in practice we expect the two equations also to give different gluon distributions, see the discussion below.

In deriving the integral equations (and also $\Delta_{ns}$ and $\Delta_s$ out of $S^2_{ne}$ and $S^2_{eik}$), the approximation that the soft emissions do not change the propagator momenta was used. However, there are also a class of hard emissions which are, in transverse momenta, as "soft" as the soft emissions. Within the approximations of CCFM, these hard emissions also do not change the propagator transverse momenta. As was realised first in [12], those hard emissions can be used to cancel the non-Sudakov, $\Delta_{ns}$, simplifying the evolution equation even more. There are, however, two caveats. The first is that this is only possible if the kinematical constraint $k^2 > zq^2$ is explicitly introduced[2]. Secondly the explicit definition of the transverse momenta conserving hard emissions requires a cut directly in momentum space, using a theta function in the momentum variables. This is contrast with the constraint put on the soft emissions, which is not directly written in momentum space. This results in a certain ambiguity since this cut can within the approximations be chosen somewhat differently. This implies that there is no unique equation to derive out of (4). The explicit choice for example used in the LDC model [12] (and this was also one of the choices in [2]) leads to the equation

$$\partial_Y \mathcal{A}(Y,k) = \bar{\alpha}_s \int \frac{\mathrm{d}k'^2}{|k^2-k'^2|}\, h(\kappa)\Big(\theta(k^2-k'^2)\mathcal{A}(Y,k') + \theta(k'^2-k^2)\theta(Y-\ln(k'^2/k^2))\mathcal{A}(Y-\ln(k'^2/k^2),k')\Big). \tag{5}$$

[1] These are not all the emissions in the final state however. There are further emissions from the $s$ channel gluons and these have be taken into account for a faithful description of the exclusive final state.

[2] As a matter of fact, the original article [13] mentions that this constraint should indeed be included.

where

$$h(\kappa) = 1 - \frac{2}{\pi} \arctan \left( \frac{1+\sqrt{\kappa}}{1-\sqrt{\kappa}} \sqrt{\frac{2\sqrt{\kappa}-1}{2\sqrt{\kappa}+1}} \right) \theta(\kappa - 1/4). \tag{6}$$

and $\kappa \equiv \min(k^2, k'^2)/\max(k^2, k'^2)$. A different choice of the explicit cut leads instead to

$$\partial_Y \mathcal{A}(Y,k) = \bar{\alpha}_s \int \frac{\mathrm{d}k'^2}{\max(k^2,k'^2)} \Big( \theta(k^2 - k'^2)\mathcal{A}(Y,k') \\ + \theta(k'^2 - k^2)\theta(Y - \ln(k'^2/k^2))\mathcal{A}(Y - \ln(k'^2/k^2), k') \Big). \tag{7}$$

One should immediately notice that $\mathcal{A}$ now only depends on $x$ and $k$, the dependence on the third variable $\bar{q}$ has dropped out. As explained in [2] this is really only true when $k \lesssim \bar{q}$. Thus strictly speaking these last equations are not valid when $k$ is larger than $\bar{q}$.

# 3 CCFM with Absorptive Boundary

Here we shall not discuss the theoretical motivations for using the absorptive boundary method in studying the non-linear effects upon the evolution (for this we refer again to [2]) but we shall rather simply explain how it is implemented.

We first define the logarithmic variables $\rho = \ln k^2/k_0^2$ and $Y = \ln 1/x$. Introduce a line of constant gluon occupancy $\rho = \rho_c(Y)$ via the condition

$$\mathcal{A}(Y, \rho = \rho_c(Y)) = c, \tag{8}$$

where the number $c$ is smaller than one, but not *much* smaller. The saturation line $\rho_s(Y)$ would correspond to $c \sim 1$, so clearly $\rho_s(Y)$ is smaller than $\rho_c(Y)$, but relatively close to it. For $\rho < \rho_c(Y)$ and sufficiently high energy, the solution $\mathcal{A}(Y, \rho)$ would become arbitrarily large. If this equation is to be solved numerically, one may think about enforcing saturation in the following way: at each step in $Y$, one first identifies the corresponding point $\rho_c(Y)$ from the condition (8), and then one requires $\mathcal{A}(Y, \rho)$ to remain finite and of $\mathcal{O}(1)$ for any $\rho$ sufficiently far below $\rho_c(Y)$ — say, for $\rho \leq \rho_c(Y) - \Delta$ with $\Delta \simeq \ln(1/c)$. When decreasing $\rho$ below $\rho_c(Y)$, the solution $\mathcal{A}(Y, \rho)$ will typically start by rising, then reach a maximum and eventually decrease to the finite value determined by hand. The saturation scale $\rho_s(Y)$ can then be conventionally identified with the position of this maximum. In this procedure, the numbers $c$ and $\Delta$ are to be viewed as free parameters, which are however correlated with each other as $\Delta \sim \ln(1/c)$. In this way the energy dependence of the saturation scale follows from the dynamics, and is thus not fitted by hand, and is guaranteed to be consistent with evolution.

In Figure 3 we plot the respective predictions of (5) and BFKL for both fixed and running coupling. To make the comparison between the solutions easier, we have scaled the solutions to BFKL so that the solutions more or less coincide at $Y = 10$. For fixed coupling, the saturation front generated by (5) is seen to progress significantly faster than the one generated by BFKL. For the running coupling case on the other hand, it is clear that the respective fronts progress with similar speeds. Thus in this case the two equations have similar predictions for the energy dependence of the saturation momentum, so that the entire difference between them can (almost) be removed by either a rescaling of the saturation momentum $Q_s$, or as in the figure, a rescaling of $\mathcal{A}$.

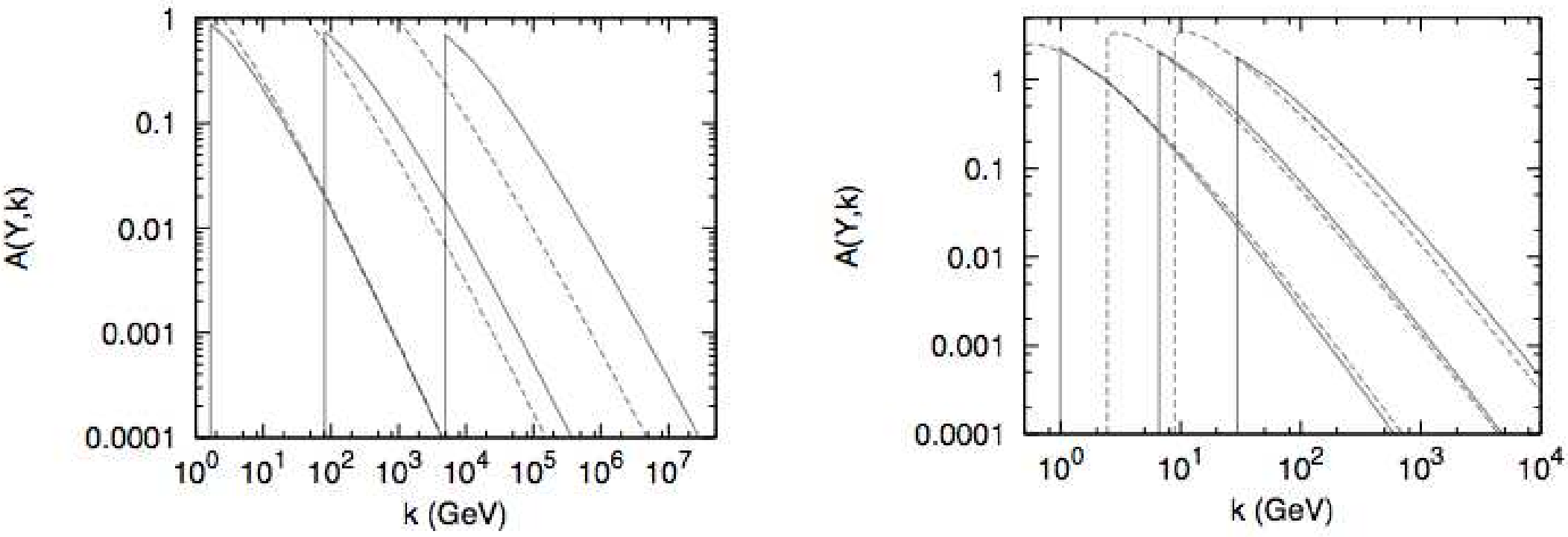

Figure 3: *The solutions to equations* (5) *(solid lines) and BFKL (dashed lines) including the absorptive boundary, for* $Y = 10, 20, 30$. *The solutions to BFKL includes the kinematical constraint and have been scaled so that the solutions match at* $Y = 10$. *Left: fixed coupling* $\alpha_s = 0.2$. *Right: running coupling.*

In Figure 4 (left) we show the respective results for the running coupling case for much higher values of $Y$. Also, in the right figure there, we compare the absorptive boundary results for (5) and (7). One clearly sees now that all the equations under consideration give approximately the same speed for the saturation front. Note that in the left figure the solutions to BFKL have been scaled to make the comparison easier. From the right figure we see that (5) and (7) roughly have the same shape and energy dependence when saturation and a running coupling is introduced. This is big contrast to the linear case (fixed or running coupling) and also to the non-linear case with a fixed coupling, where the differences between the solutions is much larger and grows with energy. Thus it seems that the inclusion of saturation and a running coupling removes much of the ambiguity inherent in the derivation of the different versions of CCFM.

From our results in [1] and [2] we also see that the solutions to the gluon distribution with saturation differs from the linear solutions quite substantially even for moderate values of $Y$ reachable at the LHC. This, however, does not necessarily imply that the effects of the absorptive boundary will be clearly visible in phenomenological applications. First one should remember that there are many more effects in such applications which slow down the growth of the gluon distribution with $Y$, most notably energy-momentum conservation which gives large corrections. Secondly, even in the application of linear equations, such as CCFM in the CASCADE event generator, there is included an absorptive boundary in $k$. This is however not a boundary in the sense of the one presented here, which has its motivation in the perturbative dynamics of non-linear evolution, but it is rather a soft cut which is fixed for all energies. It is therefore more of a non-perturbative cut. However in the limited $Y$ interval covering phenomenologically relevant energies, it may be difficult to distinguish the effects of the different boundaries.

In addition to these issues, we must say that there are further issues with CCFM not very well understood. One is the fact that non-leading modifications seem to have huge effects. This makes the whole evolution questionable as one can switch different components, within the approximations of the formalism, and get very different results. One can for example get results where the growth of the gluon density sets in very late, at $x$ values below $10^{-5}$ [14]. A notable difficulty of CCFM is also that it appears to fail in the description of forward jets, precisely due to the non-leading terms which are otherwise needed to fit the structure function. We believe that it may be premature to try seriously study CCFM with saturation for phenomenological studies before these issues are resolved. It may on the other hand be that saturation removes many of the ambiguities of the formalism, some of which we have already seen. Together with this we believe that the next step in our analysis is to study equations (3) or (4) with the absorptive boundary and compare those results to see whether the evolution is different or not. Notice that an interesting and immediate consequence of such a study will be that the saturation momentum $Q_s$ will now depend not only on $x$ but also on the momentum scale $\bar{q}$ (which in DIS is roughly $Q^2$). This is again due to the coherence and implies that the scale when the gluon fields in the proton saturates depends not only intrinsically on the proton but also on the probe, certainly an interesting result.

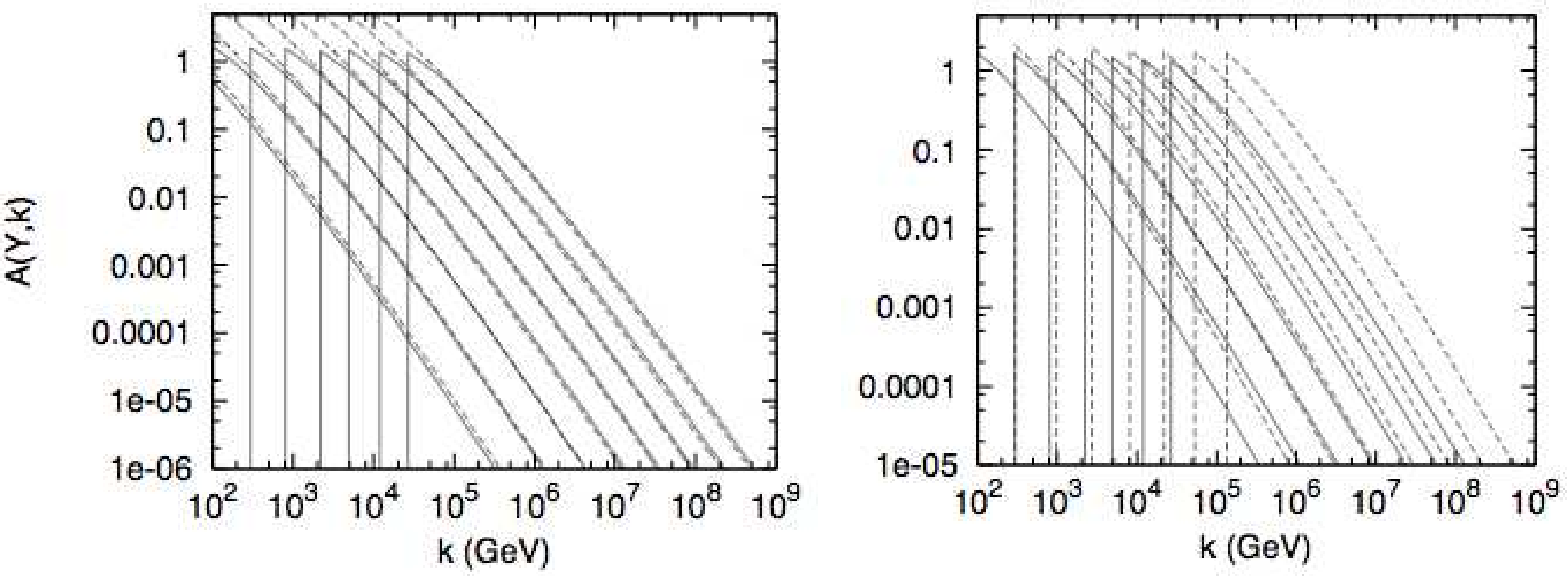

Figure 4: *Solutions to* (5), (7) *and BFKL including the saturation boundary for a running coupling. Left:* (5) *(dashed lines) and BFKL (solid lines). The solution to BFKL has been scaled by a factor 6.5. $Y$ between 50 and 140 units. Right:* (5) *(solid lines) and* (7) *(dashed lines). $Y$ between 40 and 100 units.*

# Acknowledgements

The work presented here was performed while I was a postdoc at CEA/Saclay, I would like to thank E. Iancu for this collaboration and also for all the help during my time in Saclay. I would also like to thank G. Gustafson for useful discussions. I am also grateful to H. Jung for many e-mail exchanges and also many discussions.

# References

[1] Emil Avsar and Edmond Iancu. BFKL and CCFM evolutions with saturation boundary. *Phys. Lett.*, B673:24–29, 2009.

[2] Emil Avsar and Edmond Iancu. CCFM Evolution with Unitarity Corrections. *Nucl. Phys.*, A829:31–75, 2009.

[3] L.N. Lipatov, *Sov. J. Nucl. Phys.* **23** (1976) 338;
E.A. Kuraev, L.N. Lipatov and V.S. Fadin, *Zh. Eksp. Teor. Fiz* **72**, 3 (1977) (*Sov. Phys. JETP* **45** (1977) 199);
Ya.Ya. Balitsky and L.N. Lipatov, *Sov. J. Nucl. Phys.* **28** (1978) 822.

[4] Catani, S. and Fiorani, F. and Marchesini, G. , *Nucl. Phys.* **B336** (1990) 18;
Ciafaloni, M. *Nucl. Phys.* **B296** (1988) 49;
Catani, S. and Fiorani, F. and Marchesini, G. , *Phys. Lett.* **B234** (1990) 339.

[5] Fabrizio Caola, Stefano Forte, and Juan Rojo. Deviations from NLO QCD evolution in inclusive HERA data. 2009.

[6] Edmond Iancu, Kazunori Itakura, and Larry McLerran. Geometric scaling above the saturation scale. *Nucl. Phys.*, A708:327–352, 2002.

[7] A. H. Mueller and D. N. Triantafyllopoulos. The energy dependence of the saturation momentum. *Nucl. Phys.*, B640:331–350, 2002.

[8] I. Balitsky. Operator expansion for high-energy scattering. *Nucl. Phys.*, B463:99–160, 1996.

[9] Yuri V. Kovchegov. Small-x F2 structure function of a nucleus including multiple pomeron exchanges. *Phys. Rev.*, D60:034008, 1999.

[10] H. Jung and G. P. Salam. Hadronic final state predictions from CCFM: The hadron- level Monte Carlo generator CASCADE. *Eur. Phys. J.*, C19:351–360, 2001.

[11] Hamid Kharraziha and Leif Lönnblad. The linked dipole chain Monte Carlo. *JHEP*, 03:006, 1998.

[12] Bo Andersson, G. Gustafson, and J. Samuelsson. The linked dipole chain model for dis. *Nucl. Phys.*, B467:443–478, 1996.

[13] S. Catani, F. Fiorani, and G. Marchesini. Small x Behavior of Initial State Radiation in Perturbative QCD. *Nucl. Phys.*, B336:18, 1990.

[14] G. Bottazzi, G. Marchesini, G. P. Salam, and M. Scorletti. Small-x one-particle-inclusive quantities in the CCFM approach. *JHEP*, 12:011, 1998.

# BFKL Catch Up!

*Florian Schwennsen*[1]

[1]DESY, Notketraße 85, 22607 Hamburg, Germany

I give a brief overview over the most recent progress in the field of standard dilute BFKL.

## 1 Introduction

Although there is hardly any doubt that Quantum Chromodynamics (QCD) is the proper theory to describe strong interactions, it is still far away from being completely solved. Even at large scales where the coupling constant becomes small, a perturbative calculation at a fixed order is not the end of the story. Whenever one considers kinematic regions characterized by two different large scales, logarithms of the ratio of these two scales appear at each order of perturbation theory and – being large – compensate the smallness of the coupling. A resummation of logarithms of the type $[\alpha_s \ln s]^n$ is realized by the leading logarithmic (LL) Balitsky-Fadin-Kuraev-Lipatov (BFKL) [1, 2, 3, 4] equation. It took almost twenty years to extend it to next-to-leading logarithmic accuracy (NLL) by resumming terms of the type $\alpha_s[\alpha_s \ln s]^n$ as well [5, 6].

Both the LL and the NLL BFKL equation are still being studied with great eagerness to improve the accuracy of theoretical predictions for concrete observables as well as to increase our fundamental knowledge about quantum field theory itself. The aim of my contribution to the 13th International Conference on Elastic and Diffractive Scattering is to give an overview over the most recent progresses. An exhaustive discussion of them at this place is not possible. That's why I strongly encourage the reader to have a look at the original works instead and to consider this short article just as an appetizer.

## 2 Recent Developments

The "gold-plated" process to test BFKL dynamics is the scattering of virtual photons. The virtualities of the photons can be tuned to be of equal, or at least similar, scale such that DGLAP like evolution in the transverse momenta is suppressed offering a clean environment described by the BFKL equation. In Ref. [7] an already existing NLL BFKL calculation of the production of two vector mesons [8, 9] has been collinearly improved leading to more 'sensible' energy scales. The same team has studied the total hadronic cross section of virtual photon scattering in the NLL approach [10]. With the inclusion of the LO quark box contribution an agreement with OPAL data [11] has been obtained.

Another very famous candidate for BFKL physics are forward jets in hadron collisions. In Ref. [12] such jets with a large rapidity gap in between have been investigated and the authors obtain a good agreement with Tevatron data [13, 14]. NLL BFKL calculations for Mueller-Navelet jets [15] exist already a bit longer [16, 17] and suffer from the same handicap as all previously mentioned NLL BFKL calculations: the impact factors are taken into account only at LO accuracy. However, a full NLL BFKL calculation of Mueller-Navelet jets is under way [18].

Phenomenologically very interesting is the revival of the 'discrete Pomeron' which recently has been studied [19] to describe HERA data (see also the according contribution at this conference). A phenomenological topic which became extremely popular due to the advent of the LHC is the central in- or exclusive production of Higgs particles or other objects with the appropriate quantum numbers [20, 21, 22, 23, 24, 25, 26, 27] (since at this conference a whole session was dedicated to this subject, I refer to the according contributions as well).

Also on the more theoretical side impressing progress has been made. In Ref. [28] the NLO evolution of colour dipoles has been calculated. The linear version of this evolution should reproduce the BFKL equation. While at LL this is easily verified, at NLL there were some discrepancies to the calculation of the NLL BFKL kernel in the Möbis representation [29, 30]. It was just recently [31] that the situation could be clarified. Still under investigation is the issue of conformal invariance. This feature of the LL kernel is obviously spoilt by the running of the coupling at NLL. However the approach of Ref. [28] additionally induces artificial conformal variant terms due to their regularization procedure and also the approach of Ref. [29, 30] produces conformal variant terms which are not caused by the running of the coupling. Most probably they are connected to scale ambiguities.

As another consequence of the running coupling, the LO eigen-functions $(k^2)^\gamma$ of the LL kernel are eigen-functions of the NLL kernel only if one accepts the eigen-value to become an operator. This problem has been addressed in two recent works. One idea is to shift $\gamma$ by $\alpha_s\beta_0/(4\pi)$ such that formally the eigen-equation is only violated beyond the accuracy one is working at [32]. Another idea is to numerically solve the eigen-equation [33]. Unfortunately none of these three ways is completely satisfactory.

I would also like to draw the reader's attention to two not quite pigeonholable but nevertheless very interesting publications. In Ref. [34, 35] a remarkable relation between soft gluons away from jets in electron positron annihilations and the BFKL equation in coordinate space has been identified and studied. In Ref. [36] it is shown how the correct treatment of kinematics can improve the LO colour dipole evolution in colour dipole perturbation theory, and that as a natural consequence large dipoles are suppressed.

Related to the BFKL equation is the BKP equation [37, 38] for the evolution of three instead of two gluons. It's solution is called the Odderon since it is $C$-odd while the Pomeron is $C$-even. Two solutions to the LL BKP equation are known: the so called JW-Odderon [39] with an intercept smaller than 1, and the BLV-Odderon [40] with an intercept identical to 1. Now, in Ref. [41] it has been shown that the intercept of the BLV-Odderon remains to be 1 even if the running of the coupling is taken into account, and in Ref. [42] the intercept has been shown to remain stable against complete higher order corrections. The Odderon has also been investigated in the context if gauge/string duality where the counterpart of the Odderon can be identified with the Kalb-Ramond field in Type IIB string theory [43, 44]. It should also be mentioned that there are new proposals to search for the perturbative Odderon in experiments [45, 46].

At the moment, supersymmetry is one of the most actively studied issues in high energy physics. The $\mathcal{N} = 4$ case is especially interesting because the complete vanishing of the $\beta$-function to all orders makes the theory conformal invariant. Moreover it is supposed to be related to string theory by AdS/CFT correspondence. In $\mathcal{N} = 4$ supersymmetry the asymp-totic Bethe ansatz [47] is accurate to three-loop order. The correct result for the anomalous dimension has to meet the BFKL pole structure which – due to the resummation – is in some sense an all order result. This interplay has been studied in detail in Ref. [48].

It is known that the all order BDS ansatz [49] for maximally helicity violating scattering amplitudes in $\mathcal{N} = 4$ supersymmetry is wrong for more than five external legs [50, 51]. In Refs. [52, 53, 54] the origin of this mismatch has been further investigated and could *e.g.* be traced to a class of diagram whose cut contribution is missed. Many more objects in the BFKL world are now on the way to be reinvestigated in the $\mathcal{N} = 4$ supersymmetric case, like the $R$-current impact factors [55, 56] – which can be considered to be the counterpart of the virtual photon impact factor – or like the triple Pomeron vertex [57].

# 3 Outlook

This is just a very packed presentation of the latest news in BFKL physics. Since (fortunately) there are no strict borders between the different subfields of physics, I might have missed some works which one could as well include in such a collection. Therefore, I apologize for not being literally complete. Moreover, I apologize for not summarizing because I am sure that the chapter of BFKL physics is not yet completed.

# References

[1] Victor S. Fadin, E. A. Kuraev, and L. N. Lipatov. On the Pomeranchuk singularity in asymptotically free theories. *Phys. Lett.*, B60:50–52, 1975.

[2] E. A. Kuraev, L. N. Lipatov, and Victor S. Fadin. Multi - Reggeon processes in the Yang-Mills theory. *Sov. Phys. JETP*, 44:443–450, 1976.

[3] E. A. Kuraev, L. N. Lipatov, and Victor S. Fadin. The Pomeranchuk singularity in nonabelian gauge theories. *Sov. Phys. JETP*, 45:199–204, 1977.

[4] I. I. Balitsky and L. N. Lipatov. The Pomeranchuk singularity in quantum Chromodynamics. *Sov. J. Nucl. Phys.*, 28:822–829, 1978.

[5] Victor S. Fadin and L. N. Lipatov. BFKL Pomeron in the next-to-leading approximation. *Phys. Lett.*, B429:127–134, 1998, hep-ph/9802290.

[6] Marcello Ciafaloni and Gianni Camici. Energy scale(s) and next-to-leading BFKL equation. *Phys. Lett.*, B430:349–354, 1998, hep-ph/9803389.

[7] F. Caporale, A. Papa, and A. Sabio Vera. Collinear improvement of the BFKL kernel in the electroproduction of two light vector mesons. *Eur. Phys. J.*, C53:525–532, 2008, 0707.4100.

[8] D. Yu. Ivanov and A. Papa. Electroproduction of two light vector mesons in the next-to-leading approximation. *Nucl. Phys.*, B732:183–199, 2006, hep-ph/0508162.

[9] D. Yu. Ivanov and A. Papa. Electroproduction of two light vector mesons in next-to- leading BFKL: Study of systematic effects. 2006, hep-ph/0610042.

[10] Francesco Caporale, Dmitry Yu. Ivanov, and Alessandro Papa. BFKL resummation effects in the $\gamma^*\gamma^*$ total hadronic cross section. *Eur. Phys. J.*, C58:1–7, 2008, 0807.3231.

[11] G. Abbiendi et al. Measurement of the hadronic cross-section for the scattering of two virtual photons at LEP. *Eur. Phys. J.*, C24:17–31, 2002, hep-ex/0110006.

[12] F. Chevallier, O. Kepka, C. Marquet, and C. Royon. Gaps between jets at hadron colliders in the next-to-leading BFKL framework. *Phys. Rev.*, D79:094019, 2009, 0903.4598.

[13] B. Abbott et al. Probing hard color-singlet exchange in $p\bar{p}$ collisions at $\sqrt{s} = 630$ GeV and 1800 GeV. *Phys. Lett.*, B440:189–202, 1998, hep-ex/9809016.

[14] F. Abe et al. Dijet production by color-singlet exchange at the Fermilab Tevatron. *Phys. Rev. Lett.*, 80:1156–1161, 1998.

[15] Alfred H. Mueller and H. Navelet. An inclusive minijet cross-section and the bare Pomeron in QCD. *Nucl. Phys.*, B282:727, 1987.

[16] Agustín Sabio Vera and Florian Schwennsen. The azimuthal decorrelation of jets widely separated in rapidity as a test of the BFKL kernel. *Nucl. Phys.*, B776:170–186, 2007, hep-ph/0702158.

[17] C. Marquet and C. Royon. Azimuthal decorrelation of Mueller-Navelet jets at the Tevatron and the LHC. *Phys. Rev.*, D79:034028, 2009, 0704.3409.

[18] D. Colferai, F. Schwennsen, L Szymanowski, and S. Wallon. to be published.

[19] J. Ellis, H. Kowalski, and D. A. Ross. Evidence for the Discrete Asymptotically-Free BFKL Pomeron from HERA Data. 2008, arXiv:0803.0258 [hep-ph].

[20] R. S. Pasechnik, A. Szczurek, and O. V. Teryaev. Central exclusive production of scalar $\chi_c$ meson at the Tevatron, RHIC and LHC energies. *Phys. Rev.*, D78:014007, 2008, 0709.0857.

[21] Robi Peschanski, M. Rangel, and C. Royon. Hybrid Pomeron Model of exclusive central diffractive production. 2008, 0808.1691.

[22] J. R. Cudell, A. Dechambre, O. F. Hernandez, and I. P. Ivanov. Central exclusive production of dijets at hadronic colliders. *Eur. Phys. J.*, C61:369–390, 2009, 0807.0600.

[23] S. Heinemeyer et al. Central Exclusive Diffractive MSSM Higgs-Boson Production at the LHC. *J. Phys. Conf. Ser.*, 110:072016, 2008, 0801.1974.

[24] V. A. Khoze, A. D. Martin, and M. G. Ryskin. Early LHC measurements to check predictions for central exclusive production. *Eur. Phys. J.*, C55:363–375, 2008, 0802.0177.

[25] Jeppe R. Andersen, Vittorio Del Duca, and Chris D. White. Higgs Boson Production in Association with Multiple Hard Jets. *JHEP*, 02:015, 2009, 0808.3696.

[26] M. Chaichian, P. Hoyer, K. Huitu, V. A. Khoze, and A. D. Pilkington. Searching for the triplet Higgs sector via central exclusive production at the LHC. 2009, 0901.3746.

[27] G. Chachamis, M. Hentschinski, A. Sabio Vera, and C. Salas. Exclusive central production of heavy quarks at the LHC. 2009, 0911.2662.

[28] Ian Balitsky and Giovanni A. Chirilli. Next-to-leading order evolution of color dipoles. *Phys. Rev.*, D77:014019, 2008, 0710.4330.

[29] V. S. Fadin, R. Fiore, A. V. Grabovsky, and A. Papa. The dipole form of the gluon part of the BFKL kernel. *Nucl. Phys.*, B784:49–71, 2007, 0705.1885.

[30] Victor S. Fadin, Roberto Fiore, and Alessandro Papa. The dipole form of the quark part of the BFKL kernel. *Phys. Lett.*, B647:179–184, 2007, hep-ph/0701075.

[31] V. S. Fadin, R. Fiore, and A. V. Grabovsky. On the discrepancy of the low-$x$ evolution kernels. *Nucl. Phys.*, B820:334–363, 2009, 0904.0702.

[32] Sergey Bondarenko. Conformal intercept of BFKL pomeron with NLO running coupling constant corrections. 2008, 0808.3175.

[33] D. A. Ross. Perturbative Estimates of the Eigenfunctions of the Non- forward BFKL Kernel. *Phys. Lett.*, B668:233–237, 2008, 0805.1004.

[34] Emil Avsar and Yoshitaka Hatta. Quantitative study of the transverse correlation of soft gluons in high energy QCD. *JHEP*, 09:102, 2008, 0805.0710.

[35] Emil Avsar, Yoshitaka Hatta, and Toshihiro Matsuo. Soft gluons away from jets: distribution and correlation. *JHEP*, 06:011, 2009, 0903.4285.

[36] Leszek Motyka and Anna M. Staśto. Exact kinematics in the small x evolution of the color dipole and gluon cascade. *Phys. Rev.*, D79:085016, 2009, 0901.4949.

[37] Jochen Bartels. High-energy behavior in a nonabelian gauge theory. 2. first corrections to T($n \to m$) beyond the leading ln s approximation. *Nucl. Phys.*, B175:365, 1980.

[38] J. Kwiecinski and M. Praszalowicz. Three Gluon Integral Equation and Odd $C$ Singlet Regge Singularities in QCD. *Phys. Lett.*, B94:413, 1980.

[39] R. A. Janik and J. Wosiek. Solution of the odderon problem. *Phys. Rev. Lett.*, 82:1092–1095, 1999, hep-th/9802100.

[40] Jochen Bartels, L. N. Lipatov, and G. P. Vacca. A New Odderon Solution in Perturbative QCD. *Phys. Lett.*, B477:178–186, 2000, hep-ph/9912423.

[41] M. A. Braun. Odderon with a running coupling constant. *Eur. Phys. J.*, C53:59–63, 2008, 0707.2314.

[42] Anna M. Staśto. Small $x$ resummation and the Odderon. *Phys. Lett.*, B679:288–292, 2009, 0904.4124.

[43] Richard C. Brower, Marko Djuric, and Chung-I Tan. The Kalb-Ramond Odderon in AdS/CFT. 2008, 0812.0354.

[44] Richard C. Brower, Marko Djuric, and Chung-I. Tan. Odderon in gauge/string duality. *JHEP*, 07:063, 2009.

[45] A. Bzdak, L. Motyka, L. Szymanowski, and J. R. Cudell. Exclusive J/$\psi$ and $\Upsilon$ hadroproduction and the QCD odderon. *Phys. Rev.*, D75:094023, 2007, hep-ph/0702134.

[46] B. Pire, F. Schwennsen, L. Szymanowski, and S. Wallon. Hard Pomeron-Odderon interference effects in the production of $\pi^+\pi^-$- pairs in high energy gamma-gamma collisions at the LHC. *Phys. Rev.*, D78:094009, 2008, 0810.3817.

[47] Niklas Beisert and Matthias Staudacher. Long-range PSU(2,2—4) Bethe ansaetze for gauge theory and strings. *Nucl. Phys.*, B727:1–62, 2005, hep-th/0504190.

[48] A. V. Kotikov, L. N. Lipatov, A. Rej, M. Staudacher, and V. N. Velizhanin. Dressing and Wrapping. *J. Stat. Mech.*, 0710:P10003, 2007, 0704.3586.

[49] Zvi Bern, Lance J. Dixon, and Vladimir A. Smirnov. Iteration of planar amplitudes in maximally supersymmetric Yang-Mills theory at three loops and beyond. *Phys. Rev.*, D72:085001, 2005, hep-th/0505205.

[50] Luis F. Alday and Juan Martin Maldacena. Gluon scattering amplitudes at strong coupling. *JHEP*, 06:064, 2007, 0705.0303.

[51] J. M. Drummond, J. Henn, G. P. Korchemsky, and E. Sokatchev. The hexagon Wilson loop and the BDS ansatz for the six- gluon amplitude. *Phys. Lett.*, B662:456–460, 2008, 0712.4138.

[52] J. Bartels, L. N. Lipatov, and Agustin Sabio Vera. BFKL Pomeron, Reggeized gluons and Bern-Dixon-Smirnov amplitudes. *Phys. Rev.*, D80:045002, 2009, 0802.2065.

[53] V. Del Duca, C. Duhr, and E. W. N. Glover. Iterated amplitudes in the high-energy limit. *JHEP*, 12:097, 2008, 0809.1822.

[54] Richard C. Brower, Horatiu Nastase, Howard J. Schnitzer, and Chung-I Tan. Implications of multi-Regge limits for the Bern-Dixon- Smirnov conjecture. *Nucl. Phys.*, B814:293–326, 2009, 0801.3891.

[55] J. Bartels, A. M. Mischler, and M. Salvadore. Four point function of R-currents in N=4 SYM in the Regge limit at weak coupling. *Phys. Rev.*, D78:016004, 2008, 0803.1423.

[56] J. Bartels, J. Kotanski, A. M. Mischler, and V. Schomerus. Regge limit of R-current correlators in AdS Supergravity. 2009, 0908.2301.

[57] Jochen Bartels, Martin Hentschinski, and Anna-Maria Mischler. The topology of the triple Pomeron vertex in N=4 SYM. *Phys. Lett.*, B679:460–466, 2009, 0906.3640.

# Saturation in Nuclei

*T. Lappi*

Department of Physics, P.O. Box 35, 40014 University of Jyväskylä, Finland and
Helsinki Institute of Physics, P.O. Box 64, 00014 University of Helsinki, Finland

This talk discusses some recent studies of gluon saturation in nuclei. We stress the connection between the initial condition in heavy ion collisions and observables in deep inelastic scattering (DIS). The dominant degree of freedom in the small $x$ nuclear wave-function is a nonperturbatively strong classical gluon field, which determines the initial condition for the glasma fields in the initial stages of a heavy ion collision. A correlator of Wilson lines from the same classical fields, known as the dipole cross section, can be used to compute many inclusive and exclusive observables in DIS.

## 1 Connection between Small x DIS and HIC: Wilson Line

The initial condition in a heavy ion collision (HIC) is determined by the wave-functions of the two colliding nuclei, parameterised by $Q^2$ and $x$. As in any hadronic collision, the typical magnitudes of these parameters can be estimated as $Q^2 \sim \langle p_\perp \rangle^2$ and $x \sim \langle p_\perp \rangle / \sqrt{s}$, where $\langle p_\perp \rangle$ is the typical transverse momentum of the particles being produced, and $\sqrt{s}$ is the collision energy. At relativistic energies, such as at RHIC and LHC, this means that the relevant domain for bulk particle production is at very small $x$. Gluon bremsstrahlung processes lead to an exponentially (in rapidity $y = \ln 1/x$) growing cascade of gluons in the wave-function. The number of gluons in the wave-function grows as $\mathrm{d}N/\mathrm{d}y \sim x^{-\lambda}$, where the phenomenologically observed value is $\lambda \sim 0.2 \ldots 0.3$. When the number of gluons grows large enough, eventually their phase space density becomes large, with occupation numbers $\sim 1/\alpha_\mathrm{s}$; in terms of the field strength this meas $A_\mu \sim 1/g$. At this point the nonlinear terms in the QCD Lagrangian (think of the two terms in the covariant derivative $\partial_\mu + igA_\mu$) become of the same order as the linear ones, and the dynamics becomes nonperturbative. Due to the nonlinear interactions the gluon number cannot grow indefinitely, but it must *saturate* at some $1/\alpha_\mathrm{s}$ for gluons with $p_\perp \lesssim Q_\mathrm{s}$, where $Q_\mathrm{s}$ is the *saturation scale*. When $Q_\mathrm{s}^2 \sim x^{-\lambda}$ becomes large enough, $\alpha_\mathrm{s}(Q_\mathrm{s}) \ll 1$ and the dynamics of these fields is *classical*. This situation is most conveniently described using the effective theory known as the Colour Glass Condensate (CGC, [1]), where the large $x$ degrees of freedom are described as a classical colour current $J^\mu$ and the small $x$ gluons as classical fields that this current radiates: $[D_\mu, F^{\mu\nu}] = J^\nu$.

Let us consider a hadron or a nucleus moving in the $+z$-direction. Its colour current in the CGC formalism has only one large component, the one in the +-direction (recall that $x^\pm = (t \pm z)/\sqrt{2}$). For a nucleus moving at high energy we can take the current to be independent of the light cone time $x^-$ as $J^+ = \rho(\mathbf{x}_T, x^-)$ with a very narrow, $\delta$-function-like support in $x^-$: $\rho(\mathbf{x}_T, x^-) \sim \delta(x^-)\rho(\mathbf{x}_T)$. A simple solution for the equations of motion can be found as $A^+ = \rho(\mathbf{x}_T, x^-)/\boldsymbol{\nabla}_T^2$; this is known as the covariant gauge solution. In order to have a physical partonic picture of the gluonic degrees of freedom it is necessary to gauge transform this solution to the light cone gauge $A^+ = 0$. The gauge transformation that achieves this is done with the *Wilson line* constructed from this gauge field:

$$U(\mathbf{x}_T) = \mathbb{P}\exp\left[i\int \mathrm{d}x^- A^+\right]. \tag{1}$$

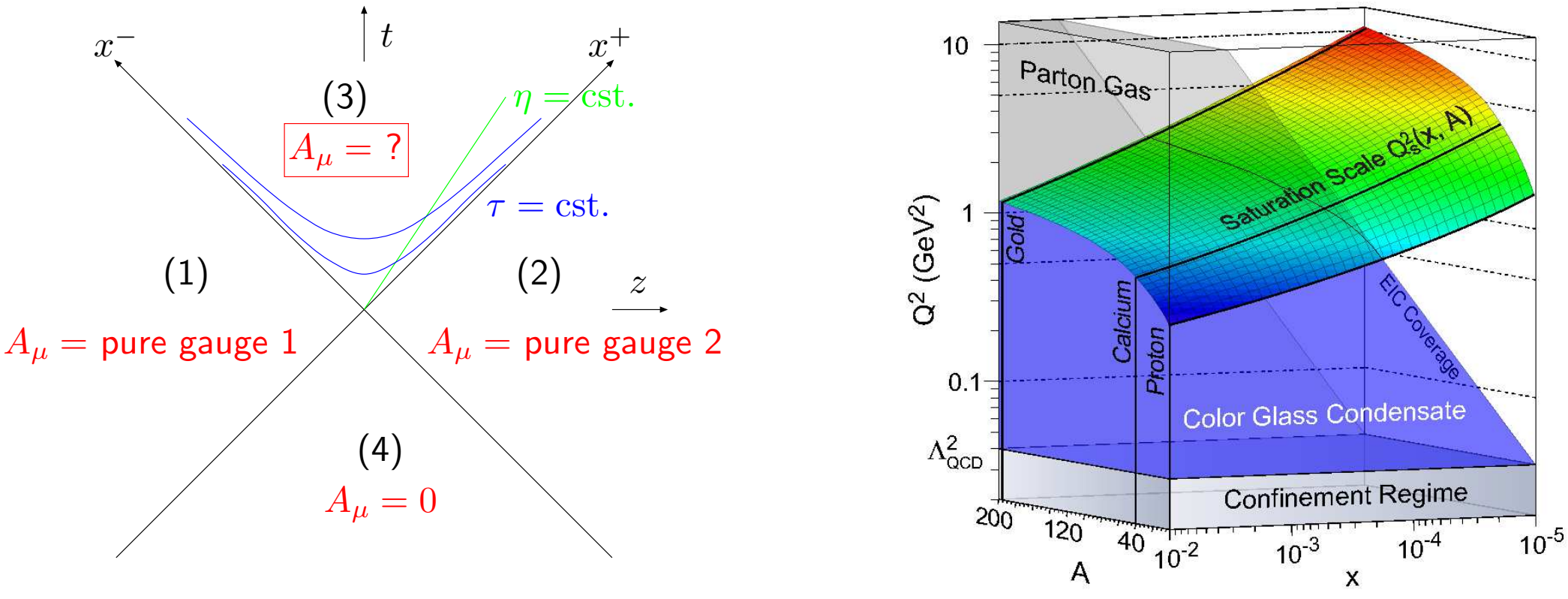

Figure 1: Left: Space-time structure of the field. In regions (1) and (2) there is a transverse pure gauge field (2) with no energy density. In region (3) after the collision there is the glasma field. Right:

This results in a field with only transverse components:

$$A^i_{(1,2)} = \frac{i}{g}U_{(1,2)}(\mathbf{x}_T)\partial_i U^\dagger_{(1,2)}(\mathbf{x}_T) \tag{2}$$

for both of the colliding nuclei $(1,2)$ separately. The initial condition for the "glasma" [2] fields at $\tau > 0$ is given in terms of these pure gauges [3].

$$\begin{aligned} A^i\big|_{\tau=0} &= A^i_{(1)} + A^i_{(2)} \\ A^\eta\big|_{\tau=0} &= \frac{ig}{2}[A^i_{(1)}, A^i_{(2)}] \end{aligned} \tag{3}$$

Inside the future light cone $\tau > 0$ the field equations must be solved either numerically or in some approximation scheme. The space-time structure described here is illustrated in Fig. 1 (left). In the rest of this talk we shall be referring to the numerical "CYM" (Classical Yang-Mills) computations [4].

To see the connection to DIS it is convenient to consider the process in a Lorentz frame where the virtual photon has a large longitudinal momentum. In the target rest frame (or more properly the "dipole frame" [5] that does not leave all the high energy evolution in the probe) the timescales of the quantum fluctuations of the virtual photon are extremely slow. In order to interact with a hadronic target it must therefore split into a quark-antiquark pair already long before the scattering. This $q\bar{q}$-dipole then interacts with the hadronic target with a scattering amplitude whose imaginary part is known as the "dipole cross section". As typical hadronic scattering amplitudes at high energy, that of the dipole is almost purely imaginary, and we shall here neglect the real part. The dipole cross section can be obtained from the quark propagator in the gluonic background field of the target, which is quite naturally given by the same Wilson line (1) [6]. The dipole cross section (which, in general, is a function of the size of the dipole $\mathbf{r}_T$, the impact parameter $\mathbf{b}_T$ and $x$) is the correlator of two Wilson lines

$$\hat{\sigma}(\mathbf{r}_T) = \int \mathrm{d}^2\mathbf{b}_T \frac{1}{N_\mathrm{c}} \left\langle 1 - U^\dagger\left(\mathbf{b}_T + \frac{\mathbf{r}_T}{2}\right) U\left(\mathbf{b}_T - \frac{\mathbf{r}_T}{2}\right)\right\rangle . \tag{4}$$

For example, the total virtual photon cross section can be obtained by convoluting the dipole cross section with the virtual photon wave-function which relates the $Q^2$ of the photon to the size of the dipole $r \sim 1/Q$:

$$\sigma^{\gamma^* p}_{L,T} = \int \mathrm{d}^2\mathbf{b}_T \int \mathrm{d}^2\mathbf{r}_T \int \mathrm{d}z \left|\Psi^\gamma_{L,T}(Q^2,\mathbf{r}_T,z)\right|^2 \sigma_\mathrm{dip}(x,\mathbf{r}_T,\mathbf{b}_T). \tag{5}$$

Fourier-transforming instead of simply integrating over the impact parameter dependence gives access to the momentum transfer to the target in diffractive scattering. The inclusive diffractive virtual photon cross section (really the elastic dipole-photon cross section) is proportional to the square of the dipole cross section

$$\frac{\sigma^{D,tot}_{L,T}}{\mathrm{d}t} = \frac{1}{16\pi} \int \mathrm{d}^2\mathbf{r}_T \int \mathrm{d}z \left|\Psi^\gamma_{L,T}(Q^2,\mathbf{r}_T,z)\right|^2 \sigma^2_\mathrm{dip}(x,\mathbf{r}_T,\mathbf{\Delta}) \tag{6}$$

and diffractive vector meson production can be obtained by projecting on the virtual photon wave-function

$$\frac{\sigma^{D,V}_{L,T}}{\mathrm{d}t} = \frac{1}{16\pi} \left|\int \mathrm{d}^2\mathbf{r}_T \int \mathrm{d}z \left(\Psi^\gamma\Psi^{*V}\right)_{L,T} \sigma_\mathrm{dip}(x,\mathbf{r}_T,\mathbf{\Delta})\right|^2 . \tag{7}$$

The exclusive cross sections are proportional to the dipole cross section (the scattering amplitude) squared, whereas the inclusive one depends on it linearly; this is due to the optical theorem and our approximation that the scattering amplitude is purely imaginary. We shall now go on to discuss some recent applications of saturation ideas to heavy ion collisions and DIS phenomenology, trying to stress the unity of the approach between the two.

## 2 Gluon Multiplicity at RHIC and LHC AA Collisions

Ideally one would like to measure the value of $Q_\mathrm{s}$ in DIS experiments and use the resulting value as an independent input in calculations of the initial state of heavy ion collisions. In practise most of the exiting CYM computations of the glasma fields have been performed in the MV model [9] in terms of the colour charge density parameter $g^2\mu$ that parameterises the fluctuations of the classical colour currents $J^\pm$. One must therefore relate the values of $g^2\mu$ and $Q_\mathrm{s}$ in a consistent way. In practise this can be done by computing the Wilson line correlator in the MV model, using exactly the same numerical implementation of the model as in the CYM calculations, and extracting the correlation length $1/Q_\mathrm{s}$ [10].

The other ingredient necessary in using the existing DIS data to calculate initial conditions for heavy ion collisions is the correct implementation of the nuclear geometry in extending

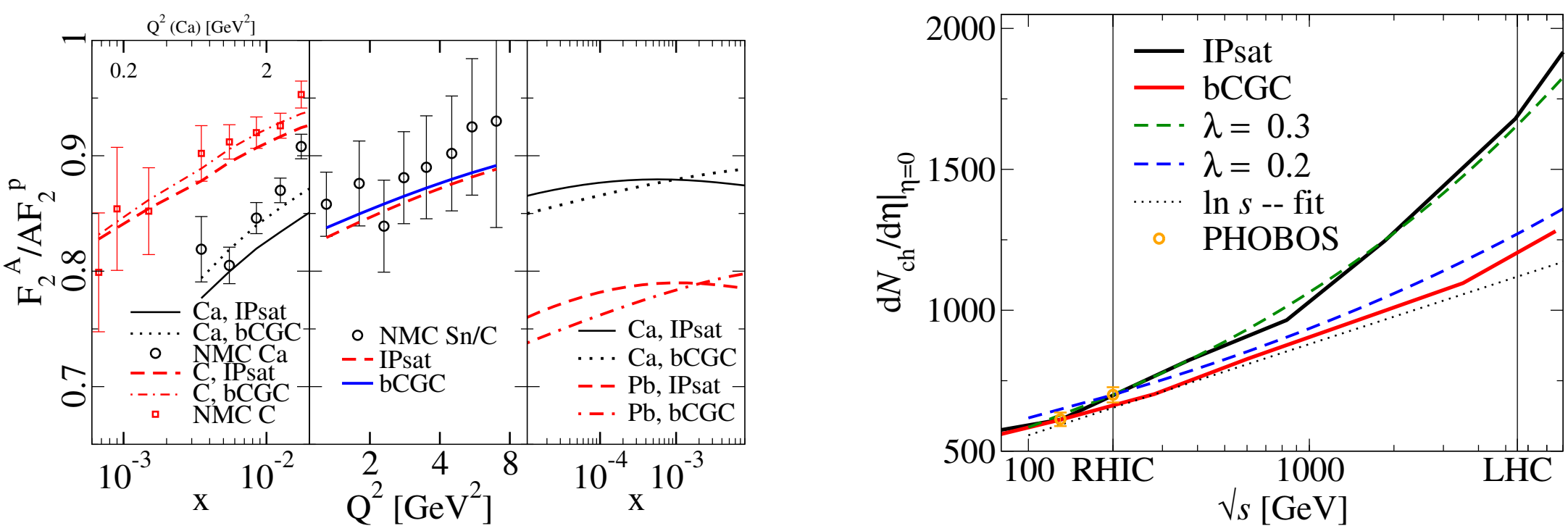

Figure 2: Left: Comparison of the fit [7] to existing nuclear DIS data from NMC. Right: Extrapolation of the gluon multiplicity to LHC energies, from [8]

the parameterisation from protons to nuclei. In a "Glauber"-like formulation of essentially independent scatterings of the dipole on each of the nucleons this is a straightforward estimate, see e.g. Refs. [7, 11]. A simple geometrical argument would give the estimate $Q^2_{\mathrm{sA}} \approx 0.5 Q^2_{sp} A^{1/3}$, where the coefficient in front follows from the inter-nucleon distance in a nucleus being smaller than the nucleon radius. The actual values in the estimate of Ref. [7] are shown in Fig. 1 (right). For other estimates of $Q_{\mathrm{s}}$ based on DIS data see Ref. [12]. Being really consistent with high energy evolution would require some further theoretical advances, since the approximation of independent dipole-nucleon scatterings will break down during the evolution. In the infinite momentum frame this can be thought of as gluons from different nucleons starting to interact with each other.

Combining these ingredients the CYM calculations [4] of gluon production paint a fairly consistent picture of gluon production at RHIC energies. The estimated value $Q_{\mathrm{s}} \approx 1.2$ GeV from HERA data [7, 11] (corresponding to the MV model parameter $g^2\mu \approx 2.1$ GeV [10]) gives a good description of existing nuclear DIS data from the NMC collaboration, see Fig. 2 (left). The same value leads to $\frac{\mathrm{d}N}{\mathrm{d}y} \approx 1100$ gluons in the initial stage. Assuming a rapid thermalization and nearly ideal hydrodynamical evolution this is consistent with the observed $\sim 700$ charged ($\sim 1100$ total) particles produced in a unit of rapidity in central collisions.

The gluon multiplicity is, across different parameterisation's, to a very good approximation proportional to $\pi R_A^2 Q_{\mathrm{s}}^2/\alpha_{\mathrm{s}}$. Thus the predictions for LHC collisions depend mostly on the energy dependence of $Q_{\mathrm{s}}$. On this front there is perhaps more uncertainty than is generally acknowledged, the estimates for $\lambda = \mathrm{d}\ln Q_{\mathrm{s}}^2/\mathrm{d}\ln 1/x$ varying between $\lambda = 0.29$ [13] and $\lambda = 0.18$ [14] in fixed coupling fits to HERA data, with a running coupling solution of the BK equation giving something in between these values [15]. This dominates the uncertainty in predictions for the LHC multiplicity, see Fig. 2 (right).

# 3 Multiplicity Distributions

One very recent application of the CGC framework has been computing the probability distribution of the number of gluons in the glasma [16]. The dominant contributions to multiparticle correlations come from diagrams that are disconnected for fixed sources and become connected only after averaging over the colour charge configurations. In other words, the dominant correlations are those arising from resummed large logarithms of the collision energy and are present already in the initial wave-functions of the colliding nuclei.

Working with the MV model Gaussian probability distribution

$$W[\rho] = \exp\left[-\int \mathrm{d}^2\mathbf{x}_T \frac{\rho^a(\mathbf{x}_T)\rho^a(\mathbf{x}_T)}{g^4\mu^2}\right] \tag{8}$$

computing the correlations in the linearised approximation is a simple combinatorial problem. The result can be expressed in terms of two parameters, the mean multiplicity $\bar{n}$, and a parameter $k$ describing the width of the distribution. The $q$'th factorial moment of the multiplicity distribution is, to leading order in $\alpha_\mathrm{s}$, proportional to $2^q(q-1)!$. Explicitly, the connected parts of the moments $m_q \equiv \langle N^q \rangle$ are

$$m_q = (q-1)!\,k\left(\frac{\bar{n}}{k}\right)^q \quad \text{with} \tag{9}$$

$$k \approx \frac{({N_\mathrm{c}}^2-1)Q_\mathrm{s}^2 S_\perp}{2\pi} \tag{10}$$

$$\bar{n} = f_N \frac{1}{\alpha_\mathrm{s}} Q_\mathrm{s}^2 S_\perp . \tag{11}$$

These moments define a *negative binomial* distribution with parameters $k$ and $\bar{n}$, which has been used as a phenomenological observation in high energy hadron and nuclear collisions already for a long time [17]. In terms of the glasma flux tube picture this result has a natural interpretation. The transverse area of a typical flux tube is $1/Q_\mathrm{s}^2$, and thus there are $Q_\mathrm{s}^2 S_\perp = N_\mathrm{FT}$ independent ones. Each of these radiates particles independently into ${N_\mathrm{c}}^2-1$ colour states in a Bose-Einstein distribution (see e.g. [18]). A sum of $k \approx N_\mathrm{FT}({N_\mathrm{c}}^2-1)$ independent Bose-Einstein-distributions is precisely equivalent to a negative binomial distribution with parameter $k$.

# 4 Inclusive Nuclear Diffraction at eRHIC and LHeC

The large fraction of diffractive events observed at HERA shows that modern colliders are approaching the nonlinear regime of QCD, where gluon saturation and unitarization effects become important. It should be possible to perform the same measurements in DIS off nuclei. There are plans for several facilities capable of high energy nuclear DIS experiments, as the EIC [19] and LHeC [20] colliders. Due to the difficulty in measuring an intact recoil nucleus deflected by a small angle, diffractive eA collisions present an experimental challenge. But if they are successful, nuclear diffractive DIS (DDIS) would provide a good test of our understanding of high energy QCD. Measuring the momentum transfer $t$ in both coherent (nucleus stays intact) and incoherent (nucleus breaks up into nucleons) would enable one to go measure directly the transverse structure of the gluonic degrees of freedom [21] instead of the electric charge distribution that is measured in low energy experiments. Figure 3 (left) demonstrates some expected results from such a measurement. The diffractive structure function can be divided into different components according to the polarisation state of the virtual photon and the inclusion of higher Fock states (e.g. $q\bar{q}g$ in addition to $q\bar{q}$) in the dipole wave-function. All of these have different dependences on the impact parameter of the dipole-target collision (see Fig. 3 right), which stresses the importance of having a detailed picture of the transverse geometry of both the proton and the nucleus.

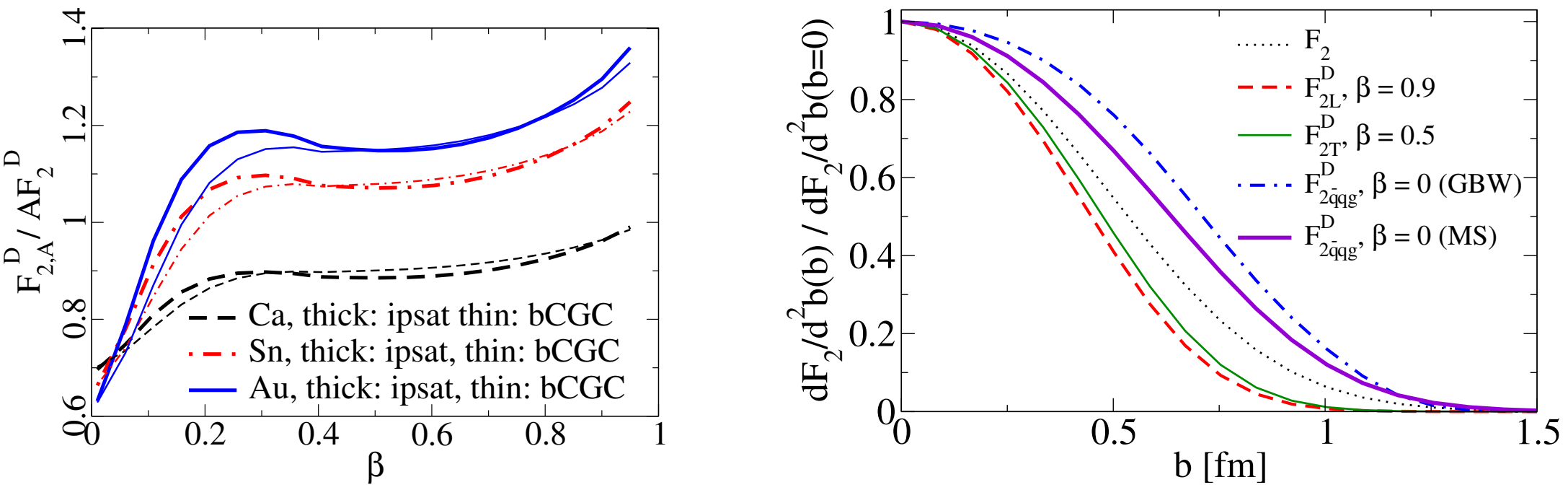

Figure 3: Left: nuclear modification of the diffractive structure function, showing characteristic suppression at small $\beta$ (large mass of the diffractive system) and enhancement at large $\beta$. From Ref. [22]. Right: dominant impact parameters for the different contributions to the proton diffractive structure function.

# Acknowledgements

Numerous conversations with F. Gelis, L. McLerran and R. Venugopalan are gratefully acknowledged. The author is supported by the Academy of Finland, contract 126604.

# References

[1] E. Iancu and R. Venugopalan, The color glass condensate and high energy scattering in QCD, in *Quark gluon plasma*, edited by R. Hwa and X. N. Wang, World Scientific, 2003, arXiv:hep-ph/0303204; H. Weigert, Prog. Part. Nucl. Phys. **55**, 461 (2005), [arXiv:hep-ph/0501087].

[2] T. Lappi and L. McLerran, Nucl. Phys. **A772**, 200 (2006), [arXiv:hep-ph/0602189].

[3] A. Kovner, L. D. McLerran and H. Weigert, Phys. Rev. **D52**, 3809 (1995), [arXiv:hep-ph/9505320]; A. Kovner, L. D. McLerran and H. Weigert, Phys. Rev. **D52**, 6231 (1995), [arXiv:hep-ph/9502289].

[4] A. Krasnitz, Y. Nara and R. Venugopalan, Phys. Rev. Lett. **87**, 192302 (2001), [arXiv:hep-ph/0108092]; T. Lappi, Phys. Rev. **C67**, 054903 (2003), [arXiv:hep-ph/0303076]; A. Krasnitz, Y. Nara and R. Venugopalan, Nucl. Phys. **A727**, 427 (2003), [arXiv:hep-ph/0305112].

[5] A. H. Mueller, arXiv:hep-ph/0111244.

[6] W. Buchmuller, M. F. McDermott and A. Hebecker, Nucl. Phys. **B487**, 283 (1997), [arXiv:hep-ph/9607290].

[7] H. Kowalski, T. Lappi and R. Venugopalan, Phys. Rev. Lett. **100**, 022303 (2008), [arXiv:0705.3047 [hep-ph]].

[8] T. Lappi, J. Phys. **G35**, 104052 (2008), [arXiv:0804.2338 [hep-ph]].

[9] L. D. McLerran and R. Venugopalan, Phys. Rev. **D49**, 2233 (1994), [arXiv:hep-ph/9309289]; L. D. McLerran and R. Venugopalan, Phys. Rev. **D49**, 3352 (1994), [arXiv:hep-ph/9311205]; L. D. McLerran and R. Venugopalan, Phys. Rev. **D50**, 2225 (1994), [arXiv:hep-ph/9402335].

[10] T. Lappi, Eur. Phys. J. **C55**, 285 (2008), [arXiv:0711.3039 [hep-ph]].

[11] H. Kowalski and D. Teaney, Phys. Rev. **D68**, 114005 (2003), [arXiv:hep-ph/0304189].

[12] A. Freund, K. Rummukainen, H. Weigert and A. Schafer, Phys. Rev. Lett. **90**, 222002 (2003), [arXiv:hep-ph/0210139]; N. Armesto, C. A. Salgado and U. A. Wiedemann, Phys. Rev. Lett. **94**, 022002 (2005), [arXiv:hep-ph/0407018].

[13] K. J. Golec-Biernat and M. Wusthoff, Phys. Rev. **D59**, 014017 (1999), [arXiv:hep-ph/9807513].

[14] H. Kowalski, L. Motyka and G. Watt, Phys. Rev. **D74**, 074016 (2006), [arXiv:hep-ph/0606272].

[15] J. L. Albacete, N. Armesto, J. G. Milhano, C. A. Salgado and U. A. Wiedemann, Phys. Rev. **D71**, 014003 (2005), [arXiv:hep-ph/0408216].

[16] F. Gelis, T. Lappi and L. McLerran, Nucl. Phys. **A828**, 149 (2009), [arXiv:0905.3234 [hep-ph]].

[17] UA1, G. Arnison *et al.*, Phys. Lett. **B123**, 108 (1983); UA5, G. J. Alner *et al.*, Phys. Lett. **B160**, 193 (1985); UA5, G. J. Alner *et al.*, Phys. Lett. **B160**, 199 (1985); UA5, R. E. Ansorge *et al.*, Z. Phys. **C37**, 191 (1988); PHENIX, S. S. Adler *et al.*, Phys. Rev. **C76**, 034903 (2007), [arXiv:0704.2894 [nucl-ex]]; PHENIX, A. Adare *et al.*, Phys. Rev. **C78**, 044902 (2008), [arXiv:0805.1521 [nucl-ex]].

[18] K. Fukushima, F. Gelis and T. Lappi, arXiv:0907.4793 [hep-ph].

[19] A. Deshpande, R. Milner, R. Venugopalan and W. Vogelsang, Ann. Rev. Nucl. Part. Sci. **55**, 165 (2005), [arXiv:hep-ph/0506148].

[20] J. B. Dainton, M. Klein, P. Newman, E. Perez and F. Willeke, JINST **1**, P10001 (2006), [arXiv:hep-ex/0603016].

[21] A. Caldwell and H. Kowalski, arXiv:0909.1254 [hep-ph].

[22] H. Kowalski, T. Lappi, C. Marquet and R. Venugopalan, Phys. Rev. **C78**, 045201 (2008), [arXiv:0805.4071 [hep-ph]].

# Parton Showering Effects in Central Heavy-Boson Hadroproduction

*M. Deak*[1], *A. Grebenyuk*[2], *F. Hautmann*[3], *H. Jung*[2,4] *and K. Kutak*[2]

[1]IFT, University of Madrid
[2]Deutsches Elektronen Synchrotron, Hamburg
[3]Theoretical Physics Department, University of Oxford
[4]Physics Department, University of Antwerp

If large-angle multigluon radiation contributes significantly to parton showers associated with heavy boson production at the LHC, appropriate parton branching methods are required for realistic Monte Carlo simulations of final states. We report on a study illustrating such effects in the case of central scalar-boson production. We comment on the possible impact of such studies on the modelling of multi-parton interactions.

## 1 Introduction

Final states containing heavy bosons and jets will be crucial in a number of experimental searches at the Large Hadron Collider. Phenomenological analyses will rely both on perturbative fixed-order calculations and on parton shower Monte Carlo generators for a realistic description of the structure of these events.

Due to the presence of multiple hard scales and the large phase space opening up at LHC energies, the treatment of these final states is potentially sensitive to complex dynamical effects in the QCD showers accompanying the events. In the case of vector bosons it has been pointed out [1] that the treatment of parton showers, and in particular of the recoils in the shower, is essential for a proper description of the W/Z $p_T$ spectrum. This affects the amount of multi-parton interactions [1, 2, 3, 4] needed to describe the events. On the other hand, parton showers which are not ordered in transverse momentum could also considerably contribute to what is typically associated with the underlying event. In the case of vector bosons this may be relevant for early phenomenology at the LHC, as the possible broadening of W and Z $p_T$ distributions [5] affects the use of these processes as luminosity monitor [6].

For scalar boson production, the role of corrections to transverse-momentum ordered showers on the structure of final states was considered in [7, 8] in terms of the heavy-top effective theory matrix elements associated with the unintegrated gluon density [9]. In this article we report on ongoing studies [10] of mini-jet radiation accompanying scalar boson production in the central region at the LHC. It is appropriate to consider this issue in view of the progress in the quantitative understanding of unintegrated gluon contributions in multi-jet final states [11]. In the case of scalar bosons as well, such studies have implications on the role of multi-parton interactions [2, 3] in the evolution of the initial state shower.

We start in Sec. 2 with a brief discussion summarising aspects of corrections to collinear-ordered showers and the role of recent jet-jet correlation measurements. In Sec. 3 we consider the application of parton showers not ordered in transverse momentum to the case of central scalar-boson hadroproduction. We conclude in Sec. 4.

## 2 Corrections to Collinear Showers and Jet Correlations

In this section we briefly discuss effects of high-energy corrections to collinear parton showers on hadronic final states with multiple jets.

Let us recall that the branching algorithms underlying the most commonly used shower Monte Carlo event generators [12, 13] are based on collinear evolution of jets developing, both "forwards" and "backwards", from the hard event [14], supplemented (in the case of certain generators) by suitable constraints for angularly-ordered phase space [15]. The angular constraints are designed to take account of coherence effects from multiple soft-gluon emission [15, 16, 17].

The main new effect one observes when trying to push this picture to higher and higher energies is that soft-gluon insertion rules [16, 17] based on eikonal emission currents [18, 19] are modified in the high-energy, multi-scale region by terms that depend on the total transverse momentum transmitted down the initial-state parton decay chain [20, 21, 22]. As a result, the physically relevant distribution to describe initial-state showers becomes the analogue not so much of an ordinary parton density but rather of an "unintegrated" parton density, dependent on both longitudinal and transverse momenta.[1]

The next observation concerns the structure of virtual corrections. Besides Sudakov form-factor effects included in standard shower algorithms [12, 13], one needs in general virtual-graph terms to be incorporated in transverse-momentum dependent (but universal) splitting functions [20, 29, 30, 31] in order to take account of gluon coherence not only for collinear-ordered emissions but also in the non-ordered region that opens up at high $\sqrt{s}/p_\perp$.

These finite-$k_\perp$ corrections to parton branching have important implications for multiplicity distributions and the structure of angular correlations in final states with high multiplicity. Refs. [11, 35] analyze examples of such effects in the case of di-jet and 3-jet production in $ep$ [32, 33] and $p\bar{p}$ [34] collisions. In particular, the accurate measurements [32] of azimuthal and transverse-momentum correlations are compared with results from collinear shower (HERWIG [36]) and $k_\perp$-shower (CASCADE [37]). The region of large azimuthal separations between the leading jets, $\Delta\phi \sim 180^o$, is dominated by soft gluon emission effects, while the region of small azimuthal separations, down to $\Delta\phi \sim 30^o$, is driven by hard parton radiation, thus offering a significant test of the quality of hard to semi-hard parton showers over the full region of phase space. The description of the angular correlation measurements by the $k_\perp$-shower is good, and provides confidence on the wider applicability of the method for multi-jet processes. Results based on collinear parton showers (HERWIG) cannot describe the shape of the $\Delta\phi$ distribution.

The $k_\perp$-shower predictions involve both transverse-momentum dependent pdfs and matrix elements. Fig. 1 [35] illustrates the relative contribution of these different components to $ep$ di-jet cross sections showing different approximations to the azimuthal dijet distribution normalised to the back-to-back cross section. The solid red curve is the full result [11]. The dashed blue curve is obtained from the same unintegrated pdf's but by taking the collinear approximation in the hard matrix element. The dashed curve drops much faster than the full result as $\Delta\phi$ decreases, indicating that the high-k$_\perp$ component in the ME [30] is necessary to describe jet correlations for small $\Delta\phi$. The dotted (violet) curve is the result obtained from the unintegrated pdf without any resolved branching. This represents the contribution of the intrinsic distribution only, corresponding to non-perturbative, predominantly low-k$_\perp$ modes. That is, in the dotted (violet) curve one retains an intrinsic k$_\perp \neq 0$ but no effects of coherence. We see that the resulting jet correlations in this case are down by an order of magnitude. The inclusion of the perturbatively computed high-k$_\perp$ correction distinguishes the calculation [11] from other shower approaches that include transverse momentum dependence in the pdfs but not in the matrix elements, see e.g. [38].

[1]See [23] for recent reviews of unintegrated pdfs. Aspects of u-pdfs from the standpoint of QCD high-energy factorisation are discussed in [24]. Associated phenomenological aspects are discussed in [8, 23, 25], and references therein; see [26, 27, 28] for recent new work. The papers in [29] contain first discussions of a more general, nonlocal operator formulation of u-pdfs applied to parton showers beyond leading order.

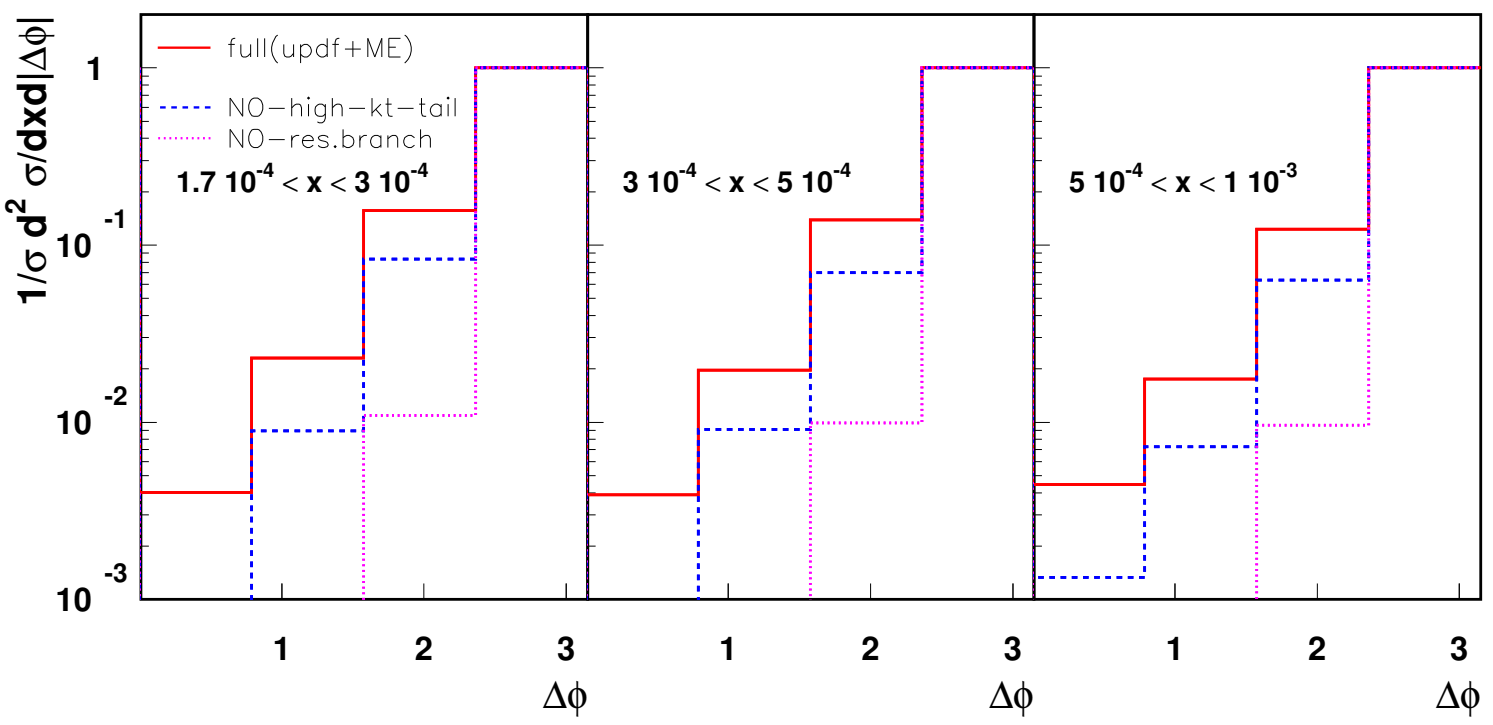

Figure 1: The dijet azimuthal distribution [35] normalised to the back-to-back cross section: (solid red) full result (u-pdf $\oplus$ ME); (dashed blue) no finite-k$_\perp$ correction in ME (u-pdf $\oplus$ ME$_{collin.}$); (dotted violet) u-pdf with no resolved branching.

The corrections to collinear showers described above embody the physics of the unintegrated gluon density and associated hard matrix elements. Besides jet-jet correlations, these corrections will affect the structure of final states associated with heavy mass production. In the next section we consider implications of the unintegrated gluon density and non-collinear contributions to showering on the jet activity accompanying production of heavy scalars in the central region at the LHC.

# 3 Central Scalar Boson Production at the LHC

To study the effect of non-collinear parton showers and its contribution to the underlying event, Ref. [10] investigates a gluon induced process which produces a colour singlet scalar system in the final state, here $gg \to h^0$. We consider radiation associated with standard model Higgs boson production, following the CDF analysis of the underlying event [39]. As shown in Fig. 2, the direction of the Higgs boson in the azimuthal plane defines the origin of the system, and four regions in the azimuthal plane are defined.

Fig. 3 [10] shows results for the average multiplicity for mini-jets with $E_t > 15$ GeV and with $E_t > 5$ GeV at LHC energies ($\sqrt{s} = 14$ TeV) in the 4 different regions of $\phi$ as a function of the Higgs transverse momentum.

The predictions of the k$_\perp$-shower Monte Carlo generator CASCADE [37] are compared with predictions from PYTHIA [1]. For comparison, CASCADE is also run in collinear mode (cascade-dglap), with the off-shell matrix element [9] replaced by the on-shell approximation and the parton showers are evolved with the one-loop splitting function and an upper restriction on the transverse momentum $p_t < \sqrt{m_h^2 + p_{t\,h}^2}$. For mini-jets with $E_T > 15$ GeV CASCADE in collinear mode reproduces the prediction of PYTHIA without multiparton interactions in both transverse regions. The full CASCADE run gives higher activities in the transverse as well as in the toward regions, and is close or larger than the prediction of PYTHIA including multiparton interactions. In the away region the slope is steeper than predicted from PYTHIA. Lowering the transverse momentum cut of the mini-jets to $E_T > 5$ GeV, CASCADE still predicts a larger multiplicity than PYTHIA without multiparton interactions, but falls clearly below the prediction including multiparton interactions. This illustrates the onset of hard perturbative contributions from the parton showers, which are simulated in PYTHIA with multiparton interactions.

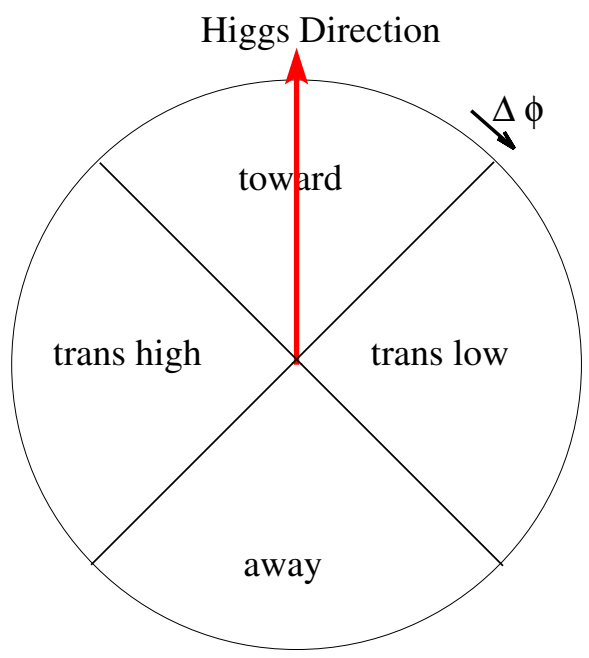

Figure 2: Different regions in $\phi$ with respect to the Higgs direction.

The phase space region where soft radiation plays a significant role is the region of minimal transverse energy in the $\phi$ plane. This is the region where multiparton interactions should be visible. We have also studied the multiplicity of charged particles (with $p_t > 150$ MeV) in the process $gg \to h^0$. The study in [10] indicates that the result from PYTHIA including multiparton interactions is above the result from CASCADE, however the multiplicity predicted from CASCADE is significantly larger than the one predicted by PYTHIA without multiparton interactions. The result using full unintegrated-pdf evolution shows significantly more activity in all regions. Thus even for the soft contribution in the charged particle multiplicity the treatment of parton showers is important.

In the region of minimum bias events, elastic and soft diffractive processes will also play a role. This is not (yet) implemented in CASCADE.

More details of this study will be reported in a forthcoming publication. We note that the effects described above can influence the description of soft underlying events and minijets [1, 4] as well as the use of exclusive scalar production channels [7].

# 4 Outlook

The production of final states containing heavy bosons and multiple jets will be characterised at the LHC by the large phase space opening up at high centre-of-mass energies, and the presence of multiple hard scales, possibly widely disparate from each other. This brings in potentially large perturbative corrections to hard-scattering events and potentially new effects in the parton-shower components of the process.

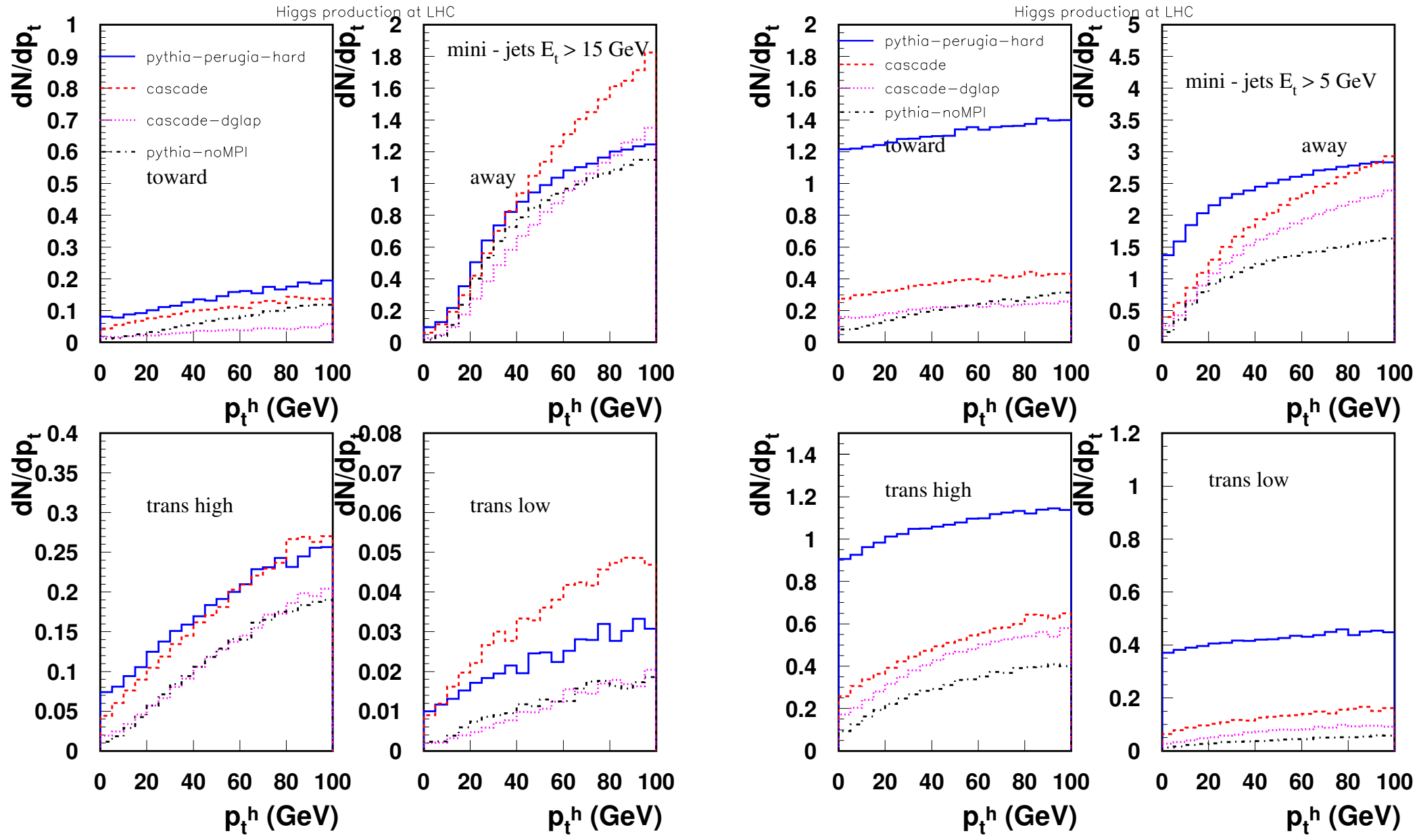

Figure 3: Multiplicity of jets as a function of the transverse momentum of the Higgs in different regions of $\phi$ as predicted from $k_\perp$-shower (CASCADE) and collinear shower (PYTHIA). Shown is also the prediction using CASCADE in collinear mode (cascade-dglap). The left figure shows results for mini-jets with $E_T > 15$ GeV, the right figure for $E_T > 5$ GeV.

If large-angle multigluon radiation gives significant contributions to the QCD showers accompanying heavy boson production at the LHC, appropriate generalisations of parton branching methods are required. In this study we have considered jet radiation associated with heavy scalars produced centrally at the LHC, and we have described applications of transverse-momentum dependent kernels [7, 8, 9, 24, 25] for parton showering. We have focused on associated minijet distributions and discussed a comparison of showering effects with multi-parton interactions effects.

This study lends itself to extensions in several directions. First, we have considered here mini-jet and effects that could be associated with the underlying event. However, the approach is much more general (see e.g. discussions in [11, 23, 26, 29]) and could be used to investigate hard radiation as well.

Next, we have considered scalar boson production which is dominated by the physics of initial-state gluonic showers, expressible in terms of unintegrated gluon densities. But treatments of quark contributions to showers at unintegrated level are also being worked on (see e.g. [26, 27, 28]). In this respect, theoretical results for splitting kernels [30] already applied to inclusive phenomenology can also be of use in calculations for exclusive final states [40]. This will have direct applications to parton showers in vector boson production.

Further, relevant areas of experimental studies will involve jet physics in the forward rapidity region [41] at the LHC. In this article we have limited ourselves to considering production processes in the central rapidity region. Note that techniques are being developed [42] to allow one to also address multi-particle hard processes at forward rapidities.

## Acknowledgements

We thank the organisers for the kind invitation and for the excellent organisation of the meeting.

## References

[1] P.Z. Skands, arXiv:0905.3418 [hep-ph] in Proc. Perugia Workshop (2008).

[2] R. Corke and T. Sjöstrand, arXiv:0911.1909 [hep-ph].

[3] M. Bähr, S. Gieseke and M. Seymour, JHEP **0807** (2008) 076.

[4] G. Gustafson, talk at Desy Workshop, Hamburg, March 2007; G. Gustafson, L. Lönnblad and G. Miu, JHEP **0209** (2002) 005.

[5] F.I. Olness, talk at HERA-LHC Workshop, CERN, May 2008; S. Berge, P.M. Nadolsky, F.I. Olness and C.P. Yuan, hep-ph/0508215.

[6] A.M. Cooper-Sarkar, arXiv:0707.1593 [hep-ph].

[7] L. Lönnblad and M. Sjödahl, JHEP **0402** (2004) 042.

[8] H. Jung, Mod. Phys. Lett. A **19** (2004) 1.

[9] F. Hautmann, Phys. Lett. B **535** (2002) 159.

[10] M. Deak, A. Grebenyuk, F. Hautmann, H. Jung and K. Kutak, in Proc. DIS09 Workshop (Madrid, 2009).

[11] F. Hautmann and H. Jung, JHEP **0810** (2008) 113.

[12] B.R. Webber, CERN Academic Training Lectures (2008).

[13] R.K. Ellis, W.J. Stirling and B.R. Webber, *QCD and collider physics*, CUP 1996; Yu.L. Dokshitzer, V.A. Khoze, A.H. Mueller and S.I. Troian, *Perturbative QCD*, Ed. Frontieres, Gif-sur-Yvette (1991).

[14] J.C. Collins, D.E. Soper and G. Sterman, Adv. Ser. Direct. High Energy Phys. 5 (1988) 1.

[15] B.R. Webber, Ann. Rev. Nucl. Part. Sci. **36** (1986) 253.

[16] A. Bassetto, M. Ciafaloni and G. Marchesini, Phys. Rept. **100** (1983) 201.

[17] Yu.L. Dokshitzer, V.A. Khoze, A.H. Mueller and S.I. Troian, Rev. Mod. Phys. **60** (1988) 373.

[18] V.N Gribov, Sov. J. Nucl. Phys. **5** (1967) 399; F.E. Low, Phys. Rev. **110** (1958) 974.

[19] J. Frenkel and J.C. Taylor, Nucl. Phys. **B246** (1984) 231; R. Doria, J. Frenkel and J.C. Taylor, Nucl. Phys. **B168** (1980) 93.

[20] M. Ciafaloni, Nucl. Phys. **B296** (1988) 49.

[21] S. Catani, M. Ciafaloni and F. Hautmann, Phys. Lett. B**242** (1990) 97.

[22] G. Marchesini and B.R. Webber, Nucl. Phys. **B386** (1992) 215.

[23] F. Hautmann, Acta Phys. Polon. B **40** (2009) 2139; F. Hautmann and H. Jung, Nucl. Phys. Proc. Suppl. **184** (2008) 64 [arXiv:0712.0568 [hep-ph]].

[24] S. Catani, M. Ciafaloni and F. Hautmann, Nucl. Phys. B**366** (1991) 135; Phys. Lett. **B307** (1993) 147.

[25] J.R. Andersen *et al.*, Eur. Phys. J. C **48** (2006) 53; B. Andersson *et al.*, Eur. Phys. J. C **25** (2002) 77.

[26] H. Jung *et al.*, Proceedings of the Workshop “HERA and the LHC”, arXiv:0903.3861 [hep-ph].

[27] S. Jadach and M. Skrzypek, arXiv:0905.1399 [hep-ph]; arXiv:0909.5588 [hep-ph].

[28] A.D. Martin, M.G. Ryskin and G. Watt, arXiv:0909.5529 [hep-ph].

[29] J.C. Collins and X. Zu, JHEP **0503** (2005) 059; J.C. Collins, Phys. Rev. D **65** (2002) 094016; J.C. Collins and F. Hautmann, JHEP **0103** (2001) 016.

[30] S. Catani and F. Hautmann, Nucl. Phys. B**427** (1994) 475; Phys. Lett. B**315** (1993) 157.

[31] F. Hautmann, Phys. Lett. B **655** (2007) 26.

[32] S. Chekanov *et al.* [ZEUS Collaboration], Nucl. Phys. B **786** (2007) 152 [arXiv:0705.1931 [hep-ex]].

[33] A. Aktas *et al.* [H1 Collaboration], Eur. Phys. J. C **33** (2004) 477 [arXiv:hep-ex/0310019].

[34] V.M. Abazov *et al.* [D0 Collaboration], Phys. Rev. Lett. **94** (2005) 221801 [arXiv:hep-ex/0409040].

[35] F. Hautmann and H. Jung, arXiv:0804.1746 [hep-ph], in Proceedings of the 8th International Symposium on Radiative Corrections RADCOR2007; arXiv:0808.0873 [hep-ph].

[36] G. Corcella *et al.*, JHEP **0101** (2001) 010.

[37] H. Jung, Comput. Phys. Commun. **143** (2002) 100.

[38] S. Höche, F. Krauss and T. Teubner, Eur. Phys. J. C**58** (2008) 17.

[39] CDF Coll., Phys. Rev. D **65** (2002) 092002.

[40] H. Jung *et al.*, in progress.

[41] D. d'Enterria, arXiv:0911.1273 [hep-ex].

[42] M. Deak *et al.*, arXiv:0908.0538 [hep-ph]; arXiv:0908.1870 [hep-ph].

# Saturation and Critical Phenomena in DIS

*L.L. Jenkovszky*[1], *S.M. Troshin*[2], *and N.E. Tyurin*[2]

[1]BITP, Kiev, 03680 Ukraine
[2]IHEP, Protvino, 142281 Russia

It is argued that the expected turn-down in $x - Q^2$ of the cross sections (structure functions $F_2(x, Q^2)$), assumed to result from the saturation of parton densities in the nucleon, is related to a phase transition from the (almost) ideal partonic gas, obeying Bjorken scaling, to a partonic "liquid". This can be quantified in the framework of statistical models, percolation and other approaches to collective phenomena of the strongly interacting matter. Similarities and differences between the case of lepton-hadron, hadron-hadron and nuclear collisions are discussed.

## 1 Introduction

Based on different observations, models and equations governing deep inelastic scattering (DIS) and related processes, a "saturation" regime is expected when certain values of low enough $x$ and relevant $Q^2$ are reached. According to the dipole model of DIS, this regime already has been achieved and it is characterised by the "saturation radius" [1] $R_0^2 = (x/x_0)^\lambda/Q_0^2$, with $Q_0^2 = 1$ GeV$^2$, $x_0 = 3 \cdot 10^{-4}$ and $\lambda = 0.29$, found from a fit to the DIS data at $x < 0.01$. On more general grounds, saturation could be expected also from unitarity: the rapid (power-like) increase with $1/x$ of the structure functions/cross sections may suggest that unitarity corrections will tamper this rise, although formally the Froissart bound has never been proven for off-mass-shell particles, thus unitarity does not provide any rigorous limitation for such amplitudes [2, 3]. One more argument is physical: the rise of the structure function $F_2(x, Q^2)$ reflects the increase of the parton density (parton number in the nucleon). Since this number increases as a power, and the nucleon radius is known to increase as $\ln s$ (or, at most $\ln^2 s$) (shrinkage of the cone), the particle number density within a nucleon increases, inevitably, reaching a critical value where the partons start to coalesce (overlap, recombine etc). The qualitative picture of this phenomenon in the $Q^2 - 1/x$ plane is well known and cited in various contexts (see, e.g., [4]).

Quantitatively, the dynamics depend on many, poorly known, details, such as the properties of the constituents and their interaction within the nucleon.

Below we propose a novel approach to the saturation phenomenon in DIS and related processes based on the collective properties of the excited nucleon. Namely, we suggest that, below the saturation reggeon, the nucleon in DIS is seen as a gas of almost free partons. With their increasing density, the constituent gradually overlap and, starting from a certain value of $x$ and $Q^2$, the gas of free partons coalesce condensing in a liquid of quarks and gluons. Saturation corresponds to the onset of the new phase.

The thermodynamic properties of such an excited nucleon are characterised by its temperature, pressure etc, and a relevant equation of state (EoS). The statistical treatment of partonic distributions is by far not new, see e.g. [5, 6, 7]. What is new in our approach, is the interpretation of the saturation in DIS as a manifestation of the transition from a dilute partonic gas to a liquid. The details (nature) of this (phase?) transition are not known. It can be of the first, of second order, or, moreover, be a smooth cross-over phenomenon. Our main argument is that the volume of the nucleon confining the partons (quarks and gluons) in the interior increases slower, at most as $\sim \ln^6 s$ (more likely, as $\sim \ln^3 s$), while the volume occupied by the interior, quarks and gluons, increases as a power, thus resulting in a limiting behaviour: a gas-to-liquid cross-over or a phase transition. The present contribution is a first step in understanding this complex process.

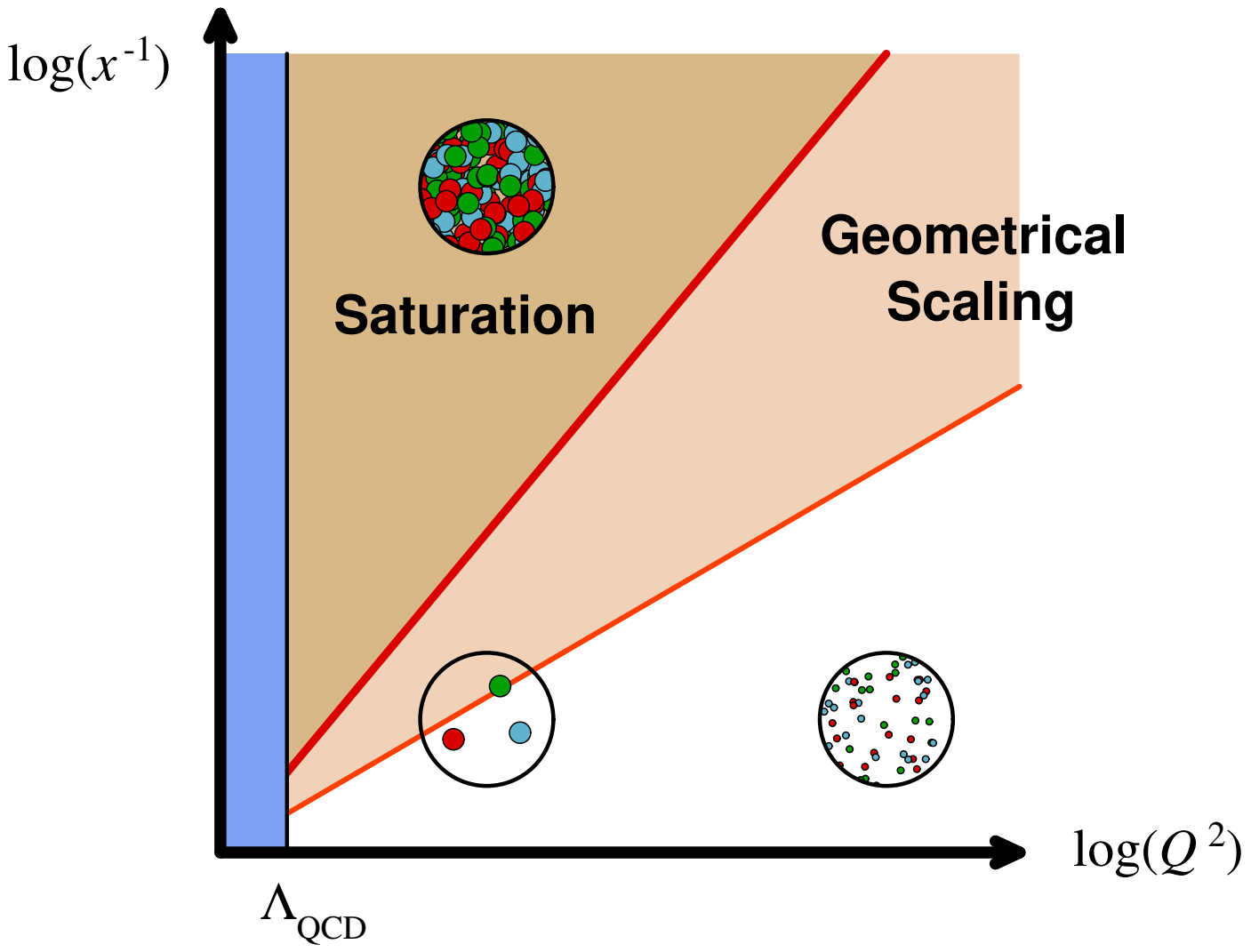

Figure 1: *Phase diagram of DIS, from Ref. [4].*

In a related paper, Ref. [8], the phase structure of the hadronic matter in terms of its temperature $T$ and its baryochemical potential $\mu$, was studied in the framework of the percolation theory. The percolation mechanism was used in [9] to obtain a limiting energy dependence of the hadronic matter at $s \to \infty$ .

## 2 Saturation

We define the saturation line (in the $x - Q^2$ plane) as the turning point (line) of the derivatives

$$B_Q(x,Q^2) = \frac{\partial F_2(x,Q^2)}{\partial(\ln Q^2)}, \quad B_x(x,Q^2) = \frac{\partial F_2(x,Q^2)}{\partial(\ln(1/x))}, \tag{1}$$

called $B_Q$ or $B_x$ slopes, where $F_2(x,Q^2)$ is a "reasonable" model for the structure function, i.e. one satisfying the basic theoretical requirements, yet fitting the data. For example, the model for $F_2(x,Q^2)$ of Ref. [10] interpolates between Regge behaviour at small $Q^2$ and the solution of the DGLAP evolution equation at asymptotically large values of $Q^2$, practically for all values of $x$. The resulting two-dimensional projection of the $Q-$slope is shown in Fig. 2. In our interpretation, the critical line (saturation = phase transition) occurs along the fold line on this figure (compare with a similar figure, Fig. 2 of Ref. [11], derived from a different model).

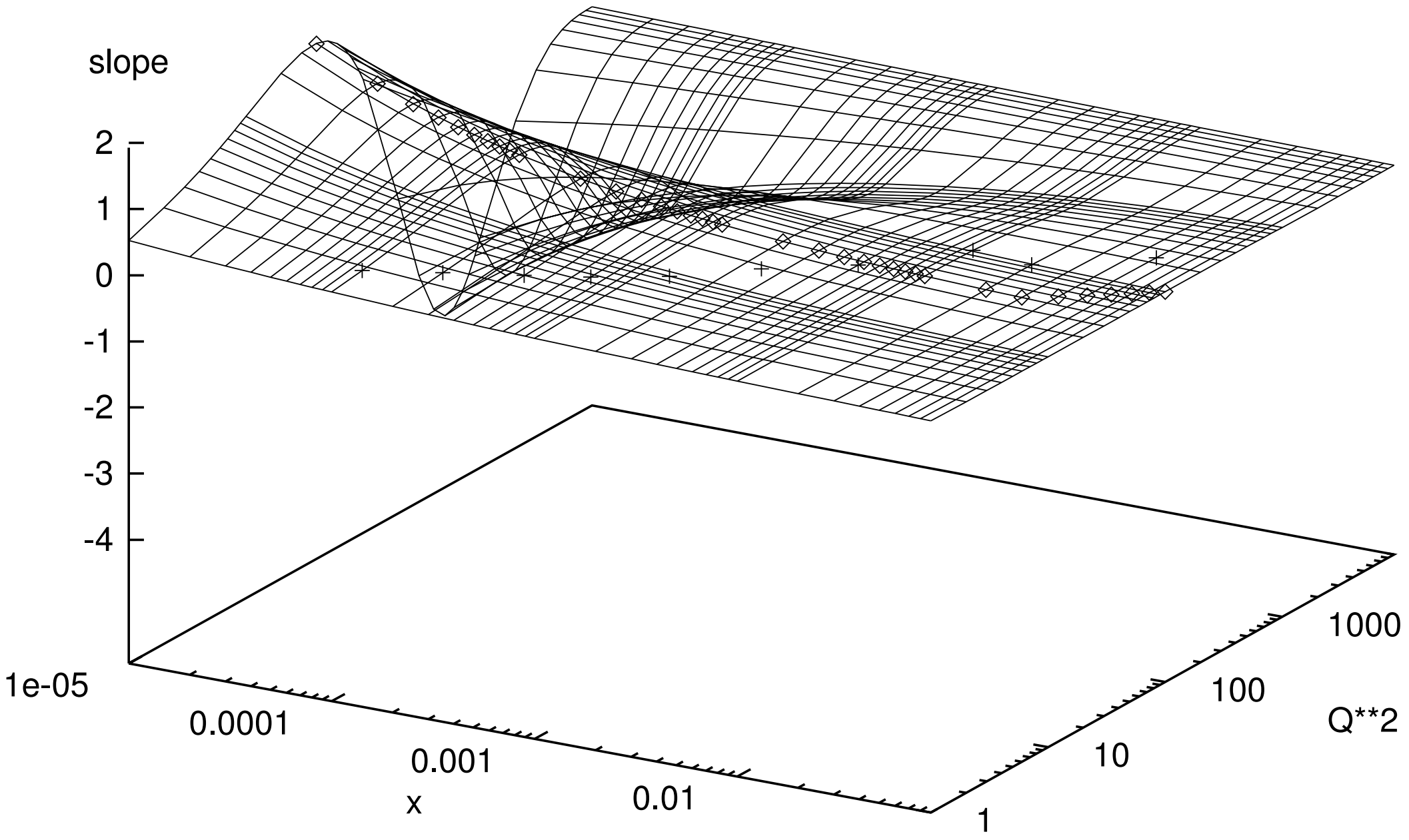

Figure 2: *The surface of $B_Q(x, Q^2)$ calculated in Ref. [10].*

The main goal of this paper is the identification of this line (point) with the critical line (point) on the $T,\ \mu$ phase diagram of an excited nucleon viewed as a thermodynamical system. The thermodynamical approach to DIS may provide a new insight to this complex phenomenon. We are aware of the limited time scales in a deep inelastic scattering from the point of view of thermalisation, a familiar problem relevant to any thermodynamical description of hadronic systems. Let us only remind that the thermodynamic approach to high-energy scattering and multiple production, originated by Fermi and Landau's papers, were applied to hadrons, rather than heavy ions.

# 3 Statistical Models of Parton Distributions

The statistical model of parton distributions in DIS was considered and developed in quite a number of papers [5, 6, 7]. In its simplest version, one assumes [5] that inside the nucleon, the valence quarks, as well as the sea quarks and antiquarks and gluons form a noninteracting gas in equilibrium. This simple picture may be further developed in two directions: 1. Introduction of the effect of the finite size (and its energy dependence!) of the nucleon on statistical expression for the number of states for the unit energy interval; 2) Account for the $Q^2$ evolution, that can be calculated either from the DGLAP equation or phenomenologically, as e.g. in Ref. [10]. A likely scenario emerging [5, 6] for this high quark density $n_q = (n_q - n_{\bar{q}})/3$ system is that quarks form Cooper pairs and new condensates develop.

The nucleon of mass $M$ consists of a gas of massless particles (quarks, antiquarks and gluons) in equilibrium at temperature $T$ in a spherical volume $V$ with radius $R(s)$ increasing with squared c.m.s. energy $s$ as $\ln s$ (or $\ln^2 s$). The invariant parton number density in phase space is given by [12]

$$\frac{dn^i}{d^3p^i\, d^3r^i} = \frac{dn}{d^3p\, d^3r} = \frac{gf(E)}{(2\pi)^3}, \tag{2}$$

where $g$ is the degeneracy ($g = 16$ for gluons and $g = 6$ for $q$ and $\bar{q}$ of a given flavour), $E, \mathbf{p}$ is the parton four-momentum and $f(E) = \Big(\exp[\beta(E-\mu)]\pm 1\Big)^{-1}$ is the Fermi or Bose distribution function with $\beta \equiv T^{-1}$. Quantities in the infinite momentum frame (IMF) are labelled by the subscript $i$.

The invariant parton density $dn^i/dx$ in the IMF is related to $dn/dE$ and $f(E)$ in the proton rest frame as follows [5]

$$\frac{dn^i}{dx} = \frac{gV(s)M^2x}{(2\pi)^2}\int_{xM/2}^{M/2} dEf(E), \tag{3}$$

and the structure function

$$F_2(x) = x\sum_q e_q^2\Big[\Big(\frac{dn^i}{dx}\Big)_q + \Big(\frac{dn^i}{dx}\Big)_{\bar{q}}\Big]. \tag{4}$$

Without any account for the finite volume of the hadron, this SF disagrees with the data. Finite volume effects can be incorporated, following R.S. Bhalerao from Ref. [5], and the result is

$$dn/dE = gf(E)(VE^2/2\pi^2 + aR^2E + bR), \tag{5}$$

where $V$ and $R$ are energy dependent and $a$, $b$, in front of the surface and curvature terms, are unknown numerical coefficients. Their values are important for the final result, but they cannot be calculated from perturbative QCD. A rather general method to calculate these important parameters can be found in Ref. [13]. We intend to come back to this point in a subsequent publication.

# 4 Percolation

Percolation as a model of phase transition from colourless hadrons to a quark gluon plasma (QGP) was studied in a number of papers, recently in Ref. [8]. Below we apply the arguments of that paper to the saturation inside a nucleon, where extended (dressed) quark and gluons, rather then mesons and baryons considered in Ref. [8] percolate into a uniform new phase of matter. Both objects are coloured particle inside a colourless nucleon. We start with a short introduction to the subject.

Consider $N$ spheres of radius $R_0$ and hence volume $V_0 = (4\pi/3)R_0^3$ in a "box" of size $V$, with $V \gg V_0$.

Percolation (*clustering*) of spheres, in three spacial directions, was studied for the case of:

a) arbitrary overlap [15] and b) for those with impenetrable hard core, allowing only partial overlap [6].

For the case a), the first percolation point (partial percolation), with 30% of occupation, occurs for the density $n = N/V$, $n_m \approx 0.35/V_0$, the largest cluster having the density of about $1.2/V_0$. A second percolation point occurs at $n_v = 1.22/V_0$, when 70% of space is covered by spheres, i.e. the vacuum disappears as a large-scale entity.

The existence of two percolation thresholds, one for the formation of the first spanning cluster of spheres and the second one for the disappearing of a spanning vacuum "cluster", is a general feature of the 3-dimensional percolation theory.

b) An impenetrable spherical core, with $R_c = R_0/2$, and the spheres can only partially overlap. Here again one has two percolation thresholds, at $\bar{n}_m \approx 0.34/V_0$ (close to the case a)), and vacuum percolation, at $\bar{n}_v = 2.0/V_0$, requiring a higher density compared to a).

## 4.1 GPD $\mathcal{H}(\mathbf{x}, \mathbf{t})$ and Impact Parameter Parton Distribution $\mathbf{q}(\mathbf{x}, \mathbf{b})$

The number of constituents in a nucleon can be found by integrating in $b$ (impact parameter) the general parton distribution, e.g. that of Ref. [14].

$$q(x, b) = \frac{1}{2\pi} \int_0^\infty \sqrt{-t} d\sqrt{-t} \mathcal{H}(x, t) J_0(b\sqrt{-t}).$$

Here, contrary to $q(x)$, $q(x, b)$ is dimensional, with a dimension of squared mass $m^2$, interpreted as the transverse size of the extended parton in the hadron, $m = R^{-1}$, $q(x, b) = m^2 \tilde{q}(x, b)$, $\tilde{q}(x, b)$ being the partons number density [14]. In paper [14], $m$ was a constant; here we choose it to depend on the photon virtuality: $m \to m_0 \ln(Q + m_\rho) = R^{-1}(Q)$, where $m$ is the mass of the lightest vector meson, $m = m_\rho$. Note the inequality $q(x, b) \leq 1/S_q$, where $S_q(Q) = \ln^{-2}(Q + m_\rho)$.

Thus, the nucleon is composed of $N = 2\pi \int_0^1 dx \int_0^\infty b\, db\, \tilde{q}(x, b, Q)$ extended partons with the transverse area $S_q(Q)$.

## 4.2 A Toy EoS

One has for mesons

$$N_M(T) = 3\frac{\pi^2}{90} T^3.$$

On the other hand, mesons percolate whenever $n_v = 1.22/V_h$, where $V_h = (4\pi/3)R_h^3$, and we use for the probe meson radius $R_h = 1.1/\ln(Q + m_\rho)$, which in the limit of a real photon, $Q \to 0$, matches the relevant value $R_h = 0.8$ fm used in Ref. [8]. Solving $n_h(T) = n_f$ yields

$$T_M(T, Q) \simeq 171 \ln(Q + m_\rho)\,\mathrm{MeV}$$

as the limiting temperature through meson fusion (cf. $T_\pi \simeq 240\,\mathrm{MeV}$ of [8]). In a similar analysis of the phase diagram of hadronic matter [8], the ($Q-$independent) limiting temperature $T_\pi \approx 240\,\mathrm{MeV}$ was obtained.

*N.B. The parameters appearing below should be rescaled with account for the replacement $m \to \ln(Q + m_\rho)$!*

The density of point-like nucleons of mass $M$ is at $T = 0$

$$n_b(\mu, T = 0) = \frac{2}{3\pi^2}(\mu^2 - M^2)^{3/2},$$

or, by using the van der Waals approach of Ref. [6]

$$n_B(\mu, T=0) = \frac{n_b(T,\mu)}{1+n_b(T,\mu)V_e}.$$

With increasing nucleon density, the empty vacuum disappears for

$$\bar{n}_v \simeq 2/V_h(Q) \approx 0.93 \ln(Q+m_{\rho)})^3 \,\mathrm{fm}^{-3},$$

which, for real photons ($Q = 0$) corresponds to about 5.5 times standard nuclear density. Solving $n_B(T = 0, \mu) = \bar{n}_v$ gives $\mu_v \simeq 1.12 \ln(Q + m_\rho)$ GeV for the limiting baryochemical potential at $T = 0$.

In the region of low or intermediate $\mu$ in the $T\mu$ diagram, one can approximate the density of point-like nucleons by the Boltzmann limit

$$n_b(\mu, T) \simeq \frac{2T^3}{\pi^2}\Big(\frac{M}{T}\Big)^2 K_2(M/T)e^{\mu/T} \simeq \frac{T^3}{2}\Big(\frac{2M}{\pi T}\Big)^{3/2} e^{(\mu-M)/T}.$$

As a result, one obtains a family of curves for the phase diagram $T(\mu)$ and for the EoS $p(T)$ for various values of $Q$ and $x$ (cf. Figs. 3 and 4 of Ref. [8]).

# 5 Conclusions

In this talk a new approach to the saturation phenomena in deeply virtual processes – DIS, DVCS, VMP – diffractive and non-diffractive – is suggested. The basic idea is a physical one: Bjorken scaling implies that the nucleon in DIS and related processes is a system of weekly interacting partonic gas, that can be described by means of Bose-Einstein or Fermi statistics. As the density increases (with decreasing $x$ and relevant $Q^2$), seen as the violation of Bjorken scaling, the system reaches a coalescence point where the gas condenses, eventually to a liquid. The thermodynamic properties of this transition can be only conjectured, and further studies are needed to quantify this phenomenon.

Of two models presented in this note, more promising seems be the first one (Sec. 3). Further studies will show its viability.

An alternative measure of the onset of saturation and expected change of phase can be related to non-linear evolution equations. Saturation and a phase transition are expected when the non-linear contribution overshoots the linear term.

The phenomenon discussed in the present note may have much in common with the colour glass condensate proposed in the context of heavy ion collisions (see, e.g. [16] and earlier references therein). Apart from similarities (condensation of quarks and gluons as $x \to 0$) there are apparent differences: for example, hydrodynamical flow is not expected in DIS. In any case, the dynamics of the strong interaction is the same in lepton-hadron, hadron-hadron and heavy ion collisions.

# Acknowledgements

We thank Mark Gorenstein, Volodymyr Magas and Francesco Paccanoni for useful discussions and criticism.

# References

[1] K. Golec-Biernat and M. Wüsthoff, Phys. Rev. D **644** 131 (2007).

[2] V.A. Petrov, A.V. Prokudin, Proceedings of the conference Elastic and diffractive scattering, Protvino 1999, p.95.

[3] S.M. Troshin, N.E. Tyurin. Eur. Phys. J. C **22** 667 (2002).

[4] Francois Gelis, *QCD at small x and nucleus-nucleus collisions*, Plenary talk given at QM2006, Shanghai, November 2006, J.Phys.G34:S421-428,2007, arXiv:hep-ph/0701225.

[5] R.S. Bhalerao, *Statistical model for the nucleon structure functions*, Phys. Lett. B **380** 1 (1996), hep-ph/9607315; R.S. Bhalerao and R.K. Bhaduri, *Droplet formation of quark-gluon plasma at low temperatures and high high densities*, hep-ph/0009333; R.S. Bhalerao, N.G. Kelkar and B. Ram *Model for polarized and unpolarized parton density functions in the nucleon*, Phys. Rev. C **63** 025208 (2001), hep-ph/9911286; K. Ganesamurtly, V. Devanathan and M. Rajasekaran, Z. Phys. C **52** 589 (1991; V. Devanathan, S. Karthiyaini and K. Ganesamurthly, Mod. Phys. Lett. A **9** 3455 (1994); V. Devanathana and J.S. McCarthy, Mod. Phys. Lett. A **11** 147 (1996); Hai Lin, hep-ph/0105050, hep-ph/0105172, hep-ph/0106100.

[6] C. Angelini and R. Pazzi, Phys. Lett. B **135** 473 (1984). E. Mac and E. Ugaz, Z. Phys. C **43** 655 (1989); J. Cleymans and R.L. Thews, Z. Phys. C **37** 315 (1988); J. Cleymans, I. Dadic, and J. Joubert, Z. Phys. **68** 275 (1994).

[7] C. Bourrely, F. Buccella, G. Miele, G. Migliore, J. Soffer, and V. Tibullo, Z.Phys. C **62** 431 (1994); C. Bourrely and J. Soffer, Phys. Rev. D **51** 2108 (1995); C. Bourrely and J. Soffer, Nucl. Phys. B **445** 341 (1995).

[8] P. Castorina, K. Redlich and H. Satz, *The Phase Diagram of Hadronic Matter*, BI-TP 2008/17 preprint, hep-ph/0807.4469.

[9] S.M. Troshin, N.E. Tyurin, Mod. Phys. Lett. A **24** 2225 (2009), see also these Proceedings.

[10] P. Desgrolard, L. Jenkovszky and F. Paccanoni, EPJ C **7** 263 (1999).

[11] R. Yoshida, *What HERA can tell us about saturation*, In the proceedings of the ISMD08 Conference, DESY, 2008,

[12] See, e.g., S.R. de Groot, W.A. van Leeuwen and Ch.G. van Weert, *Relativistic Kinetic Theory* (North-Holland, Amsterdam, 1980), p. 13.

[13] V.G. Boyko, L.L. Jenkovszky and V.M. Sysoev, Phys. Atomic Nucl. **57** 1434 (1994).

[14] S.M.Troshin and N.E. Tyurin, Mod. Phys. Lett. A **23** 3141 (2008).

[15] M.B. Ishchenko, Rev. Mod. Phys. **64**(1992)961.

[16] Larry McLerran, *Quarkyonic Matter and the Revised Phase Diagram of QCD*, arXiv:0907.4489, To appear in the conference proceedings for Quark Matter 2009, March 30 - April 4, Knoxville, Tennessee.

# Beauty Baryon Production in pp Collisions at LHC and b Quark Distribution in the Proton

*G.I. Lykasov*[1], *V.V. Lyubushkin*[1], *T.V. Lyubushkina*[1] *and V.A. Bednyakov*[1]

[1]JINR, Dubna, 141980, Moscow region, Russia

The production of charmed and beauty hadrons in proton-proton and proton-antiproton collisions at high energies are analyzed within the modified quark-gluon string model (QGSM) including the internal motion of quarks in colliding hadrons. We present some predictions for the future experiments on the beauty baryon production in $pp$ collisions at LHC energies. This analysis allows us to find interesting information on the Regge trajectories of the heavy $(b\bar{b})$ mesons and the sea beauty quark distributions in the proton.

## 1 Introduction

Various approaches of perturbative QCD including the next-to-leading order calculations (NLO QCD) have been applied to construct distributions of quarks in a proton. The theoretical analysis of the lepton deep inelastic scattering (DIS) off protons and nuclei provides rather realistic information on the distribution of light quarks like $u, d, s$ in a proton. However, to find a reliable distribution of heavy quarks like $c(\bar{c})$ and especially $b(\bar{b})$ in a proton describing the experimental data on the DIS is a non-trivial task. It is mainly due to small values of $D$ and $B$ meson yields in the DIS at existing energies. Even at the Tevatron energies the $B$- meson yield is not so large. At LHC energies the multiplicity of these mesons produced in $pp$ collisions will be significantly larger. Therefore one can try to extract a new information on the distribution of these heavy quarks in a proton. In this paper we suggest to study the distribution of heavy quarks like $c(\bar{c})$ and $b(\bar{b})$ in a proton from the analysis of the future LHC experimental data.

The multiple hadron production in hadron-nucleon collisions at high energies and large transfers is usually analyzed within the hard parton scattering model (HPSM) suggested in [1, 2]. This model was applied to the charmed meson production both in proton-proton and meson-proton interactions at high energies, see for example [3]. The HPSM is significantly improved by applying the QCD parton approach [4, 5], see details in [6] and references therein. Unfortunately the QCD including the next-to-leading order (NLO) has some uncertainties related to the renormalization parameters especially at small transverse momenta $p_t$ [6].

In [6, 7] we studied the charmed and beauty meson production in $pp$ and $p\bar{p}$ collisions at high energies within the QGSM [8] or the dual parton model (DPM) [9] based on the $1/N$ expansion in QCD [10, 11]. It was shown that this approach can be applied rather successfully at not very large values of $p_t$. In this paper we investigate the open charm and beauty baryon production in $pp$ collisions at LHC energies and very small $p_t$ within the QGSM to find new information on the Regge trajectories of the heavy $(c\bar{c})$ and $(b\bar{b})$ mesons and the sea beauty quark distributions in the proton.

## 2 General Formalism for Hadron Production in pp Collision within QGSM

Let us present briefly the scheme of the analysis of the hadron production in the $pp$ collisions within the QGSM including the transverse motion of quarks and diquarks in colliding protons [12]. As is known, the cylinder type graphs for the $pp$ collision presented in Fig. 1 make the main contribution to this process [8]. The left diagram of Fig. 1, the so-called one-cylinder graph, corresponds to the case where two colourless strings are formed between the quark/diquark ($q/qq$) and the diquark/quark ($qq/q$) in colliding protons; then, after their breakup, $q\bar{q}$ pairs are created and fragmented to a hadron, for example, $D$ meson. The right diagram of Fig. 1, the so-called multicylinder graph, corresponds to creation of the same two colourless strings and many strings between sea quarks/antiquarks $q/\bar{q}$ and sea antiquarks/quarks $\bar{q}/q$ in the colliding protons. The general form for the invariant inclusive hadron spectrum within

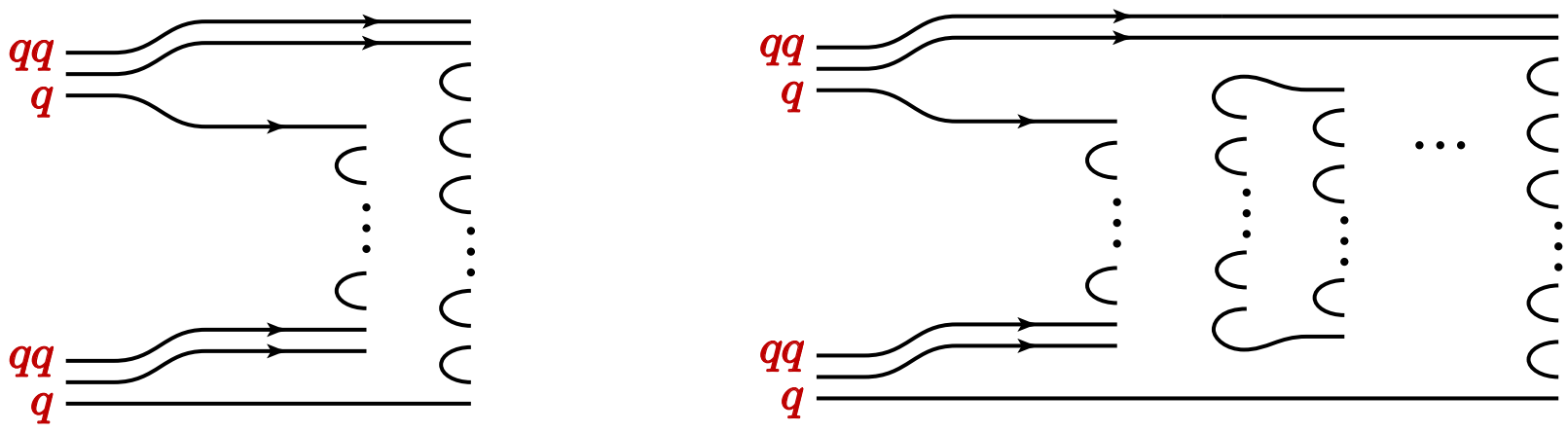

Figure 1: The one-cylinder graph (left diagram) and the multicylinder graph (right diagram) for the inclusive $pp \to hX$ process.

the QGSM is [13, 12]

$$E\frac{d\sigma}{d^3\mathbf{p}} \equiv \frac{2E^*}{\pi\sqrt{s}}\frac{d\sigma}{dxdp_t^2} = \sum_{n=1}^{\infty}\sigma_n(s)\phi_n(x,p_t) \ , \tag{1}$$

where $E, \mathbf{p}$ are the energy and the three-momentum of the produced hadron $h$ in the laboratory system (l.s.) of colliding protons; $E^*, s$ are the energy of $h$ and the square of the initial energy in the c.m.s of $pp$; $x, p_t$ are the Feynman variable and the transverse momentum of $h$; $\sigma_n$ is the cross section for production of the $n$-Pomeron chain (or $2n$ quark-antiquark strings) decaying into hadrons, calculated within the "eikonal approximation" [14]. Actually, the function $\phi_n(x,p_t)$ is the convolution of the quark (diquark) distributions in the proton and their fragmentation functions (FF), see details in [8, 9, 6, 12]. To calculate the interaction function $\phi_n(x,p_t)$ we have to know all the quark (diquark) distribution functions in the nth Pomeron chain and the FF. They are constructed within the QGSM using the knowledge of the secondary Regge trajectories, see details in [8, 13].

## 3 Heavy Baryon Production within QGSM

### 3.1 Sea Charm and Beauty Quark Distribution in the Proton

Now let us analyze the charmed and beauty baryon production in the $pp$ collision at LHC energies and very small $p_t$ within the soft QCD, e.g., the QGSM. This study can be interesting for it may allow predictions for future LHC experiments like TOTEM and ATLAS and an opportunity to find new information on the distribution of sea charmed ($c$) and beauty ($b$) quarks at very low $Q^2$. According to the QGSM, the distribution of $c(\bar{c})$ quarks in the $n$th Pomeron chain (Fig. 1, right) is, see for example [12] and references therein,

$$f^{(n)}_{c(\bar{c})}(x) = C^{(n)}_{c(\bar{c})}\delta_{c(\bar{c})}x^{a_{cn}}(1-x)^{g_{cn}} \tag{2}$$

where $a_{cn} = -\alpha_\psi(0)$, $g_{cn} = \alpha_\rho(0) - 2\alpha_B(0) + (\alpha_\rho(0) - \alpha_\psi(0)) + n - 1$; $\delta_{c(\bar{c})}$ is the weight of charmed pairs in the quark sea, $C^{(n)}_{c(\bar{c})}$ is the normalization coefficient [13], $\alpha_\psi(0)$ is the intercept of the $\psi$- Regge trajectory. Its value can be $-2.18$ assuming that this trajectory $\alpha_\psi(t)$ is linear and the intercept and the slope $\alpha'_\psi(0)$ can be determined by drawing the trajectory through the $J/\Psi$-meson mass $m_{J/\Psi} \simeq 3.1$ GeV and the $\chi$-meson mass $m_\chi = 3.554\,\text{GeV}$ [15]. Assuming that the $\psi$-Regge trajectory is nonlinear one can get $\alpha_\psi(0) \simeq 0$, which follows from perturbative QCD, as it was shown in [16]. The distribution of $b(\bar{b})$ quarks in the $n$th Pomeron chain (Fig. 1, right) has the similar form

$$f^{(n)}_{b(\bar{b})}(x) = C^{(n)}_{b(\bar{b})}\delta_{b(\bar{b})}x^{a_{bn}}(1-x)^{g_{bn}} \tag{3}$$

where $a_{bn} = -\alpha_\Upsilon(0)$, $g_{bn} = \alpha_\rho(0) - 2\alpha_B(0) + (\alpha_\rho(0) - \alpha_\Upsilon(0)) + n - 1$; $\alpha_\rho(0) = 1/2$ is the well known intercept of the $\rho$-trajectory; $\alpha_B(0) \simeq -0.5$ is the intercept of the baryon trajectory, $\alpha_\Upsilon(0))$ is the intercept of the $\Upsilon$- Regge trajectory, its value also has an uncertainty. Assuming its linearity one can get $\alpha_\Upsilon(0)) = -8, -16$, while for nonlinear ($b\bar{b}$) Regge trajectory $\alpha_\Upsilon(0) \simeq 0$, see details in [17]. Inserting these values to the form for $f^{(n)}_{c(\bar{c})}(x)$ and $f^{(n)}_{b(\bar{b})}(x)$ we get the large sensitivity for the $c$ and $b$ sea quark distributions in the $n$th Pomeron chain. Note that the FFs also depend on the parameters of these Regge trajectories. Therefore, the knowledge of the intercepts and slopes of the heavy-meson Regge trajectories is very important for the theoretical analysis of open charm and beauty production in hadron processes.

Note that all the quark distributions obtained within the QGSM are different from the parton distributions obtained within the perturbative QCD which are usually compared with the experimental data on the deep inelastic lepton scattering (DIS) off protons. To match these two kinds of quark distributions one can apply the procedure suggested in [18]. The quantities $g_{cn}$ or $g_{bn}$ entering into Eq. (2) and Eq. (3) are replaced by the following new quantities depending on $Q^2$

$$\tilde{g}_{cn} = g_{cn}\left(1 + \frac{Q^2}{Q^2 + c}\right) \; ; \; \tilde{g}_{bn} = g_{bn}\left(1 + \frac{Q^2}{Q^2 + d}\right) \tag{4}$$

The parameters $c$ and $d$ are chosen such that the structure function constructed from the valence and sea quark (antiquark) distributions in the proton should be the same as the one at the initial conditions at $Q^2 = Q_0^2$ for the perturbative QCD evolution. A similar procedure can be used to get the $Q^2$ dependence for the powers $a_{cn}$ and $a_{bn}$ entering into Eqs. (2) and (3) [18]. Then, using the DGLAP evolution equation [19], we obtain the structure functions at large $Q^2$.

## 3.2 Charmed and Beauty Baryon Production in pp Collision

The information on the charmonium ($c\bar{c}$) and bottomonium ($b\bar{b}$) Regge trajectories can be found from the experimental data on the charmed and beauty baryon production in $pp$ collisions at high energies. For example, Fig. 2 illustrates the sensitivity of the inclusive spectrum $d\sigma/dx$ of the produced charmed baryons $\Lambda_c$ to different values for $\alpha_\psi(0)$. The solid line corresponds to $\alpha_\psi(0) = 0$, whereas the dashed curve corresponds to $\alpha_\psi(0) = -2.18$. Unfortunately the experimental data presented in Fig. 2 have big uncertainties; therefore, one cannot extract the information on the $\alpha_\psi(0)$ values from the existing experimental data. A high sensitivity of the inclusive spectrum $d\sigma/dx$ of the produced beauty baryons $\Lambda_b$ to different values for $\alpha_\Upsilon(0)$ is presented in Fig. 3 (left). The $p_t$-inclusive spectrum of $\Lambda_b$ has much lower sensitivity to this quantity, according to the results presented in Fig. 3 (right). Actually, our results presented in Fig. 3 could be considered as some predictions for future experiments at LHC, see Fig. 4.

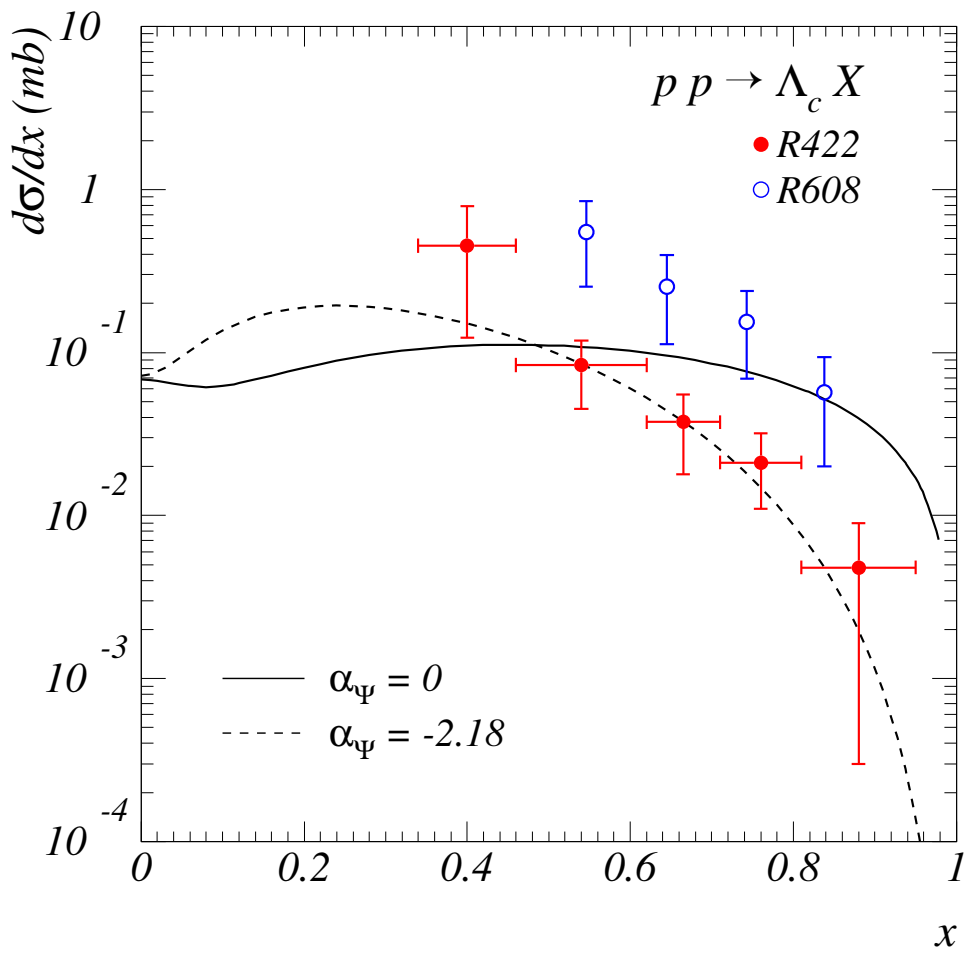

Figure 2: The differential cross section $d\sigma/dx$ for the inclusive process $pp \to \Lambda_c X$ at $\sqrt{s} =$ 62 GeV.

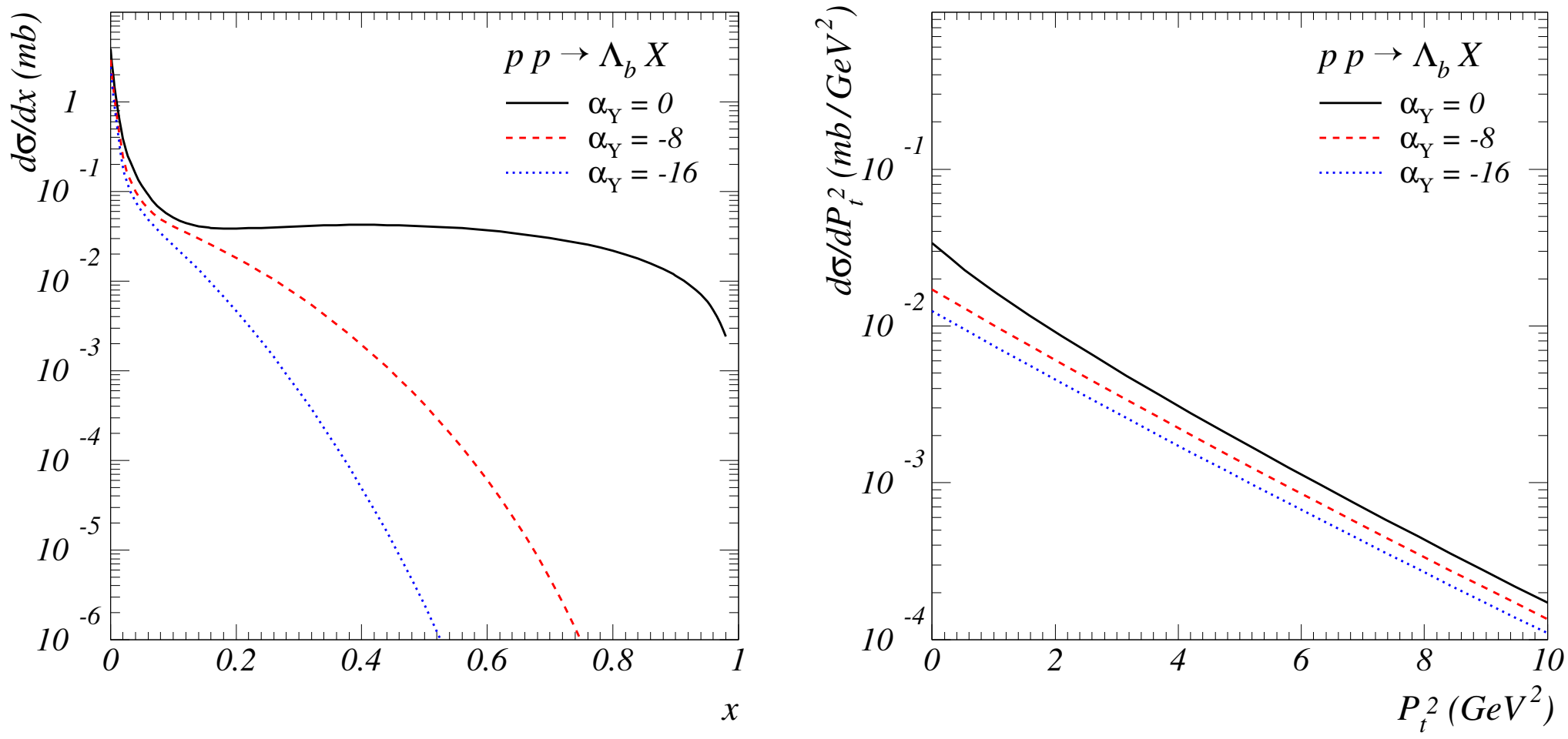

Figure 3: The differential cross section $d\sigma/dx$ (left) and $d\sigma/dP_t^2$ (right) for the inclusive process $pp \to \Lambda_b X$ at $\sqrt{s} = 4$ TeV.

Now let us analyze the production of the beauty hyperon, namely $\Lambda_b^0$, at small scattering angles $\theta_{\Lambda_b^0}$ in the $pp$ collision at LHC energies. This study would be reliable for the future forward experiments at LHC. The produced $\Lambda_b^0$ baryon can decay as $\Lambda_b^0 \to J/\Psi\Lambda^0$, and $J/\Psi$ decays into $\mu^+\mu^-$, its branching ratio ($Br = \Gamma_j/\Gamma$) is $5.93 \pm 0.06$ percent, or into $e^+e^-$ ($Br = 5.93 \pm 0.06\%$), whereas $\Lambda^0$ can decay into $p\pi^-$ ($Br = \Gamma_j/\Gamma = 63.9 \pm 05\%$), or into $n\pi^0$ ($Br = 35.8 \pm 0.5\%$), see Fig. 4.

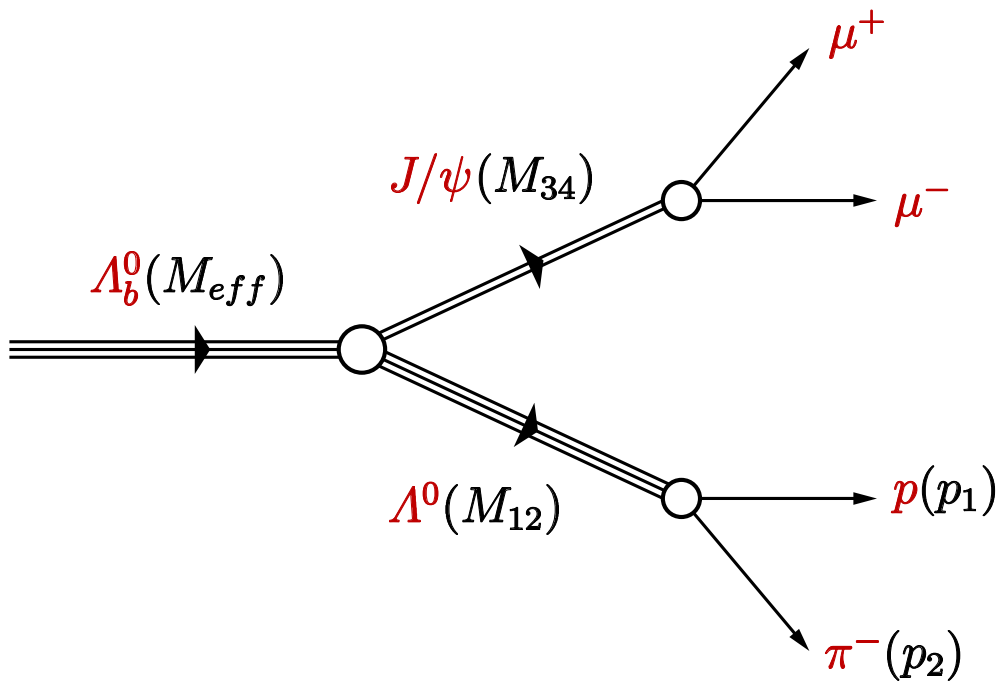

Figure 4: The decay $\Lambda_b \to J\Psi\ \Lambda^0 \to \mu^+\mu^-(e^+e^-)\ p\pi^-(n\pi^0)$.

$$\frac{d\sigma}{d^3p_1 dM_{34}} = \int \frac{d\sigma}{dM_{12}dM_{34}} \delta^{(3)}(\mathbf{p}_1 + \mathbf{p}_2 - \mathbf{p}_{12})dM_{12} \ , \tag{5}$$

where

$$\frac{d\sigma}{dM_{12}dM_{34}} = \int d^2p_{t\Lambda_b} \frac{d\sigma_{pp\to\Lambda_b X}}{dxd^2p_{t\Lambda_b}} Br_{\Lambda_b\to J/\Psi} Br_{J/\Psi\to\mu^+\mu_-} Br_{\Lambda^0\to p\pi} \frac{\pi^3}{2M_{eff}^2 M_{12}M_{34}} \lambda^{1/2}(M_{eff}^2, M_{12}^2, M_{34}^2)\lambda^{1/2}(M_{12}^2, M_1^2, M_2^2)\lambda^{1/2}(M_{34}^2, M_3^2, M_4^2) \ ,$$

$Br_{\Lambda_b\to J/\Psi} = (4.7 \pm 2.8)\cdot 10^{-4}$; $Br_{J/\Psi\to\mu^+\mu_-} = (5.93 \pm 0.06)\%$; $Br_{\Lambda^0\to p\pi} = (63.9 \pm 0.5)\%$.
Here $\lambda(x^2, y^2, z^2) = ((x^2 - (y+z)^2)((x^2 - (y-z)^2)$
One can get the following relation

$$d^3p_1 = \frac{1}{2}p\xi_p d\phi_1 d\xi_p dt_p \ , \tag{6}$$

where $\xi_p = \Delta p/p$ is the energy loss, $t_p = (p_{in} - p_1)^2$ is the four-momentum transfer, $\phi_1$ is the azimuthal angle of the final proton with the three-momentum $\mathbf{p}_1$.
Experimentally one can measure the differential cross section

$$\frac{d\sigma}{d\xi_p dt_p dM_{J/\Psi}} = \frac{1}{2}p\xi_p \int \frac{d\sigma}{d^3p_1 dM_{34}} d\phi_1 \tag{7}$$

This distribution could be reliable for the TOTEM experiment, where $J/\Psi$ decays into $\mu^+\mu^-$ and $\Lambda_b^0$ decays into $\pi^- p$ or for the ATLAS forward experiment, where $\Lambda_b^0$ decays as $\Lambda_b^0 \to J/\Psi\ \Lambda^0 \to e^+e^-\ \pi^0 n$ (Fig. 4).

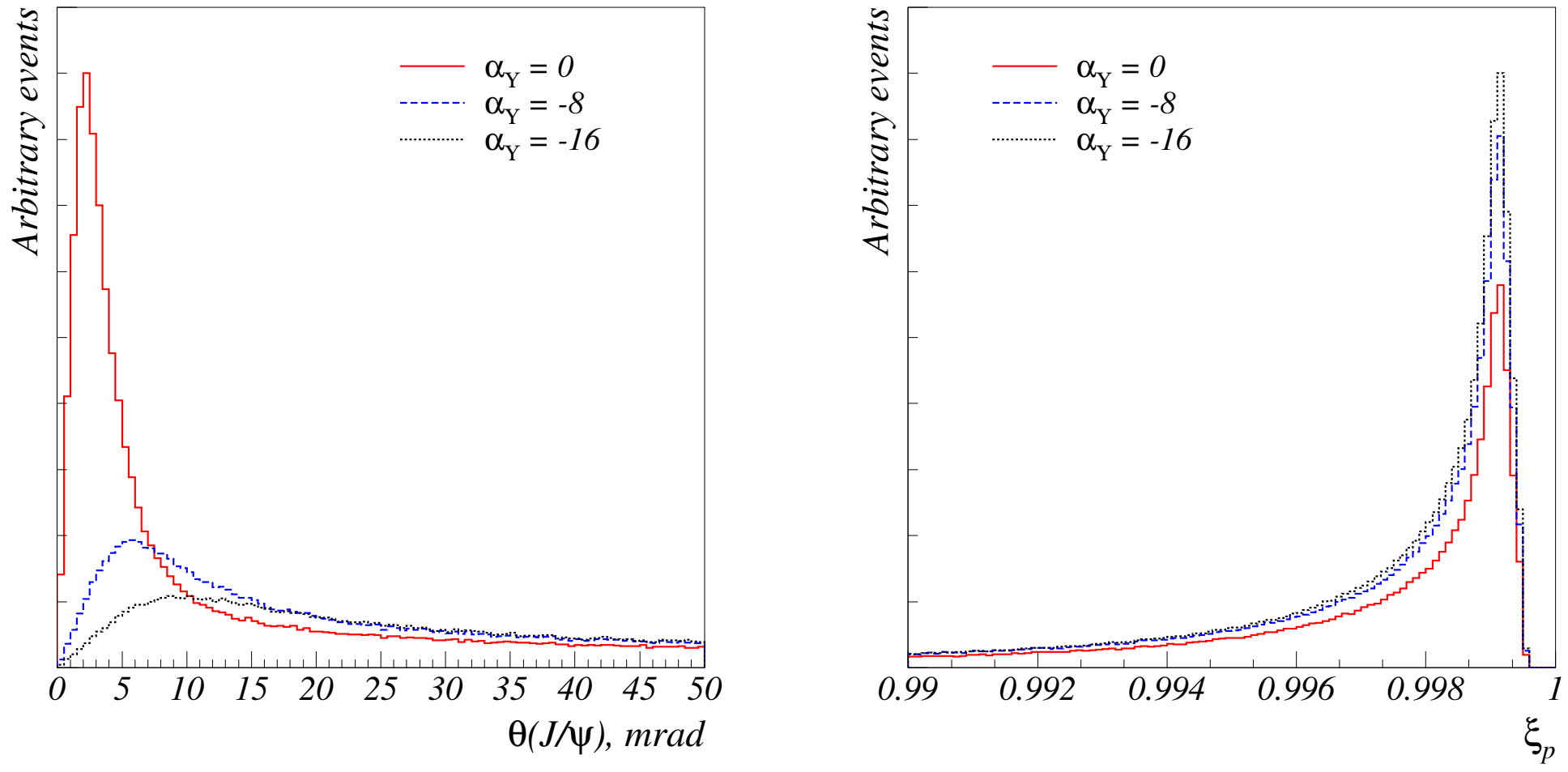

Figure 5: The distributions of $\theta_{J/\Psi}$ (left) and $\xi_p$ (right) for the inclusive process $pp \to \Lambda_b X \to \mu^+\mu^- p\pi^- X$ at $\sqrt{s} = 4\,\text{TeV}$

In Fig. 5 the distributions over $\theta_{J/\Psi}$ (left) and $\xi_p$ (right) are presented at different values of the intercept $\alpha_\Upsilon(0) = 0$ (solid line), $\alpha_\Upsilon(0) = -8$ (dashed line) and $\alpha_\Upsilon(0) = -16$ (dotted line), where $\theta_{J/\Psi}$ is the scattering angle for the final $J/\Psi$. Fig. 5 shows a sensitivity of these distributions to the intercept of the $\alpha_\Upsilon$ Regge trajectory. Actually, the result presented in Fig. 5 is a prediction for future LHC experiments on the heavy flavour baryon production at the LHC energies.

# 4 Conclusion

It was shown [6, 7] that the modified QGSM including the intrinsic longitudinal and transverse motion of quarks (antiquarks) and diquarks in colliding protons allowed us to describe rather satisfactorily the existing experimental data on inclusive spectra of heavy hadrons produced in $pp$ and $p\bar{p}$ collisions It allows us to make some predictions for future LHC forward experiments on the beauty baryon production in $pp$ collisions which can give us new information on the beauty quark distribution in the proton and very interesting information on the Regge trajectories of $(b\bar{b})$ mesons.

# Acknowledgements

We thank M. Deile, K. Eggert, D. Elia, P. Grafström, A. B. Kaidalov, A. D. Martin, M. Poghosyan and N. I. Zimin for very useful discussions. This work was supported in part by the RFBR grant N 08-02-01003.

# References

[1] A.V. Efremov, Yad. Fiz. **19** 179 (1974).

[2] R.D. Field and R.P. Feyman, Phys. Rev. **D15** 2590 (1977); R.D. Field, R.P. Feyman and G.C. Fox, Nucl. Phys. **B128** 1 (1977).

[3] V.A. Bednyakov V.A., Mod.Phys.Lett. **A10** 61 (1995).

[4] P. Nasson, S. Dawson and R.K. Ellis, Nucl. Phys. **B303** 607 (1988); *ibid.* **B327** 49 (1989); *ibid.* **B335** 260E (1989).

[5] B.A. Kniehl and G. Kramer, Phys.Rev. **D60** 014006 (1999).

[6] G.I. Lykasov, Z.M. Karpova, M.N. Sergeenko and V.A. Bednyakov, Europhys. Lett. **86** 61001 (2009); arXiv:hep-ph/0812.3220 (2009).

[7] G.I. Lykasov, V.V. Lyubushkin and V.A. Bednyakov, arXiv:hep-ph/0909.5061 (2009).

[8] A.B. Kaidalov, Phys. Lett. **B116** 459 (1982); A.B. Kaidalov and K. A. Ter-Martirosyan, Phys. Lett. **B117** 247 (1982).

[9] A. Capella, U. Sukhatme, C. I. Tan, J. Tran Than Van, Phys. Rep. **236** 225 (1994).

[10] G. t'Hooft, Nucl.Phys. **B72** 461 (1974).

[11] G. Veneziano, Phys. Lett. **B52** 220 (1974).

[12] G.I. Lykasov, G.H. Arakelian and M.N. Sergeenko, Phys. Part. Nucl. **30** 343 (1999); G.I. Lykasov, M.N. Sergeenko, Z. Phys. **C70** 455 (1996).

[13] A.B. Kaidalov and O.I. Piskunova, Z. Phys. **C30** 145 (1986).

[14] K.A. Ter-Martirosyan, Phys. Lett. B44 (1973) 377.

[15] K.G. Boreskov, A.B. Kaidalov, Sov. J. Nucl. Phys. **37** 100 (1983).

[16] A.B. Kaidalov, O.I. Piskunova, Sov. J. Nucl. Phys. **43** 994 (1986).

[17] O.I Piskunova, Yad. Fiz. **56** 176 (1993) (Phys. Atom. Nucl. **56** 1094 (1993); *ibid* **64** 392 (2001).

[18] A. Capella, A.B. Kaidalov, C. Merino and J. Tran Thanh Van, Phys. Lett. **B337** 358(1994); *ibid* **B343** 403 (1995).

[19] V.N. Gribov and L.N. Lipatov, Sov. J. Nucl. Phys. **15** 438 (1972) ; G. Altarelli and G. Parisi, Nucl. Phys. **B126** 298 (1977); Yu.L. Dokshitzer, Sov. Phys. JETP **46** 641 (1977).

# Initial-State Interactions in Drell-Yan Processes at Hadron Collisions

*Dae Sung Hwang*

Department of Physics, Sejong University, Seoul 143–747, South Korea

Single-spin asymmetry (SSA) phenomena in hadron physics are studied. The SSA in semi-inclusive DIS is understood by the final-state interactions from gluon exchange between the outgoing quark and the target spectator system. The asymmetry of the angular distribution in Drell-Yan processes is investigated by the SSA of the quark spin which is induced by the initial-state interactions.

## 1 Introduction

Since the observation of large transverse polarization of produced $\Lambda$ hyperons in the inclusive reactions $pp \to \Lambda^{\uparrow} X$ [1] and $p\,Be \to \Lambda^{\uparrow} X$ [2] in the middle of the 1970's, there have been many experimental and theoretical investigations aimed at understanding this striking polarization phenomenon [3, 4], which is called the single-spin asymmetry (SSA). The possibility of measuring the $\Lambda$ polarization at LHC was also studied [5]. SSAs in hadronic reactions have been among the most attractive phenomena to understand from basic principles in QCD. The problem has became more acute because of the observations in the semi-inclusive DIS by the HERMES [6, 7] collaboration of a strong correlation between the target proton spin $\vec{S}_p$ and the plane of the produced pion and virtual photon in semi-inclusive deep inelastic lepton scattering $\ell p^{\uparrow} \to \ell' \pi X$ at photon virtuality as large as $Q^2 = 6$ GeV$^2$. Large azimuthal single-spin asymmetries have also been seen in hadronic reactions such as $\bar{p} p^{\uparrow} \to \pi X$ [8, 9, 10], where the target antiproton is polarized normal to the pion production plane.

It was found in Ref. [11] that the final-state interaction of quark and gluon induces the single-spin asymmetry in the semi-inclusive deep inelastic scattering at the twist-two level. In Ref. [11] Brodsky, Schmidt and I calculated the single-spin asymmetry in semi-inclusive electroproduction $\gamma^* p \to HX$ induced by final-state interactions in a model of a spin-$\frac{1}{2}$ proton with mass $M$ composed of charged spin-$\frac{1}{2}$ and spin-0 constituents with respective mass $m$ and $\lambda$, which is a QCD-motivated quark-scalar diquark model of a nucleon. The basic electroproduction reaction is then $\gamma^* p \to q(qq)_0$, as illustrated in Fig. 1. Then, this time-odd twist-two effect was interpreted as the Sivers effect [12] by finding that the final-state interaction can be treated as the source of the time-odd Sivers distribution function [13, 14, 15, 16, 17, 18]. It is also often referred to as "naively $T$-odd", because the appearance of this function does not imply a violation of time-reversal invariance, since they can arise through the final-state interactions. With these developments, the existence of the Sivers distribution function has gained a firm theoretical support. The Sivers distribution function $f_{1T}^{\perp}$ describes the difference between the momentum distributions of quarks inside the nucleon transversely polarized in opposite directions. There is another quark distribution function of the nucleon induced by the final-state interaction of quark and gluon, which is called the Boer-Mulders distribution function $h_1^{\perp}$. $h_1^{\perp}$ describes the difference between the momentum distributions of the quarks transversely polarized in opposite directions inside the unpolarized nucleon [19]. The distribution functions $f_{1T}^{\perp}$ and $h_1^{\perp}$ are depicted in Figs. 2 and 3.

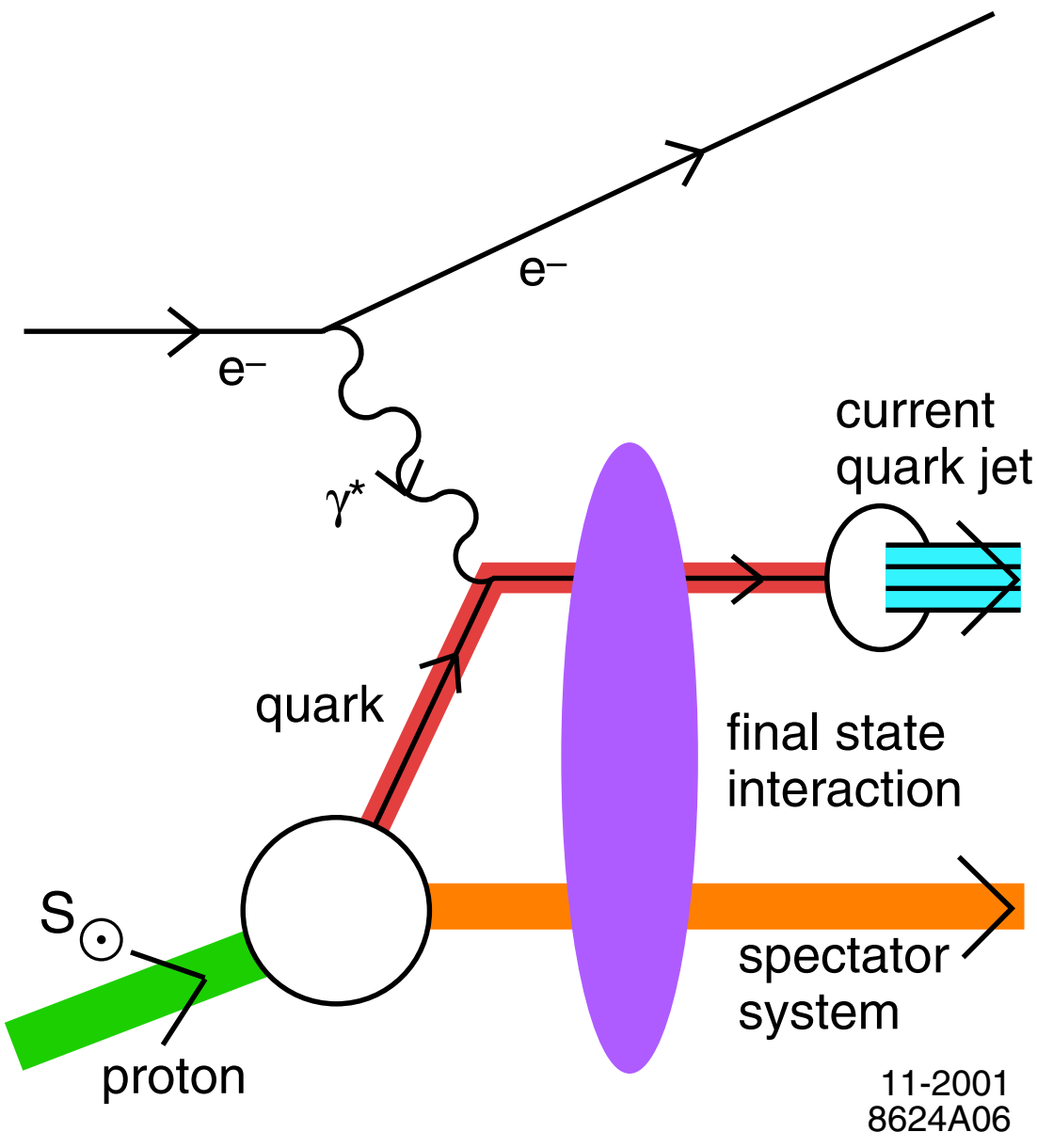

Figure 1: The final-state interaction in the semi-inclusive deep inelastic lepton scattering $\ell p^{\uparrow} \to \ell' \pi X$.

It has been shown that initial-state interactions contribute to the $\cos 2\phi$ distribution in unpolarized Drell-Yan lepton pair production $p\,p$ and $p\,\overline{p} \to \ell^+\ell^- X$, without suppression [17, 20, 21]. The asymmetry is expressed as a product of chiral-odd distributions $h_1^{\perp}(x_1, \boldsymbol{p}_\perp^2) \times \overline{h}_1^{\perp}(x_2, \boldsymbol{k}_\perp^2)$, where the quark-transversity function $h_1^{\perp}(x, \boldsymbol{p}_\perp^2)$ is the transverse momentum dependent, light-cone momentum distribution of transversely polarized quarks in an unpolarized proton. This (naive) $T$-odd and chiral-odd distribution function and the resulting $\cos 2\phi$ asymmetry were computed explicitly in a quark-scalar diquark model for the proton with initial-state gluon interaction in Ref. [17]. In this model the function $h_1^{\perp}(x, \boldsymbol{p}_\perp^2)$ equals the $T$-odd (chiral-even) Sivers effect function $f_{1T}^{\perp}(x, \boldsymbol{p}_\perp^2)$. This suggests that the single-spin asymmetries in the SIDIS and the Drell-Yan process are closely related to the $\cos 2\phi$ asymmetry of the unpolarized Drell-Yan process, since all can arise from the same underlying mechanism. This provides new insight regarding the role of quark and gluon orbital angular momentum as well as that of initial- and final-state gluon exchange interactions in hard QCD processes.

The light-cone wave functions are useful for studying the hadronic processes by treating the non-perturbative effects in a relativistically covariant way [22, 23, 24, 25, 26]. Here we calculate the Sivers and Boer-Mulders distribution functions by using their light-cone wave function representations. Then, we calculate the SSA in semi-inclusive DIS and the asymmetry of the angular distribution in the Drell-Yan process using the Sivers and the Boer-Mulders

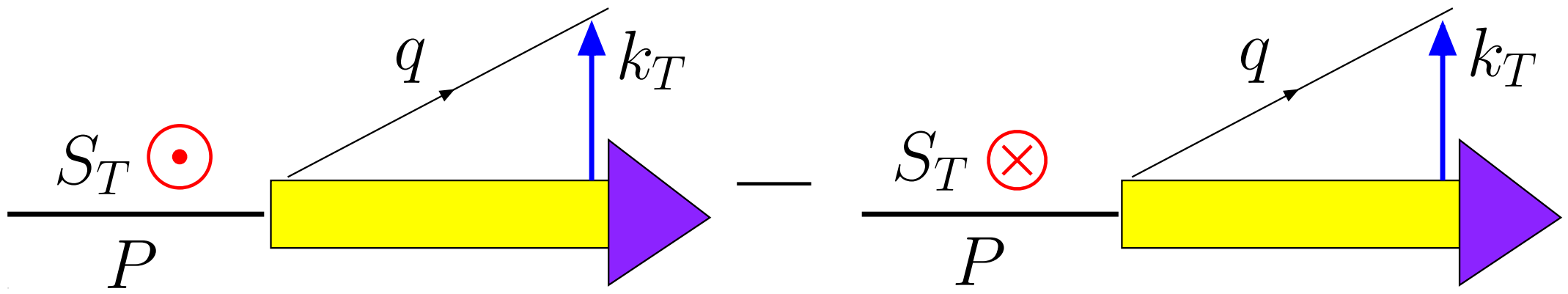

Figure 2: Schematic depiction of the Sivers distribution function $f_{1T}^{\perp}$. The spin vector $S_T$ of the nucleon points out of and into the page, respectively, and $k_T$ is the transverse momentum of the extracted quark.

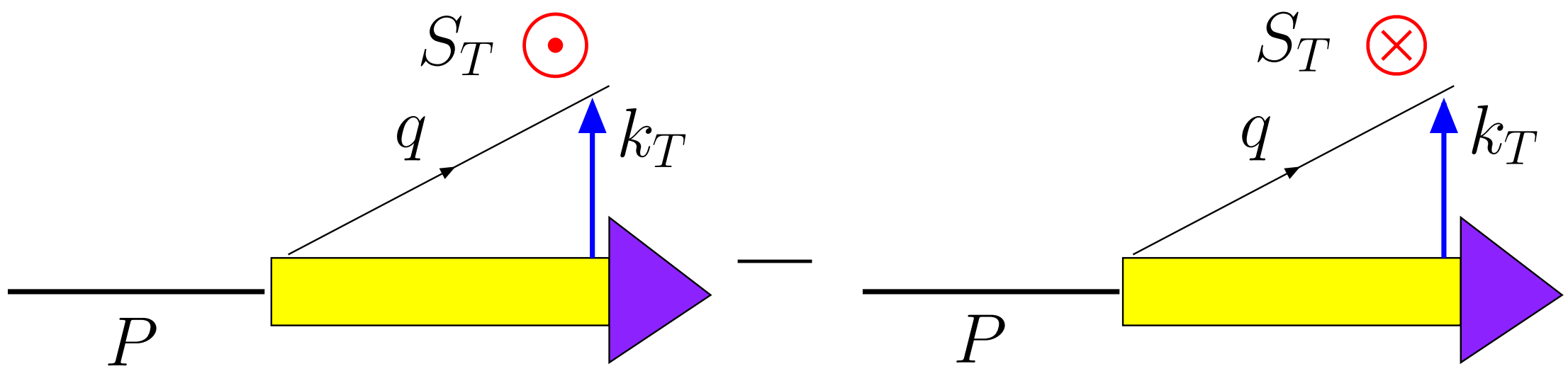

Figure 3: Schematic depiction of the Boer-Mulders distribution function $h_1^{\perp}$. The spin vector $S_T$ of the quark points out of and into the page, respectively, and $k_T$ is the transverse momentum of the extracted quark.

distribution functions.

## 2 Model Calculation of Sivers and Boer-Mulders Functions with Scalar Diquark Model

The final-state interactions in semi-inclusive deep inelastic scattering are commonly treated as a part of the proton distribution function [13, 17]. If we adopt the same treatment for the light-cone wave functions, we can consider that the final-state interactions for the scalar diquark model depicted in Fig. 4 induce the spin-dependent complex phases to the light-cone wave functions:

$$\begin{cases} \psi^{\uparrow}_{+\frac{1}{2}}(x,\vec{k}_{\perp}) = \frac{(m+xM)}{x}\left(1+ia_1\right)\varphi\ , \\ \psi^{\uparrow}_{-\frac{1}{2}}(x,\vec{k}_{\perp}) = -\frac{(+k^1+ik^2)}{x}\left(1+ia_2\right)\varphi\ , \end{cases} \tag{1}$$

$$\begin{cases} \psi^{\downarrow}_{+\frac{1}{2}}(x,\vec{k}_{\perp}) = -\frac{(-k^1+ik^2)}{x}\left(1+ia_2\right)\varphi\ , \\ \psi^{\downarrow}_{-\frac{1}{2}}(x,\vec{k}_{\perp}) = \frac{(m+xM)}{x}\left(1+ia_1\right)\varphi\ , \end{cases} \tag{2}$$

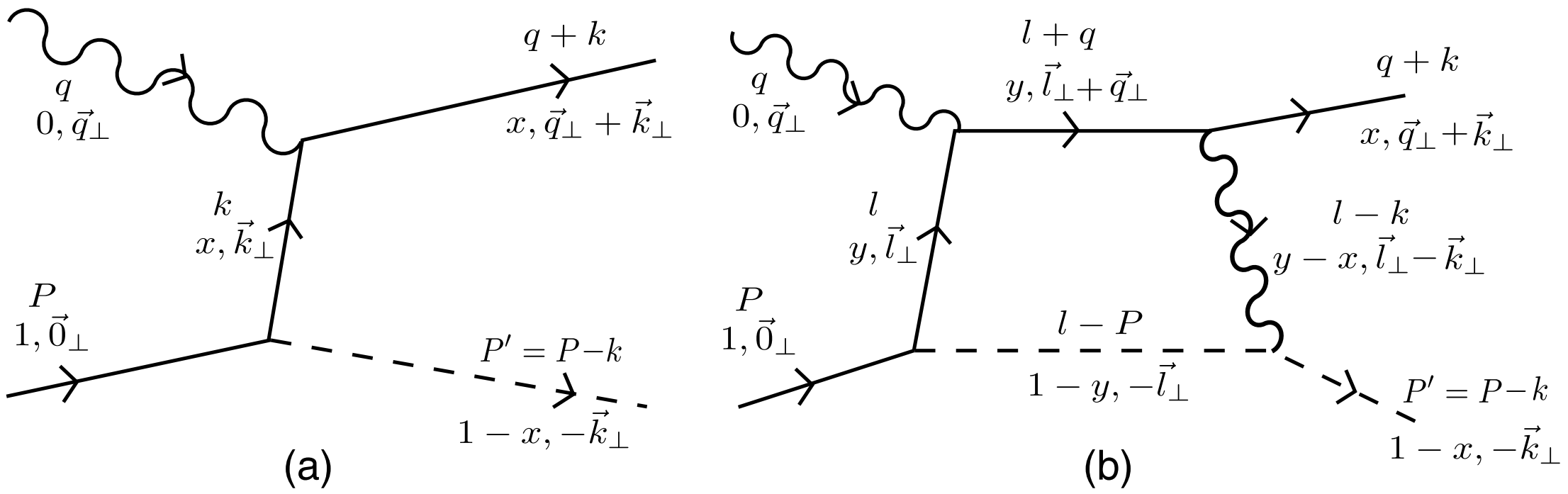

Figure 4: (a) Tree level diagram and (b) diagram with final-state interaction.

where $\varphi = \varphi(x, \vec{k}_\perp) = -g\, x\sqrt{1-x}/(\vec{k}_\perp^2 + B)$ with the nucleon-quark-diquark coupling constant $g$ and $B = -x(1-x)M^2 + (1-x)m^2 + x\lambda^2$, and $a_1$ and $a_2$ are given by

$$a_{1,2} = \frac{e_1 e_2}{8\pi}\ (\vec{k}_\perp^2 + B)\ g_{1,2} \tag{3}$$

with [11]

$$g_1 = \int_0^1 d\alpha\ \frac{-1}{\alpha(1-\alpha)\vec{k}_\perp^2 + \alpha\lambda_g^2 + (1-\alpha)B}\ , \qquad g_2 = \int_0^1 d\alpha\ \frac{-\alpha}{\alpha(1-\alpha)\vec{k}_\perp^2 + \alpha\lambda_g^2 + (1-\alpha)B}\ . \tag{4}$$

In the above, $e_1$ and $e_2$ are the quark and diquark charge, and $M$, $m$, $\lambda$ and $\lambda_g$ are the nucleon, quark, diquark and gluon mass, respectively. We take $\lambda_g = 0$ at the end of the calculation. Our analysis can be generalized to the corresponding calculation in QCD. The final-state interaction from gluon exchange has the strength $\frac{e_1 e_2}{4\pi} \to C_F \alpha_s(\mu^2)$.

Using the wave functions (1) and (2) in the light-cone wave function representations of Sivers and Boer-Mulders functions presented in Ref. [27], we obtain [11, 17, 27, 28]

$$f_1(x, \vec{k}_\perp) = \frac{1}{16\pi^3}\ \Big[(M + \frac{m}{x})^2 + \frac{\vec{k}_\perp^2}{x^2}\Big]\varphi^2\ , \tag{5}$$

$$f_{1T}^\perp(x, \vec{k}_\perp) = \frac{1}{16\pi^3}\ 2\ \frac{M}{x}\ (M + \frac{m}{x})\ \varphi^2\ \frac{e_1 e_2}{8\pi}\ (\vec{k}_\perp^2 + B)\ \frac{1}{\vec{k}_\perp^2}\ \ln\frac{(\vec{k}_\perp^2 + B)}{B}\ , \tag{6}$$

$$h_1^\perp(x, \vec{k}_\perp) = \frac{1}{16\pi^3}\ 2\ \frac{M}{x}\ (M + \frac{m}{x})\ \varphi^2\ \frac{e_1 e_2}{8\pi}\ (\vec{k}_\perp^2 + B)\ \frac{1}{\vec{k}_\perp^2}\ \ln\frac{(\vec{k}_\perp^2 + B)}{B}\ . \tag{7}$$

# 3 Semi-Inclusive DIS

The SSA in the semi-inclusive DIS (SIDIS) was calculated in Refs. [11, 17] with the light-cone wave functions given in the previous section. The formula for the SSA in the SIDIS is given by $\mathcal{P}_y = -(r_\perp^1/M)\,(f_{1T}^\perp(x, \boldsymbol{r}_\perp)/f_1(x, \boldsymbol{r}_\perp))$ [11, 17], which gives the results presented in Fig. 5 [11].

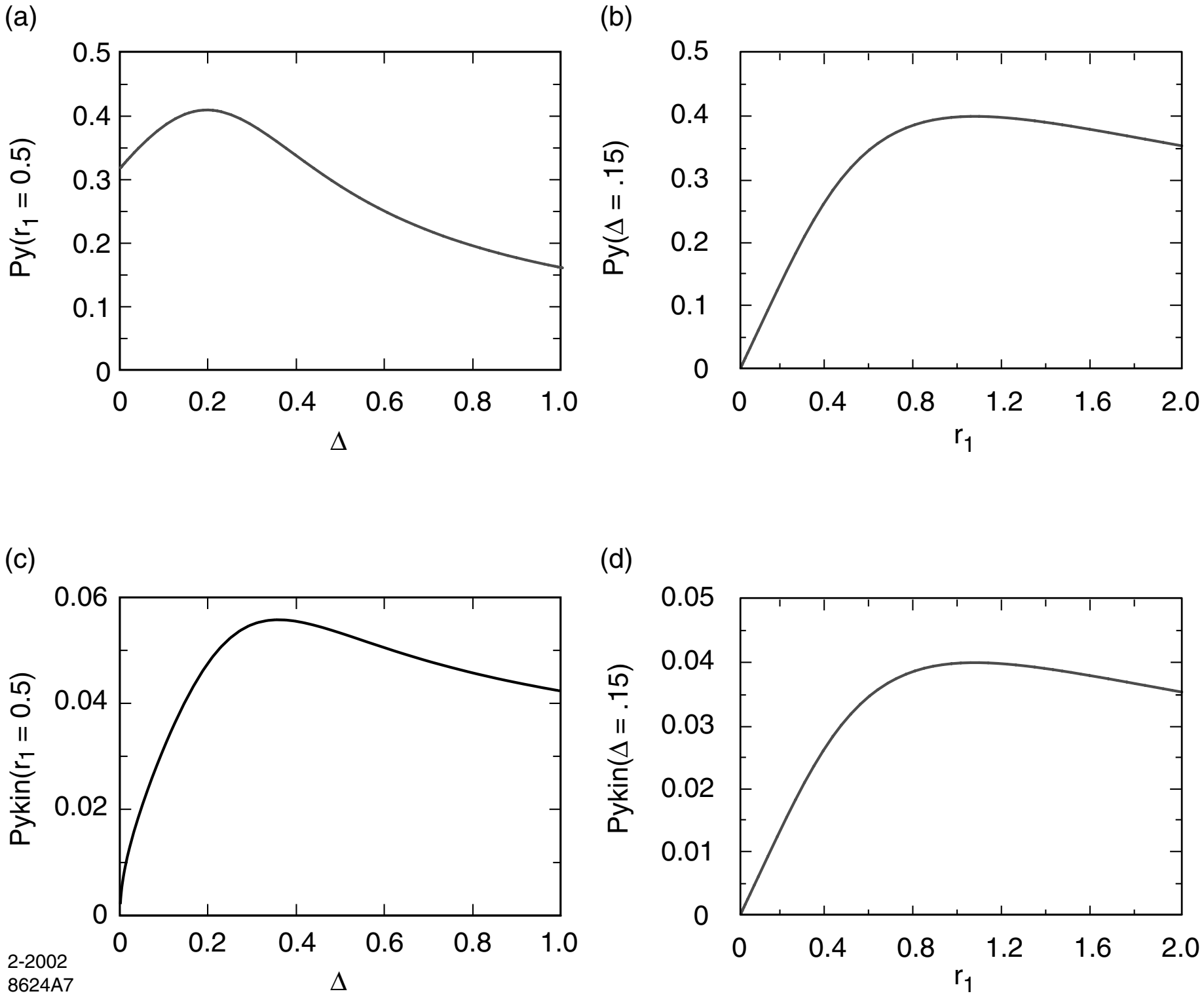

Figure 5: Model predictions for the single spin asymmetry of the proton in electroproduction resulting from gluon exchange in the final state as a function of $\Delta = x_{bj}$ and quark transverse momentum $r_\perp$. The parameters are given in the text of Ref. [11].

# 4 Drell-Yan Process

The unpolarized Drell-Yan process cross section has been measured in pion-nucleon scattering: $\pi^- N \to \mu^+ \mu^- X$, with $N$ deuterium or tungsten and a $\pi^-$-beam with energy of 140, 194, 286 GeV [29] and 252 GeV [30]. Conventionally the differential cross section is written as

$$\frac{1}{\sigma}\frac{d\sigma}{d\Omega} = \frac{3}{4\pi}\,\frac{1}{\lambda+3}\,\left(1+\lambda\cos^2\theta+\mu\sin^2\theta\cos\phi+\frac{\nu}{2}\sin^2\theta\cos 2\phi\right)\;, \tag{8}$$

where the angles $\theta$ and $\phi$ are defined, for example, in Fig. 1 of Ref. [17]. These angular dependencies can all be generated by perturbative QCD corrections, where for instance initial quarks radiate off high energy gluons into the final state. Such a perturbative QCD calculation at next-to-leading order leads to $\lambda \approx 1, \mu \approx 0, \nu \approx 0$ at very small transverse momentum of the lepton pair. More generally, the Lam-Tung relation $1-\lambda-2\nu = 0$ [31] is expected to hold at order $\alpha_s$ and the relation is hardly modified by next-to-leading order $(\alpha_s^2)$ perturbative QCD corrections [32]. However, this relation is not satisfied by the experimental data [29, 30]. The Drell-Yan data shows remarkably large values of $\nu$, reaching values of about 30% at transverse momenta of the lepton pair between 2 and 3 GeV (for $Q^2 = m^2_{\gamma^*} = (4-12\,\mathrm{GeV})^2$ and extracted in the Collins-Soper frame [33] to be discussed below). These large values of $\nu$ are not compatible with $\lambda \approx 1$ as also seen in the data.

The asymmetry given by $\nu$ in Eq. (8) is proportional to the product of chiral-odd distributions $h_1^\perp(x_1, \boldsymbol{p}_\perp^2) \times \overline{h}_1^\perp(x_2, \boldsymbol{k}_\perp^2)$ [20]. The parameter $\nu$ was estimated in Refs. [17, 20, 21], and here we present in Fig. 6 the result of Ref. [21], which was obtained with the Gaussian transverse momentum dependence.

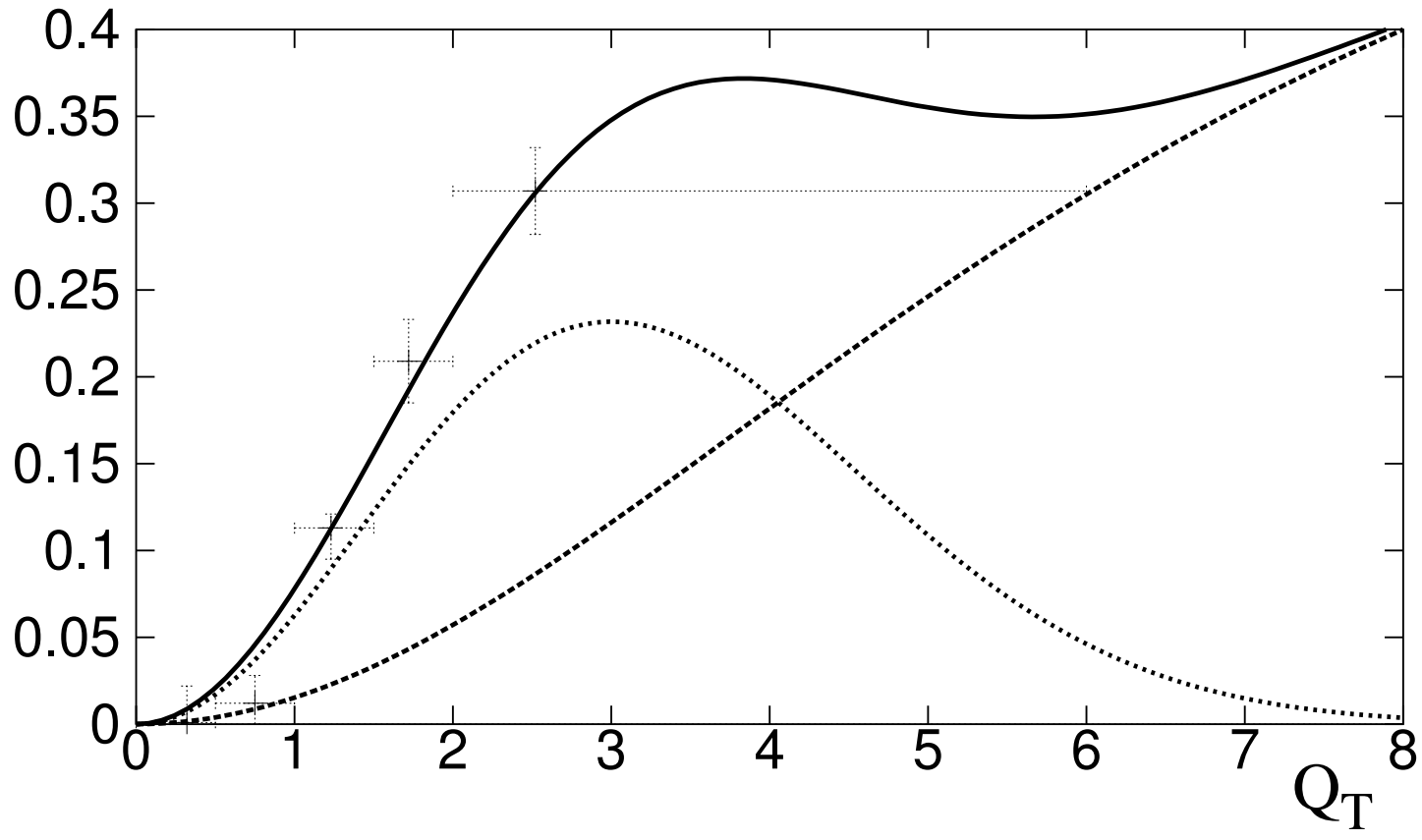

Figure 6: Possible contributions to $\nu$ as function of $Q_T$ compared to DY data of NA10 (for $Q =$ 8 GeV), which was presented in Ref. [21]. Thick dotted curve: contribution from perturbative one-gluon radiation. Thin dotted curve: contribution from a nonzero $h_1^\perp$. Solid curve: their sum.

# 5 Conclusion

Single-spin asymmetries in hadron reactions have been mysterious since the discovery of large transverse polarization of $\Lambda$ hyperons. The SSA in semi-inclusive DIS is understood by the final-state interactions from gluon exchange between the outgoing quark and the target spectator system. The asymmetry of the angular distribution in the Drell-Yan process given by the $\cos 2\phi$ distribution is investigated by the SSA of the quark spin which is induced by the initial-state interactions. This approach could explain the asymmetry of the angular distribution measured by the NA10 collaboration. It would be interesting to study this asymmetry in the Drell-Yan process further at hadron collider experiments like RHIC and LHC.

# Acknowledgements

I thank Daniel Boer, Stan Brodsky, Matthias Burkardt, Bo-Qiang Ma, Piet Mulders, and Ivan Schmidt for the collaborations on the subjects presented here. This work was supported in part by the International Cooperation Program of the KICOS (Korea Foundation for International Cooperation of Science & Technology), and in part by the Korea Research Foundation Grant funded by the Korean Government (KRF-2008-313-C00166).

# References

[1] A. Lesnik *et al.*, Phys. Rev. Lett. **35** 770 (1975).

[2] G. Bunce *et al.*, Phys. Rev. Lett. **36** 1113 (1976).

[3] For a review of experimental data, see, *e.g.*, K. Heller, in the Proceedings of Spin'96, edited by C.W. de Jager *et al.* (World Scientific, 1997); or A. D. Panagiotou, Int. J. Mod. Phys. **A5** 1197 (1990).

[4] For a review of theoretical models, see, *e.g.*, J. Felix, Mod. Phys. Lett. **A14** 827 (1999).

[5] D. Boer, C.J. Bomhof, D.S. Hwang, and P.J. Mulders, Phys. Lett. **B659** 127 (2008).

[6] HERMES Collaboration, A. Airapetian et al., Phys. Rev. Lett. **84** 4047 (2000).

[7] A. Airapetian *et al.* [HERMES Collaboration], Phys. Rev. Lett. **94** 012002 (2005).

[8] FNAL-E704 Collaboration, D.L. Adams et al., Phys. Lett. **B264** 462 (1991).

[9] STAR Collaboration, J. Adams et al., Phys. Rev. Lett. **92** 171801 (2004).

[10] STAR Collaboration, B.I. Abelev et al., Phys. Rev. Lett. **101** 222001 (2004).

[11] S.J. Brodsky, D.S. Hwang, and I. Schmidt, Phys. Lett. **B530** 99 (2002).

[12] D. Sivers, Phys. Rev. **D41** 83 (1990); Phys. Rev. **D43** 261 (1991).

[13] J.C. Collins, Phys. Lett. **B536** 43 (2002).

[14] S.J. Brodsky, D.S. Hwang and I. Schmidt, Nucl. Phys. **B642** 344 (2002).

[15] X. Ji and F. Yuan, Phys. Lett. **B543** 66 (2002).

[16] A. Belitsky, X. Ji, and F. Yuan, Nucl. Phys. **B656** 165 (2003).

[17] D. Boer, S.J. Brodsky, and D.S. Hwang, Phys. Rev. **D67** 054003 (2003).

[18] D. Boer, P.J. Mulders, and F. Pijlman, Nucl. Phys. **B667** 201 (2003).

[19] D. Boer and P. J. Mulders, Phys. Rev. **D57** 5780 (1998).

[20] D. Boer, Phys. Rev. **D60** 014012 (1999).

[21] D. Boer, in the Proceedings of "Transversity 2005", Como, Italy, 7-10 Sep. 2005; hep-ph/0511025.

[22] G.P. Lepage and S.J. Brodsky, Phys. Rev. **D22** 2157 (1980).

[23] S.J. Brodsky, H.C. Pauli, and S.S. Pinsky, Phys. Rep. **301** 299 (1998).

[24] S.J. Brodsky and S.D. Drell, Phys. Rev. **D22** 2236 (1980).

[25] S.J. Brodsky and D.S. Hwang, Nucl. Phys. **B543** 239 (1999).

[26] S.J. Brodsky, D.S. Hwang, B.-Q. Ma, and I. Schmidt, Nucl. Phys. **B593** 311 (2001).

[27] D.S. Hwang, in the Proceedings of "Light Cone 2008", Mulhouse, France, 7-11 July 2008; PoS LC2008: 033, 2008.

[28] M. Burkardt and D.S. Hwang, Phys. Rev. **D69** 074032 (2004).

[29] NA10 Collaboration, S. Falciano *et al.*, Z. Phys. **C31** 513 (1986);
NA10 Collaboration, M. Guanziroli *et al.*, Z. Phys. **C37** 545 (1988).

[30] J.S. Conway *et al.*, Phys. Rev. **D39** 92 (1989).

[31] C.S. Lam and W.K. Tung, Phys. Rev. **D21** 2712 (1980).

[32] A. Brandenburg, O. Nachtmann and E. Mirkes, Z. Phys. **C60** 697 (1993).

[33] J.C. Collins and D.E. Soper, Phys. Rev. **D16** 2219 (1977).

# Does High Energy Behaviour Depend on Quark Masses?

*Vladimir Petrov*[1]

[1]Division of Theoretical Physics, Institute for High Energy Physics, 142 281 Protvino, RF

Arguments based on the Renormalization Group invariance indicate the vanishing of hadron scattering amplitudes at infinite energies in massless confined QCD with mass gap in physical spectrum. It is shown also that if at least one quark is massive the RG arguments do not forbid increasing amplitudes.

As it is known for a long time the masslessness of fundamental–quark and gluon-fields in the non-Abelian gauge theory in no way precludes existence of physical excitations separated from the vacuum by a non-zero mass gap. This feature, called *dimensional transmutation*, goes back to the pioneering paper [1] and since then was being repeatedly discussed in various contexts (from relatively recent papers see, e.g. [2]). Moreover it is considered that QCD with three massless quarks reflects quite well the basic properties of hadron physics in the low-energy *light sector*. For instance, the nucleon mass practically does not change in the limit of massless quarks [3]. One can therefore believe that such a theory differs even less from the *genuine QCD* at very high energies when the role of mass is, to the great extent, negligible. For instance, the famous BFKL results [4] that give estimates for infinitely growing total cross-sections (due to the Pomeron intercept exceeding 1) were obtained in massless QCD.

## 1 Renormalization Group Argument

We are going to verify the possibility of infinitely growing cross- sections assuming existence of confined massless QCD [5] , i.e. of a still hypothetical but seemingly quite plausible theory with the physical state spectrum consisting of massive (due to *dimensional transmutation*) colourless hadrons and without quark and gluon asymptotic states. It means that the scattering amplitude as an analytic function does not possess any kind of zero-mass singularities related to elementary quark-gluon fields and has only singularities related to massive colourless hadrons. The only fundamental mass parameter $\Lambda_{\rm QCD}$ is *hidden* in the running coupling $\alpha_s(\mu^2)$. Now, the physical hadron masses $M_i$ are related with the coupling constant by the formula that reflects the renormalization invariance (everywhere below this also includes the scheme invariance) of physical quantities

$$M_i^2 = c_i \mu^2 \exp(-K(\alpha_s)),$$

$$dK(\alpha_s)/d\alpha_s = 1/\beta(\alpha_s),$$

where $c_i$ are fixed numerical parameters, $\beta(\alpha_s)$ is the Gell-Mann-Low function and $\mu$ stands for renormalization scale. Actually it means that (see, e.g., [6])

$$M_i^2 = c_i \Lambda^2_{\rm QCD}. \tag{1}$$

For definiteness, let us consider the scattering amplitude of two hadrons, $T$. This amplitude is a function of two independent Mandelstam variables, $s$ and $t$, and also of a (generally infinite) set of hadron masses, marked by parameters $c_i$ from Eq. (1):

$$T = T(s, t; \mu^2, \alpha_s; \{c_i\}).$$

Presence of quantities $\mu^2$ and $\alpha_s$ amongst the arguments of $T$ reflects the provenance of the amplitude from the fundamental Lagrangian and freedom of the choice of the normalization scale. In the same way as physical masses, the amplitude of a physical process must be renormalization invariant. The dependence on $\alpha_s$ is related to calculations based on the fundamental QCD Lagrangian while the very amplitude is, actually, a function of $s/\Lambda^2_{\rm QCD}$, $t/s$ and $\{c_i\}$, i.e the amplitude depends only on Mandelstam variables and physical hadron masses which define singularities of the amplitude in $s$- and $t$- planes. Nonetheless the genetic relation with the underlying theory leads to a non-trivial conclusion. Let us take for simplicity the case of forward scattering ($t = 0$). Due to our assumptions the amplitude is analytic (or at least finite) in this point. Our main argument is that the renormalization invariance, though does not define, but quite strongly restricts the functional dependence of the dimensionless amplitude on its dimensionful parameters. It is easy to verify that, in our case, the renormalization invariance implies the following general form of the amplitude:

$$T(s, 0; \mu^2, \alpha_s; \{c_i\}) = \Phi((s/\mu^2)\exp(K(\alpha_s)); \{c_i\}),$$

where the concrete form of $\Phi(Z; \{c_i\})$ is to be fixed by the dynamics.

We do not know it but we do know that, at any rate, if the coupling $\alpha_s$ goes to zero the amplitude goes there as well. As

$$K(\alpha_s) \sim 1/\beta_0\alpha_s + O(\log(1/\alpha_s)) \ \text{ at } \ \alpha_s \to 0,$$

it is equivalent to the statement that

$$\Phi(Z; \{c_i\}) \to 0 \ \text{ at } \ Z \to \infty.$$

Recalling that $Z = (s/\mu^2)\exp(K(\alpha_s))$, we see that the limit $Z \to \infty$ can be realised with $s \to \infty$ at $\mu$ and $\alpha_s$ fixed as well. In its turn it means that, in our theory, **the forward scattering amplitude vanishes with infinite energy growth**:

$$T(s, 0) \to 0 \ \text{ at } \ s \to \infty.$$

For example, the total cross-sections asymptotically decrease to zero:

$$\sigma_{\rm tot} \to 0 \ \ at \ \ s \to \infty.$$

Moreover,

$$\sigma_{\rm tot} < {\rm const}/s \ \ at \ \ s \to \infty.$$

Partial cross-sections have to drop even faster than the total one to compensate the growth of the number of open channels. It is not difficult to be convinced that the same conclusion takes place for the case of scattering at fixed non-zero angle ($t/s$ fixed). As to fixed $t$ we can rely on the results by Cornille and Martin [7] according to which if the even-signature amplitude dominates forward scattering then $|T(s,t)| \leq |T(s,0)|$, so in this case the amplitude asymptotically tends to zero at high energies and fixed momentum transfers as well. We have to note that the conclusion depends critically on the asymptotic freedom. If the number of fermion flavors were larger than 16 then our arguments would lead to quite an innocuous result that the amplitude vanishes at $s = 0$.

## 2 Discussion

From a purely theoretical viewpoint general principles of quantum field theory do not forbid such a phenomenon. For instance, if to assume the absence of oscillations, one can obtain the following high-energy lower bound [8]:

$$|T(s,0)| > \text{const}/(\log s)^{1/2}.$$

The result obtained above could mean the presence of oscillations in energy.

More liberal lower bound is [8]:

$$\sigma_{\text{tot}} > \text{const}/s^6(\log s)^2.$$

As to the fixed angle scattering the result seems to be fully consistent with the famous *quark counting rule* [9].

From the point of view of existing experimental data on total and differential small-angle cross-sections the result obtained here seems to be absolutely incredible: the total cross-sections grow at energies up to tens TeV, if to add the cosmic rays data. Differential cross sections of, say, *pp*-scattering at $t = 0$ which are proportional to $|T(s,0)|^2/s^2$ also grow. Certainly, the result deals with *infinite* energies but even so, the present-day data if to believe in the result we discuss - would mean that there exists some gigantic energy scale (no less than several tens TeV) from which the decreasing starts. However, the theory in question has, as was said, one fundamental scale $\Lambda_{\text{QCD}}$ that, in any case, does not surpass several hundreds MeV. It is very difficult to imagine, in the framework of this theory, a mechanism of generation of such a huge scale. This could be considered as an indication of invalidity of the theory for, at least, high-energy diffractive scattering. At the same time it does not contradict to the well known - both theoretically and experimentally - decrease of hadronic amplitudes at fixed angles.

## 3 Massive Case

If the fundamental QCD Lagrangian contains fermion mass terms then instead of one single RG invariant mass scale

$$\Lambda_1^2 = \mu^2 \exp(-K(\alpha_s))$$

we have another one (for one massive flavour):

$$\Lambda_2^2 = m^2 \exp(L(\alpha_s))$$

where

$$dL(\alpha_s)/d\alpha_s = \gamma_m(\alpha_s)/\beta(\alpha_s)$$

and $\gamma_m(\alpha_s) = -d(\ln m^2)/d\ln\mu^2$ is the *mass anomalous dimension* defined as in [6]. Here, taking into account the general scheme invariance of physical quantities, we use for simplicity the minimal renormalization scheme. Now the argument used above does not pass through. In fact, the amplitude now has the following general form

$$T(s,0) = F(s/\Lambda_1^2, \Lambda_2^2/\Lambda_1^2).$$

At infinite energy $T \to F(\infty, \Lambda_2^2/\Lambda_1^2)$, while at $\alpha_s \to 0$, $T > F(\infty, \infty) = 0$ because $\Lambda_2^2 \sim (\frac{1}{\alpha_s})^{\frac{\gamma_m}{\beta_0}}$ at $\alpha_s \to 0$. So the (massive) free-field limit is generally different from the high-energy limit and we cannot come to any definite conclusion concerning the latter.

The general conclusion of this paper is that massless QCD is not a good underlying approximation for high-energy diffractive scattering while it seems to be admissible for hard processes or for the low- energy sector of light hadrons.

The crucial importance of quark non-zero-masses for the rise of the total cross-sections seems a bit counterintuitive.

# Acknowledgements

I am grateful to Anton Godizov, Gia Japaridze, Alexandre Kisselev, André Martin, George Pronko, Roman Ryutin, Sergey Troshin and Tai Wu for useful discussions.

# References

[1] S. Coleman and E. Weinberg., Phys. Rev. **D7** 1888 (1973).

[2] L.D. Faddeev, Theoretical and Mathematical Physics. **148** 986 (2006).

[3] B.L. Ioffe, Physics-Uspekhi. **49** 1077 (2006).

[4] L.N. Lipatov, Phys. Atom. Nucl. **67** 83 (2004); Yad. Fiz. **67** 84 (2004).

[5] For the first time this issue was addressed in my e-print arXiv: hep-ph/0603103v1 (2006).

[6] J. Collins, Renormalization, Cambridge University Press (1984). For a scheme independent definition of $\Lambda_{\text{QCD}}$, see H. Sonoda, UCLA/94/TEP/41 (1994); arXiv: hep-th/941024v2 (1994). The most detailed outline of the renormalization group can be found in the book by N.N. Bogoliubov and D.V. Shirkov, Introduction to the Theory of Quantized Fields, New York: Wiley-Interscience (1959,1980).

[7] H. Cornille, A. Martin, Nucl. Phys. **B49** 413 (1972).

[8] Y.S. Jin and A. Martin, Phys. Rev.**135** B1369 (1964); see also other bounds in the review R.J. Eden, Rev. Mod. Phys. **43** 15 (1971).

[9] V.A. Matveev, R.M. Muradian, A.N. Tavkhelidze, Lett. Nuovo Cim. **7** 719 (1973); S.J. Brodsky, G.R. Farrar, Phys. Rev. Lett. **31** 1153 (1973).

# Part XI

# Past and Future of Forward Physics Experiments

# The Forward Detectors of CDF and DØ

*Konstantin Goulianos*

The Rockefeller University, 1230 York Avenue, New York, NY 10065-9965, USA

The forward detectors of CDF II are presented with emphasis on design aspects that proved crucial for carrying out a successful program on diffraction at the Tevatron. Alignment, calibrations and backgrounds are discussed, pointing out their relevance to the diffractive and central exclusive production physics programs planned at the LHC. The DØ forward detectors, which with forward spectrometer on both the $\bar{p}$ and $p$ sides offer the opportunity for a program complementarity to that of CDF are briefly presented for completeness.

## 1 Introduction

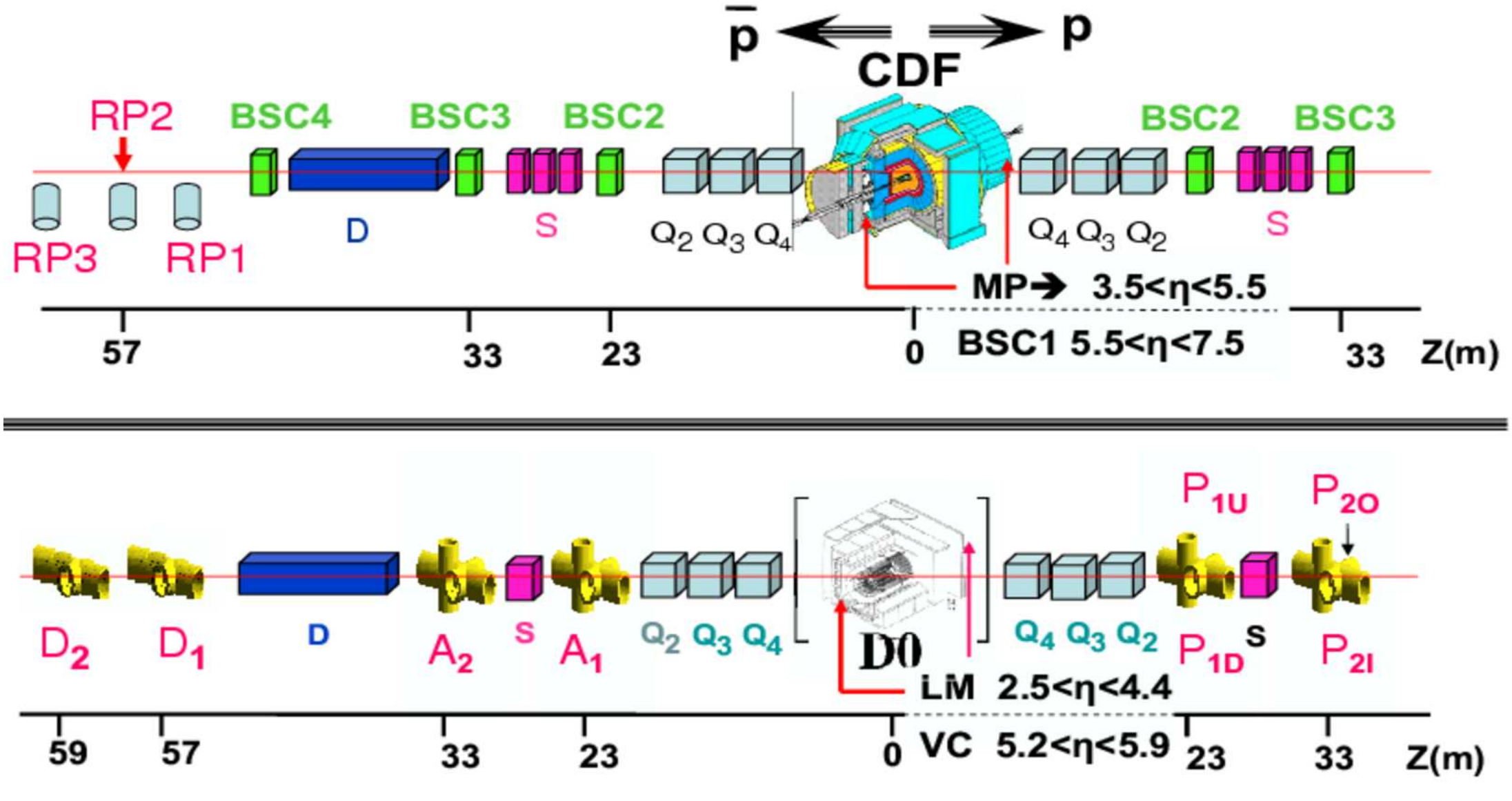

Figure 1: The CDF and D0 detectors in Run II.

The Collider Detector at Fermilab (CDF) and the DØ collaborations have been conducting studies in diffraction since the start of Tevatron operations in 1989. A plethora of results have been obtained on forward, central, and multi-gap diffraction processes, as well as on central exclusive production which is of special interest as it serves to calibrate theoretical models for exclusive Higgs boson production st the Large Hadron Collider (LHC). In this paper, we present the CDF II (CDF in Run II) and DØ forward detector configurations in Run II of the Tevatron $\bar{p}p$ collider and discuss issues of alignment, calibrations, backgrounds and physics reach.

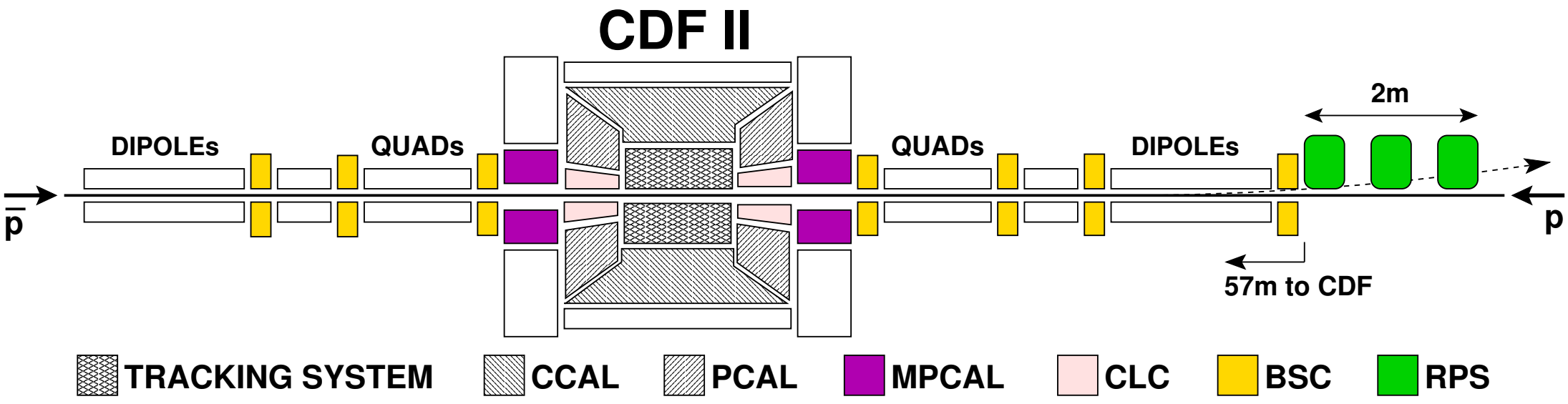

Figure 2: Schematic drawing of the CDF II detector (not to scale).

The CDF II diffractive physics data were collected with an upgraded CDF detector, which included the following forward components [1] (see Figs. 1, 2):

– **Roman Pot Spectrometer (RPS)** to detect leading **antiprotons**,
– **MiniPlug (MP)** forward calorimeters covering the region $\sim 3.5 < |\eta| < 5.5$,
– **Beam Shower Counters (BSC)** around the beam pipe at $\sim 5.5 < |\eta| < 7.5$,
– **Cerenkov Luminosity Counters (CLC)** covering the range $\sim 3.7 < |\eta| < 4.7$.

**The Roman Pot Spectrometer** was the same one that was used in Run Ic [2]. It consists of $X$-$Y$ scintillation fiber detectors [1] placed in three Roman Pot (RP) stations located at an average distance of 57 m downstream in the $\bar{p}$ direction. The detectors have a position resolution of $\pm 100\,\mu m$, which makes possible a $\sim 0.1\%$ measurement of the $\bar{p}$ momentum. In Run Ic, the $\bar{p}$-beam was behind the proton beam as viewed from the RPS side. An inverted polarity (with respect to Run I) of the electrostatic beam separators enabled moving the RPS detectors closer to the $\bar{p}$-beam and thereby gain acceptance for small $|t|$ down to $\xi \equiv 1 - x_F(\bar{p}) = 0.03$. For larger $|t|$, the values of $\xi$ that can be reached are lower.

**The MiniPlug calorimeters** were installed within the inner holes of the muon toroids. They consist of layers of lead plates immersed in liquid scintillator. The scintillation light is collected by wavelength shifting fibers strung through holes in the lead plates and read out by multi-channel photomultiplier tubes (MCPMT's). The calorimeter *tower* structure is defined by arranging fibers in groups to be read out by individual MCPMT pixels. There are 84 towers in each MiniPlug measuring energy and position for both electromagnetic (EM) and hadron showers [3].

**The Beam Shower counters** are rings of scintillation counters *hugging* the beam pipe. The BSC-1 rings are segmented into four quadrants, while the other BSC's are segmented into two halves. These counters are used to provide:

(a) rapidity gaps triggers,
(b) *exclusivity* constraints in studies of exclusive production, and
(c) beam losses, by gating the $\text{BSC2}_{\bar{p}}$ ($\text{BSC2}_{\bar{p}}$) signal to the passage of the $p$ ($\bar{p}$) beam.

**The Cerenkov Luminosity Counters** comprise a finely segmented system of gas Cerenkov counters pointing to the interaction point (I) and are normally used by CDF to measure the number of inelastic collisions per beam-beam bunch crossing and thereby the luminosity. In the diffractive program, they are used in the rapidity gap definition by detecting charged particles that might penetrate the MP calorimeters producing a signal too small to be detected above the MP tower thresholds used. Their high efficiency for detecting charged particles provides a valuable complementary input to that of the MP calorimeters.

[1] We use a coordinate system with origin at the centre of the CDF detector, $Z$ along the proton beam direction and $Y$ pointing up; the $X$ coordinate points away from the centre of the accelerator ring.

The DØ Run II forward detectors include two *state of the art* Forward Proton Detector (FPD) spectrometers on the outgoing $p$ and $\bar{p}$ directions equipped with $X$-$Y$-$V$ silicon based trackers capable of resolving ambiguities in events with multiple tracks at high luminosities. The beam-line configuration on the $\bar{p}$-side is similar to that of CDF; however, the $p$-side spectrometer has no dipole magnets between the IP and the spectrometer, which restricts access to low-$t$ recoil protons. In contrast, the MiniPlug based system of CDF allows access to low$t$ events but is not capable of measuring angular $\phi$-correlations between the outgoing $p$ and $\bar{p}$, which are important for certain physics studies.

The CDF II [4] and DØ [5] detector layouts are schematically shown and compared in Fig. 1.

## 2 Alignment

A precise alignment of the RPS detectors is crucial for an accurate determination of the $\xi$ and $t$ values of the recoil antiproton. The standard alignment method based on surveying the detectors relative to the centreline of the beam pipe and using beam position monitors (BPM) to determine the beam position relative to the centre of the pipe was hampered in Run II by several difficulties, including:

- there was no BPM in the vicinity of the RPS detectors;
- the beam position and angle could change with beam store as a result of tuning to increase the luminosity, and also during the course of a store;
- the position of the RP relative to the beam pipe could also change relative to the surveyed position due to slippage of the positioning mechanism.

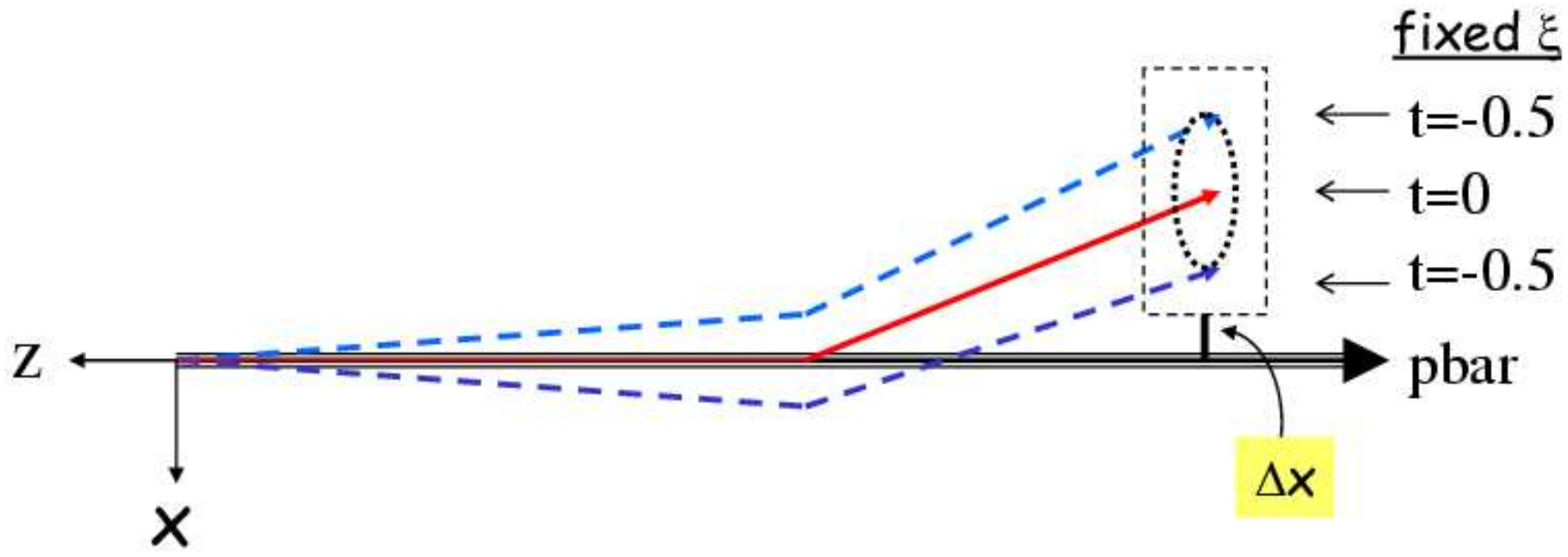

Figure 3: Schematic representation of a track detected by the RPS.

The RPS alignment was addressed in CDF by developing a *dynamic alignment* method in which the actual detector position during data taking is determined from the recorded data. The method is based on the expectation that the $t$-distribution of the recoil antiproton be maximal at $|t| = 0$. However, the measured distribution will display such a maximum only if the detectors are correctly aligned. This is illustrated in Figs. 2, 3, 4. In practice, offsets in the $X$ and $Y$ coordinates of the RPS detectors with respect to the beam line are introduced in the off-line analysis and iteratively adjusted until a maximum for $d\sigma/dt|$ is obtained at $|t| = 0$. Using the resulting values for $X_{\rm offset}$ and $Y_{\rm offset}$ corresponds to having aligned detectors. This method is very precise and only limited by the statistics of the event sample, the size of the beam, and the jitter in the beam position during data taking of the particular data sample used. Using a special data sample collected during a relatively short dedicated run an accuracy of $\pm 30$ $\mu$m in beam position was obtained [6].

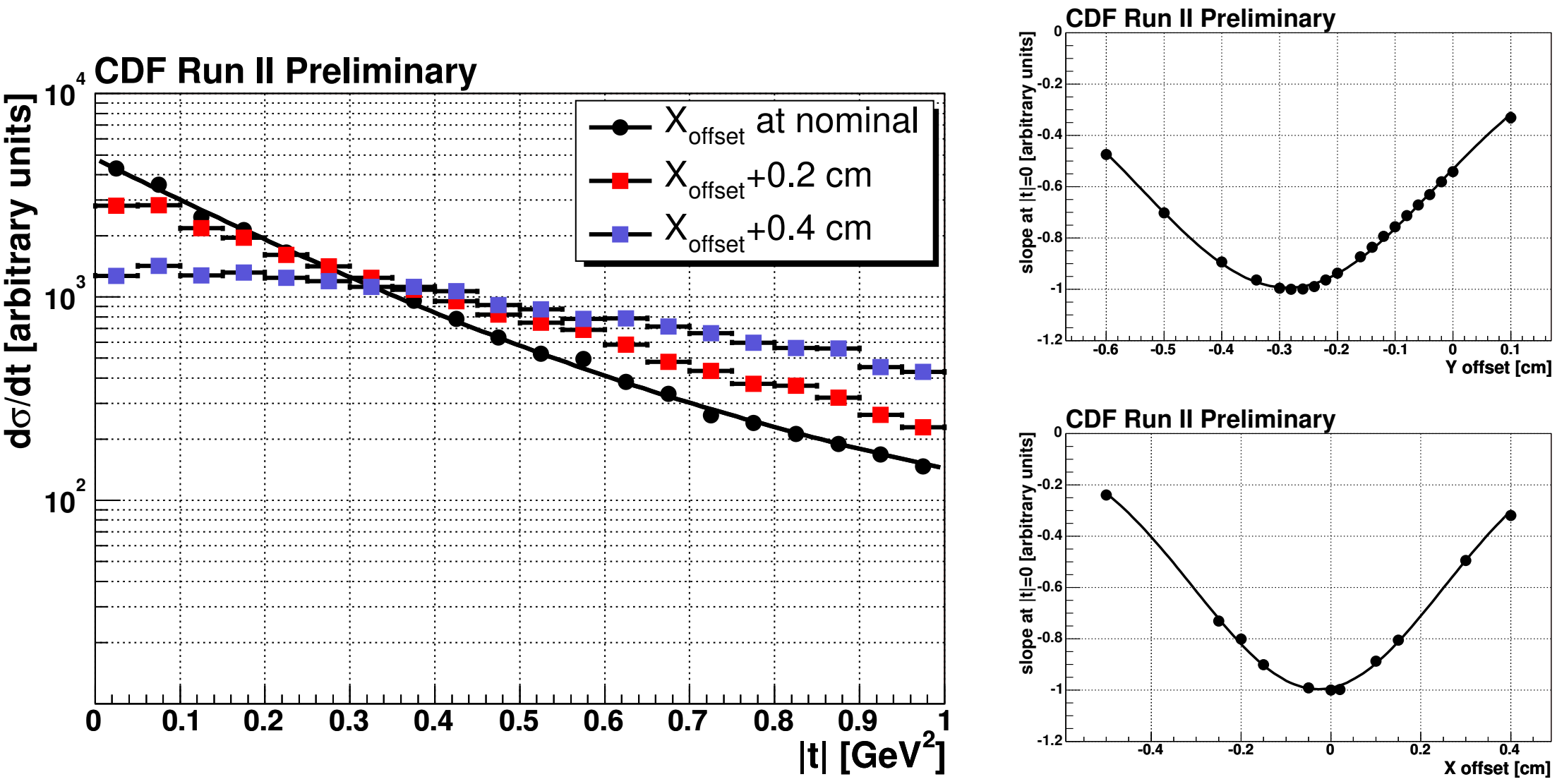

Figure 4: *Left*: $t$-distribution of reconstructed RPS tracks for positive $X_{\rm offset}$ shifts; *Right*: $|b|$ slope versus Y (*top*) and X (*bottom*) offsets.

The dynamic alignment method is quite general and can equally well be used to accurately calibrate the position of RPS detectors planned for the LHC. In principle, the offsets could be determined from small data samples and applied on-line during data taking to enrich the recorded data sample with useful events.

# 3 Calibrations and Backgrounds

## 3.1 Missing Forward Momentum and *W* Mass from Diffractive Events

A well known and frequently used data analysis tool is missing transverse energy. This tool is particularly useful in cases involving neutrinos in the final state. A good example is the determination of the mass of the $W$-boson through the $W \rightarrow e\nu$ / $\mu\nu$ decay modes. The usual technique is to use the *transverse* $W$ mass, which results is a skewed distribution that requires Monte Carlo and detector simulations that affect the accuracy of the measurement. The detection in the RPS of the forward $\bar{p}$ in diffractive $W$ production makes possible the determination of the full kinematics of the $W \rightarrow e\nu/\mu\nu$ decay.

The neutrino transverse energy $E_T^\nu$ is obtained from the missing $E_T$, as usual, and the pseudorapidity $\eta_\nu$ from the formula $\xi_{\bar{p}}^{\rm RPS}$-$\xi^{\rm cal} = (E_T/\sqrt{s})\exp[-\eta_\nu]$ , where $\xi^{\rm cal}$ is calculated from the calorimeter towers: $\xi_{\bar{p}}^{\rm cal} = \Sigma_{\rm i-towers}(E_T^{\rm i}/\sqrt{s})\cdot\exp[-\eta_i]$. This method has been applied to CDF data yielding the preliminary result shown in Fig. 5. Using a data sample of $\sim 300$ diffractive $W$ events, CDF obtained $M_W^{\rm exp} = 80.9 \pm 0.7$ GeV [7] in good agreement with the world average $W$-mass of $M_W^{\rm PDG} = 80.403 \pm 0.029$ GeV [8].

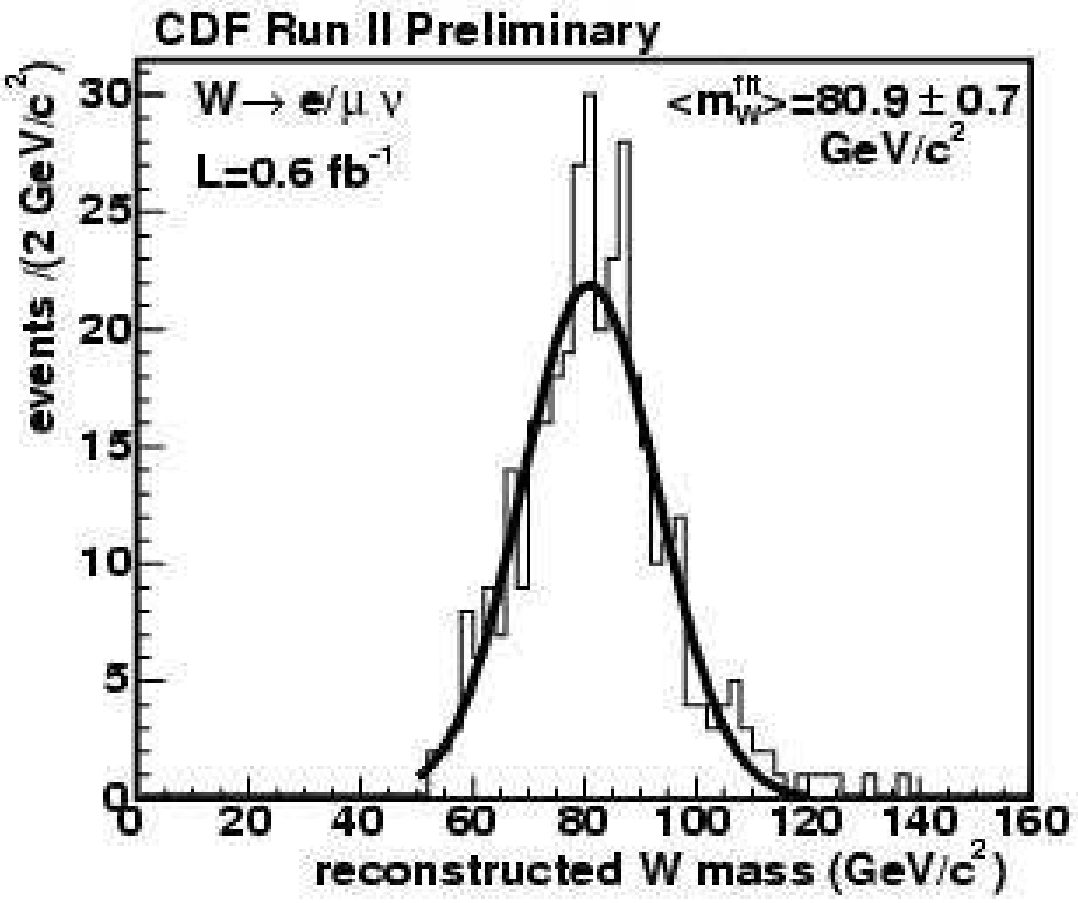

Figure 5: Diffracive $W$ mass and Gaussian fit.

## 3.2 Overlap Background and $\xi_{\bar{p}}^{\rm CAL}$ Calibration

The main event sample of diffractive events in Run II was collected without RPS tracking information. For these events, $\xi_{\bar{p}}$ was evaluated from calorimeter information using the formula:

$$\xi_{\bar{p}}^{\rm CAL} = \Sigma_{\rm i-towers}(E_T^{\rm i}/\sqrt{s})\cdot\exp[-\eta_i].$$

The calorimeter based $\xi^{\rm CAL}$ was calibrated against $\xi^{RPS}$ using a data sub-sample for which tracking was available. In implementing the calibration procedure, special care had to be taken to handle the background from overlap events at the high instantaneous luminosity ($\mathcal{L}$) of the run. Figure 6 illustrates the overlap background handling and the $\xi^{\rm CAL}$ vs. $\xi^{\rm RPS}$ calibration.

### 3.2.1 Overlap Background

The $\xi_{\bar{p}}^{\rm CAL}$ distribution is shown in Fig. 6 for three diffractive dijet data samples defined in the insert. A *single-vertex* requirement was applied to all samples. The region indicated as SD contains events which are mostly due to single-diffractive (SD) dijet production, while the events in the ND region are mainly overlaps of a non-diffractive (ND) dijet and a soft diffractive interaction that triggered the RPS but yielded no reconstructed vertex. The majority of the diffractive events are represented by the excess of events of the RPS·Jet5 over the rescaled Jet5 distribution in the SD region. The overlap background in the region $0.03 < \xi_{\bar{p}}^{\rm CAL} < 0.09$ is of $\mathcal{O}(10\%)$.

Figure 6 *(t-r)* shows a two-dimensional scatter plot of $\xi_{\bar{p}}^{\rm CAL}$ vs. $\xi_{\bar{p}}^{\rm RPS}$ for the events with a reconstructed RPS track. The mountain-like peak in the region of $\xi_{\bar{p}}^{\rm CAL} \lesssim 0.1$ is attributed to

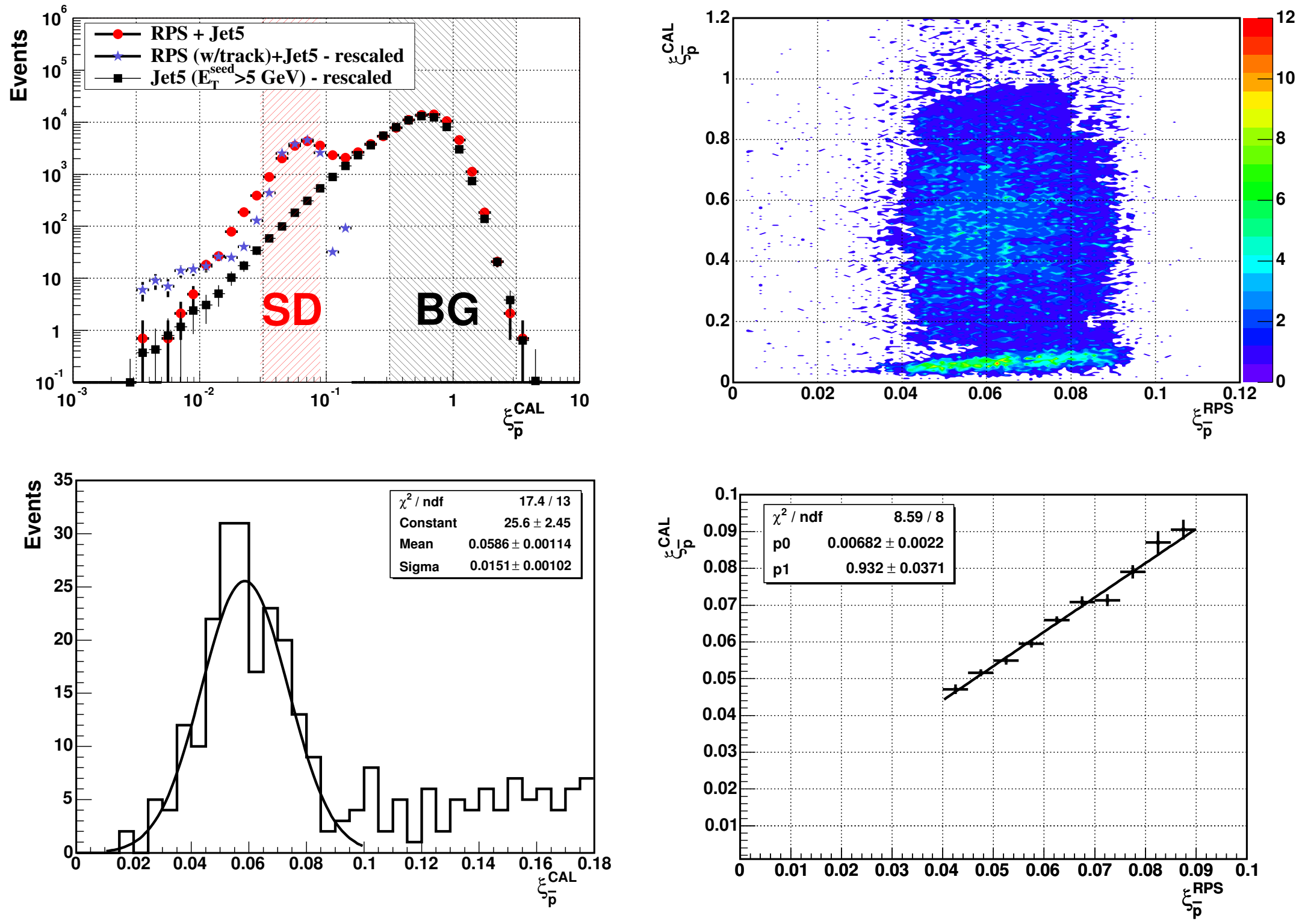

Figure 6: $\xi_{\bar{p}}^{CAL}$ vs. $\xi_{\bar{p}}^{RPS}$ (see text): *(t-l)* $\xi_{\bar{p}}^{CAL}$ distributions; *(t-r)* 2D-distribution of $\xi_{\bar{p}}^{CAL}$ vs. $\xi_{\bar{p}}^{RPS}$; *(b-l)* $\xi^{CAL}$ for a slice of the 2D-dist. of $0.055 < \xi_{\bar{p}}^{RPS} < 0.060$; *(b-r)* $\xi_{\bar{p}}^{CAL}$ vs. $\xi_{\bar{p}}^{RPS}$.

diffractive events. The events with $\xi_{\bar{p}}^{\rm CAL} > 0.1$ are mostly due to ND dijets with a superimposed soft SD overlap event.

### 3.2.2 $\xi_{\bar{p}}^{\rm CAL}$ Calibration

The calibration of of $\xi_{\bar{p}}^{\rm CAL}$ is performed by dividing the data into bins of width $\Delta\xi_{\bar{p}}^{\rm RPS} = 0.005$ and fit the $\xi_{\bar{p}}^{\rm CAL}$ values in each bin with a Gaussian distribution excluding the now well separated background, as shown in Fig. 6 *(b-l)* for $0.055 < \xi_{\bar{p}}^{\rm CAL} < 0.060$. The ratio of the half-width average of the fitted curve is $\delta\xi_{\bar{p}}^{\rm CAL}/\xi_{\bar{p}}^{\rm CAL} \approx 0.3$. Figure 6 *(b-r)* shows the results for the region $0.04 < \xi_{\bar{p}}^{\rm RPS} < 0.09$. A linear relationship is observed between $\xi_{\bar{p}}^{\rm CAL}$ and $\xi_{\bar{p}}^{\rm RPS}$ in this region.

# 4 Summary

The CDF and DØ forward detectors designed and used for studies of diffraction in Run II at the Fermilab Tevatron $\bar{p}p$ collider were presented with a focus on issues of alignment, calibrations, backgrounds and physics reach. The procedures developed are quite general can be directly adapted to the experiments planning diffractive physics studies at the LHC.

# References

[1] A. Aaltonen *et al.* (CDF Collaboration), *Observation of Exclusive Dijet Production at the Fermilab Tevatron $\bar{p}p$ Collider,* Phys. Rev. D77:052004, 2008. DOI : 10.1103/PhysRevD.77.052004. Report number: FERMILAB-PUB-07-647-E.

[2] H. Nakada, Ph.D. thesis, University of Tsukuba, January 2001.

[3] K. Goulianos, M. Gallinaro, K. Hatakeyama, S. Lami, C. Mesropian, A. Solodsky, *The CDF Miniplug Calorimeters (and references therein),* Nucl. Instrum. Methods A **496**, 333 (2003 ).

[4] D. Acosta *et al.* (CDF Collaboration, Phys. Rev. **D** 71, 232001 (2005).

[5] DØ Collaboration, *The Upgraded D0 Detector,* arXiv:physics/0507191v2.

[6] Michele Gallinaro, DIS-2006, arXiv:hep-ex/0606024.

[7] K. Goulianos, *Diffractive W and Z production at Tevatron* arXiv:0905.4281v2 [hep-ex]; paper resented for the CDF collaboration at XLIV$^{th}$ Rencontres de Moriond, QCD and High Energy Interactions, La Thuile, Aosta Valley, Italy, March 14-21, 2009.

[8] W.-M. Yao et al., Journ. Phys. G **33**, 1 (2006) and 2007 partial update for 2008, http://pdg.lbl.gov.

# The TOTEM Detector at LHC

*G. Antchev*[1], *P. Aspell*[1], *I. Atanassov*[1], *V. Avati*[1,9], *V. Berardi*[4], *M. Berretti*[5], *M. Bozzo*[2], *E. Brücken*[6], *A. Buzzo*[2], *F. Cafagna*[4], *M. Calicchio*[4], *M. G. Catanesi*[4], *M. A. Ciocci*[5], *M. Csanád*[10], *T. Csörgő*[10], *M. Deile*[1], *E. Dénes*[10], *E. Dimovasili*[1,9], *M. Doubek*[11], *K. Eggert*[9], *F. Ferro*[2], *F. Garcia*[6], *S. Giani*[1], *V. Greco*[1,5], *L. Grzanka*[1], *J. Heino*[6], *T. Hilden*[6], *M. Janda*[11], *J. Kašpar*[1,7], *J. Kopal*[1,7], *V. Kundrát*[7], *K. Kurvinen*[6], *S. Lami*[5], *G. Latino*[5], *R. Lauhakangas*[6], *E. Lippmaa*[8], *M. Lokajíček*[7], *M. Lo Vetere*[2], *F. Lucas Rodriguez*[1], *M. Macrí*[2], *G. Magazzù*[5], *S. Minutoli*[2], *H. Niewiadomski*[1,9], *G. Notarnicola*[4], *T. Novak*[10], *E. Oliveri*[5], *F. Oljemark*[6], *R. Orava*[6], *M. Oriunno*[1], *K. Österberg*[6], *E. Pedreschi*[5], *J. Petäjäjärvi*[6], *M. Quinto*[4], *E. Radermacher*[1], *E. Radicioni*[4], *F. Ravotti*[1], *E. Robutti*[2], *L. Ropelewski*[1], *G. Ruggiero* [1], *A. Rummel*[8], *H. Saarikko*[6], *G. Sanguinetti*[5], *A. Santroni*[2], *A. Scribano*[5], *G. Sette*[2], *W. Snoeys*[1], *W. Spearman*[12], *F. Spinella*[5], *A. Ster*[10], *C. Taylor*[3], *A. Trummal*[8], *N. Turini*[1,5], *V. Vacek*[11], *M. Vitek*[11], *J. Whitmore*[9], *J. Wu.*[1]

[1]CERN, 1211 Geneve 23, Switzerland
[2]Università di Genova and Sezione INFN, Genova, Italy
[3]Case Western Reserve University, Dept. of Physics, Cleveland
[4]INFN Sezione di Bari and Politecnico di Bari, Bari, Italy
[5]INFN Sezione di Pisa and Università di Siena, Italy
[6]Helsinki Institute of Physics and Department of Physics, University of Helsinki, Finland
[7]Institute of Physics of the Academy of Sciences of the Czech Republic, Prague, Czech Republic
[8]National Institute of Chemical Physics and Biophysics NICPB, Tallinn, Estonia
[9]Penn State University, Dept. of Physics, University Park, PA, USA
[10]MTA KFKI RMKI, Budapest, Hungary
[11]On leave from Czech Technical University, Prague, Czech Republic
[12]Fulbright fellow, individual participant from the University of Geneva, Switzerland.

The TOTEM experiment, small in size compared to the others at the LHC, is dedicated to the measurement of the total proton-proton cross-section with the luminosity-independent method and to the study of elastic and diffractive scattering. To achieve optimum forward coverage for charged particles emitted by the pp collisions in the interaction point IP5, two tracking telescopes, T1 and T2, are installed on each side in the pseudo-rapidity region between 3.1 and 6.5, and Roman Pot (RP) stations are placed at distances of 147 m and 220 m from IP5. The telescope closest to the interaction point (T1, centered at z = 9 m) consists of Cathode Strip Chambers (CSC), while the second one (T2, centered at 13.5 m), makes use of Gas Electron Multipliers (GEM). The proton detectors in the RPs are silicon devices designed by TOTEM with the specific objective of reducing down to a few tens of microns the insensitive area at the edge. High efficiency as close as possible to the physical detector boundary is an essential feature. It maximizes the experimental acceptance for protons scattered elastically or diffractively at polar angles down to a few micro-radians. To measure protons at the lowest possible emission angles, special beam optics have been developed, optimizing acceptance and resolution. The read-out of all TOTEM subsystems is based on the custom-designed digital VFAT chip with trigger capability.

# 1 Introduction

The Totem Experiment will measure the total pp cross-section and study elastic scattering and diffractive dissociation at LHC [1, 2]. The experimental apparatus is placed symmetrically with respect to the Interaction Point 5 (IP5) and the CMS experiment (Fig. 1). Two tracking telescopes, T1 and T2, will measure the inelastic interactions in the forward region covering an adequate acceptance over a rapidity interval of $3.2 \le \eta \le 6.5$. T1 is placed between two conical surfaces, the beam pipe and the inner envelope of the flux return yoke of the CMS end-cap, at a distance between 7.5 m and 10.5 m from the IP5, while T2 is installed at about 13.5 m in the forward shielding of CMS, between the vacuum chamber and the inner shielding of the HF calorimeter. The measurement of $\frac{d\sigma_{el}}{dt}$ down to $-t = 10^{-3}\,\text{GeV}^2$ is accomplished by silicon detectors placed in Roman Pots located at 147 m and 220 m from IP5. Since the beam of the LHC is rather thin, with a $10\,\sigma$ envelope of about 1 mm, the detectors in the Roman Pot must have a very small dead zone at the mechanical edge facing the beam. In the following sections, the different detectors will be described and their status reviewed.

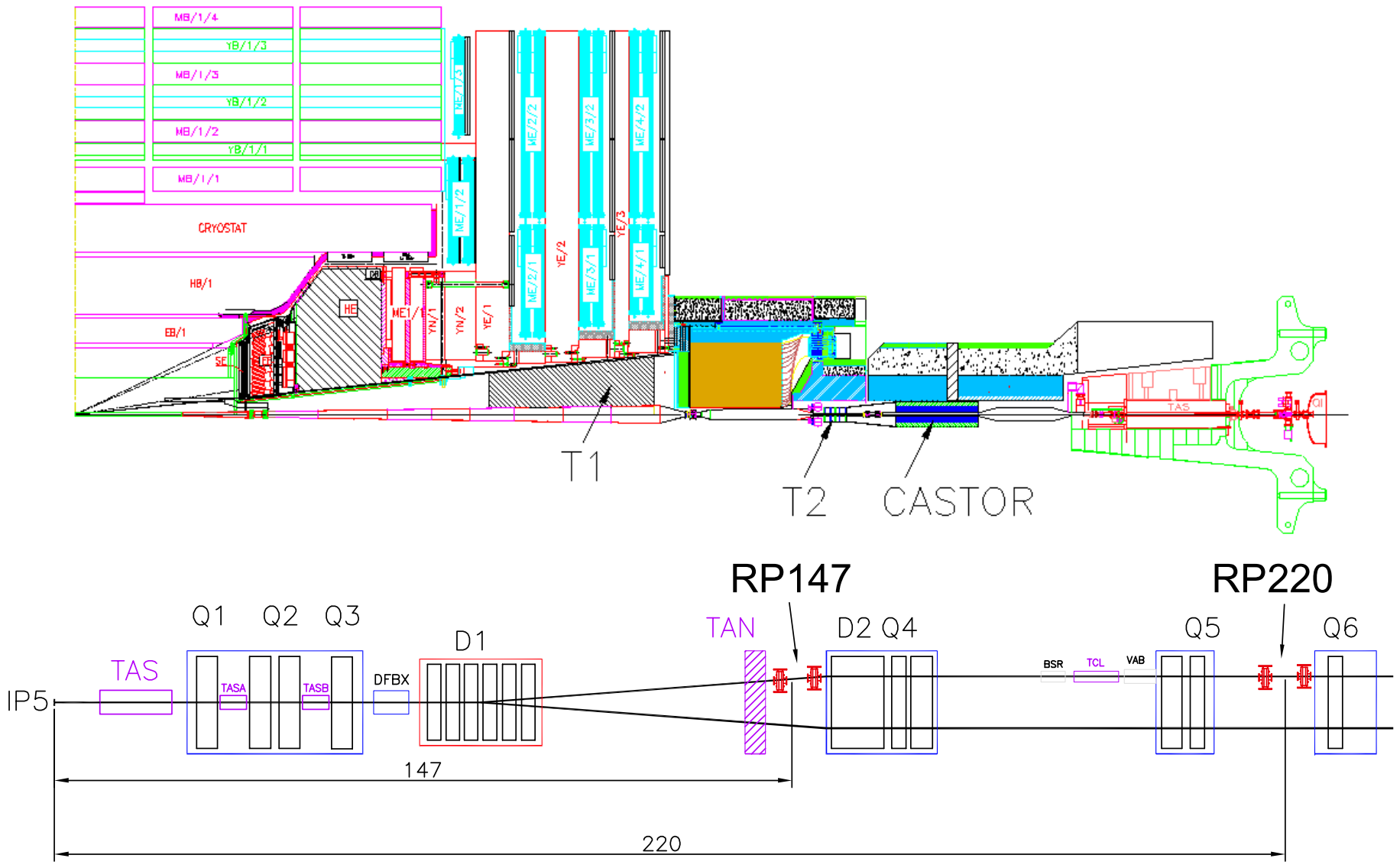

Figure 1: Top: The TOTEM forward telescopes T1 and T2 embedded in the CMS experiment together with the CMS forward calorimeter CASTOR. Bottom: The LHC beam line and the Roman Pots at 147 m (RP147) and 220 m (RP220).

# 2 The Inelastic Telescopes T1 and T2

The T1 and T2 telescopes will be employed to trigger and partially reconstruct inelastic events. Together they must provide a fully inclusive trigger for diffractive events and enable the reconstruction of the vertex of an event, in order to disentangle beam-beam events from the background. Each telescope is made of two arms, symmetrically placed with respect to IP5.

Each arm of T1 is composed of five planes of Cathode Strip Chambers, with six chambers per plane covering roughly a region of 60° in $\phi$. It is split in two halves and mounted on two different supports. A picture of one half of a T1 arm is shown in Fig. 2. In each chamber, the readout strips of the two cathode planes are oriented ±60° with respect to the anode wires. This allows the measurement of three coordinates for each particle track, which significantly helps in resolving multiple events. To improve pattern recognition, the planes are rotated by 3° with respect to each other. The production of the CSCs with their readout cards and the support structure has been completed and qualified, and the system is taking test data with cosmic rays. The T2 telescopes are made of triple Gas Electron Multipliers (GEM) [3]. GEMs are gas-filled detectors featuring the advantageous decoupling of the charge amplification structure from the charge collection and readout structure. Furthermore, they combine good spatial resolution with very high rate capability and a good resistance to radiation. Each of the two telescope arms is made of two sets of 10 aligned detector planes with almost semicircular shape, mounted on each side of the vacuum pipe. To avoid efficiency losses on the boundaries, the angular coverage of each half plane is more than 180°. The readout of the half planes has two separate layers with different patterns: one with 256 concentric rings, 80 $\mu$m wide and with a pitch of 400 $\mu$m, providing the radial coordinates of traversing tracks with a good precision, and the other with a matrix of 1536 pads varying in size from $2 \times 2\,\text{mm}^2$ to $7 \times 7\,\text{mm}^2$, used for triggering. An illustration of a GEM chamber is shown in Fig. 3 (left). Both arms of T2 have been installed in their final locations. A picture of the T2 arm during the installation, right before the insertion in the CMS HF calorimeter is shown in Fig. 3 (right).

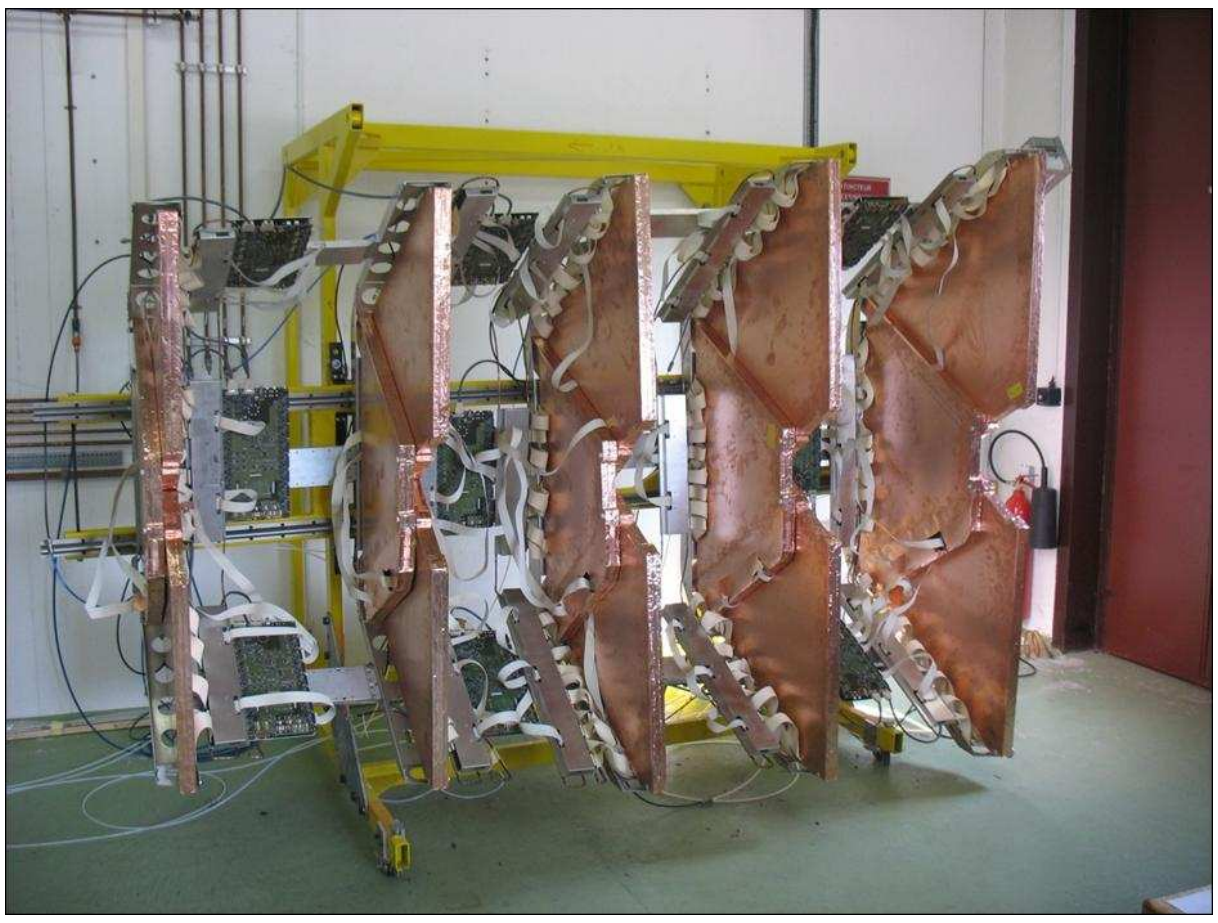

Figure 2: One quarter of the T1 telescope mounted on its support structure.

The production of the chambers has been followed by a series of acceptance tests before their assembly in half arms. Moreover, before the installation, each half arm of T2 has been extensively tested for data taking with cosmic rays. The measured and the simulated efficiency for each of the ten planes of one quarter of T2 are compared in Fig. 4.

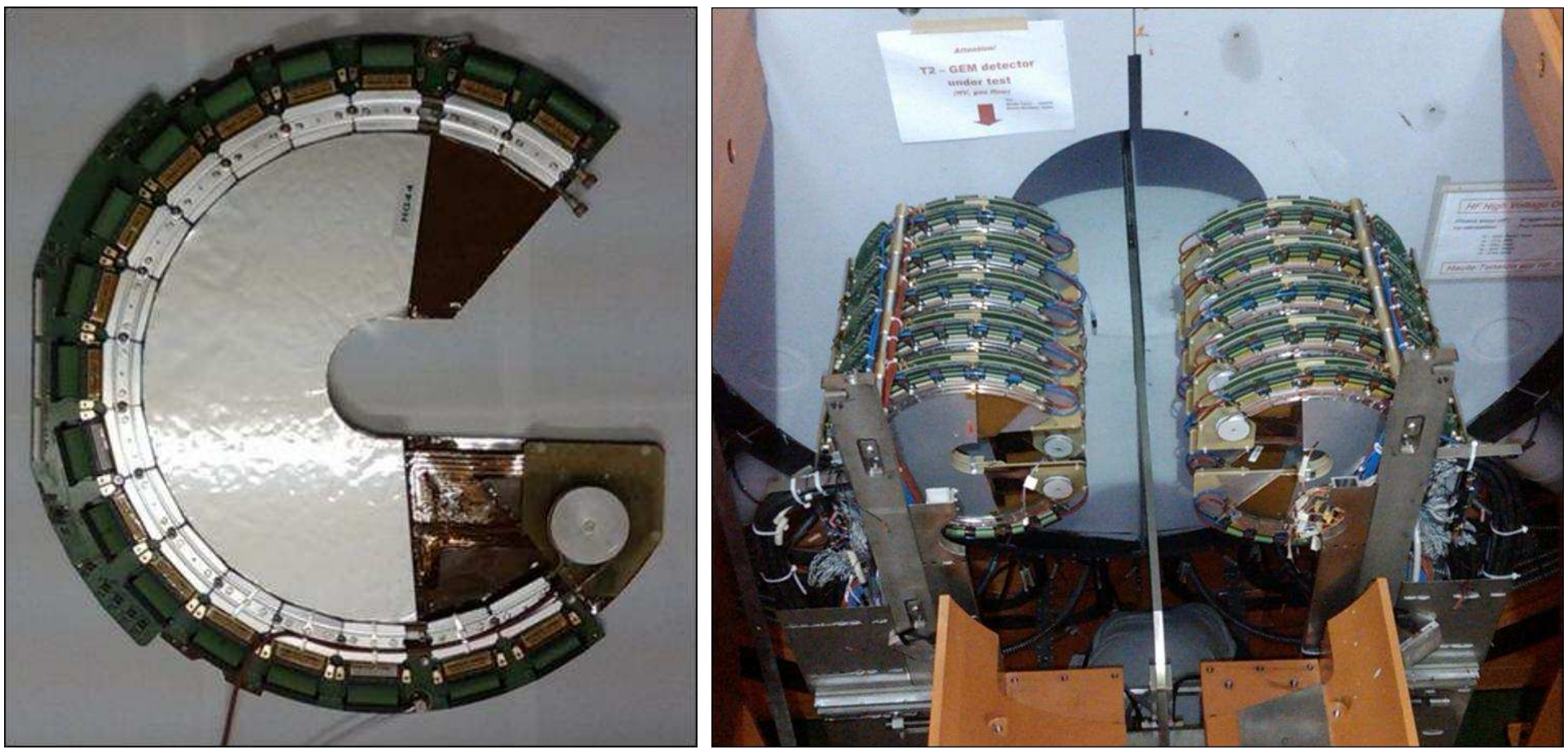

Figure 3: Left: GEM chamber assembled with its horseshoe-shaped readout card. Right: One arm of the T2 telescope just before the insertion in the CMS HF calorimeter.

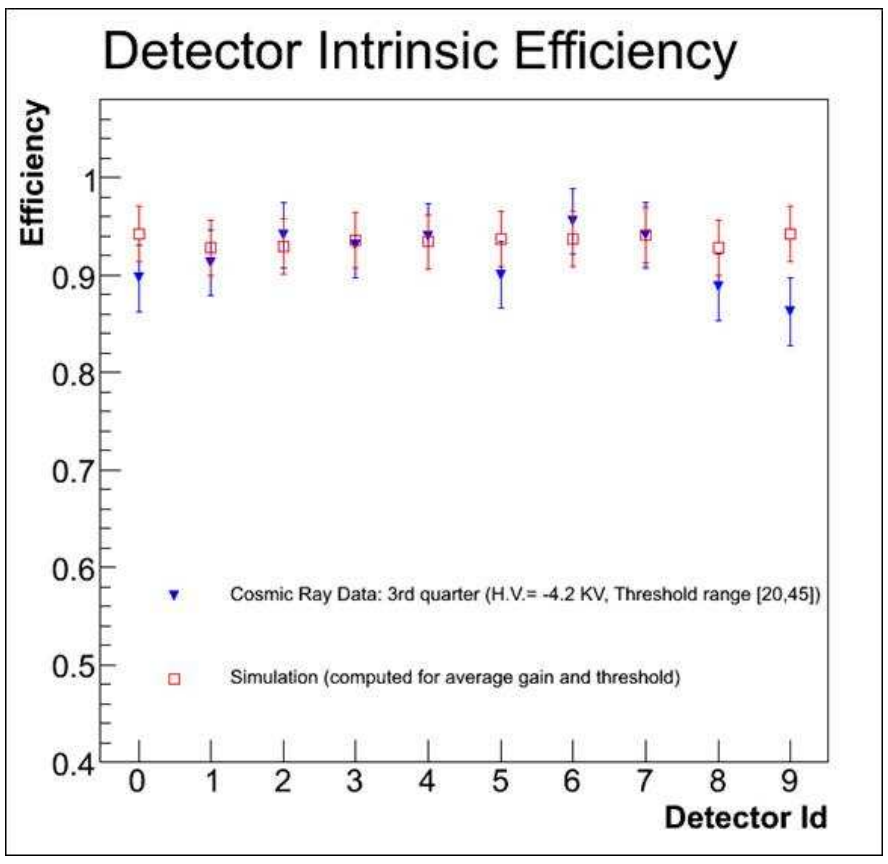

Figure 4: Comparison of the efficiency measured and simulated for each of the ten planes of the 3rd quarter of T2 with all the chambers at the nominal bias of 4.2 kV. For data taking, the threshold of the binary readout chips spanned the range from 20 to 45 DAC units, while for the simulation an average threshold was used for all the chips.

# 3 The Roman Pots

The Roman Pots are special beam pipe insertions, which allow bringing the detectors very close to the beam without interfering with the primary vacuum of the machine. Each RP station is made of two units separated by 4 m and equipped with one horizontal and two vertical pots. A photo of a fully installed unit is shown in Fig. 5 (left). Given the challenging constraints of the LHC machine, such as high beam energy, ultra high vacuum and the required physics performance of TOTEM, which needs to have active detectors at ~1 mm from the LHC beam centre, a special design has been developed. A main issue has been the welding technology employed for the thin window that separates the vacua of the machine and the Roman Pot, still minimizing the distance of the detector from the beam. As result of this development, a thickness and a planarity of less than 150 $\mu$m and 20 $\mu$m respectively have been achieved for the thin windows produced. In each pot, a detector package made of 10 planes is inserted, with the sensors approaching the thin window to few hundreds of microns. The single-sided silicon microstrip detectors have been fabricated with planar technology, with the special characteristic of reaching full sensitivity within 50 $\mu$m from the cut edge. The detector package is operated at $-30°$C by means of evaporative cooling and is kept in a controlled atmosphere with pressure between 10 mbar and 100 mbar. The microstrip sensors have a diamond-like shape and parallel strips oriented at 45° with respect to the side of the sensor facing the beam. The planes of the detector package are mounted back-to-back in pairs (flipped around the vertical axis), in order to have their strips oriented mutually orthogonal. The strip pitch of 66 $\mu$m is adequate to achieve a resolution of less than 20 $\mu$m. A photo of a detector package is shown in Fig. 5 (right).

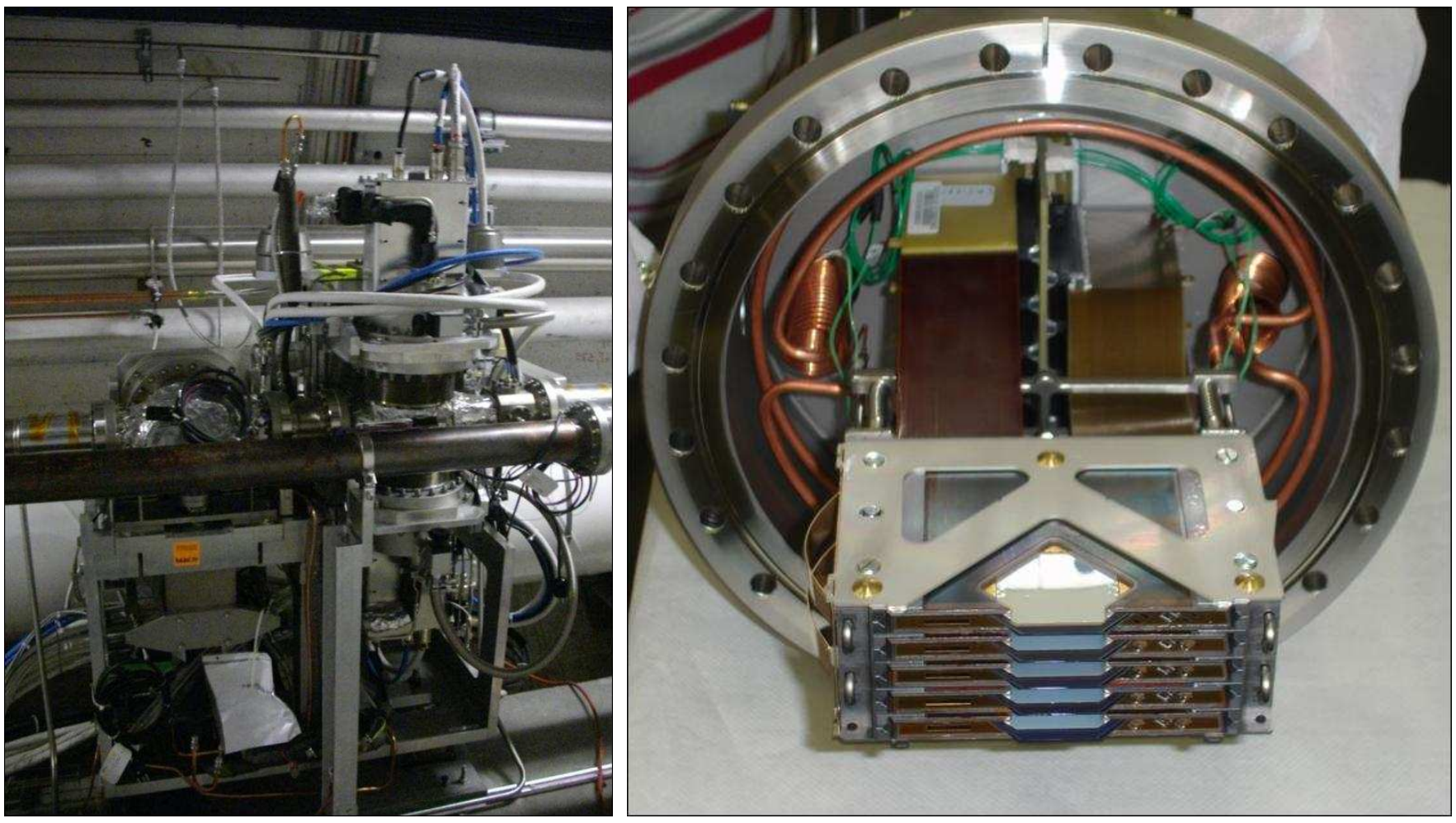

Figure 5: Left: A Roman Pot unit with two vertical and one horizontal pot installed in the LHC tunnel. Right: One of the detector packages mounted in all pots of the RP System.

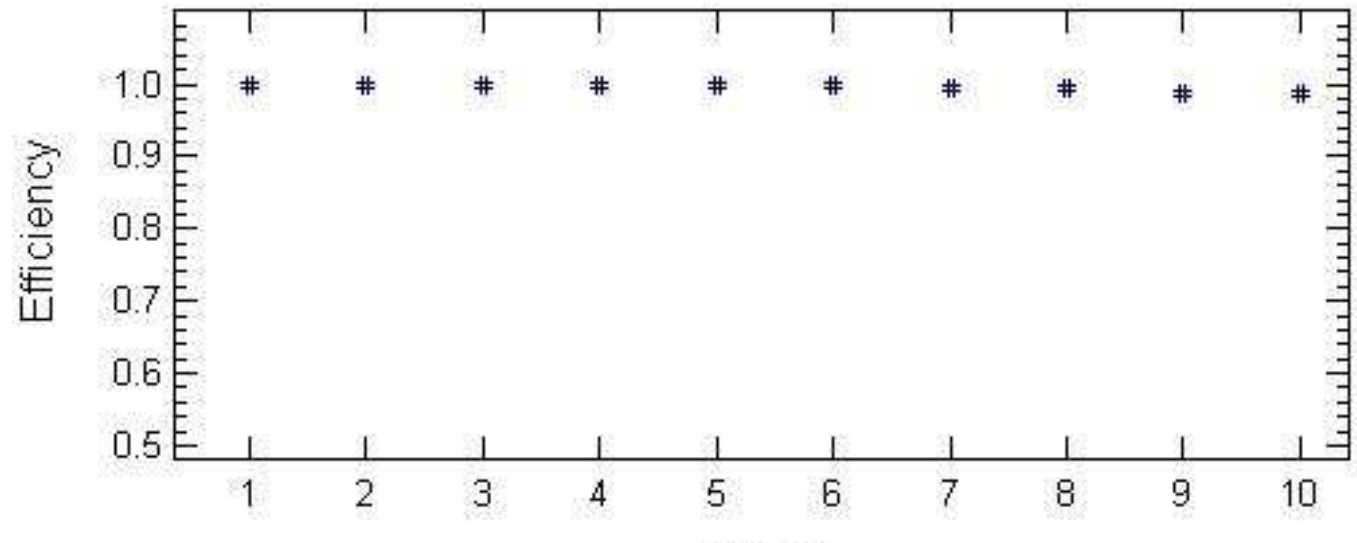

Figure 6: Efficiency measured for each of the ten planes of one detector package. The relative error on these measurements is less than 1%.

These silicon detectors, developed by the TOTEM Collaboration and also known as *planar edgeless silicon detectors with current-terminating structure (CTS)* [4], use a newly conceived design, which prevents breakdown and surface current injection at high bias, while simultaneously providing extremely reduced dead zones at the edges. Moreover, radiation hardness studies indicate that – when operated under moderate cooling – the detectors remain fully efficient up to a fluence of about $1.5 \times 10^{14}$ p cm$^{-2}$. All the Roman Pot Stations equipped with thin windows have been installed in the LHC tunnel in 2007. The detector packages, on the other hand, are being assembled and installed as they become available. A sequence of tests is performed during their assembly, including a run of data taking with particle beams or cosmic rays. The efficiency for one of the assembled detector packages measured with muons is shown in Fig. 6.

The stations at 220 m have received the first detector packages during the summer of 2008, and are now all equipped, with the last installations completed in July 2009. The installation of the detector packages in the stations at 147 m are foreseen only after the first run of LHC.

# 4 The TOTEM Electronics System

The TOTEM sub-detector systems (RP, T1, T2), based on different sensor technologies, are controlled and read out independently. However, they make use of a common electronics system architecture, based on the VFAT [5] chip. The VFAT has been designed specifically for the readout of sensors in the TOTEM experiment and has trigger and tracking capabilities. Moreover, it is able, on one side, to accommodate the considerable differences in signal properties of the three detector types and, on the other side, to be fully compatible with the CMS readout in view of common runs at a later stage. Figure 7 shows a basic block diagram of the functional components used in the system. It is subdivided into geographically separated regions and data flow. The so-called "On Detector regions" are located as close as possible to the detector. The "Local Detector regions" coincide with the readout boards in the vicinity of the detector (where the distributing control signals are grouped). A detailed description of the TOTEM Electronics System can be found in [6].

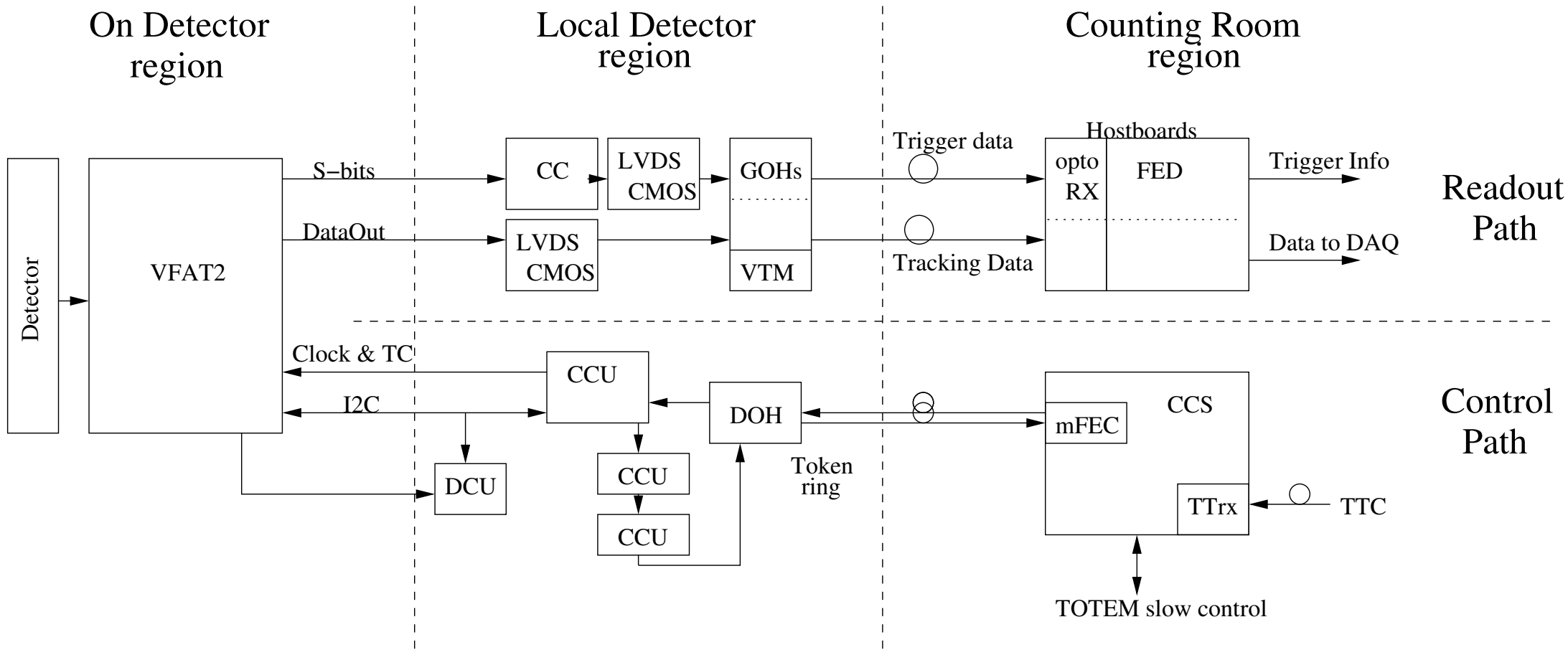

Figure 7: Functional block diagram of the TOTEM electronics system architecture.

# 5 Conclusions

The TOTEM Collaboration has installed most of its detectors in IP5. All the detectors are commissioned prior to the installation, including their electronics and readout systems. The relatively small dimension, the forward orientation and the specific placement of these detectors in the IP5 cavern and in the tunnel, have demanded a commissioning with particles on the surface with cosmic rays and – when possible – with particle beams. Today, the installation of the telescope T2 and of the Roman Pots at 220 m is complete. Both detectors have been validated after a commissioning phase done with beam particles. The commissioning of the telescope T1 is taking place now.

# Acknowledgements

This work was made possible by grants from CERN (Geneva, Switzerland), the Istituto Nazionale di Fisica Nucleare (Italy), the Institute of Physics of the ASCR, v.v.i. (Praha, Czech Republic), the Estonian Academy of Sciences (Tallinn, Estonia), the Hungarian National Science Fund OTKA (grants HA07-C 74458 and NK-73143), the Helsinki Institute of Physics HIP and Department of Physical Sciences (Helsinki, Finland), the Case Western Reserve University, Dept. of Physics, (Cleveland, OH, USA), and the Penn State University, Dept. of Physics (University Park, PA, USA). We gratefully acknowledge their support.

# References

[1] TOTEM TDR, CERN-LHCC-2004-002, TOTEM-TDR-001, 7 January 2004; addendum CERN-LHCC-2004-020.

[2] G. Anelli et al., "The TOTEM Experiment at the CERN Large Hadron Collider", 2008 JINST 3 S08007.

[3] F. Sauli, "GEM: A new concept for electron amplification in gas detectors", Nucl. Instrum.Meth. A 386 (1997) 531.

[4] G. Ruggiero et al., "Planar edgeless silicon detectors for the TOTEM experiment", IEEE Trans. Nucl. Sci. 52 (2005) 1899.

[5] P. Aspell et al., "VFAT2: a front-end system on chip providing fast trigger information...", proceedings of TWEPP-07, Topical Workshop on Electronics for Particle Physics, Prague Czech Republic (2007), http://cdsweb.cern.ch/record/1069906 .

[6] W. Snoeys et al.."The TOTEM electronics system", proceedings of TWEPP-07, Topical Workshop on Electronics for Particle Physics, Prague Czech Republic (2007), http://cdsweb.cern.ch/record/1089268 .

# Panel Discussion EDS'09 – What can we learn / expect from the LHC Experiments?

*Chaired by Karsten Eggert*[6]
*Panel members: Albert De Roeck*[1], *Konstantin Goulianos*[4], *Per Grafström*[1], *Hannes Jung*[3], *Hubert Niewiadomski*[2], *Karel Safarik*[1], *Mark Strikman*[2], *Chung-I Tan*[5]

[1]CERN, 1211 Geneve 23, Switzerland
[2]Penn State University, Dept. of Physics, University Park, PA, USA
[3]DESY, Hamburg, Physics Department, University Antwerp
[4]The Rockefeller University, 1230 York Avenue, New York, NY 10065-9965, USA
[5]Brown University, Providence, RI, USA
[6]Case Western Reserve University, Dept. of Physics, Cleveland, OH, USA

## 1 Introduction

**Author: Karsten Eggert**

The panel discussion between experimentalists (most of the LHC experiments were represented) and theorists focussed on the following subjects:

What are the most important topics on forward and diffractive physics to be addressed at the start of the LHC? How much can we learn from the experience gained during the forward detector operation at FNAL (K. Goulianos) and HERA (H. Jung) and from their latest physics results? What kind of collaborations between the LHC experiments can be envisaged to maximise synergy effects, including common trigger and run strategies, beam analysis, Monte Carlos and combination of data?

The LHC experiments benefit from their large acceptance overlaps of the very forward detectors up to the Roman Pot detectors several hundred meters upstream. As an example, the multiplicity distributions for different pseudorapidity intervals have to be corrected with the individual experimental and trigger acceptances to be able to obtain cross-sections for diffractive processes (K. Safarik). The different systematics of the experiments will help disentangle the various cross-sections.

The measurement of the elastic scattering cross-section over a large transverse momentum range $t$ ($10^{-3} < t < 10\,\mathrm{GeV}^2$) is also quite challenging, and benefits from different systematics of the two set-ups (ATLAS and TOTEM) and the way the collaborations will extract the total cross-section (Per Grafström).

New ideas about the measurements of protons with low relative momentum losses (typically $10^{-3}$) using the large dispersion of the LHC at some selected places around the LHC ring have been presented for future upgrades (A. De Roeck and H. Niewiadomski).

On the theoretical side, M. Strikman presented his ideas about probing parton correlations by studying multiparton interactions in diffractive processes, and C.-I Tan suggested the duality of diffractive scattering and Pomeron physics.

It was generally felt that the LHC will address exciting physics in diffraction and forward physics with probably some new insights, but also that close collaborations between the experiments are mandatory to fully explore the LHC potential.

## 2 Experimental Synergy – From the ATLAS Point of View

**Author: Per Grafström**

In the context of the discussion of "What can we learn /expect form the LHC experiments?" I was asked to give some examples of possible synergy between the ATLAS forward detectors and other forward detectors at the LHC.

The obvious example is the benefit that both ATLAS and TOTEM can gain from a close collaboration. Comparing the acceptance of the TOTEM Roman Pot detectors with those of ATLAS it is evident that there is a large overlap in the measured $t$-ranges between the two experiments. In addition, the overlap is in the regions which are associated with large theoretical uncertainties. Sharing experimental information of what is happening at very small angles will certainly help us to better understand this region. To reach the very small $|t|$-values will for sure be a challenge, and the possible success will to a large extent depend on detailed knowledge of the LHC halo, machine background and detailed knowledge of the optics parameters. Here clearly ATLAS and TOTEM can mutually profit from each other as the problems are close to identical. We will need to work together with the specialists from the LHC to better understand the beam conditions and share all the relevant knowledge in an efficient way.

There are also evident cross-checks of the luminosity calculation for ATLAS that can profit from early TOTEM results. ATLAS will calculate the absolute luminosity in many different ways. However one option is to also use TOTEM results on the total cross-section. The total cross-section will most likely be measured by TOTEM with higher precision than by ATLAS, and probably it will also be measured somewhat earlier. In this case, ATLAS could use the TOTEM measurement together with the Optical Theorem and data from elastic scattering in ATLAS at some moderate small $|t|$-values to estimate the luminosity for ATLAS.

There is also a case of synergy between the calorimeters of LHCf and the Zero Degree Calorimeter (ZDC) of ATLAS. Those calorimeters are installed 140 metres away from the ATLAS Interaction Point in an absorber of neutral particles (TAN), whose main function is to protect the downstream magnets from quenching. The space inside the absorber is limited, and during early data taking LHCf will occupy the space in front of the Zero Degree Calorimeter (ZDC) of ATLAS. Actually LHCf will use the space where later the electromagnetic part of the ZDC of ATLAS will be installed. During this transition phase, the ATLAS ZDC will only be equipped with the hadronic modules. Throughout this initial phase one could think of sharing energy sums between the two experiments. One could think of doing this both for the trigger and also for the actual data. Both experiments would obviously profit from such a sharing.

In a more general context, the forward detectors will contribute to the understanding of minimum bias events which in turn will be important for the understanding of the underlying event which is sort of the pedestal to the high $p_T$ events. Here again a collaboration across all forward detectors will be important. Each one covers different $\eta$ regions and has its own characteristics, and combining data will help in getting a better understanding of the global picture.

Let me just finish with what one could hope one day would be the outcome of the small angle elastic measurements at the LHC. Measurements of $\rho$ – the ratio of real to imaginary part of the forward elastic amplitude – at the ISR in the middle of the seventies were used to predict the total cross-section at energies much higher than the ISR energies. Using dispersion relations the total cross-section was correctly predicted in the energy range of the S$\bar{p}$pS collider. In the same way the results from measurements of $\rho$ at the S$\bar{p}$pS and the Tevatron have been used to predict the total cross-section at the LHC. If we succeed in measuring $\rho$ at the LHC, we could use the same method to predict the total cross section at energies well above the LHC energy. There might be many difficulties before such a programme can be realised. It may well be that the LHC halo will make it very difficult to go as close to the beam as needed to precisely measure $\rho$. In addition there might be theoretical difficulties to extract $\rho$ from the data. Maybe we will be confronted with a new regime of saturation effects and strong $t$-dependence of $\rho$ that requires extremely accurate measurements of the differential cross section in order to be able to extract the relevant parameters. Hopefully we will know in a couple of years from now.

## 3 The Alice Experiment

**Author: Karel Safarik**

The ALICE experiment at LHC was designed as the dedicated heavy-ion experiment. However, it has some unique capabilities which contribute to the interest in using the ALICE detector also for genuine pp studies, in addition to the obvious reference pp data taking. The relatively low magnetic field, 0.5 T, used in central tracking, results in a very low transverse-momentum cutoff; particles with transverse momenta down to 100 MeV/$c$ are reconstructed with a reasonable efficiency. The particle identification system in central barrel, which uses practically all known particle identification techniques (ionisation energy loss measurements in silicon detectors and TPC, time-of-flight detector, transition-radiation detection, ring-imagining Cerenkov detector), gives the possibility to identify charged-hadron species in a wide momentum range.

At the start of the LHC, ALICE will measure the charged-particle pseudorapidity density. In order to properly normalise this distribution for a given class of events (inelastic, non-single-diffractive), relative yields of non-diffractive, single-diffractive and double-diffractive processes have to be determined either by combination of measurements or by Monte-Carlo. These estimates are, however, quite model dependent, and we came to the conclusion that, taking into account the current spread of model predictions, this normalisation will be the main source of systematic uncertainty of such measurements. Therefore, we are trying to assess what we can do experimentally to constrain the relative yields of diffractive processes. For this, various detectors with different pseudorapidity coverage are used: silicon-pixel detector in central region ($-1.4 < \eta < 1.4$), two scintillating-tile arrays on two sides ($-3.7 < \eta < -1.7$ and $2.8 < \eta < 5.1$) and two sets of zero-degree calorimeters ($\eta < -6.5$ and $\eta > 6.5$). This way we cover five distinct pseudorapidity intervals and we record for each event whether or not in these intervals at least one charged particle was produced. Then we divide the event sample into $32(= 2^5)$ sets according the combination of pseudorapidity intervals which were hit. It is essential that the pseudorapidity intervals do not overlap, in order to avoid correlations between event numbers in different sets. Using a model for soft hadron-hadron collisions (usually a Monte Carlo event generator) we calculate for the three event types (non-diffractive, single-diffractive and double-diffractive events) the 32 probabilities to end-up in one of the 32 sets. We then use these probabilities to fit the relative event yields constrained to the measured event populations in the 32 sets.

In this approach the model dependence is mainly reduced to the kinematics of diffractive processes, and we are not sensitive to the relative cross sections of diffraction in the models. To study systematic uncertainties due to the diffraction kinematics we are using different event generators. At the LHC start-up we plan to record the events on bunch-bunch crossing signal (sometimes called zero-bias trigger) and the above-described procedure is perfectly adequate for selecting (offline trigger) the non-empty events from the recorded event sample. It is worth mentioning that a 'collision event' has to be defined by some selection criteria, that unavoidably introduce some bias and model dependency.

As a result of recent discussion we aim to add to the ALICE set-up another scintillating counter at higher negative pseudorapidities to enhance the rapidity-gap selection capabilities (currently covered only up to $\eta = -3.7$). Other studies under consideration concern the central diffraction production of light mesons and of charmonium states: $J/\psi$ (sensitive to odderon exchange) and $\chi_c$ (possible separation of different $\chi_c$ states is also under investigation). For central-diffractive charmonium production a selective trigger would be needed.

# 4 Beyond Inclusive Cross-Sections

**Author: Hannes Jung**

The measurement of the total proton-proton cross-section is important in its own right. This total cross-section is mainly driven by soft processes, however with hard perturbative contributions. The calculation of inclusive processes is "relatively" simple since all the final states are integrated out.

At HERA the measurement of the total deep inelastic cross-section has provided a lot of new information on the parton densities which can be used for calculating any final state process also in *pp*. However, even at HERA, a satisfactory description of dedicated final states, like the forward jet cross-section is lacking [1]. This is because the hadronic final state is sensitive to very different phenomena: higher order QCD radiation, multiparton interactions, diffraction, saturation and hadronisation. Especially at high energies or small $x$ it is expected that fixed order calculations and the DGLAP parton shower approaches are not sufficient. This has been shown with the forward jet measurements.

Approaches which go beyond the collinear factorisation and try to better describe multi-parton radiation, are available and look promising [2, 3], but are still not able to fully describe the measurements. The investigations at HERA allow to determine precisely the mechanism of multi-parton radiation and to test models on initial and final state parton showers. These tests are essential when aiming to describe final states in *pp*, since there the contribution from multi-parton scattering complicates the situation.

At high energies or at small $x$ the parton densities will become very large, and parton recombination and saturation might occur. It is essential to separate soft contributions to the taming of the parton densities from perturbative contributions. The $\Delta\phi$ dependence of the dijet cross-sections at large $E_t$ can be used to study possible saturation effects [4] in the truly perturbative region: at $\Delta\phi \sim 180^\circ$ (back-to-back jets) the transverse momentum of the incoming partons to the hard scattering is small, and is sensitive to possible saturation effects. The cross-section in the back-to-back region should be smaller than expected from standard calculations (even including resummation effects).

A still unsolved problem is the connection of the total elastic *pp* cross-section with diffractive dissociation and multiparton interaction. If there is a significant hard diffractive component, then a hard perturbative component must be also visible in multiparton interactions. Multi-parton interactions and underlying events are studied by measuring the transverse momentum spectrum and multiplicities of charged particles in jet events transverse to the jet direction [5]. However, the charged particles are sensitive to soft processes. A similar measurement using "mini-jets" with $E_t > 5(20)$ GeV could be performed, which then shows the sensitivity to a perturbative contribution.

It is important to measure not only single differential distributions, but also correlations, because they could show details on the underlying physics process, as shown in [6].

# References

[1] L. Khein, these proceedings

[2] H. Jung and G. P. Salam. Hadronic final state predictions from CCFM: The hadron- level Monte Carlo generator CASCADE. *Eur. Phys. J.*, C19:351–360, 2001.
H. Jung. The CASCADE Monte Carlo. `http://www.desy.de/~jung/cascade`, 2009.

[3] F. Hautmann, these proceedings

[4] E. Avsar, these proceedings

[5] D. E. Acosta et al. The underlying event in hard interactions at the Tevatron $\bar{p}p$ collider. *Phys. Rev.*, D70:072002, 2004.

[6] Z. Rurikova, A. Bunyatian, in "Proceedings of the workshop: HERA and the LHC workshop series on the implications of HERA for LHC physics," arXiv:0903.3861 [hep-ph]. p.675

# 5 Probing Correlations of Partons Near Nucleon Edge

**Author: Mark Strikman**

Studies of the exclusive hard processes at HERA and at fixed target energies allowed to determine the transverse spread of gluons in nucleons as a function of $x$. Using this information one can calculate the rate of the production of four-jet events originating from $4 \to 4$ hard collisions. If the transverse correlations between the partons are neglected one finds a rate which is a factor of two smaller than in experiment; for summary and references see [1]. Hence a realistic description of the $pp$ collisions at the LHC should account for such correlations.

It is important to understand how such correlations depend on the transverse distance, $\rho$ of the parton from the nucleon centre. The inclusive multijet production is dominated by impact parameters $b \leq 0.7$ fm. Hence it is predominantly sensitive to the correlations at $\rho \leq 0.5$ fm. At the same time, the presence of significant correlations at large $\rho$ may help to solve the problem with S-channel unitarity [2].

It is possible to obtain information on the correlations of partons at large $\rho$ from the study of the multiparton interactions in diffractive processes.

One could consider both cases of single and double diffraction with production of two and four jets:

$$\mathrm{pp} \to \mathrm{p} + \mathrm{X}\ (2\ \mathrm{jets} + \mathrm{Y},\ 4\ \mathrm{jets} + \mathrm{Y}) \tag{5.1}$$

$$\mathrm{pp} \to \mathrm{pp} + \mathrm{X}\ (2\ \mathrm{jets} + \mathrm{Y},\ 4\ \mathrm{jets} + \mathrm{Y}). \tag{5.2}$$

In the case of single diffraction with production of four jets, depicted in Fig. 5.1, one can study

- the rate of such events – the smaller the transverse size of the Pomeron exchange, the larger is the cross-section;

- factorization of the $(x_1, x_2)$-dependence to the product of single parton distributions as measured in the single diffraction with production of two jets;
- dependence of the $x_1 + x_2$ spectrum on $t$ – the larger $-t$, the closer is the interaction to the perturbative regime, and hence the harder is the spectrum. In particular, for large $-t$, one could look for a peak near $x_1 + x_2 = x_{I\!P}$.

It is important to study also the dependence of the cross-section on $x_3, x_4$ in production of both two and four jets. Large $x_3$ correspond to partons which are likely to be closer to the centre of the nucleon than small $x$ partons, leading to decrease of the probability of the gap survival with increase of $x_3, x_4$. Correlations between the partons should also enhance the cross-section of the exclusive channel of four jet production in the double diffraction when the light-cone fraction carried by two of the interacting partons of both nucleons are close to maximal: $(x_1 + x_2)/x_{I\!P} \sim 1$. Such a contribution should be enhanced if $-t_1, -t_2$ are large enough (few GeV$^2$) to squeeze the transverse sizes of the exchanged ladders (see Fig. 5.2) [2].

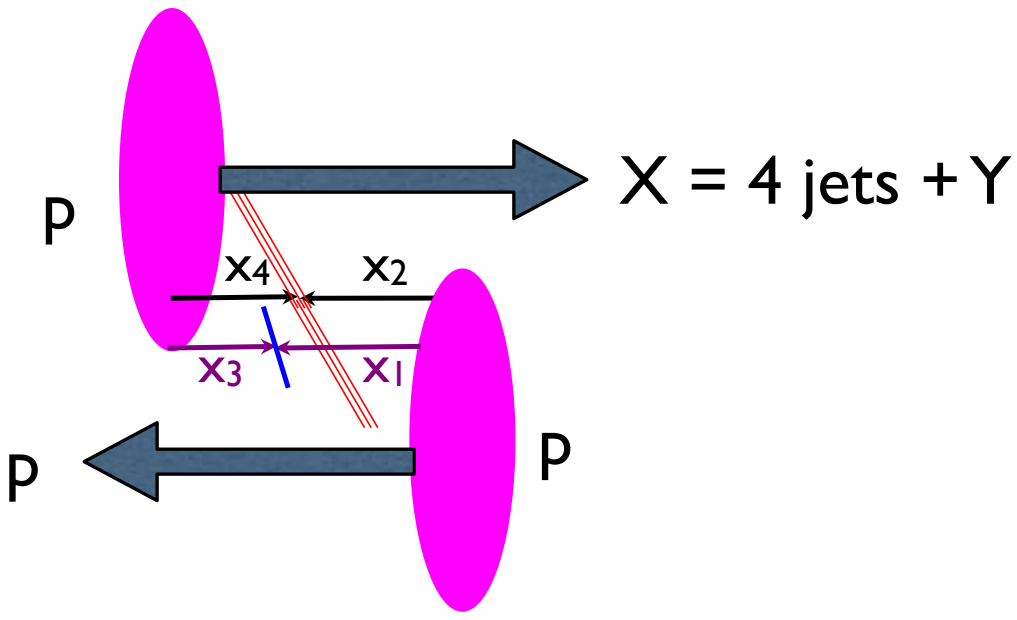

Figure 5.1: Single diffraction process with production of four jets.

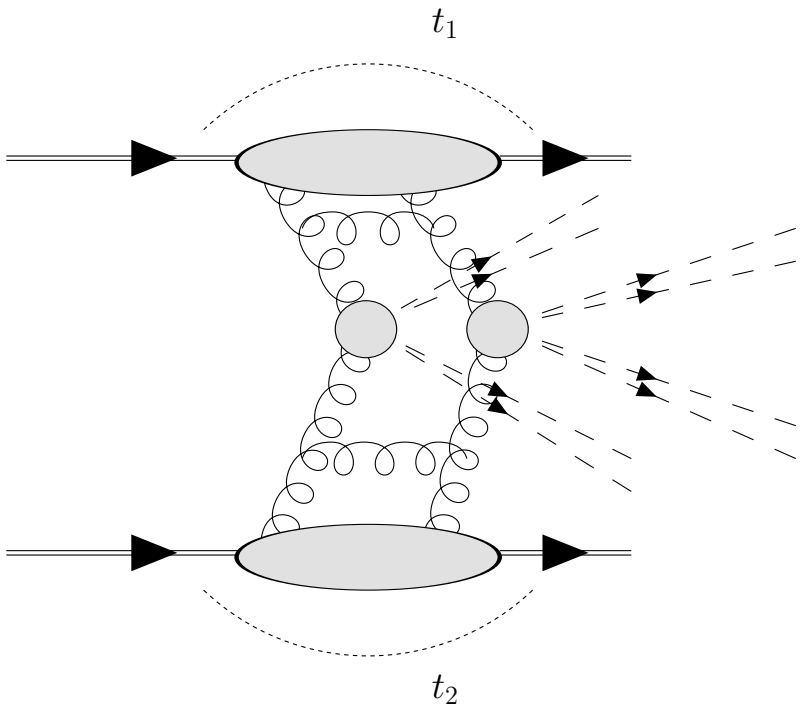

Figure 5.2: Double Pomeron process with production of two pairs of dijets.

# References

[1] L. Frankfurt, M. Strikman and C. Weiss, Ann. Rev. Nucl. Part. Sci. **55** (2005) 403 [arXiv:hep-ph/0507286].
[2] T. C. Rogers and M. Strikman, arXiv:0908.0251 [hep-ph].

# 6 What can we learn / expect *on elastic and diffractive scattering* from the LHC experiments ?

**Author: Konstantin Goulianos**

## 6.1 Introduction

Diffraction is the last frontier in the effort to harness the standard model under a computational framework that includes non-perturbative quantum electrodynamics (npQCD). Despite the success of lattice calculations in predicting the hadron mass spectrum, predictions for diffraction are still based on phenomenological models. The transition from phenomenology to theory will benefit from the larger rapidity and transverse momentum that will become available at the Large Hadron Collider (LHC). The aim should be twofold: unveil the QCD basis of diffraction, and use diffraction as a tool to discover new physics either within (dark energy?) or beyond the standard model (supersymmetry?).

The goal of conducting studies of elastic and diffractive scattering at hadron colliders should be twofold: unveil the QCD nature of the diffractive exchange, which historically is referred to as the *Pomeron*, and use diffraction as a tool in searching for new physics [1].

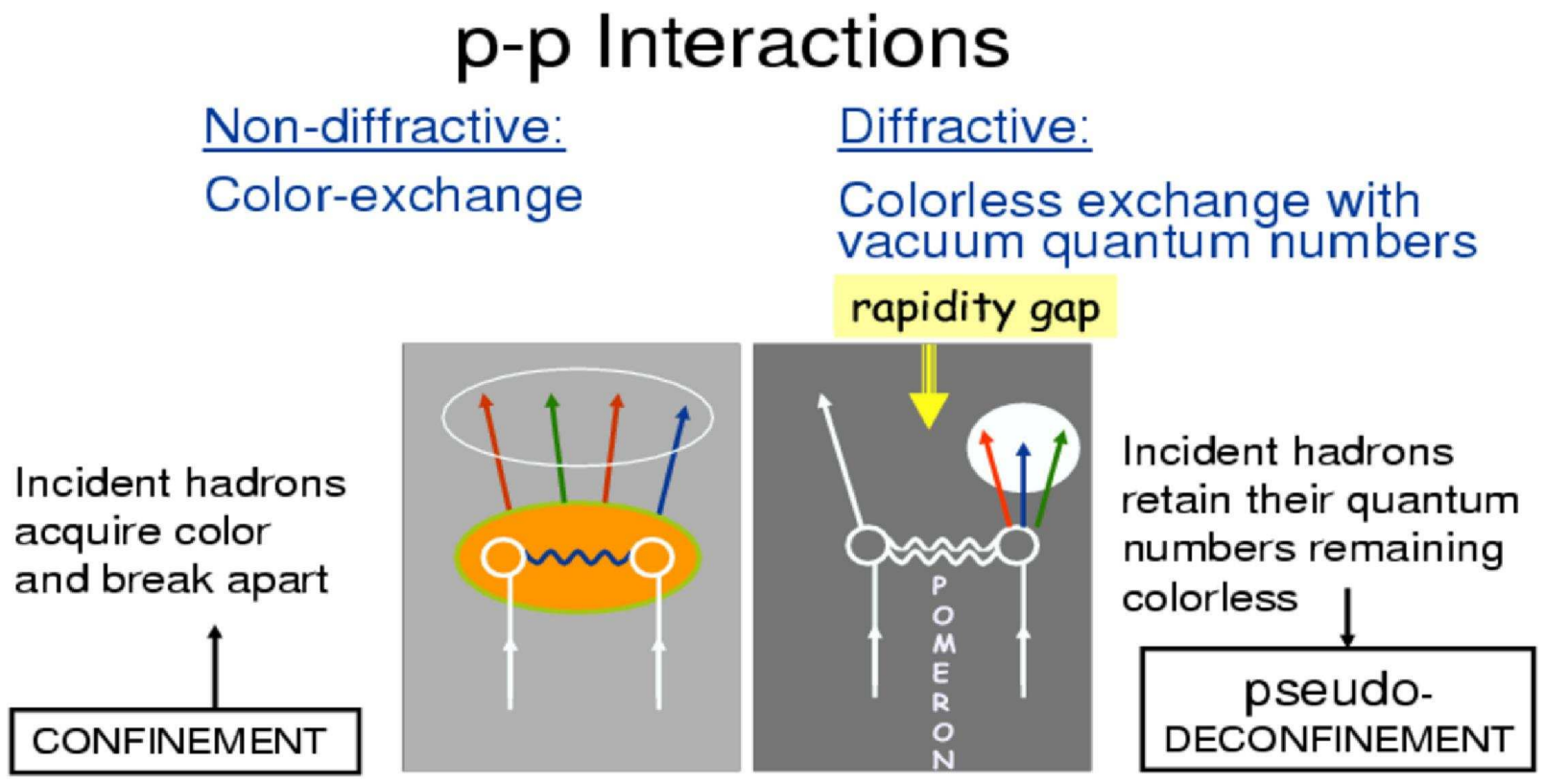

Figure 6.1: Non-diffractive and diffractive *pp* interactions.

Figure 6.1 illustrates the final-state event topologies of non-diffractive (ND) and single-diffractive (SD) *pp* interactions. A general QCD process involves a colour transfer by gluons and/or quarks. Due to colour-confinement, this is a short-range interaction. In diffraction, the exchange is a colour-singlet combination of gluons and/or quarks carrying the quantum numbers of the vacuum. As no colour is transferred, the process can be viewed as *pseudo-deconfinement*, where the prefix *pseudo* is used because the exchange has an imaginary mass and the process can proceed only if there is enough energy transferred to produce a pion. This is not unlike photon emission, in which a photon can only *deconfine* itself from the proton by interacting with an electromagnetic field, as for example in passing through matter. However, the difference is that the photon is massless, while the quantum of the strong force, the pion, has mass.

An interesting question arises: what happens if the emitted Pomeron has such low energy that it cannot produce a pion upon absorption by a nearby proton? Will it keep going in search of another hadron, or more precisely in search of a quark and be trapped in the Universe as a large wave length energy bundle in the process of being exchanged? Such an energy bundle will correspond to an imaginary mass, which brings up the next question: what are the gravitational consequences of this imaginary energy trapped in the Universe?

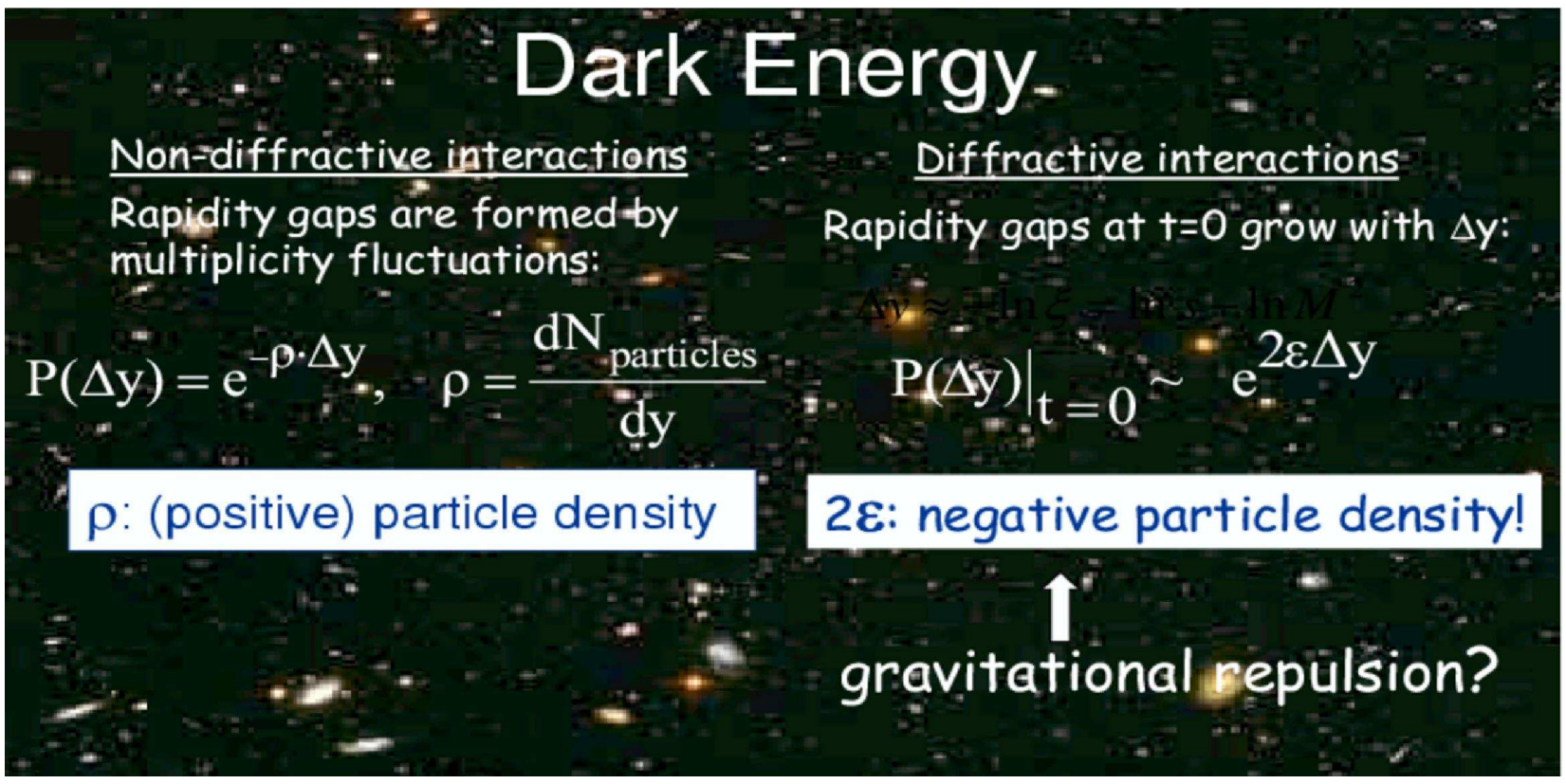

Figure 6.2: (left) In non-diffractive interactions the probability $P(\Delta y)$ for forming a gap $\Delta y$ is exponentially suppressed as $\exp[-\rho \cdot \Delta y]$, where $\rho$ is the final state particle density per unit rapidity; (right) in diffractive interactions, $P(\Delta y)$ at $|t| = 0$ increases with $\Delta y$, which corresponds to a *negative* particle density $\rho' = -2\epsilon$. Does this lead to gravitational repulsion?

The dependence of the diffractive cross-section on the size of the rapidity gap may be a clue that provides the answer. As displayed in Fig. 6.2, in writing the differential diffractive cross section in terms of the rapidity gap $\Delta y$ instead of the forward momentum loss fraction $\xi$ using $\Delta y = -\ln \xi$, the term $2\epsilon$, where $\epsilon$ is the excess above unity of the intercept of the Pomeron trajectory, appears formally as a negative particle density. Does this signify a gravitational repulsion caused by this unrealized energy permeating the Universe? If yes, can one relate the value of $\epsilon$ with the rate of gravitational expansion?

## 6.2 What to do at the LHC

| | |
|---|---|
| Goal | · understand the QCD basis of diffraction and discover new physics |
| Exploit | · large $\sqrt{s} \Rightarrow$ large $\sigma$, $\Delta\eta$, $E_T$ |
| TEV2LHC | · from Tevatron to LHC: confirm, extend, discover... |
| | $\Rightarrow$ confirm Tevatron results and extend them into the new kinematic domain |
| Specifics | · elastic, diffractive, total cross-sections, and $\rho$-value |
| | $\Rightarrow$ diffractive structure function: dijets vs. $W$-boson, ... |
| | $\Rightarrow$ multi-gap configurations |
| | $\Rightarrow$ jet·gap·jet: $d\sigma/d\Delta\eta$ vs. $E_T^{\rm jet} \Rightarrow$ BFKL, Mueller-Navelet jets |

# References

[1] V.A. Khoze, A.B. Kaidalov, A.D. Martin, M.G. Ryskin, W.J. Stirling, *Diffractive processes as a tool for searching for new physics,* arXiv:hep-ph/0507040.

[2] K. Goulianos, *Factorization Breaking in Diffraction*, in these Proceedings.

[3] K. Goulianos, *Diffractive and Total pp Cross Sections at LHC*, in these Proceedings.

[4] K. Goulianos, *The Forward Detectors of CDF and DØ*, in these Proceedings.

# 7 String Theory and the Pomeron

**Author: Chung-I Tan**

The application of the so-called Anti-de-Sitter / Conformal-Field-Theory (AdS/CFT) correspondence between strongly coupled QCD and weakly coupled gravity has recently been successfully applied to the computation of various observables in high-energy heavy-ion physics. The application of this duality to diffractive scattering and the Pomeron physics represents another area where a connection with the string-theory-based techniques can be made. Furthermore, it is now possible to extend this treatment to central diffractive production of Higgs at LHC.

The connection with the stringy aspects in a five-dimensional description is indeed very direct. In gauge theories with string-theoretical dual descriptions, the Pomeron emerges unambiguously. The Pomeron in QCD can be associated with a Reggeized Graviton, where both the IR (soft) Pomeron and the UV (BFKL) Pomeron are dealt with in a unified single step. Indeed, the Pomeron is directly related to the graviton and its higher spin partners on the leading (five-dimensional) Regge trajectory.

In AdS/CFT, confinement is associated with a deformed $AdS_5$ geometry having an effective horizon, e.g., that for a black hole. The solution to this is unknown and represents the major theoretical challenge in model-building. Each model leads to a certain unique signature. LHC data can provide guidance and direction in this endeavor.

The traditional description of high-energy small-angle scattering in QCD has two components – a soft Pomeron Regge pole associated with exchanging tensor glueballs, and a hard BFKL Pomeron at weak coupling. On the basis of gauge/string duality, a coherent treatment of the Pomeron can be achieved (BPST)[1], thus providing a firm theoretical foundation for the Pomeron in QCD. It is now possible to identify a *dual Pomeron* as a well-defined feature of the curved-space string theory. In the large 't Hooft coupling, the Pomeron can be considered as a *Reggeized Massive Graviton*, propagating in a 5-dimensional curved space, the so-called $AdS_5$.

The fact that a 5-dimensional description enters in high energy collisions can be understood as follows. In addition to the usual LC momenta, $p_\pm = p^0 \pm p^z$ (2d), and transverse impact variables, $\vec{b}$ (2d), there is one more "dimension": a "resolution" scale specified by a probe, e.g., $1/Q^2$ of the virtual photon in DIS, (see Fig. 7.1a.) Because of conformal symmetry, these 5 coordinates transform into each others, leaving the system invariant. In the strong coupling limit, conformal symmetry is realized as the $SL(2, C)$ isometries of Euclidean $AdS_3$ subspace of $AdS_5$, where the $AdS$ radius $r^2$ can be identified with $Q^2$.

The *dual Pomeron* has been identified as a well-defined feature of the curved-space string theory (BPST). In the strong coupling limit, conformal symmetry requires that the leading $C = +1$ Regge singularity is a fixed $J$-plane cut. For ultraviolet-conformal theories with confinement deformation, the spectrum exhibits a set of Regge trajectories at positive $t$, and a leading $J$-plane cut for negative $t$, the cross-over point being model-dependent (see Fig. 7.1b). For theories with logarithmically running couplings, one instead finds a discrete spectrum of poles at all $t$, with a set of slowly-varying and closely-spaced poles at negative $t$.

[1] R. Brower, J. Polchinski, M. Strassler, and C-I Tan.

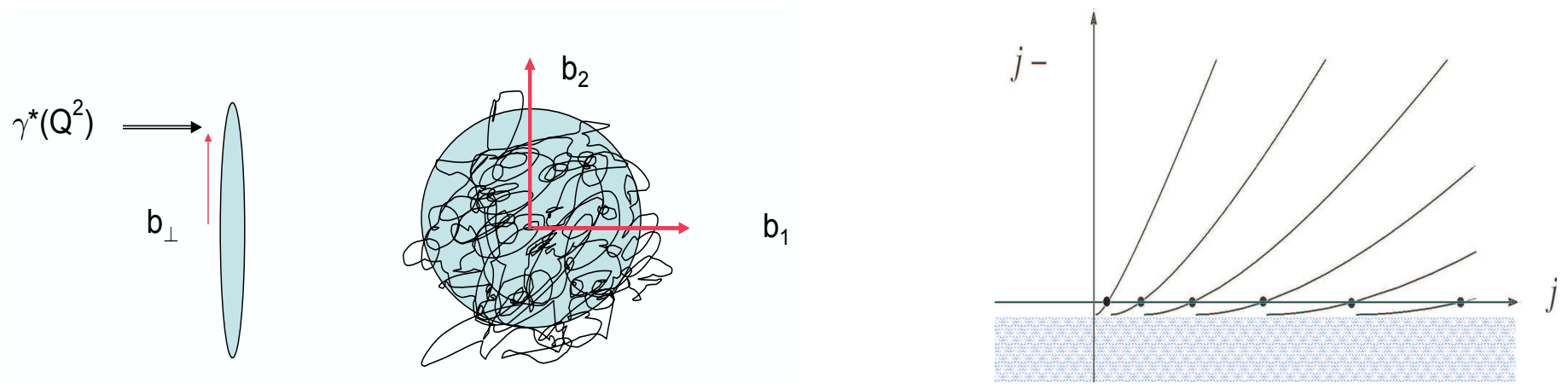

Figure 7.1: Left: Intuitive picture for $AdS_5$ kinematics. Right: Schematic representation of $J$-plane singularity structure.

This strong-coupling formalism can also be extended to diffractive central production of Higgs in forward proton-proton scattering at LHC, e.g. the double-diffractive process, $pp \to pHp$. The theoretical estimates generally involve the assumption of perturbative contribution of gluon fusion in the central rapidity region, (e.g., the Durham group.) In these estimates the Pomeron is effectively replaced by two-gluon exchange referred to in the early literature as the Low-Nussinov Pomeron. In spite of the plausibility of this approach, there are considerable uncontrolled uncertainties. The Regge description for diffractive production is well known to be intrinsically non-perturbative. An analysis in strong coupling based on the AdS/CFT correspondence and conformal strong coupling BPST Pomeron can now be carried out. While this also will have its uncertainties, a careful comparison between weak and strong coupling Pomeron should give better bounds on these uncertainties. Ultimately, the strong coupling approach calibrated by comparison with experimental numbers for double diffraction heavy quark production, can provide increasingly reliable estimates for Higgs production.

# 8 The FP420 Project

**Author: Albert De Roeck**

The physics potential of forward proton tagging at the LHC has attracted much attention in the last years. The focus of interest is the central exclusive production (CEP) process $pp \to p + \phi + p$ in which the protons remain intact and the central system is separated from the outgoing protons by a large rapidity gap. A very interesting case is the CEP process of a Higgs particle. A picture of the basic process is shown in Fig. 8.1 (left).

There are several advantages of CEP [1, 2]:

- The selection rules for CEP are such that the central system is – to a good approximation – a $0^{++}$ state. Observing CEP thus gives access to the quantum numbers of the state $\phi$.
- The three particle final state is a very constrained system. As a consequence the azimuthal correlation between the outgoing protons is directly sensitive to CP quantum numbers and is a possible way to study CP violating Higgs scenarios in detail.

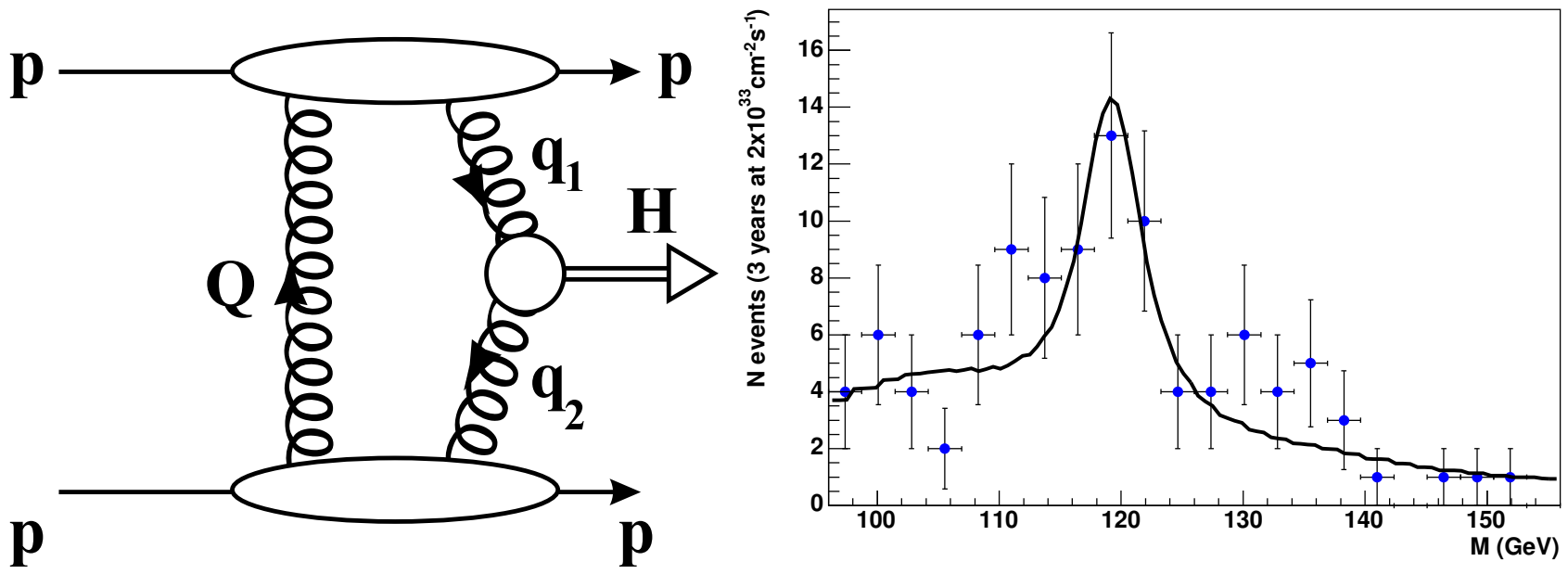

Figure 8.1: (Left) Diagram for the CEP process; (Right) A typical mass fit for 3 years of data taking at $2 \times 10^{33}$ cm$^{-2}$ s$^{-1}$ (60 fb$^{-1}$), using only events with both protons tagged at 420 m.

- The tagging of the proton allows for the measurement of the mass of the system $\phi$ with a precision of the order of 1-2 GeV, via the missing mass w.r.t. the incoming proton beams $M_{miss} = (p_1 + p_2 - p_1' - p_2')^2$ with $p_1$, $p_2$ the incoming and $p_1'$, $p_2'$ the outgoing protons. This measurement is independent of the way the central system $\phi$ decays.
- The QCD backgrounds such as $gg \to qq$ are strongly suppressed in LO.
- CEP can be a discovery channel in certain regions of the MSSM parameter space.
- CEP gives a unique access to a host of interesting QCD phenomena.

The main physics topics studied by FP420 are the Central Exclusive Production, including Higgs production and searches for new physics, QCD and diffractive studies with tagged protons and photon induced processes with tagged protons. These topics are reported in [3]. Fig. 8.1 shows an example of signal plus background estimates [4]. The cross-section can be a factor 10 or more larger than the SM model one. This has recently been explored in a systematic way in [5]. There are still some issues and concerns on the CEP soft survival probability at the LHC and the uncertainties in the PDFs. This question will be settled with the first data at the LHC.

New detectors are needed to complement the CMS and ATLAS experiments to detect these protons [6]. FP420 is an R&D collaboration that studies the feasibility to detect the protons of CEP with detectors at a distance of 420 m away from the interaction point [7]. Such detectors allow to accept protons with a fractional momentum loss (or $\xi$) of 0.1% to 1%. With these detectors the protons of CEP Higgs production in the mass range of $70 < M_\phi < 150\,\text{GeV}/c^2$ can be detected.

The FP420 project is schematically presented in Fig. 8.2. The aims of the R&D study are:

- Redesign the area of the machine around 420 m. Right now this area contains a connecting cryostat, but no magnet elements.
- Study the mechanics, stability, services for detectors at 420 m
- Design and test tracking detectors to operate close to the beam
- Design fast timing detectors (with O(10) ps resolution)

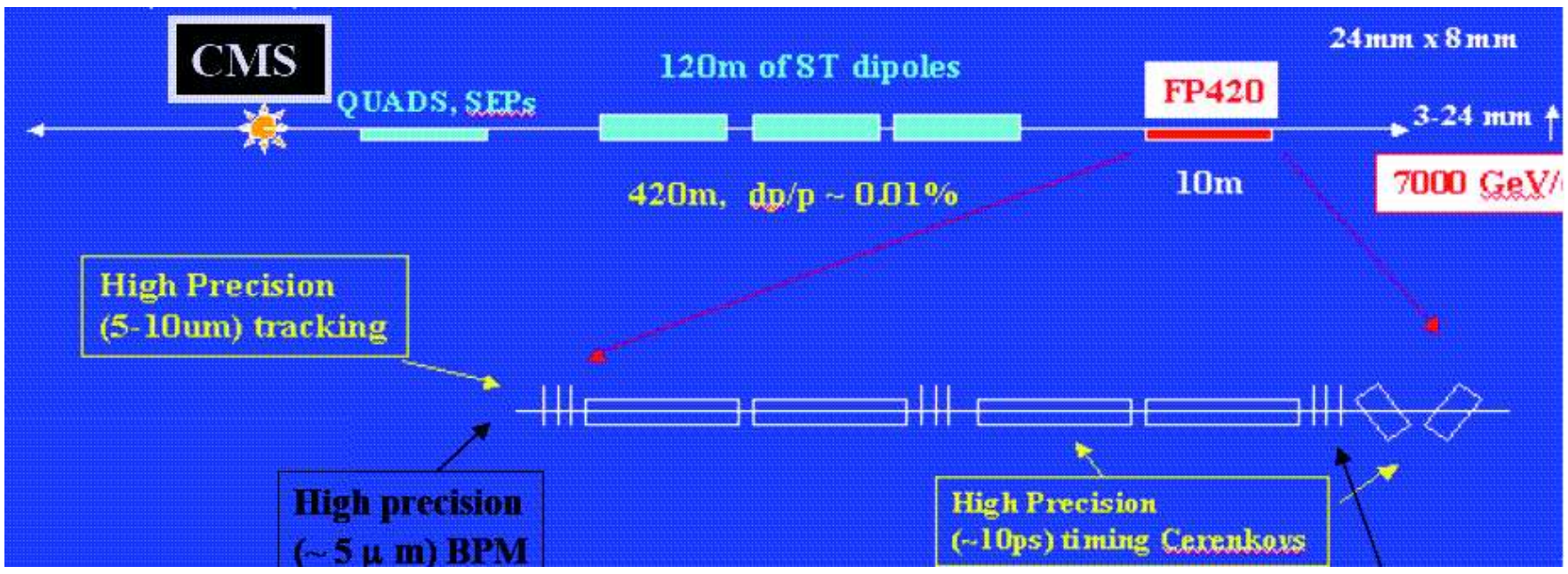

Figure 8.2: Schematic layout of the FP420 detectors.

- Study RF pickup, integration, precision alignment, radiation and resolution issues
- Study trigger, event selection, and pile-up issues.
- Study the operation of FP420 detectors at the highest LHC luminosity.

The FP420 collaboration has members from ATLAS, CMS, and 'independent' physicists, and has excellent contacts with the LHC machine group. In the emerging design the principle of FP420 is based on moving "pockets" which contain tracking and timing detectors. The tracking detectors that are developed are 3D silicon pixel detectors, which are radiation hard and can detect particles close to the edge. Timing detectors include both gas and crystal radiators. The test beam results of all these detector types have been excellent and e.g. show that the 10 ps timing can be achieved. A full pocket beam-test was performed in October 2007. A full account of the R&D results achieved so far has been published in 2008 [3] and forms the basis for the discussions on FP420 with the ATLAS and CMS experiments. Both experiments are now in the process of a review procedure.

Synergy has been a cornerstone of FP420 from the start, through the common efforts of ATLAS and CMS and externals on this project. Clearly when established to the end, this technology could also be used at other interaction points. Furthermore the exciting physics opportunities offered by FP420 have no doubt triggered the vigilant efforts at the Tevatron making measurements to check the theoretical predictions of several of the associated exclusive processes, as reported at this workshop. Recent developments include extending the FP420 techniques in the region around 220 m. For ATLAS it is already foreseen to have a common 220/420 project proposal. In the case of CMS, the TOTEM experiment is located at 220 m around IP5. So there are in principle two paths possible: either have an upgrade of the 220 m detectors with e.g. detector extensions for timing – which is absolutely essentially to control the pile-up at high luminosity – and have common readout with TOTEM/CMS, or use the 240 m area which is still free. The common readout was originally planned from the start but seemingly will not be a priority at start-up. On the other hand, the operational experience of TOTEM as the first experiment with near beam detectors will be extremely valuable and calls for a common study from all proponents interested in such type of measurements from the very beginning.

CMS and ATLAS will start their diffractive/rapidity gap programme making measurements with events which have regions void of energy and particles, at low luminosities when pile-up is absent. This will allow to measure some of the key phenomena, such as the gap survival probability, necessary to gauge the theoretical predictions for CEP processes.

In short, now that the technology is getting established for FP420-like stations, it is of interest to see where else (e.g 220 m) they could be deployed, and to use the imminent startup of the LHC in order to gain as much operation experience on near beam detectors as possible, within a collaboration across the experiments.

# References

[1] V.A Khoze, A.D. Martin and M.G. Ryskin, Eur. Phys. J. C. **23** (2002) 311.

[2] V.A Khoze, A.D. Martin and M.G. Ryskin, Eur. Phys. J. C **34** (2004) 327, hep-ph/0401078.

[3] M.G. Albrow et al., *FP420 R&D report*, JINST 4,T 10001 (2009) and arXiv:0806.0302.

[4] B.E. Cox, F.K. Loebinger, A.D. Pilkington, MAN-HEP-2007-15 (2007), arXiv:0709.3035 [hep-ph].

[5] S. Heinemeyer et al., DCPT-07-80, IPPP-07-40, Aug 2007, arXiv:0708.3052

[6] A. De Roeck, V. A. Khoze, A.D. Martin, R. Orava, and M.G. Ryskin, Eur. Phys. J. C **25** (2002) 391.

[7] M.G. Albrow et al., CERN-LHCC-2005-025, Jun 2005.

# 9 Proton Detection at IR3

**Author: Hubert Niewiadomski**

## 9.1 Introduction

As motivated in the previous chapter (A. De Roeck), TOTEM also investigated in which locations the machine dispersion is large and the beam size small, in order to optimise the proton acceptance at small momentum losses. The momentum cleaning region IR3 (Figure 9.1, left) seems to be optimal. Its optics has been optimised to absorb the protons with relative momentum deviations $\xi = \Delta p/p$ exceeding $\pm 1 \times 10^{-3}$. Such protons can be detected by near-beam insertions located in the warm region of IR3 before being intercepted by the momentum cleaning collimators. The technical aspects of the proposed RP insertions are presented in [1].

This would highly extend the diffractive mass acceptance of the TOTEM experiment. In case of the Double Pomeron Exchange process, a continuous mass acceptance from 30 GeV to 2.5 TeV would be accessible, allowing for a promising diffractive physics programme. In addition, within a certain $\xi$ range, the diffractive protons from all LHC interaction points are detected, thus making online inter-experimental luminosity calibrations possible.

## 9.2 Beam Optics and Insertion Location

By design, the IR3 region is optimised such that off-momentum protons can be intercepted by the collimators. This is achieved by maximising the ratio $D_x/\sigma_x$, i.e. exactly the beam optics property needed for a momentum measurement down to low values of $\xi$ with good resolution. The closest safe approach of a detector to the beam is given by a certain multiple – typically 10 to 15 – of the beam size $\sigma_x$, which limits the lowest detectable $\xi$-values. As a result of the large value of $D_x$, the diffractive protons are deflected further away from the beam centre and can be measured in the near-beam detectors.

Figure 9.1 (right) shows the dispersion and beam width in the IR3 region for both beams and both transverse projections, $x$ and $y$, for the nominal LHC optics configuration with $\beta^* = 0.5$ m and $\sqrt{s} = 14$ TeV. The horizontal dispersion $D_x$ at the two potential tracking detector positions, TP1 and TP2, has a magnitude in the range of 2–3 m, as compared to 8 cm at the TOTEM Roman Pot station RP220. The high ratio $D_x/\sigma_x \approx 6.7 \times 10^3$ (as compared to $\approx 1.1 \times 10^3$ at RP220) results in an acceptance down to $\xi = 1.6 \times 10^{-3}$.

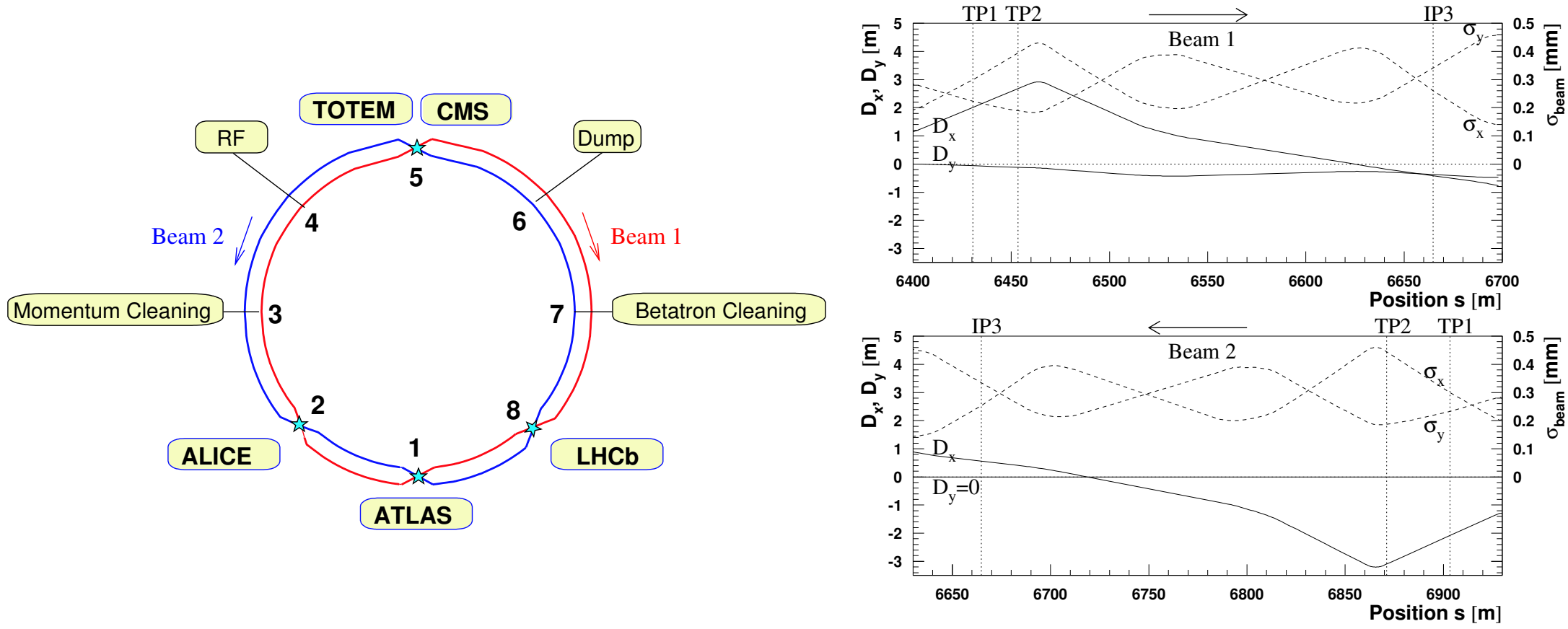

Figure 9.1: Left: Schematic drawing of the LHC with its eight "interaction" points, showing the location of the momentum cleaning insertion IR3. Right: Dispersion (left-hand axes) and beam width (right-hand axes) in $x$ and $y$ for both beams in the IR3 region. The dispersion shown is valid for protons with $\xi = 0$ and produced in IP5. The position axis $s$ follows beam 1 and has its origin in IP1. TP1 and TP2 are the two proposed tracking detector planes located in a warm region of the machine.

In addition to promising perspectives for diffractive physics, the placement of detectors in front of the momentum cleaning collimators has advantages for machine diagnostics and protection. It enables the study of beam losses at the collimators. Furthermore, all showers possibly created by the detector insertion are absorbed immediately downstream by the collimators. Finally, the insertions are proposed in a warm region and therefore their installation should not be technically too complicated.

## 9.3 Proton Acceptance and Reconstruction in IR3

The proton acceptances for both beams are shown in Figure 9.2 (left). The protons are characterised by $\xi$, integrated over all their other kinematic parametres. The IR3 acceptance for beam 1 protons originating from diffractive scattering in IP5 is reduced since these protons have to pass through the aperture limiting betatron cleaning insertion IR7. Beam 2 protons on the other hand have an almost continuous acceptance from $\xi = 1.6 \times 10^{-3}$ to 0.19 (50% acceptance limits) with only a gap between 0.01 and 0.018. This momentum acceptance gives access to diffractive masses ranging from 30 GeV to 2.5 TeV in the case of Double Pomeron Exchange events. A detailed reconstruction study, discussed in detail in [2], led to the $\xi$-resolution shown in Figure 9.2 (right). Note that the resolutions $\sigma(\xi) \sim 10^{-4}$ achieved for measurements in IR3 reach the limit imposed by the energy uncertainty of the LHC. In all cases, the relative resolution $\sigma(\xi)/\xi$ is better than 10%.

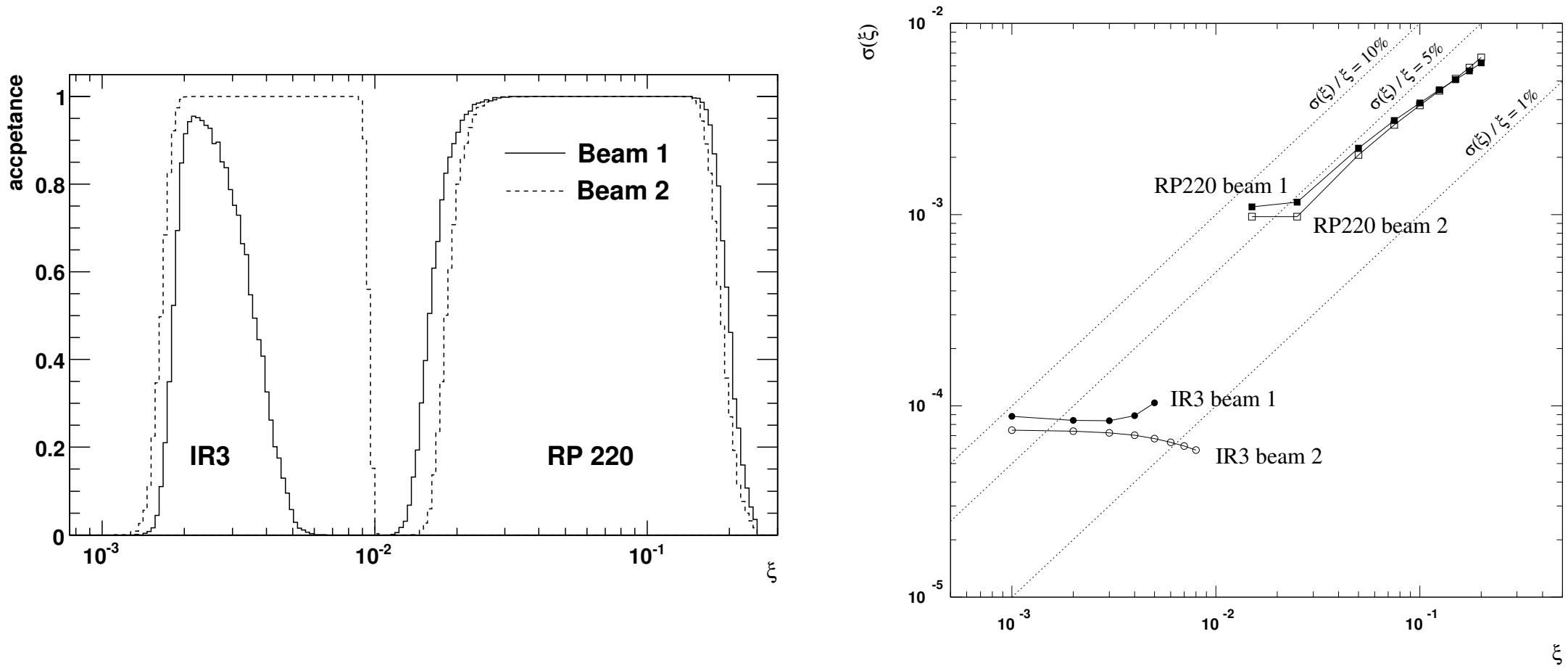

Figure 9.2: Left: Acceptance in $\xi$ at the TOTEM Roman Pots RP220 and in IR3 for both LHC beams. Right: Resolution in the reconstructed $\xi$ at the TOTEM Roman Pots RP220 and 206 m upstream of IP3 for both LHC beams. Nominal LHC optics $\beta^* = 0.5\,\mathrm{m}$ and $\sqrt{s} = 14\,\mathrm{TeV}$ was applied.

# References

[1] K. Eggert, P. Aspell, R. Assmann, V. Avati, M. Deile, H. Niewiadomski, V. Previtali, E. Radermacher, and T. Weiler, "Detection of Diffractive Protons in IR3 at the LHC", in preparation.

[2] H. Niewiadomski, "Reconstruction of Protons in the TOTEM Roman Pot Detectors at the LHC", PhD dissertation, Manchester, 2008.

# Part XII

# Summary

# Summary on Theoretical Aspects

*Jacques Soffer*

Department of Physics, Temple University, Philadelphia, Pennsylvania 19122-6082, USA

During the five days of this conference a very dense scientific program has enlightened our research fields, with the presentation of large number of interesting lectures. I will try to summarize the theoretical aspects of some of these new results.

## 1 Introduction

This meeting has confirmed once more a clear scientific evolution, that is the foundations of elastic scattering and diffraction phenomena at high energy, are now best understood in terms of the first principles of quantum chromodynamics (QCD). Since we are just at the start-up of the LHC, a great deal of the theoretical activity focus on the accessibility to this new energy regime, for the interpretation of relevant aspects of the strong interactions, by means of basic QCD mechanisms. Clearly they have to be confronted with experimental measurements, a very relevant part which will be summarized elsewhere [1].

I will essentially cover the following topics:

- Elastic and Total Cross Section
- Soft Diffraction
- Hard Diffraction and Central Production

Unfortunately, I have left out some important topics, in particular, ultra high energy cosmic rays and heavy-ion physics, because of lack of time and I apologize for that.

## 2 Elastic Scattering and Total Cross Section

This is a classical subject which was very largely discussed from different viewpoints. Let us first mention a study of the amplitudes $pp$ and $\bar{p}p$ elastic scattering in the Coulomb-Nuclear interference (CNI) region [2, 3], using a method based on derivative dispersion relations. The real and imaginary parts of the hadronic amplitude near the forward direction, whose detailed knowledge is needed, are parametrised by a single exponential, with two different hadronic slopes $B_R$ and $B_I$. The analysis of the available data, in the range from $\sqrt{s}$=19 GeV to 1800 GeV, leads to the conclusion that $B_R > B_I$, although the determination of $B_R$ is far less precise than $B_I$, for obvious reasons. Note that strictly speaking, this concept of hadronic slope is very misleading, since it is known that the derivative of the amplitude with respect to $|t|$ is a slowly decreasing function of $|t|$ as shown in Ref. [4], an approach where real and imaginary parts of the amplitude are strongly related. The relevance of the measurement of the real part of the $pp$ forward scattering amplitude at the LHC has been also emphasized in Ref. [5].
A model for $pp$ and $\bar{p}p$ elastic scattering, based on the electromagnetic and gravitational form factors related to a new set of generalized parton distributions (GPD) was used, after unitarization, to fit the data [6]. Unfortunately the quality of the fit is rather poor, with a $\chi^2/\mathrm{pt} = 3$ and it predicts a high value of the total cross section at LHC, $\sigma_{tot} = 146\,\mathrm{mb}$. One gets an even higher prediction at $\sqrt{s} = 14\,\mathrm{TeV}$, $\sigma_{tot} = 230\,\mathrm{mb}$, in another approach, which introduces the concept of reflective elastic scattering at very high energies [7]. This picture also predicts that the scattering amplitude at the LHC energy goes beyond the black disk limit.

Another phenomenological investigation of $pp$ and $\bar{p}p$ elastic scattering was carried out by considering that the proton consists of an outer region of $\bar{q}q$ condensed ground state, an inner shell of topological baryonic charge and a core where valence quarks are confined [8]. It leads to $\sigma_{tot} = 110\,\mathrm{mb}$ and for the ratio of real to imaginary parts of the forward amplitude $\rho = 0.12$ at LHC. The predicted differential cross section $d\sigma/dt$ has a smooth behaviour beyond the bump at $|t| \simeq 1\,\mathrm{GeV}^2$, with no oscillations and a much larger value, in contrast with other models.

Concerning the specific issue of the value of the $pp$ total cross section at the LHC, a highly non-perturbative quantity which cannot be predicted by QCD, we had a general presentation of different models (double poles, triple poles, cuts, etc. ...) and their experimental consequences [9]. It was stressed that the theoretical uncertainty is large, as discussed above, and therefore an accurate measurement is badly needed since it will also tell us a lot about the analytic structure of the $pp$ elastic amplitude.

The eikonal approach has been proven to be very useful in describing high energy elastic scattering. Clearly it relies on the knowledge of the impact parameter profile, which can be either more peripheral or central. The analysis of the $pp$ data at the ISR energy $\sqrt{s} = 53\,\mathrm{GeV}$ led to the conclusion that a peripheral profile is preferred in this case [10]. In another presentation [11], the validity of the optical theorem commonly used to extract the total cross section has been questioned.

A new rigorous result on the inelastic cross section was obtained recently [12] and it reads $\sigma_{inel}(s) < \frac{\pi}{4m_\pi^2}(\ln s)^2$. This bound is four times smaller than the old Froissart bound derived in 1967, $\sigma_{tot}(s) < \frac{\pi}{m_\pi^2}(\ln s)^2$ where $\frac{\pi}{m_\pi^2} = 60\,\mathrm{mb}$. This last result can be also improved by a factor two, using some reasonable assumptions and it would be nice to prove it rigorously.

A possible description of high-energy small-angle scattering in QCD can be done by means of two vacuum exchanges with $C = \pm 1$, the Pomeron and the Odderon. Recent developments in this subject, based on the weak/strong duality, relating Yang-Mills theories to string theories in Anti-de-Sitter (AdS) space, were presented in some details [13]. If the QCD Pomeron is viewed as a two-component object, soft and hard, a dual description of the Pomeron emerges unambiguously through the AdS/CFT approach and the Odderon is related to the anti-symmetric Kalb-Ramond field. Some aspects of analyticity, unitarity and confinement were also discussed.

# 3 Soft Diffraction

In an overview of soft diffraction [14], several theoretical approaches were considered for a better understanding of the relevant mechanisms of high-energy interactions and making an instructive comparison between s- and t-channel view points. Diffractive production in the s-channel is peripheral in the impact parameter and there is a strong influence of unitarity effects due to multi-Pomeron exchanges. The calculation of the survival probability for hard processes is very important, in particular for high mass diffraction, as we will see later, for example, for central

Higgs production.

Following the above ideas, a model based on Gribov's Reggeon calculus was proposed and applied to soft diffraction processes at high energy [15]. By giving a special attention to the absorptive corrections, the parameters of the model are determined following from a good description of the existing experimental data on inclusive diffraction in the energy range from ISR up to Tevatron. The model predictions for single and double diffraction at LHC energy are also given later. In another contribution [16], one was recalling the method to unitarize the Pomeron for elastic and inclusive scattering, providing as well as a comparison with data, mainly for one-particle inclusive production and some LHC predictions.

Soft scattering theory was re-visited by considering some eikonal models for simplicity and to secure s-channel unitarity [17]. After recalling the main features of two specific models [18, 19], the interplay between theory and data analysis led to some LHC predictions, in particular a total cross section of the order of 90 mb, in contrast with the prediction $\sigma_{tot} = (103.6 \pm 1.1)$ mb from Ref. [20]. Another important point from Ref. [20] to notice here, is the fact that the ratio $\sigma_{el}/\sigma_{tot}$ rises from the value 0.18 for $\sqrt{s} = 100$ GeV to 0.30 for $\sqrt{s} = 100$ TeV, whereas Refs. [18, 19] predict almost no energy dependence in this range.

Some special features of the model of soft interactions of Ref. [18] mentioned above, were discussed together with the results of the fit to determine the parameters of the model [21]. Needless to say that it is very important to estimate the survival probability for central exclusive production of the Higgs boson, which was also compared with the results of the model of Ref. [19].

This question is also related to the notion of colour fluctuations in the nucleon in high energy scattering [22], so it is legitimate to ask: how strong are fluctuations of the gluon field in the nucleon? A simple dynamical model can explain the ratio of the inelastic to the elastic cross section in vector meson production in *ep* collisions at HERA and leads to a new sum rule [23]. However it cannot explain the Tevatron CDF data and it reduces the expected survival probability in central exclusive production.

# 4 Hard Diffraction and Central Production

The standard QCD mechanism for central exclusive production for heavy systems, using the formalism of collinear generalized parton distributions has been proposed some time ago and applied for Higgs production at the LHC [24]. In this case also, it is relevant to question a possible violation of QCD factorization and some aspects of analyticity and crossing properties [25]. At the phenomenological level, the same mechanism was used to calculate the amplitudes for the central exclusive production of the $\chi_c$ mesons, using different unintegrated gluon distribution functions (UGDF) [26]. The extension of the UGDF to the non-forward case, can be obtained by saturation of positivity constraints. The resulting total cross sections for all charmonium states $\chi_c(0^+, 1^+, 2^+)$ are compared at Tevatron energy.

The present situation of theoretical predictions for central exclusive production of Higgs bosons and other heavy systems at the LHC was reviewed [27]. It was shown that the CDF dijet data can be used to reduce the uncertainty on the cross section prediction for the Higgs boson. The claim is that a cross section between 0.3 and 2 fb is expected for a standard Higgs of mass 120 GeV. Central exclusive production of vector mesons may be used as a discovery channel for the Odderon.

Some simple examples of physics beyond the Standard Model (SM), which require an ex-tended Higgs sector, were considered [28]. Assuming a central exclusive production mechanism, the sensitivity of the search for the corresponding Higgs was studied, with some experimental aspects like signal and background rates. In another presentation, along the same lines of extending the Higgs sector beyond the SM, the search for the lightest neutral Higgs boson of a model containing triplets was discussed [29]. By means of some Monte Carlo simulations, it was found that the central exclusive production mechanism is again a very powerful tool to study this new object.

Deep-inelastic scattering data in the very low-$x$ region is known to be dominated, in the Regge picture, by the Pomeron. By using a discretized version of the BFKL Pomeron, which generates discrete Regge pole solutions, an integrated positive gluon distribution was obtained [30]. It allows a good fit of the ZEUS $F_2$ data in the kinematic range $10^{-4} < x < 10^{-2}$ and $4.5 < Q^2 < 350\,\mathrm{GeV}^2$ and this gluon distribution must be tested in hadronic collisions at the LHC.

In jet production at LHC, gaps between jets is an important issue which deserves serious theoretical studies, because, it is sensitive to various QCD processes. The phenomenological impact of the Coulomb gluon contributions and super-leading logarithms on the gaps between jets cross section, has been investigated [31].

## Acknowledgements

I would like to thank the conference organizers for their warm hospitality at CERN and for making the 13th “Blois Workshop” a very successful meeting. I am also grateful to all the conference speakers for the high quality of their contributions.

## References

[1] J. Dainton, Summary on Experimental Aspects, these proceedings.

[2] E. Ferreira, these proceedings.

[3] A. Kendi Kohara, E. Ferreira and T. Kodama, arXiv:hep-ph/0905.1955 (2009).

[4] C. Bourrely, J. Soffer and T.T. Wu, Nucl. Phys. **B247** 15 (1984).

[5] C. Bourrely, N.N. Khuri, A. Martin, J. Soffer and T.T. Wu, arXiv:hep-ph/0511135 (2005).

[6] O. Selyugin, these proceedings.

[7] S. Troshin, these proceedings, arXiv:hep-ph/0909.3926 (2009).

[8] R. Luddy, these proceedings.

[9] J.R. Cudell, Total Cross Section at the LHC, these proceedings.

[10] V. Kundrat, these proceedings, arXiv:hep-ph/0909.3199(2009).

[11] M.V. Lokajicek, these proceedings, arXiv:hep-ph/0906.3961 (2009).

[12] A. Martin, these proceedings, Phys. Rev. **D80** 065013 (2009).

[13] Chung-I. Tan, these proceedings.

[14] A. Kaidalov, these proceedings.

[15] M.G. Poghosyan, these proceedings, arXiv:hep-ph/0909.5156 (2009).

[16] E. Martynov, these proceedings.

[17] U. Maor, these proceedings, arXiv:hep-ph/0910.1196 (2009).

[18] E. Gotsman, E. Levin, U. Maor and J.S. Miller, arXiv:hep-ph/0903.0247 (2009) and references therein.

[19] M.G. Ryskin, A.D. Martin and V.A. Khoze, Eur. Phys. J. **C60** 265 (2009) and references therein.

[20] C. Bourrely, J. Soffer and T.T. Wu, Eur. Phys. J. **C28** 97 (2003) and references therein.

[21] E. Gotsman, these proceedings, arXiv:hep-ph/0910.0598 (2009).

[22] M. Strikman, these proceedings.

[23] L. Frankfurt, M. Strikman, D. Treleani and C. Weiss, Phys. Rev. Lett. **101** 202003 (2008).

[24] M.A. Kaidalov, V.A. Khoze, A.D. Martin and M.G. Ryskin, Eur. Phys. J. **C33** 261 (2004) and references therein.

[25] O. Teryaev, these proceedings.

[26] O. Teryaev, R. Pasechnik and A. Szczurek, these proceedings.

[27] J.R. Cudell, Central Exclusive Production, these proceedings.

[28] S. Heinemeyer, these proceedings, arXiv:hep-ph/0909.4665 (2009).

[29] K. Huitu, these proceedings.

[30] D. Ross, these proceedings.

[31] S. Marzani, these proceedings.

# Experimental Highlights

*John Dainton*

Cockcroft Institute of Accelerator Science and Technology and the Univ. Liverpool, UK

Highlights at the 13th International Conference on Elastic & Diffractive Scattering (EDS09) of the presentations of new experimental results and developments are presented and discussed.

## 1 Pedigree and Context

Hadronic physics is the physics of colo(u)r. Colour is the degree of freedom which defines how the hadronic world which we observe and measure is actually the result of the strong interaction. Thus chromodynamics, the theory which we now have for the strong interaction, defines and determines the observable mass of the universe. This mass comes in the form of the atomic nucleus, its constituents - neutrons and protons, their constituents, quarks, and the quantum field dynamics of the non-abelian gauge freedom in chromodynamics. That we can now state all of the above with confidence is *the* monument to the triumphal progress in physics of the last 100 years since Rutherford started it all in Manchester, England, with the first experiment in search of the origin of mass.

Furthermore, the establishment of a theory of matter such as Quantum Chromodynamics (QCD) is a triumph also of human ingenuity founded on the development of the concepts of "structure", "dynamics", and "interaction". One can trace such thinking back through seminal experiments in the last two centuries, back even to ancient civilisation – to Aristotle: "By convention there is colour, by convention sweetness, by convention bitterness, but in reality there are atoms, and space."

So today, and right now at EDS09, we find ourselves concerned with progress in hadronic physics as the physics of the observable mass of the universe, and with a theory, QCD, which is established, but which is also itself complex. This complexity is such that we are a long way from understanding fully the most basic of mechanisms, confinement, by which observable mass is what it is, namely predominantly QCD field energy, gluons, and very much less the constituent mass of the fermion constituents, quarks. It is therefore the case that a cornerstone of contemporary physics continues to be to understand exactly how QCD explains the high energy interactions of hadrons. This is predominantly what we call "diffraction". In so doing, what we can learn conceptually about fundamental mechanisms in QCD, and perhaps beyond, may even lead to a deeper underlying unification.

Less prosaically, and more with the metaphorical "spanner" of the experimentalist in mind, we have at hand the experimental measurements of elastic and inelastic interactions of hadrons at high energy. Most of what we know concerning the elastic scattering of protons has defined what we mean by "diffractive scattering", namely the appealing resemblance of the angular dependence of the cross section to familiar optical diffraction patterns of apertures [1]. The observation of secondary minima and maxima consistent with the femtoscale of the hadronic diffractive aperture are governed by the convolution of the form-factors with the dynamics of the mutual interaction of that part of the constituent structure which is resolved in the interaction in each proton.

| Splitting to off-shell $t$ | Single Regge Pole splitting (momentum fraction $x_{I\!P}$) |
|---|---|
| 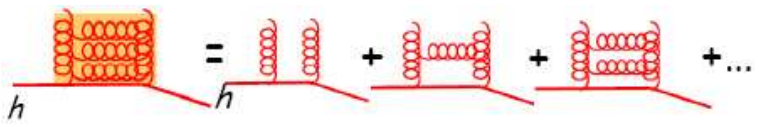 | $G_{I\!Ppp}\frac{dx}{x_{I\!P}^{2\alpha(t)-1}}f(t)\,dt$ |

Table 1: Phenomenological "splitting function" for the high energy, diffractive, interaction of a hadron in terms of the $t$-channel exchange of a single leading Regge pole; the coupling $G_{I\!Ppp}$, and the dependence $f(t)dt$ on 4-momentum transfer $t$ are not predicted; the trajectory $\alpha(t)$ can be specified by resort to the resonance structure of the crossed channel, thereby specifying the dependence on the fractional momentum $x_{I\!P}$; the Regge pole exchange is illustrated diagrammatically in the expectation that it amounts to an expansion in partonic QCD.

What (arguably) has been at the root of the importance of diffraction in hadronic physics are the experimental observations that

1. inelastic interactions at high energy are dominated by diffraction-like production: the incident hadrons can be associated with clusters of low mass particles associated in rapidity with one or other of the incident hadrons, and separated from each other and from any other hadrons by a "rapidity gap", and the production angular dependence of the clusters (4-momentum transfer squared $t$) resembles that of elastic scattering, making hadron diffraction a unique window into the diffractive shadow;

2. all such identified diffractive processes exhibit many common, and quantitatively generic, features in terms of their interaction dynamics, and

3. all diffractive processes alone survive as the only interaction mechanism for elastic and inelastic hadronic interactions at the highest energies.

These in turn have led to the phenomenology of extended, colour singlet, exchange in the $t$-channel which is related through analyticity to low energy resonance phenomena in the $s$-channel, the indisputable and indestructible triumph of Regge phenomenology [2]. Despite in many ways its crudity because it is only a leading parametrisation valid at high energy (high $s$) and small $t$, "Reggeology" often works extremely well, but sometimes fails magnificently, notably at ultra-high energy where it violates the most rigorous of theoretical requirements, unitarity! In the case of diffractive physics, it triumphs in producing a universally applicable "pomeron" trajectory $\alpha(t)$ with a well determined "intercept" $\alpha(t=0) = 1.085$ and a "slope" $\alpha' \sim 0.25$ ($\alpha(t) = \alpha(t=0) + \alpha' t$) [3], but at the cost of posing a major challenge for its "crossed" $s$-channel, namely what is the nature of the hadronic system which carries the quantum numbers of the vacuum and which defines this linear trajectory?

High (but not ultra-high) energy diffractive scattering can thus be conveniently characterised in analogy with formal field theory by a phenomenological splitting function as per Table 1, where it is clear that Reggeology is indeed a well founded, but only comparative, phenomenology. A splitting function is the probability of the diagram involving the coupling of the space-like exchange to the particle in question. Reggeology correlates phenomena in terms of a few, rigorously derived, parameters. The recipe in Table 1 gives rise (of course) to the established expectations that the leading diffractive trajectories, the pomeron $\alpha(t=0)=1.085$ and the $f^0$-meson $\alpha(t=0)=0.5$, have respectively characteristic dependences $\frac{dx}{x^{\leq 1.17}} \sim \frac{dx}{x}$ and constant in the fractional momentum variable of the Reggeon ($x = x_{I\!P}$). These are more often quoted in terms of the dependence on centre-of-mass energy squared $s$ (valid when $s$ is the only high energy scale) of the contributions of each to the diffractive production cross section, namely a gentle rise $s^{2\alpha(t)-2} \sim s^{0.17}$ and a significant decrease $s^{2\alpha(t)-2} \sim s^{-1}$ respectively, leading directly to the dominance of pomeron exchange in diffraction at high energy, that is at low $x_{I\!P}$.[1]

| Splitting to off-shell $t$ | QCD LO splitting function (momentum fraction $x$, $n_f$ flavours | High energy ($x \to 0$) |
|---|---|---|
| 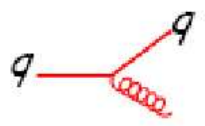 | $\frac{4}{3}\left[\frac{1+x^2}{1-x}+\frac{3}{2}\delta(1-x)\right]dx\frac{dt}{t}$ | $dx\frac{dt}{t}$ |
| 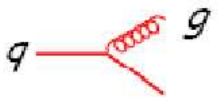 | $\frac{4}{3}\frac{1+(1-x)^2}{1-x}dx\frac{dt}{t}$ | $\frac{dx}{x}\frac{dt}{t}$ |
| 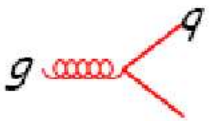 | $\frac{1}{2}\left[x^2+(1-x)^2\right]dx\frac{dt}{t}$ | $dx\frac{dt}{t}$ |
| 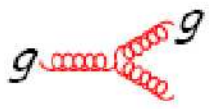 | $6\left\{\left[\frac{x}{1-x}+\frac{1-x}{x}+x(1-x)\right]+\frac{33-2n_f}{6}\delta(1-x)\right\}dx\frac{dt}{t}$ | $\frac{dx}{x}\frac{dt}{t}$ |

Table 2: QCD Splitting functions at leading order (LO); each process ($q \to q, q \to g, g \to q, g \to g$) is marked with the off-shell parton with space-like mass$^2$ $t$ and fractional momentum $x$; the high energy limit corresponds to low fractional momentum $x$.

What somewhat more recently has become an important part of the experimentalist's toolkit is the array of essential facts of QCD, which are best summarised in the "splitting functions" of the chromodynamic quanta, and which are determined from the Lagrangian of gauge field theory, QCD. They turn out to carry a set of very simple, salient, properties (experimentalists always love simplicity!). These are laid out in Table 2 and reveal the simple rules for the high energy (low x)[2] dependences, namely the probability for off-shell "emission" of a gluon $\sim \frac{dx}{x}\frac{dt}{t}$ and of a quark $\sim dx\frac{dt}{t}$. These simple outcomes are equivalent to exchange of field quanta (gluons) in interactions giving rise to no energy dependence of the cross section[3], and to exchange of fermions (quarks) in interactions giving rise to falling energy dependence. Com-paring with the Regge phenomenological expectations above, the former suggests that gluon exchange(s) must be important in high energy diffraction. In the last few years, experiment has resolved the basic QCD splitting structure in diffractive dynamics, confirming this likelihood, a major step in confronting the phenomenological "Reggeology" with the rigours of non-abelian, chromodynamic, gauge theory.

[1] The reason for explicitly laying out the Regge dynamics in the slightly unusual terms of an effective, trajectory dependent, splitting function is because the Regge analytic continuation from the low energy resonance amplitude to the crossed channel amplitude with $t/s \to 0$ is in the variable which corresponds most closely to CM scattering angle $cos\theta^*$ in the resonance amplitude and to $1/x_{I\!P}$ in the high energy amplitude. Its use in this form is also unambiguous when one or more of the initial or final state particles have sizable mass (for example virtual photon) leading to modified CM energy dependence.

[2] We here take $x$ and $t$ to be respectively the fractional momentum and the invariant mass squared of the off-shell product of the quantum splitting.

[3] The simplest manifestation of this universal property of a vector quantum field theory is of course Rutherford scattering in QED.

It is also a fact that QCD has been used in a monumental theoretical effort to calculate strong interaction dynamics in the form of multiple gluon exchanges between quarks in a calculation of diffractive quark scattering. The result is a triumph in the annals of theoretical QCD, producing from the basic splitting in Table 2 Regge behaviour with a "pomeron" intercept marginally larger than classical Reggeology in soft hadronic physics, $\alpha(t = 0)$ in Table 1 [4], and thereby also establishing new technologies in QCD concerning how to handle multiple, soft, gluons. The result, known as the "hard pomeron" or the BFKL pomeron, or the Lipatov pomeron, has become to many experimentalists something of a holy grail. Now, after a number of false "alarms", experiment has established evidence indicating the existence of a harder pomeron, and one whose intercept evolves with changes in hard (short distance) scale.

Such has been the experimentation with its associated phenomenology which has made possible quantification of diffraction in hadronic physics, and which has thereby simultaneously guided theoretical progress.

On the basis of this "potted" pedigree, since their inception, the major thrust of EDS conferences has been the synergy of simultaneous developments in both experiment and theory. At EDS09 much progress of striking significance continues to be reported. It is based on new and not so new data sets, and on new and not so new formalism and phenomenology, but this time also with the mouth-watering anticipation of the imminent new round of experiments at the CERN LHC. What follows cannot be exhaustive, so it suffers unavoidably from the author's prejudice.

## 2 Exclusive Diffraction

The aristocrat of sub-nuclear physics is proton-proton ($pp$) elastic scattering (and its partner antiproton-proton elastic scattering). Unlike its hereditary namesake, this aristocrat continues to thrive and to challenge every theoretical revolution, often leading to new insight as repeatedly it does so.

At EDS09 we have seen a snapshot of how the on-going theoretical developments, driven often by aspects of low-$x$ QCD together with the unassailable rigours of analyticity and the Optical Theorem, continue to reveal new physics insight. "Derivative Dispersion Relations" (DDR) have been brought to bear [5] on the decades of data sets, further tying down the real part of the forward elastic scattering amplitude (the $\rho$ parameter) through its interference with the electromagnetic, Rutherford scattering, amplitude (Fig. 1a). The interplay of contemporary views of proton structure with the outcome of these approaches is now at the stage that a picture of the lateral (transverse) structure of the proton emerges [6]. "Multi-pole" exchange analyses [7] of differential and total cross sections with lower energy data (lower than the LHC!) have been pushed to provide best honest estimates – and they are just that – of what to expect for the total $pp$ cross section at the Terascale. The considerable uncertainties in these expectations highlight the crucial importance of LHC measurements.

The issue facing all elastic scattering cross section measurements has been the lack of a "hard" momentum transfer scale with which to engage partonic (quark and gluon) degrees of freedom. The relationship between the "soft" Reggeon phenomenology, which has so far worked better than anything else, and the theoretical calculus of non-abelian QCD applied to strong hadronic interactions, requires measurements to elucidate when and how partonic degrees of freedom emerge from soft, extended (colour singlet), hadronic physics.

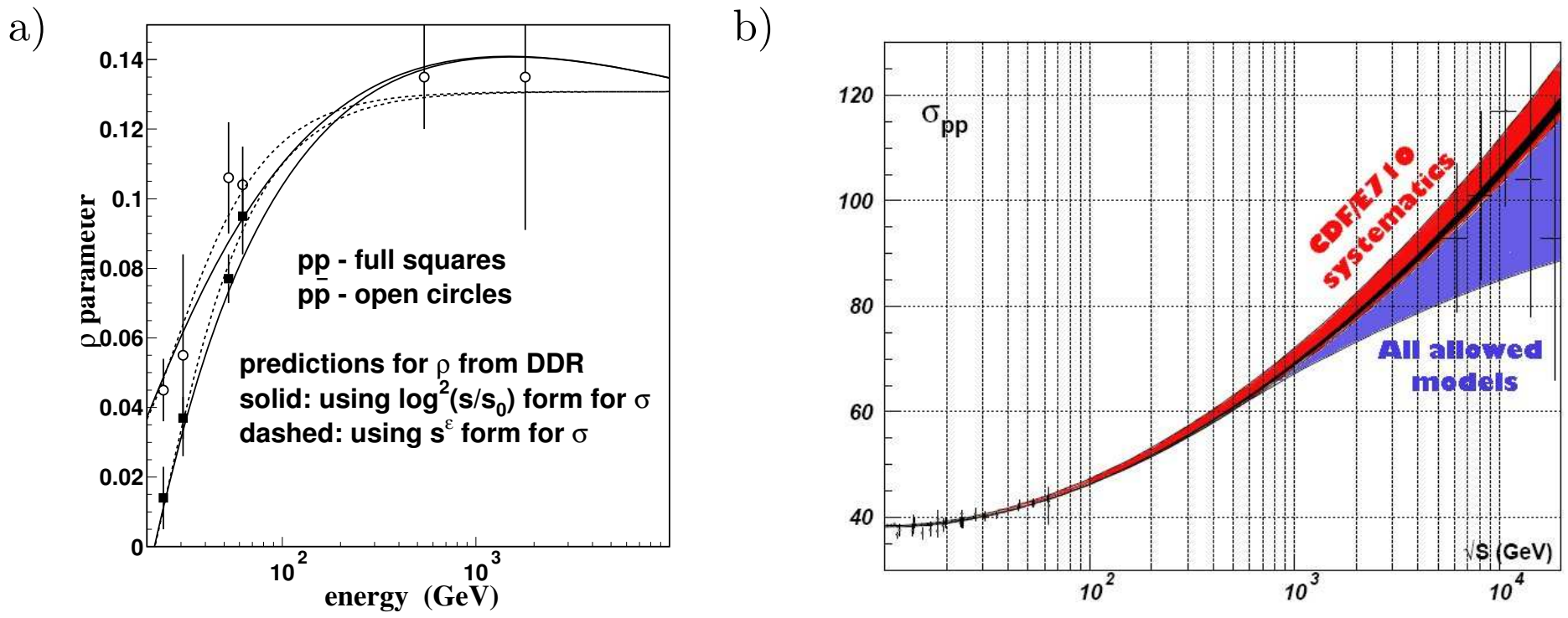

Figure 1: (a) the $\rho$-parameter, the ratio of the real to imaginary forward elastic $pp$ scattering amplitude as a function of energy showing the different predictions based on DDR (see text); (b) predictions for the total pp cross section in the LHC energy region based on a multi-pole analysis of elastic diffractive data, revealing large uncertainties in expectation.

It is here that the HERA electron-proton collider has opened completely new horizons in the last 17 years. Exclusive diffraction at HERA is measured in low-$x$ (high energy) electroproduction, where the dominant mechanism is space-like virtual photon ($\gamma^*$) exchange, but with the magnitude of the 4-momentum transfer squared $Q^2$ much less – but still sub-femtoscopic – than the virtual photon-proton interaction energy squared $W^2$. These processes are diffractive when the exclusive final state involves a photon or a vector meson (VM) $ep \to e\,\mathrm{VM}\,p$ (Fig. 2a). With (diffractive) data for the electroproduction of $\rho(770)$, $\varphi(1010)$, $\omega(770)$, $J/\Psi$ and $\Upsilon$ over a wide range of $Q^2$, it is therefore possible to probe the short distance mechanisms which, as the resolution changes, reveal the QCD view of the colour singlet exchange, the pomeron, of Regge theory.

There is a magnificently simple kinematic approach to the issue of the size of an exclusive diffractive interaction. One has to consider the interplay of spatial dimension of probe $Q^2$, 4-momentum transfer squared $t$, quark mass $m_q$, and final state meson mass squared $M^2$ in the diagram that must accompany any diffractive process involving a gauge boson (Fig. 2b). As we have appreciated for decades, indeed from the days of vector dominance to the days of HERA, any electroweak gauge boson couples to matter through a quark. Thus any interaction of a gauge boson with matter is sensitive to the "virtuality" of the inner, space-like, quark in Fig. 3a, that is to its spatial extent. The issue of how "hard" is the dynamics of a process is driven entirely by the "size" of this inner quark, in other words how large is its virtuality $v$ (taken to be its space-like 4-momentum squared). The application of simple kinematics leads to

$$v = m_q^2 - \frac{Q^2 + M^2 - t}{2}\left(1 \mp \sqrt{1 - \frac{4m_q^2}{M^2}}\right), \tag{1}$$

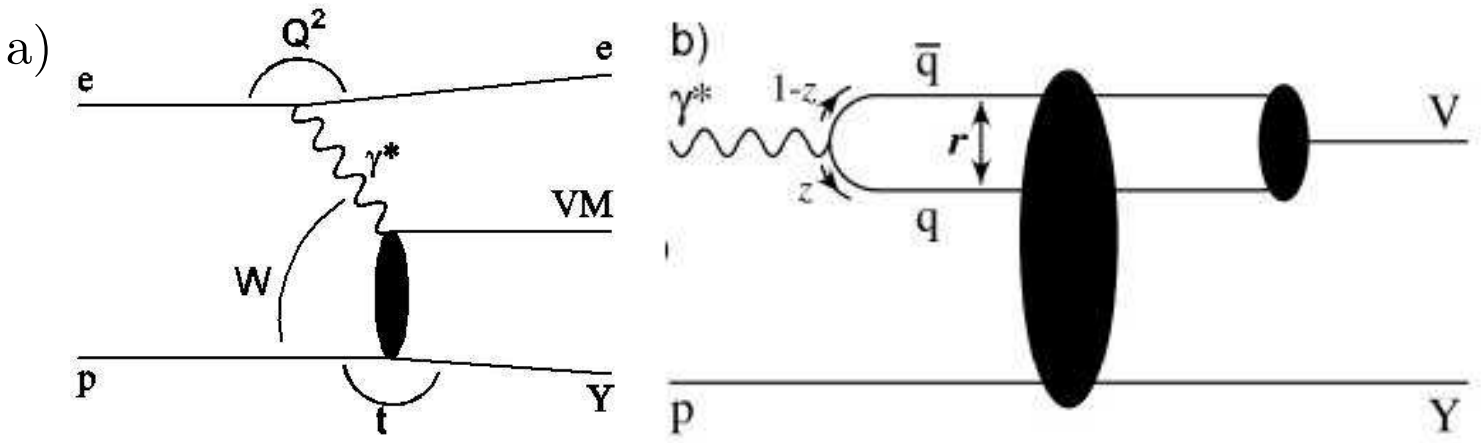

Figure 2: (a) “Elastic” (diffractive) scattering of a vector meson (VM) at HERA at high virtual photon-proton interaction energy $W$, specified by the magnitude of 4-momentum transfer squared $Q^2$ being much less than $W^2$ but still sufficient to resolve quark degrees of freedom; Y is a proton or low mass excitation of a proton ($N^*$); (b) the only way in which a gauge boson can couple to hadronic matter, namely through a quark; the dimension of the gauge boson, here a virtual photon, is specified by $Q^2$.

revealing how different possible measures, $Q^2$, $M^2$, $t$, and $m_q$ may contribute to this virtuality.

The results (only some of which are compiled in Figure 3a) are spectacular in how they reveal the evolution of the energy dependence of elastic scattering from “soft” (low $v$) to the “hard” (high $v$) domains [8]. They demonstrate unequivocally that the Regge phenomenology in the form of the trajectory $\alpha(t)$, which using Table 1 for a single leading trajectory “intercept” now drives a dependence of the cross section of the form $(Q^2+W^2)^{2\alpha(t)-2}$, changes as a function of the size of the interaction because of, one or more of, $Q^2$ and the masses of the vector meson and of its quarks. There are also measurements with sensitivity to $t$ which indicate similar dependence on the size of the interaction.

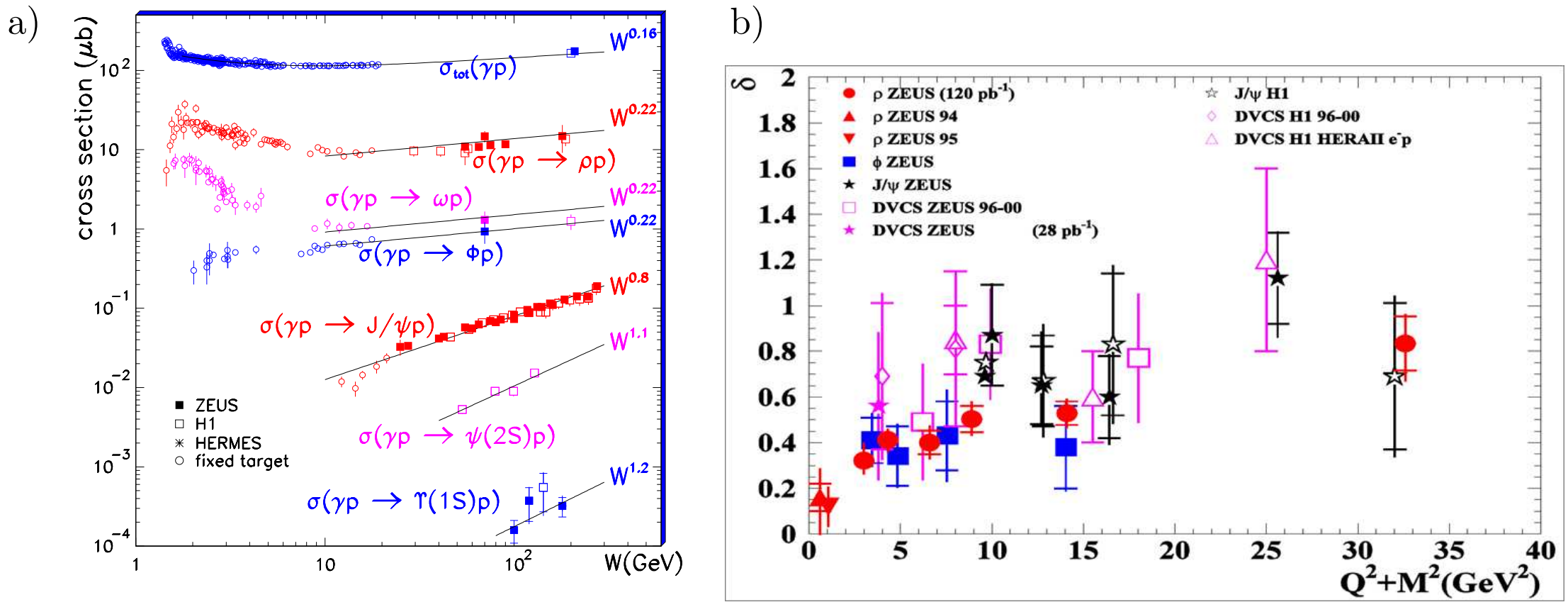

Figure 3: (a) measurements of the total cross section for “elastic” virtual photon-proton electroproduction production; (b) dependence of the $W^2$ dependence of different total cross sections for “elastic” virtual photon-proton electroproduction; for a single leading trajectory of “intercept” of $\alpha(t=0)$, the $W^2$ dependence amounts to $W^\delta$ where $\delta = 2\alpha(t=0) - 2$.

The initial significance of these and many other similar observations should not be missed. The HERA measurements produce first-time evidence, that is they discover, that the long standing, original, success of the Regge approach with universal, generic, trajectories specified by $t$-channel, colour singlet, quantum numbers and related wherever possible to cross channel resonances, fails, and the failure is because "size matters". The smaller is the size of the interaction, the steeper is the energy dependence, that is the higher is the trajectory intercept $\alpha(t = 0)$. We have therefore the evidence that the pomeron evolves through different manifestations as interaction size changes from large (soft) to small (hard). And given our dynamical picture of hard interactions in terms of field quanta in QCD, we see that these measurements are exactly what is needed to address the inter-relation of the soft pomeron with what may be the "hard" pomeron of Lipatov and colleagues [4] in QCD.

The measurements, which include exclusive electroproduction spanning different ranges of $Q^2$, $M^2$, $t$ and $m_q$, have yet to be exhaustively analysed in terms of sensitivity to the inner quark virtuality. Figure 3b summarises how the $\frac{1}{x_{I\!P}}$ – or $W^2 + Q^2$ at fixed $M^2$ – dependences, plotted as $\delta$ in a Regge motivated parametrisation $(W^2 + Q^2)^\delta$ evolve.[1] Here there is some evidence of a universal dependence, even though the kinematic analysis based on inner quark virtuality (1), if right, implies that things may be more complicated.

All of the above discussion rests on the validity of the assumption of the coupling of electroweak gauge bosons with a quark, and how the dynamics varies with the size of this coupling within the target hadron, the proton. We are addressing the phenomenology from the perspective in which somehow a quark-antiquark "dipole" defines the hadronic interaction, as Figure 2b tries to show. Many call this the new paradigm of low-$x$ deep scattering, both elastic and inelastic, and build "dipole models" to which they try then to apply QCD [9]. The dipole picture is not, however, critical to the way the data are – and hopefully will be further – analysed. For, as special relativity says, the basic electroweak gauge boson-quark coupling indeed "looks like" an incident dipole in the Lorentz frame in which the target proton is at rest and the dipole has a large Ioffe length [10]. But also as special relativity says, in the Lorentz frame in which the target proton has high momentum, we still retain the "classic" Feynman picture of the photon interacting with a frozen quark constituent of the proton, and with a similar, Lorentz invariant, lateral dimension. It is gratifying therefore to an experimentalist that these measurements are so manifestly insensitive to such paradigms, and that therefore they remain critical for fundamental QCD calculations of diffraction.

As in all important research, progress depends on the unexpected and the unanticipated. Recent, unexpected theoretical progress in QCD theory, in particular with reference to the validity of factorisation between structure and dynamics, has made it feasible to contemplate using exclusive, vector meson, electroproduction (electroweak, the photon, and chromodynamic, $\rho(770)$, $\varphi(1010)$, $\omega(770)$, $J/\Psi$ and $Y$) as the means to understand the "tomography" of the proton. In essence measurements of the differential cross sections for such processes, and in particular the exclusive electroproduction of photons $ep \to e\gamma p$ (deeply virtual Compton scattering DVCS), are sensitive to the dependence of the partonic structure of the proton on transverse proton dimension, so called "generalised parton densities" (GPDs). Measurements over a wide kinematic range are needed, which now motivates a new round of measurements to combine with the HERA results, and, after initial results from HERMES [11] at HERA, they are now also being planned for the COMPASS [12] experiment at CERN. What is truly exciting is to consider how these measurements, analysed in terms of GPDs, will tighten the understanding of diffractive elastic scattering at the energy frontier in terms of proton structure mentioned above [6].

This exemplifies the way that measurements hitherto considered too difficult, and until recently not fully appreciated for their theoretical significance, can influence the on-going challenge of understanding in QCD the original (aristocratic!) diffraction of hadrons.

I have dwelt long on the important developments in exclusive diffraction which continue to take forward the challenge of achieving a rigorous, gauge theoretic, approach to diffractive hadron scattering.

For me the highlight at EDS09 which points to where exclusive diffraction may be taking us has come with the conclusions of new measurements at the TeVatron. They are motivated by addressing experimentally whether exclusive, diffractive, Higgs-boson, ($H$) production $pp \to pHp$, that is as the afficianados say "Higgs production with no mess", is experimentally feasible at the terascale of the LHC [13]. In short, the issue has been the magnitude of the production cross section and, for experiment, the acceptance possible and the feasibility. The former involves Standard Model (SM) couplings in diagrams of the form of Fig. 4(a).

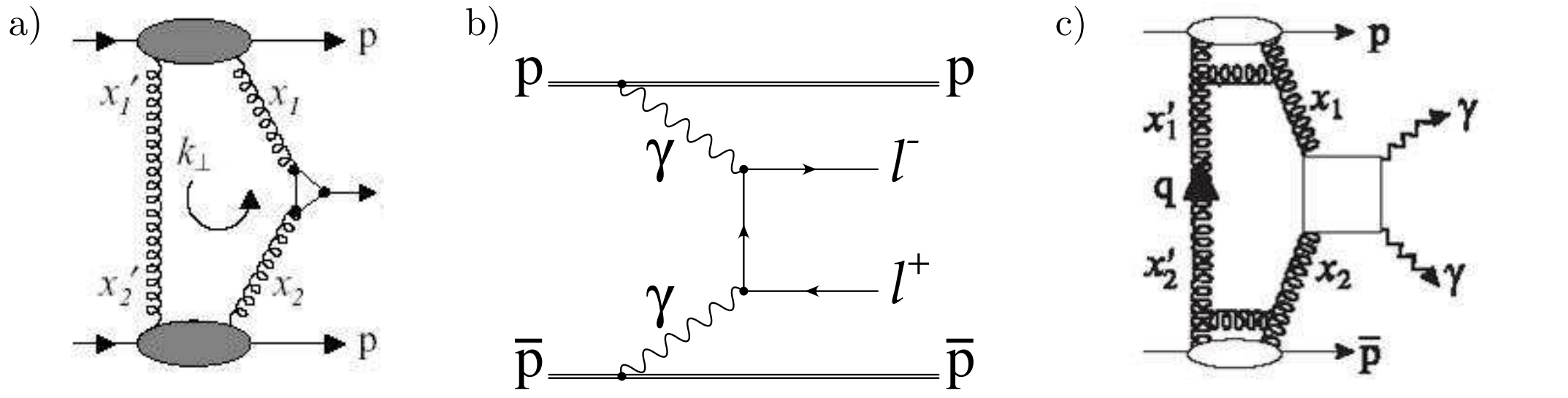

Figure 4: (a) leading order QCD diagram for the production of the Higgs in doubly-diffractive production $pp \to pHp$; and leading order QED diagrams for the production of (b) lepton pairs, and also quark pairs and thus exclusive dijets, and (c) photon pairs, all in the same "diffractive" configurations.

To this end, there has been much theoretical debate leading now to something of a consensus concerning the magnitude of the "diffractive Higgs" production cross section. There has also been much experimental R & D, based on experience with forward "pots" at TeVatron and HERA and the possibility of similar operation very close to the LHC beam with the necessary kinematic coverage.

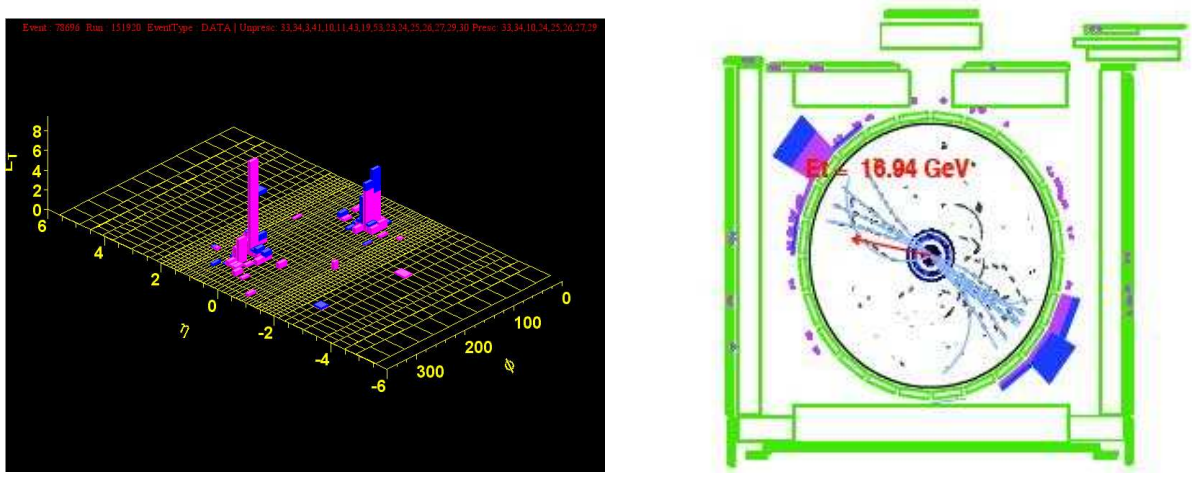

Figure 5: Event observed in CDF with the topology of only two jets in a configuration consistent with the exclusive process $pp \to p(jet + jet)p$, further illustrating the experimental feasibility of exclusive diffraction at a collider such as the LHC.

At the TeVatron, CDF has now demonstrated such exclusive diffraction in the electromagnetic sector of the SM in the processes (Fig. 4b, c) $pp \to pe^+e^-p$ and $pp \to p\gamma\gamma p$. Many of the doubts concerning the QCD predictions for Higgs production are addressed by these measurements, albeit at the TeVatron, and not the Tera, energy scale. CDF have reported signals (handfuls of events) consistent with expectation, checking QED+QCD and thereby underpinning the present estimates of the diffractive, Higgs production, cross section using electroweak+QCD at the LHC. CDF also reported here at EDS09 exclusive dijet production $pp \to p(jet+jet)p$ (Fig. 4c and Fig. 5) [13], which now piles on the pressure for more theoretical precision in diffractive QCD!

Exclusive diffraction has come along way since the days of the first high energy elastic scattering measurements. Our progress to an understanding at the level at which we can say can be applied with confidence to all strong interaction systems continues to be substantial, but far from complete. And the possibilities and horizons which are now before us at the LHC to probe further are huge.

# 3 Inclusive Diffraction

Relativistic hadron physics is remarkable for its scope. Most notably, and using the language of optical diffraction, the ability to observe and measure inelastic final states in diffractive interactions, such as $pp \to pp$ + hadrons, opens the way to understand the "diffractive shadow" as well as diffraction itself $pp \to pp$ [1]. Measurements over decades have exposed the universality of production characteristics – the pomeron at work between dissociating protons, and rules relating to quantum number flow including spin, parity, flavour isospin, and multiplicity.

At the highest energies, and pioneered at the CERN ISR, inelastic diffraction has demonstrated with the help of "Müller sub-unitarity" that the Regge approach can be extended to the concept of "triple Regge" diagrams (examples in Fig. 6a). In this context, straightforward Regge phenomenology, despite being only a leading approximation at high energy, still shows remarkable consistency and universality in the Reggeon parameters extracted from the data, including those of the (soft) pomeron. This success, for that is what it unquestionably is, forms therefore the soft physics template in analyses which pursue short distance aspects of inelastic diffraction. The best, relatively more recent, example has come with the analysis of inelastic photoproduction data from HERA. Figure 6b summarises the outcome of an exhaustive analysis [15] which also includes lower energy measurements at Fermilab [16] so as to gain the all-important "lever arm" in interaction energy alongside the substantial range which the HERA kinematics gives to the inclusive, diffractive, mass range. The fits are extremely good with a minimum number of trajectories, there is no requirement for any non-Regge "background", and the Reggeon couplings, including the pomeron, and trajectory parameters are remarkably well determined. The soft physics template is thus established for photon induced inelastic diffraction.

So it is in the face of all of the above success in recent decades that we must look at where now we are in understanding inelastic diffraction. Following in the noble footsteps of the CERN ISR and the CERN UA experiments, the CDF experiment at the TeVatron continues to pioneer what is called "single" diffraction $pp \to p +$ hadrons (Fig. 7a) and "double" diffraction $pp \to pp +$ hadrons. In all the CERN experiments and in CDF, the detector configurations have involved one of the most demanding of experimental challenges, namely the operation as close as possible to the stored beams of forward proton detectors. The results have been spectacular.

a)

$$\frac{\mathrm{d}\sigma}{\mathrm{d}t\,\mathrm{d}M_X^2} = \frac{s_0}{W^4}\sum_{i,j,k} G_{ijk}(t)\left(\frac{W^2}{M_X^2}\right)^{\alpha_i(t)+\alpha_j(t)}\left(\frac{M_X^2}{s_0}\right)^{\alpha_k(0)}\cos\left[\phi_i(t)-\phi_j(t)\right]$$

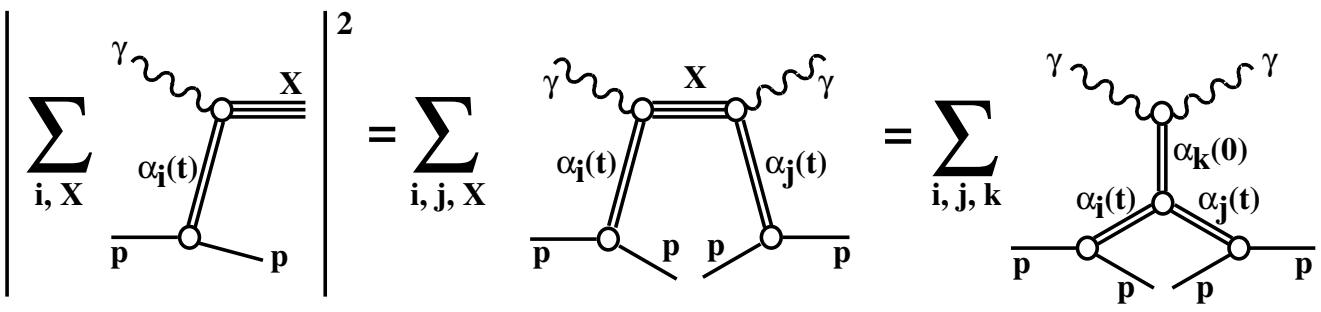

b)

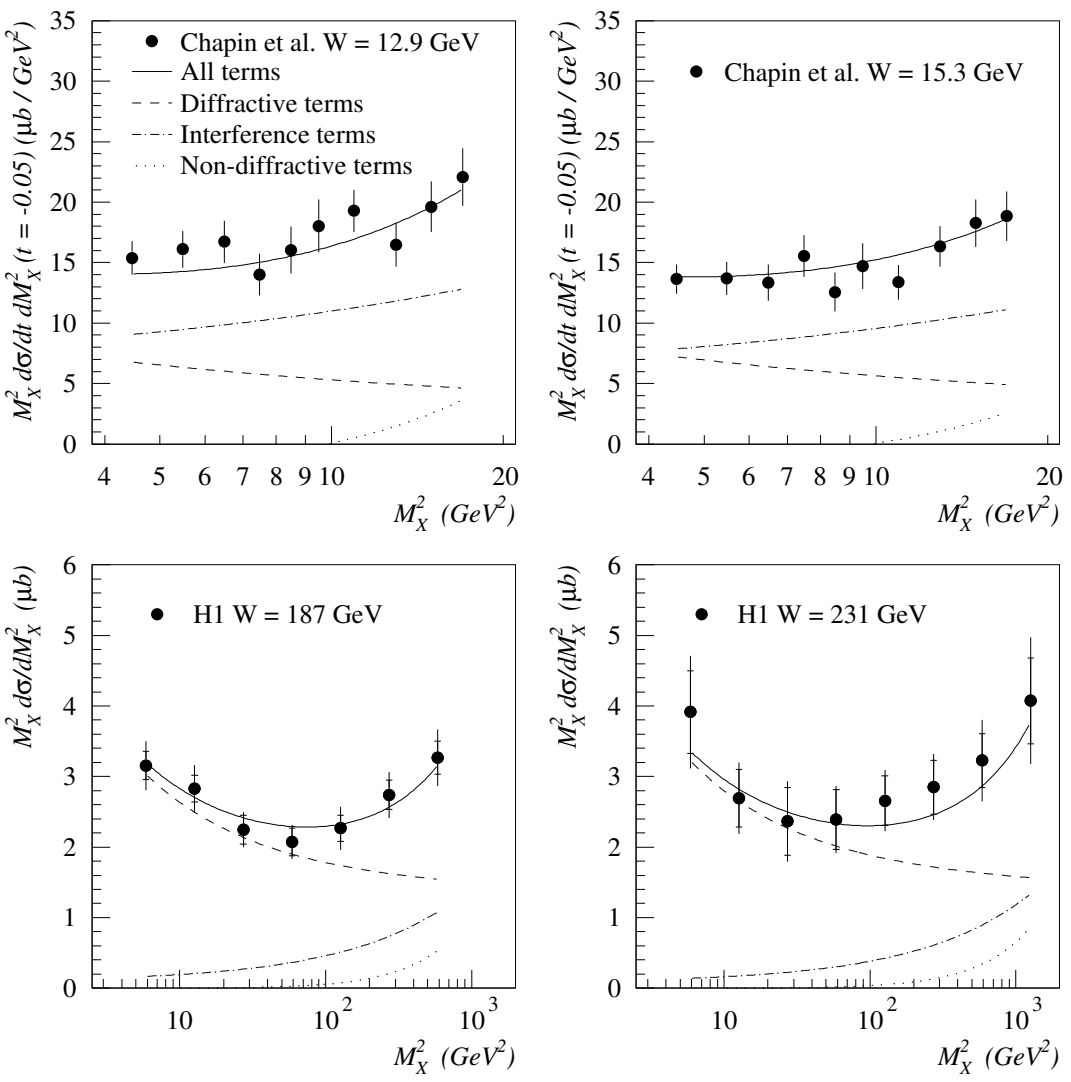

Figure 6: (a) Müller triple Regge diagrams indicating how the cross section for inclusive diffractive mass production can be replaced through sub-unitarity in this mass by the amplitude for diffractive scattering; (b) measurements of the differential cross section (multiplied by $M_X$) for the production of a diffractive mass $M_X$ at different photoproduction energies compared with a fit to a triple Regge based set of amplitudes.

For the first time jets were seen in single diffraction $pp \to p + jet + \ldots$, and, with di-jet events, even information on the fractional momentum dependence of the jet-initiating partons from the protons was obtained by UA8 [17]. The surprise was that the jets are produced with partons from the diffracting protons with surprisingly large momentum, thereby being very suggestive of "hard" pomeron dynamics. These measurements amount to the discovery of the first evidence for partonic degrees of freedom in high energy diffraction.

The status these days concerning high energy inelastic diffraction has moved on hugely thanks to the exploitation of the large diffractive component in low-$x$ deep-inelastic electron-proton scattering at HERA and the availability of good statistics for measurements of the process $ep \to eXp$ (Fig. 7b) [18]. As a result, it has been possible to probe in the classic

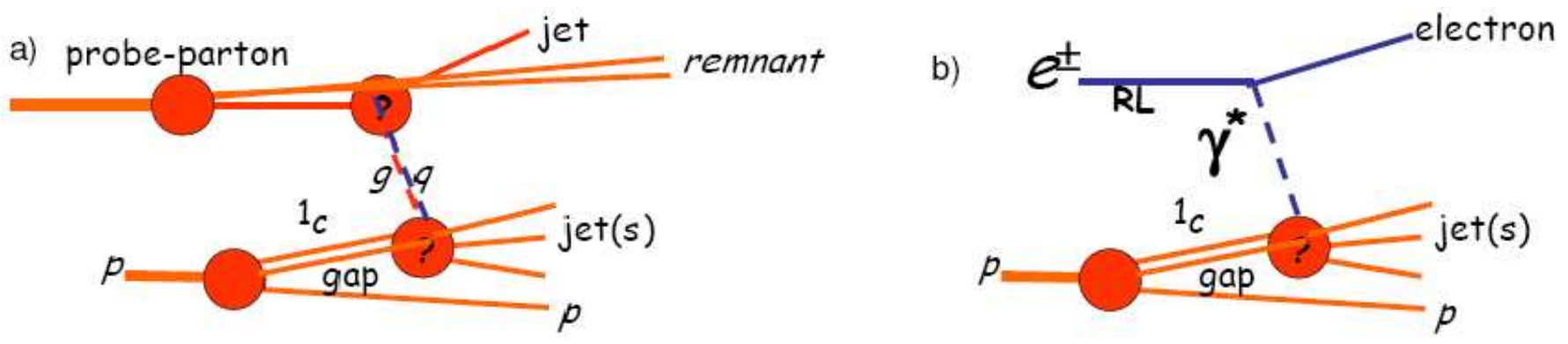

Figure 7: Diagrams illustrating (a) "single diffraction" measurements at a $pp$ collider in which a parton from one proton is used to probe the short distance structure of the other diffractively interacting proton; (b) deep-inelastic lepton-nucleon diffraction in which a lepton probes the structure of a diffractively interacting proton. In each case the rapidity "gap" is marked in which no hadrons are observed thereby defining a leading proton as a particle carrying very nearly all of its incident momentum in undergoing the inelastic diffractive interaction.

"deeply inelastic" manner the short distance structure of the diffractive interaction of a proton, rather than the proton itself. Just as for the latter where totally inclusive measurements of $ep \to eX$ are made in the form of structure functions, results have now been produced with amazing precision in the form of "diffractive structure functions" (shown because of the precision possible as "reduced cross sections"). They reveal for the first time the QCD dynamics of the diffractive $t$-channel, the pomeron.

There is no space or time here to discuss completely the wealth of detail in these measurements and how they build a picture of the QCD nature of diffractive exchange which must constrain future theoretical approaches. In Fig. 8a, $x_{I\!P}$ dependences (cf. Table 1) for different fractional momentum of the struck quark $\beta$ and different $Q^2$ are shown, revealing dominant contributions at low $x_{I\!P} \leq 0.05$ consistent with hadronic diffraction in the form of appropriate pomeron "splitting", together with sub-leading Regge splitting at larger $x_{I\!P}$ consistent with $f^0$ exchange (cf. Table 1 and discussion in text with it). In Fig. 8b the classic "scaling violation" figure shows the $Q^2$ evolution of the diffractive structure function, with rising dependence with increasing $Q^2$ for all but the highest ($\beta > 0.6$) fractional momentum $\beta$ of the quark coupling to the gauge boson (photon), the struck quark. Rising scaling violations at intermediate and larger $\beta$ are exactly what is expected if the structure is attributable to quantum fluctuation of both quarks and gluons, and not attributable to spatial extent due to a constituent bound state (cf. the scaling violations of the structure function $F_2$ of the proton for values of the appropriate variable Bjoerken-x[4] similar to the diffractive variable $\beta$). In Fig. 8c the $\beta$ dependence (note the logarithmic abscissa scale) of the reduced diffractive cross section (in essence the diffractive structure function) for different $Q^2$ reveals a dependence which is strikingly different from that of a hadron, and which is consistent with a large contribution at large fractional momentum $\beta$ from gluon, and not quark, splitting to produce the struck quarks [cf. Table 2 and the well known interpretation of the structure functions in terms of parton density functions $F_2 = \sum_q xq(x)$].

These comprehensive and beautiful measurements point unquestionably to the observation of struck quarks from gluon splitting, that is of much gluon exchange in diffractive exchange.

[4]For non-afficianados, Bjoerken-x amounts to the fractional momentum of the struck quark in the infinite-momentum frame of the target proton first invoked by Feynman.

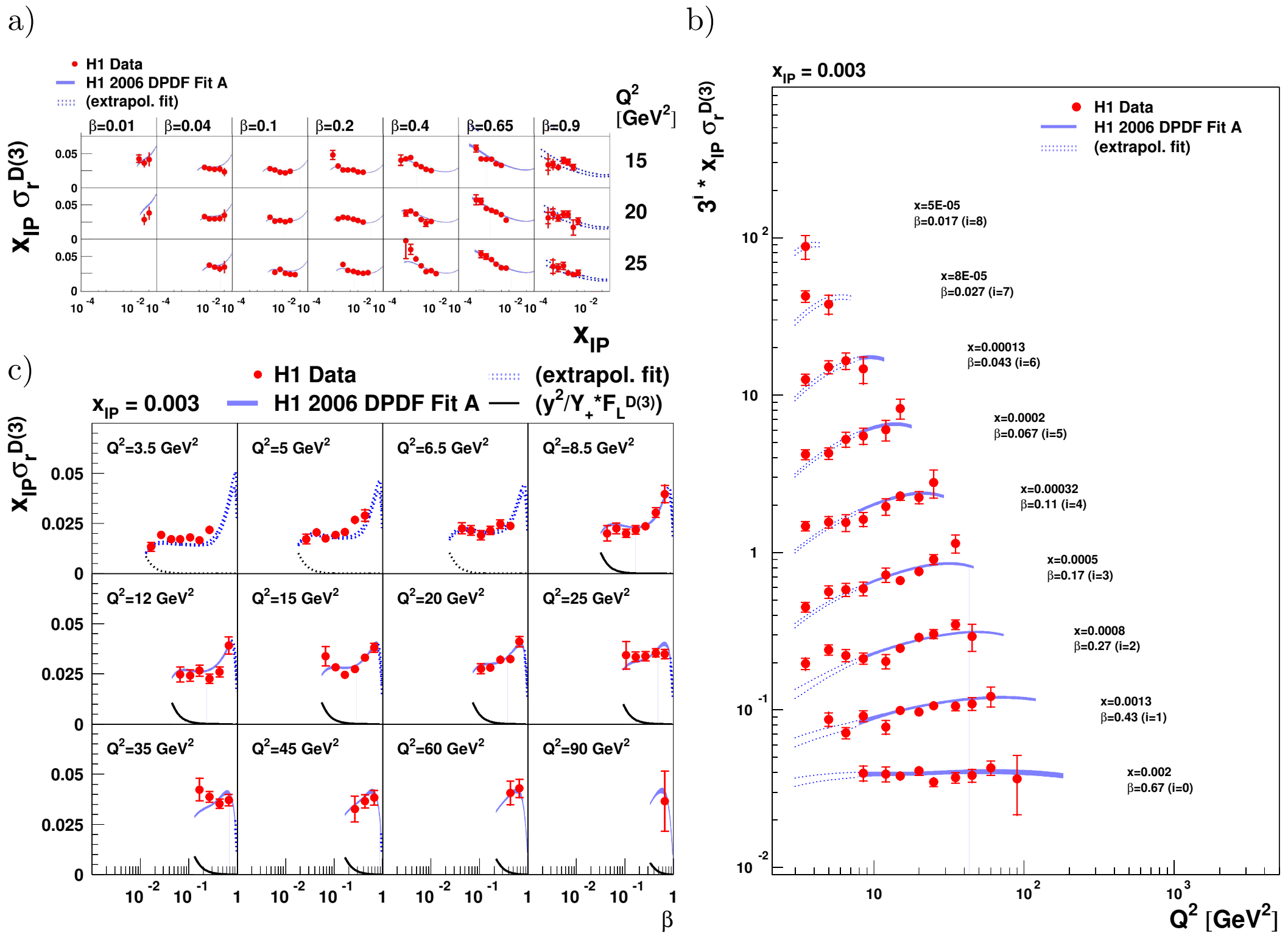

Figure 8: The reduced diffractive cross section, very nearly the diffractive structure function) multiplied by the "Regge" fractional momentum variable $x_{I\!P}$ for the diffractive $t$-channel averaged over all $t$: (a) samples of the $x_{I\!P}$ dependence for different fractional momentum of the struck quark $\beta$ and $Q^2$, revealing dominant contributions at low $x_{I\!P}$ consistent with hadronic diffraction with appropriate pomeron "splitting" together with sub-leading Regge exchange at larger $x_{I\!P}$ (Table 1); (b) The $Q^2$ evolution for $x_{I\!P} = 0.003$ showing the persistence of violations of scale invariance (no dependence on $Q^2$) of a nature which rise with increasing $Q^2$ for a wide range of values of $\beta \leq 0.6$; (c) the $\beta$ dependence (note the log abscissa scale) of the reduced diffractive cross section for different $Q^2$, revealing a dependence which is strikingly different from that of a hadron, which is consistent with a large contribution from gluon splitting (Table 2), and which amounts to the structure of the pomeron. In all figures the curves shown correspond to the results of a full QCD fit to obtain parton densities in diffractive exchange.

Quantitative fits of the data to extract factorisable diffractive parton density functions (dpdf), using QCD formalism from proton structure (DGLAP [19, 20]), quantify gluon content as being responsible for $\sim 70\%$ of diffractive exchange momentum [15]. The latest measurements of diffractive deep-inelastic scattering now include the longitudinal structure function $F_L^D$ which, like its totally inclusive partner $F_L$, has a leading contribution dependent on gluon content, so it is expected to be large. At EDS09 we have seen that there is complete consistency of this first measurement of $F_L^D$ with the above overall picture of large gluon exchanges in diffractive structure.

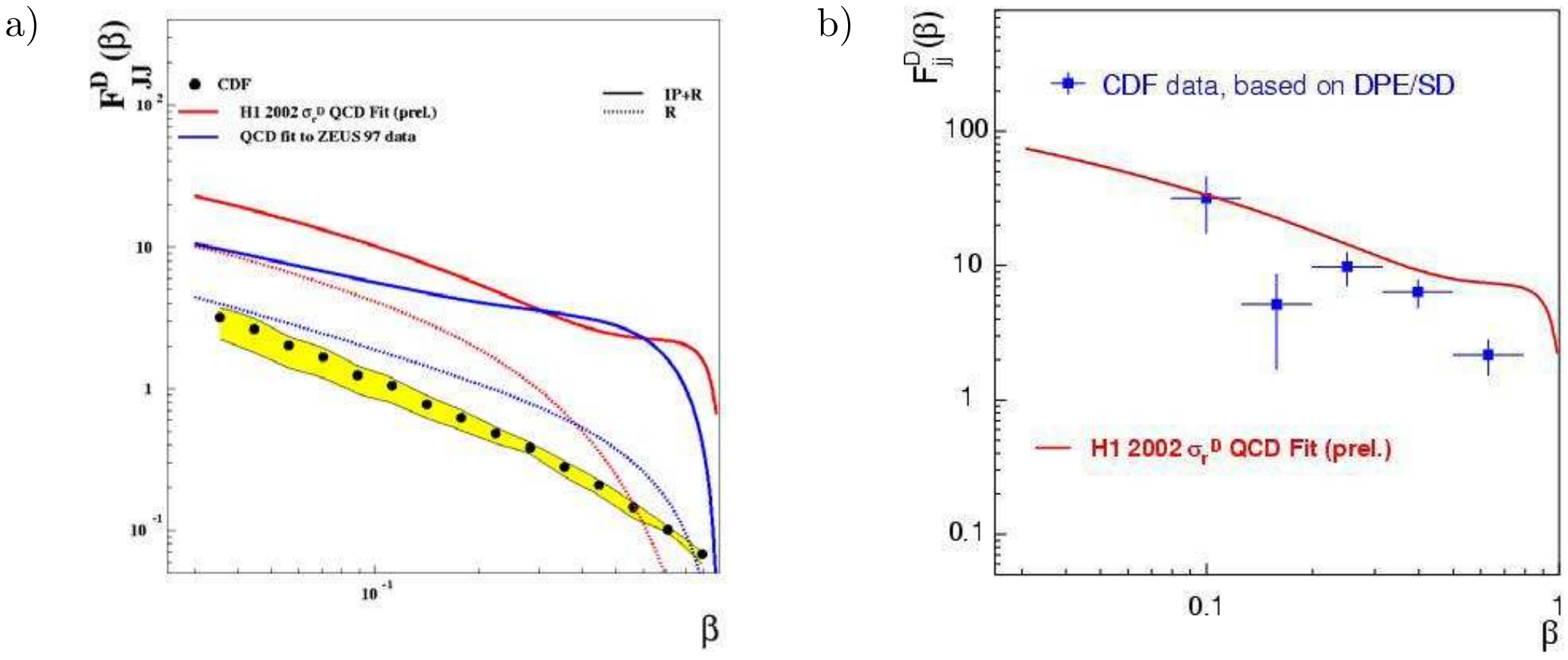

Figure 9: (a) Factorisation breakdown showing the measured cross section for inclusive jet production in single diffraction by CDF at the TeVatron, compared with the expectation using dpdfs from HERA - "gap suppression"; (b) confirmation of the universality of factorisable dpdfs from HERA comparing the measured cross section for inclusive jet production in double diffraction by CDF at the TeVatron.

The establishment theoretically of dpdfs as factorisable, and therefore portable [20], was accompanied by first comparisons of them with measurements of inclusive jet production in single diffraction (Fig. 7a) by CDF at the TeVatron. A massive discrepancy was discovered between what is taken to be a measurement of cross section, and expectation, amounting to a massive (sometimes $\times 10$) breakdown of the expectations of factorisation with the supposedly universally applicable dpdfs [14]. There then followed the invention of the paradigm of rapidity "gap suppression" to explain it (Fig. 9a) in which it is supposed that re-scattering of open colour at low momentum transfer is hypothesised to create hadrons which fill the rapidity gap (Fig. 10). Here at EDS09, CDF have followed this puzzling result with a beautiful sequel using doubly diffractive, inclusive, jet production (Fig. 7b) [14]. A similar analysis of this sample shows, in stark contrast, consistency with the universality of dpdfs, namely "gap restoration" (Fig. 9b).

In Fig. 10 are shown simple diagrammatic approaches to illustrate how the suppression of a gap with additional hadronisation due to open colour interactions is likely to be suppressed when in double *pp* diffraction the probe parton originates from a pomeron, rather than a proton. There are manifestly fewer opportunities in double diffraction for rescattering arising from open colour at small momentum transfer. Furthermore, and perhaps this is the most significant outcome of this new result from CDF, the replacement in double diffraction of a parton by a pomeron as a hard probe means exactly that, namely a remnant-free, "direct", pomeron-parton vertex. These measurements thereby expose a completely new aspect of diffractive physics and "pomeron-pomeron" interactions in which the pomeron takes on the role of a quantum phenomenon which can probe itself!

It is clear from the above that breakthroughs of major significance continue in inelastic diffractive physics. We are now at a stage where we have measurements of a sophistication and precision which must surely tie down the next steps in the direct application of QCD to high energy hadron physics. We have achieved a new level of precision in quantifying experimentally diffractive dynamics as a function of hard scale, while at the same time revealing the make-up of the partonic structure of this dynamics, conveniently, but not necessarily, referred to as the partonic structure of the pomeron. There is now a need to consider this asymptotically free, partonic, structure to see how it could further enable and liberate the theoretical QCD approaches to diffractive dynamics, whose difficulty is so manifestly illustrated in the pioneering, hugely labour intensive, Lipatov-style [4], calculations.

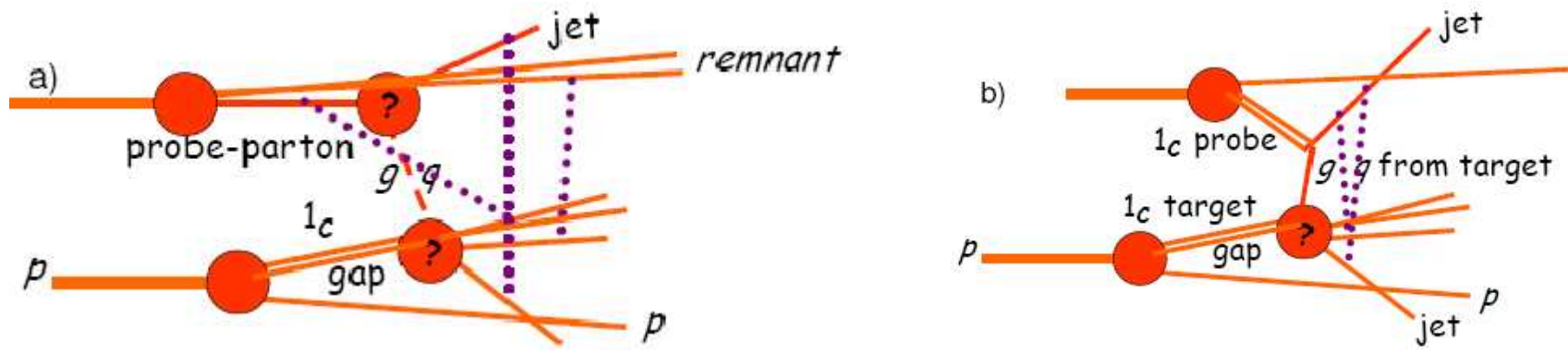

Figure 10: Diagrams illustrating (a) "single diffraction" measurements at a $pp$ collider in which a parton from one proton probes the diffractive exchange of the other; the dotted lines indicate some possible open-colour soft interactions which can occur and which would fill the rapidity gap defining the leading final proton; (b) "double diffraction" measurements in which two rapidity gaps are required with two high $p_T$ jets thereby removing the possibility of substantial open-colour soft interactions which can occur and which would mean that gap suppression was reduced.

And finally as they say, here at EDS09 we have perhaps seen a glimpse of the next steps in the detailed work which continues to be required. New results with now valuable precision are being obtained at HERA in which forward neutron and forward proton counters have operated enabling measurements of what remains at high energy of non-diffractive, colour singlet, dynamics [21]. By demanding and reconstructing a leading neutron (LN) in an interaction with a lepton $ep \to eXn$, in the language of the Regge asymptotic limit one aims to probe the structure of a colour singlet, flavour exchange, as well as leading proton (LP) production, in regions of $x_L = 1 - x_{I\!P}$ beyond just pomeron exchange (Fig. 11a).

Data have been available since the start-up of HERA, but the results have been difficult to obtain not least because the leading Reggeon is a meson with trajectory intercept $\alpha(t = 0) = 0.5$, or more likely less (for $\pi$-exchange $\alpha(t = 0) \sim 0$), and therefore with a falling dependence on interaction energy (a constant or rising dependence on $x_{I\!P}$, now written as $x_L = 1 - x_{I\!P}$ in the effective splitting function: cf. Table 1). This dependence on $x_L$, together with the small cross section, also makes the distinction between dynamics which are genuine, colour singlet, exchange (Fig. 11a), and dynamics which arise from that part of inclusive proton fragmentation with a baryon tending to lead, hazardous (Fig. 11b).

Nevertheless, the $x_L$ spectra (Fig. 11c) show a comparison of LP and LN cross sections in a kinematic region for LP in which we know diffraction is dead ($x_L < 0.95$ see Fig. 8a). The difference between the cross sections, together with the subtleties of the different dependencies, already suggest a $\pi$-exchange contribution ($\alpha(t = 0) \sim 0$) in LN at larger $x_L$, and higher trajectory meson exchanges in both LP and LN ($\alpha(t = 0) \sim 0.5$) at lower $x_L$.

The job is now to use these data samples to measure the deep-inelastic structure of the interactions with a view to revealing any effects of "hard meson" structure and dynamics, just as has been discussed above for the diffractive component and the "hard pomeron", here in Fig 11c invisible at $x_L \geq 0.95$. The salivating experimentalist is driven by the notion of discovering $Q^2$-evolving, and therefore non-universal, meson exchanges, together with structure functions of the form of a bound hadron, exhibiting scaling violations consistent with major components from $q \to q$ and $q \to g$ splitting (cf. Table 2), also constrained by appropriate QCD sum rules. For some time now initial analyses have produced results of limited precision for the structure function of the pion, which, in the light of these new data, bodes well for the future [22], and level-headed theorists continue to say that hard mesons will not be significant. So the stage is set!

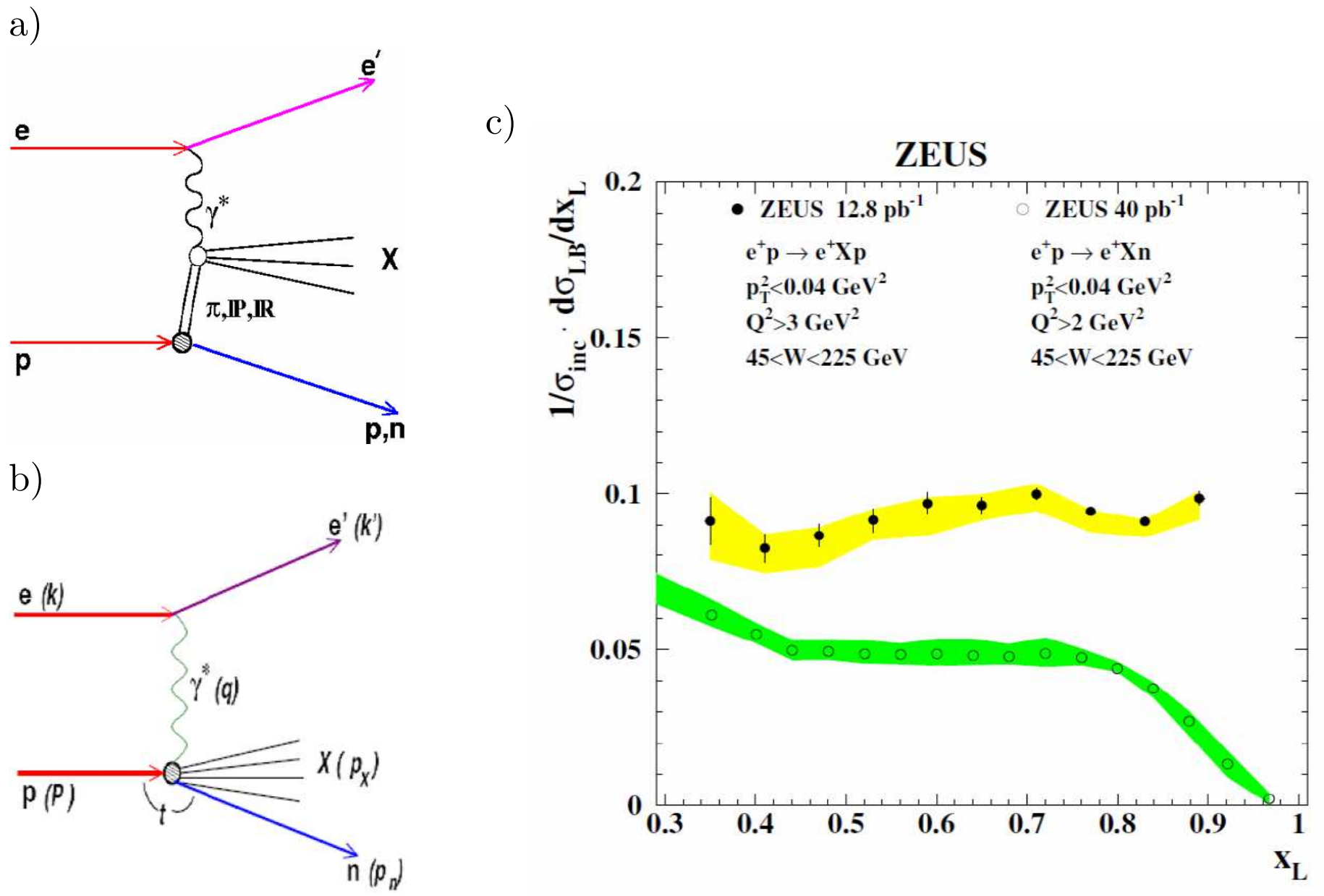

Figure 11: (a) Electron-proton, deep-inelastic, hadron production involving either a leading proton or a leading neutron; (b) electron-proton, deep-inelastic, inclusive hadron production indicating the fact that a "leading" proton or neutron is possible from the incident proton fragmentation; (c) ratio of the leading neutron and the leading proton differential cross sections to the total inclusive hadron cross sections as a function of fractional momenta of the two leading baryons. $x_L$ revealing evidence for a substantial contribution from meson exchanges with Regge intercepts $\alpha(t = 0) = 0.5$ in both leading proton and leading neutron, and a component at largest with Regge intercept $\alpha(t = 0) = 0$ consistent with $\pi$-exchange.

# 4 New Experiments

In terms of new experiments we are of course at the dawn of a new diffractive era. Already in the above we have touched on issues and challenges at the LHC Terascale.

It is now the case that substantial forward instrumentation is installed and ready to go round the CMS experiment, prosaically called TOTEM (Fig. 12a). Both experiments will run simultaneously and synchronously, so elastic and inelastic diffraction are anticipated just as soon as some luminosity is available at the LHC. Similar measurements are now possible with forward detectors at ATLAS. But also for ATLAS, a major initiative is underway to take advantage, as first pioneered in H1 at HERA, of possible modifications to the LHC cryogenic system for proton detectors 420 m either side downstream. The intention is to greatly enhance the kinematic acceptance for inelastic diffraction (with of course “Higgs production with no mess” very much in mind) [23].

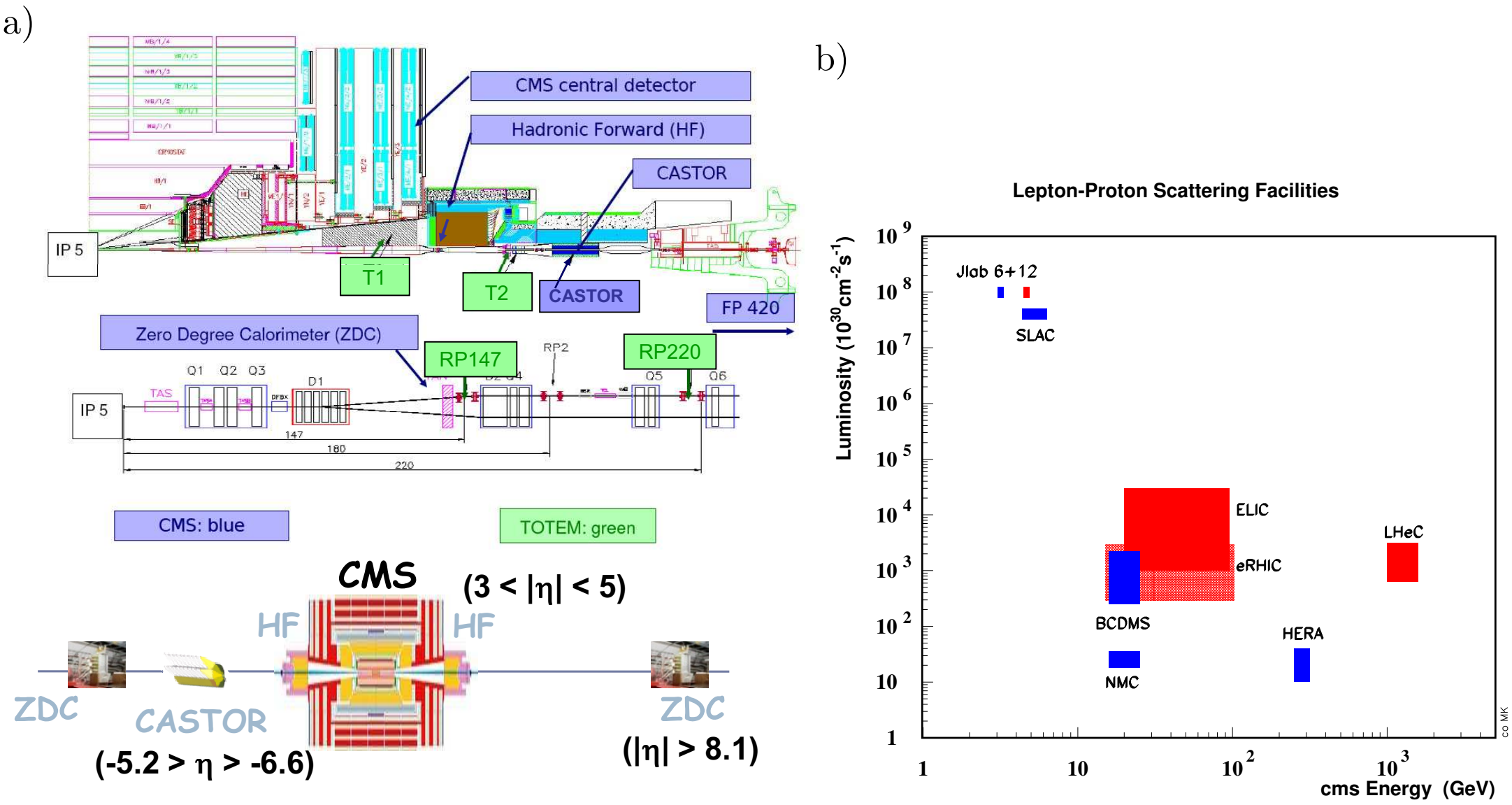

Figure 12: (a) The situation of the TOTEM experiment with respect to the CMS experiment at LHC Point 5. (b) Lepton-proton luminosity at all lepton deep-inelastic hadron experiments hitherto and all proposals now in evaluation, demonstrating the supremacy in both energy reach and luminosity of the proposed LH*e*C collider.

Another possible upgrade of the LHC, which has been kicked around in the long grass for years, indeed for as long as the LHC was proposed in the LEP tunnel, is for an intense electron (and positron) beam of 50 to 150 GeV momentum in collision with one of the LHC hadron beams, either protons or heavy ions, a Large Hadron-Electron collider LH*e*C.

First evaluations have been published [24]. It is demonstrated a) that *ep* and *e*-ion physics is possible alongside an on-going LHC programme, and b) that huge luminosity, in fact luminosity which exceeds any previous, deep-inelastic, lepton-nucleon experiment by factors except for that at the pioneering SLAC end-station in the late 1960s (Fig. 12b), is possible (power permitting) with a kinematic reach in *ep* energy of at least 1.4 TeV.

An initiative is now underway, requested by ECFA, endorsed strongly by NuPECC, the host laboratory CERN and the CERN Council Strategy group, and “blessed” by ICFA[5], to prepare and then submit a Conceptual Design Report some time in late 2010 [25]. LH$e$C will then be "on the table" alongside other future possibilities at the LHC at CERN, and hopefully will be given even more impetus to prepare a Technical Design Report. Needless to say, the opportunities for new measurements of diffraction encompassing a huge kinematic range, and with also ion targets, are "mind boggling"!

[5]The acronyms need defining: European Committee on Future Accelerators ECFA, Nuclear Physics European

Given the venue of EDS09, perhaps the most appropriate session has been that concerned with new experiments [26]. Taking advantage of the unique breadth of expertise present, the session was enlivened by a discussion of how to decide a running strategy for data taking during the first period of LHC collisions imminent later this year (2009). As a result, there cannot ever before have been such a wide representation of theoretical and experimental expertise in the immediate decision making affecting the first physics from a brand new machine at a brand new energy scale!

# 5 Conclusion

Diffraction is at the heart of hadron physics, and hadron physics remains one of the unresolved conundrums of the Standard Model, and possibly also beyond.

Despite its decades-long standing as the major feature of high energy hadron interactions, and the substantial phenomenology which grew up with it based on the analyticity of the hadronic scattering matrix, developments in only the last 15 years have revolutionised our approach and understanding of diffraction in terms of QCD. This recent progress has resulted from innovative new measurements at the $pp$ and $ep$ colliders at the Fermi scale, the TeVatron and HERA, together with intense synergy between experimentalists and theorists. With this progress comes the recognition and definition of new opportunities in the future. The programme at EDS09 reflects this state of affairs perfectly.

These opportunities span a wide energy range.

The energy frontier in the immediate future will soon pass from the TeVatron to the LHC, where we can expect diffractive physics results, elastic and inelastic, with the first $pp$ collisions.

But as has always been the case in hadronic diffraction, new understanding brings new opportunities at lower energies, where already one can see a huge program developing concerned with exclusive production leading to more precise measurements to pin down the soft-hard interface. Exclusive electroproduction will produce precision proton tomography which will take us forward in achieving better understanding of the diffractive coupling of the proton. Continuing synergy of these initiatives with the extension of the energy frontier to the Terascale cannot but influence radically our quantitative picture of diffractive scattering and dissociation.

Regrettably the HERA collider terminated before it could fully exploit the opportunities that new investment would have brought in electron-ion physics and the opportunities this brings for much progress in low-x, and therefore diffractive, physics. This loss may yet be recouped using the RHIC beam at Brookhaven (eRHIC) or with ELIC at JLab, but with nothing like the HERA kinematic reach (Fig. 12b).

So until an electron beam can be constructed at CERN, and an LH$e$C realised, elastic and diffractive scattering, and thus EDS meetings, will be concerned with classic hadron-hadron experimentation at the LHC $pp$ energy frontier, alongside what will inevitably be the on-going development of new QCD-driven phenomenology based on data we have and will still be able

Collaboration Committee NuPECC, International Committee on Future Accelerators ICFA.

to take at lower energies. This is already an important, and a wonderful, horizon. I commend it to you all.

## Acknowledgements

Accounts of all the results which I have presented here can be found in talks written up in these proceedings. They depend on the work of many, most of whom have not been present at EDS09. To all I give thanks. I have unashamedly plagiarized and copied from these EDS09 presentations in formulating this summary. I thank everyone concerned for letting me do so, and I hope that I haven't misrepresented anything.

The task of the last speaker must include calling for a vote of thanks to the organizers of EDS09, especially the conference staff here at CERN, who have worked like Trojans to ensure that EDS09 has been such a gripping, productive, and memorable meeting. On behalf of you all, thank you to all who set up and who ran this conference.

## References

[1] M L Good and W D Walker Phys. Rev. 120, (1960) 1857.

[2] T Regge, Nuov. Cim. 14 (1959) 951
G Chew, Phys. Rev. 126 (1961) 1202.

[3] A Donnachie and P Landshoff Nucl. Phys. B303 (1988) 634.

[4] L N Lipatov, Sov. Phys. JETP 63 (1986) 904
E A Kuraev et al. Sov. Phys. JETP 45 (1977) 199.

[5] E Ferreira et al., these proceedings.

[6] R Luddy et al., these proceedings.

[7] J-R Cudell et al., these proceedings.

[8] P Marage, these proceedings.

[9] K Golec-Biernat and M Wusthof
Phys. Rev. D59 (1999)14
Phys. Rev. D60 (1999) 114
A Stasto, K Golec-Biernat, J Kwiecinski, Phys. Rev. Lett. 86 (2001) 2001.

[10] L B Ioffe, V Gribov, I Pomeranchuk, Sov. Journ. Nucl. Phys. 2 (1966) 549
B L Ioffe, Phys. Lett. B30 (1969) 123.

[11] A Airapetian et al, JHEP (in press) arXiv:0909.3587 (hep-ex) and DESY-09-143.

[12] S Yaschenko, these proceedings
N d'Hose, talk at the workshop "New Opportunities in the Physics Landscape at CERN", 11-13 May 2009.

[13] M Albrow, J Pinfold et al. these proceedings.

[14] D Goulianos et al., these proceedings.

[15] H1 Collab., C. Adloff et al, Z. Phys. C74 (1997) 221.

[16] T Chapin et al. Phys. Rev. D31 (1985) 17.

[17] R Bonino et al. (UA8 collaboration)
Phys. Lett. B211 (1988) 239.

[18] A Proskuryakov, these proceedings
H1 Collab., T. Ahmed et al., Phys. Lett. B348 (1995) 681
H1 Collab., C. Adloff et al., Z. Phys. C76 (1997) 613
H1 Collab., A. Aktas et al., Eur. Phys. J. C48 (2006) 715.

[19] V N Gribov and L N Lipatov
Sov. J. Nucl. Phys. 15, 438 (1972)
Sov. J. Nucl. Phys. 15, 675 (1972)
L N Lipatov, Sov. J. Nucl. Phys. 20, 94 (1975)
Y L Dokshitzer, Sov. Phys. JETP 46, 641 (1977)
G Altarelli G Parisi, Nucl. Phys. B126, 298 (1977).

[20] J C Collins Phys. Rev. D57 (1998) 3052 Erratum-ibid 61 (1998) 019902.

[21] A Bunyatyan, these proceedings.

[22] H1 Collab., C. Adloff et al., Eur. Phys. J. C6 (1999) 587.

[23] http://www.fp420.com/ .

[24] J Dainton et al. J.Inst. 1 (2006) 10001.

[25] http://www.LHeC.org.uk/.

[26] K Eggert, these proceedings.